%% file: FLAG_master.tex
\documentclass[11pt,a4paper]{article}
\usepackage{latexsym}
\usepackage{graphicx}
\usepackage{psfrag}
\usepackage{amsmath}
\usepackage{amssymb}
\usepackage{color}
\usepackage{mathtools}
\usepackage{rotating}
\usepackage{multirow}
\usepackage{colortbl}
\usepackage{lscape,epsfig}
\usepackage{a4wide}
\usepackage[]{xcolor}
\usepackage{authblk}
\usepackage{verbatim}
\usepackage{rotating}
\usepackage{fancyheadings}
\usepackage{minibox}
\usepackage{enumitem}
\usepackage[colorlinks=true,a4paper=true,backref=false,linktocpage=true,citecolor=blue,urlcolor=blue,linkcolor=blue,pdfpagemode=UseOutlines]{hyperref}
\usepackage[squaren,binary]{SIunits}
\usepackage[sort&compress,square,comma,numbers]{natbib}
\usepackage[nottoc]{tocbibind} 
\usepackage{lineno}
\usepackage{placeins}
\usepackage[normalem]{ulem}
\usepackage{supertabular}
\usepackage{cleveref}

\title{{\bf \Huge FLAG Review 2024}\\
  \vspace{0.5cm}
  {\bf \large Flavour Lattice Averaging Group (FLAG)} }

\newcounter{affilctr} 

\addtocounter{affilctr}{1}
\newcounter{RIKEN}
\author[\theRIKEN]{Y.~Aoki}
\affil[\theRIKEN]{RIKEN Center for Computational Science, Kobe 650-0047, Japan}

\addtocounter{affilctr}{1}
\newcounter{Connecticut}
\addtocounter{affilctr}{1}
\newcounter{RIKENBNL}
\author[\theConnecticut,\theRIKENBNL]{T.~Blum}
\affil[\theConnecticut]{Physics Department, University of Connecticut, Storrs, CT 06269-3046, USA}
\affil[\theRIKENBNL]{RIKEN BNL Research Center, Brookhaven National Laboratory, Upton, NY 11973, USA}

\addtocounter{affilctr}{1}
\newcounter{Regensburg}
\author[\theRegensburg]{S.~Collins}
\affil[\theRegensburg]{Institut f\"ur Theoretische Physik, Universit\"at Regensburg, 93040 Regensburg, Germany}

\addtocounter{affilctr}{1}
\newcounter{Edinburgh}
\author[\theEdinburgh]{L.~Del Debbio}
\affil[\theEdinburgh]{Higgs Centre for Theoretical Physics, School of Physics and Astronomy,
  University of Edinburgh, Edinburgh EH9 3FD, UK}

\addtocounter{affilctr}{1}
\newcounter{Odense}
\author[\theOdense]{M.~Della~Morte}
\affil[\theOdense]{IMADA, University of Southern
  Denmark, Campusvej 55, DK-5230 Odense M, Denmark}

\addtocounter{affilctr}{1}
\newcounter{Parma}
\addtocounter{affilctr}{1}
\newcounter{ParmaINFN}
\author[\theParma,\theParmaINFN]{P.~Dimopoulos}
\affil[\theParma]{
  Dipartimento di Scienze Matematiche, Fisiche e Informatiche,
  Universit\`a di Parma, 43124 Parma, Italy}
\affil[\theParmaINFN]{INFN,
Gruppo Collegato di Parma, Parco Area delle Scienze 7/a (Campus), 43124 Parma, Italy}

\addtocounter{affilctr}{1}
\newcounter{Peking}
\addtocounter{affilctr}{1}
\newcounter{PekingCIC}
\addtocounter{affilctr}{1}
\newcounter{PekingHEP}
\addtocounter{affilctr}{1}
\newcounter{PekingNP}
\author[\thePeking,\thePekingCIC,\thePekingHEP,\thePekingNP]{X.~Feng}
\affil[\thePeking]{School of Physics, Peking University, Beijing 100871, China}
\affil[\thePekingCIC]{Collaborative Innovation Center of Quantum Matter, Beijing 100871, China}
\affil[\thePekingHEP]{Center for High Energy Physics, Peking University, Beijing 100871, China}
\affil[\thePekingNP]{State Key Laboratory of Nuclear Physics and Technology, Peking University, Beijing 100871, China}

\addtocounter{affilctr}{1}
\newcounter{SF}
\author[\theSF]{M.~Golterman}
\affil[\theSF]{Dept. of Physics and Astronomy, San Francisco State University, San Francisco, CA 94132, USA}

\addtocounter{affilctr}{1}
\newcounter{Indiana}
\author[\theIndiana]{Steven Gottlieb}
\affil[\theIndiana]{Department of Physics, Indiana University, Bloomington, IN 47405, USA}

\addtocounter{affilctr}{1}
\newcounter{LANL}
\author[\theLANL]{R.~Gupta}
\affil[\theLANL]{Los Alamos National Laboratory, Theoretical Division T-2, Los Alamos, NM 87545, USA}

\addtocounter{affilctr}{1}
\newcounter{Madrid}
\author[\theMadrid]{G.~Herdo\'iza}
\affil[\theMadrid]{Instituto de F\'{\i}sica Te\'orica UAM/CSIC and
  Departamento de F\'{\i}sica Te\'orica, Universidad Aut\'onoma de Madrid, Cantoblanco 28049 Madrid, Spain}

\addtocounter{affilctr}{1}
\newcounter{IFIC}
\author[\theIFIC]{P.~Hernandez}
\affil[\theIFIC]{IFIC (CSIC-UVEG), Parc Cient\'{\i}fic de la Universitat de Val\`encia, E-46980 Paterna, Spain}

\addtocounter{affilctr}{1}
\newcounter{Southampton}
\addtocounter{affilctr}{1}
\newcounter{STAG}
\addtocounter{affilctr}{1}
\newcounter{CERN}
\author[\theSouthampton,\theSTAG,\theCERN]{A.~J\"uttner}
\affil[\theSouthampton]{School of Physics \& Astronomy, University of Southampton, Southampton SO17 1BJ, UK}
\affil[\theSTAG]{STAG Research Center, University of Southampton, Highfield, Southampton SO17 1BJ, UK}
\affil[\theCERN]{CERN, Theoretical Physics Department, Geneva, Switzerland}

\addtocounter{affilctr}{1}
\newcounter{KEK}
\addtocounter{affilctr}{1}
\newcounter{Sokendai}
\author[\theKEK,\theSokendai]{T.~Kaneko}
\affil[\theKEK]{High Energy Accelerator Research Organization (KEK), Ibaraki 305-0801, Japan}
\affil[\theSokendai]{Graduate Institute for Advanced Studies, SOKENDAI (The Graduate University for Advanced Studies), Ibaraki 305-0801, Japan}

\author[\theIndiana]{E.~Lunghi}

\addtocounter{affilctr}{1}
\newcounter{Arizona}
\author[\theArizona]{S.~Meinel}
\affil[\theArizona]{Department  of  Physics,  University  of  Arizona,  Tucson,  AZ  85721,  USA}

\addtocounter{affilctr}{1}
\newcounter{WandM}
\addtocounter{affilctr}{1}
\newcounter{CC}
\author[\theWandM,\theCC]{C.~Monahan}
\affil[\theWandM]{Department of Physics, The College of William \& Mary, Williamsburg, VA  23187,  USA}
\affil[\theCC]{Department of Physics, Colorado College, Colorado Springs, CO 80903, USA}

\addtocounter{affilctr}{1}
\newcounter{ChapelHill}
\author[\theChapelHill]{A.~Nicholson}
\affil[\theChapelHill]{Dept.~of Physics and Astronomy, University of North Carolina, Chapel Hill, NC 27516-3255, USA}

\addtocounter{affilctr}{1}
\newcounter{Osaka}
\author[\theOsaka]{T.~Onogi}
\affil[\theOsaka]{Department of Physics, The University of Osaka, Toyonaka 560-0043, Japan}

\addtocounter{affilctr}{1}
\newcounter{BNL}
\author[\theBNL]{P.~Petreczky}
\affil[\theBNL]{Physics Department, Brookhaven National Laboratory, Upton, NY 11973, USA}

\author[\theRIKEN,\theEdinburgh,\theCERN]{A.~Portelli}

\author[\theIFIC]{A.~Ramos}

\addtocounter{affilctr}{1}
\newcounter{Seattle}
\author[\theSeattle]{S.~R.~Sharpe}
\affil[\theSeattle]{Physics Department, University of Washington, Seattle, WA 98195-1560, USA}

\addtocounter{affilctr}{1}
\newcounter{Fermilab}
\author[\theFermilab]{J.~N.~Simone}
\affil[\theFermilab]{Fermi National Accelerator Laboratory, Batavia, IL 60510 USA}

\addtocounter{affilctr}{1}
\newcounter{TrinityCollege}
\affil[\theTrinityCollege]{School of Mathematics \& Hamilton Mathematics Institute, Trinity College Dublin, Dublin 2, Ireland}
\author[\theTrinityCollege]{S.~Sint}

\addtocounter{affilctr}{1}
\newcounter{NIC}
\addtocounter{affilctr}{1}
\newcounter{Humboldt}
\author[\theNIC,\theHumboldt]{R.~Sommer}
\affil[\theNIC]{Deutsches Elektronen-Synchrotron DESY, Platanenallee~6, 15738~Zeuthen, Germany}
\affil[\theHumboldt]{Institut f\"ur Physik, Humboldt-Universit\"at zu Berlin, Newtonstr. 15, 12489 Berlin, Germany}

\addtocounter{affilctr}{1}
\newcounter{TorVergata}
\author[\theTorVergata]{N.~Tantalo}
\affil[\theTorVergata]{INFN, Sezione di Tor Vergata, c/o Dipartimento di Fisica,
 Universit\`a di Roma Tor Vergata, Via della Ricerca Scientifica 1, 00133 Rome, Italy}

\author[\theFermilab]{R.~Van de Water}

\addtocounter{affilctr}{1}
\newcounter{Zgz}
\addtocounter{affilctr}{1}
\newcounter{CAPA}
\author[\theZgz,\theCAPA]{A.~Vaquero}
\affil[\theZgz]{Departamento de Física Teórica, Universidad de Zaragoza, Calle Pedro Cerbuna 12, 50009 Zaragoza, Spain}
\affil[\theCAPA]{Center for Astroparticles and High Energy Physics (CAPA), Calle Pedro Cerbuna 12, 50009 Zaragoza, Spain}

\addtocounter{affilctr}{1}
\newcounter{Bern}
\author[\theBern]{U.~Wenger}
\affil[\theBern]{Albert Einstein Center for Fundamental Physics,
Institut f\"ur Theoretische Physik, Universit\"at Bern, Sidlerstr. 5, 3012 Bern, Switzerland}

\addtocounter{affilctr}{1}
\newcounter{Mainz}
\addtocounter{affilctr}{1}
\newcounter{Helmholtz}
\author[\theMainz,\theHelmholtz]{H.~Wittig}
\affil[\theMainz]{PRISMA Cluster of Excellence, Institut f\"ur
Kernphysik and Helmholtz Institute Mainz, University of Mainz, 55099 Mainz,
Germany}
\affil[\theHelmholtz]{Helmholtz Institute Mainz and GSI Helmholtz Center for Heavy Ion Research, 64291 Darmstadt, Germany}

\date{\today}

\begin{document}
\input{defs.tex}

 \hfill
 \begin{minipage}{0.4\textwidth}
 \begin{flushright}    
 \vspace{-1cm}
CERN-TH-2024-192
 \end{flushright}
 \end{minipage}
{\let\newpage\relax\maketitle} 

\abstract{ We review lattice results related to pion, kaon, $D$-meson,
  $B$-meson, and nucleon physics with the aim of making them easily
  accessible to the nuclear and particle physics communities. More
  specifically, we report on the determination of the light-quark
  masses, the form factor $f_+(0)$ arising in the semileptonic $K \to
  \pi$ transition at zero momentum transfer, as well as the
  decay-constant ratio $f_K/f_\pi$ and its consequences for the CKM
  matrix elements $V_{us}$ and $V_{ud}$.  We review the determination
  of the $B_K$ parameter of neutral kaon mixing as well as the
  additional four $B$ parameters that arise in theories of physics
  beyond the Standard Model.  For the heavy-quark sector, we provide
  results for $m_c$ and $m_b$ as well as those for the decay
  constants, form factors, and mixing parameters of charmed and bottom
  mesons and baryons.  These are the heavy-quark quantities most
  relevant for the determination of CKM matrix elements and the global
  CKM unitarity-triangle fit.  We review the status of lattice
  determinations of the strong coupling constant $\alpha_s$.  We
  review the determinations of nucleon charges from the matrix
  elements of both isovector and flavour-diagonal axial, scalar and
  tensor local quark bilinears, and momentum fraction, helicity moment
  and the transversity moment from one-link quark bilinears.  We also
  review determinations of scale-setting quantities. Finally, in this
  review we have added a new section on the general definition of the
  low-energy limit of the Standard Model.}

\newpage

\fancyhead{}
\fancypagestyle{firststyle}
{
\fancyhead[L]{Y.~Aoki et~al.\hfill
  \emph{{FLAG Review 2024}}\hfill
  \href{http://arxiv.org/abs/241X.XXXXX}{{\tt 241X.XXXXX}}}
}
\fancypagestyle{defaultstyle}
               {
               }
\fancypagestyle{updXXX20XXstyle}
               {
                 \fancyfoot[R]{\emph{Updated XXX.~20XX}}
               }
\fancypagestyle{draft}
               {
                 \fancyfoot[R]{\emph{Final draft 30/10/2024}}
               }

\def\reducedapptables{}


\def\noglossary{}

\tableofcontents

\clearpage
\input{introduction.tex}


\input{quality_criteria}

\clearpage
\input{./ibscheme/ediconsensus_main}

\clearpage
\input{qmass/qmass}

\clearpage
\input{Vudus/VudVus}


\clearpage
\input{BK/BK}

\clearpage
\input{HQ/macros_static.sty}
\clearpage
\input{HQ/HQD.tex}

\clearpage
\input{HQ/HQB.tex}
\clearpage
\pagestyle{plain}
\input{Alpha_s/alpha}

\clearpage
\input{NME/NME}

\clearpage
\input{ScaleSetting/scalesetting}

\clearpage
\pagestyle{plain}
\input{acknowledgment}

\appendix

\begin{appendix}
\clearpage
\input{acronyms.tex}
\ifx\noglossary\undefined  
\clearpage
\section{Glossary}\label{comm}
\input{Glossary/gloss_app}
\else
\section{Appendix}
\input{Glossary/app_qed.tex}
\input{Glossary/app_zParameterization.tex}
\fi

\clearpage
\section{Notes}

In the following Appendices we provide more detailed information on the simulations used
to calculate the quantities discussed in Secs.~\ref{sec:qmass}--\ref{sec:scalesetting}. 
\ifx\reducedapptables\undefined
\else
We present this information only for results
that are new w.r.t.~FLAG 21. For all other results the information is available
in the corresponding Appendices C.1--C.9 in FLAG 21 \cite{FlavourLatticeAveragingGroupFLAG:2021npn}, B.1--B.8 in FLAG 19 \cite{FlavourLatticeAveragingGroup:2019iem}, and B.1--B.7 in FLAG 16~\cite{Aoki:2016frl}.
\fi

\input{qmass/qmass_app}
\clearpage
\input{qmass/qmass_mc_app}

\clearpage
\input{Vudus/VudVus_app}


\clearpage
\input{BK/BK_app}

\input{BK/Kpipi_app}

\clearpage
\input{HQ/HQ_app}

\clearpage
\input{Alpha_s/Alpha_s_app}

\clearpage
\input{NME/NME_app}

\clearpage
\input{ScaleSetting/scalesetting_app}

\end{appendix}


\clearpage
\settocbibname{References}
\bibliography{FLAG}
\bibliographystyle{JHEP_FLAG}

\end{document}

%% file: defs.tex
\newcommand{\todo}[1]{~\\{\color{red}\minibox[frame,pad=0pt]{#1}}\\}
\definecolor{lightgreen}{rgb}{.85,.99,.85}
\definecolor{darkgreen}{rgb}{0,.7,0}
\definecolor{orange}{rgb}{1.0,.6,0}
\newcommand{\tbr}{\hspace{0.25mm}{\color{red}$\blacksquare$}} 
\newcommand{\tbg}{{\color{green}$\bigstar$}}
\newcommand{\tbo}{\hspace{0.25mm}{\color{orange}\Large\textbullet}}
\newcommand{\rC}{C}
\newcommand{\gA}{A}
\newcommand{\oP}{P} 
\newcommand{\red}[1]{\color[rgb]{1,0,0} #1 \color[rgb]{0,0,0}}
\newcommand{\blue}[1]{\color[rgb]{0,0,1} #1 \color[rgb]{0,0,0}}
\newcommand{\green}[1]{\color[rgb]{0,0.6,0} #1 \color[rgb]{0,0,0}}
\newcommand{\bda}{\begin{\displaymath}\begin{array}{rl}}
\newcommand{\eda}{\end{array}\end{displaymath}}
\newcommand{\be}{\begin{equation}}
\newcommand{\ee}{\end{equation}}
\newcommand{\bdm}{\begin{displaymath}}
\newcommand{\edm}{\end{displaymath}}
\newcommand{\bea}{\begin{eqnarray}}
\newcommand{\eea}{\end{eqnarray}}
\newcommand{\no}{\nonumber \\}
\newcommand{\fs}{\,.}
\newcommand{\co}{\,,}
\newcommand{\eff}{{e\hspace{-0.1em}f\hspace{-0.18em}f}}
\newcommand{\ind}{\scriptscriptstyle}
\newcommand{\indR}{\mbox{\scriptstyle R}}
\newcommand{\indL}{\mbox{\scriptstyle L}}
\newcommand{\QCD}{\mbox{\scriptstyle Q\hspace{-0.1em}CD}}
\newcommand{\qbar}{\overline{\rule[0.42em]{0.4em}{0em}}\hspace{-0.45em}q}
\newcommand{\ubar}{\overline{\rule[0.42em]{0.4em}{0em}}\hspace{-0.5em}u}
\newcommand{\dbar}{\,\overline{\rule[0.65em]{0.4em}{0em}}\hspace{-0.6em}d}
\newcommand{\sbar}{\,\overline{\rule[0.42em]{0.4em}{0em}}\hspace{-0.5em}s}
\newcommand{\lbar}{\bar{\ell}}
\newcommand{\Kbar}{\,\overline{\rule[0.62em]{0.5em}{0em}}\hspace{-0.8em}K}
\newcommand{\lsim}{\,\raisebox{-0.3em}{$\stackrel{\raisebox{-0.1em}{$<$}}{\sim}$
}\,} 
\newcommand{\gsim}{\,\raisebox{-0.3em}{$\stackrel{\raisebox{-0.1em}{$>$}}{\sim}$
}\,}
\newcommand{\lvac}{\langle 0|\,}
\newcommand{\rvac}{\,|0\rangle}
\newcommand{\rs}{\langle r^2\rangle\rule[-0.2em]{0em}{0em}_s}
\newcommand{\al}{&\!\!\!}
\newcommand{\Al}{\!\!&\hspace{-0.1cm}}
\newcommand{\Ch}{$\chi$} 
\newcommand{\mat}[1]{\begin{pmatrix}#1\end{pmatrix}}

\newcommand{\Mpibar}{\rule{0.05cm}{0cm}\overline{\hspace{-0.08cm}M}_{\hspace{-0.04cm}\pi}}
\newcommand{\MKbar}{\rule{0.05cm}{0cm}\overline{\hspace{-0.08cm}M}_{\hspace{-0.04cm}K}}
\newcommand{\ie}{i.e.}
\newcommand{\eg}{e.g.}

\newcommand{\remove}[1]{{\cblu [ remove: #1]}}
\newcommand{\off}[1]{{}}
\newcommand{\flagold}{FLAG 13}

\def\gn#1{\hspace*{3em} \pageref{#1}}%
\def\bib#1{\vspace*{-0.3cm}\bibitem{#1}}%

\newcommand{\ora}{\overset{\rightarrow}}
\newcommand{\ola}{\overset{\leftarrow}}
\newcommand{\bi}{\begin{itemize}}
\newcommand{\ei}{\end{itemize}}
\newcommand{\mycite}{\color{blue}\footnotesize\itshape}

\newcommand{\beq}{\begin{equation}}
\newcommand{\eeq}{\end{equation}}

\newcommand{\Mpi}{M_\pi}
\newcommand{\Fpi}{F_\pi}
\newcommand{\Mka}{M_K}
\newcommand{\Fka}{F_K}
\newcommand{\Met}{M_\et}
\newcommand{\Fet}{F_\et}

\newcommand{\ovr}{\over}
\newcommand{\til}{\tilde}
\newcommand{\pri}{^\prime}
\renewcommand{\dag}{^\dagger}
\newcommand{\<}{\langle}
\renewcommand{\>}{\rangle}

\newcommand{\lonebar}{\ln\frac{\Lambda_1^2}{M_\pi^2}}
\newcommand{\ltwobar}{\ln\frac{\Lambda_2^2}{M_\pi^2}}
\newcommand{\lthreebar}{\ln\frac{\Lambda_3^2}{M_\pi^2}}
\newcommand{\lfourbar}{\ln\frac{\Lambda_4^2}{M_\pi^2}}
\newcommand{\lsixbar}{\ln\frac{\Lambda_6^2}{M_\pi^2}}
\newcommand{\lMbar}{\ln\frac{\Omega_M^2}{M_\pi^2}}
\newcommand{\lFbar}{\ln\frac{\Omega_F^2}{M_\pi^2}}
\newcommand{\lSbar}{\ln\frac{\Omega_S^2}{M_\pi^2}}

\newcommand{\MeV}{\,\mathrm{MeV}}
\newcommand{\Refs}{\,\mathrm{Refs.}}
\newcommand{\eV}{\electronvolt}

\ifdefined\Ref
  \renewcommand{\Ref}{\,\mathrm{Ref.}}
\else
  \newcommand{\Ref}{\,\mathrm{Ref.}}
\fi

\newcommand{\GeV}{\,\mathrm{GeV}}
\newcommand{\fm}{\,\mathrm{fm}}

\newcommand{\ep}{\epsilon}
\newcommand{\de}{\delta}

\newcommand{\et}{\eta}


\newcommand{\kcrit}{\mbox{$\kappa_c$}}
\newcommand{\ksea}{\mbox{$\kappa^{\rm sea}$}}
\newcommand{\kval}{\mbox{$\kappa^{\rm val}$}}
\newcommand{\kstrange}{\mbox{$\kappa_s$}}
\newcommand{\keff}{\kappa_{\rm eff}}

\newcommand{\rme}{{\rm e}}
\newcommand{\rmO}{{\rm O}}
\newcommand{\msbar}{{\overline{{\rm MS}}}}
\newcommand{\lms}{\Lambda_\msbar}
\newcommand{\mut}{\multicolumn{2}{c}{\mbox{}}}
\newcommand{\mute}{\multicolumn{2}{c|}{\mbox{}}}

 
\def\mev{{\rm MeV}}
\def\gev{{\rm GeV}}
\def\tev{{\rm TeV}}
\def\fm{{\rm fm}}
 
 
\def\qbar{\bar{q}}
\def\psibar{\bar{\psi}}
\def\chibar{\bar{\chi}}
\def\ubar{\bar{u}} 

 
\def\ba{b_{\rm A}}
\def\bp{b_{\rm P}}
\def\ca{c_{\rm A}}
\def\cv{c_{\rm V}}
\def\csw{c_{\rm sw}}
\def\cs{c_{\rm s}}
\def\ct{c_{\rm t}}
\def\ctildes{\tilde{c}_{\rm s}}
\def\ctildet{\tilde{c}_{\rm t}}
\def\ctildest{\tilde{c}_{\rm s,t}}
 
\def\gbar{\bar{g}}
\def\mbar{\bar{m}}


\newcommand{\bd}{\begin{displaymath}}
\newcommand{\ed}{\end{displaymath}}


\newcommand{\eq}[1]{Eq.~(\ref{#1})}
\newcommand{\fig}[1]{Fig.~\ref{#1}}
\newcommand{\tab}[1]{Tab.~\ref{#1}}
\newcommand{\sect}[1]{Sec.~\ref{#1}}

\newcommand{\plus}{\makebox[15pt][c]{$+$}}
\newcommand{\minus}{\makebox[15pt][c]{$-$}}
\newcommand{\figurebox}[2]{\fbox{\vbox to#2in{\hbox to #1in{\hfil}\vfil}}}
%
\newcommand{\errr}[2]{\raisebox{0.08em}{\scriptsize
                                            {$\;\begin{array}{@{}l@{}}
                          \plus\makebox[0.9em][r]{#1} \\[-0.12em]
                          \minus\makebox[0.9em][r]{#2}
                        \end{array}$}}}
\newcommand{\err}[2]{\raisebox{0.08em}{\scriptsize
                          {$\;\begin{array}{@{}l@{}}
                          \plus\makebox[0.55em][r]{#1} \\[-0.12em]
                          \minus\makebox[0.55em][r]{#2}
                        \end{array}$}}}
\newcommand{\er}[2]{\raisebox{0.08em}{\scriptsize
                          {$\;\begin{array}{@{}l@{}}
                          \plus\makebox[0.15em][r]{#1} \\[-0.12em]
                          \minus\makebox[0.15em][r]{#2}
                        \end{array}$}}}
\newcommand{\Err}[2]{\raisebox{0.08em}{\scriptsize
                                            {$\;\begin{array}{@{}l@{}}
                          \plus\makebox[1.55em][r]{#1} \\[-0.12em]
                          \minus\makebox[1.55em][r]{#2}
                        \end{array}$}}}
\newcommand{\texp}{\tau^{\rm exp}}
\newcommand{\tint}{\tau^{\rm int}}
\newcommand{\tcum}{\tau^{\rm cum}}
\newcommand{\bm}[1]{\mbox{\boldmath ${#1}$}}
\newcommand{\sbm}[1]{\scriptstyle\mbox{\boldmath ${#1}$}}
\newcommand{\rvec}{\bm{r}}
\newcommand{\dvec}{\bm{d}}
\newcommand{\sdvec}{\sbm{d}}
\newcommand{\gtaeq}{\raisebox{-.6ex}{$\stackrel{\textstyle{>}}{\sim}$}}
\newcommand{\Tmin}{t_{\rm min}}
\newcommand{\Tmax}{t_{\rm max}}
\newcommand{\Nf}{N_{\hspace{-0.08 em} f}}
\newcommand{\Tr}{{\rm Tr}\,}

\newcommand{\fKfpicharged}{ \frac{f_{K^\pm}}{f_{\pi^\pm}}}
\newcommand{\fKfpichargedr}{ {f_{K^\pm}}/{f_{\pi^\pm}}}

\newcommand{\mps}{m_{\rm PS}}
\newcommand{\mpi}{M_\pi}
\newcommand{\mK}{M_{\rm K}}
\newcommand{\mv}{m_{\rm V}}
\newcommand{\FK}{F_{\rm K}}
\newcommand{\fpi}{f_\pi}
\newcommand{\fK}{f_{\rm K}}
\newcommand{\Fp}{F_{\rm PS}}
\newcommand{\fp}{f_{\rm PS}}
\newcommand{\Gp}{G_{\rm PS}}
\def\BB{B_{\rm B}}
\def\BBd{B_{\rm B_d}}
\def\BBs{B_{\rm B_s}}
\def\BBhat{\hat{B}_{\rm B}}
\def\BBdhat{\hat{B}_{\rm B_d}}
\def\BBshat{\hat{B}_{\rm B_s}}
\def\fB{f_{\rm B}}
\def\fBd{f_{\rm B_d}}
\def\fBs{f_{\rm B_s}}
\def\fD{f_{\rm D}}
\def\fDd{f_{\rm D_d}}
\def\fDs{f_{\rm D_s}}

\newcommand{\mup}{m_{\rm u}}
\newcommand{\mdo}{m_{\rm d}}
\newcommand{\mq}{m_{\rm q}}
\newcommand{\mst}{m_{\rm s}}
\newcommand{\mval}{m^{\rm val}}
\newcommand{\msea}{m^{\rm sea}}

\newcommand{\RFG}{R_{\rm M}}
\newcommand{\RF}{R_{\rm F}}
\newcommand{\mref}{m_{\rm ref}}
\newcommand{\yref}{y_{\rm ref}}

\newcommand{\avec}{\vec{a}}
\newcommand{\jvec}{\vec{j}}
\newcommand{\kvec}{\vec{k}}
\newcommand{\uvec}{\vec{u}}
\newcommand{\vvec}{\vec{v}}
\newcommand{\xvec}{\vec{x}}
\newcommand{\yvec}{\vec{y}}
\newcommand{\zvec}{\vec{z}}
\newcommand{\pvec}{\vec{p}}
\newcommand{\qvec}{\vec{q}}
\newcommand{\aop}{{\hat{a}}}
\newcommand{\aopdag}{{\hat{a}^\dagger}}
\newcommand{\xop}{\underline{\hat{x}}}
\newcommand{\pop}{\underline{\hat{p}}}
\newcommand{\qop}{\underline{\hat{q}}}
\newcommand{\Hhat}{\hat{H}}
\newcommand{\phii}{\phi_{\rm in}}
\newcommand{\phio}{\phi_{\rm out}}

\newcommand{\dbydt}{\frac{\partial}{\partial{t}}}
\newcommand{\dbydtsq}{\frac{\partial^2}{\partial{t^2}}}
\newcommand{\dxbydt}[1]{\frac{\partial{#1}}{\partial{t}}}

\newcommand{\covmeas}[1]{\frac{d^3{#1}}{(2\pi)^3\,2E(\underline{#1})}}

\newcommand{\half}{\textstyle{1\over2}}

\newcommand{\abar}{\overline{a}}

\newcommand{\zp}{Z_{\rm P}}
\newcommand{\zM}{Z_{\rm M}}
\newcommand{\za}{Z_{\rm A}}

\newcommand{\mbf}[1]{\mbox{\boldmath${#1}$}}
\newcommand{\rb}[1]{\raisebox{1.5ex}[-1.5ex]{#1}}

\newcommand{\Lo}{\stackrel{\rule[-0.1cm]{0cm}{0cm}\mbox{\tiny LO}}{=}}
\newcommand{\NLo}{\stackrel{\rule[-0.1cm]{0cm}{0cm}\mbox{\tiny NLO}}{=}} 
\newcommand{\epsilonD}{\epsilon}
\def\blue{\color{blue}}
\def\black{\color{black}}
\def\green{\color{green}}
\def\magenta{\color{magenta}}
\def\mahogany{\color{brown}}
\def\red{\color{red}}
\def\orange{\color{orange}}
\def\cyan{\color{cyan}}
\definecolor{Gray}{rgb}{0.5,0.5,0.5}
\definecolor{Black}{rgb}{0.0,0.0,0.0}
\newcommand{\gry}{\color{Gray}}
\newcommand{\bla}{\color{Black}}

\def\good{\makebox[1em]{\centering{\mbox{\color{green}$\bigstar$}}}}
\def\bad{\makebox[1em]{\centering\color{red}\tiny$\blacksquare$}}
\def\soso{\makebox[1em]{\centering{\mbox{\raisebox{-0.5mm}{\color{green}\Large$\circ$}}}}}
\def\okay{\hspace{0.25mm}\raisebox{-0.2mm}{{\color{green}\large\checkmark}}}

\newcommand{\mr}{\mathrm}

\def\half{{1\over2}}
\def\bra#1{\left\langle #1\right|}
\def\ket#1{\left| #1\right\rangle}
\def\vev#1{\left\langle #1\right\rangle}
\def\hc{\mathrm{h.c.}}
\def\su#1{{SU\left(#1\right)}}
\def\u#1{{U\left(#1\right)}}
\def\d#1#2{d\mskip 1.5mu^{#1}\mkern-1mu{#2}\,}
\def\D#1#2{{\d#1{#2} \over (2\pi)^{#1}}\,}
\def\dtilde#1{{\d3{#1} \over (2\pi)^3 \,2\omega_{#1}}\,}
\def\tr{\,\mathrm{tr}}
\def\Tr{\,\mathrm{Tr}}
\def\str{\,\mathrm{str}}
\def\det{\,\mathrm{det}}
\def\sdet{\,\mathrm{sdet}}
\def\fm{\mathrm{fm}}
\def\ev{\mathrm{e\kern-0.1em V}}
\def\kev{\mathrm{ke\kern-0.1em V}}
\def\mev{\mathrm{Me\kern-0.1em V}}
\def\gev{\mathrm{Ge\kern-0.1em V}}
\def\tev{\mathrm{Te\kern-0.1em V}}
\def\ps{\mathrm{ps}}
\def\re{\,\mathrm{Re}}
\def\im{\,\mathrm{Im}}
\let\Re=\re \let\Im=\im
\def\dotp#1#2{#1\mathord\cdot #2}
\def\n#1e#2n{{#1}\times 10^{#2}}

\newcommand{\simas}[1]{\raisebox{-.1ex}{$\stackrel{\small{#1}}{\sim}$}}

\def\ra{\rangle}
\def\la{\langle}

\def\ord#1{\mathcal{O}(#1)}
\def\bea{\begin{eqnarray}}
\def\eea{\end{eqnarray}}
\def\nn{\nonumber}
\def\On{\mathcal{O}_n}

\def\oO{\mathcal{Q}}
\def\cO{\mathcal{O}}
\def\cF{\mathcal{F}}
\def\cP{\mathcal{P}}
\def\cH{\mathcal{H}}
\def\cL{\mathcal{L}}
\def\cM{\mathcal{M}}
\def\cD{\mathcal{D}}
\def\cE{\mathcal{E}}
\def\cR{\mathcal{R}}
\def\cpv{
\math{\mathrm{CP}\hspace{-0.6cm}\slash{}\hspace{0.4cm}}}
\def\ods2{\mathcal{O}_{\Delta S=2}}
\def\zds2{Z_{\Delta S=2}}
\def\orda{$\blue O(a)$}

\def\msbar{{\overline{\mathrm{MS}}}}
\def\NDR{\mathrm{NDR}}
\def\RI{\mathrm{RI}}
\def\RGI{\mathrm{RGI}}
\def\BSM{\mathrm{BSM}}
\def\LO{\mathrm{LO}}
\def\NLO{\mathrm{NLO}}
\def\AWI{\mathrm{AWI}}
\def\lqcd{\Lambda_\mathrm{QCD}}
\def\lat{\mathrm{lat}}

\def\spose#1{\hbox to 0pt{#1\hss}}
\def\ltapprox{\mathrel{\spose{\lower 3pt\hbox{$\mathchar"218$}}
 \raise 2.0pt\hbox{$\mathchar"13C$}}}
\def\gtapprox{\mathrel{\spose{\lower 3pt\hbox{$\mathchar"218$}}
 \raise 2.0pt\hbox{$\mathchar"13E$}}}
\def\inapprox{\mathrel{\spose{\lower 3pt\hbox{$\mathchar"218$}}
 \raise 2.0pt\hbox{$\mathchar"232$}}}
 
 \def\lbabar{\mbox{{\footnotesize\sl B}\hspace{-0.4em} {\scriptsize\sl A}\hspace{-0.03em}{\footnotesize\sl B}\hspace{-0.4em} {\scriptsize\sl A\hspace{-0.02em}R}}}

\makeatletter
\def\slash#1{{\mathpalette\c@ncel{#1}}} 
\def\big#1{{\hbox{$\left#1\vbox to1.012\ht\strutbox{}\right.\n@space$}}}
\def\Big#1{{\hbox{$\left#1\vbox to1.369\ht\strutbox{}\right.\n@space$}}}
\def\bigg#1{{\hbox{$\left#1\vbox to1.726\ht\strutbox{}\right.\n@space$}}}
\def\Bigg#1{{\hbox{$\left#1\vbox
to2.083\ht\strutbox{}\right.\n@space$}}}
\makeatother

\newcommand{\nl}{\nonumber \\}
\newcommand{\mone}{M_1^{(0)}}
\newcommand{\monesub}{M_{1,sub}^{(1)}}
\newcommand{\intk}{g^2 \frac{4}{3}\int\frac{d^4k}{(2 \pi)^4}}
\newcommand{\hkt}{\hat{k}_0}
\newcommand{\hkj}{\hat{k}_j}
\newcommand{\dl}{{\delta{\cal L}}}
\newcommand{\delv}{{\bf \nabla}}
\newcommand{\delvt}{\tilde{{\bf \nabla}}}
\newcommand{\delvc}{{\bf D}}
\newcommand{\msb}{{\overline{\rm MS}}}
\newcommand{\Mbz}{{M_0}}
\newcommand{\delfour}{{\Delta^{(4)}}}
\newcommand{\delsq}{\Delta^{(2)}}
\newcommand{\delsqc}{D^{(2)}}
\newcommand{\xv}{\vec{x}}
\newcommand{\yv}{\vec{y}}
\newcommand{\nscsk}{n_{sc},n_{sk}}
\newcommand{\tminmax}{t_{min}/t_{max}}
\newcommand{\Dv}{{\bf D}}
\newcommand{\nte}{{\bf E}}
\newcommand{\Ev}{\tilde{{\bf E}}}
\newcommand{\Bv}{\tilde{{\bf B}}}
\newcommand{\sigmav}{\mbox{\boldmath$\sigma$}}
\newcommand{\kp}{k^\prime}

\def\spose#1{\hbox to 0pt{#1\hss}}
\def\ltapprox{\mathrel{\spose{\lower 3pt\hbox{$\mathchar"218$}}
\raise 2.0pt\hbox{$\mathchar"13C$}}}
\def\gtapprox{\mathrel{\spose{\lower 3pt\hbox{$\mathchar"218$}}
\raise 2.0pt\hbox{$\mathchar"13E$}}}
\def\inapprox{\mathrel{\spose{\lower 3pt\hbox{$\mathchar"218$}}
\raise 2.0pt\hbox{$\mathchar"232$}}}

\newcommand{\specialcell}[2][c]{%
  \begin{tabular}[#1]{@{}c@{}}#2\end{tabular}}
\newcommand{\specialcellthree}[3][c]{%
  \begin{tabular}[#1]{@{}c@{}}#2{}#3\end{tabular}}

\newcommand{\cred}{\color{red}}   
\newcommand{\cgre}{\color{green}}   
\newcommand{\cblu}{\color{blue}} 
\newcommand{\cmag}{\color{magenta}} 
  
\newcommand{\alphah}{\alpha_\mathrm{V'}}
\newcommand{\alphav}{\alpha_\mathrm{V}}
\newcommand{\alphap}{\alpha_\mathrm{P}}

\newcommand{\SLfnalmilcBDstar}{FNAL/MILC~21}   
\newcommand{\SLfnalmilcBD}{FNAL/MILC~15C}               
\newcommand{\SLfnalmilcBpi}{FNAL/MILC~15}                              
\newcommand{\SLjlqcdBpi}{JLQCD~22}                              
\newcommand{\SLfnalmilcBsK}{FNAL/MILC~19}                              
\newcommand{\SLfnalmilcDproc}{FNAL/MILC~14B}                            
\newcommand{\SLfnalmilcD}{FNAL/MILC~15B}                                        
\newcommand{\SLhpqcdBD}{HPQCD~15}                              
\newcommand{\SLhpqcdBpi}{HPQCD~15A}                                             
\newcommand{\SLhpqcdBsK}{HPQCD~14}                                              
\newcommand{\SLrbcukqcdBpi}{RBC/UKQCD~15}                                       
\newcommand{\SLalphaBsK}{ALPHA~14B}                                              
\newcommand{\SLLambdabp}{Detmold~15\\ $\Lambda_b \to p$}                                                
\newcommand{\SLLambdabc}{Detmold~15\\ $\Lambda_b \to \Lambda_c$}                                                
\newcommand{\SLhpqcdBK}{HPQCD~13E}                                              
\newcommand{\SLfnalmilcBK}{FNAL/MILC~15D}                              

\newcommand{\FLAGAVBEGIN}{}
\newcommand{\FLAGAVEND}{}
\newcommand{\eql}{}

\newcommand{\scal}{\mathcal{S}}
\newcommand{\rainer[1]}{{\color{red} Rainer: #1}}

\newcommand{\soutFLAG}[1]{{\red{\sout{#1}}}}
\newcommand{\newtextFLAG}[1]{{\blue #1}}

%% file: introduction.tex
\section{Introduction}
\label{sec:introduction}

Flavour physics provides an important opportunity for exploring the
limits of the Standard Model of particle physics and for constraining
possible extensions that go beyond it. 
As the LHC and its experiments continue exploring the energy frontier, and as experiments such as Belle-II, BESIII, NA62 and KOTO-2 continue extending the precision and intensity frontiers, the importance of flavour physics will grow,
both in terms of searches for signatures of new physics through
precision measurements and in terms of attempts to construct the
theoretical framework behind direct discoveries of new particles. 
Crucial to such searches for new physics is the ability to quantify
strong-interaction effects.
Large-scale numerical
calculations of lattice QCD allow for the computation of these effects
from first principles. 
The scope of the Flavour Lattice Averaging
Group (FLAG) is to review the current status of lattice results for a
variety of physical quantities that are important for flavour physics. Set up in
November 2007, it
comprises experts in Lattice Field Theory, Chiral Perturbation
Theory, and Standard Model phenomenology. 
Our aim is to provide an answer to the frequently posed
question ``What is currently the best lattice value for a particular
quantity?" in a way that is readily accessible to those who are not
expert in lattice methods.
This is generally not an easy question to answer;
different collaborations use different lattice actions
(discretizations of QCD) with a variety of lattice spacings and
volumes, and with a range of masses for the $u$ and $d$ quarks. Not
only are the systematic errors different, but also the methodology
used to estimate these uncertainties varies between collaborations. In
the present work, we summarize the main features of each of the
calculations and provide a framework for judging and combining the
different results. Sometimes, it is a single result that provides the
``best" value; more often, it is a combination of results from
different collaborations. Indeed, when consistency of values obtained using 
different formulations is found, this adds significantly to our confidence in the results. 

The first five editions of the FLAG review were made public in
2010~\cite{Colangelo:2010et}, 2013~\cite{Aoki:2013ldr},
2016~\cite{Aoki:2016frl}, 2019~\cite{FlavourLatticeAveragingGroup:2019iem}, and 2021~\cite{FlavourLatticeAveragingGroupFLAG:2021npn} 
(and will be referred to as FLAG 10, FLAG 13, FLAG 16, FLAG 19, and FLAG 21, respectively).
The fifth edition reviewed results related to both light 
($u$-, $d$- and $s$-quark), and heavy ($c$- and $b$-quark) flavours.
The quantities related to pion and kaon physics were 
light-quark masses, the form factor $f_+(0)$
arising in semileptonic $K \rightarrow \pi$ transitions 
(evaluated at zero momentum transfer), 
the decay constants $f_K$ and $f_\pi$, 
the $B_K$ parameter from neutral kaon mixing,
and the kaon mixing matrix elements of new operators that arise in 
theories of physics beyond the Standard Model.
Their implications for
the CKM matrix elements $V_{us}$ and $V_{ud}$ were also discussed.
Furthermore, results
were reported for some of the low-energy constants of SU$(2)_L \times$SU$(2)_R$ and SU$(3)_L \times$SU$(3)_R$ Chiral Perturbation Theory.
The quantities related to $D$- and $B$-meson physics that were
reviewed were the masses of the charm and bottom quarks
together with the decay constants, form factors, and mixing parameters
of $D$ and $B$ mesons.
These are the heavy-light quantities most relevant
to the determination of CKM matrix elements and the global
CKM unitarity-triangle fit. 
The current status of 
lattice results on the QCD coupling  $\alpha_s$ was reviewed.
Last but not least, we reviewed calculations of nucleon matrix elements 
of flavour nonsinglet and singlet bilinear operators, including the nucleon axial charge
$g_A$ and the nucleon sigma term.
These results are relevant for constraining $V_{ud}$, for searches for new physics
in neutron decays and other processes, and for dark matter searches.

In FLAG 21, we extended the scope of the review by adding a section on
scale setting, Sec.~\ref{sec:scalesetting}. 
The motivation for adding this section was that uncertainties in the value of
the lattice spacing $a$ are a major source of error for the calculation of
a wide range of quantities. Thus we felt that a systematic compilation of results,
comparing the different approaches to setting the scale, and summarizing the present status,
would be a useful resource for the lattice community.
An additional update was the inclusion, in Sec.~\ref{sec:Kpipi_amplitudes}, of a brief description
of the status of lattice calculations of $K\to\pi\pi$ decay amplitudes. Although some aspects of these calculations
were not yet at the stage to be included in our averages, they are approaching this stage, and
we felt that, given their phenomenological relevance, a brief review was appropriate.

In the current review, we have omitted the section on low-energy constants
in the chiral Lagrangian as progress in that area has slowed. FLAG will keep monitoring the
situation and provide updates in future editions, should new results become available.
On the other hand, we have added a new section on isospin breaking, Sec.~\ref{sec:ibscheme}.
For the most precisely determined quantities, 
isospin breaking---both from the up-down quark-mass difference and from QED---must be included.
An important issue here is that, in the context of a QED$+$QCD theory,
the separation into QED and QCD contributions to a given physical quantity
is ambiguous.  There are several ways of defining such a separation.
The new section allows a more uniform treatment 
in the sections on quark masses (see Sec.~\ref{sec:qmass}) and 
scale setting (see Sec.~\ref{sec:scalesetting}).
We stress, however, that the physical observable in QCD$+$QED is defined 
unambiguously. Any ambiguity only  arises because we are trying to separate a well-defined, physical quantity into two unphysical parts that provide useful
information for phenomenology.

Our main results are collected in Tabs.~\ref{tab:summary1}, \ref{tab:summary2},
\ref{tab:summary4}, \ref{tab:summary5}, and \ref{tab:summary_SL}.
As is clear from the tables, for most quantities there are results from ensembles with
different values for $\Nf$.  In most cases, there is reasonable agreement among
results with $\Nf=2$, $2+1$, and $2+1+1$.  As precision increases, we may
some day be able to distinguish among the different values of $\Nf$, in
which case, presumably $2+1+1$ would be the most realistic.  (If isospin
violation is critical, then $1+1+1$ or $1+1+1+1$ might be desired.)
At present, for some quantities the errors in the $\Nf=2+1$ results are smaller
than those with $\Nf=2+1+1$ (e.g., for $m_c$), while for others
the relative size of the errors is reversed. 
In most situations we expect the averages in this report for both $\Nf = 2+1$ or $\Nf = 2+1+1$ 
to provide a sufficiently accurate description of QCD. In situations where 
charm-sea-quark and/or isospin-breaking effects are expected to be subdominant systematic effects, 
both  results can be used.
We do not recommend using the $\Nf=2$ results, except for studies of
the $\Nf$-dependence of $\alpha_s$, as these have an uncontrolled
systematic error coming from quenching the strange quark.

Our plan is to continue providing FLAG updates, in the form of a peer
reviewed paper, roughly on a triennial basis. This effort is
supplemented by our more frequently updated
website \href{http://flag.itp.unibe.ch}{{\tt
http://flag.itp.unibe.ch}} \cite{FLAG:webpage}, where figures as well as pdf-files for
the individual sections can be downloaded. The papers reviewed in the
present edition have appeared before the closing date {\bf 30 April 2024}.\footnote{%
  Working groups were given the option of including papers submitted to {\tt arxiv.org}
  before the closing date but published after this date. This flexibility allows this review to
  be up-to-date at the time of submission. Two papers of this type were included, cf.~footnote \ref{footnote:deadline exceptions}.
  \label{footnote:including papers after deadline}}

\input summarytable.tex

This review is organized as follows.  In the remainder of
Sec.~\ref{sec:introduction}, we summarize the composition and rules of
FLAG and discuss general issues that arise in modern lattice
calculations.  In Sec.~\ref{sec:qualcrit}, we explain our general
methodology for evaluating the robustness of lattice results.  We also
describe the procedures followed for combining results from different
collaborations in a single average or estimate (see
Sec.~\ref{sec:averages} for our definition of these terms). The rest
of the paper consists of sections, each dedicated to a set of
closely connected physical quantities, 
or, for the final section, to the determination of the lattice scale.
Each of these
sections is accompanied by an Appendix with explicatory notes.\footnote{%
In order to keep the length of this review within reasonable bounds,
we have dropped these notes for older data, since they can be found in 
previous FLAG reviews~\cite{Colangelo:2010et,Aoki:2013ldr,Aoki:2016frl,FlavourLatticeAveragingGroup:2019iem,FlavourLatticeAveragingGroupFLAG:2021npn}.}

In previous editions, we have provided, in an appendix, a glossary summarizing
some standard lattice terminology and
describing the most commonly used lattice techniques and methodologies.
Since no significant updates in this information have occurred recently,
we have decided, in the interests of reducing the length
of this review, to omit this glossary, 
and refer the reader to FLAG 19 for this information~\cite{FlavourLatticeAveragingGroup:2019iem}.
This appendix also contained, in previous versions, 
a tabulation of the actions used in the papers that were reviewed.
Since this information is available in the discussions in the separate sections,
and is time-consuming to collect from the sections, we have dropped these tables.
In Appendix~\ref{app:acronyms}, we have added a summary and explanations of acronyms introduced in the manuscript. Collaborations referred to by an acronym can be identified through the corresponding bibliographic reference.
In Appendix~\ref{app:qed}, we provide a short review of how electromagnetic effects can be taken into account in lattice-QCD calculations.
Appendix~\ref{sec:zparam} describes the
parameterizations of semileptonic form factors that are used in Sec.~\ref{sec:BDecays}. 
A short appendix, Appendix~\ref{sec:zparam_explicit} provides all the details of the
parameters used in the form factor fits in Secs.~\ref{sec:DDecays} and \ref{sec:BDecays}.

\subsection{FLAG composition, guidelines and rules}

FLAG strives to be representative of the lattice community, both in
terms of the geographical location of its members and the lattice
collaborations to which they belong. We aspire to provide the nuclear- and
particle-physics communities with a single source of reliable
information on lattice results.

In order to work reliably and efficiently, we have adopted a formal
structure and a set of rules by which all FLAG members abide.  The
collaboration presently consists of an Advisory Board (AB), an
Editorial Board (EB), and eight Working Groups (WG). The r\^{o}le of
the Advisory Board is to provide oversight of the content, procedures, schedule
and membership of FLAG, to help resolve disputes, to serve as a source
of advice to the EB and to FLAG as a whole, and to provide a
critical assessment of drafts.
They also give their approval of the final version of the preprint before
it is released. The Editorial Board coordinates the activities
of FLAG, sets priorities and intermediate deadlines, organizes votes on
FLAG procedures, writes the introductory sections, and takes care of
the editorial work needed to integrate the sections written by the
individual working groups into a uniform and coherent review. The
working groups concentrate on writing the review of the physical
quantities for which they are responsible, which is subsequently
circulated to the whole collaboration for critical evaluation.

The current list of FLAG members and their Working Group assignments is:
\begin{itemize}
\item
Advisory Board (AB):\hfill
M.~Golterman, P.~Hernandez, T.~Onogi, S.R.~Sharpe, \\
\hbox{} \hfill and R.~Van de Water
\item
Editorial Board (EB):\hfill
S.~Gottlieb, A.~J\"uttner, and U.~Wenger
\item
Working Groups (coordinator listed first):
\begin{itemize}
\item Quark masses \hfill T.~Blum, A.~Portelli, and A.~Ramos
\item $V_{us},V_{ud}$ \hfill T.~Kaneko, J.~N.~Simone, and N.~Tantalo
\item Kaon mixing \hfill P.~Dimopoulos, X.~Feng, and G.~Herdo{\'i}za
\item $f_{B_{(s)}}$, $f_{D_{(s)}}$, $B_B$ \hfill C.~Monahan, Y.~Aoki, and M.~Della Morte
\item  $b$ and $c$ semileptonic and radiative decays
  \hfill E.~Lunghi, S.~Meinel,\\
  \hbox{} \hfill and A.~Vaquero
\item $\alpha_s$ \hfill S.~Sint, L.~Del Debbio, and P.~Petreczky
\item Nucleon matrix elements \hfill R.~Gupta, S.~Collins, A.~Nicholson, and
 H.~Wittig
 \item Scale setting \hfill R.~Sommer, N.~Tantalo, and U.~Wenger
\end{itemize}
\end{itemize}

The most important FLAG guidelines and rules are the following:
\begin{itemize}
\item
the composition of the AB reflects the main geographical areas in
which lattice collaborations are active, with members from
America, Asia/Oceania, and Europe;
\item
the mandate of regular members is not limited in time, but we expect that a
certain turnover will occur naturally;
\item
whenever a replacement becomes necessary this has to keep, and
possibly improve, the balance in FLAG, so that different collaborations, from
different geographical areas are represented;
\item
in all working groups the  
members must belong to 
different lattice collaborations;
\item
a paper is in general not reviewed (nor colour-coded, as described in
the next section) by any of its authors;
\item
lattice collaborations 
will be consulted on the colour coding
of their calculation;
\item
there are also internal rules regulating our work, such as voting procedures.
\end{itemize}
 
As for FLAG 21, for this review we sought the advice of external reviewers
once a complete draft of the review was available. For each review section, we
have asked one lattice expert (who could be a FLAG alumnus/alumna) and
one nonlattice phenomenologist for a critical assessment.\footnote{The one exception is the scale-setting section, where only a lattice
expert has been asked to provide input.}
This is similar
to the procedure followed by the Particle Data Group in the creation of the
Review of Particle Physics.  The reviewers provide comments and feedback on
scientific and stylistic matters. They are not anonymous, and enter into a discussion with
the authors of the WG. Our aim with this additional step is to make sure that a wider
array of viewpoints enter into the discussions, 
so as to make this review more useful for its intended audience.

\subsection{Citation policy}
We draw attention to this particularly important point.  As stated
above, our aim is to make lattice-QCD results easily accessible to
those without lattice expertise,
and we are well aware that it is likely that some
readers will only consult the present paper and not the original
lattice literature. It is very important that this paper not be the
only one cited when our results are quoted. We strongly suggest that
readers also cite the original sources. In order to facilitate this,
in Tabs.~\ref{tab:summary1}, \ref{tab:summary2},
\ref{tab:summary4}, \ref{tab:summary5} and \ref{tab:summary_SL}, besides
summarizing the main results of the present review, we also cite the
original references from which they have been obtained. In addition,
for each figure we make a bibtex file available on our webpage
\cite{FLAG:webpage} which contains the bibtex entries of all the
calculations contributing to the FLAG average or estimate. The
bibliography at the end of this paper should also make it easy to cite
additional papers. Indeed, we hope that the bibliography will be one of
the most widely used elements of the whole paper.

\subsection{General issues}

Several general issues concerning the present review are thoroughly
discussed in Sec.~1.1 of our initial 2010 paper~\cite{Colangelo:2010et},
and we encourage the reader to consult the relevant pages. In the
remainder of the present subsection, we focus on a few important
points. Though the discussion has been duly updated, it is similar
to that of the corresponding sections in the previous reviews~\cite{Aoki:2013ldr,Aoki:2016frl,FlavourLatticeAveragingGroup:2019iem,FlavourLatticeAveragingGroupFLAG:2021npn}.

The review aims to achieve two distinct goals:
first, to provide a {\bf description} of the relevant work done on the lattice;
and, second,
to draw {\bf conclusions} on the basis of that work,  summarizing
the results obtained for the various quantities of physical interest.

The core of the information about the work done on the lattice is
presented in the form of tables, which not only list the various
results, but also describe the quality of the data that underlie
them. We consider it important that this part of the review represents
a generally accepted description of the work done. For this reason, we
explicitly specify the quality requirements
used and provide sufficient details in appendices so that the reader
can verify the information given in the tables.\footnote{%
We also use terms
like ``quality criteria", ``rating", ``colour coding", etc., when referring to
the classification of results, as described in Sec.~\ref{sec:qualcrit}.}

On the other hand, the conclusions drawn 
on the basis of the available lattice results
are the responsibility of FLAG alone. Preferring to
err on the side of caution, in several cases we draw
conclusions that are more conservative than those resulting from
a plain weighted average of the available lattice results. This cautious
approach is usually adopted when the average is
dominated by a single lattice result, or when
only one lattice result is available for a given quantity. In such
cases, one does not have the same degree of confidence in results and
errors as when there is agreement among several different
calculations using different approaches. The reader should keep
in mind that the degree of confidence cannot be quantified, and
it is not reflected in the quoted errors. 

Each discretization has its merits, but also its shortcomings. For most
topics covered in this review we
have an increasingly broad database, and for most quantities
lattice calculations based on totally different discretizations are
now available. This is illustrated by the dense population of the
tables and figures in most parts of this review. Those
calculations that do satisfy our quality criteria indeed lead, in almost all cases, to
consistent results, confirming universality within the accuracy
reached. The consistency between independent lattice
results, obtained with different discretizations, methods, and
lattice parameters, is an important test of lattice QCD, and
observing such consistency also provides further evidence that
systematic errors are fully under control.

In the sections dealing with heavy quarks and with $\alpha_s$, the
situation is not the same. Since the $b$-quark mass can barely be resolved
with current lattice spacings, most lattice methods for treating $b$
quarks use effective field theory at some level. This introduces
additional complications not present in the light-quark sector.  An
overview of the issues specific to heavy-quark quantities is given in
the introduction of Sec.~\ref{sec:BDecays}. For $B$- and $D$-meson
leptonic decay constants, there already exist a good number of
different independent calculations that use different heavy-quark
methods, but there are only a few independent calculations of
semileptonic $B$, $\Lambda_b$, and $D$ form factors and of $B-\bar B$ mixing
parameters. 
For $\alpha_s$, most lattice methods involve a range of
scales that need to be resolved and controlling the systematic error
over a large range of scales is more demanding. The issues specific to
determinations of the strong coupling are summarized in Sec.~\ref{sec:alpha_s}.
\smallskip
\\{\it Number of sea quarks in lattice calculations:}\\
\noindent
Lattice-QCD calculations currently involve two, three or four flavours of
dynamical quarks. Most calculations set
the masses of the two lightest quarks to be equal, while the
strange and charm quarks, if present, are heavier
(and tuned to lie close to their respective physical values). 
Our notation for these calculations indicates which quarks
are nondegenerate, e.g., 
$\Nf=2+1$ if $m_u=m_d < m_s$ and $\Nf =2+1+1$ if $m_u=m_d < m_s < m_c$. 
Calculations with $\Nf =2$, i.e., two degenerate dynamical
flavours, often include strange valence quarks interacting with gluons,
so that bound states with the quantum numbers of the kaons can be
studied, albeit neglecting strange sea-quark {contributions}.  The
quenched approximation ($\Nf=0$), in which all sea-quark contributions 
are omitted, has uncontrolled systematic errors and
is no longer used in modern lattice calculations with relevance to phenomenology.
{Lattice calculations with $\Nf =2$ also have an uncontrolled systematic error,
although it is reasonable to expect it to be much smaller than for $\Nf = 0$.}    
Accordingly, we will review results obtained with $\Nf=2+1$
and $\Nf = 2+1+1$, but omit earlier results with $\Nf=0$ {and $\Nf=2$}.{\footnote{It is left to the discretion of the WG to include $\Nf = 2$ results in case they provide interesting additional information. We refer to previous editions of FLAG \cite{FLAG:webpage} for the discussion of omitted $\Nf = 2$ results.}} 
{One notable} exception concerns the QCD coupling constant $\alpha_s$.
Since this observable does not require valence light quarks,
it is theoretically well defined also in the $\Nf=0$ theory,
which is simply pure gluodynamics.
The $\Nf$-dependence of $\alpha_s$, 
or more precisely of the related quantity $r_0 \Lambda_\msbar$, 
is a theoretical issue of considerable interest; here $r_0$ is a quantity
with the dimension of length that sets the physical scale, as discussed in
 Sec.~\ref{sec:scalesetting}.
We stress, however, that only results with $\Nf \ge 3$ 
are used to determine the physical value of $\alpha_s$ at a high scale.
\smallskip
\\{\it Lattice actions, parameters, and scale setting:}\\
\noindent
The remarkable progress in the precision of lattice
calculations is due to improved algorithms, better computing resources,
and, last but not least, conceptual developments.
Examples of the latter are improved
actions that reduce lattice artifacts and actions that preserve
chiral symmetry to a very good approximation.
A concise characterization of
the various discretizations that underlie the results reported in the
present review is given in 
Appendix~A.1 of FLAG 19 \cite{FlavourLatticeAveragingGroup:2019iem}.

Physical quantities are computed in lattice calculations in units of the
lattice spacing so that they are dimensionless.
For example, the pion decay constant that is obtained from a calculation
is $f_\pi a$, where $a$ is the spacing between two neighboring lattice sites.
(All calculations with results quoted in this review use hypercubic lattices,
i.e., with the same spacing in all four Euclidean directions.)
To convert these results to physical units requires knowledge
of the lattice spacing $a$ at the fixed values of the bare QCD parameters
(quark masses and gauge coupling) used in the calculation.
This is achieved by requiring agreement between
the lattice calculation and experimental measurement of a known
quantity, which thus ``sets the scale" of a given calculation.
(See Sec.~\ref{sec:scalesetting}.)
\smallskip
\\{\it Renormalization and scheme dependence:}\\
\noindent
Several of the results covered by this review, such as quark masses,
the gauge coupling, and $B$-parameters, are for quantities defined in a
given renormalization scheme and at a specific renormalization scale. 
The schemes employed (e.g., regularization-independent MOM schemes) are often
chosen because of their specific merits when combined with the lattice
regularization. For a brief discussion of their properties, see
Appendix A.3 of FLAG 19 \cite{FlavourLatticeAveragingGroup:2019iem}.
The conversion of the results obtained in
these so-called intermediate schemes to more familiar regularization
schemes, such as the $\msbar$-scheme, is done with the aid of
perturbation theory. It must be stressed that the renormalization
scales accessible in calculations are limited, because of the presence
of an ultraviolet (UV) cutoff of $\sim \pi/a$.
To safely match to $\msbar$, a scheme defined in perturbation theory,
Renormalization Group (RG) running to higher scales is performed,
either perturbatively or nonperturbatively (the latter using
finite-size scaling techniques).
\smallskip
\\{\it Extrapolations:}\\
\noindent
Because of limited computing resources, lattice calculations are often
performed at unphysically heavy pion masses, although results at the
physical point, where all quark masses take their physical values, have become increasingly common. Further, numerical
calculations must be done at nonzero lattice spacing, and in a finite
(four-dimensional) volume.  In order to obtain physical results,
lattice data are obtained at a sequence of pion masses and a sequence
of lattice spacings, and then extrapolated to the physical pion mass
and to the continuum limit.  In principle, an extrapolation to
infinite volume is also required. However, for most quantities
discussed in this review, finite-volume effects are exponentially
small in the linear extent of the lattice in units of the pion mass,
and, in practice, one often verifies volume independence by comparing
results obtained on a few different physical volumes, holding other
parameters fixed. To control the associated systematic uncertainties,
these extrapolations are guided by effective theories.  For
light-quark actions, the lattice-spacing dependence is described by
Symanzik's effective theory~\cite{Symanzik:1983dc,Symanzik:1983gh};
for heavy quarks, this can be extended and/or supplemented by other
effective theories such as Heavy-Quark Effective Theory (HQET).  The
pion-mass dependence can be parameterized with Chiral Perturbation
Theory ($\chi$PT), which takes into account the Nambu-Goldstone nature
of the lowest excitations that occur in the presence of light
quarks. Similarly, one can use Heavy-Light Meson Chiral Perturbation
Theory (HM$\chi$PT) to extrapolate quantities involving mesons
composed of one heavy ($b$ or $c$) and one light quark.  One can
combine Symanzik's effective theory with $\chi$PT to simultaneously
extrapolate to the physical pion mass and the continuum; in this case,
the form of the effective theory depends on the discretization.  See
Appendix A.4 of FLAG 19 \cite{FlavourLatticeAveragingGroup:2019iem}
for a brief description of the different
variants in use and some useful references.  Finally, $\chi$PT can
also be used to estimate the size of finite-volume effects measured in
units of the inverse pion mass, thus providing information on the
systematic error due to finite-volume effects in addition to that
obtained by comparing calculations at different volumes.
\smallskip
\\{\it Excited-state contamination:}\\
\noindent
In all the hadronic matrix elements discussed in this review, the hadron in question
is the lightest state with the chosen quantum numbers. This implies that it dominates the
required correlation functions as their extent in Euclidean time is increased. Excited-state
contributions are suppressed by $e^{-\Delta E \Delta \tau}$, 
where $\Delta E$ is the gap between
the ground and excited states, and $\Delta \tau$ the relevant separation in Euclidean time. 
The size of $\Delta E$ depends on the hadron in question, and in general
is a multiple of the pion mass. In practice, as discussed at length in Sec.~\ref{sec:NME},
the contamination of signals due to  excited-state contributions is a much more
challenging problem for baryons than for the other particles discussed here.
This is in part due to the fact that the signal-to-noise ratio drops exponentially for
baryons, which reduces the values of $\Delta \tau$ that can be used.
\smallskip
\\{\it Critical slowing down:}\\
\noindent
The lattice spacings reached in recent calculations go down to 0.05 fm
or even smaller. In this regime, long autocorrelation times slow down
the sampling of the
configurations~\cite{Antonio:2008zz,Bazavov:2010xr,Schaefer:2010hu,Luscher:2010we,Schaefer:2010qh,Chowdhury:2013mea,Brower:2014bqa,Fukaya:2015ara,DelDebbio:2002xa,Bernard:2003gq}.
Many groups check for autocorrelations in a number of observables,
including the topological charge, for which a rapid growth of the
autocorrelation time is observed with decreasing lattice spacing.
This is often referred to as topological freezing. A solution to the
problem consists in using open boundary conditions in time~\cite{Luscher:2011kk}, 
instead of the more common periodic or antiperiodic ones. A combination of open and periodic boundary conditions have recently been employed in a parallel tempering framework \cite{Bonanno:2024zyn}.  Other approaches have been proposed, e.g., based on a multiscale
thermalization algorithm \cite{Endres:2015yca,Detmold:2018zgk}, or based on
defining QCD on a nonorientable manifold \cite{Mages:2015scv}, or using huge master fields \cite{Francis:2019muy,Fritzsch:2021klm}. Approaches using trivializing or normalizing flows \cite{Luscher:2009eq} try to solve both the problem of topological freezing and critical slowing down by finding invertible maps from simple probability distributions for the lattice configurations, which can be efficiently sampled, to the target ones. Parameterizing these flows turns out to be difficult, but can be facilitated by using  machine-learning tools \cite{Albergo:2019eim,Kanwar:2020xzo,Boyda:2020hsi,Gerdes:2022eve,Bacchio:2022vje}. So far, these attempts are restricted to simple field theories, low dimensions or, in four-dimensional SU(3) gauge theories, to very small and coarse systems \cite{Abbott:2023thq}. Reference \cite{Holland:2024muu} uses machine learning to construct RG-improved gauge actions with highly suppressed lattice artifacts, such that efficient calculations on coarse lattices suffice to yield solid continuum limits.  
The
problem of topological freezing and critical slowing down is also touched upon in Sec.~\ref{s:crit}, where it is
stressed that attention must be paid to this issue.

Few results reviewed here have been obtained with any of the above methods.
It is usually {\it  assumed} that the continuum limit can be reached by extrapolation
from the existing calculations, and that potential systematic errors due
to the long autocorrelation times have been adequately controlled.
Partially or completely frozen topology would produce a mixture of different $\theta$ vacua, and 
the difference from the desired $\theta=0$ result
may be estimated in some cases using  
 chiral perturbation theory, which gives predictions for the $\theta$-dependence of the 
physical quantity of interest \cite{Brower:2003yx,Aoki:2007ka}. These ideas have been systematically and successfully tested in various models in \cite{Bautista:2015yza,Bietenholz:2016ymo}, and a numerical test on MILC ensembles indicates that the topology dependence 
for some of the physical quantities reviewed here is small, consistent with theoretical 
expectations~\cite{Bernard:2017npd}.
\smallskip
\\ {\it Algorithms and numerical errors:}\\
\noindent
Most of the modern lattice-QCD calculations use exact algorithms such 
as those of Refs.~\cite{Duane:1987de,Clark:2006wp}, which do not produce any systematic errors when exact 
arithmetic is available. In reality, one uses numerical calculations at 
double (or in some cases even single) precision, and some errors are 
unavoidable. More importantly, the inversion of the Dirac operator is 
carried out iteratively and it is truncated once some accuracy is 
reached, which is another source of potential systematic error. In most 
cases, these errors have been confirmed to be much less than the 
statistical errors. In the following, we assume that this source of error 
is negligible. 
Some of the most recent calculations use an inexact algorithm in order to 
speed up the computation, though it may produce systematic effects. 
Currently available tests indicate that errors from the use of inexact
algorithms are under control~\cite{Bazavov:2012xda}.

%% file: summarytable.tex
\clearpage
\begin{sidewaystable}[ph!]
\vspace{-1cm}
\centering
\begin{tabular}{||l|l||l|l||l|l||l|l||l||l|l|}
\hline
Quantity \rule[-0.2cm]{0cm}{0.6cm}    & \hspace{-1.5mm}Sec.\hspace{-2mm} &$\Nf=2+1+1$ & Refs. &  $\Nf=2+1$ & Refs. &$\Nf=2$ &Refs. \\
\hline \hline
\input Tables/Summarytable2.tex
\end{tabular}\\[0.2cm]


\caption{\label{tab:summary1} Summary of the main results of this review concerning quark 
masses, light-meson decay constants, the kaon semileptonic form factor, and hadronic kaon-decay and kaon-mixing parameters.
These are grouped in terms of $\Nf$, the number of dynamical quark flavours in lattice simulations. 
Quark masses are given in the $\msbar$ scheme at running scale  $\mu=2\,\gev$ or as indicated.
BSM bag parameters $B_{2,3,4,5}$ are given in the $\msbar$ scheme at scale $\mu=3\,\gev$.
Further specifications of the quantities are given in the quoted sections.
Results for $\Nf=2$ quark masses are unchanged since FLAG~16~\cite{Aoki:2016frl},
and are not included here.
For each result, we list the references that enter the FLAG average or estimate,
and we stress again the importance of quoting these original works when referring to
FLAG results. From the entries in this column one
can also read off the number of results that enter our averages for each quantity. We emphasize that these numbers only give a very rough indication of how thoroughly the quantity in question has been explored on the lattice and recommend consulting the detailed tables and figures in the relevant section for more significant information and for explanations on the source of the quoted errors.}
\end{sidewaystable}

\clearpage
\begin{sidewaystable}[ph!]
\vspace{-1cm}
\centering
\begin{tabular}{||l|l||l|l||l|l||l|l||l|l||l|l||l|l|}
\hline
Quantity \rule[-0.2cm]{0cm}{0.6cm}    & \hspace{-1.5mm}Sec.\hspace{-2mm} &$\Nf=2+1+1$ & Refs. &  $\Nf=2+1$ & Refs. &$\Nf=2$ &Refs. \\
\hline \hline
\input Tables/Summarytable1.tex
\hline\hline
Quantity \rule[-0.2cm]{0cm}{0.6cm}    & \hspace{-1.5mm}Sec.\hspace{-2mm} &\multicolumn{3}{c|}{$\Nf=2+1$ and $\Nf=2+1+1$} & Refs. & & \\
\hline\hline
\input Tables/Summarytable1alpha.tex
\end{tabular}
\caption{\label{tab:summary2}Summary of the main results of this review concerning heavy-light 
mesons and the strong coupling constant. These are grouped in terms of $\Nf$, the number of dynamical quark flavours in lattice simulations.   The  quantities listed are specified in the quoted sections.
For each result, we list the references that enter the FLAG average or estimate,
and we stress again the importance of quoting these original works when referring to
FLAG results.
From the entries in this column one
can also read off the number of results that enter our averages for each quantity. We emphasize that these numbers only give a very rough indication of how thoroughly the quantity in question has been explored on the lattice and recommend consulting the detailed tables and figures in the relevant section for more significant information and for explanations on the source of the quoted errors. 
}
\end{sidewaystable}
\clearpage

\begin{sidewaystable}[ph!]
\vspace{-1cm}
\centering
\begin{tabular}{||l|l||l|l||l|l||l|l||l|l||l|l||l|l|}
\hline
Quantity \rule[-0.2cm]{0cm}{0.6cm}    & \hspace{-1.5mm}Sec.\hspace{-2mm} &$\Nf=2+1+1$ & Refs. &  $\Nf=2+1$ & Refs. \\
\hline \hline
\input Tables/Summarytable3.tex
\end{tabular}
\caption{\label{tab:summary4}Summary of the main results of this review concerning nuclear matrix elements, grouped in terms of $\Nf$, the number of dynamical quark flavours in lattice simulations.   The  quantities listed are specified in the quoted sections.
For each result, we list the references that enter the FLAG average or estimate,
and we stress again the importance of quoting these original works when referring to
FLAG results.
From the entries in this column one
can also read off the number of results that enter our averages for each quantity. We emphasize that these numbers only give a very rough indication of how thoroughly the quantity in question has been explored on the lattice and recommend consulting the detailed tables and figures in the relevant section for more significant information and for explanations on the source of the quoted errors. 
}

\end{sidewaystable}

\clearpage
\begin{sidewaystable}[ph!]
\footnotesize
\vspace{-1cm}
\centering
\resizebox{\textwidth}{!}{
\begin{tabular}{||l|l||l|l||l|l||l|l||l|l||l}
\hline
Quantity \rule[-0.2cm]{0cm}{0.6cm}    & \hspace{-1.5mm}Sec.\hspace{-2mm} &$\Nf=1+1+1+1$ &Refs. &$\Nf=2+1+1$ & Refs. &  $\Nf=2+1$ & Refs. &$\Nf>2+1$ &Refs. \\
\hline \hline
\input Tables/Summarytable4.tex
\end{tabular}
}
\caption{\label{tab:summary5}Summary of the main results of this review
  concerning setting of the lattice scale, grouped in terms of $\Nf$, 
the number of dynamical quark flavours in lattice simulations.   The  quantities listed are specified in the quoted section.
For each result, we list the references that enter the FLAG average or estimate,
and we stress again the importance of quoting these original works when referring to
FLAG results.
From the entries in this column one
can also read off the number of results that enter our averages for each quantity. We emphasize that these numbers only give a very rough indication of how thoroughly the quantity in question has been explored on the lattice and recommend consulting the detailed tables and figures in the relevant section for more significant information and for explanations on the source of the quoted errors. 
}

\end{sidewaystable}
\clearpage

\begin{sidewaystable}[ph!]
\footnotesize
\vspace{-1cm}
\centering
\begin{tabular}{||l|l||l|l|l||l|l||l|l|l||}
\hline
Decay \rule[-0.2cm]{0cm}{0.6cm}    &form factor& fit& \hspace{-1.5mm}Sec.\hspace{-2mm} & \hspace{-1.5mm}Fig.\hspace{-2mm} &\parbox[t]{1.5cm}{Tab.\\[-1mm] {\tiny $\Nf=2+1+1$}} &Refs. &\parbox[t]{1.5cm}{Tab.\\[-1mm]{\tiny $\Nf=2+1$}} & Refs.  \\
\hline \hline
\input Tables/Summarytable_SL.tex
\hline 
\end{tabular}
\caption{\label{tab:summary_SL}Summary of the main results of this review
  concerning $z$-parameterizations of lattice results for semileptonic meson-decay form factors and experimental data for differential decay rates (see Appendix~\ref{sec:zparam}), grouped in terms of $\Nf$, 
the number of dynamical quark flavours in lattice simulations. The entry in the column ``fit'' indicates whether the fit is to only lattice data (``lat'') or combined to both lattice and experimental data (``lat+exp''). The references listed in the columns labelled ``Tab.'' lead to the tables that list the  results for the $z$-parameterization coefficients and their correlations. 
For each result, we list the references that enter the FLAG average or estimate,
and we stress again the importance of quoting these original works when referring to
FLAG results.
From the entries in this column one
can also read off the number of results that enter our averages for each quantity. We emphasize that these numbers only give a very rough indication of how thoroughly the quantity in question has been explored on the lattice and recommend consulting the detailed tables and figures in the relevant section for more significant information and for explanations on the source of the quoted errors. 
}

\end{sidewaystable}
\clearpage


%% file: Tables/Summarytable2.tex
$ m_{ud}$[MeV]&\ref{sec:msmud}&$3.427(51)$&\cite{Alexandrou:2021gqw,Carrasco:2014cwa,Bazavov:2018omf}&$3.387(39)$&\cite{CLQCD:2023sdb,Bruno:2019vup,Blum:2014tka,Durr:2010vn,Durr:2010aw,McNeile:2010ji,Bazavov:2010yq}&&\\[1mm]
$ m_s   $[MeV]&\ref{sec:msmud}&$93.46(58)$&\cite{Alexandrou:2021gqw,Carrasco:2014cwa,Bazavov:2018omf,Lytle:2018evc,Chakraborty:2014aca}&$92.4(1.0)$&\cite{Bruno:2019vup,Bazavov:2009fk,Durr:2010vn,Durr:2010aw,McNeile:2010ji,Blum:2014tka}&&\\[1mm]
$ m_s / m_{ud} $&\ref{sec:msovermud}&$27.227(81)$&\cite{Alexandrou:2021gqw,Bazavov:2017lyh,Carrasco:2014cwa,Bazavov:2014wgs}&$27.42(12)$&\cite{Bruno:2019xed,Blum:2014tka,Bazavov:2009fk,Durr:2010vn,Durr:2010aw}&&\\[1mm]
$ m_u $[MeV]&\ref{subsec:mumd}&$2.14(8)$&\citep{Giusti:2017dmp,Bazavov:2018omf}&$2.27(9)$&\citep{Fodor:2016bgu}&&\\[1mm]
$ m_d $[MeV]&\ref{subsec:mumd}&$4.70(5)$&\citep{Giusti:2017dmp,Bazavov:2018omf}&$4.67(9)$&\citep{Fodor:2016bgu}&&\\[1mm]
$ {m_u}/{m_d} $&\ref{subsec:mumd}&$0.465(24)$&\citep{Giusti:2017dmp,Basak:2018yzz}&$0.485(19)$&\citep{Fodor:2016bgu}&&\\[1mm]
\hline
$\overline{m}_c(\mbox{3 GeV}) $[GeV]&\ref{sec:mcnf4}&$0.989(10)$&\cite{Carrasco:2014cwa,Chakraborty:2014aca,Alexandrou:2014sha,Bazavov:2018omf,Hatton:2020qhk,Alexandrou:2021gqw}&$0.991(6)$&\cite{Bussone:2023kag,McNeile:2010ji,Yang:2014sea,Nakayama:2016atf,Petreczky:2019ozv,Heitger:2021apz}&&\\[1mm]
$ m_c / m_s $&\ref{sec:mcoverms}&$11.766(30)$&\cite{Chakraborty:2014aca,Carrasco:2014cwa,Bazavov:2018omf,Alexandrou:2021gqw}&$11.82(16)$&\cite{Yang:2014sea,Davies:2009ih}&&\\[1mm]
\hline
$\overline{m}_b(\overline{m}_b) $[GeV]&\ref{s:bmass}&$4.200(14)$&\cite{Hatton:2021syc,Colquhoun:2014ica,Bussone:2016iua,Gambino:2017vkx,Bazavov:2018omf}&$4.171(20)$&\cite{McNeile:2010ji}&&\\[1mm]
\hline
$ f_+(0) $&\ref{sec:Direct}&$0.9698(17)$&\cite{Carrasco:2016kpy,Bazavov:2018kjg}&$0.9677(27)$&\cite{Bazavov:2012cd,Boyle:2015hfa}&&\\[1mm]
$ f_{K^\pm} / f_{\pi^\pm}  $&\ref{sec:Direct}&$1.1934(19)$&\cite{Dowdall:2013rya,Carrasco:2014poa,Bazavov:2017lyh,Miller:2020xhy,Alexandrou:2021bfr}&$1.1916(34)$&\cite{Follana:2007uv,Bazavov:2010hj,Durr:2010hr,Blum:2014tka,Durr:2016ulb,Bornyakov:2016dzn,CLQCD:2023sdb}&&\\[1mm]
$ f_{\pi^\pm}$[MeV]&\ref{sec:fKfpi}&&&$130.2(8)$&\cite{Follana:2007uv,Bazavov:2010hj,Blum:2014tka,CLQCD:2023sdb}&&\\[1mm]
$ f_{K^\pm}  $[MeV]&\ref{sec:fKfpi}&$155.7(3)$&\cite{Dowdall:2013rya,Bazavov:2014wgs,Carrasco:2014poa,Alexandrou:2021bfr}&$155.7(7)$&\cite{Follana:2007uv,Bazavov:2010hj,Blum:2014tka,CLQCD:2023sdb}&&\\[1mm]
\hline
$ \text{Re}(A_2) $[GeV]&\ref{sec:Kpipi_amplitudes}&&&$1.50(4)(14)\times10^{-8}$&\cite{Blum:2015ywa}&&\\[1mm]
$ \text{Im}(A_2) $[GeV]&\ref{sec:Kpipi_amplitudes}&&&$-8.34(1.03)\times10^{-13}$&\cite{Blum:2015ywa}&&\\[1mm]
$\hat{B}_{K} $&\ref{sec:BK lattice}&$0.717(18)(16)$&\cite{Carrasco:2015pra}&$0.7533(91)$&\cite{Durr:2011ap,Laiho:2011np,Blum:2014tka,Jang:2015sla,Boyle:2024gge}&$0.727(22)(12)$&\cite{Bertone:2012cu}\\[1mm]
$ B_2$&\ref{sec:Bi}&$0.46(1)(3)$&\cite{Carrasco:2015pra}&$0.488(15)$&\cite{Jang:2015sla,Garron:2016mva,Boyle:2024gge}&$0.47(2)(1)$&\cite{Bertone:2012cu}\\[1mm]
$ B_3$&\ref{sec:Bi}&$0.79(6)$&\cite{Carrasco:2015pra}&$0.757(27)$&\cite{Jang:2015sla,Garron:2016mva,Boyle:2024gge}&$0.78(4)(2)$&\cite{Bertone:2012cu}\\[1mm]
$ B_4$&\ref{sec:Bi}&$0.78(2)(4)$&\cite{Carrasco:2015pra}&$0.903(14)$&\cite{Jang:2015sla,Garron:2016mva,Boyle:2024gge}&$0.76(2)(2)$&\cite{Bertone:2012cu}\\[1mm]
$ B_5$&\ref{sec:Bi}&$0.49(3)(3)$&\cite{Carrasco:2015pra}&$0.691(14)$&\cite{Jang:2015sla,Garron:2016mva,Boyle:2024gge}&$0.58(2)(2)$&\cite{Bertone:2012cu}\\[1mm]
\hline

%% file: Tables/Summarytable1.tex
$ f_D $[MeV]&\ref{sec:fD}&$212.0(7)$&\cite{Bazavov:2017lyh,Carrasco:2014poa}&$210.4(1.5)$&\cite{Bussone:2023kag,Na:2012iu,Bazavov:2011aa,Boyle:2017jwu}&&\\[1mm]
$ f_{D_s} $[MeV]&\ref{sec:fD}&$249.9(5)$&\cite{Bazavov:2017lyh,Carrasco:2014poa}&$247.7(1.2)$&\cite{Bussone:2023kag,Davies:2010ip,Bazavov:2011aa,Boyle:2017jwu,Yang:2014sea}&&\\[1mm]
$ f_{D_s}\over{f_D} $&\ref{sec:fD}&$1.1783(16)$&\cite{Bazavov:2017lyh,Carrasco:2014poa}&$1.174(7)$&\cite{Bussone:2023kag,Na:2012iu,Bazavov:2011aa,Boyle:2017jwu}&&\\[1mm]
$ f_+^{D\to \pi}(0)$&\ref{sec:DtoPiK}&$0.6296(50)$&\cite{Lubicz:2017syv}&$0.666(29)$&\cite{Na:2011mc}&&\\[1mm]
$ f_+^{D\to K}(0)  $&\ref{sec:DtoPiK}&$0.7430(27)$&\cite{Lubicz:2017syv,Chakraborty:2021qav}&$0.747(19)$&\cite{Na:2010uf}&&\\[1mm]
$ f_{B} $[MeV]&\ref{sec:fB}&$190.0(1.3)$&\cite{Dowdall:2013tga,Bussone:2016iua,Hughes:2017spc,Bazavov:2017lyh}&$192.0(4.3)$&\cite{Bazavov:2011aa,McNeile:2011ng,Na:2012kp,Aoki:2014nga,Christ:2014uea}&$188(7)$&\cite{Carrasco:2013zta,Bernardoni:2014fva}\\[1mm]
$ f_{B_{s}} $[MeV]&\ref{sec:fB}&$230.3(1.3)$&\cite{Dowdall:2013tga,Bussone:2016iua,Hughes:2017spc,Bazavov:2017lyh}&$228.4(3.7)$&\cite{Bazavov:2011aa,McNeile:2011ng,Na:2012kp,Aoki:2014nga,Christ:2014uea}&$225.3(6.6)$&\cite{Carrasco:2013zta,Bernardoni:2014fva,Balasubramanian:2019net}\\[1mm]
$ f_{B_{s}}\over{f_B} $&\ref{sec:fB}&$1.209(5)$&\cite{Dowdall:2013tga,Bussone:2016iua,Hughes:2017spc,Bazavov:2017lyh}&$1.201(16)$&\cite{Bazavov:2011aa,Na:2012kp,Aoki:2014nga,Christ:2014uea,Boyle:2018knm}&$1.206(23)$&\cite{Carrasco:2013zta,Bernardoni:2014fva}\\[1mm]
$ f_{B_d}\sqrt{\hat{B}_{b_d}}$[MeV]&\ref{sec:BMix}&$210.6(5.5)$&\cite{Dowdall:2019bea}&$225(9)$&\cite{Gamiz:2009ku,Aoki:2014nga,Bazavov:2016nty}&$216(10)$&\cite{Carrasco:2013zta}\\[1mm]
$ f_{B_s}\sqrt{\hat{B}_{B_s}}$[MeV]&\ref{sec:BMix}&$256.1(5.7)$&\cite{Dowdall:2019bea}&$274(8)$&\cite{Gamiz:2009ku,Aoki:2014nga,Bazavov:2016nty}&$262(10)$&\cite{Carrasco:2013zta}\\[1mm]
$ \hat{B}_{B_d}$&\ref{sec:BMix}&$1.222(61)$&\cite{Dowdall:2019bea}&$1.30(10)$&\cite{Gamiz:2009ku,Aoki:2014nga,Bazavov:2016nty}&$1.30(6)$&\cite{Carrasco:2013zta}\\[1mm]
$ \hat{B}_{B_s}$&\ref{sec:BMix}&$1.232(53)$&\cite{Dowdall:2019bea}&$1.35(6)$&\cite{Gamiz:2009ku,Aoki:2014nga,Bazavov:2016nty}&$1.32(5)$&\cite{Carrasco:2013zta}\\[1mm]
$ \xi $&\ref{sec:BMix}&$1.216(16)$&\cite{Dowdall:2019bea}&$1.206(17)$&\cite{Aoki:2014nga,Bazavov:2016nty}&$1.225(31)$&\cite{Carrasco:2013zta}\\[1mm]
$ B_{B_s}/B_{B_d}  $&\ref{sec:BMix}&$1.008(25)$&\cite{Dowdall:2019bea}&$1.032(38)$&\cite{Aoki:2014nga,Bazavov:2016nty}&$1.007(21)$&\cite{Carrasco:2013zta}\\[1mm]

%% file: Tables/Summarytable1alpha.tex
$ \alpha_{\overline{\rm MS}}^{(5)}(M_Z) $&\ref{s:alpsumm}&\multicolumn{3}{c|}{$0.1183(7)$}&\cite{DallaBrida:2022eua,Petreczky:2020tky,Ayala:2020odx,Bazavov:2019qoo,Cali:2020hrj,Bruno:2017gxd,Chakraborty:2014aca,McNeile:2010ji,Aoki:2009tf,Maltman:2008bx}&&\\[1mm]
$ \Lambda_{\overline{\rm MS}}^{(5)} $[MeV]&\ref{s:alpsumm}&\multicolumn{3}{c|}{$213(8)$}&\cite{DallaBrida:2022eua,Petreczky:2020tky,Ayala:2020odx,Bazavov:2019qoo,Cali:2020hrj,Bruno:2017gxd,Chakraborty:2014aca,McNeile:2010ji,Aoki:2009tf,Maltman:2008bx}&&\\[1mm]
$ \Lambda_{\overline{\rm MS}}^{(4)} $[MeV]&\ref{s:alpsumm}&\multicolumn{3}{c|}{$295(10)$}&\cite{DallaBrida:2022eua,Petreczky:2020tky,Ayala:2020odx,Bazavov:2019qoo,Cali:2020hrj,Bruno:2017gxd,Chakraborty:2014aca,McNeile:2010ji,Aoki:2009tf,Maltman:2008bx}&&\\[1mm]
$ \Lambda_{\overline{\rm MS}}^{(3)} $[MeV]&\ref{s:alpsumm}&\multicolumn{3}{c|}{$338(10)$}&\cite{DallaBrida:2022eua,Petreczky:2020tky,Ayala:2020odx,Bazavov:2019qoo,Cali:2020hrj,Bruno:2017gxd,Chakraborty:2014aca,McNeile:2010ji,Aoki:2009tf,Maltman:2008bx}&&\\[1mm]
\hline

%% file: Tables/Summarytable3.tex
$ g_A^{u-d} $&\ref{sec:gA-S-T-IV}&$1.263(10)$&\cite{Chang:2018uxx,Walker-Loud:2019cif,Jang:2023zts,Alexandrou:2023qbg}&$1.265(20)$&\cite{Liang:2018pis,Park:2021ypf,QCDSFUKQCDCSSM:2023qlx,Bali:2023sdi,Djukanovic:2024krw}\\[1mm]
$ g_S^{u-d} $&\ref{sec:gA-S-T-IV}&$1.085(114)$&\cite{Jang:2023zts}&$1.083(69)$&\cite{Liu:2021irg,Park:2021ypf,QCDSFUKQCDCSSM:2023qlx,Bali:2023sdi,Djukanovic:2024krw}\\[1mm]
$ g_T^{u-d} $&\ref{sec:gA-S-T-IV}&$0.981(21)$&\cite{Jang:2023zts,Alexandrou:2022dtc}&$0.993(15)$&\cite{Park:2021ypf,QCDSFUKQCDCSSM:2023qlx,Bali:2023sdi,Djukanovic:2024krw}\\[1mm]
$ g_A^u  $&\ref{sec:gA-gT-FD}&$\phantom{-}0.777(25)(30)$&\cite{Lin:2018obj}&$\phantom{-}0.847(18)(32)$&\cite{Liang:2018pis}\\[1mm]
$ g_A^d  $&\ref{sec:gA-gT-FD}&$-0.438(18)(30)$&\cite{Lin:2018obj}&$-0.407(16)(18)$&\cite{Liang:2018pis}\\[1mm]
$ g_A^s  $&\ref{sec:gA-gT-FD}&$-0.053(8)$&\cite{Lin:2018obj}&$-0.035(6)(7)$&\cite{Liang:2018pis}\\[1mm]
$ g_T^u  $&\ref{sec:gA-gT-FD}&$\phantom{-}0.784(28)(10)$&\cite{Gupta:2018lvp}&&\\[1mm]
$ g_T^d  $&\ref{sec:gA-gT-FD}&$-0.204(11)(10)$&\cite{Gupta:2018lvp}&&\\[1mm]
$ g_T^s  $&\ref{sec:gA-gT-FD}&$-0.0027(16)$&\cite{Gupta:2018lvp}&&\\[1mm]
$ \sigma_{\pi N} $[MeV]&\ref{sec:gS-sum}&$60.9(6.5)$&\cite{Alexandrou:2014sha,Gupta:2021ahb}&$42.2(2.4)	$&\cite{Durr:2011mp,Durr:2015dna,Yang:2015uis,RQCD:2022xux,Agadjanov:2023efe}\\[1mm]
$ \sigma_{s} $[MeV]&\ref{sec:gS-sum}&$41.0(8.8)$&\cite{Freeman:2012ry}&$44.9(6.4)$&\cite{Durr:2011mp,Freeman:2012ry,Junnarkar:2013ac,Durr:2015dna,Yang:2015uis,RQCD:2022xux,Agadjanov:2023efe}\\[1mm]
$ \langle x \rangle_{u-d}  $&\ref{sec:moments-results}&$0.158(32)$&\cite{Mondal:2020cmt,Alexandrou:2022dtc}&$0.153(13)$&\cite{Djukanovic:2024krw,Mondal:2020ela,Yang:2018nqn}\\[1mm]
$ \langle x \rangle_{\Delta u-\Delta d}   $&\ref{sec:moments-results}&$0.213(27)$&\cite{Mondal:2020cmt}&$0.200(13)$&\cite{Djukanovic:2024krw,Mondal:2020ela}\\[1mm]
$ \langle x \rangle_{\delta u-\delta d}   $&\ref{sec:moments-results}&$0.195(25)$&\cite{Mondal:2020cmt,Alexandrou:2022dtc}&$0.206(17)$&\cite{Djukanovic:2024krw,Mondal:2020ela}\\[1mm]
\hline

%% file: Tables/Summarytable4.tex
$ \sqrt{t_0} $[fm]&\ref{sec:scale averages}&&&$0.14292(104)$&\cite{Alexandrou:2021bfr,Miller:2020evg,Bazavov:2015yea,Dowdall:2013rya}&$0.14474(57)$&\cite{RQCD:2022xux,Bruno:2016plf,Blum:2014tka,Borsanyi:2012zs}&&\\[1mm]
$ w_0 $[fm]&\ref{sec:scale averages}&$0.17236(70)$&\cite{Borsanyi:2020mff}&$0.17256(103)$&\cite{Alexandrou:2021bfr,Miller:2020evg,Bazavov:2015yea,Dowdall:2013rya}&$0.17355(92)$&\cite{Blum:2014tka,Bazavov:2014pvz,Borsanyi:2012zs}&$0.17250(70)$&\cite{Alexandrou:2021bfr,Miller:2020evg,Borsanyi:2020mff,Bazavov:2015yea,Dowdall:2013rya}\\[1mm]
$ t_0/w_0 $[fm]&\ref{sec:scale averages}&&&$0.11969(62)$&\cite{Alexandrou:2021bfr}&&&&\\[1mm]
$ r_0 $[fm]&\ref{sec:scale averages}&&&$0.4580(73)$&\cite{Brambilla:2022het,Carrasco:2014cwa}&$0.4701(36)$&\cite{Bazavov:2014pvz,Yang:2014sea,Aoki:2010dy,Gray:2005ur,Aubin:2004wf}&&\\[1mm]
$ r_1 $[fm]&\ref{sec:scale averages}&&&$0.3068(37)$&\cite{Brambilla:2022het,Dowdall:2013rya}&$0.3127(30)$&\cite{Aoki:2010dy,Bazavov:2010hj,Davies:2009tsa,Gray:2005ur,Aubin:2004wf}&&\\[1mm]
$ f_{4ps} $[MeV]&\ref{sec:scale averages}&&&$153.98(20)$&\cite{Bazavov:2017lyh}&&&&\\[1mm]
$ M_{4ps} $[MeV]&\ref{sec:scale averages}&&&$433.12(30)$&\cite{Bazavov:2017lyh}&&&&\\[1mm]
\hline

%% file: Tables/Summarytable_SL.tex
$D  \to K  	\ell\nu$&$f_+,f_0	$&lat    & \ref{sec:DtoPiK}& \ref{fig:LQCDzfitDK} & \ref{tab:FFDK}  &\cite{Lubicz:2017syv,Chakraborty:2021qav,FermilabLattice:2022gku}&&\\
$D  \to K  	\ell\nu$&$f_+,f_0	$&lat+exp&\ref{sec:Vcd}	  &\ref{fig:DtoKdGammadqsqr} &\ref{tab:FFVCSPI}&\cite{Lubicz:2017syv,Chakraborty:2021qav,FermilabLattice:2022gku}&&\\
$B  \to \pi	\ell\nu$&$f_+,f_0	$&lat    &\ref{sec:BtoPi} &\ref{fig:LQCDzfit} &          &       &\ref{tab:FFPI}&\cite{Lattice:2015tia,Flynn:2015mha,Colquhoun:2022atw}\\
$B_s\to K	\ell\nu$&$f_+,f_0	$&lat    &\ref{sec:BstoK} & \ref{fig:LQCDzfitBsK} &          &       &\ref{tab:FFBSK}&\cite{Bouchard:2014ypa,Flynn:2023nhi,Flynn:2023nhi}\\
$B  \to \pi	\ell^-\ell^+$&$f_T    	$&lat    &\ref{sec:SLBrad}&&          &       &\ref{tab:FFPIT}&\cite{Bailey:2015nbd}\\
$B  \to K  	\ell^+\ell^-(\nu\bar\nu)$&$f_+,f_0,f_T	$&lat    &\ref{sec:SLBrad}& \ref{fig:LQCDzfitBK}&          &       &\ref{tab:FFK}&\cite{Bouchard:2013pna,Bailey:2015dka}\\
$B  \to D  	\ell\nu$&$f_+,f_0	$&lat    &\ref{sec:BstoDsFFs}&\ref{fig:LQCDzfitBD}     & &       &\ref{tab:FFD}&\cite{Lattice:2015rga,Na:2015kha}\\
$B_s\to D_s	\ell\nu$&$f_+,f_0	$&lat    &\ref{sec:BstoDsFFs}&&\ref{tab:BsDs}&\cite{McLean:2019qcx}&          &       \\
$B  \to D^\ast	\ell\nu$&$g,f,F_1,F_2	$&lat    & \ref{sec:BstoDstarFFs}& \ref{fig:BDstar_latt} & \ref{tab:BDstar_latt}&\cite{Harrison:2023dzh}&\ref{tab:BDstar_latt}&\cite{FermilabLattice:2021cdg,Aoki:2023qpa}\\
$B_s\to D_s^\ast\ell\nu$&$g,f,F_1,F_2	$&lat    & \ref{sec:BstoDstarFFs}& \ref{fig:BsDsstar_latt} & &\cite{Harrison:2023dzh}&&\\
$B  \to \pi	\ell\nu$&$f_+,f_0	$&lat+exp& \ref{sec:Vub} & \ref{fig:Vub_SL_fit} & &       &\ref{tab:FFVUBPI}&\cite{Lattice:2015tia,Flynn:2015mha,Colquhoun:2022atw,delAmoSanchez:2010af,Lees:2012vv,Ha:2010rf,Sibidanov:2013rkk}\\
$B  \to D  	\ell\nu$&$f_+,f_0	$&lat+exp& \ref{sec:Vcb}& \ref{fig:Vcb_SL_fit} & &       &\ref{tab:FFVCBD}&\cite{Lattice:2015rga,Na:2015kha,Aubert:2009ac,Belle:2015pkj}\\
$B  \to D^\ast	\ell\nu$&$g,f,F_1,F_2	$&lat+exp& \ref{sec:Vcb}& \ref{fig:BDstar_FF_latt+exp}, \ref{fig:BDstar_DeltaGamma_latt+exp} & \ref{tab:BDstar_latt+exp}&\cite{Harrison:2023dzh,Belle:2018ezy,Belle:2023bwv,Belle-II:2023okj,Belle-II:2020dyp,HFLAV:2022esi}&\ref{tab:BDstar_latt+exp}&\cite{FermilabLattice:2021cdg,Aoki:2023qpa,Belle:2018ezy,Belle:2023bwv,Belle-II:2023okj,Belle-II:2020dyp,HFLAV:2022esi}\\

%% file: quality_criteria.tex
\section{Quality criteria, averaging and error estimation}
\label{sec:qualcrit}

The essential characteristics of our approach to the problem of rating
and averaging lattice quantities 
have been outlined in our first publication~\cite{Colangelo:2010et}. 
Our aim is to help the reader
assess the reliability of a particular lattice result without
necessarily studying the original article in depth. This is a delicate
issue, since the ratings may make things appear 
simpler than they are. Nevertheless,
it safeguards against the 
possibility
of using lattice results, and
drawing physics conclusions from them, without a critical assessment
of the quality of the various calculations. We believe that, despite
the risks, it is important to provide some compact information about
the quality of a calculation. We stress, however, the importance of the
accompanying detailed discussion of the results presented in the various
sections of the present review.
 
\subsection{Systematic errors and colour code}
\label{sec:color-code}

The major sources of systematic error are common to most lattice
calculations. These include, as discussed in detail below,
the chiral, continuum, and infinite-volume extrapolations.
To each such source of error for which
systematic improvement is possible we
assign one of three coloured symbols: green
star, unfilled green circle
(which replaced in Ref.~\cite{Aoki:2013ldr}
the amber disk used in the original FLAG review~\cite{Colangelo:2010et})
or red square.
These correspond to the following ratings: 
\begin{itemize}[noitemsep,nolistsep] 
\item[\good] the parameter values and ranges used 
to generate the data sets allow for a satisfactory control of the systematic uncertainties;
\item[\soso] the parameter values and ranges used to generate
the data sets allow for a reasonable attempt at estimating systematic uncertainties, which
however could be improved;
\item[\bad] the parameter values and ranges used to generate
the data sets are unlikely to allow for a reasonable control of systematic uncertainties.
\end{itemize}
The appearance of a red tag, even in a
single source of systematic error of a given lattice result,
disqualifies it from inclusion in the global average.

Note that in the first two editions~\cite{Colangelo:2010et,Aoki:2013ldr},
FLAG used the three symbols in order to rate the reliability of the systematic errors 
attributed to a given result by the paper's authors.
Starting with FLAG 16~\cite{Aoki:2016frl} the meaning of the 
symbols has changed slightly---they now rate the quality of a particular simulation, 
based on the values and range of the chosen parameters,
and its aptness to obtain well-controlled systematic uncertainties. 
They do not rate the quality of the analysis performed by the authors 
of the publication. The latter question is
deferred to the relevant sections of the present review, 
which contain detailed discussions of 
the results contributing (or not) to each FLAG average or estimate. 

For most quantities the colour-coding system refers to the following  
sources of systematic errors: (i) chiral extrapolation; 
(ii) continuum extrapolation; (iii) finite volume. 
As we will see below, renormalization is another source of systematic
uncertainties in several quantities. This we also classify using the 
three coloured symbols listed above, but now with
a different rationale:  they express how reliably these quantities are 
renormalized, from a field-theoretic point of view
(namely, nonperturbatively, or with 2-loop or 1-loop perturbation theory).

Given the sophisticated status that the field has attained,
several aspects, besides those rated by the coloured symbols,
need to be evaluated before one can conclude
whether a particular analysis leads to results that should be included in an
average or estimate. Some of these aspects are not so easily expressible
in terms of an adjustable parameter such as the lattice spacing, the pion mass
or the volume. As a result of such considerations,
it sometimes occurs, albeit rarely, that a given
result does not contribute to the FLAG average or estimate, 
despite not carrying any red tags.
This happens, for instance, whenever aspects of the analysis appear 
to be incomplete 
(e.g., an incomplete error budget), so that the presence
of inadequately controlled systematic effects cannot be excluded. 
This mostly refers to results with a statistical error only, or results
in which the quoted error budget obviously fails to account 
for an important contribution.

Of course, any colour coding has to be treated with caution; we emphasize
that the criteria are subjective and evolving. Sometimes, a single
source of systematic error dominates the systematic uncertainty and it
is more important to reduce this uncertainty than to aim for green
stars for other sources of error. In spite of these caveats, we hope
that our attempt to introduce quality measures for lattice simulations
will prove to be a useful guide. In addition, we would like to
stress that the agreement of lattice results obtained using
different actions and procedures provides further validation.

\subsubsection{Systematic effects and rating criteria}
\label{sec:Criteria}

The precise criteria used in determining the colour coding are
unavoidably time-dependent; as lattice calculations become more
accurate, the standards against which they are measured become
tighter. For this reason FLAG reassesses criteria with each edition and as a result
some of the quality criteria (the one on chiral extrapolation for instance) have been tightened up   
over time~\cite{Colangelo:2010et,Aoki:2013ldr,Aoki:2016frl,FlavourLatticeAveragingGroup:2019iem}.

In the following, we present the rating criteria used in the current report. 
While these criteria apply to most quantities without modification,
there are cases where they need to be amended or additional criteria need to be defined. 
For instance,
the discussion
of the strong coupling constant in Sec.~\ref{sec:alpha_s} requires tailored criteria
for renormalization, perturbative behaviour, and continuum extrapolation. 
Finally, in the section on nuclear matrix elements, Sec.~\ref{sec:NME},
the chiral extrapolation criterion is made slightly stronger, and a new criterion is adopted for
excited-state contributions.
In such cases,
the modified criteria are discussed in the respective sections. Apart from only a few exceptions the 
following colour code applies in the tables:

\begin{itemize}
\item Chiral extrapolation:
\begin{itemize}[noitemsep,nolistsep] 
	\item[\good] $M_{\pi,\mathrm{min}}< 200$ MeV, with three or more pion masses used in the extrapolation \\
	\underline{or} two values of $M_\pi$ with one lying within 10 MeV of 135 MeV (the physical neutral pion mass) and the other one below 200 MeV  
	\item[\soso]  200 MeV $\le M_{\pi,{\mathrm{min}}} \le$ 400 MeV, with three or more pion masses used in the extrapolation \\\underline{or} two values of $M_\pi$ with $M_{\pi,{\mathrm{min}}}<$ 200 MeV \\\underline{or} a single value of $M_\pi$, lying within 10 MeV of 135 MeV (the physical neutral pion mass)
	\item[\bad] otherwise  
	\end{itemize}
This criterion is unchanged from FLAG 19. 
In Sec.~\ref{sec:NME} the upper end of the range for $M_{\pi,{\mathrm{min}}}$
in the green circle criterion is lowered to 300 MeV, as in FLAG 19.
\item 
Continuum extrapolation:
\begin{itemize}[noitemsep,nolistsep] 
	\item[\good] at least three lattice spacings \underline{and} at least two points below 0.1 fm \underline{and} a range of lattice spacings satisfying $[a_{\mathrm{max}}/a_{\mathrm{min}}]^2 \geq 2$
	\item[\soso] at least two lattice spacings \underline{and} at least one point below 0.1 fm 
	\underline{and} a range of lattice spacings 
	satisfying $[a_{\mathrm{max}}/a_{\mathrm{min}}]^2 \geq 1.4$
	\item[\bad] otherwise
\end{itemize}
It is assumed that the lattice action is $\cO(a)$-improved (i.e., the
discretization errors vanish quadratically with the lattice spacing);
otherwise this will be explicitly mentioned. For
unimproved actions an additional lattice spacing is required.
This condition is unchanged from FLAG 19.
\item 
Finite-volume effects:\\ 
The finite-volume colour code used for a result is 
chosen to be the worse of the QCD and the QED codes, as described below. If only QCD is used the QED colour code is ignored.

\emph{-- For QCD:}
\begin{itemize}[noitemsep,nolistsep] 
	\item[\good] $[M_{\pi,\mathrm{min}} / M_{\pi,\mathrm{fid}}]^2 \exp\{4-M_{\pi,\mathrm{min}}[L(M_{\pi,\mathrm{min}})]_{\mathrm{max}}\} < 1$,
	\underline{or} at least three volumes
	\item[\soso] $[M_{\pi,\mathrm{min}} / M_{\pi,\mathrm{fid}}]^2 \exp\{3-M_{\pi,\mathrm{min}}[L(M_{\pi,\mathrm{min}})]_{\mathrm{max}}\} < 1$,
	\underline{or} at least two volumes
	\item[\bad]  otherwise 
\end{itemize}
where we have introduced $[L(M_{\pi,\mathrm{min}})]_{\mathrm{max}}$, which is the maximum box size used in 
the simulations performed at the smallest pion mass $M_{\pi,{\rm min}}$, as well as a fiducial pion mass 
$M_{\pi,{\rm fid}}$, which we set to 200
MeV (the cutoff value for a green star in the chiral extrapolation). 
It is assumed here that calculations are in the $p$-regime of chiral perturbation
theory, and that all volumes used exceed 2~fm. 
The rationale for this condition is as follows.
Finite-volume effects contain the universal factor $\exp\{- M_\pi L\}$,
and if this were the only contribution a criterion based on
the values of $M_{\pi,\textrm{min}} L$ would be appropriate. 
However, as pion masses decrease, one must also account for
the weakening of the pion couplings. In particular,
1-loop chiral perturbation theory~\cite{Colangelo:2005gd} 
reveals a behaviour proportional to
$M_\pi^2 \exp\{- M_\pi L\}$. 
Our  condition includes this weakening of the coupling, 
and ensures, for example, that simulations with
$M_{\pi,\mathrm{min}} = 135~{\rm MeV}$ and $M_{\pi,\mathrm{min}} L =
3.2$ are rated equivalently to those with $M_{\pi,\mathrm{min}} = 200~{\rm MeV}$
and $M_{\pi,\mathrm{min}} L = 4$.

\emph{-- For QED (where applicable):}
\begin{itemize}[noitemsep,nolistsep]
	\item[\good]$1/([M_{\pi,\mathrm{min}}L(M_{\pi,\mathrm{min}})]_{\mathrm{max}})^{n_{\mathrm{min}}}<0.02$,
		\underline{or} at least four volumes
	\item[\soso] $1/([M_{\pi,\mathrm{min}}L(M_{\pi,\mathrm{min}})]_{\mathrm{max}})^{n_{\mathrm{min}}}<0.04$,
		\underline{or} at least three volumes
	\item[\bad]  otherwise 
\end{itemize}
Because of the infrared-singular structure of QED, electromagnetic finite-volume effects decay only like a power of the inverse spatial extent. In several cases like mass splittings~\cite{Borsanyi:2014jba,Davoudi:2014qua} or leptonic decays~\cite{Lubicz:2016xro}, the leading corrections are known to be universal, i.e., independent of the structure of the involved hadrons. In such cases, the leading universal effects can be directly subtracted exactly from the lattice data. We denote $n_{\mathrm{min}}$ the smallest power of $\frac{1}{L}$ at which such a subtraction cannot be done. In the widely used finite-volume formulation $\mathrm{QED}_L$, one always has $n_{\mathrm{min}}\leq 3$ due to the nonlocality of the theory~\cite{Davoudi:2018qpl}.
The QED criteria are used here only in Sec.~\ref{sec:qmass}.
Both QCD and QED criteria are unchanged from FLAG 19.
\item Isospin-breaking effects (where applicable):
\begin{itemize}[noitemsep,nolistsep]
	\item[\good] all leading isospin-breaking effects are included in the lattice calculation
	\item[\soso] isospin-breaking effects are included using the electro-quenched approximation
	\item[\bad] otherwise
\end{itemize}
This criterion is used for quantities which are breaking isospin symmetry or which can be determined at the sub-percent accuracy where isospin-breaking effects, if not included, are expected to be the dominant source of uncertainty. In the current edition, this criterion is only used for the up- and down-quark masses, and related quantities ($\epsilon$, $Q^2$ and $R^2$).
The criteria for isospin-breaking effects 
are unchanged from FLAG 19.
\item Renormalization (where applicable):
\begin{itemize}[noitemsep,nolistsep]
	\item[\good]  nonperturbative
	\item[\soso]  1-loop perturbation theory or higher  with a reasonable estimate of truncation errors
	\item[\bad]  otherwise 
\end{itemize}	
In Ref.~\cite{Colangelo:2010et}, we assigned a red square to all
results which were renormalized at 1-loop in perturbation theory. In 
FLAG 13~\cite{Aoki:2013ldr}, we decided that this was too restrictive, since 
the error arising from renormalization constants, calculated in perturbation theory at
1-loop, is often estimated conservatively and reliably. 
These criteria have remained unchanged since then.

\item Renormalization Group (RG) running (where applicable): \\ 
For scale-dependent quantities, such as quark masses or $B_K$, it is
essential that contact with continuum perturbation theory can be established.
Various different methods are used for this purpose
(cf.~Appendix A.3 in FLAG 19 \cite{FlavourLatticeAveragingGroup:2019iem}): Regularization-independent Momentum
Subtraction (RI/MOM), the Schr\"odinger functional, and direct comparison with
(resummed) perturbation theory. Irrespective of the particular method used,
the uncertainty associated with the choice of intermediate
renormalization scales in the construction of physical observables
must be brought under control. This is best achieved by performing
comparisons between nonperturbative and perturbative running over a
reasonably broad range of scales. These comparisons were initially
only made in the Schr\"odinger functional approach, but are now
also being performed in RI/MOM schemes.  We mark the data for which
information about nonperturbative-running checks is available and
give some details, but do not attempt to translate this into a
colour code. 
\end{itemize}

The pion mass plays an important role in the criteria relevant for
chiral extrapolation and finite volume.  For some of the
regularizations used, however, it is not a trivial matter to identify
this mass. 
In the case of twisted-mass fermions, discretization
effects give rise to a mass difference between charged and neutral
pions even when the up- and down-quark masses are equal: the charged pion
is found to be the heavier of the two for twisted-mass Wilson fermions
(cf.~Ref.~\cite{Boucaud:2007uk}).
In early works, typically
referring to $\Nf=2$ simulations (e.g., Refs.~\cite{Boucaud:2007uk}
and~\cite{Baron:2009wt}), chiral extrapolations are based on chiral
perturbation theory formulae which do not take these regularization
effects into account. After the importance of accounting for isospin
breaking when doing chiral fits was shown in Ref.~\cite{Bar:2010jk},
later works, typically referring to $\Nf=2+1+1$ simulations, have taken
these effects into account~\cite{Carrasco:2014cwa}.
We use $M_{\pi^\pm}$ for $M_{\pi,\mathrm{min}}$
in the chiral-extrapolation rating criterion. On the
other hand, 
we identify $M_{\pi,\mathrm{min}}$ with
the root mean square (RMS) of $M_{\pi^+}$,
$M_{\pi^-}$ and $M_{\pi^0}$ in the finite-volume rating criterion.

In the case of staggered fermions,
discretization effects give rise to several light states with the
quantum numbers of the pion.\footnote{
We refer the interested reader to a number of reviews on the
subject~\cite{Durr:2005ax,Sharpe:2006re,Kronfeld:2007ek,Golterman:2008gt,Bazavov:2009bb}.}
The mass splitting among these ``taste'' partners represents a
discretization effect of $\cO(a^2)$, which can be significant at large
lattice spacings but shrinks as the spacing is reduced. In the
discussion of the results obtained with staggered quarks given in the
following sections, we assume that these artifacts are under
control. We conservatively identify $M_{\pi,\mathrm{min}}$ with the root mean
square (RMS) average of the masses of all the taste partners, 
both for chiral-extrapolation and finite-volume criteria.

In some of the simulations, the fermion formulations employed for the
valence quarks are different from those used for the sea quarks. Even
when the fermion formulations are the same, there are cases where the
sea-  and valence-quark masses differ. In such cases, we use the smaller
of the valence-valence and valence-sea $M_{\pi_{\rm min}}$ values in the
finite-volume criteria, since either of these channels may give the leading
contribution depending on the quantity of interest at the 1-loop
level of chiral perturbation theory. For the chiral-extrapolation
criteria, on the other hand, we use the unitary point, where the sea- and
valence-quark masses are the same, to define $M_{\pi_{\rm min}}$.

The strong coupling $\alpha_s$ is computed in lattice QCD with methods
differing substantially
from those used in the calculations of the other quantities 
discussed in this review. Therefore, we have established separate criteria for
$\alpha_s$ results, which will be discussed in Sec.~\ref{s:crit}.

In Sec.~\ref{sec:NME} on nuclear matrix elements, 
an additional criterion is used.
This concerns the level of control over contamination from excited states,
which is a more challenging issue for nucleons than for mesons. 
In response to an improved understanding of the impact of this
contamination, the excited-state contamination criterion has been made more stringent compared
to that in FLAG 19.

\subsubsection{Data-driven criteria}
\label{sec:DataDriven}

For some time, the FLAG working groups have been considering using a `data-driven' criterion in assessing how well the continuum limit is controlled.
The quantity $\delta(a)$ is defined as
\begin{equation}
	\delta(a) \equiv {|Q(a)-Q(0)|\over \sigma_Q}\,,
   \label{eq:delta_a}
\end{equation}
were $Q(a)$ is the quantity under consideration with lattice spacing $a$,
$Q(0)$ is the extrapolated continuum-limit value, and
$\sigma_Q$ is its error in the continuum limit.  If $a_{\rm min}$ is the 
smallest lattice spacing used, there is concern if $\delta(a_{\rm min})$
is very large.  That is, the results at the finest lattice spacing should
not be too many standard deviations from the continuum limit in order for the extrapolation to be considered reliable.

The following is adopted for the current edition of the review: (1)
Each working group attempts to determine $\delta(a_{\rm min})$ for
each calculation that contributes to a FLAG average.  However, it is
not currently used as a criterion for inclusion in the averages.  (2)
The text of the report includes these values for calculations
contributing to FLAG averages. (3) For the current edition of FLAG it
is at the discretion of each working group to decide whether they wish
to inflate the error of contributions to the average for calculations
with large values of $\delta(a_{\rm min})$.  If this is done, the
inflation factor will be
\begin{equation}
  s(\delta) = \max[1, 1 + 2 (\delta - 3) / 3].
  \label{eq:stretching_factor}
\end{equation}
The inflation of the error is not displayed in tables or plots.  
It is only used to evaluate FLAG averages.

\subsubsection{Heavy-quark actions}
\label{sec:HQCriteria}

For the $b$ quark,
the discretization of the
heavy-quark action follows a very different approach from that used for light
flavours. There are several different methods for
treating heavy quarks on the lattice, each with its own issues and
considerations.  Most of these methods use
Effective Field Theory (EFT) at some point in the computation, either
via direct simulation of the EFT, or by using EFT
as a tool to estimate the size of cutoff errors, 
or by using EFT to extrapolate from the simulated
lattice quark masses up to the physical $b$-quark mass. 
Because of the use of an EFT, truncation errors must be
considered together with discretization errors. 

The charm quark lies at an intermediate point between the heavy
and light quarks. In our earlier reviews, the calculations
involving charm quarks often treated it using one of the approaches adopted
for the $b$ quark. Since FLAG 16~\cite{Aoki:2016frl}, however, 
most calculations simulate the charm quark using light-quark actions.
This has become possible thanks to the increasing availability of
dynamical gauge field ensembles with fine lattice spacings.
But clearly, when charm quarks are treated relativistically, discretization
errors are more severe than those of the corresponding light-quark quantities.

In order to address these complications, 
the heavy-quark section adds an additional, bipartite,
treatment category to the rating system. The purpose of this
criterion is to provide a guideline for the level of action and
operator improvement needed in each approach to make reliable
calculations possible, in principle. 

A description of the different approaches to treating heavy quarks on
the lattice 
can be found in Appendix A.1.3 of FLAG 19~\cite{FlavourLatticeAveragingGroup:2019iem}.
For truncation errors we use HQET power counting throughout,
since this review is focused on heavy-quark quantities involving $B$
and $D$ mesons rather than bottomonium or charmonium quantities.  
Here we describe the criteria for how each approach
must be implemented in order to receive an acceptable rating (\okay) 
for both the heavy-quark actions and the weak operators.  Heavy-quark
implementations without the level of improvement described below are
rated not acceptable (\bad). The matching is evaluated together with
renormalization, using the renormalization criteria described in
Sec.~\ref{sec:Criteria}.  We emphasize that the heavy-quark
implementations rated as acceptable and described below have been
validated in a variety of ways, such as via phenomenological agreement
with experimental measurements, consistency between independent
lattice calculations, and numerical studies of truncation errors.
These tests are summarized in Sec.~\ref{sec:BDecays}.  \smallskip
\\ {\it Relativistic heavy-quark actions:} \\
\noindent 
\okay \hspace{0.2cm}   at least tree-level $\cO(a)$-improved action and 
weak operators  \\
This is similar to the requirements for light-quark actions. All
current implementations of relativistic heavy-quark actions satisfy
this criterion. \smallskip \\
{\it NRQCD:} \\
\noindent 
\okay \hspace{0.2cm}   tree-level matched through $\cO(1/m_h)$ 
and improved through $\cO(a^2)$ \\
The current implementations of NRQCD satisfy this criterion, and also
include tree-level corrections of $\cO(1/m_h^2)$ in the action. 
\smallskip \\
{\it HQET: }\\
\noindent 
\okay \hspace{0.2cm}  tree-level  matched through $\cO(1/m_h)$ 
with discretization errors starting at $\cO(a^2)$ \\
The current implementation of HQET by the ALPHA collaboration
satisfies this criterion, since both action and weak operators are
matched nonperturbatively through $\cO(1/m_h)$.  Calculations that
exclusively use a static-limit action do not satisfy this criterion,
since the static-limit action, by definition, does not include $1/m_h$
terms.  We therefore include static computations in our final estimates only if truncation errors (in $1/m_h$)  are discussed and included in the systematic uncertainties.\smallskip \\
{\it Light-quark actions for heavy quarks:}  \\
\noindent 
\okay \hspace{0.2cm}  discretization errors starting at $\cO(a^2)$ or higher \\
This applies to calculations that use the twisted-mass Wilson action, a
nonperturbatively improved Wilson action, domain-wall fermions or the HISQ action for charm-quark 
quantities. It also applies to calculations that use these light-quark actions in the charm region and above together with either the
static limit or with an HQET-inspired extrapolation to obtain results
at the physical $b$-quark mass. 
In these cases, the combined list of lattice spacings used for the data sets 
with $m_h > 0.5 m_{h,{\rm phys}}$ must satisfy the continuum-extrapolation 
criteria.

\subsubsection{Conventions for the figures}
\label{sec:figurecolours}

For a coherent assessment of the present situation, the quality of the
data plays a key role, but the colour coding cannot be carried over to
the figures. On the other hand, simply showing all data on equal
footing might give the misleading impression that the overall
consistency of the information available on the lattice is
questionable. Therefore, in the figures we indicate the quality of the data
in a rudimentary way, using the following symbols:
\begin{itemize}[noitemsep,nolistsep]
	\item[{\color{darkgreen}$\blacksquare$}] corresponds to results included in the average or estimate (i.e., results that contribute to the black square below);
	\item[{\color{lightgreen}$\blacksquare$\hspace{-0.3cm}\color{darkgreen}$\square$}] corresponds to results that are not included in the average but pass all quality criteria;
	\item[{\color{red}\parbox[c]{3.1mm}{$\bigcirc$}}] corresponds to all other results;
	\item[{\color{black}$\blacksquare$}]corresponds to FLAG averages or estimates; they are also highlighted by a gray vertical band.
\end{itemize} 
The reason for not including a given result in
the average is not always the same: the result may fail one of the
quality criteria; the paper may be unpublished; 
it may be superseded by newer results;
or it may not offer a complete error budget. 

Symbols other than the ones above are
used to distinguish results with specific properties and are always
explained in the caption.\footnote{%
For example, for quark-mass results we
distinguish between perturbative and nonperturbative renormalization, 
and for heavy-flavour results we distinguish between
those from leptonic and semileptonic decays.}

Often, nonlattice data are also shown in the figures for comparison. 
For these we use the following symbols:
\begin{itemize}[noitemsep,nolistsep]
	\item[\raisebox{0.15mm}{\hspace{0.65mm}\color{blue}\Large\textbullet}]
	corresponds to nonlattice results;
	\item[\raisebox{0.35mm}{\hspace{0.65mm}{\color{black}$\blacktriangle$}}] corresponds to Particle Data Group (PDG) results.
\end{itemize}
\subsection{Averages and estimates}\label{sec:averages}

FLAG results of a given quantity are denoted either as {\it averages} or as {\it estimates}. Here we clarify this distinction. To start with, both {\it averages} and {\it estimates} are based on results without any red tags in their colour coding. For many observables there are enough independent lattice calculations of good quality, with all sources of error (not merely those related to the colour-coded criteria), as analyzed in the original papers, appearing to be under control. In such cases, it makes sense to average these results and propose such an {\it average} as the best current lattice number. The averaging procedure applied to this data and the way the error is obtained is explained in detail in Sec.~\ref{sec:error_analysis}. In those cases where only a sole result passes our rating criteria (colour coding), we refer to it as our FLAG {\it average}, provided it also displays adequate control of all other sources of systematic uncertainty.

On the other hand, there are some cases in which this procedure leads to a result that, in our opinion, does not cover all uncertainties. Systematic  errors are by their nature often subjective and difficult to estimate, and may thus end up being underestimated in one or more results that receive green symbols for all explicitly tabulated criteria.   
Adopting a conservative policy, in these cases we opt for an {\it estimate} (or a range), which we consider as a fair assessment of the knowledge acquired on the lattice at present. This {\it estimate} is not obtained with a prescribed mathematical procedure, but reflects what we consider the best possible analysis of the available information. The hope is that this will encourage more detailed investigations by the lattice community.

There are two other important criteria that also play a role in this
respect, but that cannot be colour coded, because a systematic
improvement is not possible. These are: {\em i)} the publication
status, and {\em ii)} the number of sea-quark flavours $\Nf$. As far as the
former criterion is concerned, we adopt the following policy: we
average only results that have been published in peer-reviewed
journals, i.e., they have been endorsed by referee(s). The only
exception to this rule consists in straightforward updates of previously
published results, typically presented in conference proceedings. Such
updates, which supersede the corresponding results in the published
papers, are included in the averages. 
Note that updates of earlier results rely, at least partially, on the
same gauge-field-configuration ensembles. For this reason, we do not
average updates with earlier results. 
Nevertheless, all results are
listed in the tables,\footnote{%
Whenever tables and figures turn out to be overcrowded,
older, superseded results are omitted. However, all the most recent results
from each collaboration are displayed.}
and their publication status is identified by the following
symbols:
\begin{itemize}
\item Publication status:\\
\gA  \hspace{0.2cm}published or plain update of published results\\
\oP  \hspace{0.2cm}preprint\\ 
\rC  \hspace{0.2cm}conference contribution
\end{itemize}
In the present edition, the
publication status on the {\bf 30th of April 2024} is relevant.
If the paper appeared in print after that date, this is accounted for in the
bibliography, but does not affect the averages.\footnote{%
As noted above in footnote \ref{footnote:including papers after deadline}, two exceptions to this deadline were made, Refs.~\cite{Boyle:2024gge,Brambilla:2023fsi}.
\label{footnote:deadline exceptions}
}

As noted above,
in this review we present results from simulations with $\Nf=2$,
$\Nf=2+1$ and $\Nf=2+1+1$ (except for $ r_0 \Lambda_\msbar$ where we
also give the $\Nf=0$ result). We are not aware of an {\em a priori} way
to quantitatively estimate the difference between results produced in
simulations with a different number of dynamical quarks. We therefore
average results at fixed $\Nf$ separately; averages of calculations
with different $\Nf$ are not  provided.

To date, no significant differences between results with different
values of $\Nf$ have been observed in the quantities 
listed in Tabs.~\ref{tab:summary1}, \ref{tab:summary2},
\ref{tab:summary4}, and \ref{tab:summary5}.
In particular, differences between results from simulations with $\Nf = 2$ and $\Nf = 2 + 1$ 
would reflect Zweig-rule violations related to strange-quark loops. 
Although not of direct phenomenological relevance,
the size of such violations is an interesting theoretical issue {\em per se}, 
and one that can be quantitatively addressed only with lattice calculations.
It remains to be seen whether the status presented here will change in the future,
since this will require dedicated $\Nf=2$ and $\Nf=2+1$ calculations, which
are not a priority of present lattice work.

The question of differences between results with $\Nf=2+1$ and
$\Nf=2+1+1$ is more subtle.
The dominant effect of including the charm sea quark is to
shift the lattice scale, an effect that is accounted for by
fixing this scale nonperturbatively using physical quantities.
For most of the quantities discussed in this review, it is 
expected that residual effects are small in the continuum limit,
suppressed by $\alpha_s(m_c)$ and powers of $\Lambda^2/m_c^2$.
Here $\Lambda$ is a hadronic scale that can only be
roughly estimated and depends on the process under consideration.
Note that the $\Lambda^2/m_c^2$ effects have been addressed 
in~Refs.~\cite{Bruno:2014ufa,Knechtli:2017xgy,Athenodorou:2018wpk,Cali:2019enm,Cali:2021xwh},
and were found to be small for the quantities considered.
Assuming that such effects are generically small, it might be reasonable to
average the results from $\Nf=2+1$ and $\Nf=2+1+1$ simulations,
although we do not do so here.

\subsection{Averaging procedure and error analysis}
\label{sec:error_analysis}

In the present report, we repeatedly average results
obtained by different collaborations, and estimate the error on the resulting
averages. 
Here we provide details on how averages are obtained.

\subsubsection{Averaging --- generic case}
We continue to follow the procedure of FLAG 13 and FLAG 16~\cite{Aoki:2013ldr,Aoki:2016frl}
which we describe here in full detail.

One of the problems arising when forming averages is that not all
of the data sets are independent.
In particular, the same gauge-field configurations,
produced with a given fermion discretization, are often used by
different research teams with different valence-quark lattice actions,
obtaining results that are not really independent.  
Our averaging procedure takes such correlations into account. 

Consider a given measurable quantity $Q$, measured by $M$ distinct,
not necessarily uncorrelated, numerical experiments (simulations). The result
of each of these measurement is expressed as
\begin{equation}
Q_i \,\, = \,\, x_i \, \pm \, \sigma^{(1)}_i \pm \, \sigma^{(2)}_i \pm \cdots
\pm \, \sigma^{(E)}_i  \,\,\, ,
\label{eq:resultQi}
\end{equation}
where $x_i$ is the value obtained by the $i^{\rm th}$ experiment
($i = 1, \cdots , M$) and $\sigma^{(\alpha)}_i$ (for $\alpha = 1, \cdots , E$) 
are the various errors.
Typically $\sigma^{(1)}_i$ stands for the statistical error 
and $\sigma^{(\alpha)}_i$ ($\alpha \ge 2$) are the different
systematic errors from various sources. 
For each individual result, we estimate the total
error $\sigma_i $ by adding statistical and systematic errors in quadrature:
\begin{eqnarray}
Q_i \,\, &=& \,\, x_i \, \pm \, \sigma_i \,\,\, ,
\nonumber \\
\sigma_i \,\, &\equiv& \,\, \sqrt{\sum_{\alpha=1}^E \Big [\sigma^{(\alpha)}_i \Big ]^2} \,\,\, .
\label{eq:av-err-Qi}
\end{eqnarray}
With the weight factor of each total error estimated in standard fashion,
\begin{equation}
\omega_i \,\, = \,\, \dfrac{\sigma_i^{-2}}{\sum_{i=1}^M \sigma_i^{-2}} \,\,\, ,
\label{eq:weighti}
\end{equation}
the central value of the average over all simulations is given by
\begin{eqnarray}
x_{\rm av} \,\, &=& \,\, \sum_{i=1}^M x_i\, \omega_i \,\, . 
\end{eqnarray}
The above central value corresponds to a $\chi_{\rm min}^2$-weighted
average, evaluated by adding statistical and systematic errors in quadrature.
If the fit is not of good quality ($\chi_{\rm min}^2/{\rm dof} > 1$),
the statistical and systematic error bars are stretched by a factor
$S = \sqrt{\chi^2/{\rm dof}}$.

Next, we examine error budgets for
individual calculations and look for potentially correlated
uncertainties. Specific problems encountered in connection with
correlations between different data sets are described in the text
that accompanies the averaging.
If there is reason to believe that a source of error is correlated
between two calculations, a $100\%$ correlation is assumed.
The covariance matrix $C_{ij}$ for the set of correlated lattice results is
estimated by a prescription due to Schmelling~\cite{Schmelling:1994pz}.
This consists in defining
\begin{equation}
\sigma_{i;j} \,\, = \,\, \sqrt{{\sum_{\alpha}}^\prime \Big[ \sigma_i^{(\alpha)} \Big]^2 } \,\,\, ,
\label{eq:sigmaij}
\end{equation}
with $\sum_{\alpha}^\prime$ running only over those errors of $x_i$ that
are correlated with the corresponding errors of the measurement $x_j$. 
This expresses the part of the uncertainty in $x_i$
that is correlated with the uncertainty in $x_j$. 
If no such correlations are known to exist, then
we take $\sigma_{i;j} =0$. 
The diagonal and off-diagonal elements of the covariance
matrix are then taken to be
\begin{eqnarray}
C_{ii} \,\,&=& \,\, \sigma_i^2 \qquad \qquad (i = 1, \cdots , M) \,\,\, ,
\nonumber \\
C_{ij} \,\,&=& \,\, \sigma_{i;j} \, \sigma_{j;i} \qquad \qquad (i \neq j) \,\,\, .
\label{eq:Ciiij}
\end{eqnarray}
Finally, the error of the average is estimated by
\begin{equation}
\sigma^2_{\rm av} \,\, = \,\, \sum_{i=1}^M \sum_{j=1}^M \omega_i \,\omega_j \,C_{ij}\,\,,
\label{eq:sigma2av}
\end{equation}
and the FLAG average is
\begin{equation}
Q_{\rm av} \,\, = \,\, x_{\rm av} \, \pm \, \sigma_{\rm av} \,\,\, .
\end{equation}
\input{HQ/HQSubsections/nested_average.tex}

%% file: HQ/HQSubsections/nested_average.tex
 \subsubsection{Nested averaging}
\label{sec:nested_average}

We have encountered one case
where the correlations between results are more involved,
and a nested averaging scheme is required.
This concerns the $B$-meson bag parameters discussed in Sec.~\ref{sec:BMix}.
In the following, we describe the details of the nested averaging scheme.
This is an updated version of the section added in the web update of the FLAG 16 report.

The issue arises for a quantity $Q$ that is given by a ratio, $Q=Y/Z$.
In most simulations, both $Y$ and $Z$ are calculated, and the error in $Q$ can be
obtained in each simulation in the standard way.
However, in other simulations only $Y$ is calculated,
with $Z$ taken from a global average of some type.
The issue to be addressed is that this average value $\overline{Z}$ has errors
that are correlated with those in $Q$.

In the example that arises in Sec.~\ref{sec:BMix},
$Q=B_B$,  $Y=B_B f_B^2$ and $Z=f_B^2$.
In one of the simulations that contribute to the average, 
$Z$ is replaced by $\overline{Z}$, 
the PDG average for $f_B^2$~\cite{Rosner:2015wva}
(obtained with an averaging procedure similar to that used by FLAG).
This simulation is labeled with $i=1$, so that
\begin{equation}
 Q_1 = \frac{Y_1}{\overline{Z}}.
  \label{eq:FNAL_B_PDG}
\end{equation}
The other simulations have results labeled $Q_j$, with $j\ge 2$.
In this set up, the issue is that $\overline{Z}$ is correlated with the $Q_j$, $j\ge 2$.\footnote{%
There is also a small correlation between $Y_1$ and $\overline{Z}$, but we follow the
original Ref.~\cite{Bazavov:2016nty}
 and do not take this into account. Thus, the error in $Q_1$
is obtained by simple error propagation from those in $Y_1$ and $\overline{Z}$.
Ignoring this correlation is conservative, because, as in the
calculation of $B_K$, the correlations between $B_B f_B^2$ and $f_B^2$ tend to
lead to a cancellation of errors. By ignoring this effect we are making a small overestimate
of the error in $Q_1$.}

We begin by decomposing the error in $Q_1$ in the same
schematic form as above,
\begin{equation}
 Q_1 
  = x_1 
  \pm \frac{\sigma_{Y_1}^{(1)}}{\overline{Z}}
  \pm \frac{\sigma_{Y_1}^{(2)}}{\overline{Z}} \pm\cdots
  \pm \frac{\sigma_{Y_1}^{(E)}}{\overline{Z}}
  \pm \frac{Y_1 \sigma_{\overline{Z}}}{\overline{Z}^2}.
  \label{eq:Q1nested}
\end{equation}
Here the last term represents the error propagating from that in $\overline{Z}$,
while the others arise from errors in $Y_1$.
For the remaining $Q_j$ ($j\ge 2$) the decomposition is as in Eq.~(\ref{eq:resultQi}).
The total error of $Q_1$ then reads 
\begin{equation}
 \sigma_1^2 = 
  \left(\frac{\sigma_{Y_1}^{(1)}}{\overline{Z}}\right)^2
  + \left(\frac{\sigma_{Y_1}^{(2)}}{\overline{Z}}\right)^2 +\cdots
  + \left(\frac{\sigma_{Y_1}^{(E)}}{\overline{Z}}\right)^2
  + \left(\frac{Y_1}{\overline{Z}^2}\right)^2 \sigma_{\overline{Z}}^2,
  \label{eq:sigma1}
\end{equation}
while that for the $Q_j$ ($j\ge 2$) is
\begin{equation}
 \sigma_j^2 = 
  \left(\sigma_j^{(1)}\right)^2
  + \left(\sigma_j^{(2)}\right)^2 +\cdots
  + \left(\sigma_j^{(E)}\right)^2.
  \label{eq:sigmaj}
\end{equation}
Correlations between $Q_j$ and $Q_k$ ($j,k\ge 2$) are taken care of by
Schmelling's prescription, as explained above.
What is new here is how the correlations 
between $Q_1$ and $Q_j$ ($j\ge 2$) are taken into account.

To proceed, we recall from Eq.~(\ref{eq:sigma2av}) that
$\sigma_{\overline{Z}}$ is given by
\begin{equation}
 \sigma_{\overline{Z}}^2 = \sum_{{i'},{j'}=1}^{M'} \omega[Z]_{i'}
  \omega[Z]_{j'} C[Z]_{i'j'}.
\end{equation}
Here the indices
$i'$ and $j'$ run over the $M'$ simulations that contribute to $\overline{Z}$,
which, in general, are different from those contributing to the results for $Q$.
The weights $\omega[Z]$ and covariance matrix $C[Z]$ are given an explicit
argument $Z$ to emphasize that they refer to the calculation of this quantity
and not to that of $Q$.
$C[Z]$ is calculated using the Schmelling prescription
[Eqs.~(\ref{eq:sigmaij})--(\ref{eq:sigma2av})] in terms of the errors, $\sigma[Z]_{i'}^{(\alpha)}$,
taking into account the correlations between the different calculations of $Z$.

We now generalize Schmelling's prescription for $\sigma_{i;j}$, Eq.~(\ref{eq:sigmaij}),
to that for $\sigma_{1;k}$ ($k\ge 2$), i.e., the part of the error in $Q_1$ that
is correlated with $Q_k$. We take
\begin{equation}
 \sigma_{1;k} \,\, = \,\, 
  \sqrt{
  \frac{1}{\overline{Z}^2} \sum^\prime_{(\alpha)\leftrightarrow k}
  \Big[\sigma_{Y_1}^{(\alpha)} \Big]^2 
  + \frac{Y_1^2}{\overline{Z}^4} 
  \sum_{i',j'}^{M'} \omega[Z]_{i'} \omega[Z]_{j'} C[Z]_{i'j'\leftrightarrow k}
  }
  \,\,\, .
\label{eq:sigma1k}
\end{equation}
The first term under the square root sums those sources of error in $Y_1$ that
are correlated with $Q_k$. Here we are using a more explicit notation from that
in Eq.~(\ref{eq:sigmaij}), with $(\alpha) \leftrightarrow k$ indicating that the sum
is restricted to the values of $\alpha$ for which the error $\sigma_{Y_1}^{(\alpha)}$
is correlated with $Q_k$.
The second term accounts for the correlations within $\overline{Z}$ with $Q_k$,
and is the nested part of the present scheme.
The new matrix $C[Z]_{i'j'\leftrightarrow k}$ is a restriction
of the full covariance matrix $C[Z]$, and is defined as follows.
Its diagonal elements are given by
\begin{eqnarray}
C[Z]_{i'i'\leftrightarrow k} \,\,&=& \,\, (\sigma[Z]_{i'\leftrightarrow k})^2 \qquad \qquad (i' = 1, \cdots , M') \,\,\, ,
 \\
 (\sigma[Z]_{i'\leftrightarrow k})^2 & = &
  \sum^\prime_{(\alpha)\leftrightarrow k} (\sigma[Z]_{i'}^{(\alpha)})^2,
\label{eq:sigmaZipk}
\end{eqnarray}
where the summation 
$\sum^\prime_{(\alpha)\leftrightarrow k}$ 
over $(\alpha)$ is restricted to those $\sigma[Z]_{i'}^{(\alpha)}$ that are
correlated with $Q_k$.
The off-diagonal elements are
\begin{eqnarray}
C[Z]_{i'j'\leftrightarrow k} \,\,&=& \,\, \sigma[Z]_{i';j'\leftrightarrow k} \, \sigma[Z]_{j';i'\leftrightarrow k} \qquad \qquad (i' \neq j') \,\,\, ,\\
 \sigma[Z]_{i';j'\leftrightarrow k} & = &
  \sqrt{
  \sum^\prime_{(\alpha)\leftrightarrow j'k} 
  (\sigma[Z]_{i'}^{(\alpha)})^2},
\label{eq:sigmaZipjpk}
\end{eqnarray}
where the summation 
$\sum^\prime_{(\alpha)\leftrightarrow j'k}$ 
over $(\alpha)$ is restricted to $\sigma[Z]_{i'}^{(\alpha)}$ that are
correlated with {\it both} $Z_{j'}$ and $Q_k$.

The last quantity that we need to define is $\sigma_{k;1}$.
\begin{equation}
\sigma_{k;1} \,\, = \,\, \sqrt{\sum^\prime_{(\alpha)\leftrightarrow 1} \Big[ \sigma_k^{(\alpha)} \Big]^2 } \,\,\, ,
\label{eq:sigmak1}
\end{equation}
where the summation $\sum^\prime_{(\alpha)\leftrightarrow 1}$ is
restricted to those $\sigma_k^{(\alpha)}$ that are correlated with one of
the terms in Eq.~(\ref{eq:sigma1}).

In summary, we construct the covariance matrix $C_{ij}$ using
Eq.~(\ref{eq:Ciiij}), as in the generic case, except the expressions
for $\sigma_{1;k}$ and $\sigma_{k;1}$ are now given by
Eqs.~(\ref{eq:sigma1k}) and (\ref{eq:sigmak1}), respectively. All other $\sigma_{i;j}$ are given by
the original Schmelling prescription, Eq.~(\ref{eq:sigmaij}).
In this way, we extend the philosophy of Schmelling's approach while accounting
for the more involved correlations.

%% file: ibscheme/ediconsensus_main.tex
\section{General definition of the low-energy limit of the Standard
Model}\label{sec:ibscheme}
Authors: A.~Portelli, A.~Ramos, N.~Tantalo\\

\noindent This section discusses the matching of quantum chromodynamics (QCD) and quantum
electrodynamics (QED) to nature in order to obtain predictions for low-energy Standard
Model observables. In particular, we discuss the prescription dependence, i.e., the
dependence on which observables are matched, arising when one neglects electromagnetic
interactions, an approximation made in numerous lattice and phenomenological calculations.
These ambiguities need to be controlled when combining high-precision observables---typically
with less than $1\%$ of relative uncertainty---in that approximation. In order to
facilitate that, we propose here a fixed prescription for the separation of QCD and QED
contributions to any given hadronic observable. While this prescription is, in principle,
arbitrary, one has to take care not to introduce artificially large QED contributions and
to stay close to prescriptions used commonly in phenomenology.
This prescription was
discussed and agreed upon during an open workshop that took place at the Higgs Centre for
Theoretical Physics, Edinburgh, in May
2023, and therefore is referred to as the ``Edinburgh Consensus.''\footnote{\url{https://indico.ph.ed.ac.uk/event/257/}}

We note that since this consensus emerged only recently, the majority of
results in this review are averaged neglecting potential discrepancies arising from the
ambiguities. This is, on the one hand, an adequate procedure in the case of quantities
with uncertainties larger than the size of expected QED corrections. On the other hand, it
can be difficult to correct these ambiguities to a common prescription since it requires the
knowledge of derivatives of observables in quark masses and couplings, rarely communicated
in papers. We emphasize the present consensus in the hope that it will be widely adopted
in upcoming high-precision Standard Model predictions, allowing future editions of this
review to avoid uncertainties resulting from these ambiguities.
\subsection{First-order isospin-breaking expansion}
According to our present knowledge, hadronic physics is well described by the low-energy limit of the Standard Model, which is
understood as energies well below the electroweak symmetry-breaking scale
$\mathcal{S}_{\mathrm{ESB}}\approx 100~\gev$. In that limit, the Standard Model is an
$\mathrm{SU}(3)\times\mathrm{U}(1)$ gauge theory defined by the QCD+QED Lagrangian,
whose free parameters are the $e$-, $\mu$-, and $\tau$-lepton masses, the $u$-, $d$-, $s$-,
$c$-, and $b$-quark masses, and the strong and electromagnetic coupling constants,
respectively, $g_s$ and $e$. In that context, isospin symmetry is defined by assuming that
the up and down quarks are identical particles apart from their flavour. This symmetry is 
only approximate and it is broken by two effects: the small but
different masses of the two quarks, and their different electric charges. The total effect is expected to be small,
typically a $\mathcal{O}(1\%)$ perturbation of a hadronic energy or amplitude. Therefore, we 
consider only first-order perturbations in isospin-breaking effects, and we expect this
approximation to be accurate at the level of $\mathcal{O}(10^{-4})$ relative precision.

The asymptotic states of QCD are hadrons not quarks, and hadron
properties are the only unambiguous observables experimentally available. Similarly, the
strong coupling constant is not directly accessible and can be substituted through
dimensional transmutation by a dimensionful hadronic energy scale. Moreover, the running
of the electromagnetic coupling constant is a higher-order correction beyond the order
considered here. It can be fixed to its
Thomson-limit value. Finally, nature can be reproduced (up to weak and gravitational effects) by
fixing the bare parameters of the QCD+QED Lagrangian to reproduce the following inputs:
\begin{enumerate}
  \item the Thomson-limit constant $\alpha^{\phi}=\frac{e^2}{4\pi}=7.2973525693(11)\times
  10^{-3}$~\citep{ParticleDataGroup:2022pth},
  \item the experimentally observed lepton masses $m_\ell^{\phi}$,
  \item a choice of $\Nf$ known independent hadronic quantities $M^{\phi}$, setting the
  quark masses,
  \item a single known dimensionful hadronic quantity $\mathcal{S}^{\phi}$, setting the QCD  scale.
\end{enumerate}
The vectors $m_\ell$ and $M$ have three and $\Nf$ components, respectively, where $\Nf$ is the number
of  quark flavours in the calculation. In the present context, ``known'' is understood as experimentally known for measurable
quantities, or theoretically predicted for more abstract quantitities, which are not accessible experimentally, but are renormalized and gauge invariant and can be predicted by lattice gauge theory. If the dependency of a given observable
$X(\alpha, m_\ell, M, \mathcal{S})$ on the above variables is known, then its physical value
is predicted by
\begin{equation}
  X^{\phi}=(\mathcal{S}^{\phi})^{[X]}\tilde{X}(\alpha^{\phi}, m_\ell^{\phi}/\mathcal{S}^{\phi}, M^{\phi}/\mathcal{S}^{\phi})
  \equiv X(\alpha^{\phi}, m_\ell^{\phi}, M^{\phi}, \mathcal{S}^{\phi})\,,
\end{equation}
where $\tilde{X}$ is the dimensionless function describing $X$ in units of the scale $\mathcal{S}$, and $[X]$ is the energy dimension of $X$. Here $M$ and
$\mathcal{S}$ are assumed, without loss of generality, to have an energy dimension of~$1$.
Due to the renormalizability of QCD+QED, this
prediction is unambiguous, \ie, changing the variables $M^{\phi}$ and $\mathcal{S}^{\phi}$ to
other inputs with known physical values will lead to the same prediction for renormalized
observables.\footnote{Here ``renormalizability'' for QED is understood as
perturbative renormalizability, which is sufficient in this context.}

In many instances, the precision required on hadronic observables is not as small as one
percent, and isospin-breaking effects are potentially negligible. In those cases, it is
generally considerably simpler to neglect the QED contributions, both for lattice and
phenomenological calculations. Moreover, even for observables requiring isospin-breaking
corrections to be computed, it can be phenomenologically relevant to separate an
isospin-symmetric value and isospin-breaking corrections (\eg, specific parts of the HVP
contribution to the muon $g-2$, decay constants in weak decays). However, since
experimental measurements always contain isospin-breaking corrections, there are no experimental
result available to define the list of inputs above for $\alpha=0$, or in the isospin-symmetric limit. Still, one would like
to define an expansion of the form
\begin{equation}
  X^{\phi}=\bar{X}+X_{\gamma}+X_{\mathrm{SU}(2)}\,,\label{eq:ibdec}
\end{equation}
where $\bar{X}$ is the isospin-symmetric value of $X$, and $X_{\gamma}$ and
$X_{\mathrm{SU}(2)}$ are the first-order electromagnetic and strong isospin-breaking
corrections, respectively. Only the sum of these three terms is unambiguous.\footnote{Here ``unambiguous'' is used in a loose sense. Ambiguities of the order $\mathcal{O}(1/m_Z)$ and $\mathcal{O}(1/m_{\Nf+1})$, as well as higher-order isospin-breaking corrections, remain and are considered to be irrelevant.}
Defining a value for individual terms is prescription-dependent, and requires additional, in principle arbitrary,
inputs. This issue has been discussed in reviews~\citep{Aoyama:2020ynm,Tantalo:2023onv},
and both the phenomenology~\citep{Gasser:1982ap,Gasser:2003hk,Gasser:2010wz} and
lattice~\citep{Budapest-Marseille-Wuppertal:2013rtp,deDivitiis:2013xla,Tantalo:2013maa,Portelli:2015wna,Horsley:2015vla, Fodor:2016bgu,Bussone:2018ybs,Basak:2018yzz,DiCarlo:2019thl,Borsanyi:2020mff,Sachrajda:2021enz,Ce:2022kxy,Boyle:2022lsi,Boccaletti:2024guq} 
literature. If quantities defined at $\alpha=0$ are
involved in the investigation of anomalies related to new physics searches, \emph{the
associated prescriptions must be matched across predictions}. In the next section, we
propose a prescription agreed upon at the dedicated May 2023 workshop in Edinburgh. 

\subsection{Edinburgh Consensus}
\begin{table}[t]
  \centering
  \begin{tabular}{l|c|c|llcc}
    \cline{2-3} \cline{6-7}                       & QCD                                        & isoQCD               &
                                                  & \multicolumn{1}{l|}{}                      &
    \multicolumn{1}{c|}{QCD}                      &
    \multicolumn{1}{c|}{isoQCD}                                                                                         \\ \cline{1-3}
    \cline{5-7} \multicolumn{1}{|l|}{$M_{\pi^+}$} &
    $135.0~\mev$                              & $135.0~\mev$                           &
    \multicolumn{1}{l|}{~~~~~~~~~~~}              &
    \multicolumn{1}{l|}{$M_{\pi^+}/f_{\pi^+}$}    &
    \multicolumn{1}{c|}{$1.034$}                  &
    \multicolumn{1}{c|}{$1.034$}                                                                                        \\
    \multicolumn{1}{|l|}{$M_{K^+}$}               & $491.6~\mev$                           & $494.6~\mev$     &
    \multicolumn{1}{l|}{}                         & \multicolumn{1}{l|}{$M_{K^+}/f_{\pi^+}$}   &
    \multicolumn{1}{c|}{$3.767$}                  & \multicolumn{1}{c|}{$3.790$}                                        \\
    \multicolumn{1}{|l|}{$M_{K^0}$}               & $497.6~\mev$                           & $494.6~\mev$     &
    \multicolumn{1}{l|}{}                         & \multicolumn{1}{l|}{$M_{K^0}/f_{\pi^+}$}   &
    \multicolumn{1}{c|}{$3.813$}                  & \multicolumn{1}{c|}{$3.790$}                                        \\
    \multicolumn{1}{|l|}{$M_{D_s^+}$}             & $1967~\mev$                            & $1967~\mev$      &
    \multicolumn{1}{l|}{}                         & \multicolumn{1}{l|}{$M_{D_s^+}/f_{\pi^+}$} &
    \multicolumn{1}{c|}{$15.07$}                  & \multicolumn{1}{c|}{$15.07$}                                        \\
    \multicolumn{1}{|l|}{$M_{B_s^0}$}             & $5367~\mev$                            & $5367~\mev$      &
    \multicolumn{1}{l|}{}                         & \multicolumn{1}{l|}{$M_{B_s^0}/f_{\pi^+}$} &
    \multicolumn{1}{c|}{$41.13$}                  & \multicolumn{1}{c|}{$41.13$}                                        \\ \cline{5-7}
    \multicolumn{1}{|l|}{$f_{\pi^+}$}             & $130.5~\mev$                           & $130.5~\mev$     &
                                                  &                                            & \multicolumn{1}{l}{} &
    \multicolumn{1}{l}{}                                                                                                \\ \cline{1-3}
  \end{tabular}
  \caption{Edinburgh Consensus for the definition of pure QCD and isospin-symmetric QCD. The rightmost table is redundant and provided for convenience.}
  \label{tbl:edinburghc}
\end{table}
The decomposition~\eq{eq:ibdec} can be unambiguously defined given two extra sets of inputs
$(\hat{m}_\ell, \hat{M}, \hat{\mathcal{S}})$ and $(\bar{m}_\ell, \bar{M}, \bar{\mathcal{S}})$
specifying pure QCD and isospin-symmetric QCD, respectively (denoted QCD
and isoQCD). It is understood that in QCD isospin symmetry can still be broken by the
up-down quark-mass difference. The QCD and isoQCD values of an observable $X$ can then be
 defined by
\begin{equation}
  \hat{X}=X(0, \hat{m}_\ell, \hat{M}, \hat{\mathcal{S}})\qquad\text{and}\qquad
  \bar{X}=X(0, \bar{m}_\ell, \bar{M}, \bar{\mathcal{S}})\,,
\end{equation}
respectively.
The variables $\bar{M}$, $\bar{\mathcal{S}}$ must have one dimension of linear dependency to reflect the exact
isospin symmetry of this theory. This means that there are only $\Nf$ independent numbers. Finally, the corrections in~\eq{eq:ibdec} are then
defined by
\begin{equation}
  X_{\gamma}=X^{\phi}-\hat{X}\qquad\text{and}\qquad X_{\mathrm{SU}(2)}=\hat{X}-\bar{X}\,.
\end{equation}
One should notice that these definitions already constitute in themselves a prescription,
as QED has an isospin-symmetric component which is here assumed to be excluded from the
component $\bar{X}$.

The proposed prescription defines lepton masses to always be equal to their experimental
values (for which negligible experimental uncertainties are discarded), \ie,
$\hat{m}_\ell=\bar{m}_\ell=m_\ell^{\phi}$, and is based on the mass variables
$M=(M_{\pi^+},M_{K^+},M_{K^0},M_{D_s^+},M_{B_s^0})$ and the scale-setting quantity
$f_{\pi^+}$, with the values given in Tab.~\ref{tbl:edinburghc}.\footnote{For
calculations with no active $c$ and/or $b$ quarks, the $M_{D_s^+}$ and/or $M_{B_s^0}$
components should be ignored.} We will now comment on
the definition and applications of that prescription.
\subsection{Comparison to other schemes}
The hadronic quantities that define the proposed prescription, as well as their input
values, have been chosen to balance between two main constraints, on the one hand numerical
and on the other hand theoretical. Since any uncertainties on the theoretical inputs have
to be propagated to the predictions, the numerical constraint requires choosing the
matching observables among those that can be computed on the lattice with the highest
accuracy. The theoretical constraint requires choosing a definition of QCD that leads to isospin-breaking corrections which are as close as
possible to what has commonly been done in the past, in particular, in phenomenological calculations.

On the numerical side, all the chosen hadronic inputs can be extracted from the leading
exponential behaviour at large Euclidean times of two-point mesonic lattice correlators
with high numerical precision. This constraint is the main reason behind the choice of
$f_{\pi^+}$ as the scale-setting observable. From the theoretical and phenomenological
perspectives, this can be seen as an uncomfortable choice. Indeed, the physical quantity
that is measured in experiments is the leptonic decay rate of the charged pion. In the
full theory (QCD+QED) soft photons as well as nonfactorisable virtual QED corrections
have to be taken into account in the theoretical calculation in order to use the
experimental values as an input, and previous knowledge of the CKM matrix element $V_{ud}$
is required. From this perspective, for example, the choice of the  $\Omega^-$-baryon mass
used by several lattice collaborations might be more natural. However, the majority of
lattice calculations are still performed in the $\alpha=0$ limit, which makes $f_{\pi^+}$
a more accessible choice than a baryonic quantity in most cases. It is crucial to note
that our prescription defines QCD and isoQCD in the space of possible $\alpha=0$ theories,
but the choice of coordinates to define these points is arbitrary and can be changed using
standard change-of-variable algebra, while keeping the prescription fixed. In particular,
the scale setting variable can be changed, as we discuss now.

The prescription above can be implemented by using other inputs. This is possible because
QCD is renormalizable. Indeed, one can start by defining QCD using our prescription to
compute $\hat X$ and $\hat M_\Omega$, following the notation of the previous section,
namely
\begin{equation}
  \hat X=X(0,\hat{m}_\ell,\hat{M},\hat{f}_{\pi^+})\qquad\text{and}\qquad
  \hat M_{\Omega}=M_{\Omega}(0,\hat{m}_\ell,\hat{M},\hat{f}_{\pi^+})\,,
\end{equation}
where $\hat{M}$ and $\hat{f}_{\pi^+}$ are given by the ``QCD'' column in
Tab.~\ref{tbl:edinburghc}. Once this calculation has been done, the value of $\hat
M_\Omega$ that has been obtained (assuming for the moment that the errors are negligible)
can be substituted to $\hat f_{\pi^+}$ to redefine our prescription independently from the
pion decay constant. In practice, though, it will not be possible to neglect the errors on
$\hat M_\Omega$. This means that the equivalence between the two sets of coordinates,
explicitly
\begin{equation}
  \hat X=
  X(0,\hat{m}_\ell,\hat{M},\hat{f}_{\pi^+})
  =
 X(0,\hat{m}_\ell,\hat{M},\hat{M}_\Omega)\,,
\end{equation}
can be established within the errors on $\hat M_\Omega$ that will have to be propagated
on any prediction. In this respect, the choice of defining QCD by prescribing with no
errors the values appearing in Tab.~\ref{tbl:edinburghc} puts the choice of $\hat
f_{\pi^+}$ on a slightly different footing than $\hat M_\Omega$. The accuracy of this
matching will directly depend on the accuracy of the dimensionless ratio $\hat
M_{\Omega}/\hat f_{\pi^+}$. The whole discussion above can be reiterated identically for
isoQCD, replacing hatted quantities ($\hat{X}$, \dots) with barred ones ($\bar{X}$,
\dots). It is important to note that $f_{\pi^+}$ is used only to define QCD and plays no
role in defining the full QCD+QED theory. In particular, through a change of scale
variable, like that discussed above, one does not need to know the QED correction to the
$\pi^+$ leptonic decay rate to use our prescription, and one does not lose the ability to
predict this rate for high-precision determinations of the $|V_{ud}|$ CKM matrix element.

Theoretical constraints are the main reason behind the particular choice of values
prescribed in Tab.~\ref{tbl:edinburghc}. Most isospin-breaking separation schemes used in
the literature aim at keeping constant the value of a definition of the renormalized quark
masses when sending $\alpha$ to zero between the physical QCD+QED theory and QCD. Such
a class of constraints was implemented in various ways, for example by the RM123/RM123S
collaboration by computing directly quark masses in the $\overline{\mathrm{MS}}$ scheme at
$2~\mathrm{GeV}$~\citep{deDivitiis:2011eh,deDivitiis:2013xla,Giusti:2017dwk,DiCarlo:2019thl}.
Another example comes from the BMW collaboration, which used in several
calculations~\citep{Budapest-Marseille-Wuppertal:2013rtp,Fodor:2016bgu,Borsanyi:2020mff,Boccaletti:2024guq}
a scheme defined by keeping fixed the squared masses of $\bar{q}q$-connected mesons when
changing $\alpha$. Although these schemes share similar aims, they are not equivalent and
differ by the choice of renormalization scale and scheme, as well as the
contribution from higher-order chiral corrections when using squared meson masses.
However, at the level of precision of current lattice calculations, no significant
discrepancies were observed between both
approaches~\citep{DiCarlo:2019thl,Portelli:Lattice2021,Boyle:2022lsi,Boccaletti:2024guq},
and the numerical values of the pion and kaon masses in Tab.~\ref{tbl:edinburghc} are
compatible with these determinations within the current level of precision. We also note
that the mass values prescribed here are compatible with those produced from
phenomenological inputs in the first edition of FLAG~\citep{Colangelo:2010et}, which
predates the lattice references quoted above.

We end this chapter with a comment on Gasser-Rusetsky-Scimemi (GRS) type schemes
\citep{Gasser:2003hk}. These authors emphasized the importance of keeping track of the
scheme dependence of the splitting in \eq{eq:ibdec}. They furthermore proposed to keep
renormalized quark masses and the strong coupling at a particular matching scale $\mu_1$
(and a chosen renormalization scheme) fixed as one turns off the electromagnetic coupling.
In contrast to the perturbative models studied by GRS, such a scheme is hard to implement
in QCD. Even on the lattice, uncertainties are introduced which are larger than the
isospin-breaking corrections  (see the sections on quark masses and $\alpha_s$).   
The RM123S scheme~\citep{deDivitiis:2011eh} mentioned above is an
electro-quenched GRS type scheme.\footnote{The electro-quenched approximation is defined by setting the
electric charges of the sea quarks to zero.} Since there are no
electromagnetic contributions to $\alpha_s$ in the electro-quenched approximation, the
generic difficulties of a GRS type scheme are circumvented.

%% file: qmass/qmass.tex
\section{Quark masses}
\label{sec:qmass}
Authors: T.~Blum, A.~Portelli, A.~Ramos\\

Quark masses are fundamental parameters of the Standard Model. An
accurate determination of these parameters is important for both
phenomenological and theoretical applications. The bottom- and charm-quark 
masses, for instance, are important sources of parametric
uncertainties in several Higgs decay modes. The up-,
down- and strange-quark masses govern the amount of explicit chiral
symmetry breaking in QCD. From a theoretical point of view, the values
of quark masses provide information about the flavour structure of
physics beyond the Standard Model. The Review of Particle Physics of
the Particle Data Group contains a review of quark masses
\cite{Zyla:2020zbs}, which covers light as well as heavy
flavours. Here, we also consider light- and heavy-quark masses, but
focus on lattice results and discuss them in more detail. We do not
discuss the top quark, however, because it decays weakly before it can
hadronize, and the nonperturbative QCD dynamics described by present
day lattice calculations is not relevant. The lattice determination of
light- (up, down, strange), charm- and bottom-quark masses is
considered below in Secs.~\ref{sec:lqm}, \ref{s:cmass},
and \ref{s:bmass}, respectively.

Quark masses cannot be measured directly in experiment because
quarks cannot be isolated, as they are confined inside hadrons. From a
theoretical point of view, in QCD with $\Nf$ flavours, a precise
definition of quark 
masses requires one to choose a particular renormalization
scheme. This renormalization procedure introduces a
renormalization scale $\mu$, and quark masses depend on this
renormalization scale according to the Renormalization Group (RG)
equations. In mass-independent renormalization schemes the RG equations
read
\begin{equation}
  \label{eq:qmass_tau}
  \mu \frac{{\rm d} \bar m_i(\mu)}{{\rm d}{\mu}} = \bar m_i(\mu) \tau(\bar g)\,,
\end{equation}
where the function $\tau(\bar g)$ is the anomalous
dimension, which
depends only on the value of the strong coupling $\alpha_s=\bar
g^2/(4\pi)$. Note that in QCD $\tau(\bar g)$ is the same for all quark
flavours.  The anomalous 
dimension is scheme dependent, but its 
perturbative expansion  
\begin{equation}
  \label{eq:tau_asymp}
  \tau(\bar g) \raisebox{-.1ex}{
            $\stackrel{\small{\bar g \to 0}}{\sim}$} -\bar g^2\left(
    d_0 + d_1\bar g^2 + \dots
  \right) 
\end{equation}
has a leading coefficient $d_0 = 8/(4\pi)^2$,  
which is scheme independent.\footnote{We follow the conventions of
  Gasser and Leutwyler~\cite{Gasser:1982ap}.}
   Equation~(\ref{eq:qmass_tau}), being a
first order differential equation, can be solved exactly by using
Eq.~(\ref{eq:tau_asymp}) as the boundary condition. The formal
solution of the RG equation reads
\begin{equation}
  \label{eq:qmass_rgi}
  M_i = \bar m_i(\mu)[2b_0\bar g^2(\mu)]^{-d_0/(2b_0)}
  \exp\left\{
    - \int_0^{\bar g(\mu)}{\rm d} x\, \left[
      \frac{\tau(x)}{\beta(x)} - \frac{d_0}{b_0x}
    \right]
  \right\}\,,
\end{equation}
where $b_0 = (11-2\Nf/3) / (4\pi)^2$ is the universal
leading perturbative coefficient in the expansion of the
$\beta$-function
\begin{equation}
  \beta(\bar g)\equiv \mu \frac{{\rm d}\bar g}{{\rm d}\mu}  \raisebox{-.1ex}{
            $\stackrel{\small{\bar g \to 0}}{\sim}$}
  -\bar g^3\left(b_0 + b_1\bar g^2 + \dots\right)
\end{equation}
which governs the running of the strong coupling. The renormalization group invariant
(RGI) quark masses $M_i$ are formally integration constants of the
RG Eq.~(\ref{eq:qmass_tau}). They are scale independent, and due
to the universality of the coefficient $d_0$, they are also scheme
independent. Moreover, they are nonperturbatively defined by
Eq.~(\ref{eq:qmass_rgi}). They only depend on the number
of flavours $\Nf$, making them a natural candidate to quote quark
masses and compare determinations from different lattice
collaborations. Nevertheless, it is customary in the phenomenology
community to use the $\overline{\rm MS}$ scheme at a scale $\mu = 2$ GeV to compare different results for light quarks and the charm quark, and to use a
scale equal to its own mass for the charm and bottom. In this review, we will quote final averages for both
quantities.

Results for quark masses are always quoted in the four-flavour
theory unless otherwise noted. $\Nf=2+1$ results have to be converted to the four-flavour theory. Fortunately, the charm quark is heavy $(\Lambda_{\rm
  QCD}/m_c)^2<1$, and this conversion can be performed in perturbation
theory with negligible ($\sim 0.2\%$) perturbative
uncertainties. 

Nonperturbative corrections in this matching are more
difficult to estimate. Lattice determinations do not show any significant deviation between
$\Nf=2+1$ and $\Nf=2+1+1$ calculations. For example, the
difference in the final averages for the mass of the strange quark
$m_s$ between $\Nf=2+1$ and $\Nf=2+1+1$ determinations is about
1.3\%, or about one standard deviation. Since these effects are suppressed by a factor of 
$1/N_{\rm c}$, and a factor of the strong coupling at the scale of the
charm mass, naive power counting arguments would suggest that the 
effects are $\sim 1\%$, in line with the above observation. On the other hand, numerical nonperturbative
studies~\cite{Hollwieser:2020qri,Athenodorou:2018wpk,Bruno:2014ufa} 
have found this power counting argument to be an overestimate by one
order of magnitude in the determination the $\Lambda$-parameter and other quantities.  

We quote all final averages at $2$ GeV in the $\overline{\rm
  MS}$ scheme and also the RGI values (in the four-flavour theory). We
use the exact RG
 Eq.~(\ref{eq:qmass_rgi}). Note that to use this equation we
need the value of the strong coupling in the $\overline{\rm MS}$
scheme at a scale $\mu = 2$ GeV. All our results are obtained from the
RG equation in the $\overline{\rm MS}$ scheme and the 5-loop beta
function together with the 
value of the $\Lambda$-parameter in the four-flavour theory
$\Lambda^{(4)}_{\overline{\rm MS}} = 295(10)\, {\rm MeV}$ obtained in
this review (see Sec.~\ref{sec:alpha_s}). 
We use the 5-loop mass anomalous dimension as well~\cite{Baikov:2014qja}. 
In the uncertainties of the RGI masses, we separate the contributions from the determination of the
quark masses and the propagation of the uncertainty of
$\Lambda^{(4)}_{\overline{\rm MS}}$. These are identified with the
subscripts $m$ and $\Lambda$, respectively. 

Conceptually, all lattice determinations of quark masses contain three
basic ingredients:
\begin{enumerate}
\item Tuning the lattice bare-quark masses to match the experimental
  values of some quantities.  Pseudo-scalar meson masses provide 
  the most common choice, since they have a strong dependence on the
  values of quark masses.

\item Renormalization of the bare-quark masses. Bare-quark masses
  determined with
  the above-mentioned criteria have to be renormalized. Many of the
  latest determinations use some nonperturbatively defined
  scheme. One can also use perturbation theory to connect directly the
  values of the bare-quark masses to the values in the $\overline{\rm
    MS}$ scheme at $2$ GeV. Experience shows that
  1-loop calculations are unreliable for the renormalization of
  quark masses: usually at least two loops are required to have
  trustworthy results.
   
\item If quark masses have been nonperturbatively renormalized, for
  example, to some MOM or SF scheme, the values in this scheme must be
  converted to the phenomenologically useful values in the
  $\overline{\rm MS}$ scheme (or to the scheme/scale independent RGI
  masses). Either option 
  requires the use of perturbation theory. The larger the
  energy scale of this matching with perturbation theory, the better,
  and many recent computations in MOM schemes do a nonperturbative
  running up to $3$--$4$ GeV. Computations in the SF scheme allow us to
  perform this running nonperturbatively over large energy scales and
  match with perturbation theory directly at the electro-weak scale $\sim 100$
  GeV. 
\end{enumerate}
Note that many lattice determinations of quark masses make use of
perturbation theory at a scale of a few GeV.

We mention that lattice-QCD calculations of the $b$-quark mass have an
additional complication which is not present in the case of the charm
and light quarks. 
At the lattice spacings currently used in numerical calculations the
direct treatment of the $b$ quark with the fermionic actions commonly
used for light quarks is very challenging. Only two determinations of
the $b$-quark mass use this approach, reaching the physical $b$-quark
mass region at two lattice spacings with $aM\sim 1$. 
There are a few widely used approaches to treat the $b$ quark on the
lattice, which have already been discussed in the FLAG 13 review (see
Sec.~8 of Ref.~\cite{Aoki:2013ldr}). 
Those relevant for the determination of the $b$-quark mass will be
briefly described in Sec.~\ref{s:bmass}.

\medskip


\input{qmass/qmass_mudms}


\input{qmass/qmass_mc}

\newpage

\input{qmass/qmass_mb}

%% file: qmass/qmass_mudms.tex
\subsection{Masses of the light quarks}
\label{sec:lqm}

Light-quark masses are particularly difficult to determine because they are very small
(for the up and down quarks) or small (for the strange quark) compared to typical hadronic
scales. Thus, their impact on typical hadronic observables is minute, and it is difficult
to isolate their contribution accurately.

Fortunately, the spontaneous breaking of SU$(3)_L\times$SU$(3)_R$ chiral symmetry provides
observables which are particularly sensitive to the light-quark masses: the masses of the
resulting Nambu-Goldstone bosons (NGB), i.e., pions, kaons, and eta. Indeed, the
Gell-Mann-Oakes-Renner relation~\cite{GellMann:1968rz} predicts that the squared mass of a
NGB is directly proportional to the sum of the masses of the quark and antiquark which
compose it, up to higher-order mass corrections. Moreover, because these NGBs are light,
and are composed of only two valence particles, their masses have a particularly clean
statistical signal in lattice-QCD calculations. In addition, the experimental
uncertainties on these meson masses are negligible. Thus, in lattice calculations,
light-quark masses are typically obtained by renormalizing the input quark mass and tuning
them to reproduce NGB masses, as described above.
\subsubsection{Lattice determination of $m_s$ and $m_{ud}$}
\label{sec:msmud}

We now turn to a review of the lattice calculations of the light-quark masses and begin
with $m_s$, the isospin-averaged up- and down-quark mass $m_{ud}$, and their ratio. Most
groups quote only $m_{ud}$, not the individual up- and down-quark masses. We then discuss
the ratio $m_u/m_d$ and the individual determinations of $m_u$ and $m_d$.

Quark masses have been calculated on the lattice since the mid-nineties. However, early
calculations were performed in the quenched approximation, leading to unquantifiable
systematics. Thus, in the following, we only review modern, unquenched calculations, which
include the effects of light sea quarks.

Tables~\ref{tab:masses3}~and~\ref{tab:masses4} list the results of $\Nf=2+1$ and
$\Nf=2+1+1$ lattice calculations of $m_s$ and $m_{ud}$. These results are given in the
$\msbar$ scheme at $2\,\gev$, which is standard nowadays, though some groups are starting
to quote results at higher scales (e.g.,~Ref.~\cite{Arthur:2012opa}). The tables also show
the colour coding of the calculations leading to these results. As indicated earlier in
this review, we treat calculations with different values of $\Nf$ separately.

\bigskip
\noindent
{\em $\Nf=2+1$ lattice calculations}
\medskip

We begin with $\Nf=2+1$ calculations (see FLAG 19 and earlier editions for two-flavour results). These and the corresponding results for $m_{ud}$
and $m_s$ are summarized in Tab.~\ref{tab:masses3}. Given the very high precision of a
number of the results, with total errors on the order of 1\%, it is important to consider
the effects neglected in these calculations.  Isospin-breaking and electromagnetic\
effects are small on $m_{ud}$ and $m_s$, and have been approximately accounted for in the
calculations that will be retained for our averages. We have already commented that the
effect of the omission of the charm quark in the sea is expected to be small, below our
current precision, and we do not add any additional uncertainty due to these effects in
the final averages. 

\begin{table}[!ht]
\vspace{2mm}
{\footnotesize{
\begin{tabular*}{\textwidth}[l]{l@{\extracolsep{\fill}}rllllllll}
Collaboration & Ref. & \hspace{0.15cm}\begin{rotate}{60}{publication status}\end{rotate}\hspace{-0.15cm} &
 \hspace{0.15cm}\begin{rotate}{60}{chiral extrapolation}\end{rotate}\hspace{-0.15cm} &
 \hspace{0.15cm}\begin{rotate}{60}{continuum  extrapolation}\end{rotate}\hspace{-0.15cm}  &
 \hspace{0.15cm}\begin{rotate}{60}{finite volume}\end{rotate}\hspace{-0.15cm}  &  
 \hspace{0.15cm}\begin{rotate}{60}{renormalization}\end{rotate}\hspace{-0.15cm} &  
 \hspace{0.15cm}\begin{rotate}{60}{running}\end{rotate}\hspace{-0.15cm}  & 
\rule{0.6cm}{0cm}$m_{ud} $ & \rule{0.6cm}{0cm}$m_s $ \\
&&&&&&&&& \\[-0.1cm]
\hline
\hline
&&&&&&&&& \\[-0.1cm]
{CLQCD 23}& \cite{CLQCD:2023sdb} & \gA & \good & \good & \good &
\good & $e$  & 3.60(11)(15)  & 98.8(2.9)(4.7)\\

{ALPHA 19}& \cite{Bruno:2019vup} & \gA & \soso & \good & \good &
\good & $e$  & 3.54(12)(9)  & 95.7(2.5)(2.4)\\

{Maezawa 16}& \cite{Maezawa:2016vgv} & \gA & \bad & \good & \good &
\good & $d$  & --  & 92.0(1.7)\\

{RBC/UKQCD 14B$^\ominus$}& \cite{Blum:2014tka} & \gA & \good & \good & \good &
\good & $d$  & 3.31(4)(4)  & 90.3(0.9)(1.0)\\

{RBC/UKQCD 12$^\ominus$}& \cite{Arthur:2012opa} & \gA & \good & \soso & \good &
\good & $d$  &  3.37(9)(7)(1)(2) & 92.3(1.9)(0.9)(0.4)(0.8)\\

{PACS-CS 12$^\star$}& \protect{\cite{Aoki:2012st}} & \gA & \good & \bad & \bad & \good & $\,b$
&  3.12(24)(8) &  83.60(0.58)(2.23) \\

{Laiho 11} & \cite{Laiho:2011np} & \rC & \soso & \good & \good & \soso
& $-$ & 3.31(7)(20)(17)
& 94.2(1.4)(3.2)(4.7)\\

{BMW 10A, 10B$^+$} & \cite{Durr:2010vn,Durr:2010aw} & \gA & \good & \good & \good & \good &
$\,c$ & 3.469(47)(48)& 95.5(1.1)(1.5)\\

{PACS-CS 10}& \cite{Aoki:2010wm} & \gA & \good & \bad & \bad & \good & $\,b$
&  2.78(27) &  86.7(2.3) \\

{MILC 10A}& \cite{Bazavov:2010yq} & \rC & \soso  & \good & \good &
\soso  &$-$& 3.19(4)(5)(16)&\rule{0.6cm}{0cm}-- \\

{HPQCD~10$^{\ast\ast}$}&  \cite{McNeile:2010ji} &\gA & \soso & \good & \good & $-$
&$-$& 3.39(6)$ $ & 92.2(1.3) \\

{RBC/UKQCD 10A}& \cite{Aoki:2010dy} & \gA & \soso & \soso & \good &
\good & $\,a$  &  3.59(13)(14)(8) & 96.2(1.6)(0.2)(2.1)\\

{Blum~10$^\dagger$}&\cite{Blum:2010ym}& \gA & \soso & \bad & \soso & \good &
$-$ &3.44(12)(22)&97.6(2.9)(5.5)\\

{PACS-CS 09}& \cite{Aoki:2009ix}& \gA &\good   &\bad   & \bad & \good  &  $\,b$
 & 2.97(28)(3) &92.75(58)(95)\\

{HPQCD 09A$^\oplus$}&  \cite{Davies:2009ih}&\gA & \soso & \good & \good & $-$
& $-$& 3.40(7) & 92.4(1.5) \\

{MILC 09A} & \cite{Bazavov:2009fk} & \rC &  \soso & \good & \good & \soso &
$-$ & 3.25 (1)(7)(16)(0) & 89.0(0.2)(1.6)(4.5)(0.1)\\

{MILC 09} & \cite{Bazavov:2009bb} & \gA & \soso & \good & \good & \soso & $-$
& 3.2(0)(1)(2)(0) & 88(0)(3)(4)(0)\\

{PACS-CS 08} & \cite{Aoki:2008sm} &  \gA & \good & \bad & \bad  & \bad & $-$ &
2.527(47) & 72.72(78)\\

{RBC/UKQCD 08} & \cite{Allton:2008pn} & \gA & \soso & \bad & \good & \good &
$-$ &$3.72(16)(33)(18)$ & $107.3(4.4)(9.7)(4.9)$\\

\hspace{-0.2cm}{\begin{tabular}{l}CP-PACS/\\JLQCD 07\end{tabular}} 
& \cite{Ishikawa:2007nn}& \gA & \bad & \good & \good  & \bad & $-$ &
$3.55(19)(^{+56}_{-20})$ & $90.1(4.3)(^{+16.7}_{-4.3})$ \\

{HPQCD 05}
& 
\cite{Mason:2005bj}& \gA & \soso & \soso & \soso & \soso &$-$&
$3.2(0)(2)(2)(0)^\ddagger$ & $87(0)(4)(4)(0)^\ddagger$\\

\hspace{-0.2cm}{\begin{tabular}{l}MILC 04, HPQCD/\\MILC/UKQCD 04\end{tabular}} 
& \cite{Aubin:2004fs,Aubin:2004ck} & \gA & \soso & \soso & \soso & \bad & $-$ &
$2.8(0)(1)(3)(0)$ & $76(0)(3)(7)(0)$\\
&&&&&&&&& \\[-0.1cm]
\hline
\hline\\
\end{tabular*}\\[-0.2cm]
}}
\begin{minipage}{\linewidth}
{\footnotesize 
\begin{itemize}
\item[$^\ominus$] The results are given in the $\msbar$ scheme at 3
  instead of 2~GeV. We run them down to 2~GeV using numerically
  integrated 4-loop
  running~\cite{vanRitbergen:1997va,Chetyrkin:1999pq} with $\Nf=3$ and
  with the values of $\alpha_s(M_Z)$, $m_b$, and $m_c$ taken
  from~Ref.~\cite{Agashe:2014kda}. The running factor is 1.106. At
  three loops it is only 0.2\% smaller, indicating that perturbative
  running uncertainties are small. We neglect them here.\\[-5mm]
\item[$^\star$] The calculation includes electromagnetic and $m_u\ne m_d$ effects
  through reweighting.\\ [-5mm]
\item[$^+$] The fermion action used is tree-level improved.\\[-5mm]
\item[$^{\ast\ast}$] $m_s$ is obtained by combining $m_c$ and
      HPQCD 09A's $m_c/m_s=11.85(16)$~\cite{Davies:2009ih}.
      Finally, $m_{ud}$
      is determined from $m_s$ with the MILC 09 result for
      $m_s/m_{ud}$. Since $m_c/m_s$ is renormalization group invariant
      in QCD, the renormalization and running of the quark masses
      enter indirectly through that of $m_c$ (see below).\\[-5mm]
\item[$^\dagger$] The calculation includes quenched electromagnetic effects.\\[-5mm]
\item[$^\oplus$] What is calculated is $m_c/m_s=11.85(16)$. $m_s$ is then obtained by combining
      this result with the determination $m_c(m_c) = 1.268(9)$~GeV
      from~Ref.~\cite{Allison:2008xk}. Finally, $m_{ud}$
      is determined from $m_s$ with the MILC 09 result for
      $m_s/m_{ud}$.\\[-5mm]
\item[$^\ddagger$] The bare numbers are those of MILC 04. The masses are simply rescaled, using the
ratio of the 2-loop to 1-loop renormalization factors.\\[-5mm]
\item[$a$] The masses are renormalized nonperturbatively at a scale of
  2~GeV in a couple of $\Nf=3$ RI-SMOM schemes. A careful study of
  perturbative matching uncertainties has been performed by comparing
  results in the two schemes in the region of 2~GeV to 3~GeV~\cite{Aoki:2010dy}.\\[-5mm]
\item[$b$] The masses are renormalized and run nonperturbatively up to
  a scale of $40\,\gev$ in the $\Nf=3$ SF scheme. In this scheme,
  nonperturbative and NLO running for the quark masses are shown to
  agree well from 40 GeV all the way down to 3 GeV~\cite{Aoki:2010wm}.\\[-5mm]
\item[$c$] The masses are renormalized and run nonperturbatively up to
  a scale of 4 GeV in the $\Nf=3$ RI-MOM scheme.  In this scheme,
  nonperturbative and N$^3$LO running for the quark masses are shown
  to agree from 6~GeV down to 3~GeV to better than 1\%~\cite{Durr:2010aw}.  \\[-5mm]
\item[$d$] All required running is performed nonperturbatively.
\item[$e$] Running is performed nonperturbatively from 200 MeV to the
  electroweak scale $\sim 100$ GeV.
\end{itemize}
}
\end{minipage}
\caption{$\Nf=2+1$ lattice results for the masses $m_{ud}$ and $m_s$ (MeV).} 
\label{tab:masses3}
\end{table}

The only new computation since the previous FLAG edition is the determination of
light-quark masses by the CLQCD collaboration (CLQCD 23~\cite{CLQCD:2023sdb}). Using stout-smeared clover fermions, the
ensembles reach the physical point and have three lattice spacings to perform the
continuum extrapolation. These look under control, having in all cases $\delta(a_{\rm
min})<2$ (see~\ref{sec:DataDriven}). Volumes are large, and these characteristics ensure that the rating is $\good$
in all criteria. Renormalization is performed nonperturbatively in two different setups
(RI/MOM and SMOM), with the difference used as a systematic effect. 
This systematic effect, in fact, dominates their error budget.

The ALPHA collaboration~\cite{Bruno:2019xed} uses nonperturbatively $\mathcal O(a)$
improved Wilson fermions (a subset of the CLS ensembles~\cite{Bruno:2014jqa}). The
renormalization is performed nonperturbatively in the SF scheme from 200 MeV up to the
electroweak scale $\sim 100$ GeV~\cite{Campos:2018ahf}. This nonperturbative running over
such large energy scales avoids any use of perturbation theory at low energy scales, but
adds a cost in terms of uncertainty: the running alone propagates to $\approx 1\%$ of the
error in quark masses. This turns out to be one of the dominant pieces of uncertainty for
the case of $m_s$. On the other hand, for the case of $m_{ud}$, the uncertainty is
dominated by the chiral extrapolations. The ensembles used include four values of the
lattice spacing below $0.09$ fm, which qualifies for a $\good$ in the continuum
extrapolation, and pion masses down to 200 MeV. This value lies just at the boundary of
the $\good$ rating, but since the chiral extrapolation is a substantial source of
systematic uncertainty, we opted to rate the work with a $\soso$. In any case, this work
enters in the average and their results show a reasonable agreement with the FLAG average.
In all cases the data driven continuum limit criteria shows $\delta(a_{\rm min}) < 3$. 

We now comment in some detail on previous works that also contribute to the averages. 

RBC/UKQCD~14~\cite{Blum:2014tka} significantly improves on their
RBC/UKQCD~12B~\cite{Arthur:2012opa} work by adding three new domain wall fermion
ensembles to three used previously. Two of the new simulations are performed at
essentially physical pion masses ($M_\pi\simeq 139\,\mev$) on lattices of about $5.4\,\fm$
in size and with lattice spacings of $0.114\,\fm$ and $0.084\,\fm$. It is complemented by
a third simulation with $M_\pi\simeq 371\,\mev$, $a\simeq 0.063$ fm and a rather small
$L\simeq 2.0\,\fm$. Altogether, this gives them six simulations with six unitary ($m_{\rm
sea}=m_{\rm val}$) $M_\pi$'s in the range of $139$ to $371\,\mev$, and effectively three
lattice spacings from $0.063$ to $0.114\,\fm$. They perform a combined global continuum
and chiral fit to all of their results for the $\pi$ and $K$ masses and decay constants,
the $\Omega$ baryon mass and two Wilson-flow parameters.  Quark masses in these fits are
renormalized and run nonperturbatively in the RI-SMOM scheme. This is done by computing
the relevant renormalization constant for a reference ensemble, and determining those for
other simulations relative to it by adding appropriate parameters in the global fit. This
calculation passes all of our selection criteria, with $\delta (a_{\rm min}) \approx 1$.

$\Nf=2+1$ MILC results for light-quark masses go back to
2004~\cite{Aubin:2004fs,Aubin:2004ck}. They use rooted staggered fermions.  By 2009 their
simulations covered an impressive range of parameter space, with lattice spacings going
down to 0.045~fm, and valence-pion masses down to approximately
180~MeV~\cite{Bazavov:2009fk}.  The most recent MILC $\Nf=2+1$ results, i.e., MILC
10A~\cite{Bazavov:2010yq} and MILC 09A~\cite{Bazavov:2009fk}, feature large statistics and
2-loop renormalization.  Since these data sets subsume those of their previous
calculations, these latest results are the only ones that need to be kept in any world
average.

The BMW 10A, 10B~\cite{Durr:2010vn,Durr:2010aw} calculation still satisfies our stricter
selection criteria. They reach the physical up- and down-quark mass by {\it interpolation}
instead of by extrapolation. Moreover, their calculation was performed at five lattice
spacings ranging from 0.054 to 0.116~fm, with small extrapolations $\delta(a_{\rm min})< 2$.
The work uses full nonperturbative renormalization and running and in volumes of up to
(6~fm)$^3$, guaranteeing that the continuum limit, renormalization, and infinite-volume
extrapolation are controlled. It does neglect, however, isospin-breaking effects, which
are small on the scale of their error bars.

Finally, we come to another calculation which satisfies our selection criteria, HPQCD~10
\cite{McNeile:2010ji}. It updates the staggered-fermions calculation of
HPQCD~09A~\cite{Davies:2009ih}. In those papers, the renormalized mass of the strange
quark is obtained by combining the result of a precise calculation of the renormalized
charm-quark mass $m_c$ with the result of a calculation of the quark-mass ratio
$m_c/m_s$. As described in Ref.~\cite{Allison:2008xk} and in Sec.~\ref{s:cmass}, HPQCD
determines $m_c$ by fitting Euclidean-time moments of the $\bar cc$ pseudoscalar density
two-point functions, obtained numerically in lattice QCD, to fourth-order, continuum
perturbative expressions. These moments are normalized and chosen so as to require no
renormalization with staggered fermions. Since $m_c/m_s$ requires no renormalization
either, HPQCD's approach displaces the problem of lattice renormalization in the
computation of $m_s$ to one of computing continuum perturbative expressions for the
moments. To calculate $m_{ud}$ HPQCD~10~\cite{McNeile:2010ji} use the MILC 09
determination of the quark-mass ratio $m_s/m_{ud}$~\cite{Bazavov:2009bb}.

HPQCD~09A~\cite{Davies:2009ih} obtains $m_c/m_s=11.85(16)$~\cite{Davies:2009ih} fully
nonperturbatively, with a precision slightly larger than 1\%. HPQCD~10's determination of
the charm-quark mass, $m_c(m_c)=1.268(6)$,\footnote{To obtain this number, we have used
the conversion from $\mu=3\,$ GeV to $m_c$ given in Ref.~\cite{Allison:2008xk}.} is even
more precise, achieving an accuracy better than 0.5\%.

This discussion leaves us with six results for our final average for $m_s$: CLQCD
23~\cite{CLQCD:2023sdb}, ALPHA~19~\cite{Bruno:2019xed}, MILC~09A~\cite{Bazavov:2009fk},
BMW~10A, 10B~\cite{Durr:2010vn,Durr:2010aw}, HPQCD~10~\cite{McNeile:2010ji} and
RBC/UKQCD~14~\cite{Blum:2014tka}. Assuming that the result from HPQCD~10 is 100\%
correlated with that of MILC~09A, as it is based on a subset of the MILC~09A
configurations, we find $m_s=92.3(1.0)\,\mev$ with a $\chi^2/$dof = 1.60.

For the light-quark mass $m_{ud}$, the results satisfying our criteria are CLQCD 23,
ALPHA~19, RBC/UKQCD 14B, BMW 10A, 10B, HPQCD 10, and MILC 10A. For the error, we include
the same 100\% correlation between statistical errors for the latter two as for the
strange case, resulting in the following (at scale 2 GeV in the $\overline{\rm MS}$
scheme, and $\chi^2/$dof=1.4),
%
\begin{align}\label{eq:nf3msmud}
&& \FLAGAVBEGIN m_{ud}&= 3.387(39)\FLAGAVEND\;\mev&&\Refs~\mbox{\cite{Bruno:2019vup,Blum:2014tka,Durr:2010vn,Durr:2010aw,McNeile:2010ji,Bazavov:2010yq}},\,\nonumber \\[-3mm]
&\Nf=2+1 :&\\[-3mm]
&&\FLAGAVBEGIN m_s    &=92.4(1.0)\FLAGAVEND\;\mev&&\Refs~\mbox{
\cite{Bruno:2019vup,Bazavov:2009fk,Durr:2010vn,Durr:2010aw,McNeile:2010ji,Blum:2014tka}}. \nonumber
\end{align}
%
And the RGI values
\begin{align}\label{eq:nf3msmud rgi}
&&  M_{ud}^{\rm RGI}&= 4.714(55)_{m}(46)_{\Lambda}\;\mev&&\Refs~\mbox{\cite{CLQCD:2023sdb,Bruno:2019vup,Blum:2014tka,Durr:2010vn,Durr:2010aw,McNeile:2010ji,Bazavov:2010yq}},\,\nonumber \\[-3mm]
&\Nf=2+1 :&\\[-3mm]
&& M_s^{\rm RGI}    &=128.5(1.4)_{m}(1.2)_{\Lambda}\;\mev&&\Refs~\mbox{\cite{CLQCD:2023sdb,Bruno:2019vup,Bazavov:2009fk,Durr:2010vn,Durr:2010aw,McNeile:2010ji,Blum:2014tka}}. \nonumber
\end{align}

\bigskip
\noindent
{\em $\Nf=2+1+1$ lattice calculations}
\medskip

Since the previous review a new computation of $m_s, m_{ud}$ has appeared, ETM
21A~\cite{Alexandrou:2021gqw}. Using twisted-mass fermions with an added clover term to
suppress $\cO(a^2)$ effects between the neutral and charged pions, this work represents a
significant improvement over ETM 14~\cite{Carrasco:2014cwa}.  
Renormalization is performed nonperturbatively in the RI-MOM scheme. Their ensembles
comprise three lattice spacings (0.095, 0.082, and 0.069 fm), two volumes for the finest
lattice spacings with pion masses reaching down to the physical point in the two finest
lattices spacings allowing a controlled chiral extrapolation. Their volumes are large, with $m_\pi
L$ between four and five. These characteristics of their ensembles pass the most stringent
FLAG criteria in all categories. This work extracts quark masses from two different
quantities, one based on the meson spectrum and the other based on the baryon spectrum.
Results obtained with these two methods agree within errors, but the size of the continuum
extrapolation is much larger for the case of the extractions based on the meson spectrum.
In particular, we estimate that $\delta(a_{\rm min}) = 4$--$4.5$ for the individual fits that
enter the determination of $m_{\rm ud}, m_{\rm s}$ respectively. 
We note that while these values are somewhat large, the systematic errors that the authors estimate in the determinations of the light-quark masses are about the same size as the
statistical fluctuations. This will reduce the stretching factors to a value close to one,
and, therefore we do not apply any additional corrections for these cases. Nevertheless, we stress that some large continuum extrapolations are present in this work.

Determinations based on the baryon spectrum agree well with the FLAG average while the
ones based on the meson sector are high in comparison (there is good agreement with their
previous results, ETM 14~\cite{Carrasco:2014cwa}). Related with the previous point, it is important to note that the determinations that involve large continuum
extrapolations are the ones that show a larger tension. 

There are three other works that enter in light-quark mass averages.
Contributing both to the average of $m_{ud}$ and $ m_s$ is
FNAL/MILC/TUMQCD~18~\cite{Bazavov:2018omf}. 
They perform a determination of the strange-quark mass using
masses of the heavy-strange mesons as input. In this case, some very large continuum
extrapolations, with $\delta(a_{\rm min})\approx 14$ enter in a global analysis, but for
the determination of the light-quark masses, we believe that the influence of the data at
heavier masses on the determination of the fit parameter that determines $m_{\rm s}$ is
small. In the region $m_{\rm heavy} < 3$ GeV the extrapolations are much better under
control, and in fact involve up to five lattice spacing. We conclude that the large value
of $\delta(a_{\rm min})$ does not influence significantly the values of the light-quark
masses.  HPQCD 18~\cite{Lytle:2018evc} and HPQCD 14A~\cite{Chakraborty:2014aca} contribute to the
determination of $m_{ud}$, and both show $\delta(a_{\rm min}) <3$ for most of their region
of parameters.

The $\Nf=2+1+1$ results are summarized in Tab.~\ref{tab:masses4}. While the results of HPQCD 14A and HPQCD 18 agree well (using different methods), there
are several tensions in the determination of $m_s$. The most significant discrepancy is
between the results of the ETM collaboration and other results. But also the two
very precise determinations of HPQCD 18 and FNAL/MILC/TUMQCD~18 show a tension.
Note that the results of
Ref.~\cite{Chakraborty:2014aca} are reported as $m_s(2\,\gev;\Nf=3)$ and those of
Ref.~\cite{Carrasco:2014cwa} as $m_{ud(s)}(2\,\gev;\Nf=4)$. We convert the former to
$\Nf=4$ and obtain $m_s(2\,\gev;\Nf=4)=93.7(8)\mev$. The average of ETM 21A,
FNAL/MILC/TUMQCD~18, HPQCD 18, ETM 14 and HPQCD 14A is 93.46(58)$\mev$ with
$\chi^2/\mbox{dof}=1.3$. For the light-quark mass, we average ETM 21A, ETM 14 and
FNAL/MILC/TUMQCD 18 to obtain 3.427(51) with a $\chi^2/\mbox{dof}=4.5$. We note these
$\chi^2$ values are large. For the case of the light-quark masses there is a clear tension
between the ETM and FNAL/MILC/TUMQCD results. 
We also note that the 2+1-flavour values are consistent with the four-flavour ones, so in
all cases we have simply quoted averages according to FLAG rules, including
stretching factors for the errors based on $\chi^2$ values of our fits. Nevertheless it is
worth pointing out that large continuum extrapolations are present in the $\Nf=2+1+1$
determination of quark masses. Global fits that aim at describing results obtained for a wide range of quark
masses are involved in many analyses. At small quark masses many lattice spacing enter
these determinations, but how the large quark mass region influences the precision
obtained at small quark masses is something that deserves further investigation.
%
\begin{align}\label{eq:nf4msmud}
&&\FLAGAVBEGIN m_{ud}&=3.427(51)\FLAGAVEND\;\mev&& \Refs~\mbox{\cite{Alexandrou:2021gqw,Carrasco:2014cwa,Bazavov:2018omf}},\nonumber\\[-3mm]
&\Nf=2+1+1 :& \\[-3mm]
&&\FLAGAVBEGIN m_s   &= 93.46(58)\FLAGAVEND\; \mev&& \Refs~\mbox{\cite{Alexandrou:2021gqw,Carrasco:2014cwa,Bazavov:2018omf,Lytle:2018evc,Chakraborty:2014aca}},\nonumber
\end{align}
%
and the RGI values
\begin{align}\label{eq:nf4msmud rgi}
&& M_{ud}^{\rm RGI}&= 4.768(71)_{m}(46)_{\Lambda} \,\mev&& \Refs~\mbox{\cite{Alexandrou:2021gqw,Bazavov:2018omf,Carrasco:2014cwa}},\nonumber\\[-3mm]
&\Nf=2+1+1 :& \\[-3mm]
&& M_s^{\rm RGI}   &=130.0(0.8)_{m}(1.3)_{\Lambda}\,\mev&& \Refs~\mbox{\cite{Alexandrou:2021gqw,Bazavov:2018omf,Lytle:2018evc,Carrasco:2014cwa,Chakraborty:2014aca}}.\nonumber
\end{align}

\begin{table}[!htb]
\vspace{2.5cm}
{\footnotesize{
\begin{tabular*}{\textwidth}[l]{l@{\extracolsep{\fill}}rllllllll}
Collaboration & Ref. & \hspace{0.15cm}\begin{rotate}{60}{publication status}\end{rotate}\hspace{-0.15cm} &
 \hspace{0.15cm}\begin{rotate}{60}{chiral extrapolation}\end{rotate}\hspace{-0.15cm} &
 \hspace{0.15cm}\begin{rotate}{60}{continuum  extrapolation}\end{rotate}\hspace{-0.15cm}  &
 \hspace{0.15cm}\begin{rotate}{60}{finite volume}\end{rotate}\hspace{-0.15cm}  &  
 \hspace{0.15cm}\begin{rotate}{60}{renormalization}\end{rotate}\hspace{-0.15cm} &  
 \hspace{0.15cm}\begin{rotate}{60}{running}\end{rotate}\hspace{-0.15cm}  & 
\rule{0.6cm}{0cm}$m_{ud} $ & \rule{0.6cm}{0cm}$m_s $ \\
&&&&&&&&& \\[-0.1cm]
\hline
\hline
&&&&&&&&& \\[-0.1cm]
{ETM 21A}& \cite{Alexandrou:2021gqw} & \gA & \good & \good & \good &
\good & $-$  & $3.636(66)(^{+60}_{-57})$  & $98.7(2.4)(^{+4.0}_{-3.2})$\\

{HPQCD 18}$^\dagger$ & \cite{Lytle:2018evc} & \gA & \good & \good & \good &
$\good$ & $-$  &  & 94.49(96) \\

{FNAL/MILC/TUMQCD 18}& \cite{Bazavov:2018omf} & \gA & \good & \good & \good &
\good & $-$  & 3.404(14)(21)  & 92.52(40)(56)\\

{HPQCD 14A $^\oplus$} & \cite{Chakraborty:2014aca} & \gA & \good & \good & \good &
$-$ & $-$  &  & 93.7(8) \\
  
{ETM 14$^\oplus$}& \cite{Carrasco:2014cwa} & \gA & \soso & \good & \good &
\good & $-$  & 3.70(13)(11)  & 99.6(3.6)(2.3)\\

&&&&&&&&& \\[-0.1cm] 
\hline
\hline\\[-2mm]
\end{tabular*}
}}
\begin{minipage}{\linewidth}
{\footnotesize 
  \begin{itemize}
  \item[$^\dagger$] Bare-quark masses are renormalized nonperturbatively in
    the RI-SMOM scheme at scales $\mu\sim$ 2--5 GeV for different
    lattice spacings and translated to the $\overline{\rm MS}$
    scheme. Perturbative running is then used to run all results to a
    reference scale $\mu = 3$ GeV.
\item[$^\oplus$] As explained in the text, $m_s$ is obtained by combining the
        results $m_c(5\,\gev;\Nf=4)=0.8905(56)$~GeV and
        $(m_c/m_s)(\Nf=4)=11.652(65)$, determined on the same data
        set. A subsequent scale and scheme conversion, performed by
        the authors, leads to the value 93.6(8). In the table, we have converted this
        to $m_s(2\,\gev;\Nf=4)$, which makes a very small change. 
\end{itemize}
}
\end{minipage}

\caption{$\Nf=2+1+1$ lattice results for the masses $m_{ud}$ and $m_s$ (MeV).} 
\label{tab:masses4}
\end{table}

\begin{figure}[!htb]
\begin{center}
\includegraphics[width=11.5cm]{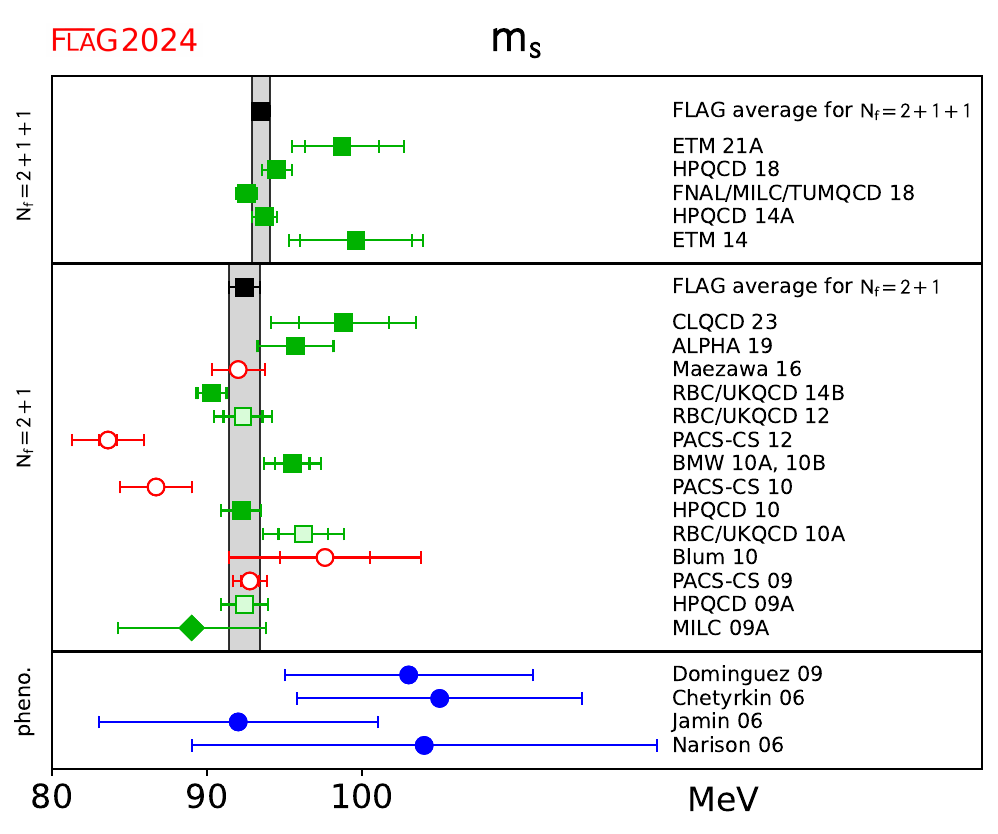}
\end{center}
\vspace{0.2cm}
\begin{center}
\caption{ \label{fig:ms} $\msbar$ mass of the strange quark (at 2 GeV scale) in MeV. The
 upper two panels show the lattice results listed in Tabs.~\ref{tab:masses3} and
 \ref{tab:masses4}, while the bottom panel collects  sum rule
 results~\cite{Dominguez:2008jz, Chetyrkin:2005kn,Jamin:2006tj, Narison:2005ny,
 Vainshtein:1978nn}. The diamond represents a result based on perturbative renormalization. The squares and open circles denote results based on nonperturbative renormalization. The black squares and the grey bands represent our averages (\ref{eq:nf3msmud}) and (\ref{eq:nf4msmud}). The significance of the colours is explained in Sec.~\ref{sec:qualcrit}.
}\end{center}

\end{figure}

\begin{figure}[!htb]

\begin{center}
\includegraphics[width=11.5cm]{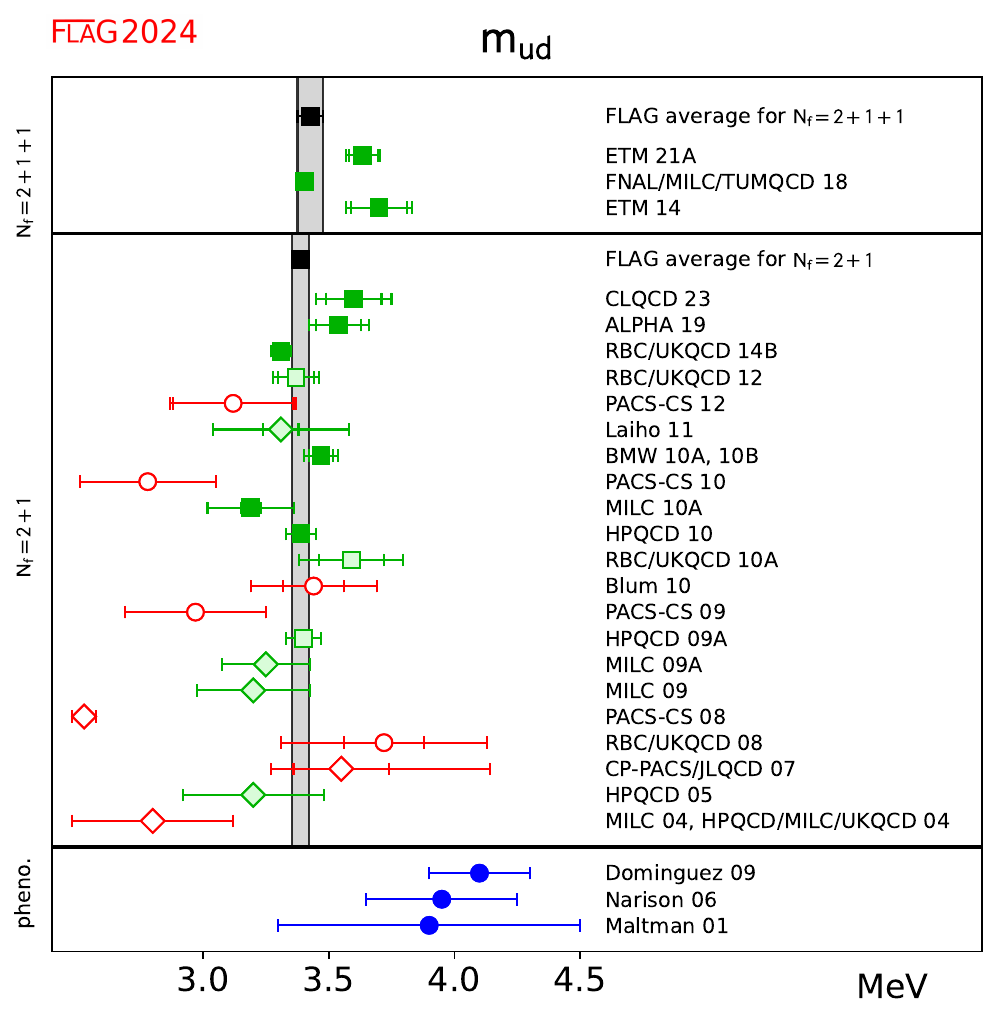}
\end{center}
\begin{center}
\caption{ \label{fig:mud} Mean mass of the two lightest quarks,
 $m_{ud}=\frac{1}{2}(m_u+m_d)$. The bottom panel shows 
 results based on sum rules~\cite{Dominguez:2008jz,Narison:2005ny,Maltman:2001nx} (for more details see Fig.~\ref{fig:ms}).}\end{center}

\end{figure}

In Figs.~\ref{fig:ms} and \ref{fig:mud} the lattice results listed in
Tabs.~\ref{tab:masses3} and \ref{tab:masses4} and the FLAG averages obtained at each value
of $\Nf$ are presented and compared with various phenomenological results. 

\subsubsection{Lattice determinations of $m_s/m_{ud}$}
\label{sec:msovermud}

\begin{table}[!htb]
\vspace{3cm}
{\footnotesize{
\begin{tabular*}{\textwidth}[l]{l@{\extracolsep{\fill}}rllllll}
Collaboration & Ref. & $\Nf$ & \hspace{0.15cm}\begin{rotate}{60}{publication status}\end{rotate}\hspace{-0.15cm}  &
 \hspace{0.15cm}\begin{rotate}{60}{chiral extrapolation}\end{rotate}\hspace{-0.15cm} &
 \hspace{0.15cm}\begin{rotate}{60}{continuum  extrapolation}\end{rotate}\hspace{-0.15cm}  &
 \hspace{0.15cm}\begin{rotate}{60}{finite volume}\end{rotate}\hspace{-0.15cm}  & \rule{0.1cm}{0cm} 
$m_s/m_{ud}$ \\
&&&&&& \\[-0.1cm]
\hline
\hline
&&&&&& \\[-0.1cm]

{ETM 21A}& \cite{Alexandrou:2021gqw} & 2+1+1 & \gA & \good & \good & \good & 27.17(32)$^{+56}_{-38}$\\

{MILC 17 $^\ddagger$} & \cite{Bazavov:2017lyh} & 2+1+1 & \gA & \good & \good & \good & $27.178(47)^{+86}_{-57}$\\

{FNAL/MILC 14A} & \cite{Bazavov:2014wgs} & 2+1+1 & \gA & \good & \good & \good & $27.35(5)^{+10}_{-7}$\\

{ETM 14}& \cite{Carrasco:2014cwa} & 2+1+1 & \gA & \soso & \good & \soso & 26.66(32)(2)\\

&&&&&& \\[-0.1cm]
\hline
&&&&&& \\[-0.1cm]
{CLQCD 23}& \cite{CLQCD:2023sdb} &2+1 & \gA & \good & \good & \good & 27.47(30)(13)\\

{ALPHA 19}& \cite{Bruno:2019xed} &2+1 & \gA & \soso & \good & \good & 27.0(1.0)(0.4)\\

{RBC/UKQCD 14B}& \cite{Blum:2014tka} &2+1  & \gA & \good & \good & \good & 27.34(21)\\

{RBC/UKQCD 12$^\ominus$}& \cite{Arthur:2012opa} &2+1  & \gA & \good & \soso & \good & 27.36(39)(31)(22)\\

{PACS-CS 12$^\star$}& \cite{Aoki:2012st}       &2+1  & \gA & \good & \bad & \bad & 26.8(2.0)\\

{Laiho 11} & \cite{Laiho:2011np}              &2+1  & \rC & \soso & \good & \good & 28.4(0.5)(1.3)\\

{BMW 10A, 10B$^+$}& \cite{Durr:2010vn,Durr:2010aw} &2+1  & \gA & \good & \good & \good & 27.53(20)(8) \\

{RBC/UKQCD 10A}& \cite{Aoki:2010dy}           &2+1  & \gA & \soso & \soso & \good & 26.8(0.8)(1.1) \\

{Blum 10$^\dagger$}&\cite{Blum:2010ym}         &2+1  & \gA & \soso & \bad & \soso & 28.31(0.29)(1.77)\\

{PACS-CS 09}  & \cite{Aoki:2009ix}            &2+1  &  \gA &\good   &\bad   & \bad & 31.2(2.7)  \\

{MILC 09A}    & \cite{Bazavov:2009fk}       &2+1  & \rC & \soso & \good & \good & 27.41(5)(22)(0)(4)  \\
{MILC 09}      & \cite{Bazavov:2009bb}      &2+1  & \gA & \soso & \good & \good & 27.2(1)(3)(0)(0)  \\

{PACS-CS 08}   & \cite{Aoki:2008sm}           &2+1  & \gA & \good & \bad  & \bad & 28.8(4)\\

{RBC/UKQCD 08} & \cite{Allton:2008pn}         &2+1  & \gA & \soso & \bad  & \good & 28.8(0.4)(1.6) \\

\hspace{-0.2cm}{\begin{tabular}{l}MILC 04, HPQCD/\\MILC/UKQCD 04\end{tabular}} 
& \cite{Aubin:2004fs,Aubin:2004ck}            &2+1  & \gA & \soso & \soso & \soso & 27.4(1)(4)(0)(1)  \\
&&&&&& \\[-0.1cm]
\hline
\hline\\
\end{tabular*}\\[-0.2cm]
}}
\begin{minipage}{\linewidth}
{\footnotesize 
\begin{itemize}
\item[$^\ddagger$] The calculation includes electromagnetic effects.\\[-5mm]
\item[$^\ominus$] The errors are statistical, chiral and finite volume.\\[-5mm]
\item[$^\star$] The calculation includes electromagnetic and $m_u\ne m_d$ effects through reweighting.\\[-5mm]
\item[$^+$] The fermion action used is tree-level improved.\\[-5mm]
\item[$^\dagger$] The calculation includes quenched electromagnetic effects.
\end{itemize}
}
\end{minipage}
\caption{Lattice results for the ratio $m_s/m_{ud}$.}
\label{tab:ratio_msmud}
\end{table}

The lattice results for $m_s/m_{ud}$ are summarized in Tab.~\ref{tab:ratio_msmud}.
In the ratio $m_s/m_{ud}$, one of the sources of systematic error---the
uncertainties in the renormalization factors---drops out. This is
especially important for the recent determination by the CLQCD
collaboration, since their error budget for the individual quark
masses was dominated by the systematic associated with the renormalization.
Also, other systematic effects (like the effect of the scale setting)
are reduced in these ratios.  
This might explain that despite the discrepancies that are present in
the individual quark mass determinations, the ratios show an overall
very good agreement. 

\medskip
\noindent
{\em $\Nf=2+1$ lattice calculations}
\medskip

CLQCD 23~\cite{CLQCD:2023sdb}, discussed already, is the
only new result for this section. 
The other works contributing to this average are
ALPHA 19, RBC/UKQCD 14B, which replaces  
RBC/UKQCD 12 (see Sec.~\ref{sec:msmud}), and the results of MILC 09A
and BMW 10A, 10B.  

The results show very good agreement with a $\chi^2/{\rm dof} = 0.14$. 
The final uncertainty ($\approx 0.5\%$) is smaller than the
ones of the quark masses themselves.
At this level of precision, the uncertainties in the electromagnetic
and strong isospin-breaking corrections might not be completely 
negligible. Nevertheless, we decided not to add any uncertainty
associated with this effect. The main reason is that most recent
determinations try to estimate this uncertainty themselves and found
an effect smaller than naive power counting estimates (see $\Nf=2+1+1$
section), 
 \be
     \label{eq:msovmud3} 
     \mbox{$\Nf = 2+1$ :} \qquad \FLAGAVBEGIN{m_s}/{m_{ud}} = 27.42 ~ (12) \FLAGAVEND\qquad\Refs~\mbox{\cite{Bruno:2019xed,Blum:2014tka,Bazavov:2009fk,Durr:2010vn,Durr:2010aw}}\,.
 \ee 

\bigskip
\noindent
{\em $\Nf=2+1+1$ lattice calculations}
\medskip

For $\Nf = 2+1+1$ there are four results, ETM
21~\cite{Alexandrou:2021gqw}, MILC 17~\cite{Bazavov:2017lyh}, ETM
14~\cite{Carrasco:2014cwa} and FNAL/MILC 14A~\cite{Bazavov:2014wgs},
all of which satisfy our selection criteria. 

All these works have been discussed in the previous FLAG
edition~\cite{FlavourLatticeAveragingGroup:2019iem}, except the new result ETM 21A, that we
have already examined.
The fit has $\chi^2/{\rm dof} \approx
1.7$, and the result shows reasonable agreement with the $N_{f}=2+1$ result.

 \be
      \label{eq:msovmud4} 
      \mbox{$\Nf = 2+1+1$ :}\qquad \FLAGAVBEGIN m_s / m_{ud} = 27.227 ~ (81)\FLAGAVEND \qquad\Refs~\mbox{\cite{Alexandrou:2021gqw,Bazavov:2017lyh,Carrasco:2014cwa,Bazavov:2014wgs}},
 \ee
which corresponds to an overall uncertainty equal to 0.4\%. It is
worth noting that Ref.~\cite{Bazavov:2017lyh} estimates the EM effects in
this quantity to be $\sim 0.18\%$ (or 0.049 which is less than the quoted error above).

\begin{figure}[!htb]
\begin{center}
\includegraphics[width=11cm]{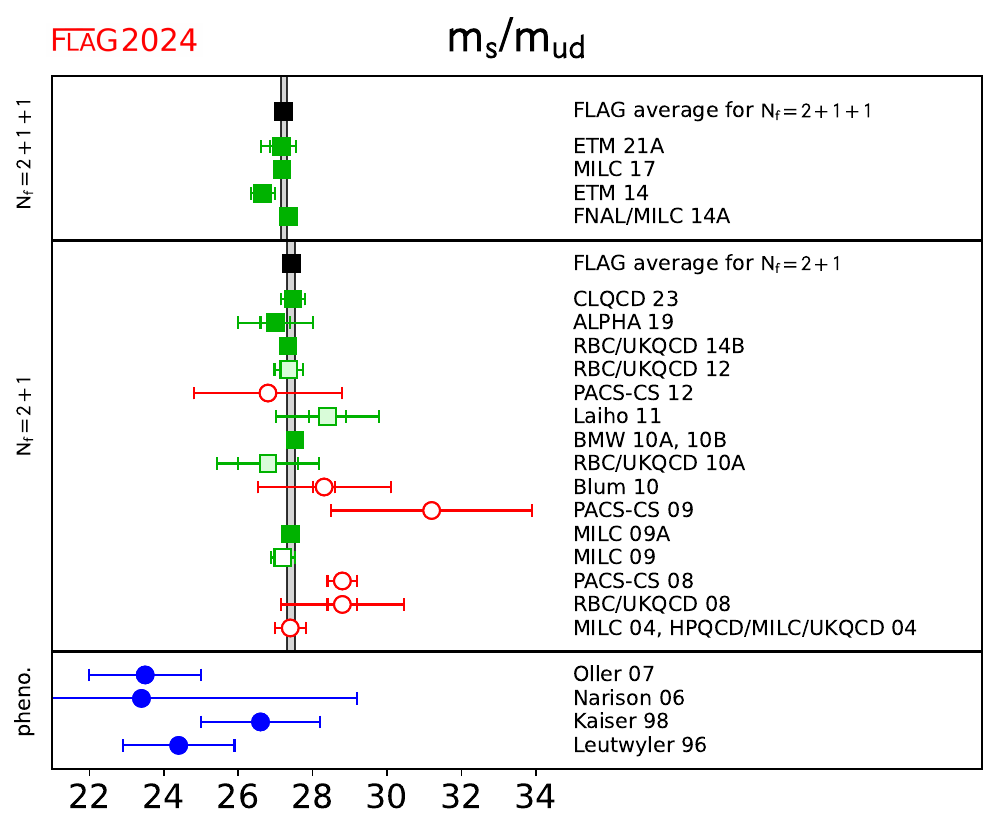}
\end{center}
\vspace{0.5cm}
\begin{center}
	\caption{ \label{fig:msovmud}Results for the ratio $m_s/m_{ud}$. The upper part
shows the lattice results listed in Tab.~\ref{tab:ratio_msmud} together with the FLAG
averages for each value of $\Nf$. The lower part shows results obtained from $\chi$PT and
sum rules~\cite{Oller:2006xb,Narison:2005ny,Kaiser,Leutwyler:1996qg,Weinberg:1977hb}. 
}
\end{center}
\end{figure}

All the lattice results listed in Tab.~\ref{tab:ratio_msmud} as well
as the FLAG averages for each value of $\Nf$ are reported in
Fig.~\ref{fig:msovmud} and compared with $\chi$PT and sum rules.

\subsubsection{Lattice determination of $m_u$ and $m_d$}
\label{subsec:mumd}
In this section, we review computations of the individual $m_u$ and $m_d$ quark
masses, as well as the parameter $\epsilon$ related to
the violations of Dashen's theorem
\be
\epsilon=\frac{(\Delta M_{K}^{2}-\Delta M_{\pi}^{2})_{\gamma}}
{\Delta M_{\pi}^{2}}\,,
\label{eq:epsdef}
\ee
where $\Delta M_{\pi}^{2}=M_{\pi^+}^{2}-M_{\pi^0}^{2}$ and $\Delta
M_{K}^{2}=M_{K^+}^{2}-M_{K^0}^{2}$ are the pion and kaon squared mass splittings,
respectively. The subscript $\gamma$, here and in the following, denotes corrections
that arise from electromagnetic effects only according to the prescription given in
Section~\ref{sec:ibscheme}. This parameter is often a crucial intermediate quantity in the
extraction of the individual light-quark masses. Indeed, it can be shown using the
$G$-parity symmetry of the pion triplet, that $\Delta M_{\pi}^{2}$ does not receive
$\cO(m_u-m_d)$ isospin-breaking corrections. In
other words
\be
\Delta M_{\pi}^{2}=(\Delta M_{\pi}^{2})_{\gamma}
\qquad
\text{and}
\qquad
\epsilon=\frac{(\Delta M_{K}^{2})_{\gamma}}
{\Delta M_{\pi}^{2}}-1\,,
\label{eq:epslo}
\ee
at leading order in the isospin-breaking expansion. Once known, $\epsilon$ allows one to
consistently subtract the electromagnetic part of the kaon-mass splitting to obtain the
QCD part of the kaon mass splitting $(\Delta M_{K}^{2})_{\mathrm{SU}(2)}$. In contrast with the pion, the kaon QCD
splitting is sensitive to $m_u-m_d$ and, in particular, proportional to it at leading
order in $\chi$PT. Therefore, the knowledge of $\epsilon$ allows for the determination of
$m_u-m_d$ from a chiral fit to lattice-QCD data. Originally introduced in another form
in~\citep{Dashen:1969eg}, $\epsilon$ vanishes in the SU(3) chiral limit, a result known
as Dashen's theorem. However, in the 1990's numerous phenomenological papers pointed out
that $\epsilon$ might be an $\cO(1)$ number, indicating a significant failure of SU(3)
$\chi$PT in the description of electromagnetic effects on light-meson masses. However, the
phenomenological determinations of $\epsilon$ feature some level of controversy, leading
to the rather imprecise estimate $\epsilon=0.7(5)$ given in the first edition of FLAG.
Starting with the FLAG 19 edition of the review, we quote more precise averages for
$\epsilon$, directly obtained from lattice-QCD+QED simulations. We refer the reader to
earlier editions of FLAG and to the review~\citep{Portelli:2015wna} for discussions of the
phenomenological determinations of $\epsilon$.

The quality criteria regarding finite-volume effects for calculations including QED are
presented in Sec.~\ref{sec:Criteria}. Due to the long-distance nature of the
electromagnetic interaction, these effects are dominated by a power law in the lattice
spatial size. The coefficients of this expansion depend on the chosen finite-volume
formulation of QED. For $\mathrm{QED}_{\mathrm{L}}$, these effects on the squared mass
$M^2$ of a charged meson are given
by~\citep{Borsanyi:2014jba,Davoudi:2014qua,Davoudi:2018qpl}
\be
  \Delta_{\mathrm{FV}}M^2=
  \alpha M^2\left\{
  \frac{c_{1}}{ML}+\frac{2c_1}{(ML)^2}+
  \cO\left[\frac{1}{(ML)^3}\right]\right\}\co
\ee
with $c_1\simeq-2.83730$. It has been shown in~\citep{Borsanyi:2014jba} that the two first
orders in this expansion are exactly known for hadrons, and are equal to the pointlike
case. However, the $\cO[1/(ML)^{3}]$ term and higher orders depend on the structure of the
hadron. The universal corrections for $\mathrm{QED}_{\mathrm{TL}}$ can also be found in
\citep{Borsanyi:2014jba}. In all this part, for all computations using such universal
formulae, the QED finite-volume quality criterion has been applied with
$n_{\mathrm{min}}=3$, otherwise $n_{\mathrm{min}}=1$ was used (see~\ref{sec:Criteria}).

Since FLAG 21, one new result has been reported for nondegenerate light-quark masses,
namely CLQCD 23~\citep{CLQCD:2023sdb}. This result is based on a new set of $\Nf=2+1$
stout-smeared clover fermion simulations, including one ensemble at the physical
light-quark mass. This calculation achieves a $\good$ rating in all criteria except the
inclusion of isospin-breaking effects. Regarding the latter, $(\Delta M_{K}^{2})^{\gamma}$
from RM123~17~\citep{Giusti:2017dmp} is used to estimate the QCD kaon-mass splitting
required to constrain $m_u$ and $m_d$. Because of the use of a result already averaged for
$\Nf=2+1+1$ up- and down-quark masses, and in application of our quality criterion, we do
not include CLQCD~23 in our average for $m_u/m_d$.

\begin{sidewaystable}[ph!]
\centering
\vspace{2.5cm}
{\footnotesize{
\begin{tabular*}{\textwidth}[l]{l@{\extracolsep{\fill}}r@{\hspace{2mm}}l@{\hspace{2mm}}l@{\hspace{1.5mm}}l@{\hspace{1.5mm}}l@{\hspace{1.5mm}}l@{\hspace{1.5mm}}l@{\hspace{1.5mm}}l@{\hspace{1.5mm}}l@{\hspace{1.5mm}}l@{\hspace{1.5mm}}l}
Collaboration \al  Ref. \al \hspace{0.15cm}\begin{rotate}{60}{publication status}\end{rotate}\hspace{-0.15cm}  \al 
\hspace{0.15cm}\begin{rotate}{60}{chiral extrapolation}\end{rotate}\hspace{-0.15cm} \al 
\hspace{0.15cm}\begin{rotate}{60}{continuum  extrapolation}\end{rotate}\hspace{-0.15cm}  \al 
\hspace{0.15cm}\begin{rotate}{60}{finite volume}\end{rotate}\hspace{-0.15cm}  \al  
\hspace{0.15cm}\begin{rotate}{60}{isospin breaking}\end{rotate}\hspace{-0.15cm} \al
\hspace{0.15cm}\begin{rotate}{60}{renormalization}\end{rotate}\hspace{-0.15cm} \al   
\hspace{0.15cm}\begin{rotate}{60}{running}\end{rotate}\hspace{-0.15cm}  \al  
\rule{0.6cm}{0cm}$m_u$\al 
\rule{0.6cm}{0cm}$m_d$ \al \rule{0.3cm}{0cm} $m_u/m_d$\\
\al \al \al \al \al \al \al \al \al \al  \\[-0.1cm]
\hline
\hline
\al \al \al \al \al \al \al \al \al \al  \\[-0.1cm]
{MILC 18} \al \cite{Basak:2018yzz} \al \gA \al \good \al \good \al  \good \al \soso
\al \good \al $-$ \al \al
\al $0.4529(48)({}^{+150}_{-67})$\\
{FNAL/MILC/TUMQCD 18$^*$} \al \cite{Bazavov:2018omf} \al \gA \al \good \al \good \al  \good \al \soso
\al \good \al $-$ \al 2.118(17)(32)(12)(03) \al 4.690(30)(36)(26)(06)
\al \\
{MILC~17$^\dagger$} \al \cite{Bazavov:2017lyh} \al \gA \al \good \al \good \al \good \al \soso 
\al \good \al $-$ \al \al \al $0.4556(55)({}^{+114}_{-67})(13)$ \\
{RM123~17} \al \cite{Giusti:2017dmp} \al \gA \al \soso \al \good \al \good
\al \soso \al \good \al $\,b$ \al 2.50(15)(8)(2) \al 4.88(18)(8)(2)
\al 0.513(18)(24)(6) \\
{ETM 14}& \cite{Carrasco:2014cwa}  \al \gA \al \good \al \good \al \good \al \bad \al \good \al 
$\,b$ \al 2.36(24) \al 5.03(26) \al 0.470(56) \\[0.5ex]
\hline
\al \al \al \al \al \al \al \al \al \al  \\[-0.2cm]
{CLQCD 23} \al \cite{CLQCD:2023sdb} \al \gA \al \good \al \good \al \good \al \bad \al \good \al $\,c$ \al 2.45(22)(20) \al 4.74(11)(09) \al 0.519(51)(34) \\
{BMW 16A} \al \cite{Fodor:2016bgu} \al \gA \al \good \al \good \al \good
\al \soso \al \good \al $-$ \al 2.27(6)(5)(4) \al 4.67(6)(5)(4) \al 
0.485(11)(8)(14)\\

{MILC 16} \al \cite{Basak:2016jnn} \al \rC \al \soso \al \good \al \good 
\al \soso \al \good \al $-$ \al \al \al 
$0.4582(38)({}^{+12}_{-82})(1)(110)$ \\

{QCDSF/UKQCD 15} \al \protect{\cite{Horsley:2015eaa}} \al \gA \al \soso \al \bad \al \bad \al \good \al $-$\al $-$
\al  \al   \al 0.52(5)\\

{PACS-CS 12} \al \protect{\cite{Aoki:2012st}} \al \gA \al \good \al \bad \al \bad \al \good \al \good \al $\,a$
\al  2.57(26)(7) \al  3.68(29)(10) \al 0.698(51)\\

{Laiho 11} \al \cite{Laiho:2011np} \al \rC \al \soso \al \good \al
\good \al \bad \al \soso \al $-$ \al 1.90(8)(21)(10) \al
4.73(9)(27)(24) \al 0.401(13)(45)\\

{HPQCD~10$^\ddagger$}\al \cite{McNeile:2010ji} \al \gA \al \soso \al \good \al \good \al \bad \al \good \al
$-$ \al 2.01(14) \al 4.77(15) \al  \\

{BMW 10A, 10B$^+$}\al \cite{Durr:2010vn,Durr:2010aw} \al \gA \al \good \al \good \al \good \al \bad \al \good \al
$\,b$ \al 2.15(03)(10) \al 4.79(07)(12) \al 0.448(06)(29) \\

{Blum~10}\al\cite{Blum:2010ym}\al \gA \al \soso \al \bad \al \soso \al \soso \al \good \al $-$ \al 2.24(10)(34)\al 4.65(15)(32)\al 0.4818(96)(860)\\

{MILC 09A} \al  \cite{Bazavov:2009fk} \al  \rC \al  \soso \al \good \al \good \al \bad \al \soso \al $-$
\al 1.96(0)(6)(10)(12)
\al  4.53(1)(8)(23)(12)  \al   0.432(1)(9)(0)(39) \\

{MILC 09} \al  \cite{Bazavov:2009bb} \al  \gA \al  \soso \al  \good \al  \good \al  \bad \al \soso \al 
$-$\al  1.9(0)(1)(1)(1)
\al  4.6(0)(2)(2)(1) \al  0.42(0)(1)(0)(4) \\

\hspace{-0.2cm}{\begin{tabular}{l}MILC 04, HPQCD/\rule{0.1cm}{0cm}\\MILC/UKQCD 04\end{tabular}} \al \cite{Aubin:2004fs}\cite{Aubin:2004ck} \al  \gA \al  \soso \al  \soso \al  \soso \al \bad \al
\bad \al$-$\al  1.7(0)(1)(2)(2)
\al  3.9(0)(1)(4)(2)  \al  0.43(0)(1)(0)(8) \\

\al \al \al \al \al \al \al \al \al \al \al  \\[-0.3cm]
\hline
\hline\\
\end{tabular*}\\[-0.2cm]
}}
\begin{minipage}{\linewidth}
{\footnotesize 
\begin{itemize}
\item[$^*$] FNAL/MILC/TUMQCD~18 uses $\epsilon$ from MILC~18 to produce the individual
$m_u$ and $m_d$ masses.
\item[$^\dagger$]MILC~17 additionally quotes an optional 0.0032 uncertainty on $m_u/m_d$
corresponding to QED and QCD separation scheme ambiguities. Because this variation is not
per se an error on the determination of $m_u/m_d$, and because it is generally not
included in other results, we choose to omit it here.
\item[$^\ddagger$]Values obtained by combining the HPQCD 10 result for $m_s$ with the MILC
  09 results for $m_s/m_{ud}$ and $m_u/m_d$.\\[-5mm]
\item[$^+$] The fermion action used is tree-level improved.\\[-5mm]
\item[$a$] The masses are renormalized and run nonperturbatively up to a scale of
  $100\,\gev$ in the $\Nf=2$ SF scheme. In this scheme, nonperturbative and NLO running
  for the quark masses are shown to agree well from 100 GeV all the way down to 2
  GeV~\cite{DellaMorte:2005kg}.\\[-5mm]
\item[$b$] The masses are renormalized and run nonperturbatively up to a scale of 4 GeV in
  the $\Nf=3$ RI-MOM scheme.  In this scheme, nonperturbative and N$^3$LO running for the
  quark masses are shown to agree from 6~GeV down to 3~GeV to better than
  1\%~\cite{Durr:2010aw}.
\item[$c$] The masses are renormalized and run nonperturbatively in both the RI/MOM and
SMOM schemes. The quoted quark-mass value is the RI/MOM one, with an assigned systematic
error coming from the difference between the two schemes.
\end{itemize}
}
\end{minipage}
\caption{Lattice results for $m_u$, $m_d$ (MeV) and for the ratio $m_u/m_d$. The values refer to the 
$\msbar$ scheme  at scale 2 GeV.  The top part of the table lists the results obtained with $\Nf=2+1+1$,  
while the lower part presents calculations with $\Nf = 2+1$.}
\label{tab:mu_md_grading}
\end{sidewaystable}
\begin{figure}[t]
\begin{center}
\includegraphics[width=13cm]{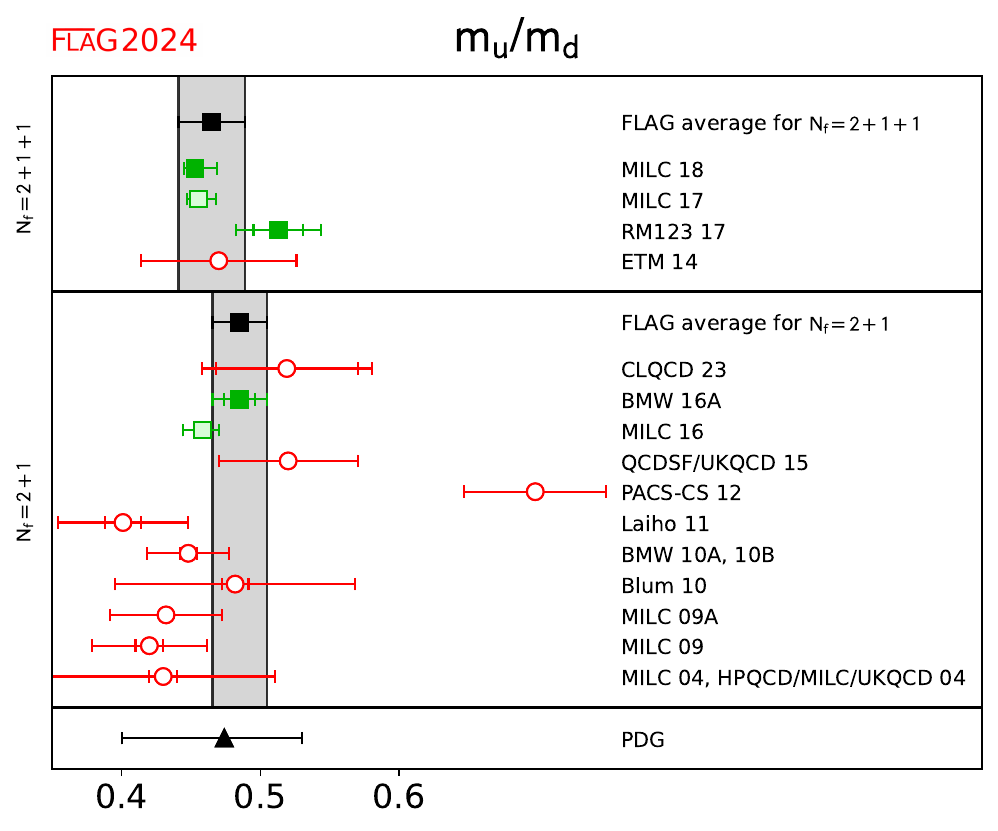}
\end{center}
\caption{\label{fig:mu over md}Lattice results and FLAG averages at $\Nf = 2+1$ and
$2+1+1$ for the up-down-quark masses ratio $m_u/m_d$, together with the current PDG
estimate.}
\end{figure}
Regarding results already presented in previous FLAG editions, we start by reviewing
predictions for the $\Nf=2+1$ sector. MILC 09A~\cite{Bazavov:2009fk} uses the mass
difference between $K^0$ and $K^+$, from which they subtract electromagnetic effects using
Dashen's theorem with corrections, as discussed in the introduction of this section.  The
up  and down sea quarks remain degenerate in their calculation, fixed to the value of
$m_{ud}$ obtained from $M_{\pi^0}$. To determine $m_u/m_d$, BMW 10A,
10B~\cite{Durr:2010vn,Durr:2010aw} follow a slightly different strategy. They obtain this
ratio from their result for $m_s/m_{ud}$ combined with a phenomenological determination of
the isospin-breaking quark-mass ratio $Q=22.3(8)$, from $\eta\to3\pi$
decays~\cite{Leutwyler:2009jg} (the decay $\eta\to3\pi$ is very sensitive to QCD isospin
breaking, but fairly insensitive to QED isospin breaking). Instead of subtracting
electromagnetic effects using phenomenology, RBC~07~\cite{Blum:2007cy} and
Blum~10~\cite{Blum:2010ym} actually include a quenched electromagnetic field in their
calculation. This means that their results include corrections to Dashen's theorem, albeit
only in the presence of quenched electromagnetism. Since the up and down quarks in the sea
are treated as degenerate, very small isospin corrections are neglected, as in MILC's
calculation. PACS-CS 12~\cite{Aoki:2012st} takes the inclusion of isospin-breaking effects
one step further. Using reweighting techniques, it also includes electromagnetic and
$m_u-m_d$ effects in the sea. However, they  do not correct for the large finite-volume
effects coming from electromagnetism in their $M_{\pi}L\sim 2$ simulations, but provide
rough estimates for their size, based on Ref.~\citep{Hayakawa:2008an}.
QCDSF/UKQCD~15~\cite{Horsley:2015eaa} uses QCD+QED dynamical simulations performed at the
SU(3)-flavour-symmetric point, but at a single lattice spacing, so they do not enter our
average. The smallest partially quenched ($m_{\rm sea}\neq m_{\rm val}$) pion mass is
greater than 200 MeV, so our chiral-extrapolation criteria require a $\soso$ rating.
Concerning finite-volume effects, this work uses three spatial extents $L$ of
$1.6~\mathrm{fm}$, $2.2~\mathrm{fm}$, and $3.3~\mathrm{fm}$. QCDSF/UKQCD~15 claims that
the volume dependence is not visible on the two largest volumes, leading them to assume
that finite-size effects are under control. As a consequence of that, the final result for
quark masses does not feature a finite-volume extrapolation or an estimation of the
finite-volume uncertainty. However, in their work on the QED corrections to the hadron
spectrum~\citep{Horsley:2015eaa} based on the same ensembles, a volume study shows some
level of compatibility with the $\mathrm{QED}_{\mathrm{L}}$ finite-volume effects derived
in~\citep{Davoudi:2014qua}. We see two issues here. First, the analytical result quoted
from~\citep{Davoudi:2014qua} predicts large, $\cO(10\%)$ finite-size effects from QED on
the meson masses at the values of $M_{\pi}L$ considered in QCDSF/UKQCD~15, which is
inconsistent with the statement made in the paper. Second, it is not known that the
zero-mode regularization scheme used here has the same volume scaling as
$\mathrm{QED}_{\mathrm{L}}$. We therefore chose to assign the \bad~rating for finite
volume to QCDSF/UKQCD~15. BMW~16A~\citep{Fodor:2016bgu} reuses the data set produced from
their determination of the light-baryon octet-mass splittings~\citep{Budapest-Marseille-Wuppertal:2013rtp}
using electro-quenched QCD+$\mathrm{QED}_{\mathrm{TL}}$ smeared clover-fermion
simulations. Finally, MILC~16~\citep{Basak:2016jnn}, which is a preliminary result for the
value of $\epsilon$ published in MILC~18~\citep{Basak:2018yzz}, also provides a $\Nf=2+1$
computation of the ratio $m_u/m_d$.

We now describe the $\Nf=2+1+1$ calculations. ETM 14~\cite{Carrasco:2014cwa} uses
simulations in pure QCD, but determines $m_u-m_d$ from the slope $\partial M_K^2/\partial
m_{ud}$ and the physical value for the QCD kaon-mass splitting taken from the
phenomenological estimate in FLAG 13. In the $\Nf=2+1+1$ sector,
MILC~18~\citep{Basak:2018yzz} computed $\epsilon$ using $\Nf=2+1$ asqtad electro-quenched
QCD+$\mathrm{QED}_{\mathrm{TL}}$ simulations and extracted the ratio $m_u/m_d$ from a new
set of $\Nf=2+1+1$ HISQ QCD simulations. Although $\epsilon$ comes from $\Nf=2+1$
simulations, $(\Delta M_{K}^{2})^\text{SU(2)}$, which is about three times larger than $(\Delta
M_{K}^{2})^{\gamma}$, has been determined in the $\Nf=2+1+1$ theory. We therefore chose to
classify this result as a four-flavour one. This result is explicitly described by the
authors as an update of MILC~17~\citep{Bazavov:2017lyh}. In
MILC~17~\citep{Bazavov:2017lyh}, $m_u/m_d$ is determined as a side-product of a global
analysis of heavy-meson decay constants, using a preliminary version of $\epsilon$ from
MILC~18~\citep{Basak:2018yzz}. In FNAL/MILC/TUMQCD~18~\citep{Bazavov:2018omf} the ratio
$m_u/m_d$ from MILC~17~\citep{Bazavov:2017lyh} is used to determine the individual masses
$m_u$ and $m_d$ from a new calculation of $m_{ud}$. The work
RM123~17~\citep{Giusti:2017dmp} is the continuation of the $\Nf=2$ work named
RM123~13~\citep{deDivitiis:2013xla} in the previous edition of FLAG. This group now uses
$\Nf=2+1+1$ ensembles from ETM~10~\citep{Baron:2010bv}, however, still with a rather large
minimum pion mass of $270~\mathrm{MeV}$, leading to the \soso~rating for chiral
extrapolations. 

Lattice results for $m_u$, $m_d$ and $m_u/m_d$ are summarized in
Tab.~\ref{tab:mu_md_grading}. 
The colour coding is specified in detail in Sec.~\ref{sec:color-code}. Considering the
important progress in the last years on including isospin-breaking effects in lattice
simulations, we are now in a position where averages for $m_u$ and $m_d$ can be made
without the need of phenomenological inputs. Therefore, lattice calculations of the
individual quark masses using phenomenological inputs for isospin-breaking effects will be
coded \bad.

We begin with $\Nf=2+1$ (for $\Nf=2$ see the 2021 edition). The only result that qualifies to enter the FLAG average is BMW~16A~\citep{Fodor:2016bgu},
%
\begin{align}
&&\FLAGAVBEGIN m_u &=2.27(9)\FLAGAVEND\,\mev&\Ref~\mbox{\cite{Fodor:2016bgu}}\,, \nonumber\\
\label{eq:mumd} \hspace{0cm}\Nf = 2+1:\hspace{0.2cm}
&&\FLAGAVBEGIN m_d &= 4.67(9) \FLAGAVEND\,\mev&\Ref~\mbox{\cite{Fodor:2016bgu}}\,,\\
&&\FLAGAVBEGIN {m_u}/{m_d} &= 0.485(19)\FLAGAVEND&\Ref~\mbox{\cite{Fodor:2016bgu}}\,,\nonumber
\end{align}
%
with errors of roughly 4\%, 2\% and 4\%, respectively. 
These numbers result in the following RGI averages
\begin{align}
&& M_u^{\rm RGI} &=3.15(12)_m(4)_\Lambda \,\mev&\Ref~\mbox{\cite{Fodor:2016bgu}}\,, \nonumber\\
\label{eq:mumd rgi} \hspace{0cm}\Nf = 2+1:\hspace{0.2cm}\\[-5mm]
&& M_d^{\rm RGI} &= 6.49(12)_m(7)_\Lambda \,\mev&\Ref~\mbox{\cite{Fodor:2016bgu}}\,.\nonumber
\end{align}

Finally, for $\Nf=2+1+1$, RM123~17~\citep{Giusti:2017dmp} and
FNAL/MILC/TUMQCD~18~\citep{Bazavov:2018omf} enter the average for the individual $m_u$ and
$m_d$ masses, and RM123~17~\citep{Giusti:2017dmp} and MILC~18~\citep{Basak:2018yzz} enter
the average for the ratio $m_u/m_d$, giving
%
\begin{align}
	&&\FLAGAVBEGIN m_u &=2.14(8)\FLAGAVEND \,\mev&\Refs~\mbox{\cite{Giusti:2017dmp,Bazavov:2018omf}}\,,\nonumber\\
\label{eq:mumd 4 flavour} \Nf = 2+1+1:\hspace{0.15cm}
        &&\FLAGAVBEGIN m_d &= 4.70(5)\FLAGAVEND \,\mev&\Refs~\mbox{\cite{Giusti:2017dmp,Bazavov:2018omf}}\,,\\
	&&\FLAGAVBEGIN {m_u}/{m_d} &= 0.465(24)\FLAGAVEND&\Refs~\mbox{\cite{Giusti:2017dmp,Basak:2018yzz}}\,,\nonumber
\end{align}
%
with errors of roughly 4\%, 1\% and 5\%, respectively. One can observe some marginal
discrepancies between results coming from the MILC collaboration and
RM123~17~\citep{Giusti:2017dmp}. More specifically, adding all sources of uncertainties in
quadrature, one obtains a 1.7$\sigma$ discrepancy between RM123~17~\citep{Giusti:2017dmp}
and MILC~18~\cite{Basak:2018yzz} for $m_u/m_d$, and a 2.2$\sigma$ discrepancy between
RM123~17~\citep{Giusti:2017dmp} and FNAL/MILC/TUMQCD~18~\citep{Bazavov:2018omf} for $m_u$.
However, the values of $m_d$ and $\epsilon$ are in very good agreement between the two
groups. These discrepancies are presently too weak to constitute evidence for concern, and
will be monitored as more lattice groups provide results for these quantities. The RGI
averages for $m_u$ and $m_d$ are

\begin{align}
	&& M_u^{\rm RGI} &=2.97(11)_m(3)_\Lambda \,\mev&\Refs~\mbox{\cite{Giusti:2017dmp,Bazavov:2018omf}}\,,\nonumber\\
\label{eq:mumd 4 flavour rgi} \Nf = 2+1+1:\hspace{0.15cm}\\[-5mm]
        && M_d^{\rm RGI} &= 6.53(7)_m(8)_\Lambda \,\mev&\Refs~\mbox{\cite{Giusti:2017dmp,Bazavov:2018omf}}\,.\nonumber
\end{align}

Every result for $m_u$ and $m_d$ used here to produce the FLAG averages relies on
electro-quenched calculations, so there is some interest to comment on the size of
quenching effects. Considering phenomenology and the lattice results presented here, it is
reasonable for a rough estimate to use the value $(\Delta M_{K}^{2})^{\gamma}\sim
2000~\mathrm{MeV}^2$ for the QED part of the kaon-mass splitting. Using the arguments
presented in Sec.~\ref{app:qed}, one can assume that the QED sea contribution
represents $\cO(10\%)$ of $(\Delta M_{K}^{2})^{\gamma}$. Using SU(3)
PQ$\chi$PT+QED~\citep{Bijnens:2006mk,Portelli:2012pn} gives a $\sim 5\%$ effect. Keeping
the more conservative $10\%$ estimate and using the experimental value of the kaon-mass
splitting, one finds that the QCD kaon-mass splitting $(\Delta M_{K}^{2})^\text{SU(2)}$ suffers
from a reduced $3\%$ quenching uncertainty. Considering that this splitting is
proportional to $m_u-m_d$ at leading order in SU(3) $\chi$PT, we can estimate that a
similar error will propagate to the quark masses. So the individual up and down masses
look mildly affected by QED quenching. However, one notices that $\sim 3\%$ is the level
of error in the new FLAG averages, and increasing significantly this accuracy will require
using fully dynamical calculations. 

In view of the fact that a {\it massless up quark} would solve the strong CP problem, many
authors have considered this an attractive possibility, but the results presented above
exclude this possibility: the value of $m_u$ in Eq.~(\ref{eq:mumd}) differs from zero by
$26$ standard deviations. We conclude that nature solves the strong CP problem
differently.

Finally, we conclude this section by giving the FLAG averages for $\epsilon$ defined in
Eq.~(\ref{eq:epsdef}). For $\Nf=2+1+1$, we average the results of
RM123~17~\citep{Giusti:2017dmp} and MILC~18~\cite{Basak:2018yzz} with the value of
$(\Delta M_{K}^{2})^{\gamma}$ from BMW~14~\citep{Borsanyi:2014jba} combined with
Eq.~(\ref{eq:epslo}), giving
\begin{align}
\label{eq:epsilon 4 flavour} \Nf = 2+1+1:\hspace{0.15cm}\\[-5mm]
        && \epsilon &= 0.79(6) & \Refs~\mbox{\cite{Giusti:2017dmp,Basak:2018yzz,Borsanyi:2014jba}}\,.\nonumber
\end{align}

Although BMW~14~\citep{Borsanyi:2014jba} focuses on hadron masses and did not extract the
light-quark masses, they are the only fully unquenched QCD+QED calculation to date that
qualifies to enter a FLAG average. With the exception of renormalization, which is not
discussed in the paper, that work has a \good~rating for every FLAG criterion considered
for the $m_u$ and $m_d$ quark masses. For $\Nf=2+1$ we use the results from
BMW~16A~\citep{Fodor:2016bgu}, 
\begin{align}
\label{eq:epsilon 3 flavour} \Nf = 2+1:\hspace{0.15cm}\\[-5mm]
        && \epsilon &= 0.73(17) & \Ref~\mbox{\cite{Fodor:2016bgu}}\,.\nonumber
\end{align}

It is important to notice that the $\epsilon$ uncertainties from BMW~16A and RM123~17 are
dominated by estimates of the QED quenching effects. Indeed, in contrast with the quark
masses, $\epsilon$ is expected to be rather sensitive to the sea-quark QED contributions.
Using the arguments presented in Sec.~\ref{app:qed}, if one conservatively assumes
that the QED sea contributions represent $\cO(10\%)$ of $(\Delta M_{K}^{2})^{\gamma}$,
then Eq.~(\ref{eq:epslo}) implies that $\epsilon$ will have a quenching error of $\sim
0.15$ for $(\Delta M_{K}^{2})^{\gamma}\sim (45~\mathrm{MeV})^2$, representing a large
$\sim 20\%$ relative error. It is interesting to observe that such a discrepancy does not
appear between BMW~14 and RM123~17, although the $\sim 10\%$ accuracy of both results
might not be sufficient to resolve these effects. On the other hand, in the context of
SU(3) chiral perturbation theory, Bijnens and Danielsson~\cite{Bijnens:2006mk} show that
the QED quenching effects on $\epsilon$ do not depend on unknown LECs at NLO in the chiral expansion and are
therefore computable at that order.  In that approach, MILC 18 finds the effect at NLO to
be only 5\%.  
To conclude, although the controversy around the value of $\epsilon$ has been
significantly reduced by lattice-QCD+QED determinations, computing this at few-percent
accuracy requires simulations with charged sea quarks.

\subsubsection{Estimates for $R$ and $Q$}\label{sec:RandQ} The quark-mass ratios
\be\label{eq:Qm} R\equiv \frac{m_s-m_{ud}}{m_d-m_u}\hspace{0.5cm}
\mbox{and}\hspace{0.5cm}Q^2\equiv\frac{m_s^2-m_{ud}^2}{m_d^2-m_u^2}
\ee
compare SU(3) breaking  with isospin breaking. Both numbers only depend on the ratios
$m_s/m_{ud}$ and $m_u/m_d$,
\be
R=\frac{1}{2}\left(\frac{m_s}{m_{ud}}-1\right)\frac{1+\frac{m_u}{m_d}}{1-\frac{m_u}{m_d}}
\qquad\text{and}\qquad Q^2=\frac{1}{2}\left(\frac{m_s}{m_{ud}}+1\right)R\,.
\ee
The quantity $Q$ is of particular interest because of a low-energy
theorem~\cite{Gasser:1984pr}, which relates it to a ratio of meson masses,  
\begin{equation}\label{eq:QM}
 Q^2_M\equiv \frac{\hat{M}_K^2}{\hat{M}_\pi^2}\frac{\hat{M}_K^2-\hat{M}_\pi^2}{\hat{M}_{K^0}^2-
   \hat{M}_{K^+}^2}\co\hspace{1cm}\hat{M}^2_\pi\equiv\mbox{$\frac{1}{2}$}( \hat{M}^2_{\pi^+}+ \hat{M}^2_{\pi^0})
 \co\hspace{0.5cm}\hat{M}^2_K\equiv\mbox{$\frac{1}{2}$}( \hat{M}^2_{K^+}+ \hat{M}^2_{K^0})\fs\end{equation}
(We remind the reader that the $\,\hat{ }\,$ denotes a quantity evaluated in the
$\alpha\to 0$ limit.) Chiral symmetry implies that the expansion of $Q_M^2$ in powers of
the quark masses (i) starts with $Q^2$ and (ii) does not receive any contributions at NLO~\cite{Gasser:1984pr}:
\be\label{eq:LET Q}Q_M\NLo Q \fs\ee
%
%
For $\Nf=2+1$, we use Eqs.~(\ref{eq:msovmud3}) and (\ref{eq:mumd}) and obtain
\be\label{eq:RQres} R=38.1(1.5)\co\hspace{2cm}Q=23.3(0.5)\ ,\ee 
and for $\Nf=2+1+1$,
\be\label{eq:RQresNf4} R=35.9(1.7)\co\hspace{2cm}Q=22.5(0.5)\ ,\ee
which are quite compatible (see the 2021 edition for the two flavour numbers which are also compatible with the above). It is interesting to
note that the most recent phenomenological determination of $R$ and $Q$ from $\eta\to
3\pi$ decay~\citep{Colangelo:2018jxw} gives the values $R=34.4(2.1)$ and $Q=22.1(0.7)$,
which are consistent with the averages presented here. The authors
of Refs.~\citep{Amoros:2001cp,Colangelo:2018jxw} point out that this discrepancy is likely due
to surprisingly large corrections to the approximation in Eq.~(\ref{eq:LET Q}) used in the
phenomenological analysis.

Our final results for the masses $m_u$, $m_d$, $m_{ud}$, $m_s$ and the mass ratios
$m_u/m_d$, $m_s/m_{ud}$, $R$, $Q$ are collected in Tabs.~\ref{tab:mudms} and
\ref{tab:mumdRQ}.

\begin{table}[!thb]\vspace{0.5cm}
{
\begin{tabular*}{\textwidth}[l]{@{\extracolsep{\fill}}cccc}
\hline\hline
$\Nf$ & $m_{ud}$ & $ m_s $ & $m_s/m_{ud}$ \\ 
&&& \\[-2ex]
\hline\rule[-0.1cm]{0cm}{0.5cm}
&&& \\[-2ex]
2+1+1 & 3.410(43) & 93.44(68) & 27.23(10)\\ 
&&& \\[-2ex]
\hline\rule[-0.1cm]{0cm}{0.5cm}
&&& \\[-2ex]
2+1 & 3.364(41) & 92.03(88) & 27.42(12)\\ 
&&& \\[-2ex]
\hline
\hline
\end{tabular*}
\caption{\label{tab:mudms} Our estimates for the average
  up-down-quark mass and the strange-quark mass in the $\msbar$ scheme at running scale
  $\mu=2\,\gev$. Mass values are given in MeV. In the
  results presented here, the error is the one which we obtain
  by applying the averaging procedure of Sec.~\ref{sec:error_analysis} to the
  relevant lattice results. 
  } }
\end{table}

\begin{table}[!thb]
{
\begin{tabular*}{\textwidth}[l]{@{\extracolsep{\fill}}cccccc}
\hline\hline
$\Nf$ & $m_u  $ & $m_d $ & $m_u/m_d$ & $R$ & $Q$\\ 
&&&&& \\[-2ex]
\hline\rule[-0.1cm]{0cm}{0.5cm}
&&&&& \\[-2ex]
2+1+1 & 2.14(8) & 4.70(5) & 0.465(24) & 35.9(1.7) & 22.5(0.5) \\ 
&&&&& \\[-2ex]
\hline\rule[-0.1cm]{0cm}{0.5cm}
&&&&& \\[-2ex]
2+1 & 2.27(9) & 4.67(9) & 0.485(19) & 38.1(1.5) & 23.3(0.5) \\ 
&&&&& \\[-2ex]
\hline
\hline
\end{tabular*}
\caption{\label{tab:mumdRQ} Our estimates for the masses of the
  two lightest quarks and related, strong isospin-breaking
  ratios. Again, the masses refer to the $\msbar$ scheme  at running
  scale $\mu=2\,\gev$. Mass values are given
  in MeV.}  }
\end{table}
\clearpage

%% file: qmass/qmass_mc.tex
\subsection{Charm-quark mass}
\label{s:cmass}

In the following, we collect and discuss the lattice determinations of the $\overline{\rm
MS}$ charm-quark mass $\overline{m}_c$. Most of the results have been obtained by
analyzing the lattice-QCD simulations of two-point heavy-light- or heavy-heavy-meson
correlation functions, using as input the experimental values of the $D$, $D_s$, and
charmonium mesons. Some groups use the moments method. The latter is based on the lattice
calculation of the Euclidean time moments of pseudoscalar-pseudoscalar correlators for
heavy-quark currents followed by an OPE expansion dominated by perturbative QCD effects,
which provides the determination of both the heavy-quark mass and the strong-coupling
constant $\alpha_s$.

The heavy-quark actions adopted by various lattice collaborations have been discussed in
previous FLAG reviews~\cite{Aoki:2013ldr,Aoki:2016frl,FlavourLatticeAveragingGroup:2019iem}, and their
descriptions can be found in Sec.~A.1.3 of FLAG 19 \cite{FlavourLatticeAveragingGroup:2019iem}. While the charm
mass determined with the moments method does not need any lattice evaluation of the
mass-renormalization constant $Z_m$, the extraction of $\overline{m}_c$  from two-point
heavy-meson correlators does require the nonperturbative calculation of $Z_m$. The lattice
scale at which $Z_m$ is obtained is usually at least of the order 2--3 GeV, and
therefore it is natural in this review to provide the values of $\overline{m}_c(\mu)$ at
the renormalization scale $\mu = 3~\gev$. Since the choice of a renormalization scale
equal to $\overline{m}_c$ is still commonly adopted (as by the PDG~\cite{Zyla:2020zbs}),
we have collected in Tab.~\ref{tab:mc} the lattice results for both
$\overline{m}_c(\overline{m}_c)$ and $\overline{m}_c(\mbox{3 GeV})$, obtained  for $\Nf
=2+1$ and $2+1+1$. For $\Nf=2$, interested readers are referred to previous
reviews~\cite{Aoki:2013ldr,Aoki:2016frl}.

When not directly available in the published work, we apply a conversion factor using
perturbative QCD evolution at five loops to run down from $\mu = 3$ GeV to the scales
$\mu = \overline{m}_c$  and 2 GeV of $0.7739(60)$ and 0.9026(23), respectively, where
the error comes from the uncertainty in $\Lambda_{\rm QCD}$.  We use $\Lambda_{\rm QCD}
= 297(12)$ MeV for $\Nf = 4$ (see Sec.~\ref{sec:alpha_s}). Perturbation theory
uncertainties, estimated as the difference between results that use 4- and 5-loop
running, are significantly smaller than the parametric uncertainty coming from
$\Lambda_{\rm QCD}$. For $\mu = \overline{m}_c$, the former is about about 2.5 times
smaller. 
\input{qmass/qmass_tab_mc}

In the next subsections, we review separately the results for $\overline{m}_c$ with three
or four flavours of quarks in the sea.

\subsubsection{$\Nf = 2+1$ results}
\label{sec:mcnf3}

Since the last review~\cite{FlavourLatticeAveragingGroupFLAG:2021npn}, there is one new
result: ALPHA 23~\cite{Bussone:2023kag}. This work uses a subset of CLS ensembles,
based on simulations of nonperturbatively $O(a)$-improved Wilson fermions. The difference with ALPHA 21 is that the valence sector uses both Wilson and twisted-mass discretizations instead of just Wilson. Renormalization is based on
previous work by the ALPHA collaboration, and is performed nonperturbatively from 100 MeV
to the electroweak scale. The subset of ensembles used have large volumes, four lattice
spacings, and reach pion masses of 200 MeV, which guarantees entering in the average.
Contrary to the extraction of light-quark masses in ALPHA 19, the chiral extrapolation
does not dominate the error budget, and being less critical in this case we decide to give
a $\good$ for the chiral extrapolation.  
The data-driven criteria quantity for the continuum extrapolation $\delta(a_{\rm min})$ (see~\ref{sec:DataDriven}) is
smaller than 3 in all cases.

Petreczky 19 employs the HISQ action on ten ensembles with ten lattice spacings down to
0.025 fm, physical strange-quark mass, and two light-quark masses, the lightest
corresponding to 161 MeV pions. Their study incorporates lattices with 11 different sizes,
ranging from 1.6 to 5.4 fm. The masses are computed from moments of pseudoscalar
quarkonium correlation functions, and $\overline{\rm MS}$ masses are extracted with 4-loop
continuum perturbation theory. Thus, that work easily rates green stars in all categories.
Continuum extrapolations are challenging, but judging the data itself the values of
$\delta(a_{\rm min})$ are not very large. It is just that the functional form of the data
is complicated.

ALPHA 21 uses the $\cO(a)$-improved Wilson-clover action with five lattice spacings from
0.087 to 0.039 fm, produced by the CLS collaboration. For each lattice spacing, several
light sea-quark masses are used in a global chiral-continuum extrapolation (the lightest
pion mass for one ensemble is 198 MeV). The authors also use nonperturbative
renormalization and running through application of step-scaling and the Schr\"odinger
functional scheme. Finite-volume effects are investigated at one lattice spacing and only
for $\sim 400$ MeV pions on the smallest two volumes where results are compatible within
statistical errors. ALPHA 21 satisfies the FLAG criteria for green-star ratings in all of
the categories listed in Tab.~\ref{tab:mc}. The values of $\delta(a_{\rm min})$ are
smaller than 3 in all continuum extrapolations. Descriptions of the other works in this
section can be found in an earlier review~\cite{FlavourLatticeAveragingGroup:2019iem}.

According to our rules on the publication status, the FLAG average for the charm-quark
mass at $\Nf = 2+1$ is obtained by combining the results HPQCD 10, $\chi$QCD 14, JLQCD 16,
Petreczky 19, ALPHA 21 and ALPHA 23,
\begin{align}
      \label{eq:mcmcnf3} 
&& \overline{m}_c(\overline{m}_c)         & = 1.278(6) ~ \gev          &&\Refs~\mbox{\cite{Bussone:2023kag,McNeile:2010ji,Yang:2014sea,Nakayama:2016atf,Petreczky:2019ozv,Heitger:2021apz}}\,, \\[-3mm]
&\mbox{$\Nf = 2+1$:}&\nonumber\\[-3mm]
&&\FLAGAVBEGIN\overline{m}_c(\mbox{3 GeV})& = 0.991(6)\FLAGAVEND ~ \gev&&\Refs~\mbox{\cite{Bussone:2023kag,McNeile:2010ji,Yang:2014sea,Nakayama:2016atf,Petreczky:2019ozv,Heitger:2021apz}}\,,
\end{align}
This result corresponds to the following RGI average
\begin{align}
      \label{eq:mcmcnf3 rgi} 
&& M_c^{\rm RGI} & = 1.526(7)_m(21)_\Lambda ~ \gev&&\Refs~\mbox{\cite{McNeile:2010ji,Yang:2014sea,Nakayama:2016atf,Petreczky:2019ozv,Heitger:2021apz}}\,.
\end{align}

\subsubsection{$\Nf = 2+1+1$ results}
\label{sec:mcnf4}

For a discussion of older results, see the previous FLAG reviews. Since FLAG 19 two groups
have produced updated values with charm quarks in the sea. 

HPQCD 20A~\cite{Hatton:2020qhk} is an update of HPQCD 18, including a new finer ensemble
($a\approx 0.045$ fm) and EM corrections computed in the quenched approximation of QED for
the first time. Besides these new items, the analysis is largely unchanged from HPQCD 18
except for an added $\alpha_s^3$ correction to the SMOM-to-$\overline{\rm MS}$ conversion
factor and tuning the bare charm mass via the $J/\psi$ mass rather than the $\eta_c$.
Their new value in pure QCD is $\overline{m}_c(3~\rm GeV)=0.9858(51)$ GeV which is quite
consistent with HPQCD 18 and the FLAG 19 average. The effects of quenched QED in both the
bare charm-quark mass and the renormalization constant are small. Both effects are
precisely determined, and the overall effect shifts the mass down slightly to
$\overline{m}_c(3~\rm GeV)=0.9841(51)$ where the uncertainty due to QED is invisible in
the final error. The shift from their pure QCD value due to quenched QED is about
$-0.2\%$. 

ETM 21A~\cite{Alexandrou:2021gqw} is a new  work that follows a similar
methodology as ETM 14, but with significant improvements. Notably, a clover-term is added
to the twisted mass fermion action which suppresses $\cO(a^2)$ effects between the neutral
and charged pions. Additional improvements include new ensembles lying very close to the
physical mass point, better control of nonperturbative renormalization systematics, and
use of both meson and baryon correlation functions to determine the quark mass. They use
the RI-MOM scheme for nonperturbative renormalization. The analysis comprises ten
ensembles in total with three lattice spacings (0.095, 0.082, and 0.069 fm), two volumes
for the finest lattice spacings and four for the other two, and pion masses down to 134
MeV for the finest ensemble. The values of $m_\pi L$ range mostly from almost four to
greater than five. According to the FLAG criteria, green stars are earned in all
categories. The authors find $m_c(3~\rm GeV)=1.036(17)(^{+15}_{-8})$ GeV. In
Tab.~\ref{tab:mc} we have applied a factor of 0.7739(60) to run from 3 GeV to
$\overline{m}_c$. As in FLAG 19, the new value is consistent with ETM 14 and ETM 14A, but
is still high compared to the FLAG average. The authors plan future improvements,
including a finer lattice spacing for better control of the continuum limit and a new
renormalization scheme, like RI-SMOM. 

Six results enter the FLAG average for $\Nf = 2+1+1$ quark flavours: ETM 14, ETM 14A,
HPQCD 14A, FNAL/MILC/TUMQCD 18, HPQCD 20A, and ETM 21A. We note that while the ETM
determinations of $\overline{m}_c$ agree well with each other, they are incompatible with
HPQCD 14A, FNAL/MILC/TUMQCD 18, and HPQCD 20A by several standard deviations. While the
ETM 14 and ETM 14A use the same configurations, the analyses are quite different and
independent, and ETM 21A is a new result on new ensembles with improved methodology. As
mentioned earlier, $m_{ud}$ and $m_s$ values by ETM are also systematically high compared
to their respective averages. Combining all six results yields yields

 \begin{align}
      \label{eq:mcmcnf4} 
&& \overline{m}_c(\overline{m}_c)           & = 1.280(13)~\gev          &&\Refs~\mbox{\cite{Carrasco:2014cwa,Chakraborty:2014aca,Alexandrou:2014sha,Bazavov:2018omf,Hatton:2020qhk,Alexandrou:2021gqw}}\,, \\[-3mm]
&\mbox{$\Nf = 2+1+1$:}& \nonumber\\[-3mm]
&&  \FLAGAVBEGIN\overline{m}_c(\mbox{3 GeV})& = 0.989(10)\FLAGAVEND ~ \gev&&\Refs~\mbox{\cite{Carrasco:2014cwa,Chakraborty:2014aca,Alexandrou:2014sha,Bazavov:2018omf,Hatton:2020qhk,Alexandrou:2021gqw}}\,,
 \end{align}
where the errors include large stretching factors $\sqrt{\chi^2/\mbox{dof}}\approx2.0$ and
$2.4$, respectively. We have assumed 100\% correlation for statistical errors between ETM
14 and ETM 14A results and the same for HPQCD 14A, HPQCD 20A, and FNAL/MILC/TUMQCD 18. 

\begin{figure}[!htb]
\begin{center}
\includegraphics[width=11.5cm]{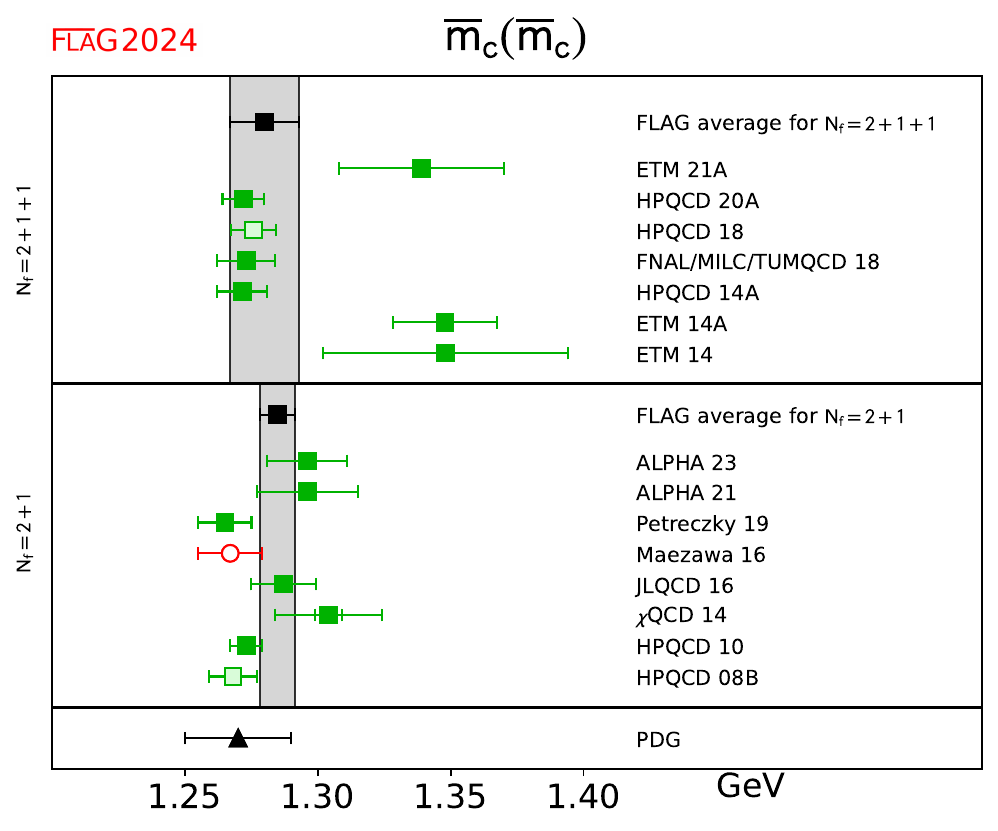}
\end{center}
\vspace{-1cm}
\caption{\label{fig:mc} The charm-quark mass for $2+1$ and $2+1+1$ flavours. For the
latter a large stretching factor is used for the FLAG average due to poor $\chi^2$ from
our fit.}
\end{figure}

These are obviously poor $\chi^2$ values, and the stretching factors are quite large.
While it may be prudent in such a case to quote a range of values covering the central
values of all results that pass the quality criteria, we believe in this case that would
obscure rather than clarify the situation.  From Fig.~\ref{fig:mc}, we note that not only
do ETM 21A, ETM 14A, and ETM 14 lie well above the other 2+1+1 results, but also above all
of the 2+1 flavour results. A similar trend is apparent for the light-quark masses (see
Figs.~\ref{fig:ms} and \ref{fig:mud}) while for mass ratios there is better agreement
(Figs.~\ref{fig:msovmud},~\ref{fig:mu over md}, and \ref{fig:mc over ms}). The latter
suggests there may be underestimated systematic uncertainties associated with scale
setting and/or renormalization which have not been detected. Finally we note the ETM
results are significantly higher than the PDG average. For these reasons, which admittedly
are not entirely satisfactory, we continue to quote an average with a stretching factor as
in previous reviews.

The RGI average reads as follows,
 \begin{align}
      \label{eq:mcmcnf4 rgi} 
&&  M_c^{\rm RGI}& = 1.528(15)_m(21)_\Lambda  ~ \gev&&\Refs~\mbox{\cite{Alexandrou:2014sha,Chakraborty:2014aca,Bazavov:2018omf,Carrasco:2014cwa,Hatton:2020qhk,Alexandrou:2021gqw}}\,.
 \end{align}

Figure~\ref{fig:mc} presents the values of $\overline{m}_c(\overline{m}_c)$ given in
Tab.~\ref{tab:mc} along with the FLAG averages obtained for $2+1$ and $2+1+1$ flavours.

\subsubsection{Lattice determinations of the ratio $m_c/m_s$}
\label{sec:mcoverms}
Because some of the results for quark masses given in this review are obtained via the
quark-mass ratio $m_c/m_s$, we review these lattice calculations, which are listed in
Tab.~\ref{tab:mcoverms}, as well.

\input{qmass/qmass_tab_mcoverms}

The $\Nf = 2+1$ results from $\chi$QCD 14 and HPQCD 09A~\cite{Davies:2009ih} are from the
same calculations that were described for the charm-quark mass in the previous review.
Maezawa 16 does not pass our chiral-limit test (see the previous review), though we note
that it is quite consistent with the other values. Combining $\chi$QCD 14 and HPQCD 09A,
we obtain the same result reported in FLAG 19, 
 \be
      \label{eq:mcmsnf3} 
      \mbox{$\Nf = 2+1$:} \qquad\FLAGAVBEGIN m_c / m_s = 11.82(16)\FLAGAVEND\qquad\Refs~\mbox{\cite{Yang:2014sea,Davies:2009ih}},
 \ee
with a $\chi^2/\mbox{dof} \simeq 0.85$.

Turning to $\Nf = 2+1+1$, there is a new result from ETM 21A (see the previous section for
details). The errors have actually increased compared to ETM 14, due to larger
uncertainties in the baryon sector which enter their average with the meson sector. 
See the earlier reviews for a discussion of previous results.

We note that some tension exists between the HPQCD 14A and FNAL/MILC/TUMQCD results.
Combining these with ETM 14 and ETM 21A yields

 \be
      \label{eq:mcmsnf4} 
      \mbox{$\Nf = 2+1+1$:} \qquad \FLAGAVBEGIN m_c / m_s = 11.766(30)\FLAGAVEND\qquad\Refs~\mbox{\cite{Chakraborty:2014aca,Carrasco:2014cwa,Bazavov:2018omf,Alexandrou:2021gqw}},
 \ee
where the error includes the stretching factor $\sqrt{\chi^2/\mbox{dof}} \simeq 1.4$. We
have assumed a 100\% correlation of statistical errors for FNAL/MILC/TUMQCD 18 and HPQCD
14A.

\begin{figure}[!htb]
\begin{center}
\includegraphics[width=11cm]{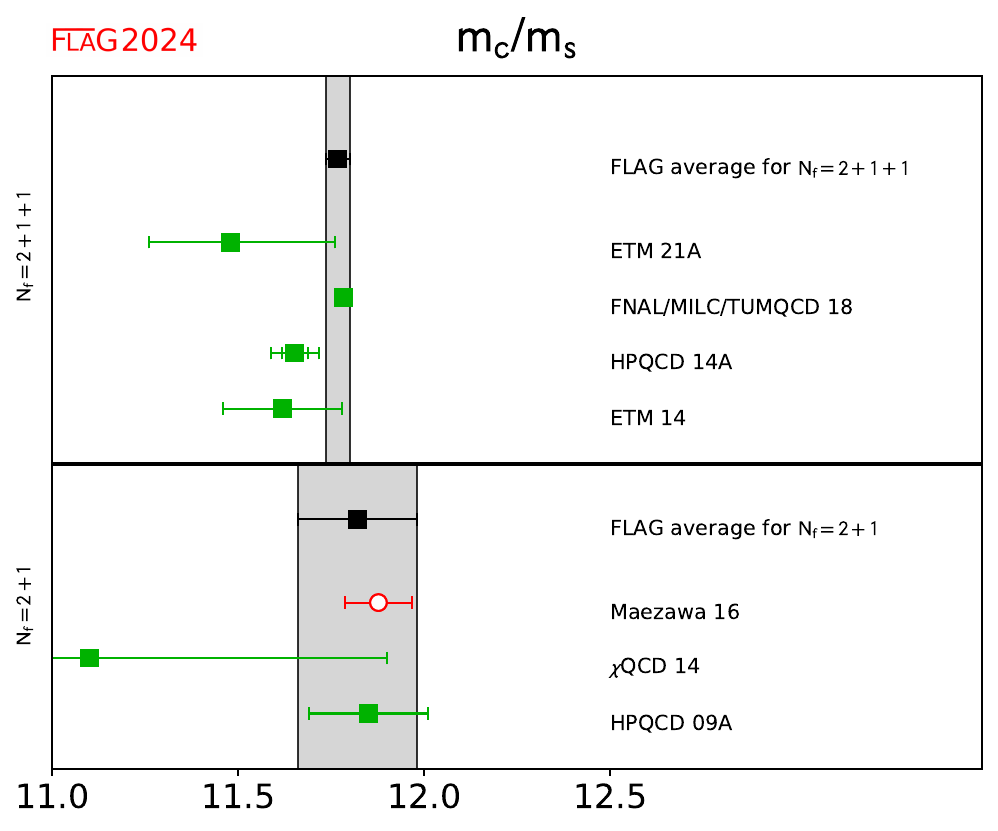}
\end{center}
\begin{center}
\caption{\label{fig:mc over ms}Lattice results for the ratio $m_c / m_s$ listed in
Tab.~\ref{tab:mcoverms} and the FLAG averages corresponding to $2+1$ and $2+1+1$ quark
flavours. The latter average includes a stretching factor of 1.4 on the error due a poor
$\chi^2$ from our fit.}
\end{center}
\end{figure}

Results for $m_c/m_s$ are shown in Fig.~\ref{fig:mc over ms} together with the FLAG
averages for $\Nf = 2+1$ and $2+1+1$ flavours. 

%% file: qmass/qmass_tab_mc.tex
\begin{table}[!htb]
\vspace{3cm}
{\footnotesize{
\begin{tabular*}{\textwidth}[l]{l@{\extracolsep{\fill}}rllllllll}
Collaboration & Ref. & $\Nf$ & \hspace{0.15cm}\begin{rotate}{60}{publication status}\end{rotate}\hspace{-0.15cm} &
 \hspace{0.15cm}\begin{rotate}{60}{chiral extrapolation}\end{rotate}\hspace{-0.15cm} &
 \hspace{0.15cm}\begin{rotate}{60}{continuum  extrapolation}\end{rotate}\hspace{-0.15cm} &
 \hspace{0.15cm}\begin{rotate}{60}{finite volume}\end{rotate}\hspace{-0.15cm} &  
 \hspace{0.15cm}\begin{rotate}{60}{renormalization}\end{rotate}\hspace{-0.15cm} & 
  \rule{0.5cm}{0cm}$\overline{m}_c(\overline{m}_c)$ & 
  \rule{0.3cm}{0cm}$\overline{m}_c(\mbox{3 GeV})$ \\
&&&&&&&&& \\[-0.1cm]
\hline
\hline
&&&&&&&&& \\[-0.1cm]
ETM 21A & \cite{Alexandrou:2021gqw} & 2+1+1 & \oP & \good & \good & \good & \good & 1.339(22)($^{+19}_{-10}$)(10)$^\dagger$ & 1.036(17)($^{+15}_{-8}$) \\ 
HPQCD 20A & \cite{Hatton:2020qhk} & 2+1+1 & \gA & \good & \good & \good & \good & 1.2719(78) & 0.9841(51) \\ 
HPQCD 18  & \cite{Lytle:2018evc} & 2+1+1 & \gA & \good & \good & \good & \good & 1.2757(84) & 0.9896(61) \\
\hspace{-0.2cm}{\begin{tabular}{l}FNAL/MILC/\rule{0.1cm}{0cm}\\TUMQCD 18\end{tabular}}
		& \cite{Bazavov:2018omf} & 2+1+1 & \gA & \good &  \good & \good & $-$ & 1.273(4)(1)(10) & 0.9837(43)(14)(33)(5)  \\ 
HPQCD 14A  & \cite{Chakraborty:2014aca} & 2+1+1 & \gA & \good & \good & \good & $-$ & 1.2715(95) & 0.9851(63) \\ 
ETM 14A & \cite{Alexandrou:2014sha} & 2+1+1 & \gA & \soso & \good & \soso & \good & 1.3478(27)(195) & 1.0557(22)(153)$^*$ \\ 
ETM 14 & \cite{Carrasco:2014cwa} & 2+1+1 & \gA & \soso & \good & \soso & \good & 1.348(46) &1.058(35)$^*$ \\ 
&&&&&&&&& \\[-0.1cm]
\hline
&&&&&&&&& \\[-0.1cm]
ALPHA 23  & \cite{Bussone:2023kag} & 2+1 & \gA$^+$ & \good & \good & \good & \good &  1.296(15) &  1.006(9) \\
ALPHA 21  & \cite{Heitger:2021apz} & 2+1 & \gA$^+$ & \good & \good & \good & \good &  1.296(19) &  1.007(16) \\
Petreczky 19  & \cite{Petreczky:2019ozv} & 2+1 & \gA$$ & \good & \good & \good & \good &  1.265(10) &  1.001(16) \\
Maezawa 16  & \cite{Maezawa:2016vgv} & 2+1 & \gA & \bad & \good & \good  & $\good$ &  1.267(12) &  \\
JLQCD 16 & \cite{Nakayama:2016atf} & 2+1 & \gA & \soso & \good & \good & $-$ & 1.2871(123) & 1.0033(96) \\
$\chi$QCD 14 & \cite{Yang:2014sea} & 2+1 & \gA& \soso & \soso & \soso & \good & 1.304(5)(20) & 1.006(5)(22) \\                  
HPQCD 10  & \cite{McNeile:2010ji} & 2+1 & \gA & \soso & \good & \soso  & $-$ & 1.273(6) & 0.986(6) \\
HPQCD 08B & \cite{Allison:2008xk} & 2+1 & \gA &  \soso & \good & \soso & $-$ & 1.268(9) & 0.986(10) \\
&&&&&&&&& \\[-0.1cm]
\hline \hline
&&&&&&&&& \\[-0.1cm]
PDG & \cite{Zyla:2020zbs} & & & & & & & 1.27(2) & \\[1.0ex]
\hline \hline
&&&&&&&&& \\
\end{tabular*}\\[-0.2cm]
}}
\begin{minipage}{\linewidth}
{\footnotesize 
\begin{itemize}
\item[$^\dagger$] We applied the running factor 0.7739(60) for $\mu=3$ GeV to $\overline{m}_c$. The errors are statistical, systematic, and the uncertainty in the running factor.\\[-5mm]
\item[$^*$] A running factor equal to 0.900 between the scales $\mu = 2$ GeV and $\mu = 3$ GeV was applied by us.  \\[-5mm]
\item[$^+$] Published after the FLAG deadline. \\[-5mm]
\end{itemize}
}
\end{minipage}
\caption{\label{tab:mc} Lattice results for the $\msbar$ charm-quark mass $\overline{m}_c(\overline{m}_c)$ and $\overline{m}_c(\mbox{3 GeV})$ in GeV, together with the colour coding of the calculations used to obtain them.}
\end{table}

%% file: qmass/qmass_tab_mcoverms.tex
\begin{table}[!htb]
\vspace{3cm}
{\footnotesize{
\begin{tabular*}{\textwidth}[l]{l@{\extracolsep{\fill}}rllllll}
Collaboration & Ref. & $\Nf$ & \hspace{0.15cm}\begin{rotate}{60}{publication status}\end{rotate}\hspace{-0.15cm}  &
 \hspace{0.15cm}\begin{rotate}{60}{chiral extrapolation}\end{rotate}\hspace{-0.15cm} &
 \hspace{0.15cm}\begin{rotate}{60}{continuum  extrapolation}\end{rotate}\hspace{-0.15cm}  &
 \hspace{0.15cm}\begin{rotate}{60}{finite volume}\end{rotate}\hspace{-0.15cm}  & \rule{0.1cm}{0cm} 
$m_c/m_s$ \\
&&&&&& \\[-0.1cm]
\hline
\hline
&&&&&& \\[-0.1cm]
ETM 21A & \cite{Alexandrou:2021gqw} & 2+1+1 & \oP & \good &  \good & \good & 11.48(12)($^{+25}_{-19}$)\\ 
FNAL/MILC/TUMQCD 18  & \cite{Bazavov:2018omf} & 2+1+1 & \gA & \good &  \good & \good & 11.784(11)(17)(00)(08) \\ 
HPQCD 14A  & \cite{Chakraborty:2014aca} & 2+1+1  & \gA & \good & \good & \good  & 11.652(35)(55) \\
ETM 14 & \cite{Carrasco:2014cwa}  & 2+1+1  & \gA & \soso & \good & \soso & 11.62(16) \\
&&&&&& \\[-0.1cm]  
\hline 
&&&&&& \\[-0.1cm]
Maezawa 16  & \cite{Maezawa:2016vgv} & 2+1 & \gA & \bad & \good & \good  & 11.877(91)  \\
$\chi$QCD 14 & \cite{Yang:2014sea} & 2+1  & \gA & \soso & \soso & \soso & 11.1(8) \\
HPQCD 09A & \cite{Davies:2009ih}  & 2+1  & \gA & \soso & \good & \good & 11.85(16) \\
&&&&&& \\[-0.1cm]  
\hline
\hline
\end{tabular*}
}}
\caption{Lattice results for the quark-mass ratio $m_c/m_s$, together with the colour coding of the calculations used to obtain them.}
\label{tab:mcoverms}
\end{table}

%% file: qmass/qmass_mb.tex
\subsection{Bottom-quark mass}
\label{s:bmass}

Now we review the lattice results for the $\overline{\rm MS}$ bottom-quark mass
$\overline{m}_b$.  Related heavy-quark actions and observables have been discussed in
previous FLAG reviews \cite{Aoki:2013ldr,Aoki:2016frl,FlavourLatticeAveragingGroup:2019iem}, and descriptions can
be found in Sec.~A.1.3 of FLAG 19 \cite{FlavourLatticeAveragingGroup:2019iem}.  In Tab.~\ref{tab:mb}, we collect
results for $\overline{m}_b(\overline{m}_b)$ obtained with $\Nf =2+1$ and $2+1+1$
sea-quark flavours.  Available results for the quark-mass ratio $m_b / m_c$ are also
reported. After discussing the new results, we evaluate the corresponding FLAG averages.

\input{qmass/qmass_tab_mb}

\subsubsection{$\Nf=2+1$}

There are no new results since the last review, so we simply quote the same average of HPQCD~10 and Petreczky~19 (both are reported for $\Nf=5$, so we simply quote the average for $\Nf=5$).
\begin{align}
&\Nf= 2+1 :  &\FLAGAVBEGIN\overline{m}_b(\overline{m}_b)& = 4.171 (20)  \FLAGAVEND ~ \gev&&\Refs ~\mbox{\cite{McNeile:2010ji,Petreczky:2019ozv}}\,.
\end{align}
The corresponding (four-flavour) RGI average is
\begin{align}
&\Nf= 2+1 :  & M_b^{\rm RGI} & = 6.888(33)_m(45)_\Lambda  ~ \gev &&\Refs ~\mbox{\cite{McNeile:2010ji,Petreczky:2019ozv}}\,.
\end{align}

\subsubsection{$\Nf=2+1+1$}

HPQCD 21 \cite{Hatton:2021syc} is an update of HPQCD 14A (and replaces it in our average. See FLAG 19 for details.),
including EM corrections for the first time for the $b$-quark mass. Four flavours of HISQ
quarks are used on MILC ensembles with lattice spacings from about 0.09 to 0.03 fm.
Ensembles with physical- and unphysical-mass sea-quarks are used. Quenched QED is used to
obtain the dominant $\cO(\alpha)$ effect. The ratio of bottom- to charm-quark masses is
computed in a completely nonperturbative formulation, and the $b$-quark mass is extracted
using the value of $\overline{m}_c(3~\rm GeV)$ from HPQCD 20A. Since EM effects are
included, the QED renormalization scale enters the ratio which is quoted for 3 GeV and
$\Nf=4$. The total error on the new result is more than two times smaller than for HPQCD
14A, but is only slightly smaller compared to the NRQCD result reported in HPQCD 14B. The
inclusion of QED shifts the ratio $m_b/m_c$ up slightly from the pure QCD value by about
one standard deviation, and the value of $\overline{m}_b(\overline{m}_b)$ is consistent, within
errors, to the other pure QCD results entering our average. Therefore, we quote a single
average. Cutoff effects are significant in that work, and are the dominant source of
uncertainty in the ratio $m_b/m_c$. It is difficult to estimate the value of
$\delta(a_{\rm min})$ from the data present in the publication, but the authors provided
extra information about their analysis with the result that $\delta(a_{\rm min})\approx
3$. Therefore, we do not inflate the errors of that computation. The work rates green stars for all FLAG criteria except for the continuum limit (see Tab.~\ref{tab:mb}) where less than three ensembles at the physical $b$-quark mass were used in the $a\to0$ extrapolation (in the previous FLAG review this was missed and is corrected here).

HPQCD 14B employs the NRQCD action~\cite{Colquhoun:2014ica} to treat the $b$ quark.  The
$b$-quark mass is computed with the moments method, that is, from Euclidean-time moments
of two-point, heavy-heavy-meson correlation functions (see also Sec.~\ref{s:curr} for a
description of the method). Due to the effective treatment of the heavy quark, continuum
extrapolations are under control since five lattice spacings are employed, with the smallest about 0.09 fm, but the requirement that $a m_b\ll 1$ is not relevant.
Their final result is $\overline{m}_b(\mu = 4.18\, \gev) = 4.207(26)$
GeV, where the error is from adding systematic uncertainties in quadrature only
(statistical errors are smaller than $0.1 \%$ and ignored). The errors arise from
renormalization, perturbation theory, lattice spacing, and NRQCD systema\-tics. The
finite-volume uncertainty is not estimated, but at the lowest pion mass they have $ m_\pi
L \simeq 4$, which leads to the tag \good\ . In this case, the continuum extrapolations
seem mild, in part, thanks to the NRQCD action used to treat the $b$ quark. The data-driven continuum-limit criterion $\delta(a_{\rm
min})<3$, so no correction factor is necessary here.

The next four-flavour result (ETM 16B~\cite{Bussone:2016iua}) is from the ETM collaboration and
updates their preliminary result appearing in a conference
proceedings~\cite{Bussone:2014cha}. The calculation is performed on a set of
ensembles generated with twisted-Wilson fermions with three lattice spacings in the
range 0.06 to 0.09 fm and with pion masses in the range 210 to 440 MeV. The $b$-quark mass
is determined from a ratio of heavy-light pseudoscalar meson masses designed to yield the
quark pole mass in the static limit. The pole mass is related to the $\overline{\rm MS}$
mass through perturbation theory at N$^3$LO. The key idea is that by taking ratios of
ratios, the $b$-quark mass is accessible through fits to heavy-light(strange)-meson
correlation functions computed on the lattice in the range $\sim 1$--$2\times m_c$ and the
static limit, the latter being exactly 1. By simulating below $\overline{m}_b$, taking the
continuum limit is easier. They find $\overline{m}_b(\overline{m}_b) = 4.26(3)(10)$ GeV,
where the first error is statistical and the second systematic. The dominant errors come
from setting the lattice scale and fit systematics.

Gambino {\it et al.}~\cite{Gambino:2017vkx} use twisted-mass-fermion ensembles from the
ETM collaboration and the ETM ratio method as in ETM 16B. Three values of the lattice
spacing are used, ranging from 0.062 to 0.089 fm. Several volumes are also used. The
light-quark masses produce pions with masses from 210 to 450 MeV. The main difference with
ETM 16 is that the authors use the kinetic mass defined in the heavy-quark expansion (HQE)
to extract the $b$-quark mass instead of the pole mass. They include an additional uncertainty stemming from the conversion between kinetic and $\overline{\rm MS}$ schemes which leads to a somewhat larger total uncertainty compared to ETM 16B.

The final $b$-quark mass result is FNAL/MILC/TUM 18~\cite{Bazavov:2018omf}. The mass is
extracted from the same fit and analysis done for the charm quark mass.
Note that relativistic HISQ valence masses reach the physical $b$ mass on the two finest
lattice spacings ($a = 0.042$ fm, 0.03 fm) with physical and $0.2\times m_s$ light-quark masses,
respectively. In lattice units, the heavy valence masses correspond to $aM^{\rm RGI} >
0.90$, making the continuum extrapolation challenging. The extrapolations have
$\delta(a_{\rm min}) \approx 14$ (taking into account only the statistical error of the
continuum extrapolation, which is a 40\% of their total error budget). According to our
policy (\ref{sec:DataDriven}) we increase the error for the average by a factor 3.5. Their results are also
consistent with an analysis dropping the finest lattice spacing from the fit. Since the $b$-quark
mass region is only reached with two lattice spacings, we rate this work with a green
circle for the continuum extrapolation (the same as HPQCD 21). Note, however, that for other values of the quark
masses they use up to five values of the lattice spacing (cf.~their charm-quark mass
determination) with small values of $\delta(a_{\rm min})$ in the continuum extrapolation.
In summary, we judge that these large scaling violations affect mainly the determination of
the $b$-quark mass.

All of the above results enter our average. We note that here the ETM 16B result is
consistent with the average and a stretching factor on the error is not used. 
\begin{align}
&\Nf= 2+1+1 :&\FLAGAVBEGIN\overline{m}_b(\overline{m}_b)& = 4.200 (14)  \FLAGAVEND ~ \gev&&\Refs~\mbox{\cite{Hatton:2021syc,Colquhoun:2014ica,Bussone:2016iua,Gambino:2017vkx,Bazavov:2018omf}}\,.
\end{align}
We have included a 100\% correlation on the statistical errors of ETM 16B and Gambino 17,
since the same ensembles are used in both. While FNAL/MILC/TUM 18 and HPQCD 21 also use
the same MILC HISQ ensembles, the statistical error in the HPQCD 21 analysis is
negligible, so we do not include a correlation between them. The average has $\chi^2/{\rm
dof}=0.02$.

The above translates to the RGI average
\begin{align}
&\Nf= 2+1+1 :& M_b^{\rm RGI} & = 6.938(23)_m(45)_\Lambda ~\gev&&\Refs~\mbox{\cite{Hatton:2021syc,Colquhoun:2014ica,Bussone:2016iua,Gambino:2017vkx,Bazavov:2018omf}}\,.
\end{align}

\begin{figure}[!htb]
\begin{center}
\includegraphics[width=11cm]{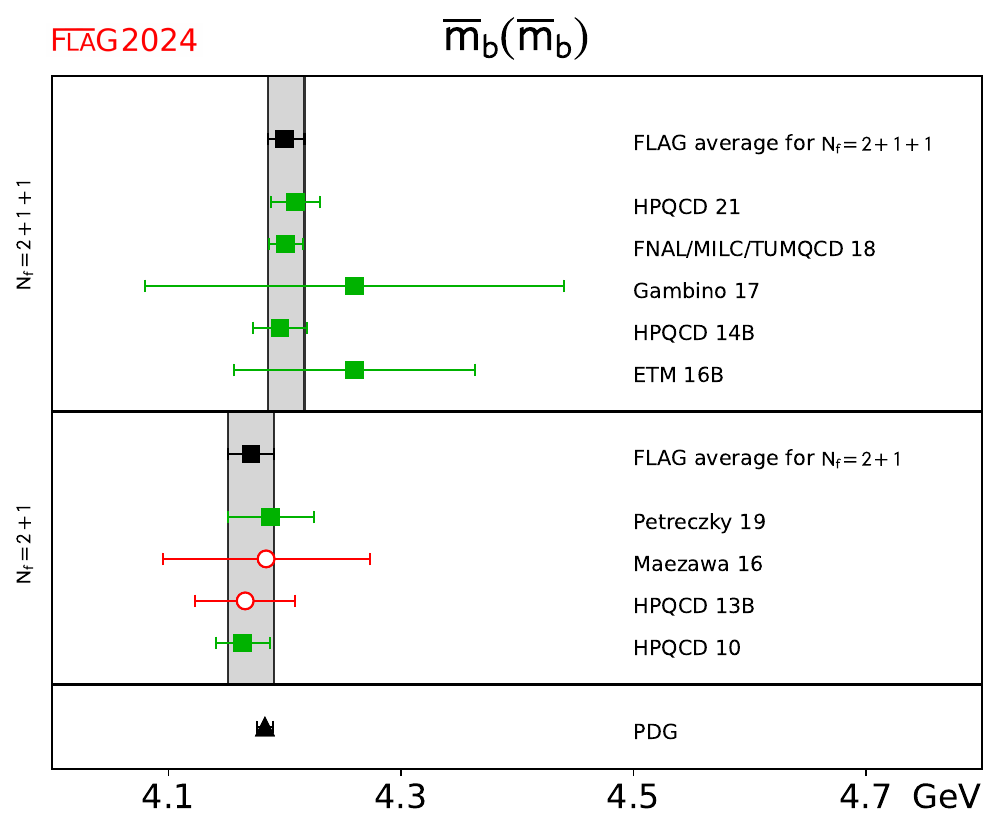}
\end{center}
\vspace{-1cm}
\caption{ \label{fig:mb} The $b$-quark mass for $\Nf =2+1$ and $2+1+1$ flavours. The
  updated PDG value from Ref.~\cite{ParticleDataGroup:2024cfk} is reported for comparison.}
\end{figure}

Results for $\overline{m}_b(\overline{m}_b)$ are shown in
Fig.~\ref{fig:mb} together with the FLAG averages corresponding to $\Nf=2+1$ and $2+1+1$
quark flavours.

%% file: qmass/qmass_tab_mb.tex
\begin{table}[!htb]
\vspace{3cm}
{\footnotesize{
\begin{tabular*}{\textwidth}[l]{l@{\extracolsep{\fill}}rl@{\hspace{0mm}}l@{\hspace{0mm}}l@{\hspace{0mm}}l@{\hspace{0mm}}l@{\hspace{0mm}}l@{\hspace{0mm}}lll}
Collaboration & Ref. & $\Nf$ & \hspace{0.15cm}\begin{rotate}{60}{publication status}\end{rotate}\hspace{-0.15cm} &
 \hspace{0.15cm}\begin{rotate}{60}{chiral extrapolation}\end{rotate}\hspace{-0.15cm} &
 \hspace{0.15cm}\begin{rotate}{60}{continuum extrapolation}\end{rotate}\hspace{-0.15cm} &
 \hspace{0.15cm}\begin{rotate}{60}{finite volume}\end{rotate}\hspace{-0.15cm} &  
 \hspace{0.15cm}\begin{rotate}{60}{renormalization}\end{rotate}\hspace{-0.15cm} &  
 \hspace{0.15cm}\begin{rotate}{60}{heavy-quark treatment}\end{rotate}\hspace{-0.15cm} & 
 \rule{0.2cm}{0cm}$\overline{m}_b(\overline{m}_b)$ & 
 \rule{0.2cm}{0cm}$m_b / m_c$ \\
&&&&&&&&&& \\[-0.1cm]
\hline
\hline
&&&&&&&&&& \\[-0.1cm]
HPQCD 21  & \cite{Hatton:2021syc} & 2+1+1 & \gA & \good &\soso& \good & $-$ & \okay & 4.209(21)$^{++}$ & 4.586(12)$^{**}$ \\ 
FNAL/MILC/TUM 18  & \cite{Bazavov:2018omf} & 2+1+1 & A & \good &  \soso & \good & $-$ & \okay & 4.201(12)(1)(8)(1)  & 4.578(5)(6)(0)(1)  \\ 
Gambino 17 & \cite{Gambino:2017vkx} & 2+1+1 & A & \soso & \good  & \soso &\good  &\okay & 4.26(18) & \\ 
ETM 16B & \cite{Bussone:2016iua} & 2+1+1 & A &\soso &\good & \soso &\good & \okay & 4.26(3)(10)$^+$ & 4.42(3)(8) \\ 
HPQCD 14B  & \cite{Colquhoun:2014ica} & 2+1+1 & \gA &\good & \good & \good & \good & \okay & 4.196(0)(23)$^\dagger$ & \\
&&&&&&&&&& \\[-0.1cm]
\hline
&&&&&&&&&& \\[-0.1cm]
Petreczky19  & \cite{Petreczky:2019ozv} & 2+1 & \gA & \good & \good & \good  & $\good$ &\okay & 4.188(37)  & 4.586(43)  \\
Maezawa 16  & \cite{Maezawa:2016vgv} & 2+1 & \gA & \bad & \good & \good  & $\good$ &\okay & 4.184(89)  & 4.528(57)  \\
HPQCD 13B  & \cite{Lee:2013mla} & 2+1 & \gA &\bad &\soso & $-$ & $-$ & \okay & 4.166(43) & \\ 
HPQCD 10 & \cite{McNeile:2010ji} & 2+1 & \gA &\good & \good & \good& $-$ & \okay & 4.164(23) & 4.51(4) \\ 
&&&&&&&&&& \\[-0.1cm]
\hline
&&&&&&&&&& \\[-0.1cm]
ETM 13B & \cite{Carrasco:2013zta} & 2 & \gA & \soso & \good & \soso & \good &\okay & 4.31(9)(8) & \\
ALPHA 13C & \cite{Bernardoni:2013xba} & 2 & \gA & \good & \good & \good & \good & \okay & 4.21(11) & \\
ETM 11A & \cite{Dimopoulos:2011gx} & 2 & \gA & \soso & \good & \soso & \good & \okay & 4.29(14) & \\[1.0ex]
\hline \hline
&&&&&&&&&& \\[-0.1cm]
PDG & \cite{Zyla:2020zbs} & & & & & & & & 4.18$^{+0.02}_{-0.03}$ & \\[1.0ex]
\hline \hline
&&&&&&&&&& \\
\end{tabular*}\\[-0.2cm]
}}
\begin{minipage}{\linewidth}
{\footnotesize 
\begin{itemize}
\item[$^{++}$] We quote the four-flavour result. For $\Nf=5$, the value is 4.202(21). \\[-5mm]
\item[$^{**}$] The ratio is quoted in the $\overline{\rm MS}$ scheme for $\mu=3$ GeV because of the different charges of the bottom and charm quarks. \\[-5mm]
\item[$^\dagger$] Only two pion points are used for chiral extrapolation. \\[-5mm]
\end{itemize}
}
\end{minipage}
\caption{\label{tab:mb} Lattice results for the $\msbar$ bottom-quark mass $\overline{m}_b(\overline{m}_b)$ in GeV,
  together with the systematic error ratings for each. Available results for the quark-mass ratio $m_b / m_c$ are also reported.}
\end{table}

%% file: Vudus/VudVus.tex
\section{Leptonic and semileptonic kaon and pion decay and $|V_{ud}|$ and $|V_{us}|$}
\label{sec:vusvud}
Authors: T.~Kaneko, J.~N.~Simone, N.~Tantalo\\

This section summarizes state-of-the-art lattice calculations of the leptonic kaon and pion decay constants, $f_K^\pm$ and $f_\pi^\pm$,
and the kaon semileptonic-decay form factor $f_+(0)$, and provides an analysis in the framework of the Standard Model.
With respect to the previous edition of the FLAG review \cite{FlavourLatticeAveragingGroupFLAG:2021npn},
there has been a new study each for the decay constants and $f_+(0)$ for $\Nf = 2+1$,
and a new entry to the average of the ratio $f_{K^\pm}/f_{\pi^\pm}$
for $\Nf = 2+1+1$.\footnote{In this edition,
we omit results for $\Nf = 2$, because there has been no new entry after 2014. We refer to the 2016 edition~\cite{Aoki:2016frl} for the $\Nf = 2$ results.}
As in Ref.~\cite{FlavourLatticeAveragingGroupFLAG:2021npn}, when combining lattice data with experimental results, we take into account the strong isospin correction, either obtained in lattice calculations or estimated by using chiral perturbation theory ({\Ch}PT), both for $f_{K^\pm}$ and $f_{K^\pm} / f_{\pi^\pm}$.

\subsection{Experimental information concerning $|V_{ud}|$, $|V_{us}|$, $f_+(0)$ and $\fKfpichargedr$}\label{sec:Exp} 

The following review relies on the fact that precision 
experimental data on kaon decays
very accurately determine the product $|V_{us}|f_+(0)$ \cite{Moulson:2017ive} and the ratio
$|V_{us}/V_{ud}|f_{K^\pm}/f_{\pi^\pm}$ \cite{Moulson:2017ive,ParticleDataGroup:2022pth}: 
\be\label{eq:products}
|V_{us}| f_+(0) = 0.21654(41)\co \hspace{1cm} \;
\left|\frac{V_{us}}{V_{ud}}\right|\frac{ f_{K^\pm}}{ f_{\pi^\pm}} \;
=0.27599(41)\fs\ee 
Here, and in the following, $f_{K^\pm}$ and $f_{\pi^\pm}$ are the isospin-broken 
decay constants in QCD. We will refer to the decay 
constants in the isospin-symmetric limit as $f_K$ and $f_\pi$ 
(the latter at leading order in the mass difference ($m_u - m_d$) coincides with $f_{\pi^\pm}$).
The parameters $|V_{ud}|$ and $|V_{us}|$ are
elements of the Cabibbo-Kobayashi-Maskawa matrix and $f_+(q^2)$ represents
one of the form factors relevant for the semileptonic decay
$K^0\rightarrow\pi^-\ell\,\nu$, which depends on the momentum transfer $q$
between the two mesons.  What matters here is the value at $q^2 = 0$:
\be
 f_+(0) \equiv f_+^{K^0\pi^-}(0) = f_0^{K^0\pi^-}(0) = q^\mu \langle \pi^-(p^\prime) | \bar{s} \gamma_\mu u | K^0(p) \rangle / (M_K^2 - M_\pi^2)
 \,\rule[-0.15cm]{0.02cm}{0.5cm}_{\;q^2\rightarrow 0}.
\ee 
  The pion and kaon decay constants are defined by\footnote{The pion
  decay constant represents a QCD matrix element---in the full Standard
  Model, the one-pion state is not a meaningful notion: the correlation
  function of the charged axial current does not have a pole at
  $p^2=M_{\pi^+}^2$, but a branch cut extending from $M_{\pi^+}^2$ to
  $\infty$. The analytic properties of the correlation function and the
  problems encountered in the determination of $f_\pi$ are thoroughly
  discussed in Ref.~\cite{Gasser:2010wz}. The ``experimental'' value of $f_\pi$
  depends on the convention used when splitting the sum ${\cal
    L}_{\mbox{\tiny QCD}}+{\cal L}_{\mbox{\tiny QED}}$ into two parts. The lattice
  determinations of $f_\pi$ do not yet reach the accuracy where this is of
  significance, but at the precision claimed by the Particle Data Group
  \cite{Agashe:2014kda,Patrignani:2016xqp}, the numerical value does depend on the convention 
  used~\cite{Gasser:2003hk,Rusetsky:2009ic,Gasser:2007de,Gasser:2010wz}. 
  }
\be
\lvac \dbar\gamma_\mu\gamma_5 \hspace{0.05cm}u|\pi^+(p)\rangle=i
\hspace{0.05cm}p_\mu f_{\pi^+}\co\hspace{1cm} \lvac \sbar\gamma_\mu\gamma_5
\hspace{0.05cm} u|K^+(p)\rangle=i \hspace{0.05cm}p_\mu f_{K^+}\fs
\ee
In this normalization, $f_{\pi^\pm} \simeq 130$~MeV, $f_{K^\pm}\simeq 155$~MeV.

 In Eq.~(\ref{eq:products}), the
electromagnetic effects have already been subtracted in the experimental
analysis using {\Ch}PT~\cite{Cirigliano:2001mk,Cirigliano:2004pv,Cirigliano:2008wn,Cirigliano:2011tm}.
In 2015, a new method~\cite{Carrasco:2015xwa} has been proposed
by the RM123-SOTON collaboration
for calculating the leptonic decay rates of hadrons including both QCD and QED on the lattice, 
and successfully applied to the case of the ratio of the leptonic decay rates of kaons and 
pions~\cite{Giusti:2017dwk,DiCarlo:2019thl}.
By employing the twisted-mass discretization,
  they simulate $\Nf = 2+1+1$ QCD at three lattice spacings $a=0.07$, 0.08, 0.09~fm
  with pion masses down to $\approx 220$~MeV on multiple lattice volumes
  to directly examine finite-volume effects.
The correction to the $K_{\mu2} / \pi_{\mu 2}$ decay rate, including both electromagnetic and strong
isospin-breaking effects, is found to be equal to $-1.26 (14) \%$~\cite{DiCarlo:2019thl}
to be compared to the estimate $-1.12 (21) \%$ based on {\Ch}PT \cite{Rosner:2015wva,Cirigliano:2011tm}.\footnote{See the discussion concerning the definition of QCD and of the isospin-breaking corrections in Sec.~\ref{sec:ibscheme}.} 
Using the experimental values of the $K_{\mu2} $ and $\pi_{\mu 2}$ decay rates the result of 
Ref.~\cite{DiCarlo:2019thl} implies
\be\label{eq:VusVud_new}
\left|\frac{V_{us}}{V_{ud}}\right|\frac{f_K}{f_\pi} = 0.27683 \, (29)_{\rm exp} \, (20)_{\rm th} \, [35] ~ , \ee 
where the last error in brackets is the sum in quadrature of the experimental and theoretical uncertainties, 
and the ratio of the decay constants is the one corresponding to isosymmetric QCD.
A large part of the theoretical uncertainty comes from the statistical error and continuum and chiral extrapolation of lattice data, which can be systematically reduced by a more realistic simulation with high statistics.

An independent study of the electromagnetic effects is carried out
by the RBC/UKQCD collaboration using the domain-wall discretization~\cite{Boyle:2022lsi}.
They simulate $\Nf=2+1$ QCD at a single lattice spacing $a = 0.11$~fm,
a pion mass close to its physical value,
and a lattice volume with $M_\pi L \sim 3.9$.
Their result $-0.86(^{+41}_{-40})$\,\%
including the strong isospin corrections
is consistent with the RM123-SOTON estimate.
The larger uncertainty is due to the possibly large finite-volume effects,
which are under active investigation in different lattice-QED prescriptions~\cite{Hermansson-Truedsson:2023krp}.

At present,
the superallowed nuclear $\beta$ transitions provide the most precise determination of $|V_{ud}|$.
Its accuracy has been limited by hadronic uncertainties in 
the universal electroweak radiative correction $\Delta_R^V$.
A 2018 analysis in terms of a dispersion relation~\cite{Seng:2018qru,Seng:2018yzq}
found $\Delta_R^V$ larger than the previous estimate~\cite{Marciano:2005ec}.
A more straightforward update~\cite{Czarnecki:2019mwq}
of Ref.~\cite{Marciano:2005ec} on the description of relevant hadronic contributions
as well as 
a lattice and perturbative-QCD calculation~\cite{Ma:2023kfr}
also reported larger $\Delta_R^V$,
which is consistent with the dispersive estimate within uncertainties.
Together with conservative estimate of nuclear corrections~\cite{Towner:2007np,Miller:2008my,Auerbach:2008ut,Liang:2009pf,Miller:2009cg,Towner:2010bx,Hardy:2014qxa,Hardy:2016vhg,Seng:2018qru,Gorchtein:2018fxl},
a recent reanalysis of twenty-three $\beta$ decays obtained~\cite{Hardy:2020qwl,ParticleDataGroup:2022pth}
\be\label{eq:Vud beta}
|V_{ud}| = 0.97373(31).
\ee

The matrix element $|V_{us}|$ can be determined from inclusive hadronic
$\tau$ decays~\cite{Gamiz:2002nu,Gamiz:2004ar,Maltman:2008na,Pich_Kass}.
Both Gamiz {\em et al.}~\cite{Gamiz:2007qs,Gamiz:2013wn}
and Maltman {\em et al.}~\cite{Maltman:2008ib,Maltman:2008na,Maltman:2009bh}
arrived at very similar values of $|V_{us}|$
by separating the inclusive decay $\tau \rightarrow X_{\{d,s\}}\nu_\tau$
into nonstrange ($X_d \nu_\tau$) and strange ($X_s \nu_\tau$) final states
and evaluating the relevant spectral integral
using the operator product expansion (OPE).
However, $|V_{us}| = 0.2195(19)$ quoted by HFLAV 18~\cite{Amhis:2019ckw} 
differs from the result one obtains from the kaon decays
by about three standard deviations (see Tab.~\ref{tab:Final results}
in Sec.~\ref{sec:SM}).
A new treatment of higher orders in the OPE
obtained a slightly larger value of $|V_{us}| = 0.2219(22)$
with a different experimental input~\cite{10.21468/SciPostPhysProc.1.006}.

Reference~\cite{Boyle:2018ilm} proposed a new method to determine $|V_{us}|$
without any recourse to the OPE
by evaluating the spectral integral from lattice-QCD data
of the hadronic vacuum polarization function
through generalized dispersion relations.
This led to an analysis~\cite{10.21468/SciPostPhysProc.1.006} yielding
$|V_{us}| = 0.2240(18)$,
which is consistent with that from the kaon decays.
However, this result mostly relies on
the $\tau \to K\nu_{\tau}$ decay channel,
which represents only $\sim 24$\,\% of the inclusive $\tau \to X_s \nu_\tau$ decay,
due to their choice of the generalized dispersion relation~\cite{Crivellin:2022rhw}.

The ETM collaboration carried out a first lattice calculation of the
{\it fully} inclusive rate of the hadronic $\tau$ decays based on ideas 
to study inclusive processes on the lattice~\cite{Hansen:2017mnd,Gambino:2020crt}.
Their study of $\tau \to X_d \nu_\tau$ led to 0.4\,\% determination of 
$|V_{ud}|=0.9752(39)$,
which is nicely consistent
with Eq.~(\ref{eq:Vud beta}) from nuclear $\beta$ decay~\cite{Evangelista:2023fmt}.
Their extension to the $\tau \to X_s \nu_\tau$ decay yields
$|V_{us}| = 0.2189(19)$ and confirms the above tension
with that from the kaon decays~\cite{Alexandrou:2024gpl}.
In Sec.~\ref{sec:SM} of this review, we quote 
\be\label{eq:Vus tau}
|V_{us}| = 0.2184(21)
\ee
from HFLAV 22~\cite{HFLAV:2022esi}
as $|V_{us}|$ from the inclusive hadronic $\tau$ decays.

The experimental results in Eq.~(\ref{eq:products}) are for the 
semileptonic decay of a neutral kaon into a negatively charged pion
and the charged pion and kaon leptonic decays, respectively, in QCD.
In the case of the semileptonic decays the corrections for strong
and electromagnetic isospin breaking in {\Ch}PT
at NLO have allowed for averaging the different experimentally
measured isospin channels~\cite{Antonelli:2010yf}. 
This is quite a convenient procedure as long as lattice-QCD calculations do not include
strong or QED isospin-breaking effects. 
Several lattice results for $f_K/f_\pi$ are quoted for QCD with (squared)
pion and kaon masses of $M_\pi^2=M_{\pi^0}^2$ and $M_K^2=\frac 12
\left(M_{K^\pm}^2+M_{K^0}^2-M_{\pi^\pm}^2+M_{\pi^0}^2\right)$
for which the leading strong and electromagnetic isospin violations cancel.
For these results,
contact with experimental results is made
by correcting leading isospin breaking 
guided either by {\Ch}PT or by lattice calculations. 
We note, however, that
the modern trend for the leptonic decays is
to include strong and electromagnetic isospin breaking in the lattice calculations
(e.g.,~Refs.~\cite{Aoki:2008sm,deDivitiis:2011eh,Ishikawa:2012ix,TakuLat12,deDivitiis:2013xla,Tantalo:2013maa,Portelli:2015wna,Carrasco:2015xwa,Giusti:2017dwk}).

This trend is being extended to the semileptonic decays.
Calculating the electromagnetic correction to the $K_{\ell 3}$ semileptonic decays
on the lattice is more involved
due to the photon exchange between $\pi^{\pm}$ and $\ell^{\mp}$ in the final state.
A framework has been proposed~\cite{Christ:2023lcc},
and its applicability to the kaon semileptonic decays has been discussed
in Ref.~\cite{Christ:2024xzj}.
References~\cite{Seng:2020jtz,Ma:2021azh,Seng:2021wcf} pursue  
an effective field theory setup supplemented by nonperturbative lattice-QCD
inputs to estimate the radiative corrections.

\subsection{Lattice results for $f_+(0)$ and $f_{K^\pm}/f_{\pi^\pm}$}

The traditional way of determining $|V_{us}|$ relies on using estimates for
the value of $f_+(0)$, invoking the Ademollo-Gatto theorem~\cite{Ademollo_Gatto}.
This theorem states that the corrections to the SU(3) symmetric limit $f_+(0)=1$
start at second order in SU(3) breaking, namely $\propto (m_s-m_{ud})^2$.
Theoretical models are used to estimate higher-order corrections.
Lattice methods have now reached the stage where quantities like
$f_+(0)$ or $f_K/f_\pi$ can be determined to good accuracy.
As a consequence, the uncertainties inherent in the
theoretical estimates for the higher order effects in the value of $f_+(0)$
do not represent a limiting factor any more, and we shall, therefore, not
invoke those estimates. Also, we will use the experimental results based on
nuclear $\beta$ decay and inclusive hadronic $\tau$ decay exclusively
for comparison---the
main aim of the present review is to assess the information gathered with
lattice methods and to use it for testing the consistency of the SM and its
potential to provide constraints for its extensions.

The database underlying the present review of the semileptonic form factor 
and the ratio of decay constants is
listed in Tabs.~\ref{tab:f+(0)}, \ref{tab:FKFpi} and \ref{tab:correctedfKfPi}. The properties of the
lattice data play a crucial role for the conclusions to be drawn from these
results: ranges of $a$, $M_\pi$ and $LM_\pi$
to control continuum extrapolation, extrapolation in the quark masses,
finite-size effects, etc.
The key features of the various data sets are characterized by means of the 
colour code specified in Sec.~\ref{sec:color-code}.  
More detailed information
on individual computations are compiled in Appendix~\ref{app:VusVud}, 
which in this edition is limited to new results and to those entering the FLAG 
averages. For other calculations the reader should refer to the Appendix B.2 
of Ref.~\cite{Aoki:2016frl}.

The quantity $f_+(0)$ represents a matrix element of a strangeness-changing
null-plane charge, $f_+(0)=\langle K|Q^{\bar{u}s}|\pi \rangle$ (see Ref.~\cite{Gasser:1984ux}). The vector charges obey the
commutation relations of the Lie algebra of SU(3), in particular
$[Q^{\bar{u}s},Q^{\bar{s}u}]=Q^{\bar{u}u-\bar{s}s}$. This relation implies the sum rule $\sum_n
|\langle K|Q^{\bar{u}s}|n \rangle|^2-\sum_n |\langle K|Q^{\bar{s}u}|n \rangle|^2=1$. Since the contribution from
the one-pion intermediate state to the first sum is given by $f_+(0)^2$,
the relation amounts to an exact representation for this quantity
\cite{Furlan}: \be \label{eq:Ademollo-Gatto} f_+(0)^2=1-\sum_{n\neq \pi}
|\langle K|Q^{\bar{u}s}|n \rangle|^2+\sum_n |\langle K |Q^{\bar{s}u}|n \rangle|^2\fs\ee While the first sum on the
right extends over nonstrange intermediate states, the second runs over
exotic states with strangeness $\pm 2$ and is expected to be small compared
to the first.

The expansion of $f_+(0)$ in SU(3) {\Ch}PT
in powers of $m_u$, $m_d$, and $m_s$ starts with
$f_+(0)=1+f_2+f_4+\ldots\,$ \cite{Gasser:1984gg}.
The NLO contribution $f_2$ is known, since it can be expressed in terms of $M_\pi$,
$M_K$, $M_\eta$ and $f_\pi$ \cite{Gasser:1984ux}.
In the language of the sum rule (\ref{eq:Ademollo-Gatto}), $f_2$
stems from nonstrange intermediate states with three mesons. Like all
other nonexotic intermediate states, it lowers the value of $f_+(0)$:
$f_2=-0.023$ when using the experimental value of $f_\pi$ as input.  
The corresponding expressions have also been derived in
quenched or partially quenched (staggered) {\Ch}PT
\cite{Bernard:2013eya,Bazavov:2012cd}.  At the same order in the SU(2) expansion
\cite{Flynn:2008tg}, $f_+(0)$ is parameterized in terms of $M_\pi$ and two
\textit{a priori} unknown parameters. The latter can be determined from the
dependence of the lattice results on the masses of the quarks.
For the SU(3) {\Ch}PT formula for $f_2$,
one may use $f_0$, that is the decay constant in the chiral limit, instead of $f_\pi$.
While this affects the result only at higher orders, it may make a significant
numerical difference in calculations where the higher-order corrections
are not explicitly accounted for. (Lattice results concerning the
value of the ratio $f_\pi/f_0$ are reviewed in Sec.~5.3 
of the previous review \cite{FlavourLatticeAveragingGroupFLAG:2021npn}.)

The lattice results shown in Fig.~\ref{fig:lattice data semileptonic}
indicate that the higher order contributions $\Delta f\equiv
f_+(0)-1-f_2$ are negative and thus amplify the effect generated by $f_2$.
This confirms the expectation that the exotic contributions are small. The
entries in the lower part represent various model
estimates for $f_4$. In Ref.~\cite{Leutwyler:1984je}, the symmetry-breaking
effects are estimated in the framework of the quark model. The more recent
calculations are more sophisticated, as they make use of the known explicit
expression for the $K_{\ell3}$ form factors to NNLO in {\Ch}PT
\cite{Post:2001si,Bijnens:2003uy}. The corresponding formula for $f_4$
accounts for the chiral logarithms occurring at NNLO and is not subject to
the ambiguity mentioned above.\footnote{Fortran programs for the
  numerical evaluation of the form factor representation in
  Ref.~\cite{Bijnens:2003uy} are available on request from Johan Bijnens.} 
 The numerical result, however, depends on
the model used to estimate the low-energy constants occurring in $f_4$
\cite{Bijnens:2003uy,Jamin:2004re,Cirigliano:2005xn,Kastner:2008ch}. The
figure indicates that the most recent numbers obtained in this way
correspond to a positive or an almost vanishing rather than a negative value for $\Delta f$.
We note that FNAL/MILC 12I~\cite{Bazavov:2012cd},
JLQCD 17~\cite{Aoki:2017spo},
FNAL/MILC 18~\cite{Bazavov:2018kjg},
and Ref.~\cite{Bernard:2007tk} have made an attempt 
at determining a combination of some of the low-energy constants appearing 
in $f_4$ from lattice data.

\subsection{Direct determination of $f_+(0)$ and $f_{K^\pm}/f_{\pi^\pm}$}\label{sec:Direct} 

Many lattice results for the form factor $f_+(0)$ and for the ratio of decay constants, which we summarize here in Tabs.~\ref{tab:f+(0)} and~\ref{tab:FKFpi}, respectively, have been computed in isospin-symmetric QCD. 
The reason for this unphysical parameter choice is that there are only  a few simulations of isospin-breaking effects in lattice QCD, which is ultimately the cleanest way for predicting these effects
~\cite{Duncan:1996xy,Basak:2008na,Blum:2010ym,Portelli:2010yn,deDivitiis:2011eh,deDivitiis:2013xla,Tantalo:2013maa,Portelli:2015wna,Carrasco:2015xwa,Giusti:2017dwk,Boyle:2022lsi}. 
In the meantime, one relies either on {\Ch}PT~\cite{Gasser:1984gg,Aubin:2004fs} to estimate the correction to the isospin limit or one calculates the breaking at leading order in $(m_u-m_d)$ in the valence quark sector by extrapolating the lattice data for the charged kaons to the physical value of the $up$($down$)-quark mass (the result for the pion decay constant is always extrapolated to the value of the average light-quark mass $\hat m$).
This defines the prediction for $f_{K^\pm}/f_{\pi^\pm}$.

Since the majority of results that qualify for inclusion into the FLAG average include the strong isospin-breaking correction, we provide in Fig.~\ref{fig:lattice data leptonic} the overview of the world data of $f_{K^\pm}/f_{\pi^\pm}$.
For all the results of Tab.~\ref{tab:FKFpi} provided only in the isospin-symmetric limit we apply individually an isospin correction that will be described later on (see Eqs.~(\ref{eq:convert})\,--\,(\ref{eq:iso})).

\begin{figure}[ht]
\centering
\psfrag{y}{\tiny $\star$}
\includegraphics[height=9.00cm]{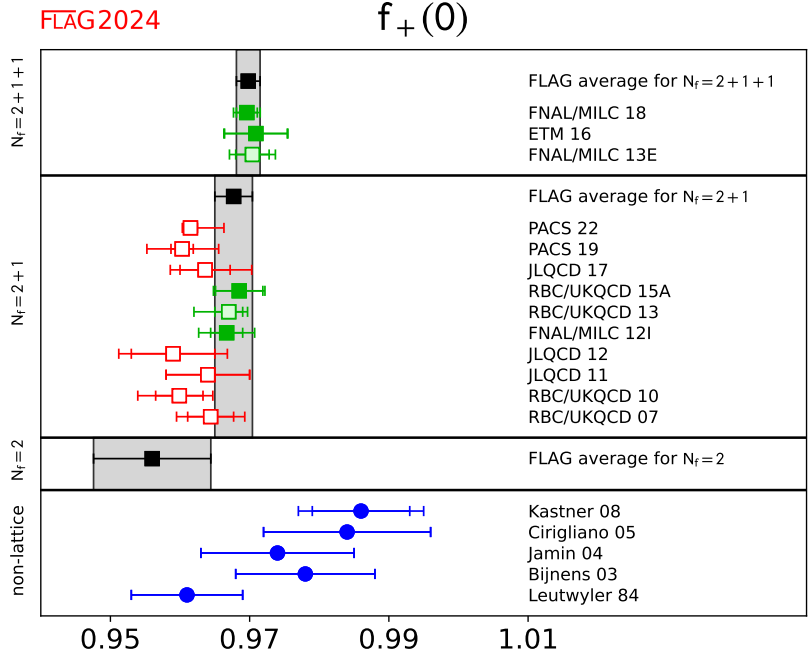}

\caption{\label{fig:lattice data semileptonic} 
Comparison of lattice results (squares and empty circles) for $f_+(0)$ with various model estimates based on {\Ch}PT~\cite{Kastner:2008ch,Cirigliano:2005xn,Jamin:2004re,Bijnens:2003uy,Leutwyler:1984je} (filled blue circles). The black squares and grey bands indicate our averages in Eqs.~(\ref{eq:fplus_direct_2p1p1}) and (\ref{eq:fplus_direct_2p1}). The significance of the colours is explained in Sec.~\ref{sec:qualcrit}.}

\end{figure}

\begin{figure}[t]
\centering
\psfrag{y}{\tiny $\star$}
\includegraphics[height=9.0cm]{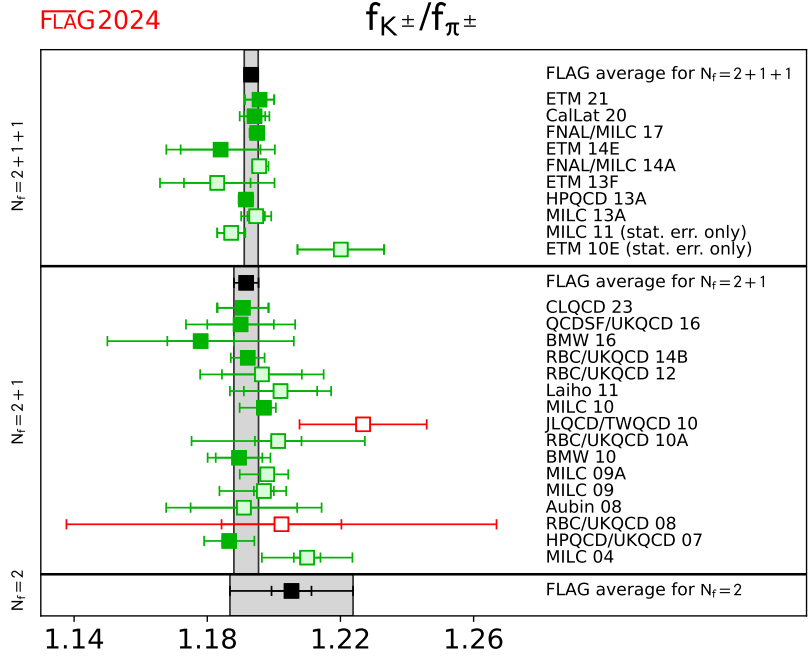}
\caption{\label{fig:lattice data leptonic}
Comparison of lattice results for $f_{K^\pm}/ f_{\pi^\pm}$.
This ratio is obtained in pure QCD including the isospin-breaking correction
(see Sec.~\ref{sec:Direct}).
The black squares and grey bands indicate our averages in Eqs.~(\ref{eq:fKfpi_direct_broken_2p1p1}) and (\ref{eq:fKfpi_direct_broken_2p1}).
}

\end{figure}

The plots in Figs.~\ref{fig:lattice data semileptonic} and \ref{fig:lattice data leptonic} illustrate our compilation of data for $f_+(0)$ and $f_{K^\pm}/f_{\pi^\pm}$.
The lattice data for the latter quantity is largely consistent even when comparing simulations with different $\Nf$.
In the case of $f_+(0)$, a slight tendency to get higher values when increasing $\Nf$ seems to be visible,
  while it does not exceed one standard deviation.
We now proceed to form the corresponding averages, separately for the data with $\Nf=2+1+1$ and $\Nf=2+1$ dynamical flavours, and in the following we will refer to these averages as the ``direct'' determinations.

\begin{table}[t]
\centering 
\vspace{2.8cm}
{\footnotesize\noindent
\begin{tabular*}{\textwidth}[l]{@{\extracolsep{\fill}}llllllll}
Collaboration & Ref. & $\Nf$ & 
\hspace{0.15cm}\begin{rotate}{60}{publication status}\end{rotate}\hspace{-0.15cm}&
\hspace{0.15cm}\begin{rotate}{60}{chiral extrapolation}\end{rotate}\hspace{-0.15cm}&
\hspace{0.15cm}\begin{rotate}{60}{continuum extrapolation}\end{rotate}\hspace{-0.15cm}&
\hspace{0.15cm}\begin{rotate}{60}{finite-volume errors}\end{rotate}\hspace{-0.15cm}&\rule{0.3cm}{0cm}
$f_+(0)$ \\
&&&&&&& \\[-0.1cm]
\hline
\hline&&&&&&& \\[-0.1cm]
FNAL/MILC 18               &\cite{Bazavov:2018kjg} &2+1+1  &\gA&\good&\good&\good& {0.9696(15)(12)}\\
ETM 16                     &\cite{Carrasco:2016kpy} &2+1+1  &\gA&\soso&\good&\soso& 0.9709(45)(9)\\
FNAL/MILC 13E               &\cite{Bazavov:2013maa} &2+1+1  &\gA&\good&\good&\good& {0.9704(24)(22)}\\
&&&&&&& \\[-0.1cm]
\hline
&&&&&&& \\[-0.1cm]
PACS 22          & \cite{Ishikawa:2022ulx} &2+1  &\gA&\soso&\tbr&\good&  0.9615(10)($^{+47}_{-6}$)\\
PACS 19             & \cite{Kakazu:2019ltq} &2+1  &\gA&\soso&\tbr&\good& 0.9603(16)($^{+50}_{-48}$)\\
JLQCD 17               & \cite{Aoki:2017spo} &2+1  &\gA&\soso&\tbr&\soso& 0.9636(36)($^{+57}_{-35}$)\\
RBC/UKQCD 15A              &\cite{Boyle:2015hfa}  &2+1  &\gA&\good&\soso&\soso& {0.9685(34)(14)}\\
RBC/UKQCD 13              & \cite{Boyle:2013gsa}  &2+1  &\gA&\good&\soso&\soso& 0.9670(20)($^{+18}_{-46}$)\\
FNAL/MILC 12I                 & \cite{Bazavov:2012cd} &2+1  &\gA&\soso&\soso&\tbg& {0.9667(23)(33)}\\
JLQCD 12                        & \cite{Kaneko:2012cta} &2+1  &\rC&\soso&\tbr&\tbg& 0.959(6)(5)\\
JLQCD 11                        & \cite{Kaneko:2011rp}  &2+1  &\rC&\soso&\tbr&\tbg& 0.964(6)\\
RBC/UKQCD 10              & \cite{Boyle:2010bh}   &2+1  &\gA&\soso&\tbr&\tbg& 0.9599(34)($^{+31}_{-47}$)(14)\rule{0cm}{0.4cm}\\ 
RBC/UKQCD 07              & \cite{Boyle:2007qe}   &2+1  &\gA&\soso&\tbr&\tbg& 0.9644(33)(34)(14)\\
&&&&&&& \\[-0.1cm]
\hline
\hline
\end{tabular*}}
\caption{Colour codes for the data on $f_+(0)$. In this and previous editions~\cite{FlavourLatticeAveragingGroup:2019iem,FlavourLatticeAveragingGroupFLAG:2021npn}, old results with two red tags have been dropped.\hfill}\label{tab:f+(0)}
\end{table}

\subsubsection{Results for $f_+(0)$}
 
For $f_+(0)$ there are currently two computational strategies: 
FNAL/MILC uses the Ward identity to relate the $K\to\pi$ form factor at zero momentum transfer to the matrix element $\langle \pi|S|K\rangle$ of the flavour-changing scalar current $S = \bar{s} u$. 
Peculiarities of the staggered fermion discretization used by FNAL/MILC (see Ref.~\cite{Bazavov:2012cd}) makes this the favoured choice. 
The other collaborations are instead computing the vector current matrix element $\langle \pi | \bar{s} \gamma_\mu u |K\rangle$. 
Apart from FNAL/MILC 13E, RBC/UKQCD 15A, FNAL/MILC 18, PACS 19 and 22, all simulations in Tab.~\ref{tab:f+(0)} involve unphysically heavy quarks and, therefore, the lattice data needs to be extrapolated to the physical pion and kaon masses corresponding to the $K^0\to\pi^-$ channel. 
We note also that the recent computations of $f_+(0)$ make use of the partially-twisted boundary conditions to determine the form-factor results directly at the relevant kinematical point $q^2=0$ \cite{Guadagnoli:2005be,Boyle:2007wg}, avoiding in this way any uncertainty due to the momentum dependence of the vector and/or scalar form factors. 
The ETM collaboration uses partially-twisted boundary conditions to compare the momentum dependence of the scalar and vector form factors with the one of the experimental data \cite{Lubicz:2010bv,Carrasco:2016kpy}, while keeping at the same time the advantage of the high-precision determination of the scalar form factor at the kinematical end-point $q_{max}^2 = (M_K - M_\pi)^2$ \cite{Becirevic:2004ya,Lubicz:2009ht} for the interpolation at $q^2 = 0$.

According to the colour codes reported in Tab.~\ref{tab:f+(0)} and to the FLAG rules of Sec.~\ref{sec:averages}, the results FNAL/MILC 12I and  RBC/UKQCD 15A with $\Nf=2+1$, and the results ETM 16 and FNAL/MILC 18 with $\Nf=2+1+1$ dynamical flavours of fermions, respectively, can enter the FLAG averages.
Therefore, there is no new entry 
to form the averages in Eqs.~(\ref{eq:fplus_direct_2p1p1}) and (\ref{eq:fplus_direct_2p1})
in this edition.

At $\Nf=2+1+1$ the result from the FNAL/MILC collaboration, $f_+(0) = 0.9704 (24) (22)$ (FNAL/MILC 13E), is based on the use of the Highly Improved Staggered Quark (HISQ) action (for both valence and sea quarks), which has been tailored to reduce staggered taste-breaking effects, and includes simulations with three lattice spacings and physical light-quark masses.
These features lead to uncertainties due to the chiral extrapolation and the discretization artifacts that are well below the statistical error.
The remaining largest systematic uncertainty comes from finite-size effects, which have been investigated in Ref.~\cite{Bernard:2017scg} using one-loop {\Ch}PT (with and without taste-violating effects).
In Ref.~\cite{Bazavov:2018kjg}, the FNAL/MILC collaboration presented a more precise determination of $f_+(0)$, $f_+(0) = 0.9696 (15) (11)$ (FNAL/MILC 18).
In this update, 
their analysis is extended to two smaller lattice spacings $a = 0.06$ and 0.042~fm.
The physical light-quark mass is simulated at four lattice spacings.
They also added a simulation at a small volume to study the finite-size effects.
The improvement of the precision with respect to FNAL/MILC 13E is obtained mainly
by an estimate of finite-size effects,
which is claimed to be controlled at the level of $\sim 0.05$\,\% by comparing
two analyses with and without the one-loop correction.
The total uncertainty is reduced to $\sim 0.2$\,\%.
An independent calculation of such high precision would be highly welcome
to solidify the lattice prediction of $f_+(0)$,
which currently suggests a tension with CKM unitarity
with the updated value of $|V_{ud}|$ (see Sec.~\ref{sec:testing}).

The result from the ETM collaboration, $f_+(0) = 0.9709 (45) (9)$ (ETM 16), makes use of the twisted-mass discretization adopting three values of the lattice spacing in the range $0.06 - 0.09$ fm and pion masses simulated in the range $210 - 450$ MeV.  
The chiral and continuum extrapolations are performed in a combined fit together with the momentum dependence, using both a SU(2)-{\Ch}PT inspired ansatz (following Ref.~\cite{Lubicz:2010bv}) and a modified z-expansion fit.
The uncertainties coming from the chiral extrapolation, the continuum extrapolation and the finite-volume effects turn out to be well below the dominant statistical error, which includes also the error due to the fitting procedure.
A set of synthetic data points, representing both the vector and the scalar semileptonic form factors at the physical point for several selected values of $q^2$, is provided together with the corresponding correlation matrix.

In ETM 16, 
a measure of the scaling violation $\delta(a)$ defined in Eq.~(\ref{eq:delta_a}) estimated
from their continuum and chiral extrapolation decreases toward the chiral limit
with the strange-quark mass kept fixed, 
because the SU(3)-breaking effects to be calculated on the lattice increases,
and more statistics are needed to keep the statistical accuracy toward this limit.
At the physical point, $\delta(a)$ is consistent with zero in their region of
the lattice spacing $a$.
This is also the case for FNAL/MILC 18, where they demonstrated that
$f_+(0)$ extrapolated to the physical point at each simulated value of $a$
is consistent with the value extrapolated to the continuum limit within 2 $\sigma$.
We note that, in contrast to the heavy-meson semileptonic decays,
relevant meson masses and momenta at zero momentum transfer are
at most ${\cal O}(M_K)$, and hence well below the cutoff $a^{-1}$.

The PACS collaboration carried out a calculation (PACS 19) for $\Nf = 2+1$
using the ${\cal O}(a)$-improved Wilson quark action
by creating an ensemble with the physical light-quark mass on a large lattice volume of $(10.9\,\mbox{fm})^4$ at a single spacing $a = 0.085$~fm~\cite{Kakazu:2019ltq}.
Such a large lattice enables them to interpolate $f_+(q^2)$ to zero momentum transfer and study the momentum-transfer dependence of the form factors 
without using partially-twisted boundary conditions.
This was extended to a smaller lattice spacing $a = 0.063$~fm
in PACS 22, which yields $f_+(0) = 0.9615(10)\left(^{+47}_{-6}\right)$.
However, their result does not enter the FLAG average,
because they simulate only two lattice spacings
using unimproved local and conserved vector currents.
That setup is the source of the largest (and very asymmetric) error in their calculation.
Further extension to an even smaller lattice spacing $a = 0.041$~fm
has been reported in Ref.~\cite{Yamazaki:2023swq},
where authors estimate the statistical error only,
and refrain from quoting a numerical value of $f_+(0)$.

For $\Nf=2+1$, the two results eligible to enter the FLAG average are the one from RBC/UKQCD 15A, $f_+(0) = 0.9685 (34) (14)$~\cite{Boyle:2015hfa}, and the one from FNAL/MILC 12I, $f_+(0)=0.9667(23)(33)$~\cite{Bazavov:2012cd}. 
These results, based on different  fermion discretizations (staggered fermions in the case of FNAL/MILC and domain wall fermions in the case of RBC/UKQCD) are in nice agreement.
Moreover, in the case of FNAL/MILC the form factor has been determined from the scalar current matrix element, while in the case of RBC/UKQCD it has been determined including also the matrix element of the vector current. 
To a certain extent, both simulations are expected to be affected by different systematic effects.

RBC/UKQCD 15A has analyzed results on ensembles with pion masses down to 140~MeV, mapping out the complete range from the SU(3)-symmetric limit to the physical point. 
No significant cut-off effects (results for two lattice spacings) were observed in the simulation results.
Ensembles with unphysical light-quark masses are weighted to work as a guide for small corrections toward the physical point, reducing in this way the model dependence in the fitting ansatz.
The systematic uncertainty turns out to be dominated by finite-volume effects, for which an estimate based on effective theory arguments is provided. 

The result FNAL/MILC 12I is from simulations reaching down to a lightest RMS pion mass of about 380~MeV (the lightest valence pion mass for one of their ensembles is about 260~MeV).
Their combined chiral and continuum extrapolation (results for two lattice spacings) is based on NLO staggered {\Ch}PT supplemented by the continuum NNLO expression~\cite{Bijnens:2003uy} and a phenomenological parameterization of the breaking of the Ademollo-Gatto theorem at finite lattice spacing inherent in their approach.
The $p^4$ low-energy constants entering the NNLO expression have been fixed in terms of external input~\cite{Amoros:2001cp}. 

Since there has been no new entry after the previous edition,
the FLAG average for $f_+(0)$ remains unchanged.
The $\Nf = 2+1+1$ average is based on the FNAL/MILC 18 and ETM 16 (uncorrelated) results, the $\Nf =2+1$ average based on FNAL/MILC 12I and RBC/UKQCD 15A, which we consider uncorrelated:
\begin{align}
&\label{eq:fplus_direct_2p1p1}
\mbox{direct},\,\Nf=2+1+1:&\FLAGAVBEGIN f_+(0) & = 0.9698(17)\FLAGAVEND  && \Refs~\mbox{\cite{Carrasco:2016kpy,Bazavov:2018kjg}},\\
&\label{eq:fplus_direct_2p1}                                                               
\mbox{direct},\,\Nf=2+1:  &\FLAGAVBEGIN f_+(0) &= 0.9677(27) \FLAGAVEND     &&\Refs~\mbox{\cite{Bazavov:2012cd,Boyle:2015hfa}}.   
\end{align}
We stress that the results (\ref{eq:fplus_direct_2p1p1}) and (\ref{eq:fplus_direct_2p1}), corresponding to $\Nf = 2+1+1$ and $\Nf = 2+1$, respectively, include simulations with physical light-quark masses.

\begin{table}[!htb]
\centering
\vspace{3.0cm}{\footnotesize\noindent
\begin{tabular*}{\textwidth}[l]{@{\extracolsep{\fill}}lrlllllll}
Collaboration & Ref. & $\Nf$ &
\hspace{0.15cm}\begin{rotate}{60}{publication status}\end{rotate}\hspace{-0.15cm}&
\hspace{0.15cm}\begin{rotate}{60}{chiral extrapolation}\end{rotate}\hspace{-0.15cm}&
\hspace{0.15cm}\begin{rotate}{60}{continuum extrapolation}\end{rotate}\hspace{-0.15cm}&
\hspace{0.15cm}\begin{rotate}{60}{finite-volume errors}\end{rotate}\hspace{-0.15cm}&
\rule{0.2cm}{0cm} $f_K/f_\pi$ &
\rule{0.2cm}{0cm} $f_{K^\pm}/f_{\pi^\pm}$ \\  
&&&&&&& \\[-0.1cm]
\hline
\hline
&&&&&&& \\[-0.1cm]
ETM 21 &\cite{Alexandrou:2021bfr}	     &2+1+1&\gA&\good &\good&\good    		&1.1995(44)(7)    & 1.1957(44)(7)\\
CalLat 20 &\cite{Miller:2020xhy}	     &2+1+1&\gA&\good &\good&\good    		&1.1964(32)(30)    & 1.1942(32)(31) \\
FNAL/MILC 17 &\cite{Bazavov:2017lyh}	     &2+1+1&\gA&\good &\good&\good    		&{1.1980(12)($_{-15}^{+5}$)}    &{1.1950(15)($_{-18}^{+6}$)} \\
ETM 14E       &\cite{Carrasco:2014poa}          &2+1+1&\gA&\soso &\good&\soso    		&	1.188(11)(11)   &{1.184(12)(11)} \\
FNAL/MILC 14A &\cite{Bazavov:2014wgs}	     &2+1+1&\gA&\good &\good&\good    		&					      &{1.1956(10)($_{-18}^{+26}$)} \\
ETM 13F       &\cite{Dimopoulos:2013qfa}      &2+1+1&\rC&\soso &\good&\soso    		&	 1.193(13)(10)    	      &1.183(14)(10)	\\
HPQCD 13A       &\cite{Dowdall:2013rya}	     &2+1+1&\gA&\good &\soso&\good    		&	 1.1948(15)(18)&{1.1916(15)(16)} \\
MILC 13A        &\cite{Bazavov:2013cp}	     &2+1+1&\gA&\good &\good&\good    		&					      &1.1947(26)(37) \\
MILC 11        &\cite{Bazavov:2011fh}	             &2+1+1&\rC&\soso &\soso&\soso    		&					      &1.1872(42)$^\dagger_{\rm stat.}$ \\
ETM 10E       &\cite{Farchioni:2010tb}            &2+1+1&\rC&\soso&\soso&\soso		        &       1.224(13)$_{\rm stat}$   &						\\
&&&&&&& \\[-0.1cm]                                                                                                              
\hline                                                                                                                          
&&&&&&& \\[-0.1cm]                                                                                                              
CLQCD 23        &\cite{CLQCD:2023sdb}  &2+1&\gA&\good&\good&\good        & 			                        &    1.1907(76)(17) \\
QCDSF/UKQCD 16  &\cite{Bornyakov:2016dzn}  &2+1&\gA&\soso&\good&\soso     & 1.192(10)(13)			        &    1.190(10)(13) \\
BMW 16          &\cite{Durr:2016ulb,Scholz:2016kcr}  &2+1&\gA&\good&\good&\good     & 1.182(10)(26)			        &    1.178(10)(26)                 \\
RBC/UKQCD 14B   &\cite{Blum:2014tka}    &2+1&\gA&\good    & \good	 &  \good  	&1.1945(45)					&					\\
RBC/UKQCD 12   &\cite{Arthur:2012opa}    &2+1&\gA&\good    & \soso	 &  \good  	&{1.199(12)(14)}				&					\\
Laiho 11       &\cite{Laiho:2011np}       &2+1&\rC&\soso    & \good   &  \soso  	&                                       	&$1.202(11)(9)(2)(5)$$^{\dagger\dagger}$	\\
MILC 10        &\cite{Bazavov:2010hj}&2+1&\rC&\soso&\good&\good			&                             			&{1.197(2)($^{+3}_{-7}$)}			\\
JLQCD/TWQCD 10 &\cite{Noaki:2010zz}&2+1&\rC&\soso&\tbr&\tbg			&1.230(19)					&                               		\\
RBC/UKQCD 10A  &\cite{Aoki:2010dy}   &2+1&\gA&\soso&\soso&\good			&1.204(7)(25)					&                               		\\
BMW 10         &\cite{Durr:2010hr}         &2+1&\gA&\good &\tbg&\tbg			&{1.192(7)(6)}					&                               		\\
MILC 09A       &\cite{Bazavov:2009fk}&2+1&\rC&\soso&\tbg&\tbg			&                                               &1.198(2)($^{\hspace{0.01cm}+6}_{-8}$)	\\
MILC 09        &\cite{Bazavov:2009bb}&2+1&\gA&\soso&\tbg&\tbg			&                                               &1.197(3)($^{\;+6}_{-13}$)		\\
Aubin 08       &\cite{Aubin:2008ie}  &2+1&\rC&\soso&\soso&\soso			&                                               &1.191(16)(17)					\\
RBC/UKQCD 08   &\cite{Allton:2008pn} &2+1&\gA&\soso&\tbr&\tbg			&1.205(18)(62)					&                                               \\
HPQCD/UKQCD 07 &\cite{Follana:2007uv}&2+1&\gA&\soso&\soso&\soso	&{1.189(2)(7)}		&                                               \\
MILC 04 &\cite{Aubin:2004fs}&2+1&\gA&\soso&\soso&\soso				&						&1.210(4)(13)				\\
&&&&&&& \\[-0.1cm]                                                                                                              
\hline
\hline
&&&&&&& \\[-0.1cm]
\end{tabular*}}\\[-2mm]
\begin{minipage}{\linewidth}
{\footnotesize 
\begin{itemize}
\item[$^\dagger$] Result with statistical error only from polynomial interpolation to the physical point.\\[-5mm]
\item[$^{\dagger\dagger}$] This work is the continuation of Aubin 08.
\end{itemize}
}
\end{minipage}
\vspace{-0.3cm}
\caption{Colour codes for the data on the ratio of decay constants: $f_K/f_\pi$ is the pure QCD isospin-symmetric ratio, while $f_{K^\pm}/f_{\pi^\pm}$ is in pure QCD including
the isospin-breaking correction. In this and previous editions~\cite{FlavourLatticeAveragingGroup:2019iem,FlavourLatticeAveragingGroupFLAG:2021npn}, old results with two red tags have been dropped.\hfill}
\label{tab:FKFpi}
\end{table}

\subsubsection{Results for $f_{K^\pm}/f_{\pi^\pm}$}

In the case of the ratio of decay constants, the data sets that meet the criteria formulated in the introduction are HPQCD 13A~\cite{Dowdall:2013rya}, ETM 14E~\cite{Carrasco:2014poa}, FNAL/MILC 17~\cite{Bazavov:2017lyh} (which updates FNAL/MILC 14A~\cite{Bazavov:2014wgs}), CalLat 20~\cite{Miller:2020xhy} and ETM 21~\cite{Alexandrou:2021bfr} with $\Nf=2+1+1$, and HPQCD/UKQCD 07~\cite{Follana:2007uv}, MILC 10~\cite{Bazavov:2010hj}, BMW 10~\cite{Durr:2010hr}, RBC/UKQCD 14B~\cite{Blum:2014tka}, BMW 16~\cite{Durr:2016ulb,Scholz:2016kcr}, QCDSF/UKQCD 16~\cite{Bornyakov:2016dzn}, and CLQCD 23~\cite{CLQCD:2023sdb} with $\Nf=2+1$ dynamical flavours.
Note that
the new entry in this edition is ETM 21 for $\Nf=2+1+1$,
which did not enter the previous FLAG average
due to its publication status,
and CLQCD 23 for $\Nf=2+1$.

CalLat 20
employs a mixed action setup with M\"obius domain-wall
valence quarks on gradient-flowed HISQ ensembles at four lattice spacings
$a = 0.06$--0.15~fm.
The valence pion mass reaches the physical point at three lattice spacings,
and the smallest valence-sea and sea pion masses are below 200~MeV.
Finite-volume corrections are studied on 
three lattice volumes at $a = 0.12$~fm and $M_\pi \sim 220$~MeV.
The extrapolation to the continuum limit and the physical point is based on
NNLO {\Ch}PT~\cite{Ananthanarayan_2018}.
A comprehensive study of systematic uncertainties is performed
by exploring several options including
the use of the mixed-action effective theory expression,
and the inclusion of N$^3$LO counter terms.
They obtain
$f_{K^\pm}/f_{\pi^\pm}=1.1942(32)_{\rm stat}(12)_{\chi}(20)_{a^2}(1)_{FV}(12)_{M}(7)_{IB}$,
where the errors are statistical, due to the extrapolation in pion and kaon masses,
extrapolation in $a^2$, finite-size effects, choice of the fitting form
and strong isospin-breaking corrections.

ETM 14E uses the twisted-mass discretization and provides a comprehensive study of the systematics by presenting results for three lattice spacings in the range 0.06--0.09 fm and for pion masses in the range 210--450 MeV.  
This makes it possible to constrain the chiral extrapolation, using both SU(2) \cite{Flynn:2008tg} {\Ch}PT and polynomial fits.
The ETM collaboration includes the spread in the central values obtained from different ans\"atze into the systematic errors.
The final result of their analysis is $\fKfpichargedr = 1.184(12)_{\rm stat+fit}(3)_{\rm Chiral}(9)_{\rm a^2}(1)_{Z_P}(3)_{FV}(3)_{IB}$ where the errors are (statistical + the error due to the fitting procedure), due to the chiral extrapolation, the continuum extrapolation, the mass-renormalization constant, the finite-volume and (strong) isospin-breaking effects.

In ETM~21~\cite{Alexandrou:2021bfr},  
the ETM collaboration presented an independent estimate of $f_K/f_\pi$
in isosymmetric QCD with 2+1+1 dynamical flavours of
the twisted-mass quarks.
Their new set of gauge ensembles reaches the physical pion mass.
The quark action includes the Sheikoleslami-Wohlert term~\cite{Sheikholeslami:1985ij}
for a better control of discretization effects.
The finite-volume effects are examined by simulating three spatial volumes,
and are corrected by SU(2) {\Ch}PT formulae~\cite{Colangelo:2005gd}.
Their new estimate
$f_K/f_\pi = 1.1995(44)_{\rm stat+fit}(7)_{\rm sys}$
is consistent with ETM 14E
with the total uncertainty reduced by a factor of $\sim$~3.5.

FNAL/MILC 17 has determined the ratio of the decay constants from a comprehensive set of HISQ ensembles with $\Nf = 2+1+1$ dynamical flavours. 
They have generated 24 ensembles for six values of the lattice spacing (0.03--0.15 fm, scale set with $f_{\pi^+}$) and with both physical and unphysical values of the light sea-quark masses, controlling in this way the systematic uncertainties due to chiral and continuum extrapolations.
With respect to FNAL/MILC 14A they have increased the statistics and added three ensembles at very fine lattice spacings, $a \simeq 0.03$ and $0.042$ fm, including for the latter case also a simulation at the physical value of the light-quark mass.
The final result of their analysis is $\fKfpichargedr=1.1950(14)_{\rm stat}($$_{-17}^{+0}$$)_{\rm a^2} (2)_{FV} (3)_{f_\pi, PDG} (3)_{EM} (2)_{Q^2}$, where the errors are statistical, due to the continuum extrapolation, finite-volume, pion decay constant from PDG, electromagnetic effects and sampling of the topological charge distribution.\footnote{To form the average in Eq.~(\ref{eq:fKfpi_direct_broken_2p1p1}), we have symmetrized the asymmetric systematic error and shifted the central value by half the difference as will be done throughout this section.}

HPQCD 13A has analyzed ensembles generated by MILC and therefore its study of $\fKfpichargedr$ is based on the same set of ensembles as FNAL/MILC 17 bar the ones at the finest lattice spacings (namely, only $a = 0.09$--0.15 fm, scale set with $f_{\pi^+}$ and relative scale set with the Wilson flow~\cite{Luscher:2010iy,Borsanyi:2012zs}) supplemented by some simulation points with heavier quark masses.
HPQCD employs a global fit based on continuum NLO SU(3) {\Ch}PT for the decay constants supplemented by a model for higher-order terms including discretization and finite-volume effects (61 parameters for 39 data points supplemented by Bayesian priors). 
Their final result is $f_{K^\pm}/f_{\pi^\pm}=1.1916(15)_{\rm stat}(12)_{\rm a^2}(1)_{FV}(10)$, where the errors are statistical, due to the continuum extrapolation, due to finite-volume effects and the last error contains the combined uncertainties from the chiral extrapolation, the scale-setting uncertainty, the experimental input in terms of $f_{\pi^+}$ and from the uncertainty in $m_u/m_d$.

Because CalLat 20,
FNAL/MILC 17 and HPQCD 13A partly share their gauge ensembles,
we assume a 100\,\% correlation among their statistical errors.
A 100\,\% correlation on the total systematic uncertainty 
is also assumed between FNAL/MILC 17 and HPQCD 13A with the HISQ valence quarks.

The discretization effects are not large, typically at the $\lesssim 1$~\% level
in HPQCD 13A, FNAL/MILC 17 and ETM21 in their simulation region of $a$.
This does not necessarily mean that $\delta(a)$ in units of the uncertainty
of the observable is small. HPQCD 13A observed that it also depends on
the choice of the input to fix the lattice scale: 
$\delta(a)$ is consistent with zero with the relative scale setting
using $r_1$ from the static potential and $w_0$ from the gradient flow,
whereas $\delta(a) \lesssim 7$ with another flow scale $\sqrt{t_0}$.\footnote{We refer to Sec.~\ref{sec:scalesetting} for detailed discussions on the scale setting and choices of the input.}
It is not surprising that CalLat 20 observed larger scaling violation of
$\lesssim 4$~\%: while they partly share gauge ensembles with HPQCD 13A
and FNAL/MILC 17, the M\"obius domain-wall action
without the tree-level $O(a^2)$ improvement is employed in their mixed action setup.

For $\Nf=2+1$,
the CLQCD collaboration obtained a new result (CLQCD 23)
by employing the tadpole-improved tree-level Symanzik gauge action
and tadpole-improved tree-level clover quark action
with one-step stout smearing for the gauge link.
They simulate three values of the lattice spacing
in a wide range $a=0.05$\,--\,0.11~fm fixed from the gradient-flow scale $w_0$.
The unitary light-quark mass reaches the physical point on the coarsest lattice.
Two additional partially-quenched masses satisfying $M_\pi L \geq 3.5$
are considered to calculate relevant two-point functions on each ensemble.
Their results for $f_K$ and $f_\pi$ are renormalized nonperturbatively
through the RI/MOM scheme.
The chiral extrapolation is based on NLO partially-quenched {\Ch}PT
for $f_\pi$ and a polynomial form for $f_K$.
Finite-volume effects are taken into account
assuming $e^{-M_\pi L}$ dependence
for data in two volumes on the two coarser lattices.
They obtain $f_K^\pm/f_\pi^\pm = 1.1907(76)_{\rm stat+fit}(17)_{\rm sys}$,
where the second error is taken from the difference
with respect to a similar analysis using the RI/SMOM scheme,
which shows larger discretization effects.

The results BMW 16 and QCDSF/UKQCD 16 are also eligible to enter the FLAG average.
BMW 16 has analyzed the decay constants evaluated for 47 gauge ensembles generated using tree-level clover-improved fermions with two HEX-smearings and the tree-level Symanzik-improved gauge action. 
The ensembles correspond to five values of the lattice spacing (0.05--0.12 fm, scale set by $\Omega$ mass), to pion masses in the range 130--680 MeV and to values of the lattice size from $1.7$ to $5.6$ fm, obtaining a good control over the interpolation to the physical mass point and the extrapolation to the continuum and infinite volume limits.

QCDSF/UKQCD 16 has used the nonperturbatively ${\cal{O}}(a)$-improved clover action for the fermions (mildly stout-smeared) and the tree-level Symanzik action for the gluons.
Four values of the lattice spacing (0.06--0.08 fm) have been simulated with pion masses down to $\sim 220$ MeV and values of the lattice size in the range 2.0--2.8 fm.
The decay constants are evaluated using an expansion around the symmetric SU(3) point $m_u = m_d = m_s = (m_u + m_d + m_s)^{phys} / 3$.

Note, that for $\Nf=2+1$ MILC 10 and HPQCD/UKQCD 07 are based on staggered fermions, BMW 10, BMW 16, QCDSF/UKQCD 16 and CLQCD 23 have used improved Wilson fermions, and RBC/UKQCD 14B's result is based on the domain-wall formulation. 
In contrast to RBC/UKQCD 14B, BMW 16 and CLQCD 23, the other simulations are for unphysical values of the light-quark masses (corresponding to smallest pion masses in the range 220--260 MeV in the case of MILC 10, HPQCD/UKQCD 07, and QCDSF/UKQCD 16) and, therefore, slightly more sophisticated extrapolations needed to be controlled.
Various ans\"atze for the mass and cutoff dependence comprising SU(2) and SU(3) {\Ch}PT or simply polynomials were used and compared in order to estimate the model dependence.
While BMW 10, RBC/UKQCD 14B, QCDSF/UKQCD 16, and CLQCD 23 are entirely independent computations, subsets of the MILC gauge ensembles used by MILC 10 and HPQCD/UKQCD 07 are the same.
MILC 10 is certainly based on a larger and more advanced set of gauge configurations than HPQCD/UKQCD 07. 
This allows them for a more reliable estimation of systematic effects. 
In this situation, we consider both statistical and systematic uncertainties to be correlated.

Before determining the average for $f_{K^\pm}/f_{\pi^\pm}$, which should be used for applications to Standard Model phenomenology, we apply the strong-isospin correction individually to all those results that have been published only in the isospin-symmetric limit, i.e.,~BMW 10, HPQCD/UKQCD 07 and RBC/UKQCD 14B at $\Nf = 2+1$.
To this end, as in the previous editions of the FLAG reviews \cite{Aoki:2013ldr,Aoki:2016frl,FlavourLatticeAveragingGroup:2019iem,FlavourLatticeAveragingGroupFLAG:2021npn}, we make use of NLO SU(3) {\Ch}PT~\cite{Gasser:1984gg,Cirigliano:2011tm}, which predicts
\begin{equation}\label{eq:convert}
	\fKfpicharged = \frac{f_K}{f_\pi} ~ \sqrt{1 + \delta_\text{SU(2)}} ~ ,
\end{equation}
where~\cite{Cirigliano:2011tm}
\begin{equation}\label{eq:iso}
 \begin{array}{rcl}
	 \delta_\text{SU(2)}& \approx&
	\sqrt{3}\,\epsilon_\text{SU(2)}
	\left[-\frac{4}{3} \left(f_K/f_\pi-1\right)+\frac 2{3 (4\pi)^2 f_0^2}
        \left(M_K^2-M_\pi^2-M_\pi^2\ln\frac{M_K^2}{M_\pi^2}\right)
        \right]\,.
  \end{array}
 \end{equation}
We use as input $\epsilon_\text{SU(2)} = \sqrt{3} / (4 R)$ with the FLAG result for $R$ of Eq.~(\ref{eq:RQres}), $F_0 = f_0 / \sqrt{2} = 80\,(20)$ MeV,
$M_\pi = 135$ MeV and $M_K = 495$ MeV (we decided to choose a conservative uncertainty on $f_0$ in order to reflect the magnitude of potential higher-order 
corrections).
The results are reported in Tab.~\ref{tab:correctedfKfPi}, where in the last column the last error is due to the isospin correction (the remaining errors are quoted in the same order as in the original data).

\begin{table}[!htb]
\begin{center}
\begin{tabular}{llll}
\hline\hline\\[-4mm]
		& $f_K/f_\pi$	& $\delta_\text{SU(2)}$ & $f_{K^\pm}/f_{\pi^\pm}$\\
\hline\\[-4mm]
HPQCD/UKQCD 07	& 1.189(2)(7)	& $-$0.0038(6) & 1.187(2)(7)(2)\\
BMW 10		& 1.192(7)(6)	& $-$0.0039(6) & 1.190(7)(6)(2)\\
RBC/UKQCD 14B	& 1.1945(45)	& $-$0.0039(6) & 1.1921(45)(24)\\
\hline\hline
\end{tabular}
\caption{Values of the isospin-breaking correction $\delta_\text{SU(2)}$ applied to the lattice data for $f_K/f_\pi$, entering the FLAG average at $\Nf=2+1$, for obtaining the corrected charged ratio $f_{K^\pm}/f_{\pi^\pm}$.
The last error in the last column is due to a 100\,\% uncertainty assumed for
$\delta_\text{SU(2)}$ from SU(3) {\Ch}PT.
}
\label{tab:correctedfKfPi}
\end{center}
\end{table}

For $\Nf=2+1+1$, HPQCD~\cite{Dowdall:2013rya}, FNAL/MILC~\cite{Bazavov:2017lyh} and ETM~\cite{Giusti:2017xrv} estimate a value for $\delta_\text{SU(2)}$ equal to $-0.0054(14)$, $-0.0052(9)$ and $-0.0073(6)$, respectively.
Note that the ETM result is obtained using the insertion of the isovector scalar current according to the expansion method of Ref.~\cite{deDivitiis:2011eh}, while the HPQCD and FNAL/MILC results correspond to the difference between the values of the decay constant ratio extrapolated to the physical $u$-quark mass $m_u$ and to the average $(m_u + m_d) / 2$ light-quark mass. 

To remain on the conservative side, we add a $100 \%$ error to the correction based on SU(3) {\Ch}PT. 
For further analyses, we add (in quadrature) such an uncertainty to the systematic error (see Tab.~\ref{tab:correctedfKfPi}).

\begin{figure}[t]
\vspace{0.2cm}
\begin{center}
\hspace{0.5cm}\includegraphics[height=9.0cm]{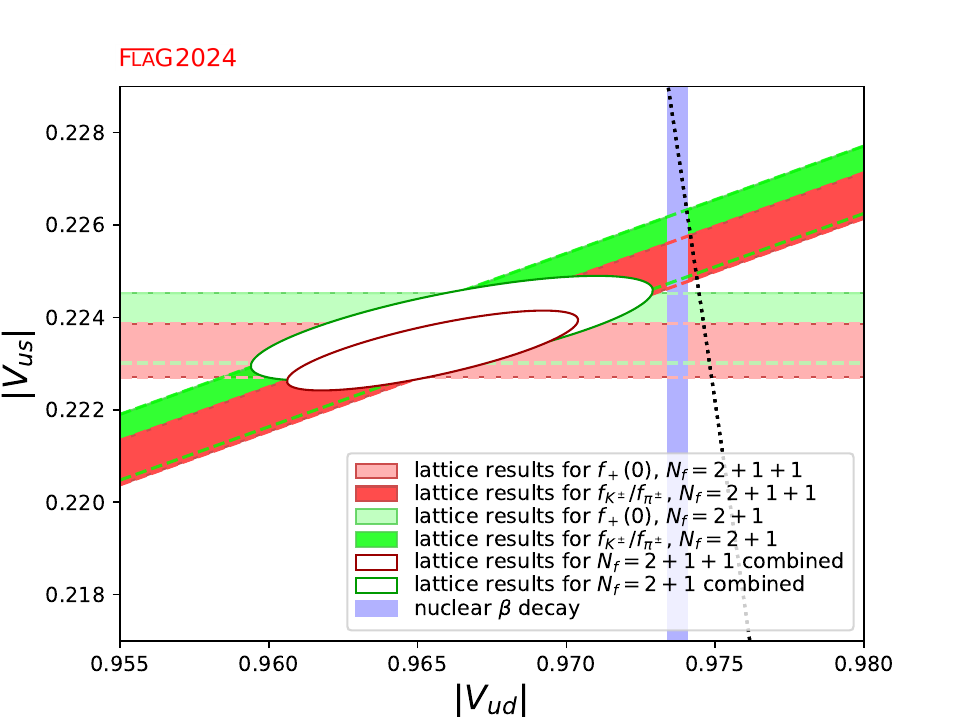}
\end{center}  
\caption{\label{fig:VusVersusVud} The plot compares the information for $|V_{ud}|$, $|V_{us}|$ obtained using lattice QCD for $\Nf = 2+1$ and $\Nf = 2+1+1$
with $|V_{ud}|$ extracted from nuclear $\beta$ transitions Eq.~(\ref{eq:Vud beta}).
The black dotted line indicates the correlation between $|V_{ud}|$ and $|V_{us}|$ that follows if the CKM-matrix is unitary.}
\end{figure}

\begin{figure}[h]
\vspace{0.2cm}
\begin{center}
\hspace{0.5cm}\includegraphics[height=9.0cm]{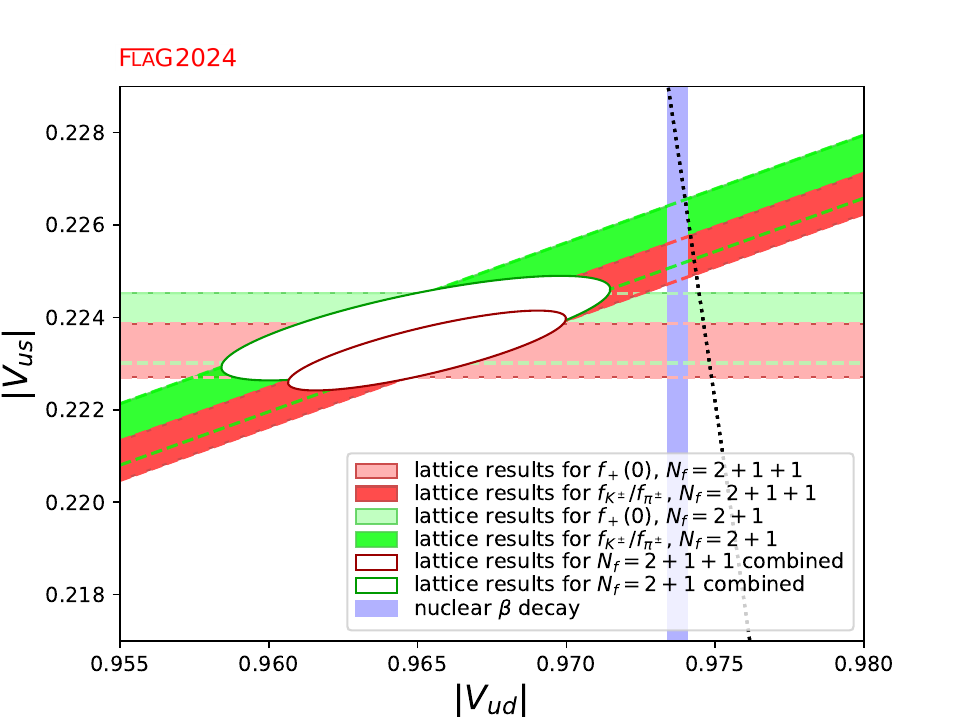}
\end{center}  

\caption{\label{fig:VusVersusVud_em-lat}
Same as Fig.~\protect\ref{fig:VusVersusVud}
but with $|V_{us}|/|V_{ud}|$ obtained using Eq.~(\ref{eq:VusVud_new}).
}
\end{figure}

Using the results of Tab.~\ref{tab:correctedfKfPi} for $\Nf = 2+1$ we obtain
\begin{align}
    \label{eq:fKfpi_direct_broken_2p1p1} 
&\mbox{direct},\,\Nf=2+1+1:&\FLAGAVBEGIN f_{K^\pm} / f_{\pi^\pm} & =  1.1934(19)\FLAGAVEND     && \Refs~\mbox{\cite{Dowdall:2013rya,Carrasco:2014poa,Bazavov:2017lyh,Miller:2020xhy,Alexandrou:2021bfr}},        \\
    \label{eq:fKfpi_direct_broken_2p1}                                                                             
&\mbox{direct},\,\Nf=2+1:  &\FLAGAVBEGIN f_{K^\pm} / f_{\pi^\pm} & =  1.1916(34)\FLAGAVEND     && \Refs~\mbox{\cite{Follana:2007uv,Bazavov:2010hj,Durr:2010hr,Blum:2014tka,Durr:2016ulb,Bornyakov:2016dzn,CLQCD:2023sdb}}, 
\end{align}
for QCD with broken isospin.

The averages obtained for $f_+(0)$ and $\fKfpichargedr$ at $\Nf=2+1$ and $\Nf=2+1+1$ [see Eqs.~(\ref{eq:fplus_direct_2p1p1}-\ref{eq:fplus_direct_2p1}) and (\ref{eq:fKfpi_direct_broken_2p1p1}-\ref{eq:fKfpi_direct_broken_2p1})] exhibit a precision better than $\sim 0.3 \%$.
At such a level of precision, QED effects cannot be ignored, and a consistent lattice treatment of both QED and QCD effects in leptonic and semileptonic decays becomes mandatory.

\subsubsection{Extraction of $|V_{ud}|$ and $|V_{us}|$}

It is instructive to convert the averages for $f_+(0)$ and $\fKfpichargedr$ into a corresponding range for the CKM matrix elements $|V_{ud}|$ and $|V_{us}|$, using the relations in Eq.~(\ref{eq:products}). 
Consider first the results for $\Nf=2+1+1$. 
The average for $f_+(0)$ in Eq.~(\ref{eq:fplus_direct_2p1p1}) is mapped into the interval $|V_{us}|=0.22328(58)$,
depicted as a horizontal red band in Fig.~\ref{fig:VusVersusVud}.
That for $\fKfpichargedr$ in Eq.~(\ref{eq:fKfpi_direct_broken_2p1p1})
is converted into $|V_{us}|/|V_{ud}|= 0.23126(50)$
using the result for $|V_{us}/V_{ud}|(\fKfpichargedr)$ in Eq.~(\ref{eq:products}),
shown as a tilted red band.
The red ellipse is the intersection of these two bands and represents the 68\% likelihood contour, obtained by treating the above two results as independent measurements. 
Repeating the exercise for $\Nf=2+1$ leads to the green ellipse.\footnote{Note that the ellipses shown in Fig.~5 of both Ref.~\cite{Colangelo:2010et} and Ref.~\cite{Aoki:2013ldr} correspond instead to the 39\% likelihood contours. Note also that in Ref.~\cite{Aoki:2013ldr} the likelihood was erroneously stated to be $68 \%$ rather than $39 \%$.}
The vertical band shows $|V_{ud}|$ from nuclear $\beta$ decay, Eq.~(\ref{eq:Vud beta}).
The PDG value (\ref{eq:Vud beta}) indicates a tension with 
both the $\Nf=2+1+1$ and $\Nf=2+1$ results from lattice QCD.
  
As we mentioned, the isospin corrections are becoming relevant
for the extraction of the CKM elements at the current precision of
lattice QCD inputs.
We obtain $|V_{us}|/|V_{ud}|=0.23131(45)$
by taking the average of $f_K/f_\pi$ in isosymmetric QCD
and combining it with the value for
$|V_{us}|f_K/|V_{ud}|f_\pi$ in Eq.~(\ref{eq:VusVud_new}).
This estimate plotted in Fig.~\ref{fig:VusVersusVud_em-lat}
is consistent with that obtained from Eq.~(\ref{eq:products})
using the isospin corrections from ChPT.
Unlike the corrections from ChPT, the accuracy of the isospin corrections
from lattice QCD can be readily improved by more realistic simulations
and higher statistics, further sharpening the comparisons shown in the figure.

\subsection{Tests of the Standard Model}\label{sec:testing}
  
In the Standard Model, the CKM matrix is unitary. In particular, the elements of the first row obey
\be
   \label{eq:CKM unitarity}
   |V_u|^2\equiv |V_{ud}|^2 + |V_{us}|^2 + |V_{ub}|^2 = 1\fs
\ee 
The tiny contribution from $|V_{ub}|$ is known much better than needed in the present context: $|V_{ub}|= 3.82(24) \times 10^{-3}$~\cite{ParticleDataGroup:2022pth}.\footnote{See also Sec.~\ref{sec:Vub} for our determination of $|V_{ub}|$.}
In the following, we test the first row unitarity Eq.~(\ref{eq:CKM unitarity}) by calculating $|V_u|^2$ and by analyzing the lattice data within the Standard Model.

In Fig.~\ref{fig:VusVersusVud}, the correlation between $|V_{ud}|$ and $|V_{us}|$ imposed by the unitarity of the CKM matrix is indicated by a dotted line (more precisely, in view of the uncertainty in $|V_{ub}|$, the correlation corresponds to a band of finite width, but the effect is too small to be seen here).
The plot shows that there is a tension with unitarity in the data for $\Nf = 2 + 1 + 1$: Numerically, the outcome for the sum of the squares of the first row of the CKM matrix reads $|V_u|^2 = 0.9820(65)$, which deviates from unity at the level of $\simeq 2.8$ standard deviations. 
Still, it is fair to say that at this level the Standard Model passes a nontrivial test
on the kaon (semi)leptonic and pion leptonic decays.

The test sharpens considerably by
combining the lattice results for $f_+(0)$
with the $\beta$ decay value of $|V_{ud}|$:
$f_+(0)$ for $\Nf = 2 + 1 +1$ in Eq.~(\ref{eq:fplus_direct_2p1p1})
and the PDG estimate of $|V_{ud}|$ in Eq.~(\ref{eq:Vud beta})
lead to $|V_u|^2 = 0.99802(66)$, which also shows 
a $\simeq 3.0~\sigma$ deviation with unitarity.
On the other hand,
unitarity is fulfilled (1.7~$\sigma$) with $\fKfpichargedr$
and $|V_{ud}|$~(\ref{eq:Vud beta}) ($|V_u|^2 = 0.99888(67)$).
Note that
the uncertainties on $|V_u|^2$ coming from the error of $|V_{ud}|$
is larger by a factor of about three than that from $|V_{us}|$.

The situation is similar for $\Nf=2+1$: with the lattice data alone one has $|V_u|^2 = 0.9835(89)$, which is consistent with unity at the level of $\simeq 1.8$ standard deviations.
The lattice results for $f_+(0)$ in Eqs.~(\ref{eq:fplus_direct_2p1}) with the PDG value of $|V_{ud}|$~(\ref{eq:Vud beta}) lead to $|V_u|^2 = 0.99824(69)$, implying a $\simeq 2.5\,\sigma$ deviation from unitarity,
whereas the deviation is reduced to 1.4\,$\sigma$
with $\fKfpichargedr$ in Eq.~(\ref{eq:fKfpi_direct_broken_2p1})
($|V_u|^2 = 0.99903(72)$).

\subsection{Analysis within the Standard Model} \label{sec:SM} 
 
The Standard Model implies that the CKM matrix is unitary. 
The precise experimental constraints quoted in Eq.~(\ref{eq:products}) and the unitarity condition Eq.~(\ref{eq:CKM unitarity}) then reduce the four quantities $|V_{ud}|,|V_{us}|,f_+(0),\fKfpichargedr$ to a single unknown: any one of these determines the other three within narrow uncertainties.

\begin{figure}[!htb]
\psfrag{y}{\tiny $\star$}
\begin{center}
\vspace{-0.5cm} 
\hspace*{-1.2cm} 
\includegraphics[height=10.0cm]{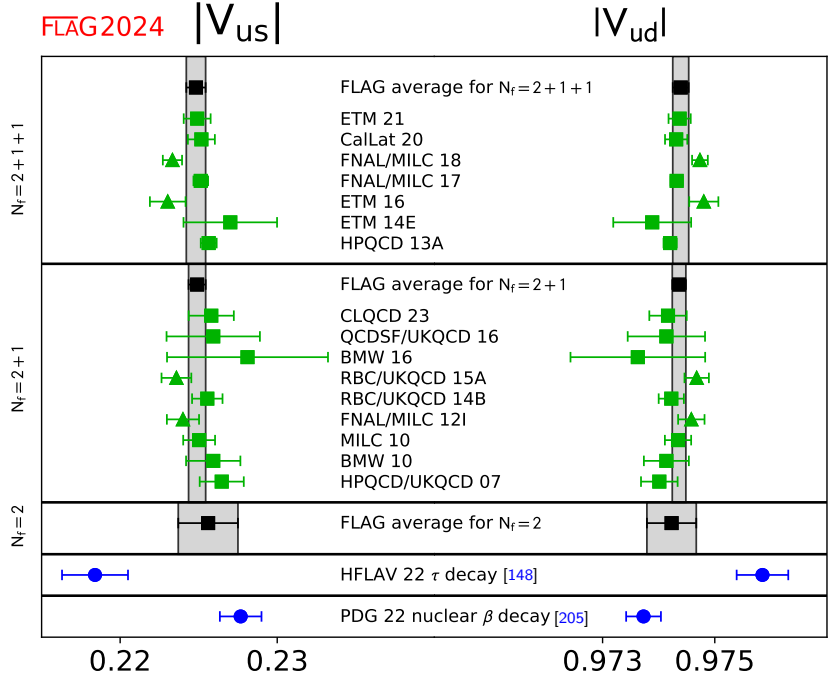}
\end{center}
\vspace{-2.47cm}\hspace{8.31cm}\parbox{6cm}{
\sffamily\tiny  \cite{HFLAV:2022esi}\\

\vspace{0.36em}\hspace{3.0em}\cite{ParticleDataGroup:2022pth}
}
\vspace{0.75cm}
\caption{\label{fig:Vus Vud} Results for $|V_{us}|$ and $|V_{ud}|$ that follow from the lattice data for $f_+(0)$ (triangles) and $\fKfpichargedr$ (squares), on the basis of the assumption that the CKM matrix is unitary. 
The black square and the grey band represent the average for each value of $\Nf$.
For comparison, the figure also indicates the results obtained if the data on nuclear $\beta$ decay and inclusive hadronic $\tau$ decay is analyzed within the Standard Model.}
\end{figure}

Numerical results for $|V_{us}|$ and $|V_{ud}|$ are listed in Tab.~\ref{tab:Vus},
where we restrict ourselves to those determinations that enter the FLAG average in
Sec.~\ref{sec:Direct}
(the error in the experimental numbers used to convert the values of $f_+(0)$
and $\fKfpichargedr$ into values for $|V_{us}|$ is included in the statistical error).
As Fig.~\ref{fig:Vus Vud} shows, the results obtained for $|V_{us}|$ and $|V_{ud}|$ from the data on $\fKfpichargedr$ (squares) are consistent with the determinations via $f_+(0)$ (triangles), while there is a tendency that 
$|V_{us}|$ ($|V_{ud}|$) from $f_+(0)$ is systematically smaller (larger)
than that from $\fKfpichargedr$.

\begin{table}[!htb]
\centering
\noindent
\begin{tabular*}{\textwidth}[l]{@{\extracolsep{\fill}}lclcll}
Collaboration & Ref. &\rule{0.5cm}{0cm}$\Nf$&from&\rule{0.6cm}{0cm}$|V_{us}|$&\rule{0.6cm}{0cm}$|V_{ud}|$\\
&&&&& \\[-2ex]
\hline \hline &&&&&\\[-2ex]
FNAL/MILC 18 & \cite{Bazavov:2018kjg}    & $2+1+1$ & $f_+(0)$ \rule{0cm}{0.45cm}         & 0.22333(55)(28) & 0.97474(13)(6)  \\
ETM 16       & \cite{Carrasco:2016kpy}   & $2+1+1$ & $f_+(0)$ \rule{0cm}{0.45cm}         & 0.2230(11)(2)   & 0.97480(26)(5)  \\
ETM 21       & \cite{Alexandrou:2021bfr} & $2+1+1$ & $\fKfpichargedr$ \rule{0cm}{0.45cm} & 0.22490(85)(13) & 0.97437(20)(3)  \\
CalLat 20    & \cite{Miller:2020xhy}     & $2+1+1$ & $\fKfpichargedr$ \rule{0cm}{0.45cm} & 0.22517(65)(56) & 0.97431(15)(13) \\
FNAL/MILC 17 & \cite{Bazavov:2017lyh}    & $2+1+1$ & $\fKfpichargedr$ \rule{0cm}{0.45cm} & 0.22513(42)(21) & 0.97432(10)(5)  \\
ETM 14E      & \cite{Carrasco:2014poa}   & $2+1+1$ & $\fKfpichargedr$ \rule{0cm}{0.45cm} & 0.2270(22)(20)  & 0.97388(51)(47) \\
HPQCD 13A    & \cite{Dowdall:2013rya}    & $2+1+1$ & $\fKfpichargedr$ \rule{0cm}{0.45cm} & 0.22564(42)(29) & 0.97420(10)(7)  \\
&&&&& \\[-2ex]
\hline
&&&&& \\[-2ex]
RBC/UKQCD 15A  & \cite{Boyle:2015hfa}      & $2+1$ & $f_+(0)$ \rule{0cm}{0.45cm}         & 0.22358(89)(32) & 0.97468(20)(7)  \\
FNAL/MILC 12I  & \cite{Bazavov:2012cd}     & $2+1$ & $f_+(0)$ \rule{0cm}{0.45cm}         & 0.22400(68)(76) & 0.97458(16)(18) \\
CLQCD 23       & \cite{CLQCD:2023sdb}      & $2+1$ & $\fKfpichargedr$ \rule{0cm}{0.45cm} & 0.2258(14)(3)   & 0.97417(33)(7)  \\ 
QCDSF/UKQCD 16 & \cite{Bornyakov:2016dzn}  & $2+1$ & $\fKfpichargedr$ \rule{0cm}{0.45cm} & 0.2259(18)(23)  & 0.97414(42)(54) \\ 
BMW 16 &\cite{Durr:2016ulb,Scholz:2016kcr} & $2+1$ & $\fKfpichargedr$ \rule{0cm}{0.45cm} & 0.2281(19)(48)  & 0.9736(4)(11)   \\
RBC/UKQCD 14B  & \cite{Blum:2014tka}       & $2+1$ & $\fKfpichargedr$ \rule{0cm}{0.45cm} & 0.22555(87)(43) & 0.97422(20)(10) \\ 
MILC 10        & \cite{Bazavov:2010hj}     & $2+1$ & $\fKfpichargedr$ \rule{0cm}{0.45cm} & 0.22503(48)(89) & 0.97434(11)(21) \\
BMW 10         & \cite{Durr:2010hr}        & $2+1$ & $\fKfpichargedr$ \rule{0cm}{0.45cm} & 0.2259(13)(11)  & 0.97414(30)(26) \\
HPQCD/UKQCD 07 & \cite{Follana:2007uv}     & $2+1$ & $\fKfpichargedr$ \rule{0cm}{0.45cm} & 0.2265(5)(13)   & 0.97401(11)(31) \\
&&&&& \\[-2ex]
\hline \hline 
\end{tabular*}
\caption{\label{tab:Vus} Values of $|V_{us}|$ and $|V_{ud}|$ obtained from the lattice determinations of either $f_+(0)$ or $\fKfpichargedr$ assuming CKM unitarity. The first number in brackets represents the statistical error including the experimental uncertainty, whereas the second is the systematic one.
}
\end{table} 

In order to calculate the average of $|V_{us}|$ for $\Nf=2+1+1$,
we consider the data both for $f_+(0)$ and $\fKfpichargedr$, treating ETM 16 and ETM 14E on the one hand and FNAL/MILC 18, CalLat 20, FNAL/MILC 17, and HPQCD 13A on the other hand, as statistically correlated according to the prescription of Sec.~\ref{sec:error_analysis}. 
We obtain $|V_{us}|=0.22483(61)$,
where the error is stretched
by a factor $\sqrt{\chi^2/{\rm dof}} \sim \sqrt{2.0}$.
This result is indicated on the left hand side of Fig.~\ref{fig:Vus Vud} by the narrow vertical band. 
In the case $\Nf = 2+1$, we consider MILC 10, FNAL/MILC 12I and HPQCD/UKQCD 07 on the one hand, and RBC/UKQCD 14B and RBC/UKQCD 15A on the other hand, as mutually statistically correlated, since the analysis in the two cases starts from partially the same set of gauge ensembles.
In this way, we arrive at $|V_{us}| = 0.22490(54)$ with $\chi^2/{\rm dof} \simeq 0.7$. 
The figure shows that the results obtained for the data with $\Nf=2+1$ and $\Nf=2+1+1$ are consistent with each other.
However, the larger error for $\Nf=2+1+1$ due to the stretch factor
$\sqrt{\chi^2/{\rm dof}}$ suggests
a slight tension between the estimates
from the semileptonic and leptonic decays.

We take the average of $|V_{ud}|$ similarly.
Again, the result $|V_{ud}|=0.97439(14)$ for $\Nf=2+1+1$
is perfectly consistent with the values $|V_{ud}|=0.97436(12)$ obtained from the data with $\Nf=2+1$.
These values are consistent with Eq.~(\ref{eq:Vud beta})
from the superallowed nuclear transitions within 2 $\sigma$.

As mentioned in Sec.~\ref{sec:Exp},
the HFLAV value of $|V_{us}|$ from the inclusive hadronic $\tau$ decays
differs from those obtained from the kaon decays
by about three standard deviations.
Assuming the first row unitarity defined in Eq.~(\ref{eq:CKM unitarity})
leads to a larger value of $|V_{ud}|$ than those
from the kaon and nuclear decays.

\begin{table}[!htb]
\centering
\begin{tabular*}{\textwidth}[l]{@{\extracolsep{\fill}}llllll}
  \rule[-0.2cm]{0cm}{0.5cm}& Ref. & \rule{0.3cm}{0cm} $|V_{us}|$&
  \rule{0.3cm}{0cm} $|V_{ud}|$ \\
\\[-2ex]
\hline \hline
\\[-2ex]
$\Nf= 2+1+1$& &\rule{0cm}{0.4cm} 0.22483(61) &  0.97439(14) \\
\\[-2ex]
\hline
$\Nf= 2+1$&   &\rule{0cm}{0.4cm} 0.22490(54) &  0.97436(12) \\
\\[-2ex]
\hline\hline
\\[-2ex]
nuclear $\beta$ decay & \cite{ParticleDataGroup:2022pth}& 0.2277(13) & 0.97373(31) \\
\\[-2ex]
inclusive $\tau$ decay &\cite{HFLAV:2022esi} & 0.2184(21) & 0.97585(47)   \\
\
\\[-2ex]
\hline\hline
\end{tabular*}
\caption{\label{tab:Final results}The upper half of the table shows the results for $|V_{us}|$ and $|V_{ud}|$
  from the analysis of the kaon and pion decays within the Standard Model. 
For comparison, the lower half lists the values that follow if the lattice results are replaced by the experimental results on nuclear $\beta$ decay and inclusive hadronic $\tau$ decay, respectively.}
\end{table}

\subsection{Direct determination of $f_{K^\pm}$ and $f_{\pi^\pm}$}\label{sec:fKfpi}

It is useful for flavour-physics studies to provide not only the lattice average of $f_{K^\pm} / f_{\pi^\pm}$, but also the average of the decay constant $f_{K^\pm}$. 
The case of the decay constant $f_{\pi^\pm}$ is different, since the 
the PDG value~\cite{Patrignani:2016xqp} of this quantity, based on the use of the value of $|V_{ud}|$ obtained from superallowed nuclear $\beta$ decays \cite{Hardy:2016vhg},
 is often used for setting the scale in lattice QCD.
However, the physical scale can be set in different ways, namely, by using as input the mass of the $\Omega$ baryon ($m_\Omega$) or the $\Upsilon$-meson spectrum ($\Delta M_\Upsilon$), which are less sensitive to the uncertainties of the chiral extrapolation in the light-quark mass with respect to $f_{\pi^\pm}$.\footnote{See Sec.~\ref{sec:scalesetting} for detailed discussions.} 
In such cases, the value of the decay constant $f_{\pi^\pm}$ becomes a direct prediction of the lattice-QCD simulations.
Therefore, it is interesting to provide also the average of the decay constant $f_{\pi^\pm}$, obtained when the physical scale is set through another hadron observable, in order to check the consistency of different scale-setting procedures.

Our compilation of the values of $f_{\pi^\pm}$ and $f_{K^\pm}$ with the corresponding colour code is presented in Tab.~\ref{tab:FK Fpi}.
The new entry in this edition is CLQCD 23 for $\Nf=2+1$.

In comparison to the case of $f_{K^\pm} / f_{\pi^\pm}$, we have added two columns indicating which quantity is used to set the physical scale and the possible use of a renormalization constant for the axial current.
For several lattice formulations, the use of the nonsinglet axial-vector Ward identity allows us to avoid the use of any renormalization constant.

One can see that the determinations of $f_{\pi^\pm}$ and $f_{K^\pm}$ suffer from larger uncertainties than those of the ratio $f_{K^\pm} / f_{\pi^\pm}$, which is less sensitive to various systematic effects (including the uncertainty of a possible renormalization constant) and, moreover, is not exposed to the uncertainties of the procedure used to set the physical scale.

According to the FLAG rules, for $\Nf = 2 + 1 + 1$ four data sets can form the average of $f_{K^\pm}$ only: ETM 21 \cite{Alexandrou:2021bfr}, ETM 14E \cite{Carrasco:2014poa}, FNAL/MILC 14A \cite{Bazavov:2014wgs}, and HPQCD 13A \cite{Dowdall:2013rya}.
Following the same procedure already adopted in Sec.~\ref{sec:Direct} for the ratio of the decay constants,
we assume 100\,\% statistical and systematic correlation between
FNAL/MILC 14A and HPQCD 13A.
For $\Nf = 2 + 1$ four data sets can form the average of $f_{\pi^\pm}$ and $f_{K^\pm}$ : CLQCD 23 \cite{CLQCD:2023sdb}, RBC/UKQCD 14B \cite{Blum:2014tka} (update of RBC/UKQCD 12), HPQCD/UKQCD 07 \cite{Follana:2007uv}, and MILC 10 \cite{Bazavov:2010hj}, which is the latest update from the MILC program.
We consider HPQCD/UKQCD 07 and MILC 10 as statistically and systematically correlated and use the prescription of Sec.~\ref{sec:error_analysis} to form an average.

\begin{figure}[!htb]
\begin{center}
\includegraphics[height=9.0cm]{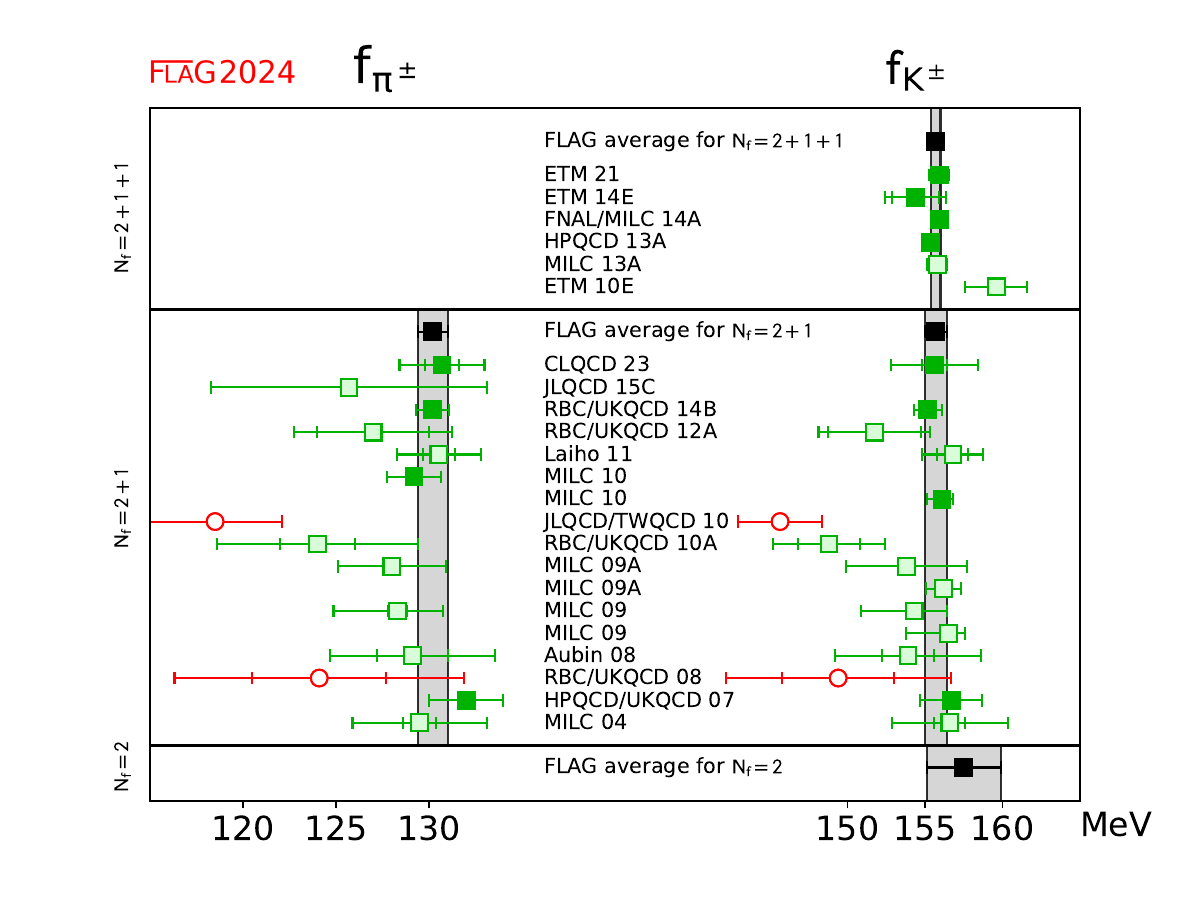}
\end{center}
\vspace{-0.75cm}
\caption{\label{fig:latticedata_decayconstants}
Values of $f_{\pi^\pm}$ and $f_{K^\pm}$.
The black squares and grey bands indicate our averages in Eqs.~(\ref{eq:fPi}) and (\ref{eq:fK}).
}
\end{figure}

Thus, our averages read
\begin{align}
  \label{eq:fPi}
&\Nf = 2 + 1:     &\FLAGAVBEGIN f_{\pi^\pm}&= 130.2 ~ (0.8)\FLAGAVEND  ~ \mbox{MeV} &&\Refs~\mbox{\cite{Follana:2007uv,Bazavov:2010hj,Blum:2014tka,CLQCD:2023sdb}},\\ \nonumber 
                \\                                                       
&\Nf = 2 + 1 + 1: &\FLAGAVBEGIN f_{K^\pm} & = 155.7 ~ (0.3)\FLAGAVEND  ~ \mbox{MeV} &&\Refs~\mbox{\cite{Dowdall:2013rya,Bazavov:2014wgs,Carrasco:2014poa,Alexandrou:2021bfr}}         ,\\ 
&\Nf = 2 + 1:     &\FLAGAVBEGIN f_{K^\pm} & = 155.7 ~ (0.7)\FLAGAVEND  ~ \mbox{MeV} &&\Refs~\mbox{\cite{Follana:2007uv,Bazavov:2010hj,Blum:2014tka,CLQCD:2023sdb}},\label{eq:fK}
\\ \nonumber
\end{align}
The lattice results of Tab.~\ref{tab:FK Fpi} and our averages in Eqs.~(\ref{eq:fPi})--(\ref{eq:fK}) are reported in Fig.~\ref{fig:latticedata_decayconstants}. 
Note that the FLAG average of $f_{K^\pm}$ for $\Nf = 2 + 1 + 1$ is based on calculations in which $f_{\pi^\pm}$ is used to set the lattice scale, while the $\Nf = 2 + 1$ average does not rely on that. 

\begin{table}[!htb]
       {\centering
\vspace{2.0cm}{\footnotesize\noindent
\begin{tabular*}{\textwidth}[l]{@{\extracolsep{\fill}}l@{\hspace{1mm}}r@{\hspace{1mm}}l@{\hspace{1mm}}l@{\hspace{1mm}}l@{\hspace{1mm}}l@{\hspace{1mm}}l@{\hspace{3mm}}l@{\hspace{1mm}}l@{\hspace{1mm}}l@{\hspace{5mm}}l@{\hspace{1mm}}l}
Collaboration & Ref. & $\Nf$ &
\hspace{0.15cm}\begin{rotate}{40}{publication status}\end{rotate}\hspace{-0.15cm}&
\hspace{0.15cm}\begin{rotate}{40}{chiral extrapolation}\end{rotate}\hspace{-0.15cm}&
\hspace{0.15cm}\begin{rotate}{40}{continuum extrapolation}\end{rotate}\hspace{-0.15cm}&
\hspace{0.15cm}\begin{rotate}{40}{finite-volume errors}\end{rotate}\hspace{-0.15cm}& 
\hspace{0.15cm}\begin{rotate}{40}{renormalization}\end{rotate}\hspace{-0.15cm}&
\hspace{0.05cm}\begin{rotate}{40}{physical scale}\end{rotate}\hspace{-0.15cm}&\rule{0cm}{0cm}
\hspace{0.0cm}\begin{rotate}{40}{isospin breaking}\end{rotate}\hspace{-0.15cm}&\rule{0.5cm}{0cm}
$f_{\pi^\pm}$&\rule{0.5cm}{0cm}$f_{K^\pm}$ \\
&&&&&&& \\[-0.1cm]
\hline
\hline
&&&&&&& \\[-0.1cm]
ETM 21 &\cite{Alexandrou:2021bfr}&2+1+1&\gA&\good&\good&\good&na&$f_\pi$&&--&{155.92(62)(9)$^{\dagger}$}\\
ETM 14E &\cite{Carrasco:2014poa}&2+1+1&\gA&\soso&\good&\soso&na&$f_\pi$&&--&{154.4(1.5)(1.3)}\\
FNAL/MILC 14A&\cite{Bazavov:2014wgs}&2+1+1&\gA&\good&\good&\good&na&$f_\pi$&&--&{155.92(13)($_{-23}^{+34}$)}\\
HPQCD 13A&\cite{Dowdall:2013rya}&2+1+1&\gA&\good&\soso&\good&na&$f_\pi$&&--&{155.37(20)(27)}\\
MILC 13A&\cite{Bazavov:2013cp}&2+1+1&\gA&\good&\soso&\good&na&$f_\pi$&&--&155.80(34)(54)\\
ETM 10E &\cite{Farchioni:2010tb}&2+1+1&\rC&\soso&\soso&\soso&na&$f_\pi$&\checkmark&--&159.6(2.0)\\
&&&&&&& \\[-0.1cm]
\hline
&&&&&&& \\[-0.1cm]
CLQCD 23        &\cite{CLQCD:2023sdb}&2+1&\gA&\good&\soso&\good& NPR& $w_0$& & 130.7(0.9)(2.1) & 155.6(0.8)(2.7)\\
JLQCD 15C       &\cite{Fahy:2015xka}&2+1&\rC&\soso&\tbg&\tbg&NPR&$t_0$& &125.7(7.4)$_{\rm stat}$&\\
RBC/UKQCD 14B   &\cite{Blum:2014tka}&2+1&\gA&\good&\good&\good&NPR&$m_\Omega$ &\checkmark& 130.19(89) & 155.18(89) \\
RBC/UKQCD 12   &\cite{Arthur:2012opa}&2+1&\gA&\tbg&\soso&\good&NPR&$m_\Omega$ &\checkmark& 127.1(2.7)(2.7)& 152.1(3.0)(1.7) \\
Laiho 11       &\cite{Laiho:2011np}   &2+1&\rC&\soso&\good&\soso&na&${}^{\dagger\dagger}$ && $130.53(87)(2.10)$&$156.8(1.0)(1.7)$\\
MILC 10 &\cite{Bazavov:2010hj}&2+1&\rC&\soso&\good&\good&na&${}^{\dagger\dagger}$ & &{129.2(4)(1.4)}&--\\
MILC 10 &\cite{Bazavov:2010hj}&2+1&\rC&\soso&\good&\good&na&$f_\pi$ &&--          &{156.1(4)($_{-9}^{+6}$)}\\
JLQCD/TWQCD 10 &\cite{Noaki:2010zz}&2+1&\rC&\soso&\tbr&\tbg&na&$m_\Omega$&\checkmark&118.5(3.6)$_{\rm stat}$&145.7(2.7)$_{\rm stat}$\\
RBC/UKQCD 10A  &\cite{Aoki:2010dy} &2+1&\gA&\soso&\soso&\good&NPR&$m_\Omega$&\checkmark&124(2)(5)&148.8(2.0)(3.0)\\
MILC 09A &\cite{Bazavov:2009fk}&2+1&\rC&\soso&\tbg&\tbg &na&$\Delta M_\Upsilon$ &&128.0(0.3)(2.9)&          153.8(0.3)(3.9)\\
MILC 09A &\cite{Bazavov:2009fk}&2+1&\rC&\soso&\tbg&\tbg &na&$f_\pi$&&--&156.2(0.3)(1.1)\\
MILC 09 &\cite{Bazavov:2009bb}&2+1&\gA&\soso&\tbg&\tbg &na&$\Delta M_\Upsilon$&&128.3(0.5)($^{+2.4}_{-3.5}$)&154.3(0.4)($^{+2.1}_{-3.4}$) \\
MILC 09 &\cite{Bazavov:2009bb}&2+1&\gA&\soso&\tbg&\tbg &na&$f_\pi$&&&156.5(0.4)($^{+1.0}_{-2.7}$)\\
Aubin 08       &\cite{Aubin:2008ie} &2+1&\rC&\soso&\soso&\soso&na&$\Delta M_\Upsilon$     && 129.1(1.9)(4.0)   & 153.9(1.7)(4.4)  \\
RBC/UKQCD 08   &\cite{Allton:2008pn} &2+1&\gA&\soso&\tbr&\tbg&NPR&$m_\Omega$&\checkmark&124.1(3.6)(6.9) &        149.4(3.6)(6.3)\\
HPQCD/UKQCD 07 &\cite{Follana:2007uv}&2+1&\gA&\soso&\soso&\soso&na&$\Delta M_\Upsilon$&\checkmark& {132(2)}                & {156.7(0.7)(1.9)}\\
MILC 04 &\cite{Aubin:2004fs}&2+1&\gA&\soso&\soso&\soso&na&$\Delta M_\Upsilon$&&129.5(0.9)(3.5)     &     156.6(1.0)(3.6)\\[-1mm]
&&&&&&& \\[-0.1cm]
\hline
\hline
&&&&&&& \\[-0.1cm]
\end{tabular*}}\\[-2mm]
}

\begin{minipage}{\linewidth}
\footnotesize The label 'na' indicates the lattice calculations that do not require the use of any renormalization constant for the axial current, while the label 'NPR' signals the use of a renormalization constant calculated nonperturbatively. 
\begin{itemize}
{\footnotesize 
\item[$^{\dagger}$] We evaluated from $f_{K^\pm}/f_{\pi^\pm}$ in Tab.~\ref{tab:FKFpi}
  and their input to fix the scale $f_\pi = 130.4(2)$.
\item[$^{\dagger\dagger}$] The ratios of lattice spacings within the ensembles were determined using the quantity $r_1$.
	The conversion to physical units was made on the basis of Ref.~\cite{Davies:2009tsa}, and we note that such a determination depends on the PDG value~\cite{Patrignani:2016xqp} of the pion decay constant.\\[-5mm]
}
\end{itemize}
\end{minipage}
\caption{Colour codes for the lattice data on $f_{\pi^\pm}$ and $f_{K^\pm}$ together with information on the way the lattice spacing was converted to physical units and on whether or not an isospin-breaking correction has been applied to the quoted result (see Sec.~\ref{sec:Direct}). The numerical values are listed in MeV units. In this and previous editions~\cite{FlavourLatticeAveragingGroup:2019iem,FlavourLatticeAveragingGroupFLAG:2021npn}, old results with two red tags have been dropped.\hfill}
\label{tab:FK Fpi}
\end{table}

\clearpage

%% file: BK/BK.tex
\section{Kaon mixing}
\label{sec:BK}
Authors: P.~Dimopoulos, X. Feng, G.~Herdo\'iza\\

The mixing of neutral pseudoscalar mesons plays an important role in the understanding of the physics of quark-flavour mixing and CP violation. In this section, we discuss $K^0 - \bar K^0$ oscillations, which probe the physics of indirect CP violation. Extensive reviews on this subject can be found in  Refs.~\cite{Branco:1999fs,Sozzi:2008bookcp,Buras2020bookcp,Buchalla:1995vs,Buras:1998raa,Lellouch:2011qw}.
 The main changes in this section with respect to the FLAG 21 edition~\cite{FlavourLatticeAveragingGroupFLAG:2021npn} are  as follows: A discussion on the $\epsilon_{K}$ calculation has been added  in Sec.~\ref{sec:indCP}. An updated discussion  
 regarding new  lattice determinations of the $K \to \pi\pi$ decay amplitudes and
  related quantities is provided in  Sec.~\ref{sec:Kpipi_amplitudes}. New FLAG averages for SM and BSM bag 
  	parameters are reported  in Secs.~\ref{sec:BK lattice} and \ref{sec:Bi}, which  concern the kaon
  mixing within the Standard Model (SM) and Beyond the Standard Model (BSM), respectively.  

\subsection{Indirect CP violation and $\epsilon_{K}$ in the
  SM \label{sec:indCP}} 

Indirect CP violation arises in $K_L \rightarrow \pi \pi$ transitions
through the decay of the $\rm CP=+1$ component of $K_L$ into two pions
(which are also in a $\rm CP=+1$ state). Its measure is defined as
\be 
\epsilon_{K} \,\, = \,\, \dfrac{{\cal A} [ K_L \rightarrow
(\pi\pi)_{I=0}]}{{\cal A} [ K_S \rightarrow (\pi\pi)_{I=0}]} \,\, ,
\ee
with the final state having total isospin zero. The parameter
$\epsilon_{K}$ may also be expressed in terms of $K^0 - \bar K^0$
oscillations.   In the Standard Model, $\epsilon_{K}$ 
is given by the following expression~\cite{Buras:1998raa,Anikeev:2001rk,Nierste:2009wg,Buras:2008nn,Buras:2010pz}
\be
\epsilon_K \,\,\, = \,\,\, \exp(i \phi_\epsilon) \, \sin(\phi_\epsilon)
\left[
\frac{\text{Im}(M_{12}^\text{SD})}{\Delta M_K}
+ \frac{\text{Im}(M_{12}^\text{LD})}{\Delta M_K}
+ \frac{\text{Im}(A_0)}{\text{Re}(A_0)}
\right] ,
\label{eq:epsK}
\ee
where the various contributions can be related to:
  (i) short-distance (SD) physics given by $\Delta
  S = 2$ ``box diagrams'' involving $W^\pm$ bosons and $u,c$ and $t$
  quarks; (ii) long-distance (LD) physics from light hadrons
  contributing to the imaginary part of the dispersive amplitude
  $M_{12}$, $\text{Im}\,(M_{12}^\text{LD})$,  used in the two-component description of $K^0-\bar{K}^0$
  mixing; (iii) the imaginary part of the absorptive amplitude
  $\Gamma_{12}$ from $K^0-\bar{K}^0$ mixing which can be related to $\text{Im}(A_0)/\text{Re}(A_0)$, where $A_0$ is the $K \to
  (\pi\pi)_{I=0}$ decay amplitude, as $(\pi\pi)_{I=0}$ states provide the dominant contribution to the absorptive part of the integral in $\Gamma_{12}$. The various factors of this decomposition may vary according to phase conventions. In terms of the
  $\Delta
  S = 2$ effective Hamiltonian, ${\cal H}_\text{eff}^{\Delta S = 2}$, it is
  common to represent contribution~(i) by
\be
 \text{Im}(M_{12}^\text{SD}) \equiv \frac{1}{2M_K}\text{Im} [ \langle
   \bar{K}^0 | {\cal H}_\text{eff}^{\Delta S = 2} | K^0 \rangle]\, .
\ee
 The phase of $\epsilon_{K}$ is
given by
\be
\phi_\epsilon \,\,\, = \,\,\, \arctan \frac{\Delta M_{K}}{\Delta
  \Gamma_{K}/2} \,\,\, . 
\ee
The quantities $\Delta M_K$ and $\Delta \Gamma_K$ are the mass and
decay width differences between long- and short-lived neutral kaons.
The experimentally known values of the above quantities
are\,\cite{ParticleDataGroup:2024cfk}:
\begin{eqnarray}
\vert \epsilon_{K} \vert \,\, &=& \,\, 2.228(11) \times 10^{-3} \,\, , \label{eq:epsilonK_exp}
 \\
\phi_\epsilon \,\, &=& \,\, 43.52(5)^\circ \,\, , \label{eq:phi_epsilonK}
 \\
\Delta M_{K} \,\, &\equiv& \,\, M_{K_{L}} - M_{K_{S}} \,\, = \,\,  3.484(6) \times 10^{-12}\, {\rm MeV} \,\, ,
 \\
\Delta \Gamma_{K}  \,\, &\equiv& \,\ \Gamma_{K_{S}} - \Gamma_{K_{L}} ~\, \,\, = \,\,  7.3382(33) \times 10^{-12} \,{\rm MeV} 
\,\,, 
\end{eqnarray}
where the latter three measurements have been obtained by imposing CPT
symmetry.

We will start by discussing the short-distance effects (i) since they
 provide the dominant contribution to $\epsilon_K$. To lowest
order in the electroweak theory, the contribution to $K^0 -
  \bar K^0$ oscillations arises from the box diagrams, in which
two $W$ bosons and two ``up-type'' quarks (i.e., up, charm, top) are
exchanged between the constituent down and strange quarks of the $K$
mesons. The loop integration of the box diagrams can be performed
exactly. In the limit of vanishing external momenta and external quark
masses, the result can be identified with an effective four-fermion
interaction, expressed in terms of the effective Hamiltonian
\be
  {\cal H}_{\rm eff}^{\Delta S = 2} \,\, = \,\,
  \frac{G_F^2 M_{\rm{W}}^2}{16\pi^2} {\cal F}^0 Q^{\Delta S=2} \,\, +
  \,\, {\rm h.c.} \,\,.
  \label{eq:HDeltaS2}
\ee
In this expression, $G_F$ is the Fermi coupling, $M_{\rm{W}}$ the
$W$-boson mass, and
\be
   Q^{\Delta S=2} =
   \left[\bar{s}\gamma_\mu(1-\gamma_5)d\right]
   \left[\bar{s}\gamma_\mu(1-\gamma_5)d\right]
   \equiv O_{\rm VV+AA}-O_{\rm VA+AV} \,\, ,
\label{eq:Q1def}
\ee
is a dimension-six, four-fermion operator. The subscripts V and A denote vector ($\bar{s}\gamma_{\mu} d$) and axial-vector ($\bar{s} \gamma_{\mu} \gamma_5 d$) bilinears, respectively. The function ${\cal F}^0$
is given by 
\be
{\cal F}^0 \,\, = \,\, \lambda_c^2 S_0(x_c) \, + \, \lambda_t^2
S_0(x_t) \, + \, 2 \lambda_c  \lambda_t S_0(x_c,x_t)  \,\, , 
\label{eq:F0InamiLin}
\ee
where $\lambda_a = V^\ast_{as} V_{ad}$, and $a=c\,,t$ denotes a
flavour index. The quantities $S_0(x_c),\,S_0(x_t)$ and $S_0(x_c,x_t)$
with $x_c=m_c^2/M_{\rm{W}}^2$, $x_t=m_t^2/M_{\rm{W}}^2$ are the
Inami-Lim functions \cite{Inami:1980fz}, which express the basic
electroweak loop contributions without QCD corrections. The
contribution of the up quark, which is taken to be massless in this
approach, has been taken into account by imposing the unitarity
constraint $\lambda_u + \lambda_c + \lambda_t = 0$. 
By substituting $\lambda_c=-\lambda_u-\lambda_t$, one can rewrite ${\cal F}^0$
as~\cite{Christ:2012se,Brod:2019rzc}
\be
{\cal F}^0 \,\, = \,\, \lambda_u^2 S_0(x_c) \, + \, \lambda_t^2
[S_0(x_t)+S_0(x_c)-2S_0(x_c,x_t)] \, + \, 2 \lambda_u  \lambda_t [S_0(x_c)-S_0(x_c,x_t)]  \,\, .
\label{eq:F0InamiLin_ut}
\ee
Equations~(\ref{eq:F0InamiLin}) and (\ref{eq:F0InamiLin_ut}) are denoted as ``$c$-$t$ unitarity'' and ``$u$-$t$ unitarity'', respectively. Since $\lambda_u^2 S_0(x_c)$ is real, it does not factor into $\epsilon_K$, even when accounting for QCD corrections.

When strong interactions are included, $\Delta{S}=2$ transitions can
no longer be discussed at the quark level. Instead, the effective
Hamiltonian must be considered between mesonic initial and final
states. Since the strong coupling is large at typical hadronic scales,
the resulting weak matrix element cannot be calculated in perturbation
theory. The operator product expansion (OPE) does, however, factorize
long- and short-distance effects. For energy scales below the charm
threshold, the $K^0-\bar K^0$ transition amplitude of the effective
Hamiltonian can be expressed in terms of the $c$-$t$ unitarity framework as follows
\begin{eqnarray}
\label{eq:Heff}
\langle \bar K^0 \vert {\cal H}_{\rm eff}^{\Delta S = 2} \vert K^0
\rangle  \,\, = \,\, \frac{G_F^2 M_{\rm{W}}^2}{16 \pi^2}  
\Big [ \lambda_c^2 S_0(x_c) \eta_1  \, + \, \lambda_t^2 S_0(x_t)
  \eta_2 \, + \, 2 \lambda_c  \lambda_t S_0(x_c,x_t) \eta_3
  \Big ]  \nn \\ 
\times 
  \left(\frac{\gbar(\mu)^2}{4\pi}\right)^{-\gamma_0/(2\beta_0)}
  \exp\bigg\{ \int_0^{\gbar(\mu)} \, dg \, \bigg(
  \frac{\gamma(g)}{\beta(g)} \, + \, \frac{\gamma_0}{\beta_0g} \bigg)
  \bigg\} 
   \langle \bar K^0 \vert  Q^{\Delta S=2}_{\rm R} (\mu) \vert K^0
   \rangle \,\, + \,\, {\rm h.c.} \,\, ,
\end{eqnarray}
where $\gbar(\mu)$ and $Q^{\Delta S=2}_{\rm R}(\mu)$ are the
renormalized gauge coupling and the four-fermion operator in some
renormalization scheme. The factors $\eta_1, \eta_2$ and $\eta_3$
depend on the renormalized coupling $\gbar$, evaluated at the various
flavour thresholds $m_t, m_b, m_c$ and $ M_{\rm{W}}$, as required by
the OPE and Renormalization-Group (RG) running procedure that separate high- and low-energy
contributions. Explicit expressions can be found
in Ref.~\cite{Buchalla:1995vs} and references therein, except that $\eta_1$
and $\eta_3$ have been  calculated to NNLO in
Refs.~\cite{Brod:2011ty} and \cite{Brod:2010mj}, respectively.
We follow the same conventions for the RG equations as in
Ref.~\cite{Buchalla:1995vs}. Thus the Callan-Symanzik function and the
anomalous dimension $\gamma(\gbar)$ of $Q^{\Delta S=2}$ are defined by
\be
\dfrac{d \gbar}{d \ln \mu} = \beta(\gbar)\,,\qquad
\dfrac{d Q^{\Delta S=2}_{\rm R}}{d \ln \mu} =
-\gamma(\gbar)\,Q^{\Delta S=2}_{\rm R} \,\,,  
\label{eq:four_quark_operator_anomalous_dimensions}
\ee
with perturbative expansions
\begin{eqnarray}
\beta(g)  &=&  -\beta_0 \dfrac{g^3}{(4\pi)^2} \,\, - \,\, \beta_1
\dfrac{g^5}{(4\pi)^4} \,\, - \,\, \cdots , 
\label{eq:four_quark_operator_anomalous_dimensions_perturbative}
\\
\gamma(g)  &=&  \gamma_0 \dfrac{g^2}{(4\pi)^2} \,\, + \,\,
\gamma_1 \dfrac{g^4}{(4\pi)^4} \,\, + \,\, \cdots \,.\nn
\end{eqnarray}
We stress that $\beta_0, \beta_1$ and $\gamma_0$ are universal,
i.e., scheme independent. As for $K^0-\bar K^0$ mixing, this is usually considered
in the naive dimensional regularization (NDR) scheme of $\msbar$, and
below we specify the perturbative coefficient $\gamma_1$ in that
scheme:
\begin{eqnarray}
& &\beta_0 = 
         \left\{\frac{11}{3}N-\frac{2}{3}\Nf\right\}, \qquad
   \beta_1 = 
         \left\{\frac{34}{3}N^2-\Nf\left(\frac{13}{3}N-\frac{1}{N}
         \right)\right\}, \label{eq:RG-coefficients}\\[0.3ex]
& &\gamma_0 = \frac{6(N-1)}{N}, \qquad
         \gamma_1 = \frac{N-1}{2N} 
         \left\{-21 + \frac{57}{N} - \frac{19}{3}N + \frac{4}{3}\Nf
         \right\}\,.\nn
\end{eqnarray}
Note that for QCD the above expressions must be evaluated for $N=3$
colours, while $\Nf$ denotes the number of active quark flavours. As
already stated, Eq.~(\ref{eq:Heff}) is valid at scales below the charm
threshold, after all heavier flavours have been integrated out,
i.e., $\Nf = 3$.

In Eq.~(\ref{eq:Heff}), the terms proportional to $\eta_1,\,\eta_2$
and $\eta_3$, multiplied by the contributions containing
$\gbar(\mu)^2$, correspond to the Wilson coefficient of the OPE,
computed in perturbation theory. Its dependence on the renormalization
scheme and scale $\mu$ is canceled by that of the weak matrix element
$\langle \bar K^0 \vert Q^{\Delta S=2}_{\rm R} (\mu) \vert K^0
\rangle$. The latter corresponds to the long-distance effects of the
effective Hamiltonian and must be computed nonperturbatively. For
historical, as well as technical reasons, it is convenient to express
it in terms of the $B$-parameter $B_{K}$, defined as
\be
   B_{K}(\mu)= \frac{{\left\langle\bar{K}^0\left|
         Q^{\Delta S=2}_{\rm R}(\mu)\right|K^0\right\rangle} }{
     {\frac{8}{3} f_{K}^2 M_{K}^2}} \,\, .
   \label{eq:defBK}
\ee
The four-quark operator $Q^{\Delta S=2}(\mu)$ is renormalized at scale $\mu$
in some regularization scheme, for instance, NDR-$\msbar$. Assuming that
$B_{K}(\mu)$ and the anomalous dimension $\gamma(g)$ are both known in
that scheme, the renormalization group independent (RGI) $B$-parameter
$\hat{B}_{K}$ is related to $B_{K}(\mu)$ by the exact formula
\be
  \hat{B}_{K} = 
  \left(\frac{\gbar(\mu)^2}{4\pi}\right)^{-\gamma_0/(2\beta_0)}
  \exp\bigg\{ \int_0^{\gbar(\mu)} \, dg \, \bigg(
  \frac{\gamma(g)}{\beta(g)} \, + \, \frac{\gamma_0}{\beta_0g} \bigg)
  \bigg\} 
\, B_{K}(\mu) \,\,\, .
\ee
At NLO in perturbation theory, the above reduces to
\be
   \hat{B}_{K} =
   \left(\frac{\gbar(\mu)^2}{4\pi}\right)^{- \gamma_0/(2\beta_0)}
   \left\{ 1+\dfrac{\gbar(\mu)^2}{(4\pi)^2}\left[
   \frac{\beta_1\gamma_0-\beta_0\gamma_1}{2\beta_0^2} \right]\right\}\,
   B_{K}(\mu) \,\,\, .
\label{eq:BKRGI_NLO}
\ee
To this order, this is the scale-independent product of all
$\mu$-dependent quantities in Eq.~(\ref{eq:Heff}).

Lattice-QCD calculations provide results for $B_K(\mu)$.
However, these
results are usually obtained in intermediate schemes other
than the continuum $\msbar$ scheme used to calculate the Wilson
coefficients appearing in Eq.~(\ref{eq:Heff}). Examples of
intermediate schemes are the RI/MOM scheme \cite{Martinelli:1994ty}
(also dubbed the ``Rome-Southampton method'') and the Schr\"odinger
functional (SF) scheme \cite{Luscher:1992an}. These schemes permit the nonperturbative renormalization of the four-fermion
operator to be conducted, using an auxiliary lattice simulation.  This allows
$B_K(\mu)$ to be calculated with percent-level accuracy, as described
below.

In order to make contact with phenomenology, however, and in
particular to use the results presented above, one must convert from
the intermediate scheme to the $\msbar$ scheme or to the RGI quantity
$\hat{B}_{K}$. This conversion relies on 1- or
2-loop
perturbative matching calculations, the truncation errors in which
are, for many  calculations, the dominant source of error in
$\hat{B}_{K}$ (see, for instance,
Refs.~\cite{Laiho:2011np,Arthur:2012opa,Bae:2014sja,Blum:2014tka,Jang:2015sla,Boyle:2024gge}).
While this scheme-conversion error is not, strictly speaking, an error
of the lattice calculation itself, it must be included in results for
the quantities of phenomenological interest, namely,
$B_K(\msbar,2\,{\rm GeV})$ and $\hat{B}_{K}$. Incidentally,
  we remark that this truncation error is estimated in different ways
  and that its relative contribution to the total error can
  considerably differ among the various lattice calculations.  We
note that this error can be minimized by matching between the
intermediate scheme and $\msbar$ at as large a scale $\mu$ as possible
(so that the coupling which determines the rate of convergence is
minimized).   The latest available calculations have pushed the matching $\mu$ up to
the range 3--3.5~GeV. This is possible because of the use of
nonperturbative RG running determined on the
lattice~\cite{Durr:2011ap,Arthur:2012opa,Blum:2014tka}. The
Schr\"odinger functional offers the possibility to run
nonperturbatively to scales $\mu\sim M_{\rm{W}}$ where the truncation
error can be safely neglected. However, so far this has been applied
only for two flavours for $B_K$ in Ref.~\cite{Dimopoulos:2007ht} and for
the case of the BSM bag parameters in Ref.~\cite{Dimopoulos:2018zef}, and in Ref.~\cite{CamposPlasencia:2024mug} for three flavours. (See
more details in Sec.~\ref{sec:Bi}.)`

Perturbative truncation errors in Eq.~(\ref{eq:Heff}) also affect the
Wilson coefficients $\eta_1$, $\eta_2$ and~$\eta_3$. It turns out that
the largest uncertainty arises from the charm quark contribution
$\eta_1=1.87(76)$~\cite{Brod:2011ty}. Although it is now calculated at
NNLO, the series shows poor convergence. 
 The net effect from the uncertainty on $\eta_1$ on the amplitude in Eq.~(\ref{eq:Heff}) is larger than that of present 
lattice  calculations of $B_K$. Exploiting an idea presented
  in Ref.~\cite{Christ:2012se}, it has been 
  shown in Ref.~\cite{Brod:2019rzc} that, by using the $u$-$t$ instead of the usual  
$c$-$t$ unitarity in the $\epsilon_{K}$ computation, the perturbative uncertainties associated   with residual  short-distance quark contributions can be reduced.
We will elaborate upon this point later.

  Returning to Eq.~(\ref{eq:epsK}), we note that an analytical estimate of the leading contribution  from $\Im(M_{12}^\text{LD})$ based on $\chi$PT, shows
  that it is approximately proportional to $\xi_0 \equiv \Im(A_0)/\Re(A_0)$
  so that Eq.~(\ref{eq:epsK}) can be written as follows~\cite{Buras:2008nn,Buras:2010pz}:
\be
    \epsilon_{K} \,\,\, = \,\,\, \exp(i \phi_\epsilon) \,\,
    \sin(\phi_\epsilon) \,\, \Big [ \frac{\text{Im}(M_{12}^{\rm SD})} {\Delta M_K }
                \,\,\, + \,\,\, \rho \,\xi_0 \,\, \Big ] \, ,
                \label{eq:epsK-phenom}
\ee
 where the deviation of $\rho$ from one parameterizes the
  long-distance effects in $\Im(M_{12})$.

 The general formula  presented in Eq.~(\ref{eq:epsK-phenom}) for the parameter $\epsilon_K$  provides one of the most valuable inputs for tests of CKM unitarity. Moreover, it holds significant potential as a probe for New Physics, provided that its precision can be enhanced. In the following, we will provide a general overview of the current status of the computation of $|\epsilon_K|$.

With a very good approximation the formula for $|\epsilon_K|$ can be written in the so-called Wolfenstein parametrization~\cite{Wolfenstein:1983yz}. 
 The determination of  $|\epsilon_K|$ requires the knowledge of more than a dozen input quantities, which can be categorized into four groups. The first group includes six quantities ($G_F, \phi_{\epsilon}, M_{K^0}, \Delta M_{K}, M_W$ and $m_t$) whose values are known from experiment with high precision. 
The second  group consists of several observables computed in lattice QCD, including the kaon decay constant $f_K$,  the charm-quark mass $m_c(m_c)$, the neutral kaon mixing bag parameter $B_K$, and the ratio $\xi_0=\Im(A_0)/\Re(A_0)$.\footnote{Furthermore, the long-distance effects owing to light hadrons can be estimated on the lattice as noted below in Sec.~\ref{sec:Kpipi_amplitudes}, c.f. Ref.~\cite{Bai:2023lkr}. However, the current accuracy of this calculation is not yet high enough to constrain the determination 
of $|\epsilon_K|$. } Moreover,  the values of the CKM  matrix elements  $|V_{ud}|$,  $|V_{us}|$ and $|V_{cb}|$ are  required\textemdash see for instance Ref.~\cite{Jwa:2023uuz}\textemdash which are based on lattice-QCD computations.   
It is worth recalling that the present FLAG report provides average values for these quantities, see Secs.~\ref{sec:vusvud} and \ref{sec:BDecays}.   
 The third   group involves the short-distance interaction factors calculated in perturbation theory. In the $c$-$t$ unitarity formula,  these factors are $\eta_1$, $\eta_2$, and $\eta_3$, as mentioned earlier in this section and shown in Eq.~(\ref{eq:Heff}). In the $u$-$t$ unitarity case, there appear only two relevant factors (see Refs.~\cite{Brod:2019rzc, Brod:2022har}). 
Finally,  the fourth  group of inputs consists of the pair of CKM triangle variables  ($\overline {\rho}, \overline{\eta}$)  whose values are estimated from the unitarity triangle analysis. In particular,  the  
Angle-only Fit (AoF) analysis used in Refs.~\cite{Lee:2014jxl, Bailey:2015tba, Bailey:2018feb} (see also Ref.~\cite{UTfit:2022hsi}) prevents any correlation of ($\overline {\rho}, \overline{\eta}$) with the rest of the inputs used in the formula for $|\epsilon_K|$. 

{Among the various inputs the value of $|V_{cb}|$ has a dominant impact on {the uncertainty of $|\epsilon_K|$ because $|V_{cb}|$ appears to the fourth power in the expression of $|\epsilon_K|$}.}
{Among the various inputs, $|V_{cb}|$  introduces the dominant contribution on the uncertainty of $|\epsilon_K |$. This effect can be attributed to the presence of terms with $|V_{cb}|$ to the fourth and second power in the expression of $|\epsilon_K |$.}

{As discussed in Sec.~\ref{sec:BDecays} of this report, to which we refer for a comprehensive review, a longstanding discrepancy persists between the exclusive and inclusive determinations of $|V_{cb}|$. Exclusive determinations, which are primarily based on the differential decay rates of processes such as $B\to D^* \ell\nu$ and $B\to D \ell\nu$, rely on nonperturbative inputs from lattice QCD. In contrast, inclusive determinations employ the total semileptonic decay rate $B\to X_c\ell\nu$ and are analysed within the framework of heavy quark effective theory and the operator product expansion. The application of these two types of theoretical approaches, in conjunction with distinct experimental inputs, yields the current sizeable tension between the exclusive and inclusive determinations of $|V_{cb}|$.} 
	
Another significant source of uncertainty, when the $c$-$t$ unitarity formula for $|\epsilon_K|$ is employed, is related to the  factor $\eta_1$ that is computed to NNLO in perturbation theory. 
For more information on the estimation of the systematic error due to perturbative truncation, see Refs.~\cite{Brod:2011ty, Buras:2013raa, Bailey:2015tba}. This  source of uncertainty can be mitigated  by adopting the $u$-$t$ unitarity formula for $|\epsilon_K|$.  In this case, it is found that the two relevant QCD perturbative factors are not subject to significant systematic uncertainties. Furthermore, this approach reduces the correlations between the individual perturbative contributions~\cite{Brod:2019rzc}. 

We close this discussion by mentioning that the use of the $u$-$t$ unitarity
 formula leads to   a total statistical error of about 8\% in $|\epsilon_K|$. In this case, when analyzing the error budget, we see that nearly half of the total error comes from the propagation of the uncertainty from $|V_{cb}|$. Furthermore, the propagated error owing to the $\overline{\eta}$ error is the second most significant source of uncertainty in $|\epsilon_K|$. It is noteworthy that the propagated error from $B_K$ is much smaller, accounting for only a few percent in the final error budget. It should also be noted that the relative uncertainties contributing to the error budget are rather sensitive to improvements in the precision of $|V_{cb}|$. 
 {The tension between the exclusive and inclusive determinations of $|V_{cb}|$ induces a corresponding difference in the predicted values of $|\epsilon_K|$ derived from these inputs. Recent detailed analyses of $|\epsilon_K|$, including the impact of different $|V_{cb}|$ determinations and the use of the $c$–$t$ and $u$–$t$ unitarity relations, can be found in Refs.~\cite{Jwa:2023uuz, Jwa:2024xjq, Jwa:2025fon}. It is also worth noting that ongoing advancements in lattice-QCD techniques may contribute in the future to reducing uncertainties in the inclusive determination of $|V_{cb}|$ {\cite{Hashimoto:2017wqo,Hansen:2017mnd,Gambino:2020crt,Gambino:2022dvu,Barone:2023tbl}}. The resolution of the longstanding discrepancy between the exclusive and inclusive determinations of $|V_{cb}|$ is highly desirable, as it would significantly enhance the precision of $|\epsilon_K|$ and its potential for probing New Physics scenarios.}\,\footnote{Note that a more precise determination of $|\epsilon_K|$ will require taking into account the effect of short-distance power corrections from dim-8 operators to the $\Delta S=2$ effective Hamiltonian.  It is estimated that their effect leads to an increase of the central value by 1\%, see Refs.~\cite{Cata:2003mn, Ciuchini:2021zgf}.}      
\vspace*{0.2cm}
  
  In order to facilitate the subsequent
  discussions about the status of the lattice studies of $K \to
  \pi\pi$ and of the current estimates of $\xi_0 \equiv \text{Im}(A_0) / \text{Re}(A_0)$, we provide a brief
  account  of the parameter $\epsilon^{\prime}$ that describes  
  direct CP-violation in the kaon sector. The definition of $\epsilon^{\prime}$ is given by: 
\be 
\epsilon^{\prime} \equiv \dfrac{1}{\sqrt{2}}\dfrac{{\cal A} [ K_S
    \rightarrow (\pi\pi)_{I=2}] }{{\cal A} [ K_S \rightarrow (\pi\pi)_{I=0}]} \left( \dfrac{{\cal A}[ K_L \rightarrow
        (\pi\pi)_{I=2}]}{{\cal A} [ K_S \rightarrow (\pi\pi)_{I=2}]} - \dfrac{{\cal A}[ K_L \rightarrow
            (\pi\pi)_{I=0}]}{{\cal A} [ K_S \rightarrow (\pi\pi)_{I=0}]} \right)\,.
\ee 
By selecting appropriate phase conventions for the mixing
parameters between $K^0$ and $\bar{K}^0$ CP-eigenstates (see,
e.g., Ref.~\cite{Sozzi:2008bookcp} for further details), the expression
of $\epsilon^{\prime}$ can be expressed in terms of the real and
imaginary parts of the isospin amplitudes as follows:
\be
     \epsilon^{\prime} \,\,\, = \,\,\, \dfrac{i \omega \, e^{i (\delta_2 - \delta_0)}}{\sqrt{2}} \,
    \Big [\xi_2  - \xi_0  \, \Big ] \, ,
                    \label{eq:epsprime-phenom}
\ee
where $\omega \equiv \text{Re}(A_2) / \text{Re}(A_0)$, 
$\xi_2 \equiv \text{Im}(A_2) / \text{Re}(A_2)$, $A_2$
denotes the $\Delta{I}=3/2$ $K\to\pi\pi$ decay amplitude, and $\delta_I$
denotes the strong scattering phase shifts in the corresponding,
$I=0,2$, $K\to(\pi\pi)_I$ decays. Given that the phase
$\phi_\epsilon^{\prime}=\delta_2 - \delta_0 + \pi/2 \approx
42.3(1.5)^\circ$~\cite{ParticleDataGroup:2024cfk} is nearly equal to
$\phi_\epsilon$ in
Eq.~(\ref{eq:phi_epsilonK}), 
the ratio of parameters 
characterizing the direct and indirect CP-violation in the kaon sector
can be approximated in the following way,
\be
\epsilon^{\prime} / \epsilon \,\,\, \approx \,\,\, \Re(\epsilon^{\prime} / \epsilon) \,\,\, = \,\,\, \dfrac{\omega}{\sqrt{2}\, |\epsilon_{K}|} \,
    \Big [ \xi_2 - \xi_0  \, \Big ] \, ,
\label{eq:epsprimeOVeps-phenom}
\ee
where  on the left hand side we have set $\epsilon \equiv \epsilon_{K}$.
The experimentally measured value reads~\cite{ParticleDataGroup:2024cfk},
\be
\Re(\epsilon^{\prime} / \epsilon) = 16.6(2.3) \times
    10^{-4}\,. 
\label{eq:epspovepsexp}
\ee
We remark that isospin breaking and electromagnetic effects (see
Refs.~\cite{Cirigliano:2003nn,Cirigliano:2019cpi}, and the discussion
in Ref.~\cite{Buras2020bookcp}) introduce additional correction terms
into Eq.~(\ref{eq:epsprimeOVeps-phenom}).

\subsection{Lattice-QCD studies of the $K\to(\pi\pi)_I$ decay amplitudes, $\xi_0$, $\xi_2$ and $\epsilon^{\prime}/\epsilon$}
\label{sec:Kpipi_amplitudes}

 As a preamble to this section, it should be noted that the study of
 $K \to \pi\pi$ decay amplitudes requires the development of
 computational strategies that are at the forefront of lattice QCD
 techniques. These studies represent a significant
   advance in the study of kaon physics. However, at present, they
   have not yet reached the same level of maturity of most of the
   quantities analyzed in the FLAG report, where, for instance,
   independent results by various lattice collaborations are being
   compared and averaged. We
   will, therefore, review the current status of $K \to \pi\pi$ lattice
   computations, but we will provide a FLAG average only for the case
   of the decay amplitude $A_2$.

We start by reviewing the determination of the parameter
$\xi_0 = \Im(A_0)/\Re(A_0)$.
An estimate of $\xi_0$ has been obtained  from a direct evaluation of the
  ratio of amplitudes $\Im(A_0)/\Re(A_0)$, where $\Im(A_0)$ is
  determined from a lattice-QCD computation  by 
    RBC/UKQCD~20~\cite{Abbott:2020hxn} employing $\Nf=2+1$ M\"obius domain-wall fermions at
  a single value of the lattice spacing,  while $\Re(A_0) \simeq |A_0|$ and   the value $|A_0| = 3.320(2) \times 10^{-7}$ GeV are used based on
  the relevant experimental input~\cite{Zyla:2020zbs} from the decay
  to two pions. This leads to a result for $\xi_0$ with a rather large
  relative error, 
\begin{equation}
   \xi_0 = -2.1(5)\times 10^{-4}.
    \label{eq:xilat1}
\end{equation}
Following a similar procedure, an estimate of $\xi_0$ was
  obtained through the use of a previous lattice QCD determination of
  $\Im(A_0)$ by RBC/UKQCD~15G~\cite{Bai:2015nea}. We refer to
Tab.~\ref{tab_A0_nf21} for further details about these computations of $\Im(A_0)$. The
comparison of the estimates of $\xi_0$ based on lattice QCD input are
collected in Tab.~\ref{tab_xi_nf21}. 

To determine the value of $\xi_0$, the expression in Eq.~(\ref{eq:epsprimeOVeps-phenom}) together with the experimental values of
$\Re(\epsilon^{\prime}/\epsilon)$, $|\epsilon_K|$ and $\omega$
can also be used. This approach has been pursued with the help of a lattice-QCD calculation of the ratio of amplitudes
$\Im(A_2)/\Re(A_2)$ by RBC/UKQCD~15F~\cite{Blum:2015ywa} where the continuum-limit
result is based on computations at two values of the lattice spacing
employing $\Nf=2+1$ M\"obius domain-wall fermions. Further details about the lattice computations of $A_2$ are collected in Tab.~\ref{tab_A2_nf21}.
 In this case, we obtain   $\xi_0 = -1.6  (2)\times 10^{-4}$.
The use of the updated value of $\text{Im}(A_2)=
  -8.34(1.03) \times 10^{-13}$\,GeV from
  Ref.~\cite{Abbott:2020hxn},
  in combination with the experimental value of $\text{Re}(A_2) =
  1.479(4) \times 10^{-8}$\,GeV, introduces a small change with
  respect to the above result.\footnote{The update in
      $\text{Im}(A_2)$ is due to a change in the value of the imaginary part of the ratio of CKM matrix elements, $\tau =
  	-V^\ast_{ts}V_{td}/V^\ast_{us}V_{ud}$, as given in
        Ref.~\cite{Tanabashi:2018oca}. The lattice-QCD input is
    therefore the one reported in Ref.~\cite{Blum:2015ywa}.} The value for $\xi_0$ reads\,\footnote{  
  The current estimates for  the
  corrections owing to isospin breaking and electromagnetic effects~\cite{Cirigliano:2019cpi}  imply a 
  relative change on the theoretical value for $\epsilon^{\prime} / \epsilon$ by about $-20$\% 
  with respect to the determination based on Eq.~(\ref{eq:epsprimeOVeps-phenom}). The size 
of these isospin breaking and electromagnetic corrections is related
to the enhancement of the decay amplitudes between the $I=0$ and the
$I=2$ channels. As a consequence, one obtains a similar reduction on $\xi_0$, leading to a value that is close
  to the result of Eq.~(\ref{eq:xilat1}).} 
\begin{equation}
   \xi_0 = -1.7 (2)\times 10^{-4}.
\label{eq:xilat2}   
\end{equation}

A phenomenological estimate can also
be obtained from the relationship of $\xi_0$ to
$\Re (\epsilon^\prime/\epsilon)$, using the experimental value of the latter and further 
assumptions concerning the estimate of hadronic contributions.
The corresponding value of $\xi_0$ reads~\cite{Buras:2008nn,Buras:2010pz}
\begin{equation}
   \xi_0 = -6.0(1.5) \times 10^{-2} \times \sqrt{2}\,|\epsilon_K| 
       = -1.9(5)\times 10^{-4}. 
       \label{eq:xipheno}
\end{equation}
We note that the use of the experimental value for $\Re(\epsilon^\prime/\epsilon)$ is based on the assumption that it is
free from New Physics contributions. The value of $\xi_0$ can then be combined with a ${\chi}\rm PT$-based
estimate for the long-range contribution,
$\rho=0.6(3)$~\cite{Buras:2010pz}. Overall, the combination $\rho\xi_0$
appearing in Eq.~(\ref{eq:epsK-phenom}) leads to a suppression of the
SM prediction of $|\epsilon_K|$ by about $3(2)\%$ relative to the
experimental measurement of $|\epsilon_K|$ given in
Eq.~(\ref{eq:epsilonK_exp}), regardless of whether the
phenomenological estimate of $\xi_0$ [see Eq.~(\ref{eq:xipheno})] or the
most precise lattice result [see Eq.~(\ref{eq:xilat1})] are used. The
uncertainty in the suppression factor is dominated by the error on
$\rho$.
Although this is a small correction, we note that its
contribution to the error of $\epsilon_K$ is larger than that arising
from the value of $B_{K}$ reported below.

The evolution of lattice-QCD methodologies has enabled recent ongoing efforts to calculate the long-distance contributions to $\epsilon_{K}$\,\cite{Bai:2016gzv,Bai:2023lkr} and the $K_L-K_S$ mass difference\,\cite{Christ:2012se,Bai:2014cva,Christ:2015pwa,Wang:2020jpi,Wang:2022lfq}.
 However,
the results are not yet precise enough to improve the accuracy in the
determination of the parameter $\rho$.

The lattice-QCD study of $K \to \pi\pi$ decays provides crucial
input to the SM prediction of $\epsilon_{K}$. 
During the last decade, the RBC/UKQCD collaboration has undertaken a series of lattice-QCD
calculations of $K \to \pi\pi$ decay amplitudes~\cite{Blum:2015ywa,Bai:2015nea,Abbott:2020hxn,Blum:2023mtn}.
In 2015, the first calculation of the $K\to(\pi\pi)_{I=0}$ decay
amplitude $A_0$ was performed using physical kinematics on a
$32^3\times64$ lattice with an inverse lattice spacing of $a^{-1}=1.3784(68)$
GeV~\cite{Bai:2015nea, Bai:2016ocm}. 
The main features of the RBC/UKQCD~15G calculation included, fixing the $I=0$
$\pi\pi$ energy very close to the kaon mass by imposing G-parity boundary conditions, a continuum-like operator mixing pattern through the use of a domain-wall fermion action with accurate chiral symmetry, and the construction of the complete set of correlation functions by computing seventy-five distinct diagrams. Results for the real and the imaginary parts of the decay amplitude $A_0$ from the RBC/UKQCD~15G computation are collected in Tab.~\ref{tab_A0_nf21}, where the first error is statistical and the second is systematic. 
\begin{table}[!t]
	\mbox{} \\[3.0cm]
	\footnotesize
	\begin{tabular*}{\textwidth}[l]{@{\extracolsep{\fill}}l@{\hspace{1mm}}r@{\hspace{1mm}}l@{\hspace{1mm}}l@{\hspace{1mm}}l@{\hspace{1mm}}l@{\hspace{1mm}}l@{\hspace{1mm}}l@{\hspace{1mm}}l@{\hspace{5mm}}l@{\hspace{1mm}}c@{\hspace{1mm}}c}
		Collaboration & Ref. & $\Nf$ & 
		\hspace{0.15cm}\begin{rotate}{60}{publication status}\end{rotate}\hspace{-0.15cm} &
		\hspace{0.15cm}\begin{rotate}{60}{continuum extrapolation}\end{rotate}\hspace{-0.15cm} &
		\hspace{0.15cm}\begin{rotate}{60}{chiral extrapolation}\end{rotate}\hspace{-0.15cm}&
		\hspace{0.15cm}\begin{rotate}{60}{finite volume}\end{rotate}\hspace{-0.15cm}&
		\hspace{0.15cm}\begin{rotate}{60}{renormalization}\end{rotate}\hspace{-0.15cm}  &
		\hspace{0.15cm}\begin{rotate}{60}{running/matching}\end{rotate}\hspace{-0.15cm} & 
		\rule{0.15cm}{0cm} $\text{Re}(A_0)$ & 
		$\text{Im}(A_0)$ \\
		&&&&&&&&&  {\scriptsize $[10^{-7}$~GeV$]$} & {\scriptsize $[10^{-11}$~GeV$]$} \\[0.5ex]
		&&&&&&&&&&\\[-0.1cm]
		\hline
		\hline
		&&&&&&&&&& \\[-0.1cm]
        RBC/UKQCD~23A & \cite{Blum:2023mtn} & 2+1 & \gA & \tbr & \soso & \good
		& \good&  $\,a$ & 2.84(0.57)(0.87)&$-8.7(1.2)(2.6)$ \\[0.5ex]
		RBC/UKQCD~20 & \cite{Abbott:2020hxn} & 2+1 & \gA & \tbr & \soso & \soso
		& \good&  $\,a$ &2.99(0.32)(0.59) &$-6.98(0.62)(1.44)$ \\[0.5ex]
		RBC/UKQCD~15G & \cite{Bai:2015nea} & 2+1 & \gA & \tbr & \soso & \soso
		& \good&  $\,b$ &4.66(1.00)(1.26) &$-1.90(1.23)(1.08)$ \\[0.5ex]

		&&&&&&&&&& \\[-0.1cm]
		\hline
		\hline\\[-0.1cm]
	\end{tabular*}
	\begin{minipage}{\linewidth}
		{\footnotesize 
			\begin{itemize}
                \item[$a$] Nonperturbative renormalization with the RI/SMOM scheme
				at a scale of 1.53\,GeV and running to 4.0\,GeV employing a nonperturbatively determined step-scaling function. Conversion to $\msbar$ at 1-loop order.   \\[-5mm]
				\item[$b$] Nonperturbative renormalization with the RI/SMOM scheme
		       at a scale of 1.53\,GeV. Conversion to $\msbar$ at 1-loop order at the same scale.     \\[-5mm]	
		\end{itemize}
		}
	\end{minipage}
	\caption{Results for the real and imaginary parts of the
           $K \to \pi\pi$  decay
		amplitude $A_0$ from lattice-QCD computations with $\Nf=2+1$
		dynamical flavours. Information about the renormalization, running
		and matching to the $\msbar$ scheme is indicated in
		the column ``running/matching'', with details given at the bottom of
		the table. We refer to the text for
                  further details about the main differences between
                  the lattice computations in Refs.~\cite{Abbott:2020hxn}~and~\cite{Bai:2015nea}.}
	\label{tab_A0_nf21}
\end{table}

\begin{table}[h]
	\mbox{} \\[3.0cm]
	\footnotesize
	\begin{tabular*}{\textwidth}[l]{@{\extracolsep{\fill}}l@{\hspace{1mm}}r@{\hspace{1mm}}l@{\hspace{1mm}}l@{\hspace{1mm}}l@{\hspace{1mm}}l@{\hspace{1mm}}l@{\hspace{1mm}}l@{\hspace{1mm}}l@{\hspace{5mm}}l@{\hspace{1mm}}c@{\hspace{1mm}}c}
		Collaboration & Ref. & $\Nf$ & 
		\hspace{0.15cm}\begin{rotate}{60}{publication status}\end{rotate}\hspace{-0.15cm} &
		\hspace{0.15cm}\begin{rotate}{60}{continuum extrapolation}\end{rotate}\hspace{-0.15cm} &
		\hspace{0.15cm}\begin{rotate}{60}{chiral extrapolation}\end{rotate}\hspace{-0.15cm}&
		\hspace{0.15cm}\begin{rotate}{60}{finite volume}\end{rotate}\hspace{-0.15cm}&
		\hspace{0.15cm}\begin{rotate}{60}{renormalization}\end{rotate}\hspace{-0.15cm}  &
		\hspace{0.15cm}\begin{rotate}{60}{running/matching}\end{rotate}\hspace{-0.15cm} & 
		\rule{0.15cm}{0cm} $\text{Re}(A_2)$ & 
		$\text{Im}(A_2)$ \\
		&&&&&&&&&  {\scriptsize $[10^{-8}$~GeV$]$} & {\scriptsize $[10^{-13}$~GeV$]$} \\[0.5ex]
		&&&&&&&&&&\\[-0.1cm]
		\hline
		\hline
		&&&&&&&&&& \\[-0.1cm]
        RBC/UKQCD~23A & \cite{Blum:2023mtn} & 2+1 & \gA & \tbr & \soso & \good
	    & \good&  $\,a$ & 1.74(0.15)(0.48)&   $-5.91(0.13)(1.75)$  \\[0.5ex]
		RBC/UKQCD~15F & \cite{Blum:2015ywa} & 2+1 & \gA & \soso & \soso & \good
		& \good&  $\,b$ & 1.50(0.04)(0.14)&  $-8.34(1.03) ^\diamondsuit$  \\[0.5ex]
		&&&&&&&&&& \\[-0.1cm]
		\hline
		\hline\\[-0.1cm]
	\end{tabular*}
	\begin{minipage}{\linewidth}
		{\footnotesize 
			\begin{itemize}
                \item[$a$] Nonperturbative renormalization with the RI/SMOM scheme
				at a scale of 1.53\,GeV and running to 4.0\,GeV employing a nonperturbatively determined step-scaling function. Conversion to $\msbar$ at 1-loop order.   \\[-5mm]
				\item[$b$] Nonperturbative renormalization with the RI/SMOM scheme
				at a scale of 3 GeV. Conversion to $\msbar$ at 1-loop order.     \\[-5mm]
				\item[$^\diamondsuit$] This value of $\text{Im}(A_2)$ is an update  
					reported in
                                        Ref.~\cite{Abbott:2020hxn}
                                        which is based on the lattice QCD computation in Ref.~\cite{Blum:2015ywa} but where a change in the
					value of the imaginary part of the ratio of CKM matrix elements $\tau =
					-V^\ast_{ts}V_{td}/V^\ast_{us}V_{ud}$ reported in Ref.~\cite{Tanabashi:2018oca} has been applied.
				\\[-5mm]
			\end{itemize}
		}
	\end{minipage}
	\caption{Results for the real and the imaginary parts of the
          $K \to \pi\pi$ decay
		amplitude $A_2$ from lattice-QCD computations with $\Nf=2+1$
		dynamical flavours. Information about the renormalization
		and matching to the $\msbar$ scheme is indicated in
		the column ``running/matching'', with details given at the bottom of
		the table. }
	\label{tab_A2_nf21}
\end{table}

\begin{table*}[t!]
	\begin{center}
		\mbox{} \\[3.0cm]
		{\footnotesize{
				\vspace*{-2cm}\begin{tabular*}{\textwidth}[l]{l
                                    @{\extracolsep{\fill}} c c c c }
					Collaboration & Ref. &
                                         $\Nf$ &  \rule{0.3cm}{0cm}$\xi_0$  		 \\
					&& \\[-0.1cm]
					\hline
					\hline
					&& \\[-0.1cm]

                     RBC/UKQCD~23A$^\circ$ &
                                        \cite{Blum:2023mtn} & 2+1 & $-2.63(37)(68)\cdot10^{-4}$  \\[0.5ex]     
					RBC/UKQCD~20$^\dagger$ &
                                        \cite{Abbott:2020hxn} & 2+1 & $-2.1(5)\cdot10^{-4}$  \\[0.5ex]
				    RBC/UKQCD~15G$^\diamond$  &
				    \cite{Bai:2015nea} & 2+1 &  $-0.6(5)\cdot10^{-4}$  \\[0.5ex]
					RBC/UKQCD~15F$^\ast$  &
                                        \cite{Blum:2015ywa} & 2+1 &  $-1.7(2)\cdot10^{-4}$  \\[0.5ex]
                  	&& \\[-0.1cm]
					\hline
					\hline\\[-0.1cm]
				\end{tabular*}
		}}
		\begin{minipage}{\linewidth}
			{\footnotesize 
				\begin{itemize}
					\item[$^\circ$] Estimate for $\xi_0$ has been provided by RBC/UKQCD (private communication with Masaaki Tomii.)
					\item[ $^\dagger$] Estimate for $\xi_0$  obtained  from a direct  evaluation of the
					ratio of amplitudes $\Im(A_0)/\Re(A_0)$ where $\Im(A_0)$ is
					determined from the
					lattice-QCD computation of
					Ref.~\cite{Abbott:2020hxn}
					while for $\Re(A_0) \simeq
					|A_0|$ is taken from  the experimental value for  $|A_0|$.  
						\item[ $^\diamond$] Estimate for $\xi_0$  obtained  from a direct  evaluation of the
					ratio of amplitudes $\Im(A_0)/\Re(A_0)$ where $\Im(A_0)$ is
					determined from the
					lattice-QCD computation of
					Ref.~\cite{Bai:2015nea}
					while for $\Re(A_0) \simeq
					|A_0|$ is taken from  the experimental value for  $|A_0|$.
					\item[ $^\ast$]  Estimate for
						$\xi_0$ based on the use of
						Eq.~(\ref{eq:epsprimeOVeps-phenom}).
						The new value of
                                                $\text{Im}(A_2)$
                                                reported in
                                                Ref.~\cite{Abbott:2020hxn}---based on the
                                                lattice-QCD
                                                computation of
                                                Ref.~\cite{Blum:2015ywa}
                                                following an update of
                                                a nonlattice input---is used in combination with
						the experimental values for
						$\text{Re}(A_2)$,
						$\Re(\epsilon^{\prime}/\epsilon)$,
						$|\epsilon_K|$ and $\omega$.

				\end{itemize}
			}
		\end{minipage}
		\caption{Results for the parameter
			$\xi_0=\Im(A_0)/\Re(A_0)$ obtained through the
			combination of lattice-QCD determinations of $K \to
			\pi\pi$ decay amplitudes with $\Nf=2+1$ dynamical
			flavours and experimental
			inputs.   \label{tab_xi_nf21}}
	\end{center}
\end{table*}

The calculation in RBC/UKQCD~20~\cite{Abbott:2020hxn} using the same lattice setup has improved upon RBC/UKQCD~15G~\cite{Bai:2015nea} in three
important aspects: (i) an increase in statistics by a factor of 3.4; (ii) the inclusion of
a scalar two-quark operator and the addition of another pion-pion
operator to isolate the ground state, and (iii) the use of step 
scaling techniques
to raise the renormalization
scale from 1.53 GeV to 4.01 GeV.
The updated determinations of  the real and the imaginary
    parts of $A_0$ in Ref.~\cite{Abbott:2020hxn} are shown in Tab.~\ref{tab_A0_nf21}.

In addition to utilizing G-parity boundary conditions to address the challenges associated with extracting excited states for achieving the correct kinematics of $K\to\pi\pi$,  the latest publications, RBC/UKQCD~23A\,\cite{Blum:2023mtn} and RBC/UKQCD~23B\,\cite{RBC:2023xqv},  also investigate alternative approaches for overcoming this issue, namely employing variational methods and periodic boundary conditions. Two-pion scattering calculations are carried out for the isospin channels $I=0$ and $I=2$ on two gauge-field ensembles with physical pion masses and inverse lattice spacings of 1.023 and 1.378 GeV\,\cite{RBC:2023xqv}  employing domain-wall fermions. The results for scattering phase shifts in both $I=0$ and $I=2$ channels exhibit consistency with the Roy equation and chiral perturbation theory, although the statistical error for $I=0$ remains relatively large. The computation of $K\to\pi\pi$ decay amplitudes and $\epsilon'$ is performed on a single ensemble with a physical pion mass and an inverse lattice spacing of 1.023 GeV\,\cite{Blum:2023mtn}. The value obtained for $\operatorname{Re}(\epsilon'/\epsilon)$ is consistent with that of the previous 2020 calculation, albeit with 1.7 times larger uncertainty.   Results from RBC/UKQCD~23A for the real and imaginary parts of $A_0$ and $A_2$ are reported in Tabs.~\ref{tab_A0_nf21} and \ref{tab_A2_nf21}, respectively.

As previously discussed, the 
determination of $\operatorname{Im}(A_0)$ from Ref.~\cite{Abbott:2020hxn}  has been used to 
  obtain the value of the parameter $\xi_0$ in Eq.~(\ref{eq:xilat1}). 
A first-principles computation of $\operatorname{Re}(A_0)$ is
  essential to address the so-called $\Delta I=1/2$ puzzle associated
  to the enhancement of $\Delta I=1/2$ over $\Delta I=3/2$ transitions
  owing, crucially, to long distance effects. Indeed, short-distance enhancements in the Wilson coefficients
are not large enough to explain the $\Delta I=1/2$
rule~\cite{Gaillard:1974nj,Altarelli:1974exa}.
Lattice-QCD calculations do provide a method to study such a
long-distance enhancement. The combination of the most precise result for $A_0$ in Tab.~\ref{tab_A0_nf21}, Ref.~\cite{Abbott:2020hxn},  with the earlier lattice calculation of $A_2$ in Ref.~\cite{Blum:2015ywa} leads to the ratio,
$\operatorname{Re}(A_0)/\operatorname{Re}(A_2)=19.9(5.0)$, which
agrees with the  value  
$\operatorname{Re}(A_0)/\operatorname{Re}(A_2)=22.45(6)$ that we obtain based solely on PDG 24~\cite{ParticleDataGroup:2024cfk} experimental input. 
In Ref.~\cite{Abbott:2020hxn}, the lattice
determination of relative size of direct CP violation was updated as follows,
\be
\operatorname{Re}(\epsilon'/\epsilon)=21.7(2.6)(6.2)(5.0)\times10^{-4},
 \label{eq:epspoveps}
\ee
where the first two errors are statistical and systematic, respectively. 
The third error arises from having omitted the strong and electromagnetic
isospin breaking effects. 
The value of
$\operatorname{Re}(\epsilon'/\epsilon)$ in Eq.~(\ref{eq:epspoveps}) uses the experimental values
of $\Re(A_0)$ and $\Re(A_2)$.  The lattice determination of $\operatorname{Re}(\epsilon'/\epsilon)$ 
is in good agreement with the experimental result in Eq.~(\ref{eq:epspovepsexp}). However,  while the result  in Eq.~(\ref{eq:epspoveps}) represents a significant step forward, it is 
important to keep in mind that the calculation of $A_0$ is currently based on a single value of the lattice spacing. 
It is expected that future work with additional values of the lattice spacing will contribute to improve the precision. 
For a description of the computation of the $\pi\pi$ scattering phase shifts entering in 
the determination of $\operatorname{Re}(\epsilon'/\epsilon)$ in Eq.~(\ref{eq:epspoveps}), we refer to Ref.~\cite{Blum:2021fcp}.

The complex amplitude $A_2$
  has been determined by RBC/UKQCD~15F~\cite{Blum:2015ywa}
  employing $\Nf=2+1$ M\"obius domain-wall fermions at two values of
  the lattice spacing, namely $a=0.114$\,fm and $0.083$\,fm, and
  performing simulations  at the physical pion mass with $M_{\pi}L
  \approx 3.8$.

A compilation of  lattice results  for the real and imaginary parts of the $K \to \pi\pi$ decay amplitudes, $A_0$ and $A_2$, with $\Nf=2+1$  flavours of
	dynamical quarks is shown in
        Tabs.~\ref{tab_A0_nf21}~and~\ref{tab_A2_nf21}. In Appendix~\ref{app-Kpipi}, we collect the corresponding information about  the lattice QCD
        simulations, including  the values of some of the most relevant
        parameters.

The determination of the real and imaginary parts
  of $A_2$ by RBC/UKQCD~15F shown in Tab.~\ref{tab_A2_nf21} is free
  of red tags. We therefore quote the following FLAG averages:
%
%
\begin{align}
&&\FLAGAVBEGIN \text{Re}(A_2) &= 1.50(0.04)(0.14) \times 10^{-8}\FLAGAVEND~\text{GeV},  &\nonumber\\[-3mm]
& \Nf=2+1: & &&\Ref~\mbox{\cite{Blum:2015ywa}}.\\[-3mm]
&&\FLAGAVBEGIN \text{Im}(A_2) &= -8.34(1.03) \times 10^{-13}\FLAGAVEND~\text{GeV},  & \nonumber
\end{align}
%
 
 Results for the parameter $\xi_0$  are
	presented  in Tab.~\ref{tab_xi_nf21}. 
	Except for the most recent calculation RBC/UKQCD~23A, which is based on the direct lattice calculation of the relevant quantities, we note that, for the other reported values of $\xi_0$, the total uncertainty depends on the specific way in which  
	lattice and experimental inputs are selected. 

Besides the RBC/UKQCD collaboration programme~\cite{Blum:2015ywa,Bai:2015nea,Abbott:2020hxn, Blum:2023mtn, RBC:2023xqv} using
domain-wall fermions, an approach based on improved Wilson
fermions~\cite{Ishizuka:2015oja, Ishizuka:2018qbn} has presented a
determination of the $K \to \pi\pi$ decay amplitudes,  $A_0$ and
$A_2$, at unphysical quark masses.
See Refs.~\cite{Donini:2016lwz,Donini:2020qfu,Baeza-Ballesteros:2022azb}
for an analysis of the scaling with the number of colours of $K \to \pi\pi$
decay amplitudes using lattice-QCD computations

Proposals aiming at the inclusion of electromagnetism in
lattice-QCD calculations of $K \to \pi\pi$ decays are being
explored~\cite{Christ:2017pze,Cai:2018why,Christ:2021guf} in order to reduce the
uncertainties associated   with isospin breaking effects.

\subsection{Lattice computation of $B_{K}$}
\label{sec:BK lattice}

Lattice calculations of $B_{K}$ are affected by the same type of
systematic effects discussed in the various sections of this review. However, the issue
of renormalization merits special attention. The reason is that the
multiplicative renormalizability of the relevant operator $Q^{\Delta
S=2}$ is lost once the regularized QCD action ceases to be invariant
under chiral transformations.  As a result, the
  renormalization pattern of $B_K$ depends on the specific choice
  of the fermionic discretization.

In the case of Wilson fermions, $Q^{\Delta S=2}$
mixes with four additional dimension-six operators, which belong to
different representations of the chiral group, with mixing
coefficients that are finite functions of the gauge coupling. This
complicated renormalization pattern was identified as the main source
of systematic error in earlier, mostly quenched calculations of
$B_{K}$ with Wilson quarks. It can be bypassed via the
implementation of specifically designed methods, which are either
based on Ward identities~\cite{Becirevic:2000cy} or on a modification
of the Wilson quark action, known as twisted-mass
QCD~\cite{Frezzotti:2000nk,Dimopoulos:2006dm,Dimopoulos:2007cn}.

An advantage of staggered fermions is the presence of a remnant $U(1)$
chiral symmetry. However, at nonvanishing lattice spacing, the
symmetry among the extra unphysical degrees of freedom (tastes) is
broken. As a result, mixing with other dimension-six operators cannot
be avoided in the staggered formulation, which complicates the
determination of the $B$-parameter.   In general, taste conserving
  mixings are implemented directly in the lattice computation of the
  matrix element.  The effects of the broken taste symmetry are
usually treated  through an effective field theory, staggered
  Chiral Perturbation Theory
  (S$\chi$PT)~\cite{VandeWater:2005uq,Bailey:2012wb},
  parameterizing the quark-mass and lattice-spacing dependences.

Fermionic lattice actions based on the Ginsparg-Wilson
relation~\cite{Ginsparg:1981bj} are invariant under the chiral group,
and hence four-quark operators such as $Q^{\Delta S=2}$ renormalize
multiplicatively. However, depending on the particular formulation of
Ginsparg-Wilson fermions, residual chiral symmetry breaking effects
may be present in actual calculations. 
For instance, in the case of domain-wall fermions, the finiteness of the extra 5th dimension implies that the decoupling of modes with different chirality is not exact, which produces a residual nonzero quark mass $m_{\rm{res}}$ in the chiral limit. The mixing with dimension-six operators of different chirality is expected to be an $\cO(m_{\rm{res}}^2)$ suppressed effect~\cite{Aoki:2005ga, Christ:2005xh} that should be investigated on a case-by-case basis.

Before describing the results for $B_K$, we would like to reiterate a discussion presented in previous FLAG reports about an issue related to the
computation of the kaon bag parameters through lattice-QCD simulations with $\Nf=2+1+1$ dynamical quarks. In practice, this only concerns the calculations of the kaon $B$-parameters including dynamical charm-quark effects in Ref.~\cite{Carrasco:2015pra}, that
were examined in the FLAG 16 report. As described in Sec.~\ref{sec:indCP}, the effective Hamiltonian in Eq.~(\ref{eq:HDeltaS2}) depends solely on the operator $Q^{\Delta
S=2}$ in Eq.~(\ref{eq:Q1def}) ---which appears in the definition of $B_K$ in Eq.~(\ref{eq:defBK})--- at energy scales below the charm threshold where charm-quark contributions are absent. As a result, a computation of $B_K$ based on $\Nf=2+1+1$ dynamical simulations will include an extra sea-quark contribution from charm-quark loop effects for which there is at present no direct evaluation in the literature.

When the matrix element of $Q^{{\Delta}S=2}$ is evaluated in a theory
that contains a dynamical
charm quark, the resulting estimate for $B_K$ must then be matched to
the three-flavour theory that underlies the effective four-quark
interaction.\footnote{We thank Martin L\"uscher for an interesting
  discussion on this issue.} In general, the matching of $2+1$-flavour
QCD with the theory containing $2+1+1$ flavours of sea quarks 
is performed around the charm threshold.  It is usually accomplished by requiring that the coupling and quark  masses are equal in the two theories at a renormalization scale $\mu$ around $m_c$. In addition, $B_K$ should be renormalized and run, in the four-flavour theory, to the value of $\mu$ at which the two theories are matched, as described in Sec.~\ref{sec:indCP}. The corrections associated with this matching are of
order $(E/m_c)^2$, where $E$ is a typical energy in the process under study, since the subleading operators have dimension eight
\cite{Cirigliano:2000ev}.

When the kaon-mixing amplitude is considered, the matching also
involves the relation between the relevant box diagrams and the
effective four-quark operator. In this case, corrections of order
$(E/m_c)^2$ arise not only from the charm quarks in the sea, but also
from the valence sector, since the charm quark propagates in the box
diagrams. We note that the original derivation of the effective
four-quark interaction is valid up to corrections of order
$(E/m_c)^2$. The kaon-mixing amplitudes evaluated in the
  $\Nf=2+1$ and $2+1+1$ theories are thus subject to corrections of
  the same order in $E/m_c$ as the derivation of the conventional
  four-quark interaction.

Regarding perturbative QCD corrections at the scale of the charm-quark mass on the amplitude in Eq.~(\ref{eq:Heff}), 
the uncertainty on  $\eta_1$ and
$\eta_3$ factors is of $\cO(\alpha_s(m_c)^3)$~\cite{Brod:2011ty,Brod:2010mj}, while that on $\eta_2$ is 
of
$\cO(\alpha_s(m_c)^2)$~\cite{Buras:1990fn}.\,\footnote{The  results of Ref.~\cite{Brod:2019rzc}, based on the use of $u$-$t$ unitarity for the two corresponding perturbative factors, also have an uncertainty of $\cO(\alpha_s(m_c)^2)$ and $\cO(\alpha_s(m_c)^3)$. The estimates for the     missing higher-order contributions are, however, expected to be reduced with respect to the more traditional case where $c$-$t$ unitarity is used;  for a discussion on the $|\epsilon_K|$ computation in the $u$-$t$ unitarity, see the relevant discussion in  Sec.~\ref{sec:indCP}.} 
On the other hand, the corrections of order $(E/m_c)^2$ 
due to dynamical charm-quark effects in the matching of the amplitudes are further suppressed by powers of $\alpha_s(m_c)$ and by a factor of
$1/N_c$, given that they arise from quark-loop diagrams.
In order to make progress in resolving this so far uncontrolled
systematic uncertainty, it is essential that any future calculation of
$B_K$ with $\Nf=2+1+1$ flavours properly addresses the size of these
residual dynamical charm effects in a quantitative way.

Another issue in this context is how the lattice scale and the
physical values of the quark masses are determined in the $2+1$ and
$2+1+1$ flavour theories. Here it is important to consider in which
way the quantities used to fix the bare parameters are affected by a
dynamical charm quark.

A recent study~\cite{Hollwieser:2020qri} using three degenerate light
quarks, together with a charm quark, indicates that the deviations
between the $\Nf=3+1$ and the $\Nf=3$ theories are considerably below the 1\%
level in dimensionless quantities constructed from ratios of gradient-flow
observables, such as $t_0$ and $w_0$, used for scale
setting. This study extends the nonperturbative investigations with
two heavy mass-degenerate
quarks~\cite{Bruno:2014ufa,Athenodorou:2018wpk} which indicate that
dynamical charm-quark effects in low-energy hadronic observables are
considerably smaller than the expectation from a naive power counting
in terms of $\alpha_s(m_c)$. For an additional discussion on this
point, we refer to Ref.~\cite{Carrasco:2015pra}.
Given the hierarchy of scales between the charm-quark
  mass and that of $B_K$, we expect these errors to be modest.  The ETM~15 $\Nf=2+1+1$ $B_K$ result does not include an estimate of this systematic uncertainty. A more quantitative understanding will be required as the statistical uncertainties in $B_K$ will be reduced. Within this review we will not discuss this issue
further. However, we wish to point out that the present discussion  also
applies to $\Nf=2+1+1$ computations of the kaon BSM $B$-parameters discussed in Sec.~\ref{sec:Bi}.

\begin{figure}[ht]
\centering
\includegraphics[width=13cm]{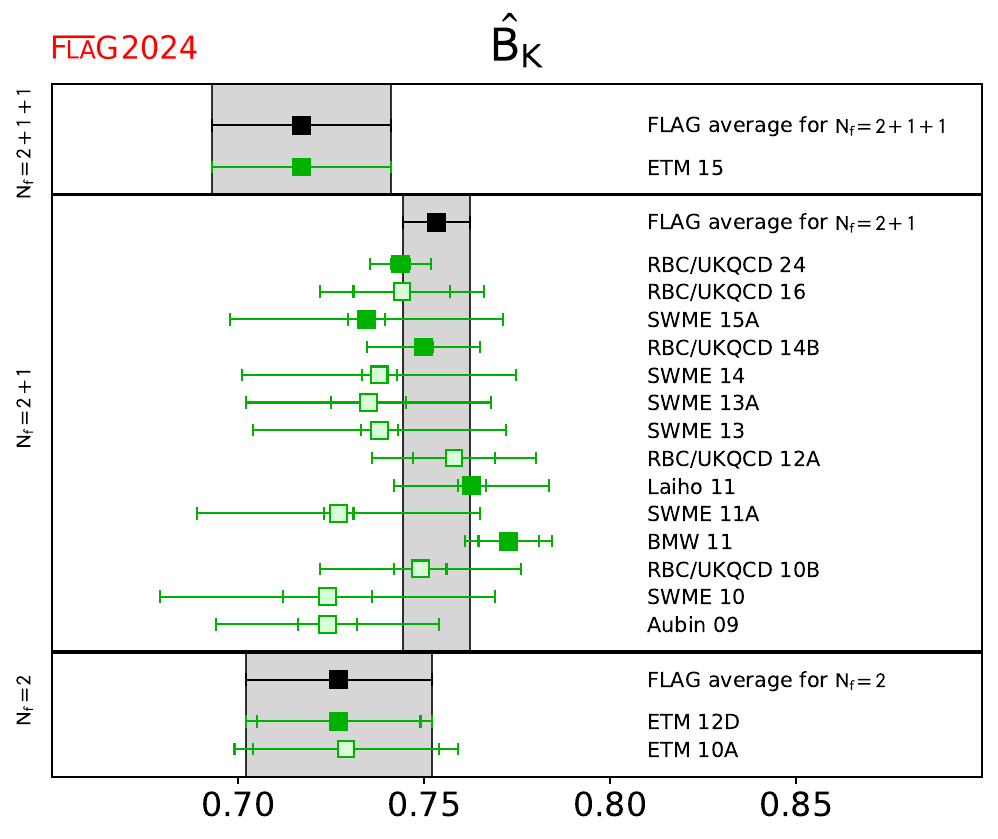}
\caption{Unquenched lattice results for the RGI $B$-parameter
  $\hat{B}_{K}$. The grey bands indicate our averages
  described in the text. For $\Nf=2+1+1$ and $\Nf=2$ the FLAG averages coincide
  with the results by ETM~15 and ETM~12D, respectively. \label{fig_BKsumm}}
\end{figure}

A compilation of  results  for $B_{K}$ with $\Nf=2+1+1, 2+1$ and $2$ flavours of
dynamical quarks is shown in Tabs.~\ref{tab_BKsumm}
and~\ref{tab_BKsumm_nf2}, as well as Fig.~\ref{fig_BKsumm}. An
overview of the quality of systematic error studies is represented by
the colour coded entries in Tabs.~\ref{tab_BKsumm}
and~\ref{tab_BKsumm_nf2}. 
The values of the most relevant lattice parameters and comparative tables on the various
estimates of systematic errors have been  collected in the
corresponding Appendices of the previous FLAG editions~\cite{FlavourLatticeAveragingGroup:2019iem,Aoki:2016frl,Aoki:2013ldr}.

 Since the  last FLAG report, one new result for $B_{K}$ appeared in RBC/UKQCD~24~\cite{Boyle:2024gge}.\,\footnote{We also mention the report of an ongoing work~\cite{Suzuki:2020zue} related to the calculation of $B_{K}$ in which the relevant operators are defined in the framework of gradient flow. A small flow time expansion method was applied in order to compute, 
to 1-loop approximation, the finite matching coefficients between the gradient flow and the
			$\overline{\rm MS}$ schemes for the operators entering the 
			$B_K$ computation.}  For the determination of $B_K$, the  RBC/UKQCD Collaboration employs domain-wall fermions at three lattice spacings spanning the range $[0.07, 0.11]~\rm{fm}$. For the two coarsest lattice spacings, simulations have been performed at the physical pion mass, whereas for the finest lattice spacing, a pion mass of  about 230 MeV has been used.  Residual  chiral symmetry breaking effects induced by the finite extent of the 5th dimension in the
	domain-wall fermion formulation have been checked and found to contribute  
  to the systematic uncertainty of the final estimate of $B_K$ at the per-mille level. Finite-volume effects are found to be negligible. The renormalization constants of the lattice operators are determined nonperturbatively    in two RI-SMOM schemes, namely $(\slash{q},\slash{q})$ and $(\gamma_{\mu}, \gamma_{\mu})$, corresponding to two different choices of renormalization conditions (see Ref.~\cite{Blum:2014tka}). The final values of the renormalization constants are obtained from the average over the results of the two schemes. The error from the $(\gamma_{\mu}, \gamma_{\mu})$ scheme is used
to quote the uncertainty arising from the lattice computation. 
The renormalization constants in the RI-SMOM schemes are computed at the renormalization
scale $\mu = 2~\rm{GeV}$. A nonperturbative step-scaling procedure is used to run them to
$\mu = 3~\rm{GeV}$ where the results are perturbatively matched to the $\overline{\rm{MS}}$
scheme.
The continuum and physical point result for $B_K$ is obtained through a combined chiral and
continuum extrapolation using NLO SU(2) chiral perturbation theory.
The spread  between the result obtained as described above and the result of a calculation performed directly at $\mu = 3~\rm{GeV}$ is taken as an estimate of the uncertainty due to discretization effects. The dominant error of the RBC/UKQCD~24 computation of $B_K$ arises from the perturbative matching of the RI-SMOM schemes used in the lattice computation to the $\overline{\rm{MS}}$ scheme.
This is estimated as half the difference of the results
obtained from the use of the two intermediate RI-SMOM schemes in the matching.
In this computation of $B_K$, a green star symbol is assigned to all FLAG quality criteria.

For a detailed description of previous $B_K$ calculations we refer the reader to FLAG 16~\cite{Aoki:2016frl}.

We now give the FLAG averages for $B_K$ for $\Nf=2+1+1, 2+1$, and 2 dynamical flavours.

 We begin with the $\Nf=2+1$ global average, which is estimated by employing five different $B_K$ results, namely BMW~11~\cite{Durr:2011ap}, Laiho~11~\cite{Laiho:2011np}, RBC/UKQCD~14B~\cite{Blum:2014tka}, 
SWME~15A~\cite{Jang:2015sla}, and RBC/UKQCD~24~\cite{Boyle:2024gge}.  Moreover, we recall that the expression of $\epsilon_K$ in terms of $B_K$ is derived in the three-flavour  theory (see Sec.~\ref{sec:indCP}). 
Our procedure is: first, we combine in quadrature the statistical and systematic errors of each individual result of the RGI $B$ parameter $\hat{B}_{\rm{K}}$.  A weighted average is then obtained from the set of results. For the final error estimate, we take correlations between different collaborations into account. Specifically, we consider the statistical and finite-volume errors of SWME~15A and Laiho~11 to be correlated, since both groups use the asqtad ensembles generated by the MILC Collaboration. Laiho~11 and RBC/UKQCD~14B both use domain-wall quarks in the valence sector and  employ similar procedures for the nonperturbative determination of matching factors. Hence, we treat  their quoted renormalization and matching uncertainties  as correlated. Moreover, we treat  the results obtained by RBC/UKQCD~14B and RBC/UKQCD~24 as fully correlated because part of the sea ensembles in the two calculations are common.\footnote{However, due to partly different methodology in the analysis and the renormalization procedure the two computations are considered as separate, and for this reason they are both included in the global average.}  In the calculation of the average, we incorporate the new  FLAG data-driven criterion (see Sec.~\ref{sec:DataDriven}) concerning the extrapolation to the continuum limit  which increases  by approximately 3.7\%  the total error of the RBC/UKQCD~24 calculation.  Following
Schmelling's procedure~\cite{Schmelling:1994pz}
to construct the global covariance matrix of the results contributing to the average, we arrive at the following value, $\hat{B}_{K} = 0.7533(85)$. Since the fit implementing the weighted average has $\chi^2/\rm{dof} = 1.142$,  according to the general FLAG rule, we stretch the error by the square root of the reduced $\chi^2$ value. This effect is mainly driven by the two most precise determinations of $\hat{B}_{\rm{K}}$, corresponding to RBC/UKQCD~24 and BMW~11, which differ at the 2$\sigma$ level. This procedure leads to the following result:
%
\begin{equation}
  \Nf=2+1:\hspace{1.5cm}\FLAGAVBEGIN \hat{B}_{K} = 0.7533(91)\FLAGAVEND\qquad\Refs~\mbox{\cite{Durr:2011ap,Laiho:2011np,Blum:2014tka,Jang:2015sla,Boyle:2024gge}},
  \label{eq:BK2+1}
\end{equation}
%
 After applying the NLO conversion
factors $\hat{B}_{K}/B_{K}^\msbar (2\,{\rm GeV})=1.369$  and  $\hat{B}_{K}/B_{K}^\msbar (3\,{\rm GeV})=1.415$ \footnote{
We refer to FLAG~19 \cite{FlavourLatticeAveragingGroup:2019iem}  for a discussion of the procedure followed in estimating the conversion factors to $\msbar$ at 2 GeV.  In addition, for the computation of  the conversion factor  from  RGI to the $\msbar$ scheme at 3 GeV, which is new here,  we have used the three-flavour $\Lambda_{\rm{QCD}}$ from FLAG~21 and the 4-loop formula for the $\beta$-function of the strong coupling constant. The propagation error owing to the error of $\Lambda_{\rm{QCD}}$ is found to be negligible compared to the total uncertainty of the $B_K$ estimate.}, this   
becomes
\begin{equation}
  \hspace*{-0.3cm}\Nf=2+1:\hspace{0.1cm} B_{K}^\msbar(2\,{\rm GeV})=0.5503(66), \hspace{0.1cm} B_{K}^\msbar(3\,{\rm GeV})=0.5324(64)\, , \hspace{0.1cm} \Refs~\mbox{\cite{Durr:2011ap,Laiho:2011np,Blum:2014tka,Jang:2015sla,Boyle:2024gge}}.
\end{equation}

Improvements in lattice calculations in recent years have led to a considerable reduction in statistical errors.  This has implied that some of the results contributing to the global average are nowadays statistically incompatible. Only by taking into account the contributions to systemic uncertainties, both from the lattice calculations themselves and, notably, from perturbative matching, can it be seen that the weighted average produces a value of $\cO(1)$ for the reduced $\chi^2$.

There is only a single  result for $\Nf=2+1+1$, computed by the ETM
collaboration\,\cite{Carrasco:2015pra}. Since it is free of red tags,
it qualifies  to the following average, 
%
\begin{equation}
\Nf=2+1+1:\hspace{1.0cm} \FLAGAVBEGIN\hat{B}_{K} = 0.717(18)(16)\FLAGAVEND\, ,
 \hspace*{0.4cm} \Ref~\mbox{\cite{Carrasco:2015pra}}.
\end{equation}
%
 Using the same conversion factors as in the three-flavour theory, this value translates into 
\begin{equation}
  \hspace*{-0.3cm}\Nf=2+1+1:\hspace{0.1cm} B_{K}^\msbar (2\,{\rm GeV}) = 0.524(13)(12),\,\,  B_{K}^\msbar (3\,{\rm GeV}) = 0.507(13)(11), \hspace*{0.1cm} \Ref~\mbox{\cite{Carrasco:2015pra}}.
\end{equation}

For $\Nf=2$ flavours the   average is based on a single result, that 
of ETM~12D\,\cite{Bertone:2012cu}: 
%
\begin{equation}
\Nf=2:\hspace{.3cm}\FLAGAVBEGIN\hat{B}_{K} = 0.727(22)(12)\FLAGAVEND ,
\hspace{.4cm} \Ref~\mbox{\cite{Bertone:2012cu}}\, ,
\end{equation}
%
which, using the same conversion factors as in the three-flavour theory, translates into 
\begin{equation}
  \hspace*{-0.3cm}\Nf=2:\hspace{0.1cm} B_{K}^\msbar (2\,{\rm GeV}) =  0.531(16)(9),\,\,  B_{K}^\msbar (3\,{\rm GeV}) = 0.514(16)(8), \hspace*{0.1cm} \Ref~\mbox{\cite{Bertone:2012cu}}.
\end{equation}

\begin{table*}[ht]
\begin{center}
\mbox{} \\[3.0cm]
{\footnotesize{
\vspace*{-2cm}\begin{tabular*}{\textwidth}[l]{l @{\extracolsep{\fill}} r@{\hspace{1mm}}l@{\hspace{1mm}}l@{\hspace{1mm}}l@{\hspace{1mm}}l@{\hspace{1mm}}l@{\hspace{1mm}}l@{\hspace{1mm}}l@{\hspace{1mm}}l@{\hspace{1mm}}l}
Collaboration & Ref. & $\Nf$ & 
\hspace{0.15cm}\begin{rotate}{60}{publication status}\end{rotate}\hspace{-0.15cm} &
\hspace{0.15cm}\begin{rotate}{60}{continuum extrapolation}\end{rotate}\hspace{-0.15cm} &
\hspace{0.15cm}\begin{rotate}{60}{chiral extrapolation}\end{rotate}\hspace{-0.15cm}&
\hspace{0.15cm}\begin{rotate}{60}{finite volume}\end{rotate}\hspace{-0.15cm}&
\hspace{0.15cm}\begin{rotate}{60}{renormalization}\end{rotate}\hspace{-0.15cm}  &
\hspace{0.15cm}\begin{rotate}{60}{running}\end{rotate}\hspace{-0.15cm} & 
\rule{0.3cm}{0cm}$B_{{K}}(\overline{\rm MS},2\,{\rm GeV})$ 
& \rule{0.3cm}{0cm}$\hat{B}_{{K}}$ \\
&&&&&&&&&& \\[-0.1cm]
\hline
\hline
&&&&&&&&&& \\[-0.1cm]

ETM~15 & \cite{Carrasco:2015pra} & 2+1+1 & \gA & \good & \soso & \soso
& \good&  $\,a$ &   0.524(13)(12)  & 0.717(18)(16)$^1$ \\[0.5ex]
&&&&&&&&&& \\[-0.1cm]
\hline
&&&&&&&&&& \\[-0.1cm]
RBC/UKQCD~24 & \cite{Boyle:2024gge} & 2+1 & \gA & \good & \good &
\good & \good & $\,b$ & 0.540(2)(20)$^2$ & 0.7436(25)(78) \\[0.5ex]

RBC/UKQCD~16 & \cite{Garron:2016mva} & 2+1 & \gA & \soso & \soso &
\soso & \good & $\,c$ & 0.543(9)(13)$^3$ & 0.744(13)(18)$^4$ \\[0.5ex]

SWME~15A & \cite{Jang:2015sla} & 2+1 & \gA & \good & \soso &
\good & \soso$^\ddagger$  & $-$ & 0.537(4)(26) & 0.735(5)(36)$^5$ \\[0.5ex]

RBC/UKQCD~14B
& \cite{Blum:2014tka} & 2+1 & \gA & \good & \good &
     \good  & \good & $\,c$  & 0.5478(18)(110)$^3$ & 0.7499(24)(150) \\[0.5ex]  

SWME~14 & \cite{Bae:2014sja} & 2+1 & \gA & \good & \soso &
\good & \soso$^\ddagger$  & $-$ & 0.5388(34)(266) & 0.7379(47)(365) \\[0.5ex]

SWME~13A & \cite{Bae:2013tca} & 2+1 & \gA & \good & \soso  &
\good & \soso$^\ddagger$  & $-$ & 0.537(7)(24) & 0.735(10)(33) \\[0.5ex]

SWME~13 & \cite{Bae:2013lja} & 2+1 & \rC & \good & \soso &
\good & \soso$^\ddagger$ & $-$ & 0.539(3)(25) & 0.738(5)(34) \\[0.5ex]

RBC/UKQCD~12A
& \cite{Arthur:2012opa} & 2+1 & \gA & \soso & \good &
     \soso & \good & $\,c$ & 0.554(8)(14)$^3$ & 0.758(11)(19) \\[0.5ex]  

Laiho~11 & \cite{Laiho:2011np} & 2+1 & \rC & \good & \soso &
     \soso & \good & $-$ & 0.5572(28)(150)& 0.7628(38)(205)$^5$ \\[0.5ex]  

SWME~11A & \cite{Bae:2011ff} & 2+1 & \gA & \good & \soso &
\soso & \soso$^\ddagger$ & $-$ & 0.531(3)(27) & 0.727(4)(38) \\[0.5ex]

BMW~11 & \cite{Durr:2011ap} & 2+1 & \gA & \good & \good & \good & \good
& $\,d$ & 0.5644(59)(58) & 0.7727(81)(84) \\[0.5ex]

RBC/UKQCD~10B & \cite{Aoki:2010pe} & 2+1 & \gA & \soso & \soso & \good &
\good & $\,e$ & 0.549(5)(26) & 0.749(7)(26) \\[0.5ex] 

SWME~10 & \cite{Bae:2010ki} & 2+1 & \gA & \good & \soso & \soso & \soso
& $-$ & 0.529(9)(32) &  0.724(12)(43) \\[0.5ex] 

Aubin~09 & \cite{Aubin:2009jh} & 2+1 & \gA & \soso & \soso &
     \soso & \tbg & $-$ & 0.527(6)(21)& 0.724(8)(29) \\[0.5ex]  

&&&&&&&&&& \\[-0.1cm]
\hline
\hline\\[-0.1cm]
\end{tabular*}
}}
\begin{minipage}{\linewidth}
{\footnotesize 
\begin{itemize}
\item[$^\ddagger$] The renormalization is performed using perturbation
        theory at 1-loop, with a conservative estimate of the uncertainty. \\[-5mm]
\item[$a$]  $B_K$ is renormalized nonperturbatively at scales $1/a \sim$ 2.2--3.3 $\gev$ in the $\Nf = 4$ RI/MOM scheme 
     using two different lattice momentum scale intervals, the first around $1/a$ while the second around  3.5 GeV. 
        The impact of the two ways to the final 
        result is taken into account  in the error budget. Conversion to $\msbar$ is at 1-loop order at 3 GeV.  \\[-5mm]
\item[$b$] $B_K$ is renormalized nonperturbatively at a scale of 2.0 GeV
in two RI/SMOM schemes for $\Nf = 3$, and 
then run to 3 GeV using a nonperturbatively determined step-scaling
function.  A direct computation at 3 GeV is also used to estimate systematic
uncertainties related to discretization effects.
Conversion to $\msbar$ is at 1-loop order at 3 GeV.\\[-5mm]
\item[$c$] $B_K$ is renormalized nonperturbatively at a scale of 1.4 GeV
        in two RI/SMOM schemes for $\Nf = 3$, and 
	then run to 3 GeV using a nonperturbatively determined step-scaling
        function. 
	Conversion to $\msbar$ is at 1-loop order at 3 GeV.\\[-5mm]
\item[$d$] $B_K$ is renormalized and run nonperturbatively to a scale of
        $3.5\,\gev$ in the RI/MOM scheme. At the same scale, conversion at 1-loop order to $\msbar$ is applied. 
	Nonperturbative and NLO
        perturbative running agrees down to scales of $1.8\,\gev$ within
        statistical
	uncertainties of about 2\%.\\[-5mm]
\item[$e$] $B_K$ is renormalized nonperturbatively at a scale of 2\,GeV
        in two RI/SMOM schemes for $\Nf = 3$, and then 
	run to 3 GeV using a nonperturbatively determined step-scaling
        function. Conversion to $\msbar$ is at 
	1-loop order at 3 GeV.\\[-5mm]
\item[$^1$] $B_{K}(\msbar, 2\,\gev)$ and $\hat{B}_{{K}}$ are related
        using the conversion factor  1.369, i.e., the one obtained
        with $\Nf=2+1$.  \\[-5mm]
\item[$^2$] $B_{K}(\msbar, 2\,\gev)$ value from a private communication with the  authors. The first error is due to lattice statistical and systematic uncertainties;  the second error is associated with the perturbative truncation uncertainty in
matching to $\msbar$ at a scale of 2~GeV.  \\[-5mm]
\item[$^3$]  $B_{K}(\msbar, 2\,\gev)$ is obtained from the  estimate for $\hat{B}_{{K}}$ using the conversion factor 1.369.            \\[-5mm]
\item[$^4$] $\hat{B}_{{K}}$ is obtained from $B_{K}(\msbar, 3\,\gev)$ using the conversion factor 
            employed in  Ref.~\cite{Blum:2014tka}.           \\[-5mm]             
\item[$^5$] $\hat{B}_{{K}}$ is obtained from the estimate for
        $B_{K}(\msbar, 2\,\gev)$ using the conversion factor 1.369. 
\end{itemize}
}
\end{minipage}
\caption{Results for the kaon $B$-parameter in QCD with $\Nf=2+1+1$
  and $\Nf=2+1$, together with a summary of
  systematic errors.
      Information about nonperturbative 
  running is indicated in the column ``running,'' with details given at
  the bottom of the table.\label{tab_BKsumm}}
\end{center}
\end{table*}

\clearpage

\begin{table*}[h]
\begin{center}
\mbox{} \\[3.0cm]
{\footnotesize{
\vspace*{-2cm}\begin{tabular*}{\textwidth}[l]{l @{\extracolsep{\fill}} r l l l l l l l l l}
Collaboration & Ref. & $\Nf$ & 
\hspace{0.15cm}\begin{rotate}{60}{publication status}\end{rotate}\hspace{-0.15cm} &
\hspace{0.15cm}\begin{rotate}{60}{continuum extrapolation}\end{rotate}\hspace{-0.15cm} &
\hspace{0.15cm}\begin{rotate}{60}{chiral extrapolation}\end{rotate}\hspace{-0.15cm}&
\hspace{0.15cm}\begin{rotate}{60}{finite volume}\end{rotate}\hspace{-0.15cm}&
\hspace{0.15cm}\begin{rotate}{60}{renormalization}\end{rotate}\hspace{-0.15cm}  &
\hspace{0.15cm}\begin{rotate}{60}{running}\end{rotate}\hspace{-0.15cm} & 
\rule{0.3cm}{0cm}$B_{{K}}(\overline{\rm MS},2\,{\rm GeV})$ 
& \rule{0.3cm}{0cm}$\hat{B}_{{K}}$ \\
&&&&&&&&&& \\[-0.1cm]
\hline
\hline
&&&&&&&&&& \\[-0.1cm]

ETM~12D & \cite{Bertone:2012cu} & 2 & \gA & \good & \soso & \soso
& \good&  $\,f$ &   0.531(16)(9)  & 0.727(22)(12)$^1$ \\[0.5ex]
ETM~10A & \cite{Constantinou:2010qv} & 2 & \gA & \good & \soso & \soso
& \good&  $\,g$ &   0.533(18)(12)$^1$  & 0.729(25)(17) \\[0.5ex]
&&&&&&&&&& \\[-0.1cm]
\hline
\hline\\[-0.1cm]
\end{tabular*}
}}
\begin{minipage}{\linewidth}
{\footnotesize 
\begin{itemize}
\item[$f$] $B_K$ is renormalized nonperturbatively at scales $1/a \sim$ 2--3.7 $\gev$ in the $\Nf = 2$ RI/MOM scheme. In this
        scheme, nonperturbative and NLO
        perturbative running are shown to agree from 4 GeV down to 2 GeV to
        better than 3\%
        \cite{Constantinou:2010gr,Constantinou:2010qv}.  \\[-5mm]
\item[$g$] $B_K$ is renormalized nonperturbatively at scales $1/a \sim$ 2--3 $\gev$ in the $\Nf = 2$ RI/MOM scheme. In this
        scheme, nonperturbative and NLO
        perturbative running are shown to agree from 4 GeV down to 2 GeV to
        better than 3\%
        \cite{Constantinou:2010gr,Constantinou:2010qv}.  \\[-5mm]
        
\item[$^1$] $B_{K}(\msbar, 2\,\gev)$ and $\hat{B}_{{K}}$ are related using the conversion factor  1.369, i.e., the one obtained with $\Nf=2+1$. 
\end{itemize}
}
\end{minipage}
\caption{Results for the kaon $B$-parameter in QCD with $\Nf=2$, 
  together with a summary of systematic
  errors.  Information about nonperturbative 
  running is
  indicated in the column ``running,'' with details given at the bottom
  of the table.\label{tab_BKsumm_nf2}}
\end{center}
\end{table*}

\subsection{Kaon BSM $B$-parameters}
\label{sec:Bi}

We now consider the matrix elements of
operators that encode the effects of physics beyond the Standard Model
(BSM) to the mixing of neutral kaons. In this theoretical framework,
both the SM and BSM contributions add up to reproduce the
experimentally observed value of $\epsilon_K$. 
As long as BSM contributions involve heavy particles with masses much larger than $\lqcd$, they will be short-distance dominated.
The effective Hamiltonian for generic ${\Delta}S=2$
processes including BSM contributions reads
\begin{equation}
  {\cal H}_{\rm eff,BSM}^{\Delta S=2} = \sum_{i=1}^5
  C_i(\mu)Q_i(\mu),
\end{equation}
where $Q_1$ is the four-quark operator of Eq.~(\ref{eq:Q1def}) that
gives rise to the SM contribution to $\epsilon_K$. In the so-called
SUSY basis introduced by Gabbiani et al.~\cite{Gabbiani:1996hi}, the
 operators $Q_2,\ldots,Q_5$ are\,\footnote{Thanks to QCD
  parity invariance lattice computations for three more dimension-six operators,
  whose parity conserving parts coincide with the corresponding parity
  conserving contributions of the operators $Q_1, Q_2$ and $Q_3$, can be ignored.}
\begin{eqnarray}
 & & Q_2 = \big(\bar{s}^a(1-\gamma_5)d^a\big)
           \big(\bar{s}^b(1-\gamma_5)d^b\big), \nonumber\\
 & & Q_3 = \big(\bar{s}^a(1-\gamma_5)d^b\big)
           \big(\bar{s}^b(1-\gamma_5)d^a\big), \nonumber\\
 & & Q_4 = \big(\bar{s}^a(1-\gamma_5)d^a\big)
           \big(\bar{s}^b(1+\gamma_5)d^b\big), \nonumber\\
 & & Q_5 = \big(\bar{s}^a(1-\gamma_5)d^b\big)
           \big(\bar{s}^b(1+\gamma_5)d^a\big),
\end{eqnarray}
where $a$ and $b$ are colour indices.  In analogy to the case of
$B_{K}$, one then defines the $B$-parameters of $Q_2,\ldots,Q_5$
according to
\be
   B_i(\mu) = \frac{\left\langle \bar{K}^0\left| Q_i(\mu)\right|K^0
     \right\rangle}{N_i\left\langle\bar{K}^0\left|\bar{s}\gamma_5
     d\right|0\right\rangle \left\langle0\left|\bar{s}\gamma_5
     d\right|K^0\right\rangle}, \quad i=2,\ldots,5.
\ee
The factors $\{N_2,\ldots,N_5\}$ are given by $\{-5/3, 1/3, 2, 2/3\}$,
and it is understood that $B_i(\mu)$ is specified in some
renormalization scheme, such as $\msbar$ or a variant of the
regularization-independent momentum subtraction (RI-MOM) scheme.

The SUSY basis has been adopted in
Refs.~\cite{Boyle:2012qb,Bertone:2012cu,Carrasco:2015pra,Garron:2016mva,Boyle:2024gge}. Alternatively,
one can employ the chiral basis of Buras, Misiak and
Urban\,\cite{Buras:2000if}. The SWME collaboration prefers the latter
since the anomalous dimension that enters the RG running has been
calculated to 2-loop order in perturbation
theory\,\cite{Buras:2000if}. Results obtained in the chiral basis can
be easily converted to the SUSY basis via
\be
   B_3^{\rm SUSY}={\textstyle\frac{1}{2}}\left( 5B_2^{\rm chiral} -
   3B_3^{\rm chiral} \right).
\ee
The remaining $B$-parameters are the same in both bases. In the
following, we adopt the SUSY basis and drop the superscript.

Older quenched results for the BSM $B$-parameters can be found in
Refs.~\cite{Allton:1998sm, Donini:1999nn, Babich:2006bh}. For a nonlattice approach 
to get estimates for the BSM $B$-parameters see Ref.~\cite{Buras:2018lgu}.  

Estimates for $B_2,\ldots,B_5$ have been reported for QCD with $\Nf=2$
(ETM~12D~\cite{Bertone:2012cu}), $\Nf=2+1$
(RBC/UKQCD~12E\,\cite{Boyle:2012qb}, SWME~13A\,\cite{Bae:2013tca},
SWME~14C\,\cite{Jang:2014aea}, SWME~15A\,\cite{Jang:2015sla}, \\
RBC/UKQCD~16 \cite{Garron:2016mva,Boyle:2017skn}, RBC/UKQCD~24 \cite{Boyle:2024gge})  and $\Nf=2+1+1$
(ETM~15\,\cite{Carrasco:2015pra}) flavours of dynamical quarks. 
Since the publication of FLAG~19~\cite{FlavourLatticeAveragingGroup:2019iem} 
 a single new work Ref.~\cite{Boyle:2024gge} has appeared. The basic characteristics of this calculation have been reported in the $B_K$ section, see Sec.~\ref{sec:BK lattice}. 
As in the case of $B_K$, the dominant
error for all the BSM $B$-parameters arises from  the systematic uncertainty associated to
the truncation error in the perturbative matching from the intermediate
schemes to the $\msbar$ scheme.
This is estimated as half the difference of the results obtained from the matching to $\msbar$ of the two
intermediate schemes. 
The ratio of the BSM to SM matrix elements are also reported in Ref.~\cite{Boyle:2024gge}.       
  
\begin{figure}[ht]
\centering
\leavevmode
\includegraphics[width=\textwidth]{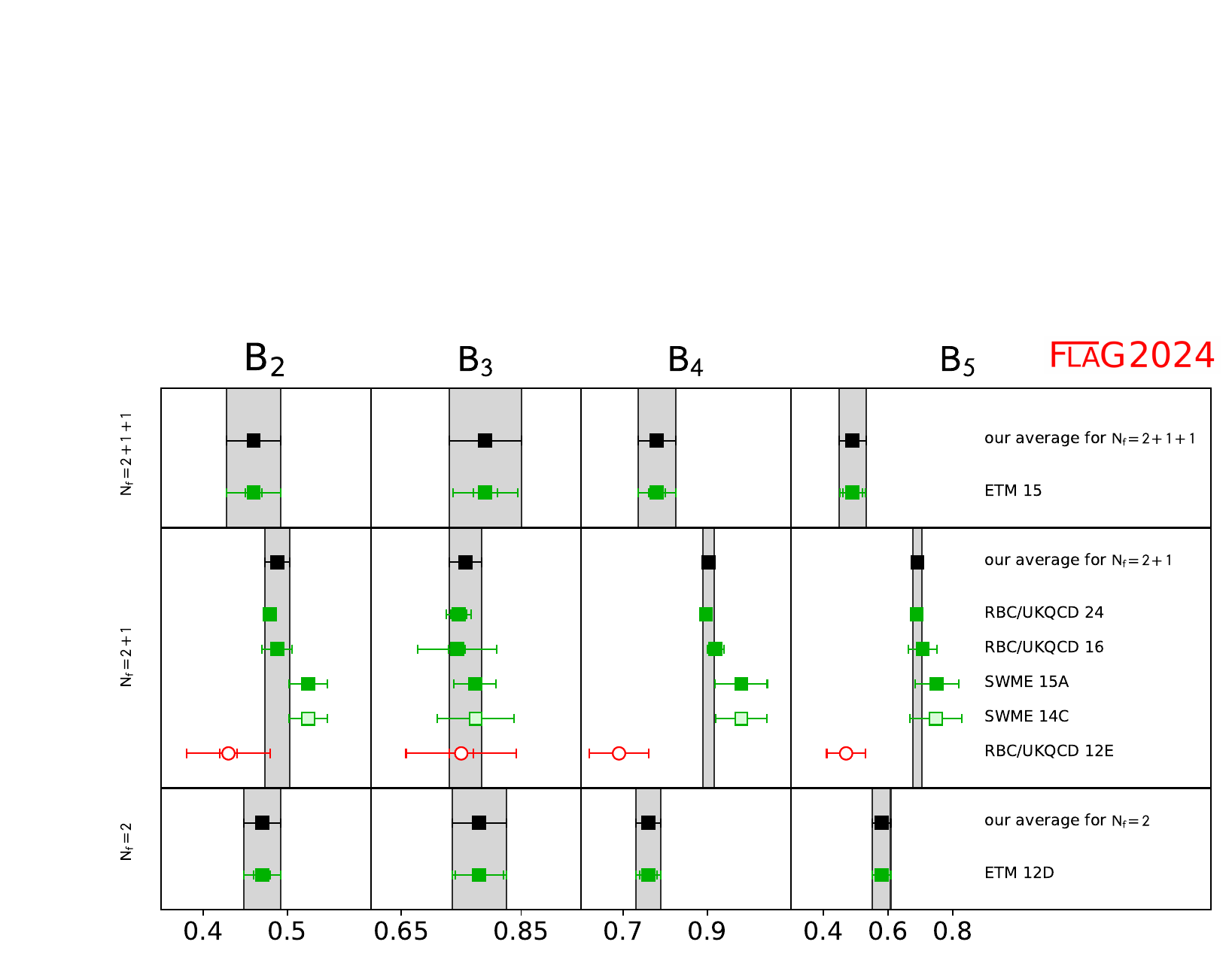}
\caption{Results for the BSM $B$-parameters defined in the
  $\msbar$ scheme at a reference scale of 3\,GeV (see Tab.~\ref{tab_Bi}).
\label{fig_Bisumm}}
\end{figure}

All the available results 
are listed and compared in Tab.~\ref{tab_Bi} and
Fig.~\ref{fig_Bisumm}. In general, one finds that the BSM $B$-parameters computed by different collaborations do not show the same
level of consistency as the SM kaon-mixing parameter $B_K$ discussed
previously. 
Control over the systematic uncertainties from chiral and continuum extrapolations as well as finite-volume effects in $B_2,\ldots,B_5$
is expected to be at a comparable level as that for $B_{K}$, as far as the
results by ETM~12D, ETM~15, SWME~15A and RBC/UKQCD~16 are concerned,
since the set of gauge ensembles employed in both kinds of computations is the same. However, the most recent  results by RBC/UKQCD~24 with $\Nf=2+1$ flavours are, in general, much more precise than the older ones. 
Notice that the calculation by RBC/UKQCD~12E has been performed at a single value of
the lattice spacing and a minimum pion mass of 290\,MeV.

As reported in  RBC/UKQCD~16~\cite{Garron:2016mva} and RBC/UKQCD~24 \cite{Boyle:2024gge}, the comparison of 
results obtained in the conventional RI-MOM and in two RI-SMOM
schemes  shows significant discrepancies for some of the BSM $B$-parameters. Tensions are observed for the cases of $B_4$ and $B_5$, where the discrepancies  between results obtained with RI-MOM and RI-SMOM are  at the level of  $2.6~ \sigma$ and $4.5~ \sigma$, respectively. The results of RBC/UKQCD~16 and RBC/UKQCD~24 lie closer to those of SWME~15A which rely on  perturbative renormalization at 1-loop order.
On the other hand, the results for $B_2$ and $B_3$ obtained by ETM~15, SWME~15A, RBC/UKQCD~16 and RBC/UKQCD~24 show a better level of compatibility.

The findings by RBC/UKQCD~16 \cite{Garron:2016mva}, RBC/UKQCD~17A \cite{Boyle:2017skn} and RBC/UKQCD~24 \cite{Boyle:2024gge} highlight the importance of carefully assessing the systematic effects on the implementation of the Rome-Southampton
method used for nonperturbative renormalization. In particular, the RI-MOM and RI-SMOM schemes differ in that the use of nonexceptional kinematics, in the RI-SMOM scheme, removes the need to subtract the pseudo-Goldstone boson pole contamination, as is required in the RI-MOM case. In addition, for each of the schemes a specific analysis of the truncation error in the perturbative matching to $\msbar$ must be carried out.

\begin{table}[!h]
\begin{center}
\mbox{} \\[3.0cm]
{\footnotesize{
\begin{tabular*}{\textwidth}[l]{l @{\extracolsep{\fill}}r@{\hspace{1mm}}l@{\hspace{1mm}}l@{\hspace{1mm}}l@{\hspace{1mm}}l@{\hspace{1mm}}l@{\hspace{1mm}}l@{\hspace{1mm}}l@{\hspace{1mm}}l@{\hspace{1mm}}l@{\hspace{1mm}}l@{\hspace{1mm}}l}
Collaboration & Ref. & $\Nf$ & 
\hspace{0.15cm}\begin{rotate}{60}{publication status}\end{rotate}\hspace{-0.15cm} &
\hspace{0.15cm}\begin{rotate}{60}{continuum extrapolation}\end{rotate}\hspace{-0.15cm} &
\hspace{0.15cm}\begin{rotate}{60}{chiral extrapolation}\end{rotate}\hspace{-0.15cm}&
\hspace{0.15cm}\begin{rotate}{60}{finite volume}\end{rotate}\hspace{-0.15cm}&
\hspace{0.15cm}\begin{rotate}{60}{renormalization}\end{rotate}\hspace{-0.15cm}  &
\hspace{0.15cm}\begin{rotate}{60}{running}\end{rotate}\hspace{-0.15cm} & 
$B_2$ & $B_3$ & $B_4$ & $B_5$ \\
&&&&&&&&& \\[-0.1cm]
\hline
\hline
&&&&&&&&& \\[-0.1cm]

ETM~15 & \cite{Carrasco:2015pra} & 2+1+1 & \gA & \good & \soso & \soso
& \good&  $\,a$ & 0.46(1)(3) & 0.79(2)(5) & 0.78(2)(4) & 0.49(3)(3)  \\[0.5ex]
&&&&&&&&& \\[-0.1cm]

\hline

&&&&&&&&&& \\[-0.1cm]
RBC/UKQCD~24 & \cite{Boyle:2024gge} & 2+1 & \gA & \good & \good &
\good & \good & $\,b$ & 0.4794(25)(35) & 0.746(13)(17) & 0.897(02)(10) &
0.6882(78)(94) \\[0.5ex]
&&&&&&&&& \\[-0.1cm]
RBC/UKQCD~16 & \cite{Garron:2016mva} & 2+1 & \gA & \soso & \soso &
\soso & \good & $\,b$ & 0.488(7)(17) & 0.743(14)(65) & 0.920(12)(16) &
0.707(8)(44) \\[0.5ex]

&&&&&&&&& \\[-0.1cm]
SWME~15A & \cite{Jang:2015sla} & 2+1 & \gA & \good & \soso &
\good & \soso$^\dagger$ & $-$ & 0.525(1)(23) & 0.773(6)(35) & 0.981(3)(62) & 0.751(7)(68)  \\[0.5ex]
&&&&&&&&& \\[-0.1cm]

SWME~14C & \cite{Jang:2014aea} & 2+1 & C & \good & \soso &
\good & \soso$^\dagger$ & $-$ & 0.525(1)(23) & 0.774(6)(64) & 0.981(3)(61) & 0.748(9)(79)  \\[0.5ex]
&&&&&&&&& \\[-0.1cm]

SWME~13A$^\ddagger$ & \cite{Bae:2013tca} & 2+1 & \gA & \good & \soso  &
\good & \soso$^\dagger$ & $-$ & 0.549(3)(28)  & 0.790(30) & 1.033(6)(46) & 0.855(6)(43)   \\[0.5ex]
&&&&&&&&& \\[-0.1cm]

RBC/
& \cite{Boyle:2012qb} & 2+1 & \gA & \tbr & \soso & \good &
\good & $\,b$ & 0.43(1)(5)  & 0.75(2)(9)  & 0.69(1)(7)  & 0.47(1)(6)
\\
UKQCD~12E & & & & & & & & & & & & \\[0.5ex]  
&&&&&&&&& \\[-0.1cm]

\hline

&&&&&&&&& \\[-0.1cm]
ETM~12D & \cite{Bertone:2012cu} & 2 & \gA & \good & \soso & \soso
& \good&  $\,c$ & 0.47(2)(1)  & 0.78(4)(2)  & 0.76(2)(2)  & 0.58(2)(2)  \\[0.5ex]

&&&&&&&&& \\[-0.1cm]
\hline
\hline\\[-0.1cm]
\end{tabular*}
}}
\begin{minipage}{\linewidth}
{\footnotesize 
\begin{itemize}
\item[$^\dagger$] The renormalization is performed using perturbation
        theory at 1-loop order, with a conservative estimate of
         the uncertainty. \\[-5mm]
\item[$a$] $B_i$ are renormalized nonperturbatively at scales $1/a
        \sim$ 2.2--3.3 $\gev$ in the $\Nf = 4$ RI/MOM scheme 
        using two different lattice momentum scale intervals, with
        values around $1/a$ for the first and around
        3.5~GeV for the second one. The impact of
        these two ways to the final result is taken into account
         in the error budget. Conversion to $\msbar$ is at 1-loop order at 3~GeV.\\[-5mm]
\item[$b$] The $B$-parameters are renormalized nonperturbatively at a scale of 3~GeV. \\[-5mm]
\item[$c$] $B_i$ are renormalized nonperturbatively at scales $1/a
        \sim$ 2--3.7 $\gev$ in the $\Nf = 2$ RI/MOM scheme using
        two different lattice momentum scale intervals,  
        with values around $1/a$ for the first and around 3~GeV
        for the second one.\\[-5mm]
\item[$^\ddagger$] The computation of $B_4$ and $B_5$ has been
        revised in Refs.~\cite{Jang:2015sla} and \cite{Jang:2014aea}. 
\end{itemize}
}
\end{minipage}
\caption{Results for the BSM $B$-parameters $B_2,\ldots,B_5$ in the
  $\msbar$ scheme at a reference scale of 3\,GeV.    Information about nonperturbative 
  running is indicated in the column
  ``running,'' with details given at the bottom of the
  table.~\label{tab_Bi}}
\end{center}
\end{table}
\clearpage

 A nonperturbative computation of the running of the four-fermion
operators contributing to the $B_2$, \dots , $B_5$ parameters has been
carried out with two dynamical flavours using the Schr\"odinger
functional renormalization
scheme~\cite{Dimopoulos:2018zef}. Renormalization matrices of the
operator basis are used to build step-scaling functions governing the
continuum-limit running between hadronic and electroweak scales. A
comparison to perturbative results using NLO (2-loop order) for the
four-fermion operator anomalous dimensions indicates that, at scales
of about 3\,GeV, nonperturbative effects can induce a sizeable
contribution to the running.  Similar conclusions are obtained on the basis of preliminary results  for the  renormalization-group running of the complete basis 
of $\Delta F = 2$ four-fermion operators  using $\Nf=3$ dynamical massless flavours in the Schr\"odinger setup~\cite{CamposPlasencia:2024mug}.

A closer look at the 
calculations reported in ETM~15
\cite{Carrasco:2015pra}, SWME~15A \cite{Jang:2015sla}, 
RBC/UKQCD~16 \cite{Garron:2016mva}, and RBC/UKQCD~24~\cite{Boyle:2024gge} reveals that cutoff effects  tend to 
be larger for the BSM $B$-parameters compared to those of $B_K$. 
In order to take into account this effect in the average analysis, we  make use of the new  FLAG data-driven  criterion (see Sec.~\ref{sec:DataDriven}) concerning the extrapolation to the continuum limit. In summary, we report that in the average procedure, (a) for $B_2$ the total errors by RBC/UKQCD~24 and RBC/UKQCD~16 have been inflated by a factor 2.6 and by 22\%,  respectively; (b) for $B_3$ the  total errors by ETM~15,  RBC/UKQCD~16 and RBC/UKQCD~24 have been inflated by 11\%, 45\% and 52\%,  respectively; (c) for $B_4$ no error inflation is required; and (d) for $B_5$ the total errors by SWME~15A and RBC/UKQCD~16 have been inflated by 3\% and 24\%, respectively.

Finally we present our estimates for the BSM $B$-parameters, quoted in
the $\msbar$-scheme at scale 3\,GeV. For
$\Nf=2+1$ our estimate is given by the average of the results
from SWME~15A, RBC/UKQCD~16, and RBC/UKQCD~24. In our analysis, the results in RBC/UKQCD~16 and RBC/UKQCD~24, though obtained through partially different analyses, are considered  as fully correlated because some gauge ensembles are common in the two computations. We find $B_2 = 0.488(12)$ ($\chi^2/\rm{dof} = 1.58$); $B_3 = 0.757(27)$ ($\chi^2/\rm{dof} = 0.17$); $B_4 = 0.903(12)$ ($\chi^2/\rm{dof} = 1.36$); $B_5 = 0.691(14)$ ($\chi^2/\rm{dof} = 0.43$). Following the FLAG rule, for cases that have a value of the reduced $\chi^2$ greater than unity, we use the square root of the latter to stretch the respective error. Hence our averages are
%
%
\begin{align}
  & \Nf=2+1: \\ 
  &\FLAGAVBEGIN B_2=0.488(15)\FLAGAVEND,\quad 
     \FLAGAVBEGIN B_3=0.757(27)\FLAGAVEND,\quad 
     \FLAGAVBEGIN B_4=0.903(14)\FLAGAVEND,\quad
      \FLAGAVBEGIN B_5=0.691(14)\FLAGAVEND,
  \quad\Refs~\mbox{\cite{Jang:2015sla,Garron:2016mva,Boyle:2024gge}}.  \nonumber
\end{align}
%
For $\Nf=2+1+1$ and $\Nf=2$, our estimates coincide---with one exception---with the ones by
ETM~15 and ETM~12D, respectively, since there is only one computation
for each case.  Only for the case of $B_3$ with $\Nf=2+1+1$, owing to the application of the  $\delta(a_{\rm min})$  criterion the error of the average estimate is inflated by about 11\% with respect to the ETM~15 reported value.   Thus we quote
\begin{align}
  & \Nf=2+1+1: \\ 
  & \FLAGAVBEGIN B_2=0.46(1)(3)\FLAGAVEND,\quad
      \FLAGAVBEGIN B_3=0.79(6)\FLAGAVEND,\quad 
      \FLAGAVBEGIN B_4=0.78(2)(4)\FLAGAVEND,\quad 
      \FLAGAVBEGIN B_5=0.49(3)(3)\FLAGAVEND, \quad\Ref~\mbox{\cite{Carrasco:2015pra}}, \nonumber\\  \nonumber\\ 
%
%
  & \Nf=2:  \\ 
  &\FLAGAVBEGIN B_2=0.47(2)(1)\FLAGAVEND,\quad
     \FLAGAVBEGIN B_3=0.78(4)(2)\FLAGAVEND,\quad
     \FLAGAVBEGIN B_4=0.76(2)(2)\FLAGAVEND,\quad 
     \FLAGAVBEGIN B_5=0.58(2)(2)\FLAGAVEND, 
  \quad\Ref~\mbox{\cite{Bertone:2012cu}}. \nonumber
\end{align}
%
 Based on the above discussion about the effects of employing different
intermediate momentum subtraction schemes in the nonperturbative
renormalization of the operators, there is evidence that the discrepancy in the  $B_4$ and $B_5$
results between $\Nf=2, 2+1+1$, and $\Nf=2+1$ calculations should not
be directly attributed to an effect of the number of dynamical
flavours. To clarify the present situation, it would be
  important to perform a direct comparison of results by the ETM
  collaboration obtained both with RI-MOM and RI-SMOM
  methods.   A calculation based an on a different nonperturbative renormalization scheme, such as the  Schr\"odinger functional, 
  would  provide valuable information to shed light on the current situation.
  
  In closing, we encourage authors to provide
the correlation matrix of the $B_i$ parameters---as done in Ref.~\cite{Boyle:2024gge}---since this information
is required in phenomenological studies of New Physics scenarios.

%% file: HQ/HQD.tex
\section{Charm-hadron decay constants and form factors}
\label{sec:DDecays}
Authors: Y.~Aoki, M.~Della~Morte, E.~Lunghi, S.~Meinel, C.~Monahan, A.~Vaquero\\

Leptonic and semileptonic decays of charmed $D$ and $D_s$ mesons or $\Lambda_c$ and other charm baryons occur
via charged $W$-boson exchange, and are sensitive probes of $c \to d$
and $c \to s$ quark flavour-changing transitions.  Given experimental
measurements of the branching fractions combined with sufficiently
precise theoretical calculations of the hadronic matrix elements, they
enable the determination of the CKM matrix elements $|V_{cd}|$ and
$|V_{cs}|$ (within the Standard Model) and a precise test of the
unitarity of the second row of the CKM matrix.  Here, we summarize the
status of lattice-QCD calculations of the charmed leptonic decay
constants.  Significant progress has
been made in charm physics on the lattice in recent years,
largely due to the availability of gauge configurations produced using
highly-improved lattice-fermion actions that enable treating the
$c$ quark with the same action as for the $u$, $d$, and $s$ quarks.

This section updates the corresponding section in the last review (FLAG~21 \cite{FlavourLatticeAveragingGroupFLAG:2021npn})
for results that appeared before April 30, 2024.
As in FLAG~19 \cite{FlavourLatticeAveragingGroup:2019iem} and FLAG~21 \cite{FlavourLatticeAveragingGroupFLAG:2021npn}, we limit our
review to results based on modern simulations with reasonably light
pion masses (below approximately 500~MeV). 
This excludes results with two flavours in the sea, even if they use light pion masses.
$\Nf=2$ results can still be checked in previous FLAG editions. 

For the heavy-meson decay constants and mixing parameters, estimates
of the quantity $\delta(a_{\rm min})$ described in Sec.~\ref{sec:DataDriven} are provided
for all computations entering the final FLAG averages or ranges.
For heavy-hadron semileptonic-decay form factors, implementing this
data-driven continuum-limit criterion was found to be not feasible.
The problem is that these quantities are functions of the momentum transfer in addition to the
other lattice parameters, and many calculations are based on global fits whose reconstruction
was not possible.

Following our review of lattice-QCD calculations of $D_{(s)}$-meson
leptonic decay constants and charm-hadron semileptonic form factors, we then
interpret our results within the context of the Standard Model.  We
combine our best-determined values of the hadronic matrix elements
with the most recent experimentally-measured branching fractions to
obtain $|V_{cd(s)}|$ and test the unitarity of the second row of the
CKM matrix.

\input{HQ/HQSubsections/DL.tex}

\input{HQ/HQSubsections/DSL.tex}

\input{HQ/HQSubsections/DCKM.tex}

%% file: HQ/HQSubsections/DL.tex
\subsection{Leptonic decay constants $f_D$ and $f_{D_s}$}
\label{sec:fD}

In the Standard Model, and up to electromagnetic corrections,
the decay constant $f_{D_{(s)}}$ of a
pseudoscalar $D$ or $D_s$ meson is related to the branching ratio for
leptonic decays mediated by a $W$ boson through the formula
\be
{\mathcal{B}}(D_{(s)} \to \ell\nu_\ell)= {{G_F^2|V_{cq}|^2 \tau_{D_{(s)}}}\over{8 \pi}}
f_{D_{(s)}}^2 m_\ell^2 
m_{D_{(s)}} \left(1-{{m_\ell^2}\over{m_{D_{(s)}}^2}}\right)^2\;,
 \label{eq:Dtoellnu}
\ee
where $q$ is $d$ or $s$ and $V_{cd}$ ($V_{cs}$) is the appropriate CKM matrix element 
for a
$D$ ($D_s$) meson.  The branching fractions have been experimentally
measured by CLEO, Belle, Babar and BES with a precision around 2.5--4.5$\%$ for
both the $D$ and the $D_s$-meson
decay modes~\cite{ParticleDataGroup:2024cfk}.  When
combined with lattice results for the decay constants, they allow for
determinations of $|V_{cs}|$ and $|V_{cd}|$.

The decay constants $f_{D_{(s)}}$ are defined through 
the matrix elements of the axial current
\be
\langle 0| A^{\mu}_{cq} | D_q(p) \rangle = if_{D_q}\;p_{D_q}^\mu  \;,
\label{eq:dkconst}
\ee
with $q=d,s$ and $A^{\mu}_{cq} =\bar{c}\gamma^\mu \gamma_5 q$.
Such matrix elements can be extracted from Euclidean two-point
functions computed on the lattice.

\begin{figure}[tb]
\hspace{-0.8cm}\includegraphics[width=0.58\linewidth]{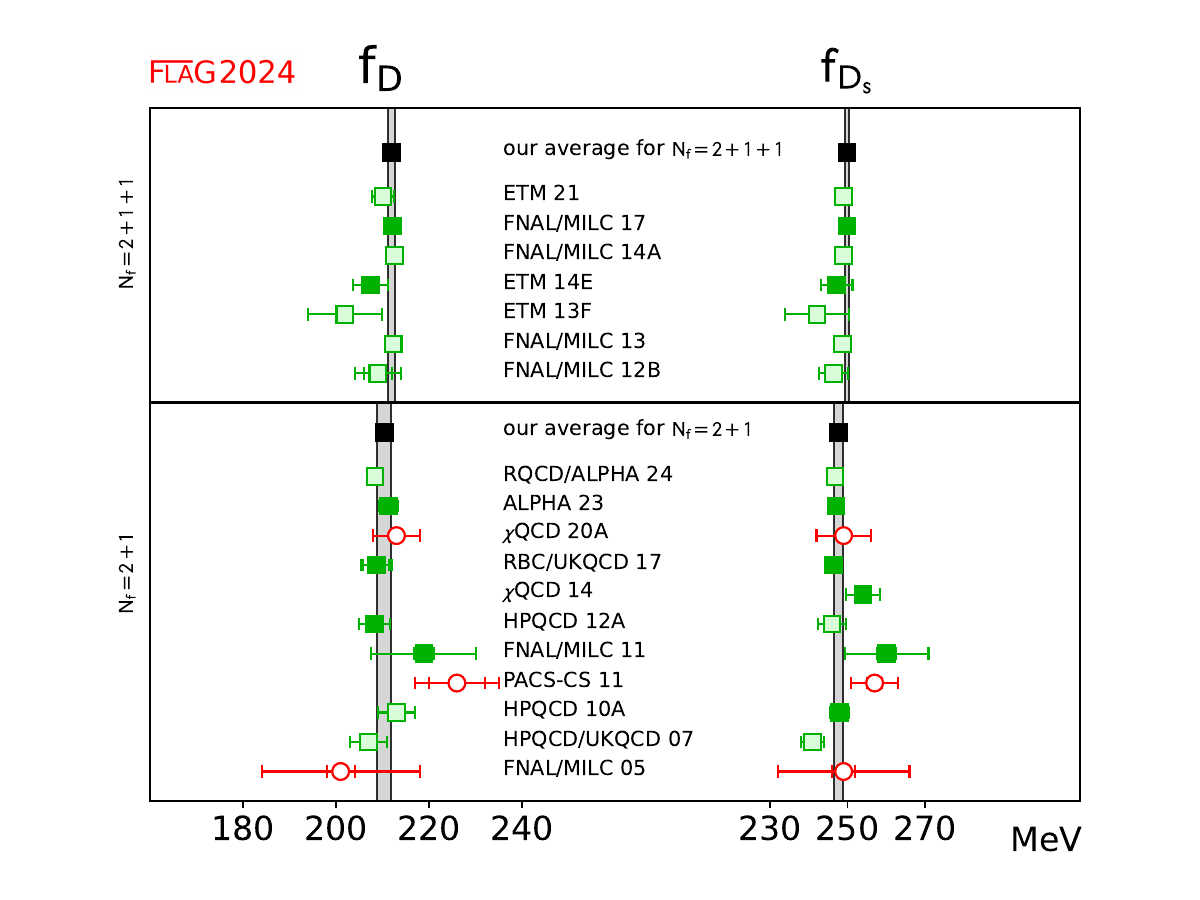} \hspace{-1cm}
\includegraphics[width=0.58\linewidth]{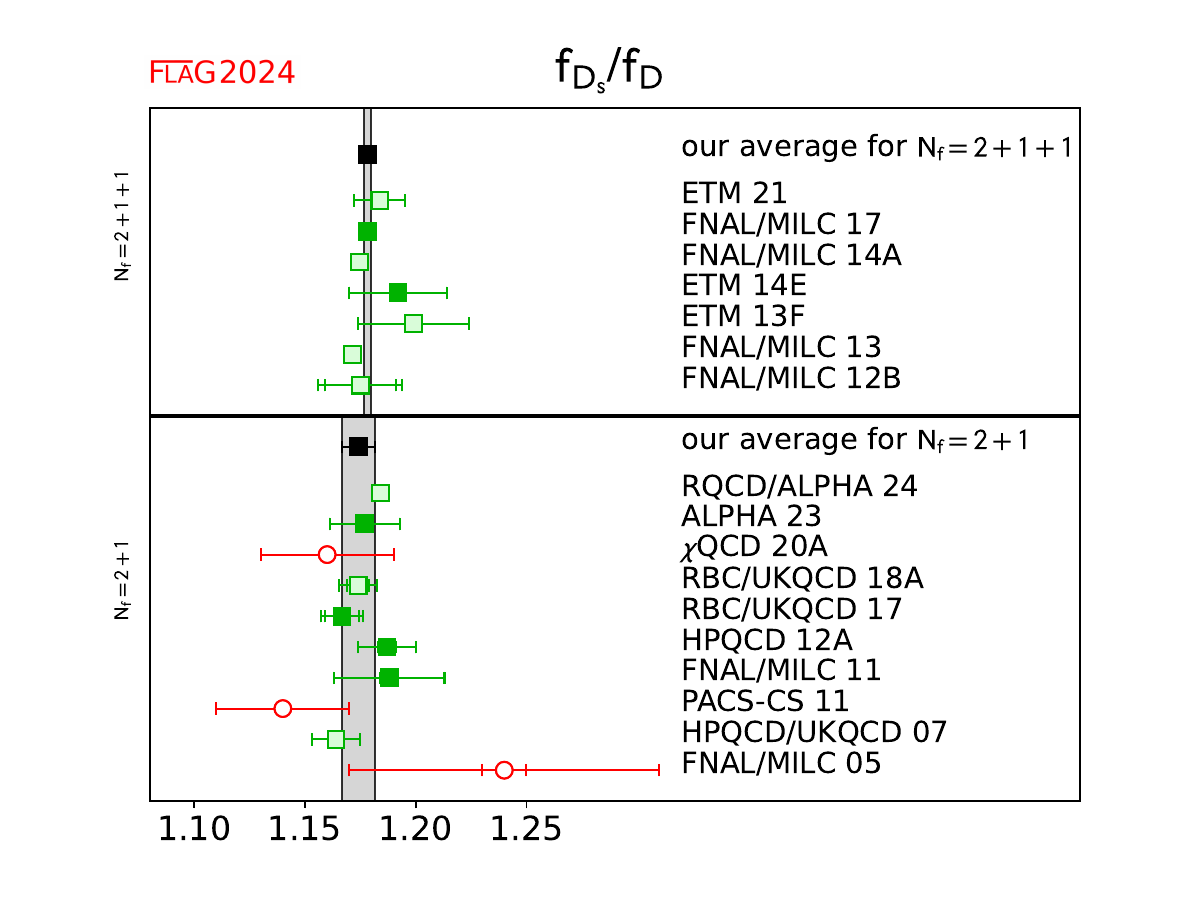}
\vspace{-2mm}
\caption{Decay constants of the $D$ and $D_s$ mesons [values in Tab.~\ref{tab_FDsummary} 
and  Eqs.~(\ref{eq:fD2+1}-\ref{eq:fDratio2+1+1})].
As usual, full green squares are used in the averaging procedure,
    pale green squares have either been superseded by later
    determinations or are only published in Proceedings or have not been published within
    the current deadline (April 30, 2024), while empty red circles do not
    satisfy the criteria.
The black squares and grey bands
  indicate our averages. }
\label{fig:fD}
\end{figure}

Results for $\Nf=2+1$ and $2+1+1$ dynamical flavours are
summarized in Tab.~\ref{tab_FDsummary} and Fig.~\ref{fig:fD}.
Since the publication of FLAG 21, a handful of results
for  $f_D$ and $f_{D_s}$ have appeared, as described below.
We consider isospin-averaged quantities, although, in a few cases, results for $f_{D^+}$ 
are quoted
(see, for example, the FNAL/MILC~11,14A and 17 computations, where
the strong-isospin-breaking effect given by the difference 
between $f_D$ and $f_{D^+}$ has been estimated to be around 0.5 MeV).

For the first time, we restrict the review here to results obtained using $\Nf=2+1$ and $2+1+1$
dynamical flavours.
No new results with $\Nf=2$ appeared since 2019 and they have been
presented in previous FLAG reviews.

Another novelty is the re-inclusion of the quantity $\delta(a_{\rm min})$ described 
in the Introduction.
Our working group introduced and applied this quantity in FLAG 13~\cite{Aoki:2013ldr}, 
but it was not applied in following reviews.
As computations have become increasingly precise and often dominated by systematic 
uncertainties, we believe that a closer scrutiny of the continuum
extrapolations is needed since such extrapolations usually produce one of the largest 
systematic errors.
Here, we provide (where possible) an estimate of $\delta(a_{\rm min})$ for all computations 
entering the final FLAG averages or ranges.
Those estimates do not need to be very precise as the natural size of the error on 
$\delta(a_{\rm min})$ is ${\mathcal{O}}(1)$.

\begin{table}[h!]
\begin{center}
\mbox{} \\[3.0cm]
\footnotesize
\begin{tabular*}{\textwidth}[l]{@{\extracolsep{\fill}}l@{\hspace{1mm}}r@{\hspace{1mm}}
	l@{\hspace{1mm}}l@{\hspace{1mm}}l@{\hspace{1mm}}l@{\hspace{1mm}}l@{\hspace{1mm}}
	l@{\hspace{1mm}}l@{\hspace{1mm}}l@{\hspace{1mm}}l@{\hspace{1mm}}l}
Collaboration & Ref. & $\Nf$ & 
\hspace{0.15cm}\begin{rotate}{60}{publication status}\end{rotate}\hspace{-0.15cm} 
&
\hspace{0.15cm}\begin{rotate}{60}{continuum extrapolation}\end{rotate}\hspace{-0.15cm} 
&
\hspace{0.15cm}\begin{rotate}{60}{chiral extrapolation}\end{rotate}\hspace{-0.15cm}&
\hspace{0.15cm}\begin{rotate}{60}{finite volume}\end{rotate}\hspace{-0.15cm}&
\hspace{0.15cm}\begin{rotate}{60}{renormalization/matching}\end{rotate}\hspace{-0.15cm} 
 &
\hspace{0.15cm}\begin{rotate}{60}{heavy-quark treatment}\end{rotate}\hspace{-0.15cm} 
& 
\rule{0.4cm}{0cm}$f_D$ & \rule{0.4cm}{0cm}$f_{D_s}$  & 
 \rule{0.3cm}{0cm}$f_{D_s}/f_D$ \\[0.2cm]
\hline
\hline
&&&&&&&&&&& \\[-0.1cm]
ETM 21B & \cite{Dimopoulos:2021qsf} & 2+1+1 & \rC & \good & \good  &  \good & \good 
 &  \okay &
210.1(2.4)   & 248.9(2.0) &  1.1838(115) \\[0.5ex]
FNAL/MILC 17 $^{\nabla\nabla}$ & \cite{Bazavov:2017lyh} & 2+1+1 & \gA & \good & \good 
&\good & \good & \okay & 
212.1(0.6)   & 249.9(0.5) &  1.1782(16) \\[0.5ex]

FNAL/MILC 14A$^{**}$ & \cite{Bazavov:2014wgs} & 2+1+1 & \gA & \good & \good &\good 
& \good & \okay & 
212.6(0.4) $+1.0 \choose -1.2$   & 249.0(0.3)$+1.1 \choose -1.5$ &  1.1745(10)$+29 
\choose -32$ \\[0.5ex]

ETM 14E & \cite{Carrasco:2014poa} & 2+1+1 & \gA & \good & \soso  &  \soso & \good 
 &  \okay &
207.4(3.8)   & 247.2(4.1) &  1.192(22) \\[0.5ex]

ETM 13F & \cite{Dimopoulos:2013qfa} & 2+1+1 & \rC & \soso & \soso  &  \soso & \good 
 &  \okay &
202(8)   & 242(8) &  1.199(25) \\[0.5ex]

FNAL/MILC 13 & \cite{Bazavov:2013nfa} & 2+1+1 & \rC & \good    & \good    & \good 
    
&\good & \okay  & 212.3(0.3)(1.0)   & 248.7(0.2)(1.0) & 1.1714(10)(25)\\[0.5ex]

FNAL/MILC 12B & \cite{Bazavov:2012dg} & 2+1+1 & \rC & \good    & \good    & \good 
    
&\good & \okay  & 209.2(3.0)(3.6)   & 246.4(0.5)(3.6) & 1.175(16)(11)\\[0.5ex]

&&&&&&&&&&& \\[-0.1cm]
\hline
&&&&&&&&&&& \\[-0.1cm]
RQCD/ALPHA 24 &\cite{Kuberski:2024pms} & 2+1 & \oP & \good & \good & \good 
& \good & \okay & 208.4(0.7)(0.7)(1.1) &246.8(0.6)(0.6)(1.0)  & 1.1842(21)(22)(19) 
\\[0.5ex]
ALPHA 23 &\cite{Bussone:2023kag} & 2+1 & \gA & \good & \soso & \good & \good & 
\okay & 211.3(1.9)(0.6) &247.0(1.9)(0.7)  & 1.177(15)(5) \\[0.5ex]
$\chi$QCD 20A$^{\dagger\dagger}$ &\cite{Chen:2020qma} & 2+1 & \gA & \tbr & \good 
& \good & \good & \okay & 213(5) & 249(7) &1.16(3) \\[0.5ex]
RBC/UKQCD 18A$^{\square\nabla}$ &\cite{Boyle:2018knm} & 2+1 & \oP & \good & \good 
& \good & \good & \okay &  &  &1.1740(51)$+68 \choose -68$ \\[0.5ex]
RBC/UKQCD 17 &\cite{Boyle:2017jwu} & 2+1 & \gA & \good & \good &\soso & \good & \okay
& 208.7(2.8)$+2.1 \choose -1.8$ & 246.4(1.3)$+1.3 \choose -1.9$    & 1.1667(77)$+57 
\choose -43$\\[0.5ex] 
$\chi$QCD~14$^{\dagger\square}$ &\cite{Yang:2014sea} & 2+1 &  \gA &\soso &\soso & 
\soso & \good & \okay
& & 254(2)(4) & \\[0.5ex]
HPQCD 12A &\cite{Na:2012iu} & 2+1 & \gA &\soso  &\soso &\soso &\good &\okay 
& 208.3(1.0)(3.3) & 246.0(0.7)(3.5) & 1.187(4)(12)\\[0.5ex]

FNAL/MILC 11& \cite{Bazavov:2011aa} & 2+1 & \gA & \soso &\soso &\soso  & 
 \soso & \okay & 218.9(11.3) & 260.1(10.8)&   1.188(25)   \\[0.5ex]  

PACS-CS 11 & \cite{Namekawa:2011wt} & 2+1 & \gA & \tbr & \good & \tbr  & 
\soso & \okay & 226(6)(1)(5) & 257(2)(1)(5)&  1.14(3)   \\[0.5ex] 

HPQCD 10A & \cite{Davies:2010ip} & 2+1 & \gA & \good  & \soso  & 
\good & \good & \okay & 213(4)$^{*}$ & 248.0(2.5)  \\[0.5ex]

HPQCD/UKQCD 07 & \cite{Follana:2007uv} & 2+1 &  \gA & \soso & \soso & 
\soso & \good  & \okay & 207(4) & 241 (3)& 1.164(11)  \\[0.5ex] 

FNAL/MILC 05 & \cite{Aubin:2005ar} & 2+1 & \gA &\soso &   \soso    &
\tbr      & \soso    &  \okay       & 201(3)(17) & 249(3)(16)  & 1.24(1)(7) \\[0.5ex]

&&&&&&&&&&& \\[-0.1cm]
\hline
\hline
\end{tabular*}
\begin{tabular*}{\textwidth}[l]{l@{\extracolsep{\fill}}lllllllll}
  \multicolumn{10}{l}{\vbox{\begin{flushleft} 
$^{*}$ This result is obtained by using the central value for $f_{D_s}/f_D$ from 
HPQCD/UKQCD~07 
and increasing the error to account for the effects from the change in the physical 
value of $r_1$. \\
$^{**}$ 
At $\beta = 5.8$, $m_{\pi, \rm min}L=3.2$ but this lattice spacing is not used in 
the final cont./chiral extrapolations.\\
$^{\nabla\nabla}$ Update of FNAL/MILC~14A. The ratio quoted is $f_{D_s}/f_{D^+}=1.1749(16)$. 
In order to compare with
results from other collaborations, we rescale the number by the ratio of central 
values for $f_{D+}$ and $f_D$.
We use the same rescaling in FNAL/MILC~14A. At the finest lattice spacing the
finite-volume criterium would produce an empty green circle, however, as checked 
by the authors, results would not significantly change by excluding this ensemble, 
which instead sharpens the continuum limit extrapolation.\\
$\square\nabla$ Update of RBC/UKQCD 17. \\
$\dagger\square$ Two values of sea pion masses.\\
$\dagger\dagger$ Four valence pion masses between 208 MeV and 114 MeV have been used 
at one value of the sea pion mass of 139 MeV.
\end{flushleft}}}
\end{tabular*}

\vspace{-0.5cm}
\caption{Decay constants of the $D$ and $D_{s}$ mesons (in MeV) and their ratio.
}
\label{tab_FDsummary}
\end{center}
\end{table}
%
%

Two new results appeared with $\Nf=2+1$.
In Ref.~\cite{Bussone:2023kag} (ALPHA~23)
maximally twisted Wilson valence fermions (for light and heavy quarks) are
implemented on a set of ensembles of configurations generated within
the CLS initiative using ${\mathcal{O}}(a)$-improved Wilson fermions.
As a consequence of the maximal twist, observables in the charm sector
are free from  ${\mathcal{O}}(am_c)$ discretisation effects.
In addition the decay constants $f_{D_{(s)}}$ are automatically normalized
and do not require the computation of normalization factors.
Four different lattice spacings have been used in the continuum
extrapolation, ranging between $0.087$ and $0.05$ fm.
Pion masses reach down to $200$ MeV and volumes are such that $3.9 \leq m_\pi L \leq 
6.4$.
The uncertainties are dominated by statistics
and the chiral-continuum fits.
Judging from the plots in Ref.~\cite{Bussone:2023kag},
the values for $\delta(a_{\rm min})$ are around 1 for $f_D$ and around 3 for $f_{D_s}$.

A second new computation with $\Nf=2+1$ has been performed by the RQCD-ALPHA
Collaboration~\cite{Kuberski:2024pms} on a set of 49 gauge ensembles generated
again within the CLS effort.
For this reason statistical errors between ALPHA~23 and RQCD/ALPHA 24 will 
be
treated as 100$\%$ correlated when performing averages. Notice, however, that
since RQCD/ALPHA 24 was not yet published in a journal by the FLAG deadline,
it is not being considered in the averages for this review.
In RQCD/ALPHA 24 nonperturbatively ${\mathcal{O}}(a)$-improved Wilson
fermions
have been used both in the valence sector and the sea.\footnote{The coefficient $\bar{b}_{\rm A}$ has been neglected
  because its nonperturbative
  value, computed in~\cite{Bali:2021qem}, turned out to be compatible with zero for the relevant range of gauge couplings.}
The simulations cover six different lattice spacings with $0.039$~fm~$\leq a\leq 
0.098$~fm,
pion masses from $420$ MeV  down to $130$ MeV and $m_\pi L$ ranging from $2.83$ to 
$6.42$.
The largest volume at $m_\pi=130$ MeV gives $m_\pi L=4.05$.
In the discussion of the final errors the uncertainty due to the scale setting is 
treated
separately. That turns out to be the largest contribution to the total error
for $f_D$ and $f_{D_s}$ (around $50\%$), while for the ratio of decay constants statistical,
systematic (chiral and continuum extrapolations) and scale-setting uncertainties 
are of
about the same size. The quantity $\delta(a_{\rm min})$, as estimated from the figures
in~\cite{Kuberski:2024pms} is around 1.

The updated $\Nf=2+1$ FLAG averages read
\begin{align}
&\label{eq:fD2+1}
\Nf=2+1:&\FLAGAVBEGIN f_D &= 210.4(1.5) \FLAGAVEND\;{\rm MeV}
&&\Refs~\mbox{\cite{Bussone:2023kag,Na:2012iu,Bazavov:2011aa,Boyle:2017jwu}},\\
&\label{eq:fDs2+1}
\Nf=2+1: &\FLAGAVBEGIN f_{D_s} &= 247.7(1.2) \FLAGAVEND\; {\rm MeV} 
&&\Refs~\mbox{\cite{Bussone:2023kag,Davies:2010ip,Bazavov:2011aa,Boyle:2017jwu,Yang:2014sea}}, 
\\
&\label{eq:fDratio2+1}
\Nf=2+1: &\FLAGAVBEGIN f_{D_s}\over{f_D} &= 1.174(0.007)\FLAGAVEND
&&\Refs~\mbox{\cite{Bussone:2023kag,Na:2012iu,Bazavov:2011aa,Boyle:2017jwu}}.
\end{align}
Those come from the results in HPQCD~12A~\cite{Na:2012iu}, FNAL/MILC~11~\cite{Bazavov:2011aa} 
as well as RBC/UKQCD~17 \cite{Boyle:2017jwu}
and ALPHA~23 \cite{Bussone:2023kag}
concerning $f_D$ while for $f_{D_s}$ 
also the $\chi$QCD~14 \cite{Yang:2014sea} result contributes, and instead of the 
value in HPQCD~12A~\cite{Na:2012iu}
the one in HPQCD~10A \cite{Davies:2010ip} is used.
In addition, the statistical errors between the results of FNAL/MILC and HPQCD have 
been everywhere treated as 100\% correlated since
the two collaborations use overlapping sets of configurations. The same procedure 
had been used in the past reviews.
Concerning the values of $\delta(a_{\rm min})$ for older computations entering those 
estimates, they are all smaller than 2 for the results
before 2013, as discussed in the second FLAG review~\cite{Aoki:2013ldr}, where that 
was used as a necessary condition to enter the averages.
For RBC/UKQCD~17 $\delta({a_{\rm min}})$ is estimated to be around 1.5, while for 
$\chi$QCD~14 it is not possible to assess the value
of $\delta({a_{\rm min}})$ from the published figures and tables.

For $\Nf=2+1+1$ only a Proceedings contribution to the 2021 Lattice Conference by 
the ETM Collaboration~\cite{Dimopoulos:2021qsf} appeared containing new results.
This ETM~21B result extends ETM~14E~\cite{Carrasco:2014poa} by including simulations 
closer to the physical point for light and heavy quarks.
Twisted-mass fermions at maximal twist are used in the sea, in order to ensure automatic 
${\mathcal{O}}(a)$ improvement.
In the valence sector Osterwalder-Seiler fermions are adopted for the strange and 
charm quarks to avoid
mixing effects at  ${\mathcal{O}}(a^2)$.
Three different lattice resolutions between $0.095$ fm  and $0.069$ fm have been 
used with $m_\pi L$ at the lightest
pion mass ($134$ MeV) being around $3.7$. Also in this case the final errors are 
dominated by statistics and the chiral-continuum
extrapolations.
Although we do not provide an estimate of $\delta(a_{\rm min})$ for results that 
do not enter the final averages, ETM~21B
makes an important observation in showing that the cutoff effects strongly depend 
on the intermediate scaling variable used.
In the case of $f_{D_s}$, when using $w_0$, $\delta(a_{\rm min})$ would turn out 
to be very large, while when using the strange-charm
meson mass cutoff effects are much reduced and $\delta(a_{\rm min})$ is around 1.

Our global averages coincide with those in FLAG 21, Ref.~\cite{FlavourLatticeAveragingGroupFLAG:2021npn}, 
namely
\begin{align}
&\label{eq:fD2+1+1}
\Nf=2+1+1:&\FLAGAVBEGIN f_D &= 212.0(0.7) \FLAGAVEND\;{\rm MeV}
&&\Refs~\mbox{\cite{Bazavov:2017lyh,Carrasco:2014poa}},\\
&\label{eq:fDs2+1+1}
\Nf=2+1+1: &\FLAGAVBEGIN f_{D_s} &= 249.9(0.5) \FLAGAVEND\; {\rm MeV} 
&&\Refs~\mbox{\cite{Bazavov:2017lyh,Carrasco:2014poa}}, \\
&\label{eq:fDratio2+1+1}
\Nf=2+1+1: &\FLAGAVBEGIN f_{D_s}\over{f_D} &= 1.1783(0.0016)\FLAGAVEND
&&\Refs~\mbox{\cite{Bazavov:2017lyh,Carrasco:2014poa}},
\end{align}
where the error on the average of $f_{D}$ has been rescaled by the factor $\sqrt{\chi^2/\mbox{dof}}=1.22$.
For the two computations entering the results above $\delta(a_{\rm min})$ is around 
2 at most.

Concerning the inclusion of QED effects, significant progress has been made in the
computation of form factors for radiative leptonic decays of $D$ mesons.\footnote{The accuracy of the estimates presented here is often
  below the percent level and a first-principles computation of isospin-breaking corrections is therefore
  very desirable. However, for the determination of the CKM matrix elements, the experimental
accuracy on the branching ratios and hence on the products $|V_{cq}|^2f^2_{D_{(q)}}$ varies between 2.2\% and 5\%, see section~\ref{sec:Vcd}.}
We do not present results in detail here since they are not yet at the level
to be reviewed according to the FLAG criteria, however, such processes are important for two reasons.
In the region of soft-photon energies
they are needed in order to compute the QED corrections to leptonic decays.
In that case they have to be combined with the contributions stemming from
virtual exchanges of photons between the meson and the charged lepton, in order
to remove infrared divergent terms.
For hard photons radiative leptonic decays become important probes of the internal
structure of hadrons and therefore of physics Beyond the Standard Model.
The form factors appear in the decomposition of the hadronic matrix element
\begin{equation}
H_W^{\alpha r}(k, {\bold{p}})=\epsilon_\mu^r(k)\int d^4y\, e^{iky}\, {\rm T}\langle0|j_W^\alpha(0) 
j_{em}^\mu(y)| P({\bold{p}})\rangle\;,
\end{equation}
with $\epsilon_\mu^r(k)$ the polarisation vector of the outgoing photon (with momentum 
$k$), $\bold{p}$ the momentum of the generic pseudoscalar meson $P$ and 
$j_W^\alpha$ and $j_{em}^\mu$ the weak and electromagnetic currents, respectively.
Such matrix elements can be extracted from suitable three-point correlation
functions that can be computed on an Euclidean lattice.
In Ref.~\cite{Giusti:2023pot} a set of numerical methods is explored with the main
goals of keeping systematic effects due to contributions from unwanted states
under control and of optimizing the signal.
The study is performed on a single ensemble with $2+1$ flavours of domain wall fermions,
$a\simeq 0.11$ fm and $m_\pi\simeq 340$ MeV.

In Ref.~\cite{Frezzotti:2023ygt}, which extends Ref.~\cite{Desiderio:2020oej}, the 
form
factors for the decay $D_s \to \ell \nu_\ell \gamma$ have been computed on four different
ensembles of $\Nf=2+1+1$ gauge configurations produced by the ETM Collaboration.
Lattice spacings span the interval $[0.056, 0.09]$ fm and quarks masses are close
to their physical values.
The full kinematical range, with a cut $E_\gamma \geq 10$ MeV, is covered by the 
results.
The structure-dependent contribution is found to dominate the amplitude for $\ell=e$, 
as opposed
to the cases with $\ell=\mu$ and $\tau$. Since the point-like contribution is
(helicity) suppressed by $(m_\ell/m_P)^2$, a nonperturbative computation
of the form factors is of paramount importance for $B$ mesons.
An analysis of the noise-to-signal ratio for the three-point functions
is presented following the Parisi-Lepage approach~\cite{Parisi:1983ae,Lepage:1989hd}
and a strategy to mitigate the problem is discussed.
That coincides with one of the methods studied, with different motivations,
in Ref.~\cite{Giusti:2023pot}.

%% file: HQ/HQSubsections/DSL.tex
\FloatBarrier

\subsection{Form factors for $D\to \pi \ell\nu$ and $D\to K \ell \nu$ semileptonic decays}
 \label{sec:DtoPiK}


The SM prediction for the differential decay rate of the semileptonic processes $D\to \pi \ell\nu$ and
$D\to K \ell \nu$ can be written as
\begin{align}
  \frac{d\Gamma(D\to P\ell\nu)}{dq^2} =
  &
    \frac{\eta_{\rm EW}^2 G_{\rm\scriptscriptstyle F}^2 |V_{cx}|^2}{24 \pi^3}
    \,\frac{(q^2-m_\ell^2)^2\sqrt{E_P^2-m_P^2}}{q^4m_{D}^2}
    \nonumber\\
  & \times
    \left[ \left(1+\frac{m_\ell^2}{2q^2}\right)
    m_{D}^2(E_P^2-m_P^2)|f_+(q^2)|^2 
    + \frac{3m_\ell^2}{8q^2}(m_{D}^2-m_P^2)^2|f_0(q^2)|^2\right]
    \label{eq:DtoPiKFull}
\end{align}
where $x = d, s$ is the daughter light quark, $P= \pi, K$ is the
daughter light-pseudoscalar meson, $\ell=e,\mu$ indicates the light charged lepton, $E_P$ is the light-pseudoscalar meson energy 
in the rest frame of the decaying $D$, and $q = (p_D - p_P)$ is the
momentum of the outgoing lepton pair. Here, we have included
the short-distance electroweak correction factor \cite{Sirlin:1981ie}, whose
value at $\mu=m_D$ is $\eta_{\rm EW}=1.009$ \cite{FermilabLattice:2022gku}.
The vector and scalar form
factors $f_+(q^2)$ and $f_0(q^2)$ parameterize the hadronic matrix
element of the heavy-to-light quark flavour-changing vector current
$V_\mu = \overline{x} \gamma_\mu c$,
\begin{equation}
\langle P| V_\mu | D \rangle  = f_+(q^2) \left( {p_D}_\mu+ {p_P}_\mu - \frac{m_D^2 - m_P^2}{q^2}\,q_\mu \right) + f_0(q^2) \frac{m_D^2 - m_P^2}{q^2}\,q_\mu \,,
\end{equation}
and satisfy the kinematic constraint $f_+(0) = f_0(0)$.  Because the contribution to the decay width from
the scalar form factor is proportional to $m_\ell^2$, within current precision standards it can be
neglected for $\ell = e$, and Eq.~(\ref{eq:DtoPiKFull})
simplifies to
\begin{equation}
\frac{d\Gamma \!\left(D \to P e \nu\right)}{d q^2} = \frac{\eta_{\rm EW}^2 G_{\rm\scriptscriptstyle F}^2}{24 \pi^3} |\vec{p}_{P}|^3 {|V_{cx}|^2 |f_+ (q^2)|^2} \,. \label{eq:DtoPiK}
\end{equation}
In models of new physics, decay rates may also receive contributions from matrix elements of other
parity-even currents. In the case of the scalar density ($\bar x c$), partial vector-current conservation allows one
to write its matrix elements in terms of $f_+$ and $f_0$, while for tensor currents $T_{\mu\nu}=\bar x\sigma_{\mu\nu}c$
a new form factor has to be introduced, viz.,
\begin{equation}
\langle P| T^{\mu\nu} | D \rangle  = \frac{2}{m_D+m_P}\left[p_P^\mu p_D^\nu - p_P^\nu p_D^\mu \right]f_T(q^2)\,.
\end{equation}
Recall that, unlike the Noether current $V_\mu$, the operator $T_{\mu\nu}$ requires
a scale-dependent renormalization.


Lattice-QCD computations of $f_{+,0}$ allow for comparisons to experiment
to ascertain whether the SM provides the correct prediction for the $q^2$-dependence of
$d\Gamma(D\to P\ell\nu)/dq^2$;
and, subsequently, to determine the CKM matrix elements $|V_{cd}|$ and $|V_{cs}|$
from Eq.~(\ref{eq:DtoPiKFull}). The inclusion of $f_T$ allows for analyses to
constrain new physics. Currently, state-of-the-art experimental results by
CLEO-c~\cite{Besson:2009uv} and BESIII~\cite{Ablikim:2017oaf,Ablikim:2018frk}
provide data for the differential rates in the whole $q^2$ range,
with a precision of order 2--3\% for the total branching fractions in both
the electron and muon final channels.


Calculations of the $D\to \pi \ell\nu$ and $D\to
K \ell \nu$ form factors typically use the same light-quark and
charm-quark actions as those of the leptonic decay constants $f_D$ and
$f_{D_s}$. Therefore, many of the same issues arise; in particular,
considerations about cutoff effects coming from the large charm-quark mass,
or the normalization of weak currents, apply.
Additional complications arise,
however, due to the necessity of covering a sizeable range of values in $q^2$:
\begin{itemize}

\item Lattice kinematics impose restrictions on the values
of the hadron momenta.
Because lattice calculations are performed
in a finite spatial volume, the pion or kaon three-momentum components can only
take discrete values in units of $2\pi/L$ when periodic boundary
conditions are used.  For typical box sizes in lattice $D$- and
$B$-meson form-factor calculations at heavier-than-physical pion masses, $L \sim 2.5$--3~fm; thus, the
smallest nonzero momentum in most of these analyses is 
$|\vec{p}_P| \sim 400$--$500$~MeV. On the other hand, the ranges relevant
for the semileptonic decays are $0\leq |\vec{p}_\pi| \lesssim 940$~MeV
and $0\leq |\vec{p}_K| \lesssim 1$~GeV, respectively. Thus, when
using periodic boundary conditions, only a small number of allowed lattice momenta fall into this range.
As a consequence, many studies have incorporated the use of
nonperiodic ``twisted'' boundary conditions (tbc)~\cite{Bedaque:2004kc,Sachrajda:2004mi}
in the valence fields used for the computation of observables,
which allows a continuous choice of momentum and hence finer resolution of the $q^2$-dependence
\cite{DiVita:2011py,Koponen:2011ev,Koponen:2012di,Koponen:2013tua,Lubicz:2017syv,Lubicz:2018rfs}.
Note that more recent calculations \cite{Chakraborty:2021qav,FermilabLattice:2022gku}
include ensembles with physical pion masses and $L\approx 5.5$--5.75~fm, so
the momentum unit when using periodic boundary conditions is correspondingly smaller,
making the use of twisted boundary conditions less relevant.

\item Final-state pions and kaons can have energies
$\gtrsim 1~{\rm GeV}$, given the available kinematical range $0 \lesssim q^2 \leq q_{\rm\scriptscriptstyle max}^2=(m_D-m_P)^2$.
This makes the use of (heavy-meson) chiral perturbation theory to extrapolate to physical
light-quark masses potentially problematic. This issue has become less relevant as
modern calculations include ensembles with physical light-quark masses.

\item Accurate comparisons to experiment, including the determination of CKM parameters,
requires good control of systematic uncertainties in the parameterization of the $q^2$-dependence of form factors. While this issue is far more important for semileptonic
$B$ decays, where it is harder to cover the kinematic range on the lattice,
the increase in experimental precision requires accurate work in the charm sector
as well. The parameterization of semileptonic form factors is discussed in detail
in Appendix \ref{sec:zparam}.

\end{itemize}


The first published $\Nf = 2+1$ lattice-QCD calculation of the $D \to
\pi \ell \nu$ and $D \to K \ell \nu$ form factors came from the
Fermilab Lattice, MILC, and HPQCD
collaborations (FNAL/MILC~04)~\cite{Aubin:2004ej}.\footnote{Because only two of the
  authors of this work are members of HPQCD, and to distinguish it
  from other more recent works on the same topic by HPQCD, we
  hereafter refer to this work as ``FNAL/MILC.''}  This work uses
asqtad-improved staggered sea quarks and light ($u,d,s$) valence
quarks and the Fermilab action for the charm quarks, with a single
lattice spacing of $a \approx 0.12$~fm, and a minimum RMS-pion
mass of $\approx 510$~MeV, dictated by the presence of fairly large
staggered taste splittings. The vector current is normalized using a
mostly nonperturbative approach, such that the perturbative truncation
error is expected to be negligible compared to other
systematics. Results for the form factors are provided over the full
kinematic range, rather than focusing just at $q^2=0$ as was customary
in most previous work, and fitted to a Be{\v{c}}irevi{\'c}-Kaidalov ansatz
(calculations in the full kinematic range had already been done earlier in the
quenched approximation \cite{Lubicz:1991bi,Abada:2000ty}).
The publication of Ref.~\cite{Aubin:2004ej} predated the precise
measurements of the $D\to K \ell\nu$ decay width by the
FOCUS~\cite{Link:2004dh} and Belle experiments~\cite{Abe:2005sh}, and
showed good agreement with the experimental determination of the shape
of $f_+^{D\to K}(q^2)$.  Progress on extending this work was reported
in~\cite{Bailey:2012sa}; efforts are aimed at reducing both the
statistical and systematic errors in $f_+^{D\to\pi}(q^2)$ and
$f_+^{D\to K}(q^2)$ by increasing the number of configurations analyzed,
simulating with lighter pions, and adding lattice spacings as fine as
$a \approx 0.045$~fm.

The most precise published calculations of the $D \to \pi \ell \nu$ and $D \to K \ell \nu$ form factors in $\Nf=2+1$ QCD
are by the HPQCD collaboration (HPQCD~11~\cite{Na:2011mc} and HPQCD~10B~\cite{Na:2010uf}, respectively).
They are also based on $\Nf = 2+1$ asqtad-improved staggered MILC configurations, but use two lattice spacings
$a \approx 0.09$ and 0.12~fm, and a HISQ action for the valence
$u,d,s$, and $c$ quarks. In these mixed-action calculations, the HISQ
valence light-quark masses are tuned so that the ratio $m_l/m_s$ is
approximately the same as for the sea quarks; the minimum RMS sea-pion
mass $\approx 390$~MeV. Form factors are determined only at $q^2=0$,
by using a Ward identity to relate matrix elements of vector
currents to matrix elements of the absolutely normalized quantity
$(m_{c} - m_{x} ) \langle P | \bar{x}c | D \rangle$ (where $x=u,d,s$),
and exploiting the kinematic identity $f_+(0) = f_0(0)$
to yield
$f_+(q^2=0) = (m_{c} - m_{x} ) \langle P | \bar{x}c | D \rangle / (m^2_D - m^2_P)$.
A modified $z$-expansion
(cf.~Appendix \ref{sec:zparam})
is employed to simultaneously extrapolate to the physical
light-quark masses and the continuum and to interpolate to $q^2 = 0$, and
allow the coefficients of the series expansion to vary with the light-
and charm-quark masses.  The form of the light-quark dependence is
inspired by $\chi$PT, and includes logarithms of the form $m_\pi^2
{\rm log} (m_\pi^2)$ as well as polynomials in the valence-, sea-, and
charm-quark masses.  Polynomials in $E_{\pi(K)}$ are also included to
parameterize momentum-dependent discretization errors.
The number of terms is increased until the result for $f_+(0)$
stabilizes, such that the quoted fit error for $f_+(0)$ not only contains
statistical uncertainties, but also reflects relevant systematics.  The
largest quoted uncertainties in these calculations are from statistics and
charm-quark discretization errors.

The most recent $\Nf=2+1$ computation of $D$ semileptonic form factors has
been carried out by the JLQCD collaboration, and so far only published in conference
proceedings; most recently in Ref.~\cite{Kaneko:2017xgg} (JLQCD~17B).
They use their own M\"obius domain-wall configurations at three values
of the lattice spacing $a=0.080, 0.055, 0.044~{\rm fm}$, with several
pion masses ranging from 226 to 501~MeV (though there is so far only one
ensemble, with $m_\pi=284~{\rm MeV}$, at the finest lattice spacing).
The vector and scalar form factors are computed at four values of the momentum transfer for each ensemble.
The computed form factors are observed to depend mildly on both the
lattice spacing and the pion mass.
The momentum dependence of the form factors is fitted to a BCL
$z$-parameterization (see Appendix~\ref{sec:zparam}) with a Blaschke factor that contains the measured
value of the $D_{(s)}^*$ mass in the vector channel,
and a trivial Blaschke factor in the scalar channel. The systematics
of this latter fit is assessed by a BCL fit with the experimental value
of the scalar resonance mass in the Blaschke factor.
Continuum and chiral extrapolations are carried out through a
linear fit in the squared lattice spacing and the squared pion and $\eta_c$ masses.
A global fit that uses hard-pion HM$\chi$PT to model the mass dependence
is furthermore used for a comparison of the form factor shapes with experimental data.\footnote{It is important
to stress the finding in Ref.~\cite{Colangelo:2012ew} that
the factorization of chiral logs in hard-pion $\chi$PT breaks down,
implying that it does not fulfill the expected requisites for a proper
effective field theory. Its use to model the mass dependence of form
factors can thus be questioned. \label{footnote:hardpion}}
Since the computation is only published in proceedings so far, it will not
enter our $\Nf=2+1$ average.\footnote{The ensemble parameters
quoted in Ref.~\cite{Kaneko:2017xgg} appear to show that the volumes
employed at the lightest pion masses are insufficient to meet our criteria
for finite-volume effects. There is, however, a typo in the table which results
in a wrong assignment of lattice sizes, whereupon the criteria are indeed met.
We thank T.~Kaneko for correspondence on this issue.} Another $\Nf=2+1$ calculation
of the $D\to \pi$, $D\to K$, and $D_s \to K$ form factors using domain-wall fermions
is currently being carried out by the RBC/UKQCD collaboration, as reported
in Ref.~\cite{Marshall:2022xbz}.

The first full computation of both the vector and scalar form factors
in $\Nf=2+1+1$ QCD was achieved by the
ETM collaboration~\cite{Lubicz:2017syv} (ETM~17D).  Furthermore, they have provided a separate
determination of the tensor form factor, relevant for new-physics analyses~\cite{Lubicz:2018rfs} (ETM~18).
Both works use the available $\Nf = 2+1+1$ twisted-mass Wilson ensembles~\cite{Baron:2010bv},
totaling three lattice spacings down to $a\approx 0.06$~fm,
and a minimum pion mass of 220~MeV.
Matrix elements are extracted from suitable double ratios of correlation functions
that avoid the need of nontrivial current normalizations. Only one source-sink separation per ensemble is
used for the three-point functions, although the authors state that this separation was optimized
to achieve a balance between excited-state contamination and statistical uncertainties.
The use of twisted boundary conditions allows both for imposing
several kinematical configurations, and considering arbitrary frames
that include moving initial mesons. 
After interpolation to the physical strange- and charm-quark masses,
the results for form factors are fitted to a modified $z$-expansion
that takes into account both the light-quark mass dependence through
hard-pion SU(2) $\chi$PT~\cite{Bijnens:2010ws}, and the
lattice-spacing dependence. In the latter  case,
a detailed study of Lorentz-breaking effects due to the breaking of
rotational invariance down to the hypercubic subgroup
is performed, leading to a nontrivial momentum-dependent parameterization
of cutoff effects.
The $z$-parameterization (see Appendix~\ref{sec:zparam}) itself includes a single-pole Blaschke factor
(save for the scalar channel in $D\to K$, where the Blaschke factor is trivial),
with pole masses treated as free parameters.
The final quoted uncertainty on the form factors
is about 5--6\% for $D\to\pi$, and 4\% for $D\to K$.
The dominant source of uncertainty is quoted as statistical+fitting procedure+input parameters ---
the latter referring to the values of quark masses, the lattice spacing (i.e., scale setting),
and the LO SU(2) LECs.

The second $\Nf=2+1+1$ computation of $f_+$ and $f_0$ in the full kinematical range for the $D\to Kl\nu$ mode,
performed by HPQCD, has been published in 2021 --- HPQCD~21A (Ref.~\cite{Chakraborty:2021qav}).
This work uses MILC's HISQ ensembles at five values of the lattice spacing, 
and pion masses reaching to the physical point for the three coarsest values
of~$a$. Vector currents are normalized nonpertubatively by imposing that form factors
satisfy Ward identities exactly at zero recoil.
Results for the form factors are fitted to a modified $z$-expansion ansatz,
with all sub-threshold poles removed by using the experimental value of the
mass shifted by a factor that matches the corresponding result at finite lattice spacing.
The accuracy of the description of the $q^2$-dependence is crosschecked by comparing
to a fit based on cubic splines.
Finite-volume effects are expected to be small, and chiral-perturbation-theory-based
estimates for them are included in the chiral fit.  The impact of frozen
topology at the finest lattice spacing is neglected (the size of this effect was later shown to be $\lesssim0.03\%$ in
a similar calculation \cite{FermilabLattice:2022gku}).
The final uncertainty from the form factors in the determination of $|V_{cs}|$ quoted
in HPQCD~21A is at the 0.5\% level, and comparable to the rest of the uncertainty
(due to the experimental error, as well as weak and electromagnetic corrections);
in particular, the precision of the form factors is around seven times higher than 
that of the earlier $\Nf=2+1+1$ determination by ETM~17D.
The work also provides an accurate prediction for the lepton-flavour-universality ratio
between the muon and electron modes, where the uncertainty is overwhelmingly dominated
by the electromagnetic corrections. An extension of the work of HPQCD~21A to heavier quark masses has also
enabled the determination of the $B\to K$ form factors \cite{Parrott:2022rgu} (HPQCD~22), and
provides the tensor form factors for both $B\to K$ and $D\to K$ in addition to the vector form factors.

In 2022, the FNAL/MILC collaboration completed another $\Nf=2+1+1$ computation of $f_+$ and $f_0$
in the full kinematic ranges for $D\to K \ell \nu$, $D \to \pi \ell \nu$, and $D_s \to K l\nu$ -- FNAL/MILC~22 \cite{FermilabLattice:2022gku}.
Like HPQCD~21A, this calculation uses the MILC HISQ ensembles and renormalization using the vector Ward identity.
This calculation does not include the 0.15~fm ensembles that were part of the HPQCD~21A analysis, and shares only one of
the two 0.12~fm ensembles used in HPQCD~21A. Compared to HPQCD~21A, FNAL/MILC~22 reaches a finer lattice spacing
at the physical pion mass, 0.057~fm, while the ensemble at the finest lattice spacing of 0.042~fm is common to
both calculations. Overall, four of the seven ensembles are shared, but FNAL/MILC~22 uses more configurations and
source positions on those ensembles. In FNAL/MILC~22, the chiral/continuum extrapolation is performed
using rooted staggered heavy-meson chiral perturbation theory prior to a continuum BCL $z$ expansion fit.
This work also corrects the effects of the frozen topology at the finest lattice spacing using chiral
perturbation theory; the correction is found to be $\lesssim0.03\%$.


\begin{table}[h]
\begin{center}
\mbox{} \\[3.0cm]
\footnotesize
\begin{tabular*}{\textwidth}[l]{l @{\extracolsep{\fill}} r ll  l @{\hspace{1mm}}l @{\hspace{1mm}}l@{\hspace{1mm}} l@{\hspace{1mm}} l c c}
Collaboration & Ref. & $\Nf$ & 
\hspace{0.15cm}\begin{rotate}{60}{publication status}\end{rotate}\hspace{-0.15cm} &
\hspace{0.15cm}\begin{rotate}{60}{continuum extrapolation}\end{rotate}\hspace{-0.15cm} &
\hspace{0.15cm}\begin{rotate}{60}{chiral extrapolation}\end{rotate}\hspace{-0.15cm}&
\hspace{0.15cm}\begin{rotate}{60}{finite volume}\end{rotate}\hspace{-0.15cm}&
\hspace{0.15cm}\begin{rotate}{60}{renormalization}\end{rotate}\hspace{-0.15cm}  &
\hspace{0.15cm}\begin{rotate}{60}{heavy-quark treatment}\end{rotate}\hspace{-0.15cm}  &
$f_+^{D\to \pi}(0)$ & $f_+^{D\to K}(0)$\\
&&&&&&&&& \\[-0.1cm]
\hline
\hline
&&&&&&&&& \\[-0.1cm]
FNAL/MILC~22 & \cite{FermilabLattice:2022gku} & 2+1+1 & \gA  & \good & \good & \good & \good & \okay & 0.6300(51) & 0.7452(31)
 \\[0.5ex]
HPQCD~22 & \cite{Parrott:2022rgu} & 2+1+1 & \gA  & \good & \good & \good & \good & \okay & n/a & 0.7441(40)
 \\[0.5ex]
HPQCD~21A & \cite{Chakraborty:2021qav} & 2+1+1 & \gA	 & \good & \good & \good & \good & \okay & n/a & 0.7380(44) \\[0.5ex]
HPQCD~20 & \cite{Cooper:2020wnj} & 2+1+1 & \gA	 & \good & \soso & \good & \good & \okay & n/a & n/a \\[0.5ex]
ETM~17D, 18 & \cite{Lubicz:2017syv,Lubicz:2018rfs} & 2+1+1 & \gA	 & \good & \soso & \soso & \good & \okay & 0.612(35) & 0.765(31) \\[0.5ex]
&&&&&&&&& \\[-0.1cm]
\hline\\[0.5ex]
JLQCD~17B & \cite{Kaneko:2017xgg} & 2+1 & \rC & \good & \good & \soso & \good & \okay & 0.615(31)($^{+17}_{-16}$)($^{+28}_{-7}$)$^*$ & 0.698(29)(18)($^{+32}_{-12}$)$^*$ \\[0.5ex]
HPQCD~11 & \cite{Na:2011mc} & 2+1 & \gA  & \soso & \soso & \soso & \good &  \okay & 0.666(29) &\\[0.5ex]
HPQCD~10B & \cite{Na:2010uf} & 2+1 & \gA  & \soso & \soso & \soso & \good &  \okay & & 0.747(19)  \\[0.5ex]
FNAL/MILC~04 & \cite{Aubin:2004ej} & 2+1 & \gA  & \tbr & \tbr & \soso & \soso & \okay & 0.64(3)(6)& 0.73(3)(7)
\\[0.5ex]
&&&&&&&&& \\[-0.1cm]
\hline
\hline
\end{tabular*}\\
\begin{minipage}{\linewidth}
{\footnotesize 
\begin{itemize}
   \item[$^*$] The first error is statistical, the second from the $q^2\to 0$ extrapolation, the third from the chiral-continuum extrapolation.
\end{itemize}
}
\end{minipage}
\caption{Summary of computations of charmed-meson semileptonic form factors.
Note that HPQCD~20 (discussed in Sec.~\protect\ref{sec:charmheavyspectator}) addresses the $B_c \to B_s$ and $B_c \to B_d$ transitions---hence
the absence of quoted values for $f_+^{D\to \pi}(0)$ and $f_+^{D\to K}(0)$---while ETM~18 and HPQCD~22 provide computations of tensor form factors.
The value for $f_+^{D\to K}(0)$ from HPQCD~22 \cite{Parrott:2022rgu} is obtained as a by-product of the $B\to K$ analysis and is not independent from HPQCD~21A \cite{Chakraborty:2021qav}.
FNAL/MILC~22 also provides results for the $D_s \to K$ form factors in addition to the $D\to K$ and $D\to\pi$ form factors \cite{FermilabLattice:2022gku}. }
\label{tab_DtoPiKsumm2}
\end{center}
\end{table}

Table \ref{tab_DtoPiKsumm2} contains our summary of the existing
calculations of the charm-meson semileptonic form factors.  Additional tables in
Appendix~\ref{app:DtoPi_Notes} provide further details on the
simulation parameters and comparisons of the error estimates. Recall
that only calculations without red tags that are published in a
refereed journal are included in the FLAG average.
For $\Nf=2+1$, only HPQCD~10B,11 qualify,
which provides our estimate for $f_+(q^2=0)=f_0(q^2=0)$.
For $\Nf=2+1+1$, we quote as the FLAG estimate for $f_+^{D\to \pi}(0)$ the
weighted average of the results by ETM~17D and FNAL/MILC~22, while for $f_+^{D\to K}(0)$ we quote
the weighted average of the values published by ETM~17D, HPQCD~21A, and FNAL/MILC~22:

%
\begin{align}
	&&\FLAGAVBEGIN f_+^{D\to \pi}(0)&=  0.666(29)\FLAGAVEND&&\Ref~\mbox{\cite{Na:2011mc}},\nonumber\\[-3mm]
&\Nf=2+1:&\label{eq:Nf=2p1Dsemi}\\[-3mm]
        &&\FLAGAVBEGIN f_+^{D\to K}(0)  &= 0.747(19)\FLAGAVEND &&\Ref~\mbox{\cite{Na:2010uf}},\nonumber
\end{align}
%

%
\begin{align}
	&&\FLAGAVBEGIN f_+^{D\to \pi}(0)&=  0.6296(50)\FLAGAVEND&&\Refs~\mbox{\cite{Lubicz:2017syv,FermilabLattice:2022gku}},\nonumber\\[-3mm]
&\Nf=2+1+1:&\label{eq:Nf=2p1p1Dsemi}\\[-3mm]
        &&\FLAGAVBEGIN f_+^{D\to K}(0)  &= 0.7430(27)\FLAGAVEND &&\Refs~\mbox{\cite{Lubicz:2017syv,Chakraborty:2021qav,FermilabLattice:2022gku}}.\nonumber
\end{align}
%

\begin{figure}[h]
\begin{center}
\includegraphics[width=0.7\linewidth]{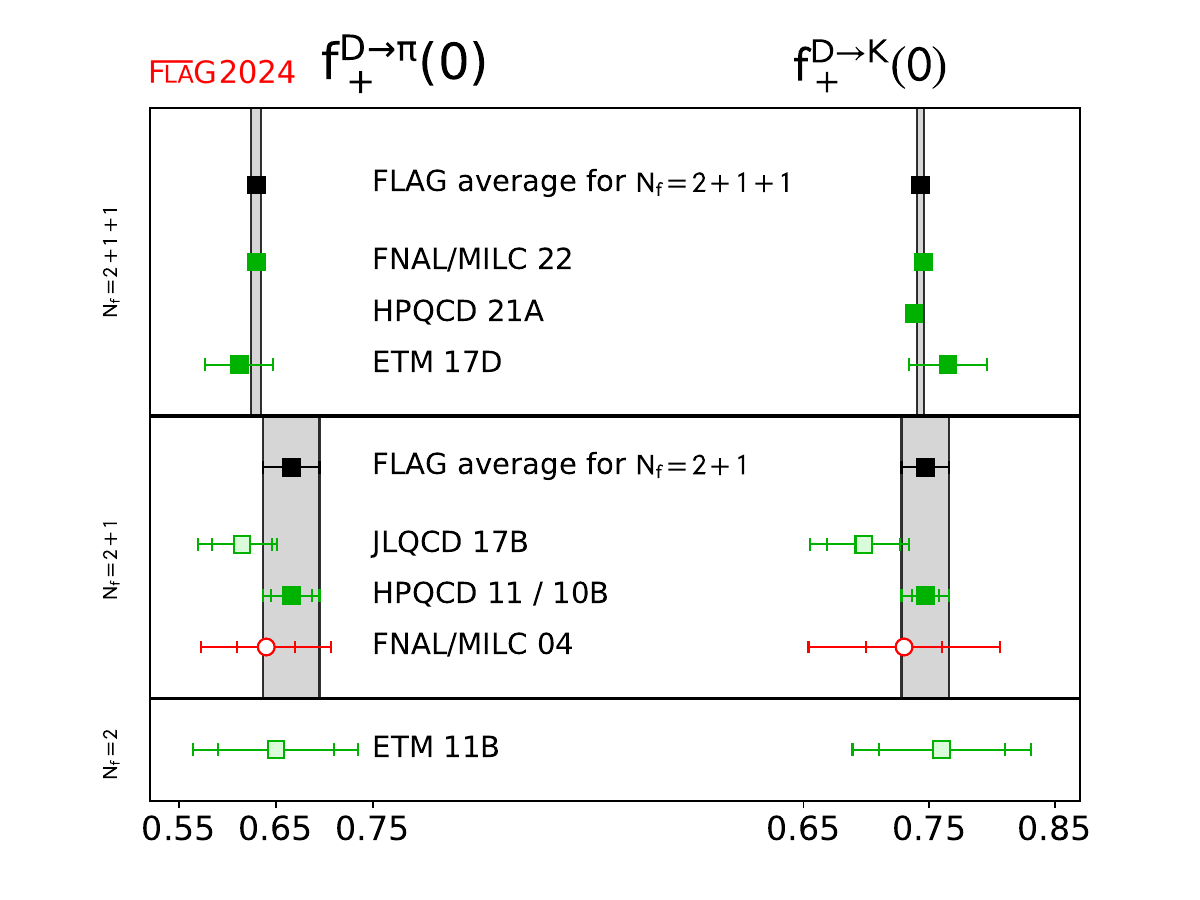}

\vspace{-2mm}
\caption{$D\to\pi \ell\nu$ and $D\to K\ell\nu$ semileptonic form
  factors at $q^2=0$. The $\Nf=2+1$ HPQCD result for
  $f_+^{D\to \pi}(0)$ is from HPQCD~11, the one for $f_+^{D\to K}(0)$
  represents HPQCD~10B (see Tab.~\ref{tab_DtoPiKsumm2}). \label{fig:DtoPiK}}
 \end{center}
\end{figure}

In Fig.~\ref{fig:DtoPiK}, we display the existing $\Nf =2$, $\Nf = 2+1$, and $\Nf=2+1+1$
results for $f_+^{D\to \pi}(0)$ and $f_+^{D\to K}(0)$; the grey bands show our
estimates of these quantities.

\begin{figure}[tbp]
\begin{center}
\includegraphics[width=0.49\textwidth]{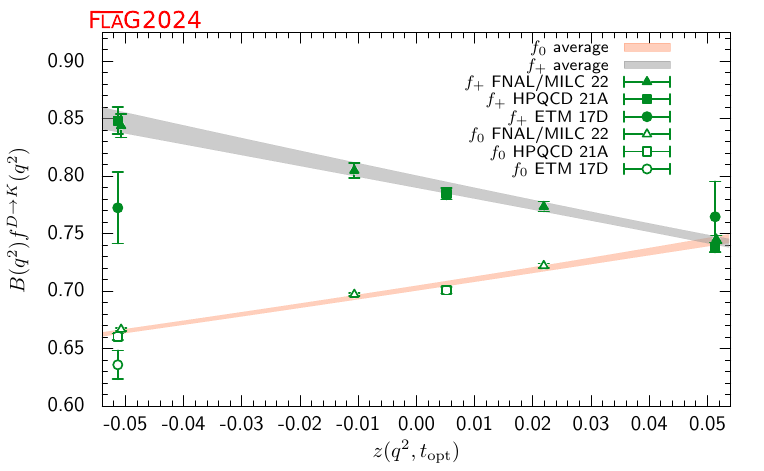}
\includegraphics[width=0.49\textwidth]{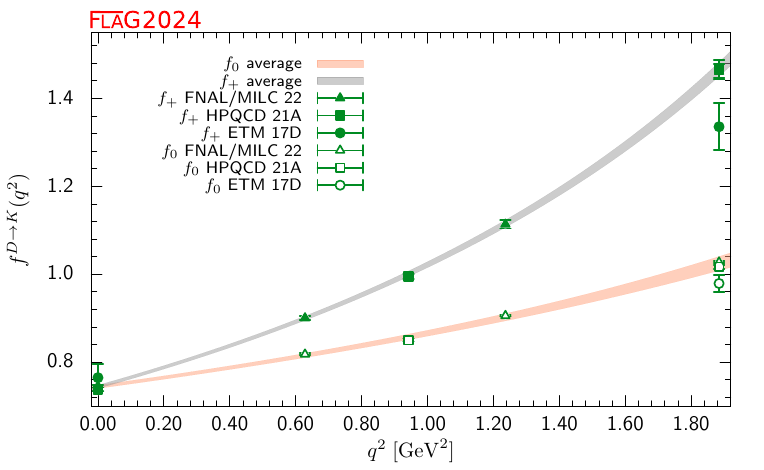}
\caption{The form factors $f_+(q^2)$ and $f_0 (q^2)$ for $D \to K\ell\nu$ plotted versus $z$ (left panel) and $q^2$ (right panel).
In the left plot, we removed the Blaschke factors.
See text for a discussion of the data set. The grey and salmon bands display our preferred $N^+=N^0=4$ BCL fit (seven parameters).}\label{fig:LQCDzfitDK}
\end{center}
\end{figure}

In the case of $\Nf=2+1+1$, we can also provide an analysis of the $q^2$-dependence of $f_+$ and $f_0$. FLAG~21 included a BCL fit to the ETM~17D and
HPQCD~21 results for the $D\to K$ form factors; this fit had a relatively poor $\chi^2/{\rm dof}=9.17/3$
due to a tension between the results from the two collaborations at large $q^2$; for $D\to \pi$, only the
ETM~17D results were available at that time. Now, the FNAL/MILC~22 calculation \cite{FermilabLattice:2022gku}
provides new high-precision $\Nf=2+1+1$ results for both $D\to K$ and $D\to \pi$ (as well as $D_s\to K$).
For $D\to K$, we update our previous BCL fit to include the FNAL/MILC~22 results. We consider the statistical correlations
between the final HPQCD~21A and FNAL/MILC~22 results to be negligible, given that there is only partial overlap among the ensembles, the source positions
for the correlation functions are different, and the analyses are performed with different fit methodologies. As in FLAG~21,
we generate synthetic data from the parameterizations provided by the collaborations. The inputs to our fit from ETM~17D and HPQCD~21A are unchanged;
for FNAL/MILC~22 we use four $q^2$ values because the parameterization used in
that reference is of higher order.  In both cases, this includes the kinematical endpoints $q^2=0$ and $q^2=(m_D-m_K)^2$ of the semileptonic interval.
We fit the resulting dataset to a BCL ansatz (cf. Eqs.~(\ref{eq:bcl_c}) and (\ref{eq:bcl_f0}));
the constraint $f_+(0)=f_0(0)$ is used to rewrite the highest-order coefficient $a^0_{N_0-1}$
in $f_0$ in terms of the other $N_++N_0-1$ coefficients. In both form factors, we include
nontrivial Blaschke factors, with pole masses set to the experimental values of the $D_s^*$
(for the vector channel) and $D_{s0}^*$ (scalar channel) masses found in the PDG~\cite{Zyla:2020zbs}.
We take flavour averages of charged and neutral states for the $D$ and $K$ masses.
Our external input is thus $m_D=1.87265~{\rm GeV}$, $m_K=495.644~{\rm MeV}$,
$m_{D_s^*}=2.1122~{\rm GeV}$, and $m_{D_{s0}^*}=2.317~{\rm GeV}$. As a result of including the new FNAL/MILC~22
data points, we found it necessary to increase the order of the $z$ expansion from $N_+=N_0=3$ (as used in FLAG~21)
to  $N_+=N_0=4$. The fit has $\chi^2/{\rm dof}\approx 2.39$ (due to the tension between the ETM~17D results at high $q^2$ and the results from the
other two collaborations, and due to a slight tension between the results from HPQCD~21A and  FNAL/MILC~22 in $f_0$) and we have scaled the uncertainties of all parameters by a factor of $\sqrt{\chi^2/{\rm dof}}\approx1.55$. The results
are quoted in full in Tab.~\ref{tab:FFDK} and illustrated in Fig.~\ref{fig:LQCDzfitDK}.

As can be seen in Fig.~19 of
Ref.~\cite{FermilabLattice:2022gku}, for $D\to \pi$ there is a very large tension between the ETM~17D and FNAL/MILC~22
 results at high $q^2$, in the same direction as the tension also seen for $D\to K$. In this case, the tension is so significant
 that attempting BCL fits to average the ETM~17D and FNAL/MILC~22 results gives values of $\chi^2/{\rm dof}$
of order 100. We are concerned about possible excited-state contamination in ETM~17D, but the authors of ETM~17D
stated that there is no evidence of an uncontrolled systematic effect; the tension remains unexplained. We therefore do not quote any results for
the $D\to \pi$ form factors away from $q^2=0$.

\begin{table}[t]\footnotesize
\begin{center}
\begin{tabular}{|c|r|rrrrrrr|}
\multicolumn{7}{l}{$D\to K\ell\nu \; (\Nf=2+1+1)$} \\[0.2em]\hline
        & \multicolumn{1}{c}{values} & \multicolumn{7}{|c|}{correlation matrix} \\[0.2em]\hline
$a_0^+$  & 0.7953(53)    &    1.       & $-$0.690759 & $-$0.051101 & $-$0.061092 &    0.501293 & 0.469810 &    0.132470 \\[0.2em]
$a_1^+$  & $-$1.0090(87) & $-$0.690759 &    1.       & $-$0.231861 &    0.133663 &    0.004097 & 0.149657 &    0.137516 \\[0.2em]
$a_2^+$  & 0.22(59)      & $-$0.051101 & $-$0.231861 &    1.       & $-$0.113075 & $-$0.095636 & 0.101738 &    0.238861 \\[0.2em]
$a_3^+$  & 0.14(10)      & $-$0.061092 &    0.133663 & $-$0.113075 &    1.       & $-$0.109883 & 0.116543 &    0.112918 \\[0.2em]
$a_0^0$  & 0.7026(21)    &    0.501293 &    0.004097 & $-$0.095636 & $-$0.109883 &    1.       & 0.339786 & $-$0.251322 \\[0.2em]
$a_1^0$  & 0.773(39)     &    0.469810 &    0.149657 &    0.101738 &    0.116543 &    0.339786 & 1.       &    0.589149  \\[0.2em]
$a_2^0$  & 0.54(40)      &    0.132470 &    0.137516 &    0.238861 &    0.112918 & $-$0.251322 & 0.589149 &    1. \\[0.2em]
\hline
\end{tabular}
\end{center}
\caption{Coefficients for the $N^+ =4, N^0=4$ $z$-expansion of the $\Nf=2+1+1$ FLAG average for the $D\to K$ form factors $f_+$ and $f_0$, and their correlation matrix. The inputs are from ETM~17D, HPQCD~21A, and FNAL/MILC~22. The form factors can be reconstructed using parameterization and inputs given in Appendix~\ref{sec:app_D2K}.
\label{tab:FFDK}}
\end{table}

\FloatBarrier

\subsection{Form factors for $\Lambda_c$ and $\Xi_c$ semileptonic decays}

The motivation for studying charm-baryon semileptonic decays is two-fold. First, these decays
allow for independent determinations of $|V_{cs}|$.
Second, given that possible new-physics contributions to the $c\to s\ell\nu$ weak effective Hamiltonian
are already constrained to be much smaller compared to $b\to u\ell\bar{\nu}$ and $b\to s \ell\ell$,
charm-baryon semileptonic decays allow testing the lattice techniques for baryons that are also
employed for bottom-baryon semileptonic decays (see Sec.~\ref{sec:Lambdab}) in a better-controlled
environment.

The amplitudes of the decays $\Lambda_c\to \Lambda\ell\nu$
receive contributions from both the vector and the axial components of the current
in the matrix element
$\langle \Lambda|\bar s\gamma^\mu(\mathbf{1}-\gamma_5)c|\Lambda_c\rangle$,
and can be parameterized in terms of six different form factors  $f_+$, $f_0$, $f_\perp$, $g_+$, $g_0$, $g_\perp$ --- see, e.g., Ref.~\cite{Feldmann:2011xf} for a complete description.

The computation in Meinel 16~\cite{Meinel:2016dqj} uses RBC/UKQCD $\Nf=2+1$ DWF ensembles,
and treats the $c$ quarks within the Columbia RHQ approach.
Two values of the lattice spacing ($a\approx 0.11,~0.085~{\rm fm}$) are considered,
with the absolute scale set from the $\Upsilon(2S)$--$\Upsilon(1S)$ splitting.
In one ensemble, the pion mass $m_\pi\approx 139~{\rm MeV}$ is at the physical point,
while for other ensembles it ranges from 295 to 352~MeV.
Results for the form factors are obtained from suitable three-point functions,
and fitted to a modified $z$-expansion ansatz that combines the $q^2$-dependence
with the chiral and continuum extrapolations. The paper 
predicts for the total rates in the $e$ and $\mu$ channels
\begin{gather}
\begin{split}
\frac{\Gamma(\Lambda_c\to \Lambda e^+\nu_e)}{|V_{cs}|^2} &= 0.2007(71)(74)~{\rm ps}^{-1}\,,\\
\frac{\Gamma(\Lambda_c\to \Lambda\mu^+\nu_\mu)}{|V_{cs}|^2} &= 0.1945(69)(72)~{\rm ps}^{-1}\,,
\end{split}
\end{gather}
where the uncertainties are statistical and systematic, respectively. In combination with the recent experimental determination of the total branching fractions
by BESIII~\cite{Ablikim:2015prg,Ablikim:2016vqd}, it is possible to extract $|V_{cs}|$
as discussed in Sec.~\ref{sec:Vcd} below.

Lattice results are also available for the $\Lambda_c \to N$ form factors, where $N$ is a neutron or proton \cite{Meinel:2017ggx}.
This calculation uses the same lattice actions but a different set of ensembles with parameters matching those used in the
2015 calculation of the $\Lambda_b \to p$ form factors in Ref.~\cite{Detmold:2015aaa} (cf. Sec.~\ref{sec:Lambdab}). Predictions
are given for the rates of the $c\to d$ semileptonic decays $\Lambda_c\to n \ell^+\nu_\ell$; these modes have not yet
been observed. Reference~\cite{Meinel:2017ggx} also studies the phenomenology of the flavour-changing neutral-current decay
$\Lambda_c \to p \mu^+\mu^-$. As is typical for rare charm decays to charged leptons, this mode is dominated by long-distance effects
that have not yet been calculated on the lattice and whose description is model-dependent.

The authors of Zhang~21 \cite{Zhang:2021oja} also performed a first lattice calculation of the $\Xi_c \to \Xi$ form factors 
and extracted $|V_{cs}|$, with still large uncertainties, from the recent Belle measurement of the $\Xi_c \to \Xi \ell^+ \nu_\ell$ branching fractions \cite{Li:2021uhk}. This calculation uses only two ensembles with $2+1$ flavours of clover fermions, with lattice spacings of $0.108$ and $0.080$~fm and nearly identical pion masses of 290 and 300 MeV. The results are extrapolated to the continuum limit but are not extrapolated to the physical pion
mass. No systematic uncertainty is estimated for the effect of the missing chiral extrapolation. A new calculation of the $\Xi_c \to \Xi$ form factors using domain-wall fermions is in progress \cite{Farrell:2023vnm}.

The calculations discussed so far in this section all have $J^P=\frac12^+$ baryons in the final state.
A first lattice calculation of the form factors for a charm-baryon semileptonic decay to a $J^P=\frac32^-$ baryon, $\Lambda_c \to \Lambda^*(1520)\ell^+\nu_\ell$, is also available:
Meinel~21B \cite{Meinel:2021mdj}.
The calculation was done using three RBC/UKQCD ensembles with $2+1$ flavours of domain-wall fermions, with $a\approx0.11,~0.08~{\rm fm}$ and pion masses in the range from approximately 300 to 430~MeV.
Chiral-continuum extrapolations linear in $m_\pi^2$ and $a^2$ were performed, with systematic uncertainties estimated using higher-order fits.
Finite-volume effects and effects associated with the strong decays of the $\Lambda^*(1520)$ are not quantified.
The calculation was done in the $\Lambda^*(1520)$ rest frame, where the cubic symmetry is sufficient to avoid mixing with unwanted lower-mass states.

A summary of the lattice calculations of charm-baryon semileptonic-decay form factors is given in Tab.~\ref{tab_CharmBaryonSLsumm2}.

\begin{table}[h]
\begin{center}
\mbox{} \\[3.0cm]
\footnotesize
\begin{tabular}{l l @{\extracolsep{\fill}} r l l l l l l l}
Process & Collaboration & Ref. & $\Nf$ & 
\hspace{0.15cm}\begin{rotate}{60}{publication status}\end{rotate}\hspace{-0.15cm} &
\hspace{0.15cm}\begin{rotate}{60}{continuum extrapolation}\end{rotate}\hspace{-0.15cm} &
\hspace{0.15cm}\begin{rotate}{60}{chiral extrapolation}\end{rotate}\hspace{-0.15cm}&
\hspace{0.15cm}\begin{rotate}{60}{finite volume}\end{rotate}\hspace{-0.15cm}&
\hspace{0.15cm}\begin{rotate}{60}{renormalization}\end{rotate}\hspace{-0.15cm}  &
\hspace{0.15cm}\begin{rotate}{60}{heavy-quark treatment}\end{rotate}\hspace{-0.15cm} \\
&&&&&&&& \\[-0.1cm]
\hline
\hline
&&&&&&&& \\[-0.1cm]
$\Lambda_c\to  \Lambda^*(1520) \ell\nu$    & Meinel 21B\:\:     & \cite{Meinel:2021mdj} & 2+1 & \gA & \soso & \soso & \tbr  & \soso & \okay \\[0.5ex]
$\Xi_c\to\Xi \ell\nu$                      & Zhang  21          & \cite{Zhang:2021oja}  & 2+1 & \gA & \tbr & \tbr  & \soso & \good & \tbr  \\[0.5ex] 
$\Lambda_c\to  n \ell\nu$                  & Meinel 17          & \cite{Meinel:2017ggx} & 2+1 & \gA & \soso & \soso & \tbr  & \soso & \okay \\[0.5ex]
$\Lambda_c\to  \Lambda \ell\nu$            & Meinel 16          & \cite{Meinel:2016dqj} & 2+1 & \gA & \soso & \good & \good & \soso & \okay \\[0.5ex]
&&&&&&&& \\[-0.1cm]
\hline
\hline
\end{tabular}
\caption{Summary of computations of charmed-baryon semileptonic form factors.  The rationale for the \tbr~ rating of finite-volume effects in Meinel 21B
(despite meeting the {\color{green}\Large$\circ$} criterion based on the minimum pion mass) is that the unstable nature of the final-state baryons was neglected in the analysis.}
\label{tab_CharmBaryonSLsumm2}
\end{center}
\end{table}

\subsection{Form factors for charm semileptonic decays with heavy spectator quarks}
\label{sec:charmheavyspectator}

Two other decays mediated by the $c\to s\ell\nu$ and $c\to d\ell\nu$ transitions are $B_c \to B_s \ell\nu$ and $B_c \to B^0 \ell\nu$, respectively. At present, there are no experimental results for these processes, but it may be possible to produce them at LHCb in the future. The HPQCD Collaboration has recently computed the form factors for both of these $B_c$ decay modes with $\Nf=2+1+1$~\cite{Cooper:2020wnj}. The calculation uses six different MILC ensembles with HISQ light, strange, and charm quarks, and employs the PCAC 
Ward identity to nonperturbatively renormalize the $c\to s$ and $c\to d$ currents. Data were generated for two different choices of the lattice action for the spectator $b$ quark: lattice NRQCD on five of the six ensembles, and HISQ on three of the six ensembles (cf.~Sec.~\ref{sec:BDecays} for a discussion of different lattice approaches used for the $b$ quark). For the NRQCD calculation, two of the ensembles have a physical light-quark mass, and the lattice spacings are 0.15~fm, 0.12~fm, and 0.09~fm. The heavy-HISQ calculation is performed only at $m_l/m_s=0.2$, and at lattice spacings of 0.12~fm, 0.09~fm, and 0.06~fm. The largest value of the heavy-HISQ mass used is 0.8 in lattice units on all three ensembles, which does not reach the physical $b$-quark mass even at the finest lattice spacing.

Form-factor fits are performed using $z$-expansions (see Appendix~\ref{sec:zparam}) modified to include a dependence on the lattice spacing and quark masses, including an expansion in the inverse heavy quark mass in the case of the heavy-HISQ approach. The parameters $t_+$ are set to $(m_{B_c}+m_{B_{(s)}})^2$ even though the branch cuts start at $(m_D+m_K)^2$ or $(m_D+m_\pi)^2$, as also noted by the authors. The variable $z$ is rescaled by a constant. The lowest charmed-meson poles are removed before the $z$-expansion, but this still leaves the branch cuts and higher poles below $t_+$. As a consequence of this structure, the good convergence properties
of the $z$-expansion are not necessarily expected to apply. Fits are performed (i) using the NRQCD data only, (ii) using the HISQ data only, and (iii) using the NRQCD data, but with priors on the continuum-limit form-factor parameters equal to the results of the HISQ fit. The results from fits (i) and (ii) are mostly consistent, with the NRQCD fit having smaller uncertainties than the HISQ fit. Case (iii) then results in the smallest uncertainties and gives the predictions (for massless leptons)
\begin{gather}
\begin{split}
\frac{\Gamma(B_c \to B_s \ell^+\nu_\ell)}{|V_{cs}|^2} &= 1.738(55)\times10^{-11}~{\rm MeV}\,,\\
\frac{\Gamma(B_c \to B^0 \ell^+\nu_\ell)}{|V_{cd}|^2} &= 2.29(12)\times10^{-11}~{\rm MeV}\,, \\
\frac{\Gamma(B_c \to B_s \ell^+\nu_\ell)|V_{cd}|^2}{\Gamma(B_c \to B^0 \ell^+\nu_\ell)|V_{cs}|^2}&= 0.759(44).
\end{split}
\end{gather}
We note that there is a discrepancy between the NRQCD and HISQ results in the case of $f_0(B_c\to B^0)$, and the uncertainty quoted for method (iii) does not cover this discrepancy. However, this form factor does not enter in the decay rate for massless leptons.

%% file: HQ/HQSubsections/DCKM.tex
\subsection{Determinations of $|V_{cd}|$ and $|V_{cs}|$ and test of  second-row CKM unitarity}
\label{sec:Vcd}

We now use the lattice-QCD results for the charm-hadron decays
to determine the CKM matrix elements $|V_{cd}|$ and $|V_{cs}|$ in the Standard Model.

For the leptonic decays, we use the latest experimental averages from
the Particle Data Group~\cite{ParticleDataGroup:2024cfk} (see Sec.~72.3.1)
\begin{equation}
f_D |V_{cd}| = 45.82(1.10)~{\rm MeV} \,, \qquad f_{D_s} |V_{cs}| = 243.5(2.7)~{\rm MeV}, \,
\end{equation}
where the errors include those from nonlattice theory, e.g., estimates of radiative corrections to 
lifetimes~\cite{Dobrescu:2008er}. Also, the values quoted by the Particle Data Group are obtained after 
applying the correction factor $\eta_{\rm EW}^2=1.018$, due to universal short-distance electroweak 
contributions~\cite{Sirlin:1981ie}, to the branching ratios. Hadronic-structure-dependent
electromagnetic corrections to the rate have not been computed on the lattice for
the case of $D_{(s)}$ mesons, while they have been calculated for pion and kaon decays~\cite{DiCarlo:2019thl,Boyle:2022lsi}. 
The errors given above include a systematic
uncertainty of 0.7\% estimated as half the size of the applied radiative corrections.

By combining these with the averaged $\Nf=2+1$ and $2+1+1$ results for $f_D$ and $f_{D_s}$ in
Eqs.~(\ref{eq:fD2+1}-\ref{eq:fDs2+1+1}),
we obtain
\begin{alignat}{3}
\Nf=2+1+1 \text{: }
&\begin{cases}
|V_{cd}| &= 0.2161(7)(52) \\
|V_{cs}| &= 0.974(2)(11)
\end{cases}
\quad\quad &&[D_{(s)}\to \ell\nu, \text{Refs.~\cite{Bazavov:2017lyh,Carrasco:2014poa}}], \label{eq:VcdsL2p1p1} \\
\Nf=2+1 \text{: }
&\begin{cases}
|V_{cd}| &= 0.2178(16)(52) \\
|V_{cs}| &= 0.983(5)(11)
\end{cases}
\quad\quad &&[D_{(s)}\to \ell\nu, \text{Refs.~\cite{Bussone:2023kag,Bazavov:2011aa,Davies:2010ip,Na:2012iu,Yang:2014sea,Boyle:2017jwu}}], \label{eq:VcdsL2p1}
\end{alignat}
where the errors shown are from the lattice calculation and experiment
(plus nonlattice theory), respectively.  For the $\Nf = 2+1$ and the $\Nf=2+1+1$
determinations, the uncertainties from the lattice-QCD calculations of
the decay constants are significantly smaller than the
experimental uncertainties in the branching fractions.

\vskip 5mm

For $D$-meson semileptonic decays, in the case of $\Nf=2+1$ there are no changes with respect to FLAG~21
other than the inclusion of the short-distance electroweak correction and a systematic uncertainty due to
missing long-distance QED corrections;
the only works entering the FLAG averages are HPQCD~10B/11~\cite{Na:2010uf,Na:2011mc}, which provide
$f_+^{D\pi}(0)$ and $f_+^{DK}(0)$. We use these results in combination with the HFLAV averages
for the combinations $f_+(0)\eta_{\rm EW}|V_{cx}|$~\cite{Amhis:2019ckw},
\begin{equation}
\label{eq:fpDtoPiandKexp}
	f_+^{D\pi}(0)\eta_{\rm EW} |V_{cd}| =  0.1426(18) \,, \qquad f_+^{DK}(0)\eta_{\rm EW} |V_{cs}| =  0.7180(33), 
\end{equation}
and obtain
\begin{alignat}{3}
\Nf=2+1 \text{: } &|V_{cd}| =  0.2121(92)(29)(21) \quad\quad &&[D\to \pi \ell \nu, \text{Ref.~\cite{Na:2011mc}}], \label{eq:Nf=2p1VcdSL} \\
\Nf=2+1 \text{: } &|V_{cs}| =  0.958(25)(5)(10) \quad\quad &&[D\to K \ell \nu, \text{Ref.~\cite{Na:2010uf}}],
\label{eq:Nf=2p1VcsSL}
\end{alignat}
where the uncertainties are lattice, experimental (plus nonlattice theory), and missing long-distance
QED corrections (estimated to be $1\%$), respectively.

\begin{figure}
\begin{center}
\includegraphics[width=0.8\linewidth]{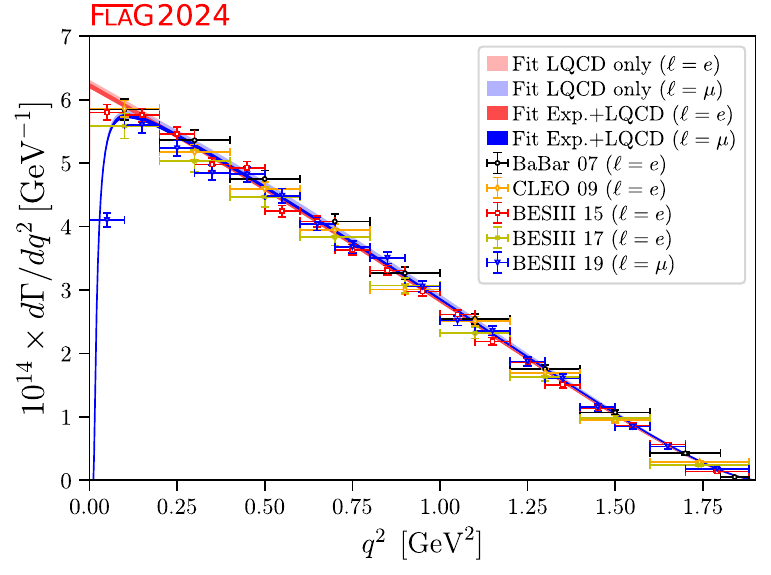}
\end{center}
\caption{Our fits to the $D\to K \ell \nu$ differential decay rates used to determine $|V_{cs}|$, with experimental inputs from Refs.~\cite{Aubert:2007wg,Besson:2009uv,Ablikim:2015ixa,Ablikim:2017lks,Ablikim:2018evp} and lattice inputs from ETM17D \cite{Lubicz:2017syv},  HPQCD 21A \cite{Chakraborty:2021qav}, and FNAL/MILC 22 \cite{FermilabLattice:2022gku}.\label{fig:DtoKdGammadqsqr}}
\end{figure}

For $\Nf=2+1+1$, we update our BCL fit to the binned $D\to K \ell \nu$ differential decay rates by
adding the FNAL/MILC 22 inputs for $f_+(q^2)$ and $f_0(q^2)$ at four $q^2$ values (the ETM 17D and HPQCD 21A inputs remain unchanged).
The experimental datasets we include are unchanged with respect to FLAG~21 and
are three different measurements of the $D^0 \to K^- e^+\nu_e$ mode by BaBar (BaBar 07, Ref.~\cite{Aubert:2007wg}),
CLEO-c (CLEO 09/$0$, Ref.~\cite{Besson:2009uv}), and BESIII (BESIII 15, Ref.~\cite{Ablikim:2015ixa});
CLEO-c (CLEO 09/$+$, Ref.~\cite{Besson:2009uv}) and BESIII measurements of the $D^+ \to \bar K^0 e^+\nu_e$ mode (BESIII 17, Ref.~\cite{Ablikim:2017lks});
and the recent first measurement of the $D^0 \to K^-\mu^+\nu_\mu$ mode by BESIII, Ref.~\cite{Ablikim:2018evp}.
There is also a Belle dataset available in Ref.~\cite{Widhalm:2006wz}, but it provides
results for parameterized form factors rather than partial widths, which implies that
reverse modelling of the $q^2$-dependence of the form factor would be needed
to add them to the fit, which involves an extra source of systematic uncertainty;
it is, furthermore, the measurement with the largest error.  Thus, we will 
drop it. The CLEO collaboration provides correlation matrices for the systematic uncertainties
across the channels in their two measurements; the latter are, however, not available for BESIII,
and, therefore, we will conservatively treat their systematics with a 100\% correlation, following
the same prescription as in the HFLAV review~\cite{Amhis:2019ckw}. Since all lattice results have been obtained in the
isospin limit, we average over the $D^0$ and $D^+$ electronic modes. The parameterization of the form factors
we use here is the same as in the lattice-only fit discussed in Sec.~\ref{sec:DtoPiK}, and we again
increase the order of the $z$ expansion (with respect to FLAG~21) to  $N^+ = N^0=4$. In contrast to
FLAG~21, we now include the short-distance electroweak correction $\eta_{\rm EW}^2$ \cite{Sirlin:1981ie} in the calculation of the differential decay rate, using $\eta_{\rm EW}=1.009$ \cite{FermilabLattice:2022gku}. The fit has $\chi^2/{\rm dof}\approx 1.66$ and we have scaled all uncertainties by a factor of $\sqrt{\chi^2/{\rm dof}}\approx 1.29$. The results for the $z$-expansion parameters and $|V_{cs}|$, as well as their correlation matrix, are given in Tab.~\ref{tab:FFVCSPI}, and a plot of the differential decay rates is shown in Fig.~\ref{fig:DtoKdGammadqsqr}.
For $D\to \pi l\nu$, we do not use the lattice results away from $q^2=0$ as discussed in Sec.~\ref{sec:DtoPiK}.
To extract $|V_{cd}|$, we instead combine the average for $f_+^{D\pi}(0)$ from ETM 17D and FNAL/MILC 22 with the HFLAV result (\ref{eq:fpDtoPiandKexp}). Thus, we obtain
\begin{alignat}{2}
\Nf=2+1+1 \text{: } |V_{cd}| =  0.2245(33)(22) &\quad\quad [D\to \pi \ell \nu, \text{Ref.~\cite{Lubicz:2017syv,FermilabLattice:2022gku}}], \label{eq:Nf=2p1p1VcdSL} \\
\Nf=2+1+1 \text{: } |V_{cs}| =  0.9592(50)(96) &\quad\quad [D\to K \ell \nu, \text{Ref.~\cite{Lubicz:2017syv,Chakraborty:2021qav,FermilabLattice:2022gku}}], \label{eq:Nf=2p1p1VcsSL}
\end{alignat}
where the two uncertainties correspond, respectively, to the combined lattice-QCD and experimental errors, and an estimate of the size of missing long-distance QED corrections, taken to be 1\% following Ref.~\cite{FermilabLattice:2022gku}. Note that FNAL/MILC 22 \cite{FermilabLattice:2022gku} also determined $|V_{cd}|$ from $D_s \to K\ell\nu$ using a BESIII measurement \cite{BESIII:2018xre}, with the result
\begin{align}
\Nf=2+1+1 \text{: } |V_{cd}| =  0.258(15)(03) \quad\quad [D_s\to K \ell \nu, \text{Ref.~\cite{FermilabLattice:2022gku}}], \label{eq:Nf=2p1p1VcdSLK}
\end{align}
where the large uncertainty is dominated by the experimental measurement.

\begin{table}\scriptsize
\begin{center}
\resizebox{1\textwidth}{!}{
\begin{tabular}{|c|r|rrrrrrrr|}
\multicolumn{10}{l}{$D\to K\ell\nu \; (\Nf=2+1+1)$} \\[0.2em]\hline
        & \multicolumn{1}{c}{values} & \multicolumn{8}{|c|}{correlation matrix} \\[0.2em]\hline
$a_0^+$    & 0.7896(38) &  1.           & $-$0.555568 & $-$0.069722 & $-$0.021610 &    0.587914 & 0.646372 &    0.247552 &    0.795354   \\[0.2em]
$a_1^+$    & $-$0.945(51) & $-$0.555568 & 1.          & $-$0.303470 &    0.102546 & $-$0.014576 & 0.043616 &    0.036587 & $-$0.280176  \\[0.2em]
$a_2^+$    & 0.29(49)   &   $-$0.069722 & $-$0.303470 & 1.          & $-$0.109799 & $-$0.092179 & 0.107676 &    0.243102 & $-$0.033821  \\[0.2em]
$a_3^+$    & 0.257(84)  &   $-$0.021610 &    0.102546 & $-$0.109799 & 1.          & $-$0.112476 & 0.104107 &    0.101692 & $-$0.003737 \\[0.2em]
$a_0^0$    & 0.7029(18) &      0.587914 & $-$0.014576 & $-$0.092179 & $-$0.112476 & 1.          & 0.341851 & $-$0.256955 & 0.554412 \\[0.2em]
$a_1^0$    & 0.748(32)  &      0.646372 &    0.043616 &    0.107676 &    0.104107 &    0.341851 & 1.       &    0.578012 & 0.651080  \\[0.2em]
$a_2^0$    & 0.11(33)   &      0.247552 &    0.036587 &    0.243102 &    0.101692 & $-$0.256955 & 0.578012 & 1.          & 0.279081 \\[0.2em]
$|V_{cs}|$ & 0.9592(50) &      0.795354 & $-$0.280176 & $-$0.033821 & $-$0.003737 &    0.554412 & 0.651080 &    0.279081 & 1. \\[0.2em]
\hline
\end{tabular}
}
\end{center}
\caption{Coefficients for the $N^+ = N^0=4$ $z$-expansion simultaneous fit of the $D\to K$ form factors $f_+$ and $f_0$ and $|V_{cs}|$ to the $D\to K\ell\nu$ differential decay rates and the ETM 17D, HPQCD 21A, and FNAL/MILC 22 lattice results. The form factors can be reconstructed using parameterization and inputs given in Appendix~\ref{sec:app_D2K}. \label{tab:FFVCSPI}}
\end{table}

For $\Lambda_c \to \Lambda \ell \nu$, there are new experimental results for the electronic and muonic branching fractions from BESIII, published in 2022 and 2023 \cite{BESIII:2022ysa,BESIII:2023jxv}. In addition, the world average of the $\Lambda_c$ lifetime has been updated in the 2024 Review of Particle Physics to $\tau_{\Lambda_c}=(202.6\pm1.0)\times 10^{-15}\:{\rm s}$, following a new precise measurement by Belle II \cite{Belle-II:2022ggx}. Using these results together with the lattice-QCD predictions of Meinel 16 for $\Gamma(\Lambda_c\to\Lambda\ell\nu)/|V_{cs}|^2$ \cite{Meinel:2016dqj}, and including the factor of $\eta_{\rm EW}^2$ (not done in Ref.~\cite{Meinel:2016dqj}), we obtain
\begin{align}
\Nf=2+1 \text{: } |V_{cs}| =  0.929(24)(16)(2)(9) \quad\quad 
[\Lambda_c\to \Lambda \ell \nu, \text{Ref.~\cite{Meinel:2016dqj}}],  \label{eq:Nf=2p1p1VcdSLLambda}
\end{align}
where the uncertainties are from the lattice calculation, from the $\Lambda_c\to \Lambda \ell \nu$ branching fractions, from the $\Lambda_c$ lifetime, and from the missing long-distance QED corrections, respectively.

\vskip 5mm

\begin{figure}[h]

\begin{center}
\includegraphics[width=0.7\linewidth]{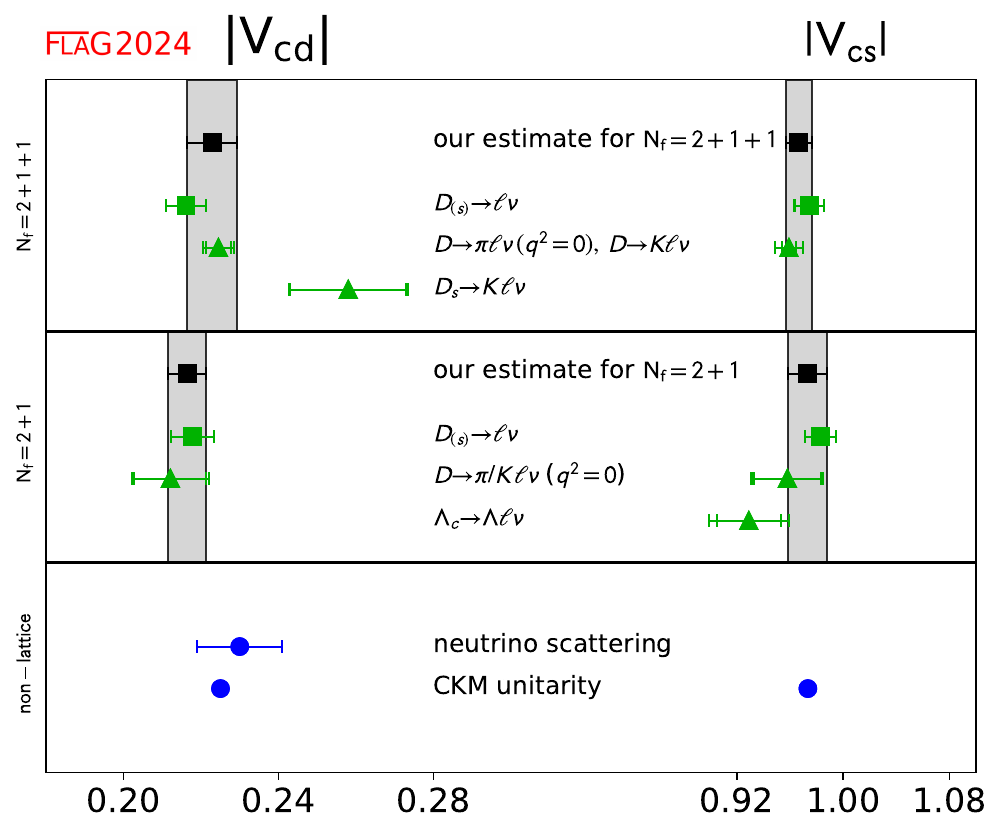}

\vspace{-2mm}
\caption{Comparison of determinations of $|V_{cd}|$ and $|V_{cs}|$
  obtained from lattice methods [Eqs.~(\ref{eq:VcdsL2p1p1}), (\ref{eq:VcdsL2p1}), (\ref{eq:Nf=2p1VcdSL}), (\ref{eq:Nf=2p1VcsSL}), (\ref{eq:Nf=2p1p1VcdSL}), (\ref{eq:Nf=2p1p1VcsSL}), (\ref{eq:Nf=2p1p1VcdSLK}), (\ref{eq:Nf=2p1p1VcdSLLambda}), (\ref{eq:Vcdsfinal2p1}), (\ref{eq:Vcdsfinal2p1p1})] with a nonlattice determination from neutrino scattering (for $|V_{cd}|$ only) \cite{Zyla:2020zbs}
and with the Standard-Model predictions from a global fit assuming CKM unitarity \cite{UTfit:2022hsi}.
\label{fig:VcdVcs}}
\end{center}
\end{figure}

In Fig.~\ref{fig:VcdVcs}, we summarize the results for $|V_{cd}|$
and $|V_{cs}|$ from leptonic 
and semileptonic
 decays, and compare
them to determinations from neutrino scattering (for $|V_{cd}|$ only)
and global fits assuming CKM unitarity (see~\cite{Zyla:2020zbs,UTfit:2022hsi}).
For both $|V_{cd}|$ and $|V_{cs}|$, the errors in the direct determinations from
leptonic  and semileptonic 
decays are approximately one order of magnitude larger
than the indirect determination from CKM unitarity.

In order to provide final estimates, we average the available results from the different processes
separately for each value of $\Nf$ and obtain
%
%
\begin{alignat}{2}
\Nf = 2+1+1 \text{:} 
\begin{cases}
\FLAGAVBEGIN |V_{cd}| = 0.2229(64) \FLAGAVEND \\
|V_{cs}| = 0.9667(96)
\end{cases}
&
[\text{FLAG average, Refs.~\cite{Bazavov:2017lyh,Carrasco:2014poa,Chakraborty:2021qav,FermilabLattice:2022gku}}]\,, \label{eq:Vcdsfinal2p1} \\
\Nf = 2+1 \text{:} 
\begin{cases}
\FLAGAVBEGIN |V_{cd}| = 0.2165(49) \FLAGAVEND \\
|V_{cs}| = 0.973(14)
\end{cases}
&
[\text{FLAG average, Refs.~\cite{Bussone:2023kag, Bazavov:2011aa, Davies:2010ip, Na:2012iu, Yang:2014sea,Boyle:2017jwu, Na:2010uf, Na:2011mc, Meinel:2016dqj}}]\,, \label{eq:Vcdsfinal2p1p1}
\end{alignat}
%
where the errors include both theoretical and experimental
uncertainties, and scale factors equal to $\sqrt{\chi^2/{\rm dof}}$ of 1.88 and 1.26 have been included for $|V_{cd}|_{\Nf=2+1+1}$ and $|V_{cs}|_{\Nf=2+1}$, respectively. These averages also appear in Fig.~\ref{fig:VcdVcs}, and are compatible with the values from the CKM global fit based on unitarity \cite{UTfit:2022hsi} within at most $1.5\sigma$. The slight increases in the uncertainties of the $\Nf=2+1+1$ averages compared to FLAG~21 are due to the inclusion of QED systematic uncertainties (treated as 100\% correlated between the different processes) and the scale factors. The large scale factor for $|V_{cd}|_{\Nf=2+1+1}$ is caused by the $D_s\to K\ell\nu$ result that has large uncertainty but also a considerably higher central value. Removing this result would change the average to  $|V_{cd}|_{\Nf=2+1+1}=0.2214(44)$.

Using the lattice determinations of $|V_{cd}|$ and $|V_{cs}|$ in
Eqs.~(\ref{eq:Vcdsfinal2p1}), (\ref{eq:Vcdsfinal2p1p1}) and $|V_{cb}|\approx 0.04$,
we can test the unitarity of the second row of the CKM matrix.  We obtain
\begin{alignat}{2}
\Nf=2+1+1 \text{: }& |V_{cd}|^2 + |V_{cs}|^2 + |V_{cb}|^2 - 1 = -0.01(2) &  \nonumber \\
&& \hskip -3.7cm
[\text{FLAG average, Refs.~\cite{Bazavov:2017lyh,Carrasco:2014poa,Chakraborty:2021qav,FermilabLattice:2022gku}}],\\  
\Nf=2+1   \text{: }& |V_{cd}|^2 + |V_{cs}|^2 + |V_{cb}|^2 - 1 = 0.00(3) &  \nonumber \\
&&  \hskip -3.7cm
[\text{FLAG average, Refs.~\cite{Bussone:2023kag, Bazavov:2011aa, Davies:2010ip, Na:2012iu, Yang:2014sea,Boyle:2017jwu, Na:2010uf, Na:2011mc, Meinel:2016dqj}}].
\end{alignat}

\clearpage

%% file: HQ/HQB.tex
\section{Bottom-hadron decays and mixings}
\label{sec:BDecays}
Authors: Y.~Aoki, M.~Della~Morte, E.~Lunghi, S.~Meinel, C.~Monahan, A.~Vaquero\\

Exclusive (semi)leptonic decays and mixing processes of $B_{(s)}$ mesons play
a crucial role in flavour physics.   In particular, they contain
important information for the investigation of the $b{-}d$ unitarity
triangle in the Cabibbo-Kobayashi-Maskawa (CKM) matrix, and provide
ideal probes of physics beyond the Standard Model.
The charged-current decay channels $B^{+} \rightarrow l^{+}
\nu_{l}$ and $B^{0} \rightarrow \pi^{-} l^{+} \nu_{l}$, where $l^{+}$
is a charged lepton with $\nu_{l}$ being the corresponding neutrino, are
essential in extracting the CKM matrix element $|V_{ub}|$.  Similarly,
the $B$ to $D^{(\ast)}$ semileptonic transitions can be used to
determine $|V_{cb}|$.   Flavour-changing neutral-current (FCNC)
processes, such as $B\to K^{(*)} \ell^+
\ell^-$ and $B_{d(s)} \to \ell^+ \ell^-$,  occur only beyond the tree level in weak interactions and are suppressed in the Standard
Model. Therefore, these processes could be sensitive to new
physics, since heavy particles can contribute to the loop diagrams.
FCNC processes are also suitable channels for the extraction of the CKM matrix
elements involving the top quark, which appears in loop contributions.
The decays $B\to D^{(*)}\ell\nu$ and $B\to K^{(*)} \ell\ell$ can also be used 
to test lepton flavour universality by comparing results for $\ell = e$, $\mu$ and $\tau$. 
In particular, anomalies have been seen in the ratios $R(D^{(*)}) = {\cal B} (B\to D^{(*)}\tau\nu) /{\cal B} (B\to D^{(*)}\ell\nu)_{\ell=e,\mu}$ and ${R}(K^{(*)}) = {\cal B} (B\to K^{(*)}\mu\mu) /{\cal B} (B\to K^{(*)}ee)$, although the latter are no longer statistically significant.
In addition, the neutral $B_{d(s)}$-meson mixings are FCNC processes and
are dominated by the 1-loop ``box'' diagrams containing the top quark
and the $W$ bosons.  Thus, using the experimentally measured neutral $B^0_{d(s)}$-meson oscillation
frequencies, $\Delta M_{d(s)}$, and the theoretical calculations for
the relevant hadronic mixing matrix elements, one can obtain
$|V_{td}|$ and $|V_{ts}|$ in the Standard Model.

At the Large Hadron Collider, decays of $b$ quarks can also be probed with $\Lambda_b$ and other bottom baryons, which
can provide complementary constraints on physics beyond the Standard Model. The most important processes are the charged-current
decays $\Lambda_b \to p \ell\bar{\nu}$ and $\Lambda_b \to \Lambda_c \ell\bar{\nu}$, and the neutral-current
decay $\Lambda_b \to \Lambda \ell^+\ell^-$.

Accommodating the light quarks and the $b$ quark simultaneously in
lattice-QCD computations is a challenging endeavour. To incorporate
the pion and the $b$ hadrons with their physical masses, the simulations have to be performed using
{a lattice that is large enough in physical units to accommodate light pions without significant finite-volume effects,
but, at the same time, fine enough to keep the heavy quark discretization errors under control.
This usually results in a} lattice size $\hat{L} = L/a \sim \cO(10^{2})$, where $a$ is the lattice spacing and $L$
is the physical (dimensionful) box size.   
The most ambitious calculations are now using such volumes; 
however, many ensembles are smaller.
Therefore, in addition to employing chiral perturbation theory for the extrapolations in the
light-quark mass, current lattice calculations for quantities involving
$b$ hadrons often make use of effective theories that allow one to
expand in inverse powers of $m_{b}$. In this regard, two general
approaches are widely adopted.  On the one hand, effective field theories
such as Heavy-Quark Effective Theory (HQET) and Nonrelativistic
QCD (NRQCD) can be directly implemented in numerical computations. On
the other hand, a relativistic quark action can be improved {\it \`{a} la}
Symanzik to suppress cutoff errors, and then re-interpreted in a manner
that is suitable for heavy-quark physics calculations.   
This latter strategy is often referred to as the method of the Relativistic
Heavy-Quark Action (RHQA).
The utilization of such effective theories inevitably introduces systematic
uncertainties that are not present in light-quark calculations.  These
uncertainties 
can arise from the truncation of the expansion in constructing the
effective theories (as in HQET and NRQCD),
or from more intricate
cutoff effects (as in NRQCD and RHQA).  They can also be introduced
through more complicated renormalization
procedures, which often lead to significant systematic effects in
matching the lattice operators to their continuum counterparts.  For
instance, due to the use of different actions for the heavy and the
light quarks, it is more difficult to construct absolutely 
normalized bottom-light currents.  

Complementary to the above ``effective theory approaches'', 
another popular method is to simulate the heavy and the light quarks
using the same (typically Symanzik-improved) lattice action at several values of
the heavy-quark mass $m_{h}$ with $a m_{h} < 1$ and $m_{h} < m_{b}$.   
This enables one to employ HQET-inspired relations to extrapolate the
computed quantities to the physical $b$ mass.  When combined with
results obtained in the static heavy-quark limit, this approach can be
rendered into an interpolation, instead of extrapolation, in
$m_{h}$. The discretization errors are the main source of the
systematic effects in this method, and very small lattice spacings are
needed to keep such errors under control.

In recent years, it has also been
possible to perform lattice simulations at very fine lattice
spacings and treat heavy quarks as
fully relativistic fermions without resorting to effective field
theories.  
Such simulations are, of course, very demanding in computing
resources.  

Because of the challenge described above, efforts to obtain reliable, accurate lattice-QCD results for the physics of the $b$ quark
have been enormous.   These efforts include significant theoretical progress in
formulating QCD with heavy quarks on the lattice. This aspect is
briefly reviewed in Appendix A.1.3 of FLAG 19 \cite{FlavourLatticeAveragingGroup:2019iem}.

In this section, we summarize the results of the $B$-meson leptonic
decay constants, the neutral $B$-mixing parameters, and the
semileptonic form factors of $B$ mesons and $\Lambda_b$ baryons, 
from lattice QCD.  To focus on the
calculations that have strong phenomenological impact, we limit the
review to results based on modern simulations containing dynamical
fermions with reasonably light pion masses (below
approximately 500~MeV).

For heavy-meson decay constants and mixing parameters, estimates
of the quantity $\delta(a_{\rm min})$ described in Sec.~\ref{sec:DataDriven} are provided, where possible,
for all computations entering the final FLAG averages or ranges.
For heavy-hadron semileptonic-decay form factors, implementing this
data-driven continuum-limit criterion was found to be not feasible.
The problem is that these quantities are functions of the momentum transfer in addition to the
other lattice parameters, and many calculations are based on global fits whose reconstruction
was not possible.

Following our review of $B_{(s)}$-meson
leptonic decay constants, the neutral $B$-meson mixing parameters, and
semileptonic form factors, we then interpret our results within the
context of the Standard Model.  We combine our best-determined values
of the hadronic matrix elements with the most recent
experimentally-measured branching fractions to obtain $|V_{ub}|$
and  $|V_{cb}|$,
and compare these results to those obtained from inclusive
semileptonic $B$ decays.

\input{HQ/HQSubsections/BL.tex}

\input{HQ/HQSubsections/BB.tex}

\input{HQ/HQSubsections/BSL.tex}

\input{HQ/HQSubsections/BCKM.tex}

%% file: HQ/HQSubsections/BL.tex
\subsection{Leptonic decay constants $f_B$ and $f_{B_s}$}
\label{sec:fB}
The $B$- and $B_s$-meson decay constants are crucial inputs for
extracting information from leptonic $B$ decays.  Charged $B$ mesons
can decay to a lepton-neutrino final state  through the
charged-current weak interaction.  On the other hand, neutral
$B_{d(s)}$ mesons can decay to a charged-lepton pair via a FCNC process.

In the Standard Model, the decay rate for $B_{(s)}^+ \to \ell^+ \nu_{\ell}$
is described by a formula identical to Eq.~(\ref{eq:Dtoellnu}), with $D_{(s)}$ replaced 
by $B_{(s)}$, $f_{D_{(s)}}$ replaced by $f_{B_{(s)}}$, and the 
relevant CKM matrix element $V_{cq}$ replaced by $V_{bq}$,
\be
\Gamma ( B_{(s)} \to \ell \nu_{\ell} ) =  \frac{ m_{B_{(s)}}}{8 \pi} G_F^2  f_{B_{(s)}}^2 |V_{bq}|^2 m_{\ell}^2 
           \left(1-\frac{ m_{\ell}^2}{m_{B_{(s)}}^2} \right)^2 \;. \label{eq:B_leptonic_rate}
\ee
The only two-body charged-current $B$-meson decay that has been observed so far is $B^{+} \to \tau^{+} \nu_{\tau}$, which has been measured by the Belle
and Babar collaborations~\cite{Lees:2012ju,Kronenbitter:2015kls}.
Both collaborations have reported results with errors around $20\%$. These measurements 
can be used to 
extract $|V_{ub}|$ when combined with lattice-QCD predictions of the corresponding
decay constant, but the experimental uncertainties currently preclude a precise determination.

Neutral $B_{d(s)}$-meson decays to a charged-lepton pair $B_{d(s)}
\rightarrow \ell^{+} \ell^{-}$ is a FCNC process, and can only occur at
1-loop in the Standard Model.  Hence these processes are expected to
be rare, and are sensitive to physics beyond the Standard Model.
The corresponding expression for the branching fraction has the form 
\be
B ( B_q \to \ell^+ \ell^-) = \tau_{B_q} \frac{G_F^2}{\pi} \, Y \,
\left(  \frac{\alpha_s}{4 \pi \sin^2 \Theta_W} \right)^2
m_{B_q} f_{B_q}^2 |V_{tb}^*V_{tq}|^2 m_{\ell}^2 
           \sqrt{1- 4 \frac{ m_{\ell}^2}{m_{B_q}^2} }\;, 
\ee
where the light quark $q=s$ or $d$, $\tau_{B_q}$ is the mean meson lifetime, and the function $Y$ includes NLO QCD and electro-weak
corrections that depend on the strong coupling $\alpha_s$
and the weak mixing angle $\Theta_W$ \cite{Inami:1980fz,Buchalla:1993bv}. 
Evidence for the $B_s \to \mu^+ 
\mu^-$ decay was first observed
by the LHCb \cite{Aaij:2012nna} and CMS collaborations, and a combined analysis was
presented in 2014 in Ref.~\cite{CMS:2014xfa}.  In 2020, the ATLAS~\cite{ATLAS:2018cur}, 
CMS~\cite{CMS:2019bbr} and LHCb~\cite{Aaij:2017vad} collaborations 
reported their measurements from a preliminary combined analysis as~\cite{ATLAS:2020acx}
\begin{eqnarray} 
   B(B \to \mu^+ \mu^-) &<& 1.9 \times \,10^{-10} \;\mathrm{at}\; 95\% \;\mathrm{CL}, 
\nonumber\\
   B(B_s \to \mu^+ \mu^-) &=& (2.69^{+0.37}_{-0.35}) \times \,10^{-9} ,
\label{eq:B_to_mumu_ATLAS_2020}
\end{eqnarray}
which are compatible with the Standard Model predictions within approximately 2 standard 
deviations~\cite{Beneke:2019slt}. More recently, updated observations have been reported 
by the LHCb collaboration~\cite{LHCb:2021awg} and the CMS collaboration~\cite{CMS:2022mgd}, but these results do not improve on 
the precision of the combined analysis.\footnote{The PDG quotes the branching fraction $B(B^0 \to \mu^+ \mu^-) < 1.5 \times \,10^{-10} \;\mathrm{at}\; 90\% \;\mathrm{CL}$~\cite{ParticleDataGroup:2024cfk}. Ref.~\cite{Altmannshofer:2021qrr} obtains $B(B^0 \to \mu^+ \mu^-) = (0.56\pm70) \times \,10^{-10}$ using a correlated global analysis.}
We note that the errors of these results are currently too large to enable a precise 
determination of $|V_{td}|$ and $|V_{ts}|$.

The related radiative leptonic decay, $B_s \to \mu^+ \mu^- \gamma$, is another FCNC 
process that is sensitive to new physics and is expected to occur at a comparable 
rate to $B_s \to \mu^+ \mu^-$. Recent searches for this decay by the LHCb collaboration 
found an upper limit of~\cite{LHCb:2021awg,LHCb:2021vsc}
\begin{equation}
B(B_s \to \mu^+ \mu^-\gamma ) < 2.0 \times \,10^{-9} \;\mathrm{at}\; 95\% \;\mathrm{CL},
\end{equation}
in the kinematic region $m_{\mu \mu} > 4.9$ GeV.
The dominant hadronic contributions are parameterized by local form factors and by 
nonlocal resonance contributions, which have been estimated using light-cone sum 
rules~\cite{Janowski:2021yvz}, QCD-inspired models~\cite{Kozachuk:2017mdk,Belov:2024vkv}, 
and from models of the transition form factors based on lattice calculations of the 
$D_s$ meson, assuming vector-meson dominance~\cite{Guadagnoli:2023zym}. The first 
lattice calculation of the local form factors were reported in \cite{Frezzotti:2024kqk}. 
The form factors provide a reasonable estimate of the decay rate for large di-muon 
invariant mass, $q^2 > (4.15\,\mathrm{GeV})^2$, where long-distance contributions 
are expected to be subdominant. Improved determinations of the branching fraction 
at lower di-muon invariant masses requires a systematic and quantitative treatment 
of the resonance region.

The rare leptonic $B^+\to \ell^+\nu_\ell\gamma$ decay is proportional to $|V_{ub}|^2$ and has been constrained by the CLEO \cite{CLEO:1996cze}, BaBar \cite{BaBar:2009pvj}, and Belle Collaborations \cite{Belle:2015mpp,Belle:2018jqd}. The most stringent constraint, in the region $E_\gamma > 1$ GeV, is \cite{Belle:2018jqd} 
\begin{equation}
B(B^+ \to \ell^+ \nu_\ell \gamma ) < 3.0 \times \,10^{-6} \;\mathrm{at}\; 90\% \;\mathrm{CL}.
\end{equation}
This branching fraction can be expressed in terms of form factors that
are yet to be directly determined on the lattice but have been
modelled using QCD sum rules and dispersive approaches combined with
an expansion in $\Lambda_{\mathrm{QCD}}/m_B$ and $\Lambda_{\mathrm{QCD}}/E_\gamma$ \cite{Beneke:2018wjp}. At leading
order in this expansion, the branching fraction depends only on the
light-cone distribution amplitude of the $B$ meson. At present, this
channel is primarily viewed as providing experimental constraints on the light-cone distribution amplitude. Direct calculations of this distribution amplitude from lattice QCD are now feasible with recent theoretical developments \cite{Zhao:2020bsx,Xu:2022krn} and, in combination with experimental data, would provide a novel method for the determination of $|V_{ub}|^2$.

The decay constants $f_{B_q}$ (with $q=u,d,s$) parameterize the matrix
elements of the corresponding axial-vector currents $A^{\mu}_{bq}
= \bar{b}\gamma^{\mu}\gamma^5q$ analogously to the definition of
$f_{D_q}$ in Sec.~\ref{sec:fD}:
\be
\langle 0| A^{\mu} | B_q(p) \rangle = i p_{B_q}^{\mu} f_{B_q} \;.
\label{eq:fB_from_ME}
\ee
For heavy-light mesons, it is convenient to define and analyse the quantity 
\be
 \Phi_{B_q} \equiv f_{B_q} \sqrt{m_{B_q}} \;,
\ee
which approaches a constant (up to logarithmic corrections) in the
$m_{B_q} \to \infty$ limit, because of heavy-quark symmetry.
In the following discussion, we denote lattice data for $\Phi$, and the corresponding 
decay constant $f$,
obtained at a heavy-quark mass $m_h$ and light valence-quark mass
$m_{\ell}$ as $\Phi_{h\ell}$ and $f_{hl}$, to differentiate them from
the corresponding quantities at the physical $b$ and light-quark
masses.

The SU(3)-breaking ratio $f_{B_s}/f_B$ is of phenomenological
interest, because many systematic effects can be partially reduced in lattice-QCD 
calculations of this ratio.  The discretization errors, heavy-quark-mass
tuning effects, and renormalization/matching errors may all be partially reduced. 

This SU(3)-breaking ratio is, however, still sensitive to the chiral
extrapolation. Provided the chiral extrapolation is under control,
one can then adopt $f_{B_s}/f_B$ as an input in extracting
phenomenologically-interesting quantities.  In addition, it often
happens to be easier to obtain lattice results for $f_{B_{s}}$ with
smaller errors than direct calculations of $f_{B}$.  Therefore, one can combine the 
$B_{s}$-meson
decay constant with the SU(3)-breaking ratio to calculate $f_{B}$.  Such
a strategy can lead to better precision in the computation of the
$B$-meson decay constant, and has been adopted by the
ETM~\cite{Carrasco:2013zta, Bussone:2016iua} and the
HPQCD collaborations~\cite{Na:2012kp}. An alternative strategy to the direct calculation of $f_{B_{s}}$, used in Ref.~\cite{Balasubramanian:2019net}, 
is to obtain the $B_s$-meson decay constant by combining the $D_{s}$-meson decay 
constant with the ratio $f_{B_s}/f_{D_s}$.

It is clear that the decay constants for charged and neutral $B$
mesons play different roles in flavour-physics phenomenology.  Knowledge of the $B^{+}$-meson 
decay constant
$f_{B^{+}}$ is essential for extracting $|V_{ub}|$ from
leptonic $B^{+}$ decays.   The neutral $B$-meson decay constants
$f_{B^{0}}$ and $f_{B_{s}}$ are inputs to searches for new physics in rare leptonic 
$B^{0}$
decays.  In
view of this, it is desirable to include isospin-breaking effects in
lattice computations for these quantities and to provide lattice
results for both $f_{B^{+}}$ and $f_{B^{0}}$.   
With the high precision of recent lattice calculations, isospin splittings 
for $B$-meson decay constants can be significant, 
and will play an important role in the foreseeable
future.   

A few collaborations have reported $f_{B^{+}}$ and $f_{B^{0}}$
separately by taking into account strong isospin effects in the
valence sector, and estimated the corrections from
electromagnetism~\cite{Bazavov:2017lyh,Bazavov:2011aa,Dowdall:2013tga,Christ:2014uea}. The $N_{f}=2+1+1$ strong isospin-breaking effect was
computed in HPQCD 13~\cite{Dowdall:2013tga} (see 
Tab.~\ref{tab:FBssumm} in this subsection).  However, since only
unitary points (with equal sea- and valence-quark masses) were considered in
HPQCD 13~\cite{Dowdall:2013tga}, this procedure only correctly accounts for the effect 
from the
valence-quark masses, while introducing a spurious sea-quark
contribution. The decay constants $f_{B^{+}}$ and $f_{B^{0}}$ are also
separately reported in FNAL/MILC 17~\cite{Bazavov:2017lyh} by
taking into account the strong-isospin effect.  The FNAL/MILC
results were obtained by keeping the averaged light sea-quark
mass fixed when varying the quark masses in their analysis procedure.
Their finding indicates that the strong isospin-breaking effects, $f_{B^+}-f_B\sim 
0.5$ MeV, could be smaller than those suggested by previous computations. One would 
have to take into
account QED effects in the $B$-meson leptonic decay rates to properly use these results 
for extracting phenomenologically relevant information.\footnote{See Ref.~\cite{Carrasco:2015xwa} 
for a strategy that
has been proposed to account for QED effects.}  Currently, errors on the
experimental measurements of these decay rates are still very large.   In this review, 
we
will therefore concentrate on the isospin-averaged result $f_{B}$ and the
$B_{s}$-meson decay constant, as well as the SU(3)-breaking ratio
$f_{B_{s}}/f_{B}$.

\begin{figure}[!htb]
\centering	
\includegraphics[width=0.48\linewidth]{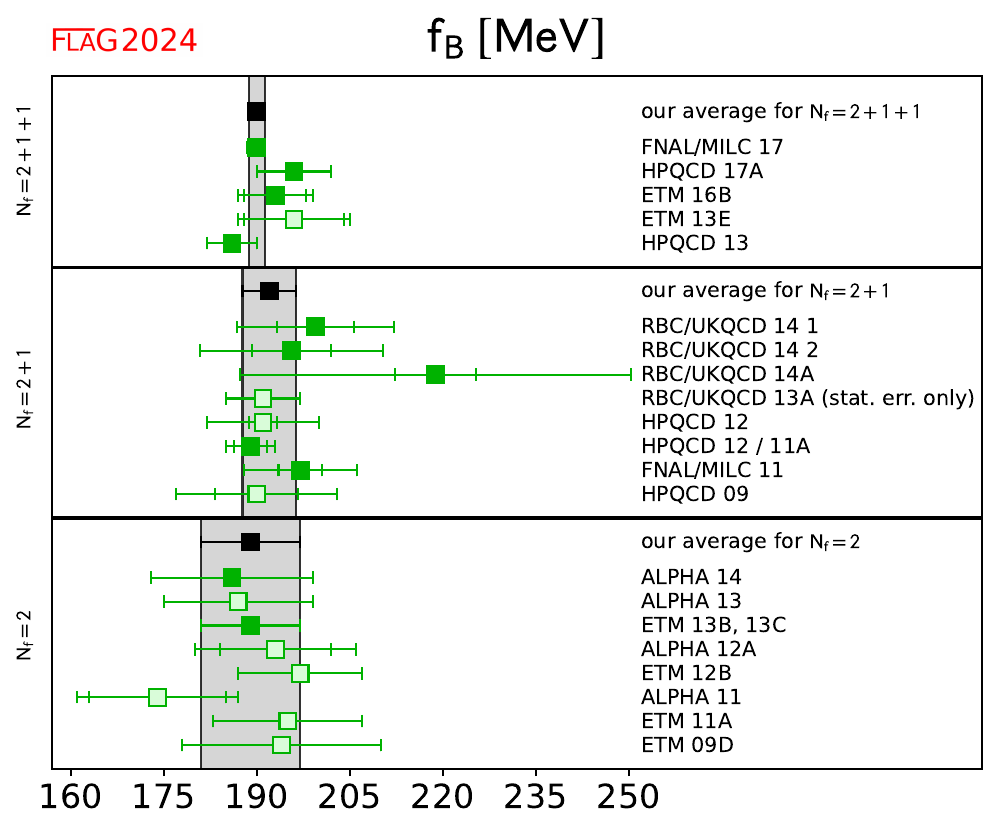}
\includegraphics[width=0.48\linewidth]{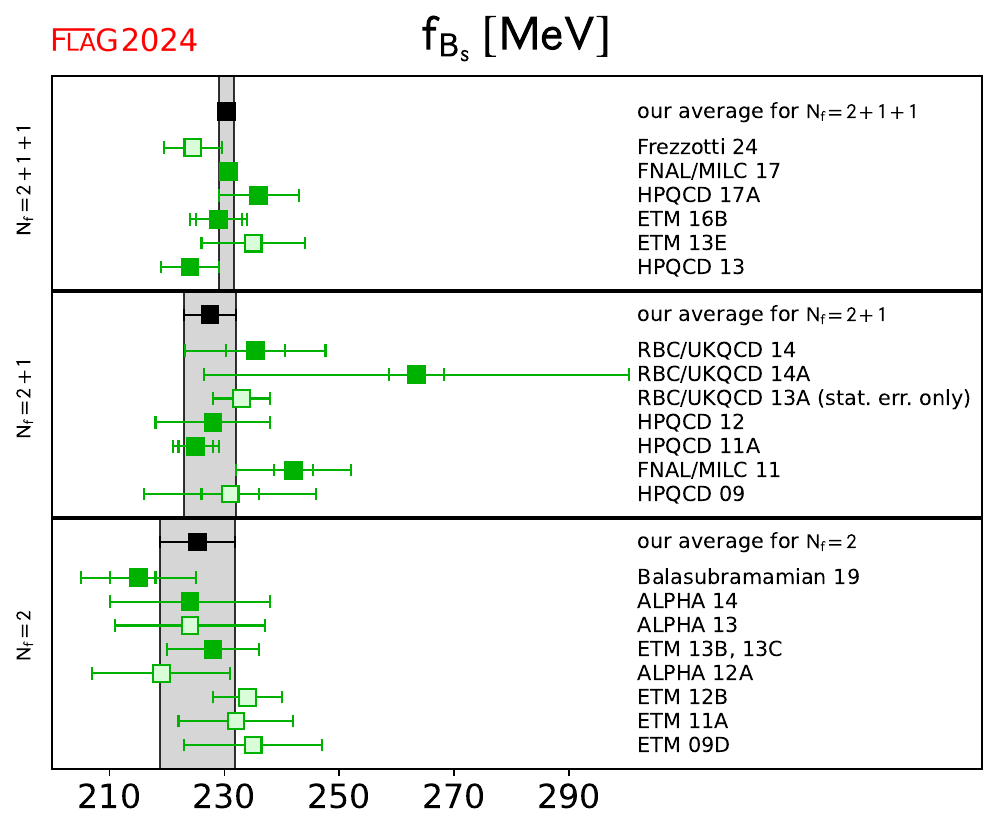}
 \vspace{-2mm}
\caption{Decay constants of the $B$ and $B_s$ mesons. The values are taken from Tab.~\ref{tab:FBssumm} 
(the $f_B$ entry for FNAL/MILC 11 represents $f_{B^+}$). The
significance of the colours is explained in Sec.~\ref{sec:qualcrit}.
The black squares and grey bands indicate
our averages in Eqs.~(\ref{eq:fB2}), (\ref{eq:fB21}),
(\ref{eq:fB211}), (\ref{eq:fBs2}), (\ref{eq:fBs21}) and
(\ref{eq:fBs211}).}
\label{fig:fB}
\end{figure}

\begin{figure}[!htb]
\begin{center}
\includegraphics[width=0.7\linewidth]{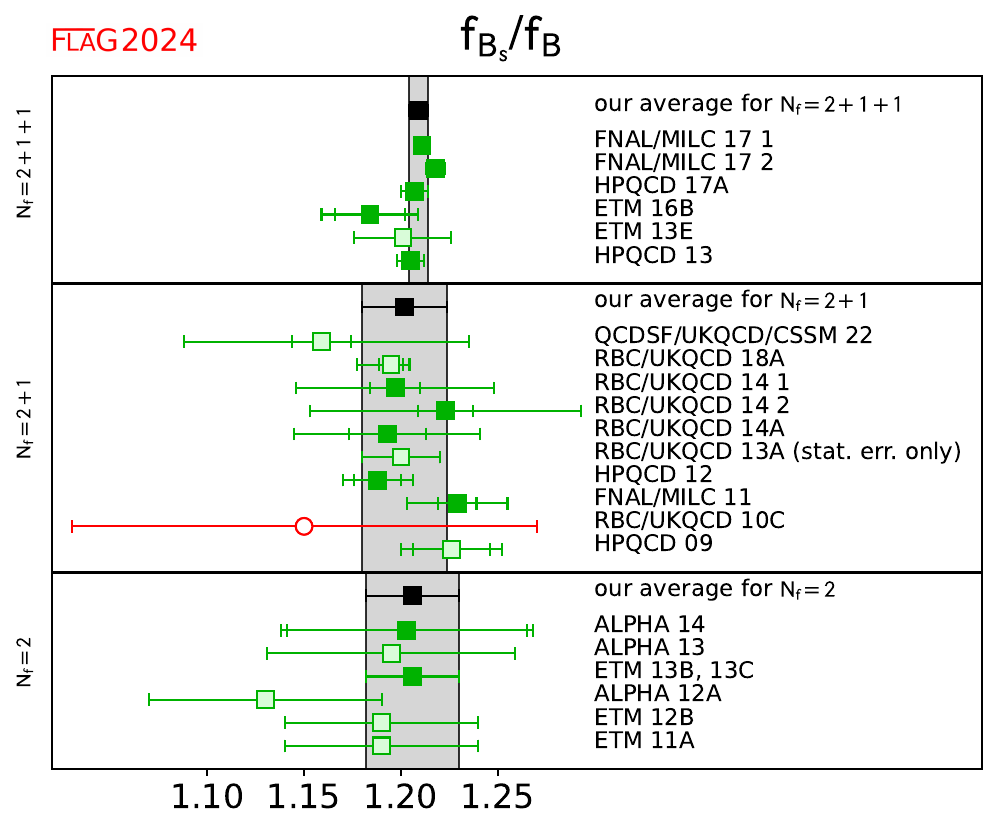}
\vspace{-2mm}
\caption{Ratio of the decay constants of the $B$ and $B_s$ mesons. The
  values are taken from Tab.~\ref{tab:FBratsumm}.  Results labelled
  as FNAL/MILC 17 1 and FNAL/MILC 17 2 correspond to
  those for $f_{B_{s}}/f_{B^{0}}$ and $f_{B_{s}}/f_{B^{+}}$ reported in FNAL/MILC
  17.  The
significance of the colours is explained in
Sec.~\ref{sec:qualcrit}.
The black squares and grey bands indicate
our averages in Eqs.~(\ref{eq:fBratio2}), (\ref{eq:fBratio21}), and
(\ref{eq:fBratio211}).}
\label{fig:fBratio}
\end{center}
\end{figure}

The status of lattice-QCD computations for $B$-meson decay constants
and the SU(3)-breaking ratio, using gauge-field ensembles
with light dynamical fermions, is summarized in Tabs.~\ref{tab:FBssumm}
and~\ref{tab:FBratsumm}. Figs.~\ref{fig:fB} and~\ref{fig:fBratio} contain the graphical
presentation of the collected results and our averages. Most results in these tables 
and plots have been reviewed in detail in
FLAG 19~\cite{FlavourLatticeAveragingGroup:2019iem} and in FLAG 21~\cite{FlavourLatticeAveragingGroupFLAG:2021npn}. 
Here, we describe the new results that have appeared since January 2021.

We also review the continuum-limit quantity, $\delta(a_{\rm min})$, described in 
Sec.~\ref{sec:qualcrit}. We estimate, where possible, $\delta(a_{\rm min})$ for results 
entering the FLAG averages of $f_{B}$, $f_{B_s}$, and $f_{B_s}/f_B$, but we do not 
use $\delta(a_{\rm min})$ for averaging. We include estimates of $\delta(a_{\rm min})$ 
for those calculations that explicitly provide the relevant data in the manuscript.

As lattice calculations of leptonic decays have become statistically more precise, 
results are often dominated by systematic uncertainties. The continuum extrapolation 
is frequently the largest source of systematic uncertainty for lattice calculations 
of heavy quarks, for which the heavy-quark discretization can introduce effects of 
the ${\cal O}(am)^n$, and a more quantitative measure of discretization effects is 
a useful guide to the quality of the continuum extrapolation. For the lattice calculations 
of leptonic decay constants of bottom hadrons that appear in this review, the continuum-limit 
quantity should be interpreted with caution, because many final results are quoted 
from combined chiral-continuum extrapolations and, typically, more recent computations 
do not quote numerical values for the leptonic decay constants at the finest lattice 
spacings. Moreover, the finest ensembles may not be at, or close to, the physical 
pion mass. Thus, we generally quote our estimations of $\delta(a_{\rm min})$ to one 
significant figure because the natural size of the uncertainty on $\delta(a_{\rm 
min})$ is ${\mathcal O}(1)$. 
\begin{table}[!htb]
\mbox{} \\[3.0cm]
\footnotesize
\begin{tabular*}{\textwidth}[l]{@{\extracolsep{\fill}}l@{\hspace{1mm}}r@{\hspace{1mm}}
	l@{\hspace{1mm}}l@{\hspace{1mm}}l@{\hspace{1mm}}l@{\hspace{1mm}}l@{\hspace{1mm}}
	l@{\hspace{1mm}}l@{\hspace{5mm}}l@{\hspace{1mm}}l@{\hspace{1mm}}l@{\hspace{1mm}}l@{\hspace{1mm}}l}
Collaboration & Ref. & $\Nf$ & 
\hspace{0.15cm}\begin{rotate}{60}{publication status}\end{rotate}\hspace{-0.15cm} 
&
\hspace{0.15cm}\begin{rotate}{60}{continuum extrapolation}\end{rotate}\hspace{-0.15cm} 
&
\hspace{0.15cm}\begin{rotate}{60}{chiral extrapolation}\end{rotate}\hspace{-0.15cm}&
\hspace{0.15cm}\begin{rotate}{60}{finite volume}\end{rotate}\hspace{-0.15cm}&
\hspace{0.15cm}\begin{rotate}{60}{renormalization/matching}\end{rotate}\hspace{-0.15cm} 
 &
\hspace{0.15cm}\begin{rotate}{60}{heavy-quark treatment}\end{rotate}\hspace{-0.15cm} 
& 
 $f_{B^+}$ & $f_{B^0}$   & $f_{B}$ & $f_{B_s}$  \\
&&&&&&&&&&&&\\[-0.1cm]
\hline
\hline
&&&&&&&&&&&& \\[-0.1cm]

Frezzotti 24  & \cite{Frezzotti:2024kqk} & 2+1+1 & \oP & \good & \good & \good
& \good & \okay & $-$ & $-$ & $-$ & 224.5(5.0) \\[0.5ex]

FNAL/MILC 17  & \cite{Bazavov:2017lyh} & 2+1+1 & \gA & \good & \good & \good 
& \good &  \okay &  189.4(1.4) & 190.5(1.3) & 189.9(1.4) & 230.7(1.2) \\[0.5ex]

HPQCD 17A & \cite{Hughes:2017spc} & 2+1+1 & \gA & \soso & \good & \good 
& \soso &  \okay & $-$ & $-$ & 196(6) & 236(7) \\[0.5ex]

ETM 16B & \cite{Bussone:2016iua} & 2+1+1 & \gA & \good & \soso & \soso 
& \soso &  \okay & $-$ & $-$ & 193(6) & 229(5) \\[0.5ex]

ETM 13E & \cite{Carrasco:2013naa} & 2+1+1 & \rC & \good & \soso & \soso 
& \soso &  \okay & $-$ & $-$ & 196(9) & 235(9) \\[0.5ex]

HPQCD 13 & \cite{Dowdall:2013tga} & 2+1+1 & \gA & \soso & \good & \good & \soso
& \okay &  184(4) & 188(4) &186(4) & 224(5)  \\[0.5ex]

&&&&&&&&&& \\[-0.1cm]
\hline
&&&&&&&&&& \\[-0.1cm]

RBC/UKQCD 14 & \cite{Christ:2014uea} & 2+1 & \gA & \soso & \soso & \soso 
  & \soso & \okay & 195.6(14.9) & 199.5(12.6) & $-$ & 235.4(12.2) \\[0.5ex]

RBC/UKQCD 14A & \cite{Aoki:2014nga} & 2+1 & \gA & \soso & \soso & \soso 
  & \soso & \okay & $-$ & $-$ & 219(31) & 264(37) \\[0.5ex]

RBC/UKQCD 13A & \cite{Witzel:2013sla} & 2+1 & \rC & \soso & \soso & \soso 
  & \soso & \okay & $-$ & $-$ &  191(6)$_{\rm stat}^\diamond$ & 233(5)$_{\rm stat}^\diamond$ 
\\[0.5ex]

HPQCD 12 & \cite{Na:2012kp} & 2+1 & \gA & \soso & \soso & \soso & \soso
& \okay & $-$ & $-$ & 191(9) & 228(10)  \\[0.5ex]

HPQCD 12 & \cite{Na:2012kp} & 2+1 & \gA & \soso & \soso & \soso & \soso
& \okay & $-$ & $-$ & 189(4)$^\triangle$ &  $-$  \\[0.5ex]

HPQCD 11A & \cite{McNeile:2011ng} & 2+1 & \gA & \good & \soso &
 \good & \good & \okay & $-$ & $-$ & $-$ & 225(4)$^\nabla$ \\[0.5ex] 

FNAL/MILC 11 & \cite{Bazavov:2011aa} & 2+1 & \gA & \soso & \soso &
     \good & \soso & \okay & 197(9) & $-$ & $-$ & 242(10) &  \\[0.5ex]  

HPQCD 09 & \cite{Gamiz:2009ku} & 2+1 & \gA & \soso & \soso & \soso &
\soso & \okay & $-$ & $-$ & 190(13)$^\bullet$ & 231(15)$^\bullet$  \\[0.5ex] 

&&&&&&&&&& \\[-0.1cm]
\hline
&&&&&&&&&& \\[-0.1cm]

Balasubramamian 19$^\dagger$ & \cite{Balasubramanian:2019net} & 2 & \gA & \good & 
\good & \good & \soso & \okay & $-$ & $-$ & $-$ & 215(10)(2)({\raisebox{0.5ex}{\tiny$\substack{+2 
\\ -5}$}})\\[0.5ex]

ALPHA 14 & \cite{Bernardoni:2014fva} & 2 & \gA & \good & \good &\good 
& \good & \okay &  $-$ & $-$ & 186(13) & 224(14) \\[0.5ex]

ALPHA 13 & \cite{Bernardoni:2013oda} & 2 & \rC  & \good   & \good   &
\good    &\good  & \okay   & $-$ & $-$ & 187(12)(2) &  224(13) &  \\[0.5ex] 

ETM 13B, 13C$^\ddagger$ & \cite{Carrasco:2013zta,Carrasco:2013iba} & 2 & \gA & \good 
& \soso & \good
& \soso &  \okay &  $-$ & $-$ & 189(8) & 228(8) \\[0.5ex]

ALPHA 12A& \cite{Bernardoni:2012ti} & 2 & \rC  & \good      & \good      &
\good          &\good  & \okay   & $-$ & $-$ & 193(9)(4) &  219(12) &  \\[0.5ex]

ETM 12B & \cite{Carrasco:2012de} & 2 & \rC & \good & \soso & \good
& \soso &  \okay &  $-$ & $-$ & 197(10) & 234(6) \\[0.5ex]

ALPHA 11& \cite{Blossier:2011dk} & 2 & \rC  & \good      & \soso      &
\good          &\good  & \okay  & $-$ & $-$ & 174(11)(2) &  $-$ &  \\[0.5ex]  

ETM 11A & \cite{Dimopoulos:2011gx} & 2 & \gA & \good & \soso & \good
& \soso &  \okay & $-$ & $-$ & 195(12) & 232(10) \\[0.5ex]

ETM 09D & \cite{Blossier:2009hg} & 2 & \gA & \good & \soso & \soso
& \soso &  \okay & $-$ & $-$ & 194(16) & 235(12) \\[0.5ex]
&&&&&&&&&& \\[-0.1cm]
\hline
\hline
\end{tabular*}
\begin{tabular*}{\textwidth}[l]{l@{\extracolsep{\fill}}lllllllll}
  \multicolumn{10}{l}{\vbox{\begin{flushleft} 
	$^\diamond$Statistical errors only. \\
        $^\triangle$Obtained by combining $f_{B_s}$ from HPQCD 11A with $f_{B_s}/f_B$ 
calculated in this work.\\
        $^\nabla$This result uses one ensemble per lattice spacing with light to 
strange sea-quark mass 
        ratio $m_{\ell}/m_s \approx 0.2$. \\
        $^\bullet$This result uses an old determination of  $r_1=0.321(5)$ fm from 
Ref.~\cite{Gray:2005ur} that 
        has since been superseded. \\
        $^\dagger$Obtained by combining $f_{D_s}$, updated in this work, with $f_{B_s}/f_{D_s}$, 
 calculated in this work.\\
        $^\ddagger$Update of ETM 11A and 12B. 
\end{flushleft}}}
\end{tabular*}
\vspace{-0.5cm}
\caption{Decay constants of the $B$, $B^+$, $B^0$ and $B_{s}$ mesons
  (in MeV). Here $f_B$ stands for the mean value of $f_{B^+}$ and
  $f_{B^0}$, extrapolated (or interpolated) in the mass of the light
  valence-quark to the physical value of $m_{ud}$.}
\label{tab:FBssumm}
\end{table}

\begin{table}[!htb]
\begin{center}
\mbox{} \\[3.0cm]
\footnotesize
\begin{tabular*}{\textwidth}[l]{@{\extracolsep{\fill}}l@{\hspace{1mm}}r@{\hspace{1mm}}
	l@{\hspace{1mm}}l@{\hspace{1mm}}l@{\hspace{1mm}}l@{\hspace{1mm}}l@{\hspace{1mm}}
	l@{\hspace{1mm}}l@{\hspace{5mm}}l@{\hspace{1mm}}l@{\hspace{1mm}}l@{\hspace{1mm}}l}
Collaboration & Ref. & $\Nf$ & 
\hspace{0.15cm}\begin{rotate}{60}{publication status}\end{rotate}\hspace{-0.15cm} 
&
\hspace{0.15cm}\begin{rotate}{60}{continuum extrapolation}\end{rotate}\hspace{-0.15cm} 
&
\hspace{0.15cm}\begin{rotate}{60}{chiral extrapolation}\end{rotate}\hspace{-0.15cm}&
\hspace{0.15cm}\begin{rotate}{60}{finite volume}\end{rotate}\hspace{-0.15cm}&
\hspace{0.15cm}\begin{rotate}{60}{renormalization/matching}\end{rotate}\hspace{-0.15cm} 
 &
\hspace{0.15cm}\begin{rotate}{60}{heavy-quark treatment}\end{rotate}\hspace{-0.15cm} 
& 
 $f_{B_s}/f_{B^+}$  & $f_{B_s}/f_{B^0}$  & $f_{B_s}/f_{B}$  \\
&&&&&&&&&& \\[-0.1cm]
\hline
\hline
&&&&&&&&&& \\[-0.1cm]

FNAL/MILC 17  & \cite{Bazavov:2017lyh} & 2+1+1 & \gA & \good & \good
                                                                                 
  & \good 
& \good &  \okay &  1.2180(49) & 1.2109(41) & $-$ \\[0.5ex]

HPQCD 17A & \cite{Hughes:2017spc} & 2+1+1 & \gA & \soso & \good & \good 
& \soso &  \okay &  $-$ & $-$ & 1.207(7) \\[0.5ex]

ETM 16B & \cite{Bussone:2016iua} & 2+1+1 & \gA & \good & \soso & \soso 
& \soso &  \okay &  $-$ & $-$& 1.184(25) \\[0.5ex]

ETM 13E & \cite{Carrasco:2013naa} & 2+1+1 & \rC & \good & \soso & \soso
& \soso &  \okay &  $-$ & $-$ & 1.201(25) \\[0.5ex]

HPQCD 13 & \cite{Dowdall:2013tga} & 2+1+1 & \gA & \soso & \good & \good & \soso
& \okay & 1.217(8) & 1.194(7) & 1.205(7)  \\[0.5ex]

&&&&&&&&&& \\[-0.1cm]
\hline
&&&&&&&&&& \\[-0.1cm]

QCDSF/UKQCD/CSSM 22 & \cite{Hollitt:2022exk} & 2+1 & \rC & \good & \good & \soso 
& \soso & \okay & $-$ & $-$ & 1.159(15)({\raisebox{0.5ex}{\tiny$\substack{+76 \\ 
-71}$}}) \\[0.5ex]

RBC/UKQCD 18A & \cite{Boyle:2018knm} & 2+1 & \oP & \good & \good & \good & \good 
& \okay & $-$ & $-$ & 1.1949(60)({\raisebox{0.5ex}{\tiny$\substack{+95 \\ -175}$}})
\\[0.5ex]

RBC/UKQCD 14 & \cite{Christ:2014uea} & 2+1 & \gA & \soso & \soso & \soso 
  & \soso & \okay & 1.223(71) & 1.197(50) & $-$ \\[0.5ex]

RBC/UKQCD 14A & \cite{Aoki:2014nga} & 2+1 & \gA & \soso & \soso & \soso 
  & \soso & \okay & $-$ & $-$ & 1.193(48) \\[0.5ex]

RBC/UKQCD 13A & \cite{Witzel:2013sla} & 2+1 & \rC & \soso & \soso & \soso 
  & \soso & \okay & $-$ & $-$ &  1.20(2)$_{\rm stat}^\diamond$ \\[0.5ex]

HPQCD 12 & \cite{Na:2012kp} & 2+1 & \gA & \soso & \soso & \soso & \soso
& \okay & $-$ & $-$ & 1.188(18) \\[0.5ex]

FNAL/MILC 11 & \cite{Bazavov:2011aa} & 2+1 & \gA & \soso & \soso &
     \good& \soso & \okay & 1.229(26) & $-$ & $-$ \\[0.5ex]  
     
RBC/UKQCD 10C & \cite{Albertus:2010nm} & 2+1 & \gA & \tbr & \tbr & \tbr 
  & \soso & \okay & $-$ & $-$ & 1.15(12) \\[0.5ex]

HPQCD 09 & \cite{Gamiz:2009ku} & 2+1 & \gA & \soso & \soso & \soso &
\soso & \okay & $-$ & $-$ & 1.226(26)  \\[0.5ex] 

&&&&&&&&&& \\[-0.1cm]
\hline
&&&&&&&&&& \\[-0.1cm]
ALPHA 14 \al \cite{Bernardoni:2014fva} & 2 & \gA & \good & \good & \good 
& \good &  \okay &  $-$ \al $-$ & 1.203(65)\\[0.5ex]

ALPHA 13 & \cite{Bernardoni:2013oda} & 2 & \rC  & \good  & \good  &
\good   &\good  & \okay   & $-$ & $-$ & 1.195(61)(20)  &  \\[0.5ex] 

ETM 13B, 13C$^\dagger$ & \cite{Carrasco:2013zta,Carrasco:2013iba} & 2 & \gA & \good 
& \soso & \good
& \soso &  \okay &  $-$ & $-$ & 1.206(24)  \\[0.5ex]

ALPHA 12A & \cite{Bernardoni:2012ti} & 2 & \rC & \good & \good & \good
& \good &  \okay & $-$ & $-$ & 1.13(6)  \\ [0.5ex]

ETM 12B & \cite{Carrasco:2012de} & 2 & \rC & \good & \soso & \good
& \soso &  \okay & $-$ & $-$ & 1.19(5) \\ [0.5ex]

ETM 11A & \cite{Dimopoulos:2011gx} & 2 & \gA & \soso & \soso & \good
& \soso &  \okay & $-$ & $-$ & 1.19(5) \\ [0.5ex]
&&&&&&&&&& \\[-0.1cm]
\hline
\hline
\end{tabular*}
\begin{tabular*}{\textwidth}[l]{l@{\extracolsep{\fill}}lllllllll}
  \multicolumn{10}{l}{\vbox{\begin{flushleft}
 	 $^\diamond$Statistical errors only. \\
          $^\dagger$Update of ETM 11A and 12B. 
\end{flushleft}}}
\end{tabular*}
\vspace{-0.5cm}
\caption{Ratios of decay constants of the $B$ and $B_s$ mesons (for details see Tab.~\ref{tab:FBssumm}).}
\label{tab:FBratsumm}
\end{center}
\end{table}

There have been no new $\Nf=2$ calculations of $f_B$, $f_{B_s}$, or $f_{B_{s}}/f_B$. 
Therefore, our averages for these quantities stay the same as those in FLAG 21~\cite{FlavourLatticeAveragingGroupFLAG:2021npn}. 
Our estimates for the continuum-limit quantity $\delta(a_{\rm min})$ are $\delta(a_{\rm 
min}) = 0.01$ for $f_{B_s}$ in Ref.~\cite{Carrasco:2013zta}. Data do not permit 
estimates of the continuum-limit quantity for $f_B$ and $f_{B_s}/f_B$ from Ref.~\cite{Carrasco:2013zta}, 
but discretization effects are generally small. 
From Ref.~\cite{Bernardoni:2014fva} we obtain $\delta(a_{\rm min}) = 0.6$ for $f_B$, 
$\delta(a_{\rm min}) = 0.3$ for $f_{B_s}$, and $\delta(a_{\rm min}) = 0.3$ for $f_{B_{s}}/f_B$. 
Finally, $\delta(a_{\rm min}) = 2.6$ for $f_{B_s}$ in \cite{Balasubramanian:2019net}. 

Our averages of the $\Nf = 2$ results are:
%
\begin{align}
&\label{eq:fB2}
\Nf=2:&\FLAGAVBEGIN f_{B} &= 188(7) \FLAGAVEND\;{\rm MeV}
&&\Refs~\mbox{\cite{Carrasco:2013zta,Bernardoni:2014fva}},\\
&\label{eq:fBs2}
\Nf=2: &\FLAGAVBEGIN f_{B_{s}} &= 225.3(6.6)\FLAGAVEND \; {\rm MeV} 
&&\Refs~\mbox{\cite{Carrasco:2013zta,Bernardoni:2014fva,Balasubramanian:2019net}}, 
\\
&\label{eq:fBratio2}
\Nf=2: &\FLAGAVBEGIN f_{B_{s}}\over{f_B} &= 1.206(0.023)\FLAGAVEND
&&\Refs~\mbox{\cite{Carrasco:2013zta,Bernardoni:2014fva}}.
\end{align}

Two new $\Nf=2+1$ calculations of $f_{B_{s}}/f_B$ were presented in conference proceedings 
after the publication of FLAG 21~\cite{FlavourLatticeAveragingGroupFLAG:2021npn}. 
Only one of these calculations, Ref.~\cite{Hollitt:2022exk}, provides a preliminary 
quantitative result. In Tab.~\ref{tab:FBratsumm}, this result is labelled QCDSF/UKQCD/CSSM 
22~\cite{Hollitt:2022exk}. The second work, Ref.~\cite{Black:2022eph}, is described 
in the text below, but not listed in Tab.~\ref{tab:FBratsumm}.

In QCDSF/UKQCD/CSSM~22~\cite{Hollitt:2022exk} the QCDSF/UKQCD/CSSM collaboration presented  
the ratio of
decay constants, $f_{B_{s}}/f_B$, using $\Nf=2+1$ dynamical ensembles generated using 
nonperturbatively ${\cal O}(a)$-improved clover-Wilson fermions. Four lattice spacings, 
of $a=0.082$, 0.074, 0.068, and 0.059 fm, were used, with pion masses ranging from 
155 to 468~MeV, and lattice sizes between 2.37 and 4.35 fm. The light-quark masses 
were tuned using the QCDSF procedure~\cite{Bietenholz:2010jr}, for fixing the light- 
and strange-quark masses. Quark masses were chosen to keep the value of the SU(3) 
flavour-singlet mass, $\overline{m} = (2m_\ell+m_s)/3$, constant. Heavy quarks were 
simulated with a relativistic heavy-quark (RHQ) action, with bare-quark masses chosen 
to keep the SU(3) flavour-singlet mass, $X_B^2 = (2M_{B_\ell}+M_{B_s})/3$, constant. 
The bare parameters of the RHQ action were chosen to ensure that the masses and hyperfine 
splitting of the $X_B$ and $X_{B^\ast}$ mesons reproduce the properties of the physical, 
spin-averaged $X_B$ and $X_{B^\ast}$~\cite{Aoki:2012xaa}.

The chiral extrapolation was performed using both linear and quadratic terms in $(M_\pi^2/M_X^2 
- 1)$ and assuming that the SU(3) flavour breaking does not depend on the lattice 
spacing. The reported value for the ratio of decay constants assumes that the renormalization 
parameters for light- and strange-quark currents are approximately equal, but this 
is only true near the SU(3)-symmetric point. Effects of the order of 1-2\% are 
expected near the physical point and calculations of the relevant parameters on near-physical 
ensembles are underway. Tests of ${\cal O}(a^2)$ discretization effects indicate 
little dependence and the final results are quoted from the subset of ensembles with 
$m_\pi L > 4$ and assuming no dependence on $a^2$. Tests of heavy-quark mistuning 
effects indicate that the ratio of decay constants are minimally affected.

The RBC/UKQCD collaboration described ongoing efforts to calculate pseudoscalar and 
vector heavy-meson decay constants in Ref.~\cite{Black:2022eph}, using $\Nf=2+1$ 
dynamical ensembles generated using Domain Wall Fermions (DWF). Four lattice spacings, 
of $a = 0.11$, 0.083, 0.071, and 0.063 fm were used,
with pion masses ranging from 267 to 433 MeV, and lattice sizes between 2.0 and 3.4 
fm.
Light and strange quarks were simulated with the Shamir DWF discretization and charm 
quarks were simulated with M{\"o}bius DWF action. These discretizations correspond 
to two different choices for the
DWF kernel. The M{\"o}bius DWF are loosely equivalent to Shamir DWF at twice the 
extension
in the fifth dimensions~\cite{Blum:2014tka}. Ref.~\cite{Black:2022eph} presents a preliminary 
analysis with a two-step procedure. The first step corrects for strange-quark-mass mistunings 
and the second applies NLO SU(2) heavy-meson chiral perturbation theory to carry 
out a chiral-continuum extrapolation using various fit Ans{\"a}tze to enable a full 
systematic error analysis. This analysis is ongoing at time of publication.

The results of Refs.~\cite{Hollitt:2022exk} and~\cite{Black:2022eph} have not been 
published and therefore neither calculation is included in our average. Thus, our 
averages remain the same as in FLAG~21~\cite{FlavourLatticeAveragingGroupFLAG:2021npn},
%
\begin{align}
&\label{eq:fB21}
\Nf=2+1:&\FLAGAVBEGIN f_{B} &= 192.0(4.3) \FLAGAVEND\;{\rm MeV}
&&\Refs~\mbox{\cite{Bazavov:2011aa,McNeile:2011ng,Na:2012kp,Aoki:2014nga,Christ:2014uea}},\\
&\label{eq:fBs21}
\Nf=2+1: &\FLAGAVBEGIN f_{B_{s}} &= 228.4(3.7)\FLAGAVEND \; {\rm MeV} 
&&\Refs~\mbox{\cite{Bazavov:2011aa,McNeile:2011ng,Na:2012kp,Aoki:2014nga,Christ:2014uea}}, 
\\
&\label{eq:fBratio21}
\Nf=2+1: &\FLAGAVBEGIN f_{B_{s}}\over{f_B} &= 1.201(0.016)\FLAGAVEND
&&\Refs~\mbox{\cite{Bazavov:2011aa,Na:2012kp,Aoki:2014nga,Christ:2014uea,Boyle:2018knm}}.
\end{align}

Our estimates for the continuum-limit quantity $\delta(a_{\rm min})$ for the results 
entering the FLAG averages for the $\Nf = 2+1$ bottom-hadron leptonic decay constants, 
and their ratio, are: $\delta(a_{\rm min}) = 5.6$ and $\delta(a_{\rm min}) = 7.4$ 
for $f_{B_s}$ and $f_B$, respectively, in Ref.~\cite{Bazavov:2011aa}; $\delta(a_{\rm 
min}) = 1.5$ for $f_B$ in Ref.~\cite{McNeile:2011ng}; $\delta(a_{\rm min}) = 0.01$ 
and $\delta(a_{\rm min}) = 0.6$ for $f_{B_s}$ and $f_B$, respectively, in Ref.~\cite{Na:2012kp}; 
$\delta(a_{\rm min}) = 1.9$ and $\delta(a_{\rm min}) = 2.3$ for $f_{B_s}$ and $f_B$, 
respectively, in Ref.~\cite{Aoki:2014nga}; and $\delta(a_{\rm min}) = 1.7$ for $f_{B_s}$ 
in Ref.~\cite{Christ:2014uea}. For $f_{B_s}/f_B$ we obtain approximately $\delta(a_{\rm min}) = 0.4$ for \cite{Bazavov:2011aa}, 
approximately 2 for \cite{Na:2012kp} and \cite{Aoki:2014nga}, 3 for \cite{Christ:2014uea}, and around 0.5 for \cite{Boyle:2018knm}.

No new $N_{f}=2+1+1$ calculations of $f_{B}$ and $f_{B_{s}}/f_{B}$ have appeared 
since FLAG 21. There has been one new calculation of $f_{B_{(s)}}$ in 
Ref.~\cite{Frezzotti:2024kqk}, labelled Frezzotti 24 in Tab.~\ref{tab:FBssumm}.

As part of the determination of the form factors for the radiative leptonic decay 
$B_s\to\mu^+\mu^-\gamma$, the decay constant $f_{B_s}$ was determined in Ref.~\cite{Frezzotti:2024kqk}. 
This work used ensembles with $\Nf=2+1+1$ clover-Wilson twisted-mass fermions at 
maximal twist. Four lattice spacings, ranging from $0.057$ to $0.091$ fm, were included 
and pion masses spanned a range from 137 to 175 MeV. The heavy-strange meson was 
simulated using  clover-Wilson twisted-mass fermions at a range of heavy-strange 
masses, extrapolated up to the physical $B_s$ mass. Ref.~\cite{Frezzotti:2024kqk} 
determined $f_{H_s}$ from both two-point functions and the spatial part of the axial 
hadronic tensor to better constrain the continuum limit because these determinations 
differ only by discretization effects. The results from both methods were simultaneously 
extrapolated to the continuum limit at fixed values of the heavy-strange meson mass 
$M_{H_s}$, with six different fit variations for each of the five values of $M_{H_s}$. 
The results of each fit were combined using the Akaike Information Criterion~\cite{Akaike:1974vps} 
and the corresponding continuum decay constants were then extrapolated to the physical 
$B_s$ mass. The extrapolation in the heavy-strange mass was carried out using a fit 
form guided by HQET, with modifications to account for the anomalous dimension of 
the axial current in HQET and and the matching between QCD and HQET. 

Ref.~\cite{Frezzotti:2024kqk} has not been published at the time of publication of 
this review.
Therefore, our averages for $f_{B}$, $f_{B_{(s)}}$ and $f_{B_{s}}/f_{B}$ remain the 
same as in FLAG 21~\cite{FlavourLatticeAveragingGroupFLAG:2021npn},
%
\begin{align}
&\label{eq:fB211}
\Nf=2+1+1:&\FLAGAVBEGIN f_{B} &= 190.0(1.3) \FLAGAVEND\;{\rm MeV}
&&\Refs~\mbox{\cite{Dowdall:2013tga,Bussone:2016iua,Hughes:2017spc,Bazavov:2017lyh}},\\
&\label{eq:fBs211}
\Nf=2+1+1: &\FLAGAVBEGIN f_{B_{s}} &= 230.3(1.3)  \FLAGAVEND\; {\rm MeV}
&&\Refs~\mbox{\cite{Dowdall:2013tga,Bussone:2016iua,Hughes:2017spc,Bazavov:2017lyh}}, 
\\
&\label{eq:fBratio211}
\Nf=2+1+1: &\FLAGAVBEGIN f_{B_{s}}\over{f_B} &= 1.209(0.005)\FLAGAVEND
&&\Refs~\mbox{\cite{Dowdall:2013tga,Bussone:2016iua,Hughes:2017spc,Bazavov:2017lyh}}.
\end{align}

The data reported in the calculations that appear in these averages do not permit 
estimates of $\delta(a_{\rm min})$.

The PDG presented averages for the $N_{f}=2+1$ and $N_{f}=2+1+1$ lattice-QCD determinations 
of the isospin-averaged $f_{B}$, $f_{B_{s}}$ and $f_{B_{s}}/f_{B}$ in 2024~\cite{ParticleDataGroup:2024cfk}. 
The $N_{f}=2+1$ and $N_{f}=2+1+1$ lattice-computation results used in Ref.~\cite{ParticleDataGroup:2024cfk} 
are identical to those included in our current work, and the averages quoted in Ref.~\cite{ParticleDataGroup:2024cfk} 
are those determined in \cite{FlavourLatticeAveragingGroup:2019iem} and \cite{FlavourLatticeAveragingGroupFLAG:2021npn}.

%% file: HQ/HQSubsections/BB.tex
\subsection{Neutral $B$-meson mixing matrix elements}
\label{sec:BMix}

Neutral $B$-meson mixing is induced in the Standard Model through
1-loop box diagrams to lowest order in the electroweak theory,
similar to those for short-distance effects in neutral kaon mixing. The effective 
Hamiltonian
is given by
\begin{equation}
  {\cal H}_{\rm eff}^{\Delta B = 2, {\rm SM}} \,\, = \,\,
  \frac{G_F^2 M_{\rm{W}}^2}{16\pi^2} ({\cal F}^0_d {\cal Q}^d_1 + {\cal F}^0_s {\cal 
Q}^s_1)\,\, +
   \,\, {\rm h.c.} \,\,,
   \label{eq:HeffB}
\end{equation}
with
\begin{equation}
 {\cal Q}^q_1 =
   \left[\bar{b}\gamma^\mu(1-\gamma_5)q\right]
   \left[\bar{b}\gamma_\mu(1-\gamma_5)q\right],
   \label{eq:Q1}
\end{equation}
where $q=d$ or $s$. The short-distance function ${\cal F}^0_q$ in
Eq.~(\ref{eq:HeffB}) is much simpler compared to the kaon mixing case
due to the hierarchy in the CKM matrix elements. Here, only one term
is relevant,
\begin{equation}
 {\cal F}^0_q = \lambda_{tq}^2 S_0(x_t)
\end{equation}
where
\begin{equation}
 \lambda_{tq} = V^*_{tq}V_{tb},
 \label{eq:lambdatq}
\end{equation}
and where $S_0(x_t)$ is an Inami-Lim function with $x_t=m_t^2/M_W^2$,
which describes the basic electroweak loop contributions without QCD
\cite{Inami:1980fz}. 

The transition amplitude for $B_q^0$ with $q=d$
or $s$ can be written as
\begin{align}
  \langle \bar B^0_q\vert{\cal H}_{\rm eff}^{\Delta B = 2}\vert B^0_q\rangle
  =&
     \frac{G_F^2 M_W^2}{16\pi^2}  
     \left[\lambda_{tq}^2 S_0(x_t) \eta_{2B} \right]
     \nn \\ 
  & \times 
    \left(\frac{\gbar(\mu)^2}{4\pi}\right)^{-\gamma_0/(2\beta_0)}
    \exp\left\{ \int_0^{\gbar(\mu)}dg \left(
    \frac{\gamma(g)}{\beta(g)}+\frac{\gamma_0}{\beta_0g} \right) \right\}
    \nn \\
   & \times
     \langle \bar B^0_q \vert  Q^q_{\rm R} (\mu) \vert B^0_q\rangle \,\, + \,\, {\rm 
h.c.} \,\, ,
   \label{eq:BBME}
\end{align}
where $Q^q_{\rm R} (\mu)$ is the renormalized four-fermion operator
(usually in the NDR scheme of $\msbar$). The running coupling
$\gbar$, the $\beta$-function $\beta(g)$, and the anomalous
dimension of the four-quark operator $\gamma(g)$ are defined in
Eqs.~(\ref{eq:four_quark_operator_anomalous_dimensions})~and~
(\ref{eq:four_quark_operator_anomalous_dimensions_perturbative}).
The product of $\mu$-dependent terms on the second line of
Eq.~(\ref{eq:BBME}) is, of course, $\mu$-independent (up to truncation
errors arising from the use of perturbation theory). The explicit expression for
the short-distance QCD correction factor $\eta_{2B}$ (calculated to
NLO) can be found in Ref.~\cite{Buchalla:1995vs}.

For historical reasons the $B$-meson-mixing matrix elements are often
parameterized in terms of bag parameters defined as
\begin{equation}
 B_{B_q}(\mu)= \frac{{\left\langle\bar{B}^0_q\left|
   Q^q_{\rm R}(\mu)\right|B^0_q\right\rangle} }{
         {\frac{8}{3}f_{B_q}^2 m_{B_q}^2}} \,\, .
         \label{eq:bagdef}
\end{equation}
The renormalization-group-independent (RGI) $B$ parameter $\hat{B}$ is defined as 
in the case of the kaon,
and expressed to 2-loop order as
\begin{equation}
 \hat{B}_{B_q} = 
   \left(\frac{\gbar(\mu)^2}{4\pi}\right)^{- \gamma_0/(2\beta_0)}
   \left\{ 1+\dfrac{\gbar(\mu)^2}{(4\pi)^2}\left[
   \frac{\beta_1\gamma_0-\beta_0\gamma_1}{2\beta_0^2} \right]\right\}\,
   B_{B_q}(\mu) \,\,\, ,
\label{eq:BBRGI_NLO}
\end{equation}
with $\beta_0$, $\beta_1$, $\gamma_0$, and $\gamma_1$ defined in
Eq.~(\ref{eq:RG-coefficients}). Note, as Eq.~(\ref{eq:BBME}) is
evaluated above the bottom threshold ($m_b<\mu<m_t$), the active number
of flavours here is $\Nf=5$.

Nonzero transition amplitudes result in a mass difference between the
CP eigenstates of the neutral $B$-meson system. Writing the mass
difference for a $B_q^0$ meson as $\Delta m_q$, its Standard Model
prediction is
\begin{equation}
 \Delta m_q = \frac{G^2_Fm^2_W m_{B_q}}{6\pi^2} \,
  |\lambda_{tq}|^2 S_0(x_t) \eta_{2B} f_{B_q}^2 \hat{B}_{B_q},
\end{equation}
where $\lambda_{tq}$ is defined in Eq.~\eqref{eq:lambdatq}.
Experimentally, the mass difference is determined from the oscillation
frequency of the CP eigenstates. The frequencies are measured
precisely with an error of less than a percent. Many different
experiments have measured $\Delta m_d$, but the current average
\cite{Zyla:2020zbs} is dominated by the 
LHC$b$ experiment. For $\Delta m_s$ the experimental average is again dominated by 
results
from LHC$b$
\cite{Zyla:2020zbs} and the precision reached is about one per mille.
With these experimental results and
lattice-QCD calculations of $f_{B_q}^2\hat{B}_{B_q}$,
$\lambda_{tq}$ can be determined.  In lattice-QCD calculations the
flavour SU(3)-breaking ratio
\begin{equation}
 \xi^2 = \frac{f_{B_s}^2B_{B_s}}{f_{B_d}^2B_{B_d}}
 \label{eq:xidef}
\end{equation} 
can be obtained more precisely than the individual $B_q$-mixing matrix
elements because statistical and systematic errors cancel in part.
From $\xi^2$, the ratio $|V_{td}/V_{ts}|$ can be determined and used
to constrain the apex of the CKM triangle.

Neutral $B$-meson mixing, being loop-induced in the Standard Model, is
also a sensitive probe of new physics. The most general $\Delta B=2$
effective Hamiltonian that describes contributions to $B$-meson mixing
in the Standard Model and beyond is given in terms of five local
four-fermion operators:
\be
  {\cal H}_{\rm eff, BSM}^{\Delta B = 2} = \sum_{q=d,s}\sum_{i=1}^5 {\cal C}_i
  {\cal Q}^q_i \;,
\ee
where ${\cal Q}_1$ is defined in Eq.~(\ref{eq:Q1}) and where
\begin{align}
{\cal Q}^q_2 & = \left[\bar{b}(1-\gamma_5)q\right]
   \left[\bar{b}(1-\gamma_5)q\right], \qquad
{\cal Q}^q_3  = \left[\bar{b}^{\alpha}(1-\gamma_5)q^{\beta}\right]
   \left[\bar{b}^{\beta}(1-\gamma_5)q^{\alpha}\right],\nonumber \\
{\cal Q}^q_4 & = \left[\bar{b}(1-\gamma_5)q\right]
   \left[\bar{b}(1+\gamma_5)q\right], \qquad
{\cal Q}^q_5 = \left[\bar{b}^{\alpha}(1-\gamma_5)q^{\beta}\right]
   \left[\bar{b}^{\beta}(1+\gamma_5)q^{\alpha}\right], 
   \label{eq:Q25}
\end{align}
with the superscripts $\alpha,\beta$ denoting colour indices, which
are shown only when they are contracted across the two bilinears.
There are three other basis operators in the $\Delta
B=2$ effective Hamiltonian. When evaluated in QCD, however, 
they give identical matrix elements to the ones already listed due to
parity invariance in QCD.
The short-distance Wilson coefficients ${\cal C}_i$ depend on the
underlying theory and can be calculated perturbatively.  In the
Standard Model only matrix elements of ${\cal Q}^q_1$ contribute to
$\Delta m_q$, while all operators do, for example, for general SUSY
extensions of the Standard Model~\cite{Gabbiani:1996hi}.
The matrix elements or bag parameters for the non-SM operators are also 
useful to estimate the width difference $\Delta \Gamma_q$ 
between the CP eigenstates of the neutral $B$ meson in the Standard Model,
where combinations of matrix elements of ${\cal Q}^q_1$,
${\cal Q}^q_2$, and ${\cal Q}^q_3$ contribute to $\Delta \Gamma_q$ 
at $\cO(1/m_b)$~\cite{Lenz:2006hd,Beneke:1996gn}.  

\begin{figure}[!htb]
\hspace{-0.8cm}\includegraphics[width=0.57\linewidth]{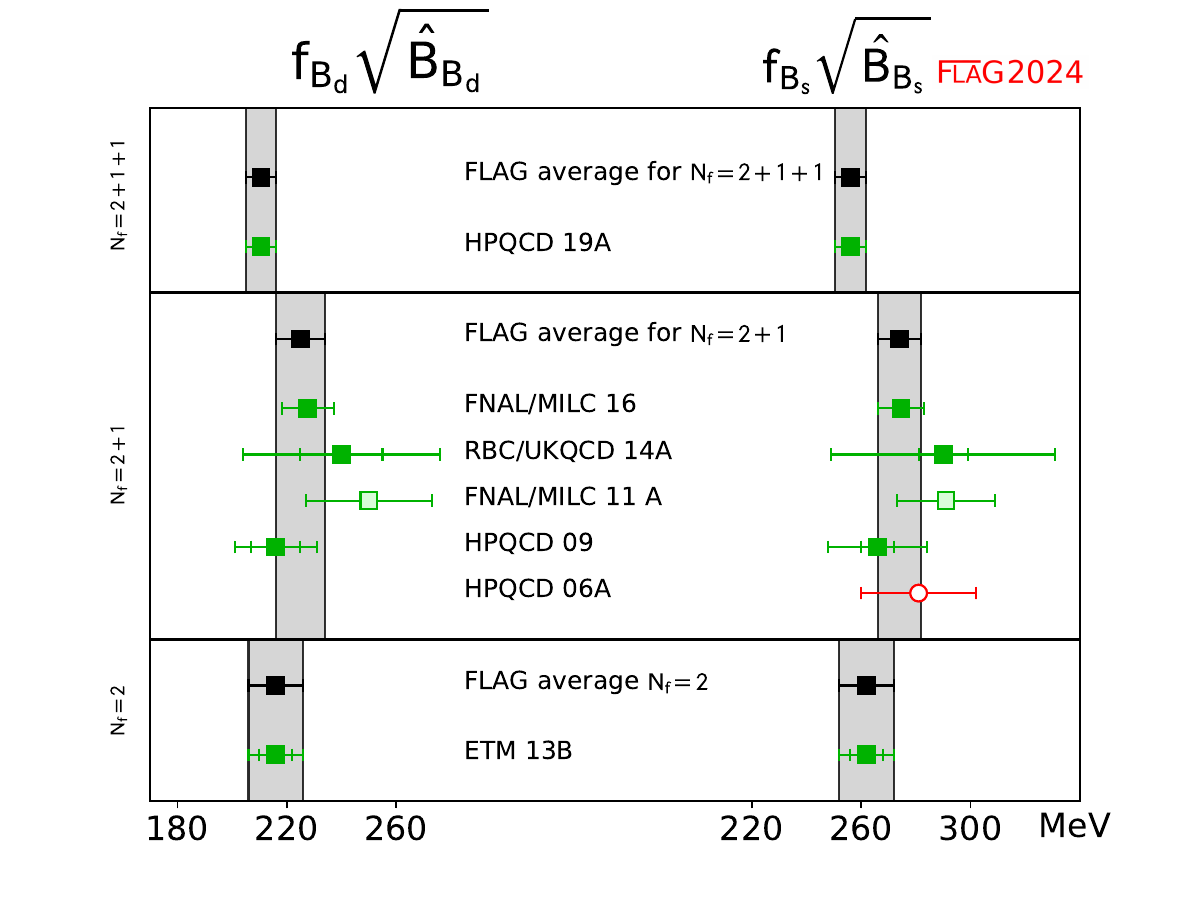}\hspace{-0.8cm}
\includegraphics[width=0.57\linewidth]{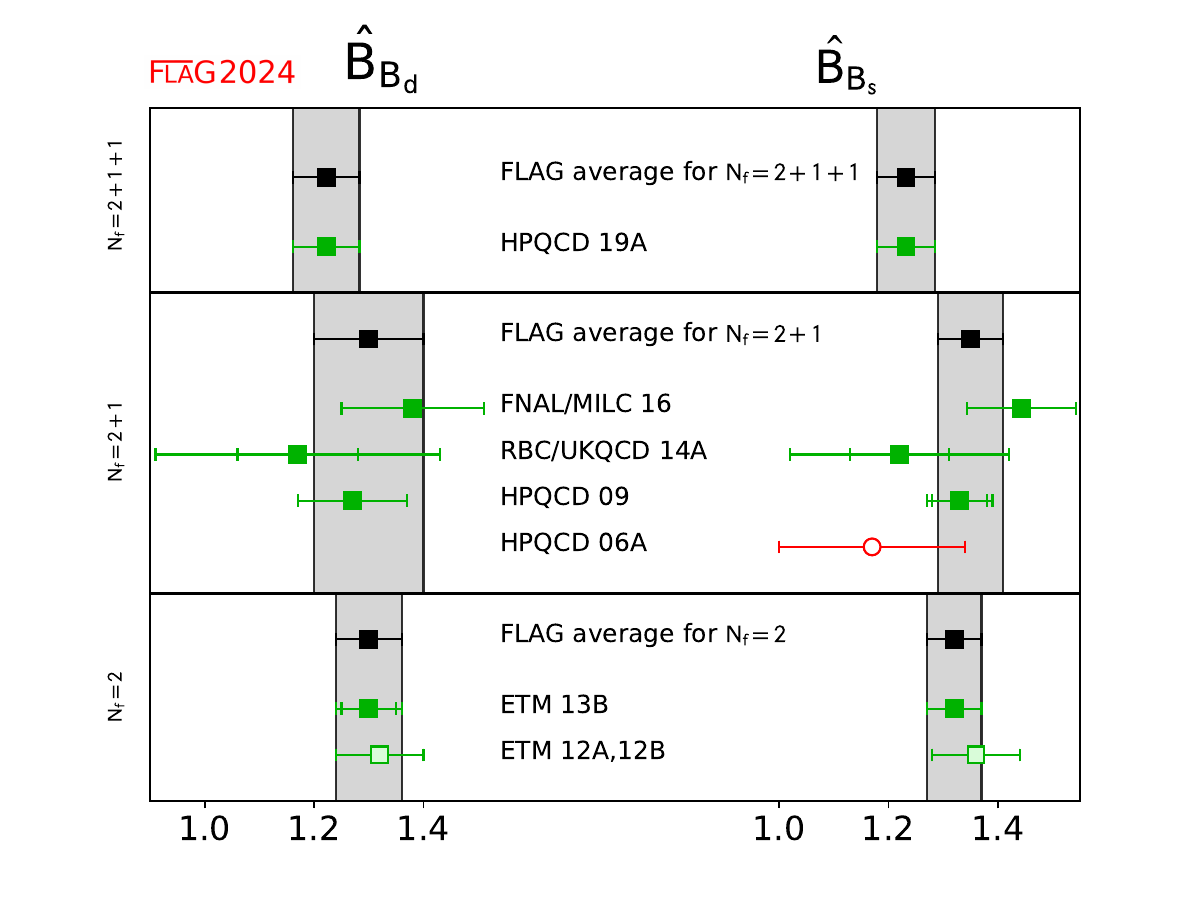}

\vspace{-5mm}
\caption{Neutral $B$- and $B_{\rm s}$-meson-mixing matrix
 elements and bag parameters [values in Tab.~\ref{tab_BBssumm} and
 Eqs.~(\ref{eq:avfBB2}), (\ref{eq:avfBB}), (\ref{eq:avfBB4}), (\ref{eq:avBB2}), (\ref{eq:avBB}), 
(\ref{eq:avBB4})].
 \label{fig:fBsqrtBB2}}
\end{figure}

In this section, we report on results from lattice-QCD calculations for
the neutral $B$-meson mixing parameters $\hat{B}_{B_d}$,
$\hat{B}_{B_s}$, $f_{B_d}\sqrt{\hat{B}_{B_d}}$,
$f_{B_s}\sqrt{\hat{B}_{B_s}}$ and the SU(3)-breaking ratios
$B_{B_s}/B_{B_d}$ and $\xi$ defined in Eqs.~(\ref{eq:bagdef}),
(\ref{eq:BBRGI_NLO}), and (\ref{eq:xidef}).  The results are
summarized in Tabs.~\ref{tab_BBssumm} and \ref{tab_BBratsumm} and in
Figs.~\ref{fig:fBsqrtBB2} and \ref{fig:xi}. Additional details about
the underlying simulations and systematic error estimates are given in
Appendix~\ref{app:BMix_Notes}.  Some collaborations do not provide the
RGI quantities $\hat{B}_{B_q}$, but quote instead
$B_B(\mu)^{\rm \overline{MS},NDR}$. In such cases, we convert the results
using Eq.~(\ref{eq:BBRGI_NLO})
to the RGI quantities quoted in Tab.~\ref{tab_BBssumm}
with a brief description for each case.
More detailed descriptions for these cases are provided in FLAG13 \cite{Aoki:2013ldr}.
We do not provide the $B$-meson-matrix elements of the other operators
${\cal Q}_{2-5}$ in this report. They have been calculated in
Ref.~\cite{Carrasco:2013zta} for the $\Nf=2$ case,
in Refs.~\cite{Bouchard:2011xj,Bazavov:2016nty} for $\Nf=2+1$,
and in Ref.~\cite{Dowdall:2019bea} for $\Nf=2+1+1$.
A discussion is provided on the comparison of these results in a recent review \cite{Tsang:2023nay}.

\begin{table}[!htb]
\begin{center}
\mbox{} \\[3.0cm]
\footnotesize
\begin{tabular*}{\textwidth}[l]{l@{\extracolsep{\fill}}@{\hspace{1mm}}r@{\hspace{1mm}}
	c@{\hspace{1mm}}l@{\hspace{1mm}}l@{\hspace{1mm}}l@{\hspace{1mm}}l@{\hspace{1mm}}
	l@{\hspace{1mm}}l@{\hspace{5mm}}l@{\hspace{1mm}}l@{\hspace{1mm}}l@{\hspace{1mm}}
	l@{\hspace{1mm}}l}
Collaboration \al Ref. \al $\Nf$ \al
\hspace{0.15cm}\begin{rotate}{60}{publication status}\end{rotate}\hspace{-0.15cm} 
\al
\hspace{0.15cm}\begin{rotate}{60}{continuum extrapolation}\end{rotate}\hspace{-0.15cm} 
\al
\hspace{0.15cm}\begin{rotate}{60}{chiral extrapolation}\end{rotate}\hspace{-0.15cm}\al
\hspace{0.15cm}\begin{rotate}{60}{finite volume}\end{rotate}\hspace{-0.15cm}\al
\hspace{0.15cm}\begin{rotate}{60}{renormalization/matching}\end{rotate}\hspace{-0.15cm} 
 \al
\hspace{0.15cm}\begin{rotate}{60}{heavy-quark treatment}\end{rotate}\hspace{-0.15cm} 
\al 
\rule{0.12cm}{0cm}
\parbox[b]{1.2cm}{$f_{\rm B_d}\sqrt{\hat{B}_{\rm B_d}}$} \al
\rule{0.12cm}{0cm}
\parbox[b]{1.2cm}{$f_{\rm B_s}\sqrt{\hat{B}_{\rm B_s}}$} \al
\rule{0.12cm}{0cm}
$\hat{B}_{\rm B_d}$ \al 
\rule{0.12cm}{0cm}
$\hat{B}_{\rm B_{\rm s}}$ \\
&&&&&&&&&& \\[-0.1cm]
\hline
\hline
&&&&&&&&&& \\[-0.1cm]

HPQCD 19A \al \cite{Dowdall:2019bea} \al 2+1+1 \al \gA \al \soso \al \soso \al \good 
\al \soso
       \al \okay & 210.6(5.5) \al 256.1(5.7) \al 1.222(61) \al 1.232(53)\\[0.5ex]
&&&&&&&&&& \\[-0.1cm]
\hline
&&&&&&&&&& \\[-0.1cm]
FNAL/MILC 16 \al \cite{Bazavov:2016nty} \al 2+1 \al \gA \al \good \al \soso \al
     \good \al \soso
	\al \okay & 227.7(9.5) \al 274.6(8.4) \al 1.38(12)(6)$^\odot$ \al 
	1.443(88)(48)$^\odot$\\[0.5ex]

        RBC/UKQCD 14A \al \cite{Aoki:2014nga} \al 2+1 \al \gA \al \soso \al \soso 
\al
     \soso \al \soso
	\al \okay & 240(15)(33) \al 290(09)(40) \al 1.17(11)(24) \al 1.22(06)(19)\\[0.5ex]

FNAL/MILC 11A \al \cite{Bouchard:2011xj} \al 2+1 \al \rC \al \good \al \soso \al
     \good \al \soso
	\al \okay & 250(23)$^\dagger$ \al 291(18)$^\dagger$ \al $-$ \al $-$\\[0.5ex]

HPQCD 09 \al \cite{Gamiz:2009ku} \al 2+1 \al \gA \al \soso \al \soso$^\nabla$ \al 
\soso \al
\soso 
\al \okay & 216(15)$^\ast$ \al 266(18)$^\ast$ \al 1.27(10)$^\ast$ \al 1.33(6)$^\ast$ 
\\[0.5ex] 

HPQCD 06A \al \cite{Dalgic:2006gp} \al 2+1 \al \gA \al \tbr \al \tbr \al \good \al 

\soso 
	\al \okay & $-$ \al  281(21) \al $-$ \al 1.17(17) \\
&&&&&&&&&& \\[-0.1cm]
\hline
&&&&&&&&&& \\[-0.1cm]
ETM 13B \al \cite{Carrasco:2013zta} \al 2 \al \gA \al \good \al \soso \al \soso \al
    \good \al \okay & 216(6)(8) \al 262(6)(8) \al  1.30(5)(3) \al 1.32(5)(2) \\[0.5ex]

ETM 12A, 12B \al \cite{Carrasco:2012dd,Carrasco:2012de} \al 2 \al \rC \al \good \al 
\soso \al \soso \al
    \good \al \okay & $-$ \al $-$ \al  1.32(8)$^\diamond$ \al 1.36(8)$^\diamond$ 
\\[0.5ex]
&&&&&&&&&& \\[-0.1cm]
\hline
\hline\\
\end{tabular*}\\[-0.2cm]
\begin{minipage}{\linewidth}
{\footnotesize 
\begin{itemize}
 \item[$^\odot$] PDG averages of decay constant $f_{B^0}$ and $f_{B_s}$ \cite{Rosner:2015wva} 
are used to obtain these values.\\[-5mm]
   \item[$^\dagger$] Reported $f_B^2B$ at $\mu=m_b$ is converted to RGI by
	multiplying the 2-loop factor
	1.517.\\[-5mm]
   \item[$^\nabla$] While wrong-spin contributions are not included in
		the HMrS$\chi$PT fits, the effect is expected to be
		small for these quantities (see description in FLAG 13
		\cite{Aoki:2013ldr}). \\[-5mm] 
        \item[$^\ast$] This result uses an old determination of
		     $r_1=0.321(5)$~fm from Ref.~\cite{Gray:2005ur} that
		     has since been superseded, which however has
		     only a small effect in the total error budget (see
		     description in FLAG 13 \cite{Aoki:2013ldr}) .\\[-5mm]
        \item[$^\diamond$] Reported $B$ at $\mu=m_b=4.35$ GeV is converted to
     RGI by multiplying the 2-loop factor 1.521.
\end{itemize}
}
\end{minipage}
\caption{Neutral $B$- and $B_{\rm s}$-meson mixing matrix
 elements (in MeV) and bag parameters.}
\label{tab_BBssumm}
\end{center}
\end{table}

\begin{table}[!htb]
\begin{center}
\mbox{} \\[3.0cm]
\footnotesize
\begin{tabular*}{\textwidth}[l]{l @{\extracolsep{\fill}} r l l l @{\hspace{-0.1mm}}l 
@{\hspace{-0.1mm}}l @{\hspace{-0.1mm}}l @{\hspace{-0.1mm}}l l l}
Collaboration & Ref. & $\Nf$ & 
\hspace{0.15cm}\begin{rotate}{60}{publication status}\end{rotate}\hspace{-0.15cm} 
&
\hspace{0.15cm}\begin{rotate}{60}{continuum extrapolation}\end{rotate}\hspace{-0.15cm} 
&
\hspace{0.15cm}\begin{rotate}{60}{chiral extrapolation}\end{rotate}\hspace{-0.15cm}&
\hspace{0.15cm}\begin{rotate}{60}{finite volume}\end{rotate}\hspace{-0.15cm}&
\hspace{0.15cm}\begin{rotate}{60}{renormalization/matching}\end{rotate}\hspace{-0.15cm} 
 &
\hspace{0.15cm}\begin{rotate}{60}{heavy-quark treatment}\end{rotate}\hspace{-0.15cm} 
& 
\rule{0.12cm}{0cm}$\xi$ &
 \rule{0.12cm}{0cm}$B_{\rm B_{\rm s}}/B_{\rm B_d}$ \\
&&&&&&&&&& \\[-0.1cm]
\hline
\hline
&&&&&&&&&& \\[-0.1cm]

HPQCD 19A \al \cite{Dowdall:2019bea} & 2+1+1 & \gA & \soso & \soso & \good & \soso
       & \okay & 1.216(16) & 1.008(25) \\[0.5ex]

&&&&&&&&&& \\[-0.1cm]

\hline

&&&&&&&&&& \\[-0.1cm]

RBC/UKQCD 18A \al \cite{Boyle:2018knm} & 2+1 & \oP & \good & \good & 
     \good & \good & \okay & 1.1939(67)($^{+95}_{-177}$) & 0.9984(45)($^{+80}_{-63}$) 
\\[0.5ex]

FNAL/MILC 16 & \cite{Bazavov:2016nty} & 2+1 & \gA & \good & \soso &
     \good & \soso & \okay & 1.206(18) & 1.033(31)(26)$^\odot$ \\[0.5ex]

RBC/UKQCD 14A & \cite{Aoki:2014nga} & 2+1 & \gA & \soso & \soso &
     \soso & \soso & \okay & 1.208(41)(52) & 1.028(60)(49) \\[0.5ex]

FNAL/MILC 12 & \cite{Bazavov:2012zs} & 2+1 & \gA & \soso & \soso &
     \good & \soso & \okay & 1.268(63) & 1.06(11) \\[0.5ex]

RBC/UKQCD 10C
 & \cite{Albertus:2010nm} & 2+1 & \gA & \tbr & \tbr & \tbr
  & \soso & \okay & 1.13(12) & $-$ \\[0.5ex]

HPQCD 09 & \cite{Gamiz:2009ku} & 2+1 & \gA & \soso & \soso$^\nabla$ & \soso &
\soso & \okay & 1.258(33) & 1.05(7) \\[0.5ex] 

&&&&&&&&&& \\[-0.1cm]

\hline

&&&&&&&&&& \\[-0.1cm]

ETM 13B & \cite{Carrasco:2013zta} & 2 & \gA & \good & \soso & \soso & \good
			     & \okay & 1.225(16)(14)(22) & 1.007(15)(14) \\

ETM 12A, 12B & \cite{Carrasco:2012dd,Carrasco:2012de} & 2 & \rC & \good & \soso & 
\soso & \good
			     & \okay & 1.21(6) & 1.03(2) \\
&&&&&&&&&& \\[-0.1cm]
\hline
\hline\\[5mm]
\end{tabular*}\\[-0.2cm]
\begin{minipage}{\linewidth}
{\footnotesize 
\begin{itemize}
 \item[$^\odot$] PDG average of the ratio of decay constants
	      $f_{B_s}/f_{B^0}$ \cite{Rosner:2015wva} is used to obtain
	      the value.\\[-5mm] 
   \item[$^\nabla$] Wrong-spin contributions are not included in the
		HMrS$\chi$PT fits. As the effect may not be negligible,
		these results are excluded from the average (see
		description in FLAG 13 \cite{Aoki:2013ldr}).
\end{itemize}
}
\end{minipage}
\caption{Results for SU(3)-breaking ratios of neutral $B_{d}$- and 
 $B_{s}$-meson-mixing matrix elements and bag parameters.}
\label{tab_BBratsumm}
\end{center}
\end{table}

\begin{figure}[!htb]
\begin{center}
\includegraphics[width=11.5cm]{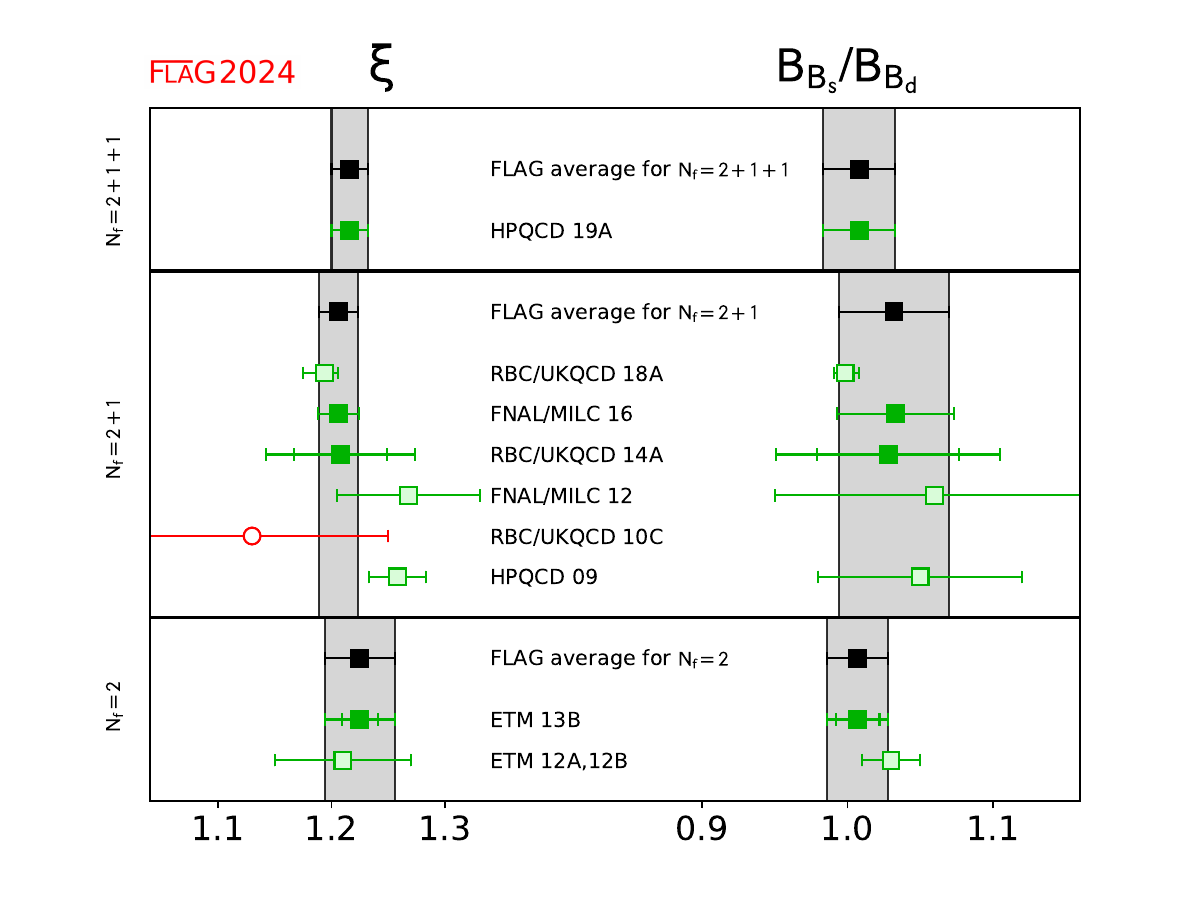}

\vspace{-2mm}
\caption{The SU(3)-breaking quantities $\xi$ and $B_{B_s}/B_{B_d}$
 [values in Tab.~\ref{tab_BBratsumm} and Eqs.~(\ref{eq:avxiBB2}), (\ref{eq:avxiBB}), 
(\ref{eq:avxiBB4})].}\label{fig:xi} 
\end{center}
\end{figure}

Let us mention that our averages here have no updates from the previous review
\cite{FlavourLatticeAveragingGroupFLAG:2021npn}. 
The new addition to this subsection is that we review a measure of continuum-limit 
quality $\delta(a_{min})$ for each result that is included in the average.
We used this quantity for the continuum-limit criterion for heavy-quark related
quantities in FLAG 13 \cite{Aoki:2013ldr}. 
This time we only quote the value for information and we do not use it when calculating averages.

There are no new results for $\Nf=2$ reported after FLAG 16 \cite{Aoki:2016frl}.
In this category, one work (ETM~13B)~\cite{Carrasco:2013zta} passes
the quality criteria. 
A description of this work can be found in FLAG 13
\cite{Aoki:2013ldr} where it did not enter the average as it had not
appeared in a journal. 
This is the only result 
available for $\Nf=2$, so we quote their values as our estimates  
\begin{align}
      &&  \FLAGAVBEGIN f_{B_d}\sqrt{\hat{B}_{b_d}}&= 216(10)\FLAGAVEND\,{\rm MeV}\,,
         &\FLAGAVBEGIN f_{B_s}\sqrt{\hat{B}_{B_s}}&= 262(10)\FLAGAVEND\,{\rm MeV}
         &\Ref~\mbox{\cite{Carrasco:2013zta}},  \label{eq:avfBB2}\\[2mm]
\Nf=2:&&\FLAGAVBEGIN \hat{B}_{B_d}&= 1.30(6)\FLAGAVEND\,, 
         &\FLAGAVBEGIN \hat{B}_{B_s}&= 1.32(5)\FLAGAVEND 
	 &\Ref~\mbox{\cite{Carrasco:2013zta}},  \label{eq:avBB2}\\[2mm]
      &&  \FLAGAVBEGIN \xi &=  1.225(31)\FLAGAVEND\,,  
  	& \FLAGAVBEGIN B_{B_s}/B_{B_d} & =  1.007(21)\FLAGAVEND
 	&\Ref~\mbox{\cite{Carrasco:2013zta}}. \label{eq:avxiBB2}
\end{align}

The continuum-limit measure, $\delta(a_{\rm min})$, cannot be estimated for the ETM~13B results for $\hat{B}_{B_d}$ because the relevant continuum-limit information is not provided.
For the other quantities of ETM~13B, $\delta(a_{\rm min})\simeq 0.1$ ($\hat{B}_{B_d}$), 
2 ($\xi$) and 0.7 ($B_{B_s}/B_{B_d}$).

For  $\Nf=2+1$ 
the results that enter our averages for $\Nf=2+1$ are 
FNAL/MILC~16~\cite{Bazavov:2016nty}, which had been included in the averages at FLAG~19~\cite{FlavourLatticeAveragingGroup:2019iem}, RBC/UKQCD~14A~\cite{Aoki:2014nga},
included in the averages at FLAG~16~\cite{Aoki:2016frl},
and HPQCD~09~\cite{Gamiz:2009ku} for which a description is available in
FLAG~13~\cite{Aoki:2013ldr}.
The work in RBC/UKQCD~18A~\cite{Boyle:2018knm} on the flavour SU(3)-breaking ratios,
whose description can be found in FLAG 21 \cite{FlavourLatticeAveragingGroupFLAG:2021npn},
has not been published yet and therefore do not enter into the averages.
Thus, the averages for $\Nf=2+1$ are unchanged:\\

$\Nf=2+1:$
\begin{align}
        && \FLAGAVBEGIN f_{B_d}\sqrt{\hat{B}_{B_d}} &=  225(9)\FLAGAVEND
	{\rm MeV}   \,,
         & \FLAGAVBEGIN f_{B_s}\sqrt{\hat{B}_{B_s}} &=  274(8)\FLAGAVEND \, 
	 {\rm MeV}
	  &\Refs~\mbox{\cite{Gamiz:2009ku,Aoki:2014nga,Bazavov:2016nty}},  \label{eq:avfBB}\\[2mm]
&& \FLAGAVBEGIN \hat{B}_{B_d}  &= 1.30(10)\FLAGAVEND\,, 
         & \FLAGAVBEGIN \hat{B}_{B_s} &=  1.35(6)\FLAGAVEND 
          &\Refs~\mbox{\cite{Gamiz:2009ku,Aoki:2014nga,Bazavov:2016nty}}, \label{eq:avBB}\\[2mm]
        && \FLAGAVBEGIN \xi  &=  1.206(17)\FLAGAVEND\,, 
        & \FLAGAVBEGIN B_{B_s}/B_{B_d}  &=  1.032(38)\FLAGAVEND 
          &\Refs~\mbox{\cite{Aoki:2014nga,Bazavov:2016nty}}. \label{eq:avxiBB} %
\end{align}
Here all the above equations have not been changed from FLAG 19. 
The averages were obtained using the nested averaging scheme described in Sec.~\ref{sec:nested_average},
due to a nested correlation structure among the results. Details are discussed in
the FLAG 19 report \cite{FlavourLatticeAveragingGroup:2019iem}.

We estimate $\delta(a_{\rm min})\simeq 2$ for both $\hat{B}_{B_s}$ and $\hat{B}_{B_d}$ 
of FNAL/MILC 16. 
Data are not available in FNAL/MILC 16 to estimate $\delta(a_{\rm min})$ for the 
ratio of the bag parameters.
Since the $f_{B_s}\sqrt{\hat{B}_{B_s}}$, $f_{B_d}\sqrt{\hat{B}_{B_d}}$ and $\xi$ 
are quantities derived using PDG estimates of the decay constants and their ratio, 
we do not provide an estimate of $\delta(a_{\rm min})$ of these quantities.
For RBC/UKQCD~14A, $\delta(a_{\rm min})\simeq 0.7$ ($f_{B_d}\sqrt{\hat{B}_{B_d}}$), 
1.3 ($f_{B_s}\sqrt{\hat{B}_{B_s}}$), 0.3 ($\xi$), 0.3 ($\hat{B}_{B_d}$), 0.4 ($\hat{B}_{B_s}$) 
and 0 ($B_{B_s}/B_{B_d}$).
For HPQCD 09, $\delta(a_{\rm min})\simeq 0.8$ ($f_{B_d}\sqrt{\hat{B}_{B_d}}$), 3 
($f_{B_s}\sqrt{\hat{B}_{B_s}}$), 0.3 ($\xi$), at most 1 ($\hat{B}_{B_d}$), 0.8 ($\hat{B}_{B_s}$) 
and 1 ($B_{B_s}/B_{B_d}$).

We note that, for $\Nf=2+1$, there is an on-going study 
involving the JLQCD and RBC/UKQCD collaborations, with 
initial results reported in the Lattice 2021 proceedings~\cite{Boyle:2021kqn}. 
These results utilize coarse lattices at the physical 
point from RBC/UKQCD along with very fine lattices from 
JLQCD (up to $a^{-1}=4.5$ GeV) with unphysical pion 
masses, both using domain-wall fermions.

The only result available for $\Nf=2+1+1$ is HPQCD 19A \cite{Dowdall:2019bea},
which uses MILC collaboration's HISQ ensembles and NRQCD for the $b$ quark.
A detailed description can be found in the previous review \cite{FlavourLatticeAveragingGroupFLAG:2021npn}.
We quote their values as the FLAG estimates \\

$\Nf=2+1+1$:
\begin{align}
  &&  \FLAGAVBEGIN f_{B_d}\sqrt{\hat{B}_{b_d}}&= 210.6(5.5)\FLAGAVEND \;{\rm MeV}\,,\;\;
         &\FLAGAVBEGIN f_{B_s}\sqrt{\hat{B}_{B_s}}&= 256.1(5.7)\FLAGAVEND\; {\rm 
MeV}
         &\Ref~\mbox{\cite{Dowdall:2019bea}},  \label{eq:avfBB4}\\
      &&\FLAGAVBEGIN \hat{B}_{B_d}&= 1.222(61)\FLAGAVEND\,, 
         &\FLAGAVBEGIN \hat{B}_{B_s}&= 1.232(53)\FLAGAVEND 
	 &\Ref~\mbox{\cite{Dowdall:2019bea}},  \label{eq:avBB4}\\
      &&  \FLAGAVBEGIN \xi &=  1.216(16)\FLAGAVEND\,,  
  	& \FLAGAVBEGIN B_{B_s}/B_{B_d} & =  1.008(25)\FLAGAVEND
 	&\Ref~\mbox{\cite{Dowdall:2019bea}}. \label{eq:avxiBB4}
\end{align}

We estimate $\delta(a_{\rm min})\simeq 0.1$ for $\hat{B}_{B_s}$, 1 for $B_{B_s}/B_{B_d}$ 
and at most 1 for $\hat{B}_{B_d}$. The other quantities are derived ones using the 
estimates of decay constants in FNAL/MILC 17.

We note that the above results 
with the same $\Nf$ 
(e.g., those in Eqs.~(\ref{eq:avfBB4}-\ref{eq:avxiBB4}))
are all correlated with each other,
due to the use of the same 
gauge-field ensembles for different quantities.
The results are also correlated with the averages obtained in 
Sec.~\ref{sec:fB} and shown in
Eqs.~(\ref{eq:fB2})--(\ref{eq:fBratio2}) for $\Nf=2$,
Eqs.~(\ref{eq:fB21})--(\ref{eq:fBratio21}) for $\Nf=2+1$ and
Eqs.~(\ref{eq:fB211})--(\ref{eq:fBratio211}) for $\Nf=2+1+1$.
This is because the calculations of $B$-meson decay constants and  
mixing quantities 
are performed on the same (or on similar) sets of ensembles, and results obtained 
by a 
given collaboration 
use the same actions and setups. These correlations must be considered when 
using our averages as inputs to unitarity triangle (UT) fits. 
For this reason, if one were for example to estimate $f_{B_s}\sqrt{\hat{B}_s}$ from 
the separate averages of $f_{B_s}$ (Eq.~(\ref{eq:fBs21})) and $\hat{B}_s$ (Eq.~(\ref{eq:avBB})) 
for $\Nf=2+1$, one would obtain a value about one standard deviation below the one 
quoted above in Eq.~(\ref{eq:avfBB}).  While these two estimates lead to compatible 
results, giving us confidence that all uncertainties have been properly addressed, 
we do not recommend combining averages this way, as many correlations would have 
to be taken into account to properly assess the errors. We recommend instead using 
the numbers quoted above.
In the future, as more independent 
calculations enter the averages, correlations between the lattice-QCD inputs to UT 
fits will become less significant.

%% file: HQ/HQSubsections/BSL.tex
\input{HQ/HQSubsections/BSL_B_to_light.tex}

\input{HQ/HQSubsections/BSL_B_to_heavy.tex}

\input{HQ/HQSubsections/BSL_other.tex}

%% file: HQ/HQSubsections/BSL_B_to_light.tex
\FloatBarrier
\subsection{Semileptonic form factors for $B$ decays to light flavours}
\label{sec:BtoPiK}
The Standard Model differential rate for the decay $B_{(s)}\to
P\ell\nu$ involving a quark-level $b\to u$ transition is given, at
leading order in the weak interaction, by a formula analogous to the
one for $D$ decays in Eq.~(\ref{eq:DtoPiKFull}), but with $D \to
B_{(s)}$ and the relevant CKM matrix element $|V_{cq}| \to |V_{ub}|$:
\begin{align}
  \frac{d\Gamma(B_{(s)}\to P\ell\nu)}{dq^2} =
  & \frac{G_F^2 |\eta_{\rm EW}|^2 |V_{ub}|^2}{24 \pi^3}
    \frac{(q^2-m_\ell^2)^2\sqrt{E_P^2-m_P^2}}{q^4m_{B_{(s)}}^2}
    \nn\\
  & \times
    \left[ \left(1+\frac{m_\ell^2}{2q^2}\right)
    m_{B_{(s)}}^2(E_P^2-m_P^2)|f_+(q^2)|^2
    \right.
    \nn\\
  & \left. \;\;\;\;\;
   + \frac{3m_\ell^2}{8q^2}(m_{B_{(s)}}^2-m_P^2)^2|f_0(q^2)|^2
    \right].
    \label{eq:B_semileptonic_rate}
\end{align}
Again, for $\ell=e,\mu$ the contribution from the scalar form factor
$f_0$ can be neglected, and one has a similar expression to
Eq.~(\ref{eq:DtoPiK}), which, in principle, allows for a direct
extraction of $|V_{ub}|$ by matching theoretical predictions to
experimental data.  However, while for $D$ (or $K$) decays the entire
physical range $0 \leq q^2 \leq q^2_{\rm max}$ can be covered with
moderate momenta accessible to lattice simulations, in
$B \to \pi \ell\nu$ decays one has $q^2_{\rm max} \sim 26~{\rm GeV}^2$
and only part of the full kinematic range is reachable.
As a consequence, obtaining $|V_{ub}|$ from $B\to\pi\ell\nu$ is more
complicated than obtaining $|V_{cd(s)}|$ from semileptonic $D$-meson
decays.

In practice, lattice computations are restricted
to large values of the momentum transfer $q^2$ (see Sec.~\ref{sec:DtoPiK})
where statistical and momentum-dependent discretization errors can be
controlled, which in existing calculations roughly cover the upper third of
the kinematically allowed $q^2$ range.\footnote{The variance of hadron correlation functions at
nonzero three-momentum is dominated at large Euclidean times by
zero-momentum multiparticle states~\cite{DellaMorte:2012xc}; therefore
the noise-to-signal grows more rapidly than for the vanishing three-momentum
case.}
Since, on the other hand, the decay rate is
suppressed by phase space at large $q^2$, most of the semileptonic $B\to
\pi$ events are observed in experiment at lower values of $q^2$, leading
to more accurate experimental results for the binned differential rate
in that region.\footnote{Upcoming data from Belle~II are expected to
significantly improve the precision of experimental results,
in particular, for larger values of $q^2$.}
It is, therefore, a challenge to find a window of
intermediate values of $q^2$ at which both the experimental and
lattice results can be reliably evaluated.

State-of-the-art determinations of CKM matrix elements, say, $|V_{ub}|$, are obtained
from joint fits to lattice and experimental results, keeping the relative normalization
$|V_{ub}|^2$ as a free parameter.
This requires, in particular, that both experimental and lattice data for the
$q^2$-dependence be parameterized by fitting data to specific ans\"atze,
with the ultimate aim of minimizing the systematic uncertainties involved.
This plays a key role in assessing the systematic uncertainties of CKM determinations,
and will be discussed extensively in this section.
A detailed discussion of the parameterization of form factors as a function of $q^2$ can be found
in Appendix \ref{sec:zparam}.

\subsubsection{Form factors for $B\to\pi\ell\nu$}
\label{sec:BtoPi}

The semileptonic decay process $B\to\pi\ell\nu$ enables the determination of the CKM matrix element $|V_{ub}|$
within the Standard Model via Eq.~(\ref{eq:B_semileptonic_rate}).
Early results for $B\to\pi\ell\nu$ form factors came from the HPQCD~\cite{Dalgic:2006dt}
and FNAL/MILC~\cite{Bailey:2008wp} collaborations (HPQCD~06 and FNAL/MILC~08A).

Our 2016 review featured a significantly extended
calculation of $B\to\pi\ell\nu$ from FNAL/MILC~\cite{Lattice:2015tia} (FNAL/MILC~15)
and a new computation from  RBC/UKQCD~\cite{Flynn:2015mha} (RBC/UKQCD~15).
In 2022, the JLQCD collaboration published
another new calculation using M\"obius Domain Wall fermions -- \SLjlqcdBpi~\cite{Colquhoun:2022atw}.
FNAL/MILC and RBC/UKQCD continue working on further new calculations of the $B\to\pi$ form factors
and have reported on their progress at the annual Lattice conferences and the 2020
Asia-Pacific Symposium for Lattice Field Theory. The results are preliminary or blinded, so
 not yet ready for inclusion in this review. FNAL/MILC is using $\Nf=2+1+1$ HISQ ensembles with $a\approx 0.15$,
0.12, 0.088~fm, 0.057~fm, with Goldstone-pion mass down to its physical
value~\cite{Gelzer:2017edb,Gelzer:2019zwx}.
The RBC/UKQCD Collaborations have added a new M\"obius-domain-wall-fermion ensemble with 
$a\approx 0.07$~fm and $m_\pi \approx 230$ MeV to their analysis \cite{Flynn:2019jbg}.
In addition, HPQCD using MILC ensembles had published the first
$\Nf=2+1+1$ results for the $B\to\pi\ell\nu$ scalar
form factor, working at zero recoil ($q^2=q^2_{\rm max}$) and pion masses down to the physical value~\cite{Colquhoun:2015mfa};
this adds to previous reports on ongoing work to upgrade their 2006
computation~\cite{Bouchard:2012tb,Bouchard:2013zda}. Since this latter
result has no immediate impact on current $|V_{ub}|$ determinations,
which come from the vector-form-factor-dominated decay channels into light leptons,
we will from now on concentrate on the $\Nf=2+1$ determinations of the
$q^2$-dependence of $B\to\pi$ form factors.

Both the HPQCD~06 and the FNAL/MILC~15 computations of $B\to\pi\ell\nu$
amplitudes use ensembles of gauge configurations with $\Nf=2+1$
flavours of rooted staggered quarks produced by the MILC collaboration;
however, FNAL/MILC~15 makes a much more extensive
use of the currently available ensembles, both in terms of
lattice spacings and light-quark masses.
HPQCD~06 has results at two values of the lattice spacing
($a\approx0.12,~0.09~{\rm fm}$), while FNAL/MILC~15 employs four values
($a\approx0.12,~0.09,~0.06,~0.045~{\rm fm}$).
Lattice-discretization
effects are estimated within heavy-meson rooted staggered chiral perturbation theory (HMrS$\chi$PT) in the FNAL/MILC~15
computation, while HPQCD~06 quotes the results at $a\approx 0.12~{\rm fm}$
as central values and uses the $a\approx 0.09~{\rm fm}$ results to quote
an uncertainty.
The relative scale is fixed in both cases through the quark-antiquark potential-derived ratio $r_1/a$.
HPQCD~06 set the absolute scale through the $\Upsilon$ $2S$--$1S$ splitting,
while FNAL/MILC~15 uses a combination of $f_\pi$ and the same $\Upsilon$
splitting, as described in Ref.~\cite{Bazavov:2011aa}.
The spatial extent of the lattices employed by HPQCD~06 is $L\simeq 2.4~{\rm fm}$,
save for the lightest-mass point (at $a\approx 0.09~{\rm fm}$) for which $L\simeq 2.9~{\rm fm}$.
FNAL/MILC~15, on the other hand, uses extents up to $L \simeq 5.8~{\rm fm}$, in order
to allow for light-pion masses while keeping finite-volume effects under
control. 

Indeed, while in the HPQCD~06 work the lightest RMS pion mass is $400~{\rm MeV}$,
the FNAL/MILC~15 work includes pions as light as $165~{\rm MeV}$---in both cases
the bound $m_\pi L \gtrsim 3.8$ is kept.
Other than the qualitatively different range of MILC ensembles used
in the two computations, the main difference between HPQCD~06 and FNAL/MILC~15 lies in the treatment of
heavy quarks. HPQCD~06 uses the NRQCD formalism, with a 1-loop matching
of the relevant currents to the ones in the relativistic
theory. FNAL/MILC~15 employs the clover action with the Fermilab
interpretation, with a mostly-nonperturbative renormalization of the
relevant currents, within which the overall renormalization factor of the heavy-light current
is written as a product of the square roots of the renormalization factors of the
light-light and heavy-heavy temporal vector currents
(which are determined nonperturbatively) and a residual factor that is computed
using 1-loop perturbation theory. (See Tab.~\ref{tab_BtoPisumm2};
full details about the computations are provided in tables in
Appendix~\ref{app:BtoPi_Notes}.)

The RBC/UKQCD~15 computation is based on $\Nf=2+1$ DWF ensembles at two
values of the lattice spacing ($a\approx0.12,~0.09~{\rm fm}$), and pion masses
in a narrow interval ranging from slightly above $400~{\rm MeV}$ to slightly below $300~{\rm MeV}$,
keeping $m_\pi L \gtrsim 4$.
The scale is set using the $\Omega^-$ baryon mass. Discretization effects
coming from the light sector
are estimated in the $1\%$ ballpark using HM$\chi$PT supplemented with effective higher-order
interactions to describe cutoff effects.
The $b$ quark is treated using the Columbia RHQ action, with
a mostly nonperturbative renormalization of the relevant currents. Discretization
effects coming from the heavy sector are estimated with power-counting
arguments to be below $2\%$. The collaboration has also reported on progress toward an
improved calculation that adds a third, finer lattice spacing \cite{Flynn:2021ttz}.

The \SLjlqcdBpi~calculation is using M\"obius Domain Wall
 fermions, including for the heavy quark, with $a\approx 0.08$, 0.055, and 0.044~fm and pion masses
 down to 230 MeV. The relative scales are set using the gradient-flow time $t_0^{1/2}/a$, with the absolute scale $t_0^{1/2}$ taken
 from Ref.~\cite{Borsanyi:2012zs}. All ensembles have $m_\pi L\gtrsim 4.0$. 
 The bare heavy-quark masses satisfy $a m_Q<0.7$ and reach from the charm mass up to 2.44 times the charm mass.
 The form factors are extrapolated linearly in $1/m_Q$ to the bottom mass. For the lower range of the quark masses,
 the vector current is renormalized using a factor $Z_{V_{qq}}$ obtained from position-space current-current correlators.
For heavier quark masses, $\sqrt{Z_{V_{QQ}}  Z_{V_{qq}}}$ is used, where $Z_{V_{QQ}}$ is the renormalization factor of the flavour-conserving temporal vector current, determined using charge conservation. This corresponds to mostly nonperturbative  renormalization with tree-level residual matching factors, but the residual matching factors are expected to be close to 1 and approach this value exactly in the continuum limit. We therefore assign a \soso rating for renormalization.

Given the large kinematical range available in the $B\to\pi$ transition,
chiral extrapolations are an important source of systematic uncertainty:
apart from the eventual need to reach physical pion masses in the extrapolation,
the applicability of $\chi$PT is not guaranteed for large values of the pion energy $E_\pi$.
Indeed, in all computations $E_\pi$ reaches values in the $1~{\rm GeV}$ ballpark,
and chiral-extrapolation systematics is the dominant source of errors.
FNAL/MILC uses SU(2) NLO HMrS$\chi$PT for the continuum-chiral extrapolation,
supplemented by NNLO analytic terms
and hard-pion $\chi$PT terms~\cite{Bijnens:2010ws};\footnote{It is important
to stress the finding in Ref.~\cite{Colangelo:2012ew} that
the factorization of chiral logs in hard-pion $\chi$PT breaks down,
implying that it does not fulfill the expected requisites for a proper
effective field theory. Its use to model the mass dependence of form
factors can thus be questioned.} systematic uncertainties
are estimated through an extensive study of the effects of varying the
specific fit ansatz and/or data range. RBC/UKQCD and JLQCD use
SU(2) hard-pion HM$\chi$PT to perform their combined continuum-chiral
extrapolations, and obtain estimates for systematic uncertainties
by varying the ans\"{a}tze and ranges used in fits. HPQCD performs chiral
extrapolations using HMrS$\chi$PT formulae, and estimates systematic
uncertainties by comparing the result with the ones from fits to a
linear behaviour in the light-quark mass, continuum HM$\chi$PT, and
partially quenched HMrS$\chi$PT formulae (including also data with
different sea and valence light-quark masses).

FNAL/MILC~15, RBC/UKQCD~15, and \SLjlqcdBpi~describe the $q^2$-dependence of
$f_+$ and $f_0$ by applying a BCL parameterization to
the form factors extrapolated to the continuum
limit, within the range of values of $q^2$ covered by data. (A discussion of the
various parameterizations can be found in Appendix~\ref{sec:zparam}.)
RBC/UKQCD~15 and \SLjlqcdBpi~generate synthetic data for the form factors at some values
of $q^2$ (evenly spaced in $z$) from the continuous function of $q^2$ obtained
from the joint chiral-continuum extrapolation,
which are then used as input for the fits. After having checked that the
kinematical constraint $f_+(0)=f_0(0)$ is satisfied within errors by the extrapolation
to $q^2=0$ of the results of separate fits, this constraint is imposed
to improve fit quality. In the case of FNAL/MILC~15, rather than producing
synthetic data a functional method is used to extract the $z$-parameterization
directly from the fit functions employed in the continuum-chiral extrapolation.
In the case of HPQCD~06, the parameterization of the $q^2$-dependence of form factors is
somewhat intertwined with chiral extrapolations: a set of fiducial
values $\{E_\pi^{(n)}\}$ is fixed for each value of the light-quark
mass, and $f_{+,0}$ are interpolated to each of the $E_\pi^{(n)}$;
chiral extrapolations are then performed at fixed $E_\pi$
(i.e., $m_\pi$ and $q^2$ are varied subject to $E_\pi$=constant). The
interpolation is performed using a Ball-Zwicky (BZ) ansatz~\cite{Ball:2005tb}.  The $q^2$-dependence of
the resulting form factors in the chiral limit is then described by
means of a BZ ansatz, which is cross-checked
against Becirevic-Kaidalov (BK) \cite{Becirevic:1999kt}, Richard Hill (RH) \cite{Hill:2005ju},
and Boyd-Grinstein-Lebed (BGL) \cite{Boyd:1994tt} parameterizations
(see Appendix \ref{sec:zparam}),
finding agreement
within the quoted uncertainties. Unfortunately, the correlation matrix for the values
of the form factors at different $q^2$ is not provided, which severely
limits the possibilities of combining them with other computations into
a global $z$-parameterization.

\begin{table}[t]
\begin{center}
\mbox{} \\[3.0cm]
\footnotesize
\begin{tabular}{l @{\extracolsep{\fill}} c @{\hspace{2mm}} c l l l l l l l }
Collaboration & Ref. & $\Nf$ &
\hspace{0.15cm}\begin{rotate}{60}{publication status}\end{rotate}\hspace{-0.15cm} &
\hspace{0.15cm}\begin{rotate}{60}{continuum extrapolation}\end{rotate}\hspace{-0.15cm} &
\hspace{0.15cm}\begin{rotate}{60}{chiral extrapolation}\end{rotate}\hspace{-0.15cm}&
\hspace{0.15cm}\begin{rotate}{60}{finite volume}\end{rotate}\hspace{-0.15cm}&
\hspace{0.15cm}\begin{rotate}{60}{renormalization}\end{rotate}\hspace{-0.15cm}  &
\hspace{0.15cm}\begin{rotate}{60}{heavy-quark treatment}\end{rotate}\hspace{-0.15cm}  &
\hspace{0.15cm}\begin{rotate}{60}{$z$-parameterization}\end{rotate}\hspace{-0.15cm} \\%
&&&&&&&&& \\[-0.0cm]
\hline
\hline
&&&&&&&&& \\[-0.0cm]
\SLjlqcdBpi& \cite{Colquhoun:2022atw} & 2+1 & \gA & \good & \soso & \good & \soso & \okay &
 BCL\\[-0.0cm]
\SLfnalmilcBpi & \cite{Lattice:2015tia} & 2+1 & \gA  & \good & \soso & \good & \soso & \okay &
 BCL\\[-0.0cm]
\SLrbcukqcdBpi & \cite{Flynn:2015mha} & 2+1 & \gA  & \soso & \soso & \soso & \soso & \okay &
 BCL \\[-0.0cm]
HPQCD~06 & \cite{Dalgic:2006dt} & 2+1 & \gA  & \soso & \soso & \soso
& \soso & \okay &
 n/a \\[-0.0cm]
&&&&&&&&& \\[-0.0cm]
\hline
\hline
\end{tabular}
\caption{Results for the $B \to \pi\ell\nu$ semileptonic form factor.
\label{tab_BtoPisumm2}}
\end{center}
\end{table}

The different ways in which the current results are presented do not
allow a straightforward averaging procedure.
RBC/UKQCD~15 only provides synthetic
values of $f_+$ and $f_0$ at a few values of $q^2$ as an illustration
of their results, and FNAL/MILC~15 does not quote synthetic values at all.
In both cases, full results for BCL $z$-parameterizations defined by
Eq.~(\ref{eq:bcl_c}) are quoted.
In the case of HPQCD~06, unfortunately,
a fit to a BCL $z$-parameterization is not possible, as discussed above.

In order to combine these form factor calculations, we start from sets
of synthetic data for several $q^2$ values. HPQCD~06, RBC/UKQCD~15, and \SLjlqcdBpi~
directly provide this information; FNAL/MILC~15 present only fits to a
BCL $z$-parameterization from which we can easily generate an
equivalent set of form factor values. It is important to note that in
both the RBC/UKQCD~15 and \SLjlqcdBpi~synthetic data and the FNAL/MILC
$z$-parameterization fits the kinematic constraint at $q^2=0$ is
automatically included (in the FNAL/MILC~15 case the constraint is
manifest in an exact degeneracy of the $(a_n^+ ,a_n^0)$ covariance
matrix). Due to these considerations, in our opinion, the most accurate
procedure is to perform a simultaneous fit to all synthetic data for
the vector and scalar form factors. Unfortunately, the absence of
information on the correlation in the HPQCD~06 result between the vector
and scalar form factors even at a single $q^2$ point makes it
impossible to include consistently this calculation in the overall
fit. In fact, the HPQCD~06 and FNAL/MILC~15 statistical uncertainties are
highly correlated (because they are based on overlapping subsets of
MILC $\Nf=2+1$ ensembles) and, without knowledge of the $f_+ - f_0$
correlation we are unable to construct the HPQCD~06-FNAL/MILC~15
off-diagonal entries of the overall covariance matrix.

In conclusion, we will present as our best result a combined vector
and scalar form factor fit to the FNAL/MILC~15, RBC/UKQCD~15, and \SLjlqcdBpi~results that
we treat as completely uncorrelated. 

\begin{figure}[tbp]
\begin{center}
\includegraphics[width=0.49\textwidth]{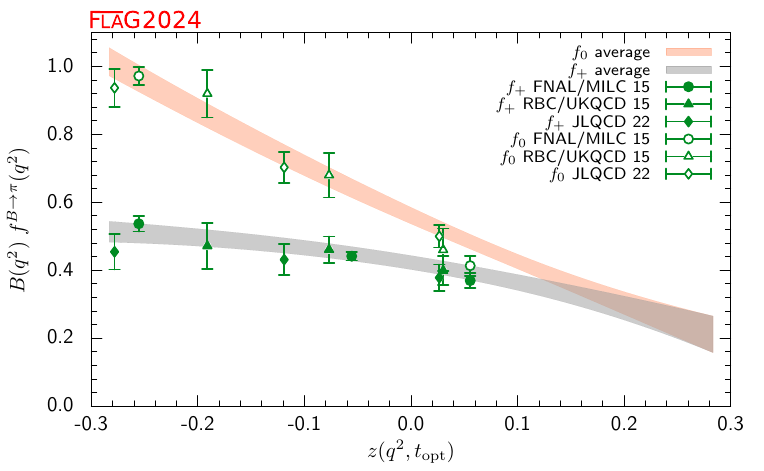}
\includegraphics[width=0.49\textwidth]{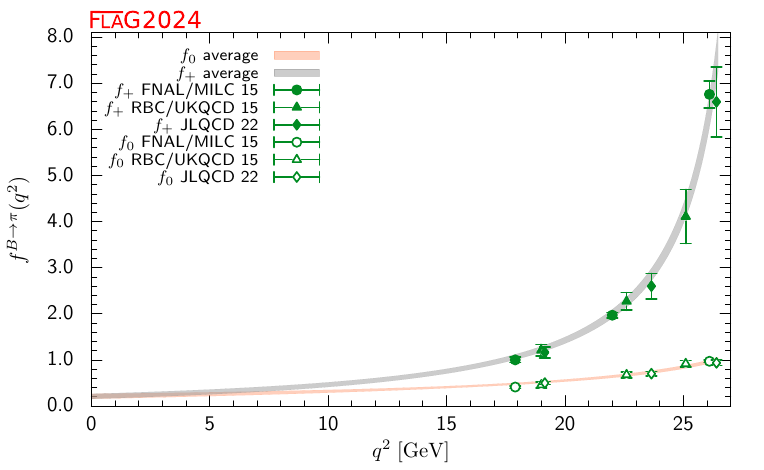}
\caption{The form factors $f_+(q^2)$ and $f_0 (q^2)$ for $B \to \pi\ell\nu$ plotted versus $z$ (left panel) and $q^2$ (right panel). In the left plot, we removed the Blaschke factors. See text for a discussion of the data set. The grey and salmon bands display our preferred $N^+=N^0=3$ BCL fit (five parameters).}\label{fig:LQCDzfit}
\end{center}
\end{figure}

The resulting data set is then fitted to the BCL parameterization in
Eqs.~(\ref{eq:bcl_c}) and (\ref{eq:bcl_f0}). We assess the systematic
uncertainty due to truncating the series expansion by considering fits
to different orders in $z$.  In Fig.~\ref{fig:LQCDzfit} (left), we show
$(1-q^2/m_{B^*}^2)
f_+(q^2)$ and $f_0 (q^2)$ versus $z$; Fig.~\ref{fig:LQCDzfit} (right) shows the full form factors versus $q^2$.
The fit has $\chi^2/{\rm dof} = 43.6/12$ with
$N^+ = N^0 = 3$. 
The poor quality of the fit is caused by tensions between the results from the different
collaborations; in particular in the slopes of $f_0$, which are very constrained due to strong correlations between data points. We have therefore
rescaled the uncertainties of the $z$ parameters by $\sqrt{\chi^2/{\rm dof}} = 1.9$.
We point out that tensions in the form factors, especially in $f_0$, might be an artifact associated with the basis of form factors employed to take the continuum limit, as explained in Appendix~\ref{sec:zparam}.
The outcome of the five-parameter $N^+ =N^0=3$ BCL fit to the FNAL/MILC~15, RBC/UKQCD~15, and \SLjlqcdBpi~calculations is shown in Tab.~\ref{tab:FFPI}. 

The fit shown in Tab.~\ref{tab:FFPI}
can therefore be used as the averaged FLAG result for the lattice-computed form
factor $f_+(q^2)$. The coefficient $a_3^+$ can be obtained from the
values for $a_0^+$--$a_2^+$ using Eq.~(\ref{eq:red_coeff}). The
coefficient $a_2^0$ can be obtained from all other coefficients
imposing the $f_+(q^2=0) = f_0(q^2=0)$ constraint.
\begin{table}[t]
\begin{center}
\begin{tabular}{|c|c|ccccc|}
\multicolumn{7}{l}{$B\to \pi \; (\Nf=2+1)$} \\[0.2em]\hline
        & Central Values & \multicolumn{5}{|c|}{Correlation Matrix} \\[0.2em]\hline
$a_0^+$ & 0.423 (21)  &   1 & $-$0.00466 & $-$0.0749 & 0.402 & 0.0920 \\[0.2em]
$a_1^+$ & $-$0.507 (93)  &   $-$0.00466 & 1 & 0.498 & $-$0.0556 & 0.659 \\[0.2em]
$a_2^+$ & $-$0.75 (34)  &  $-$0.0749 & 0.498 & 1 & $-$0.152 & 0.677 \\[0.2em]
$a_0^0$ & 0.561 (24)  &    0.402 & $-$0.0556 & $-$0.152 & 1 & $-$0.548  \\[0.2em]
$a_1^0$ & $-$1.42 (11)  &    0.0920 & 0.659 & 0.677 & $-$0.548 & 1 \\[0.2em]
\hline
\end{tabular}
\end{center}
\caption{Coefficients and correlation matrix for the $N^+ =N^0=3$ $z$-expansion fit of the $B\to \pi$ form factors $f_+$ and $f_0$. The coefficient $a_2^0$ is fixed by the $f_+(q^2=0) = f_0(q^2=0)$ constraint. The chi-square per degree of freedom is $\chi^2/{\rm dof} = 43.6/12$ and the errors on the $z$-parameters have been rescaled by  $\sqrt{\chi^2/{\rm dof}} = 1.9$. The lattice calculations that enter this fit are taken from \SLfnalmilcBpi~\cite{Lattice:2015tia}, \SLrbcukqcdBpi~\cite{Flynn:2015mha} and \SLjlqcdBpi~\cite{Colquhoun:2022atw}. The parameterizations are defined in Eqs.~(\ref{eq:bcl_c}) and (\ref{eq:bcl_f0}). The form factors can be reconstructed using parameterization and inputs given in Appendix~\ref{sec:app_B2pi}.
\label{tab:FFPI} }
\end{table}
We emphasize that future
lattice-QCD calculations of semileptonic form factors should publish
their full statistical and systematic correlation matrices to enable
others to use the data. It is also preferable to  present a
set of synthetic form-factor data equivalent to the $z$-fit results,
 since this allows for an independent
analysis that avoids further assumptions about the compatibility of
the procedures to arrive at a given $z$-parameterization.\footnote{
Note that generating synthetic data is a trivial task, but less so is choosing
the
number of required points and the $q^2$ values that lead to an optimal description of the form factors. }
 It is also preferable to present covariance/correlation matrices with enough significant digits to calculate correctly all their eigenvalues.

\subsubsection{Form factors for $B\to\rho\ell\nu$}
\label{sec:BtoRho}

Another process sensitive to $|V_{ub}|$ is $B\to\rho\ell\nu$, with experimental data available from Babar, Belle, and Belle II \cite{delAmoSanchez:2010af,Sibidanov:2013rkk,Belle-II:2022fsw}. Early lattice calculations of the $B\to \rho \ell\nu$ form factors were done in the quenched approximation and assumed the $\rho$ resonance to be stable under the strong interaction \cite{Bowler:2004zb,Flynn:2008zr}. {A proper treatment of the $\rho$ final state requires a calculation of the $B\to\pi\pi \ell\nu$ ($P$ wave) form factors as a function of both $q^2$ and $\pi\pi$ invariant mass, followed by analytic continuation to the $\rho$ resonance pole. On the lattice, this can be done using the Lellouch-L\"uscher finite-volume method} \cite{Luscher:1986pf,Luscher:1990ux,Luscher:1991cf,Lage:2009zv,Bernard:2010fp,Doring:2011vk,Hansen:2012tf,Briceno:2012yi,Dudek:2014qha,Briceno:2014uqa,Briceno:2015csa}.  Early lattice results for the $B\to\pi\pi \ell\nu$ $P$-wave vector form factor at $m_\pi\approx 320$ MeV were reported in Refs.~\cite{Leskovec:2022ubd,Leskovec:2024sfx}.

\subsubsection{Form factors for $B_s\to K\ell\nu$}
\label{sec:BstoK}
Similar to $B\to\pi\ell\nu$, measurements of $B_s\to K\ell\nu$ decay rates enable determinations
of the CKM matrix element $|V_{ub}|$
within the Standard Model via Eq.~(\ref{eq:B_semileptonic_rate}).
From the lattice point of view, the two channels are very similar.  As
a matter of fact, $B_s\to K\ell\nu$ is actually somewhat simpler,
in that the kaon mass region is easily accessed by all simulations
making the systematic uncertainties related to chiral extrapolation
smaller. Lattice calculations of the $B_s\to K$ form factors are available from \SLhpqcdBsK~\cite{Bouchard:2014ypa},
RBC/UKQCD~\cite{Flynn:2015mha,Flynn:2023nhi} (\SLrbcukqcdBpi~and RBC/UKQCD~23), and \SLfnalmilcBsK~\cite{Bazavov:2019aom}.

The \SLhpqcdBsK~computation uses ensembles of gauge configurations with $\Nf=2+1$
flavours of asqtad rooted staggered quarks produced by the MILC collaboration
at two values of the lattice spacing ($a\approx0.12,~0.09~{\rm fm}$), for three
and two different sea-pion masses, respectively, down to a value of $260~{\rm MeV}$.
The $b$ quark is treated within the NRQCD formalism, with a 1-loop matching
of the relevant currents to the ones in the relativistic theory, omitting terms
of $\cO(\alpha_s\Lambda_{\rm QCD}/m_b)$. The HISQ action
is used for the valence light and $s$ quarks. The continuum-chiral extrapolation
is combined with the description of the $q^2$-dependence of the form factors
into a modified $z$-expansion
(cf.~Appendix \ref{sec:zparam})
that formally coincides
in the continuum with the BCL ansatz. The dependence of
form factors on the pion energy and quark masses is fitted to a 1-loop ansatz
inspired by hard-pion $\chi$PT~\cite{Bijnens:2010ws},
that factorizes out the chiral logarithms describing soft physics.

The FNAL/MILC computation (\SLfnalmilcBsK) coincides with \SLhpqcdBsK~in using ensembles of gauge configurations with $\Nf=2+1$
flavours of asqtad rooted staggered quarks produced by the MILC collaboration, but
only one ensemble is shared, and a different valence regularization is employed;
we will thus treat the two results as fully independent from the statistics point of view.
\SLfnalmilcBsK~uses three values of the lattice spacing ($a\approx0.12,~0.09,~0.06~{\rm fm}$);
only one value of the sea pion mass and the volume is available at the extreme values
of the lattice spacing, while four different masses and volumes are considered at $a=0.09~{\rm fm}$.
Heavy quarks are treated within the Fermilab approach.
HMrS$\chi$PT expansion is used at next-to-leading
order in SU(2) and leading order in $1/M_B$, including next-to-next-to-leading-order
(NNLO) analytic and generic discretization terms, to perform continuum-chiral extrapolations.
Hard kaons are assumed to decouple, i.e., their effect is reabsorbed in the SU(2) LECs.
Continuum- and chiral-extrapolated values of the form factors are fitted to a $z$-parametrization
imposing the kinematical constraint $f_+(0)=f_0(0)$. See Tab.~\ref{tab_BstoKsumm} and the tables in Appendix~\ref{app:BtoPi_Notes} for full details.

The RBC/UKQCD~15 computation~\cite{Flynn:2015mha} had been published together with the $B\to\pi\ell\nu$
computation discussed in Sec.~\ref{sec:BtoPi}, all technical details being
practically identical. The RBC/UKQCD~23 computation~\cite{Flynn:2023nhi} (which considers $B_s\to K\ell\nu$ only)
differs from RBC/UKQCD~15 by the addition of one new ensemble with a third, finer lattice spacing that
also has a lower pion mass than the other ensembles, updated scale setting and updated tuning of $m_s$
and of the RHQ parameters, and a change of the form-factor basis in which the chiral-continuum extrapolation is
performed (previously: $f_\parallel$ and $f_\perp$, now: $f_+$ and $f_0$). RBC/UKQCD~23 \cite{Flynn:2023nhi} furthermore uses
a new method to perform extrapolations of the form factors to the full $q^2$ range with unitarity bounds, taking into account
that the dispersive integral ranges only over an arc of the unit circle instead of the full circle \cite{Flynn:2023qmi,Blake:2022vfl}.
However, we do not use these extrapolations in performing our average and instead use the synthetic data points
provided in RBC/UKQCD~23~\cite{Flynn:2023nhi}. This allows users of our average to impose their own
dispersive bounds in phenomenological applications if desired, since such bounds should be imposed only once.

\begin{table}[t]
\begin{center}
\mbox{} \\[3.0cm]
\footnotesize
\begin{tabular}{l  c @{\hspace{2mm}} c l l l l l l l l }
Collaboration & Ref. & $\Nf$ &
\hspace{0.15cm}\begin{rotate}{60}{publication status}\end{rotate}\hspace{-0.15cm} &
\hspace{0.15cm}\begin{rotate}{60}{continuum extrapolation}\end{rotate}\hspace{-0.15cm} &
\hspace{0.15cm}\begin{rotate}{60}{chiral extrapolation}\end{rotate}\hspace{-0.15cm}&
\hspace{0.15cm}\begin{rotate}{60}{finite volume}\end{rotate}\hspace{-0.15cm}&
\hspace{0.15cm}\begin{rotate}{60}{renormalization}\end{rotate}\hspace{-0.15cm}  &
\hspace{0.15cm}\begin{rotate}{60}{heavy-quark treatment}\end{rotate}\hspace{-0.15cm}  &
\hspace{0.15cm}\begin{rotate}{60}{$z$-parameterization}\end{rotate}\hspace{-0.15cm}\\%
&&&&&&&&& \\[-0.0cm]
\hline
\hline
&&&&&&&&& \\[-0.0cm]
RBC/UKQCD~23$^*$ & \cite{Flynn:2023nhi} & 2+1 & \gA  & \good & \soso & \good & \soso & \okay &
BGL$^{\S}$ \\[-0.0cm]
\SLfnalmilcBsK & \cite{Bazavov:2019aom} & 2+1 & \gA  & \good & \soso & \good & \soso & \okay &
BCL \\[-0.0cm]
\SLrbcukqcdBpi & \cite{Flynn:2015mha} & 2+1 & \gA  & \soso & \soso & \soso & \soso & \okay &
BCL \\[-0.0cm]
\SLhpqcdBsK & \cite{Bouchard:2014ypa} & 2+1 & \gA  & \soso & \soso & \soso & \soso & \okay &
BCL$^\dagger$   \\[-0.0cm]
&&&&&&&&& \\[-0.0cm]
\hline
\hline\\
\multicolumn{4}{l}{$^*$ Supersedes \SLrbcukqcdBpi.}\\
\multicolumn{4}{l}{$^{\S}$ Generalized as discussed in Ref.~\cite{Flynn:2023qmi}.}\\
\multicolumn{4}{l}{$^\dagger$ Results from modified $z$-expansion.}\\

\end{tabular}\\[-0.2cm]
\begin{minipage}{\linewidth}
\end{minipage}
\caption{Summary of lattice calculations of the $B_s \to K\ell\nu$ semileptonic form factors.
\label{tab_BstoKsumm}}
\end{center}
\end{table}

\begin{table}[t]
\begin{center}
\begin{tabular}{|c|c|ccccccc|}
\multicolumn{9}{l}{$B_s\to K \; (\Nf=2+1)$} \\[0.2em]\hline
        & Central Values & \multicolumn{7}{|c|}{Correlation Matrix} \\[0.2em]\hline
$a_0^+$ & 0.370(21)   & 1. & 0.2781 & $-0.3169$ & $-0.3576$ & 0.6130 & 0.3421 & 0.2826 \\[0.2em]
$a_1^+$ & $-0.68(10)$ & 0.2781 & 1. & 0.3672 & 0.1117 & 0.4733 & 0.8487 & 0.8141 \\[0.2em]
$a_2^+$ & 0.55(48)    &  $-0.3169$ & 0.3672 & 1. & 0.8195 & 0.3323 & 0.6614 & 0.6838 \\[0.2em]
$a_3^+$ & 2.11(83)    &  $-0.3576$ & 0.1117 & 0.8195 & 1. & 0.2350 & 0.4482 & 0.4877 \\[0.2em]
$a_0^0$ & 0.234(10)   & 0.6130 & 0.4733 & 0.3323 & 0.2350 & 1. & 0.6544 & 0.5189 \\[0.2em]
$a_1^0$ & 0.135(86)   & 0.3421 & 0.8487 & 0.6614 & 0.4482 & 0.6544 & 1. & 0.9440 \\ [0.2em]
$a_2^0$ & 0.20(35)    &  0.2826 & 0.8141 & 0.6838 & 0.4877 & 0.5189 & 0.9440 & 1. \\[0.2em]
\hline
\end{tabular}
\end{center}
\caption{Coefficients and correlation matrix for the $N^+ =N^0=4$ $z$-expansion of the $B_s\to K$ form factors $f_+$ and $f_0$. The coefficient $a_3^0$ is fixed by the $f_+(q^2=0) = f_0(q^2=0)$ constraint. The chi-square per degree of freedom is $\chi^2/{\rm dof} = 3.82$ and the errors on the $z$-parameters have been rescaled by  $\sqrt{\chi^2/{\rm dof}} = 1.95$. The form factors can be reconstructed using parameterization and inputs given in Appendix~\ref{sec:app_Bs2K}.
\label{tab:FFBSK}}
\end{table}

\begin{figure}[tbp]
\begin{center}
\includegraphics[width=0.49\textwidth]{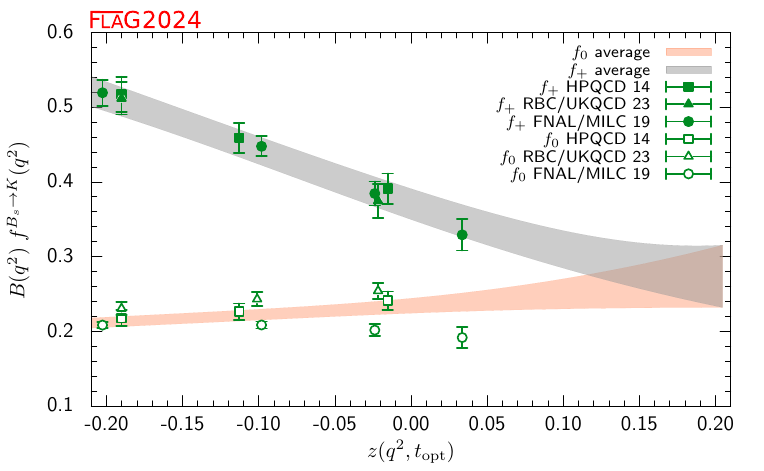}
\includegraphics[width=0.49\textwidth]{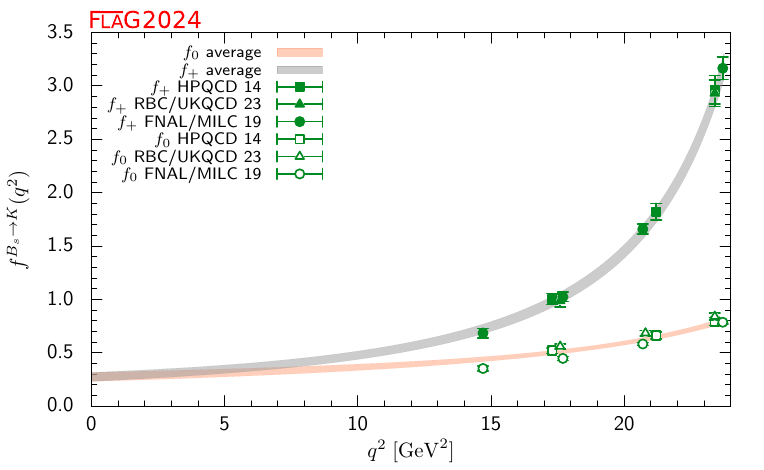}
\caption{The form factors $f_+(q^2)$ and $f_0(q^2)$ for $B_s \to K\ell\nu$ plotted versus $z$ (left panel) and $q^2$ (right panel). In the left plot, we remove the Blaschke factors. See text for a discussion of the data sets. The grey and salmon bands display our preferred $N^+=N^0=4$ BCL fit (seven parameters).}
\label{fig:LQCDzfitBsK}
\end{center}
\end{figure}

In order to combine the results for the $q^2$-dependence of the form factors from the
three collaborations, we will follow a similar approach to the one
adopted above for $B\to\pi\ell\nu$, and produce synthetic data
from the preferred fits quoted in the papers (or use the synthetic data provided in the paper),
to obtain a dataset to which a joint fit can be performed. Note that the kinematic
constraint at $q^2=0$ is included in all three cases; we will 
impose it in our fit as well, since the synthetic data will implicitly
depend on that fitting choice.  However, it is worth mentioning that
the systematic uncertainty of the resulting extrapolated value
$f_+(0)=f_0(0)$ can be fairly large, the main reason being the required long extrapolation
from the high-$q^2$ region covered by lattice data. While we stress that the
average far away from the high-$q^2$ region has to be used carefully, it is possible that 
increasing the number of $z$ coefficients beyond what is sufficient for a good description of
the lattice data and using unitarity constraints to control the size of additional terms, might 
yield fits with a more stable extrapolation at very low $q^2$. We plan to include said unitarity analysis
into the next edition of the FLAG review. It is, however, important to emphasize
that joint fits with experimental data, where the latter accurately
map the low $q^2$ region, are expected to be safe.

Our fits employ a BCL ansatz with $t_+=(M_{B}+M_{\pi})^2$ and
$t_0=t_+-\sqrt{t_+(t_+ - t_-)}$, with $t_-=(M_{B_s}-M_K)^2$.
Our pole factors will contain a single pole in both the vector and scalar
channels, for which we take the mass values $M_{B^*}=5.32465~\GeV$
and $M_{B^*(0^+)}=5.68~\GeV$.\footnote{These are the values used in the
\SLfnalmilcBsK~determination, while \SLhpqcdBsK~and \SLrbcukqcdBpi~use $M_{B^*(0+)}=5.6794(10)~\GeV$
and $M_{B^*(0+)}=5.63~\GeV$, respectively. They also employ different values of $t_+$
and $t_0$ than employed here, which again coincide with \SLfnalmilcBsK's choice.}
The constraint $f_+(0)=f_0(0)$ is imposed by expressing the coefficient $b^0_{N^0-1}$ in
terms of all others.
The outcome of the seven-parameter $N^+ = N^0 = 4$ BCL fit, which we
quote as our preferred result, is shown in Tab.~\ref{tab:FFBSK}. The fit has a chi-square per degree of freedom $\chi^2/{\rm dof} = 3.82$. Following the PDG recommendation, we rescale the whole covariance matrix by $\chi^2/{\rm dof}$: the errors on the $z$-parameters are increased by $\sqrt{\chi^2/{\rm dof}} = 1.95$ and the correlation matrix is unaffected. The parameters shown in Tab.~\ref{tab:FFBSK} provide the averaged FLAG results
for the lattice-computed form factors $f_+(q^2)$ and $f_0(q^2)$. The
coefficient $a_4^+$ can be obtained from the values for
$a_0^+$--$a_3^+$ using Eq.~(\ref{eq:red_coeff}). The fit is
illustrated in Fig.~\ref{fig:LQCDzfitBsK}.\footnote{Note that in FLAG~19 \cite{FlavourLatticeAveragingGroup:2019iem} we had adopted the threshold $t_+=(M_{B_s}+M_{K})^2$ rather than $t_+=(M_{B}+M_{\pi})^2$. This change impacted the $z$-range which the physical $q^2$ interval maps onto. We also point out that, in the FLAG~19 version of Fig.~\ref{fig:LQCDzfitBsK}, the three synthetic $f_0$ data points from HPQCD were plotted incorrectly, but this did not affect the fit.}
As can be seen in Fig.~\ref{fig:LQCDzfitBsK}, the large value of $\chi^2/{\rm dof}$ is caused by a significant tension between the lattice results from the different collaborations for $f_0$. Compared to the FLAG~21 fit that used RBC/UKQCD~15, the tension has increased as the RBC/UKQCD results for $f_0$ have shifted upward. The tension
indicates that the uncertainties have been underestimated in at least some of the calculations. One possible, at least partial, explanation was offered by the
authors of RBC/UKQCD~23 \cite{Flynn:2023nhi}, who found that the results for $f_0$ shift upward when performing the chiral/continuum extrapolation directly for $f_0$ and $f_+$ rather than $f_\parallel$ and $f_\perp$ as was done in RBC/UKQCD~15 and FNAL/MILC~19. Using $f_0$ and $f_+$ is argued to be the better choice because these form factors have definite $J^P$ quantum numbers for the bound states producing poles in $q^2$, and the chiral-continuum extrapolation fit functions include these poles.
More details on the problems associated with taking the chiral/continuum extrapolation in the $f_\parallel$ and $f_\perp$ basis can be found in Appendix~\ref{sec:zparam}.

A number of new calculations of the $B_s\to K$ form factors are underway. The JLQCD
collaboration is using a fully-relativistic approach with M\"obius domain-wall fermions \cite{Mohanta:2024hzi}.
FNAL/MILC is pursuing two new calculations with HISQ light quarks, one of which uses
Fermilab $b$ quarks \cite{Jeong:2024glv} and the other uses HISQ $b$ quarks \cite{Lytle:2024zfr}.

We will conclude by pointing out progress in the application of the npHQET method to the
extraction of semileptonic form factors, reported for $B_s \to K$ transitions in Ref.~\cite{Bahr:2019eom},
which extends the work of Ref.~\cite{Bahr:2014iqa}.
This is a methodological study based on CLS $\Nf=2$ ensembles at two different values of the
lattice spacing and pion masses, and full $1/m_b$ corrections are incorporated within the npHQET
framework. Emphasis is on the role of excited states in the extraction of the bare form factors,
which are shown to pose an impediment to reaching precisions better than a few percent.

\subsubsection{Form factors for rare and radiative $B$-semileptonic decays to light flavours}\label{sec:SLBrad}

Lattice-QCD input is also available for some exclusive semileptonic
decay channels involving neutral-current $b\to q$ transitions at the
quark level, where $q=d,s$. Being forbidden at tree level in the SM,
these processes allow for stringent tests of potential new physics;
simple examples are $B\to K^*\gamma$, $B\to K^{(\ast)}\ell^+\ell^-$, or
$B\to\pi\ell^+\ell^-$ where the $B$ meson (and therefore the light
meson in the final state) can be either neutral or charged.

The corresponding SM effective weak Hamiltonian is considerably more
complicated than the one for the tree-level processes discussed above:
after integrating out the top quark and the $W$ boson, as many as ten
dimension-six operators formed by the product of two hadronic currents
or one hadronic and one leptonic current appear.\footnote{See, e.g.,
  Ref.~\cite{Antonelli:2009ws} and references therein.}  Three of the
latter, coming from penguin and box diagrams, dominate at short
distances and have matrix elements that, up to small QED corrections,
are given entirely in terms of $B\to (\pi,K,K^*)$ form factors. The
matrix elements of the remaining seven operators can be expressed, up
to power corrections whose size is still unclear, in terms of form
factors, decay constants and light-cone distribution amplitudes (for
the $\pi$, $K$, $K^*$ and $B$ mesons) by employing OPE arguments (at
large di-lepton invariant mass)~\cite{Grinstein:2004vb, Beylich:2011aq} 
and results from QCD factorization (at small di-lepton invariant mass)~\cite{Beneke:2001at}. In conclusion,
the most important contributions to all of these decays are expected
to come from matrix elements of current operators (vector, tensor, and
axial-vector) between one-hadron states, which in turn can be
parameterized in terms of a number of form factors (see
Ref.~\cite{Liu:2009dj} for a complete description).

\begin{table}[t]
\begin{center}
\mbox{} \\[3.0cm]
\footnotesize
\begin{tabular}{l  c @{\hspace{2mm}} c l l l l l l l l }
Collaboration & Ref. & $\Nf$ &
\hspace{0.15cm}\begin{rotate}{60}{publication status}\end{rotate}\hspace{-0.15cm} &
\hspace{0.15cm}\begin{rotate}{60}{continuum extrapolation}\end{rotate}\hspace{-0.15cm} &
\hspace{0.15cm}\begin{rotate}{60}{chiral extrapolation}\end{rotate}\hspace{-0.15cm}&
\hspace{0.15cm}\begin{rotate}{60}{finite volume}\end{rotate}\hspace{-0.15cm}&
\hspace{0.15cm}\begin{rotate}{60}{renormalization}\end{rotate}\hspace{-0.15cm}  &
\hspace{0.15cm}\begin{rotate}{60}{heavy-quark treatment}\end{rotate}\hspace{-0.15cm}  &
\hspace{0.15cm}\begin{rotate}{60}{$z$-parameterization}\end{rotate}\hspace{-0.15cm}\\%
&&&&&&&&& \\[-0.0cm]
\hline
\hline
&&&&&&&&& \\[-0.0cm]
HPQCD~22 & \cite{Parrott:2022rgu} & 2+1+1 & \gA  & \good & \good & \good & \good & \okay &
 BCL \\[-0.0cm]
\SLfnalmilcBK & \cite{Bailey:2015dka} & 2+1 & \gA  & \good & \soso & \good & \soso & \okay &
 BCL \\[-0.0cm]
\SLhpqcdBK & \cite{Bouchard:2013pna} & 2+1 & \gA  & \soso & \soso & \soso & \soso & \okay &
BCL   \\[-0.0cm]
&&&&&&&&& \\[-0.0cm]
\hline
\hline
\end{tabular}
\caption{Summary of lattice calculations of the $B \to K$ semileptonic form factors.
\label{tab_BtoKsumm}}
\end{center}
\end{table}

In channels with pseudoscalar mesons in the final state, the level of
sophistication of lattice calculations is similar to the $B\to \pi$
case. Early calculations of the vector, scalar, and tensor form
factors for $B\to K\ell^+\ell^-$ by HPQCD~13E~\cite{Bouchard:2013pna} and FNAL/MILC~15D \cite{Bailey:2015dka}
were performed with $\Nf=2+1$ flavours and EFT-based heavy-quark actions. FNAL/MILC~15E also determined the form factors for $B\to\pi\ell^+\ell^-$~\cite{Bailey:2015nbd}.
Recently, HPQCD completed a new calculation of the $B\to K$ form factors with $\Nf=2+1+1$ flavours and HISQ $b$ quarks (HPQCD~22) \cite{Parrott:2022rgu}. In the following, we present an average of the two $\Nf=2+1$ calculations and a comparison with HPQCD's new $\Nf=2+1+1$ results. Details of the calculations are provided in Tab.~\ref{tab_BtoKsumm} and in Appendix~\ref{app:BtoK_Notes}.

The $\Nf=2+1$ calculations both employ MILC asqtad ensembles.  HPQCD~13E \cite{Bouchard:2013mia}
and FNAL/MILC~15D~\cite{Du:2015tda} have also companion papers in which
they calculate the Standard Model predictions for the differential
branching fractions and other observables and compare to experiment.
The HPQCD computation employs NRQCD $b$ quarks and HISQ valence light
quarks, and parameterizes the form factors over the full kinematic
range using a model-independent $z$-expansion as in
Appendix~\ref{sec:zparam},
including the covariance matrix of the fit
coefficients.  In the case of the (separate) FNAL/MILC computations,
both of them use Fermilab $b$ quarks and asqtad light quarks, and a
BCL $z$-parameterization of the form factors.

FNAL/MILC~15E \cite{Bailey:2015nbd} includes results for the tensor form factor
for $B\to\pi\ell^+\ell^-$ not included in previous publications on the
vector and scalar form factors (FNAL/MILC~15) \cite{Lattice:2015tia}.
Nineteen ensembles from four lattice
spacings are used to control continuum and chiral extrapolations.
The results for $N_z=4$ $z$-expansion of the tensor form factor and its
correlations with the expansions for the vector and scalar form factors presented in Table~II of Ref.~\cite{Bailey:2015nbd}, which we consider
the FLAG estimate, are shown in Tab.~\ref{tab:FFPIT}.
Partial decay widths for decay into light leptons or
$\tau^+\tau^-$ are presented as a function of $q^2$.  The former is
compared with results from LHCb~\cite{Aaij:2015nea}, while the
latter is a prediction.
\begin{table}[t]
\begin{center}
\begin{tabular}{|c|c|cccc|}
\multicolumn{6}{l}{$B\to \pi \; (\Nf=2+1)$} \\[0.2em]\hline
        & Central Values & \multicolumn{4}{|c|}{Correlation Matrix} \\[0.2em]\hline
$a_0^T$ & 0.393(17)   & 1.000  & 0.400 & 0.204 & 0.166 \\[0.2em]
$a_1^T$ & $-$0.65(23) & 0.400  & 1.000 & 0.862 & 0.806 \\[0.2em]
$a_2^T$ & $-$0.6(1.5) & 0.204  & 0.862 & 1.000 & 0.989 \\[0.2em]
$a_3^T$ & 0.1(2.8)    & 0.166  & 0.806 & 0.989 & 1.000 \\[0.2em]
\hline
\end{tabular}
\end{center}
\caption{Coefficients and correlation matrix for the $N^T =4$ $z$-expansion of the $B\to \pi$ form factor $f_T$. Results taken from Table~II of Ref.~\cite{Bailey:2015nbd}. \label{tab:FFPIT}}
\end{table}

\begin{figure}[tbp]
\begin{center}
\includegraphics[width=0.49\textwidth]{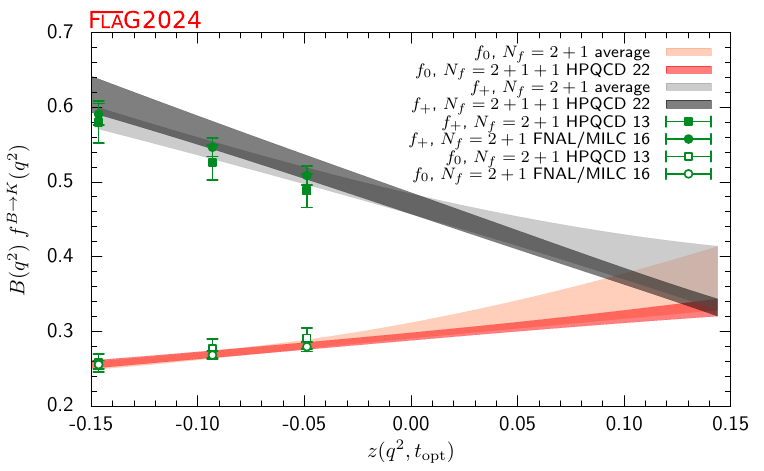}
\includegraphics[width=0.49\textwidth]{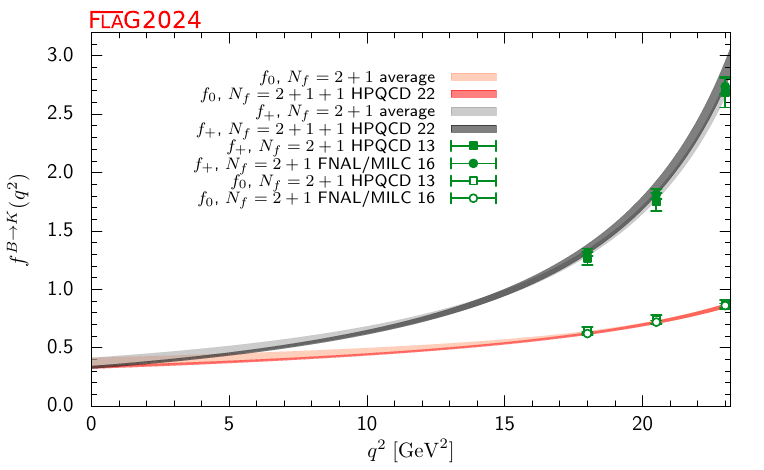}
\includegraphics[width=0.49\textwidth]{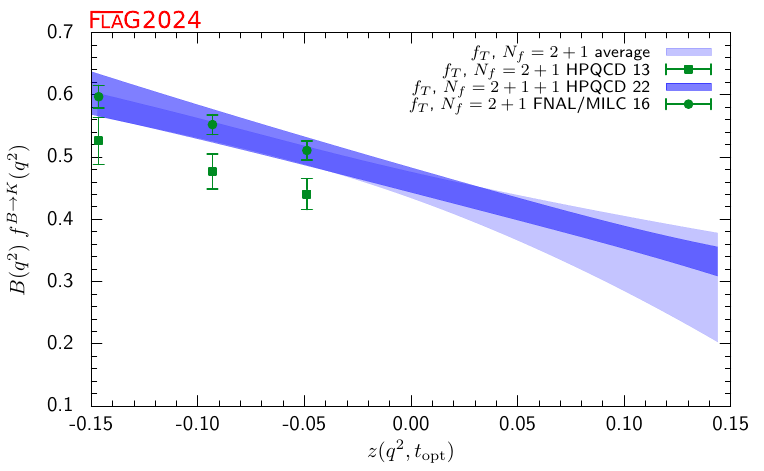}
\includegraphics[width=0.49\textwidth]{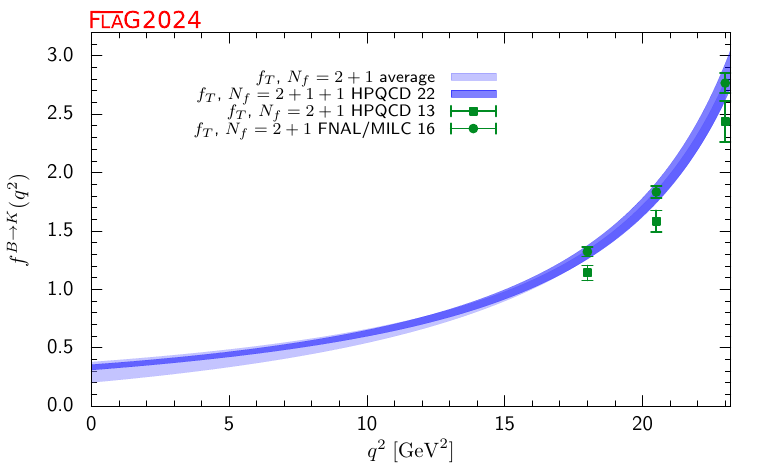}
\caption{The $B\to K$ form factors $f_+(q^2)$, $f_0(q^2)$ and $f_T(q^2)$
  plotted versus $z$ (left panels) and $q^2$ (right panels). In the plots as a function of $z$, we remove the Blaschke factors. See text for a discussion of the data sets. The light-shaded grey, salmon and blue bands display our preferred $N^+=N^0=N^T=3$ BCL fit (eight parameters) to the $\Nf=2+1$ lattice results. The dark-shaded grey, salmon and blue bands display the $\Nf=2+1+1$ HPQCD~22 results \cite{Parrott:2022rgu}.}
\label{fig:LQCDzfitBK}
\end{center}
\end{figure}

The averaging of the HPQCD~13E and FNAL/MILC~15D $\Nf=2+1$ results for the $B\to K$ form factors is similar to our
treatment of the $B\to \pi$ and $B_s\to K$ form factors. In this case,
even though the statistical uncertainties are partially correlated
because of some overlap between the adopted sets of MILC ensembles, we
choose to treat the two calculations as independent. The reason is
that, in $B\to K$, statistical uncertainties are subdominant and
cannot be easily extracted from the results presented by HPQCD~13E and
FNAL/MILC~15D. Both collaborations provide only the outcome of a
simultaneous $z$-fit to the vector, scalar and tensor form factors,
that we use to generate appropriate synthetic data. We then impose the
kinematic constraint $f_+(q^2=0) = f_0(q^2=0)$ and fit to a $(N^+ = N^0
= N^T = 3)$ BCL parameterization. The functional forms of the form
factors that we use are identical to those adopted in
Ref.~\cite{Du:2015tda}.\footnote{Note in particular that not much is
  known about the sub-threshold poles for the scalar form
  factor. FNAL/MILC~15D includes one pole at the $B_{s0}^*$ mass as taken
  from the calculation in Ref.~\cite{Lang:2015hza}.} The results of the fit are
  presented in Tab.~\ref{tab:FFK}. The fit is illustrated in Fig.~\ref{fig:LQCDzfitBK}. Note that the
average for the $f_T$ form factor appears to prefer the FNAL/MILC~15D
synthetic data. This happens because we perform a correlated fit of
the three form factors simultaneously (both FNAL/MILC~15D and HPQCD~13E
present covariance matrices that include correlations between all form
factors). We checked that the average for the $f_T$ form factor,
obtained neglecting correlations with $f_0$ and $f_+$, is a little
lower and lies in between the two data sets.
There is still a noticeable tension between the FNAL/MILC~15D and HPQCD~13E data
for the tensor form factor; indeed, a standalone fit to these data results
in $\chi^2_{\rm\scriptscriptstyle red}=7.2/3=2.4$, while a similar standalone
joint fit to $f_+$ and $f_0$ has $\chi^2_{\rm\scriptscriptstyle red}=9.2/7=1.3$.
Finally, the global fit that is shown in the figure has $\chi^2_{\rm\scriptscriptstyle red}=18.6/10=1.86$.
\begin{table}[t]
\begin{center}
\begin{tabular}{|c|c|cccccccc|}
\multicolumn{10}{l}{$B\to K \; (\Nf=2+1)$} \\[0.2em]\hline
        & Central Values & \multicolumn{8}{|c|}{Correlation Matrix} \\[0.2em]\hline
$a_0^+$ & 0.471 (14)   & 1 & 0.513 & 0.128 & 0.773 & 0.594 & 0.613 & 0.267 & 0.118   \\[0.2em]
$a_1^+$ & $-$0.74 (16) & 0.513 & 1 & 0.668 & 0.795 & 0.966 & 0.212 & 0.396 & 0.263   \\[0.2em]
$a_2^+$ & 0.32 (71)    & 0.128 & 0.668 & 1 & 0.632 & 0.768 & $-$0.104 & 0.0440 & 0.187 \\[0.2em]
$a_0^0$ & 0.301 (10)   & 0.773 & 0.795 & 0.632 & 1 & 0.864 & 0.393 & 0.244 & 0.200   \\[0.2em]
$a_1^0$ & 0.40 (15)    & 0.594 & 0.966 & 0.768 & 0.864 & 1 & 0.235 & 0.333 & 0.253   \\[0.2em]
$a_0^T$ & 0.455 (21)   & 0.613 & 0.212 & $-$0.104 & 0.393 & 0.235 & 1 & 0.711 & 0.608  \\[0.2em]
$a_1^T$ & $-$1.00 (31) & 0.267 & 0.396 & 0.0440 & 0.244 & 0.333 & 0.711 & 1 & 0.903  \\[0.2em]
$a_2^T$ & $-$0.9 (1.3) & 0.118 & 0.263 & 0.187 & 0.200 & 0.253 & 0.608 & 0.903 & 1   \\[0.2em]
\hline
\end{tabular}
\end{center}
\caption{Coefficients and correlation matrix for the $N^+ =N^0=N^T=3$ $z$-expansion of the $B\to K$ form factors $f_+$, $f_0$ and $f_T$ for $\Nf=2+1$. The coefficient $a_2^0$ is fixed by the $f_+(q^2=0) = f_0(q^2=0)$ constraint. The chi-square per degree of freedom is $\chi^2/{\rm dof} = 1.86$ and the errors on the $z$-parameters have been rescaled by  $\sqrt{\chi^2/{\rm dof}} = 1.36$. The form factors can be reconstructed using parameterization and inputs given in Appendix~\ref{sec:app_B2K}.\label{tab:FFK}}
\end{table}

The new $\Nf=2+1+1$ HPQCD~22 calculation of the $B\to K$ form factors~\cite{Parrott:2022rgu} uses the HISQ action for all quarks including the $b$ quark, which
allows the determination of the vector- and axial-vector-current renormalization factors using Ward identities. The tensor current is renormalized using RI-SMOM~\cite{Hatton:2020vzp}.
The calculation is performed for multiple lighter-than-physical values of the heavy-quark mass
and six different lattice spacings down to 0.044~fm; at the finest lattice spacing, the heavy-light pseudoscalar mass reaches
approximately $0.94 M_{B,{\rm phys}}$. Three of the eight ensembles used have an approximately physical pion mass. The form factors in the physical limit are extracted from a modified BCL $z$-expansion fit with terms incorporating dependence on the heavy-quark mass, light and strange-quark masses, lattice spacing, and cover the entire $q^2$ range. The paper \cite{Parrott:2022rgu} includes supplemental files with the form-factor parameters and a Python code that can be used to reconstruct the form factors. The form factors are shown in Fig.~\ref{fig:LQCDzfitBK} with the dark-shaded bands and are seen to be consistent with our average of the older $\Nf=2+1$ results. The $\Nf=2+1+1$ form factors are substantially more precise at low $q^2$ and somewhat less precise at high $q^2$. Standard-Model predictions  $B\to K\ell^+\ell^-$ and $B\to K\nu\bar{\nu}$ using these form factors are presented in a separate paper \cite{Parrott:2022zte}.

Lattice computations of form factors in channels with a vector meson
in the final state face extra challenges with respect to the case of a
pseudoscalar meson: the state is unstable, and the extraction of the
relevant matrix element from correlation functions is significantly
more complicated; $\chi$PT cannot be used as a guide to extrapolate
results at unphysically-heavy pion masses to the chiral limit. While
field-theory procedures to take resonance effects into account are
available~\cite{Luscher:1986pf,Luscher:1990ux,Luscher:1991cf,Lage:2009zv,Bernard:2010fp,Doring:2011vk,Hansen:2012tf,Briceno:2012yi,Dudek:2014qha,Briceno:2014uqa,Briceno:2015csa},
they have not yet been implemented in the available  
computations of $B\to K^*$ and similar form factors, which therefore suffer from uncontrolled systematic
errors (however, new calculations using these procedures are underway \cite{Leskovec:2024sfx}).\footnote{In
  cases such as $B\to D^\ast$ transitions, that will be discussed below,
  this is much less of a practical problem due to the very narrow
  nature of the resonance.}

As a consequence of the complexity of the problem, the level of maturity
of these computations is significantly below the one present for pseudoscalar form factors.
Therefore, we only provide a short guide to the existing results.
Horgan {\it et al.} have obtained the seven form factors governing $B \to K^* \ell^+
\ell^-$ (as well as those for $B_s \to \phi\, \ell^+ \ell^-$ and for the charged-current decay $B_s \to K^* \ell\nu$) in
Ref.~\cite{Horgan:2013hoa} using NRQCD $b$ quarks and asqtad staggered
light quarks.  In this work, they use a modified $z$-expansion to
simultaneously extrapolate to the physical light-quark masses and
fit the $q^2$-dependence.  As discussed above, the unstable nature of the vector mesons was not taken
into account. Horgan {\it et al.} use
their form-factor results to calculate the differential branching
fractions and angular distributions and discuss the implications for
phenomenology in a companion paper~\cite{Horgan:2013pva}. An update
of the form factor fits that enforces endpoint relations and also provides
the full correlation matrices can be found in Ref.~\cite{Horgan:2015vla}.
Finally, preliminary results on $B\to K^*\ell^+\ell^-$ and $B_s\to \phi\ell^+\ell^-$
by RBC/UKQCD have been reported in Refs.~\cite{Flynn:2015ynk,Flynn:2016vej,Lizarazo:2016myv}.

%% file: HQ/HQSubsections/BSL_B_to_heavy.tex
\subsection{Semileptonic form factors for $B_{(s)} \to D_{(s)} \ell \nu$ and $B_{(s)} \to D^\ast_{(s)}  \ell \nu$}
\label{sec:BtoD}

The semileptonic processes $ B_{(s)} \rightarrow D_{(s)} \ell \nu$ and
$B_{(s)} \rightarrow D^\ast_{(s)} \ell \nu$ have been studied
extensively by experimentalists and theorists over the years.  They
allow for the determination of the CKM matrix element $|V_{cb}|$, an
extremely important parameter of the Standard Model. The matrix
element $V_{cb}$
appears in many quantities that serve as inputs to CKM unitarity-triangle 
analyses and reducing its uncertainties is of paramount
importance.  For example, when $\epsilon_K$, the measure of indirect
CP violation in the neutral kaon system, is written in terms of the
parameters $\rho$ and $\eta$ that specify the apex of the unitarity
triangle, a factor of $|V_{cb}|^4$ multiplies the dominant term.  As a
result, the errors coming from $|V_{cb}|$ (and not those from $B_K$)
are now the dominant uncertainty in the Standard Model (SM) prediction
for this quantity.

\subsubsection{ $B_{(s)} \rightarrow D_{(s)}$ decays}
\label{sec:BstoDsFFs}

The decay rate for $B \rightarrow D\ell\nu$ can be parameterized in terms of vector and scalar form factors in the same way as, e.g., $B\to\pi\ell\nu$ (see Sec.~\ref{sec:BtoPiK}). The quantities directly studied are the form factors $h_\pm$ defined by
\begin{equation}
\frac{\langle D(p_D)| i\bar c \gamma_\mu b| B(p_B)\rangle}{\sqrt{m_D m_B}} =
h_+(w)(v_B+v_D)_\mu\,+\,h_-(w)(v_B-v_D)_\mu\,,
\label{B2D-formfactor}
\end{equation}
which are related to the standard vector and scalar form factors by
\begin{align}
  f_+(q^2) &= \frac{1+r}{2\sqrt{r}} \left[h_+(w)-\frac{1-r}{1+r} h_-(w)\right] \equiv \frac{1+r}{2\sqrt{r}} {\cal G}(q^2), \label{defG}\\
  f_0(q^2) &= \sqrt{r} \left[\frac{1+w}{1+r} h_+(w) + \frac{1-w}{1-r}h_-(w) \right],
\end{align}
where $r=m_D/m_B$, $q^2=(p_B-p_D)^2$, $v_A^\mu = p_A^\mu/m_A\; (A=D,B)$ are the four-velocities of the heavy mesons and $w = v_B \cdot v_D = (m_B^2 + m_D^2 - q^2)/(2 m_B m_D)$.

The differential decay rate can then be written as
\begin{align}
    \frac{d\Gamma_{B^-\to D^{0} \ell^-\bar{\nu}}}{dw} = &
        \frac{G^2_{\rm F} m^3_{D}}{48\pi^3}(m_B+m_{D})^2(w^2-1)^{3/2}  |\eta_\mathrm{EW}|^2|V_{cb}|^2 |\mathcal{G}(w)|^2,
    \label{eq:vxb:BtoD}
\end{align}
where $\eta_\mathrm{EW}=1.0066$ is the 1-loop electroweak correction~\cite{Sirlin:1981ie}. This formula does not include terms
 that are proportional to the lepton mass squared, which can be neglected for $\ell = e, \mu$.  
 
Until recently, most unquenched lattice calculations for $B \rightarrow D \ell \nu$ decays focused on the form factor at zero recoil ${\cal G}^{B \rightarrow D}(1)$, which can then be combined with experimental input to extract $|V_{cb}|$.
The main reasons for concentrating on the zero-recoil point are that (i) the decay rate then depends on a single form factor, and (ii) there are no $\cO(\Lambda_{QCD}/m_Q)$ contributions due to Luke's theorem~\cite{Luke:1990eg}.                
Since HQET sets $\lim_{m_Q\to\infty}{\cal G}^{B \rightarrow D}(1) = 1$~\cite{Isgur:1989ed,Isgur:1989vq,Neubert:1993mb}, high precision calculations of ${\cal G}^{B \rightarrow D}(1)$ are possible~\cite{Falk:1992wt,Neubert:1992tg,Neubert:1994qt}.
The application of these HQET developments to lattice calculations leads to a better control of the systematic errors, especially at zero recoil~\cite{Kronfeld:2000ck,Harada:2001fj}.
In particular, the zero-recoil form factor can be computed via a double ratio in which most of the current renormalization cancels and heavy-quark discretization errors are suppressed by an additional power of $\Lambda_{\rm QCD}/m_Q$~\cite{Hashimoto:1999yp}.

\begin{table}
\begin{center}
\mbox{} \\[3.0cm]
\footnotesize\hspace{-0.2cm}
\begin{tabular*}{\textwidth}[l]{l @{\extracolsep{\fill}} r l l l l l l l c l}
Collaboration & Ref. & $\Nf$ &
\hspace{0.15cm}\begin{rotate}{60}{publication status}\end{rotate}\hspace{-0.15cm} &
\hspace{0.15cm}\begin{rotate}{60}{continuum extrapolation}\end{rotate}\hspace{-0.15cm} &
\hspace{0.15cm}\begin{rotate}{60}{chiral extrapolation}\end{rotate}\hspace{-0.15cm}&
\hspace{0.15cm}\begin{rotate}{60}{finite volume}\end{rotate}\hspace{-0.15cm}&
\hspace{0.15cm}\begin{rotate}{60}{renormalization}\end{rotate}\hspace{-0.15cm}  &
\hspace{0.15cm}\begin{rotate}{60}{heavy-quark treatment}\end{rotate}\hspace{-0.15cm}  &
\multicolumn{2}{l}{\hspace{0mm} $w=1$ form factor / ratio}\\
&&&&&&&&&& \\[-0.1cm]
\hline
\hline
&&&&&&&&& \\[-0.1cm]
\SLhpqcdBD, HPQCD~17 & \cite{Na:2015kha,Monahan:2017uby} & 2+1   & \gA & \soso & \soso & \soso     & \soso & \okay & ${\mathcal G}^{B\to D}(1)$       & 1.035(40)      \\[0.5ex]
\SLfnalmilcBD        & \cite{Lattice:2015rga}            & 2+1   & \gA & \good & \soso & \good     & \soso & \okay & ${\mathcal G}^{B\to D}(1)$       & 1.054(4)(8)    \\[0.5ex]
&&&&&&&&& \\[-0.1cm]
\hline
&&&&&&&&& \\[-0.1cm]
HPQCD~19             & \cite{McLean:2019qcx}             & 2+1+1 & \gA & \good & \good & \soso$^*$ & \good & \okay & ${\mathcal G}^{B_s\to D_s}(1)$   & 1.071(37)      \\[0.5ex]
\SLhpqcdBD, HPQCD~17 & \cite{Na:2015kha,Monahan:2017uby} & 2+1   & \gA & \soso & \soso & \soso     & \soso & \okay & ${\mathcal G}^{B_s\to D_s}(1)$   & 1.068(40)      \\[0.5ex]
&&&&&&&&& \\[-0.1cm]
\hline
&&&&&&&&& \\[-0.1cm]
FNAL/MILC~21         & \cite{FermilabLattice:2021cdg}    & 2+1   & \gA & \good & \soso & \good     & \soso & \okay & ${\mathcal F}^{B\to D^\ast}(1)$  & 0.909(17)      \\[0.5ex]
JLQCD~23             & \cite{Aoki:2023qpa}               & 2+1   & \gA & \good & \soso & \good     & \soso & \okay & ${\mathcal F}^{B\to D^\ast}(1)$  & 0.887 (14)     \\[0.5ex]
HPQCD~23             & \cite{Harrison:2023dzh}           & 2+1+1 & \gA & \good & \good & \good     & \good & \okay & ${\mathcal F}^{B\to D^\ast}(1)$  & 0.903(14)      \\[0.5ex]
&&&&&&&&& \\[-0.1cm]
\hline
&&&&&&&&& \\[-0.1cm]
HPQCD~23             & \cite{Harrison:2023dzh}           & 2+1+1 & \gA & \good & \good & \good     & \good & \okay & ${\mathcal F}^{B_s\to D_s^*}(1)$ & 0.8970(92)     \\[0.5ex]
&&&&&&&&& \\[-0.1cm]
\hline
&&&&&&&&& \\[-0.1cm]
\SLhpqcdBD, HPQCD~17 & \cite{Na:2015kha,Monahan:2017uby} & 2+1   & \gA & \soso & \soso & \soso     & \soso & \okay & ${\mathcal G}^{B_s\to D_s}(1)$   & 1.068(40)      \\[0.5ex]
&&&&&&&&& \\[-0.1cm]
\hline
&&&&&&&&& \\[-0.1cm]
HPQCD~20B            & \cite{Harrison:2020gvo}           & 2+1+1 & \gA & \good & \soso & \good     & \good & \okay &                n/a               &     n/a        \\[0.5ex]
&&&&&&&&& \\[-0.1cm]
\hline
&&&&&&&&& \\[-0.1cm]
\SLhpqcdBD, HPQCD~17 & \cite{Na:2015kha,Monahan:2017uby} & 2+1   & \gA & \soso & \soso & \soso     & \soso & \okay & $R(D)$                           & 0.300(8)       \\[0.5ex]
\SLfnalmilcBD        & \cite{Lattice:2015rga}            & 2+1   & \gA & \good & \soso & \good     & \soso & \okay & $R(D)$                           & 0.299(11)      \\[0.5ex]
&&&&&&&&& \\[-0.1cm]
FNAL/MILC~21         & \cite{FermilabLattice:2021cdg}    & 2+1   & \gA & \good & \soso & \good     & \soso & \okay & $R(D^\ast)$                      & 0.265(13)      \\[0.5ex]
JLQCD~23             & \cite{Aoki:2023qpa}               & 2+1   & \gA & \good & \soso & \good     & \soso & \okay & $R(D^\ast)$                      & 0.252(22)      \\[0.5ex]
HPQCD~23             & \cite{Harrison:2023dzh}           & 2+1+1 & \gA & \good & \good & \good     & \good & \okay & $R(D^\ast)$                      & 0.273(15)      \\[0.5ex]
&&&&&&&&& \\[-0.1cm]
\hline
&&&&&&&&& \\[-0.1cm]
HPQCD~23             & \cite{Harrison:2023dzh}           & 2+1+1 & \gA & \good & \good & \good     & \good & \okay & $R(D_s^\ast)$                    & 0.266(9)       \\[0.5ex]
&&&&&&&&& \\[-0.1cm]
\hline
\hline
\end{tabular*}\\
\begin{minipage}{\linewidth}
{\footnotesize 
\begin{itemize}
   \item[$^*$] The rationale for assigning a \soso rating is discussed in the text.
\end{itemize}
}
\end{minipage}
\caption{Lattice results for mesonic processes involving $b \to c$ transitions. The form factor $\cal G$ is defined in Eqs.~(\ref{B2D-formfactor}, \ref{defG}),  the form factor $\cal F$ is defined in Eqs.~(\ref{eq:BtoDstarAxialFormFactor}, \ref{defF}), and the ratios $R$ are defined in Eq.~(\ref{defR}). Note that the results for ${\mathcal F}^{B\to D^*}(1)$, ${\mathcal F}^{B_s\to D_s^*}(1)$, $R(D^\ast)$ and $R(D_s^\ast)$ have been obtained using the results of the BGL fits described in the text and do not necessarily coincide with the results presented by the individual collaborations. \label{tab_BtoDStarsumm2}}
\end{center}
\end{table}

Early computations of the form factors for $B \rightarrow D\ell\nu$ decays include $\Nf=2+1$ results by FNAL/MILC~04A and FNAL/MILC~13B \cite{Okamoto:2004xg,Qiu:2013ofa} for ${\cal G}^{B \rightarrow D}(1)$ and the $\Nf=2$ study by Atoui {\it et al.}~\cite{Atoui:2013zza},
that in addition to providing ${\cal G}^{B \rightarrow D}(1)$ explored the $w>1$ region.
This latter work also provided the first results for $B_s \rightarrow D_s\ell\nu$ amplitudes, again including information about the momentum-transfer dependence.
In 2014 and 2015, full results for $B \rightarrow D\ell\nu$ at $w \geq 1$ were published by FNAL/MILC~15C~\cite{Lattice:2015rga} and HPQCD~15~\cite{Na:2015kha}.
These works also provided full results for the scalar form factor, allowing analysis of the decay with a final-state $\tau$.
In FLAG~19 \cite{FlavourLatticeAveragingGroup:2019iem}, we included new results for $B_s \rightarrow D_s\ell\nu$ form factors over the full kinematic range for $\Nf=2+1$ from HPQCD (HPQCD~17 \cite{Monahan:2017uby} and Ref.~\cite{Monahan:2016qxu}).
Recently, HPQCD published new calculations of the $B_s \to D_s$ form factors in the full kinematic range \cite{McLean:2019qcx} (HPQCD~19),
now using MILC's HISQ $\Nf=2+1+1$ ensembles and using the HISQ action also for the $b$ quark, reaching up to $m_b = 4m_c$ (unrenormalized mass) in their finest ensemble.\footnote{The ratio showed here is the ratio between the bare masses, which are inputs of the lattice action.
The ratio between the renormalized masses of the quarks is usually very different from the ratio of bare masses.
In order to tune the bare heavy-quark masses so they result in physical values of the renormalized quark masses, one normally tries to find out the value of the bare mass that results in a heavy meson with the right physical mass.}
This calculation has recently been used by LHCb to determine $|V_{cb}|$ \cite{LHCb:2020cyw,LHCb:2021qbv}, as discussed further in Sec.~\ref{sec:Vcb}.

In the discussion below, we mainly concentrate on the latest generation of results, which allows for an extraction of $|V_{cb}|$
that incorporates information about the $q^2$-dependence of the decay rate (cf.~Sec.~\ref{sec:Vcb}).

We will first discuss the $\Nf=2+1$ computations of $B \rightarrow D \ell \nu$ by \SLfnalmilcBD~and \SLhpqcdBD, both based on MILC asqtad ensembles.
Full details about all the computations are provided in Tab.~\ref{tab_BtoDStarsumm2} and in the tables in Appendix~\ref{app:BtoD_Notes}.

The \SLfnalmilcBD~study~\cite{Lattice:2015rga} employs ensembles at four values of the lattice spacing ranging between approximately $0.045~{\rm fm}$ and $0.12~{\rm fm}$,
and several values of the light-quark mass corresponding to pions with RMS masses ranging between $260~{\rm MeV}$ and $670~{\rm MeV}$ (with just one ensemble with $M_\pi^{\rm RMS} \simeq 330~{\rm MeV}$ at the finest lattice spacing).
The $b$ and $c$ quarks are treated using the Fermilab approach.

The hadronic form factor relevant for experiment, $\mathcal{G}(w)$, is then obtained from the relation $\mathcal{G}(w)=\sqrt{4r}f_+(q^2)/(1+r)$.
The form factors are obtained from double ratios of three-point functions in which the flavour-conserving current renormalization factors cancel.
The remaining matching factor to the flavour-changing normalized current is estimated with 1-loop lattice perturbation theory.
In order to obtain $h_\pm(w)$, a joint continuum-chiral fit is performed to an ansatz that contains the light-quark mass and lattice-spacing dependence predicted by next-to-leading order HMrS$\chi$PT,
and the leading dependence on $m_c$ predicted by the heavy-quark expansion ($1/m_c^2$ for $h_+$ and $1/m_c$ for $h_-$).
The $w$-dependence, which allows for an interpolation in $w$, is given by analytic terms up to $(1-w)^2$, as well as a contribution from the logarithm proportional to $g^2_{D^\ast D\pi}$.
The total resulting systematic error, determined as a function of $w$ and quoted at the representative point $w=1.16$ as $1.2\%$ for $f_+$ and $1.1\%$ for $f_0$, dominates the final error budget for the form factors.
After $f_+$ and $f_0$ have been determined as functions of $w$ within the interval of values of $q^2$ covered by the computation, synthetic data points are generated to be subsequently fitted to a $z$-expansion of the BGL form, cf.~Sec.~\ref{sec:BtoPiK}, with pole factors set to unity.
This in turn enables one to determine $|V_{cb}|$ from a joint fit of this $z$-expansion and experimental data. The value of the zero-recoil form factor resulting from the $z$-expansion is
\begin{equation}
{\cal G}^{B \rightarrow D}(1)= 1.054(4)_{\rm stat}(8)_{\rm sys}\,.
\end{equation}

The HPQCD computations \SLhpqcdBD~and HPQCD~17~\cite{Na:2015kha,Monahan:2017uby} use ensembles at two values of the lattice
spacing, $a=0.09,~0.12~{\rm fm}$, and two and three values of light-quark masses, respectively.
The $b$ quark is treated using NRQCD, while for the $c$ quark the HISQ action is used.
The form factors studied, extracted from suitable three-point functions, are
\begin{equation}
\langle D_{(s)}(p_{D_{(s)}})| V^0 | B_{(s)}\rangle = \sqrt{2M_{B_{(s)}}}f^{(s)}_\parallel\,,~~~~~~~~
\langle D_{(s)}(p_{D_{(s)}})| V^k | B_{(s)}\rangle = \sqrt{2M_{B_{(s)}}}p^k_{D_{(s)}} f^{(s)}_\perp\,,
\end{equation}
where $V_\mu$ is the relevant vector current and the $B_{(s)}$ rest frame is chosen.
The standard vector and scalar form factors are retrieved as
\begin{align}
  f^{(s)}_+ =& \frac{1}{\sqrt{2M_{B_{(s)}}}}
               \left[ f^{(s)}_\parallel +
               (M_{B_{(s)}}-E_{D_{(s)}})f^{(s)}_\perp \right],
  \\
  f^{(s)}_0 =& \frac{\sqrt{2M_{B_{(s)}}}}{M_{B_{(s)}}^2-M_{D_{(s)}}^2}
               \left[(M_{B_{(s)}}-E_{D_{(s)}})f^{(s)}_\parallel
               +(M_{B_{(s)}}^2-E_{D_{(s)}}^2)f^{(s)}_\perp\right].
\end{align}
The currents in the effective theory are matched at 1-loop to their continuum
counterparts. Results for the form factors are then fitted to a modified BCL $z$-expansion
ansatz~\cite{Boyd:1994tt}, that takes into account simultaneously the lattice spacing, light-quark masses,
and $q^2$-dependence. For the mass dependence, NLO chiral logarithms are included, in the
form obtained in hard-pion $\chi$PT (see footnote~\ref{footnote:hardpion}). As in the case of the \SLfnalmilcBD~computation,
once $f_+$ and $f_0$ have been determined as functions of $q^2$, $|V_{cb}|$ can
be determined from a joint fit of this $z$-expansion and experimental data.
The papers quote for the zero-recoil vector form factor the result
\begin{equation}
{\cal G}^{B \rightarrow D}(1)=1.035(40)\,~~~~{\cal G}^{B_s \rightarrow D_s}(1)=1.068(40)\,.
\end{equation}
The \SLhpqcdBD~and \SLfnalmilcBD~results for $B\to D$ differ by less than half a standard
deviation (assuming they are uncorrelated, which they are not as some of
the ensembles are common) primarily because of lower precision of the former
result.
The dominant source of errors in the $|V_{cb}|$ determination by \SLhpqcdBD~are discretization
effects and the systematic uncertainty associated with the perturbative matching.

In order to combine the form-factor determination of \SLhpqcdBD \; and the one of \SLfnalmilcBD \; into a lattice average, we proceed in a similar way as with $B\to\pi\ell\nu$
and $B_s\to K\ell\nu$ above. \SLfnalmilcBD \; quotes synthetic values for each
form factor at three values of $w$ (or, alternatively, $q^2$) with a full
correlation matrix, which we take directly as input. In the case of \SLhpqcdBD,
we use their preferred modified $z$-expansion parameterization to produce
synthetic values of the form factors at five different values of $q^2$ (three for $f_+$ and two for $f_0$).
This leaves us with a total of six (five) data points in the kinematical
range $w\in[1.00,1.11]$ for the form factor $f_+$ ($f_0$). As in the case of $B\to\pi\ell\nu$, we conservatively
assume a 100\% correlation of statistical uncertainties between \SLhpqcdBD~
and \SLfnalmilcBD. We then fit this data set to a BCL ansatz, using
$t_+=(M_{B^0}+M_{D^\pm})^2 \simeq 51.12~\GeV^2$ and
$t_0=(M_{B^0}+M_{D^\pm})(\sqrt{M_{B^0}}-\sqrt{M_{D^\pm}})^2 \simeq 6.19~\GeV^2$.
In our fits, pole factors have been set to unity, i.e., we do not
take into account the effect of sub-threshold poles, which is then
implicitly absorbed into the series coefficients. The reason for this
is our imperfect knowledge of the relevant resonance spectrum in this channel,
which does not allow us to decide the precise number of poles needed.\footnote{As noted
above, this is the same approach adopted by \SLfnalmilcBD~in their fits to a BGL
ansatz. \SLhpqcdBD, meanwhile, uses one single pole in the pole factors that
enter their modified $z$-expansion, using their spectral studies to fix
the value of the relevant resonance masses.}
This, in turn, implies that unitarity bounds do not rigorously apply,
which has to be taken into account when interpreting the results
(cf.~Appendix \ref{sec:zparam}).

With a procedure similar to what we adopted for the $B\to \pi$ and
$B_s\to K$ cases, we impose the kinematic constraint at $q^2=0$ by
expressing the $a^0_{N^0-1}$ coefficient in the $z$-expansion of $f_0$
in terms of all the other coefficients. As mentioned above, \SLfnalmilcBD \;
provides synthetic data for $f_+$ and $f_0$ including correlations;
\SLhpqcdBD~presents the result of simultaneous $z$-fits to the two form
factors including all correlations, thus enabling us to generate a
complete set of synthetic data for $f_+$ and $f_0$. Since both
calculations are based on MILC ensembles, we then reconstruct the
off-diagonal \SLhpqcdBD-\SLfnalmilcBD \; entries of the covariance matrix by
conservatively assuming that statistical uncertainties are 100\%
correlated. The \SLfnalmilcBD \; (\SLhpqcdBD) statistical error is 58\% (31\%)
of the total error for every $f_+$ value, and 64\% (49\%) for every
$f_0$ one. Using this information we can easily build the off-diagonal
block of the overall covariance matrix (e.g., the covariance between
$[f_+(q_1^2)]_{\rm FNAL}$ and $[f_0(q_2^2)]_{\rm HPQCD}$ is $(\delta
[f_+(q_1^2)]_{\rm FNAL} \times 0.58)\; (\delta [f_0(q_2^2)]_{\rm
  HPQCD} \times 0.49)$, where $\delta f$ is the total error). 

\begin{figure}[tbp]
\begin{center}
\includegraphics[width=0.49\textwidth]{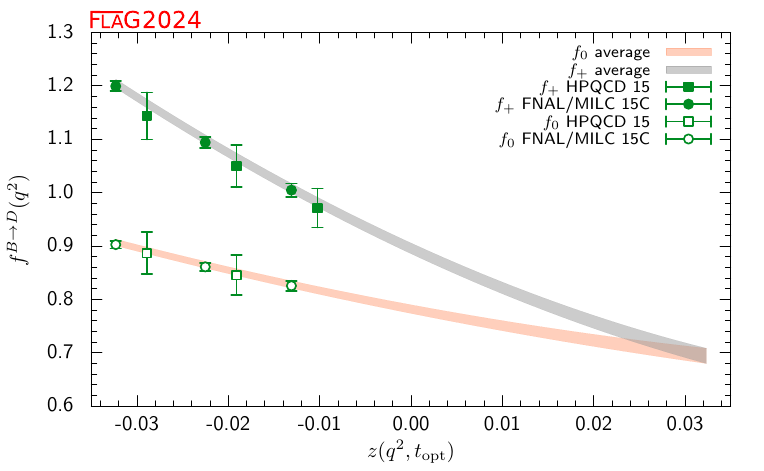}
\includegraphics[width=0.49\textwidth]{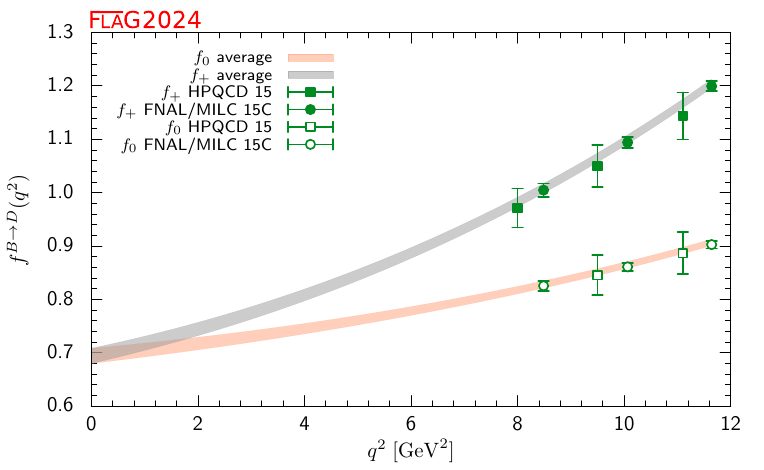}
\caption{The form factors $f_+(q^2)$ and $f_0(q^2)$ for $B \to D\ell\nu$ plotted versus $z$ (left panel) and $q^2$ (right panel). See text for a discussion of the data sets. The grey and salmon bands display our preferred $N^+=N^0=3$ BCL fit (five parameters).}
\label{fig:LQCDzfitBD}
\end{center}
\end{figure}

For our central value, we choose an $N^+ =N^0=3$ BCL fit, shown in Tab.~\ref{tab:FFD}. The coefficient $a_3^+$ can be obtained from the values for $a_0^+$--$a_2^+$ using Eq.~(\ref{eq:red_coeff}).  We find $\chi^2/{\rm dof} = 4.6/6 = 0.77$. The fit, which is dominated by the \SLfnalmilcBD~calculation, is illustrated in Fig.~\ref{fig:LQCDzfitBD}.

\begin{table}[t]
\begin{center}
\begin{tabular}{|c|c|ccccc|}
\multicolumn{7}{l}{$B\to D \; (\Nf=2+1)$} \\[0.2em]\hline
$a_n^i$ & Central Values & \multicolumn{5}{|c|}{Correlation Matrix}       \\[0.2em]\hline
$a_0^+$ &    0.896 (10)  & 1        & 0.423 & $-$0.231 & 0.958    & 0.596 \\[0.2em]
$a_1^+$ & $-$7.94 (20)   & 0.423    & 1     & 0.325    & 0.498    & 0.919 \\[0.2em]
$a_2^+$ &    51.4 (3.2)  & $-$0.231 & 0.325 & 1        & $-$0.146 & 0.317 \\[0.2em]
$a_0^0$ &    0.7821 (81) & 0.958    & 0.498 & $-$0.146 & 1        & 0.593 \\[0.2em]
$a_1^0$ & $-$3.28 (20)   & 0.596    & 0.919 & 0.317    & 0.593    & 1     \\[0.2em]
\hline
\end{tabular}
\end{center}
\caption{Coefficients and correlation matrix for the $N^+ =N^0=3$ $z$-expansion of the $B\to D$ form factors $f_+$ and $f_0$. The chi-square per degree of freedom is $\chi^2/{\rm dof} = 4.6/6=0.77$. The lattice calculations that enter this fit are taken from \SLfnalmilcBD~\cite{Lattice:2015rga} and \SLhpqcdBD~\cite{Na:2015kha}. The form factors can be reconstructed using parameterization and inputs given in Appendix~\ref{sec:app_B2D}. \label{tab:FFD}}
\end{table}

Let us finally discuss the most recent results for $B_s \to D_s$ form factors,
obtained by the HPQCD collaboration using MILC's $\Nf=2+1+1$ ensembles in
Ref.~\cite{McLean:2019qcx} (HPQCD~19). Three values of the lattice spacing are
used, including a very fine ensemble at $a \simeq 0.044~{\rm fm}$;
the pion mass is kept fixed at around 300~{\rm MeV}, and in addition at the
coarser $a \simeq 0.09~{\rm fm}$ lattice an ensemble with the physical pion
mass is included. The scalar current needs no renormalization because of the Partial Conservation of the Vector Current (PCVC) relation,
while the vector current is nonperturbatively normalized by imposing a condition
based on the PCVC relation at zero recoil.
Heavy quarks are treated in a fully relativistic fashion
through the use of the HISQ regularization, employing bare values of the quark
mass up to $am_h=0.8$ for the extrapolation to the physical $b$ point.

Results for the form factors are fitted to a modified $z$-expansion ansatz,
based on a BCL ansatz with a Blaschke factor containing one sub-threshold pole, tuned
to reproduce the lattice-spacing and heavy-quark-mass-dependent mass of the corresponding resonance.
The final error budget is equally dominated by statistics and the combined effect
of the continuum and heavy quark mass extrapolations, which correspond to 1.1\% and 1.2\%
uncertainties, respectively, for the scalar form factor at zero recoil. The total
uncertainty of {$f_0$} is thus below 2\%, which remains true in the whole $q^2$
range. The uncertainty of $f_+$ is somewhat larger, starting at around 2\% at $q^2=0$
and increasing up to around 3.5\% at zero recoil.

One important matter of concern with this computation is the use of the $a \simeq 0.044~{\rm fm}$
ensemble with periodic boundary conditions, which suffers from severe topology freezing.
Other than possible implications for statistical uncertainties, the lack of topology
fluctuations are expected to significantly enhance finite-volume effects, which are no
longer exponential in $m_\pi L$, but become power-like in the spatial volume.
The authors neglect the impact of finite-volume effects in the computation, with a
twofold argument: for the two coarser lattice spacings, the impact of pion-mass-related
corrections on the heavy-meson states involved is presumably negligible; and, for the finest
ensemble, the estimate of finite-volume effects on the $D_s$ decay constant obtained
in Ref.~\cite{Bernard:2017npd} turns out to be very small, a result which is presumed to extend
to form factors. It is however unclear whether the latter argument would really hold,
since the computation in Ref.~\cite{Bernard:2017npd} does show that the expected effect
is heavily observable-dependent, reaching, e.g., more than 1\% for $f_D$.
We have, therefore, concluded that our standard criteria for finite-volume effects
cannot be applied at the finest lattice spacing, and opted to assign $\soso$ rating
to them.

We thus proceed to quote the final result of HPQCD~19 as the FLAG estimate for the $\Nf=2+1+1$
$B_s \to D_s$ form factors. The preferred fit is a constrained BCL form with the imposition
of the kinematical constraint $f_+(0)=f_0(0)$, carried through $z^2$ for $f_0$ and $z^3$ for $f_+$.
Both form factors contain just one sub-threshold pole, to which the masses
$M_{B_c^*}= 6.329~\GeV$ and $M_{B_{c0}}= 6.704~\GeV$, respectively, have been assigned.
The fit parameters and covariance matrix, quoted in Table~VIII of Ref.~\cite{McLean:2019qcx},
are reproduced in Tab.~\ref{tab:BsDs}.

\begin{table}[t]
\begin{center}
\begin{tabular}{|c|c|cccccc|}
\multicolumn{7}{l}{$B_s\to D_s \; (\Nf=2+1+1)$} \\[0.2em]\hline
$a_n^i$ & Central Values & \multicolumn{6}{|c|}{Correlation Matrix}                        \\[0.2em]\hline
$a_0^0$ &    0.666(12)   & 1       & 0.62004 & 0.03149    & 1       & 0.03973 & 0.00122    \\[0.2em]
$a_1^0$ & $-$0.26(25)    & 0.62004 & 1       & 0.36842    & 0.62004 & 0.12945 & 0.00002    \\[0.2em]
$a_2^0$ & $-$0.1(1.8)    & 0.03149 & 0.36842 & 1          & 0.03149 & 0.22854 & $-$0.00168 \\[0.2em]
$a_0^+$ &    0.666(12)   & 1       & 0.62004 & 0.03149    & 1       & 0.03973 & 0.00122    \\[0.2em]
$a_1^+$ & $-$3.24(45)    & 0.03973 & 0.12945 & 0.22854    & 0.03973 & 1       & 0.11086    \\[0.2em]
$a_2^+$ & $-$0.1(2.0)    & 0.00122 & 0.00002 & $-$0.00168 & 0.00122 & 0.11086 & 1          \\[0.2em]
\hline
\end{tabular}
\end{center}
\caption{Coefficients and correlation matrix for the $z$-expansion of the $B_s\to D_s$ form factors $f_+$ and $f_0$. These results are a reproduction of Table~VIII of Ref.~\cite{McLean:2019qcx} (HPQCD~19). The form factors can be reconstructed using parameterization and inputs given in Appendix~\ref{sec:app_Bs2Ds}. \label{tab:BsDs}}
\end{table}
There are ongoing efforts in these channels from several collaborations. The JLQCD collaboration is working on a $B\to D$ analysis at nonzero recoil using the domain-wall action for heavy and light quarks~\cite{Kaneko:2021tlw}.
The FNAL/MILC collaborations are working on two parallel calculations of the form factors of the $B_{(s)}\to D_{(s)}$ channels sharing the same light-quark action, but with different heavy-quark actions~\cite{Lytle:2024zfr}.

\subsubsection{$B_{(s)} \rightarrow D^\ast_{(s)}$ decays}
\label{sec:BstoDstarFFs}
The community has been focusing on the decays with final vector states, $B_{(s)} \to D_{(s)}^*$, because of increasing availability of high-quality experimental data.
The decay rate for $B \rightarrow D^\ast\ell\nu$ involves a spin-1 hadron in the final-state whose vector and axial-vector current matrix elements require the introduction of four form factors:
\begin{align}
  \frac{\langle D^\ast | V_\mu | B \rangle}{\sqrt{m_B m_{D^\ast}}}  &=  h_V(w)
   \varepsilon_{\mu\nu\alpha\beta} \epsilon^{*\nu} v_{D^\ast}^\alpha v_B^\beta \,, \\
  \frac{\langle D^\ast | A_\mu | B \rangle}{i \sqrt{m_B m_{D^\ast}}}  &= 
    h_{A_1}(w) (1+w) \epsilon^{*\mu} -
    h_{A_2}(w) \epsilon^*\cdot v_B {v_B}_\mu   
  - h_{A_3}(w) \epsilon^*\cdot v_B {v_{D^\ast}}_\mu .
\label{eq:BtoDstarAxialFormFactor}
\end{align}
where $w = v_B \cdot v_{D^{(\ast)}} = (m_B^2 + m_{D^*}^2 - q^2)/(2 m_B m_{D^*})$. As has become customary, we further express the four form factors $h_{V,A_1,A_2,A_3}$ in terms of the form factors $g$, $f$, $F_1$ and $F_2$ as follows (see, for instance, Eq.~(31) of Ref.~\cite{Harrison:2023dzh}):
\begin{align}
g &= \frac{h_V}{m_B\sqrt{r}} \; , \label{eq:B2Dstar:g} \\
f &= m_B \sqrt{r} (1+w) h_{A_1} \; , \label{eq:B2Dstar:f}\\
F_1 &= m_B^2 \sqrt{r} (1+w) \Big[ (w-r) h_{A_1} - (w-1) (r h_{A_2} + h_{A_3} ) \big] \; , \label{eq:B2Dstar:F1}\\
F_2 &= \frac{1}{\sqrt{r}}\Big[ (1+w) h_{A_1} + (rw-1) h_{A_2} + (r-w) h_{A_3}  \big] \; .\label{eq:B2Dstar:F2}
\end{align}

One can then write the differential decay rate as~\cite{Korner:1989qb,Bigi:2017njr}
\begin{align}
\frac{d\Gamma_{\bar B \to D^{*}\ell \bar{\nu}}}{dw d c_v d c_l d\chi} = 
& \frac{\eta_\mathrm{EW}^2 3 m_B m_{D^*}}{4 (4\pi)^4} \sqrt{w^2-1} (1-2w r+r^2) G_F^2 |V_{cb}|^2  \nonumber \\
&  \times \Big[ (1-c_l)^2 s_v^2 H_+^2 + (1+c_l)^2 s_v^2 H_-^2 + 4 s_l^2 c_v^2 H_0^2 -2 s_l^2 s_v^2 \cos(2\chi) H_+ H_- \nonumber \\
& -4 s_l(1-c_l) s_v c_v \cos\chi H_+ H_0 + 4 s_l (1+c_l) s_v c_v \cos \chi H_- H_0 \Big] \; ,
\label{eq:BtoDstarGammaDiff}
\end{align}
where $c_v \equiv \cos \theta_v$, $s_v \equiv \sin \theta_v$, $c_l \equiv \cos \theta_l$, $s_l \equiv \sin \theta_l$. The angles $\theta_v$, $\theta_l$ and $\chi$ parameterize the kinematics of the three-body final state (see, for instance, Fig.~3 of Ref.~\cite{Belle:2018ezy}).
The helicity amplitudes $H_{\pm,0}$ have simple expressions in terms of the form factors $g$, $f$ and $F_1$ (see, for instance, Eq.~(13) of Ref.~\cite{Belle:2018ezy}):
\begin{align}
H_0   &= \frac{F_1}{\sqrt{q^2}} \; , \\
H_\pm &= f \mp m_B m_{D^*} \sqrt{w^2-1}\;  g \; .
\end{align}
For the calculation of the ratio of the semileptonic rates in the $\tau$ and $\ell = e,\mu$ channels, it is necessary to consider the differential $d\Gamma/dw$ decay rate for nonzero lepton mass:\footnote{This formula can be found, for instance, in Eq.~(7) of Ref.~\cite{FermilabLattice:2021cdg}. Note that in Ref.~\cite{FermilabLattice:2021cdg} the normalizations of the helicity amplitudes $H_{\pm,0}$ differ from those adopted here.}
\begin{align}
\frac{d\Gamma_{\bar B \to D^{*}\ell \bar{\nu}}}{dw} &= 
 |V_{cb}|^2 G_F^2 \eta_{\rm EW}^2 \frac{m_B^3}{48 \pi^3} r^2 \sqrt{w^2-1}  \left( 1- \frac{m_l^2}{q^2} \right)^2  \nonumber \\
& \times \Big[ \left( 1+ \frac{m_l^2}{2 q^2} \right) \frac{q^2}{m_B^2} ( H_+^2 + H_-^2  + m_B^2 H_0^2) 
+ \frac{3}{2} r^2 \frac{m_B^2}{q^2} m_l^2 (w^2-1) F_2^2 \Big] \; .
\label{eq:B2Dstar:dGammadwMLEP}
\end{align}
In the limit of vanishing lepton mass, Eq.~(\ref{eq:B2Dstar:dGammadwMLEP}) reduces to
\begin{align}
    \frac{d\Gamma_{B^-\to D^{0*}\ell^-\bar{\nu}}}{dw} = &
        \frac{G^2_{\rm F} m^3_{D^\ast}}{4\pi^3}(m_B-m_{D^\ast})^2(w^2-1)^{1/2}  |\eta_\mathrm{EW}|^2|V_{cb}|^2\chi(w)|\mathcal{F}(w)|^2 \; .
    \label{eq:vxb:BtoDstar_w}
\end{align}
The function $\chi(w)$ in Eq.~(\ref{eq:vxb:BtoDstar_w}) depends on the recoil $w$ and the meson masses, and reduces to unity at zero recoil~\cite{Antonelli:2009ws}.
In particular, the normalization factor $\chi(w)$~\cite{Antonelli:2009ws} is defined in such a way that at zero recoil 
\begin{align}
\mathcal{F}(1)=h_{A_1}(1) = \frac{f(1)}{2 \sqrt{m_B m_{D^*}}} \; .
\label{defF}
\end{align}

Unquenched lattice calculations for $B \rightarrow D^\ast \ell \nu$ decays have focused on the form factors at zero recoil ${\cal F}^{B \rightarrow D^\ast}(1)$ until a few years ago
(see, for instance, FNAL/MILC~08~\cite{Bernard:2008dn}, FNAL/MILC~14~\cite{Bailey:2014tva}, HPQCD~17B~\cite{Harrison:2016gup,Harrison:2017fmw}); these can then be combined with experimental input to extract $|V_{cb}|$.
The situation mirrors that of the channel $B\to D\ell\nu$: at the zero-recoil point a single form factor is enough to calculate the decay rate and Luke's theorem~\cite{Luke:1990eg} guarantees the absence of $\cO(\Lambda_{QCD}/m_Q)$ corrections.
By heavy-quark symmetry, $\lim_{m_Q\to\infty}{\cal F}^{B \rightarrow D^\ast}(1) = 1$~\cite{Isgur:1989ed,Isgur:1989vq,Neubert:1993mb}, since in that limit there is no distinction between heavy quarks.
The calculation of higher-order corrections to this value has been systematically addressed in several publications~\cite{Falk:1992wt,Neubert:1992tg,Neubert:1994qt,Neubert:1994vy}, and also applied to lattice calculations~\cite{Kronfeld:2000ck,Harada:2001fj}.
On the lattice, the zero recoil form factor of this channel can also be computed via a double ratio, cancelling most of the current renormalization and suppressing heavy-quark discretization errors by an additional power of $\Lambda_{QCD}/m_Q$~\cite{Hashimoto:2001nb}.
The situation has dramatically improved recently, and now data away from the zero-recoil region is available from several sources.
For that reason, we mainly concentrate on the latest generation of results in the discussion below, which allows for an extraction of $|V_{cb}|$ that incorporates information about the $q^2$-dependence of the decay rate (cf.~Sec.~\ref{sec:Vcb}).
 
Extraction of the form factors away from the zero-recoil point is quite challenging.
The polarization of the $D^\ast$ plays a key role in the correlation functions, as shown in Eq.~(\ref{eq:BtoDstarAxialFormFactor}).
One can build the following double ratio:
\begin{equation}
{\cal R}_{A_1}({\bf p}) = \frac{\langle D^\ast({\bf p}, \varepsilon_\bot)|\bar{c} \gamma_\bot \gamma_5 b | \overline{B}({\bf 0})
\rangle \; \langle \overline{B}({\bf 0})| \bar{b} \gamma_\bot \gamma_5 c | D^\ast({\bf p}, \varepsilon_\bot) \rangle}
{\langle D^\ast({\bf 0})|\bar{c} \gamma_4 c | D^\ast({\bf 0})
\rangle \; \langle \overline{B}({\bf 0})| \bar{b} \gamma_4 b | \overline{B}({\bf 0}) \rangle}
\propto |h_{A_1}(w)|^2,
\label{hA1NonZeroRecoil}
\end{equation}
which is proportional to $\left|h_{A_1}(w)\right|^2$, as long as the $D^\ast$ is transversally polarized (the spatial components of $\varepsilon_\bot$ are perpendicular to ${\bf p}$)
and parallel to the axial current, which displays only spatial components ($\gamma_\bot$ is parallel to the spatial components of $\varepsilon_\bot$).
At zero recoil, Eq.~\eqref{hA1NonZeroRecoil} greatly simplifies to give
\begin{equation}
{\cal R}_{A_1}({\bf 0}) = |h_{A_1}(1)|^2.
\label{hA1ZeroRecoil}
\end{equation}
Hence, an alternative to directly computing Eq.~\eqref{hA1NonZeroRecoil} is to evaluate Eq.~\eqref{hA1ZeroRecoil}, and then compute the following ratio
\begin{equation}                                                                                                                         
{\cal Q}_{A_1} = \frac{\langle D^\ast({\bf p},\varepsilon_\bot)|\bar{c} \gamma_\bot \gamma_5 b | \overline{B}({\bf 0}) \rangle}                            
                      {\langle D^\ast({\bf 0},\varepsilon)     |\bar{c} \gamma_j \gamma_5 b | \overline{B}({\bf 0})) \rangle},                           
\end{equation}                                                                                                                           
which gives $h_{A_1}(w)/h_{A_1}(1)$ times extra factors that must be removed.
Other form factors can be extracted by considering other polarizations and components of the axial current in Eq.~\eqref{eq:BtoDstarAxialFormFactor}, as well as the vector current.
Normally, all the form factors are referenced to $h_{A_1}(w)$, therefore any systematics associated to the extraction of $h_{A_1}(w)$ are carried over to the remaining form factors.

Currently, there are two $\Nf=2+1$ calculations of the $B\to D^\ast\ell\nu$ form factors.
One comes from the FNAL/MILC collaborations~\cite{FermilabLattice:2021cdg} (FNAL/MILC~21). It uses 15 MILC $\Nf = 2 + 1$ ensembles generated with asqtad staggered quarks in the sea.
The bottom and charm quarks are simulated using the clover action with the Fermilab interpretation, and they are tuned to their physical masses by using the $D_s$ and the $B_s$ mesons as references.
This implies that the renormalization cannot be fully nonperturbative. The collaboration employs a clever scheme that computes ratios where the largest component of the renormalization factors cancels out, leaving a small component that is computed perturbatively.
The MILC ensembles employed span five lattice spacings, ranging from $a \approx 0.15~{\rm fm}$ to $a \approx 0.045~{\rm fm}$, and as many as five values of the light-quark masses
per ensemble (though just one at the finest lattice spacing).
Results are then extrapolated to the physical, continuum/chiral, limit employing staggered, heavy-light meson $\chi$PT.

The $D^\ast$ meson is not a stable particle in QCD and decays predominantly into a $D$ plus a pion.
Nevertheless, heavy-light meson $\chi$PT can be applied to extrapolate lattice simulation results for the $B\to D^\ast\ell\nu$ form factor to the physical light-quark mass.
The $D^\ast$ width is quite narrow, 0.083(2) MeV for the $D^{*\pm}(2010)$ and less than 2.1 MeV for the $D^{*0}(2007)$~\cite{ParticleDataGroup:2024cfk}, making this system much more stable and long lived than the $\rho$ or the $K^*$ systems.
Therefore it is appropriate to consider the $D^\ast$ as a stable particle on the lattice, at the current level of precision.
The fact that the $D^\ast - D$ mass difference is close to the pion mass leads to the well-known ``cusp'' in ${\cal R}_{A_1}$ just above the physical pion mass~\cite{Randall:1993qg,Savage:2001jw,Hashimoto:2001nb}.
This cusp makes the chiral extrapolation sensitive to values used in the $\chi$PT formulas for the $D^\ast D\pi$ coupling $g_{D^\ast D\pi}$.
In order to take this sensitivity into account, the FNAL/MILC collaboration includes this coupling in their fits as an input prior $g_{D^\ast D\pi} = 0.53 \pm 0.08$,
but they do not analyze the impact of such a prior in the final result.
By looking at their previous calculation at zero recoil~\cite{Bailey:2014tva} (FNAL/MILC~14), which used the same ensembles and statistics, we estimate a subpercent increase in the total uncertainty for $h_{A_1}(1)$.

The final result presented in Ref.~\cite{FermilabLattice:2021cdg} (FNAL/MILC~21) is provided as synthetic data points for the four form factors in the HQET basis,
$\{h_{A_1}, h_{A_2}, h_{A_3}, h_V\}$, at three different values of the recoil parameter, and a full covariance matrix.
The result at zero recoil is
\begin{equation}
\Nf=2+1 \text{: } {\cal F}^{B \rightarrow D^\ast}(1) =  0.909(17) \quad\quad [\text{FNAL/MILC~21~\cite{FermilabLattice:2021cdg}}]                
\label{eq:BDstarFNAL}                                                             
\end{equation} 
making up a total error of $1.9$\%.
The largest systematic uncertainty comes from discretization errors followed by effects of higher-order corrections in the chiral perturbation theory ansatz.
The JLQCD collaboration has published the other $\Nf=2+1$ study of the $B\to D^\ast\ell\nu$ form factors away from the zero recoil point -- JLQCD~23 \cite{Aoki:2023qpa}.
Their calculation is based on nine $\Nf=2+1$ M\"obius domain-wall ensembles, using the same action for the valence, heavy quarks $b$ and $c$.
The ensembles cover three different lattice spacings, starting from $0.080$ fm down to $0.044$ fm, and several pion masses ranging from $\sim 230$ MeV to $\sim 500$ MeV.
The charm-quark mass is always physical, whereas the largest value of the bottom-quark mass reached is $\approx 3m_c$ (unrenormalized mass) in their finest ensemble.
Each ensemble features at least 3 different values of the bottom-quark mass, but in the coarsest ensemble only $m_Q\approx 1.5m_c$ is reached.
In terms of lattice units, the bottom-quark mass never exceeds $am_Q\lesssim 0.7$, and the final result does not significantly change if only data with $am_Q\lesssim 0.5$ (or equivalently $m_Q\lesssim 2.0m_c$) is employed.
The three-point functions leading to the form factors are evaluated for four source-sink separations 
to properly control the excited-states contamination, and also the effects of possible topological freezing are carefully analyzed to rule out finite-volume effects.
The renormalization scheme employed to renormalize the axial and vector currents is equivalent to a mostly nonperturbative renormalization scheme at tree level.
However, the properties of the Domain-Wall action establish that $Z_A\approx Z_V$ at finite lattice spacing.
Hence, we expect large cancellations of renormalization factors in ratios like Eq.~\eqref{hA1NonZeroRecoil}.
Also, discretization errors in the coefficients are expected to behave better than $O(a)$ for the same reason.

Physical data is obtained after performing combined chiral-continuum and heavy-quark-mass extrapolations, which employs an approximate estimator for the covariance matrix,
due to the low statistics of the input data and the large number of parameters involved (heavy- and light-quark masses, and lattice spacings).
The ansatz for the extrapolation is motivated by heavy-light meson $\chi$PT and HQET, and the collaboration uses the same value for the $D^\ast D\pi$ coupling $g_{D^\ast D\pi}$
as the FNAL/MILC collaboration, $g_{D^\ast D\pi} = 0.53 \pm 0.08$, but instead of including it as a prior in the fit, they estimate the systematics associated to the coupling
by shifting the central value by $\pm\sigma$. The uncertainty arising from this choice is not provided, although it is explicitly stated that it is small.

The collaboration provides three synthetic data points per form factor in the BGL basis, $\{g, f, F_1, F_2\}$ as their final result of their extrapolation, along with a full covariance matrix.
The result at zero recoil is not directly provided, but their BGL fit results in the following value,
\begin{equation}
\Nf=2+1 \text{: } {\cal F}^{B \rightarrow D^\ast}(1) =  0.887(14) \quad\quad [\text{JLQCD~23~\cite{Aoki:2023qpa}}].
\label{eq:BDstarJLQCD}
\end{equation} 

For $\Nf=2+1+1$ there is only one calculation away from the zero-recoil point, by the HPQCD collaboration~\cite{Harrison:2023dzh} -- HPQCD~23.
They use five MILC HISQ ensembles and the HISQ action for both the light and the heavy quarks, reaching up to $m_b = 4m_c$ (unrenormalized mass) in their finest ensemble.
The lattice spacings range from $0.090$ fm down to $0.044$ fm, and the pion masses are physical in two of the ensembles, whereas the rest use values $m_\pi\approx 320$ MeV.
They calculate the form factors for three or four bare values of the heavy-quark mass, depending on the ensemble, topping at $am_Q\leq 0.8$.
For the three-point functions, three different source-sink separations are evaluated, and the currents are renormalized nonperturbatively using the PCAC/PCVC relations and, for the tensor current, the RI-SMOM scheme~\cite{Hatton:2020vzp}.
The renormalization factors are interpolated for some correlators in one of the coarsest ensembles, and they are estimated for the finest ensemble with a physical pion mass, adding a conservative $1\%$ error.
As in previous analyses of HPQCD with a similar setup, the impact of fixing the topological charge in the finest ensembles is not discussed;
nonetheless, it has been pointed out that the impact on the form factors of MILC ensembles with nonequilibrated topological charge is below $0.1\%$~\cite{FermilabLattice:2022gku}.
An important difference of this analysis from the $\Nf=2+1$ ones is the inclusion of twisted boundary conditions to reach larger values of the recoil parameter.
As a result, HPQCD~23 offers data in the whole recoil range, as opposed to the other analyses, which are limited to the range $w\in [1.0, 1.2]$.
The constraint between the form factors at maximum recoil then is naturally satisfied with great precision without any need to impose it.
This feature also allows them to include higher powers of $(w-1)$ in the chiral-continuum extrapolation to model the recoil parameter dependence.
Using BGL-inspired priors, the collaboration includes terms up to $(w-1)^{10}$, steming from a $z$ expansion up to $z^4$.
 
HPQCD~23 provides five synthetic data points per form factor, of which only three are completely independent, in the HQET basis, along with the full covariance matrix.
The zero-recoil value of the decay amplitude is
{
\begin{equation}
\Nf=2+1+1 \text{: } {\cal F}^{B \rightarrow D^\ast}(1) =  0.903(14) \quad\quad [\text{FLAG average, HPQCD~23~\cite{Harrison:2023dzh}}]\,,
\label{eq:BDstarHPQCD}
\end{equation} 
}
in agreement with the value from FNAL/MILC~21, but with a slightly smaller total error, $1.6$\%.
The largest systematic uncertainty comes from the treatment of the heavy quark.

\begin{figure}[ht]
 \begin{center}
 \includegraphics[width=0.49\textwidth]{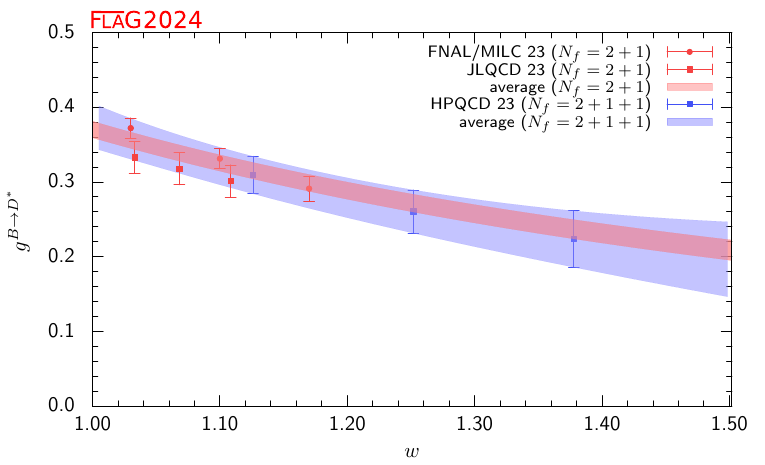}
 \includegraphics[width=0.49\textwidth]{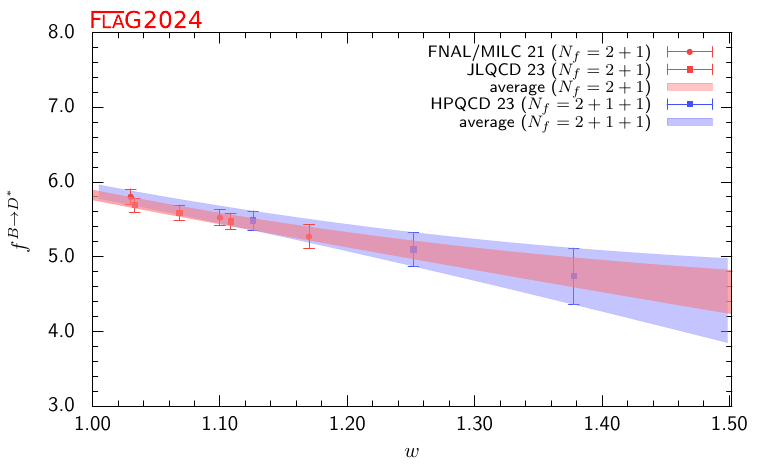}
 \includegraphics[width=0.49\textwidth]{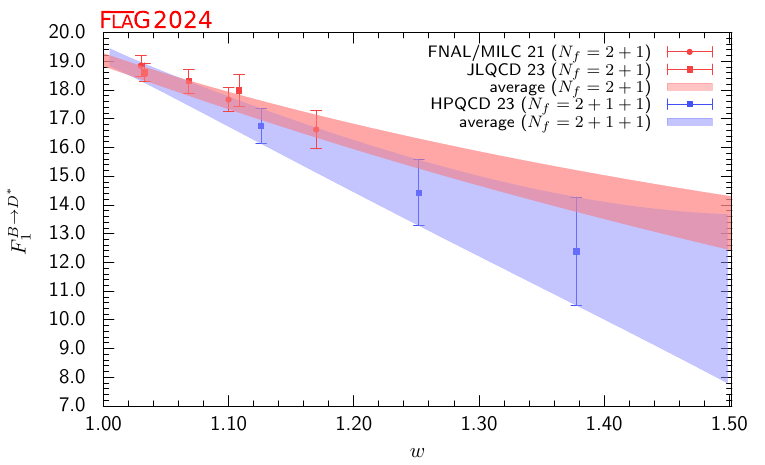}
 \includegraphics[width=0.49\textwidth]{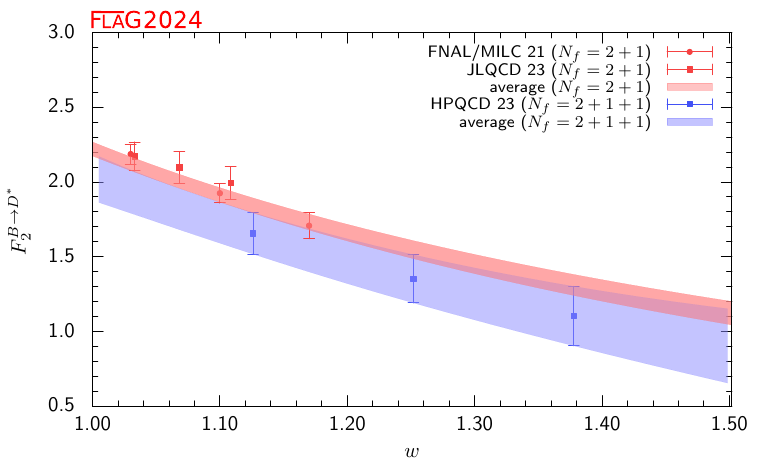}
 \caption{{The form factors $g$, $f$, $F_1$ and $F_2$ for $B \to D^* \ell\nu$ as a function of $w$. The red band displays our preferred $(N_g,\Nf,N_{F_1},N_{F_2}) = ( 2,3,3,2)$ BGL fit (eight parameters) to  $\Nf = 2+1$ lattice data. The constraints at zero and maximum recoil are imposed exactly. No use of unitarity constraints and priors has been made. The blue band is obtained using directly the $\Nf = 2+1+1$ HPQCD~23~\cite{Harrison:2023dzh} results.}}\label{fig:BDstar_latt}
 \end{center}
 \end{figure}

\begin{figure}[ht]
 \begin{center}
 \includegraphics[width=0.49\textwidth]{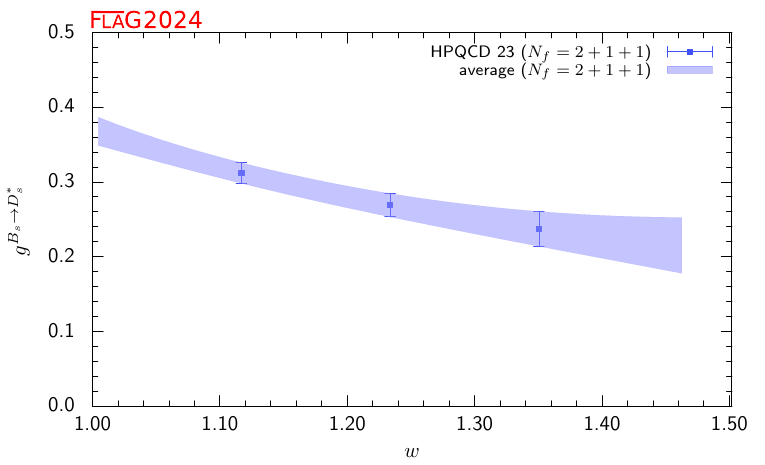}
 \includegraphics[width=0.49\textwidth]{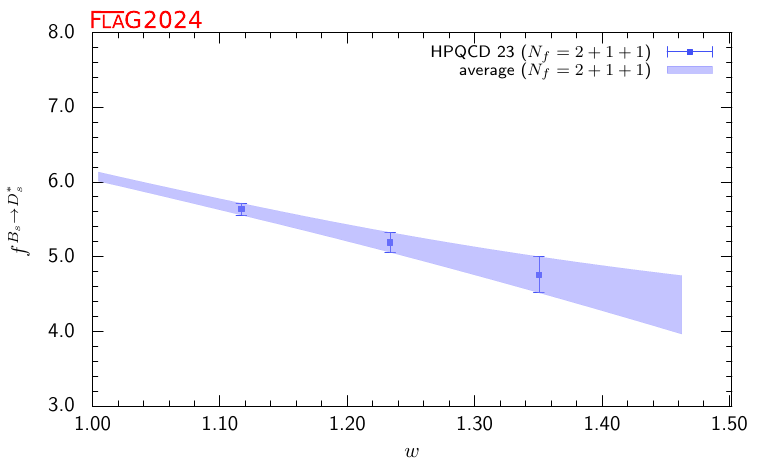}
 \includegraphics[width=0.49\textwidth]{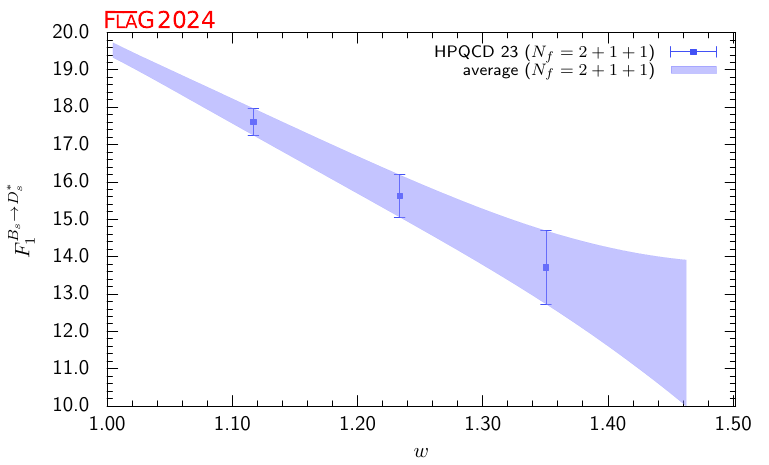}
 \includegraphics[width=0.49\textwidth]{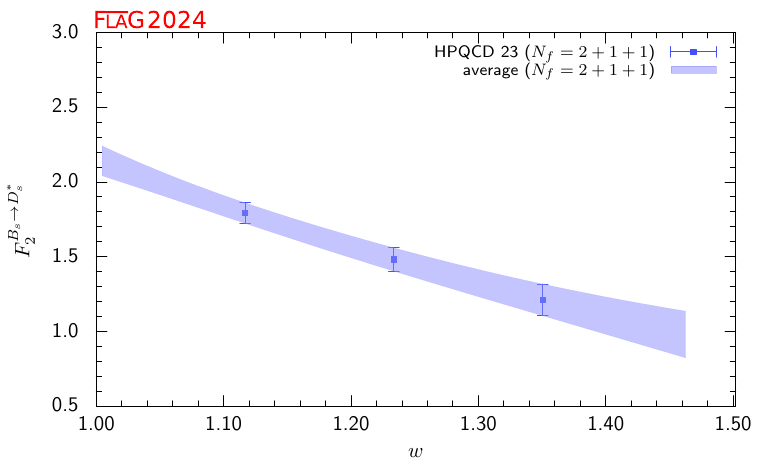}
 \caption{{ The form factors $g$, $f$, $F_1$, and $F_2$ for $B_s \to D_s^* \ell\nu$ as a function of $w$. The blue band is obtained using directly the $\Nf = 2+1+1$ HPQCD~23~\cite{Harrison:2023dzh} results.}}\label{fig:BsDsstar_latt}
 \end{center}
 \end{figure}

  \begin{table}[ht]
 \begin{center}
\tiny
  \begin{tabular}{|c|c|cccccccc|}
  \multicolumn{10}{l}{$B\to D^* \; (\Nf=2+1)$} \\[0.2em]\hline
  coeff & Central Values & \multicolumn{8}{|c|}{Correlation Matrix} \\[0.2em]\hline
  $a^g_0$ & 0.03132(93) & 1 & 0.1331 & 0.1786 & 0.03800 & 0.006578 & 0.06997 & 0.1061 & 0.03250 \\[0.2em]
$a^g_1$ & -0.057(26) & 0.1331 & 1 & 0.001304 & 0.2425 & 0.1505 & 0.1342 & 0.1966 & 0.2331 \\[0.2em]
$a^f_0$ & 0.01208(14) & 0.1786 & 0.001304 & 1 & -0.02370 & 0.09098 & 0.04710 & 0.1573 & 0.1161 \\[0.2em]
$a^f_1$ & 0.0135(72) & 0.03800 & 0.2425 & -0.02370 & 1 & -0.3968 & 0.6172 & -0.01165 & 0.5136 \\[0.2em]
$a^f_2$ & -0.08(27) & 0.006578 & 0.1505 & 0.09098 & -0.3968 & 1 & -0.2518 & 0.1880 & -0.05661 \\[0.2em]
$a^{F_1}_1$ & -0.0032(18) & 0.06997 & 0.1342 & 0.04710 & 0.6172 & -0.2518 & 1 & -0.1105 & 0.6653 \\[0.2em]
$a^{F_1}_2$ & -0.014(25) & 0.1061 & 0.1966 & 0.1573 & -0.01165 & 0.1880 & -0.1105 & 1 & 0.5974 \\[0.2em]
$a^{F_2}_1$ & -0.188(44) & 0.03250 & 0.2331 & 0.1161 & 0.5136 & -0.05661 & 0.6653 & 0.5974 & 1 \\[0.2em]
  \hline
  \end{tabular}
\end{center}
  \caption{Coefficients and correlation matrix for the $(N_g,\Nf,N_{F_1},N_{F_2}) = (2,3,3,2)$ BGL fit to the $B\to D^*$ form factors $g$, $f$, $F_1$ and $F_2$ for $\Nf=2+1$. The form factors can be reconstructed using parameterization and inputs given in Appendix~\ref{sec:app_B2D*}. \label{tab:BDstar_latt}}
\end{table}

We use synthetic data points provided by FNAL/MILC~21~\cite{FermilabLattice:2021cdg} and JLQCD~23~\cite{Aoki:2023qpa}
to fit the form factors $g$, $f$, $F_1$, and $F_2$ using a BGL parameterization.
We adopt the same outer functions, poles, and $z$ definition as in Sec.~5.1 of Ref.~\cite{FermilabLattice:2021cdg}.
In particular, we impose the kinematic constraints at zero and maximal recoil (see Eqs.(72, 73) of Ref.~\cite{FermilabLattice:2021cdg}) by eliminating the coefficients $a^{F_1}_0$ and $a^{F_2}_0$.
We also do not adopt priors for any of the coefficients and do not impose unitarity constraints.
We found that a fit with $(N_g,\Nf,N_{F_1},N_{F_2}) = (2,3,3,2)$ provides an adequate description of the lattice data.\footnote{Adequate in the sense that the coefficients do not change much when adding more terms in the $z$ expansion,
but any extra coefficient becomes unphysically large with equally large errors.
Hence, our choice is the maximum number of coefficients that can be reasonably determined with the given data without including extra information, like unitarity constraints.}
The results of the fits are presented in Tab.~\ref{tab:BDstar_latt} and in Fig.~\ref{fig:BDstar_latt}.
The two $\Nf = 2+1$ calculations of FNAL/MILC~21~\cite{FermilabLattice:2021cdg} and JLQCD~23~\cite{Aoki:2023qpa} are quite compatible and the combined fit yields $\chi^2_{\rm min}/{\rm dof} = 15.0/16$.
{
  In Fig.~\ref{fig:BDstar_latt}, we present also the $\Nf=2+1+1$ form factors obtained directly from results of the $\Nf=2+1+1$ HPQCD~23~\cite{Harrison:2023dzh} calculation in order to allow for a direct comparison between the $\Nf=2+1$ and $\Nf=2+1+1$ determinations. 
}
For completeness, we present the result for ${\cal F}^{B\to D^*}(1)$ as extracted from the fits in Tab.~\ref{tab:BDstar_latt}:
{
\begin{alignat}{2}
\Nf=2+1 \text{: }&  {\mathcal F}^{B\to D^*}(1) = 0.894 (10) \quad\quad [\text{FLAG average, Refs.~\cite{FermilabLattice:2021cdg, Aoki:2023qpa}}]. \label{eq:BDstarNf2+1FLAG}  
\end{alignat}
}

Calculations in the $B_s\to D_s^\ast$ channel are relatively recent. The first calculations at zero recoil were done by the HPQCD collaboration in 2017 and 2019~\cite{Harrison:2017fmw,McLean:2019sds} (HPQCD~17B and HPQCD~19B).
In 2021, the same collaboration published the first study of the form factors of this channel at nonzero recoil~\cite{Harrison:2021tol} (HPQCD~21B), using four $\Nf=2+1+1$ MILC ensembles and the HISQ regularization for both sea and valence quarks, including the $b$ quark.
The lattice spacings range from $0.090$ fm to $0.044$ fm, and one of the coarsest ensembles features a physical pion mass, whereas the rest are generated with $m_\pi\approx 320$ MeV.
Correlators are generated for each ensemble at three/four values of the bare-quark mass, never exceeding $am_Q \leq 0.8$, and the maximum heavy-quark mass simulated is $m_Q \approx 4m_c$ (nonrenormalized).
Momentum is injected using twisted boundary conditions, which allows them to calculate the form factors directly at large values of the recoil parameter.
This calculation was recently superseded by a combined study of the $B_{(s)}\to D_{(s)}^\ast$ channels by HPQCD~23~\cite{Harrison:2023dzh}, adding one more ensemble and increasing statistics.
The details have already been outlined earlier in this section.
{Since there is only one available result (see Fig.~\ref{fig:BsDsstar_latt}), we set our $\Nf=2+1+1$ average to the HPQCD~23 value~\cite{Harrison:2023dzh},}
\begin{align}
\Nf=2+1+1 \text{: } {\mathcal F}^{B_s\to D_s^*}(1) = 0.8970 (92) \quad\quad [\text{FLAG average, Ref.~\cite{Harrison:2023dzh}}]. 
\label{eq:BsDsstarNf2+1+1FLAG}
\end{align}

There are still ongoing efforts on both the $B\to D^\ast$ and the $B_s\to D_s^\ast$ channels, and we can expect improvements in the coming years.
The FNAL/MILC collaborations are working in two different calculations in parallel for $B\to D^\ast$, mainly differing on the heavy-quark action: one calculation uses Fermilab heavy quarks, whereas the other uses the HISQ action for the $c$ and the $b$ quarks.
Both calculations employ the HISQ action for the light sector~\cite{Lytle:2024zfr}.
The LANL-SWME collaboration is working on a different calculation, using MILC HISQ ensembles and the Oktay-Kronfeld action for the heavy sector~\cite{Bhattacharya:2023llf}.

\subsubsection{Lepton-flavour-universality ratios $R(D^{(\ast)})$ and $R(D^{(\ast)}_s)$}

The availability of results for the scalar form factor $f_0$ for $B\to D\ell\nu$ amplitudes allows us to study interesting observables that involve the decay in the $\tau$ channel.
One such quantity is the ratio
\begin{equation}
  R(D_{(s)}^{(*)}) = \frac{{\cal B}(B \rightarrow D_{(s)}^{(*)} \tau \nu)}{{\cal B}(B \rightarrow D_{(s)}^{(*)} \ell \nu)}
  \;\;\;\;\mbox{with}\;\;\;\; \ell=e,\mu\,,
  \label{defR}
\end{equation}
which, in the Standard Model, depends only on the form factors and hadron and lepton masses.
Indeed, the recent availability of experimental results for $R(D)$ has made this quantity particularly relevant in the search for possible physics beyond the Standard Model.
The most recent HFLAV average reads (see Ref.~\cite{HFLAV:2022esi} and the Moriond 2024 update):
\begin{align}
R(D)_{\rm exp} &= 0.342 (26) \; .
\end{align}

Using the FLAG average of the $B\to D$ form factors discussed above and presented in Table~\ref{tab:FFD}, we find $R(D)_{\rm lat}^{\rm FLAG} = 0.2938(38)$.
The ratio $R(D)$ requires the integral of the branching ratios for $\ell=e,\mu,\tau$ over the whole phase space.
Since lattice simulations are sensitive mostly to relatively large $q^2$ values, lattice-only calculations of $R(D)$ rely on the extrapolation of the form factors to low $q^2$ and are especially sensitive to the choice of parameterization.
In order to estimate this source of systematics, we repeated the fit using the parameterization adopted by HPQCD in Ref.~\cite{Na:2015kha}.
The main difference with respect to our default paremeterization is the inclusion of Blaschke factors for the form factors $f_+$ and $f_0$ located at $M_+ = M_{B_c^*} = 6.330(9)$ GeV and $M_0= 6.420(9)\:{\rm GeV}\sim M_{B_{c0}}$\footnote{{The value of $M_{B_{c0}}$ is not known from experiment and here we adopted the value used by HPQCD in Ref.~\cite{Na:2015kha}.}}; additionally, the parameter $t_0$ is set to $(m_B-m_D)^2$.
Using five coefficients ($a_{1,2,3}^+$ and $a_{1,2}^0$ with $a_3^0$ fixed by the $f_+(q^2=0) = f_0(q^2=0)$ condition) we find $R(D)_{\rm lat}^{\rm HPQCD} = 0.3009(38)$ which deviates from $R(D)_{\rm lat}^{\rm FLAG}$ by 1.4 $\sigma$.
To take this potential source of systematic uncertainty into account we rescale accordingly the uncertainty of our default fit and obtain:
\begin{equation}
\Nf=2+1 \text{: } R(D)_{\rm lat} = 0.2938(54) \quad\quad [\text{FLAG average, Refs.~\cite{Lattice:2015rga, Na:2015kha}}].
\label{HQeq:RDlat}
\end{equation}
This result is about 1.5$\sigma$ lower than the current experimental average \cite{HFLAV:2022esi} for this quantity.
It has to be stressed that achieving this level of precision critically depends on the reliability with which the low-$q^2$ region is controlled by the parameterizations of the form factors.

HPQCD~17 also computes values for $R(D_s)$, the analog of $R(D)$ with both heavy-light mesons containing a strange quark. The earlier calculation using NRQCD $b$ quarks gives
\begin{gather}
\Nf=2+1 \text{: } R(D_s)_{\rm lat} = 0.301(6) \quad\quad \text{\cite{Monahan:2017uby}}.
\end{gather}
The newer calculation with HISQ $b$ quarks, HPQCD~19, yields the somewhat more precise value
\begin{gather}
\Nf=2+1+1 \text{: } R(D_s)_{\rm lat} = 0.2987(46) \quad\quad \text{\cite{McLean:2019qcx}}.
\end{gather}

A similar ratio $R(D^\ast)$ can be considered for $B \rightarrow D^\ast$ transitions.
As a matter of fact, the experimental value of $R(D^\ast)$ is significantly more precise than the one of $R(D)$. 
The most recent HFLAV average reads (see Ref.~\cite{HFLAV:2022esi} and the Moriond 2024 update):
\begin{align}
R(D^*)_{\rm exp} &= 0.287 (12) \; .
\end{align}
The recent developments in decays with vector products have yielded a variety of new lattice results for this LFU ratio.
For $\Nf=2+1$ in the sea, the Fermilab lattice and MILC collaborations (FNAL/MILC~21) report the value
\begin{gather}
\Nf=2+1 \text{: } R(D^\ast)_{\rm lat} = 0.265(13) \quad\quad \text{\cite{FermilabLattice:2021cdg}},
\end{gather}
which is around 1.5$\sigma$ lower than the current experimental average~\cite{HFLAV:2022esi}.

The JLQCD collaboration has obtained the following value (JLQCD~23)
\begin{gather}
\Nf=2+1 \text{: } R(D^\ast)_{\rm lat} = 0.252(22) \quad\quad \text{\cite{Aoki:2023qpa}}\,.
\end{gather}
Their result is compatible with the FNAL/MILC~21 value, but it increases the tension with the experimental average up to 1.6$\sigma$, in spite of the larger error.

The HPQCD collaboration has also computed this ratio using $\Nf=2+1+1$ configurations, obtaining (HPQCD~23)
\begin{gather}
\Nf=2+1+1 \text{: } R(D^\ast)_{\rm lat} = 0.273(15)\quad\quad [\text{{FLAG average,} \cite{Harrison:2023dzh}}],
\end{gather}
which is closer to the current HFLAV average, but still lower by 1.3$\sigma$.

Using the results of the $\Nf=2+1$ (FNAL/MILC~21 and JLQCD~23) \cite{FermilabLattice:2021cdg, Aoki:2023qpa} fit summarized in Tab.~\ref{tab:BDstar_latt},  we calculate the following value for the ratio $R(D^\ast)$:
\begin{alignat}{2}
\Nf=2+1 \text{: }   & R(D^\ast)_{\rm lat}  = 0.2582 (51) & \quad\quad [\text{FLAG average, Refs.~\cite{FermilabLattice:2021cdg, Aoki:2023qpa}}].
\end{alignat}

The HPQCD~23 analysis also covered the $B_s\to D_s^\ast$ channel, and for the first time a result for the $R(D^\ast_s)$ ratio is provided:
\begin{gather}
\Nf=2+1+1 \text{: } R(D_s^\ast)_{\rm lat} = 0.266(9) \quad\quad [\text{FLAG average, Ref.~\cite{Harrison:2023dzh}}].
\end{gather}

\subsubsection{Fragmentation fraction ratio $f_s / f_d$}

Another area of immediate interest in searches for physics
beyond the Standard Model is the measurement of $B_s \rightarrow \mu^+ \mu^-$ decays,
recently studied at the LHC.
One of the inputs required by the LHCb analysis is the ratio
of $B_q$ meson ($q = d,s$) fragmentation fractions $f_s / f_d$,
where $f_q$ is the probability that a $q$ quark hadronizes into a $B_q$.
This ratio can be measured by writing it as a product of ratios that involve
experimentally measurable quantities, cf. Refs.~\cite{Fleischer:2010ay,LHCb:2011ldp}.
One of the factors is the ratio $f_0^{(s)}(M_\pi^2) / f_0^{(d)}(M_K^2)$
of scalar form factors for the corresponding semileptonic meson decay,
which is where lattice input becomes useful.

A dedicated $\Nf=2+1$ study, FNAL/MILC~12C \cite{Bailey:2012rr} addresses the
ratios of scalar form factors $f_0^{(q)}(q^2)$,\footnote{This work also provided
a value for $R(D)$, now superseded by FNAL/MILC~15C~\cite{Lattice:2015rga}.} and quotes:
\begin{equation}
f_0^{(s)}(M_\pi^2) / f_0^{(d)}(M_K^2) = 1.046(44)(15),
\qquad
f_0^{(s)}(M_\pi^2) / f_0^{(d)}(M_\pi^2) = 1.054(47)(17),
\end{equation}
where the first error is statistical and the second systematic.
The more recent results from HPQCD~17~\cite{Monahan:2017uby} are:
\begin{equation}
f_0^{(s)}(M_\pi^2) / f_0^{(d)}(M_K^2) = 1.000(62),
\qquad
f_0^{(s)}(M_\pi^2) / f_0^{(d)}(M_\pi^2) = 1.006(62).
\end{equation}
Results from both groups lead to fragmentation fraction
ratios $f_s/f_d$ that are
consistent with LHCb's measurements via other methods~\cite{LHCb:2011ldp}.

%% file: HQ/HQSubsections/BSL_other.tex
\subsection{Semileptonic form factors for $B_c\to (\eta_c, J/\psi)\ell\nu$ decays}
\label{sec:Bcdecays}

In a recent publication, HPQCD 20B \cite{Harrison:2020gvo} 
provided the first full determination of $B_c\to J/\psi$ form factors,
extending earlier preliminary work that also covered $B_c \to \eta_c$,
Refs.~\cite{Lytle:2016ixw, Colquhoun:2016osw}. While the latter employed
both NRQCD and HISQ actions for the valence $b$ quark,
and the HISQ action for the $c$ quark, in HPQCD 20B the HISQ action is used
throughout for all flavours. The setup is the same as for the $B_s\to D_s$ computation
discussed above, HPQCD 19; we refer to the entries for the latter paper in summary
tables for details. The flavour-singlet nature of the final state
means that there are contributions to the relevant three-point functions from disconnected
Wick contractions, which are not discussed in the paper.

Both the $J/\psi$ and the $\eta_c$ are unstable resonances, and the correct approach on the lattice would involve treating the $J/\psi$ and the $\eta_c$ as such.
However, as in the case of the $D^\ast$ meson, their widths are very narrow (93(2) keV for the $J/\psi$ and 30.5(5) keV for the $\eta_c$).
Hence, we can consider them as stable particles on the lattice.

In the $J/\psi$ case, since the hadron in the final state has vector quantum numbers, the description of the
hadronic amplitude requires four independent form factors, which in Ref.~\cite{Harrison:2020gvo} have been chosen as
\begin{gather}
\label{eq:BcJpsiFF}
\begin{split}
 \langle  J/\psi(p',\lambda)|\bar{c}\gamma^\mu  b|B_c^-(p)\rangle =&
 \frac{2i V(q^2)}{M_{B_c} + M_{J/\psi}} \varepsilon^{\mu\nu\rho\sigma}\epsilon^*_\nu(p',\lambda) p'_\rho p_\sigma\,, \\[2.0ex]
\langle  J/\psi(p',\lambda)|\bar{c}\gamma^\mu \gamma^5 b|B_c^-(p)\rangle =&
 2M_{J/\psi}A_0(q^2)\frac{\epsilon^*(p',\lambda)\cdot q}{q^2} q^\mu\\
&\quad +(M_{B_c}+M_{J/\psi})A_1(q^2)\Big[ \epsilon^{*\mu}(p',\lambda) - \frac{\epsilon^*(p',\lambda)\cdot q}{q^2} q^\mu \Big] \\
&\quad - A_2(q^2)\frac{\epsilon^*(p',\lambda)\cdot q}{M_{B_c}+M_{J/\psi}}\Big[ p^\mu + p'^\mu - \frac{M_{B_c}^2-M_{J/\psi}^2}{q^2}q^\mu \Big],
\end{split}
\end{gather}
where $\epsilon_\mu$ is the polarization vector of the $J/\psi$ state. 
The computed form factors are fitted
to a $z$-parameterization-inspired ansatz, where coefficients are modified to model
the lattice-spacing and the heavy- and light-mass dependences, for a total of 280~fit parameters.
In the continuum and at physical kinematics only 16 parameters survive, as each form factor
is parameterized by an expression of the form
\begin{gather}
\label{eq:FFpsi}
F(q^2) = \frac{1}{P(q^2)} \sum_{n=0}^3 a_n z^n\,,
\end{gather}
where the pole factor is given by
\begin{gather}
P(q^2)=\prod_k z(q^2,M_k^2)\,,
\end{gather}
with $\{M_k\}$ a different set of pole energies below the $BD^\ast$ threshold for each set of $J^P$ quantum numbers,
taken from a mixture of experimental results, lattice determinations, and model estimates.
The values used (in GeV) are
\begin{gather}
\begin{split}
&0^-:~6.275,~6.872,~7.25;\\
&1^-:~6.335,~6.926,~7.02,~7.28;\\
&1^+:~6.745,~6.75,~7.15,~7.15.
\end{split}
\end{gather}
The outcome of the fit, that we quote as a FLAG estimate, is 
\begin{center}
\begin{tabular}{ | c | c c c c | }
\hline
& $a_0$ & $a_1$ & $a_2$ & $a_3$ \\\hline
${V}$&  0.1057(55)&     $-$0.746(92)&     0.10(98)&       0.006(1.000)\\
${A0}$& 0.1006(37)&     $-$0.731(72)&     0.30(90)&       $-$0.02(1.00)\\
${A1}$& 0.0553(19)&     $-$0.266(40)&     0.31(70)&       0.11(99)\\
${A2}$& 0.0511(91)&     $-$0.22(19)&      $-$0.36(82)&      $-$0.05(1.00)\\
\hline
\end{tabular}
\end{center}
The correlation matrix for the coefficients is provided in Tables~XIX--XXVII of
Ref.~\cite{Harrison:2020gvo}. Using these form factors, the following Standard-Model prediction
for the lepton-flavour ratio $R(J/\psi)$ is obtained:
\begin{gather}
R(J/\psi)_{\rm lat} 
= \frac{\Gamma (B_c^+ \to J/\psi \: \tau^+ \nu_\tau)}{\Gamma (B_c^+ \to J/\psi\: \mu^+ \nu_\mu)} 
=  0.2582(38)\,,~~~~~\Nf=2+1+1\:\:\mbox{\cite{Harrison:2020nrv}}.
\end{gather}

\subsection{Semileptonic form factors for $\Lambda_b\to (p,\Lambda_c^{(\ast)})\ell\bar{\nu}$ decays}
\label{sec:Lambdab}

The $b\to c\ell\bar{\nu}$ and $b\to u\ell\bar{\nu}$ transitions can also be probed
in decays of $\Lambda_b$ baryons. With the LHCb experiment, the final state of $\Lambda_b\to p\mu\bar{\nu}$
is easier to identify than that of $B\to\pi\mu\bar{\nu}$ \cite{LHCbRICHGroup:2012mgd},
and the first determination of $|V_{ub}|/|V_{cb}|$ at the Large Hadron Collider
was performed using a ratio of $\Lambda_b\to p\mu\bar{\nu}$ and $\Lambda_b\to \Lambda_c\mu\bar{\nu}$
decay rates~\cite{Aaij:2015bfa} (cf.~Sec.~\ref{sec:VubVcb}).

The amplitudes of the decays $\Lambda_b\to p\ell\bar{\nu}$ and $\Lambda_b\to \Lambda_c\ell\bar{\nu}$
receive contributions from both the vector and the axial-vector components of the current
in the matrix elements $\langle p|\bar u\gamma^\mu(\mathbf{1}-\gamma_5)b|\Lambda_b\rangle$
and $\langle \Lambda_c|\bar c\gamma^\mu(\mathbf{1}-\gamma_5)b|\Lambda_b\rangle$.
The matrix elements split into three form factors $f_+$, $f_0$, $f_\perp$
mediated by the vector component of the current, and another three form factors $g_+$, $g_0$, $g_\perp$
mediated by the axial-vector component---see, e.g., Ref.~\cite{Feldmann:2011xf}
for a complete description. Given the sensitivity to all Dirac structures,
measurements of the baryonic decay rates also provides useful complementary constraints on right-handed
couplings beyond the Standard Model \cite{Aaij:2015bfa}.

To date, only one unquenched lattice-QCD computation of the  $\Lambda_b\to p$
and $\Lambda_b\to \Lambda_c$ form factors with physical heavy-quark masses has been published: Detmold 15 \cite{Detmold:2015aaa}.
This computation uses RBC/UKQCD $\Nf=2+1$ DWF ensembles,
and treats the $b$ and $c$ quarks within the Columbia RHQ approach. The renormalization of
the currents is carried out using a mostly nonperturbative method, with residual matching factors computed
at one loop.
Two values of the lattice spacing ($a\approx0.11,~0.085~{\rm fm}$) are considered,
with the absolute scale set from the $\Upsilon(2S)$--$\Upsilon(1S)$ splitting.
Sea-pion masses lie in a narrow interval ranging from slightly above
$400~{\rm MeV}$ to slightly below $300~{\rm MeV}$, keeping $m_\pi L \gtrsim 4$;
however, lighter pion masses are considered in the valence DWF action
for the $u,d$ quarks. The lowest valence-valence pion mass is 227(3) MeV,
which leads to a \tbr~ rating of finite-volume effects.
Results for the form factors are obtained from suitable three-point functions,
and fitted to a modified $z$-expansion ansatz that combines the $q^2$-dependence
with the chiral and continuum extrapolations. The main results of the paper are
the predictions (errors are statistical and systematic, respectively)
\begin{align}
\zeta_{p\mu\bar\nu}(15{\rm GeV}^2) &\equiv \frac{1}{|V_{ub}|^2}\int_{15~{\rm GeV}^2}^{q^2_{\rm max}}\frac{{\rm d}\Gamma(\Lambda_b\to p\mu^-\bar\nu_\mu)}{{\rm d}q^2}\,{\rm d}q^2 &= 12.31(76)(77)~{\rm ps}^{-1}\,,\\
\zeta_{\Lambda_c \mu\bar\nu}(7{\rm GeV}^2) &\equiv\frac{1}{|V_{cb}|^2}\int_{7~{\rm GeV}^2}^{q^2_{\rm max}}\frac{{\rm d}\Gamma(\Lambda_b\to \Lambda_c\mu^-\bar\nu_\mu)}{{\rm d}q^2}\,{\rm d}q^2 &= 8.37(16)(34)~{\rm ps}^{-1}\,,\\
\displaystyle \frac{\zeta_{p\mu\bar\nu}(15{\rm GeV}^2)}{\zeta_{\Lambda_c \mu\bar\nu}(7{\rm GeV}^2)} &= 1.471(95)(109)\,,
\end{align}
which are the input for the LHCb analysis. Predictions for the total rates in all possible
lepton channels, as well as for ratios similar to $R(D)$ (cf.~Sec.~\ref{sec:BtoD}) between the $\tau$
and light-lepton channels are also available, in particular,
\begin{equation}
 R(\Lambda_c)=\frac{\Gamma (\Lambda_b \to \Lambda_c\: \tau^- \bar{\nu}_\tau)}{\Gamma (\Lambda_b \to \Lambda_c\: \mu^- \bar{\nu}_\mu)} = 0.3328(74)(70).
\end{equation}
Datta 2017 \cite{Datta:2017aue} additionally includes results for the ${\Lambda}_b\to {\Lambda}_c$ tensor
form factors $h_+$, $h_\perp$, $\widetilde{h}_+$, $\widetilde{h}_\perp$, based on the same lattice computation
as Detmold 15 \cite{Detmold:2015aaa}. The main focus of Datta 2017 is the phenomenology of
the $ {\Lambda}_b\to {\Lambda}_c\tau {\overline{\nu}}_{\tau } $ decay
and how it can be used to constrain contributions from beyond the Standard Model
physics. Unlike in the case of the vector and axial-vector currents, the residual matching factors
of the tensor currents are set to their tree-level value. While the matching systematic uncertainty is augmented to
take this fact into account, the procedure implies that the tensor current
retains an uncanceled logarithmic divergence at $\mathcal{O}(\alpha_s)$.

Progress with next-generation lattice calculations of the $\Lambda_b \to p$ and $\Lambda_b \to \Lambda_c$ form factors was reported in Ref.~\cite{Meinel:2023wyg}.

Recently, first lattice calculations have also been completed for $\Lambda_b$ semileptonic decays to negative-parity baryons in the final state. Such calculations are substantially more challenging and have not yet reached the same level of precision. Meinel 21 \cite{Meinel:2021rbm}, which was updated in Meinel 21B \cite{Meinel:2021mdj}, considers the decays $\Lambda_b \to \Lambda_c^*(2595)\ell\bar{\nu}$ and $\Lambda_b \to \Lambda_c^*(2625)\ell\bar{\nu}$, where the $\Lambda_c^*(2595)$ and $\Lambda_c^*(2625)$ are the lightest charm baryons with isospin 0 and $J^P=\frac12^-$ and $J^P=\frac32^-$, respectively.
 These decay modes may eventually provide new opportunities to test lepton-flavour universality at the LHC, but are also very interesting from a theoretical point of view. The lattice results for the form factors may help tighten dispersive constraints in global analyses of $b\to c$ semileptonic decays \cite{Cohen:2019zev}, and may provide new insights into the internal structure of the negative-parity heavy baryons and their description in heavy-quark-effective-theory \cite{Papucci:2021pmj,DiRisi:2023npw}. The $\Lambda_c^*(2595)$ and $\Lambda_c^*(2625)$ are very narrow resonances decaying through the strong interaction into $\Lambda_c \pi\pi$. The strong decays are neglected in Meinel 21 and Meinel 21B \cite{Meinel:2021rbm,Meinel:2021mdj}. The calculation was performed using the same lattice actions as previously for $\Lambda_b \to \Lambda_c$, albeit with newly tuned RHQ parameters. Only three ensembles are used, with $a\approx0.11,~0.08~{\rm fm}$ and pion masses in the range from approximately 300 to 430 MeV, with valence-quark masses equal to the sea-quark masses. Chiral-continuum extrapolations linear in $m_\pi^2$ and $a^2$ are performed, with systematic uncertainties estimated using higher-order fits. Finite-volume effects and effects associated with the strong decays of the $\Lambda_c^*$'s are not quantified. The calculation is done in the $\Lambda_c^*$ rest frame, where the cubic symmetry is sufficient to avoid mixing with unwanted lower-mass states. As a consequence, the calculation is limited to a small kinematic region near the zero-recoil point $w=1$. On each ensemble, lattice data were produced for two values of $w-1$ of approximately 0.01 and 0.03. The final results for the form factors are parameterized as linear functions of $w-1$ and can be found in Meinel 21B \cite{Meinel:2021mdj} and associated supplemental files. 

\subsection{Semileptonic form factors for $\Lambda_b\to \Lambda^{(\ast)}\ell\ell$}
\label{sec:LambdabLambda}

The decays $\Lambda_b\to \Lambda\ell^+\ell^-$ are mediated by the same underlying $b\to s\ell^+\ell^-$ FCNC transition as, for
example, $B\to K\ell^+\ell^-$ and $B\to K^*\ell^+\ell^-$, and can therefore provide additional information on the hints for
physics beyond the Standard Model seen in the meson decays. The $\Lambda$ baryon in the final state decays through the weak
interaction into $p \pi^-$ (or $n \pi^0$), leading to a wealth of angular observables even for unpolarized $\Lambda_b$. When
including the effects of a nonzero $\Lambda_b$ polarization, $\Lambda_b\to \Lambda(\to p \pi^-)\ell^+\ell^-$ decays are
characterized by five angles leading to 34 angular observables \cite{Blake:2017une}, which have been measured by LHCb in
the bin $q^2\in[15,20]\:{\rm GeV}^2$ \cite{Aaij:2018gwm}. Given that the $\Lambda$ is stable under the strong
interactions, the $\Lambda_b \to \Lambda$ form factors parametrizing the matrix elements of local $\bar{s}\Gamma b$
currents can be calculated on the lattice with high precision using standard methods. Of course, the process
$\Lambda_b\to \Lambda\ell^+\ell^-$ also receives contributions from nonlocal matrix elements of four-quark and
quark-gluon operators in the weak effective Hamiltonian combined with the electromagnetic current. As with the
mesonic $b\to s\ell^+\ell^-$ decays, these contributions cannot easily be calculated  on the lattice and one
relies on other theoretical tools for them, including the local OPE at high $q^2$ and a light-cone OPE / QCD
factorization at low $q^2$.

Following an early calculation with static $b$ quarks \cite{Detmold:2012vy}, Detmold 16 \cite{Detmold:2016pkz} provides results for
all ten relativistic $\Lambda_b\to \Lambda$ form factors parametrizing the matrix elements of the local vector, axial-vector and
tensor $b\to s$ currents.  The lattice setup is identical to that used in the 2015 calculation of the $\Lambda_b \to p$ form
factors in Detmold 15 \cite{Detmold:2015aaa}, and similar considerations as in the previous section thus apply. The lattice data cover the
upper 60\% of the $q^2$ range, and the form factors are extrapolated to the full $q^2$ range using BCL $z$-expansion fits.
This extrapolation is done simultaneously with the chiral and continuum extrapolations. The caveat regarding the
renormalization of the tensor currents also applies here. Progress with next-generation lattice calculations of the $\Lambda_b \to \Lambda$ form factors was reported in Ref.~\cite{Meinel:2023wyg}.

Reference \cite{Blake:2019guk} uses the lattice results for the $\Lambda_b\to \Lambda$ form factors together with the
experimental results for $\Lambda_b\to \Lambda(\to p \pi^-)\mu^+\mu^-$ from LHCb \cite{Aaij:2015xza,Aaij:2018gwm} to perform
fits of the $b\to s\mu^+\mu^-$ Wilson coefficients and of the $\Lambda_b$ polarization parameter. Given the uncertainties
(which are still dominated by experiment), the results for the Wilson coefficients are presently consistent both with the
Standard-Model values and with the deviations seen in global fits that include all mesonic decays
\cite{Alguero:2019ptt,Altmannshofer:2021qrr}.		 

As with the $b\to c$ semileptonic form factors, a first lattice calculation, Meinel 2020 \cite{Meinel:2020owd} (updated in Meinel 21B \cite{Meinel:2021mdj}), 
was also completed for a $b\to s$ transition to a negative-parity baryon in the final state, in this case the $\Lambda^*(1520)$ with $J^P=\frac32^-$ 
(no calculation has yet been published for the strange $J^P=\frac12^-$ final states, which would
be the broader and even more challenging $\Lambda^*(1405)/\Lambda^*(1380)$ \cite{Zyla:2020zbs}). The $\Lambda^*(1520)$
decays primarily to $ pK^-/n \bar{K}^0$, $\Sigma \pi$, and $\Lambda\pi\pi$ with a total width of $15.6\pm 1.0$ MeV
\cite{Zyla:2020zbs} . The analysis of the lattice data again neglects the strong decays and does not quantify
finite-volume effects, and is again limited to a small kinematic region near $q^2_{\rm max}$. The results of Meinel 2020 are superseded by
Meinel 21B \cite{Meinel:2021mdj}, in which the fits to the lattice data were improved by including exact endpoint relations in the form-factor parametrizations.

\begin{table}[h]
\begin{center}
\mbox{} \\[3.0cm]
\footnotesize
\begin{tabular}{l l @{\extracolsep{\fill}} r l l l l l l l}
Process & Collaboration & Ref. & $\Nf$ & 
\hspace{0.15cm}\begin{rotate}{60}{publication status}\end{rotate}\hspace{-0.15cm} &
\hspace{0.15cm}\begin{rotate}{60}{continuum extrapolation}\end{rotate}\hspace{-0.15cm} &
\hspace{0.15cm}\begin{rotate}{60}{chiral extrapolation}\end{rotate}\hspace{-0.15cm}&
\hspace{0.15cm}\begin{rotate}{60}{finite volume}\end{rotate}\hspace{-0.15cm}&
\hspace{0.15cm}\begin{rotate}{60}{renormalization}\end{rotate}\hspace{-0.15cm}  &
\hspace{0.15cm}\begin{rotate}{60}{heavy-quark treatment}\end{rotate}\hspace{-0.15cm} \\
&&&&&&&& \\[-0.1cm]
\hline
\hline
&&&&&&&& \\[-0.1cm]
$\Lambda_b\to  \Lambda_c^*(2625) \,\ell^- \bar{\nu}_\ell$  & Meinel 21B                    & \cite{Meinel:2021mdj}                & 2+1 & \gA & \soso & \soso & \tbr  & \soso & \okay \\[0.5ex]
$\Lambda_b\to  \Lambda_c^*(2595) \,\ell^- \bar{\nu}_\ell$  & Meinel 21B                    & \cite{Meinel:2021mdj}                & 2+1 & \gA & \soso & \soso & \tbr  & \soso & \okay \\[0.5ex]
$\Lambda_b\to  \Lambda_c^*(2625) \,\ell^- \bar{\nu}_\ell$  & Meinel 21                         & \cite{Meinel:2021rbm}                & 2+1 & \gA & \soso & \soso & \tbr  & \soso & \okay \\[0.5ex]
$\Lambda_b\to  \Lambda_c^*(2595) \,\ell^- \bar{\nu}_\ell$  & Meinel 21                         & \cite{Meinel:2021rbm}                & 2+1 & \gA & \soso & \soso & \tbr  & \soso & \okay \\[0.5ex]
$\Lambda_b\to  \Lambda^*(1520) \,\ell^+\ell^-$    & Meinel 21B                    & \cite{Meinel:2021mdj}                & 2+1 & \gA & \soso & \soso & \tbr  & \soso & \okay \\[0.5ex]
$\Lambda_b\to  \Lambda^*(1520) \,\ell^+\ell^-$    & Meinel 20                         & \cite{Meinel:2020owd}                & 2+1 & \gA & \soso & \soso & \tbr  & \soso & \okay \\[0.5ex]
$\Lambda_b\to  \Lambda \,\ell^+\ell^-$            & Detmold 16                        & \cite{Detmold:2016pkz}               & 2+1 & \gA & \soso & \soso & \tbr  & \soso & \okay \\[0.5ex]
$\Lambda_b\to  p \,\ell^- \bar{\nu}_\ell$         & Detmold 15 \hspace{1ex}           & \cite{Detmold:2015aaa}               & 2+1 & \gA & \soso & \soso & \tbr  & \soso & \okay \\[0.5ex]
$\Lambda_b\to  \Lambda_c \,\ell^- \bar{\nu}_\ell$ & Detmold 15, Datta 17 \hspace{1ex} & \cite{Detmold:2015aaa,Datta:2017aue} & 2+1 & \gA & \soso & \soso & \tbr  & \soso & \okay \\[0.5ex]
&&&&&&&& \\[-0.1cm]
\hline
\hline
\end{tabular}
\caption{
Summary of computations of bottom-baryon semileptonic form factors (see also 
Refs.~\cite{Detmold:2012vy,Detmold:2013nia} 
for calculations with static $b$ quarks). The rationale for the \tbr~ rating of finite-volume effects in Meinel 20, Meinel 21, and Meinel 21B
(despite meeting the {\color{green}\Large$\circ$} criterion based on the minimum pion mass) 
is that the unstable nature of the final-state baryons was neglected in the analysis.
}
\label{tab_BottomBaryonSLsumm2}
\end{center}
\end{table}

\FloatBarrier

%% file: HQ/HQSubsections/BCKM.tex
\subsection{Determination of $|V_{ub}|$}
\label{sec:Vub}

We now use the lattice-determined Standard Model transition amplitudes
for leptonic (Sec.~\ref{sec:fB}) and semileptonic
(Sec.~\ref{sec:BtoPiK}) $B$-meson decays to obtain exclusive
determinations of the CKM matrix element $|V_{ub}|$.
In this section, we describe the aspect of our work
that involves experimental input for the relevant charged-current
exclusive decay processes.
The relevant
formulae are Eqs.~(\ref{eq:B_leptonic_rate})
and~(\ref{eq:B_semileptonic_rate}). Among leptonic channels the only
input comes from $B\to\tau\nu_\tau$, since the rates for decays to $e$
and $\mu$ have not yet been measured.  In the semileptonic case, we
only consider $B\to\pi\ell\nu$ transitions (experimentally
measured for $\ell=e,\mu$).

We first investigate the determination of $|V_{ub}|$ through the
$B\to\tau\nu_\tau$ transition.  
The experimental measurements of the branching fraction of
this channel, $B(B^{-} \to \tau^{-} \bar{\nu})$, have not been
updated since the publication of FLAG~16~\cite{Aoki:2016frl}. The status of the experimental results for this branching fraction, summarized in Tab.~\ref{tab:leptonic_B_decay_exp}, is unchanged from FLAG~16~\cite{Aoki:2016frl}. Our corresponding values of $|V_{ub}|$ are unchanged from FLAG~19~\cite{FlavourLatticeAveragingGroup:2019iem}.
\begin{table}[h]
\begin{center}
\noindent
\begin{tabular*}{\textwidth}[l]{@{\extracolsep{\fill}}lll}
Collaboration & Tagging method  & $B(B^{-}\to \tau^{-}\bar{\nu})
                                  \times 10^{4}$\\
&& \\[-2ex]
\hline \hline &&\\[-2ex]
Belle~\cite{Adachi:2012mm} &  Hadronic  & $0.72^{+0.27}_{-0.25}\pm 0.11$ \\
Belle~\cite{Kronenbitter:2015kls} &  Semileptonic & $1.25 \pm 0.28 \pm
                                                    0.27$ \\
&& \\[-2ex]
 \hline
&& \\[-2ex]
BaBar~\cite{Lees:2012ju} & Hadronic & $1.83^{+0.53}_{-0.49}\pm 0.24$\\
BaBar~\cite{Aubert:2009wt} & Semileptonic  & $1.7 \pm 0.8 \pm 0.2$\\
&& \\[-2ex]
\hline \hline && \\[-2ex]
\end{tabular*}
\caption{Experimental measurements for $B(B^{-}\to \tau^{-}\bar{\nu})$.
  The first error on each result is statistical, while the second
  error is systematic.}
\label{tab:leptonic_B_decay_exp}
\end{center}
\end{table}

It is obvious that all the measurements listed in Tab.~\ref{tab:leptonic_B_decay_exp} have significance smaller than
$5\sigma$, and the large uncertainties are dominated by statistical errors. These measurements lead to the averages
of experimental measurements for $B(B^{-}\to \tau \bar{\nu})$~\cite{Kronenbitter:2015kls,Lees:2012ju},
\begin{eqnarray}
 B(B^{-}\to \tau \bar{\nu} )\times 10^4 &=& 0.91 \pm 0.22 \mbox{ }{\rm from}\mbox{ } {\rm Belle,} \label{eq:leptonic_B_decay_exp_belle}\\
                             &=& 1.79 \pm 0.48 \mbox{ }{\rm from }\mbox{ } {\rm BaBar,} \label{eq:leptonic_B_decay_exp_babar}\\
                             &=& 1.06 \pm 0.33 \mbox{ }{\rm average,}
\label{eq:leptonic_B_decay_exp_ave}
\end{eqnarray}
where, following our standard procedure, we perform a weighted average and rescale the uncertainty by the square root of the reduced chi-squared. Note that the Particle Data Group~\cite{ParticleDataGroup:2024cfk} did not inflate the uncertainty in the calculation of the averaged branching ratio.

Combining the results in Eqs.~(\ref{eq:leptonic_B_decay_exp_belle}--\ref{eq:leptonic_B_decay_exp_ave}) with
the experimental measurements of the mass of the $\tau$-lepton and the
$B$-meson lifetime and mass we get
\begin{eqnarray}
 |V_{ub}| f_{B} &=& 0.72 \pm 0.09 \mbox{ }{\rm MeV}\mbox{ }{\rm from}\mbox{ } {\rm Belle,}\label{eq:Vub_fB_belle}\\
                &=& 1.01 \pm 0.14 \mbox{ }{\rm MeV}\mbox{ }{\rm from }\mbox{ } {\rm BaBar,}\label{eq:Vub_fB_babar} \\
                &=& 0.77 \pm 0.12 \mbox{ }{\rm MeV}\mbox{ } {\rm average,}\label{eq:Vub_fB}
\end{eqnarray}
which can be used to extract $|V_{ub}|$ using the averages in Eqs.~(\ref{eq:fB2}), (\ref{eq:fB21}) and (\ref{eq:fB211}), viz.,
\begin{align}
\phantom{\Nf = 2\text{: }} &|V_{ub}| = 3.83(14)(48) \times 10^{-3} \quad\quad [B\to\tau\nu_\tau \text{, Belle}], \\
         \Nf = 2\text{: }  &|V_{ub}| = 5.37(20)(74) \times 10^{-3} \quad\quad [B\to\tau\nu_\tau \text{, Babar}], \\
\phantom{\Nf = 2\text{: }} &|V_{ub}| = 4.10(15)(64) \times 10^{-3} \quad\quad [B\to\tau\nu_\tau \text{, average}], \\
\nonumber \\
\phantom{\Nf = 2+1\text{: }} &|V_{ub}| = 3.75(8)(47) \times 10^{-3} \quad\quad [B\to\tau\nu_\tau \text{, Belle}], \\
         \Nf = 2+1\text{: }  &|V_{ub}| = 5.26(12)(73) \times 10^{-3} \quad\quad [B\to\tau\nu_\tau \text{, Babar}], \\
\phantom{\Nf = 2+1\text{: }} &|V_{ub}| = 4.01(9)(63) \times 10^{-3} \quad\quad [B\to\tau\nu_\tau \text{, average}], \label{eq:vub_btn_2p1}\\
\nonumber \\
\phantom{\Nf = 2+1+1\text{: }} &|V_{ub}| = 3.79(3)(47) \times 10^{-3} \quad\quad [B\to\tau\nu_\tau \text{, Belle}], \\
         \Nf = 2+1+1\text{: }  &|V_{ub}| = 5.32(4)(74) \times 10^{-3} \quad\quad [B\to\tau\nu_\tau \text{, Babar}], \\
\phantom{\Nf = 2+1+1\text{: }} &|V_{ub}| = 4.05(3)(64) \times 10^{-3} \quad\quad [B\to\tau\nu_\tau \text{, average}], \label{eq:vub_btn_2p1p1}
\end{align}
where the first error comes from the uncertainty in $f_B$ and the second comes from experiment. The experimental branching fractions do not yet meet the five-sigma discovery threshold and the relative uncertainties are significantly larger than the radiative electroweak corrections. Therefore, in line with the Particle Data Group~\cite{ParticleDataGroup:2024cfk} and in contrast to the $D_{(s)}$ decays, we do not include in these results the electroweak corrections.

Let us now turn our attention to semileptonic decays. The experimental
value of $|V_{ub}|f_+(q^2)$ can be extracted from the measured
branching fractions for $B^0\to\pi^\pm\ell\nu$ or
$B^\pm\to\pi^0\ell\nu$ by applying
Eq.~(\ref{eq:B_semileptonic_rate}).\footnote{Since $\ell=e,\mu$ the
  contribution from the scalar form factor in
  Eq.~(\ref{eq:B_semileptonic_rate}) is negligible.}  
We then determine $|V_{ub}|$
by performing fits to the constrained BCL $z$-parameterization of the form factor $f_+(q^2)$ given in
Eq.~(\ref{eq:bcl_c}). This can be done in two ways: one option is to
perform separate fits to lattice and experimental results, and extract
the value of $|V_{ub}|$ from the ratio of the respective $a_0$
coefficients; a second option is to perform a simultaneous fit to
lattice and experimental data, leaving their relative normalization
$|V_{ub}|$ as a free parameter. We adopt the second strategy, because
it combines the lattice and experimental input in a more efficient
way, leading to a smaller uncertainty on $|V_{ub}|$.

The available state-of-the-art experimental input consists of five
data sets: three untagged measurements by BaBar
(6-bin~\cite{delAmoSanchez:2010af} and 12-bin~\cite{Lees:2012vv}) and
Belle~\cite{Ha:2010rf}, all of which assume isospin symmetry and
provide combined $B^0\to\pi^-$ and $B^+\to\pi^0$ data; and the two
tagged Belle measurements of $\bar B^0\to\pi^+$ (13-bin) and
$B^-\to\pi^0$ (7-bin)~\cite{Sibidanov:2013rkk}.  Including all of
them, along with the available information about cross-correlations,
will allow us to obtain a meaningful final error
estimate.\footnote{See, e.g., Sec.~V.D of Ref.~\cite{Lattice:2015tia} for
  a detailed discussion.} The lattice input data set will be that
discussed in Sec.~\ref{sec:BtoPiK}.

\begin{figure}[tbp]
\begin{center}
\includegraphics[width=0.49\textwidth]{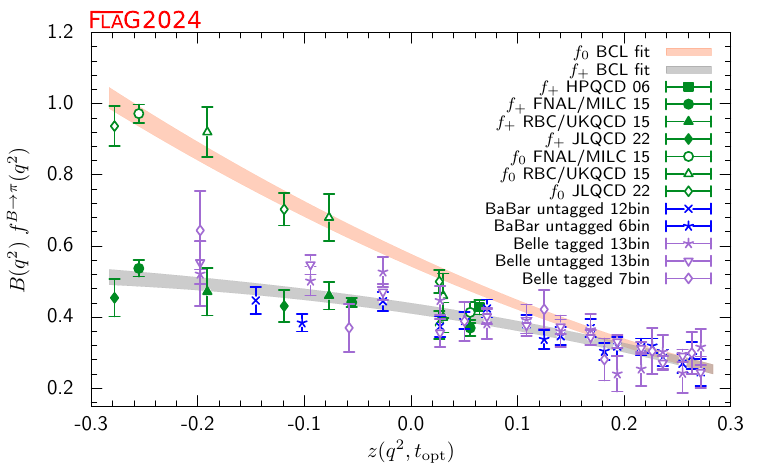}
\includegraphics[width=0.49\textwidth]{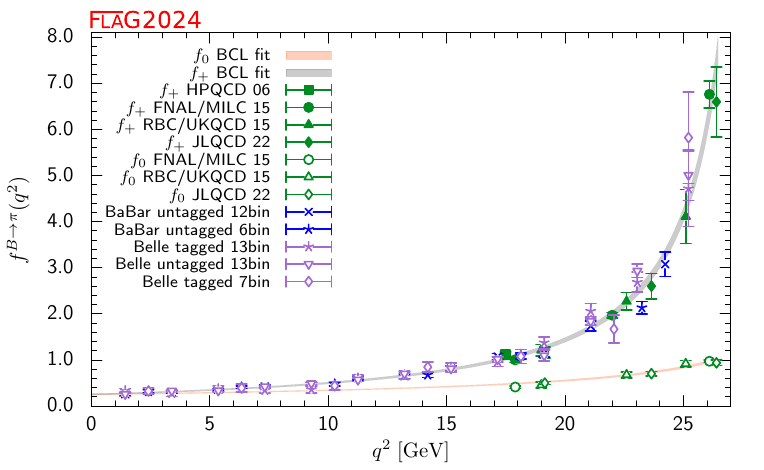}
\caption{
Lattice and experimental data for $f_+^{B\to\pi}(q^2)$ and $f_0^{B\to \pi} (q^2)$ versus $z$ (left panel) and $q^2$ (right panel). Experimental data has been rescaled by the value for $|V_{ub}|$ found from the joint fit. Green symbols denote lattice-QCD points included in the fit, while blue and indigo points show experimental data divided by the value of $|V_{ub}|$ obtained from the fit. The grey and orange bands display the preferred $N^+ = N^0 = 3$ BCL fit (five $z$-parameters and $|V_{ub}|$).}
\label{fig:Vub_SL_fit}
\end{center}
\end{figure}

We perform a constrained BCL fit of the vector and scalar form factors (this is necessary in order to take into account the $f_+(q^2=0) = f_0 (q^2=0)$ constraint) together with the combined experimental data sets. We find that the error on $|V_{ub}|$ stabilizes for $N^+ = N^0 = 3$. The result of the combined fit is presented in Tab.~\ref{tab:FFVUBPI}. The fit has a chi-square per degree of freedom $\chi^2/{\rm dof} = 116.7/62=1.88$. Following the PDG recommendation, we rescale the whole covariance matrix by $\chi^2/{\rm dof}$: the errors on the $z$-parameters are increased by $\sqrt{\chi^2/{\rm dof}} = 1.37$ and the correlation matrix is unaffected. The value of $|V_{ub}|$ which we obtain is:
\begin{align}
\Nf = 2+1 \text{: } |V_{ub}| = ( 3.61 \pm 0.16 ) \times 10^{-3} \nonumber \\
&\hskip -3.7cm [B\to \pi\ell\nu \text{, FLAG average, Refs.~\cite{Lattice:2015tia, Flynn:2015mha, Colquhoun:2022atw, delAmoSanchez:2010af, Lees:2012vv, Ha:2010rf, Sibidanov:2013rkk}}].
\label{eq:vub_b2pi}
\end{align}
\begin{table}[t]
\begin{center}
\begin{tabular}{|c|c|cccccc|}
\multicolumn{8}{l}{$B\to \pi\ell\nu \; (\Nf=2+1)$} \\[0.2em]\hline
        & Central Values & \multicolumn{6}{|c|}{Correlation Matrix} \\[0.2em]\hline
$|V_{ub}^{}| \times 10^3$ & 3.61 (16)   &  1 & $-$0.812 & $-$0.108 & 0.128 & $-$0.326 & $-$0.151 \\[0.2em]
$a_0^+$                 & 0.425 (15)  &    $-$0.812 & 1 & $-$0.188 & $-$0.309 & 0.409 & 0.00926 \\[0.2em]
$a_1^+$                 & $-$0.441 (39) &  $-$0.108 & $-$0.188 & 1 & $-$0.498 & $-$0.0343 & 0.150\\[0.2em]
$a_2^+$                 & $-$0.52 (13)  & 0.128 & $-$0.309 & $-$0.498 & 1 & $-$0.190 & 0.128 \\[0.2em]
$a_0^0$                 & 0.560 (17)  &   $-$0.326 & 0.409 & $-$0.0343 & $-$0.190 & 1 & $-$0.772  \\[0.2em]
$a_1^0$                 & $-$1.346 (53) & $-$0.151 & 0.00926 & 0.150 & 0.128 & $-$0.772 & 1 \\[0.2em]
\hline
\end{tabular} 
\end{center}
\caption{Value of $|V_{ub}|$, coefficients for the $N^+ =N^0=N^T=3$ $z$-expansion of the $B\to \pi$ form factors $f_+$ and $f_0$, and their correlation matrix. The chi-square per degree of freedom is $\chi^2/{\rm dof} = 116.7/62=1.88$ and the errors on the fit parameters have been rescaled by $\sqrt{\chi^2/{\rm dof}} =1.37$. The lattice calculations that enter this fit are taken from FNAL/MILC~15~\cite{Lattice:2015tia}, RBC/UKQCD~15~\cite{Flynn:2015mha} and \SLjlqcdBpi~\cite{Colquhoun:2022atw}. The experimental inputs are taken from BaBar~\cite{delAmoSanchez:2010af,Lees:2012vv} and Belle~\cite{Ha:2010rf,Sibidanov:2013rkk}. The form factors can be reconstructed using parameterization and inputs given in Appendix~\ref{sec:app_B2pi}.
\label{tab:FFVUBPI}}
\end{table}
In Fig.~\ref{fig:Vub_SL_fit}, we show both the lattice and experimental data for
$(1-q^2/m_{B^*}^2)f_+(q^2)$ as a function of $z(q^2)$, together with our preferred fit;
experimental data has been rescaled by the resulting value for $|V_{ub}|^2$.
It is worth noting the good consistency between the form-factor shapes from
lattice and experimental data. This can be quantified, e.g., by computing the ratio of the
two leading coefficients in the constrained BCL parameterization: the fit to lattice form
factors yields  $a_1^+/a_0^+=-1.20(23)$   (cf.~the results presented in Sec.~\ref{sec:BtoPi}),
while the above lattice+experiment fit yields  $a_1^+/a_0^+=-1.039(94)$. 

Finally we combine the $\Nf = 2+1$ determinations of $|V_{ub}|$ from $B\to \tau \nu$ and $B\to \pi\ell\nu$ in Eqs.~(\ref{eq:vub_btn_2p1}) and (\ref{eq:vub_b2pi_final}) and obtain:
\begin{align}
\Nf = 2+1 \text{: } |V_{ub}| = ( 3.63 \pm 0.16 ) \times 10^{-3} \nonumber \\
& \hskip -2.7cm  [B\to (\pi\ell\nu,\tau\nu) \text{, FLAG average,} \nonumber \\
& \hskip -2.7cm \text{\; Refs.~\cite{Lattice:2015tia, Flynn:2015mha, Colquhoun:2022atw, delAmoSanchez:2010af, Lees:2012vv, Ha:2010rf, Sibidanov:2013rkk, Bazavov:2011aa,McNeile:2011ng,Na:2012kp,Aoki:2014nga,Christ:2014uea, Kronenbitter:2015kls, Lees:2012ju}}].
\label{eq:vub_b2pi_final}
\end{align}

Our results are summarized in Tab.~\ref{tab:Vubsummary} and will be displayed after our discussion of $|V_{cb}|$. In that figure
we also show the PDG inclusive determination $|V_{ub}^{}|_{\rm incl} = (4.13 \pm 0.12_{\rm exp} \pm {{}^{+0.13}_{-0.14}}_{\rm theo} \pm 0.18_{\Delta\text{model}}) \times 10^{-3}$~\cite{ParticleDataGroup:2024cfk} (the $\Delta\text{model}$ error has been added in Ref.~\cite{ParticleDataGroup:2024cfk} to account for the spread in results obtained using different theoretical models).

\begin{table}[!t]
\begin{center}
\noindent
\begin{tabular*}{\textwidth}[l]{@{\extracolsep{\fill}}lcc}
 & from  & $|V_{ub}| \times 10^3$\\
&& \\[-2ex]
\hline \hline &&\\[-2ex]
FLAG average for $\Nf=2+1$ & $B \to \pi\ell\nu$ & $3.61(16)$ \\
FLAG average for $\Nf=2+1$ & $B \to \tau\nu$ &  $4.01(64)$  \\
FLAG average for $\Nf=2+1$ & $B \to (\pi\ell\nu,\tau\nu)$  & $3.63(16)$ \\
&& \\[-2ex]
 \hline
FLAG average for $\Nf=2+1+1$ & $B \to \tau\nu$ & $4.05(64)$ \\
&& \\[-2ex]
 \hline
 \hline
PDG 24 & $B \to X_u\ell\nu$ & $4.13(26)$ \\
&& \\[-2ex]
\hline \hline && \\[-2ex]
\end{tabular*}
\caption{Results for $|V_{ub}|$. The averages involving $B\to\pi\ell\nu$ and $B\to \tau\nu$ can be found in Eqs.~(\ref{eq:vub_b2pi}), (\ref{eq:vub_btn_2p1}), (\ref{eq:vub_b2pi_final}) and (\ref{eq:vub_btn_2p1p1}); all uncertainties have been added in quadrature. The inclusive average is taken from PDG 24~\cite{ParticleDataGroup:2024cfk}. The lattice calculations for the $B\to \pi$ form factors are taken from Refs.~\cite{Lattice:2015tia, Flynn:2015mha, Colquhoun:2022atw}, for $f_B$ at $\Nf = 2+1$ from Refs.~\cite{Bazavov:2011aa,McNeile:2011ng,Na:2012kp,Aoki:2014nga,Christ:2014uea} and for $f_B$ at $\Nf = 2+1+1$ from Refs.~\cite{Dowdall:2013tga,Bussone:2016iua,Hughes:2017spc,Bazavov:2017lyh}.}
\label{tab:Vubsummary}
\end{center}
\end{table}

\subsection{Determination of $|V_{cb}|$}
\label{sec:Vcb}
We now combine the lattice-QCD results for the $B \to D^{(*)}$ form factors with all available experimental information on $B\to D^{(*)} \ell\nu\; (\ell=e,\mu)$ semileptonic decays to obtain determinations of the CKM matrix element $|V_{cb}|$ in the Standard Model. 

For $B\to D$ we perform a joint fit to the available lattice data, i.e., the $\Nf = 2+ 1$ FNAL/MILC~15C~\cite{Lattice:2015rga} and HPQCD~15~\cite{Na:2015kha} calculations discussed in Sec.~\ref{sec:BtoD}, and state-of-the-art experimental determinations. We combine the Belle measurement~\cite{Belle:2015pkj},
which provides partial integrated decay rates in 10 bins in the recoil parameter $w$, with the 2010 BaBar data set in Ref.~\cite{Aubert:2009ac}, which quotes the value of $\mathcal{G}^{B\to D}(w)\eta_{\rm EW}|V_{cb}|$ for 10 values of $w$.\footnote{We thank Marcello Rotondo for providing the 10 bins result of the BaBar analysis.} The fit is dominated by the more precise Belle data. Given this, and the fact that only partial correlations among systematic uncertainties are to be expected, we will treat both data sets as uncorrelated.\footnote{We have checked that results using just one experimental data set are compatible within $1\sigma$. In the case of BaBar, we have taken into account the introduction of some EW corrections in the data.} The formula for the differential $B\to D \ell\nu$ branching ratio is given in Eq.~(\ref{eq:vxb:BtoD}). 

\begin{figure}[!t]
\begin{center}
\includegraphics[width=0.49\textwidth]{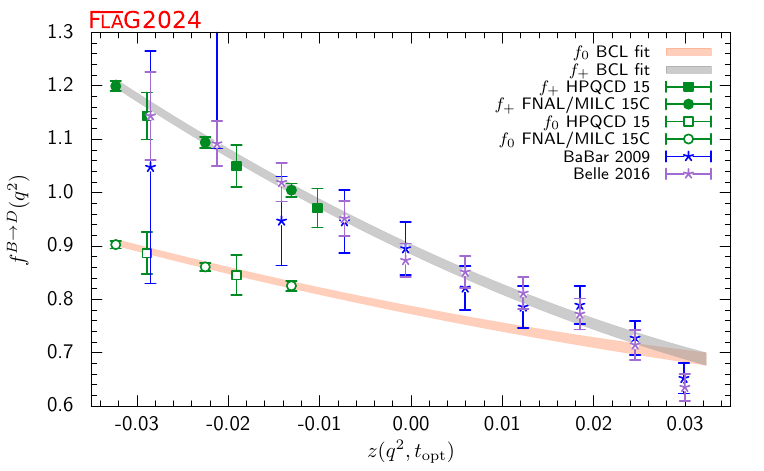}
\includegraphics[width=0.49\textwidth]{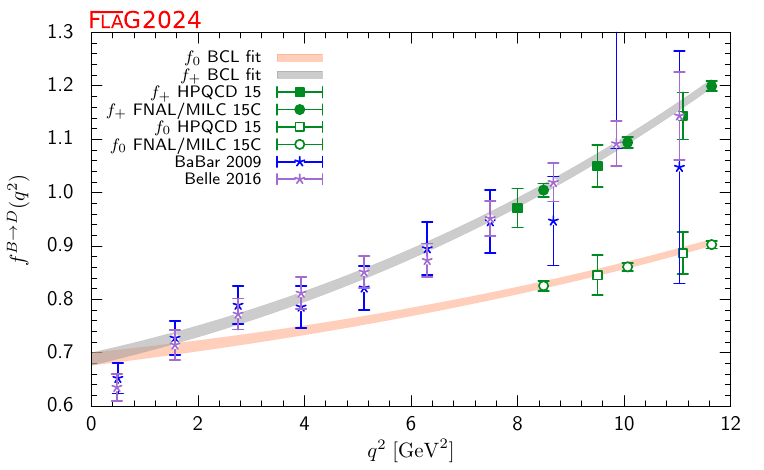}
\caption{
Lattice and experimental data for $f_+^{B\to D}(q^2)$ and $f_0^{B\to D}(q^2)$ versus $z$ (left panel) and $q^2$ (right panel). Green symbols denote lattice-QCD points included in the fit, while blue and indigo
points show experimental data divided by the value of $|V_{cb}|$ obtained from the fit. The grey and orange bands display the preferred $N^+=N^0=3$ BCL fit (five $z$-parameters and $|V_{cb}|$).
}
\label{fig:Vcb_SL_fit}
\end{center}
\end{figure}

A constrained $(N^+ = N^0 = 3)$ BCL fit using the same ansatz as for lattice-only data in Sec.~\ref{sec:BtoD} yields our average:
\begin{align}
\Nf=2+1 \text{: } |V_{cb}| &= 40.0 (1.0) \times 10^{-3} \nonumber \\
& [B\to D\ell\nu \text{, FLAG average, Refs.~\cite{Lattice:2015rga, Na:2015kha, Belle:2015pkj, Aubert:2009ac}}].
\end{align}
The fit has a chi-square per degree of freedom $\chi^2/{\rm dof} = 20.0/25=0.80$.  The result of the full fit, including the correlation matrix between $|V_{cb}|$ and the BCL coefficients is presented in Tab.~\ref{tab:FFVCBD} and illustrated in Fig.~\ref{fig:Vcb_SL_fit}. In passing, we note that, if correlations between the FNAL/MILC and HPQCD calculations are neglected, the $|V_{cb}|$ central value rises to $40.3 \times 10^{-3}$ in nice agreement with the results presented in Ref.~\cite{Bigi:2016mdz}. 

Finally, using the fit results in Tab.~\ref{tab:FFVCBD}, we extract a value for $R(D)$ which includes both lattice and experimental information:
\begin{align}
\Nf=2+1 \text{: } R(D)_{\rm lat+exp} &= 0.2955 (32) \nonumber \\
& [\text{FLAG average, Refs.~\cite{Lattice:2015rga, Na:2015kha, Belle:2015pkj, Aubert:2009ac}}].
\end{align} 
Note that we do not need to rescale the uncertainty on $R(D)_{\rm lat+exp}$ because, after the inclusion of experimental $B\to D \ell\nu \; (\ell =e,\mu)$ results, the shift in central value caused by using a different parameterization is negligible (see the discussion above Eq.~(\ref{HQeq:RDlat})). 
\begin{table}[t]
\begin{center}
\begin{tabular}{|c|c|cccccc|}
\multicolumn{8}{l}{$B\to D\ell\nu \; (\Nf=2+1)$} \\[0.2em]\hline
        & Central Values & \multicolumn{6}{|c|}{Correlation Matrix} \\[0.2em]\hline
$|V_{cb}^{}| \times 10^3$ & 40.0 (1.0) &   1.00 & $-$0.525 & $-$0.339 & 0.0487 & $-$0.521 & $-$0.433 \\[0.2em]
$a_0^+$                   & 0.8946 (94) &   $-$0.525 & 1.00 & 0.303 & $-$0.351 & 0.953 & 0.529 \\[0.2em]
$a_1^+$                   & $-$8.03 (16)  &  $-$0.339 & 0.303 & 1.00 & 0.203 & 0.375 & 0.876 \\[0.2em]
$a_2^+$                   & 50.1 (3.1)  &  0.0487 & $-$0.351 & 0.203 & 1.00 & $-$0.276 & 0.196 \\[0.2em]
$a_0^0$                   & 0.7804 (75) &   $-$0.521 & 0.953 & 0.375 & $-$0.276 & 1.0 & 0.502 \\[0.2em]
$a_1^0$                   & $-$3.38 (16)  &   $-$0.433 & 0.529 & 0.876 & 0.196 & 0.502 & 1.0 \\[0.2em]
\hline
\end{tabular}
\end{center}
\caption{Value of $|V_{cb}|$, coefficients for the $N^+ =N^0$ $z$-expansion of the $B\to D$ form factors $f_+$ and $f_0$, and their correlation matrix. The coefficient $a_2^0$ is fixed by the $f_+(q^2=0) = f_0(q^2=0)$ constrain. The chi-square per degree of freedom is $\chi^2/{\rm dof} = 20.0/25=0.80$. The lattice calculations that enter this fit are taken from FNAL/MILC~15C~\cite{Lattice:2015rga} and HPQCD~15~\cite{Na:2015kha}. The experimental inputs are taken from BaBar~\cite{Aubert:2009ac} and Belle~\cite{Belle:2015pkj}. The form factors can be reconstructed using parameterization and inputs given in Appendix~\ref{sec:app_B2D}.
\label{tab:FFVCBD}}
\end{table}
For $B\to D^*$, we perform a joint fit to all available lattice and experimental data. On the lattice side, we consider separately the two $\Nf=2+1$ calculations FNAL/MILC~21~\cite{FermilabLattice:2021cdg} and JLQCD~23~\cite{Aoki:2023qpa}  and the single $\Nf = 2+1+1$ HPQCD~23~\cite{Harrison:2023dzh} calculation. On the experimental side, the situation is more complicated because we need to combine the following results.
\begin{itemize}
\item The Belle untagged measurement~\cite{Belle:2018ezy} of the differential $B^0\to D^{*-} \ell^+ \nu_\ell$ partial width.
\item The Belle tagged measurement~\cite{Belle:2023bwv} of the \underline{\it normalized} differential $B\to D^* \ell\nu_\ell$ partial width (averaged over the $B^-$ and $\bar B^0$ modes).
\item The Belle II tagged measurement~\cite{Belle-II:2023okj} of the \underline{\it normalized} differential $\bar B^0 \to D^{*+} \ell^- \bar\nu_\ell$ partial width.
\item The Belle II tagged branching ratio measurement ${\rm BR} (\bar B^0 \to D^{*+} \ell^- \bar\nu_\ell) = (4.922 \pm 0.023\pm  0.220)\%$~\cite{Belle-II:2023okj}.
\item A modified HFLAV world average for the branching ratio of $\bar B^0 \to D^{*+} \ell^- \bar\nu_\ell$ mode in which the contributions from the Belle untagged~\cite{Belle:2018ezy} (already included in the differential results we use) and Belle II tagged~\cite{Belle-II:2020dyp} (superseded by the Belle II tagged result~\cite{Belle-II:2023okj} which we include separately) measurements have been removed. Using the results from Table 69 of Ref.~\cite{HFLAV:2022esi}, we calculate ${\rm BR}(\bar B^0 \to D^{*+} \ell^- \bar\nu_\ell) = (5.12 \pm 0.19)\%$ where a PDG rescaling factor 1.36 has been applied.
\item The HFLAV world average ${\rm BR} ( B^- \to D^{*0} \ell^- \bar\nu_\ell) = (5.58 \pm 0.07_{\rm stat} \pm 0.21_{\rm syst})\%$~\cite{HFLAV:2022esi} (which is not included in the Belle tagged shape-only measurement).
\end{itemize}
The theoretical predictions for the differential $B\to D^*\ell\nu$ rate binned over the variables $w$, $\cos\theta_v$, $\cos\theta_l$ and $\chi$ are obtained easily via direct integration of Eq.~(\ref{eq:B2Dstar:dGammadwMLEP}). One small subtlety is the inclusion of the so-called Coulomb factor $(1+\alpha \pi)$ for final states involving two charged particles, i.e., only for ${\rm BR}(\bar B^0\to D^{*+} \ell^- \bar\nu_\ell)$. Regarding the fit methodology, we chose not to use any prior nor to impose unitarity constraints on the BGL coefficients. The Belle untagged analysis~\cite{Belle:2018ezy} presents the data in 10 bins of each kinematical variable; since the integral over the bins in each of the four distributions are identical, we remove the last bin in each of the three angular distributions. Moreover, we marginalize over $N_{B^0}$, the number of $B^0$ mesons in the data sample, thus properly correlating its impact over all the distributions and over the electron and muon modes. 

\begin{figure}[tbp]
\begin{center}
\includegraphics[width=0.49\textwidth]{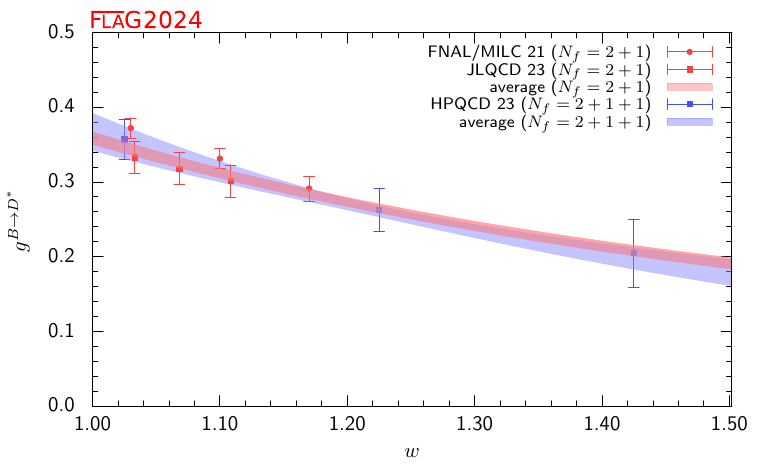}
\includegraphics[width=0.49\textwidth]{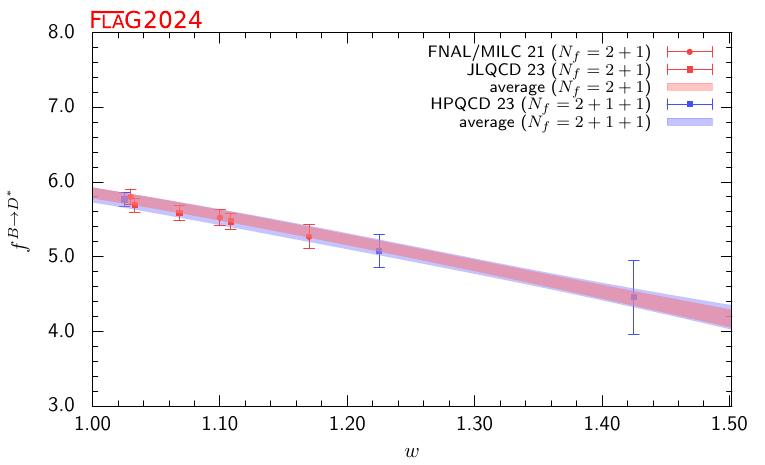}
\includegraphics[width=0.49\textwidth]{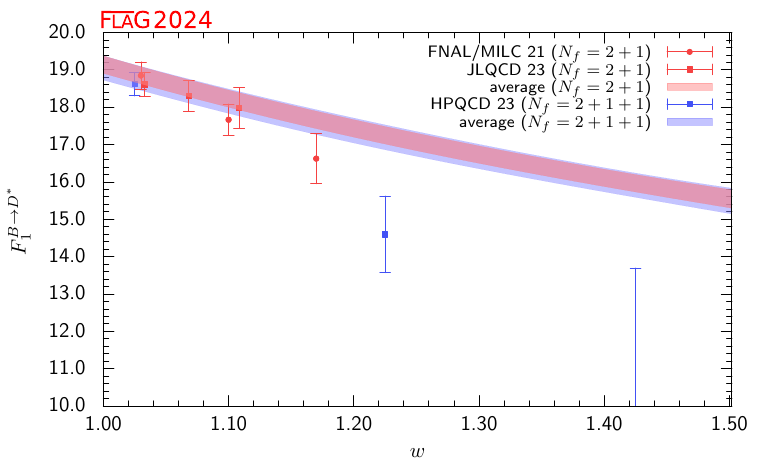}
\includegraphics[width=0.49\textwidth]{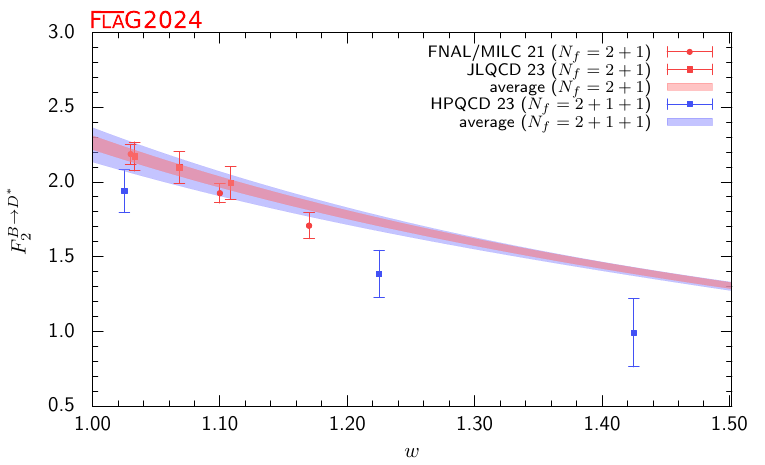}
\caption{The form factors $g(q^2)$, $f (q^2)$, $F_1(q^2)$, and $F_2(q^2)$ for $B \to D^* \ell\nu$ plotted as a function of $w$. The red (blue) band displays our preferred $(N_g,\Nf,N_{F_1},N_{F_2}) = ( 2,3,3,2)$ BGL fit (eight parameters) to experimental and $\Nf = 2+1 \; (2+1+1)$ lattice data. The constraints at zero and maximum recoil are imposed exactly. No use of unitarity constraints and priors has been made.}\label{fig:BDstar_FF_latt+exp}
\end{center}
\end{figure}

\begin{figure}[tbp]
\begin{center}
\includegraphics[width=0.49\textwidth]{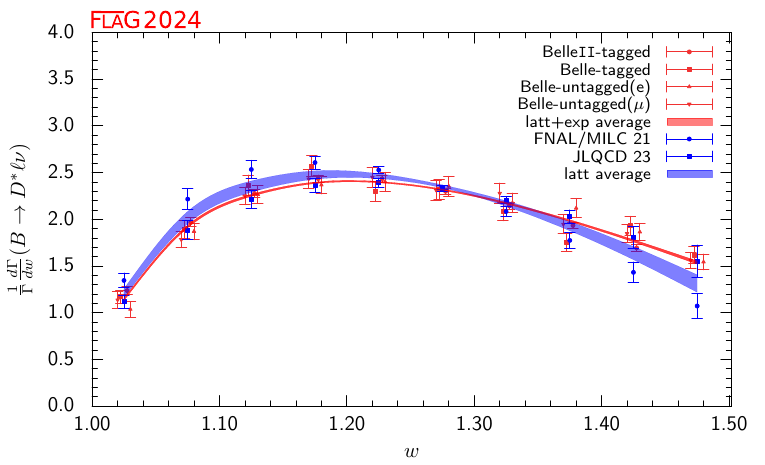}
\includegraphics[width=0.49\textwidth]{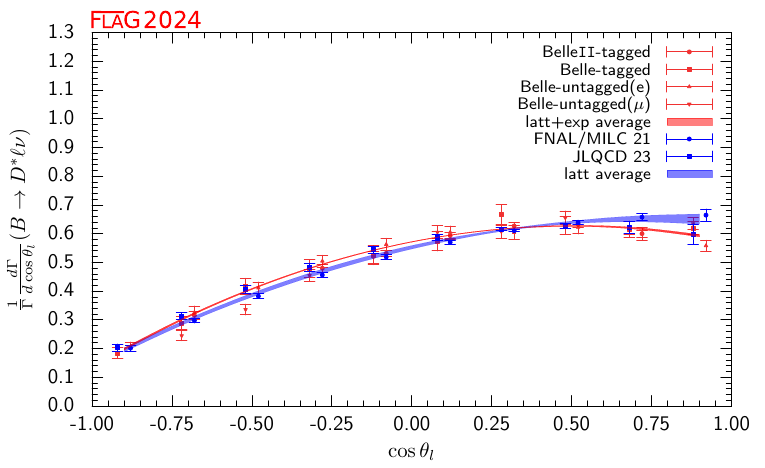}
\includegraphics[width=0.49\textwidth]{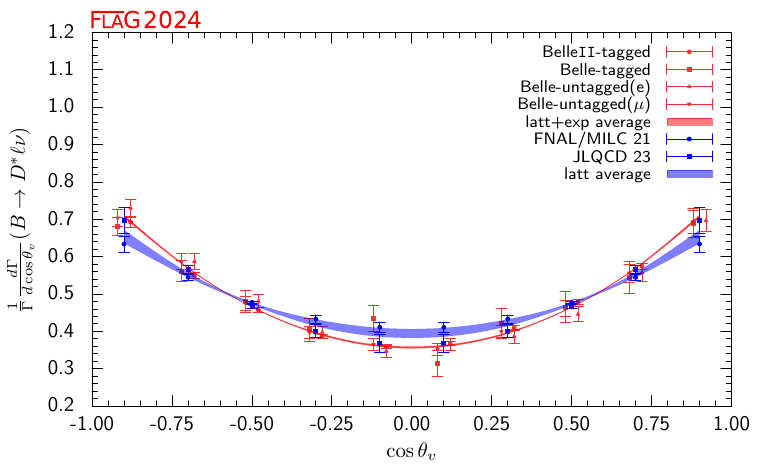}
\includegraphics[width=0.49\textwidth]{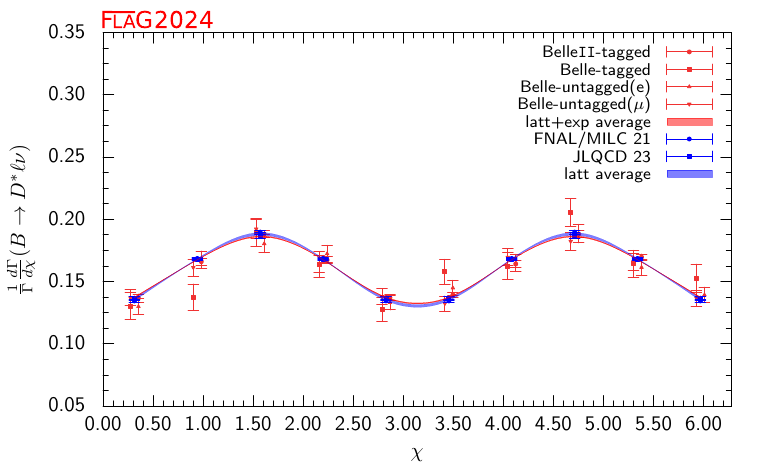}
\includegraphics[width=0.49\textwidth]{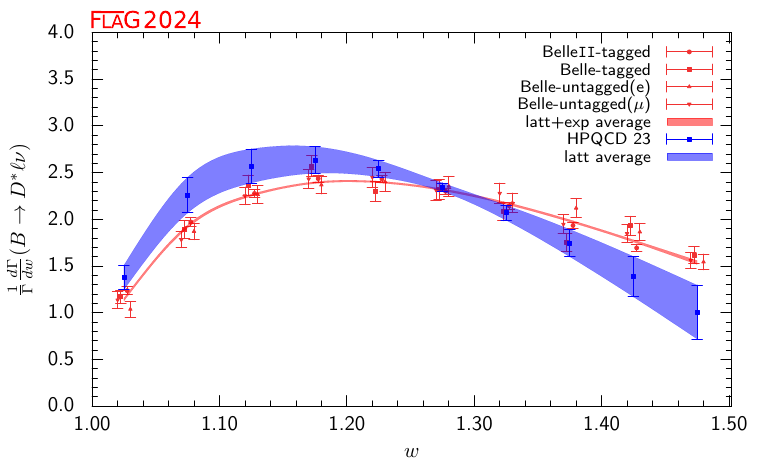}
\includegraphics[width=0.49\textwidth]{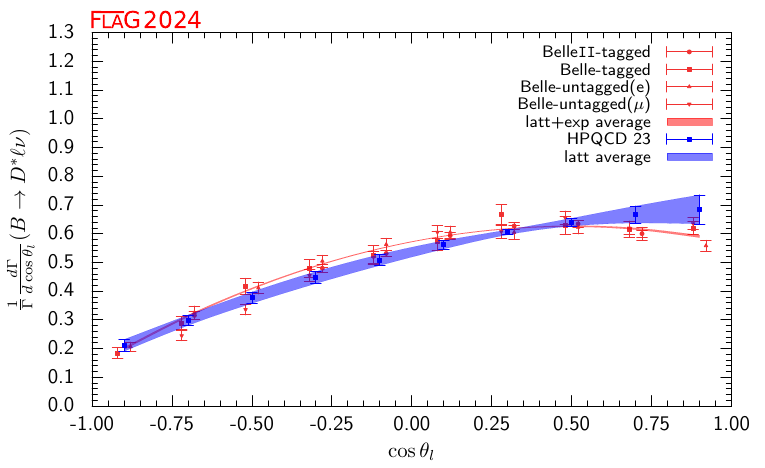}
\includegraphics[width=0.49\textwidth]{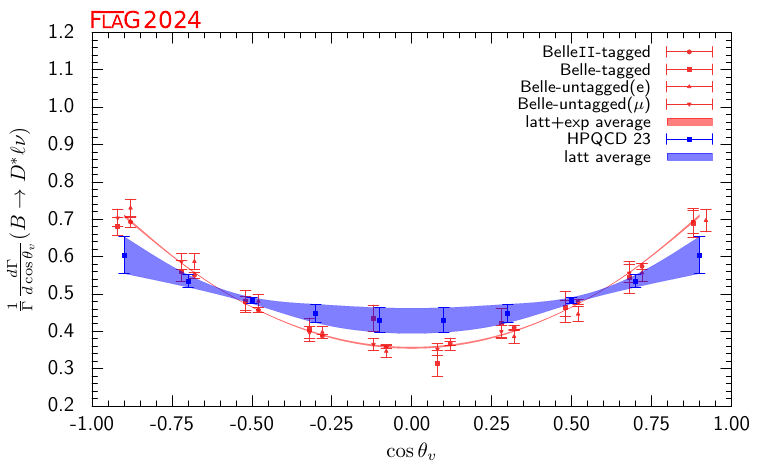}
\includegraphics[width=0.49\textwidth]{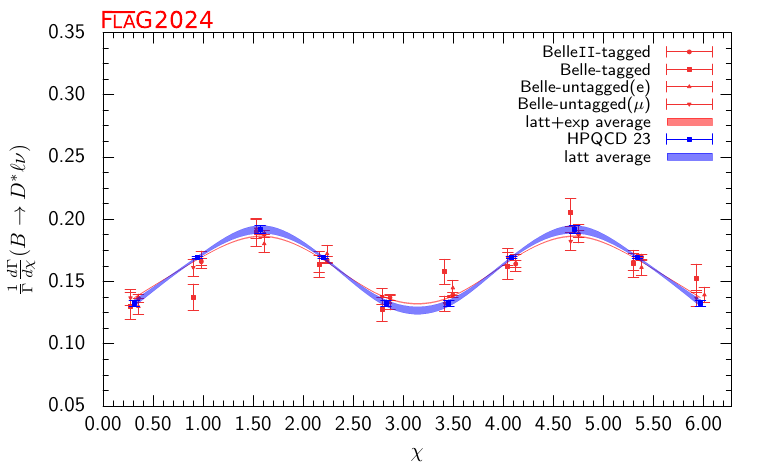}
\caption{Normalized differential decay rates with respect to the variables $w$, $\cos\theta_l$, $\cos\theta_{v}$ and $\chi$. The red (blue) band displays our preferred $(N_g,\Nf,N_{F_1},N_{F_2}) = ( 2,3,3,2)$ BGL fit (eight parameters) obtained from lattice calculations with (without) the inclusion of experimental data. The constraints at zero and maximum recoil are imposed exactly. No use of unitarity constraints and priors has been made. The top and bottom four distributions are obtained using $\Nf=2+1$ and $\Nf=2+1+1$ lattice calculations, respectively.}\label{fig:BDstar_DeltaGamma_latt+exp}
\end{center}
\end{figure}

The results of this global fit are presented in Tab.~\ref{tab:BDstar_latt+exp}. The chi-square per degree of freedom of the two fits are $\chi^2/{\rm dof} = 216/160 = 1.35$ for $\Nf = 2+1$ and $\chi^2/{\rm dof} = 200/148 = 1.35$ for $\Nf = 2+1+1$ (the difference in the degrees of freedom is simply due to the presence of two sets of lattice synthetic data, each comprised of 12 points, for the $\Nf = 2+1$ case). Note that we have rescaled all the errors by $\sqrt{\chi^2/{\rm dof}}$ following the standard PDG recipe. In particular, we find:
\begin{align}
\Nf=2+1\text{: }&   |V_{cb}| = 39.23 (65) \times 10^{-3} \nonumber\\
& \hskip -0.2cm
 [B\to D^*\ell\nu \text{, FLAG average, Refs.~\cite{FermilabLattice:2021cdg, Aoki:2023qpa, Belle:2018ezy, Belle:2023bwv, Belle-II:2023okj, HFLAV:2022esi}}]  \, ,\\
\Nf=2+1+1\text{: }& |V_{cb}| = 39.44 (89) \times 10^{-3} \nonumber \\
& \hskip -0.2cm
[B\to D^*\ell\nu \text{, FLAG average, Refs.~\cite{Harrison:2023dzh, Belle:2018ezy, Belle:2023bwv, Belle-II:2023okj, HFLAV:2022esi}}]  \,.
\end{align}
In Fig.~\ref{fig:BDstar_FF_latt+exp}, we show the form factors obtained from combining lattice and experimental results. In Fig.~\ref{fig:BDstar_DeltaGamma_latt+exp}, we present a comparison of the four \underline{\it normalized} differential distributions extracted from the experimental data, from the individual lattice results and from the combined lattice plus experiment fit.\footnote{For the Belle untagged case~\cite{Belle:2018ezy} we produce the normalized binned distributions by inverting the electron and muon response matrices and averaging over the leptons. Note that these distributions are presented for illustrative purpose only.} The top (bottom) four panels correspond to $\Nf = 2+1$ $(2+1+1)$. Direct inspection of these distributions shows quite a good agreement (as already evidenced by the relatively good chi-square per degree of freedom of the fits) albeit with some tensions in some of the shapes. In particular, the normalized distributions extracted from $\Nf = 2+1$ and $\Nf = 2+1+1$ results tend to deviate from the measured ones along similar patterns: deficit at large $w$, excess at large $\cos\theta_v$, flatter distribution in $\cos \theta_\ell$. The tensions in the $\Nf = 2+1+1$ are only apparently more pronounced because of the larger lattice uncertainties.  

Finally, using the fit results in Tab.~\ref{tab:BDstar_latt+exp}, we extract a value for $R(D^*)$ which includes both lattice and experimental information:
\begin{align}
\Nf=2+1\text{: }& R(D^*)_{\rm lat+exp} = 0.2505 (11)  \nonumber\\
& \quad\quad
 [\text{FLAG average, Refs.~\cite{FermilabLattice:2021cdg, Aoki:2023qpa, Belle:2018ezy, Belle:2023bwv, Belle-II:2023okj, HFLAV:2022esi}}]  \, ,\\
\Nf=2+1+1\text{: }& R(D^*)_{\rm lat+exp} = 0.2506 (17)  \nonumber \\
& \quad\quad
[\text{FLAG average, Refs.~\cite{Harrison:2023dzh, Belle:2018ezy, Belle:2023bwv, Belle-II:2023okj, HFLAV:2022esi}}]  \,.
\end{align} 

\begin{table}[t]
\tiny
    \begin{center}
 \resizebox{1\textwidth}{!}{  
  \begin{tabular}{|c|c|ccccccccc|}
  \multicolumn{11}{l}{$B\to D^* \; (\Nf=2+1)$} \\[0.2em]\hline
  coeff & Central Values & \multicolumn{9}{|c|}{Correlation Matrix} \\[0.2em]\hline
$|V_{cb}|\times 10^3$ & 39.23(65) & 1 & $-$0.3552 & $-$0.1269 & $-$0.6672 & $-$0.3260 & 0.2331 & $-$0.2412 & 0.1118 & $-$0.08658 \\[0.2em]
$a^g_0$ & 0.03036(72) & $-$0.3552 & 1 & $-$0.4976 & 0.3645 & $-$0.0009317 & $-$0.02169 & 0.1026 & $-$0.02327 & $-$0.09817 \\[0.2em]
$a^g_1$ & $-$0.083(21) & $-$0.1269 & $-$0.4976 & 1 & 0.02961 & 0.1874 & $-$0.2543 & 0.08161 & $-$0.03930 & 0.1177 \\[0.2em]
$a^f_0$ & 0.01213(15) & $-$0.6672 & 0.3645 & 0.02961 & 1 & $-$0.08990 & 0.07897 & $-$0.08767 & 0.07594 & $-$0.09589 \\[0.2em]
$a^f_1$ & 0.0234(64) & $-$0.3260 & $-$0.0009317 & 0.1874 & $-$0.08990 & 1 & $-$0.8384 & 0.4660 & $-$0.2491 & 0.3552 \\[0.2em]
$a^f_2$ & $-$0.59(16) & 0.2331 & $-$0.02169 & $-$0.2543 & 0.07897 & $-$0.8384 & 1 & $-$0.2414 & 0.07961 & $-$0.2880 \\[0.2em]
$a^{F_1}_1$ & 0.00141(97) & $-$0.2412 & 0.1026 & 0.08161 & $-$0.08767 & 0.4660 & $-$0.2414 & 1 & $-$0.9135 & $-$0.06385 \\[0.2em]
$a^{F_1}_2$ & $-$0.005(17) & 0.1118 & $-$0.02327 & $-$0.03930 & 0.07594 & $-$0.2491 & 0.07961 & $-$0.9135 & 1 & 0.2820 \\[0.2em]
$a^{F_1}_1$ & $-$0.093(17) & $-$0.08658 & $-$0.09817 & 0.1177 & $-$0.09589 & 0.3552 & $-$0.2880 & $-$0.06385 & 0.2820 & 1 \\[0.2em]
\hline
  \end{tabular}
}
 \resizebox{1\textwidth}{!}{  
  \begin{tabular}{|c|c|ccccccccc|}
    \multicolumn{11}{l}{$B\to D^* \; (\Nf=2+1+1)$} \\[0.2em]\hline
    coeff & Central Values & \multicolumn{9}{|c|}{Correlation Matrix} \\[0.2em]\hline
$|V_{cb}|\times 10^3$ & 39.44(89) & 1 & $-$0.1717 & $-$0.06581 & $-$0.7257 & $-$0.4981 & 0.4426 & $-$0.2473 & 0.08156 & $-$0.2155 \\[0.2em]
$a^g_0$ & 0.0311(21) & $-$0.1717 & 1 & $-$0.9267 & 0.1121 & $-$0.004683 & 0.1735 & 0.1230 & $-$0.003372 & 0.07094 \\[0.2em]
$a^g_1$ & $-$0.125(75) & $-$0.06581 & $-$0.9267 & 1 & 0.09615 & 0.1018 & $-$0.2899 & $-$0.03844 & $-$0.03789 & $-$0.03009 \\[0.2em]
$a^f_0$ & 0.01207(21) & $-$0.7257 & 0.1121 & 0.09615 & 1 & 0.01430 & $-$0.04137 & $-$0.03342 & 0.02486 & 0.07847 \\[0.2em]
$a^f_1$ & 0.023(12) & $-$0.4981 & $-$0.004683 & 0.1018 & 0.01430 & 1 & $-$0.9267 & 0.2522 & 0.03052 & 0.3601 \\[0.2em]
$a^f_2$ & $-$0.55(31) & 0.4426 & 0.1735 & $-$0.2899 & $-$0.04137 & $-$0.9267 & 1 & $-$0.06981 & $-$0.1655 & $-$0.3503 \\[0.2em]
$a^{F_1}_1$ & 0.0016(14) & $-$0.2473 & 0.1230 & $-$0.03844 & $-$0.03342 & 0.2522 & $-$0.06981 & 1 & $-$0.9270 & $-$0.1678 \\[0.2em]
$a^{F_1}_2$ & $-$0.008(27) & 0.08156 & $-$0.003372 & $-$0.03789 & 0.02486 & 0.03052 & $-$0.1655 & $-$0.9270 & 1 & 0.3148 \\[0.2em]
$a^{F_1}_1$ & $-$0.090(48) & $-$0.2155 & 0.07094 & $-$0.03009 & 0.07847 & 0.3601 & $-$0.3503 & $-$0.1678 & 0.3148 & 1 \\[0.2em]
       \hline
    \end{tabular}   
}
\end{center}
  \caption{$|V_{cb}|$, coefficients and correlation matrix for the $(N_g,\Nf,N_{F_1},N_{F_2}) = ( 2,3,3,2)$ BGL fit to the $B\to D^*$ form factors $g$, $f$, $F_1$ and $F_2$ for $\Nf=2+1$ and $\Nf = 2+1+1$. The form factors can be reconstructed using parameterization and inputs given in Appendix~\ref{sec:app_B2D*}. \label{tab:BDstar_latt+exp}}
\end{table}
  
Before discussing the combination of the above $|V_{cb}|$ results, we note that the LHCb Collaboration recently reported
the first determination of $|V_{cb}|$ at the Large Hadron Collider
using $B_s\to D_s^- \mu^+\nu_\mu$ and $B_s\to D^{*-}_s \mu^+\nu_\mu$ decays \cite{LHCb:2020cyw,LHCb:2021qbv}. The differential decay
rates, in combination with the $\Nf=2+1+1$ HPQCD~19 \cite{McLean:2019qcx} and HPQCD~19B \cite{McLean:2019sds} lattice results for $f_+^{B_s\to D_s}$
and ${\mathcal F}^{B_s\to D_s^*}(1)$, were analyzed using either the CLN or BGL form-factor parameterizations. The result for $|V_{cb}|$
from the BGL fit is \cite{LHCb:2021qbv}
\begin{align}
\Nf=2+1+1\text{: } |V_{cb}^{}| &= (41.7 \pm 0.8 \pm 0.9 \pm 1.1)\times 10^{-3} \quad\quad [B_s\to D^{(*)-}_s \mu^+\nu_\mu \text{, LHCb}]\;, \label{eq:VcbLHCb}
\end{align}
where the first two uncertainties are the statistical and systematic experimental uncertainties, and the third is due to the external inputs used, including the
lattice inputs.

The LHCb analysis used ratios to the reference decay modes $B^0\to D^- \mu^+\nu_\mu$ and $B^0\to D^{*-} \mu^+\nu_\mu$, whose branching fractions
are used as input in the form of the Particle Data Group averages of measurements by other experiments \cite{Tanabashi:2018oca}. The result (\ref{eq:VcbLHCb})
is therefore correlated with the determinations of $|V_{cb}|$ from $B\to D$ and $B\to D^*$ semileptonic decays.
Given the challenges
involved in performing our own fit to the LHCb data, we do not, at present, include the LHCb results for $B_s\to D_s^- \mu^+\nu_\mu$ and $B_s\to D^{*-}_s \mu^+\nu_\mu$ in our combination of $|V_{cb}|$.

\begin{figure}[!h]
\begin{center}
\includegraphics[width=0.49\linewidth]{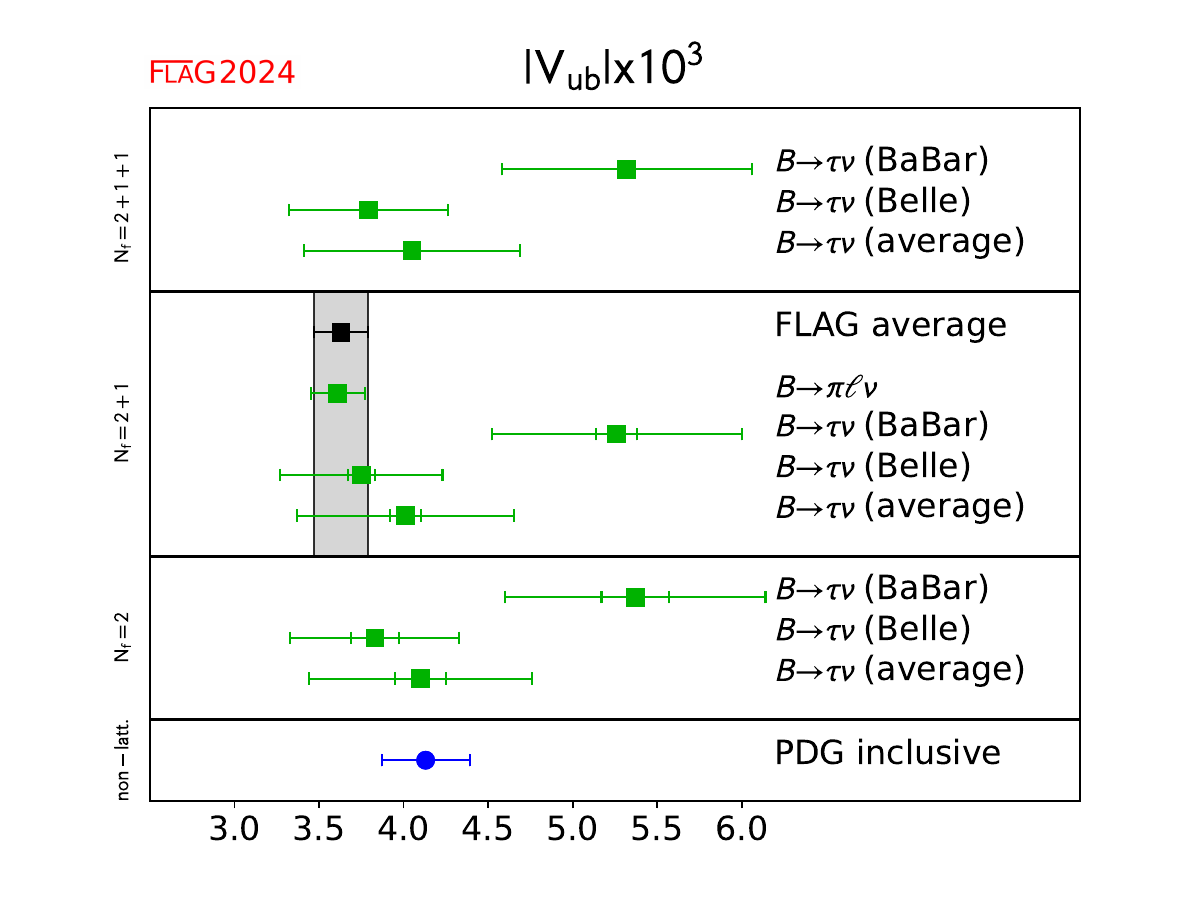}
\includegraphics[width=0.49\linewidth]{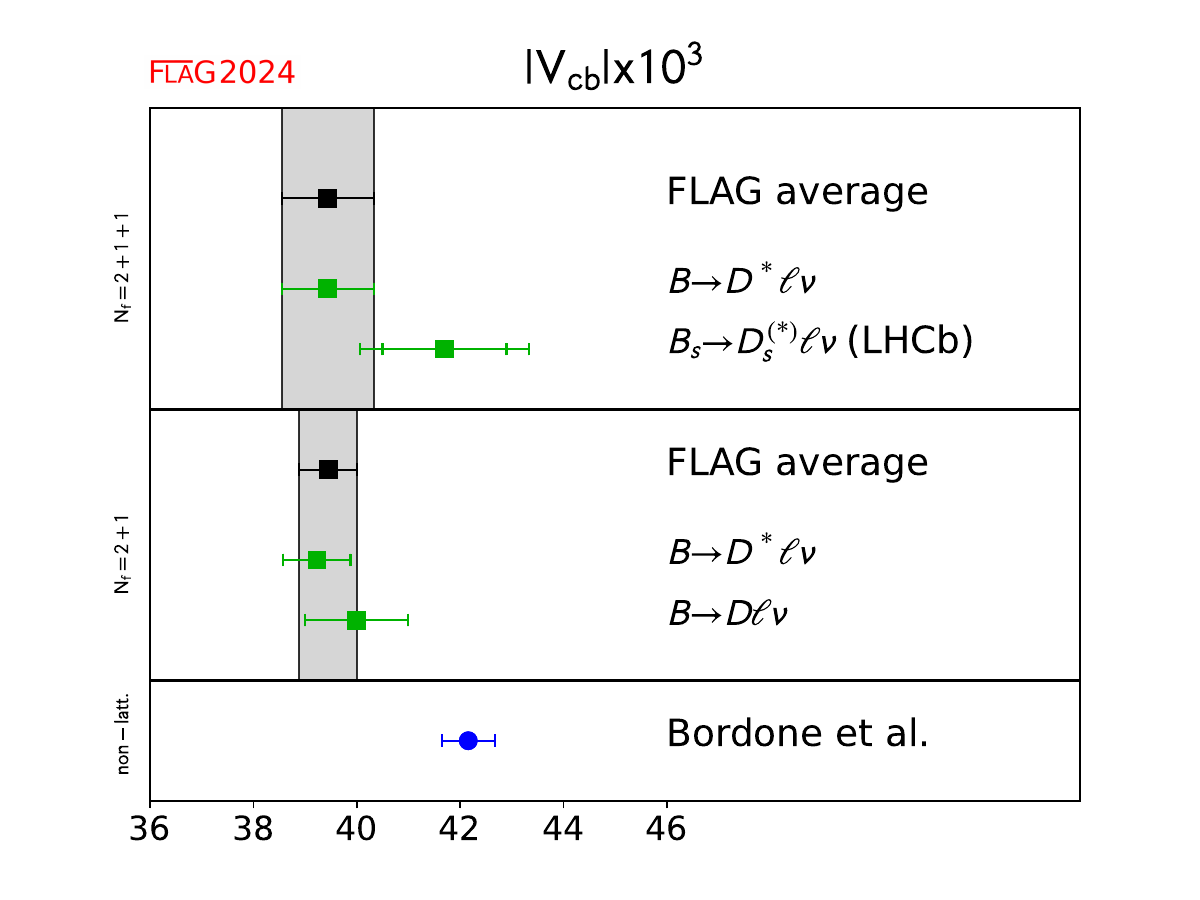}
\caption{Left: Summary of $|V_{ub}|$ determined using: i) the $B$-meson leptonic
decay branching fraction, $B(B^{-} \to \tau^{-} \bar{\nu})$, measured
at the Belle and BaBar experiments, and our averages for $f_{B}$ from
lattice QCD; and ii) the various measurements of the $B\to\pi\ell\nu$
decay rates by Belle and BaBar, and our averages for lattice determinations
of the relevant vector form factor $f_+(q^2)$. The inclusive result is taken from PDG~\cite{ParticleDataGroup:2024cfk}.
Right: Same for determinations of $|V_{cb}|$ using semileptonic decays.
The inclusive result is taken from Ref.~\cite{Bordone:2021oof}.}
\label{fig:Vxbsummary}
\end{center}
\end{figure}

We now proceed to combine the two $\Nf = 2+1$ determinations of $|V_{cb}|$ from exclusive $B\to D$ and $B\to D^*$ semileptonic decays. To this end, we include an estimate of the correlation between the statistical lattice uncertainties on $|V_{cb}|_{B\to D}^{\Nf = 2+1}$ (FNAL/MILC and HPQCD) and $|V_{cb}|_{B\to D^*}^{\Nf = 2+1}$ (FNAL/MILC), because they are based on the same MILC configurations (albeit on different subsets). An estimate of this correlation is complicated due to the difficulty of disentangling lattice and experimental sources of uncertainties in a global BGL fit. Here we follow an approximate procedure which relies on estimating these correlations by looking at the $B\to D$ and $B\to D^*$ form factors at zero recoil, ${\cal G}^{B\to D} (1)$ and ${\cal F}^{B\to D^*}(1)$. The inclusion of these correlations has a very small impact on the average, thus providing an {\em a posteriori} justification for this approximate method.  
We obtain:
\begin{align}
\Nf=2+1\text{: }& |V_{cb}| = 39.45(56) \times 10^{-3} \nonumber \\
& \quad\quad 
[B\to (D,D^*)\ell\nu \text{, FLAG average,} \nonumber \\
& \quad\quad
\text{\; Refs.~\cite{Lattice:2015rga, Na:2015kha, Belle:2015pkj, Aubert:2009ac, FermilabLattice:2021cdg, Aoki:2023qpa, Belle:2018ezy, Belle:2023bwv, Belle-II:2023okj, HFLAV:2022esi}}].
\end{align}
Our results are summarized in Tab.~\ref{tab:Vcbsummary}, which also shows the inclusive determination of $|V_{cb}|=42.16(51) \times 10^{-3}$~\cite{Bordone:2021oof} for comparison, and are illustrated in Fig.~\ref{fig:Vxbsummary}.\footnote{This determination of $|V_{cb}|$ is also adopted by the PDG~\cite{ParticleDataGroup:2024cfk}.}

\begin{table}[!t]
\begin{center}
\noindent
\begin{tabular*}{\textwidth}[l]{@{\extracolsep{\fill}}lcc}
 & from  & $|V_{cb}| \times 10^3$\\
&& \\[-2ex]
\hline \hline &&\\[-2ex]
FLAG average for $\Nf=2+1$ & $B \to D^*\ell\nu$ & $39.23(65)$ \\
FLAG average for $\Nf=2+1$ & $B \to D\ell\nu$ &  $40.0(1.0)$  \\
FLAG average for $\Nf=2+1$ & $B \to (D,D^*)\ell\nu$  & $39.45 (56)$ \\
&& \\[-2ex]
 \hline
FLAG average for $\Nf=2+1+1$ & $B \to D^*\ell\nu$ & $39.44 (89) $ \\
&& \\[-2ex]
 \hline
LHCb result for $\Nf=2+1+1$ (BGL) & $B_s \to D_s^{(*)}\ell\nu$ & $41.7(0.8)(0.9)(1.1)$ \\
&& \\[-2ex]
 \hline
Bordone et al. & $B \to X_c\ell\nu$ & $42.16(51)$ \\
&& \\[-2ex]
\hline \hline && \\[-2ex]
\end{tabular*}
\caption{Results for $|V_{cb}|$. The lattice calculations for the $B\to D$ form factors at $N_f=2+1$ are taken from \SLfnalmilcBpi~\cite{Lattice:2015tia}, \SLrbcukqcdBpi~\cite{Flynn:2015mha} and \SLjlqcdBpi~\cite{Colquhoun:2022atw}; for the $B\to D^*$ form factors at $\Nf=2+1$ from FNAL/MILC~21~\cite{FermilabLattice:2021cdg} and JLQCD~23~\cite{Aoki:2023qpa}; for the $B\to D^*$ form factors at $\Nf=2+1+1$ from HPQCD~23~\cite{Harrison:2023dzh}. The LHCb result using $B_s \to D_s^{(*)}\ell\nu$ decays~\cite{LHCb:2020cyw,LHCb:2021qbv,McLean:2019qcx,McLean:2019sds}, as well as the inclusive average obtained in the kinetic scheme from Ref.~\cite{Bordone:2021oof} are shown for comparison. In the LHCb result, the first two uncertainties are the statistical and systematic experimental uncertainties, and the third is due to the external inputs used, including the
lattice inputs.}
\label{tab:Vcbsummary}
\end{center}
\end{table}

\subsection{Determination of $|V_{ub}/V_{cb}|$ from $\Lambda_b$ decays}
\label{sec:VubVcb}

In 2015, the LHCb Collaboration reported a measurement of the ratio \cite{Aaij:2015bfa}
\begin{equation}
R_{\rm BF}(\Lambda_b)=\frac{\displaystyle\int_{15~{\rm GeV}^2}^{q^2_{\rm max}}\frac{{\rm d}\mathcal{B}(\Lambda_b\to p\mu^-\bar\nu_\mu)}{{\rm d}q^2}\,{\rm d}q^2}{\displaystyle\int_{7~{\rm GeV}^2}^{q^2_{\rm max}}\frac{{\rm d}\mathcal{B}(\Lambda_b\to \Lambda_c\mu^-\bar\nu_\mu)}{{\rm d}q^2}\,{\rm d}q^2}, 
\end{equation}
which, combined with the lattice QCD prediction from Ref.~\cite{Detmold:2015aaa} (Detmold 15) discussed in Sec.~\ref{sec:Lambdab} yields a determination of $|V_{ub}/V_{cb}|$. The LHCb analysis uses the decay $\Lambda_c \to p K \pi$ to reconstruct the $\Lambda_c$ and requires the branching fraction $\mathcal{B}(\Lambda_c \to p K \pi)$ of this decay as an external input. Using the latest world average of $\mathcal{B}(\Lambda_c \to p K \pi)=(6.28\pm0.32)\%$ \cite{Zyla:2020zbs} to update the LHCb measurement gives \cite{Amhis:2019ckw}
\begin{equation}
R_{\rm BF}(\Lambda_b) = (0.92 \pm 0.04 \pm 0.07) \times 10^{-2},
\end{equation}
and, combined with the lattice QCD prediction for $\frac{\zeta_{p\mu\bar\nu}(15{\rm GeV}^2)}{\zeta_{\Lambda_c \mu\bar\nu}(7{\rm GeV}^2)}$ discussed in Sec.~\ref{sec:Lambdab},
\begin{equation}
 |V_{ub}/V_{cb}| = 0.079 \pm 0.004_{\rm\, lat.} \pm 0.004_{\rm\, exp.}.
 \label{eq:vubOvcb_Lambda}
\end{equation}
We remind the reader that the lattice calculation for the form factor ratio currently has a \tbr~ rating; thus we will not use the result in Eq.~(\ref{eq:vubOvcb_Lambda}) in the global $[V_{ub},V_{cb}]$ fit. 

\subsection{Determination of $|V_{ub}/V_{cb}|$ from $B_s$ decays}

More recently, LHCb reported the measurements \cite{Aaij:2020nvo}
\begin{eqnarray}
\nonumber R_{\rm BF}(B_s, \text{low}) &=& \frac{\displaystyle\int_{q^2_{\rm min}=m_\mu^2}^{7\:{\rm GeV}^2}\frac{{\rm d}\mathcal{B}(B_s\to K^- \mu^+\nu_\mu)}{{\rm d}q^2}\,{\rm d}q^2}{\mathcal{B}(B_s\to D_s^-\mu^+\nu_\mu)} \\
 &=& (1.66\pm0.12)\times10^{-3}, \\
\nonumber  && \\
\nonumber R_{\rm BF}(B_s, \text{high}) &=& \frac{\displaystyle\int_{7\:{\rm GeV}^2}^{q^2_{\rm max}=(m_{B_s}-m_{K})^2}\frac{{\rm d}\mathcal{B}(B_s\to K^- \mu^+\nu_\mu)}{{\rm d}q^2}\,{\rm d}q^2}{\mathcal{B}(B_s\to D_s^-\mu^+\nu_\mu)} \\
&=& (3.25\pm0.28)\times10^{-3}, \\
 \nonumber && \\
\nonumber R_{\rm BF}(B_s, \text{all}) &=& \frac{\mathcal{B}(B_s\to K^- \mu^+\nu_\mu)}{\mathcal{B}(B_s\to D_s^-\mu^+\nu_\mu)}\\
&=& (4.89\pm0.33)\times10^{-3}.
\end{eqnarray}
Using our average of the $B_s \to K$ form factors from lattice QCD as discussed in Sec.~\ref{sec:BstoK}, we obtain the Standard-Model predictions
\begin{eqnarray}
 \frac{1}{|V_{ub}|^2}\int_{q^2_{\rm min}=m_\mu^2}^{7\:{\rm GeV}^2}\frac{{\rm d}\Gamma(B_s\to K^- \mu^+\nu_\mu)}{{\rm d}q^2} &=& (2.51 \pm 0.62) \:{\rm ps}^{-1}, \\
 \frac{1}{|V_{ub}|^2}\int_{7\:{\rm GeV}^2}^{q^2_{\rm max}=(m_{B_s}-m_{K})^2}\frac{{\rm d}\Gamma(B_s\to K^- \mu^+\nu_\mu)}{{\rm d}q^2} &=& (4.02 \pm 0.51) \:{\rm ps}^{-1}, \\
 \frac{1}{|V_{ub}|^2}\Gamma(B_s\to K^- \mu^+\nu_\mu) &=& (6.5 \pm 1.1) \:{\rm ps}^{-1}.
\end{eqnarray}
For the denominator, we use the $B_s\to D_s$ form factors from Ref.~\cite{McLean:2019qcx} (HPQCD~19), which yields
\begin{eqnarray}
  \frac{1}{|V_{cb}|^2}\Gamma(B_s\to D_s^- \mu^+\nu_\mu) &=& (9.15 \pm 0.37) \:{\rm ps}^{-1}.
\end{eqnarray}
Since the form factor shape is most reliably constrained by the lattice data only at high-$q^2$, the most reliable determination of the ratio $|V_{ub}/V_{cb}|$ is obtained by using LHCb measurements limited to the high-$q^2$ region. The result which we obtain and which is used in the combination presented in Sec.~\ref{sec:VubVcbsummary}, reads:
\begin{eqnarray}
 \frac{|V_{ub}|}{|V_{cb}|}(\text{high}) &=& 0.0861 \pm 0.0057_{\rm\, lat.} \pm 0.0038_{\rm\, exp.} \; .
\end{eqnarray}
For reference, the corresponding CKM ratio obtained at low-$q^2$ and in the whole $q^2$ regions are, $|V_{ub}|/|V_{cb}|(\text{low})  = 0.0779 \pm 0.0098_{\rm\, lat.} \pm 0.0028_{\rm\, exp.}$ and $|V_{ub}|/|V_{cb}|(\text{all})  = 0.0828 \pm 0.0070_{\rm\, lat.} \pm 0.0028_{\rm\, exp.}$, respectively.

\subsection{Summary: $|V_{ub}|$ and $|V_{cb}|$}
\label{sec:VubVcbsummary}

\begin{figure}[!h]
\begin{center}
\includegraphics[width=0.6\linewidth]{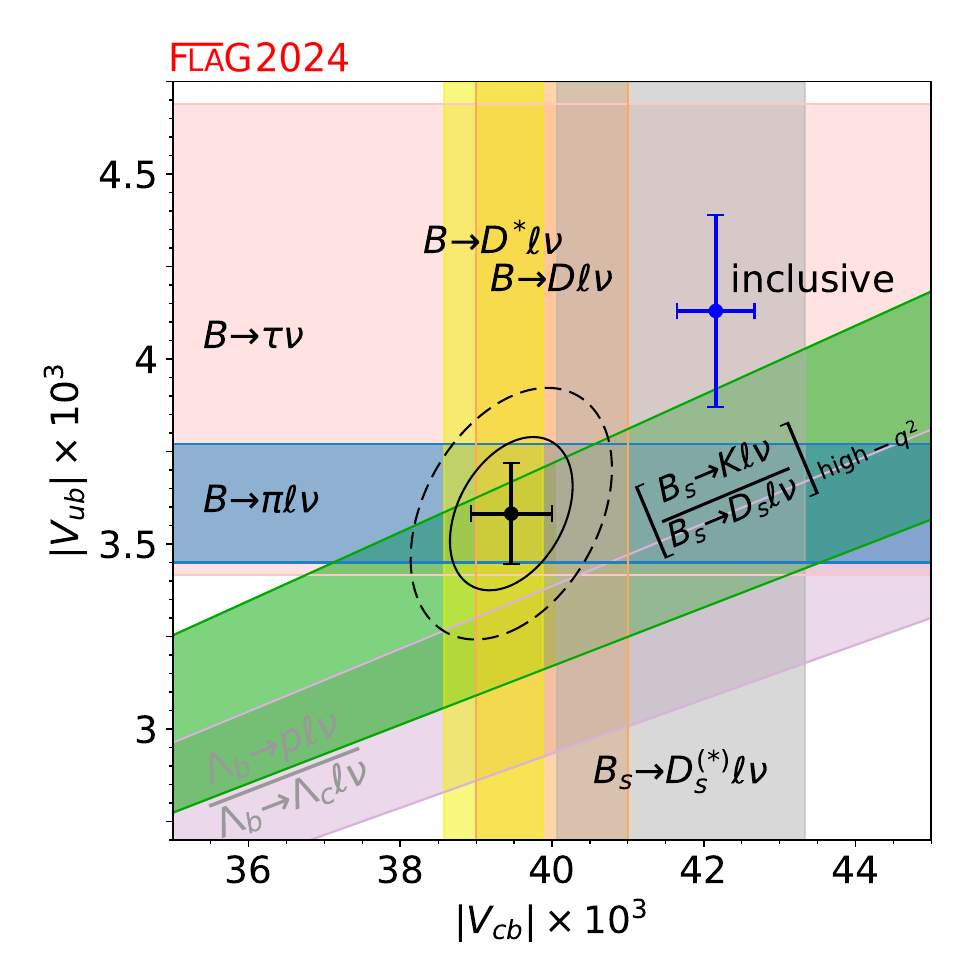}
\caption{Summary of $|V_{ub}|$ and $|V_{cb}|$ determinations. The black solid and dashed lines correspond to 68\% and 95\% C.L. contours, respectively. The result of the global fit (which does not include $\left| V_{ub}^{}/V_{cb}^{} \right|$ from baryon modes nor $|V_{cb}|$ from $B_s \to D_s^{(*)}\ell\nu$) is 
$(|V_{cb}^{}|,|V_{ub}^{}|) = (39.46\pm 0.53, 3.60 \pm 0.14) \times 10^{-3}$ with a $p$-value of 0.66.
The lattice and experimental results that contribute to the various contours are the following. 
$B\to\pi\ell\nu$: lattice (FNAL/MILC~15~\cite{Lattice:2015tia}, RBC/UKQCD~15~\cite{Flynn:2015mha}, and \SLjlqcdBpi~\cite{Colquhoun:2022atw}) and experiment (BaBar~\cite{delAmoSanchez:2010af,Lees:2012vv} and Belle~\cite{Ha:2010rf,Sibidanov:2013rkk}). 
$B\to D\ell\nu$: lattice (FNAL/MILC~15C~\cite{Lattice:2015rga} and HPQCD~15~\cite{Na:2015kha}) and experiment (BaBar~\cite{Aubert:2009ac} and Belle~\cite{Belle:2015pkj}).
$B\to D^*\ell\nu$: lattice (FNAL/MILC~21~\cite{FermilabLattice:2021cdg}, JLQCD~23~\cite{Aoki:2023qpa}, HPQCD~23~\cite{Harrison:2023dzh}) and experiment (Belle~\cite{Belle:2018ezy, Belle:2023bwv}, Belle II~\cite{Belle-II:2023okj}, HFLAV~\cite{HFLAV:2022esi}). 
$B\to \tau\nu$: lattice ($\Nf=2+1+1$ determination of $f_B$ in Eq.~(\ref{eq:fB211})~\cite{Dowdall:2013tga,Bussone:2016iua,Hughes:2017spc,Bazavov:2017lyh}) and experiment (BaBar~\cite{Kronenbitter:2015kls} and Belle~\cite{Lees:2012ju}).
$B_s\to K\ell\nu/B_s\to D_s\ell\nu$: lattice (HPQCD~14~\cite{Bouchard:2014ypa}, RBC/UKQCD~23~\cite{Flynn:2023nhi}, FNAL/MILC~19~\cite{Bazavov:2019aom}, HPQCD~19~\cite{McLean:2019qcx}) and experiment (LHCb~\cite{Aaij:2020nvo}).
$\Lambda_b\to p\ell\nu/\Lambda_b\to \Lambda_c \ell\nu$: lattice (Detmold~15~\cite{Detmold:2015aaa}) and experiment (LHCb~\cite{Aaij:2015bfa}).
$B_s \to D_s^* \ell\nu / B_s \to D_s \ell\nu$: lattice (HPQCD~19~\cite{McLean:2019qcx} and HPQCD~19B~\cite{McLean:2019sds}) and experiment (LHCb~\cite{LHCb:2020cyw,LHCb:2021qbv}).
The inclusive determinations are taken from Refs.~\cite{Bordone:2021oof, Amhis:2019ckw, Zyla:2020zbs} and read  $(|V_{cb}^{}|,|V_{ub}^{}|)_{\rm incl}  = (42.16\pm 0.51, 4.13 \pm 0.26) \times 10^{-3}$. \label{fig:VubVcb}}
\end{center}
\end{figure}

In Fig.~\ref{fig:VubVcb}, we present a summary of determinations of $|V_{ub}|$ and $|V_{cb}|$ from $B\to (\pi,D^{(*)})\ell\nu$, $B_s\to (K,D_s) \ell\nu$ (high $q^2$ only), $B\to \tau \nu$ and $\Lambda_b\to (p,\Lambda_c)\ell\nu$, as well as the results from inclusive $B\to X_{u,c} \ell\nu$ decays. Currently, the determinations of $V_{cb}$ from $B\to D^*$ and $B\to D$ decays are quite compatible; however, a sizeable tension involving the extraction of $V_{cb}$ from inclusive decays remains. Note that constraints on $\left| V_{ub}^{}/V_{cb}^{} \right|$ from baryon modes are displayed but, in view of the rating in Tab.~\ref{tab_BottomBaryonSLsumm2}, are not included in the global fit. As discussed in Sec.~\ref{sec:Vcb}, experimental inputs used in the extraction of $|V_{cb}|$ from $B_s \to D_s^{(*)} \ell\nu$ decays~\cite{LHCb:2020cyw,LHCb:2021qbv} given in Eq.~(\ref{eq:VcbLHCb}) are highly correlated with those entering the global $(|V_{ub}|,|V_{cb}|)$ fit described in this section. Given these correlations and the challenges in reproducing the LHCb analysis, for the time being we do not include the result Eq.~(\ref{eq:VcbLHCb}) into the global fit. 

In the global fit we include an estimate of the correlations between the $|V_{ub}|$ and $|V_{cb}|$ determinations from semileptonic $B$ decays. We conservatively assume 100\% correlation between the statistical lattice uncertainties on (1) $|V_{ub}|$ (FNAL/MILC), $|V_{cb}|_{B\to D}$ (FNAL/MILC and HPQCD) and $|V_{cb}|_{B\to D^*}$ (FNAL/MILC) and (2) $|V_{ub}|$ (JLQCD) and $|V_{cb}|_{B\to D}$ (JLQCD). Due to the difficulty of disentangling statistical lattice uncertainties in the three BGL fits for $B\to (\pi,D,D^*)$, we follow the same approximate procedure described at the end of Sec.~\ref{sec:Vcb} and estimate the correlations by looking at the zero-recoil form factors $f_+(0)$, ${\cal F}^{B\to D}(1)$ and ${\cal F}^{B\to D^*}(1)$. The results of the fit are
\begin{align}
|V_{cb}^{}| & =  39.46 (53) \times 10^{-3}\;, \\
|V_{ub}^{}| & = 3.60 (14) \times 10^{-3}  \;, \\
p{\rm -value} & = 0.66 \; ,
\end{align}
with a 0.36 correlation coefficient. For reference, the fit without the inclusion of any correlation between the various lattice calculations yield $|V_{cb}^{}| =  39.50 (51) \times 10^{-3}$,  $|V_{ub}^{}| = 3.60 (13) \times 10^{-3}$ with a 0.09 correlation coefficient (the latter does not vanish because of the inclusion of $|V_{ub}/V_{cb}|$ from $B_s\to (K,D_s) \ell\nu$ decays).

The inclusive determinations read $|V_{cb}^{}|_{\rm incl} = (42.16 \pm 0.51 ) \times 10^{-3}$~\cite{Gambino:2016jkc} and $|V_{ub}^{}|_{\rm incl} = (4.13 \pm 0.12_{\rm exp} \pm {{}^{+0.13}_{-0.14}}_{\rm theo} \pm 0.18_{\Delta\text{model}}) \times 10^{-3}$~\cite{ParticleDataGroup:2024cfk}.

%% file: Alpha_s/alpha.tex
\section{The strong coupling $\alpha_{\rm s}$}
\label{sec:alpha_s}

Authors
: L.~Del Debbio, P.~Petreczky, S.~Sint\\

\newcommand{\todoalpha}[1]{{\color{magenta} \bf [ todo: #1]}}

\input{Alpha_s/s12.tex} 
\input{Alpha_s/s3.tex} 
\input{Alpha_s/s0.tex}  
\input{Alpha_s/s4.tex} 
\input{Alpha_s/s5.tex} 
\input{Alpha_s/s6.tex} 
\input{Alpha_s/s7.tex} 
\input{Alpha_s/s_gf.tex} 

\input{Alpha_s/s10.tex} 

%% file: Alpha_s/s12.tex
\subsection{Introduction}

\label{s:intro}


The strong coupling $\alpha_s(\mu)=\gbar_s^2(\mu)/(4\pi)$ defined at scale $\mu$, is 
the parameter that determines the strength of strong interactions in the Standard Model. 
It plays a key
role in the understanding of QCD and in its application to collider
physics, where it is ubiquitous in calculations of physical processes, e.g., at the LHC. 
For example, the parametric uncertainty from $\alpha_s$ is one of
the dominant sources of uncertainty in the Standard Model predictions for
the Higgs boson~\cite{LHCHiggsCrossSectionWorkingGroup:2016ypw} 
and top quark cross sections, see, e.g., Ref.~\cite{Salam:2017qdl}. In order to fully exploit the 
experimental results that will be collected during the high-luminosity run of the LHC in the near future, 
it is mandatory to reduce the 
uncertainty on $\alpha_s$ below 1\%. Similarly, high-accuracy determinations of this coupling 
will help in understanding the stability of the vacuum of the Standard 
Model and will yield one of the essential boundary conditions
for completions of the Standard Model at high 
energies~\cite{Dittmaier:2012vm,Heinemeyer:2013tqa,Adams:2013qkq,Dawson:2013bba,Accardi:2016ndt,Lepage:2014fla,Buttazzo:2013uya,Espinosa:2013lma}.
At this level of precision, it becomes imperative to have a robust understanding of
systematic errors. Lattice simulations are ideally placed to play a central role in this quest.
In the following we try to summarize the main features of the lattice approach in a way
that we hope is understandable by nonexperts. For recent, complementary review articles, we refer the reader 
to  Refs.~\cite{DelDebbio:2021ryq,DallaBrida:2020pag}.

In order to determine the running coupling at scale $\mu$
\begin{eqnarray}
   \alpha_s(\mu) = { \gbar^2_{s}(\mu) \over 4\pi} \,,
\end{eqnarray}
we should first ``measure'' a short-distance quantity ${\oO}$ at scale
$\mu$ either experimentally or by lattice calculations, and then 
match it to a perturbative expansion in terms of a running coupling,
conventionally taken as $\alpha_{\overline{\rm MS}}(\mu)$,
\begin{eqnarray}
   {\oO}(\mu) = c_1 \alpha_{\overline{\rm MS}}(\mu)
              +  c_2 \alpha_{\overline{\rm MS}}^2(\mu) + \cdots \,.
\label{eq:alpha_MSbar}
\end{eqnarray}
We note that in some cases also a lowest-order constant term, $c_0$, may be present; 
in the following, we always assume that such a term has been subtracted on both sides and absorbed in a re-definition of $\oO(\mu)$.
We distinguish between phenomenological and lattice determinations of $\alpha_s$,
the essential difference being the origin of the values of
$\oO$ in \eq{eq:alpha_MSbar}. 
The basis of phenomenological determinations are 
experimentally measurable cross sections or decay widths from which $\oO$ is
defined. These cross sections have to be sufficiently inclusive 
and at sufficiently high scales such that perturbation theory 
can be applied. Often hadronization corrections have to be used
to connect the observed hadronic cross sections to the perturbative
ones. Experimental data at high $\mu$, where perturbation theory
is progressively more precise, usually have increasing experimental errors,
not least due to the very smallness of $\alpha_s(\mu)$.
Hence, it is not easy to find processes that allow one
to follow the $\mu$-dependence of a single $\oO(\mu)$ over
a range where $\alpha_s(\mu)$ changes significantly and precision is 
maintained. Note also that determinations of $\alpha_s$ from experimental data 
at hadron colliders necessarily require a simultaneous fit of the Parton Distribution Functions 
(PDFs)~\cite{Forte:2020pyp}, 
making the whole procedure more complicated and prone to systematic errors.

In contrast, in lattice gauge theory, one can design $\oO(\mu)$ 
Euclidean short-distance quantities that are not directly related to
experimental observables. This allows us to follow the $\mu$-dependence 
until the perturbative regime is reached and nonperturbative ``corrections'' are negligible.  The only
experimental input for lattice computations of $\alpha_s$ are the masses or decay constants 
of hadrons, which fixes the overall energy scale of the theory and
the quark masses. Therefore, experimental errors are completely
negligible and issues such as hadronization do not occur.  We can
construct many short-distance quantities that are easy to calculate
nonperturbatively in lattice simulations with small statistical
uncertainties.  We can also simulate at parameter values that do not
exist in nature (for example, with unphysical quark masses between
bottom and charm) to help control systematic uncertainties.  These
features mean that precise results for $\alpha_s$ can be achieved
with lattice-gauge-theory computations.  Further, as in phenomenological determinations,
the different methods available to determine $\alpha_s$ in
lattice calculations with different associated systematic
uncertainties enable valuable cross-checks.  Practical limitations are
discussed in the next section, but a simple one is worth mentioning
here. Experimental results (and therefore the phenomenological
determinations) of course have all quarks present, while in lattice
gauge theories in practice only the lighter ones are included and one
is then forced to use the matching at thresholds, as discussed in the following
subsection.

It is important to keep in mind that the dominant source of uncertainty
in most present day lattice-QCD calculations of $\alpha_s$ are from
the truncation of continuum/lattice perturbation theory and from
discretization errors. Perturbative truncation errors are of 
particular concern because they often cannot easily be estimated
from studying the data itself. Perturbation theory provides an asymptotic series
and the size of higher-order coefficients can sometimes
turn out to be larger than suggested by naive expectations 
based on power counting from the behaviour of lower-order terms.  
We note that perturbative truncation errors are also the dominant
source of uncertainty in several of the phenomenological
determinations of $\alpha_s$.  

The various phenomenological approaches to determining the running
coupling constant, $\alpha^{(5)}_{\overline{\rm MS}}(M_Z)$ are summarized by the
Particle Data Group \cite{Zyla:2020zbs}.
The PDG review lists five categories of phenomenological results 
used to obtain the running coupling: using hadronic
$\tau$ decays, hadronic final states of $e^+e^-$ annihilation,
deep inelastic lepton--nucleon scattering, electroweak precision data, and high-energy hadron-collider data.
Excluding lattice results, the PDG, in their most recent update~\cite{ParticleDataGroup:2024cfk}, 
quotes the weighted average as
\begin{eqnarray}
   \alpha^{(5)}_{\overline{\rm MS}}(M_Z) &=& 0.1175(10) \,, 
   \quad 
   \mbox{PDG 24 \cite{ParticleDataGroup:2024cfk}}
\label{PDG_nolat}
\end{eqnarray}
compared to 
$
   \alpha^{(5)}_{\overline{\rm MS}}(M_Z) = 0.1176(11) 
$
of the older PDG 2020 \cite{Zyla:2020zbs}.
For a general overview of the various phenomenological
and lattice approaches 
see, e.g., Ref.~
 \cite{Salam:2017qdl}. 
The extraction of $\alpha_s$ from $\tau$ data, which is one of the most precise and thus
has a large impact on the nonlattice average in Eq.~(\ref{PDG_nolat}), is
especially sensitive to the treatment of higher-order perturbative
terms as well as the treatment of nonperturbative effects.  
This is important to keep in mind when comparing our chosen
range for $\alpha^{(5)}_{\overline{\rm MS}}(M_Z)$ from lattice
determinations in Eq.~(\ref{eq:alpmz}) with the nonlattice average
from the PDG.

\subsubsection{Scheme and scale dependence of $\alpha_s$ and $\Lambda_{\rm QCD}$}

Despite the fact that the notion of the QCD coupling is 
initially a perturbative concept, the associated $\Lambda$-parameter
is nonperturbatively defined,
\begin{eqnarray}
   \Lambda 
      &\equiv& \mu\,\varphi_s(\gbar_s(\mu)),\nonumber\\
      \varphi_s(\gbar_s) &=&
      (b_0\gbar_s^2)^{-b_1/(2b_0^2)} 
              e^{-1/(2b_0\gbar_s^2)}
             \exp\left[ -\int_0^{\gbar_s}\,dx 
                        \left( {1\over \beta(x)} + {1 \over b_0x^3} 
                                                 - {b_1 \over b_0^2x}
                        \right) \right] \,,
\nonumber\\ \label{eq:Lambda}
\end{eqnarray}
provided that $\beta(\gbar_s) = \mu\frac{\partial\gbar_s(\mu)}{\partial\mu}$ is the full renormalization group function 
in the (mass-independent) scheme which defines $\gbar_s$. The first two coefficients, $b_0$ and $b_1$, in 
the perturbative expansion
\begin{eqnarray}
   \beta(x) \sim -b_0 x^3 -b_1 x^5 + \ldots \,,
\label{eq:beta_pert}
\end{eqnarray}
are scheme-independent (``universal'') and given by
\begin{eqnarray}
   b_0 = {1\over (4\pi)^2}
           \left( 11 - {2\over 3}\Nf \right) \,, \qquad
   b_1 = {1\over (4\pi)^4}
           \left( 102 - {38 \over 3} \Nf \right) \,.
\label{b0+b1}
\end{eqnarray}
In the $\overline{\rm MS}$ scheme, the coefficients of the $\beta$-function have
been calculated up to 5-loop order, i.e., $b_2$, $b_3$ and $b_4$ 
are known~\cite{vanRitbergen:1997va,Czakon:2004bu,Luthe:2016ima,Herzog:2017ohr,Baikov:2016tgj}.

As a renormalization-group-invariant quantity, the $\Lambda$-parameter is $\mu$-independent. 
However, it does depend on the renormalization scheme albeit in an exactly computable way:
A perturbative change of the coupling from one mass-independent scheme $S$ to another (taken here
to be the $\overline{\rm MS}$ scheme) takes the form
\begin{eqnarray}
   g_{\overline{\rm MS}}^2(\mu) 
      = g_{\rm S}^2(\mu) (1 + c^{(1)}_g g_{\rm S}^2(\mu) + \ldots ) \,,
\label{eq:g_conversion}
\end{eqnarray}
where $c^{(i)}_g, \, i\geq 1$ are finite coefficients. Performing this change
in the expression for the $\Lambda$-parameter at a large scale $\mu$,
so that higher-order terms can be neglected, one obtains the
exact relation between the respective $\Lambda$-parameters of the two schemes,
\begin{eqnarray}
   \Lambda_{\overline{\rm MS}} 
      = \Lambda_{\rm S} \exp\left[ c_g^{(1)}/(2b_0)\right] \,.
      \label{eq:Lambdaconversion}
\end{eqnarray}
Note that this exact relation allows us to nonperturbatively define $\Lambda_{\overline{\rm MS}}$,
by starting from any nonperturbatively defined scheme $S$ for which $c^{(1)}_g$ is known. 
Given the high-order knowledge (5-loop by now) of
$\beta_\msbar$ then means that the errors in $\alpha_\msbar(m_\mathrm{Z})$
correspond almost completely with the  errors of $\Lambda_S$. We will therefore
mostly discuss them in that way.  Starting from \eq{eq:Lambda}, we have to consider (i) the
error of $\gbar_S^2(\mu)$ (denoted as $\left(\frac{\Delta \Lambda}{\Lambda}\right)_{\Delta \alpha_S}$ ) 
and (ii) the truncation error in $\beta_S$ (denoted as $\left( \frac{\Delta \Lambda}{\Lambda}\right)_{\rm trunc}$).
Concerning (ii), note that knowledge of $c_g^{(n_{\mathrm{l}})}$ for the scheme $S$ means that $\beta_S$ is known to $n_{\mathrm{l}}+1$ loop order; $b_{n_{\mathrm{l}}}$ is known. 
We thus see that in the region where perturbation theory can be applied, 
the following errors of $\Lambda_S$ (or consequently $\Lambda_{\overline{\rm MS}}$) have to be considered
\begin{eqnarray}
  \left(\frac{\Delta \Lambda}{\Lambda}\right)_{\Delta \alpha_S} 
  &=& \frac{\Delta \alpha_{S}(\mu)}{ 8\pi b_0 \alpha_{S}^2(\mu)} \times \left[1 + \cO(\alpha_S(\mu))\right]\,,
  \label{eq:i}\\
 \left( \frac{\Delta \Lambda}{\Lambda}\right)_{\rm trunc} 
  &=& k \alpha_{S}^{n_\mathrm{l}}(\mu) + \cO(\alpha_S^{n_\mathrm{l}+1}(\mu))\,,
  \label{eq:ii}  
\end{eqnarray}
where the pre-factor $k$ depends on $b_{n_\mathrm{l}+1}$ and in typical 
good schemes such as $\msbar$ it is numerically of order one. 
Statistical and systematic errors such as discretization
effects contribute to  $\Delta \alpha_{S}(\mu)$.  In the above we dropped a 
scheme subscript for the $\Lambda$-parameters because of
      \eq{eq:Lambdaconversion}.

By convention $\alpha_\msbar$ is usually quoted at a scale $\mu=M_Z$
where the appropriate effective coupling is the one in the
five-flavour theory: $\alpha^{(5)}_{\overline{\rm MS}}(M_Z)$.  In
order to obtain it from a result with fewer flavours, one connects effective
theories with different number of flavours as discussed by Bernreuther
and Wetzel~\cite{Bernreuther:1981sg}.  For example, one considers the
$\msbar$ scheme, matches the three-flavour theory to the four-flavour
theory at a scale given by the charm-quark mass~\cite{Chetyrkin:2005ia,Schroder:2005hy,Kniehl:2006bg}, runs with the
5-loop $\beta$-function~\cite{vanRitbergen:1997va,Czakon:2004bu,Luthe:2016ima,Herzog:2017ohr,Baikov:2016tgj} 
of the four-flavour theory to a scale given by
the $b$-quark mass, and there matches to the five-flavour theory, after
which one runs up to $\mu=M_Z$ with the five-loop $\beta$-function.
For the matching relation at a given
quark threshold we use the mass $m_\star$ which satisfies $m_\star=
\overline{m}_\msbar^{(\Nf)}(m_\star)$, where $\overline{m}$ is the running
mass (analogous to the running coupling). Then
\begin{eqnarray}
\label{e:convnfm1}
 \gbar^2_{\Nf-1}(m_\star) =  \gbar^2_{\Nf}(m_\star)\times 
      [1+ 0\times\gbar^{2}_{\Nf}(m_\star) + \sum_{n\geq 2}t_n\,\gbar^{2n}_{\Nf}(m_\star)]
\label{e:grelation}
\end{eqnarray}
{with  \cite{Grozin:2011nk,Kniehl:2006bg,Chetyrkin:2005ia} }
\def\nli{(\Nf-1)}
\begin{eqnarray}
  t_2 &=&  {1 \over (4\pi^2)^2} {11\over72}\,,\\
  t_3 &=&  {1 \over (4\pi^2)^3} \left[- {82043\over27648}\zeta_3 + 
                     {564731\over124416}-{2633\over31104}(\Nf-1)\right]\,, \\
  t_4 &=& {1 \over (4\pi^2)^4} \big[5.170347 - 1.009932 \nli - 0.021978 \,\nli^2\big]\,,
\end{eqnarray}
(where $\zeta_3$ is the Riemann zeta-function) provides the matching
at the thresholds in the $\msbar$ scheme.  Often the software packages {\tt RunDec} \cite{Chetyrkin:2000yt,Herren:2017osy}
or the more recent one, {\tt REvolver}~\cite{Hoang:2021fhn}, 
are used for quark-threshold matching and running in the $\msbar$-scheme. 

While $t_2,\,t_3,\,t_4$ are
numerically small coefficients, the charm-threshold scale is also
relatively low and so there are nonperturbative
uncertainties in the matching procedure, which are difficult to
estimate but which we assume here to be negligible. This is supported
by nonperturbative tests~\cite{Athenodorou:2018wpk}, where perturbative decoupling relations
in the $\msbar$ scheme were shown to quantitatively describe
decoupling at the few permille level, down to the charm-quark region.
Obviously there is no perturbative matching formula across
the strange ``threshold''; here matching is entirely nonperturbative.
Model-dependent extrapolations of $\gbar^2_{\Nf}$ from $\Nf=0,2$ to
$\Nf=3$ were done in the early days of lattice gauge theory. We will
include these in our listings of results but not in our estimates,
since such extrapolations are based on untestable assumptions.

\subsubsection{Overview of the review of $\alpha_s$}

We begin by explaining lattice-specific difficulties in \sect{s:crit}
and the FLAG criteria designed to assess whether the
associated systematic uncertainties can be controlled and estimated in
a reasonable manner. These criteria remain unchanged since  
the FLAG 19 report, as there has still not been sufficiently broad progress
to make them more stringent. However, in this report we have
implemented a systematic scale variation to help assess systematic
errors due to the truncation of the perturbative series. Scale variations are
widely used in phenomenology and its application to lattice determinations
has been advocated in Ref.~\cite{DelDebbio:2021ryq}. We explain
the procedure at the end of this introduction and, where possible, 
we will quote corresponding results.

We then discuss, in \sect{s:SF} -- \sect{s:gf},
the various lattice approaches and results 
from calculations with $\Nf = 0$, 2, 2+1, and 2+1+1 flavours.
For lattice approaches with neither a new result nor a result passing all FLAG criteria,
we refer to the discussion in previous FLAG reports. 
In particular, this regards determinations of $\alpha_s$ from QCD vertices
and from the eigenvalue spectrum of the Dirac operator.

Since FLAG 21, the strategy of nonperturbative renormalization by decoupling, 
as introduced by the ALPHA collaboration in Ref.~\cite{DallaBrida:2019mqg},
produced a new result for $\alpha_s$.  It is important to realize
that this method shifts the perspective on results 
for the $\Lambda$-parameter with unphysical flavour
numbers, in particular for $\Nf=0$: Such results can be
related to $\Nf>0$ results by a nonperturbative matching calculation.
We therefore made an effort to review $\Nf=0$ results, some of which are
now over 20 years old. In particular, we also included a new section on the gradient-flow (GF) coupling 
in infinite space-time volume, even though only results for $\Nf=0$ exist at the moment.

After the discussion of the various lattice methods, we proceed, 
in Sec.~\ref{s:alpsumm}, with the averages together with our best
estimates for $\alpha_{\overline{\rm MS}}^{(5)}$. These are currently determined
from three- and four-flavour QCD simulations only, however, with the decoupling result also relying on 
the $\Nf=0$ $\Lambda$-parameter as input. 
Therefore, we discuss results for the $\Nf=0$ $\Lambda$-parameter in some detail, in addition to the physical cases with $\Nf=3$, 4 and 5, where the latter is derived from $\Nf=3$ and 4 results by the standard perturbative evolution across the bottom-quark threshold.

\subsubsection{Additions with respect to the FLAG 21 report}
\label{sec:npapers}
Since the FLAG 21 report there were two new papers on $\Nf=3$:
 \begin{itemize} 
\item[] 
   Petreczky 20 \cite{Petreczky:2020tky}
   from heavy-quark current two-point functions (\sect{s:curr}).
\item[]
   ALPHA~22 \cite{DallaBrida:2022eua}
   from the decoupling method (\sect{s:dec}).
 \end{itemize}
In $\Nf=0$ QCD, there are a number of additional works:
\begin{itemize}
\item[]
   Bribian~21 \cite{Bribian:2021cmg}, from step-scaling with the twisted
   periodic gradient-flow coupling (\sect{s:SF}).
\item[]
   Hasenfratz~23 \cite{Hasenfratz:2023bok} and
   Wong~23 \cite{Wong:2023jvr}
   from the GF scheme in infinite volume (\sect{s:gf})
\item[]
  Chimirri~23 \cite{Chimirri:2023iro}
  from heavy-quark current two-point functions (\sect{s:curr})
\item[] 
  Brambilla~23 \cite{Brambilla:2023fsi},
  from the force between static quarks (\sect{s:qq})
\end{itemize}
 
\subsection{General issues}

\subsubsection{Discussion of criteria for computations entering the averages}


\label{s:crit}


As in the PDG review, we only use calculations of $\alpha_s$ published
in peer-reviewed journals, and that use NNLO or higher-order
perturbative expansions, to obtain our final range in
Sec.~\ref{s:alpsumm}.  We also, however, introduce further
criteria designed to assess the ability to control important
systematics, which we describe here.  Some of these criteria, 
e.g., that for the continuum extrapolation, are associated with
lattice-specific systematics and have no continuum analogue.  Other
criteria, e.g., that for the renormalization scale, could in
principle be applied to nonlattice determinations.
Expecting that lattice calculations will continue to improve significantly 
in the near future, our goal in reviewing the state-of-the-art here 
is to be conservative and avoid prematurely choosing an overly small range.

In lattice calculations, we generally take ${\oO}$ to be some
combination of physical amplitudes or Euclidean correlation functions
which are free from UV and IR divergences and have a well-defined
continuum limit.  Examples include the force between static quarks and
two-point functions of quark-bilinear currents.

In comparison to values of observables ${\oO}$ determined
experimentally, those from lattice calculations require two more
steps. The first step concerns obtaining the scale $\mu$ in physical units (\mbox{GeV}), given its value, 
$a\mu$, in lattice units. Ideally one compares the lattice result for some hadron mass
$a M_\text{had}$ with the known experimental result for $M_\text{had}$ to determine $a$ and thus
$\mu$ in physical units. Alternatively, convenient intermediate scales such as  $\sqrt{t_0}$, $w_0$, $r_0$, $r_1$,
\cite{Luscher:2010iy,Borsanyi:2012zs,Sommer:1993ce,Bernard:2000gd} can be used 
if their relation to an experimental dimensionful observable is established. For more details 
we refer to Sec.~\ref{sec:scalesetting} on scale setting in this FLAG report.
The low-energy scale $\mu$ needs to be computed at the same
lattice spacings (i.e., the same bare couplings) where ${\oO}$ is determined, at least as long as
one does not use the step-scaling method (see below).  This induces a
practical difficulty given present computing resources.  In the
determination of the low-energy reference scale the volume needs to be
large enough to avoid finite-size effects. On the other hand, in order
for the perturbative expansion of Eq.~(\ref{eq:alpha_MSbar}) to be
reliable, one has to reach sufficiently high values of $\mu$,
i.e., short enough distances. To avoid uncontrollable discretization
effects the lattice spacing $a$ has to be accordingly small.  This
means
\begin{eqnarray}
   L \gg \mbox{hadron size}\sim \Lambda_{\rm QCD}^{-1}\quad 
   \mbox{and} \quad  1/a \gg \mu \,,
   \label{eq:scaleproblem}
\end{eqnarray}
(where $L$ is the box size) and therefore
\begin{eqnarray} 
   L/a \ggg \mu/\Lambda_{\rm QCD} \,.
   \label{eq:scaleproblem2}
\end{eqnarray}
The currently available computer power, however, limits $L/a$, 
typically to
$L/a = 32-96$. 
Unless one accepts compromises in controlling  discretization errors
or finite-size effects, this means one needs to set 
the scale $\mu$ according to
\begin{eqnarray}
   \mu \lll L/a \times \Lambda_{\rm QCD} & \sim 10-30\, \mbox{GeV} \,.
\end{eqnarray}
(Here $\lll$ or $\ggg$ means at least one order of magnitude smaller or larger.) 
Therefore, $\mu$ can be $1-3\, \mbox{GeV}$ at most.
This raises the concern whether the asymptotic perturbative expansion
truncated at $1$-loop, $2$-loop, or $3$-loop in Eq.~(\ref{eq:alpha_MSbar})
is sufficiently accurate. There is a finite-size scaling method,
usually called step-scaling method, which solves this problem by identifying 
$\mu=1/L$ in the definition of ${\oO}(\mu)$, see \sect{s:SF}. 

For the second step after setting the scale $\mu$ in physical units
($\mbox{GeV}$), one should compute ${\oO}$ on the lattice,
${\oO}_{\rm lat}(a,\mu)$ for several lattice spacings and take the continuum
limit to obtain the left hand side of Eq.~(\ref{eq:alpha_MSbar}) as
\begin{eqnarray}
   {\oO}(\mu) \equiv \lim_{a\rightarrow 0} {\oO}_{\rm lat}(a,\mu) 
              \mbox{  with $\mu$ fixed}\,.
\end{eqnarray}
This is necessary to remove the discretization error.

Here it is assumed that the quantity ${\oO}$ has a continuum limit,
which is regularization-independent. 
The method discussed in \sect{s:WL}, which is based on the perturbative
expansion of a lattice-regulated, divergent short-distance quantity
$W_{\rm lat}(a)$ differs in this respect and must be
treated separately.

In summary, a controlled determination of $\alpha_s$ 
needs to satisfy the following:
\begin{enumerate}

   \item The determination of $\alpha_s$ is based on a
         comparison of a
         short-distance quantity ${\oO}$ at scale $\mu$ with a well-defined
         continuum limit without UV and IR divergences to a perturbative
         expansion formula in Eq.~(\ref{eq:alpha_MSbar}).

   \item The scale $\mu$ is large enough so that the perturbative expansion
         in Eq.~(\ref{eq:alpha_MSbar}) is precise 
         to the order at which it is truncated,
         i.e., it has good {\em asymptotic} convergence.
         \label{pt_converg}

   \item If ${\oO}$ is defined by physical quantities in infinite volume,  
         one needs to satisfy \eq{eq:scaleproblem2}.
         \label{constraints}

   \item Nonuniversal quantities, i.e., quantities that depend on the chosen lattice regularization 
            and do not have a nontrivial continuum limit need a separate discussion, see
        \sect{s:WL}.

\end{enumerate}

Conditions \ref{pt_converg}. and \ref{constraints}. give approximate lower and
upper bounds for $\mu$ respectively. It is important to see whether there is a
window to satisfy \ref{pt_converg}. and \ref{constraints}. at the same time.
If it exists, it remains to examine whether a particular lattice
calculation is done inside the window or not. 

Obviously, an important issue for the reliability of a calculation is
whether the scale $\mu$ that can be reached lies in a regime where
perturbation theory can be applied with confidence. However, the value
of $\mu$ does not provide an unambiguous criterion. For instance, the
Schr\"odinger Functional, or SF coupling (Sec.~\ref{s:SF}) is
conventionally taken at the scale $\mu=1/L$, but one could also choose
$\mu=2/L$. Instead of $\mu$ we therefore define an effective
$\alpha_{\rm eff}$.  For schemes such as SF (see Sec.~\ref{s:SF}) or
$qq$ (see Sec.~\ref{s:qq}) this is directly the coupling  of
the scheme. For other schemes such as the vacuum polarization we use
the perturbative expansion \eq{eq:alpha_MSbar} for the observable
${\oO}$ to define
\begin{eqnarray}
   \alpha_{\rm eff} =  {\oO}/c_1 \,.
   \label{eq:alpeff}
\end{eqnarray}
As mentioned earlier, if there is an $\alpha_s$-independent term it should first be subtracted.
Note that this is nothing but defining an effective,
regularization-independent coupling, a physical renormalization scheme. 
For ease of notation, here and in what follows we denote
by $\alpha_s$ the coupling $\alpha_{\overline{\rm MS}}(\mu)$ that appears in Eq.~(\ref{eq:alpha_MSbar}).
 
Let us now comment further on the use of the perturbative series.
Since it is only an asymptotic expansion, the remainder $R_n({\oO})={\oO}-\sum_{i\leq n}c_i \alpha_s^i$ of a truncated
perturbative expression ${\oO}\sim\sum_{i\leq n}c_i \alpha_s^i$
cannot just be estimated as a perturbative error $k\,\alpha_s^{n+1}$.
The error is nonperturbative. Often one speaks of ``nonperturbative
contributions'', but nonperturbative and perturbative cannot be
strictly separated due to the asymptotic nature of the series (see,
e.g., Ref.~\cite{Martinelli:1996pk}).

Still, we do have some general ideas concerning the 
size of nonperturbative effects. The known ones such as instantons
or renormalons decay for large $\mu$ like inverse powers of $\mu$
and are thus roughly of the form 
\begin{eqnarray}
   \exp(-\gamma/\alpha_s) \,,
\end{eqnarray}
with some positive constant $\gamma$. Thus we have,
loosely speaking,
\begin{eqnarray}
   {\oO} = c_1 \alpha_s + c_2 \alpha_s^2 + \ldots + c_n\alpha_s^n
                  + \cO(\alpha_s^{n+1}) 
                  + \cO(\exp(-\gamma/\alpha_s)) \,.
   \label{eq:Owitherr}
\end{eqnarray}
For small $\alpha_s$, the $\exp(-\gamma/\alpha_s)$ is negligible.
Similarly the perturbative estimate for the magnitude of
relative errors in \eq{eq:Owitherr} is small; as an
illustration for $n=3$ and $\alpha_s = 0.2$ the relative error
is $\sim 0.8\%$ (assuming coefficients $|c_{n+1} /c_1 | \sim 1$).

For larger values of $\alpha_s$ nonperturbative effects can become
significant in Eq.~(\ref{eq:Owitherr}). An instructive example comes
from the values obtained from $\tau$
decays, for which $\alpha_s\approx 0.3$. Here, different applications
of perturbation theory (fixed order and contour improved)
each look reasonably asymptotically 
convergent but the difference does not seem to decrease much 
with the order (see, e.g., the contribution by Pich to Ref.~\cite{Bethke:2011tr}; see, however, also the discussion in Refs.~\cite{Hoang:2021nlz,Benitez-Rathgeb:2022hfj}). In addition, nonperturbative terms in the spectral function may be 
nonnegligible even after the integration up to $m_\tau$ (see, e.g., Refs.~\cite{Boito:2014sta}, \cite{Boito:2016oam}). 
All of this is because $\alpha_s$ is not really small.

Since the size of the nonperturbative effects is very hard to
estimate one should try to avoid such regions of the coupling.  In a
fully controlled computation one would like to verify the perturbative
behaviour by changing $\alpha_s$ over a significant range instead of
estimating the errors as $\sim \alpha_s^{n+1}$ .  Some computations
try to take nonperturbative power `corrections' to the perturbative
series into account by including such terms in a fit to the $\mu$-dependence. 
We note that this is a delicate procedure, as a term like, e.g.,
$\alpha_s(\mu)^3$ is hard to distinguish from a $1/\mu^2$ term when
the $\mu$-range is restricted and statistical and systematic errors
are present. We consider it safer to restrict the fit range to the
region where the power corrections are negligible compared to the
estimated perturbative error.

The above considerations lead us to the following special
criteria for the determination of $\alpha_s$: 

\begin{itemize}
   \item Renormalization scale         
         \begin{itemize}
            \item[\good] all data points relevant in the analysis have
             $\alpha_\mathrm{eff} < 0.2$
            \item[\soso] all data points have $\alpha_\mathrm{eff} < 0.4$
                         and at least one 
                         $\alpha_\mathrm{eff} \le 0.25$
            \item[\bad]  otherwise                                   
         \end{itemize}

   \item Perturbative behaviour 
        \begin{itemize}
           \item[\good] verified over a range of a factor $4$ change
                        in $\alpha_\mathrm{eff}^{n_\mathrm{l}}$ without power
                        corrections  or alternatively 
                        $\alpha_\mathrm{eff}^{n_\mathrm{l}} \le \frac12 \Delta \alpha_\mathrm{eff} / (8\pi b_0 \alpha_\mathrm{eff}^2) $ is reached
           \item[\soso] agreement with perturbation theory 
                        over a range of a factor
                        $(3/2)^2$ in $\alpha_\mathrm{eff}^{n_\mathrm{l}}$ 
                        possibly fitting with power corrections or
                        alternatively 
                        $\alpha_\mathrm{eff}^{n_\mathrm{l}} \le \Delta \alpha_\mathrm{eff} / (8\pi b_0 \alpha_\mathrm{eff}^2)$
                        is reached
           \item[\bad]  otherwise
       \end{itemize}
        Here {$\Delta \alpha_\mathrm{eff}$ is the accuracy cited for the determination of 
        $\alpha_\mathrm{eff}$}
        and $n_\mathrm{l}$ is the loop order to which the 
        connection of $\alpha_\mathrm{eff}$ to the $\msbar$ scheme is known.
        Recall the discussion around Eqs.~(\ref{eq:i},\ref{eq:ii});
        the $\beta$-function of $\alpha_\mathrm{eff}$ is then known to 
        $(n_\mathrm{l}+1)$-loop order.%
        \footnote{Once one is in the perturbative region with 
        $\alpha_{\rm eff}$, the error in 
        extracting the $\Lambda$-parameter due to the truncation of 
        perturbation theory scales like  $\alpha_{\rm eff}^{n_\mathrm{l}}$,
        as discussed around Eq.~(\ref{eq:ii}). In order to 
        detect/control such corrections properly, one needs to change
        the correction term significantly; 
        we require a factor of four for a $\good$ and a factor $(3/2)^2$
        for a $\soso$. 
        An exception to the above is the situation 
        where the correction terms are small anyway, i.e.,  
        $\alpha_{\rm eff}^{n_\mathrm{l}} \approx (\Delta \Lambda/\Lambda)_\text{trunc} 
        < (\Delta \Lambda/\Lambda)_{\Delta \alpha} 
        \approx \Delta \alpha_{\rm eff} / (8\pi b_0 \alpha_{\rm eff}^2)$ is reached.}
         
   \item Continuum extrapolation 
        
        At a reference point of $\alpha_{\rm eff} = 0.3$ (or less) we require
         \begin{itemize}
            \item[\good] three lattice spacings with
                         $\mu a < 1/2$ and full $\cO(a)$
                         improvement, \\
                         or three lattice spacings with
                         $\mu a \leq 1/4$ and $2$-loop $\cO(a)$
                         improvement, \\
                         or $\mu a \leq 1/8$ and $1$-loop $\cO(a)$
                         improvement 
            \item[\soso] three lattice spacings with $\mu a < 3/2$
                         reaching down to $\mu a =1$ and full
                         $\cO(a)$ improvement, \\
                         or three lattice spacings with
                         $\mu a \leq 1/4$ and 1-loop $\cO(a)$
                         improvement        
            \item[\bad]  otherwise 
         \end{itemize}
        
\end{itemize}  
In addition to the above criteria we have looked at scale variations as a general mean to assess perturbative behaviour
(cf.~subsection below). Continuum extrapolations are often not the primary concern in determinations
of $\alpha_s$. Where appropriate we will evaluate the new FLAG data-driven criterion,
by which the distance of the data to the continuum-extrapolated value is measured in units of the quoted error.
If the observable is $Q(a)$ with an extrapolated continum value $Q(0)\pm \Delta Q$ we look at the size of 
\begin{equation}
   \delta_\text{min} =  \dfrac{|Q(0) -Q(a_\text{min})|}{\Delta Q}.
\end{equation}
Some scepticism is warranted if $\delta_\text{min}$ exceeds $3$ or so, although there may be cases where this can be justified.
While we keep the core FLAG criteria unchanged, our general assessment will be informed by these measures.

We also need to specify what is meant by $\mu$. Here are our choices:
\begin{eqnarray}
   \text{step scaling} &:& \mu=1/L\,,
   \nonumber  \\
   \text{heavy quark-antiquark potential} &:& \mu=2/r\,,
   \nonumber  \\
   \text{observables in position space}  &:& \mu=1/|x|\,,
   \nonumber \\
   \text{observables in momentum space} &:& \mu =q \,,
   \nonumber   \\ 
    \text{moments of heavy-quark currents} 
                                        &:& \mu=2\overline{m}_\mathrm{c} \,,
   \nonumber   \\ 
   \text{Gradient-Flow (GF) scheme in infinite volume} 
                                        &:& \mu = 1/\sqrt{8t}\,,
\label{mu_def}
\end{eqnarray}
where $|x|$ is the Euclidean norm of the four-vector $x$, 
$q$ is the magnitude of the momentum, $\overline{m}_\mathrm{c}$ is
the heavy-quark mass (in the $\msbar$ scheme with $\Nf$ quarks, including the heavy-quark flavour) 
and usually taken around the charm-quark mass. The parameter $t$ denotes the gradient-flow time. 
We note again that the above criteria cannot
be applied when regularization-dependent quantities
$W_\mathrm{lat}(a)$ are used instead of ${\oO}(\mu)$. These cases
are specifically discussed in \sect{s:WL}.

In principle one should also
account for electro-weak radiative corrections. However, both in the
determination of $\alpha_{s}$ at intermediate scales $\mu$ and
in the running to high scales, we expect electro-weak effects to be
much smaller than the presently reached precision. Such effects are
therefore not further discussed.

The attentive reader will have noticed that bounds such as $\mu a <
3/2$ or at least one value of $\alpha_\mathrm{eff}\leq 0.25$
which we require for a $\soso$ are
not very stringent. There is a considerable difference between
$\soso$ and $\good$. We have chosen the above bounds, unchanged since
FLAG 16, as not too many current computations would satisfy more stringent 
ones. Nevertheless, we believe that the \soso\ criteria already give
reasonable bases for estimates of systematic errors. An exception 
may be Cali~20~\cite{Cali:2020hrj}, which is discussed in detail in Sec.~\ref{s:vac}.

In anticipation of future changes of the criteria, we 
expect that we will be able to tighten our criteria for inclusion in
the average, and that many more computations will reach the present
\good\ rating in one or more categories. 

In addition to our explicit criteria, the following effects may influence
the precision of results: 

{\em Topology sampling:}
    In principle a good way to improve the quality 
    of determinations of $\alpha_s$ is to push to very small lattice 
    spacings thus enabling large $\mu$. It is known  
    that the sampling of field space becomes very difficult for the 
    HMC algorithm when the lattice spacing is small and one has the
    standard periodic boundary conditions. In practice, for all known
    discretizations the topological charge slows down dramatically for
    $a\approx 0.05\,\fm$ and smaller 
    \cite{DelDebbio:2002xa,Bernard:2003gq,Schaefer:2010hu,Chowdhury:2013mea,Brower:2014bqa,Bazavov:2014pvz,Fukaya:2015ara}. 
    Open boundary conditions solve the problem 
    \cite{Luscher:2011kk} but are not frequently used. Since the effect of
    the freezing on short-distance observables is not known, we also do need to pay
    attention to this issue. Remarks are added in the text when appropriate.
      
{\em Quark-mass effects:} 
    We assume that effects
    of the finite masses of the light quarks (including strange) 
    are negligible in the effective
    coupling itself where large, perturbative, $\mu$ is considered.

{\em Scale setting:}
    The scale does not need
    to be very precise, since using the lowest-order $\beta$-function
    shows that a 3\% error in the scale determination corresponds to a
    $\sim 0.5\%$ error in $\alpha_s(M_Z)$.  Since the errors of scale
    determinations are now typically at the 1-2 percent level or better,
    the corresponding error in $\alpha_s(M_Z)$ will remain subdominant for the foreseeable
    future.

{\em Other limits/extrapolations:}
     Besides the continuum limit and the infinite-volume extrapolation of hadronic
     observables, further limits may be required, depending on 
     the method employed. An obvious case is the large-mass extrapolation in 
     the decoupling method. While in this case, an effective theory can be deployed to derive
     plausible fit functions, this is less clear in other cases.
     An example is the infinite space-time volume extrapolation
     in the GF scheme, which is needed to make contact with the available perturbative calculations.
     One would expect the volume dependence to be quite different at low and high energies, and
     there may be a complicated intermediate regime. 
     Systematic uncertainties are then much harder to quantify and
     our approach necessarily is on a case-by-case basis. Data-driven criteria like
     the new FLAG continuum-limit criterion are considered, however, these may fail if the data does
     not sufficiently overlap with the true (and possibly unknown) asymptotic regime.

\subsubsection{Physical scale}


Since FLAG 19, a new FLAG working group on scale setting has
been established. We refer to Sec.~\ref{sec:scalesetting}
for definitions and the current status. Note that the error from scale setting is
sub-dominant for current $\alpha_s$ determinations.

A popular scale choice has been the intermediate $r_0$ scale, 
and its variant $r_1$, which both derive from the force between static quarks, 
see Eq.~(\ref{eq:r01}). 
One should bear in mind that their determination from physical
observables also has to be taken into account.  The phenomenological
value of $r_0$ was originally determined as $r_0 \approx
0.49\,\mbox{fm}$ through potential models describing quarkonia
\cite{Sommer:1993ce}. Of course the quantity is precisely defined,
independently of such model considerations.
But a lattice computation with the correct sea-quark content is 
needed to determine a completely sharp value. When the quark 
content is not quite realistic, the value of $r_0$ may depend to
some extent on which experimental input is used to determine 
(actually define) it. 
 
The latest determinations from two-flavour QCD are
$r_0$ = 0.420(14)--0.450(14)~fm by the ETM collaboration
\cite{Baron:2009wt,Blossier:2009bx}, using as input $f_\pi$ and $f_K$
and carrying out various continuum extrapolations. On the other hand,
the ALPHA collaboration \cite{Fritzsch:2012wq} determined $r_0$ =
0.503(10)~fm with input from $f_K$, and the QCDSF
collaboration \cite{Bali:2012qs} cites 0.501(10)(11)~fm from
the mass of the nucleon (no continuum limit).  Recent determinations
from three-flavour QCD are consistent with $r_1$ = 0.313(3)~fm
and $r_0$ = 0.472(5)~fm
\cite{Davies:2009tsa,Bazavov:2010hj,Bazavov:2011nk}. Due to the
uncertainty in these estimates, and as many results are based directly
on $r_0$ to set the scale, we shall often give both the dimensionless
number $r_0 \Lambda_{\overline{\rm MS}}$, as well as $\Lambda_{\overline{\rm MS}}$.
In the cases where no physical $r_0$ scale is given in
the original papers or we convert to
the $r_0$ scale, we use the value $r_0$ = 0.472~fm. In case
$r_1 \Lambda_{\overline{\rm MS}}$ is given in the publications,
we use $r_0 /r_1 = 1.508$ \cite{Bazavov:2011nk}, to convert,
which remains well consistent with the update \cite{Bazavov:2014pvz} 
neglecting the error on this ratio. In some, mostly early,
computations the string tension, $\sqrt{\sigma}$ was used.
We convert to $r_0$ using $r_0^2\sigma = 1.65-\pi/12$,
which has been shown to be an excellent approximation 
in the relevant pure gauge theory \cite{Necco:2001xg,Luscher:2002qv}.

The more recent gradient-flow scales $t_0,w_0$ are very attractive
alternatives to $r_0$, as their determination is much simpler within
a given simulation and most collaborations quote their values. 
The main downside  are potentially large cutoff effects.
We intend to transition from $r_0$ to $t_0$. In this report we
start by reporting $\Nf=0$ results both with $r_0$ and  with $\sqrt{8t_0}$, where we use as 
conversion factor the central value of $\sqrt{8t_0}/r_0 = 0.9435(97)$ 
from Dalla Brida~19~\cite{DallaBrida:2019mqg}.
A general discussion of the various scales is given in 
the scale-setting section of this FLAG report, cf.~Sec.~\ref{sec:scalesetting}.


{
\subsubsection{Studies of truncation errors of perturbation theory}
\label{s:trunc}

As discussed previously, we have to determine $\alpha_s$ in a region
where the perturbative expansion for the $\beta$-function, 
Eq.~(\ref{eq:beta_pert}) in the integral Eq.~(\ref{eq:Lambda}),
is reliable. In principle this must be checked, and  is
difficult to achieve as we need to reach up to a sufficiently high scale.
A recipe routinely used to estimate the size of truncation errors
of the perturbative series is to study the dependence on the renormalization scale
of an observable
evaluated at a fixed order in the coupling, as the renormalization scale  
is varied around some `optimal' scale $\mu_*$,  from $\mu=\mu_*/2$ to $2\mu_*$. 
For examples, see Ref.~\cite{DelDebbio:2021ryq}.

Alternatively, or in addition, the renormalization scheme chosen
can be varied, which investigates the perturbative
conversion of the chosen scheme to the perturbatively defined
$\overline{\rm MS}$ scheme and in particular `fastest apparent
convergence' when the `optimal' scale is chosen so that the
$\cO(\alpha_s^2)$ coefficient vanishes.

The ALPHA collaboration in Ref.~\cite{Brida:2016flw} and 
ALPHA 17~\cite{DallaBrida:2018rfy}, within the SF approach defined
a set of $\nu$-schemes for which the 3-loop (scheme-dependent)
coefficient of the $\beta$-function for $\Nf = 2+1$ flavours was
computed to be $b_2^\nu = -(0.064(27)+1.259(1)\nu)/(4\pi)^3$. 
The standard SF scheme has $\nu = 0$. For comparison, $b_2^\msbar = 0.324/(4\pi)^3$.
A range of scales from about 
$4\,\mbox{GeV}$ to $128\,\mbox{GeV}$ was investigated.
It was found that while the procedure of varying the
scale by a factor 2 up and down gave a correct estimate
of the residual perturbative error for $\nu \approx 0 \ldots  0.3$,  
for negative values, e.g.,  $\nu = -0.5$, the estimated perturbative
error is much too small to account for the mismatch in the 
$\Lambda$-parameter of $\approx 8\%$ at $\alpha_s=0.15$.
This mismatch, however, did, as expected, still scale with $\alpha_s^{n_{\mathrm{l}}}$ with $n_{\mathrm{l}}=2$. In the schemes
with negative $\nu$, the coupling $\alpha_s$ has to be quite small for
scale variations of a factor 2 to correctly signal the perturbative errors. 

For a systematic study of renormalization-scale variations as a measure of
perturbative truncation errors in various lattice determinations
of $\alpha_s$, we implement scale variations following the proposal in 
Ref.~\cite{DelDebbio:2021ryq}. Scale variations are commonly used in phenomenology 
as a tool to investigate truncations errors. 
While they cannot give a precise
estimate of the truncation errors, they provide a simple, quantitative test that
can be uniformly applied to all observables. Furthermore, the implementation 
proposed in Ref.~\cite{DelDebbio:2021ryq} does not rely on lattice data. The only
inputs are the coefficients of the perturbative expansion of $\alpha_{\mathrm{eff}}$, 
so that, in principle, an estimate of the truncation errors can be done {\em before}
embarking in a numerical simulation. Here we shall summarize briefly the methodology, 
provide the coefficients of the perturbative expansions for the observables of interest
in this review, and compute the corresponding truncation errors. 

\paragraph*{Methodology}
The use of scale variations for the determination of the missing higher-order uncertainties
relies on a simple observation, namely that the scale $\mu$ that appears on the 
left-hand side of Eq.~(\ref{eq:alpha_MSbar}) does not need to match the scale at which the 
running coupling constant is computed on the right-hand side of the same equation. 
Eq.~(\ref{eq:alpha_MSbar}) can be rewritten, with the same level of precision, as
\begin{eqnarray}
   {\oO}(\mu) = c_1 \alpha_{\overline{\rm MS}}(\mu')
              +  \sum_{k=2}^{n} c'_k(s) \alpha^k_{\overline{\rm MS}}(\mu') 
              + \mathcal{O}\left(\alpha^{n+1}_{\overline{\rm MS}}(\mu')\right)\, ,
              \quad (s=\mu'/\mu)\, .
\label{eq:alpha_MSbar_prime}
\end{eqnarray}
The coefficients 
\begin{eqnarray}
   \label{eq:ckslogs}
   c'_k(s) = \sum_{\ell=0}^{k-1} c'_{k,\ell} \log^\ell(s)\, ,
\end{eqnarray}
for $k\geq 2$, are determined from the coefficients $c_k$ in Eq.~(\ref{eq:alpha_MSbar}) using the recursion
\begin{eqnarray}
   c'_{k,0} &=& c_k\, , \\
   c'_{k,\ell} &=& \frac{2}{\ell} \sum_{j=1}^{k-1} j (4\pi)^{k-j} b_{k-j-1} c'_{j,\ell-1}\, ,
\end{eqnarray}
where $b_n$ are the coefficients of the beta function defined in Eq.~(\ref{eq:beta_pert}). The 
dependence on $s$, and therefore on the scale $\mu'$, is entirely due to the truncation
of the perturbative expansion. Denoting the truncated series by
\begin{eqnarray}
   {\oO}^{(n)}(\mu,\mu') = c_1 \alpha_{\overline{\rm MS}}(\mu')
              +  \sum_{k=2}^{n} c'_k(s) \alpha^k_{\overline{\rm MS}}(\mu') 
   \, ,
\label{eq:alpha_MSbar_prime_truncated}
\end{eqnarray}
it is possible to show that the scale-variation procedure described below yields a sensible estimate of
the truncation error
\begin{eqnarray}
   \delta_n=\left|{\oO}(\mu)-{\oO}^{(n)}(\mu,\mu')\right|\, ,
\end{eqnarray} 
see, e.g., the discussion in Ref.~\cite{Cacciari:2011ze}. 
Formally, 
\begin{eqnarray}
   \label{eq:scaleorder}
   \mu' \frac{\partial}{\partial \mu'} {\oO}^{(n)}(\mu,\mu') \propto \alpha^{n+1}_{\overline{\rm MS}}(\mu')\, ,
\end{eqnarray}
showing that scale variations capture the correct size of the truncation error, at least parametrically. 

\paragraph*{Implementation}
The implementation of the scale variations proceeds as follows. 

\begin{enumerate}
   \item We assume a value for $\Lambda_{\overline{\rm MS}}^{(3)}$, e.g., the current best estimate in
   FLAG. Given this value, we compute the corresponding value of 
   $\alpha_{\overline{\mathrm{MS}}}(s_{\mathrm{ref}} \mu)$ (at fixed $\Nf=3$)  where $\mu$ is 
   the scale associated to the observable ${\oO}$. Typical choices are $s_{\mathrm{ref}}=1$ or $s_{\mathrm{ref}}=s^*$, the 
   latter being the scale of fastest apparent convergence. 
   Similarly, we also compute the value of $\alpha_{\overline{\mathrm{MS}}}^{(5)}(M_Z)$. All these
   values are computed using the running of the strong coupling, the value of $\Lambda_{\overline{\rm MS}}^{(3)}$
   as the unique input, in addition to the $\msbar$ charm- and bottom-quark masses at their own scale, 
   $\bar{m}_{c}^{(4)}(\bar{m}_{c})$ and $\bar{m}_{b}^{(5)}(\bar{m}_{b})$, respectively and $m_Z$.
   
   \item Using Eq.~(\ref{eq:alpha_MSbar_prime_truncated}), we compute the value $\oO_{\mathrm{ref}}$ 
   of the observable by imposing that it
   coincides with its truncated expansion,
   \begin{eqnarray}
      {\oO}_{\mathrm{ref}} = {\oO}^{(n)}(\mu,s_{\mathrm{ref}}\mu)\, ,
   \end{eqnarray}
   where $s_{\mathrm{ref}}\mu$ is the scale associated to the observable as shown explicitly in 
   Eq.~(\ref{eq:alpha_MSbar}). 
   By construction, using the value ${\oO}_{\mathrm{ref}}$, setting $s=s_{\mathrm{ref}}$, and solving  
   Eq.~(\ref{eq:alpha_MSbar_prime_truncated}), we recover for 
   $\alpha_{\overline{\mathrm{MS}}}(s_{\mathrm{ref}}\mu)$ the value obtained in step 1. Hence, we interpret 
   ${\oO}_{\mathrm{ref}}$ as the value of the observable that yields the value of 
   $\alpha_{\overline{\mathrm{MS}}}^{(5)}(M_Z)$ in step 1, when performing the {\em usual}\ extraction 
   of the strong coupling. 
   
   \item We use Eq.~(\ref{eq:alpha_MSbar_prime_truncated}) again, but this time set 
   $s=s_{\mathrm{ref}}/2,2s_{\mathrm{ref}}$, 
   to extract
   $\alpha_{\overline{\mathrm{MS}}}(s\mu)$ by solving
   \begin{eqnarray}
      {\oO}_{\mathrm{ref}} = {\oO}^{(n)}(\mu,s\mu)\, .
   \end{eqnarray} 
   Because the expansion is truncated, the value obtained here for $\alpha_{\overline{\mathrm{MS}}}(s\mu)$
   is different from the one obtained by running the coupling from the value of 
   $\alpha_{\overline{\mathrm{MS}}}(s_{\mathrm{ref}}\mu)$ computed in step 2. 

   \item Using $\alpha_{\overline{\mathrm{MS}}}(s\mu)$ as the initial condition, we run the strong coupling
   constant and compute $\alpha_{\overline{\mathrm{MS}}}^{(5)}(M_Z)$. The difference between this value and
   the value computed in step 1 is used as an estimate of the uncertainty due to the truncation of the
   perturbative expansion.
\end{enumerate}

Typically scale variations are performed by multiplying and dividing the reference scale by a factor 2. 
For some determinations, where the perturbative matching is done at a few GeV, dividing the scale by a factor of 2
yields a low scale where perturbation theory is clearly no longer applicable and therefore the scale variation yields 
an artificially large error. In these cases, we consider only the variation obtained by multiplying the reference scale
by a factor 2. To be more specific, we define the following quantities. 
\begin{description}
   \item [$\delta_{(4)}(s_{\rm ref})$:] The renormalization scale $s_{\rm ref}\mu$ is multiplied and divided by a factor
     two. We quote a symmetric
     error by averaging the difference between the results obtained with the scales $s_{\rm ref}\mu$ and
     $2s_{\rm ref}\mu$, and the difference between the results obtained with scales $s_{\rm ref}/2 \times \mu$ and
     $s_{\rm ref}\mu$. Note however that in some cases the error is markedly
     asymmetric. We will quote the differences as a percentage deviation from the reference value of $\alpha_s(m_Z)$.
        
   \item[$\delta_{(2)}(s_{\rm ref})$:] The renormalization scale is multipied by a factor
     two only. The error $\delta_{(2)}(s_{\rm ref})$ is simply the difference between the two results obtained
     with the two scales, again taken as a percentage deviation from the reference value of  $\alpha_s(m_Z)$.
   \end{description}
   We also explore two common choices, namely $s_{\rm ref}=1$ and $s_{\rm ref}=s^*$, the scale of fastest apparent convergence, 
   i.e., the scale at which $c'_2(s^*)=0$. 

\paragraph*{Perturbative coefficients} 
The coefficients of the perturbative expansion for the observables of interest in this review 
are summarized in Tab.~\ref{tab:scale_truncation}. For each observable we report the number of
coefficients that are available for the perturbative expansion, the scale at which the 
perturbative matching is done, the list of coefficients and the relevant references. 

\begin{table}[!htb]
   \footnotesize
   \begin{tabular*}{\textwidth}{lllcr}
   Observable & $n_{\mathrm{l}}$ (loops) &\(\mu\) [GeV] & perturbative coefficients & References \\
   \hline\hline
   Step-scaling & 2 & 80 & $-1.37520970$, $0.57120172$  & \cite{Bode:1998hd,Bode:1999sm}  \\
   \hline
             & 3 & 1.5 & $-0.0485502$, $0.687447$, $0.818808$ & \cite{Fischler:1977yf,Peter:1996ig,Peter:1997me,Smirnov:2009fh,Smirnov:2008pn}     \\
   Potential &   & 2.5 & same as line above, $\mu$ changed  &      \\
             &   & 5.0 & same as line above, $\mu$ changed  &      \\
   \hline
                       & 3 & 2.0 & $-1.4346$, $0.16979$, $3.21120$ & \cite{Shintani:2010ph}  \\
   Vacuum polarization &   & 4.0 &  &  \cite{Hudspith:2018bpz}  \\
                       &   & 1.3 &  &  \cite{Cali:2020hrj} \\
   \hline
      $- \log W_{11}$      & 2 & 4.4 & $-0.87811924$, $4.20161085$ & \cite{Davies:2008sw,Lepage:1992xa}  \\
      $-\log W_{12}/u_0^6$ &   & 4.4 & $0.79128076$, $3.18658638$ &  \\
   \hline
      HQ \(r_4\) & 2 & \(m_{\rm c}\)  & $-0.07762325$, $0.07957445$  & \cite{Chetyrkin:2006xg,Chetyrkin:1997mb,Broadhurst:1991fi}   \\
      HQ \(r_4\) &  & \(2m_{\rm c}\) &  same as line above, $\mu$ changed&    \\
      HQ \(r_6\) &   & \(2m_{\rm c}\) & $0.77386542$, $-0.08560363$  &    \\
      HQ \(r_8\) &   & \(2m_{\rm c}\) & $1.08917060$, $0.20034888$ &    \\
   \hline
      GF coupling & 2 & $1/\sqrt{8t}$& \begin{tabular}{@{}c@{}}1.09778674 + 0.007555192 $\Nf$ \\ 
      $-0.98225 - 0.069913 \Nf + 0.001872234 \Nf^2$\end{tabular}  & \cite{Luscher:2010iy,Harlander:2016vzb} \\
   \hline\hline
   \end{tabular*}
   \caption{Summary of the coefficients of the perturbative expansion of the observables discussed in this review
   as a power series in $\alpha_{\overline{\mathrm{MS}}}$. We assume that the observables are normalized so 
   that $c_1=1$ and we only quote the coefficients starting from $c_2$. The coefficients are computed for 
   $\Nf=3$, unless the explicit dependence on the number of flavours is given. 
   For each observable, we quote the number of coefficients that are known analytically and the scale of 
   perturbative matching to the $\overline{\mathrm{MS}}$ scheme. Note that for the GF coupling
   there are two coefficients, reported as functions of $\Nf$, over two separate lines.}
   \label{tab:scale_truncation}
\end{table}


%% file: Alpha_s/s3.tex
\subsection{$\alpha_s$ from Step-Scaling Methods}
\label{s:SF}

\subsubsection{General considerations}


The method of step-scaling functions avoids the scale problem,
\eq{eq:scaleproblem}. It is in principle independent of the particular
boundary conditions used and was first developed with periodic
boundary conditions in a two-dimensional model~\cite{Luscher:1991wu}.

The essential idea of the step-scaling strategy
is to split the determination of the running coupling at large
$\mu$ and of a hadronic scale into two lattice calculations and
connect them by `step scaling'. In the former part, we determine the
running coupling constant in a finite-volume scheme
in which the renormalization scale is set by the inverse lattice size
$\mu = 1/L$. In this calculation, one takes a high renormalization scale
while keeping the lattice spacing sufficiently small as
\begin{eqnarray}
   \mu \equiv 1/L \sim 10\,\ldots\, 100\,\mbox{GeV}\,, \qquad a/L \ll 1 \,.
\end{eqnarray}
In the latter part, one chooses a certain 
$\gbar^2_\mathrm{max}=\gbar^2(1/L_\mathrm{max})$, 
typically such that $L_\mathrm{max}$ is around $0.5$--$1$~fm. With a 
common discretization, one then determines $L_\mathrm{max}/a$ and
(in a large volume $L \ge$ 2--3~fm) a hadronic scale
such as a hadron mass, $\sqrt{t_0}/a$ or $r_0/a$ at the same bare
parameters. In this way one gets numbers for, e.g., $L_\mathrm{max}/r_0$
and by changing the lattice spacing $a$ carries out a continuum-limit extrapolation of that ratio. 
 
In order to connect $\gbar^2(1/L_\mathrm{max})$ to $\gbar^2(\mu)$ at
high $\mu$, one determines the change of the coupling in the continuum
limit when the scale changes from $L$ to $L/s$, where $s$ is a scale factor,
set to $s=2$ in most applications. Then, starting from
$L=L_{\rm max}$ one iteratively performs $k$ steps 
to arrive at $\mu = s^k /L_{\rm max}$. This part of
the strategy is called step scaling. Combining these results yields
$\gbar^2(\mu)$ at $\mu = s^k \,(r_0 / L_\mathrm{max})\, r_0^{-1}$,
where $r_0$ stands for the particular chosen hadronic scale.

At present most applications in QCD use Schr\"odinger 
functional boundary conditions~\cite{Luscher:1992an,Sint:1993un}
and we discuss this below in a little more detail.
(However, other boundary conditions are also possible, such as
twisted periodic boundary conditions for the gauge fields 
and the discussion also applies to them.)
An important reason is that these boundary conditions avoid zero modes
for the quark fields and quartic modes \cite{Coste:1985mn} in the
perturbative expansion in the gauge fields. Furthermore the corresponding
renormalization scheme is well studied in perturbation
theory~\cite{Luscher:1993gh,Sint:1995ch,Bode:1999sm} with the
3-loop $\beta$-function and 2-loop cutoff effects (for the
standard Wilson regularization) known.

In order to have a perturbatively well-defined scheme,
the SF scheme uses Dirichlet boundary conditions at time 
$t = 0$ and $t = T$. These break translation invariance and permit
${\cO}(a)$ counter terms at the boundary through quantum corrections. 
Therefore, the leading discretization error is ${\cO}(a)$.
Improving the lattice action is achieved by adding
counter terms at the boundaries whose coefficients are denoted
as $c_t,\tilde c_t$. In practice, these coefficients are computed
with $1$-loop or $2$-loop perturbative accuracy.
A better precision in this step yields a better 
control over discretization errors, which is important, as can be
seen, e.g., in Refs.~\cite{Takeda:2004xha,Necco:2001xg}.

Also computations with Dirichlet boundary conditions do in principle
suffer from the insufficient change of topology in the HMC algorithm
at small lattice spacing. However, in a small volume the weight of
nonzero charge sectors in the path integral is exponentially
suppressed~\cite{Luscher:1981zf} and in a Monte Carlo run of
  typical length very few configurations
  with nontrivial topology should appear.\footnote{We simplify here and assume
  that the classical solution associated with the used boundary
  conditions has charge zero.  In practice this is the case.} Considering
the issue quantitatively Ref.~\cite{Fritzsch:2013yxa} finds a
strong suppression below $L\approx 0.8\,\fm$. Therefore the lack of
topology change of the HMC is not a serious issue for the high-energy regime
in step-scaling studies. However, the matching to hadronic observables
requires volumes where the problem cannot be ignored. Therefore,
Ref.~\cite{DallaBrida:2016kgh} includes a projection to zero topology 
into the {\em definition} of the coupling. 
A very interesting comparison of the step-scaling approach for a $(Q=0)$-projected coupling
and its unprojected version was recently carried out in Ref.~\cite{Bonanno:2024nba}, with $\Nf=0$ and
twisted periodic boundary conditions for the gauge field.
A new parallel-tempering approach to relate systems with different 
boundary conditions was used. The results validate the $Q=0$ approach, in that step scaling
in large volume (where contributions from $Q\ne 0$ configurations are sizeable)
leads, within errors, to indistinguishable results, once the couplings are properly matched.
We note also that a mix of Dirichlet and open boundary conditions is
expected to remove the topology issue entirely \cite{Luscher:2014kea}
and may be considered in the future.

Apart from the boundary conditions, the very definition
of the coupling needs to be chosen. 
We briefly discuss in turn, the two schemes used at present, 
namely, the `Schr\"odinger
Functional' (SF) and `Gradient-Flow' (GF) schemes.

The SF scheme is the first one, which was used in step-scaling studies
in gauge theories \cite{Luscher:1992an}. Inhomogeneous
Dirichlet boundary conditions are imposed in time,
\begin{eqnarray}
    A_k(x)|_{x_0=0} = C_k\,,
    \quad
    A_k(x)|_{x_0=L} = C_k'\,,    
\end{eqnarray}
for $k=1,2,3$.
Periodic boundary conditions (up to a phase for the fermion fields)  with period $L$ are imposed in space.
The matrices 
\begin{align}
LC_k &= i \,{\rm diag}\big( \eta- \pi/3, -\eta/2 , -\eta/2  + \pi/3 \big) \,,
\nonumber \\
LC^\prime_k &= i \,{\rm diag}\big( -(\eta+\pi), \eta/2 + \pi/3,\eta/2 + 2\pi/3 \big)\,,
\nonumber
\end{align}
just depend on the dimensionless parameter $\eta$.
The coupling $\bar{g}_\mathrm{SF}$ is obtained from
the $\eta$-derivative of the effective action,
\begin{eqnarray}
  \langle \partial_\eta S|_{\eta=0} \rangle = \frac{12\pi}{\gbar^2_\mathrm{SF}}\,.
\end{eqnarray}
For this scheme, the finite $c^{(i)}_g$, \eq{eq:g_conversion}, are 
known for $i=1,2$ 
\cite{Sint:1995ch,Bode:1999sm}.

More recently, gradient-flow couplings have been used frequently
because of their small statistical errors at large couplings (in contrast to 
$\gbar_\mathrm{SF}$, which has small statistical errors at small couplings). 
The gradient flow is introduced as follows \cite{Narayanan:2006rf,Luscher:2010iy}.
Consider the flow gauge field $B_\mu(t,x)$ with the flow time $t$, 
which is a one-parameter deformation of the bare gauge field 
$A_\mu(x)$, where $B_\mu(t,x)$ is the solution to the 
gradient-flow equation
\begin{eqnarray}
   \partial_t B_\mu(t,x) 
            &=& D_\nu G_{\nu\mu}(t,x)\,,
                                                      \nonumber \\
   G_{\mu\nu} &=& \partial_\mu B_\nu - \partial_\nu B_\mu + [B_\mu,B_\nu] \,,
   \label{eq:def-GF}
\end{eqnarray}
with initial condition $B_\mu(0,x) = A_\mu(x)$.
The renormalized coupling is defined by \cite{Luscher:2010iy}
\begin{eqnarray}
   \bar{g}^2_{\rm GF}(\mu) 
      = \left. {\cal N} t^2 \langle E(t,x)\rangle
                                        \right|_{\mu=1/\sqrt{8t}} \,,
\end{eqnarray}
with ${\cal N} = 16\pi^2/3 + \cO((a/L)^2)$
and where $E(t,x)$ is the action density given by
\begin{eqnarray}
   E(t,x) = \frac{1}{4} G^a_{\mu\nu}(t,x) G^a_{\mu\nu}(t,x). 
                                        \label{eq:Et}
\end{eqnarray}
In a finite volume, one needs to specify additional conditions.
In order not to introduce two independent scales one sets 
\begin{eqnarray}
   \sqrt{8t} = cL \,,
\end{eqnarray}
for some fixed number $c$ \cite{Fodor:2012td}. 
Schr\"odinger functional boundary conditions~\cite{Fritzsch:2013je}
or twisted periodic boundary conditions \cite{Ramos:2014kla,Ishikawa:2017xam,Bribian:2021cmg}  
have been employed.
Matching of the GF coupling to the $\overline{\rm MS}$-scheme coupling
is known to 1-loop for twisted boundary conditions with zero
quark flavours and SU(3) group \cite{Ishikawa:2017xam} and to 2-loop with SF boundary conditions with zero
quark flavours \cite{DallaBrida:2017tru}.
The former is based on a MC evaluation at small couplings and the 
latter on numerical stochastic perturbation theory.\footnote{For a variant of the 
twisted periodic finite volume scheme the 1-loop matching has been computed analytically~\cite{Bribian:2019ybc}.}



\subsubsection{Discussion of computations}


In Tab.~\ref{tab_SF3} we give results from various determinations
\begin{table}[!htb]
   \vspace{3.0cm}
   \footnotesize
   \begin{tabular*}{\textwidth}[l]{l@{\extracolsep{\fill}}rlllllllll}
      Collaboration & Ref. & $\Nf$ &
      \hspace{0.15cm}\begin{rotate}{60}{publication status}\end{rotate}
                                                       \hspace{-0.15cm} &
      \hspace{0.15cm}\begin{rotate}{60}{renormalization scale}\end{rotate}
                                                       \hspace{-0.15cm} &
      \hspace{0.15cm}\begin{rotate}{60}{perturbative behaviour}\end{rotate}
                                                       \hspace{-0.15cm} &
      \hspace{0.15cm}\begin{rotate}{60}{continuum extrapolation}\end{rotate}
                               \hspace{-0.25cm} & 
                         scale & $\Lambda_\msbar[\MeV]$ & $r_0\Lambda_\msbar$ \\
      & & & & & & & & \\[-0.1cm]
      \hline
      \hline
      & & & & & & & & \\
      ALPHA 10A & \cite{Tekin:2010mm} & 4 
                    & \gA &\good & \good & \good 
                    & \multicolumn{3}{l}{only running of $\alpha_s$ in Fig.~4}
                    \\  
      Perez 10 & \cite{PerezRubio:2010ke} & 4 
                    & \rC &\good & \good & \soso  
                    & \multicolumn{3}{l}{only step-scaling function in Fig.~4}
                    \\           
      & & & & & & & & & \\[-0.1cm]
      \hline
      & & & & & & & & & \\[-0.1cm]
      ALPHA 17   &  \cite{Bruno:2017gxd} &2+1 
                    & \gA & \good & \good & \good 
                    & $\sqrt{8t_0}= 0.415\,\mbox{fm}$ & 341(12) & 0.816(29)
                    \\  
      PACS-CS 09A& \cite{Aoki:2009tf} & 2+1 
                    & \gA &\good &\good &\soso
                    & $m_\rho$ & $371(13)(8)(^{+0}_{-27})$$^{\#}$
                    & $0.888(30)(18)(^{+0}_{-65})$$^\dagger$
                    \\ 
                    &&&\gA &\good &\good &\soso 
                    & $m_\rho$  & $345(59)$$^{\#\#}$
                    & $0.824(141)$$^\dagger$
                    \\ 
      & & & & & & & & \\[-0.1cm]
      \hline  \\[-1.0ex]
      & & & & & & & & \\[-0.1cm]
      ALPHA 12$^*$  & \cite{Fritzsch:2012wq} & 2 
                    & \gA &\good &\good &\good
                    &  $f_{\rm K}$ & $310(20)$ &  $0.789(52)$
                    \\
      ALPHA 04 & \cite{DellaMorte:2004bc} & 2 
                    & \gA &\bad &\good &\good
                    & $r_0 = 0.5\,\mbox{fm}$$^\S$  & $245(16)(16)^\S$ 
                                                   & $0.62(2)(2)^\S$
                    \\
      ALPHA 01A & \cite{Bode:2001jv} & 2 
                    &\gA & \good & \good & \good 
                    &\multicolumn{3}{l}{only running of $\alpha_s$  in Fig.~5}
                    \\
      & & & & & & & & \\[-0.1cm]
      \hline  \\[-1.0ex]
      & & & & & & & & \\[-0.1cm]
       Bribian~21   & \cite{Bribian:2021cmg} & 0
                    & \gA & \good & \good & \good
                    & $r_0=0.5\fm$ & 249.4(8.0)  & 0.632(20) 
                    \\
      Nada 20    & \cite{Nada:2020jay} & 0 
                    & \gA & \good & \good & \good
                    &\multicolumn{3}{l}{consistency checks for \cite{DallaBrida:2019wur}, same gauge configurations}
                    \\
 Dalla Brida 19 & \cite{DallaBrida:2019wur} & 0 
                    & \gA & \good & \good & \good
                    & $r_0=0.5\fm$ & 260.5(4.4)  & 0.660(11) 
                    \\
      Ishikawa 17   & \cite{Ishikawa:2017xam} & 0 
                    & \gA & \good & \good & \good
                    & $r_0$, $[\sqrt{\sigma}]$ & $253(4)(^{+13}_{-2})$$^\dagger$
                                              & $0.606(9)(^{+31}_{-5})^+$
                    \\
      CP-PACS 04$^\&$  & \cite{Takeda:2004xha} & 0 
                    & \gA & \good & \good & \soso  
                    & \multicolumn{3}{l}{only tables of $g^2_{\rm SF}$}
                    \\
      ALPHA 98$^{\dagger\dagger}$ & \cite{Capitani:1998mq} & 0 
                    & \gA & \good & \good & \soso 
                    &  $r_0=0.5\fm$ & $238(19)$ & 0.602(48) 
                    \\
      L\"uscher 93  & \cite{Luscher:1993gh} & 0 
                    & \gA & \good & \soso & \soso
                    & $r_0=0.5\fm$ & 233(23)  & 0.590(60)$^{\S\S}$ 
                    \\
      &&&&&&& \\[-0.1cm]
      \hline
      \hline\\
\end{tabular*}\\[-0.2cm]
\begin{minipage}{\linewidth}
{\footnotesize 
\begin{itemize}
\item[$^{\#}$] Result with a constant (in $a$) continuum extrapolation
              of the combination $L_\mathrm{max}m_\rho$.             \\[-5mm]
\item[$^\dagger$] In conversion from $\Lambda_\msbar$ to
                 $r_0\Lambda_{\overline{\rm MS}}$ and vice versa, $r_0$ is
                 taken to be $0.472\,\mbox{fm}$.                   \\[-5mm]
\item[$^{\#\#}$] Result with a linear continuum extrapolation
             in $a$ of the combination $L_\mathrm{max}m_\rho$.        \\[-5mm]
\item[$^*$]  Supersedes ALPHA 04.                                   \\[-5mm]
\item[$^\S$] The $\Nf=2$ results were based on values for $r_0/a$
             which have later been found to be too small by
             \cite{Fritzsch:2012wq}. The effect will be of the order of
             10--15\%, presumably an increase in $\Lambda r_0$.
             We have taken this into account by a $\bad$ in the 
             renormalization scale.                                  \\[-5mm]
\item[$^\&$] This investigation was a precursor for PACS-CS 09A
          and confirmed two step-scaling functions as well as the
          scale setting of ALPHA~98.                              \\[-5mm]
\item[$^{\dagger\dagger}$] Uses data of L\"uscher~93 and therefore supersedes it.
                                                                  \\[-5mm]
\item[$^{\S\S}$] Converted from $\alpha_\msbar(37r_0^{-1})=0.1108(25)$.
\item[$^+$] Also $\Lambda_\msbar/\sqrt{\sigma} = 0.532(8)(^{+27}_{-5})$ is quoted.

\end{itemize}
}
\end{minipage}
\caption{Results for the $\Lambda$-parameter from computations using 
         step scaling of the SF-coupling. Entries without values for $\Lambda$
         computed the running and established perturbative behaviour
         at large $\mu$. 
         }
\label{tab_SF3}
\end{table}
of the $\Lambda$-parameter. For a clear assessment of the $\Nf$-dependence, the last column also shows results that refer to a common
hadronic scale, $r_0$. As discussed above, the renormalization scale
can be chosen large enough such that $\alpha_s < 0.2$ and the
perturbative behaviour can be verified.  Consequently only $\good$ is
present for these criteria except for early work
where the $n_{\mathrm{l}}=2$ loop correction to $\msbar$ was not yet known and we assigned a $\bad$ concerning the renormalization scale.
With dynamical fermions, results for the
step-scaling functions are always available for at least $a/L = \mu a
=1/4,1/6, 1/8$.  All calculations have a nonperturbatively
$\cO(a)$ improved action in the bulk. For the discussed
boundary $\cO(a)$ terms this is not so. In most recent
calculations 2-loop $\cO(a)$ improvement is employed together
with at least three lattice spacings.\footnote{With 2-loop
  $\cO(a)$ improvement we here mean $c_t$ including
  the $g_0^4$ term and $\tilde c_\mathrm{t}$ with the $g_0^2$
  term. For gluonic observables such as the running coupling this is
  sufficient for cutoff effects being suppressed to $\cO(g^6
  a)$.} This means a \good\ for the continuum extrapolation.  In 
other computations only 1-loop $c_t$ was available and we arrive at \soso. We
note that the discretization errors in the step-scaling functions 
of the SF coupling are
usually found to be very small, at the percent level or
below. However, the overall desired precision is very high as well,
and the results in CP-PACS 04~\cite{Takeda:2004xha} show that
discretization errors at the below percent level cannot be taken for
granted.  In particular with staggered fermions (unimproved except for
boundary terms) few-percent effects are seen in
Perez~10~\cite{PerezRubio:2010ke}.

In the work by PACS-CS 09A~\cite{Aoki:2009tf}, the continuum
extrapolation in the scale setting is performed using a constant
function in $a$ and with a linear function.
Potentially the former leaves a considerable residual discretization 
error. We here use, as discussed with the collaboration, 
the continuum extrapolation linear in $a$,
as given in the second line of PACS-CS 09A \cite{Aoki:2009tf}
results in Tab.~\ref{tab_SF3}.
After perturbative conversion from a three-flavour result to five flavours
(see \sect{s:crit}), they obtain
\begin{eqnarray}
 \alpha_\msbar^{(5)}(M_Z)=0.118(3)\,. 
\end{eqnarray}

In Ref.~\cite{Bruno:2017gxd}, the ALPHA collaboration determined 
$\Lambda^{(3)}_{\msbar}$ combining step scaling in $\gbar^2_{\rm GF}$
in the lower-scale region $\mu_{\rm had} \leq \mu \leq \mu_0$, and 
step scaling in $\gbar^2_{\rm SF}$ for higher scales  
$\mu_0 \leq \mu \leq \mu_{\rm PT}$. 
Both schemes are defined with SF boundary conditions. For $\gbar^2_{\rm GF}$ a projection to the sector of zero 
topological charge is included, \eq{eq:Et} is restricted to the 
magnetic components, and $c=0.3$.
The scales $\mu_{\rm had}$, $\mu_0$, and 
$\mu_{\rm PT}$ are defined by $\gbar^2_{\rm GF} (\mu_{\rm had})= 11.3$,
$\gbar^2_{\rm SF}(\mu_0) = 2.012$, and $\mu_{\rm PT} = 16 \mu_0$ which
are roughly estimated as
\begin{eqnarray}
   1/L_\mathrm{max}\equiv \mu_{\rm had} \approx 0.2 \mbox{ GeV}, & \mu_0 \approx 4 \mbox{ GeV} \,, 
      & \mu_{\rm PT}\approx 70 \mbox{ GeV} \,.
\end{eqnarray}
Step scaling is carried out with an $\cO(a)$-improved Wilson quark action
\cite{Bulava:2013cta}
and L\"uscher-Weisz gauge action \cite{Luscher:1984xn} in the low-scale region
and an $\cO(a)$-improved Wilson quark action
\cite{Yamada:2004ja}
and Wilson gauge action in the high-energy part. 
For the step scaling using steps of
$L/a \,\to\,2L/a$, three lattice sizes $L/a=8,12,16$ were simulated for
$\gbar^2_{\rm GF}$ and four lattice sizes $L/a=(4,)\, 6, 8, 12$ for 
$\gbar^2_{\rm SF}$. The final results do not use the small lattices given
in parenthesis. The parameter $\Lambda^{(3)}_{\msbar}$ is then obtained via 
\begin{eqnarray}
   \Lambda^{(3)}_{\msbar} 
      = \underbrace{\frac{\Lambda^{(3)}_{\msbar}}{\mu_{\rm PT}}}_{\rm perturbation  ~ theory}
           \times \underbrace{\frac{\mu_{\rm PT}}{\mu_{\rm had}}}_{\rm step scaling}
           \times \underbrace{\frac{\mu_{\rm had}}{f_{\pi K}}}_{\rm large ~ volume~ simulation}
           \times \underbrace{f_{\pi K}}_{\rm experimental ~data} \,, 
\label{eq:Lambda3}
\end{eqnarray}
where the hadronic scale $f_{\pi K}$ is 
$f_{\pi K}= \frac{1}{3}(2 f_K + f_\pi) = 147.6 (5)\mbox{ MeV}$.
The first factor on the right-hand side of Eq.~(\ref{eq:Lambda3}) is 
obtained from $\alpha_{\rm SF}(\mu_{\rm PT})$ which is the output from
SF step scaling using Eq.~(\ref{eq:Lambda}) with 
$\alpha_{\rm SF}(\mu_{\rm PT})\approx 0.1$ and
the 3-loop $\beta$-function 
and the exact conversion to the $\msbar$-scheme.
The second factor is essentially obtained
from step scaling in the GF scheme and the measurement of 
$\gbar^2_{\rm SF}(\mu_0)$ (except for the trivial scaling factor of 16 
in the SF running). The third factor is obtained from a measurement
of the hadronic quantity at large volume.

A large-volume simulation is done for three lattice spacings with 
sufficiently large volume and reasonable control over the chiral
extrapolation so that the scale determination is precise enough.  
The step scaling results in both schemes
satisfy renormalization criteria, perturbation theory criteria,
and continuum-limit criteria just as previous studies using step scaling.
So we assign green stars for these criteria.

The dependence of $\Lambda$, Eq.~(\ref{eq:Lambda}) with 3-loop $\beta$-function, on $\alpha_s$ and on the chosen scheme is discussed
in \cite{Brida:2016flw}. This investigation provides a warning on estimating the 
truncation error of perturbative series. Details are explained in \sect{s:trunc}.

The result for the $\Lambda$-parameter is  
$\Lambda^{(3)}_{\overline{\rm MS}} = 341(12)~\mbox{MeV}$, 
where the dominant error comes from the error of 
$\alpha_{\rm SF}(\mu_{\rm PT})$ after step scaling in the SF scheme.
Using 4-loop matching at the charm and bottom thresholds 
and 5-loop running one finally obtains
\begin{eqnarray}
   \alpha^{(5)}_{\overline{\rm MS}}(M_Z) = 0.11852(84)\,.
\end{eqnarray}
Several other results do not have a sufficient number of
quark flavours  or do not yet contain the conversion
of the scale to physical units (ALPHA~10A \cite{Tekin:2010mm}, 
Perez~10 \cite{PerezRubio:2010ke}). Thus no value for $\alpha_\msbar^{(5)}(M_Z)$
is quoted.

The computation of Ishikawa et al.~\cite{Ishikawa:2017xam} 
is based on the gradient-flow coupling with twisted boundary conditions
\cite{Ramos:2014kla} (TGF coupling)
in the pure gauge theory. Again they use 
$c=0.3$. Step scaling with a scale factor $s=3/2$ is employed,
covering a large range of couplings from $\alpha_s\approx 0.5$ to
$\alpha_s\approx 0.1$ and taking the continuum limit through global
fits to the step-scaling function on $L/a=12,16,18$ lattices with between 6 and 
8 parameters. Systematic errors due to variations of the fit functions
are estimated. Two physical scales are considered:
$r_0/a$ is taken from \cite{Necco:2001xg} and $\sigma a^2$ from 
\cite{Allton:2008pn} and \cite{GonzalezArroyo:2012fx}.  
As the ratio $\Lambda_\mathrm{TGF}/\Lambda_\mathrm{\msbar}$   
has not yet been computed analytically, Ref.~\cite{Ishikawa:2017xam}
determines the 1-loop relation between $\gbar_\mathrm{SF}$ and 
$\gbar_\mathrm{TGF}$ from  MC simulations performed
in the weak coupling region and then uses the known
$\Lambda_\mathrm{SF}/\Lambda_\mathrm{\msbar}$. Systematic errors 
due to variations of the fit functions dominate the overall uncertainty.
\par
Two extensive $\Nf=0$ step-scaling studies have been carried out in Dalla Brida 19~\cite{DallaBrida:2019wur}
and by Nada and Ramos~\cite{Nada:2020jay}. They use different strategies for the running from mid to high energies, 
but use the same gauge configurations and share the running at low energies and matching to the hadronic scales.
These results are therefore correlated. However, given the comparatively high value for $r_0\lms$, 
it is re-assuring that these conceptually different approaches yield perfectly compatible results within errors of
similar size of around 1.5\% for $\sqrt{8t_0}\lms=0.6227(98)$, or, alternatively $r_0\lms = 0.660(11)$.

In Dalla Brida 19 \cite{DallaBrida:2019wur} two GF-coupling definitions with SF-boundary conditions are considered, corresponding to (colour-) magnetic and
electric components of the action density respectively. The coupling definitions include the projection to 
$Q=0$, as was also done in~\cite{Bruno:2017gxd}. The flow-time parameter is set to $c=0.3$, 
and both Zeuthen and Wilson flow are measured. Lattice sizes
range from $L/a=8$ to $L/a=48$, covering up to a factor of 3 in lattice spacings 
for the step-scaling function, where both $L/a$ and $2L/a$ are needed. Lattice effects in the 
step-scaling function are visible but can be extrapolated using global fits with $a^2$ errors.
Some remnant $\cO(a)$ effects from the boundaries are expected, as their perturbative
cancellation is incomplete. These $\cO(a)$ contaminations are treated as a systematic error on the data, 
following \cite{Bruno:2017gxd}, and are found to be subdominant. An intermediate reference scale $\mu_\mathrm{ref}$ is defined where $\alpha=0.2$,
and the scales above and below are analyzed separately. Again this is similar to \cite{Bruno:2017gxd}, except
that here GF-coupling data is available also at high energy scales.
The GF $\beta$-functions are then obtained by fitting to the continuum extrapolated data 
for the step-scaling functions. In addition, a nonperturbative
matching to the standard SF coupling is performed above $\mu_\mathrm{ref}$ 
for a range of couplings covering a factor of 2. The nonperturbative $\beta$-function for the SF scheme 
can thus be inferred from the GF $\beta$-function. It turns out that GF schemes are very slow 
to reach the perturbative regime.  Particularly the $\Lambda$-parameter for the magnetic GF coupling
shows a large slope in $\alpha^2$, which is the parametric uncertainty with known 3-loop $\beta$-function.
Also, convincing contact with the 3-loop $\beta$-function is barely seen down to $\alpha = 0.08$.
This is likely to be related to the rather large 3-loop $\beta$-function coefficients, 
especially for the magnetic GF scheme~\cite{DallaBrida:2017tru}.
In contrast, once the GF couplings are matched nonperturbatively to the SF scheme the contact to 
perturbative running can be safely made. It is also re-assuring that in all
cases the extrapolations (linear in $\alpha^2$) to $\alpha=0$ for the $\Lambda$-parameters agree very well,
and the authors argue in favour of such extrapolations. Their data confirms that this procedure
yields consistent results with the SF scheme for $\nu=0$, where such an extrapolation is not required.

The low-energy regime between $\mu_\mathrm{ref}$ and a hadronic scale $\mu_\mathrm{had}$ is
covered again using the nonperturbative step-scaling function and the derived $\beta$-function.
Finally, contact between $\mu_\mathrm{had}$ and hadronic scales $t_0$ and $r_0$ is established using five lattice spacings
covering a factor up to 2.7. The multitude of cross checks of both continuum limit
and perturbative truncation errors make this a study which passes all current FLAG criteria by some margin.
The comparatively high value for $r_0\lms$ found in this study must therefore be taken very seriously.

In Nada 20 \cite{Nada:2020jay}, Nada and Ramos provide further consistency checks of \cite{DallaBrida:2019wur}
for scales larger than $\mu_\mathrm{ref}$. The step-scaling function for $c=0.2$ is constructed in two steps, by determining
first the relation between couplings for $c=0.2$ and $c=0.4$ at the same $L$ and
then increasing $L$ to $2L$ keeping the flow time fixed (in units of the lattice spacing), 
so that one arrives again at $c=0.2$ on the $2L$ volume.
The authors demonstrate  that the direct construction of the step-scaling function for $c=0.2$ would require
much larger lattices in order to control the continuum limit at the same level of precision. 
The consistency with \cite{DallaBrida:2019wur} for the $\Lambda$-parameter is therefore a highly nontrivial check on the systematic effects of
the continuum extrapolations. The study obtains results for the
$\Lambda$-parameter (again extrapolating to $\alpha=0$) with a similar error as in \cite{DallaBrida:2019wur} using the low-energy running and matching to the hadronic scale from that reference.
For this reason and since gauge configurations are shared between both papers, 
these results are not independent of \cite{DallaBrida:2019wur},
so Dalla Brida 19 will be taken as representative for both works.

Since FLAG 21 a new step-scaling result with $\Nf=0$ has appeared in Bribian~21~\cite{Bribian:2021cmg}. It uses the gradient
flow in a volume with twisted periodic boundary conditions for the gauge field. The volume has two shorter directions by a factor of 3; 
however, a re-interpretation as a symmetric physical volume is possible using internal degrees of freedom of the gauge field.
This is a state-of-the-art step-scaling result, the main problem being the poor perturbative behaviour
of the gradient-flow coupling. Since the 3-loop $\beta$-function is not known, the parametric uncertainty
in estimates of the $\Lambda$-parameter is of O($\alpha$) and is quite large. 
The problem is by-passed by matching nonperturbatively to the SF scheme, which leads to stable estimates vs.~$\alpha^2$,
and the result is $\sqrt{8t_0}\Lambda_\msbar = 0.603(17)$, or, in units of the Sommer scale,  $r_0\Lambda_\msbar = 0.632(20)$.
All FLAG criteria are passed with \good, and the data-driven criterion for the continuum limit is irrelevant in this case.

\paragraph{Scale variations.} 
With a perturbative matching at $\mu\approx 80~\mathrm{GeV}$, we have computed the change in the 
determination of 
$\alpha_{\overline{\mathrm{MS}}}(M_Z)$ under scale variations as explained above. 
The systematic errors obtained from scale variations are
\begin{eqnarray}
   \delta^*_{(4)} = 0.1\%\, , \quad 
   \delta_{(2)} = 0.2\%\, \quad
   \delta^*_{(2)} = 0.2\%\, .
\end{eqnarray}
Because the perturbative matching is performed at a high-energy scale, the systematic error obtained from scale 
variations is negligible.


%% file: Alpha_s/s0.tex
\subsection{The decoupling method}

\label{s:dec}

\newcommand{\mudec}{\mu_\text{dec}}

The ALPHA collaboration  has proposed and pursued a
new strategy to compute the $\Lambda$ parameter in QCD with $\Nf \ge 3$ flavours 
based on the simultaneous decoupling of $\Nf\ge 3$ heavy quarks with RGI mass $M$~\cite{DallaBrida:2019mqg}.
We refer to \cite{DallaBrida:2020pag} for a pedagogical introduction. 
Generically, for large quark mass $M$, a running coupling in a mass-dependent renormalization scheme
\begin{equation}
\label{eq:dec-start}
   \gbar^2(\mu, M)^{(\Nf)} = \gbar^2(\mu)^{(\Nf=0)} + O\left(1/M^{k}\right)
\end{equation}
can be represented by the corresponding $\Nf=0$ coupling, up to power corrections in
$1/M$. The leading power is usually $k=2$, however renormalization schemes 
in finite volume may have $k=1$, depending on the set-up. For example, 
this is the case with standard SF or open boundary conditions in combination 
with a standard mass term. In practice such boundary
contributions can be made numerically small by a careful choice of the scheme's parameters.
In principle, power corrections can be either $(\mu/M)^k$ or $(\Lambda/M)^k$.
Fixing $\mu=\mudec$, e.g., by prescribing a value for the mass-independent coupling,
such that $\mudec/\Lambda = \cO(1)$ thus helps to reduce the need for very large $M$. 
Defining  $\gbar^2(\mudec,M) = u_M$ at fixed $\gbar^2(\mudec,M=0)$,
Eq.~(\ref{eq:dec-start}) translates to a relation 
between $\Lambda$-parameters, which can be cast in the form,
\begin{eqnarray}
 \frac{\lms^{(\Nf)}}{\mudec}\;P\left(\frac{M}\mudec \frac{\mudec}{\lms^{(\Nf)}}\right) = 
 \frac{\lms^{(0)}}{\Lambda_s^{(0)}}\, \varphi_s^{(\Nf=0)}\left( \sqrt{u_\mathrm{M}}\right) + \cO(M^{-k})\,, \nonumber \\[-2ex]
\label{eq:basic}
\end{eqnarray}
with the function $\varphi_s$  as defined in Eq.~(\ref{eq:Lambda}), for scheme $s$ and $\Nf=0$.
A crucial observation is that the function $P$, which gives the ratios of $\Lambda$-parameters $\lms^{(0)}/\lms^{(\Nf)}$,
can be evaluated perturbatively to a very good approximation~\cite{Bruno:2014ufa,Athenodorou:2018wpk}.
Equation~(\ref{eq:dec-start}) also implies a relation between the couplings in mass-independent schemes, 
in the theories with $\Nf$ and zero flavours, respectively. In the $\msbar$ scheme
this relation is analogous to Eq.~(\ref{e:grelation}),
\begin{eqnarray}
 \left[\gbar^{(\Nf=0)}_\msbar(m_\star)\right]^2 
 =  \left[\gbar^{(\Nf)}_\msbar(m_\star)\right]^2\times 
 C\left(\gbar^{(\Nf)}_\msbar(m_\star)\right),
\end{eqnarray}
where the evaluation of the coupling is done at the scale $m_\star=m^{(\Nf)}_\msbar(\mu=m_\star)$.
This removes the leading 1-loop correction of O($g^2$) in the expansion of the function, $C(g) = 1 +  c_2g^4 +\rmO(g^6)$, 
which is known up to 4-loop order~\cite{Grozin:2011nk,Chetyrkin:2005ia,Schroder:2005hy,Kniehl:2006bg,Gerlach:2018hen}. 
The mass scale $m_\star$ is in one-to-one correspondence with the RGI mass $M$, and $g^\star(y)= \gbar_\msbar^{(\Nf)}(m_\star)$
can thus be considered  a function of $y\equiv M^{(\Nf)}/\lms^{(\Nf)}$.
The function $P(y)$ can be evaluated perturbatively in the $\msbar$ scheme, as the ratio,
\begin{equation}
 P(y) = \dfrac{\varphi_\msbar^{(\Nf=0)}\left(g^\star(y) \sqrt{C(g^\star(y))}\right) }
               {\varphi_\msbar^{(\Nf)}(g^\star(y))}\,.
\end{equation}
Note that perturbation theory is only required at the scale set by the heavy-quark mass, which works better 
the larger $M$ can be chosen. 
Given the function $P(y)$, the LHS of Eq.~(\ref{eq:basic}) can be inferred from a $\Nf=0$ computation of the RHS in 
the scheme $s$, if the argument $\sqrt{u_M}$ of $\varphi_s^{(0)}$ is known (and the ratio $\lms/\Lambda_s$ for the scheme $s$).
The main challenge then consists in the computation of the mass-dependent coupling $u_M$ for large masses.


\subsubsection{Discussion of computations}

\begin{figure}[!htb]\hspace{-2cm}\begin{center}
      \includegraphics[width=13.5cm]{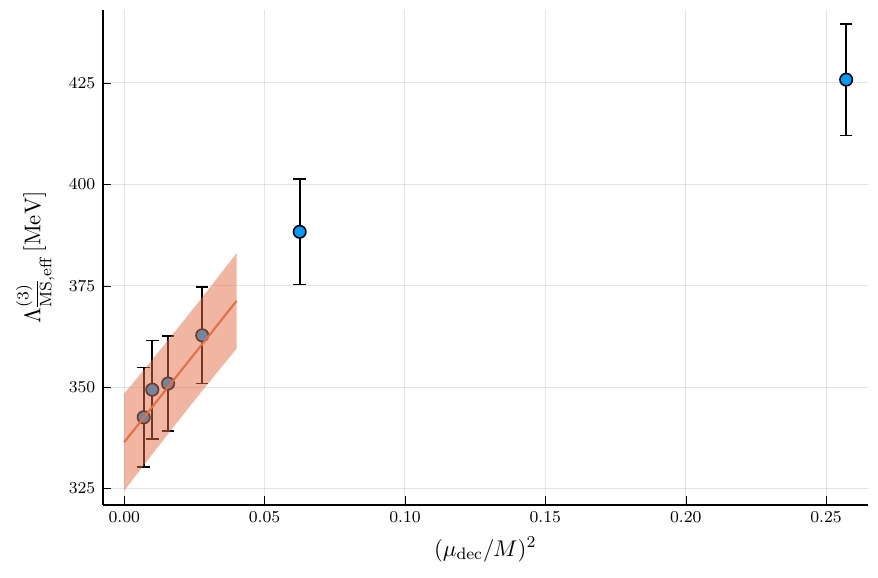}
      \end{center}
      \label{fig:decpl}
\vspace{-0.5cm}
\caption{The decoupling limit $M\rightarrow \infty$ in ALPHA~22, Ref.~\cite{DallaBrida:2022eua}.}
\label{fig:decoup}
\end{figure}

To put the decoupling strategy to work, ALPHA~22~\cite{DallaBrida:2022eua} uses $\Nf=3$, 
so that information from \cite{Bruno:2017gxd} can be used. Using the massless GF coupling in finite
volume from this project, $\mudec$ is defined through $\gbar^2_\text{GF}(\mudec) = 3.949$, and thus known in physical
units, $\mudec = 789(15)\MeV$. Imposing this condition for lattice sizes between $L/a=12$ to $L/a=48$,
a corresponding sequence of $\beta$-values between $4.302$ and $5.174$ is obtained (the lattice action is the same as 
used by CLS, there for much coarser lattice spacings at $\beta\le 3.85$). 
Using the available information on nonperturbative mass renormalization~\cite{Campos:2018ahf}, 
six values for the O($a$)-improved RGI quark masses are considered 
at each of these $\beta$-values, such that the ratios $z=M/\mudec$ are close to 2, 4, 6, 8, 10, and 12.
While great care is taken to implement nonperturbative O($a$) improvement, there is only perturbative 1-loop
information on $b_\text{g}$, which parameterizes a mass-dependent rescaling of the bare coupling,
$$
  \tilde{g}_0^2 = g_0^2(1+b_\text{g}(g_0)a\mq),\qquad b_\text{g}(g_0) = 0.012\times\Nf g_0^2 + \rmO(g_0^4).
$$
Here, $\mq$ denotes the subtracted bare quark mass, related to $M$ by a renormalization factor of $O(1)$
at the relevant lattice spacings.
Consistent O($a$) improvement requires that $\tilde{\beta}=6/\tilde{g}_0^2$ be kept fixed as the
quark mass is varied. The authors of ALPHA~22 here assume a $100\%$ uncertainty 
of the perturbative $b_\text{g}$-estimate, which is treated as a systematic error (cf.~below). 
At the  chosen quark-mass parameters, the GF coupling with doubled time extent, $T=2L$, is
measured. This GFT coupling is used in order to minimize effects from the time boundaries, which introduce
linear effects in $1/M$ in the decoupling relation, and also residual lattice effects linear in $a$.
Both of these effects are monitored and found to be negligible.
The continuum limit is then taken, either separately for each $z$-value, or using a global fit to all 
$z$-values $z>2$, which turns out too small to be useful in  the large-$M$ limit (cf.~Fig.~\ref{fig:decpl}).
The lattice effects are fitted
to O($a^2$), including an $[\alpha_\msbar(1/a)]^{\hat\Gamma}$ term, as expected from Symanzik's effective theory
with RG improvement~\cite{Balog:2009np,Balog:2009yj,Husung:2019ytz,Husung:2021mfl,Husung:2022kvi}. The global fit uses the combined arguments from heavy-quark and Symanzik 
effective theories to separate the leading $(aM)^2$ effects with yet another logarithmic correction term.
Cuts in the data are considered for $(aM)^2 < 0.25$ and $(aM)^2<0.16$.
The continuum-extrapolated values include a systematic error due to the uncertainty in $b_\text{g}$.
 The fits are repeated for different choices of
$\hat\Gamma$ and $\hat\Gamma'$ in intervals constrained by the effective heavy-quark and Symanzik theories,
and the variation is used as an estimate of systematic effects due to the possible presence of 
such non-power-like cutoff effects.
The continuum extrapolated GFT coupling defines the starting point for the $\Nf=0$ running. Before
the GF running can be used, a matching from the GFT to GF scheme is done to high precision in the $\Nf=0$ theory.
The running in $\Nf=0$ is taken from Dalla Brida 19~\cite{DallaBrida:2019wur} and the results are then
inserted into the Eq.~(\ref{eq:basic}), for each of the available $M$-values. This defines ``effective''
$\Lambda$-parameters, equal to the asymptotic value up to $1/M^2$ effects. 
Taking the $z\rightarrow \infty$ limit (again allowing for a logarithmic correction with exponent $\Gamma_m$) 
then yields the final result, with the scale set using $\sqrt{t_0}$ from Ref.~\cite{Bruno:2016plf},
\begin{equation}
     \lms^{(3)} = 336(10)(6)_{b_\text{g}} (3)_{\Gamma_m} \,\MeV = 336(12)\, \MeV
\end{equation}
which translates to $\alpha_s(m_Z) = 0.11823(84)$.
Despite some common elements with ALPHA 17, the authors emphasize that the decoupling method is largely independent, with the 
overlap in squared error amounting to 28 percent. This is due to the fact that the error in ALPHA 17 is dominated by the $\Nf=3$  step-scaling procedure at {\em high} energy, 
and this part is completely replaced by the $\Nf=0$ result by Dalla Brida 19 \cite{DallaBrida:2019wur}.
ALPHA~22 also give the covariance matrix between both results
which allows for combining both results with correlations taken into account.

The FLAG criteria are only indirectly applicable; decoupling relies on the step-scaling analysis with $\Nf=0$ in Dalla Brida~19~\cite{DallaBrida:2019wur}, which passes all FLAG criteria (cf. Sect.~\ref{s:SF}). Except for
the (well-established, cf.~Refs.~\cite{Bruno:2014ufa,Athenodorou:2018wpk}) 
perturbative evaluation of the function $P(y)$, 
perturbation theory is only applied in the $\Nf=0$ theory at very high energy, which yields a $\good$ for perturbative behaviour and
renormalization scale.  Using the FLAG criterion for continuum extrapolations 
(the constraint on values of $\alpha_\text{eff}$ is not applicable here)
the relevant scale is $M$, and the continuum extrapolations are based on data cut at $aM < 0.5$ or $aM<0.4$,
which leaves 3--4 values satisfying this cut even at the largest mass of O($10\,\GeV$). 
A remaining uncertainty of O($aM$) due to a perturbative estimate of $b_\text{g}$ is treated as
a systematic uncertainty, so that full O($a$) improvement is expected to be realized within the errors.
This is confirmed by---now available---nonperturbative data on $b_\text{g}$~\cite{DallaBrida:2023fpl},
and we use $\good$ for continuum extrapolations.
With these errors the distance of the extrapolated result is less than one sigma away from the 
last data point, i.e., $\delta(min)\approx 1$ for the data-driven criterion.

Final remark: The decoupling method offers scope for a further error reduction, 
by using the result for $b_\text{g}$ and both, improved scale setting and improved 
$\Nf=0$ step-scaling results.

In Tab.~\ref{tab:dcpl} we list the result.

\begin{table}[!htb]
   \vspace{3.0cm}
   \footnotesize
   \begin{tabular*}{\textwidth}[l]{l@{\extracolsep{\fill}}rllllllll}
   Collaboration & Ref. & $\Nf$ &
   \hspace{0.15cm}\begin{rotate}{60}{publication status}\end{rotate}
                                                    \hspace{-0.15cm} &
   \hspace{0.15cm}\begin{rotate}{60}{renormalization scale}\end{rotate}
                                                    \hspace{-0.15cm} &
   \hspace{0.15cm}\begin{rotate}{60}{perturbative behaviour}\end{rotate}
                                                    \hspace{-0.15cm} &
   \hspace{0.15cm}\begin{rotate}{60}{continuum extrapolation}\end{rotate}
      \hspace{-0.25cm} & 
                         scale & $\Lambda_\msbar[\MeV]$ & $r_0\Lambda_\msbar$ \\
   & & & & & & & & & \\[-0.1cm]
   \hline
   \hline
   & & & & & & & & & \\[-0.1cm] 
   ALPHA~22 & \cite{DallaBrida:2022eua} & 2+1 & \gA
            &     \good   &  \good      & \good  
            & $\sqrt{t_0}$ \cite{Bruno:2016plf}
            & $336(12)$\,$^*$   
            & $0.804(29)$\,$^*$                  \\ 
   & & & & & & & & & \\[-0.1cm]
   \hline
   \hline
\end{tabular*}
\begin{tabular*}{\textwidth}[l]{l@{\extracolsep{\fill}}llllllll}
\multicolumn{8}{l}{\vbox{\begin{flushleft}
   $^*$ $\alpha_\msbar^{(5)}(M_Z)=0.11823(84)$; $r_0\Lambda_\msbar$ determined 
        using $r_0 = 0.472\,\mbox{fm}$   \\
\end{flushleft}}}
\end{tabular*}
\vspace{-0.3cm}
\normalsize
\caption{Decoupling result.}
\label{tab:dcpl}
\end{table}

%% file: Alpha_s/s4.tex
\subsection{$\alpha_s$ from the potential at short distances}
\label{s:qq}


\subsubsection{General considerations}


The basic method was introduced in Ref.~\cite{Michael:1992nj} and developed in
Ref.~\cite{Booth:1992bm}. The force or potential between an infinitely
massive quark and antiquark pair defines an effective coupling
constant via
\begin{eqnarray}
   F(r) = {d V(r) \over dr} 
        = C_F {\alpha_\mathrm{qq}(r) \over r^2} \,.
\label{force_alpha}
\end{eqnarray}
The coupling can be evaluated nonperturbatively from the potential
through a numerical differentiation, see below. In perturbation theory
one also defines couplings in different schemes $\alpha_{\bar{V}}$,
$\alpha_V$ via 
\begin{eqnarray}
   V(r) = - C_F {\alpha_{\bar{V}}(r) \over r} \,, 
   \qquad \mbox{or} \quad
   \tilde{V}(Q) = - C_F {\alpha_V(Q) \over Q^2} \,,
\label{pot_alpha}
\end{eqnarray}
where one fixes the unphysical constant in the potential
by $\lim_{r\to\infty}V(r)=0$, which is compatible with fixed-order perturbation theory.
$\tilde{V}(Q)$ is the Fourier transform of $V(r)$. Nonperturbatively, the subtraction
of a constant in the potential introduces an additional 
renormalization constant, the value of $V(r_\mathrm{ref})$ at some 
distance $r_\mathrm{ref}$.  Perturbatively, it is believed to entail a 
renormalon ambiguity. In perturbation theory, the different definitions
are all simply related to each other, and their perturbative
expansions are known including the $\alpha_s^4,\,\alpha_s^4 \log\alpha_s$ 
and $\alpha_s^5 \log\alpha_s ,\,\alpha_s^5 (\log\alpha_s)^2$  terms
\cite{Fischler:1977yf,Billoire:1979ih,Peter:1997me,Schroder:1998vy,Brambilla:1999qa,Smirnov:2009fh,Anzai:2009tm,Brambilla:2009bi,Hoang:2001rr,Hoang:2002yy}.
 
The potential $V(r)$ is determined from ratios of Wilson loops,
$W(r,t)$, which behave as
\begin{eqnarray}
   \langle W(r, t) \rangle 
      = |c_0|^2 e^{-V(r)t} + \sum_{n\not= 0} |c_n|^2 e^{-V_n(r)t} \,,
      \label{e:vfromw}
\end{eqnarray}
where $t$ is taken as the temporal extension of the loop, $r$ is the
spatial one and $V_n$ are excited-state potentials.  To improve the
overlap with the ground state, and to suppress the effects of excited
states, $t$ is taken large. Also various additional techniques are
used, such as a variational basis of operators (spatial paths) to help
in projecting out the ground state.  Furthermore some
lattice-discretization effects can be reduced by averaging over Wilson
loops related by rotational symmetry in the continuum.

In order to reduce discretization errors it is of advantage 
to define the numerical derivative giving the force as
\begin{eqnarray}
   F(r_\mathrm{I}) = { V(r) - V(r-a) \over a } \,,
\end{eqnarray}
where $r_\mathrm{I}$ is chosen so that at tree level the force is the
continuum force. $F(r_\mathrm{I})$ is then a `tree-level improved' quantity
and similarly the tree-level improved potential can be defined
\cite{Necco:2001gh}.

Lattice potential results are in position space,
while perturbation theory is naturally computed in momentum space at
large momentum.
Usually, the Fourier transform 
of the perturbative expansion is then matched to  lattice data.

Finally, as was noted in Sec.~\ref{s:crit}, a determination
of the force can also be used to determine the scales $r_0,\,r_1$,
by defining them from the static force by
\begin{eqnarray}
   r_0^2 F(r_0) = {1.65} \,, \quad r_1^2 F(r_1) = 1\,.
   \label{eq:r01}
\end{eqnarray}


\subsubsection{Discussion of computations}
\label{short_dist_discuss}


\begin{table}[!htb]
   \vspace{3.0cm}
   \footnotesize
   \begin{tabular*}{\textwidth}[l]{l@{\extracolsep{\fill}}rllllll@{\hspace{1mm}}l@{\hspace{1mm}}ll}
      Collaboration & Ref. & $\Nf$ &
      \hspace{0.15cm}\begin{rotate}{60}{publication status}\end{rotate}
                                                       \hspace{-0.15cm} &
      \hspace{0.15cm}\begin{rotate}{60}{renormalization scale}\end{rotate}
                                                       \hspace{-0.15cm} &
      \hspace{0.15cm}\begin{rotate}{60}{perturbative behaviour}\end{rotate}
                                                       \hspace{-0.15cm} &
      \hspace{0.15cm}\begin{rotate}{60}{continuum extrapolation}\end{rotate}
                               \hspace{-0.25cm} & 
                         scale & $\Lambda_\msbar[\MeV]$ & $r_0\Lambda_\msbar$ \\
      & & & & & & & & & \\[-0.1cm]
      \hline
      \hline
      & & & & & & & & & \\[-0.1cm]
    Ayala 20
                   & \cite{Ayala:2020odx} & 2+1  & \gA 
                   &  \soso     &  \good     & \soso 
                   & $r_1 = 0.3106(17)\,\mbox{fm}^c$ 
                   & $338(13)$
                   & $0.802(31)$                                    \\
    TUMQCD 19
                   & \cite{Bazavov:2019qoo} & 2+1  & \gA 
                   &  \soso     &  \good     & \soso 
                   & $r_1 = 0.3106(17)\,\mbox{fm}^c$ 
                   & $314\left(^{+16}_{-8}\right)$
                   & $0.745(^{+38}_{-19})$                                \\
  
     Takaura 18
                   & \cite{Takaura:2018lpw,Takaura:2018vcy} & 2+1  & \gA 
                   &  \bad     &  \soso     & \soso 
                   & $\sqrt{t_0}=0.1465(25)\fm^a$
                   & $334(10)(^{+20}_{-18})^b$
                   & $0.799(51)$$^+$                                 \\

      {Bazavov 14}
                   & \cite{Bazavov:2014soa}  & 2+1       & \gA & \soso
                   & \good  & \soso
                   & $r_1 = 0.3106(17)\,\mbox{fm}^c$
                   & $315(^{+18}_{-12})^d$
                   & $0.746(^{+42}_{-27})$                              \\

      {Bazavov 12}
                   & \cite{Bazavov:2012ka}   & 2+1       & \gA & \soso$^\dagger$
                   & \soso   & \soso$^\#$
                   & $r_0 = 0.468\,\mbox{fm}$ 
                   & $295(30)$\,$^\star$ 
                   & $0.70(7)$$^{\star\star}$                                   \\
      & & & & & & & & & \\[-0.1cm]
      \hline
      & & & & & & & & & \\[-0.1cm]

     Karbstein 18 
                   & \cite{Karbstein:2018mzo} & 2        & \gA
                   & \soso        &  \soso    & \soso
                   & $r_0 = 0.420(14)\,\mbox{fm}$$^e$
                   & $302(16)$
                   & $0.643(34)$                                       \\

     Karbstein 14 
                   & \cite{Karbstein:2014bsa} & 2        & \gA & \soso
                   & \soso & \soso
                   & $r_0 = 0.42\,\mbox{fm}$
                   & $331(21)$
                   & 0.692(31)                                         \\

      ETM 11C      & \cite{Jansen:2011vv}    & 2         & \gA & \soso  
                   & \soso  & \soso
                   & $r_0 = 0.42\,\mbox{fm}$
                   & $315(30)^\S$ 
                   & $0.658(55)$                                        \\
      & & & & & & & & & \\[-0.1cm]
      \hline
      & & & & & & & & & \\[-0.1cm]

      Brambilla~23 & \cite{Brambilla:2023fsi} & 0 & A 
                   & \soso & \soso & \good 
                   & $\sqrt{8t_0} = 0.9569(66) r_0$
                   & 275$\left(^{+9}_{-12}\right)^{+}$ & $0.657^{+23}_{-28}$ \\
      Husung 20    & \cite{Husung:2020pxg}  & 0  & C
                   & \soso & \good   & \good
                   &  \multicolumn{3}{c}{\text{no quoted value for} $\lms$}  \\
      
      Husung 17    & \cite{Husung:2017qjz}   & 0         & C
                   & \soso & \good   & \good
                   &  $r_0 = 0.50\,\mbox{fm}$ 
                   & 232(6) & $0.590(16)$  \\

      Brambilla 10 & \cite{Brambilla:2010pp} & 0         & \gA & \soso 
                   & \good\ & \soso$^{\dagger\dagger}$ &  & $266(13)$$^{+}$&
                   $0.637(^{+32}_{-30})$$^{\dagger\dagger}$                  \\
      UKQCD 92     & \cite{Booth:1992bm}    & 0         & \gA & \good 
                                  & \soso$^{++}$   & \bad   
                                  & $\sqrt{\sigma}=0.44\,\GeV$ 
                                  & $256(20)$
                                  & 0.686(54)                             \\
      Bali 92     & \cite{Bali:1992ru}    & 0         & \gA & \good 
                                  & \soso$^{++}$   & \bad 
                                  & $\sqrt{\sigma}=0.44\,\GeV$
                                  & $247(10)$                             
                                  & 0.661(27)                             \\
      & & & & & & & & & \\[-0.1cm]
      \hline
      \hline\\
\end{tabular*}\\[-0.2cm]
\begin{minipage}{\linewidth}
{\footnotesize 
\begin{itemize}
   \item[$^a$] Scale determined from $t_0$ in
               Ref.~\cite{Borsanyi:2012zs}.
   \item[$^b$]             
               $\alpha^{(5)}_{\overline{\rm MS}}(M_Z) = 0.1179(7)(^{+13}_{-12})$.  
   \item[$^c$]
   Determination on lattices with $m_\pi L=2.2 - 2.6$. 
   Scale from $r_1$ \cite{Bazavov:2014pvz}
   as determined from  $f_\pi$ in Ref.~\cite{Bazavov:2010hj}.      \\[-5mm]
   \item[$^d$]
         $\alpha^{(3)}_{\overline{\rm MS}}(1.5\,\mbox{GeV}) = 0.336(^{+12}_{-8})$, 
         $\alpha^{(5)}_{\overline{\rm MS}}(M_Z) = 0.1166(^{+12}_{-8})$.
         \\[-5mm]
   \item[$^e$] 
         Scale determined from $f_\pi$, see \cite{Baron:2009wt}.   \\[-5mm]
   \item[$^\dagger$]
   Since values of $\alpha_\mathrm{eff}$ within our designated range are used,
   we assign a \soso\ despite
   values of $\alpha_\mathrm{eff}$ up to $\alpha_\mathrm{eff}=0.5$ being used.  
   \\[-5mm]
   \item[$^\#$]Since values of $2a/r$ within our designated range are used,
   we assign a \soso\ although
   only values of $2a/r\geq1.14$ are used at $\alpha_\mathrm{eff}=0.3$.
   \\[-5mm]
   \item[$^\star$] Using results from Ref.~\cite{Bazavov:2011nk}.  \\[-5mm]
   \item[$^{\star\star}$]
         $\alpha^{(3)}_{\overline{\rm MS}}(1.5\,\mbox{GeV}) = 0.326(19)$, 
         $\alpha^{(5)}_{\overline{\rm MS}}(M_Z) = 0.1156(^{+21}_{-22})$.  \\[-5mm]
   \item[$^\S$] Both potential and $r_0/a$ are determined on a small 
   ($L=3.2r_0$) lattice.   \\[-5mm]
   \item[$^{\dagger\dagger}$] Uses lattice results of Ref.~\cite{Necco:2001xg}, 
   some of which have very small lattice spacings where 
   according to more recent investigations a bias due to the freezing of
   topology may be present.  \\[-5mm] 
   \item[$^+$] Our conversion using $r_0 = 0.472\,\mbox{fm}$.   \\[-5mm]
   \item[$^{++}$] We give a $\soso$ because only a NLO formula is used and
       the error bars are very large; our criterion does not apply 
       well to these very early calculations.           
\end{itemize}
}
\end{minipage}
\normalsize
\caption{Short-distance potential results.}
\label{tab_short_dist}
\end{table}

In Tab.~\ref{tab_short_dist}, we list results of determinations
of $r_0\Lambda_{\msbar}$ (together with $\Lambda_{\msbar}$
using the scale determination of the authors).

The first determinations in the three-colour Yang Mills theory are by
UKQCD 92 \cite{Booth:1992bm} and Bali 92 \cite{Bali:1992ru} who used
$\alpha_\mathrm{qq}$, Eq.~(\ref{force_alpha}), as explained above, but not in the tree-level
improved form. Rather a phenomenologically determined lattice-artifact
correction was subtracted from the lattice potentials.  The comparison
with perturbation theory was on a more qualitative level on the basis
of a 2-loop $\beta$-function ($n_{\mathrm{l}}=1$) and a continuum extrapolation
could not be performed as yet. A much more precise computation of
$\alpha_\mathrm{qq}$ with continuum extrapolation was performed in
Refs.~\cite{Necco:2001xg,Necco:2001gh}. Satisfactory agreement with
perturbation theory was found \cite{Necco:2001gh} but the stability of
the perturbative prediction was not considered sufficient to be able
to extract a $\Lambda$ parameter.

In Brambilla 10 \cite{Brambilla:2010pp} the same quenched lattice
results of Ref.~\cite{Necco:2001gh} were used and a fit was performed to
the continuum potential, instead of the force. Perturbation theory to
$n_{\mathrm{l}}=3$ loop
was used including a resummation of terms $\alpha_s^3 (\alpha_s \ln\alpha_s)^n $ 
and $\alpha_s^4 (\alpha_s \ln\alpha_s)^n $. Close
agreement with perturbation theory was found when a renormalon
subtraction was performed. Note that the renormalon subtraction
introduces a second scale into the perturbative formula which is
absent when the force is considered.

Bazavov 14 \cite{Bazavov:2014soa} updates 
Bazavov 12 \cite{Bazavov:2012ka} and modifies this procedure
somewhat. They consider the 
perturbative expansion
for the force. 
They set $\mu = 1/r$
to eliminate logarithms and then integrate the force to obtain an
expression for the potential. 
The resulting integration constant is fixed by requiring
the perturbative potential to be equal to the nonperturbative 
one exactly at a reference distance $r_{\rm ref}$ and the two are then
compared at other values of $r$. As a further check,
the force is also used directly.

For the quenched calculation of Brambilla 10 \cite{Brambilla:2010pp}
very small lattice spacings,
$a \sim 0.025\,\mbox{fm}$, were available from Ref.~\cite{Necco:2001gh}.
For ETM 11C \cite{Jansen:2011vv}, Bazavov 12 \cite{Bazavov:2012ka},
Karbstein 14 \cite{Karbstein:2014bsa}
and Bazavov 14 \cite{Bazavov:2014soa} using dynamical
fermions such small lattice spacings are not yet realized 
(Bazavov 14 reaches down to $a \sim 0.041\,\mbox{fm}$). They
all use the tree-level improved potential as described above. 
We note that the value of $\Lambda_\msbar$ in physical units by
ETM 11C \cite{Jansen:2011vv} is based on a value of $r_0=0.42$~fm. 
This is at least 10\% smaller than the large majority of
other values of $r_0$. Also the values of $r_0/a$ 
on the finest lattices in ETM 11C \cite{Jansen:2011vv}
and $r_1/a$ for Bazavov 14 \cite{Bazavov:2014soa} come from
rather small lattices with $m_\pi L \approx 2.4$, $2.2$ respectively.

Instead of the procedure discussed previously, Karbstein 14 
\cite{Karbstein:2014bsa} reanalyzes the data of ETM 11C 
\cite{Jansen:2011vv} by first estimating
the Fourier transform $\tilde V(p)$ of $V(r)$ and then fitting  
the perturbative expansion of $\tilde V(p)$ in terms of 
$\alpha_\msbar(p)$. Of course, the Fourier transform requires
some modelling of the $r$-dependence of $V(r)$
at short and at large distances. The authors fit a linearly rising
potential at large distances together with string-like
corrections of order $r^{-n}$ and define the potential at large 
distances by this fit.\footnote{Note that at large distances,
where string breaking is known to occur, this is not 
any more the ground-state potential defined by \eq{e:vfromw}.}
Recall that for observables in momentum space
we take the renormalization scale entering our criteria as $\mu=q$,
Eq.~(\ref{mu_def}). The analysis (as in ETM 11C \cite{Jansen:2011vv})
is dominated by the data at the smallest lattice spacing, where
a controlled determination of the overall scale  is difficult due to 
possible finite-size effects.
Karbstein 18  \cite{Karbstein:2018mzo} is a
reanalysis of Karbstein 14 and supersedes it. Some data with a different
discretization of the static quark is added (on the same configurations)
and the discrete lattice results for the static potential in position
space are first parameterized by a continuous function, which then
allows for an analytical Fourier transformation to momentum space.

Similarly also for Takaura 18~\cite{Takaura:2018lpw,Takaura:2018vcy}
the momentum space potential $\tilde{V}(Q)$ is the central object. 
Namely, they assume that renormalon/power-law effects are absent in
$\tilde{V}(Q)$ and only come in through the Fourier transformation.
They provide evidence that renormalon effects (both $u=1/2$ and $u=3/2$) can be 
subtracted and arrive at a nonperturbative term $k\,\Lambda_\msbar^3 r^2$. 
Two different analyses are carried out with the final result 
taken from ``Analysis II''. Our numbers including the evaluation of
the criteria refer to it. Together with the 
perturbative 3-loop (including the $ \alpha_s^4\log \alpha_s$ term) 
expression, this term is fitted to the 
nonperturbative results for the potential in the region
$0.04\,\fm \, \leq \, r \,\leq 0.35\,\fm$, where $0.04\,\fm$ is 
$r=a$ on the finest lattice.
The nonperturbative potential data originates from JLQCD ensembles (Symanzik-improved
gauge action and M\"obius domain-wall quarks) at three lattice spacings with 
a pion mass around $300\,\MeV$. Since at the maximal distance in the analysis
we find $\alpha_\msbar(2/r) = 0.43$, the renormalization-scale 
criterion yields a \bad. 
The perturbative
behaviour is \soso\ because of the high orders in perturbation theory known. The 
continuum-limit criterion yields a $\soso$.

One of the main issues for all these computations is whether the
perturbative running of the coupling constant
has been reached.
While for $\Nf=0$ fermions Brambilla~10
\cite{Brambilla:2010pp} reports agreement with perturbative behaviour
at the smallest distances, Husung~17 (which goes to
shorter distances) finds relatively large corrections beyond the 3-loop
$\alpha_\mathrm{qq}$.
For dynamical fermions,  
 Bazavov 12 \cite{Bazavov:2012ka}
and Bazavov 14 \cite{Bazavov:2014soa} report good agreement with perturbation
theory after the renormalon is subtracted or eliminated. 

A second issue is the coverage of configuration space in some of the
simulations, which use very small lattice spacings with periodic
boundary conditions. Affected are the smallest two lattice spacings
of Bazavov 14 \cite{Bazavov:2014soa} where very few tunnelings of
the topological charge occur \cite{Bazavov:2014pvz}.
With present knowledge, it also seems  possible that the older data
by Refs.~\cite{Necco:2001xg,Necco:2001gh} used by Brambilla 10 
\cite{Brambilla:2010pp} are partially obtained with (close to) frozen topology.

The computation in Husung 17~\cite{Husung:2017qjz},
for $\Nf = 0$ flavours, first determines the coupling 
$\gbar_{\rm qq}^2(r,a)$ from the force and then performs a continuum extrapolation
on lattices down to $a \approx 0.015\,\mbox{fm}$, using a step-scaling method at short distances, $r/r_0 \lsim 0.5$. 
Using the $4$-loop $\beta^{\rm qq}$ function this allows $r_0\Lambda_{\rm qq}$
to be estimated, which is then converted to the $\overline{\rm MS}$ scheme.
$\alpha_{\rm eff} = \alpha_{\rm qq}$ ranges from $\sim 0.17$ to large 
values; we 
give $\soso$ for renormalization scale and \good\ for perturbative behaviour. The range
$a\mu = 2a/r \approx$ 0.37--0.14 leads to a $\good$
in the continuum extrapolation.
Recently these calculations have been extended in Husung 20~\cite{Husung:2020pxg}.
A finer lattice spacing of $a=0.01\,\fm$ (scale from $r_0=0.5~\fm$) is reached and lattice volumes up to $L/a=192$ are
simulated (in Ref.~\cite{Husung:2017qjz} the smallest lattice spacing is $0.015\,\fm$).
The Wilson action is used despite its significantly larger cutoff effects compared to Symanzik-improved actions;
this avoids unitarity violations, thus allowing for a clean ground-state extraction via a generalized eigenvalue problem.
Open boundary conditions are used to avoid the topology-freezing problem.
Furthermore, new results for the continuum approach are employed, which determine the
cutoff dependence at $\cO(a^2)$ including the exact coupling-dependent terms, in the asymptotic region where
the Symanzik effective theory is applicable~\cite{Husung:2019ytz}. An ansatz for the remaining
higher-order cutoff effects at $\cO(a^4)$ is propagated as a systematic error to the data, which
effectively discards data for $r/a < 3.5$. The large-volume step-scaling function with step factor $3/4$ is
computed and compared to perturbation theory. For $\alpha_{qq}>0.2$ there is a noticeable difference between the 2-loop and 3-loop results. 
Furthermore, the ultra-soft contributions at 4-loop level give a significant
contribution to the static $Q\bar Q$ force. While this study is for $\Nf=0$ flavours 
it does raise the question whether the weak-coupling expansion for the range of $r$-values 
used in present analyses of $\alpha_s$ is sufficiently reliable.
Around $\alpha_{\rm qq} \approx 0.21$ the differences get smaller but the error increases significantly, 
mainly due to the propagated lattice artifacts. 
The dependence of $\Lambda_{\overline{\rm MS}}^{(N_f=0)} \sqrt{8 t_0}$ on $\alpha_{\rm qq}^3$ is very similar to
the one observed in the previous study but no value for its $\alpha_{qq}\rightarrow 0$ limit
is quoted. Husung 20 \cite{Husung:2020pxg} is more pessimistic about the error on the $\Lambda$ parameter stating
the relative error has to be $5\%$ or larger, while Husung 17 quotes a relative error of $3 \%$.

In 2+1-flavour QCD two new papers appeared on the determination of the strong coupling constant
from the static quark anti-quark potential after the FLAG 19 report
\cite{Bazavov:2019qoo,Ayala:2020odx}. 
In TUMQCD 19~\cite{Bazavov:2019qoo}\footnote{The majority of authors are the same as in ~\cite{Bazavov:2014soa}.} 
the 2014 analysis of Bazavov~14~\cite{Bazavov:2014soa} 
has been extended by including three finer lattices with lattice
spacing $a=$ 0.035, 0.030 and 0.025~fm as well as lattice results
on the free energy of static quark anti-quark pair at nonzero temperature.
On the new fine lattices the effect of freezing topology has been observed, however,
it was verified that this does not affect the potential within the estimated errors
\cite{Bazavov:2017dsy,Weber:2018bam}. The comparison of the lattice result on
the static potential has been performed in the interval $[r_{\mathrm{min}},r_{\mathrm{max}}]$, with 
$r_{\mathrm{max}}$ = 0.131, 0.121, 0.098, 0.073 and 0.055~fm. The main result quoted in
the paper is based on the analysis with $r_{\mathrm{max}}=$ 0.073~fm \cite{Bazavov:2019qoo}. 
Since the new study employs a much wider range in $r$ than the previous one \cite{Bazavov:2014soa}
we give it a $\good$  for the perturbative behaviour. 
Since $\alpha_\text{eff}=\alpha_{qq}$ varies in the range 0.2--0.4 for the $r$ values used in
the main analysis we give $\soso$ for the renormalization scale.
Several values of $r_{\mathrm{min}}$ have been used
in the analysis, the largest being $r_{\mathrm{min}}/a=\sqrt{8}\simeq$ 2.82, which corresponds
to $a \mu \simeq$ 0.71. Therefore, we give a $\soso$ for continuum extrapolation
in this case. An important difference compared to the previous study \cite{Bazavov:2014soa} is
the variation of the renormalization scale. In Ref. \cite{Bazavov:2014soa} the renormalization
scale was varied by a factor of $\sqrt{2}$ around the nominal value of $\mu=1/r$, in order to
exclude very low scales, for which the running of the strong coupling constant is no longer
perturbative. In the new analysis the renormalization scale was varied by a factor of two.
As the result, despite the extended data set and shorter distances used in the new study
the perturbative error did not decrease \cite{Bazavov:2019qoo}. 
We also note that the scale dependence turned out to be nonmonotonic in the range $\mu=1/(2r)$--$2/r$ \cite{Bazavov:2019qoo}.
The final result
reads (``us" stands for ``ultra-soft"),
\begin{eqnarray}
\lms^{(\Nf=3)}&=&314.0 \pm 5.8 (\text{stat}) \pm 3.0(\text{lat}) 
    \pm 1.7 (\text{scale})\left(^{+13.4}_{-1.8}\right) (\text{pert}) \pm 4.0 (\text{pert. us}) ~{\MeV} \nonumber\\
    &=& 314\left(^{+16}_{-08}\right)\,{\MeV}\,,
\label{eq:Lam-tumqcd}
\end{eqnarray}
where all errors were combined in quadrature. This is
in very good agreement with the previous determination \cite{Bazavov:2014soa}.

The analysis was also applied to the singlet static quark anti-quark free energy
at short distances. At short distances the free energy is expected to be the same
as the static potential. This is verified numerically in the lattice calculations
TUMQCD 19~\cite{Bazavov:2019qoo} for $rT < 1/4$ with $T$ being the temperature. Furthermore,
this is confirmed by the perturbative calculations at $T>0$ at NLO {\cite{Berwein:2017thy}.
The advantage of using the free energy is that it gives access to much shorter distances.
On the other hand, one has fewer data points because the condition $rT < 1/4$
has to be satisfied. The analysis based on the free energy gives
\begin{eqnarray}
\lms^{(\Nf=3)}&=&310.9 \pm 11.3 (\text{stat}) \pm 3.0(\text{lat}) 
\pm 1.7 (\text{scale})\left(^{+5.6}_{-0.8}\right)(\text{pert}) \pm 2.1 (\text{pert. us}) ~{\MeV} \nonumber \\
 &=& 311(13) \,\MeV,
\end{eqnarray}
in good agreement with the above result and thus, providing additional confirmation of it.

The analysis of Ayala 20~\cite{Ayala:2020odx} uses a subset of data presented in 
TUMQCD~19~\cite{Bazavov:2019qoo} with the same correction of the lattice effects. 
For this reason the continuum extrapolation gets $\soso$, too.
They match to perturbation theory for $1/r >2$ GeV, which corresponds to 
$\alpha_\text{eff}=\alpha_{qq}=$ 0.2--0.4. 
Therefore, we give $\soso$ for the renormalization scale.
They verify the perturbative behaviour in the region $1~{\rm GeV}<1/r<2.9~{\rm GeV}$, which
corresponds to variation of $\alpha_\text{eff}^3$ by a factor of 3.34. However, the relative
error on the final result has $\delta \Lambda/\Lambda \simeq 0.035$ which is larger
than $\alpha_\text{eff}^3=0.011$. Therefore, we give a $\good$ for the perturbative behaviour
in this case. The final result for the $\Lambda$-parameter reads:
\begin{equation}
\Lambda_\msbar^{(\Nf=3)}=338 \pm 2 (\text{stat}) \pm 8 (\text{matching}) \pm 10 (\text{pert})~{\MeV} = 338(13)\MeV\,.
\label{eq:Lam-ayala}
\end{equation}
This is quite different from the above result. This difference is mostly due to the organization
of the perturbative series. The authors use ultra-soft (log) resummation, i.e., they resum the terms
$\alpha_s^{3+n} \ln^n\alpha_s$ to all orders instead of using fixed-order perturbation theory.
They also include what is called the terminant of the perturbative series associated to the leading renormalon 
of the force \cite{Ayala:2020odx}.
When they use fixed-order perturbation theory they obtain very similar 
results to Refs.~\cite{Bazavov:2014soa,Bazavov:2019qoo}. 
It has been argued that log resummation
cannot be justified since for the distance range available in the lattice studies $\alpha_s$
is not small enough and the logarithmic and nonlogarithmic higher-order terms are of
a similar size \cite{Bazavov:2014soa}. On the other hand, the resummation of ultra-soft logs does not lead
to any anomalous behaviour of the perturbative expansion like large scale dependence or bad convergence \cite{Ayala:2020odx}.

To obtain the value of $\lms^{(\Nf=3)}$ from the static potential
we combine the results in Eqs.~(\ref{eq:Lam-tumqcd}) and (\ref{eq:Lam-ayala}) using the weighted
average with the weight given by the perturbative error and using the difference in
the central value as the error estimate. This leads to
\begin{equation}
\lms^{(\Nf=3)}=330(24) ~{\MeV}\,,
\end{equation}
from the static potential determination. In the case of TUMQCD 19, where the perturbative error
is very asymmetric we used the larger upper error for the calculation of the corresponding weight.

A new analysis with $\Nf=0$ has been presented in Brambilla~23~\cite{Brambilla:2023fsi} where gradient
flow is used to study the static force. The use of gradient flow allows an improved determination of
the static force while adding to the problem a new scale, the gradient-flow time $\tau_F$. The lattice 
volumes used are $40\times 20^3$, $52\times 26^3$, $60\times 30^2$ and $80\times 40^3$, with 
corresponding lattice spacings ranging from 0.06 to 0.03 fm, using the Wilson action. 
On the finest lattice an increase in the autocorrelation of the topological charge is observed and 
taken into account by increasing the Monte Carlo time in-between measurements. 
The reference scale $t_0$, used throughout the analysis, 
is obtained from a measurement of the action density by imposing
\begin{eqnarray}
   \tau_F \left.\langle \frac{1}{4} G_{\mu\nu} G^{\mu\nu}\rangle\right|_{\tau_F=t_0} = 0.3\, .
\end{eqnarray}
The static force is computed from the insertion of the chromoelectric field in the expectation
value of the Wilson loop,
\begin{eqnarray}
   F(r) = -i \lim_{T\to \infty} \frac{\langle \mathrm{Tr}\, \left[W_{r\times T}\, \hat{\mathbf{r}}
   \cdot g\mathbf{E(\mathbf{r},t)}\right]\rangle}{\langle\mathrm{Tr}\, 
   W_{r\times T}\rangle}\, ,
\end{eqnarray}
and tree-level improvement is used to improve the extrapolation to the continuum limit. The 
dimensionless product $r^2 F(r)$ yields the observable used for the extraction of $\alpha_s$.

Results extrapolated to $\tau_F=0$ are used for a conventional analysis along the lines of previous 
publications using the static force. The fit uses the perturbative expansion of the force including 
3-loop contributions and leading ultrasoft logarithms. Data points with 
$r/\sqrt{t_0} \in [0.80,1.15]$ are included in the fit, which yields
\begin{eqnarray}
   \sqrt{8t_0} \Lambda_\msbar^{(\Nf=0)} = 0.6353 \pm 0.0032 (\mathrm{stat}) \pm 0.0013 (\mathrm{AIC})\, ,
\end{eqnarray} 
where the label AIC refers to the Bayesian procedure for combining results from different fit ranges 
based on Akaike's information criterion, as proposed in Ref.~\cite{Jay:2020jkz}. Note that the 
error on this result is still dominated by statistics rather than the systematics related to the 
choice of fitting range. The matching scale in these fits is the usual scale $\mu=1/r$.

Measurements at $\tau_F\neq 0$ allow an alternative way to extract the strong coupling constant by 
fitting to the perturbative expression for the force at finite flow time. The latter
perturbative expansion is only known at 1-loop, which is used as a correction of the higher-order
result at $\tau_F=0$. The best result is obtained by fitting the $r$-dependence at fixed values of 
$\tau_F$, which yields
\begin{eqnarray}
   \sqrt{8 t_0} \Lambda_\msbar^{(\Nf=0)}= 0.629^{+22}_{-26}\, .
\end{eqnarray}
The scale of perturbative matching is defined as
\begin{eqnarray}
   \mu = \frac{1}{\sqrt{sr^2 + 8b\tau_F}}\, .
\end{eqnarray}
The uncertainty related to the truncation of the perturbative expansion is estimated by scale 
variations, where $b=0$ and $s$ is varied by a factor $\sqrt{2}$ in the zero-flow-time part of the 
perturbative expansion, while $s=1$ and $b=0,1,-0.5$ in the finite-flow-time part. The central value 
corresponds to $s=1, b=0$. The error on the result above is dominated by the $s$-scale variation. 
The ratio $\sqrt{t_0}/r_0$ is computed in Brambilla~23 and allows 
to quote a final result in units of $r_0$, 
\begin{eqnarray}
   r_0 \Lambda_\msbar^{(\Nf=0)} = 0.657\left(^{+23}_{-28}\right)\, .
\end{eqnarray}
The continuum extrapolation is based on four lattice spacings. From the data reported in the figures, 
we see that for $r=0.7323 \sqrt{t_0}$, the effective 
coupling is below the requested threshold of 0.03, while the lattice spacing is such that
$0.2321 \leq \mu a \leq 0.4916$. Therefore, we can give a $\good$\ for the continuum extrapolation. 
Fits to the perturbative behaviour are performed for $0.27 \leq \alpha_{\mathrm{eff}} \leq 0.36$ and 
$n_\ell=3$ in the perturbative expansion. Hence, $\alpha_{\mathrm{eff}}^{n_\ell}$ changes by a factor 
of 2.37, which is 5\% above the threshold of $(3/2)^2$. We feel in this case we can award a $\soso$\ for the 
perturbative behaviour. Finally, given the range of values for $\alpha_{\mathrm{eff}}$ quoted above, 
we give a $\soso$\ for the renormalization scale.  

\paragraph{Scale variations.} 
The perturbative matching for the static potential is done at lower scales, $\mu = 1.5, 2.5, 5.0~\mathrm{GeV}$. 
We have computed the 
change in the determination of 
$\alpha_{\overline{\mathrm{MS}}}(M_Z)$ as explained in Sec.~\ref{s:intro}. The systematic errors 
depend on the value of the perturbative matching scale. We obtain
\begin{description}
   \item[$Q=1.5~\mathrm{GeV}$] 
   \begin{eqnarray}
      \delta_{(2)} = 2.6\%\, \quad
      \delta^*_{(2)} = 2.7\%\, .
   \end{eqnarray}
   The value of $\delta^*_{(4)}$ cannot be computed in this case, because the matching scale is low, already at the boundary
   of the region where the perturbative expansion can be trusted. 
   \item[$Q=2.5~\mathrm{GeV}$]
   \begin{eqnarray}
    \delta^*_{(4)} = 0.9\%\, , \quad 
    \delta_{(2)} = 1.5\%\, \quad
    \delta^*_{(2)} = 1.5\%\, .
    \end{eqnarray}
   \item[$Q=5.0~\mathrm{GeV}$]
   \begin{eqnarray}  
    \delta^*_{(4)} = 0.4\%\, , \quad 
    \delta_{(2)} = 0.8\%\, \quad
    \delta^*_{(2)} = 0.8\%\, .
   \end{eqnarray}
   Note that in the last two cases it was possible to compute $\delta^*_{(4)}$.
\end{description}
For the larger values of $Q$, the error obtained from scale variations is very similar to the error 
quoted in previous editions of FLAG, where scale variations were not performed systematically. 
For $Q=1.5~\mathrm{GeV}$
the error is larger, as expected since the matching of perturbation theory happens at lower energy. 

%% file: Alpha_s/s5.tex
\subsection{$\alpha_s$ from the light-quark vacuum polarization in momentum/po\-sition space}


\label{s:vac}


\subsubsection{General considerations}


Except for the calculation Cali 20 \cite{Cali:2020hrj}, where position space is used (see below),
the light-flavour-current two-point function is usually evaluated in momentum 
space, in terms of the vacuum-polarization function. Assuming $\Nf=3$ 
flavours in the isospin limit, with flavour nonsinglet currents consisting of up and down quarks, $J^a_\mu$ ($a=1,\ldots,3$),
the momentum representation takes the form
\begin{eqnarray}
   \langle J^a_\mu J^b_\nu \rangle 
      =\delta^{ab} [(\delta_{\mu\nu}Q^2 - Q_\mu Q_\nu) \Pi_J^{(1)}(Q) 
                                     - Q_\mu Q_\nu\Pi_J^{(0)}(Q)] \,,
\end{eqnarray}
where $Q_\mu$ is a space-like momentum and $J_\mu\equiv V_\mu$
for a vector current and $J_\mu\equiv A_\mu$ for an axial-vector current.\footnote{For the general mass-nondegenerate case with SU(3) flavour nonsinglet currents 
see, for example, Ref.~\cite{Bruser:2024zyg}.} 
Defining $\Pi_J(Q)\equiv \Pi_J^{(0)}(Q)+\Pi_J^{(1)}(Q)$,
the operator product expansion (OPE) of  $\Pi_{V/A}(Q)$ is given by
\begin{eqnarray}
   \lefteqn{\Pi_{V/A}|_{\rm OPE}(Q^2,\alpha_s)}
      & &                                             \nonumber  \\
      &=& c + C_1^{V/A}(Q^2) + C_m^{V/A}(Q^2)
                       \frac{\bar{m}^2(Q)}{Q^2}
            + \sum_{q=u,d,s}C_{\bar{q}q}^{V/A}(Q^2)
                        \frac{\langle m_q\bar{q}q \rangle}{Q^4}
                                                      \nonumber  \\
      & &   + C_{GG}^{V/A}(Q^2) 
                \frac{\langle \alpha_s GG\rangle}{Q^4}+{\cO}(Q^{-6}) \,,
\label{eq:vacpol}
\end{eqnarray}
for large $Q^2$. The perturbative coefficient functions $C_X^{V/A}(Q^2)$ for the
operators $X$ ($X=1$, $\bar{q}q$, $GG$) are given as $C_X^{V/A}(Q^2)=\sum_{i\geq0}\left( C_X^{V/A}\right)^{(i)}\alpha_s^i(Q^2)$  and $\bar m$ is the running 
mass of the mass-degenerate up and down quarks.
$C_1^{V/A}$ is known including $\alpha_s^4$
in a continuum renormalization scheme such as the
$\overline{\rm MS}$ scheme
\cite{Chetyrkin:1979bj,Surguladze:1990tg,Gorishnii:1990vf,Baikov:2008jh}.
Nonperturbatively, there are terms in $C_X^{V/A}$ that do not have a 
series expansion in $\alpha_s$. For an example for the unit
operator see Ref.~\cite{Balitsky:1993ki}.
The term $c$ is $Q$-independent and divergent in the limit of infinite
ultraviolet cutoff. However the Adler function defined as 
\begin{eqnarray}
   D(Q^2) \equiv - Q^2 { d\Pi(Q^2) \over dQ^2} \,,
\label{eq:adler}
\end{eqnarray}
is a scheme-independent finite quantity, which gives rise to an effective coupling. 
Therefore, one can determine the running coupling constant in the $\overline{\rm MS}$ scheme
from the vacuum-polarization function computed by a lattice-QCD
simulation. Of course, there is the choice whether to use the vector or the axial-vector channel,
or both, the canonical choice being $\Pi_{V+A} = \Pi_V+\Pi_A$. 
While perturbation theory does not distinguish between
these channels, the nonperturbative contributions are different, and the 
quality of lattice data may differ, too.
For a given choice, the lattice data of the vacuum polarization is fitted with the 
perturbative formula Eq.~(\ref{eq:vacpol}) with fit parameter 
$\Lambda_{\overline{\rm MS}}$ parameterizing the running coupling 
$\alpha_{\overline{\rm MS}}(Q^2)$.  

While there is no problem in discussing the OPE at the
nonperturbative level, the `condensates' such as ${\langle \alpha_s
  GG\rangle}$ are ambiguous, since they mix with lower-dimensional
operators including the unity operator.  Therefore, one should work in
the high-$Q^2$ regime where power corrections are negligible within
the given accuracy. Thus setting the renormalization scale as
$\mu\equiv \sqrt{Q^2}$, one should seek, as always, the window
$\Lambda_{\rm QCD} \ll \mu \ll a^{-1}$.

\subsubsection{Definitions in position space}

The two-point current correlation functions in position space
contain the same physical information as in momentum space, but the technical 
details are sufficiently different to warrant a separate discussion.
The (Euclidean) current-current correlation function for $J^\mu_{ff'}$ (with flavour indices $f,f'$)
is taken to be either the flavour nondiagonal vector or axial-vector current,
with the Lorentz indices contracted,
\begin{equation}
    C_\text{A,V}(x) = -\sum_{\mu}\left\langle J^\mu_{ff'\text{A,V}}(x)J^\mu_{f'f\text{A,V}}
(0)\right\rangle = \dfrac{6}{\pi^4(x^2)^3}\left( 1 + \frac{\alpha_s}{\pi} + \cO(\alpha^2)\right)\,.
\end{equation}
In the chiral limit, the perturbative expansion is known to $\alpha_s^4$~\cite{Chetyrkin:2010dx}, and is identical
for vector and axial-vector correlators. The only scale is set by the Euclidean distance $\mu=1/|x|$ and 
the effective coupling can thus be defined as
\begin{equation}
  \alpha_\text{eff}(\mu=1/|x|) = \pi \left[(x^2)^3(\pi^4/6) C_\text{A,V}(x) - 1\right]\,.
\end{equation}
As communicated to us by the authors of \cite{Cali:2020hrj}, 
there is a typo in Eq.~(35) of~\cite{Chetyrkin:2010dx}.
For future reference, the numerical coefficients for the 3-loop conversion
\begin{equation}
   \alpha_\text{eff}(\mu) = \alpha_\msbar(\mu) +  c_1 \alpha^2_\msbar(\mu) + c_2\alpha^3_\msbar(\mu) + c_3\alpha^4_\msbar(\mu),
  \label{eq:CVAMSbar}
   \end{equation}
should read
\begin{equation}
   c_1 = -1.4346,\qquad c_2=0.16979,\qquad   c_3= 3.21120\,. 
\end{equation}


\subsubsection{Discussion of computations}


Results using this method in momentum space are, to date, only available using
overlap fermions or domain-wall fermions. Cali 20~\cite{Cali:2020hrj} consider vacuum polarization in position space using $\cO(a)$-improved Wilson fermions.
The results are collected in Tab.~\ref{tab_vac} for
$\Nf=2$, JLQCD/TWQCD 08C \cite{Shintani:2008ga} and for $\Nf = 2+1$, JLQCD 10
\cite{Shintani:2010ph}, Hudspith 18~ \cite{Hudspith:2018bpz} and Cali 20 \cite{Cali:2020hrj}.

\begin{table}[!htb]
   \vspace{3.0cm}
   \footnotesize
   \begin{tabular*}{\textwidth}[l]{l@{\extracolsep{\fill}}rllllllll}
   Collaboration & Ref. & $\Nf$ &
   \hspace{0.15cm}\begin{rotate}{60}{publication status}\end{rotate}
                                                    \hspace{-0.15cm} &
   \hspace{0.15cm}\begin{rotate}{60}{renormalization scale}\end{rotate}
                                                    \hspace{-0.15cm} &
   \hspace{0.15cm}\begin{rotate}{60}{perturbative behaviour}\end{rotate}
                                                    \hspace{-0.15cm} &
   \hspace{0.15cm}\begin{rotate}{60}{continuum extrapolation}\end{rotate}
      \hspace{-0.25cm} & 
                         scale & $\Lambda_\msbar[\MeV]$ & $r_0\Lambda_\msbar$ \\
   & & & & & & & & & \\[-0.1cm]
   \hline
   \hline
   & & & & & & & & & \\[-0.1cm]
   Cali 20 & \cite{Cali:2020hrj} & 2+1 & \gA
            & \soso  & \good   & \good  
            &  $m_\Upsilon$$^\S$
            & $342(17)$
            & $0.818(41)$$^a$               \\
   & & & & & & & & & \\[-0.1cm]
   \hline
   & & & & & & & & & \\[-0.1cm]

   Hudspith 18 & \cite{Hudspith:2018bpz} & 2+1 & \oP
            & \soso  & \soso   & \bad  
            &  $m_\Omega$$^\star$
            & $337(40)$
            & $0.806(96)$$^b$               \\
   & & & & & & & & & \\[-0.1cm]
   \hline
   & & & & & & & & & \\[-0.1cm]

   Hudspith 15 & \cite{Hudspith:2015xoa} & 2+1 &\rC 
            & \soso  & \soso   & \bad  
            & $m_\Omega$$^\star$
            & $300(24)^+$
            & $0.717(58)$              \\
   & & & & & & & & & \\[-0.1cm]
   \hline
   & & & & & & & & & \\[-0.1cm]
   JLQCD 10 & \cite{Shintani:2010ph} & 2+1 &\gA & \bad 
            & \bad & \bad
            & $r_0 = 0.472\,\mbox{fm}$
            & $247(5)$$^\dagger$
            & $0.591(12)$              \\
   & & & & & & & & & \\[-0.1cm]
   \hline
   & & & & & & & & & \\[-0.1cm]
   JLQCD/TWQCD 08C & \cite{Shintani:2008ga} & 2 & \gA & \soso 
            & \bad & \bad
            & $r_0 = 0.49\,\mbox{fm}$
            & $234(9)(^{+16}_{-0})$
            & $0.581(22)(^{+40}_{-0})$    \\
            
   & & & & & & & & & \\[-0.1cm]
   \hline
   \hline
\end{tabular*}
\begin{tabular*}{\textwidth}[l]{l@{\extracolsep{\fill}}llllllll}
\multicolumn{8}{l}{\vbox{\begin{flushleft}
   $^\S$ via $t_0/a^2$, still unpublished. We use $r_0=0.472\,\fm$\\
   $^\star$ Determined in \cite{Blum:2014tka}.  \\
   $^a$ Evaluates to $\alpha_\msbar^{(5)}(M_Z)= 0.11864(114)$\\
   In conversion to $r_0\Lambda$ we used
        $r_0 = 0.472\,\mbox{fm}$.  \\
   $^b$ 
        $\alpha_\msbar^{(5)}(M_Z)=0.1181(27)(^{+8}_{-22})$. $\Lambda_\msbar$
        determined by us from $\alpha_\msbar^{(3)}(2\,\mbox{GeV})=0.2961(185)$.
        In conversion to $r_0\Lambda$ we used
        $r_0 = 0.472\,\mbox{fm}$.  \\
        $^+$ Determined by us from $\alpha_\msbar^{(3)}(2\,\GeV)=0.279(11)$. 
       Evaluates to $\alpha_\msbar^{(5)}(M_Z)=0.1155(18)$. \\
   $^\dagger$  $\alpha_\msbar^{(5)}(M_Z)=0.1118(3)(^{+16}_{-17})$. \\
   
\end{flushleft}}}
\end{tabular*}
\vspace{-0.3cm}
\normalsize
\caption{Results from the  vacuum polarization in both momentum and position space.}
\label{tab_vac}
\end{table}

We first discuss the results of JLQCD/TWQCD 08C 
\cite{Shintani:2008ga} and JLQCD 10 \cite{Shintani:2010ph}.
The fit to \eq{eq:vacpol} is done with the 4-loop relation between
the running coupling and $\lms$. It is found that without introducing fit parameters for
condensate contributions, the momentum scale where the perturbative
formula gives good agreement with the lattice results is very narrow,
$aQ \simeq$ 0.8--1.0. When fit parameters for condensate contributions are included the
perturbative formula gives good agreement with the lattice results for
the extended range $aQ \simeq$ 0.6--1.0. Since there is only a single
lattice spacing $a \approx$ 0.11~fm there is a 
\bad\ for the continuum limit. The renormalization scale $\mu$ is in
the range of $Q=$ 1.6--2~GeV. Approximating 
$\alpha_{\rm eff}\approx \alpha_{\overline{\rm MS}}(Q)$, we estimate that
$\alpha_{\rm eff}=$ 0.25--0.30 for $\Nf=2$ and $\alpha_{\rm  eff}=$ 0.29--0.33
for $\Nf=2+1$. Thus we give a \soso\ and \bad\ for $\Nf=2$ and 
$\Nf=2+1$, respectively, for the renormalization scale and a \bad\ for
the perturbative behaviour.

A further investigation of this method was initiated in
Hudspith 15 \cite{Hudspith:2015xoa} and completed by Hudspith 18 
\cite{Hudspith:2018bpz} (see also \cite{Hudspith:2018zlq}) based
on domain-wall fermion configurations at three lattice spacings, 
$a^{-1} =$ 1.78, 2.38, 3.15~GeV, with three different 
light-quark masses on the two coarser lattices and one on the fine lattice.
An extensive discussion of condensates, using continuum
finite-energy sum rules was employed to estimate where their contributions
might be negligible. It was found that even up to terms
of $O((1/Q^2)^8)$ (a higher order than depicted in Eq.~(\ref{eq:vacpol})
but with constant coefficients) 
no single condensate dominates
and apparent convergence was poor for low $Q^2$
due to cancellations between contributions of similar size
with alternating signs. (See, e.g., \ the list given by Hudspith 15
\cite{Hudspith:2015xoa}.) Choosing $Q^2$ to be at least
$\sim 3.8\,\mbox{GeV}^2$ mitigated the problem, but then the coarsest
lattice had to be discarded, due to large lattice artefacts.
So this gives a $\bad$ for continuum extrapolation.
With the higher $Q^2$ the quark-mass dependence of the
results was negligible, so ensembles with different quark masses were
averaged over.
A range of $Q^2$ from 3.8--16~GeV$^2$ gives 
$\alpha_{\rm eff}$ = 0.31--0.22,  so there is a $\soso$ for the
renormalization scale.
The value of $\alpha_{\rm eff}^3$ reaches
$\Delta \alpha_{\rm eff}/(8\pi b_0 \alpha_{\rm eff})$ and thus
gives a $\soso$ for perturbative behaviour.
In Hudspith 15 \cite{Hudspith:2015xoa} (superseded by Hudspith 18 
\cite{Hudspith:2018bpz}) about a 20\% difference in 
$\Pi_V(Q^2)$ was seen between the two lattice spacings and a
result is quoted only for the smaller $a$.

\subsubsection{Vacuum polarization in position space}

Cali 20~\cite{Cali:2020hrj} evaluate the light-current two-point function in position space.
The two-point functions for the nonperturbatively renormalized (nonsinglet) flavour currents is computed for
distances $|x|$ between $0.1$ and $0.25$~fm and extrapolated to the chiral limit. 
The available CLS configurations are used for this work, with lattice spacings between $0.039$ and $0.086\,\fm$. 
Despite fully nonperturbative renormalization and $\cO(a)$ improvement,
the remaining $\cO(a^2)$ effects, as measured by $O(4)$ symmetry violations, are very large, 
even after subtraction of tree-level lattice effects. 
Therefore the authors performed a numerical stochastic perturbation theory (NSPT) simulation in order to determine the lattice artifacts at
$\cO(g^2)$. Only after subtraction of these effects the constrained continuum extrapolations from three different lattice directions
to the same continuum limit are characterized by reasonable $\chi^2$-values, so the feasibility of the study crucially depends
on this step. Interestingly, there is no subtraction performed of nonperturbative effects. For instance, 
chiral symmetry breaking would manifest itself in a difference between the vector and the axial-vector two-point functions, 
and is invisible to perturbation theory, where these two-point functions are known to $\alpha_s^4$~\cite{Chetyrkin:2010dx}. 
According to the authors, phenomenological estimates suggest that a difference of 1.5\% between 
the continuum correlators would occur around 0.3~fm and this difference would not be resolvable by their lattice data. 
Equality within their errors is confirmed for shorter distances. We note, however, that chiral symmetry breaking effects
are but one class of nonperturbative effects, and their smallness does not allow for the conclusion that
such effects are generally small. In fact, the need for explicit subtractions in momentum space analyses may
lead one to suspect that such effects are not negligible at the available distance scales.
For the determination of $\lms^{(\Nf=3)}$ the authors limit the range of distances to 0.13--0.19~fm,
where $\alpha_\text{eff} \in [0.2354,0.3075]$ (private communication by the authors).
These effective couplings are converted to $\msbar$ couplings at the same scales $\mu=1/|x|$ 
by solving Eq.~(\ref{eq:CVAMSbar}) numerically. Central values for the $\Lambda$-parameter thus obtained 
are in the range 325--370~MeV (using the $\beta$-function at 5-loop order) and a weighted average
yields the quoted result 342(17)~MeV, where the average emphasizes the data  around $|x|=$ 0.16~fm, 
or $\mu=$ 1.3~GeV. 

Applying the FLAG criteria the range of lattice spacings yields $\good$ for the continuum extrapolation. However, the FLAG criterion
implicitly assumes that the remaining cutoff effects after nonperturbative $\cO(a)$ improvement are small, which is 
not the case here. Some hypercubic lattice artefacts are still rather large even after 1-loop subtraction, but these are
not used for the analysis.
As for the renormalization scale, the lowest effective coupling entering the analysis is $0.235<0.25$, so we give $\soso$.
As for perturbative behaviour, for the range of couplings in the above interval $\alpha_\text{eff}^3$ 
changes by $(0.308/0.235)^3\approx 2.2$, marginally reaching $(3/2)^2=2.25$.
The errors $\Delta \alpha_\text{eff}$ after continuum and chiral extrapolations are 4--6\% (private
communication by the authors) and the induced uncertainty in $\Lambda$ is comfortably 
above $2\alpha_\text{eff}^3$, which gives a $\good$ according to FLAG criteria.

Although the current FLAG criteria are formally passed by this result, the quoted error of 5\% for $\Lambda$ 
seems very optimistic. We have performed a simple test, converting to the $\msbar$ scheme 
by inverting Eq.~(\ref{eq:CVAMSbar}) perturbatively (instead of solving the fixed-order equation numerically).
The differences between the couplings are of order $\alpha_s^5$ and thus indicative 
of the sensitivity to perturbative truncation
errors. The resulting $\Lambda$-parameter estimates are now in the range 409--468~MeV, i.e., ca.~15--30\%
larger than before. While the difference between both estimates decreases proportionally to the expected 
$\alpha_\text{eff}^3$, an extraction of the $\Lambda$-parameter in this energy range is a priori affected by
systematic uncertainties corresponding to such differences. 
The FLAG criterion might fail to capture this, e.g., if the assumption of an $\cO(1)$ coefficient for
the asymptotic $\alpha_\text{eff}^3$ behaviour is not correct. Some indication
for a problematic behaviour is indeed seen when perturbatively inverting Eq.~(\ref{eq:CVAMSbar}) 
to order $\alpha_s^3$. The resulting $\msbar$ couplings are then closer to the values used in Cali~20,
although the difference is formally  $O(\alpha_s^4)$  rather than $O(\alpha_s^5)$. 

\paragraph{Scale variations.}
Using scale variations to determine the systematic error due to the truncation of the perturbative 
series only makes sense when the extrapolation of the observable to the continuum limit is under 
control. Therefore, we apply our common procedure only to the results in Cali 20~\cite{Cali:2020hrj}.
Using $\mu\approx 1.3~\mathrm{GeV}$ as the typical scale set by the inverse of the distance, yields
\begin{eqnarray}
   \delta^*_{(4)} = 1.0\%\, , \quad 
   \delta_{(2)} = 11.6\%\, \quad
   \delta^*_{(2)} = 0.6\%\, .
\end{eqnarray}
The discrepancy between the variation around $Q$, $\delta_{(2)} = 11.6\%$, and the variation
around the scale of fastest apparent convergence, $\delta^*_{(2)} = 0.6\%$, is due to 
the large value of the factor $s^*_{\mathrm{ref}}=2.72$. As a consequence the scale of fastest apparent
convergence is artificially large compared to the actual scale where the lattice observables is computed.  
The large value of $\delta_{(2)}$, obtained for $s_{\mathrm{ref}}=1$, shows that the scale of the 
lattice observable is too low to keep the systematic errors under control. 

%% file: Alpha_s/s6.tex
\subsection{$\alpha_s$ from observables at the lattice spacing scale}
\label{s:WL}


\subsubsection{General considerations}


The general method is to evaluate a short-distance quantity ${\oO}$
at the scale of the lattice spacing $\sim 1/a$ and then determine
its relationship to $\alpha_{\overline{\rm MS}}$ via a 
perturbative 
expansion.

This is epitomized by the strategy of the HPQCD collaboration
\cite{Mason:2005zx,Davies:2008sw}, discussed here for illustration,
which computes and then fits to a variety of short-distance quantities
\begin{eqnarray}
   Y = \sum_{n=1}^{n_{\rm max}} c_n \alphah^n(q^*) \,.
\label{Ydef}
\end{eqnarray}
The quantity $Y$ is taken as the logarithm of small Wilson loops (including some
nonplanar ones), Creutz ratios, `tadpole-improved' Wilson loops and
the tadpole-improved or `boosted' bare coupling ($\cO(20)$ quantities in
total). The perturbative coefficients $c_n$  (each depending on the
choice of $Y$) are known to $n = 3$ with additional coefficients up to
$n_{\rm max}$ being fitted numerically.   The running
coupling $\alphah$ is related to $\alphav$ from the static-quark potential
(see Sec.~\ref{s:qq}).\footnote{ $\alphah$ is defined by
  $\Lambda_\mathrm{V'}=\Lambda_\mathrm{V}$ and
  $b_i^\mathrm{V'}=b_i^\mathrm{V}$ for $i=0,1,2$ but $b_i^\mathrm{V'}=0$ for
  $i\geq3$. }

 The coupling
constant is fixed at a scale $q^* = d/a$.
The latter  is chosen as the mean value of $\ln q$ with the one-gluon loop
as measure
\cite{Lepage:1992xa,Hornbostel:2002af}. (Thus a different result
for $d$ is found for every short-distance quantity.)
A rough estimate yields $d \approx \pi$, and in general the
renormalization scale is always found to lie in this region.

For example, for the Wilson loop $W_{mn} \equiv \langle W(ma,na) \rangle$
we have
\begin{eqnarray}
   \ln\left( \frac{W_{mn}}{u_0^{2(m+n)}}\right)
      = c_1 \alphah(q^*) +  c_2 \alphah^2(q^*)  + c_3 \alphah^3(q^*)
        + \cdots \,,
\label{short-cut}
\end{eqnarray}
for the tadpole-improved version, where $c_1$, $c_2\,, \ldots$
are the appropriate perturbative coefficients and $u_0 = W_{11}^{1/4}$.
Substituting the nonperturbative simulation value in the left hand side,
we can determine $\alphah(q^*)$, at the scale $q^*$.
Note that one finds empirically that perturbation theory for these
tadpole-improved quantities have smaller $c_n$ coefficients and so
the series has a faster apparent convergence compared to the case
without tadpole improvement.

Using the $\beta$-function in the $\rm V'$ scheme,
results can be run to a reference value, chosen as
$\alpha_0 \equiv \alphah(q_0)$, $q_0 = 7.5\,\mbox{GeV}$.
This is then converted perturbatively to the continuum
$\msbar$ scheme
\begin{eqnarray}
   \alpha_{\overline{\rm MS}}(q_0)
      = \alpha_0 + d_1 \alpha_0^2 + d_2 \alpha_0^3 + \cdots \,,
\end{eqnarray}
where $d_1, d_2$ are known 1-and 2-loop coefficients.

Other collaborations have focused more on the bare `boosted'
coupling constant and directly determined its relationship to
$\alpha_{\overline{\rm MS}}$. Specifically, the boosted coupling is
defined by 
\begin{eqnarray}
   \alphap(1/a) = {1\over 4\pi} {g_0^2 \over u_0^4} \,,
\end{eqnarray}
again determined at a scale $\sim 1/a$. As discussed previously, 
since the plaquette expectation value in the boosted coupling
contains the tadpole-diagram contributions to all orders, which
are dominant contributions in perturbation theory,
there is an expectation that the perturbation theory using
the boosted coupling has 
smaller perturbative coefficients \cite{Lepage:1992xa}, and hence smaller 
perturbative errors.
 

\subsubsection{Continuum limit}


Lattice results always come along with discretization errors,
which one needs to remove by a continuum extrapolation.
As mentioned previously, in this respect the present
method differs in principle from those in which $\alpha_s$ is determined
from physical observables. In the general case, the numerical
results of the lattice simulations at a value of $\mu$ fixed in physical 
units can be extrapolated to the continuum limit, and the result can be 
analyzed as to whether it shows perturbative running as a function of 
$\mu$ in the continuum. For observables at the cutoff-scale ($q^*=d/a$),  
discretization effects cannot easily be separated out
from perturbation theory, as the scale for the coupling
comes from the lattice spacing. 
Therefore the restriction  $a\mu  \ll 1$ (the `continuum-extrapolation'
criterion) is not applicable here. Discretization errors of 
order $a^2$ are, however, present. Since 
$a\sim \exp(-1/(2b_0 g_0^2)) \sim \exp(-1/(8\pi b_0 \alpha(q^*))$, 
these errors now appear as power corrections to the perturbative 
running, and have to be taken into account in the study of the 
perturbative behaviour, which is to be verified by changing $a$. 
One thus usually fits with power corrections in this method.

In order to keep a symmetry with the `continuum-extrapolation' 
criterion for physical observables and to remember that discretization 
errors are, of course, relevant, 
we replace it here by one for the lattice spacings used:
\begin{itemize}
   \item Lattice spacings
         \begin{itemize}
            \item[\good] 
               3 or more lattice spacings, at least 2 points below
               $a = 0.1\,\mbox{fm}$
            \item[\soso]
               2 lattice spacings, at least 1 point below
               $a = 0.1\,\mbox{fm}$
            \item[\bad]
               otherwise 
         \end{itemize}
\end{itemize}


\subsubsection{Discussion of computations}

\begin{table}[!p]
   \vspace{3.0cm}
   \footnotesize
   \begin{tabular*}{\textwidth}[l]{l@{\extracolsep{\fill}}rllllllll}
   Collaboration & Ref. & $\Nf$ &
   \hspace{0.15cm}\begin{rotate}{60}{publication status}\end{rotate}
                                                    \hspace{-0.15cm} &
   \hspace{0.15cm}\begin{rotate}{60}{renormalization scale}\end{rotate}
                                                    \hspace{-0.15cm} &
   \hspace{0.15cm}\begin{rotate}{60}{perturbative behaviour}\end{rotate}
                                                    \hspace{-0.15cm} &
   \hspace{0.15cm}\begin{rotate}{60}{lattice spacings}\end{rotate}
      \hspace{-0.25cm} & 
                         scale & $\Lambda_\msbar[\MeV]$ & $r_0\Lambda_\msbar$ \\
   & & & & & & & & \\[-0.1cm]
   \hline
   \hline
   & & & & & & & & \\[-0.1cm]
   HPQCD 10$^a$$^\S$& \cite{McNeile:2010ji}& 2+1 & \gA & \soso
            & \good & \good
            & $r_1 = 0.3133(23)\, \mbox{fm}$
            & 340(9) 
            & 0.812(22)                                   \\ 
   HPQCD 08A$^a$& \cite{Davies:2008sw} & 2+1 & \gA & \soso
            & \good & \good
            & $r_1 = 0.321(5)\,\mbox{fm}$$^{\dagger\dagger}$
            & 338(12)$^\star$
            & 0.809(29)                                   \\
   Maltman 08$^a$& \cite{Maltman:2008bx}& 2+1 & \gA & \soso
            & \soso & \good
            & $r_1 = 0.318\, \mbox{fm}$
            & 352(17)$^\dagger$
            & 0.841(40)                                   \\ 
   HPQCD 05A$^a$ & \cite{Mason:2005zx} & 2+1 & \gA & \soso
            & \soso & \soso
            & $r_1$$^{\dagger\dagger}$
            & 319(17)$^{\star\star}$
            & 0.763(42)                                   \\
   & & & & & & & & &  \\[-0.1cm]
   \hline
   & & & & & & & & &  \\[-0.1cm]
   QCDSF/UKQCD 05 & \cite{Gockeler:2005rv}  & 2 & \gA & \good
            & \bad  & \good
            & $r_0 = 0.467(33)\,\mbox{fm}$
            & 261(17)(26)
            & 0.617(40)(21)$^b$                           \\
   SESAM 99$^c$ & \cite{Spitz:1999tu} & 2 & \gA & \soso
            & \bad  & \bad
            & $c\bar{c}$(1S-1P)
            & 
            &                                             \\
   Wingate 95$^d$ & \cite{Wingate:1995fd} & 2 & \gA & \good
            & \bad  & \bad
            & $c\bar{c}$(1S-1P)
            & 
            &                                             \\
   Davies 94$^e$ & \cite{Davies:1994ei} & 2 & \gA & \good
            & \bad & \bad
            & $\Upsilon$
            & 
            &                                             \\
   Aoki 94$^f$ & \cite{Aoki:1994pc} & 2 & \gA & \good
            & \bad & \bad
            & $c\bar{c}$(1S-1P)
            & 
            &                                             \\
   & & & & & & & & &  \\[-0.1cm]
   \hline
   & & & & & & & & &  \\[-0.1cm]

   {Kitazawa 16}
            & \cite{Kitazawa:2016dsl}        & 0 & \gA
            & \good  & \good   & \good
            & $w_0$
            & $260(5)$$^j$
            & $0.621(11)$$^j$                            \\

   FlowQCD 15
            & \cite{Asakawa:2015vta}        & 0 & \oP 
            & \good  & \good   & \good
            & $w_{0.4}$$^i$
            & $258(6)$$^i$
            & 0.618(11)$^i$                             \\

   QCDSF/UKQCD 05 & \cite{Gockeler:2005rv}  & 0 & \gA & \good
            & \soso & \good
            & $r_0 = 0.467(33)\,\mbox{fm}$
            & 259(1)(20)
            & 0.614(2)(5)$^b$                              \\
   SESAM 99$^c$ & \cite{Spitz:1999tu} & 0 & \gA & \good
            & \bad  & \bad
            & $c\bar{c}$(1S-1P)
            & 
            &                                             \\
   Wingate 95$^d$ & \cite{Wingate:1995fd} & 0 & \gA & \good
            & \bad  & \bad
            & $c\bar{c}$(1S-1P)
            & 
            &                                             \\
   Davies 94$^e$ & \cite{Davies:1994ei}  & 0 & \gA & \good
            & \bad & \bad
            & $\Upsilon$
            & 
            &                                             \\
   El-Khadra 92$^g$ & \cite{ElKhadra:1992vn} & 0 & \gA & \good
            & \bad    & \soso
            & $c\bar{c}$(1S-1P)
            & 234(10)
            & 0.560(24)$^h$                               \\
   & & & & & & & & &  \\[-0.1cm]
   \hline
   \hline\\
\end{tabular*}\\[-0.2cm]
\begin{minipage}{\linewidth}
{\footnotesize 
\begin{itemize}
   \item[$^a$]The numbers for $\Lambda$ have been converted from the values for 
              $\alpha_s^{(5)}(M_Z)$. \\[-5mm]
   \item[$^{\S}$]     $\alpha_{\overline{\rm MS}}^{(3)}(5\ \mbox{GeV})=0.2034(21)$,
              $\alpha^{(5)}_{\overline{\rm MS}}(M_Z)=0.1184(6)$,
              only update of intermediate scale and $c$-, $b$-quark masses,
              supersedes HPQCD 08A.\\[-5mm]
   \item[$^\dagger$] $\alpha^{(5)}_{\overline{\rm MS}}(M_Z)=0.1192(11)$. \\[-4mm]
   \item[$^\star$]    $\alpha_V^{(3)}(7.5\,\mbox{GeV})=0.2120(28)$, 
              $\alpha^{(5)}_{\overline{\rm MS}}(M_Z)=0.1183(8)$,
              supersedes HPQCD 05. \\[-5mm]
   \item[$^{\dagger\dagger}$] Scale is originally determined from $\Upsilon$
              mass splitting. $r_1$ is used as an intermediate scale.
              In conversion to $r_0\Lambda_{\overline{\rm MS}}$, $r_0$ is
              taken to be $0.472\,\mbox{fm}$. \\[-5mm]
   \item[$^{\star\star}$] $\alpha_V^{(3)}(7.5\,\mbox{GeV})=0.2082(40)$,
              $\alpha^{(5)}_{\overline{\rm MS}}(M_Z)=0.1170(12)$. \\[-5mm]
   \item[$^b$]       This supersedes 
              Refs.~\cite{Gockeler:2004ad,Booth:2001uy,Booth:2001qp}.
              $\alpha^{(5)}_{\overline{\rm MS}}(M_Z)=0.112(1)(2)$.
              The $\Nf=2$ results were based on values for $r_0 /a$
              which have later been found to be too 
              small~\cite{Fritzsch:2012wq}. The effect will  
              be of the order of 10--15\%, presumably an increase in 
              $\Lambda r_0$. \\[-5mm]
   \item[$^c$]       $\alpha^{(5)}_{\overline{\rm MS}}(M_Z)=0.1118(17)$. \\[-4mm]
   \item[$^d$]    
   $\alpha_V^{(3)}(6.48\,\mbox{GeV})=0.194(7)$ extrapolated from $\Nf=0,2$.
              $\alpha^{(5)}_{\overline{\rm MS}}(M_Z)=0.107(5)$.   \\[-4mm]
   \item[$^e$]  
              $\alpha_P^{(3)}(8.2\,\mbox{GeV})=0.1959(34)$ extrapolated
              from $\Nf=0,2$. $\alpha^{(5)}_{\overline{\rm MS}}(M_Z)=0.115(2)$.
              \\[-5mm]
   \item[ $^f$]Estimated $\alpha^{(5)}_{\overline{\rm MS}}(M_Z)=0.108(5)(4)$. \\[-5mm]
   \item[$^g$]       This early computation violates our requirement that
              scheme conversions are done at the 2-loop level.
              $\Lambda_{\overline{\rm MS}}^{(4)}=160(^{+47}_{-37})\mbox{MeV}$, 
              $\alpha^{(4)}_{\overline{\rm MS}}(5\mbox{GeV})=0.174(12)$.
              We converted this number to give
              $\alpha^{(5)}_{\overline{\rm MS}}(M_Z)=0.106(4)$.  \\[-5mm]
   \item[$^h$]We used $r_0=0.472\,\mbox{fm}$ to convert to $r_0 \lms$. \\[-5mm]
   \item[$^i$]       Reference scale $w_{0.4}$ where $w_x$ is defined 
              by $\left. t\partial_t[t^2 \langle E(t)\rangle]\right|_{t=w_x^2}=x$
              in terms of the action density $E(t)$ at positive flow time $t$ 
              \cite{Asakawa:2015vta}. Our conversion to $r_0$ scale
              using \cite{Asakawa:2015vta} $r_0/w_{0.4}=2.587(45)$ and
              $r_0=0.472\,\mbox{fm}$. 
   \item[$^j$]{Our conversion from $w_0\Lambda_\msbar=0.2154(12)$ to $r_0$ scale
              using  $r_0/w_0=(r_0/w_{0.4}) \cdot   (w_{0.4}/w_0) = 2.885(50)$ 
              with the factors cited by the collaboration \cite{Asakawa:2015vta} 
              and with $r_0=0.472\,\mbox{fm}$. }
\end{itemize}
}
\end{minipage}
\normalsize
\caption{Wilson loop results. Some early results for $\Nf=0, 2$ did not determine  $\Lambda_\msbar$.
}
\label{tab_wloops}
\end{table}

Note that due to $\mu \sim 1/a$ being relatively large the
results easily have a $\good$ or $\soso$ in the rating on 
renormalization scale.

The work of El-Khadra 92 \cite{ElKhadra:1992vn} employs a 1-loop
formula to relate $\alpha^{(0)}_{\overline{\rm MS}}(\pi/a)$
to the boosted coupling for three lattice spacings
$a^{-1} = 1.15$, $1.78$, $2.43\,\mbox{GeV}$. (The lattice spacing
is determined from the charmonium 1S-1P splitting.) They obtain
$\Lambda^{(0)}_{\overline{\rm MS}}=234\,\mbox{MeV}$, corresponding
to $\alpha_{\rm eff} = \alpha^{(0)}_{\overline{\rm MS}}(\pi/a)
\approx$ 0.15--0.2. The work of Aoki 94 \cite{Aoki:1994pc}
calculates $\alpha^{(2)}_V$ and $\alpha^{(2)}_{\overline{\rm MS}}$
for a single lattice spacing $a^{-1}\sim 2\,\mbox{GeV}$,  again
determined from charmonium 1S-1P splitting in two-flavour QCD.
Using 1-loop perturbation theory with boosted coupling,
they obtain $\alpha^{(2)}_V=0.169$ and $\alpha^{(2)}_{\overline{\rm MS}}=0.142$.
Davies 94 \cite{Davies:1994ei} gives a determination of $\alphav$
from the expansion 
\begin{equation}
   -\ln W_{11} \equiv \frac{4\pi}{3}\alphav^{(\Nf)}(3.41/a)
        \times [1 - (1.185+0.070\Nf)\alphav^{(\Nf)} ]\,,
\end{equation}
neglecting higher-order terms.  They compute the $\Upsilon$ spectrum
in $\Nf=0$, $2$ QCD for single lattice spacings at $a^{-1} = 2.57$,
$2.47\,\mbox{GeV}$ and obtain $\alphav(3.41/a)\simeq$ 0.15, 0.18, respectively.  Extrapolating the inverse coupling linearly in $\Nf$, a
value of $\alphav^{(3)}(8.3\,\mbox{GeV}) = 0.196(3)$ is obtained.
SESAM 99 \cite{Spitz:1999tu} follows a similar strategy, again for a
single lattice spacing. They linearly extrapolated results for
$1/\alphav^{(0)}$, $1/\alphav^{(2)}$ at a fixed scale of
$9\,\mbox{GeV}$ to give $\alphav^{(3)}$, which is then perturbatively
converted to $\alpha_{\overline{\rm MS}}^{(3)}$. This finally gave
$\alpha_{\overline{\rm MS}}^{(5)}(M_Z) = 0.1118(17)$.  Wingate 95
\cite{Wingate:1995fd} also follows this method.  With the scale
determined from the charmonium 1S-1P splitting for single lattice
spacings in $\Nf = 0$, $2$ giving $a^{-1}\simeq 1.80\,\mbox{GeV}$ for
$\Nf=0$ and $a^{-1}\simeq 1.66\,\mbox{GeV}$ for $\Nf=2$, they obtain
$\alphav^{(0)}(3.41/a)\simeq 0.15$ and $\alphav^{(2)}\simeq 0.18$, 
respectively. Extrapolating the inverse coupling linearly in $\Nf$, they
obtain $\alphav^{(3)}(6.48\,\mbox{GeV})=0.194(17)$.

The QCDSF/UKQCD collaboration, QCDSF/UKQCD 05
\cite{Gockeler:2005rv}, \cite{Gockeler:2004ad,Booth:2001uy,Booth:2001qp},
use the 2-loop relation (re-written here in terms of $\alpha$)
\begin{eqnarray}
   {1 \over \alpha_{\overline{\rm MS}}(\mu)} 
      = {1 \over \alphap(1/a)} 
        + 4\pi(2b_0\ln a\mu - t_1^{\rm P}) 
        + (4\pi)^2(2b_1\ln a\mu - t_2^{\rm P})\alphap(1/a) \,,
\label{gPtoMSbar}
\end{eqnarray}
where $t_1^{\rm P}$ and $t_2^{\rm P}$ are known. (A 2-loop relation corresponds
to a 3-loop lattice $\beta$-function.)  This was used to
directly compute $\alpha_{\rm \overline{\rm MS}}$, and the scale was
chosen so that the $\cO(\alphap^0)$ term vanishes, i.e., \
\begin{eqnarray}
   \mu^* = {1 \over a} \exp{[t_1^{\rm P}/(2b_0)] } 
        \approx \left\{ \begin{array}{cc}
                           2.63/a  & \Nf = 0 \\
                           1.4/a   & \Nf = 2 \\
                        \end{array}
                 \right. \,.
\label{amustar}
\end{eqnarray}
The method is to first compute $\alphap(1/a)$ and from this,  using
Eq.~(\ref{gPtoMSbar}) to find $\alpha_{\overline{\rm MS}}(\mu^*)$.
The RG equation, Eq.~(\ref{eq:Lambda}), then determines
$\mu^*/\Lambda_{\overline{\rm MS}}$ and hence using
Eq.~(\ref{amustar}) leads to the result for $r_0\Lambda_{\overline{\rm MS}}$.
This avoids giving the scale in $\mbox{MeV}$ until the end.
In the $\Nf=0$ case seven lattice spacings were used
\cite{Necco:2001xg}, giving a range $\mu^*/\Lambda_{\overline{\rm MS}}
\approx$ 24--72 (or $a^{-1} \approx$ 2--7~GeV) and
$\alpha_{\rm eff} = \alpha_{\overline{\rm MS}}(\mu^*) \approx$ 0.15--0.10. Neglecting higher-order perturbative terms (see discussion
after Eq.~(\ref{qcdsf:ouruncert}) below) in Eq.~(\ref{gPtoMSbar}) this
is sufficient to allow a continuum extrapolation of
$r_0\Lambda_{\overline{\rm MS}}$.
A similar computation for $\Nf = 2$ by QCDSF/UKQCD~05 \cite{Gockeler:2005rv}
gave $\mu^*/\Lambda_{\overline{\rm MS}} \approx$ 12--17
(or roughly $a^{-1} \approx$ 2--3~GeV) 
and $\alpha_{\rm eff} = \alpha_{\overline{\rm MS}}(\mu^*)
\approx$ 0.20--0.18.
The $\Nf=2$ results of QCDSF/UKQCD~05 \cite{Gockeler:2005rv} are affected by an 
uncertainty which was not known at the time of publication: 
It has been realized that the values of $r_0/a$ of Ref.~\cite{Gockeler:2005rv}
were significantly too low~\cite{Fritzsch:2012wq}. 
As this effect is expected to depend on $a$, it
influences the perturbative behaviour leading us to assign 
a \bad\ for that criterion. 

Results for the $\Nf = 0$ $\Lambda$-parameter 
by FlowQCD 15 \cite{Asakawa:2015vta}, later 
updated and published in Kitazawa 16 \cite{Kitazawa:2016dsl}, are obtained
following the same strategy,
cf.~Eqs.~(\ref{gPtoMSbar}), (\ref{amustar}), except that the scale $r_0$ is replaced by the gradient-flow scale $w_0$, 
leading to a determination of $w_0\Lambda_{\overline{\rm MS}}$.
The continuum limit is estimated by extrapolating the data at six
lattice spacings linearly in $a^2$. The data range used is
$\mu^*/\Lambda_{\overline{\rm MS}} \approx$ 50--120 (or 
$a^{-1} \approx$ 5--11~GeV) and
$\alpha_{\overline{\rm MS}}(\mu^*) \approx$ 0.12--0.095.
Since a very small value of $\alpha_\msbar$ is reached, there is a $\good$ 
in the perturbative behaviour. Note that our conversion to the common
$r_0$ scale unfortunately leads to a significant increase of the error of the
$\Lambda$ parameter compared to using $w_0$ directly \cite{Sommer:2014mea}. 
Again we note that the results of QCDSF/UKQCD 05
\cite{Gockeler:2005rv} ($\Nf = 0$) and Kitazawa 16 \cite{Kitazawa:2016dsl}
may be affected by frozen topology as they have
lattice spacings significantly below $a = 0.05\,\mbox{fm}$.
Kitazawa 16 \cite{Kitazawa:2016dsl} investigate this by evaluating
$w_0/a$ in a fixed topology and estimate any effect at about $\sim 1\%$.

The work of HPQCD 05A \cite{Mason:2005zx} (which supersedes
the original work \cite{Davies:2003ik}) uses three lattice spacings
$a^{-1} \approx 1.2$, $1.6$, $2.3\,\mbox{GeV}$ for $2+1$
flavour QCD. Typically the renormalization scale
$q \approx \pi/a \approx$ 3.50--7.10~GeV, corresponding to
$\alpha_\mathrm{V'} \approx$ 0.22--0.28. 

In the later update HPQCD 08A \cite{Davies:2008sw} twelve data sets
(with six lattice spacings) are now used reaching up to $a^{-1}
\approx 4.4\,\mbox{GeV}$,  corresponding to $\alpha_\mathrm{V'}\approx
0.18$. The values used for the scale $r_1$ were further updated in
HPQCD 10 \cite{McNeile:2010ji}. Maltman 08 \cite{Maltman:2008bx}
uses most of the same lattice ensembles as HPQCD
08A~\cite{Davies:2008sw}, but not the one at the smallest lattice spacing, $a\approx0.045$~fm. Maltman 08 \cite{Maltman:2008bx} also
considers a much smaller set of
quantities (three versus 22) that are less sensitive to condensates.
They also use different strategies for evaluating the condensates and
for the perturbative expansion, and a slightly different value for the
scale $r_1$. The central values of the final results from 
Maltman 08 \cite{Maltman:2008bx} and HPQCD 08A \cite{Davies:2008sw}
differ by 0.0009 (which would be decreased to 0.0007
taking into account a reduction of 0.0002 in the value of the $r_1$
scale used by Maltman 08 \cite{Maltman:2008bx}).
 
As mentioned before, the perturbative coefficients are computed
through $3$-loop order~\cite{Mason:2004zt}, while the higher-order
perturbative coefficients $c_n$ with $ n_{\rm max} \ge n > 3$ (with
$n_{\rm max} = 10$) are numerically fitted using the
lattice-simulation data for the lattice spacings with the help of
Bayesian methods.  It turns out that corrections in \eq{short-cut} are
of order $|c_i/c_1|\alpha^i=$ 5--15\% and 3--10\% for $i$ = 2, 3,
respectively.  The inclusion of a fourth-order term is necessary to
obtain a good fit to the data, and leads to a shift of the result by
$1$--$2$ sigma. For all but one of the 22 quantities, central values
of $|c_4/c_1|\approx$ 2--4 were found, with errors from the fits of
$\approx 2$. 
It should be pointed out that the description of lattice results for the short-distance quantities 
does not require Bayesian priors, once the term proportional to $c_4$ is included \cite{Maltman:2008bx}. 
We also stress that different short-distance quantities have quite different nonperturbative contributions \cite{Maltman:2010zza}.
Hence the fact that different observables lead to consistent $\alpha_s$ values is a nontrivial check of the
approach.

An important source of uncertainty is the truncation 
of perturbation theory. In HPQCD 08A \cite{Davies:2008sw}, HPQCD 10
\cite{McNeile:2010ji} it
is estimated to be about $0.4$\% of $\alpha_\msbar(M_Z)$.  In \flagold\
we included a rather detailed discussion of the issue with the result
that we prefer for the time being a more conservative error
based on the above estimate $|c_4/c_1| = 2$. 
From Eq.~(\ref{Ydef}) this gives an estimate of the uncertainty
in $\alpha_{\rm eff}$ of
\begin{eqnarray}
  \Delta \alpha_{\rm eff}(\mu_1) = 
          \left|{c_4 \over c_1}\right|\alpha_{\rm eff}^4(\mu_1) \,,
\label{qcdsf:ouruncert}
\end{eqnarray}
at the scale $\mu_1$ where $\alpha_{\rm eff}$ is computed from
the Wilson loops. This can be used with a variation
in $\Lambda$ at lowest order of perturbation theory and also
applied to $\alpha_s$ evolved to a different scale $\mu_2$,%
\footnote{From Eq.~(\ref{e:grelation}) we see that at low order in PT
the coupling $\alpha_s$ is continuous and differentiable across
the mass thresholds (at the same scale). Therefore 
to leading order $\alpha_s$ and $\Delta \alpha_s$
are independent of $\Nf$.}
\begin{eqnarray}
   {\Delta\Lambda \over \Lambda} 
      = {1\over 8\pi b_0 \alpha_s} 
                  {\Delta \alpha_s \over \alpha_s}
                                                         \,, \qquad
   {\Delta \alpha_s(\mu_2) \over \Delta \alpha_s(\mu_1)}
      = {\alpha_s^2(\mu_2) \over \alpha_s^2(\mu_1)} \,.
   \label{e:dLL}   
\end{eqnarray}
With $\mu_2 = M_Z$
and $\alpha_s(\mu_1)=0.2$ (a typical value extracted 
from Wilson loops in HPQCD 10 \cite{McNeile:2010ji}, HPQCD 08A
\cite{Davies:2008sw} at $\mu = 5\,\mbox{GeV}$) we have 
\begin{eqnarray}
  \Delta \alpha_\msbar(m_Z) = 0.0012 \,,
\label{hpqcd:ouruncert}
\end{eqnarray}
which we shall later use as the typical perturbative uncertainty 
of the method with $2+1$ fermions.
}

Table~\ref{tab_wloops} summarizes the results. Within the errors of 3--5\% $\Nf=3$ determinations of $r_0 \Lambda$ nicely agree.

\paragraph{Scale variations.} 
As discussed above, the short-distance observables are fitted to a perturbative expansion where 
the higher-order coefficients are actual parameters in the fit. Here instead we follow the exact 
same procedure introduced for all the observables, and we describe the observables using only the 
known perturbative coefficients. For illustration, we report the result of the scale variations 
for two observables, namely the simple $1\times 1$ plaquette and the $2\times 1$ Wilson loop. 
The perturbative coefficients are reported in Tab.~\ref{tab:scale_truncation} and the typical scale
is $\mu \approx 2.4/a \approx 4.4~\mathrm{GeV}$. With these values we obtain the following results.
\begin{description}
   \item[$-\log W_{11}$]
   \begin{eqnarray}
      \delta^*_{(4)} = 2.8\%\, , \quad 
      \delta_{(2)} = 3.3\%\, \quad
      \delta^*_{(2)} = 2.5\%\, .
   \end{eqnarray}
   \item[$-\log W_{112}/u_0^6$]  
   \begin{eqnarray}
      \delta^*_{(4)} = 3.5\%\, , \quad 
      \delta_{(2)} = 3.2\%\, \quad
      \delta^*_{(2)} = 3.1\%\, .
   \end{eqnarray}
\end{description}
This analysis suggests a systematic error around $3\%$ for these kind of analyses on the available 
ensembles. 


%% file: Alpha_s/s7.tex

\subsection{$\alpha_s$ from heavy-quark current two-point functions}


\label{s:curr}


\subsubsection{General considerations}


The method has been introduced in HPQCD 08, Ref.~\cite{Allison:2008xk},
and updated in HPQCD 10, Ref.~\cite{McNeile:2010ji}, see also
Ref.~\cite{Bochkarev:1995ai}.  In addition
there is a 2+1+1-flavour result, HPQCD 14A \cite{Chakraborty:2014aca}.

The basic observable is constructed from a current,
\begin{eqnarray}
  J(x) = i am_c\overline\psi_c(x)\gamma_5\psi_{c'}(x)\,,
  \label{e:Jx}
\end{eqnarray}
of two mass-degenerate heavy-valence quarks, $c$, $c^\prime$,
usually taken to be at or around the charm-quark mass.
The pre-factor $m_c$ denotes the bare mass of the quark.
When the lattice discretization respects chiral symmetry, 
$J(x)$ is a renormalization group
invariant local field, i.e., it requires no renormalization.
Staggered fermions and twisted-mass fermions have such a residual
chiral symmetry. The (Euclidean) time-slice correlation function
\begin{eqnarray}
   G(x_0) = a^6 \sum_{\vec{x}} \langle J^\dagger(x) J(0) \rangle \,,
\end{eqnarray}
($J^\dagger(x) = i am_c\overline\psi_{c'}(x)\gamma_5\psi_c(x)$)
has a $\sim x_0^{-3}$  singularity at short distances and moments
\begin{eqnarray}
   G_n = a \sum_{x_0=-(T/2-a)}^{T/2-a} x_0^n \,G(x_0) \,
\label{Gn_smu}
\end{eqnarray}
are nonvanishing for even $n$ and furthermore finite for $n \ge 4$ in the $a \rightarrow 0$ limit.
Here $T$ is the time extent of the lattice.
The moments are dominated by contributions at $x_0$ of order $1/m_c$.
For large mass $m_c$ these are short distances and the moments
become increasingly perturbative for decreasing $n$.
Denoting the lowest-order perturbation theory moments by $G_n^{(0)}$,
one defines the
normalized moments 
\begin{eqnarray}
    R_n = \left\{ \begin{array}{cc}
          G_4/G_4^{(0)}          & \mbox{for $n=4$} \,, \\[0.5em]
          {am_{\eta_c}\over 2am_c} 
                \left( { G_n \over G_n^{(0)}} \right)^{1/(n-4)}
                               & \mbox{for $n \ge 6$} \,, \\
                 \end{array}
         \right.
\label{Rn}
\end{eqnarray}
of even order $n$. Note that \eq{e:Jx} contains the variable
(bare) heavy-quark mass $m_c$. 
The normalization $G_n^{(0)}$ is introduced to help in
reducing lattice artifacts.
In addition, one can also define moments with different normalizations,
\begin{eqnarray}
   \tilde R_n = 2 R_n / m_{\eta_c} \qquad \mbox{for $n \ge 6$}\,.
\end{eqnarray}
While $\tilde R_n$ also remains renormalization-group invariant,
it now also has a scale which might introduce an
additional ambiguity \cite{Nakayama:2016atf}.

The normalized moments can then be parameterized in terms of functions
\begin{eqnarray}
   R_n \equiv \left\{ \begin{array}{cc}
                         r_4(\alpha_s(\mu))
                                        & \mbox{for $n=4$} \,,     \\[0.5em]
                         \frac{m_{\eta_c}}{2 \bar{m}_c(\mu_m)} r_n(\alpha_s(\mu))
                                        & \mbox{for $n \ge 6$} \,, \\
                      \end{array}
              \right.
              \label{e:Rn}
\end{eqnarray}
with $\bar{m}_c(\mu_m)$ being the renormalized heavy-quark  mass. 
The scale $\mu_m$ at which the heavy-quark mass is defined could be different from the scale $\mu$ at which $\alpha_s$ is defined \cite{Dehnadi:2015fra}.
The HPQCD collaboration, however, used the choice $\mu=\mu_m=3 m_c(\mu)$. This ensures that the renormalization scale is never too small.
The reduced moments $r_n$ have a perturbative expansion
\begin{eqnarray}
   r_n = 1 + r_{n,1}\alpha_s + r_{n,2}\alpha_s^2 + r_{n,3}\alpha_s^3 + \ldots\,,
\label{rn_expan}
\end{eqnarray}
where the written terms $r_{n,i}(\mu/\bar{m}_c(\mu))$, $i \le 3$ are known
for low $n$ from Refs.~\cite{Chetyrkin:2006xg,Boughezal:2006px,Maier:2008he,
Maier:2009fz,Kiyo:2009gb}. In practice, the expansion is performed in
the $\overline{\rm MS}$ scheme. Matching nonperturbative lattice results
for the moments to the perturbative expansion, one determines an
approximation to $\alpha_{\overline{\rm MS}}(\mu)$ as well as $\bar m_c(\mu)$.
With the lattice spacing (scale) determined from some extra physical input,
this calibrates $\mu$. As usual suitable pseudoscalar masses
determine the bare-quark masses, here in particular the charm-quark mass, 
and then through \eq{e:Rn} the renormalized charm-quark mass.

A difficulty with this approach is that large masses are needed to enter
the perturbative domain. Lattice artifacts can then be sizeable and
have a complicated form. The ratios in Eq.~(\ref{Rn}) use the
tree-level lattice results in the usual way for normalization.
This results in unity as the leading term in Eq.~(\ref{rn_expan}),
suppressing some of the kinematical lattice artifacts.
We note that in contrast to, e.g., the definition of $\alpha_\mathrm{qq}$,
here the cutoff effects are of order $a^k\alpha_s$, while there the
tree-level term defines $\alpha_s$ and therefore the cutoff effects
after tree-level improvement are of order $a^k\alpha_s^2$.
To obtain the continuum results for the moments it is important to perform
fits with high powers of $a$. This implies many fit parameters.
To deal with this problem the HPQCD collaboration
used Bayesian fits of their lattice results. More recent analyses of
the moments, however,  did not rely on Bayesian fits \cite{Nakayama:2016atf,Maezawa:2016vgv,Petreczky:2019ozv,Petreczky:2020tky}.

Finite-size effects (FSE) due to the omission of
$|x_0| > T /2$ in Eq.~(\ref{Gn_smu}) grow with $n$ as 
$(m_{\eta_c}T/2)^n\, \exp{(-m_{\eta_c} T/2)}$. 
In practice, however, since the (lower) moments
are short-distance dominated, the FSE are expected to be small
at the present level of precision. Possible exception could be the
ratio $R_8/R_{10}$, where the finite-volume effects could be significant
as discussed below.

Moments of correlation functions of the quark's electromagnetic
current can also be obtained from experimental data for $e^+e^-$
annihilation~\cite{Kuhn:2007vp,Chetyrkin:2009fv}.  This enables a
nonlattice determination of $\alpha_s$ using a similar analysis
method.  In particular, the same continuum perturbation-theory
computation enters both the lattice and the phenomenological determinations.


\subsubsection{Discussion of computations}

\begin{table}[!htb]
   \vspace{3.0cm}
   \footnotesize
   \begin{tabular*}{\textwidth}[l]{l@{\extracolsep{\fill}}rl@{\hspace{1mm}}l@{\hspace{1mm}}l@{\hspace{1mm}}l@{\hspace{1mm}}lll@{\hspace{1mm}}l}
      Collaboration & Ref. & $\Nf$ &
      \hspace{0.15cm}\begin{rotate}{60}{publication status}\end{rotate}
                                                       \hspace{-0.15cm} &
      \hspace{0.15cm}\begin{rotate}{60}{renormalization scale}\end{rotate}
                                                       \hspace{-0.15cm} &
      \hspace{0.15cm}\begin{rotate}{60}{perturbative behaviour}\end{rotate}
                                                       \hspace{-0.15cm} &
      \hspace{0.15cm}\begin{rotate}{60}{continuum extrapolation}\end{rotate}
      \hspace{-0.25cm} & 
                         scale & $\Lambda_\msbar[\MeV]$ 
                       & $r_0\Lambda_\msbar$ \\
      &&&&&&&&& \\[-0.1cm]
      \hline
      \hline
      &&&&&&&&& \\[-0.1cm]
      HPQCD 14A   &  \cite{Chakraborty:2014aca}
                                              & 2+1+1   & \gA
                   & \soso & \good   & \soso
                   & $w_0=0.1715(9)\,\mbox{fm}^a$
                   & 294(11)$^{bc}$
                   & 0.703(26)             \\
      &&&&&&&&& \\[-0.1cm]
      \hline
      &&&&&&&&& \\[-0.1cm]

      {Petreczky 20}
                   & \cite{Petreczky:2020tky}  & 2+1   & \gA
                   & \soso & \soso  & \good   
                   & $r_1 = 0.3106(18)$ fm                   
                   & 332(17)$^h$             & 0.792(41)$^g$  \\

        {Boito 20}
                   & \cite{Boito:2020lyp}  & 2+1   & \gA 
                   & \bad & \bad  & \soso         
                   & $m_c(m_c)=1.28(2)$ GeV  
                   & 328(30)$^h$             & 0.785(72)  \\ 
    {Petrezcky 19, $\scriptstyle m_h=m_c$} 
                   & \textcolor{blue}{\cite{Petreczky:2019ozv}}  & 2+1   & \gA 
                   & \bad & \bad  & \good          
                   & {$r_1 = 0.3106(18)\,\mbox{fm}^{g}$}  
                   &  314(10)            &    0.751(24)$^g$  \\
     {Petrezcky 19, $\scriptstyle \frac{m_h}{m_c}=1.5$} 
                   & \textcolor{blue}{\cite{Petreczky:2019ozv}}  & 2+1   & \gA
                   & \bad & \bad  & \soso  
                   & {$r_1 = 0.3106(18)\,\mbox{fm}^{g}$}
                   &  310(10)            &    0.742(24)$^g$  \\
      {Maezawa 16}
                   & \textcolor{blue}{\cite{Maezawa:2016vgv}}  & 2+1   & \gA & \bad
                   & \bad  & \soso          
                   & {$r_1 = 0.3106(18)\,\mbox{fm}$$^{d}$}  
                   & 309(10)$^{e}$             & 0.739(24)$^{e}$  \\
       {JLQCD 16}   & \cite{Nakayama:2016atf}  
                   & 2+1     & \gA 
                   & \bad & \soso  & \soso           
                   & {$\sqrt{t_0} = 0.1465(25)\,\mbox{fm}$}
                   & {331(38)$^{f}$}  &  0.792(89)$^{f}$ \\
      HPQCD 10     & \cite{McNeile:2010ji}  & 2+1       & \gA 
                   & \soso & \good   & \soso           
                   & $r_1 = 0.3133(23)\, \mbox{fm}$$^\dagger$
                   & 338(10)$^\star$           &  0.809(25)           \\
      HPQCD 08B    & \cite{Allison:2008xk}  & 2+1       & \gA 
                   & \bad  & \bad  & \bad           
                   & $r_1 = 0.321(5)\,\mbox{fm}$$^\dagger$  
                   & 325(18)$^+$             &  0.777(42)            \\
      &&&&&&&&& \\[-0.1cm]
      \hline
      \hline\\
\end{tabular*}\\[-0.2cm]
\begin{minipage}{\linewidth}
{\footnotesize 
\begin{itemize}
   \item[$^a$]  Scale determined in \cite{Dowdall:2013rya} using $f_\pi$. \\[-5mm]
   \item[$^b$]  $\alpha^{(4)}_\msbar(5\,\mbox{GeV}) = 0.2128(25)$, 
         $\alpha^{(5)}_{\overline{\rm MS}}(M_Z) = 0.11822(74)$.         \\[-5mm]
   \item[$^c$] We evaluated $\Lambda_{\overline{\rm MS}}^{(4)}$ from $\alpha^{(4)}_\msbar$. 
         We also used $r_0 = 0.472\,\mbox{fm}$.\\[-5mm]
   \item[$^{d}$] 
   Scale is determined from $f_\pi$ . 
    \\[-5mm]
   \item[$^{e}$]       $\alpha^{(3)}_\msbar(m_c=1.267\,\mbox{GeV}) = 0.3697(85)$,
               $\alpha^{(5)}_\msbar(M_Z) = 0.11622(84)$. Our conversion with $r_0 = 0.472\,\mbox{fm}$.         
               \\[-5mm]
    \item[$^{f}$]  We evaluated $\Lambda_{\overline{\rm MS}}^{(3)}$ from the given $\alpha^{(4)}_\msbar(3\,\mbox{GeV}) = 0.2528(127)$.
          $\alpha^{(5)}_{\overline{\rm MS}}(M_Z) = 0.1177(26)$.
          We also used $r_0 = 0.472\,\mbox{fm}$ to convert. \\[-5mm]
    \item[$^{g}$] We used $r_0 = 0.472\,\mbox{fm}$ to convert. \\[-5mm]
     \item[$^{h}$] We back-engineered from $\alpha^{(5)}_\msbar(M_Z) = 0.1177(20)$. We used $r_0 = 0.472\,\mbox{fm}$ to convert. 
  \\[-5mm]
   \item[$^\star$]  $\alpha^{(3)}_\msbar(5\,\mbox{GeV}) = 0.2034(21)$,
            $\alpha^{(5)}_\msbar(M_Z) = 0.1183(7)$.         \\[-4mm]
   \item[$^\dagger$] Scale is determined from $\Upsilon$ mass splitting.    \\[-5mm]
   \item[$^+$]  We evaluated $\Lambda_{\overline{\rm MS}}^{(3)}$ from the given $\alpha^{(4)}_\msbar(3\,\mbox{GeV}) = 0.251(6)$. $\alpha^{(5)}_\msbar(M_Z) = 0.1174(12)$.       

\end{itemize}
}
\end{minipage}
\normalsize
\caption{Heavy-quark current two-point function results. Note that all analysis using $2+1$
  flavour simulations perturbatively add a dynamical charm quark.
  Partially they then quote results in four-flavour 
  QCD, which we converted back to $\Nf=3$, corresponding to the
  nonperturbative sea quark content.}
\label{tab_current_2pt}
\end{table}

The determination of the strong coupling constant from the moments of quarkonium correlators
by HPQCD collaboration have been discussed in detail in the FLAG 16 and 19 reports.
Therefore, we only give the summary of these determinations in Tab.~\ref{tab_current_2pt}.
There were no new determinations of the strong coupling constant in 2+1 flavour QCD by
other groups since the FLAG 21 report. The only new development was that Petreczky 20 \cite{Petreczky:2020tky} is now published and therefore this determination enters the FLAG average.
The determinations of $\alpha_s$ by Maezawa 16, JLQCD16, Petreczky 19 and Boito 20 have been
discussed in detail in the FLAG 21 report, so we do not discuss them here again and only give
the summary of these determinations in Tab.~\ref{tab_current_2pt}. We will only discuss
the results of Petreczky 20 \cite{Petreczky:2020tky} here.

Petreczky 20 is based on the same lattice data as Petreczky 19 \cite{Petreczky:2019ozv}. Here the pseudo-scalar correlation functions
have been computed using HISQ ensembles from HotQCD Collaboration \cite{Bazavov:2014pvz} for physical strange-quark mass 
and light-quark masses corresponding to the pion mass
of $160$ MeV in the continuum limit, and lattice spacings $a^{-1}=1.81,~2.07,~2.39,~2.67,~3.01,~3.28,~4.00$ and $4.89$ GeV. Additional
calculations have been performed for light-quark mass corresponding to the pion mass of 300 MeV and lattice spacings
$a^{-1}=2.39,~4.89,~5.58,~6.62$ and $7.85$ GeV using the gauge configurations from the study of QCD equation of state at high temperatures
\cite{Bazavov:2017dsy}.
No significant light-quark-mass dependence of heavy pseudo-scalar correlators have 
been observed \cite{Petreczky:2019ozv}. Therefore, the results for the two light-quark masses have been combined into a single analysis.
Calculations have been performed at four values of the heavy-quark mass equal to the physical charm-quark mass,
one and half times the charm-quark mass, two times the charm-quark mass and three times the charm-quark mass.
In this study random-colour wall sources which greatly reduced the statistical errors were used.
In fact, the statistical errors on the moments were completely negligible compared to other sources of errors.
The strong coupling constant was extracted from $R_4$ \cite{Petreczky:2020tky}. To obtain the continuum limit
the lattice-spacing dependence of the results of $R_4$
at different quark masses was fitted simultaneously in a similar manner as in the HPQCD 10 and HPQCD 14 analyses, but
without using Bayesian priors. In extracting $\alpha_s$ several choices of the renormalization scale $\mu$ in the range 
$2/3 m_h$--$3 m_h$ have been considered. The perturbative truncation error was estimated by varying the coefficient of the unknown
4-loop term in Eq. (\ref{rn_expan}) between $-1.6 r_3$ and $+1.6 r_3$.
However, the uncertainty 
of the results due to the scale variation was larger than the estimated perturbative truncation error. The final error of the
result $\Lambda_{\overline{MS}}^{\Nf=3}=331(17)$ MeV comes mostly from the scale variation \cite{Petreczky:2020tky}.
Since there are three lattice spacing available with $a \mu<0.5$ we give $\good$ for continuum extrapolation.
Because $\alpha_\text{eff}=0.22-0.38$ we give $\soso$ for the renormalization scale. Finally, since 
$(\Delta \Lambda/\Lambda)_{\Delta \alpha}>\alpha_\text{eff}^2$ for the smallest $\alpha_\text{eff}$ value 
we give $\soso$ for the perturbative behaviour.
In addition to $R_4$ Petreczky 20 also considered using $R_6/R_8$ and $R_8/R_{10}$ for the $\alpha_s$ determination.
It was pointed out that the lattice spacing dependence of $R_6/R_8$ is quite subtle and therefore reliable continuum
extrapolations for this ratio are not possible for $m_h \ge 2 m_c$ \cite{Petreczky:2020tky}. 
For $m_h=m_c$ and $1.5m_c$ the ratio $R_6/R_8$ leads to $\alpha_s$ values that are consistent with the ones from $R_4$.
Furthermore, it was argued
that finite-volume effects in the case of $R_8/R_{10}$ are large for $m_h=m_c$ and therefore the corresponding
data are not suitable for extracting $\alpha_s$. 
This observation may explain why the central values of $\alpha_s$ extracted from $R_8/R_{10}$ in some previous studies
were systematically lower \cite{Allison:2008xk,Maezawa:2016vgv,Petreczky:2019ozv}.
On the other hand for $m_h \ge 1.5 m_c$ the finite-volume effects are sufficiently
small in the continuum extrapolated results if some small-volume lattice data are excluded from the analysis \cite{Petreczky:2020tky}.
The $\alpha_s$ obtained from $R_8/R_{10}$ with $m_h\ge 1.5 m_c$ were consistent with the ones obtained from $R_4$.

Aside from the final results for $\alpha_s(m_Z)$ obtained by matching with perturbation theory, it is 
interesting to make a comparison of the short distance quantities 
in the continuum limit $R_n$ which are available from 
HPQCD 08~\cite{Allison:2008xk}, 
JLQCD 16 \cite{Nakayama:2016atf}, Maezawa 16 \cite{Maezawa:2016vgv}, Petreczky 19 \cite{Petreczky:2019ozv} 
and Petreczky 20 \cite{Petreczky:2020tky} (all using $2+1$ flavours). 
This comparison is shown in 
Tab.~\ref{Rn_moments}.
\begin{table}[h]
\begin{center}
\begin{tabular}{c|cccccc}
\hline
             & HPQCD 08    &  HPQCD 10  & Maezawa 16 & JLQCD 16  & Petreczky 19 & Petreczky 20      \\
\hline
$R_4$        & 1.272(5)    & 1.282(4)   & 1.265(7)  & -          & 1.279(4)     & 1.278(2)          \\
$R_6$        & 1.528(11)   & 1.527(4)   & 1.520(4)  & 1.509(7)   & 1.521(3)     & 1.522(2)          \\
$R_8$        & 1.370(10)   & 1.373(3)   & 1.367(8)  & 1.359(4)   & 1.369(3)     & 1.368(3)          \\
$R_{10}$     & 1.304(9)    & 1.304(2)   & 1.302(8)  & 1.297(4)   & 1.311(7)     & 1.301(3)          \\
$R_6/R_8$    & 1.113(2)    & -          & 1.114(2)  & 1.111(2)   & 1.1092(6)    & 1.10895(32)       \\
$R_8/R_{10}$ & 1.049(2)    & -          & 1.0495(7) & 1.0481(9)  & 1.0485(8)    & -                 \\
\hline
\end{tabular}
\end{center}
\caption{Moments and the ratios of the moments from $\Nf=3$ simulations at the charm-quark mass.}
\label{Rn_moments}
\end{table} 
The results are in quite good agreement with each other.
For future studies it is of course interesting to check
agreement of these numbers before turning to the more
involved determination of $\alpha_s$.


While there have been no new determinations of $\alpha_s$ from the moments of the heavy-quark current two-point functions in 2+1+1 flavour
or 2+1 flavour QCD since the FLAG 21 report, this method has been scrutinized in quneched QCD ($\Nf=0$) 
in three conference proceedings \cite{Chimirri:2022bsu,Chimirri:2023iro,Sommer:2022wac}. 
In these works the Wilson gauge action was used for several values of the lattices spacings, down to lattice
spacing of $a=0.01$ fm, which is 2.5 times smaller than the smallest lattice spacing used in 2+1 flavour QCD.
The box size was sufficiently large for the heavy-quark current two-point functions, namely $L=2$ fm was used. In the temporal
direction open boundary conditions have been used, and the extent in the time direction was $6$ fm.
For heavy-quark twisted-mass fermion formulation was used at the maximal twist.
Five different heavy-quark masses have been used in these studies, namely $0.77M_c,~1.16M_c,~1.55 M_c,~2.32M_c$ and $3.48 M_c$, with
$M_c$ being the physical charm-quark RGI mass \cite{Chimirri:2022bsu,Chimirri:2023iro,Sommer:2022wac}. The continuum extrapolation of
$R_4$ has been performed and from it the value of $\Lambda_{\overline{MS}}^{(\Nf=0)}$ was obtained for different heavy-quark masses and different choices of $\mu$. It turned out, however, that the results obtained for different heavy-quark masses and
values of $\mu$ are not consistent with each other and often are not compatible with the value determined from step scaling
\cite{DallaBrida:2019wur}. It was argued that this is due to the log-enhanced discretization errors in $R_4$, i.e., discretization errors
that are proportional to $a^2 \log(a m_c)$ \cite{Sommer:2022wac}, and that reliable continuum extrapolation of 
$R_4$ is not possible for this lattice setup. A practical way to circument this problem was also proposed in Ref.~\cite{Sommer:2022wac}
and relies on considering a special combination of $R_4$ evaluated at two heavy-quark masses. 
The ratios $R_6/R_8$ and $R_8/R_{10}$ do not have log-enhanced discretization effects \cite{Chimirri:2023iro,Sommer:2022wac}
and therefore, can be used to obtain $\Lambda_\msbar^{(\Nf=0)}$. Such an analysis was performed in Ref.~\cite{Chimirri:2023iro}.
Here to deal with perturbative error it was assumed that $\Lambda_\msbar^{(\Nf=0)} \sqrt{8 t_0}$ obtained at different
renormalization scales $\mu$ is linear in $\alpha_s^2(\mu)$ as expected from 3-loop perturbative calculations. Performing linear
extrapolations in $\alpha_s^2(\mu)$ the final values of $\Lambda_\msbar^{(\Nf=0)} \sqrt{8 t_0}$ have been obtained.
The corresponding results for the $\Lambda$-parameter agree with the result of the step-scaling analysis but have much larger
errors, and thus are not competitive \cite{Chimirri:2023iro}.

\paragraph{Scale variations.} 
Moments of heavy-quark correlators are computed at scales that are set by the mass of the charm quark. 
We compute scale variations for the moments $r_4$, $r_6$ and $r_8$ at different values of the 
matching scale. 
\begin{description}
   \item[HQ $r_4$, $\mu=m_c$]
      \begin{eqnarray}
         \delta_{(2)} = 2.7\%\, \quad
         \delta^*_{(2)} = 2.8\%\, .   
      \end{eqnarray} 
   \item[HQ $r_4$, $\mu=2 m_c$]
      \begin{eqnarray}
         \delta^*_{(4)} = 1.2\%\, , \quad 
         \delta_{(2)} = 1.5\%\, \quad
         \delta^*_{(2)} = 1.6\%\, .   
      \end{eqnarray} 
   \item[HQ $r_6$, $\mu=2 m_c$]
      \begin{eqnarray}
         \delta_{(2)} = 2.3\%\, \quad
         \delta^*_{(2)} = 1.2\%\, .   
      \end{eqnarray} 
   \item[HQ $r_8$, $\mu=2 m_c$]
      \begin{eqnarray}
         \delta_{(2)} = 2.8\%\, \quad
         \delta^*_{(2)} = 4.8\%\, .   
      \end{eqnarray} 
\end{description}
We note here that the errors from the scale variations are in the same ballpark as previous 
estimates published in FLAG reviews. The moment $r_4$ computed at the scale $Q=2 m_c$ 
happens to have a systematic error in the range $1-2\%$.  

%% file: Alpha_s/s_gf.tex
\subsection{Gradient-flow schemes}
\label{s:gf}

\subsubsection{General considerations}
\label{sec:gf-general}

The gradient flow~\cite{Narayanan:2006rf,Luscher:2010iy} (cf.~the paragraph around Eq.~(\ref{eq:def-GF}) for the basic equations) 
allows for the definition of many new observables, both in pure gauge theory
and QCD, which are gauge invariant and automatically renormalized after the standard QCD renormalizations
of parameters and composite fields have been carried out. This has been established perturbatively 
to all orders in Ref.~\cite{Luscher:2011bx} and confirmed up to 2-loop level in practical calculations~\cite{Harlander:2016vzb}.
It is generally assumed to be valid beyond perturbation theory and many simulation results 
corroborate this assumption. 

The gradient flow comes with the flow-time parameter, $t$, which has dimensions of
length squared and thus introduces a new energy scale which is, by analogy with the diffusion equation,
naturally identified as $\mu=1/\sqrt{8t}$ (in four dimensions).
The most widely used observable is the action density at finite flow
time, 
\begin{equation}
   E(t,x) = -\frac12 \tr\{G_{\mu\nu}(t,x) G_{\mu\nu}(x)\} \, .
\end{equation}
Its expectation value has a perturbative expansion starting at $O(\alpha)$, 
which gives rise to the definition of the coupling in the GF scheme,
\begin{equation}
  \alpha_\text{GF}(\mu) \equiv \dfrac{\bar{g}_\text{GF}^2(\mu)}{4\pi} = \dfrac{4\pi t^2}{3} \langle E(t,x) \rangle
\end{equation}
and is known to 3-loop order,
\begin{equation}
   \alpha_\text{GF}(\mu=1/\sqrt{8t}) = \alpha_\msbar(\mu) + k_1 \alpha_\msbar(\mu)^2 + k_2 \alpha_\msbar(\mu)^3 + \ldots 
  \label{eq:GF-PT}
\end{equation}
with $k_1$ and $k_2$ computed in Refs.~\cite{Luscher:2010iy} and \cite{Harlander:2016vzb}, 
respectively (cf.~Tab.~\ref{tab:scale_truncation}).
Note that the GF coupling directly relates to the scale $t_0$;
its definition is equivalent to $\bar{g}_\text{GF}^2(1/\sqrt{8t_0}) = 15.8$.
With the flow time setting the renormalization scale, the $\beta$-function is 
readily obtained during the numerical integration of the flow equation, 
by also tracking the flow-time derivative of $\langle E(t,x) \rangle$,
\begin{equation}
\beta_\text{GF}(\bar{g}_\text{GF})= -2t \dfrac{d}{d t} \bar{g}_\text{GF}(1/\sqrt{8t})\,,
\end{equation}
and the 3-loop $\beta$-function coefficient $b_2$ is known. In the pure gauge
theory is is given by
\begin{equation}
   b^\text{GF}_2 = -1.90395(4)/(4\pi)^3, \quad (\Nf=0). 
\end{equation}
This is almost three times larger in magnitude than in the $\msbar$ scheme and of opposite sign.
One naturally worries about higher-order corrections being large, too.
As a result, making contact with perturbation theory requires very small couplings.
To quantify the problem, we have done the following exercises (all for $\Nf=0$): First one may
evaluate the difference in $\sqrt{8t} \Lambda_\text{GF}$ obtained by integrating
the perturbative $\beta$-function at 2- vs.~3-loop order 
from zero coupling to a reference value $\bar{g}^2_\text{GF}(1/\sqrt{8t})=1.2$, which
corresponds to the smallest coupling reached in the works discussed below.
We find that this difference is about 11 percent, again about three times larger than with the $\msbar$ scheme.
In order to vary the scale we convert to the $\msbar$ scheme,
\begin{equation}
  \alpha_\text{GF}(\mu) = \alpha_\msbar(s\mu) + k_1(s) \alpha^2_\msbar(s\mu) + k_2(s) \alpha^3_\msbar(s\mu) + \rmO(\alpha_\msbar^4),
  \label{eq:GF-MSbar-conversion}
\end{equation}
where the $s$-dependence of the coefficients is given as
\begin{equation}
 k_1(s)-k_1(1) = 8\pi b_0 \ln(s), \quad k_2(s)-k_2(1) = 32\pi^2 b_1 \ln(s) + k^2_1(s) - k^2_1(1). 
\end{equation}
In order to obtain the $\msbar$ coupling in terms of the reference coupling 
one needs to invert Eq.~(\ref{eq:GF-MSbar-conversion}),
which we do either perturbatively or numerically for the truncated equation.
We then compute,
\begin{equation}
   \sqrt{8t}\Lambda_\msbar = s\times\varphi_\msbar\left(\bar{g}_\msbar(s/\sqrt{8t})\right),
\end{equation}
for scale factors $s=1/2,1,2$, using the 5-loop $\beta$-function in the $\msbar$ scheme.
We find that the resulting variation in the $\Lambda$-parameter depends on how the 
$\msbar$-coupling is obtained: With perturbative inversion, the variation is plus $7.5$ 
and minus $4$ percent, with numerical inversion, one obtains plus $2.5$ and plus $3.3$ percent,
i.e., even monotony is lost. The central values for $s=1$ differ by 5 percent.
As an alternative, we consider the scale factor $s^\ast=0.534$ which implies $k_1(s^\ast)=0$.
Varying by a factor two around $s^\ast$ one finds that the difference in central values reduces to $1.3$ percent,
and the $\Lambda$-parameter changes by minus $6$ percent and plus $4$ percent for perturbative inversion, and 
by plus 9.7\% and minus 2.7\% for numerical inversion.

We conclude that at this reference coupling a determination of the $\Lambda$-parameter to better
than five percent seems impossible.

\subsubsection{Discussion of computations}

A determination of the $\beta$-function directly from the flow-time dependence of the GF coupling
requires  a controlled infinite-volume extrapolation. This was first suggested in Ref.~\cite{Fodor:2017die},
where the strategy was applied to a BSM model. Since then, two works have applied this scheme 
to the pure gauge theory (QCD with $\Nf=0$), namely Hasenfratz 23~\cite{Hasenfratz:2023bok} and 
Wong 23~\cite{Wong:2023jvr}, in a proceedings contribution. We mainly discuss Hasenfratz~23 who provide 
more details: the data produced for the GF coupling ranges
from $15.8$ down to $1.2$, lattice sizes vary between $L/a=20$
and $L/a=48$, depending on the $\beta$-value, and periodic boundary conditions are imposed on the gauge field.
Wong~23 do have data for larger lattices up to $L/a=64$ and even $L/a=80,96$ at selected bare couplings.
The data for both $\bar{g}_\text{GF}^2$ and $\beta_\text{GF}$ are extrapolated 
to the infinite-volume limit at fixed lattice spacing,
assuming corrections $\propto (a/L)^4$, with Wong 23 also allowing for a subleading $(a/L)^6$ term.  
Then the continuum limit is taken for $a^2/t$-values in the range
0.25--0.5, corresponding to $a\mu$-values in the range 0.177--0.25, and a somewhat wider range in Wong 23.
The continuum extrapolation data for the $\beta$-function
at fixed GF coupling are shown in plots. For Hasenfratz 23 these extrapolations look fine and would
pass any reasonable data-driven criterion. Wong 23 only show the extrapolation at the largest GF coupling
which looks fine, too.  Hence we give \good\ for the continuum extrapolation and also for 
renormalization scale, given that $\alpha_\text{eff}$ reaches down to below $0.1$. Regarding the 
formal FLAG criterion for perturbative behaviour, Hasenfratz 23 give an overall error of $0.6\%$ for $\alpha_\text{GF}$.
Using this error we have, at the smallest couplings reached, $\alpha_\text{eff}^{n_{\mathrm{l}}} = (0.1)^2  < 0.006 \times 2.85 = 0.017$,
which satisfies the criterion comfortably. This warrants a \good\ for Hasenfratz~23.
For Wong 23 the accuracy of $\alpha_\text{GF}$ is not given but they quote a per-mille accuracy for the beta function at $\bar{g}_\text{GF}^2=15.8$; 
we assign a \soso, which assumes their coupling data is perhaps a factor 2 but still less than a factor $3-4$ more 
accurate relative to the 0.6\% of Hasenfratz~23.

Unfortunately, the formal FLAG criteria do not capture the anomalously bad behaviour of the GF scheme.
As discussed above, even at $\alpha_\text{eff} =0.1$ the estimate of the $\Lambda$-parameter is
ambiguous at the level of about 5 percent.

Contact to perturbation theory is not really established, as the obtained $\beta$-function 
seems to show a slope that is different from the perturbative
expectation.  Imposing perturbative asymptotics and evaluating the integral over the beta function
numerically leads to the estimate $\sqrt{8t_0}\Lambda_\msbar = 0.622(10)$. Wong~23 
obtain an even smaller error, $\sqrt{8t_0}\Lambda_\msbar = 0.632(7)$.\footnote{Wong 23 write $t_0\Lambda_\msbar$, instead of $\sqrt{8t_0}\Lambda_\msbar$, 
which we interpret as a typo.} 
Note that both values are in agreement with each other and with Dalla Brida~19 
(who obtained $0.623(10)$) and would lend support to the high central value
compared to older results in the literature. Despite this consistency, the claimed high accuracy
seems at odds with the bad perturbative behaviour of this scheme.


Regarding the infinite-volume limit, the main problem is the lack of
guidance from theory regarding the fit ansatz. With $\Nf=0$ and in the hadronic regime, one may expect
an exponential approach to the infinite-volume limit $\propto \exp(-m_G L)$, with $m_G$ the 
$0^{++}$ glueball mass. At high energies one is necessarily in small volumes where hadrons cannot
form, and leading effects $\propto (a/L)^4$ are used as a plausible ansatz by both groups of authors.
However, there is an intermediate regime where the situation is quite unclear, and even at
high energies, once the volume is large enough to contain hadrons, the large-volume asymptotics
should be expected to change. The situation may be even more complicated in full QCD,
where massless pions are expected at low energies. Chiral perturbation theory may help
but only as long as pions are relevant degrees of freedom.

Note that boundary conditions should not matter in the infinite-volume limit, so that any of the
GF finite-volume couplings that have been used in step-scaling studies (cf.~Sec.~\ref{s:SF}) 
could be used to improve our understanding of it. In fact, the first discussion 
can be found in Ref.~\cite{Luscher:2014kea}, there with open-SF boundary conditions. 
In Dalla Brida 19, two different finite-volume schemes are considered
which should both converge to the infinite-volume GF scheme.

In step-scaling studies, the gradient-flow scale is fixed in units of $L$ to a 
constant $c=\sqrt{8t}/L$, with typical values around $c=0.3$. 
This means that the $\beta$-function cannot be obtained directly and
a detour via the step-scaling function is used in practice~\cite{DallaBrida:2019wur}
Since the schemes are defined in a finite volume, $c$ becomes an integral part of the scheme definition
as do the boundary conditions (SF, twisted periodic, etc.). In particular, the perturbative 2-loop result 
in Eq.~(\ref{eq:GF-PT}) cannot be used. For $\Nf=0$ and twisted periodic boundary conditions
there is a 1-loop computation~\cite{Bribian:2019ybc} while for $\Nf=0$ and SF boundary conditions 
there is a 2-loop result obtained using a stochastic
perturbative approach~\cite{DallaBrida:2017tru}. As in infinite volume, the perturbative behaviour of the finite-volume gradient-flow schemes  is quite bad~\cite{DallaBrida:2019wur}. This problem was circumvented in  
Refs.~\cite{DallaBrida:2019wur,Bribian:2021cmg,Ishikawa:2017xam} by matching nonperturbatively to the SF scheme, 
in order to benefit from its good perturbative behaviour. The option of such a matching is also mentioned
in Hasenfratz~23 where it is left to future work.

In Tab.~\ref{tab_GFinfinite} we list these results.
\begin{table}[!htb]
   \vspace{3.0cm}
   \footnotesize
   \begin{tabular*}{\textwidth}[l]{l@{\extracolsep{\fill}}rllllllll}
   Collaboration & Ref. & $\Nf$ &
   \hspace{0.15cm}\begin{rotate}{60}{publication status}\end{rotate}
                                                    \hspace{-0.15cm} &
   \hspace{0.15cm}\begin{rotate}{60}{renormalization scale}\end{rotate}
                                                    \hspace{-0.15cm} &
   \hspace{0.15cm}\begin{rotate}{60}{perturbative behaviour}\end{rotate}
                                                    \hspace{-0.15cm} &
   \hspace{0.15cm}\begin{rotate}{60}{continuum extrapolation}\end{rotate}
      \hspace{-0.25cm} & 
                         scale & $\sqrt{8t_0}\Lambda_\msbar$ & $r_0\Lambda_\msbar$$^*$ \\
   & & & & & & & & & \\[-0.1cm]
   \hline
   \hline
   & & & & & & & & & \\[-0.1cm] 
   Hasenfratz 23& \cite{Hasenfratz:2023bok} & 0 & \gA
                &     \good   &  \good      & \good  
                & $\sqrt{t_0}$ 
                & $0.622(10)$  
                & $0.659(11)$                   \\ 
   & & & & & & & & & \\[-0.1cm]
   Wong 23   & \cite{Wong:2023jvr}       & 0 & \rC
                &     \good   &  \soso      & \good  
                & $\sqrt{t_0}$ 
                & $0.632(7)$  
                & $0.670(8)$                   \\ 
   & & & & & & & & & \\[-0.1cm]
   \hline
   \hline
\end{tabular*}
\begin{tabular*}{\textwidth}[l]{l@{\extracolsep{\fill}}llllllll}
\multicolumn{8}{l}{\vbox{\begin{flushleft}
   $^*$ $r_0\Lambda_\msbar$ determined 
        by us using  $\sqrt{8t_0}/r_0 = 0.9435(97)$ from Dalla Brida 19~\cite{DallaBrida:2019wur} without propagating the error. \\
\end{flushleft}}}
\end{tabular*}
\vspace{-0.3cm}
\normalsize
\caption{Results for the GF scheme in infinite volume.}
\label{tab_GFinfinite}
\end{table}

\paragraph{Scale variations.}
As discussed in the general considerations of the previous subsection, the matching with 
perturbation theory is performed for $\bar{g}^2_\text{GF}(1/\sqrt{8t})=1.2$. The corresponding
energy scale $\mu=1/\sqrt{8t}$ is not given in the publications, preventing us from using the 
generic procedure that we used for the majority of the observables. Instead, we defined an alternative 
procedure to estimate the effect of scale variations directly on the ratio of $\Lambda$-parameters, 
as discussed in Sec.~\ref{sec:gf-general}.

%

%% file: Alpha_s/s10.tex
\subsection{Summary}
\label{s:alpsumm}

\newcommand{\pp}{\phantom{0}}


Having reviewed the individual computations, we are now in a position
to discuss the overall result. We first look at the current results
of the $\Lambda$-parameter for QCD with  $\Nf = 0, 2, 3, 4$ flavours in
units of the scale $r_0$ (and $\sqrt{8t_0}$ for $\Nf=0$).
These results are directly obtained from lattice simulations of 
QCD with given $\Nf$. For the $\Lambda$-parameter with $\Nf=0$ we present 
a more in depth discussion. As emphasized in our last report,
even though $\Nf=0$ is unphysical, the $\Lambda$-parameter
enters into the decoupling result, which is one of the
most accurate lattice determinations of $\alpha_\msbar^{(5)}(m_Z)$. 
Fortunately, this has motivated several collaborations to help 
clarify the situation, which is characterized by many historical 
results, with a large spread of central values, 
that are mutually incompatible due to the smallness of some error estimates.
We have decided to estimate ranges for different methods and give 
a corresponding FLAG estimate.

Then we discuss the central $\alpha_{\overline{\rm MS}}(m_Z)$ results
in five-flavour QCD. We give ranges for each sub-group discussed previously, 
and give a final FLAG average as well as an overall 
average together with the current PDG nonlattice numbers. 
In the end, we return to the $\Lambda$-parameter; for $\Nf=3,4,5$ we
derive their values from the FLAG estimate of  $\alpha_{\overline{\rm MS}}(m_Z)$.

We end with an outlook and some concluding remarks.


\subsubsection{Ranges for $[r_0 \lms]^{(\Nf)}$ and $\lms^{(\Nf)}$}

In the present situation, we give ranges for $[r_0 \lms]^{(\Nf)}$
and $\lms$, discussing their determination case by case.  We include
results with $\Nf<3$ because it is interesting to see the
$\Nf$-dependence of the connection of low- and high-energy QCD.  This
aids our understanding of the field theory and helps in finding
possible ways to tackle it beyond the lattice approach. It is also of
interest in providing an impression on the size of the 
vacuum-polarization effects of quarks, in particular with an eye on the still
difficult-to-treat heavier charm and bottom quarks. 
Most importantly, however, the decoupling strategy described in subsection~\ref{s:dec}
means that $\Lambda$-parameters at different $\Nf$ can be connected
by a nonperturbative matching computation. Thus, even results at unphysical flavour
numbers, in particular $\Nf=0$, may enter results for the physically
interesting case. Rather than phasing out results for ``unphysical flavour numbers",
continued scrutiny by FLAG will be necessary.
Having said this, we emphasize that results for $[r_0 \lms]^{(0)}$
and $[r_0 \lms]^{(2)}$ are {\em not}\/ meant to be used directly for
phenomenology. 

For the ranges we obtain: 
\begin{eqnarray}
   [r_0 \lms]^{(4)}   &=& 0.70(3)\,,               \label{eq:lms4} \\[0pt]
   [r_0 \lms]^{(3)}   &=& 0.809(23) \,,            \label{eq:lms3} \\[0pt]
   [r_0 \lms]^{(2)}   &=& 0.79(^{+~5}_{-{15}})\,,  \label{eq:lms2} \\[0pt]
   [r_0 \lms]^{(0)}   &=& 0.647(11) \,.            \label{eq:lms0}  
\end{eqnarray}  
No change has occurred since FLAG 21 for $\Nf=2,4$, so we refer to the respective discussions in earlier FLAG reports.

 For $\Nf=2+1$, we take as a central value the weighted average of 
ALPHA~22 \cite{DallaBrida:2022eua},
Petreczky~20 \cite{Petreczky:2020tky},
Cali 20 \cite{Cali:2020hrj}, 
Ayala 20 \cite{Ayala:2020odx},  
TUMQCD 19 \cite{Bazavov:2019qoo}, 
ALPHA 17 \cite{Bruno:2017gxd},
HPQCD~10 \cite{McNeile:2010ji}, 
PACS-CS~09A \cite{Aoki:2009tf} and 
Maltman~08 \cite{Maltman:2008bx},
and arrive at our range,
\begin{eqnarray}
   [r_0 \lms]^{(3)} = 0.809(23) \, ,
 \end{eqnarray}
where the error is the one from the weighted average of those results, which are statistics-dominated,
namely PACS-CS~09A, ALPHA~17 and ALPHA~22, and the known correlation between the latter two is taken into
account. This is to be compared with the much smaller 
error of $0.010$, as obtained from the weighted average.
There is good agreement with all 2+1 results without red tags. 
In physical units, using $r_0=$ 0.472~fm and neglecting
its error, we get
\begin{eqnarray}
   \lms^{(3)} = 338(10)\,\mbox{MeV}\,,
 \label{e:lms3}
\end{eqnarray}
whereas the error of the straight weighted average is around $4\MeV$.

For $\Nf=0$ there are now 12 results which pass the FLAG criteria,
four of which are new since FLAG~21. Instead of averaging individual
results we will group them by method, produce pre-ranges
and a final estimate for the range from combining the pre-ranges.
There are four different methods used:
\begin{itemize}
\item Step scaling: Combining Dalla~Brida~19 with Bribian~21, Ishikawa~17 (with symmetrized larger error) 
and ALPHA~98  in a weighted average, we obtain
\begin{equation}
   [r_0 \lms]^{(0)} = 0.648(11)\,.
\end{equation}
Leaving out Ishikawa~17 with its asymmetric error, this would change  to $0.651(11)$. 
For the error we take the statistics-dominated one from Dalla~Brida~19.

\item Static potential/force: We combine Brambilla~10 with Brambilla~23 (both with symmetrized error, using the larger ones) 
in a weighted average,
\begin{equation}
   [r_0 \lms]^{(0)} = 0.648(28)\,,
\end{equation}
where we use the error of the newer result for our estimate of the range.
\item
There are two new determinations with the GF scheme in infinite volume and continuous $\beta$-function,
by Wong~23 and Hasenfratz~23. We use the central value of the published paper by Hasenfratz~23 and 
include a perturbative uncertainty of five percent as discussed in Sec.~\ref{s:gf}, and obtain,
\begin{equation}
 [r_0 \lms]^{(0)} = 0.659(33)\,.
\end{equation}
\item Wilson loops: There are two results which are, due to their tiny errors, causing the tension noticed
in our previous FLAG report. We performed a scale-variation analysis, similar to the one explained in
Sec.~\ref{s:gf} for the GF scheme. Variations around the scale of fastest apparent convergence (cf.~Sec.~\ref{s:trunc})  
result in changes of up to 13 percent even at the finest available lattice spacings. Another way to look at the data is to note that both works perform continuum
extrapolations of the $\Lambda$-parameter assuming an $a^2$-behaviour. On the other hand, there is a parametric
uncertainty of O($\alpha_P^2(1/a)$) which is neglected. If included as a second term in a fit, 
the error gets much larger, and central values tend to increase. 
Stopping short of changing central values, we take the (unweighted) average central value and
include a symmetric range of $\pm 7$ percent as perturbative uncertainty,
\begin{equation}
 [r_0 \lms]^{(0)} = 0.618(43)\,.
\end{equation}
\end{itemize}
With these pre-ranges we perform a weighted average to obtain the central value, and then take
the statistics-dominated Dalla~Brida~19 step-scaling error as our estimate of the range,
\begin{equation}
 [r_0 \lms]^{(0)} = 0.647(11)  \quad \Rightarrow \quad [\sqrt{8t_0}\lms]^{(0)} = 0.610(10)\,.
\end{equation}

\begin{figure}[!htb]\hspace{-2cm}\begin{center}
      \includegraphics[width=11.5cm]{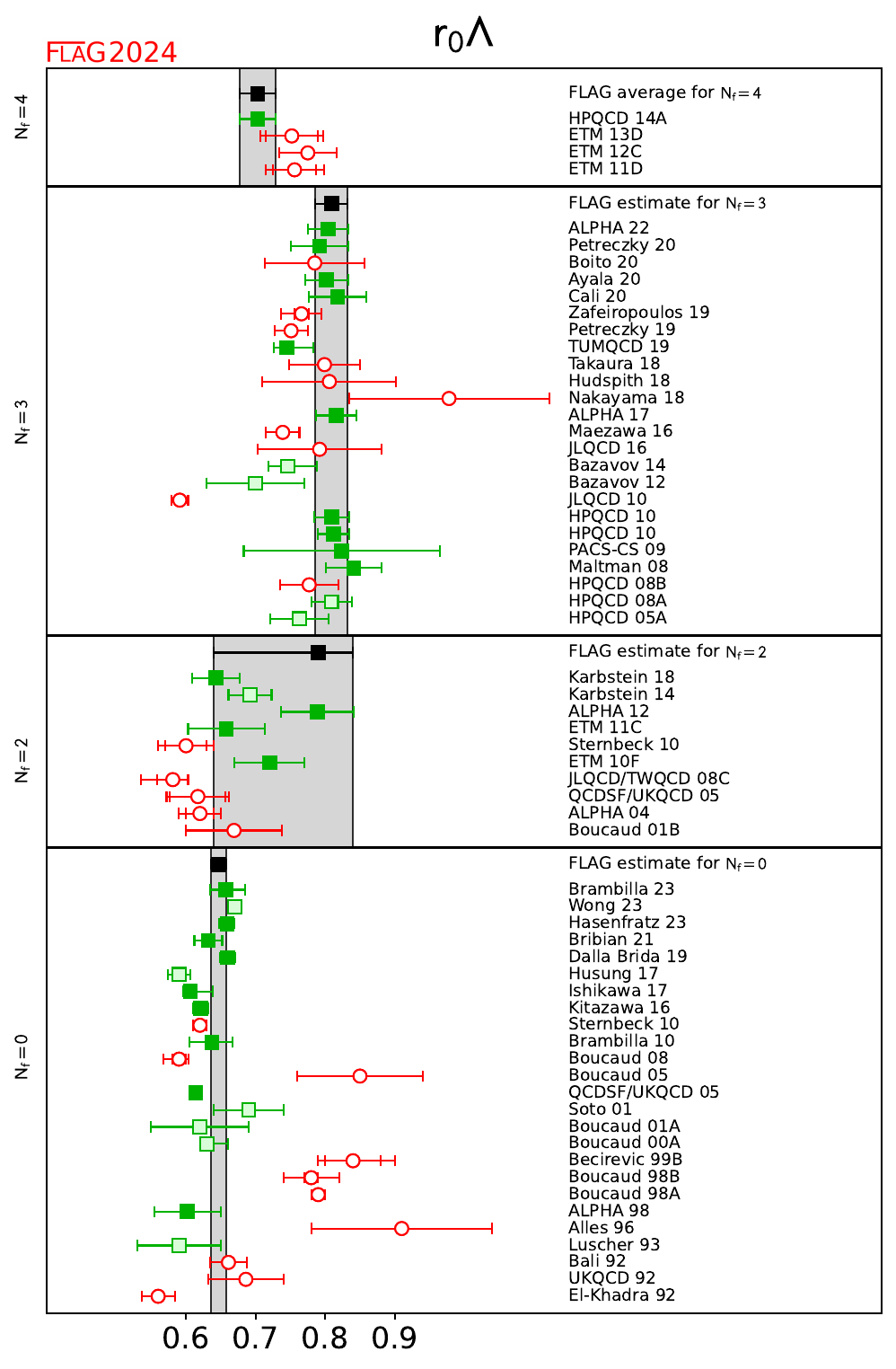}
      \end{center}
\vspace{-0.6cm}
\caption{$r_0\Lambda_{\overline{\rm MS}}$ estimates for
         $\Nf = 0$, $2$, $3$, $4$ flavours.
         Full green squares are used in our final
         ranges, pale green squares also indicate that there are no
         red squares in the colour coding but the computations were
         superseded by later more complete ones or not
         published, while red open circles mean that there is at
         least one red square in the colour coding.}
\label{fig:r0LamMSbar}
\end{figure}

\begin{figure}[!htb]\hspace{-2cm}\begin{center}
      \includegraphics[width=11.5cm]{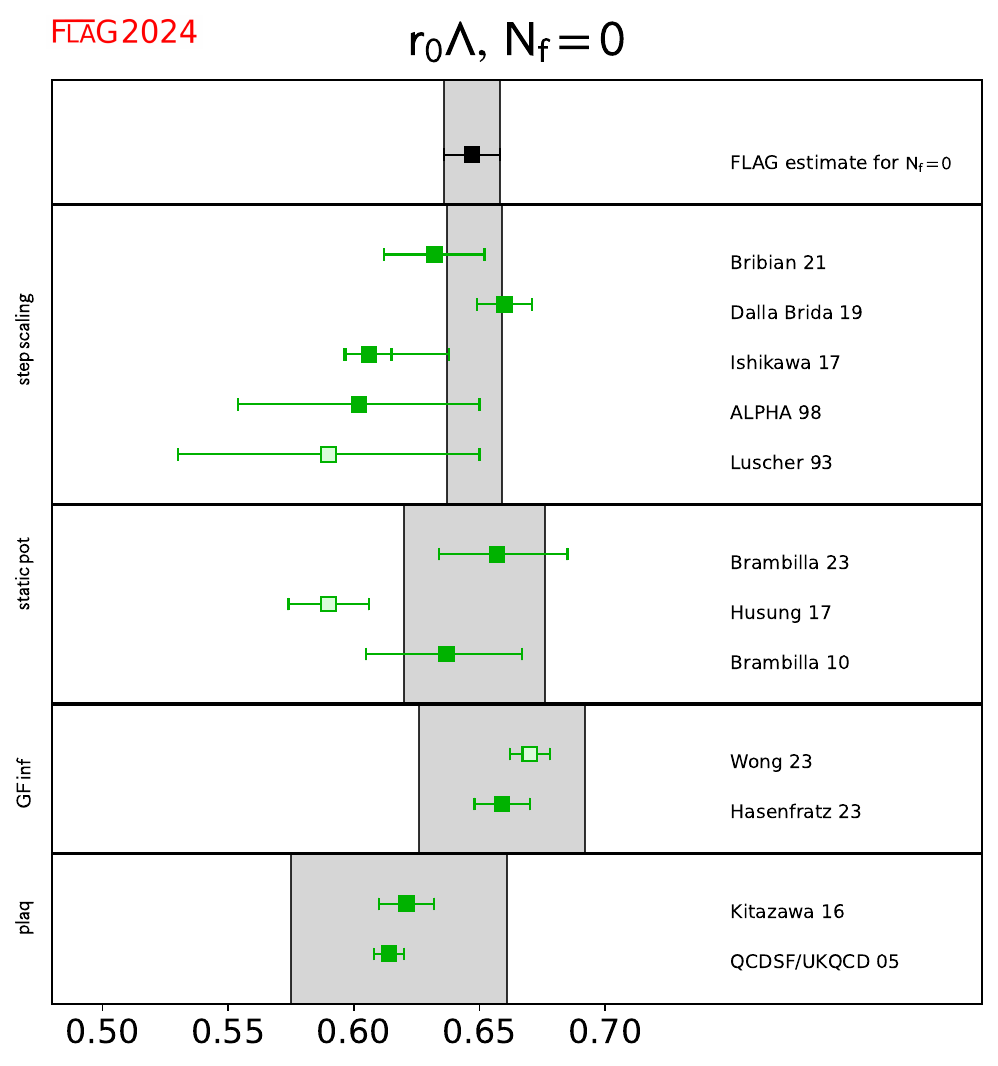}
      \end{center}
\caption{$r_0\Lambda_{\overline{\rm MS}}$ estimates for
         $\Nf = 0$ flavours. As discussed in the text, we
         group the results by method and estimate pre-ranges.
         Only full green squares are used in our final
         ranges, pale green squares indicate that the computations were
         not published or superseded by later more complete ones.}
\label{fig:r0LamMSbar_Nf0}
\end{figure}

All results are shown in Fig.~\ref{fig:r0LamMSbar} and the $\Nf=0$ results, with our pre-range by
method are shown in Fig.~\ref{fig:r0LamMSbar_Nf0}.


\begin{table}[!htb]
   \vspace{3.0cm}
   \tiny
   \begin{tabular*}
   {\textwidth}[l]{l@{\extracolsep{\fill}}rl@{\hspace{-1mm}}l@{\hspace{-1mm}}l@{\hspace{-1mm}}l@{\hspace{-1mm}}l@{\hspace{-1mm}}l@{\hspace{-1mm}}l@{\hspace{-3mm}}r}
   Collaboration & Ref. & $\Nf$ &
   \hspace{0.15cm}\begin{rotate}{60}{publication status}\end{rotate}
                                                    \hspace{-0.15cm} &
   \hspace{0.15cm}\begin{rotate}{60}{renormalization scale}\end{rotate}
                                                    \hspace{-0.15cm} &
   \hspace{0.15cm}\begin{rotate}{60}{perturbative behaviour}\end{rotate}
                                                    \hspace{-0.15cm} &
   \hspace{0.15cm}\begin{rotate}{60}{continuum extrapolation}\end{rotate}
      \hspace{-0.25cm} & 
       $\alpha_\msbar(M_\mathrm{Z})$ & Remark  & Tab. \\
   & & & & & & & & & \\[-0.1cm]
   \hline
   \hline
   & & & & & & & & & \\[-0.1cm]
   {ALPHA 17}
            & \cite{Bruno:2017gxd}    & 2+1       & \gA 
            & \good   & \good    & \good 
            & $0.11852(\pp84)$
            & step scaling
            & \ref{tab_SF3}                                   \\
  PACS-CS 09A& \cite{Aoki:2009tf} & 2+1 
            & \gA &\good &\good &\soso
            & $0.11800(300)$
            & step scaling \hspace{-0.5cm}
            & \ref{tab_SF3}                                        \\[1ex]
  \multicolumn{3}{l}{pre-range (average)}  & & & & & 0.11848(\pp81)             & &     
  \\[1ex] \hline & & & & & & & & & \\[-0.1cm]
  {AlPHA~22}
            & \cite{DallaBrida:2022eua}  & 2+1       & \gA 
            & \good  & \good   & \good
            & $0.11823(84)$
            & decoupling $\Nf=3$ to $\Nf=0$ \& step scaling
            & \ref{tab:dcpl}                            \\[1ex]
   \multicolumn{3}{l}{pre-range (average)}  & & & & & 0.11823(84)             & &    
   \\[1ex] \hline & & & & & & & & & \\[-0.1cm]
   {Ayala 20}
            & \cite{Ayala:2020odx}    & 2+1       & \gA 
            & \soso & \good   & \soso
            & $0.11836(88)$
            & $Q$-$\bar{Q}$ potential
            & \ref{tab_short_dist}                            \\[1ex]
 
 {TUMQCD 19}
            & \cite{Bazavov:2019qoo}    & 2+1       & \gA 
            & \soso & \good   & \soso
            & $0.11671(^{+110}_{-57})$
            & $Q$-$\bar{Q}$ potential (and free energy)
            & \ref{tab_short_dist}                            \\[1ex]
   
  {Takaura 18}
            & \cite{Takaura:2018lpw,Takaura:2018vcy} & 2+1  & \gA 
            & \bad  & \soso  & \soso
            & $0.11790(70)(^{+130}_{-120})$
            & $Q$-$\bar{Q}$ potential
            & \ref{tab_short_dist}                            \\[1ex]
   {Bazavov 14}
            & \cite{Bazavov:2014soa}    & 2+1       & \gA 
            & \soso & \good   & \soso
            & $0.11660(100)$
            & $Q$-$\bar{Q}$ potential
            & \ref{tab_short_dist}                            \\[1ex]
   {Bazavov 12}
            & \cite{Bazavov:2012ka}   & 2+1       & \gA 
            & \soso & \soso  & \soso
            & $0.11560(^{+210}_{-220})$ 
            & $Q$-$\bar{Q}$ potential
            & \ref{tab_short_dist}                            \\[1ex]
   \multicolumn{5}{l}{pre-range with estimated pert. error}    & & & 0.11782(165)  &      
 &      
   \\[1ex] \hline & & & & & & & & & \\[-0.1cm]
    {Cali 20} 
            & \cite{Cali:2020hrj}    & 2+1       & \gA
            & \soso  & \good  & \good
            & $0.11863(114)$
            & vacuum pol. (position space)
            & \ref{tab_vac}      \\
  {Hudspith 18} 
            & \cite{Hudspith:2018bpz}    & 2+1       & P
            & \soso  & \good     & \bad
            & $0.11810(270)(^{\pp+80}_{-220})$
            & vacuum polarization
            & \ref{tab_vac}      \\      
   JLQCD 10 & \cite{Shintani:2010ph} & 2+1 &\gA & \bad 
            & \soso & \bad
            & $0.11180(30)(^{+160}_{-170})$    
            & vacuum polarization  
            & \ref{tab_vac} \\[1ex]
  \multicolumn{5}{l}{pre-range with estimated pert. error}    & & & 0.11863(360)  &      
&      
  \\[1ex] \hline & & & & & & & & & \\[-0.1cm]

   HPQCD 10& \cite{McNeile:2010ji}& 2+1 & \gA & \soso
            & \good & \good
            & {0.11840(\pp60)}    
            & Wilson loops
            & \ref{tab_wloops}  
            \\
   Maltman 08& \cite{Maltman:2008bx}& 2+1 & \gA & \soso
            & \soso & \good
            & {$0.11920(110)$}
            & Wilson loops
            & \ref{tab_wloops}                               \\[1ex]
  \multicolumn{5}{l}{pre-range with estimated pert. error}    & & & 0.11871(128)  &      
&      
  \\[1ex] \hline & & & & & & & & & \\[-0.1cm]
{Petreczky 20} & \cite{Petreczky:2020tky}       & 2+1 & \gA &
           \soso & \soso & \good     
            & $0.11773(119)$
            & heavy current two points
             & \ref{tab_current_2pt}        \\
{Boito 20}
            & \cite{Boito:2019pqp,Boito:2020lyp}  & 2+1       & \gA 
            & \bad  &  \bad  & \soso
            & $0.1177(20)$
            & use published lattice data
            & \ref{tab_current_2pt}        \\
{Petreczky 19}
            & \cite{Petreczky:2019ozv}    & 2+1       & \gA 
            & \bad  &  \bad  & \good
            & $0.1159(12)$
            & heavy current two points
            & \ref{tab_current_2pt}      \\
{JLQCD 16}
            & \cite{Nakayama:2016atf}    & 2+1       & \gA 
            & \bad  &  \soso  & \soso
            & $0.11770(260)$
            & heavy current two points
            & \ref{tab_current_2pt}                                \\
{Maezawa 16}
            & \cite{Maezawa:2016vgv}    & 2+1       & \gA 
            & \bad  &  \bad   & \soso
            & $0.11622(\pp84)$
            & heavy current two points
            & \ref{tab_current_2pt}                            \\
{HPQCD 14A} 
            &  \cite{Chakraborty:2014aca} & 2+1+1 & \gA 
            & \soso & \good   & \soso
            & 0.11822(\pp74)
            & heavy current two points
            & \ref{tab_current_2pt}                    \\

{HPQCD 10}  & \cite{McNeile:2010ji}  & 2+1       & \gA 
            & \soso & \good  & \soso          
            & 0.11830(\pp70)          
            & heavy current two points
            & \ref{tab_current_2pt} \\
{HPQCD 08B} & \cite{Allison:2008xk}  & 2+1       & \gA 
            & \bad & \bad  & \bad
            & 0.11740(120) 
            & heavy current two points
            & \ref{tab_current_2pt}                                \\[1ex]
  \multicolumn{5}{l}{pre-range with estimated pert. error}    & & & $0.11818(119)$ &      
&      
  \\[1ex] \hline & & & & & & & & & \\[-0.1cm]
   Zafeiropoulos 19 &  \cite{Zafeiropoulos:2019flq}   & 2+1& \gA
                    & \bad & \bad  & \bad 
                    & 0.1172(11)
                    & gluon-ghost vertex
                    & 66 in \cite{FlavourLatticeAveragingGroupFLAG:2021npn}                      \\
   ETM 13D    &  \cite{Blossier:2013ioa}   & 2+1+1& \gA
                    & \soso & \soso  & \bad 
                    & 0.11960(40)(80)(60)
                    & gluon-ghost vertex
                    & 66 in \cite{FlavourLatticeAveragingGroupFLAG:2021npn}                      \\
   ETM 12C    & \cite{Blossier:2012ef}   & 2+1+1 & \gA 
                    & \soso & \soso  & \bad  
                    & 0.12000(140)
		 & gluon-ghost vertex
                    & 66 in \cite{FlavourLatticeAveragingGroupFLAG:2021npn}                      \\
   ETM 11D   & \cite{Blossier:2011tf}   & 2+1+1 & \gA 
             & \soso & \soso & \bad  
                    & $0.11980(90)(50)(^{\pp+0}_{-50})$
                    & gluon-ghost vertex
                    & 66 in \cite{FlavourLatticeAveragingGroupFLAG:2021npn}                         
\\[1ex] \hline & & & & & & & & & \\[-0.1cm]
  
  {Nakayama 18}
            & \cite{Nakayama:2018ubk}    & 2+1       & \gA
            &     \good  &  \soso      & \bad
            & $0.12260(360)$
            & Dirac eigenvalues
            & 67 in \cite{FlavourLatticeAveragingGroupFLAG:2021npn}                              \\[1ex]
   & & & & & & & & & \\[-0.1cm]
   \hline
   \hline
\end{tabular*}
\begin{tabular*}{\textwidth}[l]{l@{\extracolsep{\fill}}lllllll}
\multicolumn{8}{l}{\vbox{\begin{flushleft} 
\end{flushleft}}}
\end{tabular*}
\vspace{-0.8cm}
\caption{Results for $\alpha_\msbar(M_\mathrm{Z})$.
Different methods are listed separately and they are combined to a pre-range 
when computations are available without any \protect\bad.
The FLAG estimate is given by $0.11833(67)$, where the error
is the statistics-dominated error of the combined decoupling and step-scaling results.
}
\label{tab_alphamsbar}
\end{table} 


\subsubsection{Our range for $\alpha_{\overline{\rm MS}}^{(5)}$}
\label{subsubsect:Our range}

We now turn to the status of the essential result for phenomenology,
$\alpha_{\overline{\rm MS}}^{(5)}(M_Z)$. We only consider 
lattice results with $\Nf=3$ or $\Nf=4$ sea quarks.
Converting a $\Lambda$-parameter to $\alpha_{\overline{\rm MS}}^{(5)}(M_Z)$ 
involves the perturbative matching of the coupling across
the charm- and bottom-quark thresholds, which is available up to 4-loop order~\cite{Chetyrkin:2005ia,Schroder:2005hy}.
Note that perturbative matching at 4-loops is consistent with using the $\beta$-function at 5-loop order,
which is also available in the $\msbar$ scheme~\cite{Herzog:2017ohr,Chetyrkin:2017bjc}.
One then needs the $Z$-boson mass and the charm- and bottom-quark masses as additional
input. For definiteness, we use $m_Z=91.1876\,\GeV$, and, for the $\msbar$ quark masses at their own scale,
$m_\text{c} = 1.275(13)\,\GeV$ and $m_\text{b}=4.203(11)\,\GeV$~\cite{FlavourLatticeAveragingGroupFLAG:2021npn}.
Fortunately, the exact choices are almost irrelevant at the current accuracy: A change in
the charm-quark mass by one percent shifts the value of $\alpha_s(m_Z)$ by $3\times 10^{-5}$,
and the effect for the bottom-quark mass is even smaller. This is down by over a factor of 20 compared to the current
best total errors on $\alpha_s$. The combined perturbative uncertainty of decoupling across both
the charm- and the bottom-quark threshold is around $25\times 10^{-5}$, if one takes the difference
between 3-loop and 5-loop order as estimate, as was done, for example, 
in ALPHA 17~\cite{Bruno:2017gxd}. Even this generous estimate is still a factor 2--3 below
the best total errors. Incidentally we also note that perturbative decoupling has been
tested nonperturbatively~\cite{Athenodorou:2018wpk}. It was found that the decoupling of a heavy quark
in gluonic observables (such as the ones used to define $\alpha_\text{eff}$), 
is well described by perturbation theory. Even for the charm quark the
nonperturbative effects are expected to be at the few per-mille level. This
result justifies the use of $\Nf=3$ QCD to obtain $\alpha_{\overline{\rm MS}}^{(5)}(M_Z)$,
and it motivated the development of the decoupling method used in ALPHA~22~\cite{DallaBrida:2022eua}.

As can be seen from the tables and figures, several 
computations satisfy the FLAG criteria for inclusion in
the FLAG average. Since FLAG 21 the contribution by Petreczky~20~\cite{Petreczky:2020tky}
has been published and is now included in the average and there
is the first result from the decoupling method by the ALPHA collaboration,
ALPHA~22~\cite{DallaBrida:2022eua}.

We now explain the determination of our range.  We only include those
results without a red tag and that are published in a refereed journal.

A general issue with most determinations of $\alpha_\msbar$,
both lattice and nonlattice, is that they are dominated by
perturbative truncation errors, which are difficult to estimate.
Further, all results discussed here except for
those of Secs.~\ref{s:SF},~\ref{s:WL},~\ref{s:dec} are based on extractions of
$\alpha_\msbar$ that are largely influenced by data with
$\alpha_\mathrm{eff}\geq 0.3$.  At smaller $\alpha_s$ the momentum scale
$\mu$ quickly gets at or above $a^{-1}$. We have included computations
using $a\mu$ up to $1.5$ and $\alpha_\mathrm{eff}$ up to 0.4, but one
would ideally like to be significantly below that. Accordingly, we
choose to not simply perform weighted averages with the individual
errors estimated by each group. 
Rather, we use our own more conservative estimates of the perturbative truncation errors 
in the weighted average.  In order to improve our assessment we have also performed scale variations
as is commonly done in phenomenology. In Tab.~\ref{tab:ScaleVarsSummary}, we provide a summary of the 
variations in $\alpha_{\overline{\mathrm{MS}}}^{(5)}(M_Z)$ obtained from the procedure
explained in Sec.~\ref{s:intro} and suggested in Ref.~\cite{DelDebbio:2021ryq}. 

\begin{table}[!htb]
  \footnotesize
  \begin{tabular*}{\textwidth}{l@{\extracolsep{\fill}}cccccl}
  Observable & loops &\(Q\) [GeV] & $\delta_{(4)}^*[\%]$ & $\delta_{(2)} [\%]$ & $\delta_{(2)}^*[\%]$ Refs.\\
  \hline\hline
  Step scaling & 2 & 80 & 0.1 & 0.2 & 0.2 & \cite{Bode:1998hd,Bode:1999sm} \\
  \hline
            & 3 & 1.5 &     & 2.6 & 2.7 & 
            \cite{Fischler:1977yf,Peter:1996ig,Peter:1997me,Smirnov:2009fh,Smirnov:2008pn}\\
  Potential &   & 2.5 & 0.9 & 1.5 & 1.5 & \\
            &   & 5.0 & 0.4 & 0.8 & 0.8 & \\
  \hline
  Vacuum polarization & 3 & 1.3 & 1.0 & 11.6 & 0.6 & \cite{Cali:2020hrj}\\                    
  \hline
     $- \log W_{11}$      & 2 & 4.4 & 2.8 & 3.3 & 2.5 & \cite{Davies:2008sw,Lepage:1992xa}\\
     $-\log W_{12}/u_0^6$ &   & 4.4 & 3.5 & 3.2 & 3.1 & \\
  \hline
     HQ \(r_4\) & 2 & \(m_{\rm c}\)  & & 2.7 & 2.8 & 
      \cite{Chetyrkin:2006xg,Chetyrkin:1997mb,Broadhurst:1991fi}\\
     HQ \(r_4\) &   & \(2m_{\rm c}\) & 1.2 & 1.5 & 1.6 & \\
     HQ \(r_6\) &   & \(2m_{\rm c}\) & & 2.3 & 1.2 & \\
     HQ \(r_8\) &   & \(2m_{\rm c}\) & & 2.8 & 4.8 & \\
  \hline\hline
  \end{tabular*}
  \caption{Summary of the results of scale variations. We report results 
  for those observables for which we could use the common procedure introduced earlier.}
  \label{tab:ScaleVarsSummary}
\end{table}
In the following we first obtain separate estimates for  $\alpha_s$ from each of
the six methods with results that pass the FLAG criteria: step scaling, decoupling,
the heavy-quark potential, Wilson loops, heavy-quark current 
two-point functions and vacuum polarization. In a second step we combine them to obtain
the overall FLAG estimate. All results are collected in Tab.~\ref{tab_alphamsbar}.

\begin{itemize}
\item
{\em Step scaling\\} 
The step-scaling computations of PACS-CS~09A~\cite{Aoki:2009tf}
and ALPHA~17~\cite{Bruno:2017gxd} reach energies around the
$Z$-mass where perturbative uncertainties in the three-flavour theory 
are negligible.  We form a weighted average of the two results and obtain 
$\alpha_{\msbar}=0.11848(81)$, where the error is dominated by the statistical 
error from the simulations.
\item
{\em Decoupling\\} 
There is a single result which has been discussed in Sec.~\ref{s:dec}.
The result is $\alpha_{\msbar}=0.11823(84)$ with a statistics-dominated error. 
\item
{\em Static-quark potential computations\\} 
Brambilla 10 \cite{Brambilla:2010pp}, ETM 11C \cite{Jansen:2011vv} and Bazavov 12 \cite{Bazavov:2012ka}  give
evidence that they have reached distances where perturbation theory
can be used. However, in addition to $\Lambda$, a scale
is introduced into the perturbative prediction by the process of
subtracting the renormalon contribution. 
This subtraction is avoided in Bazavov 14 \cite{Bazavov:2014soa} by using the force and again 
agreement with perturbative running is reported.
Husung 17 \cite{Husung:2017qjz} (unpublished) studied the reliability of perturbation theory 
in the pure gauge theory with lattice spacings down to $0.015\,\fm$ and found that at weak coupling there is a downwards
trend in the $\Lambda$-parameter with a slope  $\Delta \Lambda / \Lambda \approx 9 \alpha_s^3$. 
The downward trend is broadly confirmed in Husung 20~\cite{Husung:2020pxg} albeit with larger errors.

Bazavov 14 \cite{Bazavov:2014soa} satisfies all of the criteria to enter the FLAG average for $\alpha_s$
but has been superseded by TUMQCD 19~\cite{Bazavov:2019qoo}.
Moreover, there is another study, Ayala 20~\cite{Ayala:2020odx} who use the very
same data as TUMQCD 19, but treat perturbation theory differently, resulting
in a rather different central value. This shows that perturbative truncation 
errors are the main source of errors. We combine the results
for $\lms^{\Nf=3}$ from both groups as a weighted average (with the 
larger upward error of TUMQCD 19) and take
the difference of the central values as the uncertainty of the average.
We obtain $\lms^{\Nf=3} =$ 330(24)~MeV, which translates to $\alpha_s(m_Z) = 0.11782(165)$.
This uncertainty of $1.4$ percent is in line with estimates from scale variations.

\item 
{\em Small Wilson loops\\} 
Here the situation is unchanged since FLAG~16.  In the determination of $\alpha_s$ from
observables at the lattice spacing scale, there is an interplay
of higher-order perturbative terms and lattice artifacts.
In HPQCD 05A \cite{Mason:2005zx}, HPQCD 08A \cite{Davies:2008sw}
and Maltman 08 \cite{Maltman:2008bx} both lattice artifacts (which are
power corrections in this approach) and higher-order perturbative
terms are fitted.  We note that Maltman 08~\cite{Maltman:2008bx} and
HPQCD 08A~\cite{Davies:2008sw} analyze largely the same data set but
use different versions of the perturbative expansion and treatments of
nonperturbative terms.  After adjusting for the slightly different
lattice scales used, the values of $\alpha_\msbar(M_Z)$ differ by
$0.0004$ to $0.0008$ for the three quantities considered.  In fact the
largest of these differences ($0.0008$) comes from a tadpole-improved
loop, which is expected to be best behaved perturbatively.
 We therefore replace the perturbative-truncation errors from \cite{Maltman:2008bx} 
and \cite{McNeile:2010ji} with our estimate of the perturbative uncertainty 
\eq{hpqcd:ouruncert}. Taking the perturbative errors to be 100\% correlated between the results, we obtain for the weighted average 
$\alpha_{\msbar}=0.11871(128)$. 
We note that this assessment, taken over from FLAG~21, seems optimistic in the light of the uncertainty induced
by scale variations, which are at the level of three percent for the plaquette and rectangle Wilson loops. 
One may expect that simultaneous consideration of many quantities stabilizes the estimates
as do terms of higher order in $\alpha$. It would be interesting to see a new study of this kind,
possibly with a different action. We may have to revise our range in the future.

\item 
{\em Heavy-quark current two-point functions\\}
Further computations with small errors are 
HPQCD~10 \cite{McNeile:2010ji} and HPQCD~14A~\cite{Chakraborty:2014aca},
where correlation functions of heavy valence quarks are
used to construct short-distance quantities. Due to the large quark
masses needed to reach the region of small coupling, considerable
discretization errors are present, see Fig.~30 of FLAG~16. These
are treated by fits to the perturbative running (a 5-loop running
$\alpha_{\overline{\rm MS}}$ with a fitted 5-loop coefficient in
the $\beta$-function is used) with high-order terms in a double expansion
in $a^2\Lambda^2$ and $a^2 m_\mathrm{c}^2$ supplemented by priors
which limit the size of the coefficients.  The priors play an
especially important role in these fits given the much larger number
of fit parameters than data points.  We note, however, that the size
of the coefficients does not prevent high-order terms from
contributing significantly, since the data includes values of
$am_\text{c}$ that are rather close to one.  

From a physics perspective it seems natural to use the renormalization scale set by the charm-quark mass; however, this
implies $\alpha_\text{eff} \simeq 0.38$, which is the reason why JLQCD 16, Petreczky 19~\cite{Petreczky:2019ozv} 
and Boito~20~\cite{Boito:2020lyp} do not pass the FLAG criteria. 
Still some valuable insight can be gained from these works.
While Petreczky 19/Petreczky 20 share the same lattice data for heavy quark masses in the range $m_h=m_c$--$4m_c$ 
they use a different strategy for continuum extrapolations and a different treatment of perturbative uncertainties. 
Petreczky 19 \cite{Petreczky:2019ozv} perform continuum extrapolation separately for each value of the valence-quark
mass, while Petreczky 20 rely on joint continuum extrapolations of the lattice data at different heavy-quark masses,
similar to the analysis of HPQCD, but without Bayesian priors. It is concluded that reliable continuum extrapolations 
for $m_h \ge 2 m_c$ require a joint fit to the data. This limits the eligible $\alpha_s$ 
determinations in Petreczky 19 \cite{Petreczky:2019ozv} to $m_h=m_c$ and $1.5m_c$, for which, however, 
the FLAG criteria are not satisfied.  There is also a difference in the choice of renormalization scale between both analyses: 
Petreczky 19 \cite{Petreczky:2019ozv} uses $\mu= m_h$, while Petreczky 20 \cite{Petreczky:2020tky} considers 
several choices of $\mu$ in the range $\mu=2/3 m_h$--$3 m_h$, which leads to larger perturbative uncertainties 
in the determination of $\alpha_s$ \cite{Petreczky:2020tky}. 
Boito 20 \cite{Boito:2020lyp} use published continuum extrapolated lattice results
for $m_h=m_c$ and performs their own extraction of $\alpha_s$. Limiting the choice of $m_h$ to the charm-quark mass means
that the FLAG criteria are not met ($\alpha_\text{eff} \simeq 0.38$). However, their analysis gives valuable insight
into the perturbative error. In addition to the renormalization scale $\mu$, Boito 20 also vary
the renormalization scale $\mu_m$ at which the charm-quark mass is defined. The corresponding result
$\alpha_s(M_Z)=0.1177(20)$ agrees well with previous lattice determination but has a larger error, which is dominated
by the perturbative uncertainty due to the variation of both scales. 

Since the FLAG~21 report the results of Petreczky 20 have been published and pass all FLAG criteria. There are now three determinations of $\alpha_s$ from the heavy-quark current two-point functions
that satisfy all the FLAG criteria 
and enter the FLAG average: $\alpha_{\overline{MS}}(M_Z)=0.11773(119)$ from
Petreczky 20 \cite{Petreczky:2020tky}, $\alpha_{\overline{MS}}(M_Z)=0.11822(74)$
from HPQCD 14 \cite{Chakraborty:2014aca} and $\alpha_{\overline{MS}}(M_Z)=0.11830(70)$ from
HPQCD 10 \cite{McNeile:2010ji}.
All three determinations agree well with each other within errors. 
Since these determinations are uncorrelated we take the weighted average of these
results as an estimate for the strong coupling constants from the heavy-quark current two-point functions.
The analysis in Petreczky 20 does not use Bayesian priors and considers five different
choices of the renormalization scale, while HPQCD 10 and HPQCD 14 analyses use $\mu=3 m_c$. Therefore,
the error of Petreczky 20 can be considered to be more conservative and we take it as the range for
$\alpha_{\overline{MS}}(m_Z)$. With this we arrive at $\alpha_{\overline{MS}}(M_Z)=0.11818(119)$ from
the method of the heavy-quark current two-point functions. Comparing with the scale variations, the 
perturbative uncertainty is estimated to be 1-2 percent so a one percent range is roughly in line.

\item
{\em Light-quark vacuum polarization \\}
Cali 20~\cite{Cali:2020hrj} use the light-quark current two-point functions in position space,
evaluated on a subset of CLS configurations for lattice spacings in the range 0.038--0.076~fm, and for Euclidean 
distances 0.13--0.19~fm, corresponding to renormalization scales $\mu=$ 1--1.5~GeV.
Both flavour-nonsinglet vector and axial-vector currents are considered and their difference is shown to vanish within errors.
After continuum and chiral limits are taken, the effective coupling from the axial-vector two-point function is converted
at 3-loop order to $\alpha_\msbar(\mu)$. The authors do this by numerical solution for $\alpha_\msbar$ and then
perform a weighted average of the $\Lambda$-parameter estimates for the available energy range, which yields $\lms^{\Nf=3}= 342(17)$~MeV. 
Note that this is the first calculation in the vacuum polarization category that passes the current FLAG criteria.
Yet the renormalization scales are rather low and one might suspect that nonperturbative effects that do not break chiral symmetry  
may still be sizeable.  Our main issue is a rather optimistic estimate 
of perturbative truncation errors, based only on the variation of the $\Lambda$-parameter from the range of effective couplings considered. 
If the solution for the $\msbar$ coupling is done by series expansion in $\alpha_\text{eff}$, the differences in $\alpha_\msbar$, formally of order $\alpha_\text{eff}^5$, are still 
large at the scales considered. Hence, as a measure of the systematic uncertainty we take the difference $409-355$~MeV 
between $\lms^{\Nf=3}$ estimates at $\mu=1.5$~GeV as a proxy for the total error, i.e., $\lms^{\Nf=3} =$ 342(54)~MeV, 
which translates to our pre-range, $\alpha_s(m_Z)=0.11863(360)$, from vacuum polarization.
Looking at scale variation it appears that these are of O(10) percent if the scale is identified as done by the authors.
The scale is simply too low for perturbation theory. It is an interesting observation that a variation around the scale of fastest apparent convergence, cf.~Sec.~\ref{s:trunc},
yields much smaller ambiguities of the order of one percent. A reanalysis of the data might be warranted.

\item 
{\em Other methods}
\\
Computations using other methods do not qualify for an average yet,
predominantly due to a lacking \soso\ in the continuum extrapolation. 
\end{itemize}

\begin{figure}[h]
   \begin{center}
      \includegraphics[width=11.0cm]{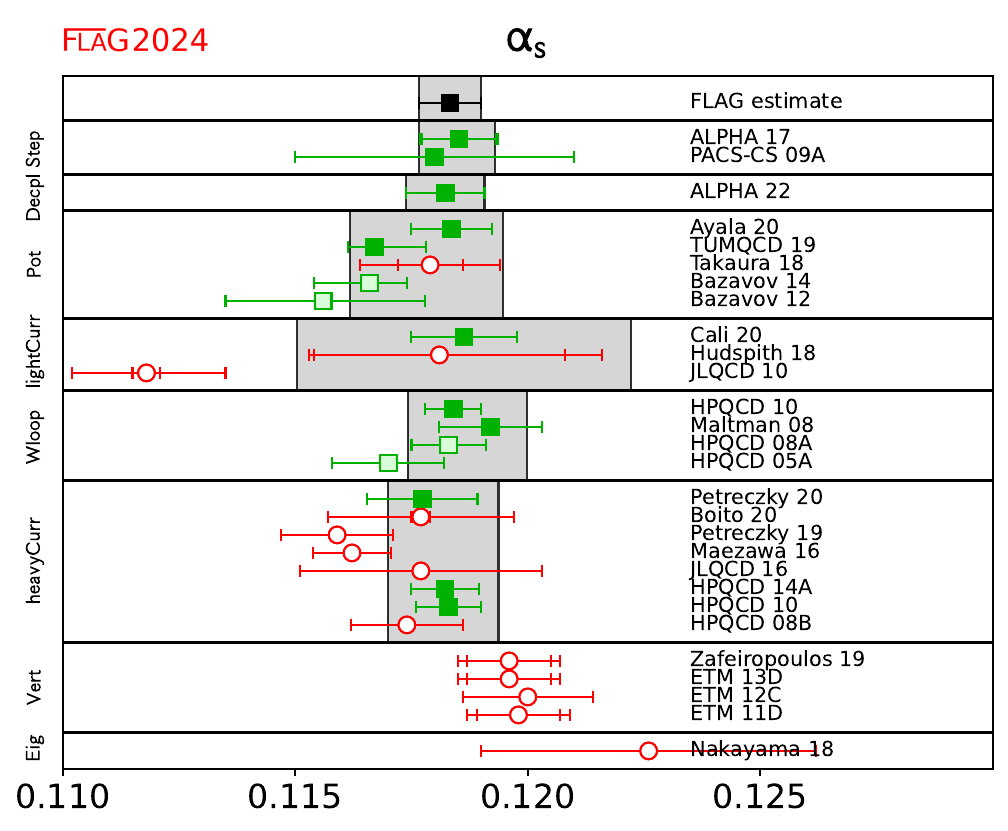}\\
      \includegraphics[width=11.0cm]{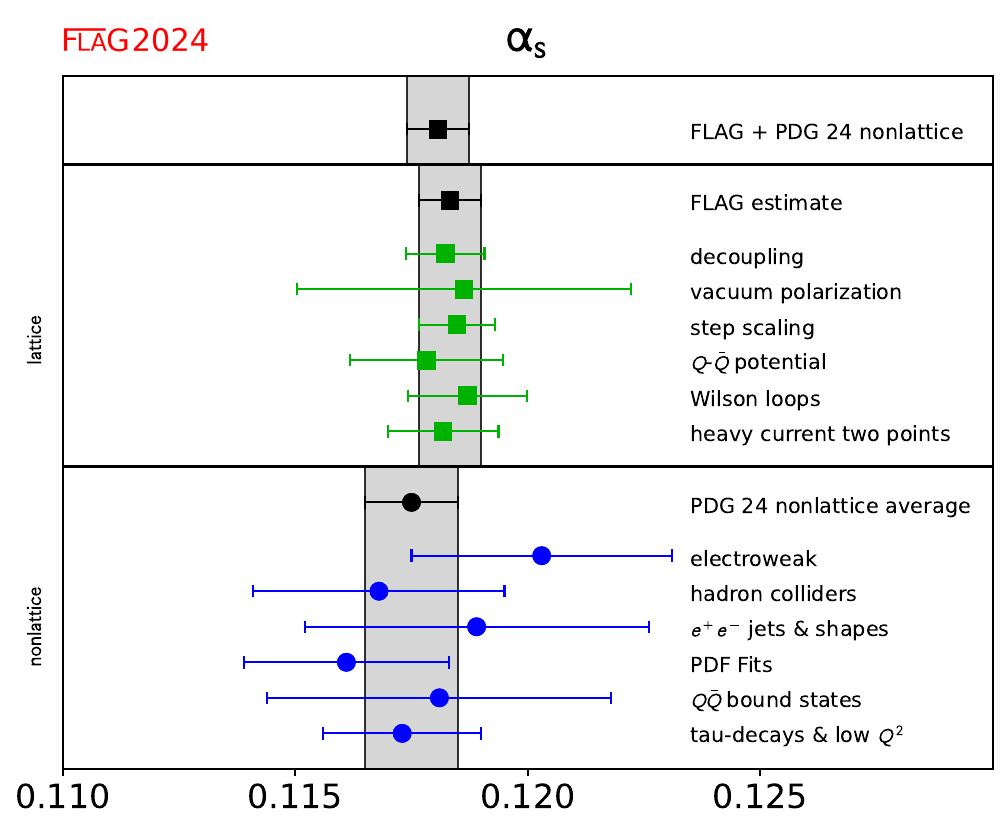}
   \end{center}
\caption{$\alpha_{\overline{\rm MS}}^{(5)}(M_Z)$, the coupling
  constant in the $\overline{\rm MS}$ scheme at the $Z$-boson mass. 
Top: lattice  results, pre-ranges from different calculation methods, and final average. 
Bottom: Comparison of the lattice pre-ranges and average with the nonlattice ranges and average.
The first PDG~24 entry gives the outcome of their analysis excluding lattice results.
At the very top we display the weighted average of PDG~24 nonlattice and FLAG lattice estimates,
with the error taken from the FLAG estimate (statistics dominated),
  see Sec.~\ref{subsubsec:alpha_s_Conclusions}.
\label{Fig:alphasMSbarZ}}
\end{figure}

We form the average in two steps, due to the known correlation between ALPHA~17 and ALPHA~22. We thus first combine these two results
by combining the respective $\Lambda$-parameters and then obtain $\alpha_s(m_Z)= 0.11836(69)$. 
Next we combine with the step-scaling result by PACS~CS-09A,
and get $\alpha_\msbar^{(5)}(m_Z)= 0.11834(67)$. This average is interesting as it combines 
the three results where the error is dominated by statistics.
A weighted average with the remaining pre-ranges yields the central value,
we quote as the new FLAG estimate,
\begin{eqnarray}
  \alpha_{\overline{\rm MS}}^{(5)}(M_Z) = 0.11833(67) = 0.1183(7) \,.
 \label{eq:alpmz}
\end{eqnarray}
where we have used the above statistics-dominated error as our range, rather than the 
25 percent smaller error from the weighted average.
All central values are remarkably consistent, as can also be seen in
Figure~\ref{Fig:alphasMSbarZ}.


\subsubsection{Conclusions}
\label{subsubsec:alpha_s_Conclusions}


With the present results our range for the strong coupling is
(repeating Eq.~(\ref{eq:alpmz}))
\begin{eqnarray*}
 \FLAGAVBEGIN \alpha_{\overline{\rm MS}}^{(5)}(M_Z) = 0.1183(7)\FLAGAVEND\qquad
 \Refs~\mbox{\cite{DallaBrida:2022eua,Petreczky:2020tky,Ayala:2020odx,Bazavov:2019qoo,Cali:2020hrj,Bruno:2017gxd,Chakraborty:2014aca,McNeile:2010ji,Aoki:2009tf,Maltman:2008bx}}, 
\end{eqnarray*}
and the associated $\Lambda$-parameters
\begin{eqnarray}
  \FLAGAVBEGIN \Lambda_{\overline{\rm MS}}^{(5)} = 213(8)\FLAGAVEND\,\MeV\hspace{5mm}\qquad
  \Refs~\mbox{\cite{DallaBrida:2022eua,Petreczky:2020tky,Ayala:2020odx,Bazavov:2019qoo,Cali:2020hrj,Bruno:2017gxd,Chakraborty:2014aca,McNeile:2010ji,Aoki:2009tf,Maltman:2008bx}}, 
  \\
  \FLAGAVBEGIN \Lambda_{\overline{\rm MS}}^{(4)} = 295(10)\FLAGAVEND\,\MeV\hspace{5mm}\qquad
  \Refs~\mbox{\cite{DallaBrida:2022eua,Petreczky:2020tky,Ayala:2020odx,Bazavov:2019qoo,Cali:2020hrj,Bruno:2017gxd,Chakraborty:2014aca,McNeile:2010ji,Aoki:2009tf,Maltman:2008bx}}, 
  \\
  \FLAGAVBEGIN \Lambda_{\overline{\rm MS}}^{(3)} = 338(10)\FLAGAVEND\,\MeV\hspace{5mm}\qquad
  \Refs~\mbox{\cite{DallaBrida:2022eua,Petreczky:2020tky,Ayala:2020odx,Bazavov:2019qoo,Cali:2020hrj,Bruno:2017gxd,Chakraborty:2014aca,McNeile:2010ji,Aoki:2009tf,Maltman:2008bx}}, 
\end{eqnarray} 
%
Compared with FLAG 21, the central values have only moved slightly and the 
errors have been reduced by ca.~15-20 percent.
Overall we find excellent agreement between all published results that pass the FLAG criteria.
The error for the reference value $\alpha_{\overline{\rm MS}}(M_Z)$ has reached the level of 
$0.6$ percent, and, as we emphasize again, is dominated by the statistical errors originating
from the stochastic process inherent in lattice simulations.
The same cannot be said about nonlattice determinations, for which PDG~24 quote the value
$\alpha^{(5)}_{\overline{\rm MS}}(M_Z) = 0.1175(10)$. Combining FLAG and PDG nonlattice
estimates, we obtain
\begin{eqnarray}
    \alpha^{(5)}_{\overline{\rm MS}}(M_Z) &=&   0.1181(7) \,, \quad 
    \mbox{FLAG 24 + PDG 24},
 \label{PDG_FLAG_alpha}  
 \end{eqnarray}
where we assign the error of the FLAG estimate as our range. In Fig.~\ref{Fig:alphasMSbarZ},
we have collected and summarized the results that go into the FLAG estimate and the PDG~24 average.
The agreement with nonlattice results is very good. Despite our conservative error estimate
the FLAG lattice estimate has an error that is 30\% smaller than the PDG~24 nonlattice result.
Compared to high-energy experiments, lattice QCD has the advantage that the complicated 
transition between hadronic and quark and gluon degrees of freedom never needs 
to be dealt with explicitly. All hadronic input quantities are very
well measured properties of hadrons, such as their masses and decay widths.
We would like to encourage experimentalists and phenomenologists at collider experiments to make
use of the FLAG lattice estimate. The higher accuracy and precision, with improvements still possible and expected in the near future, 
may help our understanding of other important physics aspects at the LHC and in other experiments. 
Currently, many experiments attempt their own determination of $\alpha_s$, and the spread of the results 
is then taken as indication of the size of systematic effects. While this provides valuable information, one may ask whether
one can learn more from the data about the origin of the systematic uncertainties, by 
using the precise lattice result for $\alpha_s$ as input for the analysis.
This may clarify where tensions or inconsistencies arise and help our understanding of
nonperturbative effects, e.g., in hadronization processes, or in some corners of parameter space.
There is also the theoretical possibility that QCD does not provide the full picture of the strong interactions.
While experimental data would be affected by any new physics, lattice QCD, by design, excludes such effects. Hence, any 
inconsistencies encountered in the analysis might also point to such new effects.

We finish by commenting on perspectives for the future. This edition of the FLAG report has seen the  first result
from the decoupling strategy, which complements the step-scaling result.
In fact, the decoupling result also relies on the step-scaling technique, however, here it is applied in
the $\Nf=0$ theory and therefore technically simpler, and with different systematics.
The nice agreement between $\Nf=3$ step-scaling and decoupling results is therefore a very strong
consistency check. Of course, further results with different schemes and systematics
would be very welcome. For step scaling with $\Nf=0$, Dalla~Brida~19 have used two different finite-volume
schemes with SF boundary conditions, and there is now a new result by Bribian~21 with twisted periodic b.c.'s.
There are also results with the GF scheme in infinite volume, where the $\beta$-function
can be measured directly, by Hasenfratz~23 and Wong~23. In some sense, the case of $\Nf=0$ flavours
is more difficult than full QCD, in that the asymptotic regime is often harder to reach. Of course,
part of the problem lies in the smallness of statistical errors, which means that even moderate 
systematic errors easily stand out. In particular, in GF schemes, both in finite and infinite volume, 
the parametric uncertainties in the $\Lambda$-parameter of order $\alpha^{n_{\mathrm{l}}}$, Eq.~(\ref{eq:ii}), 
can be still quite large at the largest scales reached while showing the expected asymptotic behaviour
$\propto \alpha^{n_{\mathrm{l}}}$ over a wide range. Rather than assigning a large systematic uncertainty at
the highest scale reached, one might be inclined  to allow for an extrapolation in $\alpha^{n_{\mathrm{l}}}$, 
together with a data-driven criterion to assess its quality. 
We will reconsider this issue in the next edition of the FLAG report.

Finally we emphasize the importance that errors remain dominated by statistics. Only
in this case a probabilistic interpretation is obvious. This is currently not the case
for the majority of lattice calculations, the exception being the step-scaling and decoupling approaches.
For those determinations, further improvements will require access to higher energy scales,
for instance, by implementing some elements of the step-scaling approach.

%% file: NME/NME.tex

\section{Nucleon matrix elements \label{sec:NME}}
Authors: S.~Collins, R.~Gupta, A.~Nicholson, H.~Wittig\\

A large number of experiments testing the Standard Model (SM) and
searching for physics Beyond the Standard Model (BSM) involve either
free nucleons (proton and neutron beams) or the scattering of
electrons, muons, neutrinos and dark matter off nuclear
targets. Necessary ingredients in the analysis of the experimental
results are the matrix elements of various probes (fundamental
currents or operators in a low-energy effective theory) between
nucleon or nuclear states. The goal of lattice-QCD calculations in
this context is to provide high-precision predictions of these matrix
elements, the simplest of which give the nucleon charges and form
factors. Determinations of the charges, the first Mellin moments of
parton distribution functions, are the most mature and in this review
we update results for twelve quantities, the isovector and flavour-diagonal axial vector, scalar and tensor charges, given in the two
previous FLAG reports in 2019 and
2021~\cite{FlavourLatticeAveragingGroupFLAG:2021npn,FlavourLatticeAveragingGroup:2019iem}.
In this edition in Sec.~\ref{sec:moments}, we also add a review of the
second Mellin moments for the vector, axial and tensor currents that
give the momentum fraction, the helicity moment and the transversity
moment as a sufficient number of calculations have been performed and
the results are considered robust.

Other quantities that are not being reviewed but for which significant
progress has been made in the last five years are the nucleon axial
vector and electromagnetic form
factors~\cite{RQCD:2019jai,Jang:2019jkn,Hasan:2019noy,Alexandrou:2020okk,Park:2021ypf,Meyer:2021vfq,Djukanovic:2021cgp,Djukanovic:2022wru,Tsuji:2023llh,Djukanovic:2023beb,Djukanovic:2023jag,Djukanovic:2023cqe,Jang:2023zts,Alexandrou:2023qbg} 
and parton distribution functions from matrix elements of nonlocal
operators~\cite{Lin:2017snn,Constantinou:2020pek,Constantinou:2020hdm,Cichy:2018mum,Monahan:2018euv,Constantinou:2022yye,DelDebbio:2022tsj}.
The more challenging calculations of nuclear matrix elements that are
needed, for example, to calculate the cross-sections of neutrinos or
dark matter scattering off nuclear targets, are proceeding along three
paths. The first is based on direct evaluations of matrix elements calculated with
initial and final states consisting of multiple
nucleons~\cite{Savage:2011xk,Chang:2017eiq}. 
The second proceeds by matching few-nucleon observables
  computed in lattice QCD to nuclear effective field theories and
  extrapolating in the mass number $A$, while the third strategy uses
  the HAL~QCD method \cite{Iritani:2017wvu} or the direct method
  \cite{Wagman:2017tmp} to extract nuclear forces and currents from
  lattice calculations as input for {\it ab initio} many-body methods.
We expect future FLAG
reviews to include results on these quantities once a sufficient level
of control over all the systematics is reached.

\subsection{Isovector and flavour-diagonal charges of the nucleon\label{sec:intro}}

The simplest nucleon matrix elements are composed of local quark-bilinear operators, $\overline{q_i} \Gamma_\alpha q_j$, where
$\Gamma_\alpha$ can be any of the sixteen Dirac matrices. In this
report, we consider two types of flavour structures: (a) when $i = u$
and $j = d$. These $\overline{u} \Gamma_\alpha d$ operators arise in
$W^\pm$ mediated weak interactions such as in neutron or pion decay.
We restrict the discussion to the matrix elements of the axial-vector~($A$),
scalar~($S$) and tensor~($T$) currents, which give the isovector charges,
$g_{A,S,T}^{u-d}$.\footnote{In the isospin-symmetric limit $\langle
  p|\bar{u}\Gamma d|n\rangle=\langle p|\bar{u}\Gamma u-\bar{d}\Gamma
  d|p\rangle=\langle n|\bar{d}\Gamma d-\bar{u}\Gamma u|n\rangle$ for
  nucleon and proton states $|p\rangle$ and $|n\rangle$,
  respectively. The latter two~(equivalent) isovector matrix elements are computed
  on the lattice.  } (b) When $i = j $ for $j \in \{u, d, s\}$,
there is no change of flavour, e.g., in processes mediated via the
electromagnetic or weak neutral interaction or dark matter.  These
$\gamma$ or $Z^0$ or possible dark matter mediated processes couple to all
flavours with their corresponding charges. Since these probes interact
with nucleons within nuclear targets, one has to include the effects
of QCD (to go from the couplings defined at the quark and gluon level
to those for nucleons) and nuclear forces in order to make contact with 
experiments. The isovector and flavour-diagonal charges, given by the
matrix elements of the corresponding operators calculated between nucleon states,
are these nucleon level couplings. Here we review results for the
light and strange flavours, $g_{A,S,T}^{u}$, $g_{A,S,T}^{d}$, and
$g_{A,S,T}^{s}$ and the isovector charges $g_{A,S,T}^{u-d}$.

The isovector and flavour-diagonal operators also arise in BSM
theories due to the exchange of novel force carriers or as effective
interactions due to loop effects.  The associated couplings are
defined at the energy scale $\Lambda_{\rm BSM}$,
while lattice-QCD calculations of matrix elements are carried out at a hadronic
scale, $\mu$, of a few GeV. The tool for connecting the couplings at
the two scales is the renormalization group. Since the operators of
interest are composed of quark fields~(and more generally also of gluon
fields), the predominant change in the corresponding couplings under a
scale transformation is due to QCD.  To define the operators and their
couplings at the hadronic scale $\mu$, one constructs renormalized
operators, whose matrix elements are finite in the continuum limit. This requires
calculating both multiplicative renormalization factors, including the
anomalous dimensions and finite terms, and the mixing with other
operators. We discuss the details of the renormalization factors
needed for each of the six operators reviewed in this report in
Sec.~\ref{sec:renorm}.

\begin{figure}[tpb] 
\centerline{
\includegraphics[width=0.32\linewidth]{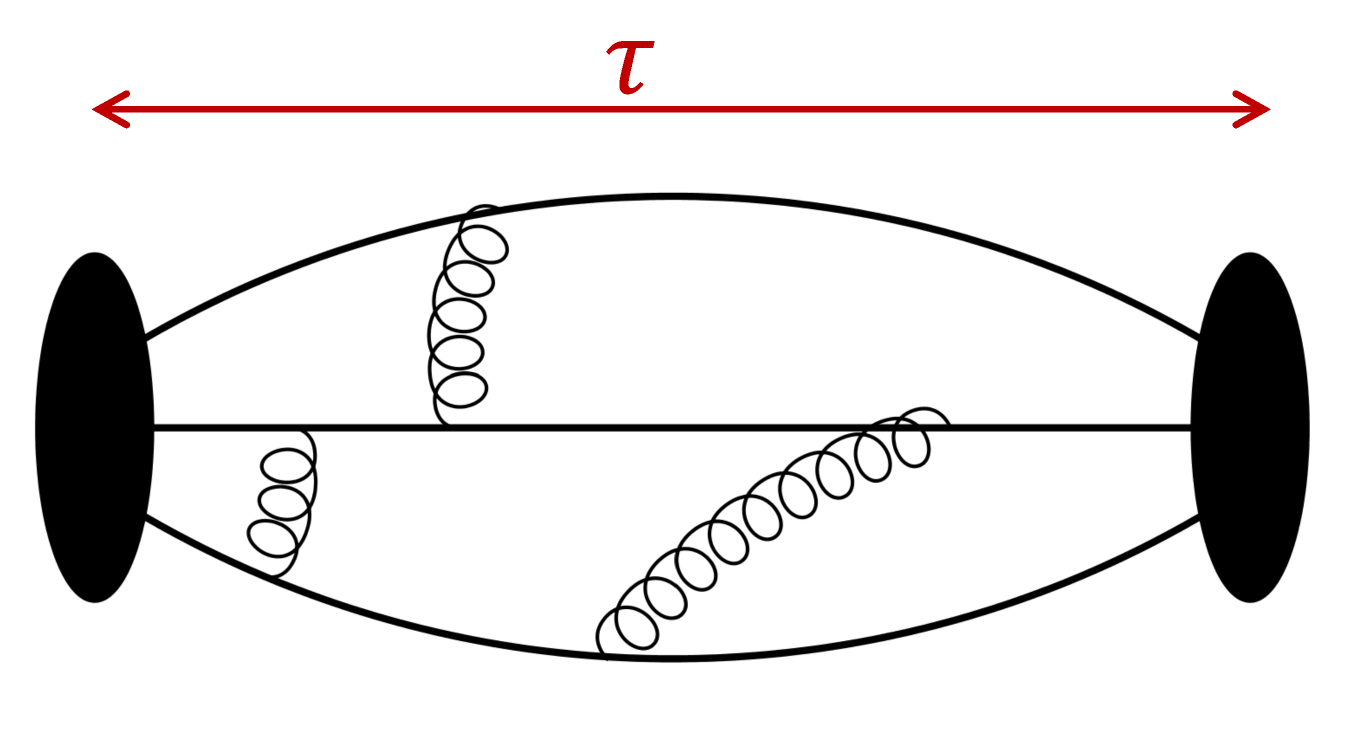} 
\includegraphics[width=0.32\linewidth]{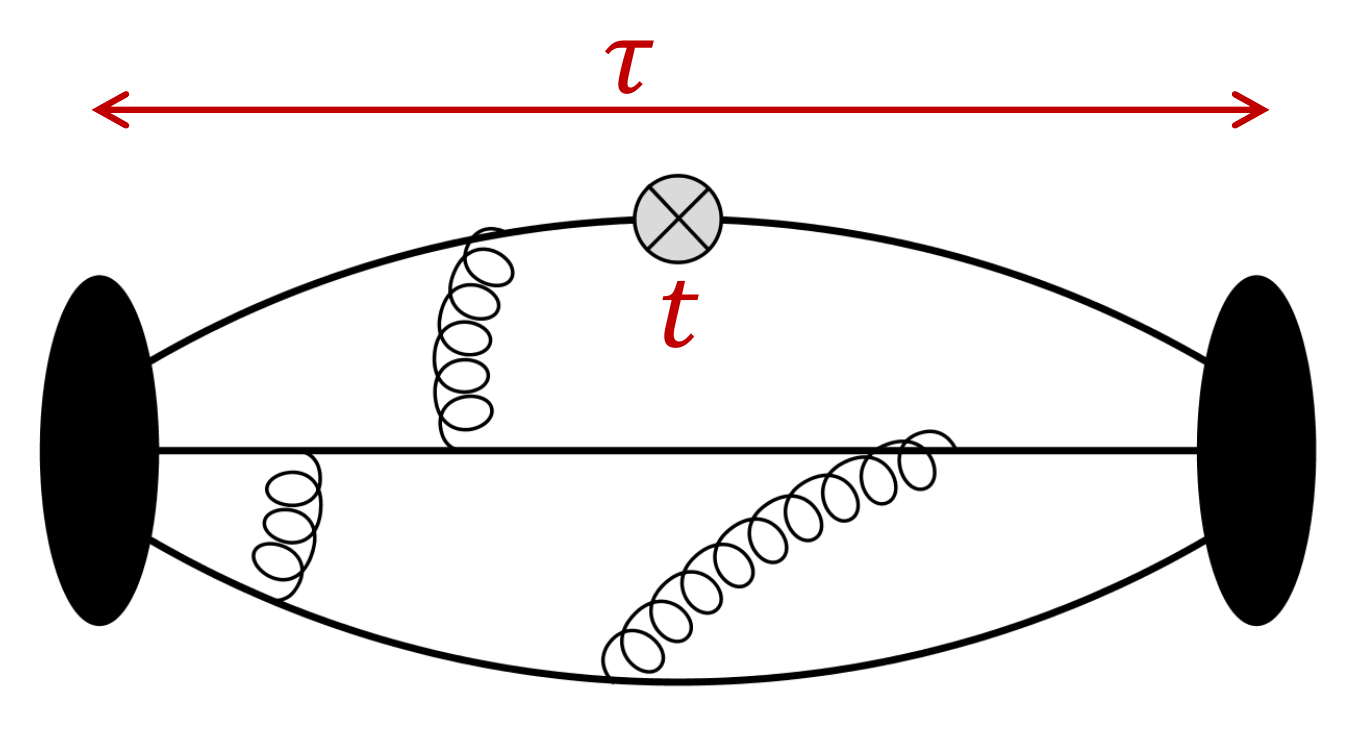}
\includegraphics[width=0.32\linewidth]{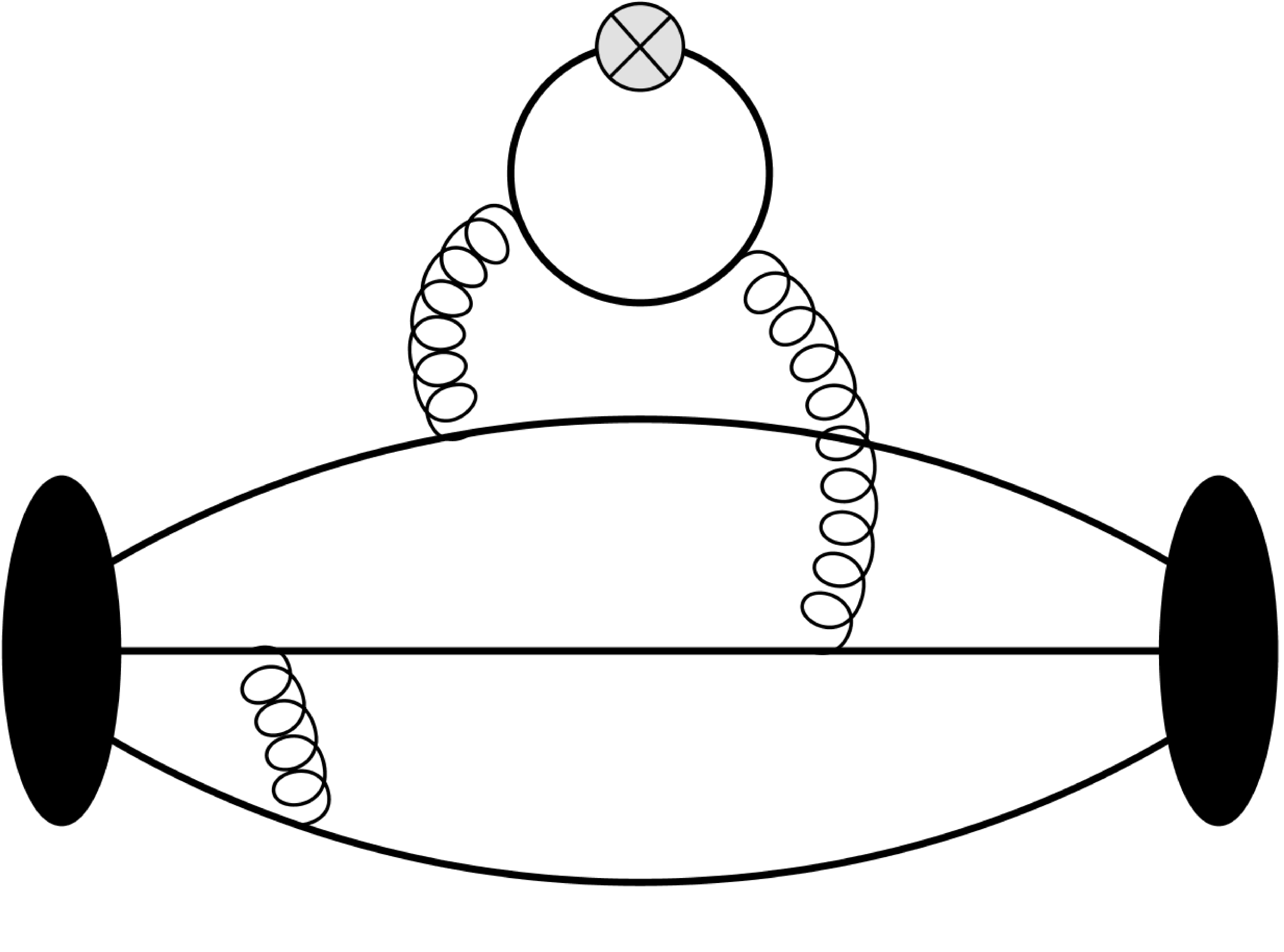}}
\caption{The two- and three-point correlation functions (illustrated by Feynman diagrams) that need to be
  calculated to extract the ground-state nucleon matrix
  elements. (Left) the nucleon two-point function. (Middle) the
  connected three-point function with source-sink separation $\tau$
  and operator-insertion time slice $t$. (Right) the quark-disconnected
  three-point function with operator insertion at $t$. }
\label{fig:feynman}
\end{figure}

Once renormalized operators are defined, the nucleon matrix elements of
interest are extracted using expectation values of two-point and
three-point correlation functions illustrated in
Fig.~\ref{fig:feynman}, where the latter can have both quark-line connected 
and disconnected contributions. In order to isolate the
ground-state matrix element, these correlation functions are analyzed
using their spectral decomposition. The current practice is to fit the
$n$-point correlation functions (or ratios involving three- and
two-point functions) including contributions from one or two excited
states.  In some cases, such as axial and vector operators, Ward
identities provide relations between correlation functions, or ground-state matrix elements, or facilitate the calculation of renormalization
factors.  It is important to ensure that all such Ward identities
are satisfied in lattice calculations, especially as in the case of
axial form factors where they provide checks of whether excited-state
contamination has been removed in obtaining matrix elements within
ground-state nucleons~\cite{Liang:2018pis,Jang:2019vkm,RQCD:2019jai}.

The ideal situation occurs if the time separation $\tau$ between the
nucleon source and sink positions, and the distance of the operator-insertion time from the source and the sink, $t$ and $\tau - t$,
respectively, are large enough such that the contribution of all
excited states is negligible. In the limit of large $\tau$, the ratio
of noise to signal in the nucleon two- and three-point correlation
functions grows exponentially as $e^{(M_N -
  \frac{3}{2}M_\pi)\tau}$~\cite{Hamber:1983vu,Lepage:1989hd}, where
$M_N$ and $M_\pi$ are the masses of the nucleon and the pion,
respectively. Therefore, in particular at small pion masses,
maintaining reasonable errors for large $\tau$ is challenging, with
most current calculations limited to $\tau \lesssim 1.5$~fm. In addition,
the mass gap between the ground and excited (including multi-particle)
states is smaller than in the meson sector and at these separations,
excited-state effects can be significant. The approach commonly taken
is to first obtain results with high statistics at multiple values of
$\tau$, using the methods described in Sec.~\ref{sec:technical}. Then,
as mentioned above, excited-state contamination is removed by fitting
the data using a fit form involving one or two excited states. The
different strategies that have been employed to minimize excited-state
contamination are discussed in Sec.~\ref{sec:ESC}.

Usually, the quark-connected part of the three-point function
(corresponding to the plot in the centre of Fig.~\ref{fig:feynman}) is
computed via the so-called ``sequential propagator method'', which
uses the product of two quark propagators between the positions of the
initial and the final nucleons as a source term for another inversion
of the lattice Dirac operator. This implies that the position of the
sink timeslice is fixed at some chosen value. Varying the value of the
source-sink separation $\tau$ then requires the calculation of another
sequential propagator.

The evaluation of quark-disconnected contributions is computationally
more challenging as the disconnected loop~(which contains the operator
insertion, as illustrated in Fig.~\ref{fig:feynman} right) is needed
at all points on a particular timeslice or, in general, over the whole
lattice. The quark loop is computed stochastically and then correlated
with the nucleon two-point function before averaging this three-point
function over the ensemble of gauge configurations. The associated
statistical error, therefore, is a combination of that due to the
stochastic evaluation~(on each configuration) and that from the gauge
average. The number of stochastic sources employed on each
configuration is, typically, optimized to reduce the overall error for
a given computational cost. The statistical errors of the connected
contributions, in contrast, usually come only from the ensemble
average since they are often evaluated exactly on each configuration,
for a small number of source positions. If these positions are
well-separated in space and time, then each measurement is
statistically independent. The methodology applied for these
calculations and the variance reduction techniques are summarized in
Sec.~\ref{sec:technical}. By construction, arbitrary values of $\tau$
across the entire temporal extent of the lattice can be realized when
computing the quark-disconnected contribution, since the source-sink
separation is determined by the part of the diagram that corresponds
to the two-point nucleon correlator. However, in practice, statistical
fluctuations of both the connected and disconnected contributions
increase sharply, so that the signal is lost in the statistical noise
for $\tau\gtrsim1.5$\,fm.

The lattice calculation is performed for a given number of quark
flavours and at a number of values of the lattice spacing $a$, the pion
mass $M_\pi$, and the lattice size, represented by $M_\pi L$. The
results need to be extrapolated to the physical point defined by
$a=0$, $M_\pi = 135$~MeV and $M_\pi L \to \infty$. This is done by
fitting the data simultaneously in these three variables using
a theoretically motivated ansatz. The ans\"atze used and the fitting
strategy are described in Sec.~\ref{sec:extrap}.

The procedure for rating the various calculations and the
criteria specific to this chapter are discussed in
Sec.~\ref{sec:rating}, which also includes a brief description of how
the final averages are constructed. The physics motivation for
computing the isovector charges, $g_{A,S,T}^{u-d}$, and the review of
the lattice results are presented in Sec.~\ref{sec:isovector}. This is
followed by a discussion of the relevance of the flavour-diagonal
charges, $g_{A,S,T}^{u,d,s}$, and a presentation of the lattice results in Sec.~\ref{sec:FDcharges}.

\subsubsection{Technical aspects of the calculations of nucleon matrix elements\label{sec:technical}}

The calculation of $n$-point functions needed to extract nucleon matrix
elements requires making four essential choices. The first involves choosing between 
the suite of background gauge field ensembles one has access to. The range of lattice parameters 
should be large enough to facilitate the
extrapolation to the continuum and infinite-volume limits, and, ideally, the
evaluation at the physical pion mass taken to be~
$M_\pi=135$~MeV. Such ensembles have been generated with a variety of
discretization schemes for the gauge and fermion actions that have
different levels of improvement and preservation of continuum
symmetries. 
The actions employed at present
include (i) Wilson gauge with nonperturbatively improved
Sheikholeslami-Wohlert fermions~(nonperturbatively improved clover
fermions)~\cite{Khan:2006de,Bali:2012qs,Horsley:2013ayv,Capitani:2012gj,Bali:2014nma,Bali:2016lvx,Capitani:2017qpc}, (ii) Iwasaki gauge with nonperturbatively improved
clover fermions~\cite{Ishikawa:2009vc,Ishikawa:2018rew}, (iii) Iwasaki gauge with twisted-mass
fermions with a clover term~\cite{Abdel-Rehim:2015owa,Abdel-Rehim:2016won,Alexandrou:2017hac,Alexandrou:2017oeh,Alexandrou:2017qyt}, (iv) tadpole Symanzik improved
gauge with highly improved staggered quarks
(HISQ)~\cite{Bhattacharya:2013ehc,Bhattacharya:2015wna,Bhattacharya:2015esa,Bhattacharya:2016zcn,Berkowitz:2017gql,Gupta:2018lvp,Lin:2018obj,Gupta:2018qil,Chang:2018uxx}, (v) Iwasaki gauge with domain-wall fermions
(DW)~\cite{Yamazaki:2008py,Yamazaki:2009zq,Aoki:2010xg,Gong:2013vja,Yang:2015uis,Gong:2015iir,Liang:2018pis} and (vi) Iwasaki gauge with overlap fermions~\cite{Ohki:2008ff,Oksuzian:2012rzb,Yamanaka:2018uud}. For details of
the lattice actions, see the glossary in the Appendix A.1 of FLAG 19 \cite{FlavourLatticeAveragingGroup:2019iem}.

The second choice is of the valence-quark action. Here there are two
choices, to maintain a unitary formulation by choosing exactly the
same action as is used in the generation of gauge configurations or to
choose a different action and tune the quark masses to match the
pseudoscalar meson spectrum in the two theories.  Such mixed-action
formulations are nonunitary but are expected to have the same
continuum limit as QCD. The reason for choosing a mixed-action
approach is expediency. For example, the generation of 2+1+1 flavour
HISQ and 2+1 flavour DW ensembles with physical quark masses has been
possible even at the coarse lattice spacing of $a=0.15$~fm and
there are indications that cut-off effects are reasonably small. These
ensembles have been analyzed using clover-improved Wilson fermions, DW
and overlap fermions since the construction of baryon correlation
functions with definite spin and parity is much simpler compared to
staggered fermions.

The third choice is the combination of the algorithm for inverting the
Dirac matrix and variance reduction techniques. Efficient inversion
and variance reduction techniques are needed for the calculation of
nucleon correlation functions with high precision because the signal-to-noise ratio degrades exponentially as $e^{({\frac{3}{2}M_\pi-M_N}) \tau}$
with the source-sink separation $\tau$. Thus, the number of
measurements needed for high precision is much larger than in the
meson sector. Commonly used inversion algorithms include the
multigrid~\cite{Babich:2010qb} and the deflation-accelerated Krylov
solvers~\cite{Luscher:2007es}, which can handle linear systems with
large condition numbers very efficiently, thereby enabling
calculations of correlation functions at the physical pion mass.

The sampling of the path integral is limited by the number $N_{\rm
  conf}$ of gauge configurations generated. One requires sufficiently
large $N_{\rm conf}$ such that the phase space (for example, different
topological sectors) has been adequately sampled and all the
correlation functions satisfy the expected lattice symmetries such as
$C$, $P$, $T$, momentum and translation invariance. Thus, one needs
gauge field generation algorithms that give decorrelated large-volume
configurations cost-effectively. On such large lattices, to reduce
errors one can exploit the fact that the volume is large enough to
allow multiple measurements of nucleon correlation functions that are
essentially statistically independent. Two other common variance
reduction techniques that reduce the cost of multiple measurements on
each configuration are: the truncated solver with bias correction
method~\cite{Bali:2009hu} and deflation of the Dirac matrix for
the low-lying modes followed by sloppy solution with bias correction
for the residual matrix consisting predominately of the high-frequency
modes~\cite{Bali:2009hu,Blum:2012uh}.

A number of other variance reduction methods are also being used and
developed. These include deflation with hierarchical probing for
disconnected diagrams~\cite{Stathopoulos:2013aci,Gambhir:2016jul}, the
coherent source sequential propagator
method~\cite{Bratt:2010jn,Yoon:2016dij}, low-mode
averaging~\cite{DeGrand:2004qw,Giusti:2004yp}, the hopping-parameter
expansion~\cite{Gupta:1989kx,Thron:1997iy} and
partitioning~\cite{Bernardson:1993he}~(also known as
dilution~\cite{Foley:2005ac}).

The final choice is of the interpolating operator used to create and
annihilate the nucleon state, and of the operator used to calculate
the matrix element. Along with the choice of the interpolating
operator (or operators if a variational method is used) one also
chooses a ``smearing'' of the source used to construct the quark
propagator. By tuning the width of the smearing, one can optimize the
spatial extent of the nucleon interpolating operator to reduce the
overlap with the excited states. Two common smearing algorithms that
are equally performant are Gaussian (Wuppertal)~\cite{Gusken:1989ad}
and Jacobi~\cite{Alexandrou:1990dq} smearing. Specific smearing
techniques for hadrons boosted to (large) nonzero momentum have also
been designed \cite{Roberts:2012tp,DellaMorte:2012xc,Bali:2016lva}.

Having made all the above choices, for which a reasonable recipe
exists, one calculates a statistical sample of correlation functions
from which the desired ground-state nucleon matrix element is
extracted. Excited states, unfortunately, contribute significantly to
nucleon correlation functions in present studies.  To remove their
contributions, calculations are performed with multiple source-sink
separations $\tau$ and fits are made to the correlation functions using
their spectral decomposition as discussed in the next section.

\subsubsection{Controlling excited-state contamination\label{sec:ESC}}

Nucleon matrix elements are determined from a combination of two- and
three-point correlation functions. To be more specific, let
$B^\alpha(\vec{x},t)$ denote an interpolating operator for the
nucleon. Placing the initial state at time slice $t=0$, the two-point
correlation function of a nucleon with momentum $\vec{p}$ reads
\begin{equation}
\label{eq:nucl2pt}
   C_2(\vec{p};\tau) =
   \sum_{\vec{x},\vec{y}}\,e^{i\vec{p}\cdot(\vec{x}-\vec{y})}\,
   \mathbb{P}_{\beta\alpha}\,\left\langle 
   B^\alpha(\vec{x},\tau)\,\overline{B}^\beta(\vec{y},0) \right\rangle,
\end{equation}
where the projector $\mathbb{P}$ selects the polarization, and
$\alpha, \beta$ denote Dirac indices. The three-point function of two
nucleons and a quark-bilinear operator $O_\Gamma$ is defined as
\begin{equation}
\label{eq:nucl3pt}
   C_3^\Gamma(\vec{q};t,\tau) = \sum_{\vec{x},\vec{y},\vec{z}}\,
   e^{ i\vec{p\,}^\prime\cdot(\vec{x}-\vec{z})}\,
   e^{-i\vec{p}\cdot(\vec{y}-\vec{z})}\,
   \mathbb{P}_{\beta\alpha}\,\left\langle 
   B^\alpha(\vec{x},\tau)\,O_\Gamma(\vec{z},t)\,
   \overline{B}^\beta(\vec{y},0) \right\rangle,
\end{equation}
where $\vec{p},\ \vec{p\,}^\prime$ denote the momenta of the nucleons at
the source and sink, respectively, and
$\vec{q}\equiv\vec{p\,}^\prime-\vec{p}$ is the momentum transfer. The
bilinear operator is inserted at time slice $t$, and $\tau$ denotes the
source-sink separation. The corresponding quark-line diagrams
for both $C_2$ and $C_3^\Gamma$,
in terms of the nonperturbative quark propagators, $D^{-1}(y,x)$
where $D$ denotes the lattice Dirac operator, are shown in Fig.~\ref{fig:feynman}.

The framework for the analysis of excited-state contamination is based
on spectral decomposition. After inserting complete sets of
eigenstates of the transfer matrix, the expressions for the
correlators $C_2$ and $C_3^\Gamma$ read
\begin{eqnarray}
  \label{eq:specdec2pt}
  C_2(\vec{p};\tau) &=& \frac{1}{L^3}
  \sum_{n}\,\mathbb{P}_{\beta\alpha}\,\langle\Omega|B^\alpha|n\rangle
  \langle n|\overline{B}^\beta|\Omega\rangle\,
  e^{-E_n\tau}, \\ 
  \label{eq:specdec3pt}
  C_3^\Gamma(\vec{q};t,\tau) &=& \frac{1}{L^3}\sum_{n,m}\,
  \mathbb{P}_{\beta\alpha}\,
  \langle\Omega|B^\alpha|n\rangle\,
  \langle n|O_\Gamma|m\rangle\,
  \langle m|\overline{B}^\beta|\Omega\rangle\,
  e^{-E_n(\tau-t)}\,e^{-E_m t},
\end{eqnarray}
where $|\Omega\rangle$ denotes the vacuum state, and $E_n$ represents
the energy of the $n^{\rm th}$ eigenstate $|n\rangle$ in the nucleon
channel. Here we restrict the discussion to vanishing momentum
transfer, i.e., the forward limit $\vec{q}=0$, and label the ground state by $n=0$. The matrix
element of interest $g_\Gamma\equiv\langle0|O_\Gamma|0\rangle$ can,
for instance, be obtained from the asymptotic behaviour of the ratio
\begin{equation}
\label{eq:3ptratio}
  R_\Gamma(t,\tau) \equiv
  \frac{C_3^\Gamma(\vec{q}=0;t,\tau)}{C_2(\vec{p}=0;\tau)}
  \stackrel{t,(\tau-t)\to\infty}{\longrightarrow} g_{\Gamma} +
        {\rm O}(e^{-\Delta t},\,e^{-\Delta(\tau-t)},\,e^{-\Delta\tau}),
\end{equation}
where $\Delta\equiv E_1-E_0$ denotes the energy gap between the 
ground state and the first excitation. We also assume that the
bilinear operator $O_\Gamma$ is appropriately renormalized (see
Sec.~\ref{sec:renorm}).

Excited states with the same quantum numbers as the nucleon include
resonances such as a Roper-like state with a mass of about 1.5\,GeV,
or multi-particle states consisting of a nucleon and one or more pions
\cite{Tiburzi:2009zp,Bar:2017gqh}.  The latter can provide significant
contributions to the two- and three-point correlators in
Eqs.~(\ref{eq:nucl2pt}) and~(\ref{eq:nucl3pt}) or their
ratios~(\ref{eq:3ptratio}) as the pion mass approaches its physical
value. Ignoring the interactions between the individual hadrons, one
can easily identify the lowest-lying multi-particle states: they
include the $N\pi\pi$ state with all three particles at rest at
$\sim1.2$\,GeV, as well as $N\pi$ states with both hadrons having
nonzero and opposite momentum. Depending on the spatial box size $L$
in physical units (with the smallest nonzero momentum equal to $2\pi
/L$), there may be a dense spectrum of $N\pi$ states before the first
nucleon resonance is encountered. Corrections to nucleon correlation
functions due to the pion continuum have been studied using chiral
effective theory \cite{Tiburzi:2009zp, Bar:2017gqh, Bar:2016uoj, Bar:2016jof} 
and L\"uscher's finite-volume quantization condition
\cite{Hansen:2016qoz}.

The well-known noise problem of baryonic correlation functions implies
that the long-distance regime, $t, (\tau-t)\to\infty$, where the
correlators are dominated by the ground state, is difficult to
reach. Current lattice calculations of baryonic three-point functions
are typically confined to source-sink separations of
$\tau\lesssim1.5$\,fm, despite the availability of efficient noise
reduction methods. In view of the dense excitation spectrum
encountered in the nucleon channel, one has to demonstrate that the
contributions from excited states are sufficiently suppressed to
guarantee an unbiased determination of nucleon matrix elements. There
are several strategies to address this problem:
\begin{itemize}
\item Multi-state fits to correlator ratios or individual two- and
  three-point functions;
\item Three-point correlation functions summed over the operator-insertion time $t$;
\item Increasing the projection of the interpolator $B^\alpha$ onto
  the ground state.
\end{itemize}
The first of the above methods includes excited state contributions
explicitly when fitting to the spectral decomposition of the
correlation functions, Eqs.~(\ref{eq:specdec2pt}, \ref{eq:specdec3pt})
or, alternatively, their ratio (see Eq.~(\ref{eq:3ptratio})). In its
simplest form, the resulting expression for $R_\Gamma$ includes the
contributions from the first excited state, i.e.,
\begin{equation}
\label{eq:multistate}
  R_\Gamma(t,\tau) = g_\Gamma +c_{01}\,e^{-{\Delta}t}
  +c_{10}\,e^{-{\Delta}(\tau-t)} +c_{11}\,e^{-{\Delta}\tau}+\ldots,
\end{equation}
where $c_{01}, c_{10}, c_{11}$ and $\Delta$ are treated as additional
parameters when fitting $R_\Gamma(t,\tau)$ simultaneously over
intervals in the source-sink separation $\tau$ and the operator-insertion timeslice~$t$. Multi-exponential fits become more difficult
to stabilize for a growing number of excited states, since an
increasing number of free parameters must be sufficiently constrained
by the data. Therefore, a high level of comparable statistical
precision over several source-sink separations is required. One common
way to address this issue is to introduce Bayesian constraints, as
described in \cite{Yoon:2016jzj}. Alternatively, one may try to reduce
the number of free parameters, for instance, by determining the energy
gap $\Delta$ from nucleon two-point function and/or using a common gap
for several different nucleon matrix elements~\cite{Harris:2019bih}.

Ignoring the explicit contributions from excited states and fitting
$R_\Gamma(t,\tau)$ to a constant in $t$ for fixed $\tau$ amounts to
applying what is called the ``plateau method''. The name derives from
the ideal situation that sufficiently large source-sink separations $\tau$ 
can be realized, which would cause $R_\Gamma(t,\tau)$ to exhibit a
plateau in $t$ independent of $\tau$. The ability to control
excited-state contamination is rather limited in this approach, since
the only option is to check for consistency in the estimate of the
plateau as $\tau$ is varied. In view of the exponential degradation of
the statistical signal for increasing $\tau$, such stability checks
are difficult to perform reliably.

Summed operator insertions, originally proposed in
Ref.~\cite{Maiani:1987by}, have also emerged as a widely used method to
address the problem of excited-state contamination. One way to
implement this method \cite{Dong:1997xr,Capitani:2010sg} proceeds by
summing $R_\Gamma(t,\tau)$ over the insertion time $t$, resulting in the
correlator ratio $S_\Gamma(\tau)$,
\begin{equation}
  S_\Gamma(\tau) \equiv \sum_{t=a}^{\tau-a}\,R_\Gamma(t,\tau).
\end{equation}
The asymptotic behaviour of $S_\Gamma(\tau)$, including sub-leading
terms, for large source-sink separations $\tau$ can be easily derived
from the spectral decomposition of the correlators and is given by
\cite{Bulava:2011yz}
\begin{equation}
\label{eq:summation}
  S_\Gamma(\tau)\;\stackrel{\tau\gg1/\Delta}{\longrightarrow}\;
  K_\Gamma+(\tau-a)\,g_\Gamma+(\tau-a)\,e^{-\Delta\tau}d_\Gamma
  +e^{-\Delta\tau}f_\Gamma +\ldots,
\end{equation}
where $K_\Gamma$ is a constant, and the coefficients $d_\Gamma$ and
$f_\Gamma$ contain linear combinations of transition matrix elements
involving the ground and first excited states. Thus, the matrix
element of interest $g_\Gamma$ is obtained from the linear slope of
$S_\Gamma(\tau)$ with respect to the source-sink separation
$\tau$. While the leading corrections from excited states $e^{-\Delta\tau}$ are
smaller than those of the original ratio
$R_\Gamma(t,\tau)$ (see Eq.~(\ref{eq:3ptratio})), extracting the slope
from a linear fit to $S_\Gamma(\tau)$ typically results in relatively
large statistical errors. In principle, one could include the
contributions from excited states explicitly in the expression for
$S_\Gamma(\tau)$. However, in practice it is often difficult to
constrain an enlarged set of parameters reliably, in particular if one
cannot afford to determine $S_\Gamma(\tau)$ except for a handful of
source-sink separations.

The original summed operator-insertion technique described in
Refs.~\cite{Maiani:1987by,Gusken:1988yi,Gusken:1989ad,Sommer:1989rf} avoids
the explicit summation over the operator-insertion time $t$ at every
fixed value of $\tau$. Instead,
one replaces one of the quark propagators that appear in the
representation of the two-point correlation function $C_2(t)$ by a
``sequential'' propagator, according to
\begin{equation}
  D^{-1}(y,x) \to D_\Gamma^{-1}(y,x) = \sum_z D^{-1}(y,z)\Gamma
    D^{-1}(z,x).
\end{equation}
In this expression, the position $z\equiv(\vec{z},t)$ of the insertion
of the quark-bilinear operator is implicitly summed over, by inverting
the lattice Dirac operator $D$ on the source field $\Gamma
D^{-1}(z,x)$. While this gives access to all source-sink separations
$0\leq\tau\leq T$, where $T$ is the temporal extent of the lattice,
the resulting correlator also contains contact terms, as well as
contributions from $\tau<t<T$ that must be controlled. This
method has been adopted
recently by the CalLat collaboration in their calculation of the
isovector axial charge \cite{Berkowitz:2017gql,Chang:2018uxx}.\footnote{In Ref.\,\cite{Bouchard:2016heu} it is shown that the
  method can be linked to the Feynman-Hellmann theorem. A direct
  implementation of the Feynman-Hellmann theorem by means of a
  modification of the lattice action is discussed and applied
  in Refs.~\cite{Chambers:2014qaa,Chambers:2015bka}.}

As in the case of explicitly summing over the operator-insertion time,
the matrix element of interest is determined from the slope of the
summed correlator. For instance, in Ref.~\cite{Chang:2018uxx}, the
axial charge was determined from the summed three-point correlation
function, by fitting to its asymptotic
behaviour~\cite{Bouchard:2016heu} including sub-leading terms.

In practice, one often uses several methods simultaneously, e.g.,
  multi-state fits and the summation method based on
  Eq.~(\ref{eq:summation}), in order to check the robustness of the
result. All of the approaches for controlling excited-state
contributions proceed by fitting data obtained in a finite interval in
$\tau$ to a function that describes the approach to the asymptotic
behaviour derived from the spectral decomposition. Obviously, the
accessible values of $\tau$ must be large enough so that the model
function provides a good representation of the data that enter such a
fit. It is then reasonable to impose a lower threshold on $\tau$ above
which the fit model is deemed reliable. We will return to this issue
when explaining our quality criteria in Sec.~\ref{sec:rating}.

The third method for controlling excited-state contamination aims at
optimizing the projection onto the ground state in the two-point and
three-point correlation functions
\cite{Owen:2012ts,Bali:2014nma,Yoon:2016dij,Egerer:2018xgu}. The RQCD collaboration
has chosen to optimize the parameters in the Gaussian smearing
procedure, so that the overlap of the nucleon interpolating operator
onto the ground state is maximized \cite{Bali:2014nma}. In this way it
may be possible to use shorter source-sink separations without
incurring a bias due to excited states. 

The variational method, originally designed to provide detailed
information on energy levels of the ground and excited states in a
given channel \cite{Fox:1981xz,Michael:1985ne, Luscher:1990ck, Blossier:2009kd},
has also been adapted to the determination of hadron-to-hadron
transition elements \cite{Bulava:2011yz}. In the case of nucleon
matrix elements, the authors of Ref.\,\cite{Owen:2012ts} have employed
a basis of operators to construct interpolators that couple to
individual eigenstates in the nucleon channel. The method has produced
promising results when applied to calculations of the axial and other
forward matrix elements at a fixed value of the pion mass
\cite{Owen:2012ts,Dragos:2016rtx,Yoon:2016dij,Egerer:2018xgu}. However, a more
comprehensive study aimed at providing an estimate at the physical
point has, until now, not been performed.

The investigation of excited-state effects is an active subfield in
calculations of nucleon matrix elements, and many refinements and extensions have been
implemented since the first edition of the FLAG report. For instance,
it has been shown that the previously observed failure of the axial
and pseudoscalar form factors to satisfy the PCAC relation linking
them could be avoided by including the enhanced contribution of $N\pi$
excitations, either by including additional information on the nucleon
excitation spectrum extracted from the three-point function of the
axial current~\cite{Jang:2019vkm}, or with guidance from chiral
effective field theory analyses of nucleon three-point
functions~\cite{RQCD:2019jai}.  Following this, in
Refs.~\cite{Barca:2022uhi,Alexandrou:2024tin} it has been demonstrated
that this enhanced $N\pi$ contribution can be significantly reduced
when performing a GEVP analysis with a basis that includes a five-quark/antiquark interpolator with the quantum numbers of the nucleon
in addition to a three-quark interpolator.  For the flavour-diagonal
$u$- and $d$-quark scalar operators, a $\chi$PT study of excited-state
corrections~\cite{Gupta:2021ahb} suggests that there is a significant
enhancement of the disconnected contribution, which impacts the
calculation of the pion-nucleon sigma term $\sigma_{\pi N}$ as
discussed in Sec.~\ref{sec:gS-FD}.

The variety of methods that are employed to address the problem of excited-state
contamination has greatly improved our understanding of and control
over excited-state effects in calculations of nucleon matrix elements. However, there is still room for further improvement:
For instance, dedicated calculations of the excitation spectrum using
the variational method could replace the often rudimentary spectral
information gained from multi-state fits to the two- and three-point
functions used primarily for the determination of the matrix elements.
In general, the development of methods to explicitly include
multi-particle states, such as $N\pi$ and $N\pi\pi$ with appropriate
momentum configurations, coupled with the determination of the
associated (transition) matrix elements, is needed to significantly
enhance the precision of a variety of nucleon matrix elements.  Such
approaches would, to some extent, eliminate the need to extend the
source-sink separation $\tau$ into a regime that is currently
inaccessible due to the signal-to-noise problem.

Since the ongoing efforts to study excited-state contamination are
producing deeper insights, we have decided to follow a more cautious
approach in the assessment of available calculations of nucleon matrix
elements. This is
reflected in a modification of the quality criterion for excited-state
contamination that is described and discussed in
Sec.~\ref{sec:rating}.

\subsubsection{Renormalization and Symanzik improvement of local currents\label{sec:renorm}}

In this section, we discuss the renormalization of lattice operators
and their matching to a continuum reference scheme such as $\msbar$,
and the application of Symanzik improvement to remove $\cO(a)$
contributions.  For the charges, the relevant operators are the
axial~($A_\mu$), tensor~($T_{\mu\nu}$) and scalar~($S$) local
operators of the form ${\cal O}_\Gamma=\overline{q}\Gamma q$, with
$\Gamma=\gamma_\mu\gamma_5$, $i\sigma_{\mu\nu}$ and $\mathbf{1}$,
respectively, whose matrix elements are evaluated in the forward
limit. The steps in the renormalization of the 1-link operators,
defined in Section~\ref{sec:moments}, used to calculate the second
Mellin moments of distribution functions are similar to those for the
charges and we refer readers to
Refs.~\cite{Gockeler:1995wg,Harris:2019bih}.

For the charges, the general form for renormalized operators in the isovector
flavour combination, at a scale $\mu$, reads
\begin{equation}
{\cal O}_\Gamma^{\msbar}(\mu) = Z_{\cal O}^{\msbar,{\rm Latt}}(\mu a,g^2)\left[{\cal O}_\Gamma(a) +ab_{\cal O}m{\cal O}_\Gamma(a)+ac_{\cal O}{\cal O}_\Gamma^{\rm imp}(a)\right] +\cO(a^2),\label{eq_op_improv}
\end{equation}
where $Z_{\cal O}^{\msbar,{\rm Latt}}(\mu a,g^2)$ denotes the
multiplicative renormalization factor determined in the chiral limit, $m \to 0$, 
and the second and third terms represent all possible quark-mass-dependent
and -independent Symanzik improvement terms,
respectively, at $\cO(a)$.\footnote{Here, $a(g^2)$ refers to the lattice spacing in
  the chiral limit, however, lattice simulations are usually carried
  out by fixing the value of $g^2$ while varying the quark
  masses. This means $a=a(\tilde{g}^2)$ where $\tilde{g}^2=g^2(1+b_g
  am_q)$~\cite{Jansen:1995ck,Luscher:1996sc} is the improved coupling
  that varies with the average sea-quark mass $m_q$. The difference
  between the renormalization factors calculated with
  respect to $g^2$ and $\tilde{g}^2$ can effectively be absorbed into
  the $b_{\mathcal{O}}$
  coefficients~\cite{Bali:2016umi,Gerardin:2018kpy}.
}  The chiral properties of overlap,
domain-wall fermions~(with improvement up to $\cO(m_{\rm res}^n)$ where $m_{\rm res}$ is the residual mass) and twisted-mass
fermions~(at maximal twist~\cite{Frezzotti:2003ni,Frezzotti:2003xj})
mean that the $\cO(a)$-improvement terms are absent, while for
nonperturbatively improved Sheikholeslami-Wohlert-Wilson
(nonperturbatively improved clover) fermions all terms appear in principle.  For
the operators of interest here there are several mass-dependent terms
but at most one dimension-four ${\cal O}_\Gamma^{\rm imp}$; see,
e.g., Refs.~\cite{Capitani:2000xi,Bhattacharya:2005rb}.
However, the latter involve external derivatives whose 
corresponding matrix elements vanish in the forward limit.  Note that
no mention is made of staggered fermions as they are not, currently,
widely employed as valence quarks in nucleon matrix element calculations.

In order to illustrate the above remarks we consider the
renormalization and improvement of the isovector axial current. This
current has no anomalous dimension and hence the renormalization
factor, $Z_A=Z_A^{\msbar,{\rm Latt}}(g^2)$, is independent of the
scale.  The factor is usually computed nonperturbatively via the
axial Ward identity~\cite{Bochicchio:1985xa} or the Rome-Southampton
method~\cite{Martinelli:1994ty}~(see Sec.~A.3 of FLAG 19 \cite{FlavourLatticeAveragingGroup:2019iem}
for details). In some studies, the ratio with the corresponding vector
renormalization factor, $Z_A/Z_V$, is determined for which some of the
systematics cancel. In this case, one constructs the combination $Z_A
g_A/(Z_V g_V)$, where $Z_V g_V=1$ and $g_A$ and $g_V$ are the lattice
forward matrix elements, to arrive at the renormalized axial
charge~\cite{Bhattacharya:2016zcn}. For domain-wall fermions the ratio is
employed in order to remove $\cO(am_{\rm res})$ terms and achieve
leading discretization effects starting at
$\cO(a^2)$~\cite{Blum:2014tka}. Thus, as mentioned above,
$\cO(a)$-improvement terms are only present for
nonperturbatively improved clover fermions. For the axial current, Eq.~(\ref{eq_op_improv})
takes the explicit form,
\begin{equation}
A_\mu^{\msbar}(\mu) = Z_A^{\msbar,{\rm Latt}}(g^2)\left[
  \left(1+ ab_A m_{\rm val}+ 3a\tilde{b}_A m_{\rm sea}\right) A_\mu(a)+ac_A
  \partial_\mu P(a)\right] +\cO(a^2),
\end{equation}
where $m_{\rm val}$ and $m_{\rm sea}$ are the average valence- and sea-quark masses derived from the vector Ward identity~\cite{Bochicchio:1985xa,Luscher:1996sc,Bhattacharya:2005rb}, and $P$ is the
pseudoscalar operator $\overline{q}\gamma_5 q$. The matrix element of
the derivative term is equivalent to $q_\mu \langle
N(p^\prime)|P|N(p)\rangle$ and hence vanishes in the forward limit
when the momentum transfer $q_\mu=0$.  The improvement coefficients
$b_A$ and $\tilde{b}_A$ are known perturbatively for a variety of
gauge actions~\cite{Sint:1997jx,Taniguchi:1998pf,Capitani:2000xi} and
nonperturbatively for the tree-level Symanzik-improved gauge action for
$\Nf=2+1$~\cite{Korcyl:2016ugy}.

Turning to operators for individual quark flavours, these can mix 
under renormalization and the singlet and nonsinglet renormalization 
factors can differ.
For the axial current, such mixing occurs for all
fermion formulations just like in the continuum, where the singlet
combination acquires an anomalous dimension due to the $U_A(1)$
anomaly. The ratio of singlet to nonsinglet renormalization
factors, $r_{\cal O}=Z^{\rm s.}_{\cal O}/Z^{\rm n.s.}_{\cal O}$ for
${\cal O}=A$ differs from 1 at $\cO(\alpha_s^2)$ in perturbation
theory~(due to quark loops), suggesting that the mixing is a small
effect.  The nonperturbative determinations performed so far find $r_A\approx
1$~\cite{Alexandrou:2017hac,Green:2017keo}, supporting this.  For the
tensor current the disconnected diagram vanishes in the continuum due
to chirality and consequently on the lattice $r_T=1$ holds for overlap
and DW fermions~(assuming $m_{\rm res}=0$ for the latter). For
twisted-mass and clover fermions the mixing is expected to be small
with $r_T=1+\cO(\alpha_s^3)$~\cite{Constantinou:2016ieh} and this is
confirmed by the nonperturbative studies of
Refs.~\cite{Alexandrou:2017qyt,Bali:2017jyw}.

The scalar operators for the individual quark flavours,
$\overline{q}q$, are relevant not only for the corresponding scalar
charges, but also for the sigma terms $\sigma_q=m_q\langle
N|\overline{q}q|N\rangle$ when combined with the quark
masses~$m_q$. For overlap and DW fermions $r_S=1$, like in the
continuum and all $\overline{q}q$ renormalize multiplicatively with
the isovector $Z_S$. The latter is equal to the inverse of the mass
renormalization and hence $m_q\overline{q}q$ is renormalization
group~(RG) invariant. For twisted-mass fermions, through the use of
Osterwalder-Seiler valence fermions, the operators
$m_{ud}(\overline{u}u+\overline{d}d)$ and $m_s\overline{s}s$ are also
invariant~\cite{Dinter:2012tt}.\footnote{Note that for twisted-mass
  fermions the pseudoscalar renormalization factor is the relevant
  factor for the scalar operator. The isovector~(isosinglet) scalar
  current in the physical basis becomes the isosinglet~(isovector)
  pseudoscalar current in the twisted basis. Perturbatively
  $r_P=1+\cO(\alpha_s^3)$ and nonperturbative determinations have found
  $r_P\approx 1$~\cite{Alexandrou:2017qyt}.}  In contrast, the lack of
good chiral properties leads to significant mixing between quark
flavours for clover fermions. 
Nonperturbative determinations via the axial Ward
identity~\cite{Fritzsch:2012wq,Bali:2016lvx} have found the ratio
$r_S$ to be much larger than the perturbative expectation
$1+\cO(\alpha_s^2)$~\cite{Constantinou:2016ieh} may suggest.
While the sum
over the quark flavours which appear in the action $\sum^{\Nf}_q m_q
\overline{q}q$ is RG invariant, large cancellations between the
contributions from individual flavours can occur when evaluating,
e.g., the strange sigma term. Note that for twisted-mass and clover
fermions there is also an additive contribution $\propto
a^{-3}\mathbf{1}$~(or $\propto \mu a^{-2}\mathbf{1}$) to the scalar
operator.  This contribution is removed from the nucleon scalar matrix
elements by working with the subtracted current, $\overline{q}q -
\langle \overline{q}q\rangle$, where $\langle \overline{q}q\rangle$ is
the vacuum expectation value of the
current~\cite{Bhattacharya:2005rb}.

Symanzik improvement for the singlet currents follows the same pattern
as in the isovector case with $\cO(a)$ terms only appearing for
nonperturbatively improved clover fermions. For the axial and tensor operators only
mass-dependent terms are relevant in the forward limit while for the
scalar there is an additional gluonic operator ${\cal O}_S^{\rm
  imp}=\text{Tr}(F_{\mu\nu}F_{\mu\nu})$ with a coefficient of
$\cO(\alpha_s)$ in perturbation theory. When constructing the sigma terms
from the quark masses and the scalar operator, the improvement terms
remain and they must be included to remove all $\cO(a)$ effects for
nonperturbatively improved clover fermions, see Ref.~\cite{Bhattacharya:2005rb} for a
discussion.

\subsubsection{Extrapolations in $a$, $M_\pi$ and $M_\pi L$\label{sec:extrap}}

To obtain physical results that can be used to compare to or make
predictions for experiment, all quantities must be extrapolated to the
continuum and infinite-volume limits. In general, either a chiral
extrapolation or interpolation must also be made to the physical pion
mass. These extrapolations need to be performed simultaneously since
discretization and finite-volume effects are themselves dependent upon
the pion mass. Furthermore, in practice it is not possible  to hold the
pion mass fixed while the lattice spacing is varied, as some variation
in $a$ occurs when tuning the quark masses at fixed gauge coupling. Thus, one 
performs a simultaneous extrapolation in all three 
variables using a theoretically motivated formula of the form,
\begin{eqnarray}
g(M_{\pi},a,L) = g_{\mathrm{phys}} + \delta_{M_{\pi}} + \delta_a + \delta_L \ ,
\end{eqnarray}
where $g_{\mathrm{phys}}$ is the desired extrapolated result, and
$\delta_{M_{\pi}}$, $\delta_a$, $\delta_L$ are the deviations due to the 
pion mass, the lattice spacing, and the volume, respectively. 
Below we outline the forms for each of these terms.

All observables discussed in this section are dimensionless, therefore
the extrapolation formulae may be parameterized by a set of
dimensionless variables:
\begin{eqnarray}
\epsilon_{\pi} = \frac{M_{\pi}}{\Lambda_{\chi}} \ , \qquad M_{\pi} L \ , \qquad \epsilon_a = \Lambda_a a \ .
\end{eqnarray}
Here, $\Lambda_{\chi} \sim 1$~GeV is a chiral symmetry breaking scale,
which, for example, can be set to $\Lambda_{\chi} = 4 \pi F_{\pi}$,
where $F_{\pi} = 92.2$~MeV is the pion decay constant, and $\Lambda_a$ is a
discretization scale, e.g., $\Lambda_a = \frac{1}{4\pi w_0}$,
where $w_0$ is a gradient-flow scale~\cite{Borsanyi:2012zs}.

Effective field theory methods may be used to determine the form of
each of these extrapolations. For the single nucleon charges, Heavy-Baryon $\chi$PT (HB$\chi$PT) is a common
choice~\cite{Jenkins:1990jv,Bernard:1995dp}, however, other variants, such as
unitarized~\cite{Truong:1988zp} or covariant~$\chi$PT~\cite{Becher:1999he,Fuchs:2003qc}, are also employed. Various 
formulations of HB$\chi$PT exist, including those
for two and three flavours, as well as with and without explicit
$\Delta$ baryon degrees of freedom. Two-flavour HB$\chi$PT is typically used
due to issues with convergence of the three-flavour 
theory~\cite{Walker-Loud:2008rui,Torok:2009dg,Ishikawa:2009vc,Jenkins:2009wv,WalkerLoud:2011ab}.
The convergence properties of all known formulations for baryon $\chi$PT,
even at the physical pion mass, have not been well-established.

To $\cO(\epsilon_{\pi}^2)$, the two-flavour chiral expansion for the nucleon charges is known to be of the form~\cite{Bernard:1992qa},
\begin{eqnarray}
\label{eq:chi}
g = g_0 + g_1 \epsilon_{\pi} + g_2 \epsilon_{\pi}^2 + \tilde{g}_2 \epsilon_{\pi}^2 \ln \left(\epsilon_{\pi}^2\right) \ ,
\end{eqnarray}
where $g_1=0$ for all charges $g$ except $g_S^{u,d}$. The
dimensionless coefficients $g_{0,1,2}, \tilde{g}_2$ are assumed to be
different for each of the different charges. The coefficients in front
of the logarithms, $\tilde{g}_2$, are known functions of the low-energy 
constants (LECs), and do not represent new, independent
LECs. Mixed-action calculations will have further dependence upon the
mixed valence-sea pion mass, $m_{vs}$.

Given the potential difficulties with convergence of the chiral
expansion, known values of the $\tilde{g}_2$ in terms of LECs are not
typically used, but are left as free fit parameters. Furthermore, many
quantities have been found to display mild pion-mass dependence, such
that Taylor expansions, i.e., neglecting logarithms in the above
expressions, are also often employed. The lack of a rigorously
established theoretical basis for the extrapolation in the pion mass
thus requires data close to the physical pion mass for obtaining high-precision extrapolated/interpolated results.

Discretization effects depend upon the lattice action used in a particular
calculation, and their form may be determined using the standard Symanzik
power counting. In general, for an unimproved action, the
corrections due to discretization effects $\delta_a$ include terms
of the form,
\begin{eqnarray}
\delta_a = c_1 \epsilon_a + c_2 \epsilon_a^2 + \cdots \ ,
\end{eqnarray}
where $c_{1,2}$ are dimensionless coefficients. Additional terms of
the form $\tilde{c}_n \left(\epsilon_{\pi} \epsilon_a\right)^n$, where
$n$ is an integer whose lowest value depends on the combined
discretization and chiral properties, will also appear. Improved
actions systematically remove correction terms, e.g., an
$\cO(a)$-improved action, combined with a similarly 
improved operator, will contain terms in the extrapolation ansatz beginning
at $\epsilon_a^2$ (see Sec.~\ref{sec:renorm}).

Finite volume corrections $\delta_L$ may be determined in the usual
way from effective field theory, by replacing loop integrals over
continuous momenta with discrete sums. Finite volume effects therefore
introduce no new undetermined parameters to the extrapolation. For
example, at next-to-leading order, and neglecting contributions from
intermediate $\Delta$ baryons, the finite-volume corrections for the
axial charge in two-flavour HB$\chi$PT take the
form~\cite{Beane:2004rf},
\begin{eqnarray}
\delta_L &\equiv& g_{A}(L) - g_{A}(\infty) = \frac{8}{3} \epsilon_{\pi}^2 \left[ g_{A,\,0}^3 F_1\left(M_{\pi}L\right) + g_{A,\,0} F_3\left(M_{\pi} L\right)\right] \ ,
\label{eq:FVdeltaL}
\end{eqnarray}
where
\begin{eqnarray}
F_1\left(mL\right) &=& \sum_{\mathbf{n\neq 0}}\left[K_0\left(mL|\mathbf{n}|\right) - \frac{K_1\left(mL|\mathbf{n}|\right)}{mL|\mathbf{n}|}\right] \,, \cr
F_3\left(mL\right) &=& -\frac{3}{2} \sum_{\mathbf{n} \neq 0} \frac{K_1\left(mL|\mathbf{n}|\right)}{mL|\mathbf{n}|} \ ,
\end{eqnarray}
and $K_{\nu}(z)$ are the modified Bessel functions of the second
kind. Some extrapolations are performed using the form for
asymptotically large $M_{\pi} L$,
\begin{eqnarray}\label{eq:Vasymp}
K_0(z) \to \frac{e^{-z}}{\sqrt{z}} \ ,
\end{eqnarray}
and neglecting contributions due to $K_1$. Care must, however, be
taken to establish that these corrections are negligible for all
included values of $M_{\pi} L$. The numerical coefficients, for
example, $8/3$ in Eq.~\eqref{eq:FVdeltaL}, are often taken to be
additional free fit parameters, due to the question of convergence of
the theory discussed above.

Given the lack of knowledge about the convergence of the expansions and
the resulting plethora of possibilities for extrapolation models at
differing orders, it is important to include statistical tests of model selection 
for a given set of data. Bayesian model averaging
\cite{doi:10.1080/01621459.1995.10476572} or use of the Akaike
Information Criterion \cite{1100705} are common choices which penalize
over-parameterized models.

\subsection{Quality criteria for nucleon matrix elements and
  averaging procedure \label{sec:rating}}

There are two specific issues that call for a modification and
extension of the FLAG quality criteria listed in
Sec.~\ref{sec:qualcrit}. The first concerns the rating of the
chiral extrapolation: The FLAG criteria reflect the ability of
$\chi$PT to provide accurate descriptions of the pion-mass dependence
of observables. Clearly, this ability is linked to the convergence
properties of $\chi$PT in a particular mass range. Quantities
extracted from nucleon matrix elements are extrapolated to the
physical pion mass using some variant of baryonic $\chi$PT, whose
convergence is not well established as compared to the mesonic
sector. Therefore, we have opted for stricter quality criteria, 
200 MeV $\le M_{\pi,\mathrm{min}} \le 300$ MeV, for a 
green circle in the chiral extrapolation of nucleon matrix elements, i.e.,\\

\noindent
\good \hspace{0.2cm} $M_{\pi,\mathrm{min}}< 200$ MeV with three or more pion masses used in the extrapolation\\ 
\noindent \rule{0.05em}{0em}\hspace{0.45cm} \underline{or}
two values of $M_\pi$ with one lying within 10 MeV of 135 MeV (the physical neutral\\
\noindent \rule{0.05em}{0em}\hspace{0.45cm} pion mass) and the other one below 200 MeV\\
\rule{0.05em}{0em}\soso \hspace{0.2cm} 200 MeV $\le M_{\pi,\mathrm{min}}
\le 300$ MeV with three or more pion masses used in the extrapolation;\\
\noindent \rule{0.05em}{0em}\hspace{0.45cm} \underline{or}
two values of $M_\pi$ with $M_{\pi,\mathrm{min}}< 200$ MeV;\\
\noindent \rule{0.05em}{0em}\hspace{0.45cm} \underline{or} a single
value of $M_\pi$ lying within 10 MeV of 135\,MeV  (the physical neutral pion mass)\\
\rule{0.05em}{0em}\bad \hspace{0.2cm} Otherwise \\

In Sec.~\ref{sec:ESC} we have discussed that insufficient control
over excited-state contributions, arising from the noise problem in 
baryonic correlation functions, may lead to a systematic bias in the
determination of nucleon matrix elements. We therefore introduce an
additional criterion that rates the efforts to suppress excited-state
contamination in the final result.
As described in Sec.~\ref{sec:ESC}, the applied methodology to control
excited-state contamination is quite diverse. Since a broad consensus
on the question which procedures should be followed has yet to emerge,
our criterion is expressed in terms of simulation parameters that can
be straightforwardly extracted on the basis of publications.
Furthermore, the criterion must also be readily applicable to a
variety of different local operators whose matrix elements are
discussed in this chapter. These requirements are satisfied by the
source-sink separation $\tau$, i.e., the
Euclidean distance between the initial and final nucleons.
The discussion at the end of Sec.~\ref{sec:ESC} shows that there is
room for improvement in the ability to control excited-state
contamination. Hence, we have reverted to a binary system, based on
the range of source-sink separations of a given calculations. While we
do not award the highest category---a green star---in this edition,
we stress that the adoption of the modified criterion for
excited-state contamination has not led
to a situation where calculations that were previously rated with a
green star are now excluded from FLAG averages. The rating scale
concerning control over excited-state contributions is thus \\

\noindent
\soso \hspace{0.2cm} Three or more source-sink separations $\tau$, at
least two of which must be above 1.0 fm. \\ 
\rule{0.05em}{0em}\bad \hspace{0.2cm} Otherwise \\

We will continue to monitor the situation concerning excited-state
contamination and, if necessary, adapt the criteria further in future
editions of the FLAG report.

As explained in Sec.~\ref{sec:qualcrit}, FLAG averages are
distinguished by the sea-quark content. Hence, for a given
configuration of the quark sea (i.e., for $\Nf=2$, $2+1$, $2+1+1$, or $1+1+1+1$), we
first identify those calculations that pass the FLAG and the additional quality criteria
defined in this section, i.e., excluding any calculation that has a red tag
in one or more of the categories. We then add statistical and
systematic errors in quadrature and perform a weighted average. If the
fit is of bad quality (i.e., if $\chi^2_{\rm min}/{\rm dof}>1$), the
errors of the input quantities are scaled by $\sqrt{\chi^2/{\rm dof}}$. In the following step, correlations among different
calculations are taken into account in the error estimate by applying
Schmelling's procedure~\cite{Schmelling:1994pz}.

\subsection{Isovector charges\label{sec:isovector}}

The axial, scalar and tensor isovector charges are needed to interpret
the results of many experiments and phenomena mediated by weak
interactions, including probes of new physics.  The most natural process from
which isovector charges can be measured is neutron beta decay ($n \to
p^{+} e^{-} \overline{\nu}_e$).  At the quark level, this process
occurs when a down quark in a neutron transforms into an up quark due
to weak interactions, in particular due to the axial-current
interaction. While scalar and tensor currents have not been observed
in nature, effective scalar and tensor interactions arise in the
SM due to loop effects. At the TeV and higher scales,
contributions to these three currents could arise due to new
interactions and/or loop effects in BSM theories. These super-weak
corrections to standard weak decays can be probed through high-precision measurements of the neutron decay distribution by examining
deviations from SM predictions as described in
Ref.~\cite{Bhattacharya:2011qm}. The lattice-QCD methodology for the
calculation of isovector charges is well established, and the control
over statistical and systematic uncertainties has become quite robust
since the first edition of the FLAG report that featured nucleon
matrix elements \cite{FlavourLatticeAveragingGroup:2019iem}.

The axial charge $g_A^{u-d}$ is an important parameter that
encapsulates the strength of weak interactions of nucleons. It enters
in many analyses of nucleon structure and of SM and BSM 
physics. For example, it enters in (i) the
extraction of $V_{ud}$ and tests of the unitarity of the
Cabibbo-Kobayashi-Maskawa (CKM) matrix; (ii) the analysis of
neutrinoless double-beta decay, (iii) neutrino-nucleus quasi-elastic 
scattering cross-section; (iv) the rate of proton-proton fusion,
the first step in the thermonuclear reaction chains that power
low-mass hydrogen-burning stars like the Sun; (v) solar and reactor
neutrino fluxes; (vi) muon capture rates, etc. Currently the best
determination of the ratio of the axial to the vector charge,
$g_A/g_V$, comes from measurement of neutron beta decay using
polarized ultracold neutrons by the UCNA collaboration,
$1.2772(20)$~\cite{Mendenhall:2012tz,Brown:2017mhw}, and by PERKEO II,
$1.2761{}^{+14}_{-17}$~\cite{Mund:2012fq}. Note that, in the SM,
$g_V=1$ up to second-order corrections in isospin
breaking~\cite{Ademollo:1964sr,Donoghue:1990ti} as a result of the
conservation of the vector current.  The percent-level contributions of radiative corrections
discussed in Ref.~\cite{Cirigliano:2022hob} will need to be considered once the
accuracy of the lattice-QCD calculations reaches that of 
$g_A^{u-d}$ measured in experiments. The current goal 
is to calculate it directly with $\cO(1\%)$ accuracy using lattice QCD.

Isovector scalar or tensor interactions contribute to the
helicity-flip parameters, called $b$ and $B$, in the neutron decay
distribution. By combining the calculation of the scalar and
tensor charges with the measurements of $b$ and $B$, one can put 
constraints on novel scalar and tensor interactions at the TeV scale
as described in Ref.~\cite{Bhattacharya:2011qm}.  To optimally
bound such scalar and tensor interactions using measurements of
$b$ and $B$ parameters in planned experiments targeting $10^{-3}$ 
precision~\cite{abBA,WilburnUCNB,Pocanic:2008pu}, 
we need to determine $g_S^{u-d}$ and $g_T^{u-d}$ at the $10\%$ level as 
explained in Refs.~\cite{Bhattacharya:2011qm,Bhattacharya:2016zcn}. 
Future higher-precision measurements of $b$ and $B$ would require
correspondingly higher-precision calculations of the matrix elements
to place even more stringent bounds on these couplings at the TeV-scale.  

One can estimate $g_S^{u-d}$ via the conserved vector current (CVC)
relation, $g_S/g_V = (M_{\mathrm{neutron}}-M_{\mathrm{proton}})^{\rm
  QCD}/ (m_d-m_u)^{\rm QCD}$, as done by Gonzalez-Alonso {\it et
  al.}~\cite{Gonzalez-Alonso:2013ura}. In their analysis, they took
estimates of the two mass differences on the right-hand side from the
global lattice-QCD data~\cite{Aoki:2013ldr} and obtained
$g_S^{u-d}=1.02(8)(7)$.

The tensor charge $g_T^{u-d}$ can be extracted experimentally from
semi-inclusive deep-inelastic scattering (SIDIS)
data~\cite{Dudek:2012vr,Ye:2016prn,Lin:2017stx,Radici:2018iag}. A sample of these 
phenomenological estimates is shown in Fig.~\ref{fig:gt}, and the noteworthy feature is 
that the current uncertainty in these phenomenological estimates is large. More recent
phenomenological analyses include lattice values for the tensor charges as constraints
in their global fits~\cite{Lin:2017stx,Gamberg:2022kdb,Cocuzza:2023oam}.

\subsubsection{Results for $g_A^{u-d},\,g_S^{u-d}$ and $g_T^{u-d}$ \label{sec:gA-S-T-IV}}

\input{NME/Tables/table_ga.tex}

Results for the isovector axial, scalar and tensor charges are
presented in Tabs.\,\ref{tab:ga}, \ref{tab:gs} and \ref{tab:gt},
respectively. Compared with previous editions of the FLAG report, we
have made two changes: First, we have stopped listing results for
isovector charges from simulations in two-flavour QCD, since no new
results have been reported since 2018. Secondly, for simulations using
$2+1$ or $2+1+1$ flavours of dynamical quarks, we have imposed a
cutoff to focus on results published after 2014. For full listings,
including results obtained in two-flavour 
QCD~\cite{Khan:2006de,Lin:2008uz,Capitani:2012gj,Horsley:2013ayv,Bali:2014nma,Abdel-Rehim:2015owa,Alexandrou:2017hac,Capitani:2017qpc,Alexandrou:2017qyt} or published prior to our cutoff date
\cite{Edwards:2005ym,Yamazaki:2008py,Yamazaki:2009zq,Aoki:2010xg,Bratt:2010jn,Green:2012ud,Green:2012ej,Bhattacharya:2013ehc}, we
refer to earlier editions of the FLAG report.

For the sake of brevity, only calculations completed after FLAG\,21
and calculations that meet the criteria for inclusion in averages are
described below. For detailed descriptions of past calculations and
those that do not meet the criteria, the reader is again referred to
earlier editions of FLAG. The final results for the scalar and tensor
charges, $g_S^{u-d}$ and $g_T^{u-d}$, are presented in the
$\msbar$-scheme at a reference scale of 2~GeV by all collaborations.

The $2+1$-flavour calculation of the scalar and tensor charges by
$\chi$QCD 21A~\cite{Liu:2021irg} was performed using a mixed-action
approach with domain-wall fermion gauge configurations generated by
the RBC/UKQCD collaboration and overlap valence quarks. They include
five pion masses ranging from $m_{\pi}\sim$ 140~MeV to 370~MeV, four
lattice spacings ($a \sim$ 0.06, 0.08, 0.11, and 0.14~fm), thereby
considerably extending the parameter range in their earlier
calculation of the axial charge in $\chi$QCD\,18~\cite{Liang:2018pis}.
Matrix elements are computed for three to six different valence-quark
masses on each ensemble. The extrapolation to the physical pion mass,
continuum and infinite-volume limits is obtained by a global fit of
all data to a partially quenched chiral perturbation theory
ansatz. Excited-state contamination is assessed using three to five
sink-source separations and multi-state fits. Renormalization factors
were determined nonperturbatively using the RI/MOM prescription.

\input{NME/Tables/table_gs.tex}

The NME 21~\cite{Park:2021ypf} $2+1$-flavour calculation utilized
seven ensembles of $\cO(a)$-improved Wilson fermions. Three lattice
spacings, ranging from $a\sim 0.07$~fm to $0.13$~fm, several pion
masses, $m_{\pi}\sim$ 165~MeV to 285~MeV, and volumes corresponding to
$m_{\pi}L\sim$ 3.75 to 6.15 were used. Combined continuum, chiral, and
infinite-volume extrapolations were performed to the physical point
using leading-order fit functions. Several fitting strategies were
explored using four to six source-sink separations ranging from
0.7--1.8~fm. Final results are quoted by averaging results from two of
these fitting strategies, in which the excited-state energy for the
three-point function is fixed using two alternative choices of
priors. Renormalization is nonperturbative (RI-SMOM) using two
strategies.

PACS 22B~\cite{Tsuji:2022ric} reports estimates for the scalar and
tensor charges, computed on two ensembles with nonperturbatively
improved Wilson quark and Iwasaki gauge action at a single lattice
spacing of 0.085~fm, pion mass near physical value, and two volumes
with $m_{\pi} L\sim 3.7$ and 7.4. Two to four source-sink separations
ranging from 0.85--1.36~fm were used to estimate contributions from
excited states. They employ the RI-SMOM$_{\gamma_{\mu}}$
renormalization procedure. Due to the use of only a single lattice
spacing, this calculation does not meet the criteria for inclusion in
the average. In PACS 23~\cite{Tsuji:2023llh}, another ensemble was
considered for the calculation of the axial charge and form factors,
which features a smaller lattice spacing of 0.063~fm, a 10~fm spatial
box size and a near-physical pion mass of 138~MeV. The range of
source-sink separations matches the choice in PACS~22B. The size of
discretization effects is estimated by the difference between results
at fine and coarser lattice spacings. Since these results are based
on only two lattice spacings, they do not
qualify for an average.

The calculation of all three isovector charges by QCDSF/UKQCD/CSSM
23~\cite{QCDSFUKQCDCSSM:2023qlx} used a Feynman-Hellmann approach
to determine matrix elements from derivatives of energies produced via
a variation of the action. These energies were determined from fits to
two-point correlation functions, where a weighted average is taken of
the results obtained when varying the fitting range.
The computations utilized the
$2+1$-flavour stout-link nonperturbative clover action with Wilson-clover
valence quarks. Pion masses range from 220--468~MeV, using a
flavour-breaking expansion around the flavour SU(3) point to
extrapolate to physical pion mass. Combined pion-mass, lattice-spacing, 
and volume extrapolations were performed, using multiple
volumes ranging from $m_{\pi} L \sim$~3.2--9, and five lattice
spacings, 0.052--0.082~fm. Only the leading discretization effects and
asymptotic form of the volume extrapolation, Eq.~(\ref{eq:Vasymp}),
were included. They employ the RI'-MOM prescription for
nonperturbative renormalization.

\input{NME/Tables/table_gt.tex}

The calculations of $g_A^{u-d},\, g_S^{u-d}$ and $g_T^{u-d}$ published
by RQCD 23~\cite{Bali:2023sdi} and Mainz
24~\cite{Djukanovic:2024krw} are both based on $2+1$-flavour
ensembles generated by the CLS effort using nonperturbatively
improved Wilson fermions. The subsets of ensembles used in the two
calculations partly overlap. The 48 ensembles used by RQCD\,23
\cite{Bali:2023sdi} span six values of the lattice spacing, from
0.039--0.098~fm, pion masses from 130~MeV up to 430~MeV, and volumes
corresponding to $m_{\pi} L\sim$ 3--6.5. Excited states are controlled
using simultaneous two- and three-state fits of up to four different
observables using four time separations, $t \approx 0.7$--1.2~fm, with
a number of fit strategies employed. Extrapolations to the physical
point were performed using leading-order chiral expressions for the
pion mass, the leading asymptotic form for finite-volume corrections,
and terms up to $a^2$ in the lattice spacing. Renormalization uses the
nonperturbative RI'-SMOM scheme. In an earlier paper (RQCD\,19
\cite{RQCD:2019jai}), the Regensburg group computed the axial form
factor on a subset of the ensembles that enter RQCD\,23. The estimate
for $g_A^{u-d}$ from an analysis including matrix elements for
nonforward kinematics is also listed in Tab.\,\ref{tab:ga} but has
been superseded by the result in RQCD\,23.

The Mainz\,24 \cite{Djukanovic:2024krw} calculation, which
supersedes Mainz 19~\cite{Harris:2019bih}, uses four lattice spacings
($a\sim 0.05$~fm to $0.086$~fm) from the CLS set of ensembles, pion
masses ranging from $\sim 130$~MeV to $\sim 350$~MeV, and volumes
corresponding to $m_{\pi}L\sim 3$--5.4. Physical-point extrapolations
were performed simultaneously in the lattice spacing, pion mass, and
volume. In Mainz 24, the range of source-sink separations used was
enlarged to 0.2--1.4~fm, 
which allowed for the inclusion of sub-leading terms in the summation
method for improved control over excited-state
effects. Renormalization was performed nonperturbatively using the
RI-SMOM scheme. The Mainz group has also performed a calculation of
the axial form factor (Mainz~22 \cite{Djukanovic:2022wru}) on the
same set of ensembles, by incorporating the summation method directly
into the $z$-expansion used to describe the $Q^2$-dependence. The
corresponding estimate for $g_A^{u-d}$ from an analysis including
nonforward matrix elements has larger errors than the most recent
result~\cite{Djukanovic:2024krw}.

New results for $\Nf=2+1+1$ flavours of dynamical fermions have been
published by PNDME \cite{Jang:2023zts} and ETM
\cite{Alexandrou:2022dtc,Alexandrou:2023qbg}. The mixed-action
calculation by PNDME~23 \cite{Jang:2023zts}, which supersedes PNDME~18
\cite{Gupta:2018qil} and PNDME~16 \cite{Bhattacharya:2016zcn}, was
performed using the MILC HISQ ensembles, with a clover valence action.
As in PNDME~18 \cite{Gupta:2018qil}, the 11 ensembles used include
three pion-mass values, $M_{\pi} \sim$ 135, 225, 320~MeV, and four
lattice spacings, $a \sim$ 0.06, 0.09, 0.12, 0.15~fm. Note that four
lattice spacings are required to meet the green star criteria, as this
calculation is not fully $\cO(a)$-improved. Lattice size ranges
between $3.3 \lesssim M_{\pi} L \lesssim 5.5$. Physical-point
extrapolations were performed simultaneously, keeping only the
leading-order terms in the various expansion parameters. For the
finite-volume extrapolation, the asymptotic limit of the $\chi$PT
prediction, Eq.~(\ref{eq:Vasymp}), was used.  PNDME~23
\cite{Jang:2023zts} adds a study of sensitivity to excited-state
contamination using between three and five source-sink time
separations from $0.72 \lesssim \tau \lesssim 1.68$~fm, and several
strategies, including removing $N\pi$ contributions. Renormalization
was performed nonperturbatively using the RI-SMOM scheme.

The ETM collaboration has presented new results for the tensor charge
(ETM 22 \cite{Alexandrou:2022dtc}) and for the axial charge (ETM 23
\cite{Alexandrou:2023qbg}). Both calculations use three ensembles with
$2+1+1$-flavour twisted-mass fermions with close-to-physical pion
masses at $a = 0.057$, 0.069 and 0.080~fm, with volumes corresponding
to $m_{\pi}L\sim3.6$--3.9. These results supersede those in
\cite{Alexandrou:2019brg} based on the single ensemble at $a =
0.080$~fm. To control excited-state effects, they compared results
from the plateau, summation method and two-state fits.  After applying
nonperturbative renormalization via the RI'-MOM method supplemented
by a perturbative subtraction of lattice artefacts
\cite{Constantinou:2014fka,Alexandrou:2015sea}, they perform the
extrapolation to the continuum limit via a fit which is linear in
$a^2$.

\begin{figure}[!t]
\begin{center}
\includegraphics[width=11.5cm]{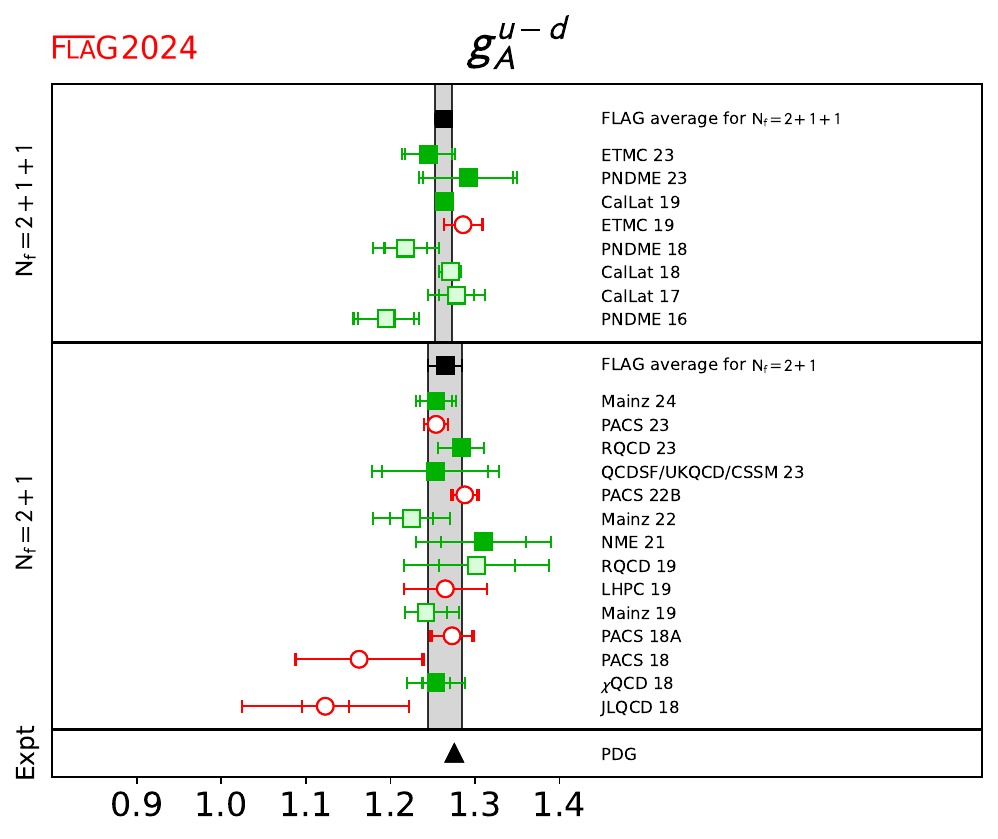}
\end{center}
\vspace{-1cm}
\caption{\label{fig:ga} Lattice results and FLAG averages for the
  isovector axial charge $g_A^{u-d}$ for $2+1$ and $2+1+1$ flavour
calculations. Also shown is the experimental result as quoted in the
PDG \cite{ParticleDataGroup:2022pth}.}
\end{figure}

We now proceed to discussing global averages for the isovector
charges. The compilation of results for the axial charge $g_A^{u-d}$,
plotted in Fig.~\ref{fig:ga}, shows that the situation has greatly
improved in terms of stability and precision thanks to several new
calculations that have been added since FLAG\,21. For QCD with
$\Nf=2+1+1$ dynamical quarks, the latest calculations by
ETM\,23\,\cite{Alexandrou:2023qbg},
PNDME\,23~\cite{Jang:2023zts} and
CalLat\,19~\cite{Walker-Loud:2019cif} pass all quality criteria. Since
PNDME and CalLat both use gauge ensembles produced by MILC, we assume
that the quoted errors are 100\% correlated, even though the range of
pion masses and lattice spacings explored in Refs.~\cite{Jang:2023zts}
and \cite{Chang:2018uxx,Walker-Loud:2019cif} is not exactly
identical. The two results are fully consistent within errors, which
is an improvement, since FLAG\,21 reported a slight tension between
CalLat\,19 \cite{Walker-Loud:2019cif} and PNDME\,18
\cite{Gupta:2018qil}. The calculation by ETM\,23
\cite{Alexandrou:2023qbg} uses an independent set of
ensembles. Performing a weighted average yields $g_A^{u-d} =
1.2633(100)$ with $\chi^2/{\rm dof}=0.30$. The result by CalLat
dominates the $2+1+1$ weighted average due to its smaller error.
Values for $\delta(a_{\rm min})$ for the two $\Nf=2+1+1$ calculations
that enter the averages vary between 1.0--1.5 (PNDME\,23: 1.0,
CalLat 19: 1.5). 

For QCD with $\Nf=2+1$ dynamical quarks, we compute a weighted average
from the results $\chi$QCD~18 \cite{Liang:2018pis}, NME~21
\cite{Park:2021ypf}, QCDSF/UKQCD/CSSM\,23
\cite{QCDSFUKQCDCSSM:2023qlx}, RQCD\,23 \cite{Bali:2023sdi} and
Mainz\,24 \cite{Djukanovic:2024krw}. Since the calculations by the
Mainz group and RQCD were both performed on ensembles generated by the
CLS effort, we treat the results RQCD\,23 \cite{Bali:2023sdi} and
Mainz\,24 \cite{Djukanovic:2024krw} as 100\% correlated. This
yields $g_A^{u-d}=1.265(20)$ with $\chi^2/\rm dof=0.28$.  Values for
$\delta(a_{\rm min})$ for the qualified calculations for $\Nf=2+1$
suggest that discretization effects are under good control (NME\,21:
0.15, QCDSF/UKQCD/CSSM 23: 0.6, RQCD\,23: 2.0, Mainz\,24:
2.3). From the information provided in the paper, it is not possible
to infer $\delta(a_{\mathrm{min}})$ for $\chi$QCD 18.

To summarize, the FLAG averages for the axial charge read
\begin{align}
&\label{eq:ga_2p1p1}
	\Nf=2+1+1:&\FLAGAVBEGIN g_A^{u-d} &= 1.263(10) \FLAGAVEND
  &&\Refs~\mbox{\cite{Chang:2018uxx,Walker-Loud:2019cif,Jang:2023zts,Alexandrou:2023qbg}},\\
&\label{eq:ga_2p1}
	\Nf=2+1:&\FLAGAVBEGIN g_A^{u-d} &= 1.265(20) \FLAGAVEND
  &&\Refs~\mbox{\cite{Liang:2018pis,Park:2021ypf,QCDSFUKQCDCSSM:2023qlx,Bali:2023sdi,Djukanovic:2024krw}}.
\end{align}
The averages computed for QCD with $\Nf=2+1+1$ and $\Nf=2+1$ flavours
are in excellent agreement, with a relative precision of 0.8\% and
1.5\%, respectively. The average for $2+1+1$ flavours exhibits a mild
tension of $1.25\sigma$ with the experimental value of
$g_A^{u-d}=1.2756(13)$ quoted by the PDG. While lattice QCD is able to
determine the axial charge with a total relative uncertainty at the
percent level, the experimental result is more precise by an order of
magnitude. We conclude with the remark that there has been enormous progress in
calculating this important benchmark quantity in lattice QCD over the
course of the past 10--15 years, owing to a variety of methods to
control excited-state effects, higher statistical precision, as well
as much better control over the extrapolation to the physical point.

\begin{figure}[t!]
\begin{center}
\includegraphics[width=11.5cm]{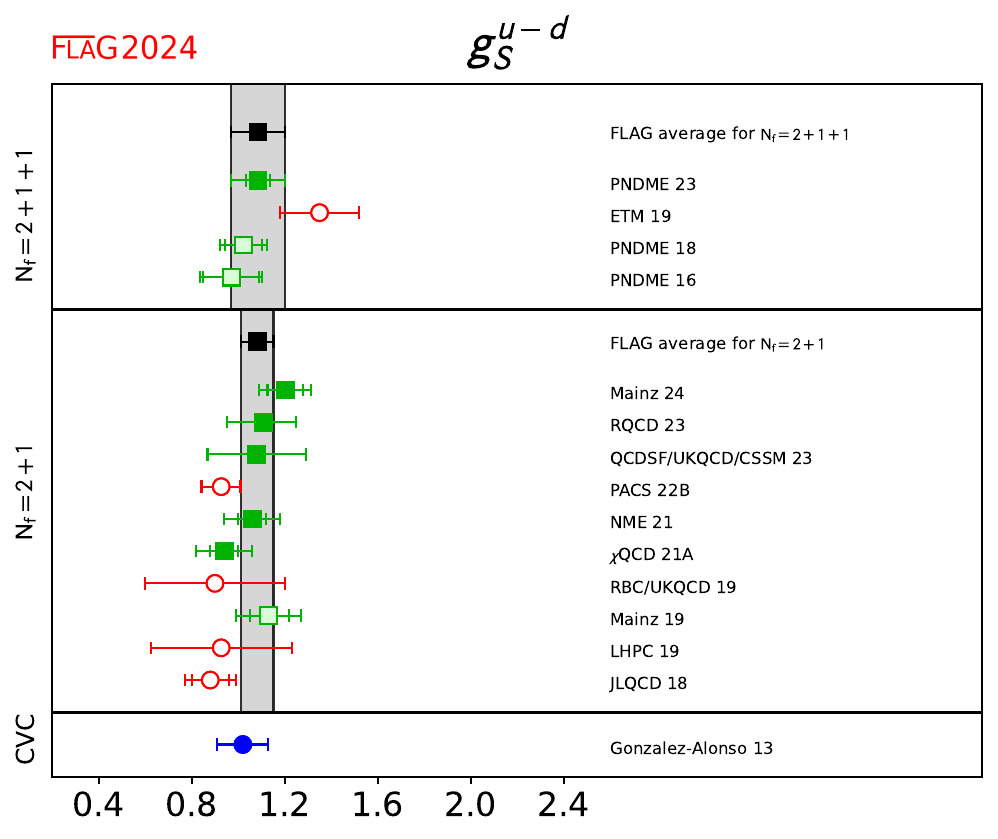}
\end{center}
\vspace{-1cm}
\caption{\label{fig:gs} Lattice results and FLAG averages for the isovector scalar charge $g^{u-d}_S$ 
for $\Nf = 2+1$ and $2+1+1$ flavour calculations. Also shown is
a phenomenological result obtained using the conserved vector
current~(CVC) relation~\cite{Gonzalez-Alonso:2013ura} (circle).}
\end{figure}

Turning now to the isovector scalar charge, we note that---in
addition to the direct three-point method---its value can also be
determined indirectly via the conserved vector current~(CVC) relation
from results for the neutron-proton mass
difference~\cite{Walker-Loud:2012ift,Shanahan:2012wa,Beane:2006fk,Horsley:2012fw,deDivitiis:2013xla,Budapest-Marseille-Wuppertal:2013rtp,Borsanyi:2014jba,Horsley:2015eaa,Brantley:2016our} and the down- and up-quark-mass
difference~(see Sec.~\ref{subsec:mumd}). For comparison, the
compilation in Fig.~\ref{fig:gs} also shows the indirect determination
by Gonzalez-Alonso {\it et al.}~\cite{Gonzalez-Alonso:2013ura}
obtained using lattice and phenomenological input.

For $2+1+1$ flavours, only PNDME~23 \cite{Jang:2023zts}, which
supersedes PNDME~18 \cite{Gupta:2018qil} and
PNDME~16 \cite{Bhattacharya:2016zcn}, meets all the criteria for
inclusion in the average. Consequently we identify the result from
PNDME~23 with the global average.

There are five $2+1$-flavour calculations which satisfy all criteria
required for inclusion in the average, i.e., $\chi$QCD
21A~\cite{Liu:2021irg}, NME\,21 \cite{Park:2021ypf},
QCDSF/UKQCD/CSSM\,23 \cite{QCDSFUKQCDCSSM:2023qlx}, RQCD\,23
\cite{Bali:2023sdi} and Mainz\,24 \cite{Djukanovic:2024krw}. The
calculations by PACS\,22B \cite{Tsuji:2022ric} and LHP\,19
\cite{Hasan:2019noy} have been performed at fewer than three lattice
spacings and therefore do not meet the criteria. As in the case of the
isovector axial charge, we assume 100\% correlation between the results
reported by Mainz\,24 and RQCD\,23, since the calculations were both
performed on the CLS set of ensembles. Values of
$\delta(a_{\mathrm{min}})$ for the qualified calculations range from
0.4--2.4 (PNDME 23: 1.6, NME 21: 2.4, RQCD 23: 0.4, Mainz 24:
0.5). It is not possible based on the information given to determine
$\delta(a_{\mathrm{min}})$ for $\chi$QCD 21 or QCDSF/UKQCD/CSSM 23,
however, in the former calculation it is noted that all data on the
finest lattice spacing is within one sigma of the quoted final result,
while for the latter extrapolations performed without accounting for
discretization effects give results within one sigma of the final quoted
result. Thus it is likely that in these cases
$\delta(a_{\mathrm{min}})$ is within a reasonable range.

The final FLAG values for $g_S^{u-d}$ are
\begin{align}
&\label{eq:gs_2p1p1}
  \Nf=2+1+1:&\FLAGAVBEGIN g_S^{u-d} &=  1.085(114) \FLAGAVEND
  &&\Ref~\mbox{\cite{Jang:2023zts}}, \\
&\label{eq:gs_2p1}    \Nf=2+1:&\FLAGAVBEGIN g_S^{u-d} &=  1.083(69)\FLAGAVEND
  &&\Refs~\mbox{\cite{Liu:2021irg,Park:2021ypf,QCDSFUKQCDCSSM:2023qlx,Bali:2023sdi,Djukanovic:2024krw}},
\end{align}
so that the total relative error for $\Nf=2+1+1$ and $2+1$ is about
10.5\% and 6.4\%, respectively. This implies that the relevant
precision target for current experimental searches for new scalar interactions
has been met.

\begin{figure}[t!]
\begin{center}
\includegraphics[width=11.5cm]{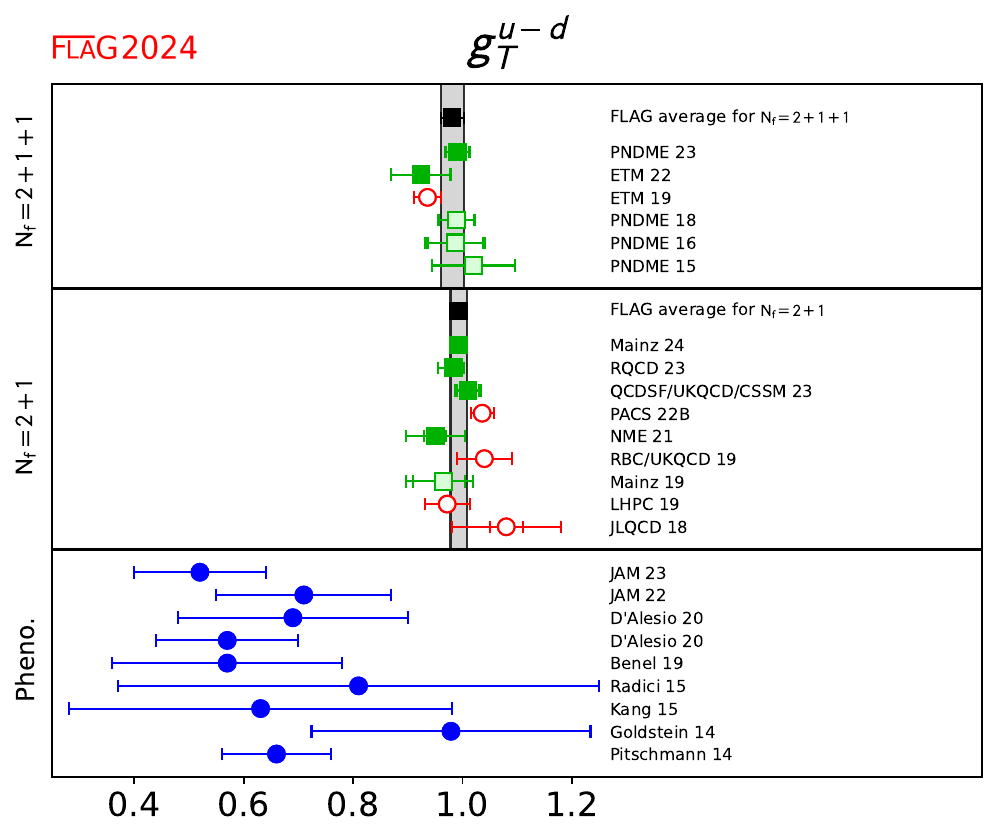}
\end{center}
\vspace{-1cm}
\caption{\label{fig:gt} Lattice results and FLAG averages for the isovector tensor charge $g^{u-d}_T$
for $\Nf = 2+1$ and 2+1+1 flavour calculations. Also shown are phenomenological results using measures of transversity~\cite{Radici:2015mwa,Kang:2015msa,Kang:pc2015,Goldstein:2014aja,Pitschmann:2014jxa,Benel:2019mcq,DAlesio:2020vtw,Gamberg:2022kdb,Cocuzza:2023oam} (circles). To maintain the independence of lattice QCD and phenomenological results, the JAM22 (private communication supplementing~\cite{Gamberg:2022kdb}) and JAM23~\cite{Cocuzza:2023oam} values shown are those obtained without including input from lattice QCD.}
\end{figure}

Estimates of the isovector tensor charge are generally at a high level of precision,
with values that are stable over time, as can be seen from the
compilation given in Tab.\,\ref{tab:gt} and plotted in
Fig.~\ref{fig:gt}. This is a consequence of the smaller statistical
fluctuations in the raw data and the very mild dependence on $a$,
$M_\pi$, and the lattice size $M_\pi L$. As a result, the uncertainty
due to the various extrapolations is small. Also shown for comparison
in Fig.~\ref{fig:gt} are phenomenological results using measures of
transversity~\cite{Radici:2015mwa,Kang:2015msa,Kang:pc2015,Goldstein:2014aja,Pitschmann:2014jxa,Benel:2019mcq,DAlesio:2020vtw,Gamberg:2022kdb,Cocuzza:2023oam}.

For $\Nf=2+1+1$ flavours, two calculations meet all the criteria for
inclusion in the average: PNDME~23 \cite{Jang:2023zts}, which
supersedes PNDME~18 \cite{Gupta:2018qil} and
PNDME~16 \cite{Bhattacharya:2016zcn}, and
ETM~22 \cite{Alexandrou:2022dtc}. Computational details for
PNDME~23 and ETM~22 have already been described above.

Using $\Nf=2+1$ flavours, four calculations meet all criteria for inclusion
in the average: NME 21\cite{Park:2021ypf}, QCDSF/UKQCD/CSSM\,23
\cite{QCDSFUKQCDCSSM:2023qlx}, RQCD 23~\cite{Bali:2023sdi}, and
Mainz\,24 \cite{Djukanovic:2024krw} calculation, which supersedes
Mainz\,19 \cite{Harris:2019bih}. Details of these calculations, as
well as the PACS~22B \cite{Tsuji:2022ric} calculation which does not
meet all criteria for inclusion in the average, have been described
above. As in the cases of the axial and scalar charge, we assume 100\%
correlation between the Mainz 24 and RQCD 23 calculations. Values of
$\delta(a_{\mathrm{min}})$ for the qualified calculations range from
0.03--2 (PNDME 23: 2, NME 21: 0.5, RQCD 23: 0.03, Mainz 24:
0.5). Similarly to the case for $g_S$, it is not possible based on the
information given to determine $\delta(a_{\mathrm{min}})$ for
$\chi$QCD 21 or QCDSF/UKQCD/CSSM 23. However, in the former
calculation it is noted that all data on the finest lattice spacing is
within one sigma of the quoted final result, while for the latter
extrapolations performed without accounting for discretization give
results within one sigma of the final quoted result. Thus, it is likely
that in these cases $\delta(a_{\mathrm{min}})$ is within a reasonable
range.

The final FLAG values for $g_T^{u-d}$ are
\begin{align}
&\label{eq:gt_2p1p1}
  \Nf=2+1+1:&\FLAGAVBEGIN g_T^{u-d} &=  0.981(21)\FLAGAVEND
  &&\Ref~\mbox{\cite{Jang:2023zts,Alexandrou:2022dtc}}, \\
 &\label{eq:gt_2p1}   \Nf=2+1:&\FLAGAVBEGIN g_T^{u-d} &=  0.993(15)\FLAGAVEND
  &&\Refs~\mbox{\cite{Park:2021ypf,QCDSFUKQCDCSSM:2023qlx,Bali:2023sdi,Djukanovic:2024krw}},
\end{align}
which implies that the isovector tensor charge is determined at the
level of 1.5--2.0\% relative precision.

\subsection{Flavour-diagonal charges\label{sec:FDcharges}}

Three examples of interactions for which matrix elements of
flavour-diagonal operators ($q \Gamma q$ where $\Gamma$ defines the
Lorentz structure of the bilinear quark operator) are needed are the
neutral-current interactions of neutrinos, elastic scattering of
electrons off nuclei, and the scattering of dark matter off
nuclei. In addition, these
matrix elements also probe intrinsic properties of nucleons (the spin,
the nucleon sigma term and strangeness content, and the contribution of the 
electric dipole moment (EDM) of the quarks to the nucleon EDM) as
explained below. For brevity, all operators are assumed to be appropriately
renormalized as discussed in Sec.~\ref{sec:renorm}. 

The matrix elements of the scalar operator $\overline{q} q$ with
flavour $q$ give the rate of change in the nucleon mass due to
nonzero values of the corresponding quark mass. This relationship is
given by the Feynman-Hellmann theorem. The quantities of interest are
the nucleon $\sigma$-term, $\sigma_{\pi N}$, and the strange and charm
content of the nucleon, $\sigma_{s}$ and $\sigma_{c}$,
\begin{align}
\sigma_{\pi N} &= m_{ud} \langle N| \overline{u} u +  \overline{d} d | N \rangle  \,, \\
\sigma_{s}     &= m_s \langle N| \overline{s} s | N \rangle  \,, \\
\sigma_{c}     &= m_c \langle N| \overline{c} c | N \rangle  \,.
\label{eq:gSdef}
\end{align}
Here, $m_{ud}$ is the average of the up- and down-quark masses and $m_s$, $m_c$ are the strange- and charm-quark masses.  The $\sigma_{\pi N, s, c}$
give the shift in $M_N$ due to nonzero light-, strange- and charm-quark
masses.  The same matrix elements are also needed to quantify the spin-independent interaction of dark matter with nucleons. Note that, while
$\sigma_b$ and $\sigma_t$ are also phenomenologically interesting,
they are unlikely to be calculated on the lattice due to the expected tiny signal in the matrix elements. In principle, the
heavy sigma terms can be estimated using $\sigma_{u,d,s}$ by exploiting
the heavy-quark
limit~\cite{Shifman:1978zn,Chetyrkin:1997un,Hill:2014yxa}.

The matrix elements of the axial operator $\overline{q} \gamma_\mu
\gamma_5 q$ give the contribution $\Delta q$ of quarks of flavour
$q$ to the spin of the nucleon:
\begin{align}
\langle N| \overline{q} \gamma_\mu \gamma_5 q | N \rangle  &= g_A^q \overline{u}_N \gamma_\mu \gamma_5 u_N,  \nonumber \\ 
g_A^q \equiv \Delta q &= \int_0^1 dx (\Delta q(x) + \Delta \overline{q} (x) )  \,.
\label{eq:gAdefnme}
\end{align}
The charge $g_A^q$ is thus the contribution of the spin of a quark of
flavour $q$ to the spin of the nucleon.  It is also related to the
first Mellin moment of the polarized parton distribution function
(PDF) $\Delta q$ as shown in the second line in
Eq.~\eqref{eq:gAdefnme}.  Measurements by the European Muon
collaboration in 1987 of the spin asymmetry in polarized deep
inelastic scattering showed that the sum of the spins of the quarks
contributes less than half of the total spin of the
proton~\cite{Ashman:1987hv}.  To understand this unexpected result,
called the ``proton spin crisis'', it is common to start with Ji's sum
rule~\cite{Ji:1996ek}, which provides a gauge invariant decomposition of
the nucleon's total spin, as
\begin{equation}
\frac{1}{2} =  \sum_{q=u,d,s,c,\cdot} \left(\frac{1}{2} \Delta q + L_q\right) + J_g \,,
\label{eq:Ji}
\end{equation}
where $\Delta q /2 \equiv g_A^q /2 $ is the contribution of the
intrinsic spin of a quark with flavour $q$; $L_q$ is the orbital
angular momentum of that quark; and $J_g$ is the total angular
momentum of the gluons.  Thus, to obtain the spin of the proton
starting from QCD requires calculating the contributions of the three
terms: the spin and orbital angular momentum of the quarks, and the
angular momentum of the gluons. Lattice-QCD calculations of the
various matrix elements needed to extract the three contributions are
underway. An alternate decomposition of the spin of the proton has
been provided by Jaffe and Manohar~\cite{Jaffe:1989jz}. The two
formulations differ in the decomposition of the contributions of the
quark orbital angular momentum and of the gluons. The contribution of
the quark spin, which is the subject of this review and given in
Eq.~\eqref{eq:gAdefnme}, is the same in both formulations.

The tensor charges are defined as the matrix elements of the tensor
operator $\overline{q} \sigma^{\mu\nu} q$ with $\sigma^{\mu\nu} = 
\{\gamma_\mu,\gamma_\nu\}/2$:
\begin{align}
g_T^q \overline{u}_N \sigma_{\mu \nu} u_N &= \langle N| \overline{q} \sigma_{\mu \nu} q | N \rangle  \,.
\label{eq:gTdef}
\end{align}
These flavour-diagonal tensor charges $g_T^{u,d,s,c}$ quantify the
contributions of the $u$, $d$, $s$, $c$ quark EDM to the neutron electric dipole moment
(nEDM)~\cite{Bhattacharya:2015wna,Pospelov:2005pr}. Since 
particles can have an EDM only due to P- and T- (or CP- assuming CPT is a good
symmetry) violating interactions, the nEDM is a very sensitive probe of
new sources of CP violation that arise in most extensions of the
SM designed to explain nature at the TeV scale. The
current experimental bound on the nEDM is $d_n < 1.8 \times
10^{-26}\ e$~cm~\cite{Abel:2020pzs,Baker:2006ts}, while the known CP violation in the SM
implies  $d_n < 10^{-31}\ e$~cm~\cite{Seng:2014lea}. A nonzero result over the
intervening five orders of magnitude would signal new physics.
Planned experiments aim to reduce the bound to around $
10^{-28}\ e$~cm. A discovery or reduction in the bound from these
experiments will put stringent constraints on many BSM theories,
provided the matrix elements of novel CP-violating interactions, of
which the quark EDM is one, are calculated with the required
precision.

One can also extract these tensor charges from the zeroth moment of the
transversity distributions that are measured in many experiments
including Drell-Yan and semi-inclusive deep inelastic scattering
(SIDIS). Of particular importance is the active program at Jefferson Lab (JLab) to measure
them~\cite{Dudek:2012vr,Ye:2016prn}. 
Transversity distributions describe the net transverse
polarization of quarks in a transversely polarized nucleon. Their 
extraction from the data taken over a limited range of $Q^2$ and
Bjorken $x$ is, however, not straightforward and requires additional
phenomenological modeling. 
At present, lattice-QCD estimates of $g_T^{u,d,s}$, presented in the next section, are more accurate than these phenomenological estimates~\cite{Radici:2015mwa,Kang:2015msa,Kang:pc2015,Goldstein:2014aja,Pitschmann:2014jxa,Benel:2019mcq,DAlesio:2020vtw,Cocuzza:2023oam}.
Future experiments will
significantly improve the extraction of the transversity
distributions.  Thus, accurate calculations of the tensor charges
using lattice QCD will continue to help elucidate the structure of the
nucleon in terms of quarks and gluons and provide a benchmark against
which phenomenological estimates utilizing measurements at JLab and
other experimental facilities worldwide can be compared.

The methodology for the calculation of flavour-diagonal charges is
well-established. The major challenges are the much larger statistical
errors in the disconnected contributions for the same computational
cost and the need for the additional calculations of the isosinglet
renormalization factors.  In this report, we present
    results for the axial and tensor charges in the same
    section~\ref{sec:gA-gT-FD} since they are mostly calculated
    together and because the statistical and systematic uncertainties
    are similar. The calculation of the scalar charges can, however,
    be done in two ways and the results are therefore presented
    separately in section~\ref{sec:gS-FD}.

\subsubsection{Results for $g_A^{u,d,s}$ and $g_T^{u,d,s}$\label{sec:gA-gT-FD}}

\input{NME/Tables/table_ga_singlet.tex}
\input{NME/Tables/table_gt_singlet.tex}

\begin{figure}[!t]
\begin{center}
\includegraphics[width=7.5cm]{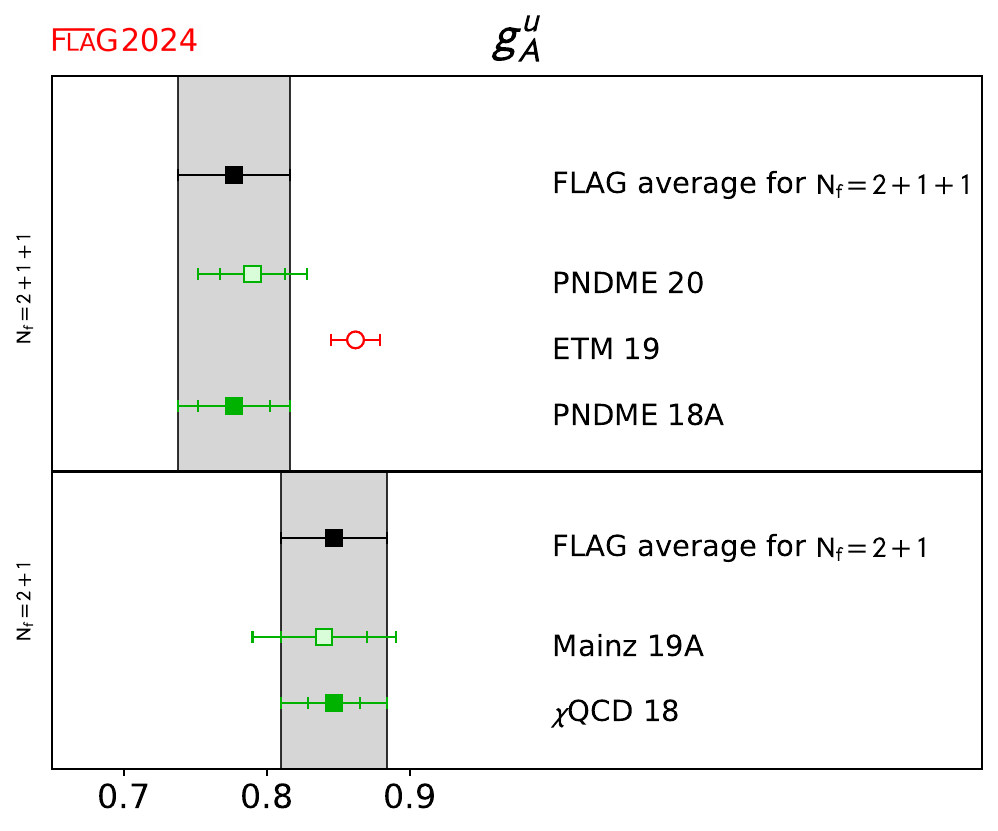}
\includegraphics[width=7.5cm]{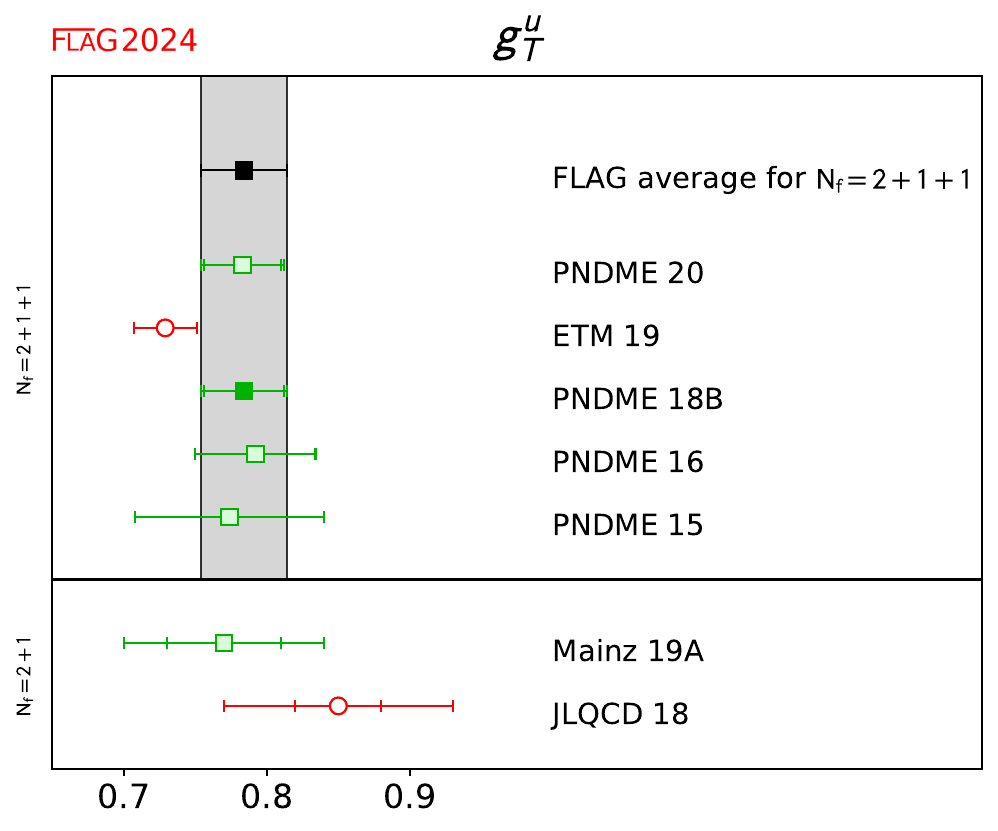}
\includegraphics[width=7.5cm]{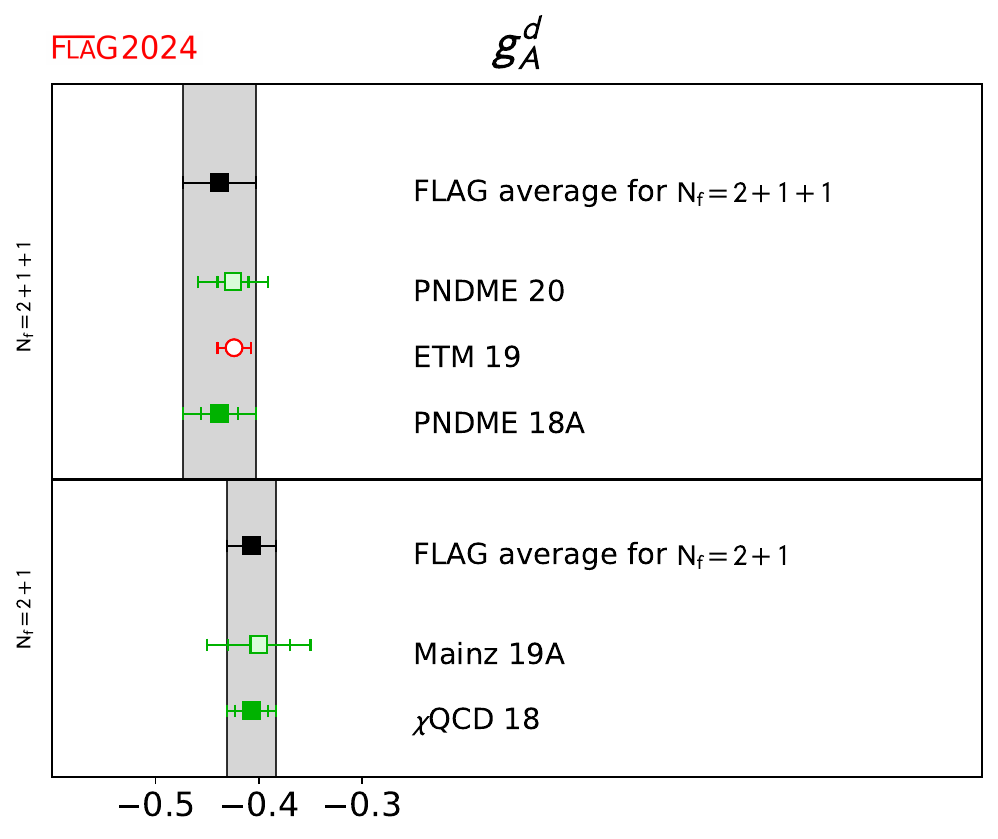}
\includegraphics[width=7.5cm]{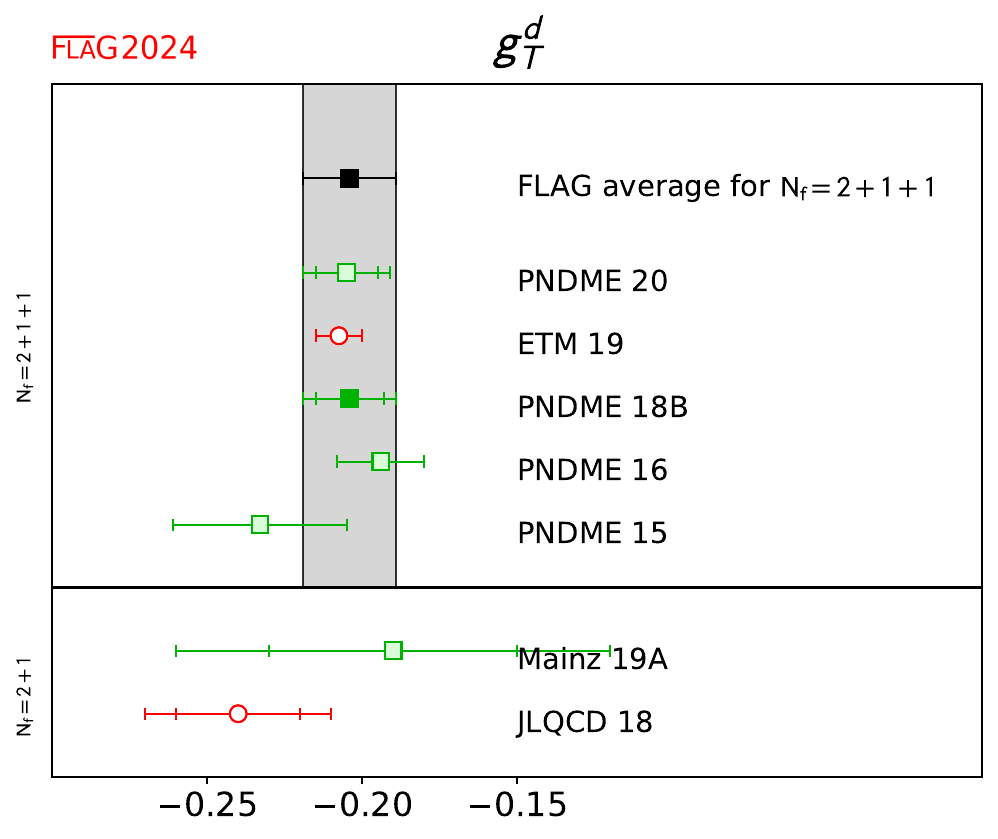}
\includegraphics[width=7.5cm]{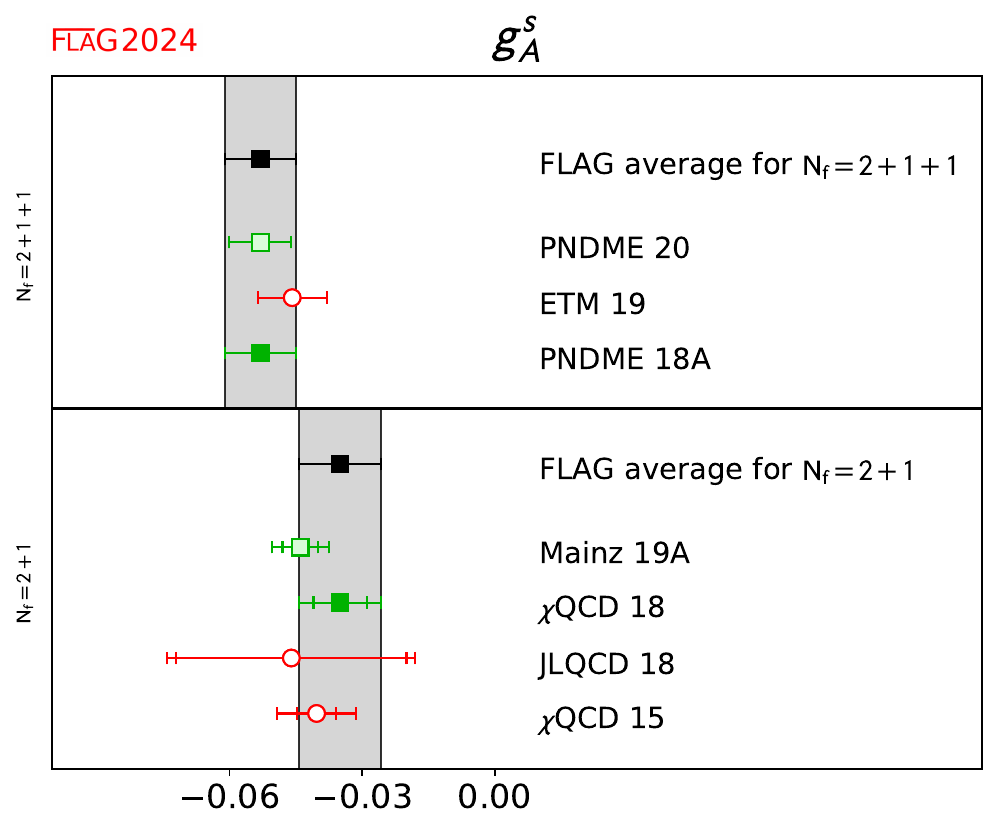}
\includegraphics[width=7.5cm]{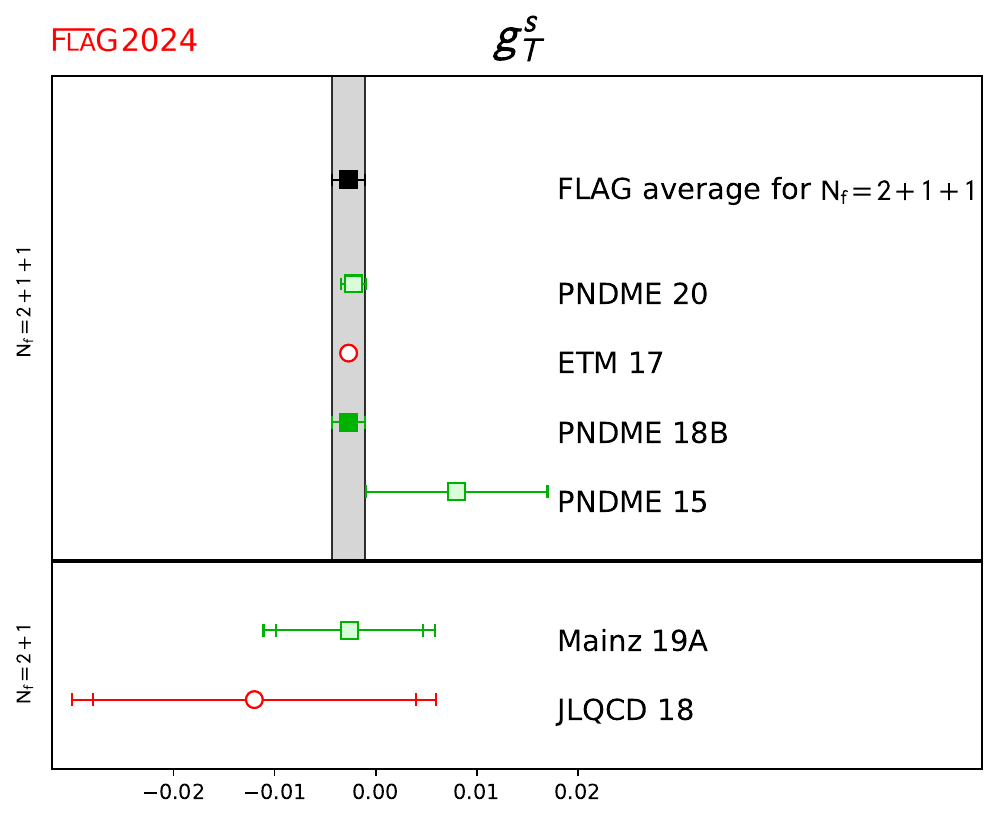}
\end{center}
\vspace{-1cm}
\caption{\label{fig:ga-gt-singlet} Lattice results and FLAG averages for 
  $g_A^{u,d,s}$ (left) and $g_T^{u,d,s}$ (right) for the $\Nf =2+1$ and $2+1+1$-flavour calculations.  }
\end{figure}

A compilation of results for the flavour-diagonal axial (tensor)
charges for the proton is given in Tab.\,\ref{tab:ga-singlet}
(Tab.\,\ref{tab:gt-singlet}), and are plotted in
Fig.~\ref{fig:ga-gt-singlet}.  Results for the neutron can be obtained
by interchanging the $u$- and $d$-flavour indices. To keep the report
current, publications from before 2014 that do not satisfy one or more
of the FLAG criteria and the $\Nf = 2$ results have been removed. They
can be obtained from the FLAG 19 ~\cite{FlavourLatticeAveragingGroup:2019iem} and
FLAG 21~\cite{FlavourLatticeAveragingGroupFLAG:2021npn} reports.

There are no new results that qualify
for FLAG averages, so the FLAG values for the proton in the $\msbar$ scheme at 2~GeV 
remain the same as in FLAG 19 ~\cite{FlavourLatticeAveragingGroup:2019iem} and
FLAG 21~\cite{FlavourLatticeAveragingGroupFLAG:2021npn}.
For $g_A^{u,d,s}$, only the PNDME 18A~\cite{Lin:2018obj} calculation qualifies 
for the 2+1+1-flavour theory, and only the $\chi$QCD 18
\cite{Liang:2018pis} for 2+1 flavours. 

The PNDME 18A~\cite{Lin:2018obj} results were obtained using the 2+1+1-flavour 
clover-on-HISQ formulation. The connected contributions were obtained
on 11 HISQ ensembles generated by the MILC collaboration with $a
\approx 0.057$, 0.87, 0.12 and 0.15~fm, $ M_\pi \approx 135$, 220 and
320~MeV, and $3.3 < M_\pi L < 5.5$. The light
disconnected contributions were obtained on six of these ensembles with the
lowest pion mass $M_\pi \approx 220$~MeV, while the strange disconnected
contributions were obtained on seven ensembles, i.e., including an additional one at $a \approx 0.087$~fm 
and $M_\pi \approx 135$~MeV. The excited-state and the
chiral-continuum fits were done separately for the connected and
disconnected contributions, which introduces a systematic that is 
hypothesised  to be small as explained in Ref.~\cite{Lin:2018obj}. 
The analysis of the excited-state contamination, 
discussed in Sec.~\ref{sec:ESC}, was done using
three-state fits for the connected contribution and two-state fits for the
disconnected contributions. Isovector renormalization factors,  
calculated on the lattice in the RI-SMOM scheme and converted to $\msbar$, 
are used for the flavour-diagonal operators, i.e., assuming
$Z_{A,S,T}^{u-d} = Z_{A,S,T}^{u,d,s} $. 
The chiral-continuum extrapolation was
done keeping the leading correction terms proportional to $M_\pi^2$ and $a$, and 
the leading finite-volume correction in $M_\pi L$ was included in the analysis of the connected contributions.
The continuum-limit criteria, $\delta(a_{\rm min})$,
can only be extracted for $g_A^s$ from PNDME 18A and is 0.3.

The PNDME 20~\cite{Park:2020axe} and the more recent conference
proceedings,~\cite{Park:2023tsj} and~\cite{Park:2024vjp}, are updates.
They extend the disconnected
calculations to eight ensembles, perform fits to the sum of the
connected and disconnected contributions, and also show, through
explicit calculations, that flavour mixing in the calculation of
renormalization factors in the RI-sMOM scheme is small, and the
isovector renormalization factor is a good approximation for
renormalizing flavour-diagonal axial and tensor charges as discussed
in Sec.~\ref{sec:renorm}.  These updates are, however, not included in
Tab.\,\ref{tab:gt-singlet} as they are preliminary.

The ETM 19~\cite{Alexandrou:2019brg} results for
$g_A^{u,d,s,c}$ are from a single ensemble with 2+1+1-flavour twisted-mass
fermions with a clover term at $a=0.0801(4)$~fm and $M_\pi=
139.3(7)$~MeV. These are not considered for the averages as they do
not satisfy the criteria for the continuum extrapolation.

The 2+1+1-flavour FLAG values for the axial charges $g_A^{u,d,s}$ of
the proton are the PNDME 18A results given
in Tab.~\ref{tab:ga-singlet}: 
\begin{align}
&  \mbox{}\Nf=2+1+1: &\FLAGAVBEGIN g_A^u  &= \phantom{-}0.777(25)(30)  \FLAGAVEND   &&\Ref~\mbox{\cite{Lin:2018obj}}, \\
&  \mbox{}\Nf=2+1+1: &\FLAGAVBEGIN g_A^d  &=           -0.438(18)(30)  \FLAGAVEND   &&\Ref~\mbox{\cite{Lin:2018obj}}, \\
&  \mbox{}\Nf=2+1+1: &\FLAGAVBEGIN g_A^s  &=           -0.053(8)   \FLAGAVEND   &&\Ref~\mbox{\cite{Lin:2018obj}}. 
\end{align}

The 2+1-flavour FLAG results from $\chi$QCD 18 \cite{Liang:2018pis}
were obtained using the overlap-on-domain-wall formalism. Three
domain-wall ensembles with lattice spacings 0.143, 0.11 and 0.083~fm
and sea-quark pion masses $M_\pi = 171$, 337 and 302~MeV,
respectively, were analyzed.  In addition to the three approximately
unitary points, the paper presents data for an additional 4--5 valence-quark masses on each ensemble, i.e., partially quenched data. Separate
excited-state fits were done for the connected and disconnected
contributions.  The continuum, chiral and volume extrapolation to the
combined unitary and nonunitary data is made including terms
proportional to both $M_{\pi,{\rm valence}}^2$ and $M_{\pi,{\rm
    sea}}^2$, and two $\cO(a^2)$ discretization terms for the two
different domain-wall actions. With just three unitary points, not all
the coefficients are well constrained. The $M_{\pi,sea}$-dependence is
omitted and considered as a systematic, and a prior is used for the
coefficients of the $a^2$-terms to stabilize the fit.
The continuum-limit criteria, $\delta(a_{\rm min})$,
could not be extracted for these results from $\chi$QCD 18.

These
$\chi$QCD~18 2+1-flavour results for the proton, which supersede the
$\chi$QCD~15 \cite{Gong:2015iir} analysis, are
\begin{align}
&  \mbox{}\Nf=2+1: &\FLAGAVBEGIN g_A^u  &= \phantom{-}0.847(18)(32)  \FLAGAVEND   &&\Ref~\mbox{\cite{Liang:2018pis}}, 
\\
&  \mbox{}\Nf=2+1: &\FLAGAVBEGIN g_A^d  &=           -0.407(16)(18)  \FLAGAVEND   &&\Ref~\mbox{\cite{Liang:2018pis}}, 
\\
&  \mbox{}\Nf=2+1: &\FLAGAVBEGIN g_A^s  &=           -0.035(6)(7)    \FLAGAVEND   &&\Ref~\mbox{\cite{Liang:2018pis}}. 
\end{align}

The results for $g_A^{u,d,s}$ from Mainz 19A \cite{Djukanovic:2019gvi} 
satisfy all the criteria, however, they are not included in the 
averages as~\cite{Djukanovic:2019gvi} is a conference proceeding. 
The JLQCD~18 \cite{Yamanaka:2018uud},
ETM~17C \cite{Alexandrou:2017oeh} and
Engelhardt~12 \cite{Engelhardt:2012gd} calculations were not
considered for the averages as they did not satisfy the criteria for
the continuum extrapolation. All three calculations were done at a
single lattice spacing. The JLQCD~18 calculation used overlap
fermions and the Iwasaki gauge action. They perform a chiral fit using
data at four pion masses in the range 290--540~MeV. Finite-volume
corrections are assumed to be negligible since each of the two pairs
of points on different lattice volumes satisfy $M_\pi L \geq 4$.  The
ETM~17C calculation is based on a single twisted-mass ensemble with
$M_\pi=130$~MeV, $a=0.094$ and a relatively small $M_\pi L = 2.98$.
Engelhardt~12 \cite{Engelhardt:2012gd} calculation was done on three asqtad ensembles with
$M_\pi = 293$, 356 and 495~MeV, but all at a single lattice spacing
$a=0.124$~fm.

Results for $g_A^s$ are also presented by LHPC in
Ref.~\cite{Green:2017keo}.  However, this calculation is not included in 
Tab.\,\ref{tab:ga-singlet} 
as it has been performed on a single ensemble with $a=$ 0.114~fm and a
heavy pion mass, $M_\pi \approx 317$~MeV.


Switching to the tensor charges, $g_T^{u,d,s}$, only one calculation,
the PNDME 18B \cite{Gupta:2018lvp}, qualifies for the FLAG averaging.
These 2+1+1-flavour theory results, which use the same
ensembles already discussed for $g_A^{u,d,s}$, supersede those in PNDME
16 \cite{Bhattacharya:2016zcn} and PNDME 15
\cite{Bhattacharya:2015wna}.  The continuum-limit criteria, $\delta(a_{\rm min})$, 
can only be extracted for $g_T^s$ from PNDME 18B and is 0.5. 
Again, results in the more recent conference proceedings,~\cite{Park:2023tsj}
and~\cite{Park:2024vjp}, are not discussed here as they are preliminary.

The FLAG values for the proton in the $\msbar$ scheme at 2~GeV, which 
remain the same as in FLAG 19 and FLAG 21, are:

\begin{align}
&  \mbox{}\Nf=2+1+1: &\FLAGAVBEGIN g_T^u  &= \phantom{-}0.784(28)(10)  \FLAGAVEND   &&\Ref~\mbox{\cite{Gupta:2018lvp}},  \\
&  \mbox{}\Nf=2+1+1: &\FLAGAVBEGIN g_T^d  &=           -0.204(11)(10)  \FLAGAVEND   &&\Ref~\mbox{\cite{Gupta:2018lvp}}, \\
&  \mbox{}\Nf=2+1+1: &\FLAGAVBEGIN g_T^s  &=           -0.0027(16)  \FLAGAVEND   &&\Ref~\mbox{\cite{Gupta:2018lvp}}. 
\end{align}

The ensembles and the analysis strategy used in PNDME~18B is the same
as described in PNDME~18A for $g_A^{u,d,s}$. The only difference for
the tensor charges was that a one-state (constant) fit was used for
the disconnected contributions as the data did not show significant
excited-state contamination. The isovector renormalization factors,
used for all three flavour-diagonal tensor operators, were calculated
on the lattice in the RI-SMOM scheme and converted to $\msbar$ at
2~GeV using 2-loop perturbation theory~\cite{Kniehl:2020sgo}. The 
proceeding~\cite{Park:2024vjp} extends the calculation to eight ensembles and 
reports that flavour mixing in the calculation of
renormalization factors is small, and the isovector renormalization
factor, which was used for renormalizing the flavour-diagonal tensor charges
in PNDME~18B, is a good approximation.

The ETM 19~\cite{Alexandrou:2019brg} results for
$g_T^{u,d,s,c}$ are from a single ensemble with 2+1+1-flavour twisted-mass
fermions with a clover term at $a=$ 0.0801(4)~fm and $M_\pi=
139.3(7)$~MeV. It was not considered for the final averages because it
did not satisfy the criteria for the continuum extrapolation. 
The same applies to the
JLQCD 18 \cite{Yamanaka:2018uud} and ETM 17 \cite{Alexandrou:2017qyt}
calculations.  The Mainz 19A \cite{Djukanovic:2019gvi} results with
2+1-flavour ensembles of clover fermions are not included in the
averages as Ref.~\cite{Djukanovic:2019gvi} is a conference proceeding.


\subsubsection{Results for $g_S^{u,d,s}$ from direct and hybrid calculations of the matrix elements\label{sec:gS-FD}}
\input{NME/Tables/table_sigmaud_s_direct.tex}

The sigma terms $\sigma_q=m_q\langle N|\bar{q}q|N\rangle=m_q g_S^q$ or
the quark-mass fractions $f_{T_q}=\sigma_q/M_N$ are normally computed
rather than $g_S^q$.  These combinations have the advantage of being
renormalization group invariant in the continuum, and this holds on
the lattice for actions with good chiral properties, see
Sec.~\ref{sec:renorm} for a discussion. In order to aid comparison
with phenomenological estimates, e.g., from $\pi$-$N$
scattering~\cite{Alarcon:2011zs,Chen:2012nx,Hoferichter:2015dsa}, the
light-quark sigma terms are usually added to give the $\pi N$ sigma
term, $\sigma_{\pi N}=\sigma_u+\sigma_d$. The direct evaluation of the
sigma terms involves the calculation of the corresponding three-point
correlation functions for different source-sink separations
$\tau$. For $\sigma_{\pi N}$ there are both connected and disconnected
contributions, while for most lattice fermion formulations only
disconnected contributions are needed for $\sigma_s$.  The techniques
typically employed lead to the availability of a wider range of $\tau$
for the disconnected contributions compared to the connected
ones~(both, however, suffer from signal-to-noise problems for large
$\tau$, as discussed in Sec.~\ref{sec:intro}) and we only comment on
the range of $\tau$ computed for the latter in the following.

Recent $\Nf=2+1$ and $\Nf=2+1+1$ results for $\sigma_{\pi N}$ and
$\sigma_s$ from the direct approach are compiled in
Tab.~\ref{tab:gs-ud-s-direct}. In the following, we summarize new
results that have appeared since the last FLAG report and previous
studies that enter the averages. Details of ETM
19~\cite{Alexandrou:2019brg} and JLQCD 18~\cite{Yamanaka:2018uud} can
be found in the FLAG 21 report. As there have been no new $\Nf=2$
studies of the sigma terms since the introduction of the
section on nucleon matrix elements~\cite{Bali:2016lvx,Abdel-Rehim:2016won}, we also refer the
reader to the previous report for a discussion of these results
and other early three- and four-flavour works with at least one red
square~\cite{Engelhardt:2012gd,Oksuzian:2012rzb,Gong:2013vja}.

Starting with $\Nf=2+1+1$, there is a new study from
PNDME~\cite{Gupta:2021ahb}.  This calculation is based on a mixed-action set-up of $\cO(a)$-improved Wilson valence fermions on top of
staggered~(HISQ) gauge ensembles generated by the MILC collaboration.
Six ensembles are utilized with lattice spacings, $a \approx 0.12,
0.09$ and 0.06~fm and pion masses $M_\pi \approx 315, 230$ and
138~MeV. The two-point and three-point correlation functions are
fitted simultaneously including contributions from four and three
states, respectively, where wide-width priors are used for the excited-state masses entering the fits. Four to five values of the source-sink
separation are utilized with the largest $\tau \approx 1.5$~fm.  The
fitting procedure is repeated using a narrow-width prior for the first
excited state which is set to the energy of the lowest multi-hadron state~($N\pi$ or
$N\pi\pi$, see Sec.~\ref{sec:ESC}). This choice is motivated by a
$\chi$PT analysis~\cite{Gupta:2021ahb}, which indicates that excited-state contributions
arising from low-lying $N\pi$ and
$N\pi\pi$ states can be significant on close-to-physical pion mass ensembles. In particular,
    there is a significant enhancement of the disconnected contribution
due to the large QCD condensate.
The quality of the
fits is, however, similar for both a narrow- and wide-width  prior for the first excited
state. Combined continuum- and chiral-limit fits are performed with a
parameterization composed of a term linear in the lattice spacing and
the NNLO SU(2) baryon $\chi$PT expression for the pion-mass dependence.  
Finite-volume effects are not resolved. The result from the narrow-width
first-excited-state prior analysis is chosen as the final value,
while the wide-width prior analysis (which has a first-excited-state energy significantly above the lowest $N \pi$ or $N \pi \pi$ noninteracting level) gives $\sigma_{\pi N}\approx 42$~MeV.

Moving on to the $\Nf=2+1$ results, Mainz
23~\cite{Agadjanov:2023efe} is a new study employing 16
nonperturbatively $\cO(a)$-improved Wilson fermion ensembles from the
CLS consortium. The flavour average of the light- and strange-quark
mass is held constant in the simulations as the pion mass varies in
the range $350\gtrsim M_\pi \gtrsim 130$~MeV.  Four lattice spacings
are realized, with $a=0.050$--0.086~fm. The connected three-point
functions are computed for a large number of source-sink
separations~(between 9 and 17 values of $\tau$, depending on the
ensemble) where the largest $\tau=1.4$--1.5~fm. The ground-state matrix
elements are extracted employing two analysis strategies: one
employing the summation method~(with only the ground-state terms) and
the other performing two-state fits to correlator ratios. For the
latter, the mass gap to the first excited state is set with a prior 
equal to twice the pion mass. As both the light- and strange-quark masses
vary in the simulations, $\sigma_{\pi N}$ and $\sigma_s$ are fitted
simultaneously with the quark-mass dependence parameterized by SU(3)
$\cO(p^3)$ covariant baryon $\chi$PT.
Combined continuum, chiral and finite-volume fits are performed, where
cuts are made on the data set entering the fit which depend on the
lattice spacing, finite volume and pion mass. 
Akaike-information-criterion~\cite{1100705} averages of the results are computed
for the two analysis choices separately. The two results are then
combined to form the final values.

The $\chi$QCD~15A~\cite{Yang:2015uis} study also qualifies for global
averaging.  In this mixed-action study, three RBC/UKQCD $\Nf=2+1$
domain-wall ensembles are analyzed comprising two lattice spacings,
$a=0.08$~fm with $M_{\pi,\rm sea}=300$~MeV and $a=0.11$~fm with
$M_{\pi,\rm sea}=330$~MeV and $139$~MeV. Overlap fermions are employed
with a number of nonunitary valence-quark masses. The connected
three-point functions are measured with three values of $\tau$ in the
range 0.9--1.4~fm. A combined chiral, continuum and volume
extrapolation is performed for all data with $M_\pi<350$~MeV. The
leading-order expressions are taken for the lattice-spacing and volume
dependence while partially quenched SU(2) HB$\chi$PT up to $M_\pi^3$-terms models the chiral behaviour for $\sigma_{\pi N}$. The
strange-quark sigma term has a milder dependence on the pion mass and
only the leading-order quadratic terms are included in this case.

MILC has also computed $\sigma_s$ using a hybrid
method~\cite{Toussaint:2009pz} which makes use of the
Feynman-Hellmann~(FH) theorem and involves evaluating the nucleon matrix
element $\langle N|\int\! d^4\!x\, \bar{s}s|N\rangle$.\footnote{Note
  that in the direct method the matrix element $\langle N|\int\!
  d^3\!x\, \bar{s}s|N\rangle$, involving the spatial-volume sum, is
  evaluated for a fixed timeslice.} This method is applied in
MILC~09D~\cite{Toussaint:2009pz} to the $\Nf=2+1$ asqtad ensembles
with lattice spacings $a=$ 0.06, 0.09, 0.12~fm and values of $M_\pi$
ranging down to 224~MeV. A continuum and chiral extrapolation is
performed including terms linear in the light-quark mass and quadratic
in $a$.  As the coefficient of the discretization term is poorly
determined, a Bayesian prior is used, with a width corresponding to a
10\% discretization effect between the continuum limit and the
coarsest lattice spacing.\footnote{This is consistent with
  discretization effects observed in other quantities at $a=0.12$~fm.}
A similar updated analysis is presented in
MILC~12C~\cite{Freeman:2012ry}, with an improved evaluation of
$\langle N|\int\! d^4\!x\, \bar{s}s|N\rangle$ on a subset of the $\Nf=2+1$
asqtad ensembles. The study is also extended to HISQ $\Nf=2+1+1$
ensembles comprising four lattice spacings with $a=$ 0.06--0.15~fm and a
minimum pion mass of 131~MeV.  Results are presented for
$g_S^s=\langle N|\bar{s}s|N\rangle$~(in the $\msbar$ scheme at 2~GeV)
rather than for $\sigma_s$. The scalar matrix element is renormalized
for both three and four flavours using the 2-loop factor for the
asqtad action~\cite{Mason:2005bj}. The error incurred by applying the
same factor to the HISQ results is expected to be small.\footnote{At
  least at 1-loop the renormalization factors for HISQ and asqtad are very
  similar, cf.  Ref.~\cite{McNeile:2012xh}.}

Both MILC~09D and MILC~12C achieve green tags for all the criteria,
see Tab.~\ref{tab:gs-ud-s-direct}. As the same set of asqtad ensembles is
utilized in both studies we take MILC~12C as superseding MILC~09D
for the three-flavour case. The global averaging is discussed in
Sec.~\ref{sec:gS-sum}.

\subsubsection{Results for $g_S^{u,d,s}$ using the Feynman-Hellmann theorem\label{sec:gS-FD-FH}}

An alternative approach for accessing the sigma terms is to determine
the slope of the nucleon mass as a function of the quark masses, or
equivalently, the squared pseudoscalar meson masses. The Feynman-Hellman~(FH)
theorem gives
\begin{equation}
\sigma_{\pi N}=m_u\frac{\partial M_N}{\partial m_u}+ m_d\frac{\partial M_N}{\partial m_d}\approx M_\pi^2 \frac{\partial M_N}{\partial M_\pi^2},\hspace{0.7cm} \sigma_s = m_s \frac{\partial M_N}{\partial m_s}\approx M_{\bar{s}s}^2 \frac{\partial M_N}{\partial M_{\bar{s}s}^2},\label{eq:fheq1}
\end{equation}
where the fictitious $\bar{s}s$ meson has a mass squared
$M^2_{\bar{s}s}=2M_K^2-M_\pi^2$.  In principle this is a
straightforward method as the nucleon mass can be extracted from fits
to two-point correlation functions, and a further fit to $M_N$ as a
function of $M_\pi$~(and also $M_K$ for $\sigma_s$) provides the
slope. Nonetheless, this approach presents its own challenges: a
functional form for the chiral behaviour of the nucleon mass is
needed, and while baryonic $\chi$PT~(B$\chi$PT) is the natural choice,
the convergence properties of the different formulations are not well
established. Results are sensitive to the formulation chosen and the
order of the expansion employed. If there is an insufficient number of
data points when implementing higher-order terms, the coefficients are
sometimes fixed using additional input, e.g., from analyses of
experimental data. This may influence the slope extracted. Simulations
with pion masses close to or bracketing the physical point can
alleviate these difficulties. In some studies the nucleon mass is used
to set the lattice spacing. This naturally forces the fit to reproduce
the physical nucleon mass at the physical point and may affect the
extracted slope. Note that, if the nucleon mass is fitted as
a  function of the pion and kaon masses, the dependence of the meson masses on
  the quark masses also, in principle, needs to be considered in order to extract the sigma terms.

\input{NME/Tables/table_sigmaud_fh.tex}

An overview of recent three- and four-flavour determinations of
$\sigma_{\pi N}$ and $\sigma_s$ is given in
Tab.~\ref{tab:gs-singlet-fh}.
All the results are eligible for global
averaging, with RQCD 22~\cite{RQCD:2022xux} being the sole new work.
For details of earlier works~(published before 2014) with at least one red
square~\cite{Walker-Loud:2008rui,Ishikawa:2009vc,MartinCamalich:2010fp,Horsley:2011wr,Oksuzian:2012rzb,Shanahan:2012wh} and all $\Nf=2$~\cite{Bali:2012qs,Ohki:2008ff} works we refer the reader to the FLAG 21 report. 
Note that the renormalization criterion is
not included in Tab.~\ref{tab:gs-singlet-fh} as renormalization is not
normally required when computing the sigma terms in the
Feynman-Hellmann approach.\footnote{An exception to this is when
clover fermions are employed. In this case one must take care of the
mixing between quark flavours when renormalizing the quark masses that
appear in Eq.~\eqref{eq:fheq1}.}  At present, a rating indicating
control over excited-state contamination is also not considered since
a wide range of source-sink separations are available for nucleon
two-point functions and ground-state dominance is normally achieved.
This issue may be revisited in the future as statistical precision
improves and this systematic is further investigated.

We first summarize the determinations of $\sigma_{\pi N}$.  BMW have
performed a $\Nf=1+1+1+1$ study BMW 20A~\cite{Borsanyi:2020bpd} which
follows a two-step analysis procedure: the dependence of the nucleon
mass on the pion and kaon masses is determined on HEX-smeared clover
ensembles with $a=$ 0.06--0.1~fm and pion masses in the range $M_\pi=$
195--420~MeV.  The meson masses as a function of the quark masses are
evaluated on stout-staggered ensembles with a similar range in $a$ and
quark masses which bracket their physical values.  As
\cite{Borsanyi:2020bpd} is a preprint, their results (for both sigma
terms) are not considered for global averaging.

Regarding $\Nf=2+1+1$, there is only one recent study. In
ETM~14A~\cite{Alexandrou:2014sha}, fits are performed to the nucleon
mass utilizing SU(2) $\chi$PT for data with $M_\pi \ge 213$~MeV as
part of an analysis to set the lattice spacing. The expansion is
considered to $\cO(p^3)$ and $\cO(p^4)$, with two and three of the coefficients
as free parameters, respectively.  The difference between the two fits
is taken as the systematic error. No discernable discretization or
finite-volume effects are observed where the lattice spacing is varied
over the range $a$ = 0.06--0.09~fm and the spatial volumes cover 
$M_\pi L=3.4$ up to $M_\pi L>5$. The results are unchanged when a near-physical-point $\Nf=2$ ensemble is added to the analysis in
Ref.~\cite{Alexandrou:2017xwd}.

Turning to $\Nf=2+1$, RQCD 22~\cite{RQCD:2022xux} utilizes 49
nonperturbatively $\cO(a)$-improved Wilson fermion CLS ensembles, with six
lattice spacings in the range $0.04\le a \le 0.1$~fm and
$M_\pi\sim 130$--410~MeV. The ensembles lie on three trajectories in the 
quark-mass plane, two of which meet at the physical point. Simultaneous fits
to the bayon octet are performed, employing SU(3) $\cO(p^3)$ covariant
baryon $\chi$PT, heavy baryon $\chi$PT and Taylor-expansion fit forms
for the quark-mass dependence.
The final values at the physical point in the continuum and infinite-volume limits are obtained by performing an Akaike-information-criterion~\cite{1100705} average of the covariant baryon $\chi$PT fits
to various reduced data sets. 
These fits include finite-volume terms to $\cO(p^3)$ as well as
terms quadratic in the lattice spacing in order to model cut-off effects.

In BMW~11A~\cite{Durr:2011mp}, stout-smeared
tree-level clover fermions are employed on 15 ensembles with
simulation parameters encompassing $a$ = 0.06--0.12~fm, $M_\pi \sim$
190--550~MeV and $M_\pi L \gsim 4$.  Taylor, Pad\'{e} and covariant
SU(3) B$\chi$PT fit forms are considered. Due to the use of smeared
gauge links, discretization effects are found to be mild even though
the fermion action is not fully $\cO(a)$-improved. Fits are performed
including an $\cO(a)$ or $\cO(a^2)$ term and also without a
lattice-spacing-dependent term.  Finite-volume effects were assessed
to be small in an earlier work~\cite{Durr:2008zz}. The final results
are computed considering all combinations of the fit ansatz weighted
by the quality of the fit.  In BMW~15~\cite{Durr:2015dna}, a more
extensive analysis on 47 ensembles is presented for HEX-smeared clover
fermions involving five lattice spacings and pion masses reaching down
to 120~MeV. Bracketing the physical point reduces the reliance on a
chiral extrapolation.  Joint continuum, chiral and infinite-volume
extrapolations are carried out for a number of fit parameterizations
with the final results determined via the Akaike-information-criterion
procedure. Although only $\sigma_{\pi N}$ is accessible
in the FH approach in the isospin limit, the individual quark
fractions $f_{T_q}=\sigma_q/M_N$ for $q=u,d$ for the proton and the
neutron are also quoted in BMW~15, using isospin
relations.\footnote{These isospin relations were also derived in
Ref.~\cite{Crivellin:2013ipa}.}

With one exception, all of the above studies have also determined the
strange-quark sigma term, while Junnarkar~13~\cite{Junnarkar:2013ac}
only presents results for $\sigma_s$. This quantity is difficult to
access via the Feynman-Hellmann method since in most simulations the
physical point is approached by varying the light-quark mass, keeping
$m_s$ approximately constant. While additional ensembles can be
generated, it is hard to resolve a small slope with respect to
$m_s$. Such problems are illustrated by the large uncertainties in the
results from BMW~11A and BMW~15. Alternative approaches have been
pursued where the physical point is approached along a trajectory
keeping the average of the light- and strange-quark masses
fixed~\cite{Horsley:2011wr}, and where quark-mass reweighting is
applied~\cite{Oksuzian:2012rzb}.  One can also fit to the whole baryon
octet and apply SU(3) flavour-symmetry constraints as investigated
in RQCD 22~\cite{RQCD:2022xux} and
Refs.~\cite{MartinCamalich:2010fp,Durr:2011mp,Horsley:2011wr,Shanahan:2012wh}.

Junnarkar~13~\cite{Junnarkar:2013ac} is a mixed-action study which
utilizes domain-wall valence fermions on MILC $\Nf=2+1$ asqtad
ensembles. The derivative $\partial M_N/\partial m_s$ is determined
from simulations above and below the physical strange-quark mass for
$M_\pi$ around 240--675~MeV. The resulting values of $\sigma_s$ are
extrapolated quadratically in $M_\pi$. The quark fraction
$f_{T_s}=\sigma_s/M_N$ exhibits a milder pion-mass dependence and
extrapolations of this quantity were also performed using ans\"atze
linear and quadratic in $M_\pi$. A weighted average of all three fits
was used to form the final result. Two lattice spacings were analyzed,
with $a$ around $0.09$~fm and $0.12$~fm, however, discretization
effects could not be resolved.

The global averaging of the results is discussed in the next section.

\subsubsection{Summary of Results for $g_S^{u,d,s}$\label{sec:gS-sum}}

We consider computing global averages of results determined via the
direct, hybrid and Feynman-Hellmann (FH) methods.
Beginning with $\sigma_{\pi N}$,
Tabs.~\ref{tab:gs-ud-s-direct} and \ref{tab:gs-singlet-fh} show that
for $\Nf=2+1+1$ ETM~14A~(FH) and PNDME 21~(direct) satisfy the selection criteria. 
The FLAG average for the four-flavour case reads
\begin{align}
  &\label{eq:sigmaud_2p1p1} \mbox{}\Nf=2+1+1: &\FLAGAVBEGIN \sigma_{\pi N} &= 60.9(6.5) \FLAGAVEND ~\mbox{MeV} &&\Refs~\mbox{\cite{Alexandrou:2014sha,Gupta:2021ahb}}.
\end{align}
We remark that although the $\Nf=1+1+1+1$ BMW 20A study~\cite{Borsanyi:2020bpd}
     also satisfies the criteria, it is not considered for averaging as it is a preprint.
 For $\Nf=2+1$ we form an average from the BMW~11A~(FH), BMW~15~(FH), $\chi$QCD~15A~(direct),
RQCD 22~(FH) and Mainz 23~(direct) results, yielding
\begin{align}
&  \mbox{}\Nf=2+1: &\FLAGAVBEGIN \sigma_{\pi N} &=  42.2(2.4)	 \FLAGAVEND ~\mbox{MeV} 
  &&\Refs~\mbox{\cite{Durr:2011mp,Durr:2015dna,Yang:2015uis,RQCD:2022xux,Agadjanov:2023efe}}.
\end{align}
Note that both BMW results are included as they were obtained on
independent sets of ensembles~(employing different fermion
actions). The RQCD 22 
and Mainz 23 studies both utilize CLS $\Nf=2+1$ ensembles~(the latter utilizes a subset of
the ensembles employed by the former). To be conservative we take the
statistical errors for these two studies to be 100\% correlated.
The FLAG result for $\Nf=2$ can be found in the FLAG 21
report~\cite{FlavourLatticeAveragingGroupFLAG:2021npn}.

Moving on to $\sigma_s$ and the calculations detailed in
Tab.~\ref{tab:gs-ud-s-direct}, for $\Nf=2+1+1$ MILC~12C~(hybrid) 
and BMW 20A satisfy the quality criteria, however,
the latter is a preprint and is not considered for averaging. In order to convert
the result for $\langle N|\bar{s}s|N\rangle$ given in MILC~12C to a
value for $\sigma_s$, we multiply by the appropriate FLAG average for
$m_s$ given in Eq.~(35) 
of FLAG 19.  This gives our
result for four flavours, which is unchanged since the last FLAG report,
\begin{align}
  &\label{eq:sigmas_2p1p1}
  \mbox{}\Nf=2+1+1:&\FLAGAVBEGIN \sigma_{s} &= 41.0(8.8)\FLAGAVEND ~\mbox{MeV} 
                       &&\Ref~\mbox{\cite{Freeman:2012ry}}.
\end{align}
For $\Nf=2+1$ we perform a weighted average of BMW~11A~(FH),
MILC~12C~(hybrid), Junnarkar~13~(FH), BMW~15~(FH), $\chi$QCD~15A
(direct), RQCD 22~(FH) and Mainz 23~(direct).
MILC~09D~\cite{Toussaint:2009pz} also passes the FLAG selection rules,
however, this calculation is superseded by MILC~12C. As for
Eq.~\eqref{eq:sigmas_2p1p1}, the strangeness scalar matrix element
determined in the latter study is multiplied by the three-flavour FLAG
average for $m_s$ given in Eq.~(33) of FLAG 19. 
There are correlations between the MILC~12C and Junnarkar~13 results
as there is some overlap between the sets of asqtad ensembles used in
both cases. We take the statistical errors for these two studies to be
100\% correlated and, similarly, for the Mainz 23 and RQCD 22 studies~(as
for $\sigma_{\pi N}$). The global average is
\begin{align}
  &\label{eq:sigmas_2p1}
  \mbox{}\Nf=2+1:  &\FLAGAVBEGIN \sigma_{s} &= 44.9(6.4)  \FLAGAVEND ~\mbox{MeV} 
                       &&\Refs~\mbox{\cite{Durr:2011mp,Freeman:2012ry,Junnarkar:2013ac,Durr:2015dna,Yang:2015uis,RQCD:2022xux,Agadjanov:2023efe}},
\end{align}
where the error has been increased by around 10\% because
$\chi^2/dof=1.2317$ for the weighted average.  For all the other
averages presented above, the $\chi^2/dof$ is less than one and no
rescaling of the error is applied. There are no $\Nf=2$ studies of
$\sigma_s$ which pass the FLAG quality criteria, see the FLAG 21 report for
further details.

We remark that  it was not possible to determine
$\delta(a_{\rm min})$  for the above works based on the information
provided.

\begin{figure}[!t]
\begin{center}
\includegraphics[width=11.5cm]{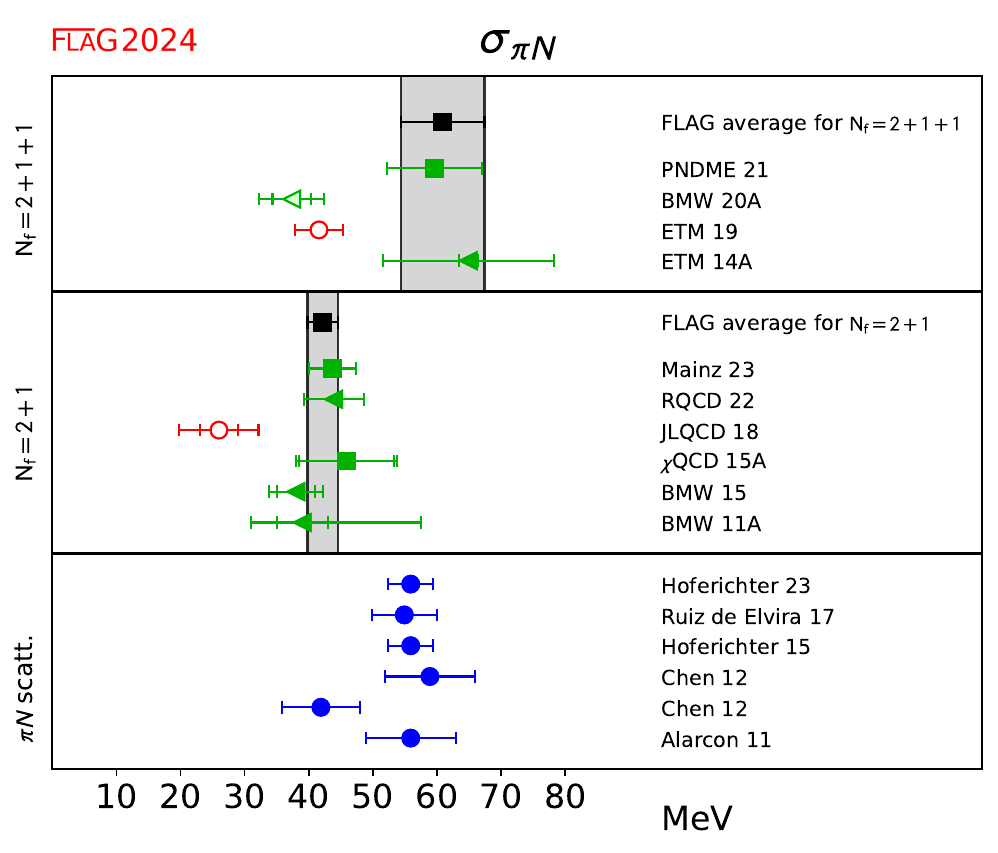}
\end{center}
\vspace{-1cm}
\caption{\label{fig:gs-sum-light} Lattice results and FLAG averages
  for the nucleon sigma term, $\sigma_{\pi N}$, for the $\Nf = 2+1$, and $2+1+1$ flavour calculations. Determinations via the
  direct approach are indicated by squares and circles, and the Feynman-Hellmann
  method by triangles.
  Results from recent   analyses of $\pi$-$N$   scattering~\cite{Alarcon:2011zs,Chen:2012nx,Hoferichter:2015dsa,RuizdeElvira:2017stg,Hoferichter:2023ptl}~(filled blue circles) are shown for comparison. Note that the charged pion is used to define the isospin limit in these phenomenological analyses,
  while the neutral pion with $M_\pi\sim 135$~MeV is usually used to define the physical point in lattice simulations. We adjust the results to be consistent with the
  latter, applying the correction for the different conventions determined in Ref.~\cite{Hoferichter:2023ptl}. }
\end{figure}

\begin{figure}[!t]
\begin{center}
\includegraphics[width=11.5cm]{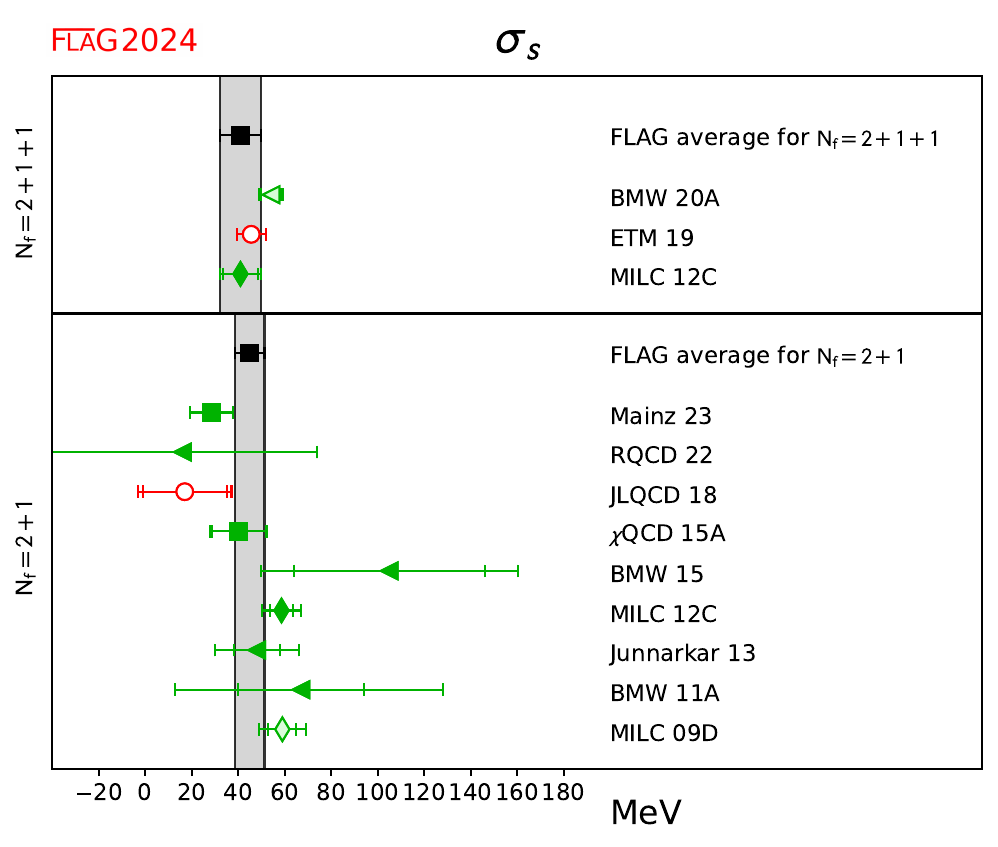}
\end{center}
\vspace{-1cm}
\caption{\label{fig:gs-sum-strange} Lattice results and FLAG averages
  for $\sigma_s$ for the $\Nf = 2+1$, and $2+1+1$ flavour
  calculations. Determinations via the direct approach are indicated
  by squares and circles, the Feynman-Hellmann method by triangles and the hybrid
  approach by diamonds.
}
\end{figure}

All the results for $\sigma_{\pi N}$ and $\sigma_s$ are displayed in
Figs.~\ref{fig:gs-sum-light} and~\ref{fig:gs-sum-strange} along with
the averages given above. Note that where
$f_{T_{ud}}=f_{T_{u}}+f_{T_{d}}$ or $f_{T_s}$ is quoted in
Tabs.~\ref{tab:gs-ud-s-direct} and \ref{tab:gs-singlet-fh}, we
multiply by the experimental proton mass in order to include the
results in the figures. For $\sigma_{\pi N}$, the averages are
consistent with the respective FLAG 21 values, however, the errors are
significantly reduced. For four flavours, this is due to the PNDME~21 direct result, which dominates the average. The results that
enter the average for three flavours, are all consistent with each
other and the addition of the RQCD~22 and Mainz~23 studies
reduces the uncertainty. The latter is the most precise result to date
which passes all the FLAG quality criteria. Notably, there is now a
2.7$\sigma$ difference between the $\Nf=2+1$ and $\Nf=2+1+1$ FLAG
averages. This is unlikely to be due to the inclusion of charm quarks
in the sea. The control of excited-state contributions remains an
issue. In particular, the PNDME~21 study utilizes a narrow-width
prior in their fitting analysis set to the lowest multi-hadron~($N\pi$
or $N\pi\pi$) excited-state energy. This is motivated by a $\chi$PT
analysis which indicates that these multi-hadron contributions are
significant at physical pion masses. If this constraint is relaxed
then a sigma term of around 42~MeV is obtained. Mainz~23 also find
an increase in the sigma term if such a prior is included in the
fitting procedure; however, the shift is much less
pronounced. Although progress is being made in terms of improving the
statistical precision of the correlation functions and realising more
source-sink separations (with the maximum separation currently around 1.5~fm),
more work needs to be done in order to control excited-state
contributions at close-to-physical pion masses. We caution the reader
that as more results for both $\sigma_{\pi N}$ and $\sigma_s$ become
available the averages may change.

Also shown for comparison in the figures are determinations of
$\sigma_{\pi N}$ from recent analyses of $\pi$-$N$
scattering~\cite{Alarcon:2011zs,Chen:2012nx,Hoferichter:2015dsa,RuizdeElvira:2017stg,Hoferichter:2023ptl}.
The $\Nf=2+1+1$ lattice average is in agreement with Hoferichter et
al.~\cite{Hoferichter:2023ptl}~(Hoferichter 23 in
Fig.~\ref{fig:gs-sum-light}), while
there is some tension, at the level of around three standard
deviations, with the lattice average for $\Nf=2+1$.\footnote{We adjust the result of
Ref.~\cite{Hoferichter:2023ptl} such that it is consistent with
defining the isospin limit using the mass of the neutral pion.}

For the strangeness sigma term, the four-flavour average is unchanged
from the previous FLAG report, while the three-flavour average has
decreased by 1$\sigma$ and there is a small reduction in the
error. There is a slight tension between the Mainz 23 and MILC 12C
$\Nf=2+1$ results, however, both FLAG averages are consistent with
each other.

Finally we remark that, by exploiting the heavy-quark limit, the
light- and strange-quark sigma terms can be used to estimate
$\sigma_q$ for the charm, bottom and top
quarks~\cite{Shifman:1978zn,Chetyrkin:1997un,Hill:2014yxa}. The
resulting estimate for the charm quark, see, e.g., the RQCD 16 $\Nf=2$
analysis of Ref.~\cite{Bali:2016lvx} that reports $f_{T_c}=0.075(4)$
or $\sigma_c=70(4)$~MeV, is consistent with the direct determinations
of ETM 19~\cite{Alexandrou:2019brg} for $\Nf=2+1+1$
  of $\sigma_c=107(22)$~MeV, ETM 16A~\cite{Abdel-Rehim:2016won} for
$\Nf=2$ of $\sigma_c=79(21)(^{12}_{8})$~MeV and $\chi$QCD
13A~\cite{Gong:2013vja} for $\Nf=2+1$ of $\sigma_c=94(31)$~MeV.  
  BMW in  BMW 20A~\cite{Borsanyi:2020bpd} employing the Feynman-Hellmann approach obtain
  $f_{T_c}=\sigma_c/m_N=0.0734(45)(55)$ for $\Nf=1+1+1+1$. MILC in MILC
12C~\cite{Freeman:2012ry} find $\langle N|\bar{c}c|N\rangle=0.056(27)$
in the $\msbar$ scheme at a scale of 2~GeV for $\Nf=2+1+1$ via the
hybrid method. Considering the large uncertainty, this is consistent
with the other results once multiplied by the charm-quark mass.

\subsection{Isovector second Mellin moments $\langle x \rangle_{u-d}$,
  $\langle x \rangle_{\Delta u - \Delta d}$
  and $\langle x \rangle_{\delta u - \delta d}$}
\label{sec:moments}

\begin{figure*}[!h]                                                                                                                                                   
\centering
\includegraphics[angle=0,width=0.92\textwidth]{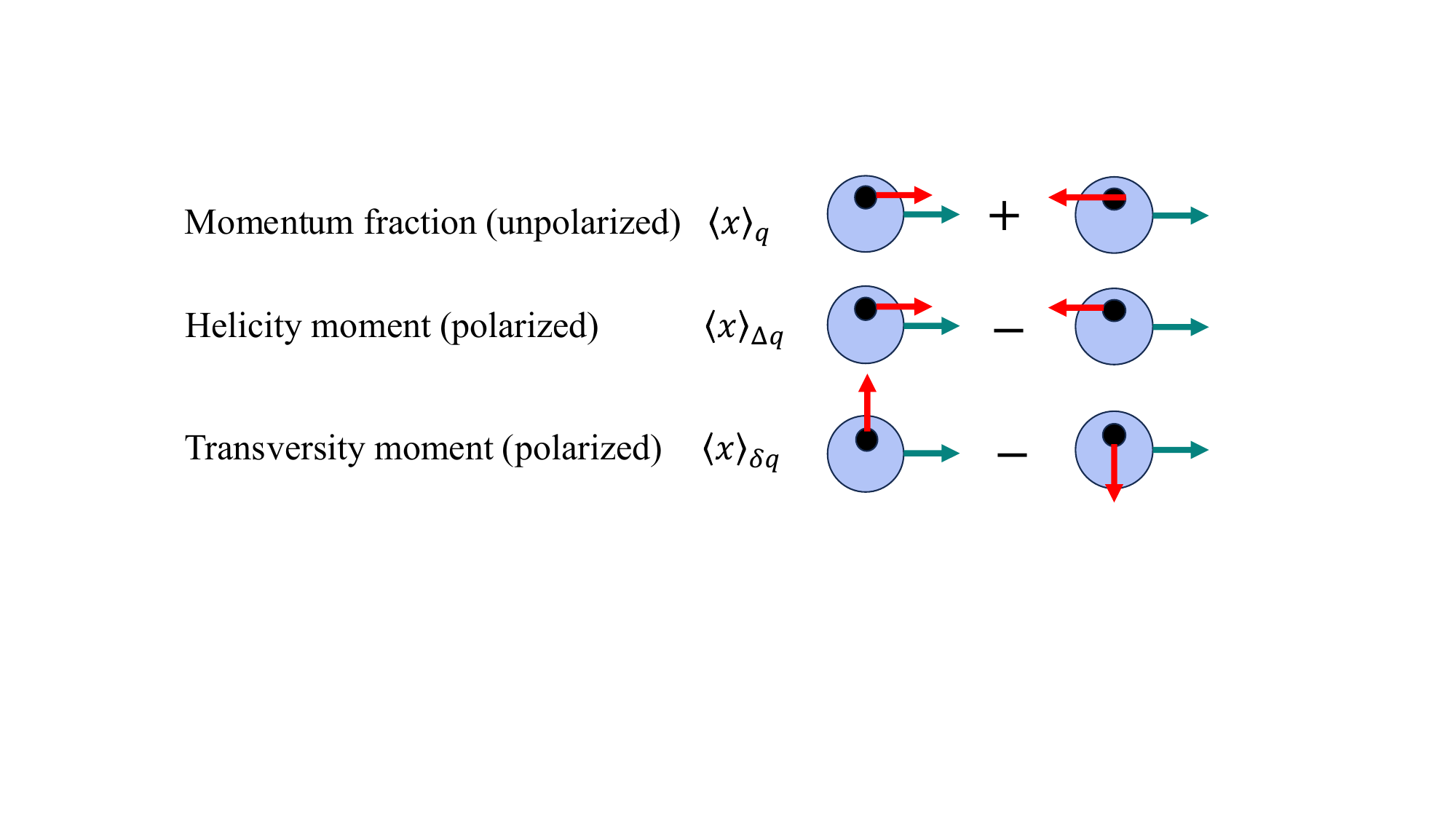}
\vspace{-0.08in}
\caption{A pictorial description of the three moments showing the direction of the spin of the quark (red arrow)
with respect to the nucleon momentum (green arrow). }
\label{fig:Moments_diag}
\end{figure*}

This section introduces the basics of the calculation of the momentum fraction
carried by the quarks and the transversity and helicity moments in the
isovector channel. These  moments of spin-independent (i.e.,
unpolarized), $q=q_\uparrow+q_\downarrow$, helicity (i.e., polarized),
$\Delta q=q_\uparrow-q_\downarrow$, and transversity,
$\delta q =q_\top+q_{\perp}$ distributions, are defined as
\begin{eqnarray}
\langle x \rangle_q &=& \int_0^1~x~[q(x)+\overline{q}(x)]~dx \,, \\
\langle x \rangle_{\Delta q} &=& \int_0^1~x~[\Delta q(x)+\Delta \overline{q}(x)]~dx \,, \\
\langle x \rangle_{\delta q} &=& \int_0^1~x~[\delta q(x)+\delta \overline{q}(x)]~dx \,,
\end{eqnarray}
where $q_{\uparrow(\downarrow)}$ corresponds to quarks with helicity
aligned (anti-aligned) with that of a longitudinally polarized target,
and $q_{\top(\perp)}$ corresponds to quarks with spin aligned
(anti-aligned) with that of a transversely polarized target. These
alignments are shown pictorially in Fig.~\ref{fig:Moments_diag}.

At leading twist, these moments can be extracted from the forward matrix
elements of one-derivative vector, axial-vector and tensor operators
within ground-state nucleons.  The complete
set of the relevant twist-two operators are
\begin{eqnarray}
   {\cal O}^{\mu \nu}_{V^a}&=&\overline{q} \gamma^{\{\mu}\overleftrightarrow{D}^{\nu\}} \tau^a q\nn \,, \\
   {\cal O}^{\mu \nu}_{A^a}&=&\overline{q}\gamma^{\{\mu}  \overleftrightarrow{D}^{\nu\}} \gamma^5 \tau^a q\nn \,, \\
   {\cal O}^{\mu \nu \rho}_{T^a}&=&\overline{q} \sigma^{[\mu\{\nu]} \overleftrightarrow{D}^{\rho\}} \tau^a q \,,
   \label{operators}
\end{eqnarray}
where $q=\{u,d\}$ is the isodoublet of light quarks and
$\sigma^{\mu\nu} = (\gamma^\mu\gamma^\nu - \gamma^\nu\gamma^\mu)/2$.
The derivative $\overleftrightarrow{D}^{\nu}\equiv\frac{1}{2}(\overrightarrow{D}^\nu-\overleftarrow{D}^\nu)$
consists of four terms defined in Ref.~\cite{Mondal:2020cmt}.
Lorentz indices within $\{ ~\}$ in Eq.~\eqref{operators} are
symmetrized and within $[\, ]$ are antisymmetrized. It is also
implicit that, where relevant, the traceless part of the above
operators is taken.

The methodology for nonperturbative renormalization of these
operators is very similar to that for the charges. Details of
these twist-two operators and their renormalization can be found in
Refs.~\cite{Gockeler:1995wg} and~\cite{Harris:2019bih}.

In numerical calculations, it is typical to set the spin of the nucleon in a given
direction. Choosing the spin to be in the ``3'' direction and restricting to the isovector case,
$\tau^a = \tau^3$, the explicit operators become
\begin{eqnarray}
 {\cal O}^{44}_{V^3} &=&  \overline{q} (\gamma^{4}\overleftrightarrow{D}^{4}  -\frac{1}{3}
 {\bm \gamma} \cdot \overleftrightarrow{\bf D}) \tau^3 q \,,
 \label{eq:finaloperatorV} \\
 {\cal O}^{34}_{A^3} &=&\overline{q} \gamma^{\{3}\overleftrightarrow{D}^{4\}} \gamma^5 \tau^3 q \,,
 \label{eq:finaloperatorA} \\
{\cal O}^{124}_{T^3} &=& \overline{q} \sigma^{[1\{2]}\overleftrightarrow{D}^{4\}} \tau^3 q \,.
\label{eq:finaloperatorT}
\end{eqnarray}
The isovector moments are then obtained from their forward matrix elements within the
nucleon ground state using the following relations:
\begin{eqnarray}
\langle 0 | {\cal O}^{44}_{V^3}| 0 \rangle &=&  -  M_N\, \langle x \rangle_{u-d} \,,
\label{eq:me2momentV} \\
\langle 0 | {\cal O}^{34}_{A^3}| 0 \rangle &=&  - \frac{i  M_N}{2} \, \langle x \rangle_{\Delta u-\Delta d} \,,
\label{eq:me2momentA} \\
\langle 0 | {\cal O}^{124}_{T^3}| 0 \rangle &=& - \frac{i M_N}{2} \, \langle x \rangle_{\delta u-\delta d} \,.
\label{eq:me2momentT}
\end{eqnarray}

\subsubsection{Results for the isovector moments 
$\langle x \rangle_{u-d}$, $\langle x \rangle_{\Delta u - \Delta d}$                              
  and $\langle x \rangle_{\delta u - \delta d}$}
\label{sec:moments-results}

\begin{figure}[!t]
\begin{center}
\includegraphics[width=7.5cm]{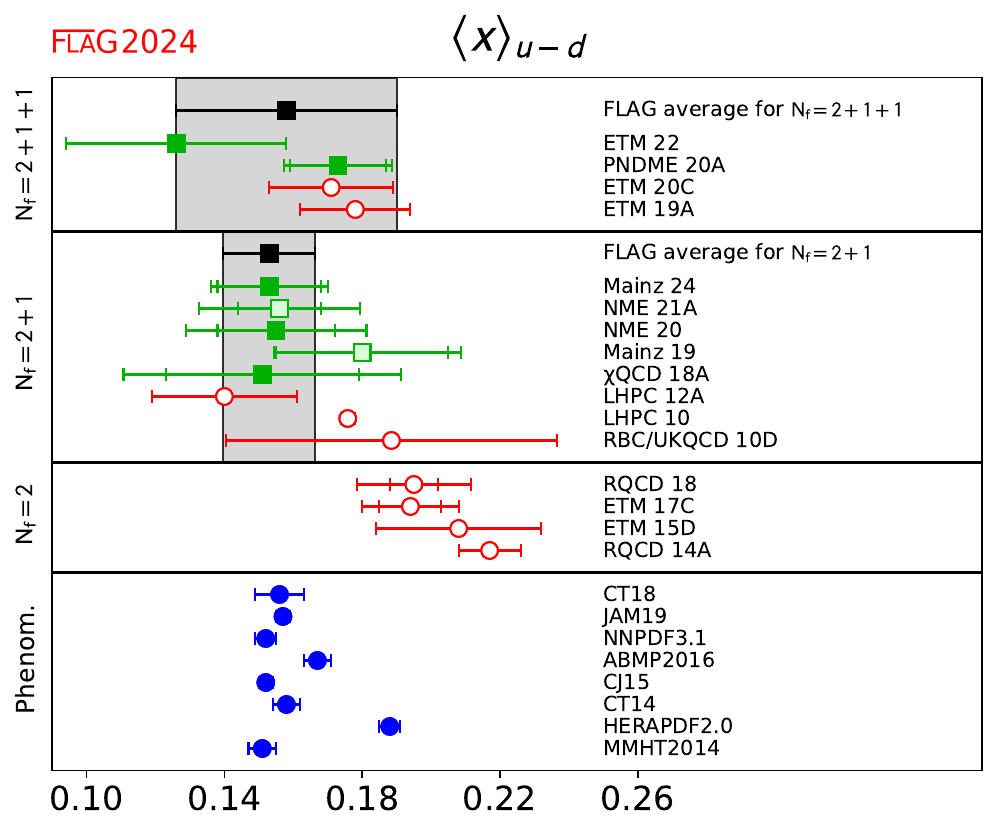}
\includegraphics[width=7.5cm]{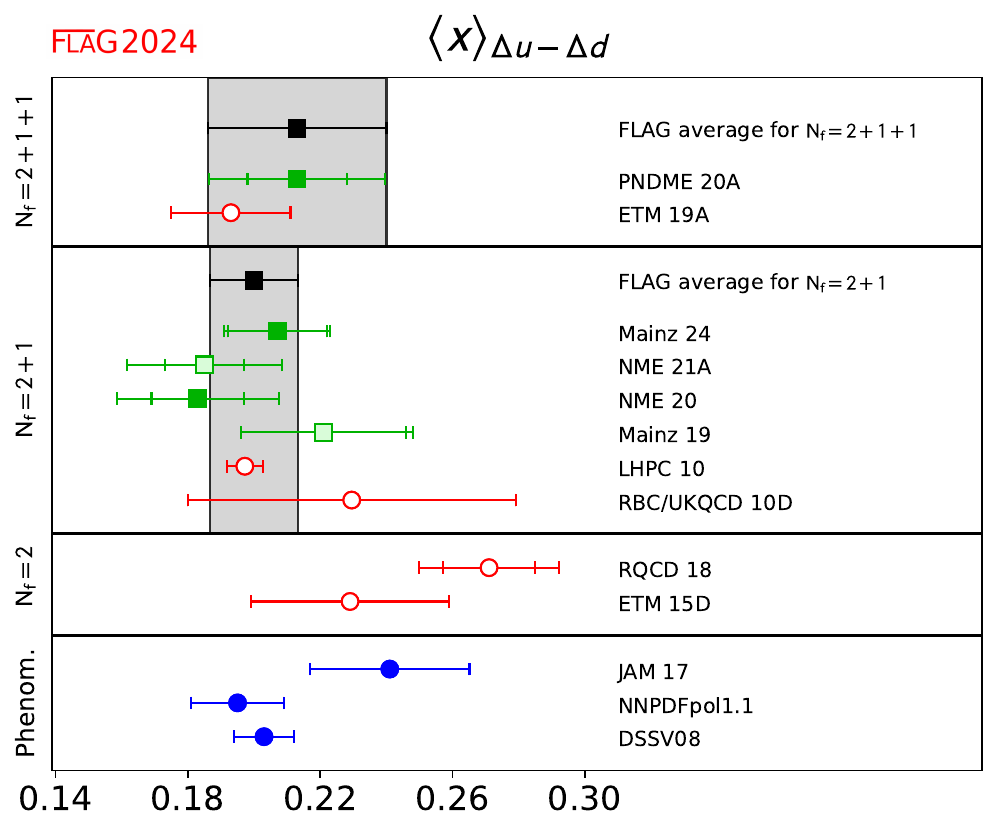}
\includegraphics[width=7.5cm]{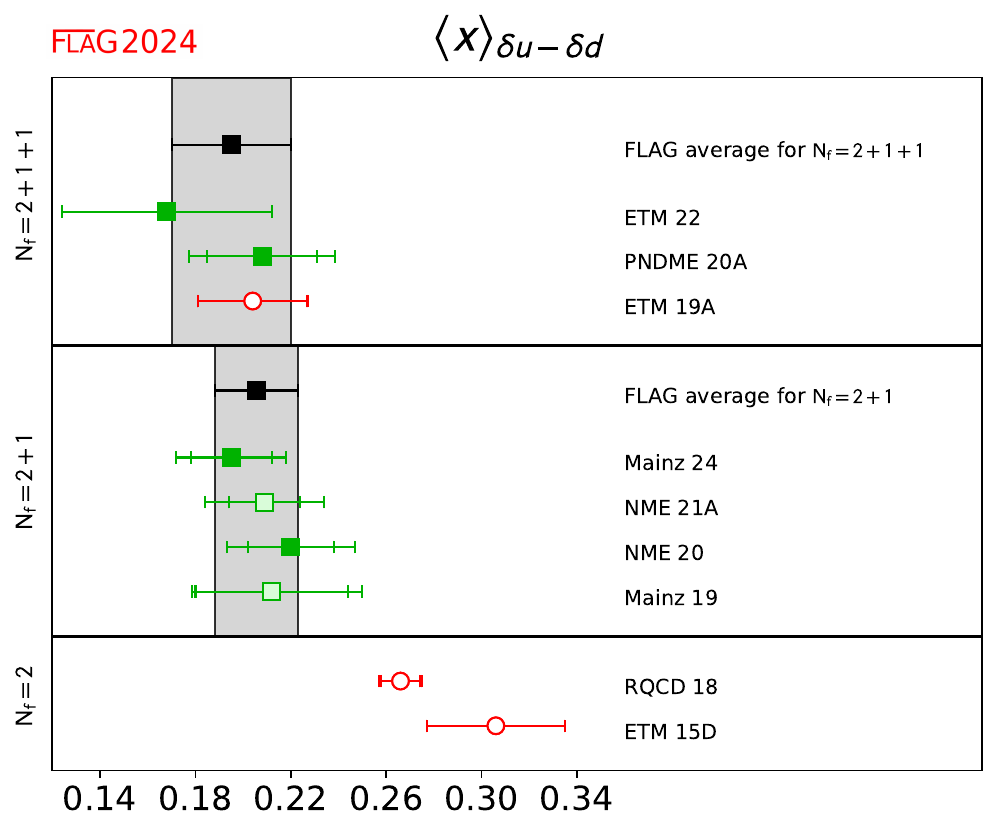}
\end{center}
\vspace{-1cm}
\caption{\label{fig:moments} Lattice-QCD results for the second Mellin moments
$\langle x \rangle_{u-d}$, $\langle x \rangle_{\Delta u - \Delta d}$
  and $\langle x \rangle_{\delta u - \delta d}$. Results from 
  $\Nf=2$ simulations and publications prior to 2014 have been 
  included as this is the first review of these quantities. For the
  momentum-fraction and helicity moment, we have also included
  phenomenological estimates~\cite{Harland-Lang:2014zoa,H1:2015ubc,Dulat:2015mca,Accardi:2016qay,Alekhin:2017kpj,NNPDF:2017mvq,Sato:2019yez,Hou:2019efy,deFlorian:2009vb,Nocera:2014gqa,Ethier:2017zbq}. 
  For reference, a recent analysis~\cite{Li:2023yda}, using a combined set of World DIS data, found $\langle x \rangle_{u-d}=0.143(5)$ at $Q^2=4$~GeV${}^2$ (not shown).
}
\end{figure}

A summary of results for these three moments is given in
Tabs.~\ref{tab:moments1} and~\ref{tab:moments2} and the values including the FLAG averages are
shown in Fig.~\ref{fig:moments}.  Results from $\Nf=2$ simulations and
publications prior to 2014 have been included as this is the first
review of these quantities. For the momentum fraction and helicity
moment, we have also included phenomenological estimates. Lattice
values for the momentum fraction are consistent with phenomenology but
have larger errors.  Results for the helicity moment, $\langle x
\rangle_{\Delta u - \Delta d}$, are consistent and have similar
uncertainties. Lattice results for the transversity moment are a
prediction.

We discuss results for these three
moments together as the methodology for their calculations and the
analysis is the same, and the systematics are similar.  All results
presented in this section are in the $\overline{\rm MS}$ scheme at
$2$~GeV.

\input{NME/Tables/table_x.tex}

For the 2+1+1-theory, the results PNDME 20A~\cite{Mondal:2020cmt} 
and ETM 22~\cite{Alexandrou:2022dtc} results in qualify for the
averages. The PNDME 20A results are from nine HISQ
ensembles analyzed using clover fermions. The operators are
renormalized nonperturbatively using the RI'-MOM scheme, and the
chiral-continuum-finite-volume extrapolation is done keeping the
leading-order corrections in each of the three variables. Analyses of
excited-state contamination are done using three strategies that differ in the selection of
the first excited-state mass. The final results are from a three-state
fit to the three-point function with the spectrum taken from the two-point
function, i.e., assuming no enhanced contribution from multihadron
excited states. An additional systematic uncertainty is assigned to
cover the spread of these three estimates.

The ETM collaboration has presented new results from three ensembles
with 2+1+1-flavour twisted-mass fermions with close-to-physical pion
masses at $a = 0.057$, 0.069 and 0.80~fm in
\cite{Alexandrou:2022dtc}. These results supersede those in
\cite{Alexandrou:2020sml,Alexandrou:2019ali} based on the single
ensemble at $a = 0.080$~fm for the momentum fraction and the
transversity moment. To control excited-state contamination, they
compare results from the plateau, summation and two-state methods with
the final values taken from the two-state fit. Operators are
renormalized nonperturbatively via the RI'-MOM scheme supplemented by
perturbative subtraction of lattice artefacts.  The continuum
extrapolation, which keeps the leading correction $\propto a^2$, shows
a significant slope for $\langle x \rangle_{u-d}$, which reduces the
continuum-limit value.

When determining the final results to quote for the $2+1+1$ theory, we
note the large difference between the results from
Refs.~\cite{Mondal:2020cmt,Alexandrou:2022dtc} for the momentum
fraction. Our conservative approach is to construct the interval defined by
the PNDME 20A value plus error and the ETM 22 value, i.e., 0.126--0.189, 
and then take the mean of the interval for the central value
and half the spread for the error as shown in Fig.~\ref{fig:moments}.
For the transversity moment we
perform the FLAG averaging assuming no correlations between the two
calculations. For the helicity fraction we quote the PNDME
20A~\cite{Mondal:2020cmt} values.
The values of $\delta(a_{\rm min})$ for the three moments for the PNDME 20A
calculation~\cite{Mondal:2020cmt} are 0.6, 0.3 and 0.13, and those for ETM 22
are roughly 0.8 (momentum fraction) and 0.0 (transversity). The FLAG averages are

\begin{align}
&   \mbox{}\Nf=2+1+1: &\FLAGAVBEGIN \langle x \rangle_{u-d}  &= 0.158(32)  \FLAGAVEND   &&\Refs~\mbox{\cite{Mondal:2020cmt,Alexandrou:2022dtc}},  \\
&  \mbox{}\Nf=2+1+1: &\FLAGAVBEGIN \langle x \rangle_{\Delta u-\Delta d}   &=    0.213(27)  \FLAGAVEND   &&\Ref~\mbox{\cite{Mondal:2020cmt}}, \\
&  \mbox{}\Nf=2+1+1: &\FLAGAVBEGIN \langle x \rangle_{\delta u-\delta d}   &=    0.195(25)  \FLAGAVEND   &&\Refs~\mbox{\cite{Mondal:2020cmt,Alexandrou:2022dtc}}.
\end{align}

Five calculations qualify for averages for the 2+1-flavour theory: the
Mainz~\cite{Djukanovic:2024krw,Harris:2019bih}, the NME~\cite{Mondal:2020ela,Mondal:2021oot}, and $\chi$QCD~\cite{Yang:2018nqn}. Of these, the Mainz
24~\cite{Djukanovic:2024krw} supersedes the Mainz
19~\cite{Harris:2019bih}, and while the NME 21A~\cite{Mondal:2021oot}
is an update of NME 20~\cite{Mondal:2020ela}, it is a conference proceeding.

The Mainz 24 results are based on fifteen $\Nf=2+1$ ensembles
produced by the CLS collaboration covering the ranges
$0.05 \le a \le 0.09$~fm and $130 \le M_\pi \le 360$~MeV.
A two-state summation method is used to control
excited-state contamination. In the continuum-chiral-finite-volume extrapolation, leading-order corrections are used for the continuum and finite-volume
corrections and up to NNLO results from SU(2) baryon chiral
perturbation theory for the chiral part.

The NME 20~\cite{Mondal:2020ela} results are based on seven
$\Nf=2+1$ clover ensembles produced by the JLab/W\&M/LANL/MIT
collaborations.  They cover the range $0.07 \le a \le 0.13$~fm and
$170 \le M_\pi \le 280$~MeV.  The analysis methodology is the same as in
Ref.~\cite{Mondal:2020cmt} already discussed above.

The $\chi$QCD~\cite{Yang:2018nqn} calculation uses four domain-wall
ensembles that have been ge\-ne\-ra\-ted by the RBC/UKQCD col\-labo\-ra\-tion that cover the
range $0.08 \le a \le 0.14$~fm and $139 \le M_\pi \le 330$~MeV. A
number of values of overlap-valence-quark masses, in addition to those
close to the unitary point $M_\pi^{\rm sea} = M_\pi^{\rm valence}$,
are used. The renormalization is carried out nonperturbatively. The
continuum-chiral-finite-volume extrapolation is carried out using the
leading corrections plus terms accounting for partial quenching, i.e.,
the leading terms in the difference $M_\pi^{\rm sea} - M_\pi^{\rm
  valence}$.

The three older calculations, LHPC
12A~\cite{Green:2012ud}, LHPC 10~\cite{Bratt:2010jn} and
RBC/UKQCD~\cite{Aoki:2010xg}, do not meet the criteria of
control over the continuum limit. Similarly, the $\Nf = 2$
calculations  fail to satisfy one or more of the FLAG criteria. 

The 2+1-flavour FLAG averages for the momentum fraction,
$\langle x \rangle_{ u-d}$, are constructed using the Mainz
24~\cite{Djukanovic:2024krw}, NME 20~\cite{Mondal:2020ela} and
$\chi$QCD 18A~\cite{Yang:2018nqn} values assuming zero correlations
between them. The results for the helicity and transversity moments
are the FLAG averages of the Mainz 24~\cite{Djukanovic:2024krw} and
NME 20~\cite{Mondal:2020ela} values again assuming zero
correlations.  The values of $\delta(a_{\rm min})$ for
the Mainz 24~\cite{Djukanovic:2024krw} for the three moments are
1.5, 0.2, 0.1 and those for the NME 20 are 0.5, 1.0 and 0.2.
The $\chi$QCD 18A work does not provide enough information to determine $\delta(a_{\rm min})$. The FLAG averages are

\begin{align}
  &  \mbox{}\Nf=2+1: &\FLAGAVBEGIN \langle x \rangle_{u-d}  &= 0.153(13)  \FLAGAVEND   &&\Refs~\mbox{\cite{Djukanovic:2024krw,Mondal:2020ela,Yang:2018nqn}},  \\
&  \mbox{}\Nf=2+1: &\FLAGAVBEGIN \langle x \rangle_{\Delta u-\Delta d}   &=    0.200(13)  \FLAGAVEND   &&\Refs~\mbox{\cite{Djukanovic:2024krw,Mondal:2020ela}}, \\
&  \mbox{}\Nf=2+1: &\FLAGAVBEGIN \langle x \rangle_{\delta u-\delta d}   &=    0.206(17)  \FLAGAVEND   &&\Refs~\mbox{\cite{Djukanovic:2024krw,Mondal:2020ela}}.
\end{align}

%% file: NME/Tables/table_ga.tex
\begin{table}[t!]
\begin{center}
\mbox{} \\[3.0cm]
\footnotesize
\begin{tabular*}{\textwidth}[l]{l @{\extracolsep{\fill}} r l l l l l l l l }
Collaboration & Ref. & $\Nf$ & 
\hspace{0.15cm}\begin{rotate}{60}{publication status}\end{rotate}\hspace{-0.15cm} &
\hspace{0.15cm}\begin{rotate}{60}{continuum extrapolation}\end{rotate}\hspace{-0.15cm} &
\hspace{0.15cm}\begin{rotate}{60}{chiral extrapolation}\end{rotate}\hspace{-0.15cm}&
\hspace{0.15cm}\begin{rotate}{60}{finite volume}\end{rotate}\hspace{-0.15cm}&
\hspace{0.15cm}\begin{rotate}{60}{renormalization}\end{rotate}\hspace{-0.15cm}  &
\hspace{0.15cm}\begin{rotate}{60}{excited states}\end{rotate}\hspace{-0.15cm}  &
$g^{u-d}_A$\\
&&&&&&&&& \\[-0.1cm]
\hline
\hline
&&&&&&&& \\[-0.1cm]

ETM 23 & \cite{Alexandrou:2023qbg} & 2+1+1 & \gA & \good & \good & \good & \good & \soso & 1.245(28)(14)$^c$ \\[0.5ex]
PNDME 23$^a$ & \cite{Jang:2023zts} & 2+1+1 & \gA & \good$^\ddag$ & \good & \good & \good & \soso & 1.292(53)(24)$^c$\\[0.5ex]
CalLat 19 & \cite{Walker-Loud:2019cif} & 2+1+1 & \rC & \soso & \good & \good & \good & \soso & 1.2642(93) \\[0.5ex]
ETM 19 & \cite{Alexandrou:2019brg} & 2+1+1 & \gA & \bad & \soso & \good & \good & \soso & 1.286(23) \\[0.5ex]
PNDME 18$^a$ & \cite{Gupta:2018qil} & 2+1+1 & \gA & \good$^\ddag$ & \good & \good & \good & \soso & 1.218(25)(30) \\[0.5ex]
CalLat 18 & \cite{Chang:2018uxx} & 2+1+1 & \gA & \soso & \good & \good & \good & \soso & 1.271(10)(7) \\[0.5ex]
CalLat 17 & \cite{Berkowitz:2017gql} & 2+1+1 & \oP & \soso & \good & \good & \good & \soso & 1.278(21)(26) \\[0.5ex]
PNDME 16$^a$ & \cite{Bhattacharya:2016zcn} & 2+1+1 & \gA & \soso$^\ddag$ & \good & \good & \good & \soso & 1.195(33)(20) \\[0.5ex]
\\[-0.1ex]\hline\\[0.2ex]
Mainz 24 & \cite{Djukanovic:2024krw} & 2+1 & \gA & \good & \good & \good & \good & \soso & 1.254(19)(15) \\[0.5ex]
PACS 23 & \cite{Tsuji:2023llh} & 2+1 & \gA & \bad & \soso & \good & \good & \soso & 1.264(14)(3) \\[0.5ex]
RQCD 23 & \cite{Bali:2023sdi} & 2+1 & \gA & \good & \good & \good & \good & \soso & 1.284($^{+0.028}_{-0.027}$) \\[0.5ex]
QCDSF/UKQCD/CSSM 23 & \cite{QCDSFUKQCDCSSM:2023qlx} & 2+1 & \gA & \good & \soso & \good & \good & \soso & 1.253(63)(41)$^d$ \\[0.5ex]
PACS 22B & \cite{Tsuji:2022ric} & 2+1 & \gA  &  \bad  & \soso & \good & \good & \soso  &  1.288(14)(9)  \\[0.5ex]
Mainz 22 & \cite{Djukanovic:2022wru} & 2+1 & \gA & \good & \good & \good & \good & \soso & 1.225(39)(25)$^c$ \\[0.5ex]
NME 21$^a$ & \cite{Park:2021ypf} & 2+1 & \gA & \soso$^\ddag$ & \good & \good & \good & \soso & 1.31(6)(5) \\[0.5ex]
RQCD 19 & \cite{RQCD:2019jai} & 2+1 & \gA & \good & \good & \good & \good & \soso & 1.302(45)(73)$^c$ \\[0.5ex]
LHPC 19 & \cite{Hasan:2019noy} & 2+1 & \gA & \bad$^\ddag$ & \good & \good & \good & \soso & 1.265(49) \\[0.5ex]
Mainz 19 & \cite{Harris:2019bih} & 2+1 & \gA & \good & \soso & \good & \good & \soso & 1.242(25)($^{+0}_{-0.030}$) \\[0.5ex]
PACS 18A & \cite{Shintani:2018ozy} & 2+1 & \gA & \bad & \good & \good & \good & \soso & 1.273(24)(5)(9) \\[0.5ex]
PACS 18 & \cite{Ishikawa:2018rew} & 2+1 & \gA & \bad & \bad & \good & \good & \bad & 1.163(75)(14) \\[0.5ex]
$\chi$QCD 18 & \cite{Liang:2018pis} & 2+1 & \gA & \soso & \good & \good & \good & \soso & 1.254(16)(30)$^\$$ \\[0.5ex]
JLQCD 18 & \cite{Yamanaka:2018uud} & 2+1 & \gA & \bad & \soso & \soso & \good & \soso & 1.123(28)(29)(90) \\[0.5ex]
&&&&&&&& \\[-0.1cm]
\hline
\hline
\end{tabular*}
\begin{minipage}{\linewidth}
{\footnotesize 
\begin{itemize}
\item[$^a$] The improvement coefficient in the valence-quark action is
  set to its tadpole-improved tree-level value. \\[-5mm]
\item[$^b$] The quark action is tree-level improved. \\[-5mm]
\item[$^c$] Determination includes data for nonforward matrix elements. \\[-5mm]
\item[$^d$] Feynman-Hellmann theorem is used to determine the matrix element. \\[-5mm]
\item[$^\ddag$]The rating takes into account that the action is not fully $\cO(a)$-improved by requiring an additional lattice spacing. \\[-5mm]\item[$^\$$] For this partially quenched analysis the criteria are applied to the unitary points.
\end{itemize}
}
\end{minipage}
\caption{Overview of results for $ g^{u-d}_A$. \label{tab:ga}}
\end{center}
\end{table}

%% file: NME/Tables/table_gs.tex
\begin{table}[t!]
\begin{center}
\mbox{} \\[3.0cm]
\footnotesize
\begin{tabular*}{\textwidth}[l]{l @{\extracolsep{\fill}} r l l l l l l l l }
Collaboration & Ref. & $\Nf$ & 
\hspace{0.15cm}\begin{rotate}{60}{publication status}\end{rotate}\hspace{-0.15cm} &
\hspace{0.15cm}\begin{rotate}{60}{continuum extrapolation}\end{rotate}\hspace{-0.15cm} &
\hspace{0.15cm}\begin{rotate}{60}{chiral extrapolation}\end{rotate}\hspace{-0.15cm}&
\hspace{0.15cm}\begin{rotate}{60}{finite volume}\end{rotate}\hspace{-0.15cm}&
\hspace{0.15cm}\begin{rotate}{60}{renormalization}\end{rotate}\hspace{-0.15cm}  &
\hspace{0.15cm}\begin{rotate}{60}{excited states}\end{rotate}\hspace{-0.15cm}  &
$g^{u-d}_S$\\
&&&&&&&&& \\[-0.1cm]
\hline
\hline
&&&&&&&& \\[-0.1cm]
PNDME 23 & \cite{Jang:2023zts} & 2+1+1 & \gA &   \good$^\ddag$ & \good & \good &  \good& \soso &   1.085(50)(103) \\[0.5ex]
ETM 19 & \cite{Alexandrou:2019brg} & 2+1+1 & \gA & \bad & \soso & \good & \good & \soso & 1.35(17) \\[0.5ex]
PNDME 18 & \cite{Gupta:2018qil} & 2+1+1 & \gA & \good$^\ddag$ & \good & \good & \good & \soso & 1.022(80)(60) \\[0.5ex]
PNDME 16 & \cite{Bhattacharya:2016zcn} & 2+1+1 & \gA & \soso$^\ddag$ & \good & \good & \good & \soso & 0.97(12)(6) \\[0.5ex]
\\[-0.1ex]\hline\\[0.2ex]
Mainz 24 & \cite{Djukanovic:2024krw} & 2+1 & \gA & \good & \good & \good & \good & \soso & 1.203(77)(81) \\[0.5ex]
RQCD 23  & \cite{Bali:2023sdi} & 2+1 & \gA & \good & \good & \good & \good & \soso & 1.11$\left(^{+14}_{-16}\right)$ \\[0.5ex]
QCDSF/UKQCD/CSSM 23 & \cite{QCDSFUKQCDCSSM:2023qlx} & 2+1 & \gA &  \good & \soso & \good & \good & \soso$^d$  &   1.08(21)(03)$^d$ \\[0.5ex]
PACS 22B & \cite{Tsuji:2022ric} & 2+1 & \gA  &  \bad  & \soso & \good & \good & \soso  & 0.927(83)(22)  \\[0.5ex]
NME 21 & \cite{Park:2021ypf} & 2+1 & \gA & \soso$^\ddag$ & \good & \good & \good & \soso & 1.06(9)(6) \\[0.5ex]
$\chi$QCD 21A & \cite{Liu:2021irg} & 2+1 & \gA & \good & \good & \good & \good & \soso & 0.94(10)(08)$^\$$ \\[0.5ex]
RBC/UKQCD 19 & \cite{Abramczyk:2019fnf} & 2+1 & \gA & \bad & \soso & \good & \good & \bad & 0.9(3) \\[0.5ex]
Mainz 19 & \cite{Harris:2019bih} & 2+1 & \gA & \good & \soso & \good & \good & \soso & 1.13(11)$\left(^{+7}_{-6}\right)$ \\[0.5ex]
LHPC 19 & \cite{Hasan:2019noy} & 2+1 & \gA & \bad$^\ddag$ & \good & \good & \good & \soso & 0.927(303) \\[0.5ex]
JLQCD 18 & \cite{Yamanaka:2018uud} & 2+1 & \gA & \bad & \soso & \soso & \good & \soso & 0.88(8)(3)(7) \\[0.5ex]
&&&&&&&& \\[-0.1cm]
\hline
\hline
\end{tabular*}
\begin{minipage}{\linewidth}
{\footnotesize 
\begin{itemize}
\item[$^d$] Feynman-Hellmann theorem is used. \\[-5mm]
\item[$^\ddag$]The rating takes into account that the action is not fully $\cO(a)$-improved by requiring an additional lattice spacing. \\[-5mm]\item[$^\$$] For this partially quenched analysis the criteria are applied to the unitary points.
\end{itemize}
}
\end{minipage}
\caption{Overview of results for $ g^{u-d}_S$. \label{tab:gs}}
\end{center}
\end{table}

%% file: NME/Tables/table_gt.tex
\begin{table}[t!]
\begin{center}
\mbox{} \\[3.0cm]
\footnotesize
\begin{tabular*}{\textwidth}[l]{l @{\extracolsep{\fill}} r l l l l l l l l }
Collaboration & Ref. & $\Nf$ & 
\hspace{0.15cm}\begin{rotate}{60}{publication status}\end{rotate}\hspace{-0.15cm} &
\hspace{0.15cm}\begin{rotate}{60}{continuum extrapolation}\end{rotate}\hspace{-0.15cm} &
\hspace{0.15cm}\begin{rotate}{60}{chiral extrapolation}\end{rotate}\hspace{-0.15cm}&
\hspace{0.15cm}\begin{rotate}{60}{finite volume}\end{rotate}\hspace{-0.15cm}&
\hspace{0.15cm}\begin{rotate}{60}{renormalization}\end{rotate}\hspace{-0.15cm}  &
\hspace{0.15cm}\begin{rotate}{60}{excited states}\end{rotate}\hspace{-0.15cm}  &
$g^{u-d}_T$\\
&&&&&&&&& \\[-0.1cm]
\hline
\hline
&&&&&&&& \\[-0.1cm]
PNDME 23 & \cite{Jang:2023zts} & 2+1+1 & \gA &  \good$^\ddag$ & \good & \good &  \good& \soso &   0.991(21)(10) \\[0.5ex]
ETM 22       & \cite{Alexandrou:2022dtc} & 2+1+1 &\gA & \good & \good & \good & \good &\soso  &0.924(54) \\[0.5ex]
ETM 19 & \cite{Alexandrou:2019brg} & 2+1+1 & \gA & \bad & \soso & \good & \good & \soso & 0.936(25) \\[0.5ex]
PNDME 18 & \cite{Gupta:2018qil} & 2+1+1 & \gA & \good$^\ddag$ & \good & \good & \good & \soso & 0.989(32)(10) \\[0.5ex]
PNDME 16 & \cite{Bhattacharya:2016zcn} & 2+1+1 & \gA & \soso$^\ddag$ & \good & \good & \good & \soso & 0.987(51)(20) \\[0.5ex]
PNDME 15, 15A & \cite{Bhattacharya:2015esa,Bhattacharya:2015wna} & 2+1+1 & \gA & \soso$^\ddag$ & \good & \good & \good & \soso & 1.020(76) \\[0.5ex]
\\[-0.1ex]\hline\\[0.2ex]
Mainz 24 & \cite{Djukanovic:2024krw} & 2+1 & \gA & \good & \good & \good & \good & \soso & 0.993(15)(05) \\[0.5ex]
RQCD 23  & \cite{Bali:2023sdi} & 2+1 & \gA & \good & \good & \good &\good  &\soso  & 0.984$\left(^{+19}_{-29}\right)$ \\[0.5ex]
QCDSF/UKQCD/CSSM 23 & \cite{QCDSFUKQCDCSSM:2023qlx} & 2+1 & \gA &  \good & \soso & \good & \good & \soso$^d$  &   1.010(21)(12) \\[0.5ex]
PACS 22B & \cite{Tsuji:2022ric} & 2+1 & \gA  &  \bad  & \soso & \good & \good & \soso & 1.036(6)(20)  \\[0.5ex]
NME 21 & \cite{Park:2021ypf} & 2+1 & \gA & \soso$^\ddag$ & \good & \good & \good & \soso & 0.95(5)(2) \\[0.5ex]
RBC/UKQCD 19 & \cite{Abramczyk:2019fnf} & 2+1 & \gA & \bad & \soso & \good & \good & \bad & 1.04(5) \\[0.5ex]
Mainz 19 & \cite{Harris:2019bih} & 2+1 & \gA & \good & \soso & \good & \good & \soso & 0.965(38)$\left(^{+13}_{-41}\right)$ \\[0.5ex]
LHPC 19 & \cite{Hasan:2019noy} & 2+1 & \gA & \bad$^\ddag$ & \good & \good & \good & \soso & 0.972(41) \\[0.5ex]
JLQCD 18 & \cite{Yamanaka:2018uud} & 2+1 & \gA & \bad & \soso & \soso & \good & \soso & 1.08(3)(3)(9) \\[0.5ex]
&&&&&&&& \\[-0.1cm]
\hline
\hline
\end{tabular*}
\begin{minipage}{\linewidth}
{\footnotesize 
\begin{itemize}
\item[$^d$] Feynman-Hellmann theorem is used. \\[-5mm]
\item[$^\ddag$]The rating takes into account that the action is not fully $\cO(a)$-improved by requiring an additional lattice spacing.
\end{itemize}
}
\end{minipage}
\caption{Overview of results for $ g^{u-d}_T$. \label{tab:gt}}
\end{center}
\end{table}

%% file: NME/Tables/table_ga_singlet.tex
\begin{table}[t!]
\begin{center}
\mbox{} \\[3.0cm]
\footnotesize
\begin{tabular*}{\textwidth}[l]{l @{\extracolsep{\fill}} r l l l l l l l l l}
Collaboration & Ref. & $\Nf$ & 
\hspace{0.15cm}\begin{rotate}{60}{publication status}\end{rotate}\hspace{-0.15cm} &
\hspace{0.15cm}\begin{rotate}{60}{continuum extrapolation}\end{rotate}\hspace{-0.15cm} &
\hspace{0.15cm}\begin{rotate}{60}{chiral extrapolation}\end{rotate}\hspace{-0.15cm}&
\hspace{0.15cm}\begin{rotate}{60}{finite volume}\end{rotate}\hspace{-0.15cm}&
\hspace{0.15cm}\begin{rotate}{60}{renormalization}\end{rotate}\hspace{-0.15cm}  &
\hspace{0.15cm}\begin{rotate}{60}{excited states}\end{rotate}\hspace{-0.15cm}  &
$g_A^u$ & $g_A^d$ \\
&&&&&&&&& & \\[-0.1cm]
\hline
\hline
&&&&&&&& &  \\[-0.1cm]

PNDME 20 & \cite{Park:2020axe} & 2+1+1 & \rC & \good$^\ddag$ & \good & \good & \good & \soso & 0.790(23)(30) & $-$0.425(15)(30)  \\[0.5ex]
ETM 19 & \cite{Alexandrou:2019brg} & 2+1+1 & \gA & \bad & \soso & \good & \good & \soso & 0.862(17) & $-$0.424(16) \\[0.5ex]
PNDME 18A & \cite{Lin:2018obj} & 2+1+1 & \gA & \good$^\ddag$ & \good & \good & \good & \soso & 0.777(25)(30)$^\#$ & $-$0.438(18)(30)$^\#$ \\[0.5ex]
\\[-0.1ex]\hline\\[0.2ex]
Mainz 19A & \cite{Djukanovic:2019gvi} & 2+1 & \rC & \good & \soso & \good & \good & \soso & 0.84(3)(4) & $-$0.40(3)(4)   \\[0.5ex]
$\chi$QCD 18 & \cite{Liang:2018pis} & 2+1 & \gA & \soso & \good & \good & \good & \soso & 0.847(18)(32)$^\$$ & $-$0.407(16)(18)$^\$$ \\[0.5ex]
 & & & & & & & & & & \\[-0.1cm]
\hline
\hline
 & & & & & & & & & & \\[-0.1cm]
 & & & & & & & & & $g_A^s$& \\[-0.1cm]
 & & & & & & & & & &\\[-0.1cm]
\hline
\hline
     & & & & & & & & & &\\[-0.1cm]
PNDME 20 & \cite{Park:2020axe} & 2+1+1 & \rC & \good$^\ddag$ & \good & \good & \good & \soso &  $-$0.053(7) &  \\[0.5ex]
ETM 19 & \cite{Alexandrou:2019brg} & 2+1+1 & \gA & \bad & \soso & \good & \good & \soso & $-0.0458(73)$ & \\[0.5ex]
PNDME 18A & \cite{Lin:2018obj} & 2+1+1 & \gA & \good$^\ddag$ & \good & \good & \good & \soso &  $-$0.053(8)$^\#$ & \\[0.5ex]
\\[-0.1ex]\hline\\[0.2ex]
Mainz 19A & \cite{Djukanovic:2019gvi} & 2+1 & \rC & \good & \soso & \good & \good & \soso & $-$0.044(4)(5) &    \\[0.5ex]
$\chi$QCD 18 & \cite{Liang:2018pis} & 2+1 & \gA & \soso & \good & \good & \good & \soso &  $-$0.035(6)(7)$^\$$ & \\[0.5ex]
JLQCD 18 & \cite{Yamanaka:2018uud} & 2+1 & \gA & \bad & \soso & \soso & \good & \soso &  $-$0.046(26)(9)$^{\#}$ & \\[0.5ex]
$\chi$QCD 15 & \cite{Gong:2015iir} & 2+1 & \gA & \bad & \soso & \bad & \good & \soso &  $-$0.0403(44)(78)$^\#$ & \\[0.5ex]
 &&&&&&&& \\[-0.1cm]
\hline
\hline
\end{tabular*}
\begin{minipage}{\linewidth}
{\footnotesize 
\begin{itemize}
\item[$^\#$] Assumed that $Z_A^{n.s.}=Z_A^{s}$. \\[-5mm]
\item[$^\ddag$] The rating takes into account that the action is not fully $\cO(a)$-improved by requiring an additional lattice spacing. \\[-5mm]\item[$^\$$] For this partially quenched analysis the criteria are applied to the unitary points.
\end{itemize}
}
\end{minipage}
\caption{Overview of results for $g^q_A$.\label{tab:ga-singlet}}
\end{center}
\end{table}

%% file: NME/Tables/table_gt_singlet.tex
\begin{table}[t!]
\begin{center}
\mbox{} \\[3.0cm]
\footnotesize
\begin{tabular*}{\textwidth}[l]{l @{\extracolsep{\fill}} r l l l l l l l l@{\hspace{1mm}} l}
Collaboration & Ref. & $\Nf$ & 
\hspace{0.15cm}\begin{rotate}{60}{publication status}\end{rotate}\hspace{-0.15cm} &
\hspace{0.15cm}\begin{rotate}{60}{continuum extrapolation}\end{rotate}\hspace{-0.15cm} &
\hspace{0.15cm}\begin{rotate}{60}{chiral extrapolation}\end{rotate}\hspace{-0.15cm}&
\hspace{0.15cm}\begin{rotate}{60}{finite volume}\end{rotate}\hspace{-0.15cm}&
\hspace{0.15cm}\begin{rotate}{60}{renormalization}\end{rotate}\hspace{-0.15cm}  &
\hspace{0.15cm}\begin{rotate}{60}{excited states}\end{rotate}\hspace{-0.15cm}  &
$g_T^u$ & $g_T^d$ \\
&&&&&&&&& & \\[-0.1cm]
\hline
\hline
&&&&&&&& &  \\[-0.1cm]
PNDME 20 & \cite{Park:2020axe} & 2+1+1 & \rC & \good$^\ddag$ & \good & \good & \good & \soso & 0.783(27)(10) & $-$0.205(10)(10)   \\[0.5ex]
ETM 19 & \cite{Alexandrou:2019brg} & 2+1+1 & \gA & \bad & \soso & \good & \good & \soso & 0.729(22) & $-$0.2075(75) \\[0.5ex]
PNDME 18B & \cite{Gupta:2018lvp} & 2+1+1 & \gA & \good$^\ddag$ & \good & \good & \good & \soso & 0.784(28)(10)$^\#$ & $-$0.204(11)(10)$^\#$ \\[0.5ex]
PNDME 16 & \cite{Bhattacharya:2016zcn} & 2+1+1 & \gA & \soso$^\ddag$ & \good & \good & \good & \soso & 0.792(42)$^{\#\&}$ & $-$0.194(14)$^{\#\&}$ \\[0.5ex]
PNDME 15 & \cite{Bhattacharya:2015wna,Bhattacharya:2015esa} & 2+1+1 & \gA & \soso$^\ddag$ & \good & \good & \good & \soso & 0.774(66)$^\#$ & $-$0.233(28)$^\#$ \\[0.5ex]
\\[-0.1ex]\hline\\[0.2ex]
Mainz 19A & \cite{Djukanovic:2019gvi} & 2+1 & \rC & \good & \soso & \good & \good & \soso & 0.77(4)(6) & $-$0.19(4)(6)   \\[0.5ex]
JLQCD 18 & \cite{Yamanaka:2018uud} & 2+1 & \gA & \bad & \soso & \soso & \good & \soso & 0.85(3)(2)(7) & $-$0.24(2)(0)(2) \\[0.5ex]
 & & & & & & & & & & \\[-0.1cm]
\hline
\hline
 & & & & & & & & & & \\[-0.1cm]
 & & & & & & & & & $ g_T^s $& \\[-0.1cm]
 & & & & & & & & & &\\[-0.1cm]
\hline
\hline
     & & & & & & & & & &\\[-0.1cm]
PNDME 20 & \cite{Park:2020axe} & 2+1+1 & \rC & \good$^\ddag$ & \good & \good & \good & \soso &  $-$0.0022(12)  & \\[0.5ex]
ETM 19 & \cite{Alexandrou:2019brg} & 2+1+1 & \gA & \bad & \soso & \good & \good & \soso & $-0.00268(58)$ & \\[0.5ex]
PNDME 18B & \cite{Gupta:2018lvp} & 2+1+1 & \gA & \good$^\ddag$ & \good & \good & \good & \soso &  $-$0.0027(16)$^\#$ & \\[0.5ex]
PNDME 15 & \cite{Bhattacharya:2015wna,Bhattacharya:2015esa} & 2+1+1 & \gA & \soso$^\ddag$ & \good & \good & \good & \soso &  0.008(9)$^\#$ & \\[0.5ex]
\\[-0.1ex]\hline\\[0.2ex]
Mainz 19A & \cite{Djukanovic:2019gvi} & 2+1 & \rC & \good & \soso & \good & \good & \soso & $-$0.0026(73)(42) &   \\[0.5ex]
JLQCD 18 & \cite{Yamanaka:2018uud} & 2+1 & \gA & \bad & \soso & \soso & \good & \soso &  $-$0.012(16)(8) & \\[0.5ex]
&&&&&&&& \\[-0.1cm]
\hline
\hline
\end{tabular*}
\begin{minipage}{\linewidth}
{\footnotesize 
\begin{itemize}
\item[$^\ddag$]The rating takes into account that the action is not fully $\cO(a)$-improved by requiring an additional lattice spacing.\\[-5mm]\item[$^\#$] Assumed that $Z_T^{n.s.}=Z_T^{s}$. \\[-5mm] \item[$^\&$] Disconnected terms omitted.
\end{itemize}
}
\end{minipage}
\caption{Overview of results for $g^q_T$.\label{tab:gt-singlet} }
\end{center}
\end{table}

%% file: NME/Tables/table_sigmaud_s_direct.tex
\section*{}
\begin{table}[t!]
\begin{center}
\mbox{} \\[3.0cm]
\footnotesize
\begin{tabular*}{\textwidth}[l]{l @{\extracolsep{\fill}} r l l l l l l l l l}
Collaboration & Ref. & $\Nf$ & 
\hspace{0.15cm}\begin{rotate}{60}{publication status}\end{rotate}\hspace{-0.15cm} &
\hspace{0.15cm}\begin{rotate}{60}{continuum extrapolation}\end{rotate}\hspace{-0.15cm} &
\hspace{0.15cm}\begin{rotate}{60}{chiral extrapolation}\end{rotate}\hspace{-0.15cm}&
\hspace{0.15cm}\begin{rotate}{60}{finite volume}\end{rotate}\hspace{-0.15cm}&
\hspace{0.15cm}\begin{rotate}{60}{renormalization}\end{rotate}\hspace{-0.15cm}  &
\hspace{0.15cm}\begin{rotate}{60}{excited states}\end{rotate}\hspace{-0.15cm}&
$\sigma_{\pi N}$~[MeV]  &
$\sigma_s$~[MeV]\\
&&&&&&&&& \\[-0.1cm]
\hline
\hline
&&&&&&&& \\[-0.1cm]
%
PNDME 21 & \cite{Gupta:2021ahb} & 2+1+1 & \gA & \soso$^\ddag$ & \good & \good & $^a$/$-$ & \soso & 59.6(7.4)   &  $-$ \\[0.5ex]
ETM 19 & \cite{Alexandrou:2019brg} & 2+1+1 & \gA & \bad & \soso & \good & na/na & \soso & 41.6(3.8) & 45.6(6.2) \\[0.5ex]
\\[-0.1ex]\hline\\[0.2ex]
Mainz 23& \cite{Agadjanov:2023efe} & 2+1 & \gA & \good$^{b}$ & \good & \good & \good/\good & \soso & 43.7(3.6) & 28.6(9.3) \\[0.5ex]
JLQCD 18 & \cite{Yamanaka:2018uud} & 2+1 & \gA & \bad & \soso & \soso & na/na & \soso & 26(3)(5)(2) & 17(18)(9) \\[0.5ex]
$\chi$QCD 15A & \cite{Yang:2015uis} & 2+1 & \gA & \soso & \good & \good & na/na & \soso & 45.9(7.4)(2.8)$^\$$ & 40.2(11.7)(3.5)$^\$$ \\[0.5ex]
&&&&&&&& \\[-0.1cm]
\hline
\hline
&&&&&&&& \\[-0.1cm]
MILC 12C & \cite{Freeman:2012ry} & 2+1+1 & \gA & \good & \good & \good & $-$/\soso & \soso & $-$ & 0.44(8)(5)$\times m_s$$^{\P\S}$ \\[0.5ex]
\\[-0.1ex]\hline\\[0.2ex]
MILC 12C & \cite{Freeman:2012ry} & 2+1 & \gA & \good & \soso & \good & $-$/\soso & \soso & $-$ &0.637(55)(74)$\times m_s$$^{\P\S}$ \\[0.5ex]
MILC 09D & \cite{Toussaint:2009pz} & 2+1 & \gA & \good & \soso & \good & $-$/na & \soso & $-$ &59(6)(8)$^\S$ \\[0.5ex]
&&&&&&&& \\[-0.1cm]
\hline
\hline
\end{tabular*}
\begin{minipage}{\linewidth}
{\footnotesize The renormalization criteria is given for $\sigma_{\pi N}$~(first) and $\sigma_s$~(second). The
  label ’na’ indicates that no renormalization is required.
\begin{itemize}
\item[$a$] Mixing between quark flavours is found to be small and is neglected. \\[-5mm]
\item[$^\ddag$]The rating takes into account that the action is not fully $\cO(a)$-improved by requiring an additional lattice spacing. \\[-5mm]
\item[$b$] The rating takes into account that the scalar current is not fully $\cO(a)$-improved by requiring an additional lattice spacing. The gluonic operator that appears in the $\cO(a)$ improvement for Wilson fermions is not implemented. The effect of this term is expected to be small. \\[-5mm]
\item[$^\$$] For this partially quenched analysis the criteria are applied to the unitary points. \\[-5mm] 
\item[$^\S$] This study employs a hybrid method, see Ref.~\cite{Toussaint:2009pz}. \\[-5mm] 
\item[$^\P$] The matrix element $\langle N|\bar{s}s|N\rangle$ at the scale $\mu=2$~GeV in the $\msbar$ scheme is computed.
\end{itemize}
}
\end{minipage}
\caption{Overview of results for $\sigma_{\pi N}$ and $\sigma_s$ from the direct approach~(above) and $\sigma_s$ from the hybrid approach~(below).  \label{tab:gs-ud-s-direct}}
\end{center}
\end{table}

%% file: NME/Tables/table_sigmaud_fh.tex
\begin{table}[t!]
\begin{center}
\mbox{} \\[3.0cm]
\footnotesize
\begin{tabular*}{\textwidth}[l]{l @{\extracolsep{\fill}} r l l l l l l l }
Collaboration & Ref. & $\Nf$ & 
\hspace{0.15cm}\begin{rotate}{60}{publication status}\end{rotate}\hspace{-0.15cm} &
\hspace{0.15cm}\begin{rotate}{60}{continuum extrapolation}\end{rotate}\hspace{-0.15cm} &
\hspace{0.15cm}\begin{rotate}{60}{chiral extrapolation}\end{rotate}\hspace{-0.15cm}&
\hspace{0.15cm}\begin{rotate}{60}{finite volume}\end{rotate}\hspace{-0.15cm}&
$\sigma_{\pi N}$~[MeV] & $\sigma_s$~[MeV] \\
 & & & & & & & &\\[-0.1cm]
\hline
\hline
 & & & & & & & &\\[-0.1cm]
BMW 20A & \cite{Borsanyi:2020bpd} & 1+1+1+1 & \oP & \good$^\ddag$ & \good & \good & 0.0398(32)(44)$\times m_N$$^\dagger$ & 0.0577(46)(33)$\times m_N$$^\dagger$ \\[0.5ex]
ETM 14A & \cite{Alexandrou:2014sha} & 2+1+1 & \gA & \good & \soso & \soso & 64.9(1.5)(13.2)$^\triangle$ & $-$ \\[0.5ex]
\\[-0.1ex]\hline\\[0.2ex]
RQCD 22& \cite{RQCD:2022xux} & 2+1 & \gA & \good & \good & \good & 43.9(4.7) & 16$\left(^{+58}_{-68}\right)$ \\[0.5ex]
BMW 15 & \cite{Durr:2015dna} & 2+1 & \gA & \good$^\ddag$ & \good & \good & 38(3)(3) & 105(41)(37) \\[0.5ex]
Junnarkar 13 & \cite{Junnarkar:2013ac} & 2+1 & \gA & \soso & \soso & \soso & $-$ & 48(10)(15) \\[0.5ex]
BMW 11A & \cite{Durr:2011mp} & 2+1 & \gA & \soso$^\ddag$ & \good & \soso & 39(4)$\left(^{+18}_{-7}\right)$ & 67(27)$\left(^{+55}_{-47}\right)$ \\[0.5ex]
&&&&&&&& \\[-0.1cm]
\hline
\hline
\end{tabular*}
\begin{minipage}{\linewidth}
{\footnotesize 
\begin{itemize}
 \item[$^\triangle$]  Two results for $\sigma_{\pi N}$ are quoted arising from different fit ans\"atze to the nucleon mass. The systematic error  is the same as in  Ref.~\cite{Alexandrou:2017xwd} for a combined $\Nf=2$ and $\Nf=2+1+1$ analysis~\cite{Kallidonis:pc2018}.
\\[-5mm] \item[$^\ddag$]The rating takes into account that the action is not fully O(a) improved by requiring an additional lattice spacing.
\\[-5mm] \item[$^\dagger$] The quark fractions $f_{T_{ud}}=f_{T_{u}}+f_{T_{d}}=\sigma_{\pi N}/m_N$ and $f_{T_s}=\sigma_s/m_N$ are computed.
\end{itemize}
}
\end{minipage}
\caption{Overview of results for $\sigma_{\pi N}$ and $\sigma_s$ from the Feynman-Hellmann approach.  \label{tab:gs-singlet-fh}}
\end{center}
\end{table}

%% file: NME/Tables/table_x.tex
\begin{table}[t!]
\begin{center}
\mbox{} \\[3.0cm]
\footnotesize
\begin{tabular*}{\textwidth}[l]{l @{\extracolsep{\fill}} r l l l l l l l l l l}
Collaboration & Ref. & $\Nf$ & 
\hspace{0.15cm}\begin{rotate}{60}{publication status}\end{rotate}\hspace{-0.15cm} &
\hspace{0.15cm}\begin{rotate}{60}{continuum extrapolation}\end{rotate}\hspace{-0.15cm} &
\hspace{0.15cm}\begin{rotate}{60}{chiral extrapolation}\end{rotate}\hspace{-0.15cm}&
\hspace{0.15cm}\begin{rotate}{60}{finite volume}\end{rotate}\hspace{-0.15cm}&
\hspace{0.15cm}\begin{rotate}{60}{renormalization}\end{rotate}\hspace{-0.15cm}  &
\hspace{0.15cm}\begin{rotate}{60}{excited states}\end{rotate}\hspace{-0.15cm}  &
$\langle x\rangle_{u-d}$ &$\langle x\rangle_{\Delta u-\Delta d}$ \\
&&&&&&&&&& \\[-0.1cm]
\hline
\hline
&&&&&&&&&& \\[-0.1cm]
ETM 22       & \cite{Alexandrou:2022dtc} & 2+1+1 &\gA & \good & \good & \good & \good &\soso  &0.126(32) & \\[0.5ex]
PNDME 20A     & \cite{Mondal:2020cmt} & 2+1+1 & \gA & \good$^\ddag$ & \good & \good & \good & \soso & 0.173(14)(07) &0.213(15)(22)\\[0.5ex]
ETM 20C       & \cite{Alexandrou:2020sml} & 2+1+1 &\gA & \bad & \soso & \good & \good &\soso  &0.171(18) & \\[0.5ex]
ETM 19A       & \cite{Alexandrou:2019ali} & 2+1+1 &\gA & \bad & \soso & \good &\good  &\soso  &0.178(16)&0.193(18)  \\[0.5ex]
\\[-0.1ex]\hline\\[0.2ex]
Mainz 24     &\cite{Djukanovic:2024krw}  & 2+1   & \gA & \good$^\ddag$ & \good & \good & \good & \soso &0.153(15)(10)& 0.207(15)(06)\\[0.5ex]
LHPC  24     & \cite{Rodekamp:2023wpe}   & 2+1 & \gA & \bad$^\ddag$ & \good & \good & \good & \soso &
0.200(17) & 0.213(16)  \\[0.5ex]
NME 21A       & \cite{Mondal:2021oot} & 2+1 & \rC   & \good$^\ddag$ & \good & \good & \good &\soso &0.156(12)(20)  &0.185(12)(20) \\[0.5ex]
NME 20       & \cite{Mondal:2020ela} & 2+1 & \gA   & \soso$^\ddag$& \good & \good & \good & \soso & 0.155(17)(20) & 0.183(14)(20)  \\[0.5ex]
Mainz 19        &\cite{Harris:2019bih}  & 2+1  & \gA & \good$^\ddag$ & \soso & \good & \good & \soso  &0.180(25)$\left(^{+14}_{-6}\right)$&0.221(25)$\left(^{+10}_{-0}\right)$   \\[0.5ex]
$\chi$QCD 18A & \cite{Yang:2018nqn}   & 2+1   & \gA & \soso & \good & \good & \good & \soso & 0.151(28)(29) &  \\[0.5ex]
LHPC 12A        & \cite{Green:2012ud}   & 2+1  & \gA & \bad$^\ddag$ & \good & \good & \good & \soso & 0.140(21) &  \\[0.5ex]
LHPC 10         & \cite{Bratt:2010jn}   & 2+1 & \gA & \bad$^\ddag$ & \soso & \bad & \good & \bad   & 0.1758(20)&0.1972(55) \\[0.5ex]
RBC/UKQCD 10D   & \cite{Aoki:2010xg}    & 2+1 & \gA & \bad & \bad & \soso & \good & \bad  & 0.140--0.237 & 0.180--0.279 \\[0.5ex]
\\[-0.1ex]\hline\\[0.2ex]
RQCD 18      & \cite{Bali:2018zgl}   & 2   & \gA & \soso$^\ddag$ & \good & \good & \good &\bad  & 0.195(7)(15) &0.271(14)(16) \\[0.5ex]
ETM 17C         & \cite{Alexandrou:2017oeh} & 2&  \gA & \bad & \soso & \soso & \good & \soso & 0.194(9)(11) &  \\[0.5ex]
ETM 15D         & \cite{Abdel-Rehim:2015owa} & 2 & \gA & \bad & \soso & \soso & \good & \soso  & 0.208(24) &0.229(30) \\[0.5ex]
RQCD 14A      & \cite{Bali:2014gha}        & 2 & \gA & \soso$^\ddag$ & \good & \good & \good & \bad & 0.217(9) & \\[0.5ex]
&&&&&&&&&& \\[-0.1cm]
\hline
\hline
\end{tabular*}
\begin{minipage}{\linewidth}
{\footnotesize 
\begin{itemize}
\item[$^\ddag$]The rating takes into account that the moments are not fully $\cO(a)$-improved by requiring an additional lattice spacing.
\end{itemize}
}
\end{minipage}
\caption{Overview of results for $\langle x\rangle_{u-d}$ and $\langle x\rangle_{\Delta u-\Delta d}$.  The $\Nf=2$ results and publications prior to 2014   are included as this is the first review of these quantities. 
\label{tab:moments1}}
\end{center}
\end{table}

\begin{table}[t!]
\begin{center}
\mbox{} \\[3.0cm]
\footnotesize
\begin{tabular*}{\textwidth}[l]{l @{\extracolsep{\fill}} r l l l l l l l l l}
Collaboration & Ref. & $\Nf$ & 
\hspace{0.15cm}\begin{rotate}{60}{publication status}\end{rotate}\hspace{-0.15cm} &
\hspace{0.15cm}\begin{rotate}{60}{continuum extrapolation}\end{rotate}\hspace{-0.15cm} &
\hspace{0.15cm}\begin{rotate}{60}{chiral extrapolation}\end{rotate}\hspace{-0.15cm}&
\hspace{0.15cm}\begin{rotate}{60}{finite volume}\end{rotate}\hspace{-0.15cm}&
\hspace{0.15cm}\begin{rotate}{60}{renormalization}\end{rotate}\hspace{-0.15cm}  &
\hspace{0.15cm}\begin{rotate}{60}{excited states}\end{rotate}\hspace{-0.15cm}  &
$\langle x\rangle_{\delta u-\delta d}$  \\
&&&&&&&&& \\[-0.1cm]
\hline
\hline
&&&&&&&&& \\[-0.1cm]
ETM 22       & \cite{Alexandrou:2022dtc} & 2+1+1 &\gA & \good & \good & \good & \good &\soso  & 0.168(44)\\[0.5ex]
PNDME 20A     & \cite{Mondal:2020cmt} & 2+1+1 & \gA & \good$^\ddag$ & \good & \good & \good & \soso &0.208(19)(24)\\[0.5ex]
ETM 19A       & \cite{Alexandrou:2019ali} & 2+1+1 &\gA & \bad & \soso & \good &\good  &\soso   & 0.204(23)   \\[0.5ex]
\\[-0.1ex]\hline\\[0.2ex]
Mainz 24     &\cite{Djukanovic:2024krw}  & 2+1   & \gA & \good$^\ddag$ & \good & \good & \good & \soso & 0.195(17)(15)\\[0.5ex]
LHPC  24     & \cite{Rodekamp:2023wpe}   & 2+1 & \gA & \bad$^\ddag$ & \good & \good & \good & \soso  & 0.219(21) \\[0.5ex]
NME 21A       & \cite{Mondal:2021oot} & 2+1 & \rC   & \good$^\ddag$ & \good & \good & \good &\soso  &0.209(15)(20) \\[0.5ex]
NME 20       & \cite{Mondal:2020ela} & 2+1 & \gA   & \soso$^\ddag$& \good & \good & \good & \soso & 0.220(18)(20) \\[0.5ex]
Mainz 19        &\cite{Harris:2019bih}  & 2+1  & \gA & \good$^\ddag$ & \soso & \good & \good & \soso  &0.212(32)($^{+20}_{-10}$) \\[0.5ex]
\\[-0.1ex]\hline\\[0.2ex]
RQCD 18      & \cite{Bali:2018zgl}   & 2   & \gA & \soso$^\ddag$ & \good & \good & \good &\bad  &0.266(8)(4) \\[0.5ex]
ETM 15D         & \cite{Abdel-Rehim:2015owa} & 2 & \gA & \bad & \soso & \soso & \good & \soso   &0.306(29) \\[0.5ex]
&&&&&&&&& \\[-0.1cm]
\hline
\hline
\end{tabular*}
\begin{minipage}{\linewidth}
{\footnotesize 
\begin{itemize}
\item[$^\ddag$]The rating takes into account that the moments are not fully $\cO(a)$-improved by requiring an additional lattice spacing.
\end{itemize}
}
\end{minipage}
\caption{Overview of results for $\langle x\rangle_{\delta u-\delta d}$.  The $\Nf=2$ results and publications prior to 2014
  are included as this is the first review of these quantities. 
\label{tab:moments2}}
\end{center}
\end{table}

%% file: ScaleSetting/scalesetting.tex

\section{Scale setting\label{sec:scalesetting}}
Authors: R.~Sommer, N.~Tantalo, U.~Wenger\\

\def\qedl{$\rm QED_L$}
\def\U{\hfill {\cred Urs}}
\def\R{\hfill{\cred Rainer}}
\def\N{\hfill{\cred Nazario}}
\newcommand{\makered}{\textcolor{red}}  
\newcommand{\todors}[1]{{\color{red}{\bf todo} #1 {\bf endtodo}}}

\ifx\versionforCollabs\undefine

\renewcommand{\cred}{}

Matching QCD to nature requires fixing the quark masses
and matching an overall scale to experiment. That overall
energy scale $\scal$ may be taken, for example, as the nucleon mass. 
This process is referred to as scale setting. 

\subsection{Impact}

The scale-setting procedure, described in some detail below, is a rather technical step necessary to obtain predictions from QCD. What may easily be overlooked is that the exact predictions obtained from the theory, including many in this review, may depend rather sensitively on the scale. 

As long as the theory is incomplete, e.g., because we have predictions from $\Nf=2+1$ QCD, results will depend on which physics scale is used. Whenever a theory scale (see Sec.~\ref{s:theoryscales}) is used, it matters which value one imposes. Thus, to know whether computations of a particular quantity agree or not, one should check which (value for a) scale was used. 

The sensitivity of predictions to the scale varies with the observable. For example, the $\Lambda$ parameter of the theory has a linear dependence,
\begin{equation}
 \frac{\delta\Lambda}{\Lambda}	\approx \frac{\delta\scal}{\scal}\,,
\end{equation}
because $\Lambda$ has mass dimension one and other hidden dependencies on the scale are (usually) suppressed. Let us preview the results. The present precision  on the most popular theory scale, $w_0$ in \eq{eq:w0_2p1p1_all} is about 0.4\% and for $\sqrt{t_0}$ it is 0.6\%. On the $\Lambda$ parameter it is about 3\%. Thus, we would think that the scale uncertainty is irrelevant. However, in Sec.~\ref{s:Scalconcl} we will discuss that differences between $\Nf=2+1$ and 2+1+1 numbers for $\sqrt{t_0}$ are at around 2\%, which {\em does matter}.

Also, light-quark masses  have an approximatively  linear dependence on the scale (roughly speaking one determines, e.g., $m_{ud} = \frac1{\scal}\times[M_\pi^2]_\mathrm{exp}\times[\frac{m_{ud}\, \scal}{M_\pi^2} ]_\mathrm{lat}$) and scale uncertainties may play an important r\^ole in the discussion of agreement vs.~disagreement of computations within their error budget.

The list of quantities where scale setting is very important may be continued;  we just want to mention an observable very much discussed at present, the hadronic vacuum-polarisation contribution to the anomalous magnetic moment of the muon \cite{Aoyama:2020ynm}. It is easily seen that the dependence on the scale is about quadratic in that case \cite{DellaMorte:2017dyu},
\begin{equation}
 \frac{\delta a_\mu^\mathrm{HVP}}{a_\mu^\mathrm{HVP}}	\approx 2 \frac{\delta\scal}{\scal}\,.
\end{equation}
This fact means that scale setting has to be precise at the few per-mille precision for the $a_\mu^\mathrm{HVP}$ lattice determination 
to be relevant in the comparison with experiment.

\subsection{Scale setting as part of hadronic renormalization schemes}
\label{sec:QCDhadRen}
We consider QCD with $\Nf$ quarks and without a $\theta$-parameter. This theory is  completely defined by its coupling constant as well as $\Nf$ quark masses. After these parameters are specified all other properties of the theory are predictions. Coupling and  quark masses depend on a renormalization scale $\mu$ as well as on a renormalization scheme. The most popular scheme in the framework of perturbative computations is the $\msbar$ scheme, but one may also define nonperturbative renormalization schemes, see Secs.~\ref{sec:qmass} and \ref{sec:alpha_s}.

In principle, a lattice computation may, therefore, use these $\Nf+1$ parameters as input together with the renormalization scale $\mu$  to fix the bare quark masses and coupling of the discretized Lagrangian, perform continuum and infinite-volume limit and obtain desired results, e.g., for decay rates.\footnote{At first sight this seems like too many inputs,  but note that it is the scale $\mu$, at which $\alpha(\mu)$ has a  particular value, which is the input. The coupling $\alpha$ by itself can have any (small) value as it runs.} However, there are various reasons why this strategy is inefficient. The most relevant one is that unless one uses lattice gauge theory to compute them, coupling and quark masses cannot be obtained from experiments without invoking perturbation theory and thus necessarily truncation errors. Moreover, these parameters are naturally short-distance quantities, since this is where perturbation theory applies. Lattice QCD on the other hand is most effective at long distances, where the lattice spacing plays a minor role. Therefore, it is  more natural to proceed differently.

Namely, we may fix $\Nf+1$ nonperturbative, long-distance observables to have the values found in nature. An obvious choice are $\Nf+1$ hadron masses that are stable in the absence of weak interactions. This hadronic renormalization scheme is defined by
\begin{equation}
	\frac{M_i(g_0,\{a m_\mathrm{0,j}\})}{
	M_1(g_0,\{a m_\mathrm{0,j}\})} 
	=\frac{M_i^\mathrm{exp}}{M_1^\mathrm{exp}}\,, \quad 
	i=2\ldots\Nf+1\,,
	\quad
	j=1\ldots\Nf\,.
	\label{e:hadscheme}
\end{equation}
Here, $M_i$ are the chosen hadron masses, $g_0$ is the bare coupling, and $am_{0,j}$ are the bare quark masses in lattice units. The ratio $M_i/M_1$ is, precisely speaking, defined through the hadron masses in lattice units, but in infinite volume. In QCD (without QED), all particles are massive. Therefore, the infinite-volume limit of the properties of stable particles is approached with exponentially small corrections, which are assumed to be estimated reliably. The power-like finite-volume corrections in QCD$+$QED are discussed in Sec.~\ref{sec:isobreak}.  For fixed $g_0$,
Eq.~\eqref{e:hadscheme} needs to be solved for the bare quark masses,
\begin{equation}
	a m_\mathrm{0,j} = \mu_j(g_0)\,. \label{e:lcp}
\end{equation}
The functions $\mu_j$ define a line in the bare parameter space, called the line of constant physics. Its dependence on the set of masses $\{M_i\}$ is suppressed. The continuum limit is obtained as $g_0\to0$ with the lattice
spacing shrinking 
roughly as $aM_1 \sim \rme^{-1/(2b_0 g_0^2)}$. More precisely, consider 
observables $\cal O$ with mass dimension $d_{\cal O}$. One defines their
dimensionless ratio
\begin{equation}
 	\hat {\cal O}(aM_1)=\left. 
 	\frac{{\cal O}}{M_1^{d_{\cal O}}}\right|_{am_{0,j}=\mu_j(g_0)}\,,
\end{equation}
and obtains the continuum prediction as
\begin{equation}
  {\cal O}^\mathrm{cont} = \left(M_1^{\mathrm{exp}}\right)^{d_{\cal
      O}} \, \lim_{aM_1\to 0} \hat {\cal O}(aM_1)
\end{equation}
which explains why the determination and use of $aM_1$ is referred to as scale setting.

Equation (\ref{e:lcp}) has to be obtained from numerical results. Therefore, it 
is easiest and most transparent if the $i$-th mass ratio depends 
predominantly on the $i$-th quark mass.
Remaining for a while in the isospin-symmetric theory with $m_{0,1} = m_{0,2}$
(we enumerate the quark masses in the order up, down, strange, charm, bottom and ignore the top quark),
we have natural candidates for the numerators as the pseudoscalar masses in the
associated flavour sectors, i.e., $\pi,\,K,\,D,\,B$. The desired strong
dependence on light- (strange-)quark masses of $\pi$- ($K$-)meson masses derives
from their pseudo-Goldstone nature of the approximate SU$(3)_\mathrm{L}\times$SU$(3)_\mathrm{R}$ symmetry of the massless QCD Lagrangian,  which
predicts that $M_\pi^2$ is roughly proportional to the light-quark mass and $M_{\mathrm K}^2$  to  the sum of light- and strange-quark masses. For $D$ and $B$ mesons approximate heavy-quark symmetry
predicts $M_{D}$ and $M_{B}$ to be proportional to charm- and bottom-quark masses.
Also other heavy-light bound states have this
property. There is another  important feature, which singles out pseudoscalar masses. Because they are the lightest particles with the given flavour quantum numbers, their correlation functions have the least signal/noise problem in the
Monte Carlo evaluation of the path integral \cite{Lepage:1989hd,Luscher:2010ae}.

Still restricting ourselves to isospin-symmetric QCD (isoQCD, see Sec.~\ref{sec:ibscheme}), we thus take it for granted that the choice $M_i,\, i\geq 2$ is easy, and we do not need to discuss it in detail: the pseudoscalar meson masses are very good choices, and some variations for 
heavy quarks may provide further improvements.

The choice of $M_1$ is more difficult. 
From the point of view of physics, a natural choice is the nucleon mass, $M_1=M_{\textrm{nucl}}$. Unfortunately it has a rather bad signal/noise problem 
when quark masses are close to their physical values. 
The ratio of signal to noise of the correlation function
at time $x_0$ from $N$ measurements behaves as \cite{Lepage:1989hd}
\begin{eqnarray}
	 R_{S/N}^\mathrm{nucl} &\simas{x_0\; \mathrm{large}}& \sqrt{N}\, \exp(-(M_\mathrm{nucl}-\frac32\mpi)\,x_0) \approx \sqrt{N}\,\exp(-x_0 / 0.27\,\fm) \,,
\end{eqnarray}
where the numerical value of $0.27\,$fm uses the experimental
masses. The behaviour in practice, but at still favourably large quark
masses, is illustrated in Fig.~\ref{f:plateaux}.
\begin{figure}[ht!]
\centering
   \includegraphics[width=0.8\textwidth]{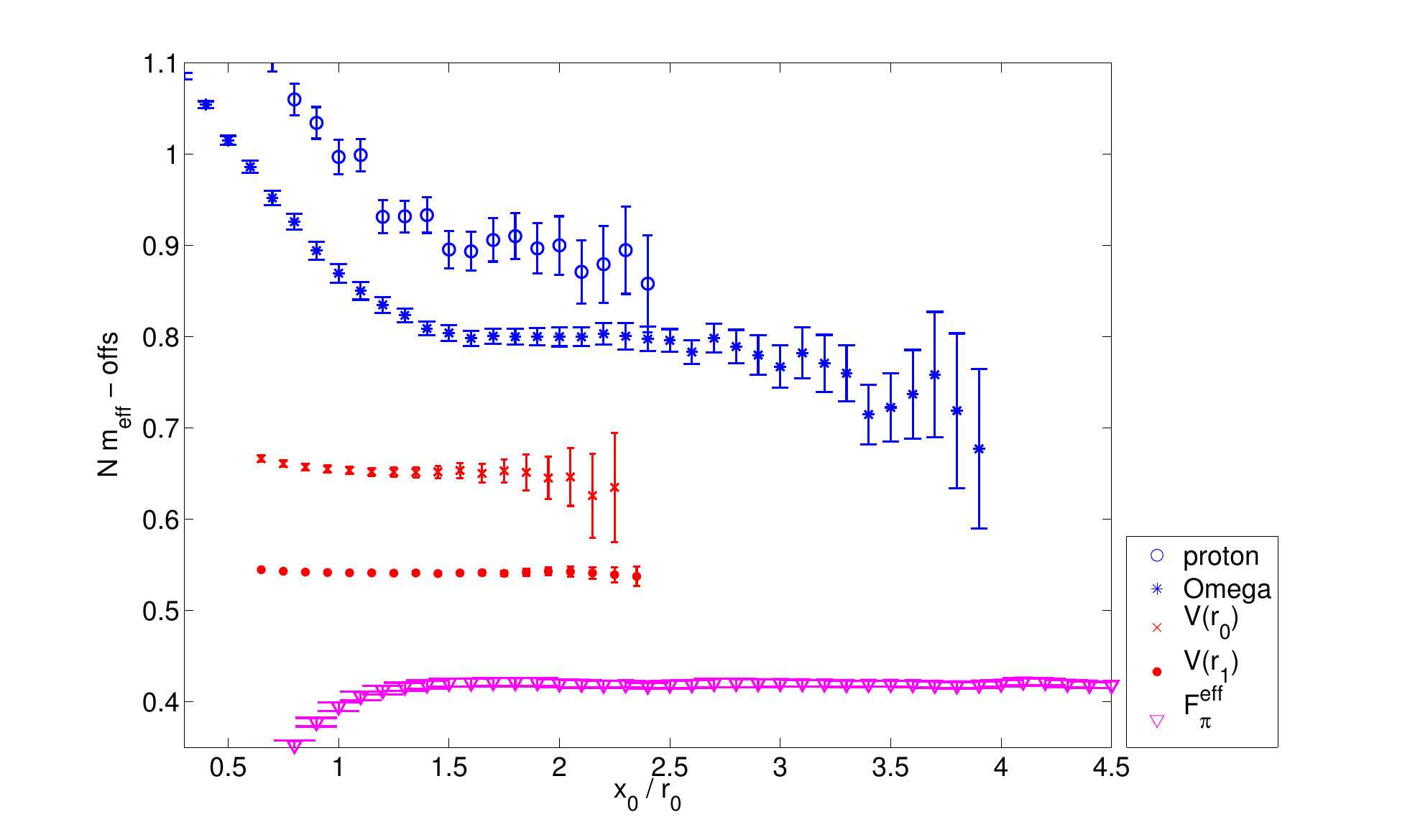}
   \vspace*{-2mm}
\caption{Effective masses for 
$M_\mathrm{proton}$ \cite{Jager:2013kha},
$M_\Omega$ \cite{Capitani:2011fg},
$V(\approx r_0)$, $V(\approx r_1)$~\cite{Fritzsch:2012wq}
and $\fpi$~\cite{Lottini:2013rfa} on $\Nf=2$ CLS ensemble N6 with 
$a=0.045\,\fm, M_\pi=340\,\MeV$ on a $48^3\times 96$ lattice \cite{Lottini:2013rfa}. 
All effective ``masses'' have been scaled 
such that the errors in the graph reflect
directly the errors of the determined scales. They are shifted vertically
by arbitrary amounts. Figure from Ref.~\cite{Sommer:2014mea}. Note that this example is at still favourably large quark masses. The situation for
$M_\mathrm{proton}$ becomes worse closer to the physical point, but may be changed by algorithmic improvements.
\label{f:plateaux}
}
\end{figure}
Because
this property leads to large statistical errors and it is further difficult to control excited-state contaminations when statistical errors are large, it is useful to search for alternative 
physics scales. The community has gone this way, and we discuss some of them below. For illustration, here we just  give one example:
the decay constants of leptonic $\pi$ or $K$ decays have mass dimension one 
and can directly replace $M_1$ above. Figure \ref{f:plateaux} demonstrates their long and precise plateaux as a function of the Euclidean time. Advantages and disadvantages of this choice
and others are discussed more systematically in Sec.~\ref{s:physscal}. 

\subsubsection{Theory scales}
Since the signal/noise problem of physics scales is rather severe,
they were already replaced by theory scales  in the very first days
of lattice QCD.  These scales cannot be determined from experiment alone. Rather, their values have to be computed by lattice QCD using a physics scale as input.

Creutz already used the string tension in his seminal paper
on SU(2) Yang Mills theory \cite{Creutz:1980zw}, because it is far easier to determine
than glueball masses. A further step was made by the potential scale
 $r_0$, defined in terms of the static force $F(r)$ as \cite{Sommer:1993ce}
 \begin{equation}
 	r_0^2 F(r_0) =1.65\,.
 \end{equation}
Even though $r_0$ can vaguely be related to the phenomenology of
charmonium and bottomonium states, its precise definition is in terms
of $F(r)$ which can be obtained accurately 
from Monte Carlo lattice computations with (improvable) control
over the uncertainties, but not from experiment. In that sense, it is a prototype of a theory scale. \\[1ex]
Useful properties of a good theory scale are high statistical precision, easy to control systematics, e.g., weak volume dependence, quark-mass dependence only due to the fermion determinant, and low numerical cost for its evaluation.
These properties are realized to varying degrees by the different theory scales 
covered in this section and, in this respect, they are much preferred
compared to physics scales. Consequently, the physics scale $M_1$ has often been replaced by a theory scale as, e.g., 
$\scal=r_0^{-1}$ in the form
\begin{equation}
  {\cal O}^\mathrm{cont} = \left(\scal^{\mathrm{phys}}\right)^{d_{\cal O}} \,
  \lim_{a\scal\to0} \hat {\cal O}_{\scal}(a\scal)\; \text{ with }\; \hat {\cal O}_{\scal}(a\scal)= \left[\scal^{-d_{\cal O}}\, \cal O\right]_{am_{0,j}=\mu_j(g_0)}\,,
\end{equation}
and
\begin{equation}
  {\scal}^\mathrm{phys} = \left(M_1^{\mathrm{exp}}\right) \, \lim_{aM_1\to0} \hat {\scal}_{M_1}(aM_1)\,. \label{e:scalesettbasic}
\end{equation}

In this section, we review the determination of numerical results for the values
of various theory scales in physical units,
\eq{e:scalesettbasic}. The main difficulty is that a physics scale $M_1$ has to
be determined first in order to connect to nature and, in particular, that the continuum limit of the theory scale in units of the physics scale has to be taken.

\subsubsection{Isospin breaking
} 
\label{sec:isobreak}

For simplicity and because it is a very good approximation, 
we have assumed above that all other interactions except for QCD can be ignored when hadron masses and many other properties of hadrons are considered. This is a
natural point of view because QCD is a renormalizable field theory and thus provides unique results. 

However, we must be aware that while it is true that the predictions
(e.g., for hadron masses $M_i, \, i>\Nf+1$) are unique once
\eq{e:hadscheme} is specified, they will change when we change the
inputs $M_i^\mathrm{exp}$. These ambiguities are due to the neglected
electroweak and gravitational interactions, namely because QCD is only an
approximate---even if precise---theory of hadrons. At the sub-percent level, QED effects and isospin violations due to $m_{u} \ne m_{d}$ must be included. At that level one has 
a very precise description of nature, where weak decays 
or weak effects, in general, can be included perturbatively and systematically
in an effective-field-theory description through the 
weak-effective-interaction Hamiltonian, while gravity may be ignored.

Scale setting is then part of
the renormalization of QCD$+$QED, and in principle it is quite analogous to the previous discussion.  
Triviality of QED does not play a r\^ole at small enough $\alpha$: we may think of replacing the continuum limit $a\to0$ by a limit $a\to a_\mathrm{w}$ with $a_\mathrm{w}$ nonzero but very far below all physical QCD$+$QED scales treated.
The definition and implementation of a hadronic renormalization scheme of QCD+QED defined on the lattice is discussed in Sec.~\ref{sec:ibscheme}. The electric charge appears as a new parameter and is conveniently fixed in the Thomson limit. Care needs to be taken in the separate definition of QED effects
 and strong isospin-breaking effects due to the up/down quark-mass difference.
Here, we repeat Eq.~\eqref{eq:ibdec}, 
\begin{equation}
  X^{\phi}=\bar{X}+X_{\gamma}+X_{\mathrm{SU}(2)}\,, 
  \label{eq:ibdecss}
\end{equation}
and again emphasize that the split of physical observables 
$X^{\phi}$ into their isoQCD part, $\bar{X}$,
the QED contributions,  $X_{\gamma}$, and 
the strong IB effects, $X_{\mathrm{SU}(2)}$, 
is scheme dependent. In order to hopefully avoid 
confusion and to make it possible to average results
also when they have a precision where the small IB-breaking effects matter, a particular scheme 
has been defined in Sec.~\ref{sec:ibscheme}.
For quantities that enter in the averages, the schemes used in the computations are listed in Tab.~\ref{tab_schemes}. 
In this way, we can, to some degree,  
judge whether differences of results may also be due to the scheme used.

\begin{table}[h!]
\footnotesize 
\centering
\begin{tabular*}{0.8\textwidth}[l]{l@{\extracolsep{\fill}}rl@{\hspace{1mm}}l@{\hspace{1mm}}l@{\hspace{1mm}}l@{\hspace{1mm}}}

Collaboration & Ref. & $\Nf$ & $M_K$ [MeV] & scale & scale [MeV]\\
\hline\hline\\[-2ex]

ETM 21 & \cite{Alexandrou:2021bfr} & 2+1+1 & 494.2 & $f_\pi$ & 130.4 \\

CalLat 20A& \cite{Miller:2020evg} & 2+1+1 & 494.2 & $M_\Omega$ & 1672.5 \\

MILC 15 &\cite{Bazavov:2015yea} & 2+1+1 & 494.5 & ${F_{p4s}(f_\pi)}$ & 153.90(9)($^{+21}_{-28}$) \\

HPQCD 13A & \cite{Dowdall:2013rya} & 2+1+1 & 494.6 & $f_\pi$ & 130.4 \\
\hline
Hudspith 24 & \cite{Hudspith:2024kzk} & 2+1 & 494.2 & $M_\Omega$ & 1672.5
\\
RQCD 22 & \cite{RQCD:2022xux} & 2+1 & 494.2 & $M_\Xi$ & 1316.9 \\

CLS 21 & \cite{Strassberger:2021tsu} & 2+1 & 497.6 & $\frac13(f_\pi +2f_K)$ & 148.3 \\

CLS 16 & \cite{Bruno:2016plf} & 2+1 & 494.2 & $\frac13(f_\pi +2f_K)$ & 147.6 \\

RBC/UKQCD 14B & \cite{Blum:2014tka} & 2+1 & 495.7 & $M_\Omega$ & 1672.5 \\

HotQCD 14 & \cite{Bazavov:2014pvz} & 2+1 & n/a$\phantom{.}^\#$ & ${r_1(f_\pi)}$ & 0.3106 fm\\

BMW 12A & \cite{Borsanyi:2012zs} & 2+1 & 494.2 & $M_\Omega$ & 1672.5 \\
\hline\\[-2ex]
Edinburgh 
consensus & & & 494.6 & $f_\pi$ & 130.5\\[1ex]
\hline\hline\\
\end{tabular*}\\[-0.2cm]
\begin{minipage}{0.8\linewidth}
  {\footnotesize
    ${}^\#$ The scheme uses 
    $M_{\eta_{s\bar s}}\approx 695~\mathrm{MeV}$ instead of fixing $M_\mathrm{K}$.
  }
\end{minipage}
\caption{isoQCD schemes used in different computations as well as the Edinburgh consensus (see Sec.~\ref{sec:ibscheme}).
We do not list the choice for $M_\pi$. It is $M_\pi=135.0~\mev$ throughout. As all quantities 
refer to the  light sector of QCD only, charm quarks 
only enter through sea-quark effects. We therefore do not list which quantity is used to fix the charm-quark mass at the present stage. 
\newline
}
\label{tab_schemes}
\end{table}

As a matter of fact, many existing lattice calculations have been performed in the isospin-symmetric limit, but not all the results considered in this review correspond to the very same definition of QCD. 
The different choices of experimental inputs are perfectly legitimate if QED radiative corrections are neglected, but
in principle, predictions of isoQCD 
do depend on these choices, and it is not meaningful to average numbers obtained with different inputs. However,
at the present level of precision the sub-percent differences in the
inputs are most likely not relevant, and we will average and compare isoQCD results irrespective of these differences. The issue will become important when results become significantly more precise. Of course, the different inputs may not be ignored, when radiative corrections, Eq.~\eqref{eq:ibdec}, 
from various collaborations are directly compared. In this case, we strongly
suggest to compare results for the unambiguous full theory observable
or sticking to a standard.

\subsection{Physical scales}
\label{s:physscal}
The purpose of this short section is to summarize the most popular
scales and give a short discussion of their advantages and
disadvantages. We restrict ourselves to those used in more recent
computations and,  thus, the list is short.

\subsubsection{The mass of the $\Omega$ baryon}
As already discussed, masses of hadrons that are stable in QCD$+$QED
and have a small width, in general, are very good candidates for
physical scales since there are no QED infrared divergences to be discussed. Furthermore, 
remaining within this class, the radiative corrections
are expected to be small. Furthermore, the $\Omega$ baryon has a significantly better noise/signal ratio than the nucleon (see \fig{f:plateaux}). It also has little dependence 
on up- and down-quark masses, since it is composed entirely of strange
valence quarks. 

Still, one has to be aware that
the mass is not extracted from the plateau
 region but from a modelling of the approach to a plateau in the form of fits \cite{Miller:2020evg,Borsanyi:2020mff,Blum:2014tka,Borsanyi:2012zs,Aoki:2010dy,Aoki:2008sm}. In this sense, the noise/signal ratio problem may persist.
The use of various interpolating fields for the 
$\Omega$ helps in constraining such analyses, but it
would be desirable to have a theoretical understanding of multi-hadron (or in QCD$+$QED multi-hadron $+$ photon) contributions as for the nucleon~\cite{Bar:2017kxh} discussed in \sect{sec:NME}. In the present review, we take the estimates of the collaborations at face value and do not try to apply a rating or an estimate of systematic error due to excited-state contributions.

\subsubsection{Pion and kaon leptonic decay rates}

These decay rates play a prominent r\^ole in scale setting 
in (pure) QCD because excited-state contaminations 
can simply be avoided by going to sufficiently large Euclidean time.
As a downside, QED radiative corrections need to  be taken into account in the values assigned to the associated decay constants. Therefore, we briefly summarize the knowledge of 
QED radiative corrections and the definition of decay constants. More details are found in the previous 
edition of this review.  

The physical observable is the decay rate $\Gamma^\mathrm{QCD+QED}[\pi^-\mapsto \mu\bar\nu_\mu(\gamma),E_\gamma]
$ of a pion at rest. It depends on the maximum energy $E_\gamma$ of photons emitted in the 
decay and registered in the experimental measurement. These soft and hard photons cannot be avoided since the cross-section vanishes as $E_\gamma \to 0$ and, e.g., the fixed-order cross-section without final-state photons is infrared divergent.
However, apart from the dependence on $E_\gamma$, there are no ambiguities in the definition of $\Gamma^\mathrm{QCD+QED}$.

In QCD,
the leptonic decay rate is,
\begin{eqnarray}
\Gamma^\mathrm{QCD}[\pi\mapsto \mu\bar\nu_\mu]
=
\frac{G_F^2}{8\pi}\vert V_{ud}\vert^2\, M_{\pi^-}^\mathrm{exp}\left(m_\mu^\mathrm{exp}\right)^2\left[
1-\frac{\left(m_\mu^\mathrm{exp}\right)^2}{\left(M_{\pi^-}^\mathrm{exp}\right)^2}
\right]\, \left(f_\pi^\mathrm{QCD}\right)^2\,,
\label{eq:piellnuQCD}
\end{eqnarray} 
where one naturally introduces the decay constant,
\begin{eqnarray}
f_\pi^\mathrm{QCD}=\frac{\bra{0}\bar u \gamma^0 \gamma^5 d\ket{\pi}^\mathrm{QCD}}{M_\pi^\mathrm{QCD}}\;.
\end{eqnarray}
of the pion. 
Radiative corrections to $f_\pi^\mathrm{QCD}$ are then defined by
\begin{eqnarray}
\delta f_\pi^\mathrm{QCD}(E_\gamma) = 
\sqrt{\frac{\Gamma^\mathrm{QCD+QED}[\pi^-\mapsto \mu\bar\nu_\mu(\gamma),E_\gamma]}
{\Gamma^\mathrm{QCD}[\pi\mapsto \mu\bar\nu_\mu]}}-1\;,
\end{eqnarray}
such that
\begin{eqnarray}
\Gamma^\mathrm{QCD+QED}[\pi^-\mapsto \mu\bar\nu_\mu(\gamma),E_\gamma]
=
\Gamma^\mathrm{QCD}[\pi\mapsto \mu\bar\nu_\mu]\left[
1+\delta f_\pi^\mathrm{QCD}(E_\gamma)
\right]^2\,.
\end{eqnarray}
 Common practice  is  to set
\begin{eqnarray}
E_\gamma=E_\gamma^\mathrm{max}=\frac{M_{\pi^-}^\mathrm{exp}}{2}\left[
1-\frac{\left(m_\mu^\mathrm{exp}\right)^2}{\left(M_{\pi^-}^\mathrm{exp}\right)^2}
\right]\;,
\end{eqnarray}
the maximum energy allowed for a single photon in the case of negligible $\cO(\alpha_\mathrm{em}^2)$ corrections.

As discussed in Sec.~\ref{sec:ibscheme}, $\delta f_\pi^\mathrm{QCD}(E_\gamma)$ depends on the scheme used to define QCD. However, the RM123 lattice determination in the electro-quenched approximation~\cite{DiCarlo:2019thl}
found the  scheme dependence to be irrelevant at the 
level of their result, $\delta f_\pi^\mathrm{isoQCD}(E_\gamma^\mathrm{max}) = 0.0076(9)$.\footnote{More precisely,
  both a hadronic scheme
  and a so-called GRS scheme were tested, where as a simplification one may replace constant $\alpha_s(\mu_\mathrm{ref})$ across theories by constant lattice spacing in the electro-quenched approximation.}  Additionally this
agrees well with the estimate, $\delta f_\pi^\mathrm{isoQCD}(E_\gamma^\mathrm{max}) = 0.0088(11)$ from ChPT~\cite{Cirigliano:2007ga,Ananthanarayan:2004qk,Cirigliano:2011tm}.
 Taking $V_{ud}$ from the PDG  \cite{Zyla:2020zbs} (beta decays) 
 and the ChPT number for $\delta f_\pi$, one has 
 $$
   f_\pi^\mathrm{isoQCD} = 130.56(2)_\mathrm{exp}(13)_\mathrm{QED}(2)_{V_{ud}}\,\mev  \,.
 $$
 With the Edinburgh consensus Sec.~\ref{sec:ibscheme},   the scale of isoQCD is {\em defined} by
 \begin{equation}
 	f_\pi^\mathrm{isoQCD} \equiv 130.5\,\mev\,. \label{eq:fpiedi}
 	\quad 
 \end{equation}
 At the present level of accuracy the difference between 
 the determined value  (with a scheme uncertainty of around $1$ permille) and the defining value \eqref{eq:fpiedi} is
 irrelevant.
 
 Some scale determinations use also the Kaon decay constant.
 There the understanding of QED radiative corrections is not yet as good as for pion decays. The ChPT estimate is $\delta f_K^\mathrm{isoQCD}(E_\gamma^\mathrm{max}) = 0.0053(11)$~\cite{Cirigliano:2007ga,Ananthanarayan:2004qk,Cirigliano:2011tm}, while the electro-quenched lattice computation yielded $\delta f_K^\mathrm{isoQCD}(E_\gamma^\mathrm{max}) = 0.0012(5)$~\cite{DiCarlo:2019thl}. As a slight update of the previous review, here we opt for a more conservative number of
 \begin{equation}
 	\delta f_K^\mathrm{isoQCD}(E_\gamma^\mathrm{max}) = 0.003(3)\,,
 \end{equation} 
 encompassing both estimates. 
 Together with $ V_{us}=0.2232(6)$  from Sec.~\ref{sec:vusvud} ($f_+(0)$ for $\Nf=2+1+1$) and the PDG decay rate,
 we have
 \begin{equation}
    f_K^\mathrm{isoQCD} =157.4(2)_\mathrm{exp}(4)_\mathrm{QED}(4)_{V_{us}}\,\mev \,.	
 \end{equation}

Depending on the lattice formulation, there is also a
nontrivial renormalization of the axial current. Since it is easily determined from a chiral Ward identity, it does not play an important r\^ole. 
When it is present, it is assumed to be accounted for in the statistical errors.

\subsubsection{Other physics scales}
Scales derived from bottomonium have been used in the past, in particular, the splitting $\Delta M_{\Upsilon} =  M_{\Upsilon(2s)}-M_{\Upsilon(1s)}$.
They have very little dependence on the light-quark masses, but need
an input for the $b$-quark mass. In all relevant cases, the $b$ quark is treated by NRQCD.

\subsection{Theory scales}
\label{s:theoryscales}
 
In the following, we consider in more detail the two classes of theory
scales that are most commonly used in typical lattice
computations. The first class consists of scales related to the static
quark-antiquark potential \cite{Sommer:1993ce}. The second class is related to the action density renormalized through the gradient flow~\cite{Luscher:2010iy}.

\subsubsection{Potential scales}
In this approach, lattice scales are derived from the properties of
the static quark-antiquark potential. In particular, a scale can be
defined by fixing the force $F(r)$ between a static quark and
antiquark separated by the distance $r$ in
physical units \cite{Sommer:1993ce}. Advantages of using the
potential include the ease and accuracy of its computation,
and its mild dependence on the valence-quark mass. In general, a
potential scale $r_c$ can be fixed through
the condition that the static force takes a predescribed value, i.e., 
\begin{equation}
\label{eq:rc definition}
r_c^2 F(r_c) = X_c\,,
\end{equation}
where $X_c$ is a suitably chosen number. Phenomenological and
computational considerations
suggest that the optimal choice for $X_c$ is in the region where the
static force turns over from Coulomb-like to linear behaviour and
before string breaking occurs. In the original work
\cite{Sommer:1993ce}, it was suggested to use $X_0=1.65$ leading to
the condition
\begin{equation}
\label{eq:r0 definition}
r_0^2 F(r_0) = 1.65 \, .
\end{equation}
In Ref.~\cite{Bernard:2000gd}, the value $X_1=1.0$ was proposed yielding the
scale $r_1$.

The static force is the derivative of the static quark-antiquark potential $V(r)$ which can be determined  through the calculation of Wilson loops. 
More specifically, the potential at distance $r$ is extracted from the asymptotic time dependence of the $r \times t$-sized Wilson loops $W(r,t)$,
\begin{equation}
\label{eq:V(r) Wilson loops}
V(r)=-\lim_{t\rightarrow \infty} \frac{d}{d t} \log \langle W(r,t) \rangle \,.
\end{equation}
The derivative of the potential needed for the force is then determined through the derivative of a suitable local parameterization of the potential as a function of $r$, e.g., 
\begin{equation}
V(r) =  C_-\frac{1}{r} + C_0 + C_+ r \,,
\label{eq:V parametrization}
\end{equation}
and estimating uncertainties due to the parameterization.
In some calculations, the gauge field is fixed to Coulomb or temporal
gauge in order to ease the computation of the potential at arbitrary
distances.

In order to optimize the overlap of the Wilson loops with the ground
state of the potential, one can use different types and levels of
spatial gauge-field smearing and extract the ground-state energy from
the corresponding correlation matrix by solving
a generalized eigenvalue problem~\cite{Michael:1985ne,Luscher:1990ck,Niedermayer:2000yx}. 
Finally, one can also make use of the noise reduction proposed in Refs.~\cite{DellaMorte:2005nwx,Donnellan:2010mx}. It includes
a smearing of the temporal parallel transporter~\cite{Hasenfratz:2001hp}
in the lattice definition of the discretized loops and thus yields a different discretization of the continuum force.

\subsubsection{Gradient-flow scales}
\label{s:flowscales}
The gradient flow $B_\mu(t,x)$ of gauge fields is defined in the continuum by the flow equation
\begin{align}
\label{eq:GF equation continuum 1}
\dot B_\mu &= D_\nu G_{\nu\mu}, \quad \left. B_\mu \right|_{t=0} =
A_\mu\, ,\\
G_{\mu\nu} &= \partial_\mu B_\nu - \partial_\nu B_\mu + [B_\mu,B_\nu],
\quad D_\mu = \partial_\mu + [B_\mu, \cdot\,]\,,
\label{eq:GF equation continuum 2}
\end{align}
where $A_\mu$ is the fundamental gauge field, $G_{\mu\nu}$ the field-strength tensor, and $D_\mu$ the covariant derivative~\cite{Luscher:2010iy}.
At finite lattice spacing, a possible form of Eqs.~(\ref{eq:GF equation continuum 1}) and (\ref{eq:GF equation continuum 2}) is 
\begin{equation}
  a^2\frac{d}{dt} V_t(x,\mu) = -g_0^2 \cdot {\partial_{x,\mu} S_G(V_t)} \cdot V_t(x,\mu)\,,
\label{eq:GF equation lattice}
\end{equation}
where $V_t(x,\mu)$ is the flow of the original gauge field $U(x,\mu)$
at flow time $t$, $S_G$ is an arbitrary lattice discretization of the
gauge action, and $\partial_{x,\mu}$ denotes the su$(3)$-valued
differential operator with respect to $V_t(x,\mu)$. An important point to
note is that the flow time $t$ has the dimension of a length squared,
i.e., $t \sim a^2$, and hence provides a means for setting the scale. 

One crucial property of the gradient flow is that any function of the
gauge fields evaluated at flow times $t>0$ is
renormalized~\cite{Luscher:2011bx} by just renormalizing the gauge
coupling. Therefore, one can define a  scale by keeping a suitable
gluonic observable defined at constant flow time $t$, e.g., the action
density $E=-\frac1{2} \Tr
\,G_{\mu\nu}G_{\mu\nu}$~\cite{Luscher:2010iy}, fixed in physical
units. This can, for example, be achieved through the condition
\begin{equation}
t_c^2 \langle E(t_c,x)\rangle = c \,, \quad E(t,x)=-\frac1{2} \Tr\, G_{\mu\nu}(t,x)G_{\mu\nu}(t,x)\,,
\label{eq:GF t definition}
\end{equation}
where $ G_{\mu\nu}(t,x)$ is the field-strength tensor evaluated on the
flowed gauge field $V_t$.
Then, the lattice scale $a$ can be determined from the dimensionless
flow time in lattice units, $\hat t_c = a^2 t_c$. The original proposal in \cite{Luscher:2010iy} was to use $c = 0.3$
yielding the scale $t_0$,
\begin{equation}
t_0^2 \langle E(t_0)\rangle = 0.3 \, .
\label{eq:GF t0 definition}
\end{equation}
 For convenience one sometimes also defines
$s_0 = \sqrt{t_0}$.

An alternative scale $w_0$ has been introduced in
Ref.~\cite{Borsanyi:2012zs}. It is defined by fixing a suitable
derivative of the action density,
\begin{equation}
\label{eq:GF w0 definition}
W(t_c) = t_c \cdot \partial_t \left(t^2 \langle
  E(t)\rangle\right)_{t=t_c}  = c\, .
\end{equation}
Setting $c=0.3$ yields the scale $w_0$ through
\begin{equation}
W(w_0^2) = 0.3 \, .
\end{equation}

In addition to the lattice scales from $t_0$ and $w_0$, one can also consider the scale from the dimensionful combination $t_0/w_0$. 
This combination is observed to have a very weak dependence on the quark mass~\cite{Deuzeman:2012jw,Abdel-Rehim:2015pwa,Alexandrou:2021bfr}.

A useful property of the gradient-flow scales is the fact that their quark-mass dependence is known from $\chi$PT \cite{Bar:2013ora}.

Since the action density at $t \sim t_0 \sim w_0^2$ usually suffers from large autocorrelation~\cite{Deuzeman:2012jw,Schaefer:2012tq}, the calculation of the statistical error needs special care.

Lattice artefacts in the
gradient-flow scales originate from different sources \cite{Ramos:2015baa}, which are systematically discussed by considering $t$ as a coordinate in a fifth dimension. First, there is the choice of the action $S_G$ for $t>0$.  Second, there is the discretization of $E(t,x)$. Third, due to the discretization of the four-dimensional quantum action, and fourth, contributions of terms localized at the boundary $t=0_+$. 
The interplay
between the different sources of lattice artefacts  turns out to be rather
subtle~\cite{Ramos:2015baa}. 

Removing discretization errors due to the first two sources requires only classical ($g_0$-independent) improvement. 
Those due to the quantum
action are common to all $t=0$ observables, but the effects of the boundary terms are not easily removed in practice. At tree level, the Zeuthen flow~\cite{Ramos:2015baa} removes these effects completely, but none of the computations reviewed here have used it. 
Discretization effects due to $S_G$
can be removed by using an improved action such as the tree-level
Symanzik-improved gauge action
\cite{Borsanyi:2012zs,Bazavov:2013gca}. More phenomenological attempts of improving the gradient-flow scales consist of
applying a
$t$-shift \cite{Cheng:2014jba}, or tree-level improvement \cite{Fodor:2014cpa}.

\subsubsection{Other theory scales}
\label{subsubsec:other scales}
The MILC collaboration has been using another set of scales,  
the partially quenched pseudoscalar decay constant $f_{p4s}$ with
degenerate valence quarks with a mass $m_q=0.4 \cdot
m_\mathrm{strange}$,  and the corresponding partially quenched pseudoscalar mass $M_{p4s}$. So far it has been a quantity only used by the
MILC collaboration \cite{Bazavov:2014wgs,Bazavov:2017lyh,Bazavov:2012xda}. We  do not perform an in-depth discussion or an average but will list numbers in the results section.

Yet another scale that has been used is the leptonic decay constant of the $\eta_s$. This fictitious particle is 
a pseudoscalar made of a valence quark-antiquark pair with different
(fictitious) flavours which are mass-degenerate
with the strange quark \cite{Dowdall:2011wh,Davies:2009tsa,Gray:2005ur}. 

\else
\fi

\subsection{List of computations and results}

\subsubsection{Gradient-flow scales}
\begin{figure}
  \includegraphics[width=0.5\textwidth]{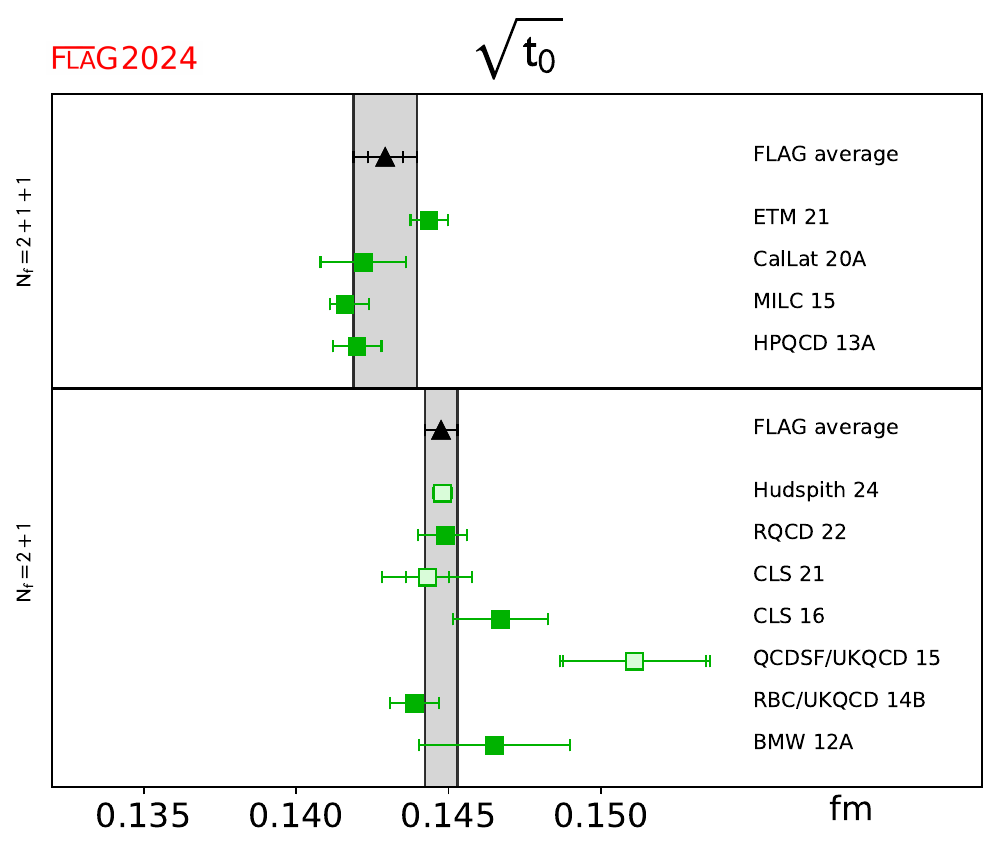}
  \includegraphics[width=0.5\textwidth]{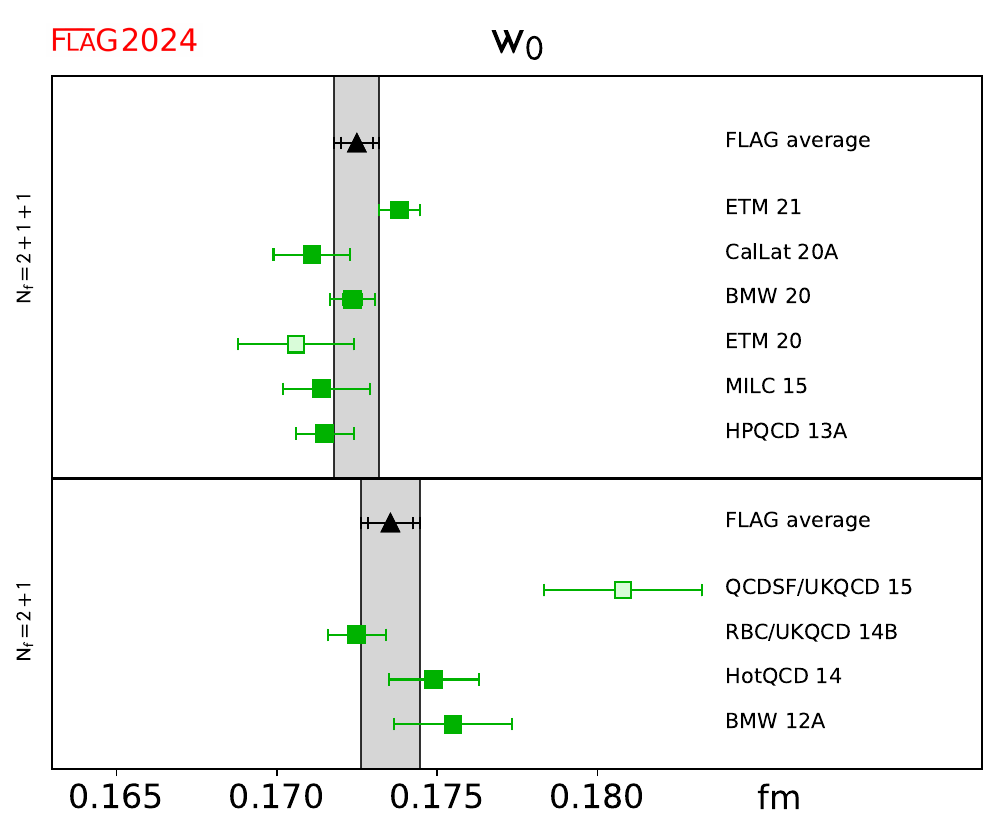}
\caption{\label{fig_GFscales} Results for gradient-flow scales.}
\end{figure}

We now turn to a review of the calculations of the gradient-flow scales $\sqrt{t_0}$ and $w_0$. The results are compiled in Tab.~\ref{tab_GFscales} and shown in Fig.~\ref{fig_GFscales}. In the following, we briefly discuss the calculations in the order that they appear in the table and figure.
\begin{table}[!h]
  \vspace*{3cm}
\footnotesize
\begin{tabular*}{1.0\textwidth}[l]{l@{\extracolsep{\fill}}rl@{\hspace{1mm}}l@{\hspace{1mm}}l@{\hspace{1mm}}l@{\hspace{1mm}}lll@{\hspace{1mm}}l}

Collaboration & Ref. & $\Nf$ & \begin{rotate}{60}{publication status}\end{rotate} &  \begin{rotate}{60}{chiral extrapolation}\end{rotate} & \begin{rotate}{60}{continuum extrapolation}\end{rotate} & \begin{rotate}{60}{finite volume}\end{rotate} &\begin{rotate}{60}{physical scale}\end{rotate}  & $\sqrt{t_0}$ [fm] & $w_0$ [fm] \\
\hline\hline

ETM 21 & \cite{Alexandrou:2021bfr} & 2+1+1 & \gA &\good&\good&\good& $f_\pi$  & 0.14436(61) & 0.17383(63) \\

CalLat 20A& \cite{Miller:2020evg} & 2+1+1 &  \gA &\good&\good&\good& $M_\Omega$ & 0.1422(14) & 0.1709(11) \\

BMW 20 & \cite{Borsanyi:2020mff} & 1+1+1+1 &\gA &\good&\good&\good& $M_\Omega$ &  &  0.17236(29)(63)[70] \\

ETM 20 & \cite{Dimopoulos:2020eqd} & 2+1+1 & \rC &\good&\good&\good& $f_\pi$  &  & 0.1706(18) \\
  
MILC 15 &\cite{Bazavov:2015yea} & 2+1+1 & \gA &\good&\good&\good& ${F_{p4s}(f_\pi)}^\#$ &  0.1416(+8/-5) &  0.1714(+15/-12) \\
  
HPQCD 13A & \cite{Dowdall:2013rya} &  2+1+1 & \gA & \good &\soso & \good & $f_\pi$ & 0.1420(8) & 0.1715(9) \\
\hline
Hudspith 24 & \cite{Hudspith:2024kzk} & 2+1 & P & \good&\good&\good & ${}^\&$  & 0.14480(32)(6)\\

RQCD 22 & \cite{RQCD:2022xux}        & 2+1 &  \gA   &  \good  & \good  &  \good &  $M_\Xi$      &  0.1449(+7/-9) & \\

CLS 21 & \cite{Strassberger:2021tsu} & 2+1 &  \rC   &  \good  & \good  &  \good &  $f_\pi, f_K$ &  0.1443(7)(13) & \\

CLS 16 & \cite{Bruno:2016plf} & 2+1 & \gA &\soso&\good&\good& $f_\pi,f_K$  & 0.1467(14)(7)  & \\

QCDSF/UKQCD 15B & \cite{Bornyakov:2015eaa} & 2+1 & \oP & \soso & \soso & \soso & $M_P^\text{SU(3)}$  & 0.1511(22)(6)(5)(3) & 0.1808(23)(5)(6)(4) \\

RBC/UKQCD 14B & \cite{Blum:2014tka} & 2+1 & \gA &\good&\good&\good& $M_\Omega$  & 0.14389(81) & 0.17250(91) \\

HotQCD 14 & \cite{Bazavov:2014pvz} &  2+1  & \gA &\good&\good&\good& ${r_1(f_\pi)}^\#$  & & 0.1749(14) \\

BMW 12A & \cite{Borsanyi:2012zs} & 2+1 &\gA & \good & \good & \good & $M_\Omega$ &  0.1465(21)(13)  &  0.1755(18)(4) \\

\hline\hline
\end{tabular*}\\
\begin{minipage}{1.0\linewidth}
  {\footnotesize
    $^\#$ These scales are not physical scales and have been determined from $f_\pi$. 
    \newline
    $^\&$ There is no physical scale as such. The input is the quark-mass dependence of
$M_\Omega$.
  }
  \end{minipage}
\caption{Results for gradient-flow scales at the physical point, cf.~\eq{e:scalesettbasic}.
Note that BMW 20 \cite{Borsanyi:2020mff} take IB and QED corrections into account. 
   An additional result for the ratio of scales is:\newline
     ETM 21 \cite{Alexandrou:2021bfr}: $t_0/w_0 = 0.11969(62)$ fm.
\label{tab_GFscales}
}\end{table}



ETM 21 \cite{Alexandrou:2021bfr} finalizes and supersedes ETM 20 discussed below. It determines the scales $\sqrt{t_0}, w_0$, also $t_0/w_0 = 0.11969(62)$ fm, and the ratio $\sqrt{t_0}/w_0 = 0.82930(65)$, cf.~also HPQCD 13A \cite{Dowdall:2013rya}. Since ETM 21 is now published, the values replace the ones of ETM 20 in the previous FLAG average.   

  CalLat 20A \cite{Miller:2020evg} use M\"obius Domain-Wall valence fermions on HISQ ensembles generated by the MILC and CalLat collaborations. The gauge fields entering the M\"obius Domain-Wall operator are gradient-flow smeared with $t=a^2$. They compute the $\Omega$ mass and the scales  $w_0,\,t_0$ and perform global fits to determine $w_0 M_\Omega$ and $\sqrt{t_0} M_\Omega$ at the physical point. The flow is discretized with the Symanzik tree-level improved action and the clover discretization of $E(t)$ is used. A global fit with Bayesian priors is performed including terms derived from  {\Ch}PT for finite-volume and quark-mass dependencies, as well as $a^2$ and $a^2 \alpha_s(1.5/a)$ terms for discretization errors. Also, a tree-level improved definition of the GF scales is used where the leading-in-$g^2$ cutoff effects are removed up to and including
  ${\cal O}(a^8/t^4)$.

  BMW 20 \cite{Borsanyi:2020mff}
     presents a result for $w_0$ in the context of their staggered-fermion 
     calculation of the muon anomalous magnetic moment. It is
     the first computation that takes QED and isospin-breaking
     corrections into account.
  The simulations are performed by using staggered fermions with stout gauge-field 
  smearing with six lattice spacings
  and several pion masses around the physical
  point with $M_\pi$ between $110$ and $140$ MeV. Volumes are around
  $L=6$ fm. 
  At the largest lattice spacing, it is demonstrated how the effective masses of the $\Omega$ correlator almost reach the plateau value extracted from a four-state fit (two states per parity). Within the range where the data are fit, the deviation 
  of data points from the estimated plateau is less than a percent. Isospin-breaking  corrections are computed by Taylor expansion around isoQCD with QED treated as \qedl~\cite{Hayakawa:2008an}. Finite-volume effects in QED are taken from the $1/L, 1/L^2$ universal corrections and $\cO(1/L^3)$ effects are neglected. The results for $M_\Omega w_0$ are extrapolated to the continuum by a fit with $a^2$ and $a^4$ terms. 

   ETM 20 \cite{Dimopoulos:2020eqd} presents in their proceedings contribution a preliminary analysis of their $\Nf=2+1+1$ Wilson twisted-mass fermion simulations at maximal twist (i.e., automatic ${\cal O}(a)$ improved), at three lattice spacings and pion masses at the physical point. Their determination of $w_0=0.1706(18)$ fm from  $f_\pi$  using an analysis in terms of $M_\pi$ is the value quoted above. They obtain the consistent value $w_0=0.1703(18)$ fm from an analysis in terms of the renormalized light quark mass.

   MILC 15 \cite{Bazavov:2015yea} sets the physical scale using the fictitious pseudoscalar decay constant ${F_{p4s}=153.90(9)(+21/-28)}$ MeV with degenerate valence quarks of mass $m_v = 0.4 m_s$ and physical sea-quark masses \cite{Bazavov:2012xda}. ($F_{p4s}$ has strong dependence on the valence-quark mass and is determined from $f_\pi$.) They use a definition of the flow scales where the tree-level lattice artefacts up to ${\cal O}(a^4/t^2)$ are divided out. Charm-quark mass mistunings are between 1\% and 11\%. They are taken into account at leading order in $1/m_c$  through $\Lambda^{(3)}_\text{QCD}$ applied directly to $F_{p4s}$ and $1/m_c$ corrections are included as terms in the fits. They use elaborate variations of fits in order to estimate extrapolation errors (both in GF scales and $F_{p4s}$). They include errors from   FV effects and experimental errors in $f_\pi$ in $F_{p4s}$.

  HPQCD~13A \cite{Dowdall:2013rya} uses
  eight MILC-HISQ ensembles with  lattice spacings $a$ = 0.088, 0.121, 0.151 fm.
Values of $L$ are between 2.5~fm and 5.8~fm with $M_\pi L =$ 3.3--4.6.
Pion masses range between $128$ and $306$~MeV.
QCD is defined by using the inputs 
$M_{\pi}=134.98(32)$ MeV,
$M_{K}=494.6(3)$ MeV,
$f_{\pi^+}=130.4(2)$ MeV
derived by model subtractions
of IB effects.
Additional scale ratios are given: $\sqrt{t_0}/w_0 = 0.835(8)$, $r_1/w_0=1.789(26)$.
%

Hudspith 24 \cite{Hudspith:2024kzk} computes the 
mass of the Omega baryon on CLS $\Nf=2+1$ configurations
along the trajectories with approximately constant trace of the bare quark-mass
matrix. They use 27 ensembles with six different values of the lattice
spacing from
$a=0.09$ fm to $a=0.04$ fm.
They compute the (nonpositive) correlation function $C_\Omega(x_0)$ of a local field with a gauge-fixed wall-source,
which results in a very good statistical precision. It is analyzed directly with a two-state fit describing the data over a large range. In addition they also extract $M_\Omega$ by constructing a 2x2 generalized Pencil-of-Functions matrix correlator from $C_\Omega(x_0),\, C_\Omega(x_0+a),\, C_\Omega(x_0+2a)$. Projecting with a GEVP eigenvector (from a fixed-time GEVP) a correlation function with a long plateau of the effective mass is found. 
Precisions for the Omega mass  on various ensembles
range from a few per mille to below a per mille.
\\
These masses, together with the scale $t_0$ are subsequently fit using a 
phenomenology- and ChPT-motivated form where a few parameters are taken from previous ChPT fits~\cite{Lutz:2023xpi} to baryon masses computed on CLS ensembles by RQCD \cite{RQCD:2022xux}. The dependence on $t_0$ is in the higher-order chiral-correction terms which include NNNLO. There is no 
term in the fit which allows for discretization effects in the chiral corrections. Their absence is justified by the results of previous fits in \cite{Lutz:2023xpi}. 
Given the unconventional analysis carried out in this work, the WG hopes that additional technical information will be provided in the published version of the paper (in particular concerning the direction in parameter space of the global fit of the $\Omega$, kaon and pion masses from which the continuum value of $t_0$ is extracted)
and may reconsider the $\good$ assigned in this review, on the basis of  the standard continuum-limit criterion,  once the paper is published and eligible to enter the average.
Once the precision for the raw values of $am_\Omega$ is independently confirmed, this paper \cite{Hudspith:2024kzk}, possibly with a new analysis, may lead to very high-precision determinations of the theory scales.

RQCD 22 \cite{RQCD:2022xux} is an independent analysis of CLS
ensembles employing $\Nf=2+1$ nonperturbatively improved Wilson
fermions and the tree-level Symanzik improved gauge action. It uses a multitude of quark-mass combinations at six different values of the lattice spacing, ranging from $a \lesssim 0.098$ fm
down to $a < 0.039$ fm. Near-physical quark masses are realized at $a
= 0.064$ fm and $a = 0.085$ fm. The input quantities used to fix
the physical point and to set the scale are $M_\pi=134.8(3)$~MeV, $M_K=494.2(3)$~MeV, and
$m_\Xi=1316.9(3)$~MeV (last line of pg.~33 in
\cite{RQCD:2022xux}). As RQCD 22 has been published since
  the last update, the result for $\sqrt{t_0}$ is now included in the FLAG
average.

CLS 21 \cite{Strassberger:2021tsu} is a proceedings contribution describing a
preliminary analysis following the one in CLS 16
\cite{Bruno:2016plf}, cf.~the description below. CLS 21
includes about twice the number of ensembles as compared to CLS 16, in
particular, ensembles at two more lattice spacings and two ensembles at
the physical point. As a consequence, this analysis is not considered
a straightforward update and hence does not supersede the result of
CLS 16.

CLS~16 \cite{Bruno:2016plf} uses CLS
   configurations of 2+1 nonperturbatively ${\cal O}(a)$-improved Wilson fermions. There are a few pion masses with the strange mass adjusted  along a line of $m_u+m_d+m_s=\mathrm{const}$. Three different lattice spacings are used. They determine $t_0$ at the physical point defined by  $\pi$ and $K$ masses and the linear combination $f_K+\frac12 f_\pi$. They use the Wilson flow with the clover definition of $E(t)$.

QCDSF 15B \cite{Bornyakov:2015eaa,Bornyakov:2015plz} 
  results, unpublished, are obtained by simulating $\Nf=2+1$ QCD with the tree-level Symanzik-improved gauge action and clover Wilson fermions with single-level stout smearing for the hopping terms together with unsmeared links for the clover term (SLiNC action). Simulations are performed at four different lattice spacings, in the range $[0.06,0.08]$~fm, with $M_{\pi,\text{min}}=228$~MeV and $M_{\pi,\text{min}}L=4.1$. The results for the gradient-flow scales have been obtained by relying on the observation that flavour-symmetric quantities get corrections of $\cO((\Delta m_q)^2)$ where $\Delta m_q$ is the difference of the quark mass from the SU(3)-symmetric value. The $\cO(\Delta m_q^2)$ terms are not detected in the data and subsequently neglected.

    RBC/UKQCD 14B \cite{Blum:2014tka} presents results for $\sqrt{t_0}$ and $w_0$ obtained in QCD with $2+1$ dynamical flavours. The simulations are performed by using domain-wall fermions on six ensembles with lattice spacing $a^{-1}=1.38, 1.73, 1.78, 2.36, 2.38$, and 3.15 GeV, pion masses in the range $M_\pi^{unitary}\in[139,360]$~MeV. The simulated volumes are such that $M_\pi L> 3.9$. The effective masses of the $\Omega$ correlator are extracted with two-state fits and it is shown, by using two different nonlocal interpolating operators at the source, that the correlators almost reach a pleateau. In the calculation of $\sqrt{t_0}$ and $w_0$, the clover definition of $E(t)$ is used. The values given are $\sqrt{t_0}=0.7292(41) \GeV^{-1}$ and $w_0=0.8742(46) \GeV^{-1}$ which we converted to the values in Tab.~\ref{tab_GFscales}.  

    HotQCD 14 \cite{Bazavov:2014pvz}
  determines the equation of state with $\Nf=2+1$ flavours using highly
  improved staggered quarks (HISQ/tree).
   As a byproduct, they update the results of HotQCD 11 \cite{Bazavov:2011nk} by adding simulations at four new values of $\beta$, for a total of 24 ensembles, with lattice spacings in the range $[0.04,0.25]$~fm and volumes in the range $[2.6,6.1]$~fm with $M_\pi= 160$ MeV. They obtain values for  the scale parameters $r_0$ and $w_0$, via the ratios $r_0/r_1, w_0/r_1$ and using $r_1 = 0.3106(14)(8)(4)$ fm  from MILC 10 \cite{Bazavov:2010hj}. They obtain for the ratios $(r_0/r_1)_{cont}=1.5092(39)$ and $(w_0/r_1)_{cont}  = 0.5619(21)$ in the continuum.  They crosscheck their   determination of the scale $r_1$ using the hadronic quantities $f_K$, $f_\eta$ from HPQCD 09B \cite{Davies:2009tsa} and the experimental value of $M_\varphi$, and find good agreement.

  BMW 12A  \cite{Borsanyi:2012zs} is the work in which $w_0$ was introduced.  Simulations with 2HEX smeared Wilson fermions and two-level stout-smeared rooted staggered fermions are done. The Wilson flow with clover $E(t)$ is used, and a test of the Symanzik flow is carried out. They take the results with Wilson fermions as their central value, because those ``do not rely on the `rooting' of the fermion determinant''. 
Staggered fermion results agree within uncertainties.

\subsubsection{Potential scales}

\begin{figure}[htb]
  \includegraphics[width=0.5\textwidth]{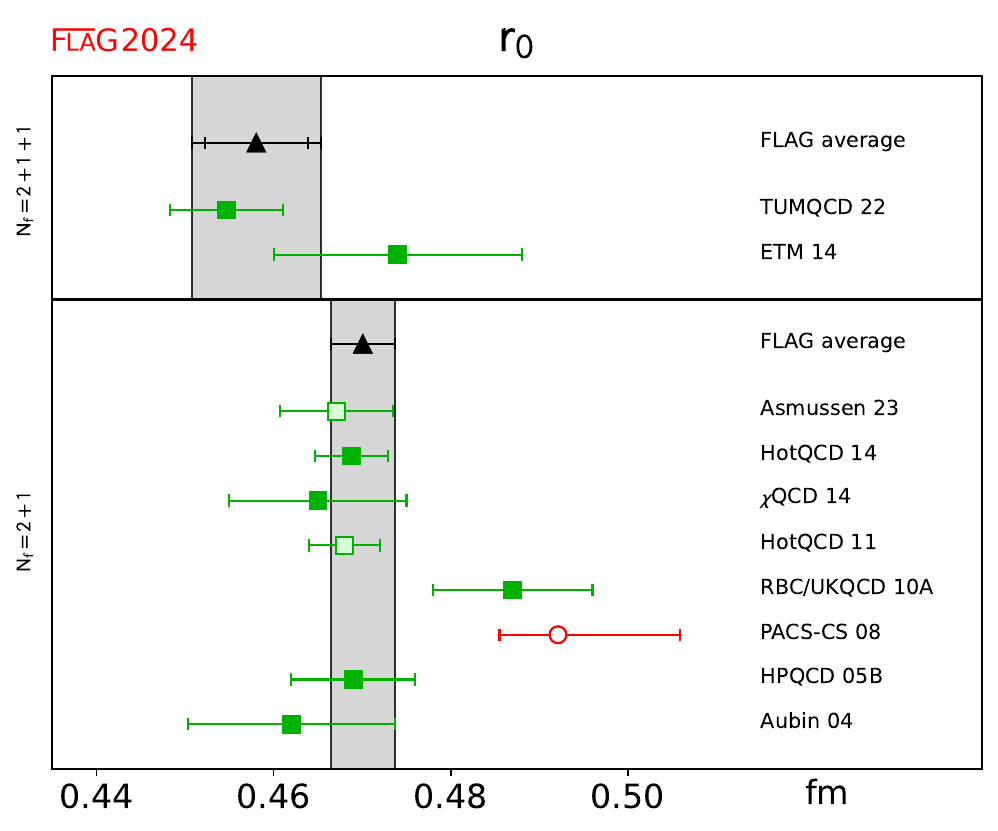}
  \includegraphics[width=0.5\textwidth]{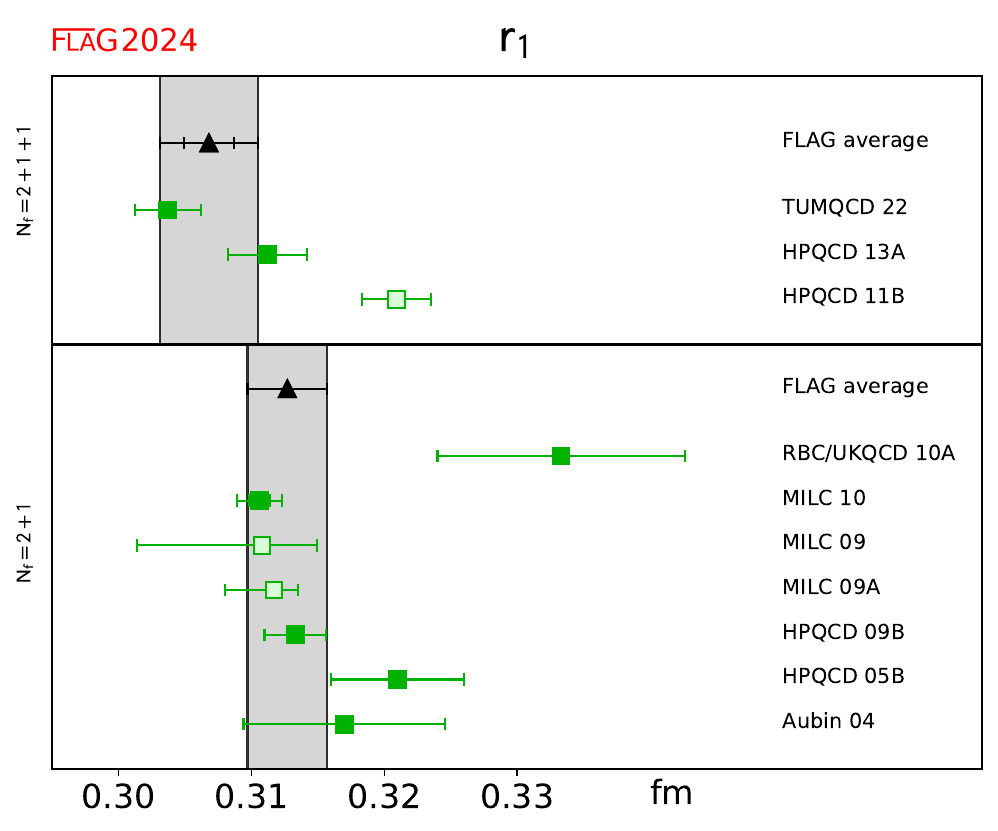}
\caption{\label{fig_potentialscales} Results for potential scales.}
\end{figure}

We now turn to a review of the calculations of the potential scales $r_0$ and $r_1$. The results are compiled in Tab.~\ref{tab_potentialscales} and shown in Fig.~\ref{fig_potentialscales}. 
With the exception of TUMQCD 22 \cite{Brambilla:2022het}, the most recent calculations date back to 2014, and we discuss them in the order that they appear in the table and the figure.
\begin{table}[!h]
  \vspace*{3cm}
\footnotesize{
\begin{tabular*}{1.0\textwidth}[l]{l@{\extracolsep{\fill}}rlll@{\hspace{1mm}}l@{\hspace{1mm}}llll}

Collaboration & Ref. & $\Nf$ & \begin{rotate}{60}{publication status}\end{rotate} &  \begin{rotate}{60}{chiral extrapolation}\end{rotate} & \begin{rotate}{60}{continuum extrapolation}\end{rotate} & \begin{rotate}{60}{finite volume}\end{rotate} &\begin{rotate}{60}{physical scale}\end{rotate} & $r_0$ [fm] & $r_1$ [fm] \\
\hline\hline  
TUMQCD 22 & \cite{Brambilla:2022het} & 2+1+1 & \gA 
&\good &\good&\good& $f_{p4s}$ (\cite{Bazavov:2017lyh}) $^\$ $ & 0.4547(64) & 0.3037(25)\\

ETM 14 & \cite{Carrasco:2014cwa} & 2+1+1 & \gA &\soso&\good&\good& $f_\pi$ & 0.474(14) & \\
  
HPQCD 13A & \cite{Dowdall:2013rya} & 2+1+1 & \gA &\good&\soso&\good& $f_\pi$ & & 0.3112(30) \\

HPQCD 11B &  \cite{Dowdall:2011wh} & 2+1+1 & \gA &\soso&\soso&\soso& $\Delta M_{\Upsilon}$, $f_{\eta_s}$ & & 0.3209(26) \\
\hline
Asmussen 23 & \cite{Asmussen:2023pia} & 2+1 & C &\good&\good&\good&
 $f_\pi,f_K$ & 0.4671(64) \\
 
HotQCD 14 & \cite{Bazavov:2014pvz} & 2+1 & \gA &\good&\good&\good& $r_1$(\cite{Bazavov:2010hj})$^\#$ & 0.4671(41) & \\

$\chi$QCD 14 & \cite{Yang:2014sea} & 2+1 & \gA & \soso    & \soso & \soso    &  three inputs\footnote{$M_{D_s^*}, M_{D_s^*}-M_{D_s}, M_{J/\psi}$}     &  0.465(4)(9) & \\

HotQCD 11 & \cite{Bazavov:2011nk} & 2+1 & \gA   &\good&\good&\good& $f_\pi$ & 0.468(4) & \\
RBC/UKQCD 10A &  \cite{Aoki:2010dy} & 2+1 & \gA & \soso & \soso & \soso & $M_\Omega$  & 0.487(9) & 0.333(9) \\
MILC 10 & \cite{Bazavov:2010hj} & 2+1 & \rC &\soso &\good &\good &$f_\pi$ & &0.3106(8)(14)(4) \\

MILC 09 & \cite{Bazavov:2009bb}  & 2+1 & \gA &\soso&\good&\good& $f_\pi$ & & 0.3108(15)($^{+26}_{-79}$) \\

MILC 09A & \cite{Bazavov:2009fk} & 2+1 & \rC &\soso&\good&\good& $f_\pi$ &  & 0.3117(6)($^{+12}_{-31}$) \\

HPQCD 09B & \cite{Davies:2009tsa} & 2+1 &  \gA  &\soso&\good&\soso& three inputs &  & 0.3133(23)(3) \\

PACS-CS 08 & \cite{Aoki:2008sm} & 2+1 & \gA &\good&\bad&\bad& $M_\Omega$ & 0.4921(64)($^{+74}_{-2}$) &  \\

HPQCD 05B & \cite{Gray:2005ur} & 2+1 & \gA &\soso&\soso&\soso&$\Delta M_{\Upsilon}$ & 0.469(7) & 0.321(5) \\

Aubin 04 & \cite{Aubin:2004wf}  & 2+1  &\gA &\soso&\soso&\soso& $\Delta M_{\Upsilon}$ & 0.462(11)(4) & 0.317(7)(3) \\
\hline\hline
\end{tabular*}\\
\begin{minipage}{1.0\linewidth}
  {\footnotesize
$^\#$ This theory scale was determined in turn from 
$r_1$ \cite{Bazavov:2010hj}.
\newline $^\$\,$ This theory scale was determined in turn from 
$f_\pi$.
  }
\end{minipage}
}
\caption{Results for potential scales at the physical point, cf.~\eq{e:scalesettbasic}. $\Delta M_{\Upsilon} =  M_{\Upsilon(2s)}-M_{\Upsilon(1s)}$.
}    
\label{tab_potentialscales}
\end{table}

Asmussen 23   \cite{Asmussen:2023pia} perform a computation of the potential at five lattice spacings down to $a=0.04$~fm on CLS ensembles. 
The ground-state level is extracted from a GEVP, starting from smeared Wilson loops with different levels of smearing. The results are thus far only available as a conference proceedings. The final result for $r_0$ originates from a global fit incorporating the pion-mass dependence and the lattice-spacing dependence.

TUMQCD 22 \cite{Brambilla:2022het}
uses HISQ ensembles generated by MILC at six lattice spacings ranging from $a=0.15$~fm to $a=0.03$~fm to compute the potential. Scale setting is performed through $f_{p4s}$ \cite{Bazavov:2017lyh}. In contrast to other determinations, the static potential is extracted using Coulomb-gauge fixing on two time-slices and the Wilson lines connecting the two time-slices. Thus, there is no variational method but fits are performed with up to three energy levels. Both continuum extrapolations with $a^2$ corrections and $\alpha^2(1/a) \,a^2$ are performed, where 
there is a preselection of the direction $\vec{r}/r$ and direction-dependent discretization effects are 
assumed to be sufficiently reduced by the use of the tree-level improved $r_\mathrm{I}$ \cite{Sommer:1993ce}.
The final results come from a Bayesian model average.

ETM 14 \cite{Carrasco:2014cwa} uses $\Nf=2+1+1$ Wilson twisted-mass fermions at maximal twist (i.e., automatic ${\cal O}(a)$-improved), three lattice spacings and pion masses reaching down to $M_\pi = 211$ MeV. They determine the scale $r_0$ through $f_\pi=f_{\pi^+}=130.41$ MeV.
  A crosscheck of the so-obtained lattice spacings with the ones obtained via the fictitious pseudoscalar meson $M_{s's'}$ made of two strange-like quarks gives consistent results. The crosscheck is done using the  dimensionless combinations $r_0 M_{s's'}$ (with $r_0$ in the chiral limit) and $f_\pi/M_{s's'}$ determined in the continuum, and then using $r_0/a$ and the value of $M_{s's'}$ obtained from the experimental value of $f_\pi$. We also note that in Ref.~\cite{Deuzeman:2012jw}  using the same ensembles the preliminary value $w_0=0.1782$ fm is determined, however, without error due to the missing or incomplete investigation of the systematic effects.  
  
  HPQCD~13A \cite{Dowdall:2013rya} was already discussed
  above in connection with the gradient-flow scales. 

HPQCD~11B \cite{Dowdall:2011wh} uses five MILC-HISQ ensembles and determines $r_1$ from  $M_{\Upsilon(2s)}-M_{\Upsilon(1s)}$ and the decay constant
$f_{\eta_s}$ (see HPQCD~09B).
The valence $b$ quark is treated by NRQCD, while the light valence quarks have the HISQ discretization, identical to the sea quarks.

HotQCD 14 \cite{Bazavov:2014pvz} was already discussed
in connection with the gradient-flow scales. 

$\chi$QCD 14 \cite{Yang:2014sea} uses  overlap fermions as valence quarks on $\Nf=2+1$ domain-wall fermion  gauge configurations generated by the RBC/UKQCD collaboration  \cite{Aoki:2010dy}. Using the physical masses of $D_s, D_s^*$ and  $J/\psi$ as inputs, the strange- and charm-quark masses and the decay  contant $f_{D_s}$ are determined as well as the scale $r_0$.

HotQCD 11 \cite{Bazavov:2011nk}  uses configurations with tree-level improved Symanzik gauge action and HISQ staggered quarks in addition to previously generated ensembles with p4 and asqtad staggered quarks. In this calculation, QCD is defined by generating lines of constant physics with $m_l/m_s=\{0.2,0.1,0.05, 0.025 \}$ and setting the strange-quark mass by requiring that the mass of a fictitious $\eta_{s\bar s}$ meson is $M_{\eta_{s\bar s}}=\sqrt{2M_K^2-M_\pi^2}$. The physical point is taken to be at $m_l/m_s=0.037$. The physical scale is set by using the value $r_1=0.3106(8)(18)(4)$~fm obtained in Ref.~\cite{Bazavov:2010hj} by using $f_\pi$ as physical input. In the paper, this result is shown to be consistent within the statistical and systematic errors with the choice of $f_K$ as physical input. {The result $r_0/r_1=1.508(5)$ is obtained by averaging over 12 ensembles at $m_l/m_s=0.05$ with lattice spacings in the range $[0.066,0.14]$~fm. This result is then used to get $r_0=0.468(4)$~fm. Finite-volume effects have been monitored with 20 ensembles in the range $[3.2,6.1]$fm with $M_\pi L>2.6$.}

RBC/UKQCD 10A \cite{Aoki:2010dy} uses $\Nf=2+1$ flavours of domain-wall quarks and the Iwasaki gauge action at two values of the lattice spacing with unitary pion masses in the approximate range $[290,420]$ MeV. They use the masses of $\pi$ and $K$ meson and of the $\Omega$ baryon to determine the physical quark masses and the lattice spacings, and so obtain estimates of the scales $r_0, r_1$ and the ratio $r_1/r_0$ from a combined chiral and continuum extrapolation.  

MILC 10  \cite{Bazavov:2010hj} presents a further update of $r_1$ with asqtad-staggered-quark ensembles with $a\in\{0.045,0.06,0.09\}$~fm. It supersedes MILC 09  \cite{Bazavov:2009bb,Bazavov:2009fk,Bazavov:2009tw}.

MILC 09  \cite{Bazavov:2009bb} presents an $\Nf=2+1$ calculation of the potential scales on asqtad-staggered-quark ensembles with $a\in\{0.045,0.06,0.09,0.12,0.15,0.18\}$~fm. The continuum extrapolation is performed by using Goldstone-boson pions as light as $M_\pi=224$~MeV (RMS pion mass of 258~MeV). The physical scale is set from $f_\pi$. The result for $r_1$ obtained in the published paper  \cite{Bazavov:2009bb} is then updated and, therefore, superseded by the conference proceedings  MILC 09A and 09B \cite{Bazavov:2009fk,Bazavov:2009tw}. 

HPQCD~09B \cite{Davies:2009tsa} is an extension of HPQCD~05B \cite{Gray:2005ur} and uses HISQ valence quarks instead of asqtad quarks. The scale $r_1$ is obtained from three different inputs. First $r_1=0.309(4)$~fm from the splitting of 2S and 1S $\Upsilon$ states as in Ref.~\cite{Gray:2005ur}, second $r_1=0.316(5)$~fm from $M_{D_s}- M_{\eta_s}/2$ and third $r_1=0.315(3)$~fm from the decay constant of the $\eta_s$. The fictitious $\eta_s$ state is operationally defined by setting quark masses to the s-quark mass and dropping disconnected diagrams. Its mass and decay constant are obtained from a partially quenched chiral-perturbation-theory analysis using the pion and kaon states from experiment together with various partially quenched lattice data. The three results are combined to $r_1=0.3133(23)(3)$~fm.

PACS-CS 08 \cite{Aoki:2008sm} presents a calculation of $r_0$ in $\Nf=2+1$ QCD by using NP ${\cal O}(a)$-improved clover Wilson quarks and Iwasaki gauge action. The calculation is done at fixed lattice spacing  $a=0.09$~fm and is extrapolated to the physical point from (unitary) pion masses in the range $[156,702]$~MeV. The $\Nf=2+1$ theory is defined by fixing $M_\pi$, $M_K$, and $M_{\Omega}$ to  $135.0,\; 497.6$, and $1672.25~\mev$, respectively. The effective masses of smeared-local $\Omega$ correlators   averaged over the four spin polarizations show quite good plateaux. 

RBC/Bielefeld 07 \cite{Cheng:2007jq} performed calculations of the equation of state with two light-quark flavours
and a heavier strange quark
using improved staggered fermions.
Zero-temperature calculations including the static-quark potential were used to set the temperature scale for the thermodynamic observables. The lattice cut-off changes by a factor 6 from $a \simeq 0.3$~fm down to $a \simeq 0.05$~fm while the pion mass is kept fixed at $M_\pi \simeq 220(4)$~MeV. Apart from the dimensionless ratio $r_0/r_1 = 1.4636(60)$ they also provide a result for the ratio $r_0 \sqrt{\sigma} = 1.1034(40)$

HPQCD~05B \cite{Gray:2005ur} performed the first bottomonium spectrum calculation in full QCD with $N_f=2+1$ on MILC asqtad configurations and the $b$ quark treated by NRQCD. They find agreement of the low lying $\Upsilon$ states with experiment and also compare to quenched and $\Nf=2$ results. They determined $r_0$ and $r_1$ from the splitting of 2S and 1S states.

Aubin 04 \cite{Aubin:2004wf} presents an $\Nf=2+1$ calculation of the potential scales by using asqtad staggered quark ensembles with $a=0.09$ and $0.12\}$~fm. The continuum extrapolation is performed by using Goldstone-boson pions as light as $M_\pi=250$~MeV. The physical scale is set from the $\Upsilon$ 2S-1S and 1P-1S splittings computed with NRQCD by HPQCD \cite{Wingate:2003gm}.

\off{
\begin{table}[!h]
\begin{tabular*}{\textwidth}[l]{l@{\extracolsep{\fill}}lllllllll}
Ref. & $\Nf$ & $M_\pi$  & $M_K$ & $M_\Omega$ & $f_\pi$ & $f_K$ & Description \\
\hline\hline \\ 

\cite{Miller:2020evg} & 2+1+1 & 134.8(3)$_*$ & 494.2(3)$_*$ & 1672.43(32)$_*$ &  & & \parbox[t]{2.5cm}{$D_s$ for charm} \\

\cite{Carrasco:2014cwa} & 2+1+1 & 134.98$_*$ & 494.2(4)$_*$$^\dagger$ & & 130.41$_*$ & & \parbox[t]{2.5cm}{$D$ and $D_s$ for charm} \\
  
\cite{Blum:2014tka} & 2+1 & 135.0$_*$ & 495.7$_*$ & 1672.25$_*$ & 130.2(9) & 155.5(8)\\

\cite{Bazavov:2009fk} & 1+1+1 & 135.0$_*$ & 494.4$_*$ &  & 130.4(2)$_*$ & 156.2(11) & \parbox[t]{2.5cm}{$m_{ud}$ defined using $M_K$ and $m_u$ fixed with $M_{K^+}$ with a partially quenched analysis}\\

\cite{Aoki:2008sm} & 2+1 & 135.0$_*$ & 497.6$_*$ & 1672.25$_*$ & 134.0(42) & 159.4(31) & \\

\end{tabular*}\\[-0.2cm]
\caption{Values in MeV for the hadronic quantities that can be used to identify the renormalization prescription adopted to define QCD in absence of QED corrections. In each line, the entries marked with the $_*$ subscript have been used by the collaboration as hadronic inputs in the renormalization procedure. Those without the $_*$ mark have been obtained as results of the simulations. \newline
${}^\dagger$Corrected for leading strong and electromagnetic isospin-breaking effects.
}
\label{tab_Hscheme}
\end{table}
}

\subsubsection{Ratios of scales}
\begin{figure}[htb]
  \centering
  \includegraphics[width=0.5\textwidth]{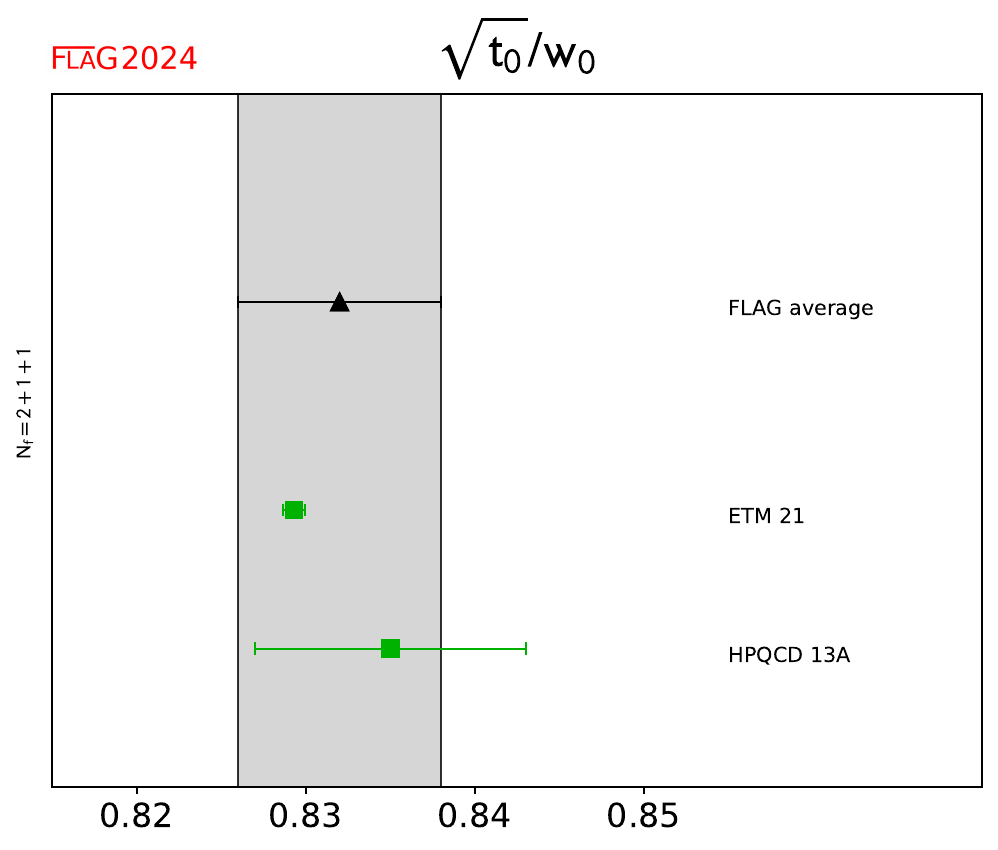}\\
  \includegraphics[width=0.49\textwidth]{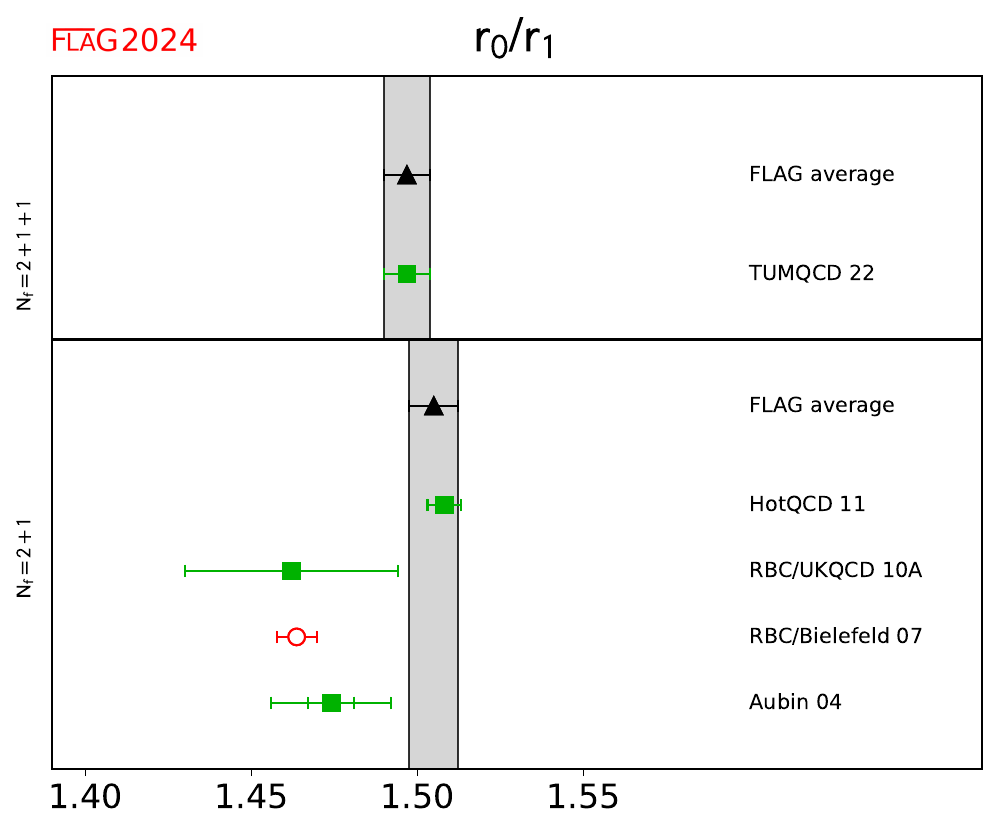}
  \includegraphics[width=0.49\textwidth]{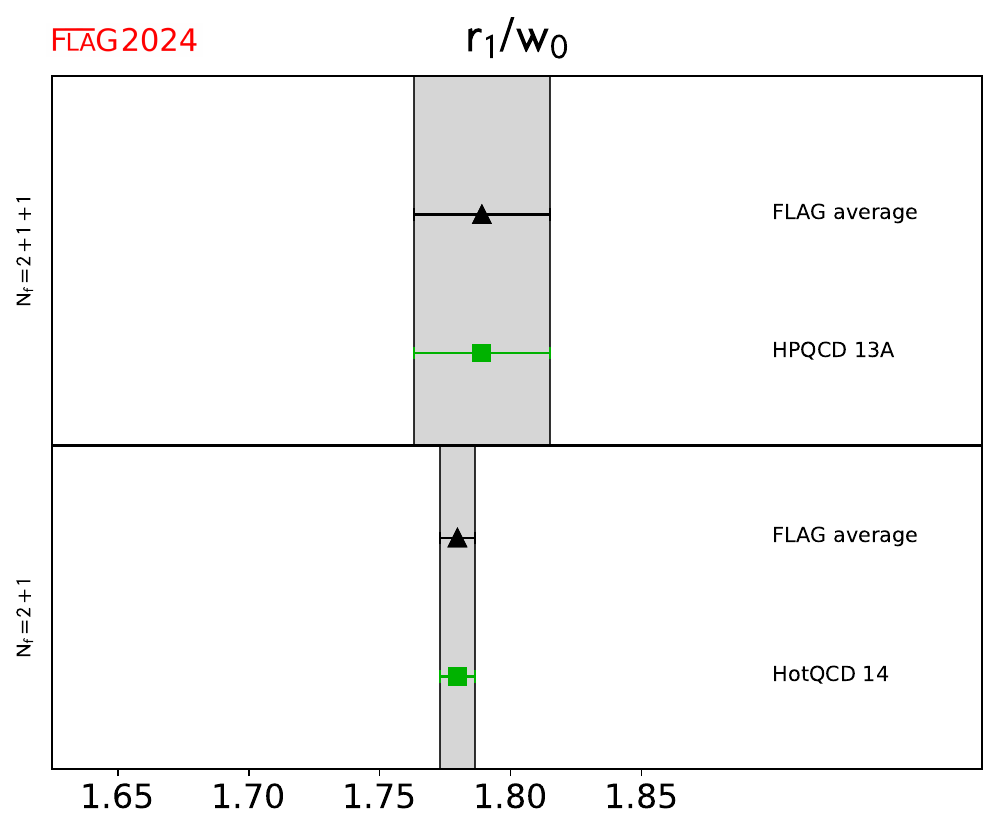}
\caption{\label{fig:scale_ratios} Results for dimensionless ratios of scales.}
\end{figure}

It is convenient in many cases to also have ratios of scales at hand. In addition to translating from one scale to another, the ratios provide important crosschecks between different determinations. Results on ratios provided by the collaborations are compiled in Tab.~\ref{tab:scale_ratios} and Fig.~\ref{fig:scale_ratios}. The details of the computations were already discussed in the previous sections.
\begin{table}[!h]
  \vspace*{3cm}
\footnotesize{
\begin{tabular*}{1.0\textwidth}[l]{l@{\extracolsep{\fill}}rllllllll}

Collaboration & Ref. & $\Nf$ & \begin{rotate}{60}{publication status}\end{rotate} &  \begin{rotate}{60}{chiral extrapolation}\end{rotate} & \begin{rotate}{60}{continuum extrapolation}\end{rotate} & \begin{rotate}{60}{finite volume}\end{rotate} & $\sqrt{t_0}/w_0$ & $r_0/r_1$  & $r_1/w_0$\\
\hline\hline
TUMQCD 22 & \cite{Brambilla:2022het} & 2+1+1 & \gA 
&\good &\good&\good& & 1.4968(69) \\

ETM 21 & \cite{Alexandrou:2021bfr} & 2+1+1 & \gA & \good & \good & \good & 0.82930(65) &  &\\
HPQCD 13A & \cite{Dowdall:2013rya} &  2+1+1 & \gA & \good &\soso & \good & 0.835(8) &  & 1.789(26) \\
\hline
HotQCD 14 & \cite{Bazavov:2014pvz} & 2+1 & \gA & \good & \good & \good & & & 1.7797(67) \\
HotQCD 11 & \cite{Bazavov:2011nk} & 2+1 & \gA   & \good & \good & \good &  & 1.508(5) & \\
RBC/UKQCD 10A &  \cite{Aoki:2010dy} & 2+1 & \gA & \soso & \soso & \soso  & & 1.462(32)$^\#$ & \\
RBC/Bielefeld 07 & \cite{Cheng:2007jq} & 2+1 & \gA & \bad & \good & \good &  & 1.4636(60) & \\
Aubin 04 & \cite{Aubin:2004wf}  & 2+1  &\gA & \soso & \soso & \soso &  & 1.474(7)(18) & \\
\hline\hline
\end{tabular*}\\
\begin{minipage}{1.0\linewidth}
  {\footnotesize
    \;\;$^\#$This value is obtained from $r_1/r_0=0.684(15)(0)(0)$.
  }
  \end{minipage}
}
\caption{Results for dimensionless ratios of scales.
}
\label{tab:scale_ratios} 
\end{table}

\clearpage

\subsection{Averages}
\label{sec:scale averages}
\noindent
{\it Data-driven continuum-limit criterion}

As discussed in Sec.~\ref{sec:DataDriven}, we evaluate the inflation 
factor
\begin{equation}
	s(a)= \max[1, 1 + 2 (\delta(a) - 3) / 3]\,,\quad 
	\delta(a)=\frac{|Q(a)-Q(0)|}{\sigma_Q}\,,
\end{equation}
where $Q$ is the quantity for which we perform an average, 
and $\sigma_Q$ is the uncertainty estimated by the collaboration for its continuum limit. 
If $s(a_\mathrm{min})$ exceeds one, i.e., if the continuum limit is more than three $\sigma_Q$ from the result at
smallest lattice spacing, $a_\mathrm{min}$, the error of the computation is inflated  by $s(a_\mathrm{min})$ before taking the average. For our quantities $s(a_\mathrm{min})=1$ except for few cases. We therefore report explicitly values of $s(a_\mathrm{min})$ only where $s(a_\mathrm{min})>1$.
\\

\noindent
{\it Gradient-flow scale $\sqrt{t_0}$}

For $\Nf=2+1+1$, we have two recent calculations from ETM 21 \cite{Alexandrou:2021bfr} and CalLat 20A \cite{Miller:2020evg}, and two less recent ones from MILC 15 \cite{Bazavov:2015yea} and HPQCD 13A \cite{Dowdall:2013rya} fulfilling the FLAG criteria to enter the average. The latter two and CalLat 20A are based on the same MILC-HISQ gauge-field ensembles, hence we consider their statistical errors to be 100\% correlated. 

 For $\Nf=2+1$, we have four calculations from RQCD 22
\cite{RQCD:2022xux}, CLS 16 \cite{Bruno:2016plf}, RBC/UKQCD 14B
\cite{Blum:2014tka}, and BMW 12A \cite{Borsanyi:2012zs} which enter
the FLAG average. RQCD 22 and CLS 16 are based on the same
  gauge-field ensembles, hence we consider their statistical errors to
  be 100\% correlated. The other two are independent computations, so there is no correlation to be taken into account. QCDSF/UKQCD 15B \cite{Bornyakov:2015eaa} does not contribute to the average, because it is not published.  CLS 21 \cite{Strassberger:2021tsu} is a proceedings contribution  based on double the number of ensembles. It is therefore not a straightforward update and does not supersede CLS 16 \cite{Bruno:2016plf}. 
Performing the weighted and correlated average we obtain 
\begin{align}
&\label{eq:sqrtt0_2p1p1}
\Nf=2+1+1:&\FLAGAVBEGIN \sqrt{t_0} &= 0.14292(104) \FLAGAVEND\text{ fm}
&&\Refs~\mbox{\cite{Alexandrou:2021bfr,Miller:2020evg,Bazavov:2015yea,Dowdall:2013rya}}, \\&\label{eq:sqrtt0_2p1}
\Nf=2+1: &\FLAGAVBEGIN \sqrt{t_0} &= 0.14474(57) \FLAGAVEND \text{ fm} &&\Refs~\mbox{\cite{RQCD:2022xux,Bruno:2016plf,Blum:2014tka,Borsanyi:2012zs}}.
\end{align}
We note that  the $\Nf=2+1+1$ results of staggered fermions and the twisted-mass result are not well compatible. The resulting stretching factor based on the $\chi^2$ value from the weighted average for $\Nf=2+1+1$ is 1.81. It causes the error to be increased compared to FLAG 21. For the $\Nf=2+1$ average the stretching factor is 1.04. We hope that the differences for $\Nf=2+1+1$ get resolved in the near future and the uncertainty of the average decreases.
\\

\noindent
{\it Gradient-flow scale $w_0$}

For $\Nf=1+1+1+1$, including QED,  there is a single calculation, BMW 20 \cite{Borsanyi:2020mff} with the result
\begin{align}
&\label{eq:w0_1p1p1p1}
\Nf=1+1+1+1+\mathrm{QED}:&\FLAGAVBEGIN w_0 &= 0.17236(70)\FLAGAVEND \text{ fm}
&&\Ref~\mbox{\cite{Borsanyi:2020mff}}.
\end{align}

For $\Nf=2+1+1$ we now have four calculations ETM 21 \cite{Alexandrou:2021bfr},  CalLat 20A \cite{Miller:2020evg}, MILC 15 \cite{Bazavov:2015yea}, and HPQCD 13A \cite{Dowdall:2013rya} entering the FLAG average. The proceedings ETM 20 is superseded by ETM 21. As discussed above in connection with $\sqrt{t_0}$, we assume 100\% correlation between the statistical errors of CalLat 20A, MILC 15, and HPQCD 13A. 

For $\Nf=2+1$, we have three calculations RBC/UKQCD 14B \cite{Blum:2014tka}, HotQCD 14  \cite{Bazavov:2014pvz}, and BMW 12A \cite{Borsanyi:2012zs} that enter the FLAG average. These calculations are independent, and no correlation is used. QCDSF/UKQCD 15B \cite{Bornyakov:2015eaa} does not contribute to the average, because it is not published.

Performing the weighted and correlated average, we obtain
\begin{align}
&\label{eq:w0_2p1p1}
\Nf=2+1+1:&\FLAGAVBEGIN w_0 &= 0.17256(103) \FLAGAVEND\text{ fm}
&&\Refs~\mbox{\cite{Alexandrou:2021bfr,Miller:2020evg,Bazavov:2015yea,Dowdall:2013rya}},\\ &\label{eq:w0_2p1}
\Nf=2+1: &\FLAGAVBEGIN w_0 &= 0.17355(92) \FLAGAVEND\text{ fm}  &&\Refs~\mbox{\cite{Blum:2014tka,Bazavov:2014pvz,Borsanyi:2012zs}}.
\end{align}
As above, $\Nf=2+1+1$ results of staggered fermions and the twisted-mass result are not well compatible. The resulting stretching factor based on the $\chi^2$ value from the weighted average is 1.67. It causes the error to be slightly increased compared to FLAG 21.  For the $\Nf=2+1$ average the stretching factor is 1.23. We hope that the differences for $\Nf=2+1+1$ get resolved in the near future and the uncertainty of the average decreases.

Isospin-breaking and electromagnetic corrections are expected to be small at the level of present uncertainties. This is also confirmed by the explicit computation by BMW~12A. Therefore, we also 
perform an average over all $\Nf>2+1$ computations and obtain
\begin{align}
&\label{eq:w0_2p1p1_all}
\Nf>2+1:&\FLAGAVBEGIN w_0 &= 0.17250(70) \FLAGAVEND\text{ fm}
&&\Refs~\mbox{\cite{Alexandrou:2021bfr,Miller:2020evg,Borsanyi:2020mff,Bazavov:2015yea,Dowdall:2013rya}}.
\end{align}
For the $\Nf>2+1$ average the rescaling factor is 1.45.
\\

\noindent
{\it Gradient-flow scale $t_0/w_0$}

Currently, there is only one calculation of the scale $t_0/w_0$ available from ETM 21 \cite{Alexandrou:2021bfr} which forms the FLAG average
\begin{align}
&\label{eq:t0w0_2p1p1}
\Nf=2+1+1:&\FLAGAVBEGIN t_0/w_0 &= 0.11969(62)\FLAGAVEND \text{ fm}
&&\Ref~\mbox{\cite{Alexandrou:2021bfr}}.
\end{align}

\noindent
{\it Potential scale $r_0$}

For $\Nf=2+1+1$, there are two determinations of $r_0$ from
  ETM 14 \cite{Carrasco:2014cwa} and TUMQCD 22
  \cite{Brambilla:2022het}, which contribute to the FLAG average and
  these 
  are uncorrelated.

For $\Nf=2+1$, all but one calculation fulfill all the criteria to enter the FLAG average. HotQCD 14 \cite{Bazavov:2014pvz} is essentially an update of HotQCD 11 \cite{Bazavov:2011nk} by enlarging the set of ensembles used in the computation. Therefore, the result from HotQCD 14 supersedes the one from HotQCD 11 and, hence, we only use the former in the average. The computation of $\chi$QCD \cite{Yang:2014sea} is based on the configurations produced by RBC/UKQCD 10A \cite{Aoki:2010dy}, and we, therefore, assume a 100\% correlation between the statistical errors of the two calculations. HPQCD 05B \cite{Gray:2005ur} enhances the calculation of Aubin 04 \cite{Aubin:2004wf} by adding ensembles at a coarser lattice spacing and using the same discretization for the valence fermion. Therefore, we consider the full errors (statistical and systematic) on the results from Aubin 04 and HPQCD 05B to be 100\% correlated.

Performing the weighted (and correlated) average, we obtain 
\begin{align}
&\label{eq:r0_2p1p1}
\Nf=2+1+1:&\FLAGAVBEGIN r_0 &= 0.4580(73) \FLAGAVEND\text{ fm}
&&\Refs~\mbox{\cite{Brambilla:2022het,Carrasco:2014cwa}},\\ &\label{eq:r0_2p1}
\Nf=2+1: &\FLAGAVBEGIN r_0 &= 0.4701(36) \FLAGAVEND\text{ fm}  &&\Refs~\mbox{\cite{Bazavov:2014pvz,Yang:2014sea,Aoki:2010dy,Gray:2005ur,Aubin:2004wf}}.
\end{align}
We note that for the $\Nf=2+1+1$ average, the stretching factor based
on the $\chi^2$-value from the weighted average is 1.25, while for the
$\Nf=2+1$ average it is 1.14.\\

\noindent
{\it Potential scale $r_1$}
    
For $\Nf=2+1+1$, there are three works that fulfill the criteria to enter the FLAG average, namely TUMQCD 22
  \cite{Brambilla:2022het}, HPQCD 13A \cite{Dowdall:2013rya} and HPQCD
  11B \cite{Dowdall:2011wh}. They are all based on a varying number of
  MILC-HISQ ensembles and we therefore assume 100\% correlation
  between the statistical errors. The result from HPQCD
  13A supersedes the result from HPQCD 11B (in line with a
  corresponding statement in HPQCD 13A), hence TUMQCD 22 and HPQCD 13A form the FLAG average.

For $\Nf=2+1$, all the results quoted in Tab.~\ref{tab_potentialscales} fulfill the FLAG criteria, but not all of them enter the average. The published result from MILC 09 \cite{Bazavov:2009bb} is superseded by the result in the proceedings MILC 10 \cite{Bazavov:2010hj}, while MILC 09A \cite{Bazavov:2009fk} is a proceedings contribution and does not enter the average. HPQCD 09B \cite{Davies:2009tsa} uses HISQ valence quarks instead of asqtad valence quarks as in HPQCD 05B \cite{Gray:2005ur}.
Therefore, we have RBC/UKQCD 10A \cite{Aoki:2010dy}, MILC 10, HPQCD 09B, HPQCD 05B, and Aubin 04 entering the average. However, since the latter four calculations are based on the aqtad MILC ensembles, we attribute 100\% correlation on the statistical error between them and 100\% correlation on the systematic error between HPQCD 05B and Aubin 04 as discussed above in connection with $r_0$.

Performing the weighted and correlated average, we obtain 
\begin{align}
&\label{eq:r1_2p1p1}
\Nf=2+1+1:&\FLAGAVBEGIN r_1 &= 0.3068(37) \FLAGAVEND\text{ fm}
&&\Refs~\mbox{\cite{Brambilla:2022het,Dowdall:2013rya}},\\ &\label{eq:r1_2p1}
\Nf=2+1: &\FLAGAVBEGIN r_1 &= 0.3127(30) \FLAGAVEND\text{ fm}  &&\Refs~\mbox{\cite{Aoki:2010dy,Bazavov:2010hj,Davies:2009tsa,Gray:2005ur,Aubin:2004wf}}. 
\end{align}
We note that for the $\Nf=2+1+1$ average the stretching
  factor based on the $\chi^2$-value from the weighted average is 1.92,
  while for the $\Nf=2+1$ average it is 1.57.
While it is not entirely clear what the reasons are
for the discrepancies encoded in these stretching factors, excited-state contaminations are likely to play a role. Also for the potential, states with additional pions will play an increasingly important role at small pion masses and are not easily captured. \\

\noindent
{\it The scales $M_{p4s}$ and $f_{p4s}$}

    As mentioned in Sec.~\ref{subsubsec:other scales}, these scales have been used only by the MILC and FNAL/MILC collaborations \cite{Bazavov:2014wgs,Bazavov:2017lyh,Bazavov:2012xda}. The latest numbers from Ref.~\cite{Bazavov:2017lyh} are $f_{4ps} = 153.98(11)(^{+2}_{-12})(12)[4]$ MeV and $M_{p4s}=433.12(14)(^{+17}_{-6})(4)[40]$ MeV and, hence, we have 
\begin{align}
  &\Nf=2+1+1: & \FLAGAVBEGIN f_{4ps} &= 153.98(20) \FLAGAVEND\text{ MeV} && \Ref~\mbox{\cite{Bazavov:2017lyh}}, \\
  &\Nf=2+1+1: & \FLAGAVBEGIN M_{4ps} &= 433.12(30) \FLAGAVEND\text{ MeV} && \Ref~\mbox{\cite{Bazavov:2017lyh}}.
\end{align}

\noindent
    {\it Dimensionless ratios of scales}
    
We start with the ratio $\sqrt{t_0}/w_0$ for which two $\Nf=2+1+1$ calculations from ETM 21 \cite{Alexandrou:2021bfr} and HPQCD 13A \cite{Dowdall:2013rya} are available and form the FLAG average 
\begin{align}
  &\Nf=2+1+1: & \sqrt{t_0}/w_0 &= 0.832(6) && \Refs~\mbox{\cite{Dowdall:2013rya,Alexandrou:2021bfr}}.
\end{align}
Here we found a large stretching factor $s(a_\mathrm{min})=12.3$
for \cite{Alexandrou:2021bfr}. 
It was applied to the uncertainty before performing the weighted average and has a large effect.
In fact, in the web-update after FLAG 21 the error was an order of magnitude 
smaller due to the very small error of ETM~21. This is now compensated by the large stretching factor.

For the ratio $r_0/r_1$ there is only one $\Nf=2+1+1$ calculation
available from TUMQCD 22 \cite{Brambilla:2022het}, which fulfills
the FLAG criteria and therefore
forms the FLAG average. For $\Nf=2+1$ there are three calculations from HotQCD 11 \cite{Bazavov:2011nk}, RBC/UKQCD 10A \cite{Aoki:2010dy}, and Aubin 04 \cite{Aubin:2004wf} available. They all fulfill the FLAG criteria and enter the FLAG average of this ratio,
\begin{align}
  &\Nf=2+1+1: & r_0/r_1 &= 1.4968(69) && \Ref~\mbox{\cite{Brambilla:2022het}},   \\
  &\Nf=2+1: & r_0/r_1 &= 1.5049(74) && \Refs~\mbox{\cite{Bazavov:2011nk,Aoki:2010dy,Aubin:2004wf}}. 
\end{align}
We note that for $\Nf=2+1$, the stretching factor based on the $\chi^2$-value from the weighted average is 1.54.

Finally, for the ratio $r_1/w_0$ there is one computation from HotQCD 14  \cite{Bazavov:2014pvz} for $\Nf=2+1+1$, and one from HPQCD 13A \cite{Dowdall:2013rya} for $\Nf=2+1$ fulfilling the FLAG criteria, and, hence, forming the FLAG values
\begin{align}
  &\Nf=2+1+1: & r_1/w_0 &= 1.789(26) && \Ref~\mbox{\cite{Dowdall:2013rya}},\\
  &\Nf=2+1: & r_1/w_0 &= 1.7797(67) && \Ref~\mbox{\cite{Bazavov:2014pvz}}. 
\end{align}

\subsection{Observations and conclusions}
\label{s:Scalconcl}
Unfortunately the different computations for theory scales reported
here are generally not in good agreement within each set of
$\Nf=2+1+1$ and $2+1$ flavour content. As a measure we list here the
stretching factors above one. We remind the reader that their squares
are equal to the $\chi^2/$dof of the weighted averages.
Quantitatively, the stretching factors are for $\Nf=2+1$:
1.2 ($w_0$), 1.1 ($r_0$), 1.6 ($r_1$), 1.5 ($r_0/r_1$).  For
$\Nf=2+1+1$ the numbers are larger: 1.8 ($\sqrt{t_0}$), 1.7 ($w_0$)
1.3 ($r_0$) 1.9 ($r_1$), and due to differences which exist between
present-days twisted-mass QCD results and staggered results.  Of
course, the limited number of large-scale QCD simulations that are
available means that there are only a small number of truly
independent determinations of the scales. For example, three out of
the five computations entering our average for $w_0$ are based on the
same HISQ rooted staggered fermion configurations and thus their
differences are only due to the choice of the physical scale
($M_\Omega$ vs.~$f_\pi$), the valence quark action (M\"obius
domain-wall valence fermions vs.~staggered fermions) employed to
compute it and different analysis of continuum limit, etc.

Due to the publication of ETM~21, differences between $\Nf=2+1$ and
2+1+1 QCD are now smaller and (within their errors) in agreement with
expectations \cite{Bruno:2014ufa,Knechtli:2017xgy}.  The effect of the
charm quark is $-0.6(8)$\% on $w_0$ and $-1.2(9)$\% on $\sqrt{t_0}$ as
computed from the FLAG averages, while precision studies of the decoupling
of charm quarks predicted generic effects of a magnitude of only
$\approx 0.2\%$ \cite{Bruno:2014ufa,Knechtli:2017xgy} for low-energy
quantities. However, the agreement within errors is due to
  large stretching factors. Taking just the individual results, they
  do not agree. The differences are between $\Nf=2+1$ calculations and $2+1+1$
  calculations, but one can also interpret them as a difference
  between staggered fermion simulations and Wilson-type ones. Since
the FLAG averages have changed quite a bit due to one more computation
entering the averages, we are looking forward to further and more
precise results to see whether the numbers hold up over time. In this
respect, it is highly desirable for future computations to also
publish ratios such as $\sqrt{t_0}/w_0$ for which there are 
few numbers so far.

Such ratios of gradient-flow scales are also of high interest in order
to better understand the specific discretization errors of gradient-flow observables. So far, systematic studies and information on the
different contributions (see Sec.~\ref{s:flowscales} and
Ref.~\cite{Ramos:2015baa}) are missing. A worrying result is, for
example, the scale-setting study of Ref.~\cite{Hollwieser:2020qri} on
ratios of scales. The authors find indications that the asymptotic
$\sim a^2$ scaling does not set in before $a\approx0.05~\fm$ and the
$a=0.04~\fm$ data has a relevant influence on their continuum
extrapolations.

A final word concerns the physics scales that all results depend
on. While the mass of the $\Omega$ baryon is more popular than the
leptonic decay rate of the pion, both have systematics which are
difficult to estimate. For the $\Omega$ baryon it is the
contaminations by excited states and for the decay rates it is the QED
effects $\delta f_\pi^\mathrm{isoQCD}$. The uncertainty in $V_{ud}$ is
{\em not} relevant at this stage, but means that one is relying more
on the standard model being an accurate low-energy theory than in the
case of the $\Omega$ mass. In principle, excited-state effects are
controlled by just going to large Euclidean time, but, in practice,
this yields errors that are too large. One, therefore, performs fits
with a very small number of excitations while theoretically there is a
multitude of multi-hadron states that are expected to contribute. For
the leptonic decay rate of the pion, the situation is quite reversed,
namely, the problematic QED contributions have a well-motivated
theory: chiral perturbation theory. The needed combination of
low-energy constants is not accessible from experiment but its
large-$N$ estimate \cite{Cirigliano:2007ga} has been (indirectly)
confirmed by the recent computation of $\delta
f_\pi^\mathrm{isoQCD}$~\cite{DiCarlo:2019thl}. Unfortunately the same
comparison is not so favourable for the leptonic Kaon decay.
\newpage

%% file: acknowledgment.tex
\section*{Acknowledgments}
\addcontentsline{toc}{section}{Acknowledgments}
We are very grateful to the external reviewers for providing detailed comments and suggestions on the draft of this review. These reviewers were
Martin Beneke, Aoife Bharucha, Chris Bouchard, Vincenzo Cirigliano, Martha Constantinou, Evgeny Epelbaum, Martin Gorbahn, Andr\'e Hoang, Christian Hoelbling, Weonjong Lee, Laurent Lellouch, Zoltan Ligeti, Agostino Patella, Chris Sachrajda, and Takeshi Yamazaki.

We are grateful to Anthony Grebe for help in making the figures more suitable for readers with color blindness.
We also wish to thank Silvano Simula for his valuable contributions in the early stages of preparing the report, and Takumi Doi for correspondence and helpful comments.

The kick-off meeting for the present review was held at CERN in January 2023.
The mid-review meeting was held in January 2024 at the Johannes Gutenberg University in Mainz and was supported by the Mainz Institute for Theoretical Physics.
We thank our hosts for their hospitality and financial support is gratefully acknowledged.

Members of FLAG were supported by funding agencies; in particular:
\begin{itemize}

\item X.F. is supported in part by NSFC of China under Grant No. 12125501 and No. 12141501, and National Key Research and Development Program of China under No. 2020YFA0406400.

\item M.G.~was supported in part by the United States Department of Energy grant No.~DE-SC0013682.
    
\item S.G.~acknowledges support from the United States Department of Energy through grant DE-SC0010120.  
  
\item G.H.~acknowledges support from the grant PID2021-127526NB-I00, funded by\linebreak MCIN/AEI/10.13039/501100011033 and by ``ERDF A way of making Europe'', and by the Spanish Research Agency (Agencia Estatal de Investigaci\'on) through grant IFT Centro de Excelencia Severo Ochoa No.~CEX2020-001007-S, funded by MCIN/AEI/10.13039/501100011033.
  
\item C.J.M. was supported in part by the United States Department of Energy under Grant DE-SC0023047 and DE-SC0025908.
  
\item T.K.~was supported in part by JSPS KAKENHI Grant Numbers JP22K21347 and JP23K20846,
and by ``Program for Promoting Researches on the Supercomputer Fugaku''
(Simulation for basic science: approaching the new quantum era, JPMXP1020230411)
through the Joint Institute for Computational Fundamental Science (JICFuS).

\item S.M.~was supported by the United States Department of Energy, Office of Science, Office of High Energy Physics under Award Number DE-SC0009913.

\item A.P.~was supported in part by funding from the European Research Council (ERC) under the European Union’s Horizon 2020 research and innovation programme under grant agreements No.~757646 \& 813942, UK STFC grants ST/X000494/1, and long-term Invitational Fellowship L23530 from the Japan Society for the Promotion of Science.

\item P.P. was supported by the U.S. Department of Energy under Contract No. DE-SC0012704.
  
\item S.R.S.~was supported in part by the United States Department of Energy grant No.~DE-SC0011637.   
\item S.S.~acknowledges support from the EU H2020-MSCA-ITN-2018-813942 (EuroPLEx).  

\item A.V.~was supported by the Spanish Research Agency (Agencia Estatal de Investigación) under Grant No. RYC2020-030244-I / AEI / 10.13039/501100011033.
  
\item U.W.~was supported by the Swiss National Science Foundation (SNSF) project No.~200020\_208222. 
\end{itemize}

%% file: acronyms.tex
\twocolumn
\section{List of acronyms}\label{app:acronyms}
\begin{supertabular}[ht]{p{.21\linewidth}@{\hspace{.02\linewidth}}p{.72\linewidth}}
B$\chi$PT&	  baryonic chiral perturbation theory\\
BCL&	          Bourrely-Caprini-Lellouch\\
BGL&	          Boyd-Grinstein-Lebed\\
BK&	          Becirevic-Kaidalov\\
BSM&	          beyond standard model\\
BZ&	          Ball-Zwicky\\
$\chi$PT&	  chiral perturbation theory\\
CKM&	          Cabibbo-Kobayashi-Maskawa\\
CLN&	          Caprini-Lellouch-Neubert\\
CP&	          charge-parity\\
CPT&	          charge-parity-time reversal\\
CVC&	          conserved vector current\\
DSDR&		  dislocation suppressing determinant ratio\\
DW&	          domain wall\\
DWF&	          domain wall fermion\\
EDM&	          electric dipole moment\\
EFT&	          effective field theory\\
EM&	 	  electromagnetic\\
ESC&	          excited state contributions\\
EW&		  electroweak\\
FCNC&	          flavour-changing neutral current\\
FH&	          Feynman-Hellman\\
FSE&	 	  finite-size effects\\
FV&	          finite volume\\
GF&	          gradient flow\\
GGOU&	          Gambino-Giordano-Ossola-Uraltsev\\
GRS&	          Gasser-Rusetsky-Scimemi\\
HEX&	          hypercubic stout\\
HISQ&	          highly-improved staggered quarks\\
HM$\chi$PT&	  heavy-meson chiral perturbation theory\\
HMC&	          hybrid Monte Carlo\\
HMrS$\chi$PT&	  heavy-meson rooted staggered chiral perturbation theory\\
HQET&	          heavy-quark effective theory\\
IR&	          infrared\\
isoQCD&	          isospin-symmetric QCD\\
LD&	          long distance\\
LEC&	          low-energy constant\\
LO&		  leading order\\
LW&	 	  L\"uscher-Weisz\\
MC&	          Monte Carlo\\
MM&	          minimal MOM\\
MOM&		  momentum subtraction\\
$\msbar$&	  modified  minimal substraction scheme \\
NDR&	          naive dimensional regularization\\
nEDM&	          nucleon electric dipole moment\\
NGB&	          Nambu-Goldstone bosons\\
NLO&	          next-to-leading order\\
NME&	          nucleon matrix elements\\
NNLO&	          next-to-next-to-leading order\\
NP&	          nonperturbative\\
npHQET&		  nonperturbative heavy-quark effective theory\\
NRQCD&	          nonrelativistic QCD\\
NSPT&	          numerical stochastic perturbation theory\\
OPE&	          operator product expansion\\
PCAC&	          partially-conserved axial current\\
PDF&	          parton distribution function\\
PDG&	          particle data group\\
QCD&	          quantum chromodynamics\\
QED&	          quantum electrodynamics\\
QED$_{\rm L}$&	  formulation of QED in finite volume (see~\cite{Hayakawa:2008an}) \\
QED$_{\rm TL}$&	  formulation of QED in finite volume (see~\cite{Duncan:1996be})\\
RG&	          renormalization group\\
RGI&	          renormalization group invariant\\
RH&	          R. Hill\\
RHQ&	          relativistic heavy-quark\\
RHQA&	          relativistic heavy-quark action\\
RI-MOM&		  regularization-independent momentum subtraction (also RI/MOM)\\
RI-SMOM&	  regularization-independent symmetric momentum (also RI/SMOM) \\
RMS&	          root mean square\\
S$\chi$PT&	  staggered chiral perturbation theory\\
SD&	          short distance\\
SF&	          Schr\"odinger functional\\
SIDIS&	          semi-inclusive deep-inelastic scattering\\
SM&	          standard model\\
SSF&		  step-scaling function\\
SUSY&	 	  supersymmetric\\
SW&		  Sheikholeslami-Wohlert\\
UT&	          unitarity triangle\\
UV&	          ultraviolet\\
\end{supertabular}
\onecolumn

%% file: Glossary/gloss_app.tex
\subsection{Lattice actions}\label{sec_lattice_actions}
In this appendix we give brief descriptions of the lattice actions
used in the simulations and summarize their main features.

\subsubsection{Gauge actions \label{sec_gauge_actions}}

The simplest and most widely used discretization of the Yang-Mills
part of the QCD action is the Wilson plaquette action\,\cite{Wilson:1974sk}:
\be
 S_{\rm G} = \beta\sum_{x} \sum_{\mu<\nu}\Big(
  1-\frac{1}{3}{\rm Re\,\Tr}\,W_{\mu\nu}^{1\times1}(x)\Big),
\label{eq_plaquette}
\ee
where $\beta \equiv 6/g_0^2$ (with $g_0$ the bare gauge coupling) and
the plaquette $W_{\mu\nu}^{1\times1}(x)$ is the product of
link variables around an elementary square of the lattice, i.e.,
\be
  W_{\mu\nu}^{1\times1}(x) \equiv U_\mu(x)U_\nu(x+a\hat{\mu})
   U_\mu(x+a\hat{\nu})^{-1} U_\nu(x)^{-1}.
\ee
This expression reproduces the Euclidean Yang-Mills action in the
continuum up to corrections of order~$a^2$.  There is a general
formalism, known as the ``Symanzik improvement programme''
\cite{Symanzik:1983dc,Symanzik:1983gh}, which is designed to cancel
the leading lattice artifacts, such that observables have an
accelerated rate of convergence to the continuum limit.  The
improvement programme is implemented by adding higher-dimensional
operators, whose coefficients must be tuned appropriately in order to
cancel the leading lattice artifacts. The effectiveness of this
procedure depends largely on the method with which the coefficients
are determined. The most widely applied methods (in ascending order of
effectiveness) include perturbation theory, tadpole-improved
(partially resummed) perturbation theory, renormalization group
methods, and the nonperturbative evaluation of improvement
conditions.

In the case of Yang-Mills theory, the simplest version of an improved
lattice action is obtained by adding rectangular $1\times2$ loops to
the plaquette action, i.e.,
\be
   S_{\rm G}^{\rm imp} = \beta\sum_{x}\left\{ c_0\sum_{\mu<\nu}\Big(
  1-\frac{1}{3}{\rm Re\,\Tr}\,W_{\mu\nu}^{1\times1}(x)\Big) +
   c_1\sum_{\mu,\nu} \Big(
  1-\frac{1}{3}{\rm Re\,\Tr}\,W_{\mu\nu}^{1\times2}(x)\Big) \right\},
\label{eq_Sym}
\ee
where the coefficients $c_0, c_1$ satisfy the normalization condition
$c_0+8c_1=1$. The {\sl Symanzik-improved} \cite{Luscher:1984xn},
{\sl Iwasaki} \cite{Iwasaki:1985we}, and {\sl DBW2}
\cite{Takaishi:1996xj,deForcrand:1999bi} actions are all defined
through \eq{eq_Sym} via particular choices for $c_0, c_1$. Details are
listed in Tab.~\ref{tab_gaugeactions} together with the
abbreviations used in the summary tables. Another widely used variant is the {\sl tadpole Symanzik-improved} \cite{Lepage:1992xa,Alford:1995hw} action which is obtained by adding additional 6-link parallelogram loops $W_{\mu\nu\sigma}^{1\times 1\times 1}(x)$ to the action in Eq.~(\ref{eq_Sym}), i.e.,
\be
S_{\rm G}^{\rm tadSym} = S_{\rm G}^{\rm imp} + \beta \sum_x c_2 \sum_{\mu<\nu<\sigma}\Big(1-\frac{1}{3} {\rm Re\,\Tr}\,W_{\mu\nu\sigma}^{1\times1\times1}(x)\Big),
\ee
where
\be
  W_{\mu\nu\sigma}^{1\times1\times1}(x) \equiv U_\mu(x)U_\nu(x+a\hat{\mu})U_\sigma(x+a\hat\mu+a\hat\nu)
   U_\mu(x+a\hat\sigma+a\hat{\nu})^{-1} U_\nu(x+a\hat\sigma)^{-1} U_\sigma(x)^{-1}
\ee
allows for 1-loop improvement \cite{Luscher:1984xn}.

\vspace{-0.07cm}
\begin{table}[!h]
\begin{center}
{\footnotesize
\begin{tabular*}{\textwidth}[l]{l @{\extracolsep{\fill}} c l}
\hline\hline \\[-1.0ex]
Abbrev. & $c_1$ & Description 
\\[1.0ex] \hline \hline \\[-1.0ex]
Wilson    & 0 & Wilson plaquette action \\[1.0ex] \hline \\[-1.0ex]
tlSym   & $-1/12$ & tree-level Symanzik-improved gauge action \\[1.0ex] \hline \\[-1.0ex]
tadSym  & variable & tadpole Symanzik-improved gauge action
 \\[1.0ex] \hline \\[-1.0ex]
Iwasaki & $-0.331$ & Renormalization group improved (``Iwasaki'')
action \\[1.0ex] \hline \\[-1.0ex]
DBW2 & $-1.4088$ & Renormalization group improved (``DBW2'') action 
\\ [1.0ex] 
\hline\hline
\end{tabular*}
}
\caption{Summary of lattice gauge actions. The leading lattice
 artifacts are $\cO(a^2)$ or better for all
  discretizations. \label{tab_gaugeactions}} 
\end{center}
\end{table}


\subsubsection{Light-quark actions \label{sec_quark_actions}}

If one attempts to discretize the quark action, one is faced with the
fermion doubling problem: the naive lattice transcription produces a
16-fold degeneracy of the fermion spectrum. \\

\noindent
{\it Wilson fermions}\\
\noindent

Wilson's solution to the fermion doubling problem is based on adding a
dimension-5 (irrelevant) operator to the lattice action. The
Wilson-Dirac operator for the massless case reads
\cite{Wilson:1974sk,Wilson:1975id}
\be
     D_{\rm w} = \half\gamma_\mu(\nabla_\mu+\nabla_\mu^*)
   +a\nabla_\mu^*\nabla_\mu,
\ee
where $\nabla_\mu,\,\nabla_\mu^*$ denote the covariant forward and
backward lattice derivatives, respectively.  The addition of the
Wilson term $a\nabla_\mu^*\nabla_\mu$, results in fermion doublers
acquiring a mass proportional to the inverse lattice spacing; close to
the continuum limit these extra degrees of freedom are removed from
the low-energy spectrum. However, the Wilson term also results in an
explicit breaking of chiral symmetry even at zero bare quark mass.
Consequently, it also generates divergences proportional to the UV
cutoff (inverse lattice spacing), besides the usual logarithmic
ones. Therefore the chiral limit of the regularized theory is not
defined simply by the vanishing of the bare quark mass but must be
appropriately tuned. As a consequence quark-mass renormalization
requires a power subtraction on top of the standard multiplicative
logarithmic renormalization.  The breaking of chiral symmetry also
implies that the nonrenormalization theorem has to be applied with
care~\cite{Karsten:1980wd,Bochicchio:1985xa}, resulting in a
normalization factor for the axial current which is a regular function
of the bare coupling.  On the other hand, vector symmetry is
unaffected by the Wilson term and thus a lattice (point split) vector
current is conserved and obeys the usual nonrenormalization theorem
with a trivial (unity) normalization factor. Thus, compared to lattice
fermion actions which preserve chiral symmetry, or a subgroup of it,
the Wilson regularization typically results in more complicated
renormalization patterns.

Furthermore, the leading-order lattice artifacts are of order~$a$.
With the help of the Symanzik improvement programme, the leading
artifacts can be cancelled in the action by adding the so-called
``Clover'' or Sheikholeslami-Wohlert (SW) term~\cite{Luscher:1996sc}.
The resulting expression in the massless case reads
\be
   D_{\rm sw} = D_{\rm w}
   +\frac{ia}{4}\,\csw\sigma_{\mu\nu}\widehat{F}_{\mu\nu},
\label{eq_DSW}
\ee
where $\sigma_{\mu\nu}=\frac{i}{2}[\gamma_\mu,\gamma_\nu]$, and
$\widehat{F}_{\mu\nu}$ is a lattice transcription of the gluon field
strength tensor $F_{\mu\nu}$. The coefficient $\csw$ can be determined
perturbatively at tree-level ($\csw = 1$; tree-level improvement or
tlSW for short), via a mean field approach \cite{Lepage:1992xa}
(mean-field improvement or mfSW) or via a nonperturbative approach
\cite{Luscher:1996ug} (nonperturbatively improved or npSW).
Hadron masses, computed using $D_{\rm sw}$, with the coefficient
$\csw$ determined nonperturbatively, will approach the continuum
limit with a rate proportional to~$a^2$; with tlSW for $\csw$ the rate
is proportional to~$g_0^2 a$.

Other observables require additional improvement
coefficients~\cite{Luscher:1996sc}.  A common example consists in the
computation of the matrix element $\langle \alpha \vert Q \vert \beta
\rangle$ of a composite field $Q$ of dimension-$d$ with external
states $\vert \alpha \rangle$ and $\vert \beta \rangle$. In the
simplest cases, the above bare matrix element diverges logarithmically
and a single renormalization parameter $Z_Q$ is adequate to render it
finite. It then approaches the continuum limit with a rate
proportional to the lattice spacing $a$, even when the lattice action
contains the Clover term. In order to reduce discretization errors to
${\cO}(a^2)$, the lattice definition of the composite operator $Q$
must be modified (or ``improved''), by the addition of all
dimension-$(d+1)$ operators with the same lattice symmetries as $Q$.
Each of these terms is accompanied by a coefficient which must be
tuned in a way analogous to that of $\csw$. Once these coefficients
are determined nonperturbatively, the renormalized matrix element of
the improved operator, computed with a npSW action, converges to the
continuum limit with a rate proportional to~$a^2$. A tlSW improvement
of these coefficients and $\csw$ will result in a rate proportional
to~$g_0^2 a$.

It is important to stress that the improvement procedure does not
affect the chiral properties of Wilson fermions; chiral symmetry
remains broken.

Finally, we mention ``twisted-mass QCD'' as a method which was
originally designed to address another problem of Wilson's
discretization: the Wilson-Dirac operator is not protected against the
occurrence of unphysical zero modes, which manifest themselves as
``exceptional'' configurations. They occur with a certain frequency in
numerical simulations with Wilson quarks and can lead to strong
statistical fluctuations. The problem can be cured by introducing a
so-called ``chirally twisted'' mass term. The most common formulation
applies to a flavour doublet $\bar \psi = ( u \quad d)$ of
mass-degenerate quarks, with the fermionic part of the QCD action in
the continuum assuming the form \cite{Frezzotti:2000nk}
\be
   S_{\rm F}^{\rm tm;cont} = \int d^4{x}\, \psibar(x)(\gamma_\mu
   D_\mu +
   m + i\mu_{\rm q}\gamma_5\tau^3)\psi(x).
\ee
Here, $\mu_{\rm q}$ is the twisted-mass parameter, and $\tau^3$ is a
Pauli matrix in flavour space. The standard action in the continuum
can be recovered via a global chiral field rotation. The physical
quark mass is obtained as a function of the two mass parameters $m$
and $\mu_{\rm q}$. The corresponding lattice regularization of twisted-mass QCD (tmWil) for $\Nf=2$ flavours is defined through the fermion
matrix
\be
   D_{\rm w}+m_0+i\mu_{\rm q}\gamma_5\tau^3 \,\, .
\label{eq_tmQCD}
\ee
Although this formulation breaks physical parity and flavour
symmetries, resulting in nondegenerate neutral and charged pions,
is has a number of advantages over standard Wilson
fermions. Firstly, the presence of the twisted-mass parameter
$\mu_{\rm q}$ protects the discretized theory against unphysical zero
modes. A second attractive feature of twisted-mass lattice QCD is the
fact that, once the bare mass parameter $m_0$ is tuned to its ``critical value''
(corresponding to massless pions in the standard Wilson formulation),
the leading lattice artifacts are of order $a^2$ without the
need to add the Sheikholeslami-Wohlert term in the action, or other
improving coefficients~\cite{Frezzotti:2003ni}. A third important advantage
is that, although the problem of explicit chiral
symmetry breaking remains, quantities computed with twisted fermions
with a suitable tuning of the mass parameter $\mu_{\rm q}$,
are subject to renormalization patterns which are simpler than the ones with
standard Wilson fermions. Well known examples are the pseudoscalar decay
constant  and $B_{\rm K}$.\\

\noindent
{\it Staggered Fermions}\\
\noindent

An alternative procedure to deal with the doubling problem is based on Kogut-Susskind fermions \cite{Kogut:1974ag,Banks:1975gq} and is now known under the name ``staggered''
 fermion formulation \cite{Susskind:1976jm, Kawamoto:1981hw,Sharatchandra:1981si}.
Here the degeneracy is only lifted partially, from 16 down to 4.  It has become customary
to refer to these residual doublers as ``tastes'' in order to distinguish them from physical
flavours.  Taste changing interactions 
can occur via the exchange of gluons with one or more components
  of momentum near the cutoff $\pi/a$.  This leads to the breaking of the SU(4) vector symmetry among 
  tastes, thereby generating order $a^2$ lattice artifacts.

The residual doubling of staggered quarks (four tastes per
flavour) is removed by taking a fractional power of the fermion determinant \cite{Marinari:1981qf} --- the ``fourth-root 
procedure,'' or, sometimes, the ``fourth root trick.''  
This procedure would be unproblematic if
the action had full SU(4) taste symmetry, which would give a
Dirac operator that was block-diagonal in taste space.  
However, the breaking of taste symmetry at nonzero lattice spacing leads to a
variety of problems. In fact, the fourth root of the determinant is not equivalent
to the determinant of any local lattice Dirac operator \cite{Bernard:2006ee}.
This in turn leads 
to violations of unitarity 
on the lattice \cite{Prelovsek:2005rf,Bernard:2006zw,Bernard:2007qf,Aubin:2008wk}.

According to standard renormalization group lore, the taste
violations, which are associated with lattice operators of dimension
greater than four, might be expected to go away in the continuum limit,
resulting in the restoration of locality and unitarity.  However,
there is a problem with applying the standard lore to this nonstandard
situation: the usual renormalization group reasoning assumes that the
lattice action is local.  Nevertheless, Shamir
\cite{Shamir:2004zc,Shamir:2006nj} shows that one may apply the
renormalization group to a ``nearby'' local theory, and thereby gives
a strong argument that that the desired local, unitary theory of QCD
is reproduced by the rooted staggered lattice theory in the continuum
limit.

A version of chiral perturbation that includes the lattice artifacts
due to taste violations and rooting (``rooted staggered chiral
perturbation theory'') can also be worked out
\cite{Lee:1999zxa,Aubin:2003mg,Sharpe:2004is} and shown to correctly
describe the unitarity-violating lattice artifacts in the pion sector
\cite{Bernard:2006zw,Bernard:2007ma}.  This provides additional
evidence that the desired continuum limit can be obtained. Further, it
gives a practical method for removing the lattice artifacts from
simulation results. Versions of rooted staggered chiral perturbation
theory exist for heavy-light mesons with staggered light quarks but
nonstaggered heavy quarks \cite{Aubin:2005aq}, heavy-light mesons with
staggered light and heavy quarks
\cite{Komijani:2012fq,Bernard:2013qwa}, staggered baryons
\cite{Bailey:2007iq}, and mixed actions with a staggered sea
\cite{Bar:2005tu,Bae:2010ki}, as well as the pion-only version
referenced above.

There is also considerable numerical evidence that the rooting
procedure works as desired.  This includes investigations in the
Schwinger model \cite{Durr:2003xs,Durr:2004ta,Durr:2006ze}, studies of
the eigenvalues of the Dirac operator in QCD
\cite{Follana:2004sz,Durr:2004as,Wong:2004nk,Donald:2011if}, and
evidence for taste restoration in the pion spectrum as $a\to0$
\cite{Aubin:2004fs,Bazavov:2009bb}.

Issues with the rooting procedure have led Creutz
\cite{Creutz:2006ys,Creutz:2006wv,Creutz:2007yg,Creutz:2007pr,Creutz:2007rk,Creutz:2008kb,Creutz:2008nk}
to argue that the continuum limit of the rooted staggered theory
cannot be QCD.  These objections have however been answered in
Refs.~\cite{Bernard:2006vv,Sharpe:2006re,Bernard:2007eh,Kronfeld:2007ek,Bernard:2008gr,Adams:2008db,Golterman:2008gt,Donald:2011if}. 
In particular, a claim that the continuum 't Hooft
vertex \cite{'tHooft:1976up,'tHooft:1976fv} could not be properly
reproduced by the rooted theory has been refuted
\cite{Bernard:2007eh,Donald:2011if}.

Overall, despite the lack of rigorous proof of the correctness of the
rooting procedure, we think the evidence is strong enough to consider staggered
QCD simulations on a par with simulations using other actions.
See the following reviews for further evidence and discussion:
Refs.~\cite{Durr:2005ax,Sharpe:2006re,Kronfeld:2007ek,Golterman:2008gt,Bazavov:2009bb}.
\\

\noindent
{\it Improved Staggered Fermions}\\
\noindent

An improvement program can be used to suppress taste-changing
interactions, leading to ``improved staggered fermions,'' with the
so-called ``Asqtad'' \cite{Orginos:1999cr}, ``HISQ''
\cite{Follana:2006rc}, ``Stout-smeared'' \cite{Aoki:2005vt}, and
``HYP'' \cite{Hasenfratz:2001hp} actions as the most common versions.
All these actions smear the gauge links in order to reduce the
coupling of high-momentum gluons to the quarks, with the main goal of
decreasing taste-violating interactions. In the Asqtad case, this is
accomplished by replacing the gluon links in the derivatives by
averages over 1-, 3-, 5-, and 7-link paths.  The other actions reduce
taste changing even further by smearing more.  In addition to the
smearing, the Asqtad and HISQ actions include a three-hop term in the
action (the ``Naik term'' \cite{Naik:1986bn}) to remove order $a^2$
errors in the dispersion relation, as well as a ``Lepage term''
\cite{Lepage:1998vj} to cancel other order $a^2$ artifacts introduced
by the smearing.  In both the Asqtad and HISQ actions, the leading
taste violations are of order $\alpha_S^2 a^2$, and ``generic''
lattices artifacts (those associated with discretization errors other
than taste violations) are of order $\alpha_S a^2$.  The overall
coefficients of these errors are, however, significantly smaller with
HISQ than with Asqtad.  With the Stout-smeared and HYP actions, the
errors are formally larger (order $\alpha_S a^2$ for taste violations
and order $a^2$ for generic lattices artifacts).  Nevertheless, the
smearing seems to be very efficient, and the actual size of errors at
accessible lattice spacings appears to be at least as small as with
HISQ.

Although logically distinct from the light-quark improvement program
for these actions, it is customary with the HISQ action to include an
additional correction designed to reduce discretization errors for
heavy quarks (in practice, usually charm quarks)
\cite{Follana:2006rc}. The Naik term is adjusted to remove leading
$(am_c)^4$ and $\alpha_S(am_c)^2$ errors, where $m_c$ is the
charm-quark mass and ``leading'' in this context means leading in
powers of the heavy-quark velocity $v$ ($v/c\sim 1/3$ for $D_s$).
With these improvements, the claim is that one can use the staggered
action for charm quarks, although it must be emphasized that it is not
obvious {\it a priori}\/ how large a value of $am_c$ may be tolerated
for a given desired accuracy, and this must be studied in the
simulations.  \\

\noindent
{\it Ginsparg-Wilson fermions}\\
\noindent

Fermionic lattice actions, which do not suffer from the doubling
problem whilst preserving chiral symmetry go under the name of
``Ginsparg-Wilson fermions''. In the continuum the massless Dirac
operator ($D$) anti-commutes with $\gamma_5$. At nonzero lattice spacing a 
chiral symmetry can be realized if this condition is relaxed
to \cite{Hasenfratz:1998jp,Hasenfratz:1998ri,Luscher:1998pqa}
\be
   \left\{D,\gamma_5\right\} = aD\gamma_5 D,
\label{eq_GWrelation}
\ee
which is now known as the Ginsparg-Wilson relation
\cite{Ginsparg:1981bj}. The Nielsen-Ninomiya
theorem~\cite{Nielsen:1981hk}, which states that any lattice
formulation for which $D$ anticommutes with $\gamma_5$ necessarily has
doubler fermions, is circumvented since $\{D,\gamma_5\}\neq 0$.

A lattice Dirac operator which satisfies \eq{eq_GWrelation} can be
constructed in several ways. The so-called ``overlap'' or
Neuberger-Dirac operator~\cite{Neuberger:1997fp} acts in four
space-time dimensions and is, in its simplest form, defined by
\be
   D_{\rm N} = \frac{1}{\abar} \left( 1-\epsilon(A)
   \right),\quad\mathrm{where}\quad\epsilon(A)\equiv A (A^\dagger A)^{-1/2}, \quad A=1+s-aD_{\rm w},\quad \abar=\frac{a}{1+s},
\label{eq_overlap}
\ee
$D_{\rm w}$ is the massless Wilson-Dirac operator and $|s|<1$
is a tunable parameter. The overlap operator $D_{\rm N}$ removes all
doublers from the spectrum, and can readily be shown to satisfy the
Ginsparg-Wilson relation. The occurrence of the sign function $\epsilon(A)$ in
$D_{\rm N}$ renders the application of $D_{\rm N}$ in a computer
program potentially very costly, since it must be implemented using,
for instance, a polynomial approximation.

The most widely used approach to satisfying the Ginsparg-Wilson
relation \eq{eq_GWrelation} in large-scale numerical simulations is
provided by \textit{Domain Wall Fermions}
(DWF)~\cite{Kaplan:1992bt,Shamir:1993zy,Furman:1994ky} and we
therefore describe this in some more detail. Following early
exploratory studies~\cite{Blum:1996jf}. this approach has been
developed into a practical formulation of lattice QCD with good chiral
and flavour symmetries leading to results which contribute
significantly to this review. In this formulation, the fermion fields
$\psi(x,s)$ depend on a discrete fifth coordinate $s=1,\ldots,N$ as well as
the physical 4-dimensional space-time coordinates $x_\mu,\,\mu=1\cdots
4$ (the gluon fields do not depend on $s$). The lattice on which the
simulations are performed, is therefore a five-dimensional one of size
$L^3\times T\times N$, where $L,\,T$ and $N$ represent the number of
points in the spatial, temporal and fifth dimensions respectively.
The remarkable feature of DWF is that for each flavour there exists a
physical light mode corresponding to the field $q(x)$:
\begin{eqnarray}
q(x)&=&\frac{1+\gamma^5}{2}\psi(x,1)+\frac{1-\gamma^5}{2}\psi(x,N)\\
\bar{q}(x)&=&\overline{\psi}(x,N)\frac{1+\gamma^5}{2} + \overline{\psi}(x,1)\frac{1-\gamma^5}{2}\,.
\end{eqnarray}
The left and right-handed modes of the physical field are located on
opposite boundaries in the 5th dimensional space which, for
$N\to\infty$, allows for independent transformations of the left and
right components of the quark fields, that is for chiral
transformations. Unlike Wilson fermions, where for each flavour the
quark-mass parameter in the action is fine-tuned requiring a
subtraction of contributions of $\cO(1/a)$ where $a$ is the lattice
spacing, with DWF no such subtraction is necessary for the physical
modes, whereas the unphysical modes have masses of $\cO(1/a)$ and
decouple.

In actual simulations $N$ is finite and there are small violations of
chiral symmetry which must be accounted for. The theoretical framework
for the study of the residual breaking of chiral symmetry has been a
subject of intensive investigation (for a review and references to the
original literature see, e.g., \cite{Sharpe:2007yd}). The breaking
requires one or more \emph{crossings} of the fifth dimension to couple
the left and right-handed modes; the more crossings that are required
the smaller the effect.  For many physical quantities the leading
effects of chiral symmetry breaking due to finite $N$ are parameterized
by a \emph{residual} mass, $m_{\mathrm{res}}$.  For example, the PCAC
relation (for degenerate quarks of mass $m$) $\partial_\mu A_\mu(x) =
2m P(x)$, where $A_\mu$ and $P$ represent the axial current and
pseudoscalar density respectively, is satisfied with
$m=m^\mathrm{DWF}+m_\mathrm{res}$, where $m^\mathrm{DWF}$ is the bare
mass in the DWF action. The mixing of operators which transform under
different representations of chiral symmetry is found to be negligibly
small in current simulations. The important thing to note is that the
chiral symmetry breaking effects are small and that there are
techniques to mitigate their consequences.

The main price which has to be paid for the good chiral symmetry is
that the simulations are performed in 5 dimensions, requiring
approximately a factor of $N$ in computing resources and resulting in
practice in ensembles at fewer values of the lattice spacing and quark
masses than is possible with other formulations. The current
generation of DWF simulations is being performed at physical quark
masses so that ensembles with good chiral and flavour symmetries are
being generated and analyzed~\cite{Arthur:2012opa}. For a discussion
of the equivalence of DWF and overlap fermions
see Refs.~\cite{Borici:1999zw,Borici:1999da}.

A third example of an operator which satisfies the Ginsparg-Wilson
relation is the so-called fixed-point action
\cite{Bietenholz:1995cy,Hasenfratz:2000xz,Hasenfratz:2002rp}. This
construction proceeds via a renormalization group approach. A related
formalism are the so-called ``chirally improved'' fermions
\cite{Gattringer:2000js}.\\

\begin{table}
\begin{center}
{\footnotesize
\begin{tabular*}{\textwidth}[l]{l @{\extracolsep{\fill}} l l l l}
\hline \hline  \\[-1.0ex]
\parbox[t]{1.5cm}{Abbrev.} & Discretization & \parbox[t]{2.2cm}{Leading lattice \\artifacts} & Chiral symmetry &  Remarks
\\[4.0ex] \hline \hline \\[-1.0ex]
Wilson     & Wilson & $\cO(a)$ & broken & 
\\[1.0ex] \hline \\[-1.0ex]
tmWil   & twisted-mass Wilson &  \parbox[t]{2.2cm}{$\cO(a^2)$
at\\ maximal twist} & broken & \parbox[t]{5cm}{flavour-symmetry breaking:\\ $(M_\text{PS}^{0})^2-(M_\text{PS}^\pm)^2\sim \cO(a^2)$}
\\[4.0ex] \hline \\[-1.0ex]
tlSW      & Sheikholeslami-Wohlert & $\cO(g^2 a)$ & broken & tree-level
impr., $\csw=1$
\\[1.0ex] \hline \\[-1.0ex]
\parbox[t]{1.0cm}{n-HYP tlSW}      & Sheikholeslami-Wohlert & $\cO(g^2 a)$ & broken & \parbox[t]{5cm}{tree-level
impr., $\csw=1$,\\
n-HYP smeared gauge links
}
\\[4.0ex] \hline \\[-1.0ex]
\parbox[t]{1.2cm}{stout tlSW}      & Sheikholeslami-Wohlert & $\cO(g^2 a)$ & broken & \parbox[t]{5cm}{tree-level
impr., $\csw=1$,\\
stout smeared gauge links
}
\\[4.0ex] \hline \\[-1.0ex]
\parbox[t]{1.2cm}{HEX tlSW}      & Sheikholeslami-Wohlert & $\cO(g^2 a)$ & broken & \parbox[t]{5cm}{tree-level
impr., $\csw=1$,\\
HEX smeared gauge links
}
\\[4.0ex] \hline \\[-1.0ex]
mfSW      & Sheikholeslami-Wohlert & $\cO(g^2 a)$ & broken & mean-field impr.
\\[1.0ex] \hline \\[-1.0ex]
npSW      & Sheikholeslami-Wohlert & $\cO(a^2)$ & broken & nonperturbatively impr.
\\[1.0ex] \hline \\[-1.0ex]
KS      & Staggered & $\cO(a^2)$ & \parbox[t]{3cm}{$\rm
  U(1)\times U(1)$ subgr.\\ unbroken} & rooting for $\Nf<4$
\\[4.0ex] \hline \\[-1.0ex]
Asqtad  & Staggered & $\cO(g^2a^2)$ & \parbox[t]{3cm}{$\rm
  U(1)\times U(1)$ subgr.\\ unbroken}  & \parbox[t]{5cm}{Asqtad
  smeared gauge links, \\rooting for $\Nf<4$}  
\\[4.0ex] \hline \\[-1.0ex]
HISQ  & Staggered & $\cO(g^2a^2)$ & \parbox[t]{3cm}{$\rm
  U(1)\times U(1)$ subgr.\\ unbroken}  & \parbox[t]{5cm}{HISQ
  smeared gauge links, \\rooting for $\Nf<4$}  
\\[4.0ex] \hline \\[-1.0ex]
DW      & Domain Wall & \parbox[t]{2.2cm}{asymptotically \\$\cO(a^2)$} & \parbox[t]{3cm}{remnant
  breaking \\exponentially suppr.} & \parbox[t]{5cm}{exact chiral symmetry and\\$\cO(a)$ impr. only in the limit \\
 $N\rightarrow \infty$}
\\[7.0ex] \hline \\[-1.0ex]
oDW      & optimal Domain Wall & \parbox[t]{2.2cm}{asymptotically \\$\cO(a^2)$} & \parbox[t]{3cm}{remnant
  breaking \\exponentially suppr.} & \parbox[t]{5cm}{exact chiral symmetry and\\$\cO(a)$ impr. only in the limit \\
 $N\rightarrow \infty$}
\\[7.0ex] \hline \\[-1.0ex]
M-DW      & Moebius Domain Wall & \parbox[t]{2.2cm}{asymptotically \\$\cO(a^2)$} & \parbox[t]{3cm}{remnant
  breaking \\exponentially suppr.} & \parbox[t]{5cm}{exact chiral symmetry and\\$\cO(a)$ impr. only in the limit \\
 $N\rightarrow \infty$}
\\[7.0ex] \hline \\[-1.0ex]
overlap    & Neuberger & $\cO(a^2)$ & exact
\\[1.0ex] 
\hline\hline
\end{tabular*}
}
\caption{The most widely used discretizations of the quark action
  and some of their properties. Note that in order to maintain the
  leading lattice artifacts of the action in nonspectral observables
  (like operator matrix elements)
  the corresponding nonspectral operators need to be improved as well. 
\label{tab_quarkactions}}
\end{center}
\end{table}

\noindent
{\it Smearing}\\
\noindent

A simple modification which can help improve the action as well as the
computational performance is the use of smeared gauge fields in the
covariant derivatives of the fermionic action. Any smearing procedure
is acceptable as long as it consists of only adding irrelevant (local)
operators. Moreover, it can be combined with any discretization of the
quark action.  The ``Asqtad'' staggered quark action mentioned above
\cite{Orginos:1999cr} is an example which makes use of so-called
``Asqtad'' smeared (or ``fat'') links. Another example is the use of
n-HYP smeared \cite{Hasenfratz:2001hp,Hasenfratz:2007rf}, stout smeared
\cite{Morningstar:2003gk,Durr:2008rw} or HEX (hypercubic stout) smeared \cite{Capitani:2006ni} gauge links in the tree-level clover improved
discretization of the quark action, denoted by ``n-HYP tlSW'',
``stout tlSW'' and ``HEX tlSW'' in the following.\\

\noindent
In Tab.~\ref{tab_quarkactions} we summarize the most widely used
discretizations of the quark action and their main properties together
with the abbreviations used in the summary tables. Note that in order
to maintain the leading lattice artifacts of the actions as given in
the table in nonspectral observables (like operator matrix elements)
the corresponding nonspectral operators need to be improved as well.

\subsubsection{Heavy-quark actions}
\label{app:HQactions}

Charm and bottom quarks are often simulated with different
lattice-quark actions than up, down, and strange quarks because their
masses are large relative to typical lattice spacings in current
simulations; for example, $a m_c \sim 0.4$ and $am_b \sim 1.3$ at
$a=0.06$~fm.  Therefore, for the actions described in the previous
section, using a sufficiently small lattice spacing to control generic
$(am_h)^n$ discretization errors at the physical $b$-quark mass is
computationally demanding and has so far not been possible, with the
first exception being the calculation of FNAL/MILC in
\cite{Bazavov:2017lyh} which uses the HISQ action and a lattice spacing of $a \approx 0.03$\,fm.

One alternative approach for lattice heavy quarks is direct application of
effective theory.  In this case the lattice heavy-quark action only
correctly describes phenomena in a specific kinematic regime, such as
Heavy-Quark Effective Theory
(HQET)~\cite{Isgur:1989vq,Eichten:1989zv,Isgur:1989ed} or
Nonrelativistic QCD (NRQCD)~\cite{Caswell:1985ui,Bodwin:1994jh}.  One
can discretize the effective Lagrangian to obtain, for example,
Lattice HQET~\cite{Heitger:2003nj} or Lattice
NRQCD~\cite{Thacker:1990bm,Lepage:1992tx}, and then simulate the
effective theory numerically.  The coefficients of the operators in
the lattice-HQET and lattice-NRQCD actions are free parameters that
must be determined by matching to the underlying theory (QCD) through
the chosen order in $1/m_h$ or $v_h^2$, where $m_h$ is the heavy-quark
mass and $v_h$ is the heavy-quark velocity in the the heavy-light
meson rest frame.

Another approach is to interpret a relativistic quark action such as
those described in the previous section in a manner suitable for heavy
quarks.  One can extend the standard Symanzik improvement program,
which allows one to systematically remove lattice cutoff effects by
adding higher-dimension operators to the action, by allowing the
coefficients of the dimension 4 and higher operators to depend
explicitly upon the heavy-quark mass.  Different prescriptions for
tuning the parameters correspond to different implementations: those
in common use are often called the Fermilab
action~\cite{ElKhadra:1996mp}, the relativistic heavy-quark action
(RHQ)~\cite{Christ:2006us}, and the Tsukuba
formulation~\cite{Aoki:2001ra}.  In the Fermilab approach, HQET is
used to match the lattice theory to continuum QCD at the desired order
in $1/m_h$.

More generally, effective theory can be used to estimate the size of
cutoff errors from the various lattice heavy-quark actions.  The power
counting for the sizes of operators with heavy quarks depends on the
typical momenta of the heavy quarks in the system.  Bound-state
dynamics differ considerably between heavy-heavy and heavy-light
systems.  In heavy-light systems, the heavy quark provides an
approximately static source for the attractive binding force, like the
proton in a hydrogen atom.  The typical heavy-quark momentum in the
bound-state rest frame is $|\vec{p}_h| \sim \Lambda_{\rm QCD}$, and
heavy-light operators scale as powers of $(\Lambda_{\rm QCD}/m_h)^n$.
This is often called ``HQET power-counting'', although it applies to
heavy-light operators in HQET, NRQCD, and even relativistic
heavy-quark actions described below.  Heavy-heavy systems are similar
to positronium or the deuteron, with the typical heavy-quark momentum
$|\vec{p}_h| \sim \alpha_S m_h$.  Therefore motion of the heavy quarks
in the bound state rest frame cannot be neglected.  Heavy-heavy
operators have complicated power counting rules in terms of
$v_h^2$~\cite{Lepage:1992tx}; this is often called ``NRQCD power
counting.''

Alternatively, one can simulate bottom or charm quarks with the same
action as up, down, and strange quarks provided that (1) the action is
sufficiently improved, and (2) the lattice spacing is sufficiently
fine.  These qualitative criteria do not specify precisely how large a
numerical value of $am_h$ can be allowed while obtaining a given
precision for physical quantities; this must be established
empirically in numerical simulations.  At present, both the HISQ and
twisted-mass Wilson actions discussed previously are being used to
simulate charm quarks.
Simulations with HISQ quarks have employed heavier-quark masses than
those with twisted-mass Wilson quarks because the action is more
highly improved, but neither action has been used to simulate at the
physical $am_b$ until the recent calculation of FNAL/MILC in
\cite{Bazavov:2017lyh}, where a lattice spacing of $a \approx 0.03$\,fm is available.
All other calculations
of heavy-light decay constants with these actions still rely on
effective theories: the ETM collaboration
interpolates between twisted-mass Wilson data generated near $am_c$
and the static point~\cite{Dimopoulos:2011gx}, while the HPQCD
collaboration, for the coarser lattice spacings,  
extrapolates HISQ data generated below $am_b$ up to the
physical point using an HQET-inspired series expansion in
$(1/m_h)^n$~\cite{McNeile:2011ng}.
\\


\noindent
{\it Heavy-quark effective theory}\\
\noindent

HQET was introduced by Eichten and Hill in
Ref.~\cite{Eichten:1989zv}. It provides the correct asymptotic
description of QCD correlation functions in the static limit
$m_{h}/|\vec{p}_h| \!\to\! \infty$. Subleading effects are described
by higher dimensional operators whose coupling constants are formally
of ${\cO}((1/m_{h})^n)$.  The HQET expansion works well for
heavy-light systems in which the heavy-quark momentum is small
compared to the mass.

The HQET Lagrangian density at the leading (static) order in the rest
frame of the heavy quark is given by
\be
{\mathcal L}^{\rm stat}(x) = \overline{\psi}_{h}(x) \,D_0\, \psi_{h}(x)\;,
\ee
with
\be
P_+ \psi_{h} = \psi_{h} \; , \quad\quad \overline{\psi}_{h} P_+=\overline{\psi}_{h} \;,  
\quad\quad P_+={{1+\gamma_0}\over{2}} \;.
\ee
A bare quark mass $m_{\rm bare}^{\rm stat}$ has to be added to the energy  
levels $E^{\rm stat}$ computed with this Lagrangian to obtain the physical ones.
 For example, the mass of the $B$ meson in the static approximation is given by
\be
m_{B} = E^{\rm stat} + m_{\rm bare}^{\rm stat} \;.
\ee
At tree-level $m_{\rm bare}^{\rm stat}$ is simply the (static approximation of
the) $b$-quark mass, but in the quantized lattice formulation it has
to further compensate a divergence linear in the inverse lattice spacing.
Weak composite fields  are also rewritten in terms of the static fields, e.g.,
\begin{equation}
A_0(x)^{\rm stat}=Z_{\rm A}^{\rm stat} \left( \overline{\psi}(x) \gamma_0\gamma_5\psi_h(x)\right)\;,
\end{equation}
where the renormalization factor of the axial current in the static
theory $Z_{\rm A}^{\rm stat}$ is scale-dependent.  Recent lattice-QCD
calculations using static $b$ quarks and dynamical light
quarks \cite{Albertus:2010nm,Dimopoulos:2011gx} perform the operator
matching at 1-loop in mean-field improved lattice perturbation
theory~\cite{Ishikawa:2011dd,Blossier:2011dg}.  Therefore the
heavy-quark discretization, truncation, and matching errors in these
results are of ${\cO}(a^2 \Lambda_{\rm QCD}^2)$, ${\cO}
(\Lambda_{\rm QCD}/m_h)$, and ${\cO}(\alpha_s^2, \alpha_s^2
a \Lambda_{\rm QCD})$.

In order to reduce heavy-quark truncation errors in $B$-meson masses
and matrix elements to the few-percent level, state-of-the-art
lattice-HQET computations now include corrections of ${\cO}(1/m_h)$.  Adding the $1/m_{h}$ terms, the HQET Lagrangian reads
\begin{eqnarray}
{\mathcal L}^{\rm HQET}(x) &=&  {\mathcal L}^{\rm stat}(x) - \omegakin{\mathcal{O}}_{\rm kin}(x)
        - \omegaspin{\mathcal{O}}_{\rm spin}(x)  \,, \\[2.0ex]
  \mathcal{O}_{\rm kin}(x) &=& \overline{\psi}_{h}(x){\bf D}^2\psi_{h}(x) \,,\quad
  \mathcal{O}_{\rm spin}(x) = \overline{\psi}_{h}(x){\boldsymbol\sigma}\!\cdot\!{\bf B}\psi_{h}(x)\,.
\end{eqnarray}
At this order, two other parameters appear in the Lagrangian,
$\omegakin$ and $\omegaspin$. The normalization is such that the
tree-level values of the coefficients are
$\omegakin=\omegaspin=1/(2m_{h})$.  Similarly the operators are
formally expanded in inverse powers of the heavy-quark mass.  The time
component of the axial current, relevant for the computation of
mesonic decay constants is given by
\begin{eqnarray}
A_0^{\rm HQET}(x) &=& Z_{\rm A}^{\rm HQET}\left(A_0^{\rm stat}(x) +\sum_{i=1}^2 c_{\rm A}^{(i)} A_0^{(i)}(x)\right)\;, \\
A_0^{(1)}(x)&=&\overline{\psi}\frac{1}{2}\gamma_5 \gamma_k  (\nabla_k-\overleftarrow{\nabla}_k)\psi_h(x), \qquad k=1,2,3\\
A_0^{(2)} &=& -\partial_kA_k^{\rm stat}(x)\;, \quad A_k^{\rm stat}=\overline{\psi}(x) \gamma_k\gamma_5\psi_h(x)\;,
\end{eqnarray}
and depends on two additional parameters $c_{\rm A}^{(1)}$ and $c_{\rm A}^{(2)}$.

A framework for nonperturbative HQET on the lattice has been
introduced in Refs.~\cite{Heitger:2003nj,Blossier:2010jk}.  As pointed out
in Refs.~\cite{Sommer:2006sj,DellaMorte:2007ny}, since $\alpha_s(m_h)$
decreases logarithmically with $m_h$, whereas corrections in the
effective theory are power-like in $\Lambda/m_h$, it is possible that
the leading errors in a calculation will be due to the perturbative
matching of the action and the currents at a given order
$(\Lambda/m_h)^l$ rather than to the missing ${\cO}((\Lambda/m_h)^{l+1})$ terms.  Thus, in order to keep matching
errors below the uncertainty due to truncating the HQET expansion, the
matching is performed nonperturbatively beyond leading order in
$1/m_{h}$. The asymptotic convergence of HQET in the limit
$m_h \to \infty$ indeed holds only in that case.

The higher dimensional interaction terms in the effective Lagrangian
are treated as space-time volume insertions into static correlation
functions.  For correlators of some multi-local fields ${\oO}$
and up to the $1/m_h$ corrections to the operator, this means
\begin{equation}
\langle {\oO} \rangle =\langle {\oO} \rangle_{\rm stat} +\omegakin a^4 \sum_x
\langle {\oO\mathcal{O}}_{\rm kin}(x) \rangle_{\rm stat} + \omegaspin a^4 \sum_x
\langle {\oO\mathcal{O}}_{\rm spin}(x) \rangle_{\rm stat} \;, 
\end{equation}
where $\langle {\oO} \rangle_{\rm stat}$ denotes the static
expectation value with ${\mathcal{L}}^{\rm stat}(x)
+{\mathcal{L}}^{\rm light}(x)$.  Nonperturbative renormalization of
these correlators guarantees the existence of a well-defined continuum
limit to any order in $1/m_h$.  The parameters of the effective action
and operators are then determined by matching a suitable number of
observables calculated in HQET (to a given order in $1/m_{h}$) and in
QCD in a small volume (typically with $L\simeq 0.5$ fm), where the
full relativistic dynamics of the $b$-quark can be simulated and the
parameters can be computed with good accuracy.
In Refs.~\cite{Blossier:2010jk,Blossier:2012qu} the Schr\"odinger Functional
(SF) setup has been adopted to define a set of quantities, given by
the small volume equivalent of decay constants, pseudoscalar-vector
splittings, effective masses and ratio of correlation functions for
different kinematics, that can be used to implement the matching
conditions.  The kinematical conditions are usually modified by
changing the periodicity in space of the fermions, i.e., by directly
exploiting a finite-volume effect.  The new scale $L$, which is
introduced in this way, is chosen such that higher orders in $1/m_hL$
and in $\Lambda_{\rm QCD}/m_h$ are of about the same size. At the end
of the matching step the parameters are known at lattice spacings
which are of the order of $0.01$ fm, significantly smaller than the
resolutions used for large volume, phenomenological, applications. For
this reason a set of SF-step scaling functions is introduced in the
effective theory to evolve the parameters to larger lattice spacings.
The whole procedure yields the nonperturbative parameters with an
accuracy which allows to compute phenomenological quantities with a
precision of a few percent
(see Refs.~\cite{Blossier:2010mk,Bernardoni:2012ti} for the case of the
$B_{(s)}$ decay constants).  Such an accuracy can not be achieved by
performing the nonperturbative matching in large volume against
experimental measurements, which in addition would reduce the
predictivity of the theory.  For the lattice-HQET action matched
nonperturbatively through ${\cO}(1/m_h)$, discretization and
truncation errors are of ${\cO}(a \Lambda^2_{\rm QCD}/m_h,
a^2 \Lambda^2_{\rm QCD})$ and ${\cO}((\Lambda_{\rm QCD}/m_h )^2)$.

The noise-to-signal ratio of static-light correlation functions grows
exponentially in Euclidean time, $\propto e^{\mu x_0}$ . The rate
$\mu$ is nonuniversal but diverges as $1/a$ as one approaches the
continuum limit. By changing the discretization of the covariant
derivative in the static action one may achieve an exponential
reduction of the noise to signal ratio. Such a strategy led to the
introduction of the $S^{\rm stat}_{\rm HYP1,2}$
actions~\cite{DellaMorte:2005nwx}, where the thin links in $D_0$ are
replaced by HYP-smeared links~\cite{Hasenfratz:2001hp}.  These actions
are now used in all lattice applications of HQET.
\\


\noindent
{\it Nonrelativistic QCD}\\
\noindent

Nonrelativistic QCD (NRQCD) \cite{Thacker:1990bm,Lepage:1992tx} is an
 effective theory that can be matched to full QCD order by order in
 the heavy-quark velocity $v_h^2$ (for heavy-heavy systems) or in
 $\Lambda_{\rm QCD}/m_h$ (for heavy-light systems) and in powers of
 $\alpha_s$.  Relativistic corrections appear as higher-dimensional
 operators in the Hamiltonian.
 
 As an effective field theory, NRQCD is only useful with an
 ultraviolet cutoff of order $m_h$ or less. On the lattice this means
 that it can be used only for $am_h>1$, which means that $\cO(a^n)$
 errors cannot be removed by taking $a\to0$ at fixed $m_h$. Instead
 heavy-quark discretization errors are systematically removed by
 adding additional operators to the lattice Hamiltonian.  Thus, while
 strictly speaking no continuum limit exists at fixed $m_h$, continuum
 physics can be obtained at finite lattice spacing to arbitrarily high
 precision provided enough terms are included, and provided that the
 coefficients of these terms are calculated with sufficient accuracy.
 Residual discretization errors can be parameterized as corrections to
 the coefficients in the nonrelativistic expansion, as shown in
 Eq.~(\ref{deltaH}).  Typically they are of the form
 $(a|\vec{p}_h|)^n$ multiplied by a function of $am_h$ that is smooth
 over the limited range of heavy-quark masses (with $am_h > 1$) used
 in simulations, and can therefore can be represented by a low-order
 polynomial in $am_h$ by Taylor's theorem (see
 Ref.~\cite{Gregory:2010gm} for further discussion).  Power-counting
 estimates of these effects can be compared to the observed lattice-spacing dependence in simulations. Provided that these effects are
 small, such comparisons can be used to estimate and correct the
 residual discretization effects.

An important feature of the NRQCD approach is that the same action can
be applied to both heavy-heavy and heavy-light systems. This allows,
for instance, the bare $b$-quark mass to be fixed via experimental
input from $\Upsilon$ so that simulations carried out in the $B$ or
$B_s$ systems have no adjustable parameters left.  Precision
calculations of the $B_s$-meson mass (or of the mass splitting
$M_{B_s} - M_\Upsilon/2$) can then be used to test the reliability of
the method before turning to quantities one is trying to predict, such
as decay constants $f_B$ and $f_{B_s}$, semileptonic form factors or
neutral $B$ mixing parameters.

Given the same lattice-NRQCD heavy-quark action, simulation results
will not be as accurate for charm quarks as for bottom ($1/m_b <
1/m_c$, and $v_b < v_c$ in heavy-heavy systems).  For charm, however,
a more serious concern is the restriction that $am_h$ must be greater
than one.  This limits lattice-NRQCD simulations at the physical
$am_c$ to relatively coarse lattice spacings for which light-quark and
gluon discretization errors could be large.  Thus recent lattice-NRQCD
simulations have focused on bottom quarks because $am_b > 1$ in the
range of typical lattice spacings between $\approx$ 0.06 and 0.15~fm.

In most simulations with NRQCD $b$-quarks during the past decade one
has worked with an NRQCD action that includes tree-level relativistic
corrections through ${\cO}(v_h^4)$ and discretization corrections
through ${\cO}(a^2)$,
 \begin{eqnarray}
 \label{nrqcdact}
&&  S_{\rm NRQCD}  =
a^4 \sum_x \Bigg\{  {\Psi}^\dagger_t \Psi_t -
 {\Psi}^\dagger_t
\left(1 \!-\!\frac{a \delta H}{2}\right)_t
 \left(1\!-\!\frac{aH_0}{2n}\right)^{n}_t \nonumber \\
& \times &
 U^\dagger_t(t-a)
 \left(1\!-\!\frac{aH_0}{2n}\right)^{n}_{t-a}
\left(1\!-\!\frac{a\delta H}{2}\right)_{t-a} \Psi_{t-a} \Bigg\} \, ,
 \end{eqnarray}
where the subscripts $``t''$ and $``t-a''$ denote that the heavy-quark, gauge, $\bf{E}$,  and $\bf{B}$-fields are on time slices $t$ or $t-a$, respectively.
 $H_0$ is the nonrelativistic kinetic energy operator,
 \be
 H_0 = - {\delsq\over2m_h} \, ,
 \ee
and $\delta H$ includes relativistic and finite-lattice-spacing
corrections,
 \begin{eqnarray}
\delta H
&=& - c_1\,\frac{(\delsq)^2}{8m_h^3}
+ c_2\,\frac{i g}{8m_h^2}\left(\delv\cdot\Ev - \Ev\cdot\delv\right) \nl
& &
 - c_3\,\frac{g}{8m_h^2} \sigmav\cdot(\delvt\times\Ev - \Ev\times\delvt)\nl
& & - c_4\,\frac{g}{2m_h}\,\sigmav\cdot\Bv
  + c_5\,\frac{a^2\delfour}{24m_h}  - c_6\,\frac{a(\delsq)^2}
{16nm_h^2} \, .
\label{deltaH}
\end{eqnarray}
 $m_h$ is the bare heavy-quark mass, $\delsq$ the lattice Laplacian,
$\delv$ the symmetric lattice derivative and $\delfour$ the lattice
discretization of the continuum $\sum_i D^4_i$.  $\delvt$ is the
improved symmetric lattice derivative and the $\Ev$ and $\Bv$ fields
have been improved beyond the usual clover leaf construction. The
stability parameter $n$ is discussed in Ref.~\cite{Lepage:1992tx}.  In most
cases the $c_i$'s have been set equal to their tree-level values $c_i
= 1$.  With this implementation of the NRQCD action, errors in
heavy-light-meson masses and splittings are of ${\cO}(\alpha_S \Lambda_{\rm QCD}/m_h )$, ${\cO}(\alpha_S (\Lambda_{\rm
QCD}/m_h)^2 )$, ${\cO}((\Lambda_{\rm QCD}/m_h )^3)$, and ${\cO}(\alpha_s a^2 \Lambda_{\rm QCD}^2)$, with coefficients that are
functions of $am_h$.  1-loop corrections to many of the coefficients
in Eq.~(\ref{deltaH}) have now been calculated, and are starting to be
included in
simulations \cite{Morningstar:1994qe,Hammant:2011bt,Dowdall:2011wh}.

Most of the operator matchings involving heavy-light currents or
four-fermion operators with NRQCD $b$-quarks and Asqtad or HISQ light
quarks have been carried out at 1-loop order in lattice perturbation
theory.  In calculations published to date of electroweak matrix
elements, heavy-light currents with massless light quarks have been
matched through ${\cO}(\alpha_s, \Lambda_{\rm QCD}/m_h, \alpha_s/(a
m_h),
\alpha_s \Lambda_{\rm QCD}/m_h)$, and four-fermion operators through \\
 ${\cO}(\alpha_s, \Lambda_{\rm QCD}/m_h, 
\alpha_s/(a m_h))$.
NRQCD/HISQ currents with massive HISQ quarks are also of interest,
e.g.,  for the bottom-charm currents in $B \rightarrow D^{(*)} l \nu$
semileptonic decays and the relevant matching calculations have been
performed at 1-loop order in Ref.~\cite{Monahan:2012dq}.  Taking all
the above into account, the most significant systematic error in
electroweak matrix elements published to date with NRQCD $b$-quarks is
the ${\cO}(\alpha_s^2)$ perturbative matching uncertainty.  Work is
therefore underway to use current-current correlator methods combined
with very high order continuum perturbation theory to do current
matchings nonperturbatively~\cite{Koponen:2010jy}.
\\


\noindent
{\it Relativistic heavy quarks}\\
\noindent

An approach for relativistic heavy-quark lattice formulations was
first introduced by El-Khadra, Kronfeld, and Mackenzie in
Ref.~\cite{ElKhadra:1996mp}.  Here they showed that, for a general
lattice action with massive quarks and non-Abelian gauge fields,
discretization errors can be factorized into the form $f(m_h
a)(a|\vec{p}_h|)^n$, and that the function $f(m_h a)$ is bounded to be
of ${\cO}(1)$ or less for all values of the quark mass $m_h$.
Therefore cutoff effects are of ${\cO}(a \Lambda_{\rm QCD})^n$
and ${\cO}((a|\vec{p}_h|)^n)$, even for $am_h \gtapprox 1$, and
can be controlled using a Symanzik-like procedure.  As in the standard
Symanzik improvement program, cutoff effects are systematically
removed by introducing higher-dimension operators to the lattice
action and suitably tuning their coefficients.  In the relativistic
heavy-quark approach, however, the operator coefficients are allowed
to depend explicitly on the quark mass.  By including lattice
operators through dimension $n$ and adjusting their coefficients
$c_{n,i}(m_h a)$ correctly, one enforces that matrix elements in the
lattice theory are equal to the analogous matrix elements in continuum
QCD through $(a|\vec{p}_h|)^n$, such that residual heavy-quark
discretization errors are of ${\cO}(a|\vec{p}_h|)^{n+1}$.

The relativistic heavy-quark approach can be used to compute the
matrix elements of states containing heavy quarks for which the
heavy-quark spatial momentum $|\vec{p}_h|$ is small compared to the
lattice spacing.  Thus it is suitable to describe bottom and charm
quarks in both heavy-light and heavy-heavy systems.  Calculations of
bottomonium and charmonium spectra serve as nontrivial tests of the
method and its accuracy.

At fixed lattice spacing, relativistic heavy-quark formulations
recover the massless limit when $(am_h) \ll 1$, recover the static
limit when $(am_h) \gg 1$, and smoothy interpolate between the two;
thus they can be used for any value of the quark mass, and, in
particular, for both charm and bottom.  Discretization errors for
relativistic heavy-quark formulations are generically of the form
$\alpha_s^k f(am_h)(a |\vec{p}_h|)^n$, where $k$ reflects the order of
the perturbative matching for operators of ${\cO}((a |\vec{p}_h|)^n)$.
For each $n$, such errors are removed completely if the operator
matching is nonperturbative. When $(am_h) \sim 1$, this gives rise to
nontrivial lattice-spacing dependence in physical quantities, and it
is prudent to compare estimates based on power-counting with a direct
study of scaling behaviour using a range of lattice spacings.  At
fixed quark mass, relativistic heavy-quark actions possess a smooth
continuum limit without power-divergences.  Of course, as $m_h \to
\infty$ at fixed lattice spacing, the static limit is recovered and by
then taking the continuum limit the corresponding power divergences
are reproduced (see, e.g., Ref.~\cite{Harada:2001fi}).

The relativistic heavy-quark formulations in use all begin with the
asymmetric (or anisotropic) Sheikholeslami-Wohlert (``clover'')
action~\cite{Sheikholeslami:1985ij}:
\begin{equation}
S_\textrm{lat} = a^4 \sum_{x,x'} \bar{\psi}(x') \left( m_0 + \gamma_0 D_0 + \zeta \vec{\gamma} \cdot \vec{D} - \frac{a}{2} (D^0)^2 - \frac{a}{2} \zeta (\vec{D})^2+ \sum_{\mu,\nu} \frac{ia}{4} c_{\rm SW} \sigma_{\mu\nu} F_{\mu\nu} \right)_{x' x} \psi(x) \,,
\label{eq:HQAct}
\end{equation}
where $D_\mu$ is the lattice covariant derivative and $F_{\mu\nu}$ is
the lattice field-strength tensor.  Here we show the form of the
action given in Ref.~\cite{Christ:2006us}.  The introduction of a
space-time asymmetry, parameterized by $\zeta$ in
Eq.~(\ref{eq:HQAct}), is convenient for heavy-quark systems because
the characteristic heavy-quark four-momenta do not respect space-time
axis exchange ($\vec{p}_h < m_h$ in the bound-state rest frame).
Further, the Sheikoleslami-Wohlert action respects the continuum
heavy-quark spin and flavour symmetries, so HQET can be used to
interpret and estimate lattice discretization
effects~\cite{Kronfeld:2000ck,Harada:2001fi,Harada:2001fj}.  We
discuss three different prescriptions for tuning the parameters of the
action in common use below.  In particular, we focus on aspects of the
action and operator improvement and matching relevant for evaluating
the quality of the calculations discussed in the main text.

The meson energy-momentum dispersion relation plays an important role
in relativistic heavy-quark formulations:
\begin{equation}
	E(\vec{p}) = M_1 + \frac{\vec{p}^2}{2M_2} + {\cO}(\vec{p}^4) \,,
\end{equation}
where $M_1$ and $M_2$ are known as the rest and kinetic masses,
respectively.  Because the lattice breaks Lorentz invariance, there
are corrections proportional to powers of the momentum.  Further, the
lattice rest masses and kinetic masses are not equal ($M_1 \neq M_2$),
and only become equal in the continuum limit.

The Fermilab interpretation~\cite{ElKhadra:1996mp} is suitable for
calculations of mass splittings and matrix elements of systems with
heavy quarks.  The Fermilab action is based on the hopping-parameter
form of the Wilson action, in which $\kappa_h$ parameterizes the
heavy-quark mass.  In practice, $\kappa_h$ is tuned such that the the
kinetic meson mass equals the experimentally-measured heavy-strange
meson mass ($m_{B_s}$ for bottom and $m_{D_s}$ for charm).  In
principle, one could also tune the anisotropy parameter such that $M_1
= M_2$.  This is not necessary, however, to obtain mass splittings and
matrix elements, which are not affected by
$M_1$~\cite{Kronfeld:2000ck}.  Therefore in the Fermilab action the
anisotropy parameter is set equal to unity.  The clover coefficient in
the Fermilab action is fixed to the value $c_{\rm SW} = 1/u_0^3$ from
mean-field improved lattice perturbation theory~\cite{Lepage:1992xa}.
With this prescription, discretization effects are of ${\cO}(\alpha_sa|\vec{p}_h|, (a|\vec{p}_h|)^2)$.  Calculations of
electroweak matrix elements also require improving the lattice current
and four-fermion operators to the same order, and matching them to the
continuum.  Calculations with the Fermilab action remove tree-level
${\cO}(a)$ errors in electroweak operators by rotating the
heavy-quark field used in the matrix element and setting the rotation
coefficient to its tadpole-improved tree-level value (see, e.g.,
Eqs.~(7.8) and (7.10) of Ref.~\cite{ElKhadra:1996mp}).  Finally,
electroweak operators are typically renormalized using a mostly
nonperturbative approach in which the flavour-conserving light-light
and heavy-heavy current renormalization factors $Z_V^{ll}$ and
$Z_V^{hh}$ are computed nonperturbatively~\cite{El-Khadra:2001wco}.  The
flavour-conserving factors account for most of the heavy-light current
renormalization.  The remaining correction is expected to be close to
unity due to the cancellation of most of the radiative corrections
including tadpole graphs~\cite{Harada:2001fi}; therefore it can be
reliably computed at 1-loop in mean-field improved lattice
perturbation theory with truncation errors at the percent to
few-percent level.

The relativistic heavy-quark (RHQ) formulation developed by Li, Lin,
and Christ builds upon the Fermilab approach, but tunes all the
parameters of the action in Eq.~(\ref{eq:HQAct})
nonperturbatively~\cite{Christ:2006us}.  In practice, the three
parameters $\{m_0a, c_{\rm SW}, \zeta\}$ are fixed to reproduce the
experimentally-measured $B_s$ meson mass and hyperfine splitting
($m_{B_s^*}-m_{B_s}$), and to make the kinetic and rest masses of the
lattice $B_s$ meson equal~\cite{Aoki:2012xaa}.  This is done by
computing the heavy-strange meson mass, hyperfine splitting, and ratio
$M_1/M_2$ for several sets of bare parameters $\{m_0a, c_{\rm
SW}, \zeta\}$ and interpolating linearly to the physical $B_s$ point.
By fixing the $B_s$-meson hyperfine splitting, one loses a potential
experimental prediction with respect to the Fermilab formulation.
However, by requiring that $M_1 = M_2$, one gains the ability to use
the meson rest masses, which are generally more precise than the
kinetic masses, in the RHQ approach.  The nonperturbative
parameter-tuning procedure eliminates ${\cO}(a)$ errors from
the RHQ action, such that discretization errors are of ${\cO}((a|\vec{p}_h|)^2)$.  Calculations of $B$-meson decay constants and
semileptonic form factors with the RHQ action are in
progress~\cite{Witzel:2012pr,Kawanai:2012id}, as is the corresponding
1-loop mean-field improved lattice perturbation
theory~\cite{Lehner:2012bt}.  For these works, cutoff effects in the
electroweak vector and axial-vector currents will be removed through
${\cO}(\alpha_s a)$, such that the remaining discretization
errors are of ${\cO}(\alpha_s^2a|\vec{p}_h|,
(a|\vec{p}_h|)^2)$.  Matching the lattice operators to the continuum
will be done following the mostly nonperturbative approach described
above.

The Tsukuba heavy-quark action is also based on the
Sheikholeslami-Wohlert action in Eq.~(\ref{eq:HQAct}), but allows for
further anisotropies and hence has additional parameters: specifically
the clover coefficients in the spatial $(c_B)$ and temporal $(c_E)$
directions differ, as do the anisotropy coefficients of the $\vec{D}$
and $\vec{D}^2$ operators~\cite{Aoki:2001ra}.  In practice, the
contribution to the clover coefficient in the massless limit is
computed nonperturbatively~\cite{Aoki:2005et}, while the
mass-dependent contributions, which differ for $c_B$ and $c_E$, are
calculated at 1-loop in mean-field improved lattice perturbation
theory~\cite{Aoki:2003dg}.  The hopping parameter is fixed
nonperturbatively to reproduce the experimentally-measured
spin-averaged $1S$ charmonium mass~\cite{Namekawa:2011wt}.  One of the
anisotropy parameters ($r_t$ in Ref.~\cite{Namekawa:2011wt}) is also
set to its 1-loop perturbative value, while the other ($\nu$ in
Ref.~\cite{Namekawa:2011wt}) is fixed noperturbatively to obtain the
continuum dispersion relation for the spin-averaged charmonium $1S$
states (such that $M_1 = M_2$).  For the renormalization and
improvement coefficients of weak current operators, the contributions
in the chiral limit are obtained
nonperturbatively~\cite{Kaneko:2007wh,Aoki:2010wm}, while the
mass-dependent contributions are estimated using 1-loop lattice
perturbation theory~\cite{Aoki:2004th}.  With these choices, lattice
cutoff effects from the action and operators are of ${\cO}(\alpha_s^2 a|\vec{p}|, (a|\vec{p}_h|)^2)$.
\\


\noindent
{\it Light-quark actions combined with HQET}\\
\noindent

The heavy-quark formulations discussed in the previous sections use
effective field theory to avoid the occurence of discretization errors
of the form $(am_h)^n$.  In this section we describe methods that use
improved actions that were originally designed for light-quark systems
for $B$ physics calculations. Such actions unavoidably contain
discretization errors that grow as a power of the heavy-quark mass. In
order to use them for heavy-quark physics, they must be improved to at
least ${\cO}(am_h)^2$.  However, since $am_b > 1$ at the smallest
lattice spacings available in current simulations, these methods also
require input from HQET to guide the simulation results to the
physical $b$-quark mass.

The ETM collaboration has developed two methods, the ``ratio
method'' \cite{Blossier:2009hg} and the ``interpolation
method'' \cite{Guazzini:2006bn,Blossier:2009gd}. They use these
methods together with simulations with twisted-mass Wilson fermions,
which have discretization errors of $\cO(am_h)^2$.  In the interpolation
method $\Phi_{hs}$ and $\Phi_{h\ell}$ (or $\Phi_{hs}/\Phi_{h\ell}$)
are calculated for a range of heavy-quark masses in the charm region
and above, while roughly keeping $am_h \ltsim 0.5 $. The relativistic
results are combined with a separate calculation of the decay
constants in the static limit, and then interpolated to the physical
$b$ quark mass. In ETM's implementation of this method, the heavy
Wilson decay constants are matched to HQET using NLO in continuum
perturbation theory. The static limit result is renormalized using
1-loop mean-field improved lattice perturbation theory, while for
the relativistic data PCAC is used to calculate absolutely normalized
matrix elements. Both, the relativistic and static limit data are then
run to the common reference scale $\mu_b = 4.5 \GeV$ at NLO in
continuum perturbation theory.  In the ratio method, one constructs
physical quantities $P(m_h)$ from the relativistic data that have a
well-defined static limit ($P(m_h) \to$ const.~for $m_h \to \infty$)
and evaluates them at the heavy-quark masses used in the simulations.
Ratios of these quantities are then formed at a fixed ratio of heavy-quark masses, $z = P(m_h) / P(m_h/\lambda)$ (where $1 < \lambda \lsim
1.3$), which ensures that $z$ is equal to unity in the static limit.
Hence, a separate static limit calculation is not needed with this
method.  In ETM's implementation of the ratio method for the $B$-meson
decay constant, $P(m_h)$ is constructed from the decay constants and
the heavy-quark pole mass as $P(m_h) = f_{h\ell}(m_h) \cdot (m^{\rm
pole}_h)^{1/2}$. The corresponding $z$-ratio therefore also includes
ratios of perturbative matching factors for the pole mass to $\msbar$
conversion.  For the interpolation to the physical $b$-quark mass,
ratios of perturbative matching factors converting the data from QCD
to HQET are also included. The QCD-to-HQET matching factors improve
the approach to the static limit by removing the leading logarithmic
corrections. In ETM's implementation of this method (ETM 11 and 12)
both conversion factors are evaluated at NLO in continuum perturbation
theory. The ratios are then simply fit to a polynomial in $1/m_h$ and
interpolated to the physical $b$-quark mass.  The ratios constructed
from $f_{h\ell}$ ($f_{hs}$) are called $z$ ($z_s$).  In order to
obtain the $B$ meson decay constants, the ratios are combined with
relativistic decay constant data evaluated at the smallest reference
mass.

The HPQCD collaboration has introduced a method in
Ref.~\cite{McNeile:2011ng} which we shall refer to as the ``heavy
HISQ'' method.  The first key ingredient is the use of the HISQ action
for the heavy and light valence quarks, which has leading
discretization errors of ${\cO} \left(\alpha_s (v/c) (am_h)^2,
(v/c)^2 (am_h)^4\right)$.  With the same action for the heavy- and
light-valence quarks it is possible to use PCAC to avoid
renormalization uncertainties.  Another key ingredient at the time of formulation was the
availability of gauge ensembles over a large range of lattice
spacings, in this case the library of $\Nf = 2+1$
asqtad ensembles made public by the MILC collaboration which include
lattice spacings as small as $a \approx 0.045$~fm.
Since the HISQ
action is so highly improved and with lattice spacings as small as
$0.045$~fm, HPQCD is able to use a large range of heavy-quark masses,
from below the charm region to almost up to the physical $b$-quark
mass with $am_h \ltsim 0.85$. They then fit their data in a combined
continuum and HQET fit (i.e., using a fit function that is motivated by
HQET) to a polynomial in $1/m_H$ (the heavy pseudoscalar-meson mass
of a meson containing a heavy ($h$) quark).

This approach has been extended in recent work by the HPQCD and
FNAL/MILC collaborations using the MILC-generated $\Nf=2+1+1$ HISQ
ensembles with lattice spacings down to
$0.03$~fm~\cite{Bazavov:2017lyh}.  These are being used by the HPQCD
and the FNAL/MILC collaborations for their B-physics programmes and
the corresponding analyses include heavy-quark masses at the physical
$b$ quark mass.

\bigskip

In Tab.~\ref{tab_heavy_quarkactions} we list the discretizations of
the quark action most widely used for heavy $c$ and $b$ quarks
together with the abbreviations used in the summary tables.  We also
summarize the main properties of these actions and the leading lattice
discretization errors for calculations of heavy-light meson matrix
quantities with them.  Note that in order to maintain the leading
lattice artifacts of the actions as given in the table in nonspectral
observables (like operator matrix elements) the corresponding
nonspectral operators need to be improved as well.

\begin{table}
\begin{center}
{\footnotesize
\begin{tabular*}{\textwidth}[l]{l @{\extracolsep{\fill}} l l l}
\hline \hline  \\[-1.0ex]
\parbox[t]{1.5cm}{Abbrev.} & Discretization & 
\parbox[t]{4cm}{Leading lattice artifacts\\and truncation errors\\for heavy-light mesons} &  
Remarks
\\[7.0ex] \hline \hline \\[-1.0ex]
tmWil   & twisted-mass Wilson &  ${\cO}\big((am_h)^2\big)$ & \parbox[t]{4.cm}{PCAC relation for axial-vector current}  
\\[3.0ex] \hline \\[-1.0ex]
HISQ  & Staggered & \parbox[t]{4cm}{${\cO}\big (\alpha_S (am_h)^2 (v/c), \\(am_h)^4 (v/c)^2 \big)$}  & \parbox[t]{4.cm}{PCAC relation for axial-vector current; Ward identity for vector current}  
\\[6.0ex] \hline \\[-1.0ex]
static  & static effective action &  \parbox[t]{4cm}{${\cO}\big( a^2 \Lambda_{\rm QCD}^2, \Lambda_{\rm QCD}/m_h, \\ \alpha_s^2, \alpha_s^2 a \Lambda_{\rm QCD} \big)$}  & \parbox[t]{4.5cm}{implementations use APE, HYP1, and HYP2 smearing}  
\\[4.0ex] \hline \\[-1.0ex]
HQET  & Heavy-Quark Effective Theory &  \parbox[t]{4cm}{${\cO}\big( a \Lambda^2_{\rm QCD}/m_h,  a^2 \Lambda^2_{\rm QCD},\\
 (\Lambda_{\rm QCD}/m_h)^2 \big)$}  & \parbox[t]{4.5cm}{Nonperturbative matching through ${\cO}(1/m_h)$}  
\\[4.0ex] \hline \\[-1.0ex]
NRQCD  & Nonrelativistic QCD & \parbox[t]{4cm}{${\cO}\big(\alpha_S \Lambda_{\rm QCD}/m_h, \\ \alpha_S (\Lambda_{\rm QCD}/m_h)^2 , \\ (\Lambda_{\rm QCD}/m_h )^3,  \alpha_s a^2 \Lambda_{\rm QCD}^2 \big)$}  & \parbox[t]{4.5cm}{Tree-level relativistic corrections through 
${\cO}(v_h^4)$ and discretization corrections through ${\cO}(a^2)$}  
\\[9.5ex] \hline \\[-1.0ex] 
Fermilab  & Sheikholeslami-Wohlert & ${\cO}\big(\alpha_sa\Lambda_{\rm QCD}, (a\Lambda_{\rm QCD})^2\big)$  & \parbox[t]{4.5cm}{Hopping parameter tuned nonperturbatively; clover coefficient computed at tree-level in mean-field-improved lattice perturbation theory}  
\\[12.0ex] \hline \\[-1.0ex] 

RHQ       & Sheikholeslami-Wohlert & ${\cO}\big( \alpha_s^2 a\Lambda_{\rm QCD}, (a\Lambda_{\rm QCD})^2 \big)$  & \parbox[t]{4.5cm}{Hopping parameter, anisoptropy and clover coefficient tuned nonperturbatively by fixing the $B_s$-meson hyperfine splitting} \\[12.0ex] \hline \\[-1.0ex] 

Tsukuba  & Sheikholeslami-Wohlert & ${\cO}\big( \alpha_s^2 a\Lambda_{\rm QCD}, (a\Lambda_{\rm QCD})^2 \big)$  & \parbox[t]{4.5cm}{NP clover coefficient at $ma=0$ plus mass-dependent corrections calculated at 1-loop in lattice perturbation theory; $\nu$ calculated NP from dispersion relation; $r_s$ calculated at 1-loop in lattice perturbation theory}  
\\[20.0ex]
\hline\hline
\end{tabular*}
}
\caption{Discretizations of the quark action most widely used for heavy $c$ and $b$ quarks  and some of their properties.
\label{tab_heavy_quarkactions}}
\end{center}
\end{table}

\subsection{Matching and running \label{sec_match}}

The lattice formulation of QCD amounts to introducing a particular
regularization scheme. Thus, in order to be useful for phenomenology,
hadronic matrix elements computed in lattice simulations must be
related to some continuum reference scheme, such as the
$\msbar$-scheme of dimensional regularization. The matching to the
continuum scheme usually involves running to some reference scale
using the renormalization group. 

In principle, the matching factors which relate lattice matrix
elements to the $\msbar$-scheme, can be computed in perturbation
theory formulated in terms of the bare coupling. It has been known for
a long time, though, that the perturbative expansion is not under good 
control. Several techniques have been developed which allow for a
nonperturbative matching between lattice regularization and continuum
schemes, and are briefly introduced here.\\


\noindent
{\sl Regularization-independent Momentum Subtraction}\\
\noindent

In the {\sl Regularization-independent Momentum Subtraction}
(``RI/MOM'' or ``RI'') scheme \cite{Martinelli:1994ty} a
nonperturbative renormalization condition is formulated in terms of
Green functions involving quark states in a fixed gauge (usually
Landau gauge) at nonzero virtuality. In this way one relates operators
in lattice regularization nonperturbatively to the RI scheme. In a
second step one matches the operator in the RI scheme to its
counterpart in the $\msbar$-scheme. The advantage of this procedure is
that the latter relation involves perturbation theory formulated in
the continuum theory. The use of lattice perturbation theory can thus
be avoided, and the continuum perturbation theory, which is
technically more feasible for higher order calculations, could be
applied if more precision is required. A technical complication is
associated with the accessible momentum scales (i.e., virtualities),
which must be large enough (typically several $\gev$) in order for the
perturbative relation to $\msbar$ to be reliable. The momentum scales
in simulations must stay well below the cutoff scale (i.e., $2\pi$
over the lattice spacing), since otherwise large lattice artifacts are
incurred. Thus, the applicability of the RI scheme traditionally
relies on the existence of a ``window'' of momentum scales, which
satisfy
\be
   \Lambda_{\rm QCD} \;\lesssim\; p \;\lesssim\; 2\pi a^{-1}.
\ee
However, solutions for mitigating this limitation, which involve
continuum limit, nonperturbative running to higher scales in the
RI/MOM scheme, have recently been proposed and implemented
\cite{Arthur:2010ht,Durr:2010vn,Durr:2010aw,Aoki:2010pe}.

Within the RI/MOM framework one has some freedom in the choice of the
external momenta used in the Green functions. In the choice made in
the original work, the virtuality of each external leg is nonzero, but
that of the momentum transfer between different legs can
vanish~\cite{Martinelli:1994ty}.  This leads to enhanced
nonperturbative contributions that fall as powers of $p^2$.  An
alternative choice that reduces these issues is the symmetric MOM scheme, in which virtualities in all channels are
nonzero~\cite{Sturm:2009kb}.  This scheme is now widely used. To
distinguish it from the original choice of virtualities, it is
referred to as the RI/SMOM (or RI-SMOM) scheme, while the original choice is called
the RI/MON (or RI-MOM) scheme.\\

\noindent
{\it Schr\"odinger functional}\\
\noindent

Another example of a nonperturbative matching procedure is provided
by the Schr\"odinger functional (SF) scheme \cite{Luscher:1992an}. It
is based on the formulation of QCD in a finite volume. If all quark
masses are set to zero the box length remains the only scale in the
theory, such that observables like the coupling constant run with the
box size~$L$. The great advantage is that the RG running of
scale-dependent quantities can be computed nonperturbatively using
recursive finite-size scaling techniques. It is thus possible to run
nonperturbatively up to scales of, say, $100\,\gev$, where one is
sure that the perturbative relation between the SF and
$\msbar$-schemes is controlled.\\

\noindent
{\sl Perturbation theory}\\
\noindent

The third matching procedure is based on perturbation theory in which
higher order are effectively resummed \cite{Lepage:1992xa}. Although
this procedure is easier to implement, it is hard to estimate the
uncertainty associated with it.\\

\noindent
{\sl Mostly nonperturbative renormalization}\\
\noindent

Some calculations of heavy-light and heavy-heavy matrix elements adopt a mostly nonperturbative matching approach.  Let us consider a weak 
decay process mediated by a current with quark flavours $h$ and $q$, where $h$ is the initial heavy quark (either bottom or charm) and 
$q$ can be a light ($\ell = u,d$), strange, or charm quark. The matrix elements of lattice current  $J_{hq}$ are matched to the 
corresponding continuum matrix elements with continuum current ${\cal J}_{hq}$ by calculating the renormalization factor $Z_{J_{hq}}$. 
The mostly nonperturbative renormalization method takes advantage of rewriting the current renormalization factor as the following product:
\begin{align}
Z_{J_{hq}} = \rho_{J_{hq}} \sqrt{Z_{V^4_{hh}}Z_{V^4_{qq}}} \,
\label{eq:Zvbl}
\end{align}
The flavour-conserving renormalization factors $Z_{V^4_{hh}}$ and $Z_{V^4_{qq}}$ can be obtained nonperturbatively from standard heavy-light 
and light-light meson charge normalization conditions.  $Z_{V^4_{hh}}$ and $Z_{V^4_{qq}}$ account for  the bulk of the renormalization. The remaining 
correction $\rho_{J_{hq}}$ is expected to be close to unity because most of the radiative corrections, including self-energy corrections and 
contributions from tadpole graphs, cancel in the ratio~\cite{Harada:2001fj,Harada:2001fi}.  The 1-loop coefficients of $\rho_{J_{hq}}$  have been calculated for
heavy-light and heavy-heavy currents for Fermilab heavy and both (improved) Wilson light \cite{Harada:2001fj,Harada:2001fi} and 
asqtad light  \cite{ElKhadra:2007qe} quarks. In all cases the 1-loop coefficients are found to be very small, yielding sub-percent to few percent level corrections.

\bigskip
\noindent
In Tab.~\ref{tab_match} we list the abbreviations used in the
compilation of results together with a short description.

\begin{table}[ht]
{\footnotesize
\begin{tabular*}{\textwidth}[l]{l @{\extracolsep{\fill}} l}
\hline \hline \\[-1.0ex]
Abbrev. & Description
\\[1.0ex] \hline \hline \\[-1.0ex]
RI  &  regularization-independent momentum subtraction scheme 
\\[1.0ex] \hline \\[-1.0ex]
SF  &  Schr\"odinger functional scheme
\\[1.0ex] \hline \\[-1.0ex]
PT1$\ell$ & matching/running computed in perturbation theory at one loop
\\[1.0ex] \hline \\[-1.0ex]
PT2$\ell$ & matching/running computed in perturbation theory at two loops 
\\[1.0ex] \hline \\[-1.0ex]
mNPR & mostly nonperturbative renormalization 
%
\\[1.0ex]
\hline\hline
\end{tabular*}
}
\caption{The most widely used matching and running
  techniques. \label{tab_match} 
}
\end{table}

\subsection{Chiral extrapolation\label{sec_ChiPT}}
As mentioned in the introduction, Symanzik's framework can be combined 
with Chiral Perturbation Theory. The well-known terms occurring in the
chiral effective Lagrangian are then supplemented by contributions 
proportional to powers of the lattice spacing $a$. The additional terms are 
constrained by the symmetries of the lattice action and therefore 
depend on the specific choice of the discretization. 
The resulting effective theory can be used to analyze the $a$-dependence of 
the various quantities of interest -- provided the quark masses and the momenta
considered are in the range where the truncated chiral perturbation series yields 
an adequate approximation. Understanding the dependence on the lattice spacing 
is of central importance for a controlled extrapolation to the continuum limit.
 
For staggered fermions, this program has first been carried out for a
single staggered flavour (a single staggered field) \cite{Lee:1999zxa}
at $\cO(a^2)$. In the following, this effective theory is denoted by
S{\Ch}PT. It was later generalized to an arbitrary number of flavours
\cite{Aubin:2003mg,Aubin:2003uc}, and to next-to-leading order
\cite{Sharpe:2004is}. The corresponding theory is commonly called
Rooted Staggered chiral perturbation theory and is denoted by
RS{\Ch}PT.

For Wilson fermions, the effective theory has been developed in
\cite{Sharpe:1998xm,Rupak:2002sm,Aoki:2003yv}
and is called W{\Ch}PT, while the theory for Wilson twisted-mass
fermions \cite{Sharpe:2004ny,Aoki:2004ta,Bar:2010jk} is termed tmW{\Ch}PT.

Another important approach is to consider theories in which the
valence and sea quark masses are chosen to be different. These
theories are called {\it partially quenched}. The acronym for the
corresponding chiral effective theory is PQ{\Ch}PT
\cite{Bernard:1993sv,Golterman:1997st,Sharpe:1997by,Sharpe:2000bc}.

Finally, one can also consider theories where the fermion
discretizations used for the sea and the valence quarks are different. The
effective chiral theories for these ``mixed action'' theories are
referred to as MA{\Ch}PT  \cite{Bar:2002nr,Bar:2003mh,Bar:2005tu,Golterman:2005xa,Chen:2006wf,Chen:2007ug,Chen:2009su}. \\

\noindent
{\sl Finite-Volume Regimes of QCD}\\
\noindent

Once QCD with $\Nf$ nondegenerate flavours is regulated both in the UV and
in the IR, there are $3+\Nf$ scales in play: The scale
$\Lambda_\mathrm{QCD}$ that reflects ``dimensional transmutation''
(alternatively, one could use the pion decay constant or the nucleon mass,
in the chiral limit), the inverse lattice spacing $1/a$, the inverse box
size $1/L$, as well as $\Nf$ meson masses (or functions of meson masses)
that are sensitive to the $\Nf$ quark masses, e.g., $\Mpi^2$,
$2\Mka^2-\Mpi^2$ and the spin-averaged masses of ${}^1S$ states of
quarkonia.

Ultimately, we are interested in results with the two regulators
removed, i.e., physical quantities for which the limits $a \to 0$ and
$L \to \infty$ have been carried out. In both cases there is an
effective field theory (EFT) which guides the extrapolation.  For the
$a \to 0$ limit, this is a version of the Symanzik EFT which depends,
in its details, on the lattice action that is used, as outlined in
Sec.~\ref{sec_lattice_actions}.  The finite-volume effects are
dominated by the lightest particles, the pions.  Therefore, a chiral
EFT, also known as {\Ch}PT, is appropriate to parameterize the
finite-volume effects, i.e., the deviation of masses and other
observables, such as matrix elements, in a finite-volume from their
infinite volume, physical values.  Most simulations of
phenomenological interest are carried out in boxes of size $L \gg
1/M_\pi$, that is in boxes whose diameter is large compared to the
Compton wavelength that the pion would have, at the given quark mass,
in infinite volume. In this situation the finite-volume corrections
are small, and in many cases the ratio $M_\mathrm{had}(L) /
M_\mathrm{had}$ or $f(L) / f$, where $f$ denotes some generic matrix
element, can be calculated in {\Ch}PT, such that the leading
finite-volume effects can be taken out analytically. In the
terminology of {\Ch}PT this setting is referred to as the $p$-regime,
as the typical contributing momenta $p \sim M_\pi \gg 1/L$.  A
peculiar situation occurs if the condition $L \gg 1/\Mpi$ is violated
(while $L\Lambda_\mathrm{QCD}\gg1$ still holds), in other words if the
quark mass is taken so light that the Compton wavelength that the pion
would have (at the given $m_q$) in infinite volume, is as large or
even larger than the actual box size. Then the pion zero-momentum mode
dominates and needs to be treated separately. While this setup is
unlikely to be useful for standard phenomenological computations, the
low-energy constants of {\Ch}PT can still be calculated, by matching
to a re-ordered version of the chiral series, and following the
details of the reordering such an extreme regime is called the
$\epsilon$- or $\delta$-regime, respectively.  Accordingly, further
particulars of these regimes are discussed in Sec.~\ref{sec:chPT} of this report.

\input{Glossary/app_zParameterization.tex}

\ifx\nosimulatedlatticeactiontables\undefined

\subsection{Summary of simulated lattice actions}
In the following Tabs.~\ref{tab:simulated Nf2 actions}--\ref{tab:simulated Nf4 bc actions}
we summarize the gauge and quark actions used
in the various calculations with $\Nf=2, 2+1$ and $2+1+1$ quark
flavours. The calculations with $\Nf=0$ quark flavours mentioned in
Sec.~\ref{sec:alpha_s} all used the Wilson gauge action and are not
listed. Abbreviations are explained in Secs.~\ref{sec_gauge_actions}, \ref{sec_quark_actions} and
\ref{app:HQactions}, and summarized in Tabs.~\ref{tab_gaugeactions},
\ref{tab_quarkactions} and \ref{tab_heavy_quarkactions}.

%
\hspace{-1.5cm}
\begin{table}[h]
{\footnotesize
\begin{tabular*}{\textwidth}[l]{l @{\extracolsep{\fill}} c c c c}
\hline \hline \\[-1.0ex]
Collab. & Ref. & $\Nf$ & \parbox{1cm}{gauge\\action} & \parbox{1cm}{quark\\action} 
\\[2.0ex] \hline \hline \\[-1.0ex]
ALPHA 01A, 04, 05, 12, 13A & \cite{Bode:2001jv,DellaMorte:2004bc,DellaMorte:2005kg,Fritzsch:2012wq,Lottini:2013rfa} & 2 & Wilson & npSW \\
[2.0ex] \hline \\[-1.0ex]
Aoki 94 & \cite{Aoki:1994pc} & 2 & Wilson &  KS \\
[2.0ex] \hline \\[-1.0ex]
Bernardoni 10 & \cite{Bernardoni:2010nf} & 2 & Wilson & npSW ${}^\dagger$ \\
[2.0ex] \hline \\[-1.0ex]
Bernardoni 11 & \cite{Bernardoni:2011kd} & 2 & Wilson & npSW \\
[2.0ex] \hline \\[-1.0ex]
Brandt 13 & \cite{Brandt:2013dua} & 2 & Wilson & npSW \\
[2.0ex] \hline \\[-1.0ex]
Boucaud 01B & \cite{Boucaud:2001qz} & 2 & Wilson &  Wilson \\
[2.0ex] \hline \\[-1.0ex]
CERN-TOV 06 & \cite{DelDebbio:2006cn} & 2 & Wilson & Wilson/npSW \\
[2.0ex] \hline \\[-1.0ex]
CERN 08 & \cite{Giusti:2008vb} & 2 & Wilson & npSW \\
[2.0ex] \hline \\[-1.0ex]
{CP-PACS 01, 04} & \cite{AliKhan:2001tx,Takeda:2004xha} & 2 & Iwasaki & mfSW  \\
[2.0ex] \hline \\[-1.0ex]
Davies 94  & \cite{Davies:1994ei} & 2 & Wilson  & KS \\
[2.0ex] \hline \\[-1.0ex]
D\"urr 11 & \cite{Durr:2011ed} & 2 & Wilson & npSW \\
[2.0ex] \hline \\[-1.0ex]
Engel 14 & \cite{Engel:2014eea} & 2 & Wilson & npSW \\
[2.0ex] 
\hline\hline \\
\end{tabular*}\\[-0.2cm]
\begin{minipage}{\linewidth}
{\footnotesize 
\begin{itemize}
   \item[${}^\dagger$] The calculation uses overlap fermions in the valence-quark sector.\\[-5mm]
\end{itemize}
}
\end{minipage}
}
\caption{Summary of simulated lattice actions with $\Nf=2$ quark
  flavours.
\label{tab:simulated Nf2 actions}}
\end{table}

\begin{table}[h]
\addtocounter{table}{-1}
{\footnotesize
\begin{tabular*}{\textwidth}[l]{l @{\extracolsep{\fill}} c c c c}
\hline \hline \\[-1.0ex]
Collab. & Ref. & $\Nf$ & \parbox{1cm}{gauge\\action} & \parbox{1cm}{quark\\action} 
\\[2.0ex] \hline \hline \\[-1.0ex]
\parbox[t]{4.0cm}{ETM 07, 07A, 08, 09, 09A-D, 09G 10B, 10D, 10F, 11C, 12, 13, 13A} & \parbox[t]{2.5cm}{
\cite{Blossier:2007vv,Boucaud:2007uk,Frezzotti:2008dr,Blossier:2009bx,Lubicz:2009ht,Jansen:2009tt,Baron:2009wt,Blossier:2009hg,Feng:2009ij,Blossier:2010cr,Lubicz:2010bv,Blossier:2010ky,Jansen:2011vv,Burger:2012ti,Cichy:2013gja,Herdoiza:2013sla}} & 2 &  tlSym & tmWil \\
[13.0ex] \hline \\[-1.0ex]
{ETM 10A, 12D} & \cite{Constantinou:2010qv,Bertone:2012cu} & 2 &  tlSym & tmWil ${}^*$ \\
[2.0ex] \hline \\[-1.0ex]
\parbox[t]{4.0cm}{ETM 14D, 15A, 16C} & \parbox[t]{2.5cm}{\cite{Abdel-Rehim:2014nka,Abdel-Rehim:2015pwa,Liu:2016cba}} & 2 &  Iwasaki & tmWil with npSW \\
[2.0ex] \hline \\[-1.0ex]
\parbox[t]{4.0cm}{ETM 15D, 16A, 17, 17B, 17C} & \parbox[t]{2.5cm}{\cite{Abdel-Rehim:2015owa,Abdel-Rehim:2016won,Alexandrou:2017hac,Alexandrou:2017oeh,Alexandrou:2017qyt}} & 2 &  Iwasaki & tmWil with npSW ${}^*$\\
[4.0ex] \hline \\[-1.0ex]
G\"ulpers 13, 15 & \cite{Gulpers:2013uca,Gulpers:2015bba} & 2 & Wilson & npSW \\
[2.0ex] \hline \\[-1.0ex]
Hasenfratz 08 & \cite{Hasenfratz:2008ce} & 2 & tadSym &
n-HYP tlSW\\
[2.0ex] \hline \\[-1.0ex]
JLQCD 08, 08B & \cite{Aoki:2008ss,Ohki:2008ff} & 2 & Iwasaki & overlap \\
[2.0ex] \hline \\[-1.0ex]
{JLQCD 02, 05} & \cite{Aoki:2002uc,Tsutsui:2005cj} & 2 & Wilson & npSW \\
[2.0ex] \hline \\[-1.0ex]
JLQCD/TWQCD 07, 08A, 08C, 10 & \cite{Fukaya:2007pn,Noaki:2008iy,Shintani:2008ga,Fukaya:2010na} & 2 & Iwasaki & overlap \\
[2.0ex] \hline \\[-1.0ex]
Mainz 12, 17 & \cite{Capitani:2012gj,Capitani:2017qpc} & 2 & Wilson & npSW \\
[2.0ex] \hline \\[-1.0ex]
QCDSF 06, 07, 12, 13 & \cite{Khan:2006de,Brommel:2007wn,Bali:2012qs,Horsley:2013ayv} & 2 & Wilson &  npSW \\
[2.0ex] \hline \\[-1.0ex]
QCDSF/UKQCD 04, 05, 06, 06A, 07 & \parbox[t]{2.5cm}{\cite{Gockeler:2004rp,Gockeler:2005rv,Gockeler:2006jt,Brommel:2006ww,QCDSFUKQCD}} & 2 & Wilson &  npSW \\
[5.0ex] \hline \\[-1.0ex]
{RBC 04, 06, 07, 08} & \cite{Aoki:2004ht,Dawson:2006qc,Blum:2007cy,Lin:2008uz} & 2 & DBW2 & DW \\
[2.0ex] \hline \\[-1.0ex]
RBC/UKQCD 07 & \cite{Boyle:2007qe} & 2 & Wilson & npSW \\ 
[2.0ex]\hline \\ [-1.0ex]
RM123 11, 13 & \cite{deDivitiis:2011eh,deDivitiis:2013xla}& 2 &  tlSym & tmWil  \\
[2.0ex] \hline \\[-1.0ex]
RQCD 14, 16 & \cite{Bali:2014nma,Bali:2016lvx} & 2 & Wilson & npSW \\
[2.0ex] \hline \\[-1.0ex]
SESAM 99 & \cite{Spitz:1999tu} & 2 & Wilson &  Wilson \\
[2.0ex] 
\hline\hline \\
\end{tabular*}\\[-0.2cm]
\begin{minipage}{\linewidth}
{\footnotesize 
\begin{itemize}
   \item[${}^*$] The calculation uses Osterwalder-Seiler fermions \cite{Osterwalder:1977pc} in the valence quark sector to treat strange and
 charm quarks.
\end{itemize}
}
\end{minipage}
}
\caption{(cntd.) Summary of simulated lattice actions with $\Nf=2$ quark
  flavours.}
\end{table}

\begin{table}[h]
\addtocounter{table}{-1}
{\footnotesize
\begin{tabular*}{\textwidth}[l]{l @{\extracolsep{\fill}} c c c c}
\hline \hline \\[-1.0ex]
Collab. & Ref. & $\Nf$ & \parbox{1cm}{gauge\\action} & \parbox{1cm}{quark\\action} 
\\[2.0ex] \hline \hline \\[-1.0ex]
Sternbeck 10, 12 & \cite{Sternbeck:2010xu,Sternbeck:2012qs}& 2 &  Wilson & npSW  \\
[2.0ex] \hline \\[-1.0ex]
{SPQcdR 05} & \cite{Becirevic:2005ta} & 2 & Wilson & Wilson \\
[2.0ex] \hline \\ [-1.0ex]
TWQCD 11, 11A  & \cite{Chiu:2011bm,Chiu:2011dz} & 2 & Wilson & optimal DW \\
[2.0ex] \hline \\ [-1.0ex]
UKQCD 04 & \cite{Flynn:2004au,Boyle:2007qe} & 2 & Wilson & npSW \\ 
[2.0ex]\hline \\ [-1.0ex]
Wingate 95 & \cite{Wingate:1995fd} & 2 & Wilson &  KS \\
[2.0ex] 
\hline\hline \\
\end{tabular*}\\[-0.2cm]
}
\caption{(cntd.) Summary of simulated lattice actions with $\Nf=2$ quark
  flavours.}
\end{table}
%
%
\begin{table}[h]
{\footnotesize
\begin{tabular*}{\textwidth}[l]{l @{\extracolsep{\fill}} c c c c}
\hline \hline \\[-1.0ex]
Collab. & Ref. & $\Nf$ & \parbox{1cm}{gauge\\action} & \parbox{1cm}{quark\\action} 
\\[2.0ex] \hline \hline \\[-1.0ex]
ALPHA 17 & \cite{Bruno:2017gxd} & $2+1$ & tlSym/Wilson & npSW\\
[2.0ex] \hline \\[-1.0ex]
Aubin 08, 09 & \cite{Aubin:2008ie,Aubin:2009jh} & $2+1$ & tadSym & Asqtad ${}^\dagger$\\
[2.0ex] \hline \\[-1.0ex]
Bazavov 12, 14 & \cite{Bazavov:2012ka,Bazavov:2014soa} & $2+1$ & tlSym & HISQ \\
[2.0ex] \hline \\[-1.0ex] 
Blum 10 & \cite{Blum:2010ym} & $2+1$ & Iwasaki & DW\\
[2.0ex] \hline \\[-1.0ex]
%
%
\parbox[t]{4.0cm}{BMW 10A-C, 11, 13, 15, 16, 16A} & \parbox[t]{2.5cm}{\cite{Durr:2010vn,Durr:2010aw,Portelli:2010yn,Durr:2011ap,Durr:2013goa,Durr:2015dna,Durr:2016ulb,Fodor:2016bgu}} & $2+1$ &  tlSym & 2-level HEX tlSW \\
[4.0ex] \hline \\[-1.0ex]
BMW 10, 11A & \cite{Durr:2010hr,Durr:2011mp} & $2+1$ & tlSym & 6-level stout tlSW\\
[2.0ex] \hline \\[-1.0ex]
Boyle 14 & \cite{Boyle:2014pja} & $2+1$ & \parbox[t]{1.5cm}{Iwasaki, Iwasaki+DSDR${}^*$} &  DW \\
[4.0ex] \hline \\[-1.0ex]
$\chi$QCD 13A, 15& \cite{Gong:2013vja,Gong:2015iir} & $2+1$ & Iwasaki & DW ${}^+$\\
[2.0ex] \hline \\[-1.0ex]
$\chi$QCD 15A& \cite{Yang:2015uis} & $2+1$ & Iwasaki & M-DW ${}^+$\\
[2.0ex] \hline \\[-1.0ex]
$\chi$QCD 18 & \cite{Liang:2018pis} & $2+1$ & Iwasaki & DW, M-DW ${}^+$ \\
[2.0ex] \hline \\[-1.0ex]
CP-PACS/JLQCD 07&  \cite{Ishikawa:2007nn} & $2+1$ & Iwasaki & npSW \\
[2.0ex] \hline \hline\\
\end{tabular*}\\[-0.2cm]
\begin{minipage}{\linewidth}
{\footnotesize 
\begin{itemize}
\item[${}^\dagger$] The calculation uses domain wall fermions in
  the valence-quark sector.\\[-5mm]
\item[${}^*$] An additional
weighting factor known as the dislocation suppressing determinant ratio (DSDR) is added to the gauge action \cite{Arthur:2012opa}.\\[-5mm]
\item[${}^+$] The calculation uses overlap fermions in the
  valence-quark sector.\\[-5mm]
\end{itemize}
}
\end{minipage}
}
\caption{Summary of simulated lattice actions with $\Nf=2+1$ or $\Nf=3$ quark flavours.
\label{tab:simulated Nf3 actions}}
\end{table}

\begin{table}[h]
\addtocounter{table}{-1}
{\footnotesize
\begin{tabular*}{\textwidth}[l]{l @{\extracolsep{\fill}} c c c c}
\hline \hline \\[-1.0ex]
Collab. & Ref. & $\Nf$ & \parbox{1cm}{gauge\\action} & \parbox{1cm}{quark\\action} 
\\[2.0ex] \hline \hline \\[-1.0ex]
Engelhardt 12 & \cite{Engelhardt:2012gd} & $2+1$ & tadSym & Asqtad $^\dagger$ \\
[2.0ex] \hline \\[-1.0ex]
FNAL/MILC 12, 12I  & \cite{Bazavov:2012zs,Bazavov:2012cd} & $2+1$ & tadSym & Asqtad \\
[2.0ex] \hline \\[-1.0ex]
HPQCD 05, 05A, 08A, 13A&  \cite{Mason:2005bj,Mason:2005zx,Davies:2008sw,Dowdall:2013rya}& $2+1$ & tadSym &  Asqtad \\
[2.0ex] \hline \\[-1.0ex]
HPQCD 10 & \cite{McNeile:2010ji} & $2+1$ & tadSym & Asqtad ${}^*$\\ 
[2.0ex] \hline \\[-1.0ex]
HPQCD/UKQCD 06 & \cite{Gamiz:2006sq} & $2+1$ & tadSym & Asqtad \\ 
[2.0ex] \hline \\[-1.0ex]
HPQCD/UKQCD 07 & \cite{Follana:2007uv} & $2+1$ & tadSym & Asqtad ${}^*$\\ 
[2.0ex] \hline \\[-1.0ex]
HPQCD/MILC/UKQCD 04&  \cite{Aubin:2004ck}& $2+1$ & tadSym &  Asqtad \\
[2.0ex] \hline \\[-1.0ex]
\parbox[t]{4.0cm}{Hudspith 15, 18}  &
\cite{Hudspith:2015xoa,Hudspith:2018bpz} & $2+1$ & \parbox[t]{2.0cm}{Iwasaki, Iwasaki+DSDR${}^+$ } &  DW, \mbox{M-DW} \\
[4.0ex] \hline \\[-1.0ex]
JLQCD 09, 10 & \cite{Fukaya:2009fh,Shintani:2010ph} & $2+1$ & Iwasaki & overlap \\
[2.0ex] \hline \\[-1.0ex]
\parbox[t]{4.0cm}{JLQCD 11, 12, 12A, 14, 15A, 17, 18} & \parbox[t]{2.5cm}{\cite{Kaneko:2011rp,Kaneko:2012cta,Oksuzian:2012rzb,Fukaya:2014jka,Aoki:2015pba,Aoki:2017spo,Yamanaka:2018uud}}          & $2+1$ & \parbox[t]{3.5cm}{Iwasaki (fixed topology)} & overlap\\
[6.0ex] \hline \\[-1.0ex]
\parbox[t]{4.0cm}{JLQCD 15B-C, 16, 16B, 17A} &
\parbox[t]{2.5cm}{\cite{Nakayama:2015hrn,Fahy:2015xka,Nakayama:2016atf,Cossu:2016eqs,Aoki:2017paw}}
& $2+1$ & tlSym & M-DW \\ 
[4.0ex] \hline \\[-1.0ex]
JLQCD/TWQCD 08B, 09A & \cite{Chiu:2008kt,JLQCD:2009sk} & $2+1$ & Iwasaki & overlap \\
[2.0ex] \hline \\[-1.0ex]
JLQCD/TWQCD 10 & \cite{Fukaya:2010na} & $2+1, 3$ & Iwasaki & overlap \\
[2.0ex] \hline \\[-1.0ex]
Junnarkar 13 & \cite{Junnarkar:2013ac} & $2+1$ & tadSym & Asqtad ${}^\dagger$\\
[2.0ex] \hline \\[-1.0ex]
Laiho 11 & \cite{Laiho:2011np} & $2+1$ & tadSym & Asqtad ${}^\dagger$\\
[2.0ex] \hline \\[-1.0ex]
LHP 04, LHPC 05, 10 & \parbox[t]{2.5cm}{\cite{Bonnet:2004fr,Edwards:2005ym,Bratt:2010jn}} & $2+1$ & tadSym & Asqtad $^\dagger$ \\
[4.0ex] \hline \\[-1.0ex]
LHPC 12, 12A & \cite{Green:2012ej,Green:2012ud} & $2+1$ & tlSym & 2-level HEX tlSW \\
[2.0ex] \hline \hline\\
\end{tabular*}\\[-0.2cm]
\begin{minipage}{\linewidth}
{\footnotesize 
\begin{itemize}
\item[${}^\dagger$] The calculation uses domain wall fermions in
  the valence-quark sector.\\[-5mm]
\item[${}^+$] An additional
weighting factor known as the dislocation suppressing determinant ratio (DSDR) is added to the gauge action \cite{Arthur:2012opa}.\\[-5mm]
\item[${}^*$] The calculation uses HISQ staggered fermions in the valence-quark sector.
\end{itemize}
}
\end{minipage}
}
\caption{(cntd.) Summary of simulated lattice actions with $\Nf=2+1$ or $\Nf=3$ quark flavours.
}
\end{table}

\begin{table}[h]
\addtocounter{table}{-1}
{\footnotesize
\begin{tabular*}{\textwidth}[l]{l @{\extracolsep{\fill}} c c c c}
\hline \hline \\[-1.0ex]
Collab. & Ref. & $\Nf$ & \parbox{1cm}{gauge\\action} & \parbox{1cm}{quark\\action} 
\\[2.0ex] \hline \hline \\[-1.0ex]
Mainz 18 & \cite{Ottnad:2018fri} & $2+1$ & tlSym & npSW \\
[2.0ex] \hline \\[-1.0ex]
Maltman 08 & \cite{Maltman:2008bx} & $2+1$ & tadSym &  Asqtad \\
[2.0ex] \hline \\[-1.0ex]
Martin Camalich 10 & \cite{MartinCamalich:2010fp} & $2+1$ & Iwasaki & npSW \\
[2.0ex] \hline \\[-1.0ex]
\parbox[t]{4.0cm}{MILC 04, 07, 09, 09A, 09D, 10, 10A, 12C, 16} & \parbox[t]{3.0cm}{\cite{Aubin:2004ck,Aubin:2004fs,Bernard:2007ps,Bazavov:2009bb,Toussaint:2009pz,Bazavov:2010hj,Bazavov:2010yq,Freeman:2012ry,Basak:2016jnn}}& $2+1$ &
tadSym & Asqtad \\
[4.0ex] \hline \\[-1.0ex]
\parbox[t]{4.0cm}{Nakayama 18} &
\cite{Nakayama:2018ubk}
& $2+1$ & tlSym & M-DW \\ 
[2.0ex] \hline \\[-1.0ex]
NPLQCD 06& \cite{Beane:2006kx}& $2+1$ & tadSym & Asqtad $^\dagger$ \\
[2.0ex] \hline \\[-1.0ex]
PACS 18 & \cite{Ishikawa:2018rew} & $2+1$ & Iwasaki & npSW \\
[2.0ex] \hline \\[-1.0ex]
\parbox[t]{4.0cm}{PACS-CS 08, 08A, 09, 09A, 10, 11A, 12, 13} & \parbox[t]{2.5cm}{\cite{Aoki:2008sm,Kuramashi:2008tb,Ishikawa:2009vc,Aoki:2009ix,Aoki:2009tf,Nguyen:2011ek,Aoki:2010wm,Sasaki:2013vxa}} & $2+1$ & Iwasaki & npSW \\
[4.0ex] \hline \\[-1.0ex]
QCDSF 11 & \cite{Bali:2011ks} & $2+1$ & tlSym & npSW \\
[2.0ex] \hline \\[-1.0ex]
QCDSF/UKQCD 15, 16 & \cite{Horsley:2015eaa,Bornyakov:2016dzn} & $2+1$ & tlSym & npSW \\
[2.0ex] \hline \\[-1.0ex]
\parbox[t]{4.0cm}{RBC/UKQCD 07, 08, 08A, 10, 10A-B, 11, 12, 13, 16}& \parbox[t]{2.5cm}{\cite{Antonio:2007pb,Allton:2008pn,Boyle:2008yd,Boyle:2010bh,Aoki:2010dy,Aoki:2010pe,Kelly:2012uy,Arthur:2012opa,Boyle:2013gsa,Garron:2016mva}} & $2+1$ & \parbox[t]{2.0cm}{Iwasaki, Iwasaki+DSDR${}^*$ } &  DW \\
[7.0ex] \hline \\[-1.0ex]
RBC/UKQCD 08B, 09B, 10D, 12E  & \cite{Yamazaki:2008py,Yamazaki:2009zq,Aoki:2010xg,Boyle:2012qb} & $2+1$ & Iwasaki &  DW \\
[2.0ex] \hline \\[-1.0ex]
\parbox[t]{4.0cm}{RBC/UKQCD 14B, 15A, 15E}  &
\parbox[t]{2.5cm}{\cite{Blum:2014tka,Boyle:2015hfa,Boyle:2015exm,Hudspith:2015xoa,Hudspith:2018bpz}} & $2+1$ & \parbox[t]{2.0cm}{Iwasaki, Iwasaki+DSDR${}^*$} &  DW, \mbox{M-DW} \\
[4.0ex] \hline \\[-1.0ex]
Shanahan 12 & \cite{Shanahan:2012wh} & $2+1$ & Iwasaki & npSW \\
[2.0ex] \hline \\[-1.0ex]
Sternbeck 12 & \cite{Sternbeck:2012qs} & $2+1$ & tlSym & npSW \\
[2.0ex] \hline \\[-1.0ex]
\parbox[t]{4.0cm}{SWME 10, 11, 11A, 13, 13A, 14A, 14C, 15A} & \parbox[t]{3.0cm}{\cite{Bae:2010ki,Kim:2011qg,Bae:2011ff,Bae:2013lja,Bae:2013tca,Bae:2014sja,Jang:2014aea,Jang:2015sla}} & $2+1$ & tadSym & Asqtad${}^+$ \\
[4.0ex] \hline \\[-1.0ex]
\parbox[t]{4.0cm}{Takaura 18} &
\cite{Takaura:2018lpw,Takaura:2018vcy}
& $2+1$ & tlSym & M-DW \\ 
[2.0ex] \hline \\[-1.0ex]
TWQCD 08 & \cite{Chiu:2008jq}  & $2+1$ & Iwasaki &  DW \\
[2.0ex] \hline\hline\\
\end{tabular*}\\[-0.2cm]
\begin{minipage}{\linewidth}
{\footnotesize 
\begin{itemize}
   \item[${}^\dagger$] The calculation uses domain wall fermions in the valence-quark sector.\\[-5mm]
\item[${}^*$] An additional
weighting factor known as the dislocation suppressing determinant ratio (DSDR) is added to the gauge action \cite{Arthur:2012opa}.\\[-5mm]
\item[${}^+$] The calculation uses HYP smeared improved staggered fermions in the valence-quark sector.
\end{itemize}
}
\end{minipage}
}
\caption{(cntd.) Summary of simulated lattice actions with $\Nf=2+1$ or $\Nf=3$ quark flavours.}
\end{table}

\begin{table}[h]
{\footnotesize
\begin{tabular*}{\textwidth}[l]{l @{\extracolsep{\fill}} c c c c}
\hline \hline \\[-1.0ex]
Collab. & Ref. & $\Nf$ & \parbox{1cm}{gauge\\action} & \parbox{1cm}{quark\\action} 
\\[2.0ex] \hline \hline \\[-1.0ex]
ALPHA 10A & \cite{Tekin:2010mm} & $4$ & Wilson &  npSW \\
[2.0ex] \hline \\[-1.0ex]
CalLat 17, 18 & \cite{Berkowitz:2017gql,Chang:2018uxx} & $2+1+1$ 
 & tadSym & HISQ ${}^*$\\
[2.0ex] \hline \\[-1.0ex]
\parbox[t]{4.0cm}{ETM 10, 10E, 11, 11D, 12C, 13, 13A, 13D, 15E, 16} &  \parbox[t]{3.0cm}{\cite{Baron:2010bv,Farchioni:2010tb,Baron:2011sf,Blossier:2011tf,Blossier:2012ef,Cichy:2013gja,Blossier:2013ioa,Herdoiza:2013sla,Helmes:2015gla,Carrasco:2016kpy}} & $2+1+1$ & Iwasaki & tmWil \\
[7.0ex] \hline \\[-1.0ex]
\parbox[t]{4.0cm}{ETM 14A, 14B, 14E, 15, 15C, 17E} &  \parbox[t]{3.0cm}{\cite{Alexandrou:2014sha,Bussone:2014cha,Carrasco:2014poa,Carrasco:2015pra,Carrasco:2016kpy,Lubicz:2017asp}} & $2+1+1$ & Iwasaki & tmWil ${}^+$ \\
[4.0ex] \hline \\[-1.0ex]
\parbox[t]{4.0cm}{FNAL/MILC 12B, 12C, 13, 13C, 13E, 14A, 17, 18} &\parbox[t]{3.0cm}{ \cite{Bazavov:2012dg,Bailey:2012rr,Bazavov:2013nfa,Gamiz:2013xxa,Bazavov:2013maa,Bazavov:2014wgs,Bazavov:2017lyh,Bazavov:2018kjg}} & $2+1+1$ & tadSym & HISQ \\
[4.0ex] \hline \\[-1.0ex] 
HPQCD 14A, 15B, 18  & \cite{Chakraborty:2014aca,Koponen:2015tkr,Lytle:2018evc} & $2+1+1$ & tadSym & HISQ \\
[2.0ex] \hline \\[-1.0ex] 
MILC 12C, 13A, 18 & \cite{Freeman:2012ry,Bazavov:2013cp,Basak:2018yzz} & $2+1+1$ & tadSym & HISQ \\
[2.0ex] \hline \\[-1.0ex] 
Perez 10 & \cite{PerezRubio:2010ke} & $4$ & Wilson &  npSW \\
[2.0ex] \hline \\[-1.0ex]
\parbox[t]{4.0cm}{PNDME 13, 15, 15A, 16, 18, 18A, 18B} & \cite{Bhattacharya:2013ehc,Bhattacharya:2015wna,Bhattacharya:2015esa,Bhattacharya:2016zcn,Gupta:2018qil,Lin:2018obj,Gupta:2018lvp} & $2+1+1$ 
 & tadSym & HISQ ${}^\dagger$\\
 [4.0ex] \hline \hline\\[-1.0ex]
\end{tabular*}\\[-0.2cm]
\begin{minipage}{\linewidth}
{\footnotesize 
\begin{itemize}
  \item[${}^*$] The calculation uses M\"obius domain-wall fermions
    (M-DW) in the valence sector.\\[-5mm]
   \item[${}^+$] The calculation uses Osterwalder-Seiler fermions
     \cite{Osterwalder:1977pc} in the valence-quark sector.\\[-5mm]
   \item[${}^\dagger$] The calculation uses mean-field improved clover
     fermions (mfSW) in the valence-quark sector.
\end{itemize}
}
\end{minipage}
}
\caption{Summary of simulated lattice actions with $\Nf=4$ or $\Nf=2+1+1$ quark flavours.\label{tab:simulated Nf4 actions}}
\end{table}

\begin{table}[!ht]
{\footnotesize
\begin{tabular*}{\textwidth}[l]{l @{\extracolsep{\fill}} c c c c c c}
\hline\hline \\[-1.0ex]
Collab. & Ref. & $\Nf$ & Gauge & \multicolumn{3}{c}{Quark actions}  \\
& & & action & sea & light valence & heavy \\[1.0ex] \hline \hline \\[-1.0ex]
\parbox[t]{3.5cm}{ALPHA 11, 12A, 13, 14, 14B} & \parbox[t]{2.5cm}{\cite{Blossier:2011dk,Bernardoni:2012ti,Bernardoni:2013oda,Bernardoni:2014fva,Bahr:2014iqa}} & 2 &  plaquette & npSW  & npSW & HQET
\\[4.0ex] \hline \\[-1.0ex]
ALPHA 13C & \cite{Heitger:2013oaa} & 2 &  plaquette & npSW  & npSW & npSW
\\[2.0ex] \hline \\[-1.0ex]
Blossier 18 & \cite{Blossier:2018jol} & 2 &  plaquette & npSW  & npSW & npSW
\\[2.0ex] \hline \\[-1.0ex]
\parbox[t]{2.5cm}{Atoui 13} & \cite{Atoui:2013zza} & 2 &  tlSym & tmWil & tmWil & tmWil
\\[2.0ex] \hline \\[-1.0ex]
\parbox[t]{3.5cm}{ETM 09, 09D, 11B, 12A, 12B, 13B, 13C} & \parbox[t]{2.5cm}{\cite{Blossier:2009bx,Blossier:2009hg,DiVita:2011py,Carrasco:2012dd,Carrasco:2012de,Carrasco:2013zta,Carrasco:2013iba}} & 2 &  tlSym & tmWil & tmWil & tmWil
\\[4.0ex] \hline \\[-1.0ex]
ETM  11A & \cite{Dimopoulos:2011gx} & 2 &  tlSym & tmWil & tmWil & tmWil, static
\\[4.0ex] \hline \\[-1.0ex]
TWQCD 14 & \cite{Chen:2014hva} & 2 & plaquette & oDW & oDW & oDW \\[2.0ex] \hline\hline
\end{tabular*}
\caption{Summary of lattice simulations $\Nf=2$ sea-quark flavours and with $b$ and $c$ valence quarks.\label{tab:simulated Nf2 bc actions}}
}
\end{table}

\begin{table}[!ht]
{\footnotesize
\begin{tabular*}{\textwidth}[l]{l @{\extracolsep{\fill}} c c c c c c}
\hline\hline \\[-1.0ex]
Collab. & Ref. & $\Nf$ & Gauge & \multicolumn{3}{c}{Quark actions}  \\
& & & action & sea & light valence & heavy \\[1.0ex] \hline \hline \\[-2.0ex]
$\chi$QCD 14 & \cite{Yang:2014sea} & 2+1 & Iwasaki & DW & overlap & overlap \\[1.0ex] \hline \\[-2.0ex]
Datta 17 & \cite{Datta:2017aue} & 2+1 & \parbox[t]{1.0cm}{Iwasaki, Iwasaki +DSDR${}^+$} & DW     & DW     & RHQ 
\\[6.0ex] \hline \\[-2.0ex]
Detmold 16 & \cite{Detmold:2016pkz} & 2+1 & \parbox[t]{1.0cm}{Iwasaki, Iwasaki +DSDR${}^+$} & DW     & DW     & RHQ 
\\[6.0ex] \hline \\[-2.0ex]
\parbox[t]{3.5cm}{FNAL/MILC 04, 04A, 05, 08, 08A, 10, 11, 11A, 12, 13B} &  \parbox[t]{2.0cm}{\cite{Aubin:2004ej,Okamoto:2004xg,Aubin:2005ar,Bernard:2008dn,Bailey:2008wp,Bailey:2010gb,Bazavov:2011aa,Bouchard:2011xj,Bazavov:2012zs,Qiu:2013ofa}}  & 2+1 & tadSym & Asqtad & Asqtad & Fermilab
\\[9.0ex] \hline \\[-2.0ex]
FNAL/MILC 14, 15C, 16      &   \cite{Bailey:2014tva,Lattice:2015rga,Bazavov:2016nty}   & 2+1  & tadSym	&	Asqtad	& Asqtad${}^*$ &	Fermilab${}^*$\\[1.0ex] \hline \\[-2.0ex]
FNAL/MILC 15, 15D, 15E    &   \cite{Lattice:2015tia,Bailey:2015dka,Bailey:2015nbd}   & 2+1  & tadSym	&	Asqtad	& Asqtad &	Fermilab\\[1.0ex] \hline \\[-2.0ex]
\parbox[t]{3.5cm}{HPQCD 06, 06A, 08B, 09, 13B}& \parbox[t]{2.0cm}{\cite{Dalgic:2006dt,Dalgic:2006gp,Allison:2008xk,Gamiz:2009ku,Lee:2013mla}} & 2+1 &  tadSym & Asqtad & Asqtad & NRQCD
\\[4.0ex] \hline \\[-2.0ex]
HPQCD 12, 13E & \cite{Na:2012kp,Bouchard:2013pna} & 2+1 &  tadSym & Asqtad & HISQ & NRQCD
\\[1.0ex] \hline \\[-2.0ex]
HPQCD 15 & \cite{Na:2015kha} & 2+1 &  tadSym & Asqtad & HISQ${}^\dagger$ & NRQCD${}^\dagger$
\\[1.0ex] \hline \\[-2.0ex]
HPQCD 17 & \cite{Monahan:2017uby} & 2+1 &  tadSym & Asqtad & HISQ & \parbox[t]{1.2cm}{HISQ, NRQCD}
\\[4.0ex] \hline \\[-2.0ex]
\parbox[t]{3.5cm}{HPQCD/UKQCD 07, HPQCD 10A,  10B, 11, 11A, 12A, 13C}  & \parbox[t]{2.0cm}{\cite{Follana:2007uv,Davies:2010ip,Na:2010uf,Na:2011mc,McNeile:2011ng,Na:2012iu,Koponen:2013tua}} & 2+1 &  tadSym & Asqtad & HISQ & HISQ
\\[6.0ex] \hline \\[-2.0ex]
JLQCD 16 & \cite{Nakayama:2016atf} & 2+1 & tlSym & M-DW & M-DW & M-DW
\\[1.0ex] \hline \\[-2.0ex]
JLQCD 17B & \cite{Kaneko:2017xgg} & 2+1 & tlSym & DW & DW & DW
\\[1.0ex] \hline \\[-2.0ex]
Maezawa 16 & \cite{Maezawa:2016vgv} & 2+1 & tlSym & HISQ & HISQ & HISQ
\\[1.0ex] \hline \\[-2.0ex]
Meinel 16 & \cite{Meinel:2016dqj} & 2+1 & \parbox[t]{1.0cm}{Iwasaki, Iwasaki + DSDR${}^+$} & DW     & DW     & RHQ 
\\[6.0ex] \hline \\[-2.0ex]
PACS-CS 11 & \cite{Namekawa:2011wt} & 2+1 & Iwasaki & npSW & npSW & Tsukuba
\\[1.0ex] \hline \\[-2.0ex]  
RBC/UKQCD 10C, 14A & \cite{Albertus:2010nm,Aoki:2014nga} & 2+1 & Iwasaki & DW & DW & static
\\[1.0ex] \hline \\[-2.0ex]
RBC/UKQCD 13A, 14, 15 & \cite{Witzel:2013sla,Christ:2014uea,Flynn:2015mha} & 2+1 & Iwasaki & DW & DW & RHQ
\\[1.0ex] \hline \\[-2.0ex]
RBC/UKQCD 17 & \cite{Boyle:2017jwu} & 2+1 & Iwasaki & DW/M-DW & M-DW & M-DW
\\[1.0ex] \hline \\[-2.0ex]
\parbox[t]{3.5cm}{ETM 13E, 13F, 14E, 17D, 18}   & \parbox[t]{2.0cm}{\cite{Carrasco:2013naa,Dimopoulos:2013qfa,Carrasco:2014poa,Lubicz:2017syv,Lubicz:2018rfs}} & 2+1+1 &  Iwasaki & tmWil & tmWil & tmWil
\\[4.0ex] \hline \hline\\
\end{tabular*}\\[-0.2cm]
\begin{minipage}{\linewidth}
{\footnotesize 
\begin{itemize}
   \item[$^*$] Asqtad for $u$, $d$ and $s$ quark; Fermilab for $b$ and $c$ quark.\\[-5mm]
\item[${}^+$] An additional
weighting factor known as the dislocation suppressing determinant ratio (DSDR) is added to the gauge action \cite{Arthur:2012opa}.\\[-5mm]
   \item[$^\dagger$] HISQ for $u$, $d$, $s$  and $c$ quark; NRQCD for $b$ quark.
\end{itemize}
}
\end{minipage}
\caption{Summary of lattice simulations with $\Nf=2+1$ sea-quark flavours and $b$ and $c$ valence quarks.  \label{tab:simulated Nf3 bc actions}
}
}
\end{table}

\begin{table}[!ht]
{\footnotesize
\begin{tabular*}{\textwidth}[l]{l @{\extracolsep{\fill}} c c c c c c}
\hline\hline \\[-1.0ex]
Collab. & Ref. & $\Nf$ & Gauge & \multicolumn{3}{c}{Quark actions}  \\
& & & action & sea & light valence & heavy \\[1.0ex] \hline \hline \\[-2.0ex]
ETM 16B   & \cite{Bussone:2016iua} & 2+1+1 &  Iwasaki & tmWil & tmWil & tmWil$^+$
\\[1.0ex] \hline \\[-2.0ex]
FNAL/MILC 12B, 13, 14A & \cite{Bazavov:2012dg,Bazavov:2013nfa,Bazavov:2014wgs} & 2+1+1 & tadSym & HISQ & HISQ & HISQ
\\[1.0ex] \hline \\[-2.0ex]
FNAL/MILC 17 & \cite{Bazavov:2017lyh} & 2+1+1 & tadSym & HISQ & HISQ & HISQ \\
[1.0ex] \hline \\[-2.0ex] 
FNAL/MILC/TUMQCD 18& \cite{Bazavov:2018omf} & 2+1+1 & tadSym & HISQ & HISQ & HISQ \\
[1.0ex] \hline \\[-2.0ex] 
Gambino 17& \cite{Gambino:2017vkx} & 2+1+1 & Iwasaki & tmWil&tmWil&tmWil$^+$ \\
[1.0ex] \hline \\[-2.0ex] 
\parbox[t]{2.5cm}{HPQCD 13, 17A} & \cite{Dowdall:2013tga,Hughes:2017spc} &  2+1+1 & tadSym & HISQ & HISQ & NRQCD
\\[1.0ex] \hline\\[-2.0ex]
\parbox[t]{2.5cm}{HPQCD 17B} & \cite{Harrison:2017fmw} &  2+1+1 & tadSym & HISQ & HISQ & HISQ, NRQCD
\\[1.0ex] \hline\\[-2.0ex]
 RM123 17 & \cite{Giusti:2017dmp} & 2+1+1 & Iwasaki &  tmWil&tmWil&tmWil$^+$\\
[1.0ex] \hline\hline\\
\end{tabular*}\\[-0.2cm]
\begin{minipage}{\linewidth}
{\footnotesize 
\begin{itemize}
   \item[${}^+$] The calculation uses Osterwalder-Seiler fermions \cite{Osterwalder:1977pc} in the valence-quark sector.\\[-5mm]
\end{itemize}
}
\end{minipage}
\caption{Summary of lattice simulations with  $\Nf=2+1+1$ sea-quark flavours and $b$ and $c$ valence quarks.  \label{tab:simulated Nf4 bc actions}
}
}
\end{table}

\fi

%% file: Glossary/app_zParameterization.tex
\subsection{Parameterizations of semileptonic form factors}\label{sec:zparam}

In this section, we discuss the description of the $q^2$-dependence of
form factors, using the vector form factor $f_+$ of $B\to\pi\ell\nu$ decays
as a benchmark case. Since in this channel the parameterization of the
$q^2$-dependence is crucial for the extraction of $|V_{ub}|$ from the existing
measurements (involving decays to light leptons), as explained
above, it has been studied in great detail in the literature. Some comments
about the generalization of the techniques involved will follow.

\paragraph{The vector form factor for $B\to\pi\ell\nu$}

All form factors are analytic functions of $q^2$ outside physical
poles and inelastic threshold branch points; in the case of
$B\to\pi\ell\nu$, the only pole expected below the $B\pi$ production
region, starting at $q^2 = t_+ = (m_B+m_\pi)^2$, is the $B^*$.  A
simple ansatz for the $q^2$-dependence of the $B\to\pi\ell\nu$
semileptonic form factors that incorporates vector-meson dominance is
the Be\v{c}irevi\'c-Kaidalov (BK)
parameterization~\cite{Becirevic:1999kt},
which for the vector form factor reads:
\begin{gather}
f_+(q^2) = \frac{f(0)}{(1-q^2/m_{B^*}^2)(1-\alpha q^2/m_{B^*}^2)}\,.
\label{eq:BKparam}
\end{gather}
Because the BK ansatz has few free parameters, it has been used
extensively to parameterize the shape of experimental
branching-fraction measurements and theoretical form-factor
calculations.  A variant of this parameterization proposed by Ball and
Zwicky (BZ) adds extra pole factors to the expressions in
Eq.~(\ref{eq:BKparam}) in order to mimic the effect of multiparticle
states~\cite{Ball:2004ye}. A similar idea, extending the use of effective
poles also to $D\to\pi\ell\nu$ decays, is explored in Ref.~\cite{Becirevic:2014kaa}.
Finally, yet another variant (RH) has been proposed by
Hill in Ref.~\cite{Hill:2005ju}. Although all of these parameterizations
capture some known properties of form factors, they do not manifestly
satisfy others.  For example,
perturbative QCD scaling constrains the
behaviour of $f_+$ in the deep Euclidean region~\cite{Lepage:1980fj,Akhoury:1993uw,Lellouch:1995yv}, and
angular momentum conservation constrains the asymptotic behaviour near
thresholds---e.g., ${\rm Im}\,f_+(q^2) \sim (q^2-t_+)^{3/2}$ (see, e.g., Ref.~\cite{Bourrely:2008za}).  Most importantly, these parameterizations do not allow for an easy
quantification of systematic uncertainties.

A more systematic approach that improves upon the use of simple models
for the $q^2$ behaviour exploits the positivity and analyticity
properties of two-point functions of vector currents to obtain optimal
parameterizations of form
factors~\cite{Bourrely:1980gp,Boyd:1994tt,Lellouch:1995yv,Boyd:1997kz,Boyd:1997qw,Arnesen:2005ez,Becher:2005bg}.
Any form factor $f$ can be shown to admit a series expansion of the
form
\begin{gather}
f(q^2) = \frac{1}{B(q^2)\phi(q^2,t_0)}\,\sum_{n=0}^\infty a_n(t_0)\,z(q^2,t_0)^n\,,
\end{gather}
where the squared momentum transfer is replaced by the variable
\begin{gather}
z(q^2,t_0) = \frac{\sqrt{t_+-q^2}-\sqrt{t_+-t_0}}{\sqrt{t_+-q^2}+\sqrt{t_+-t_0}}\,.
\end{gather}
This is a conformal transformation, depending on an arbitrary real
parameter $t_0<t_+$, that maps the $q^2$ plane cut for $q^2 \geq t_+$
onto the disk $|z(q^2,t_0)|<1$ in the $z$ complex plane. The function
$B(q^2)$ is called the {\it Blaschke factor}, and contains poles and
cuts below $t_+$ --- for instance, in the case of $B\to\pi$ decays,
\begin{gather}
B(q^2)=\frac{z(q^2,t_0)-z(m_{B^*}^2,t_0)}{1-z(q^2,t_0)z(m_{B^*}^2,t_0)}=z(q^2,m_{B^*}^2)\,.
\end{gather}
Finally, the quantity $\phi(q^2,t_0)$, called the {\em outer
function}, is some otherwise arbitrary function that does not introduce further
poles or branch cuts.  The crucial property of this series expansion
is that the sum of the squares of the coefficients
\begin{gather}
\sum_{n=0}^\infty a_n^{2} = \frac{1}{2\pi i}\oint \frac{dz}{z}\,|B(z)\phi(z)f(z)|^2\,,
\end{gather}
is a finite quantity. Therefore, by using this parameterization an
absolute bound to the uncertainty induced by truncating the series can
be obtained.  The aim in choosing $\phi$ is to obtain
a bound that is useful in practice, while
(ideally) preserving the correct behaviour of the form factor at high
$q^2$ and around thresholds.

The simplest form of the bound would correspond to $\sum_{n=0}^\infty
a_n^{2}=1$.  {\it Imposing} this bound yields the following ``standard''
choice for the outer function
\begin{gather}
\label{eq:comp_of}
\begin{split}
\phi(q^2,t_0)=&\sqrt{\frac{1}{32\pi\chi_{1^-}(0)}}\,
\left(\sqrt{t_+-q^2}+\sqrt{t_+-t_0}\right)\\
&\times\,\left(\sqrt{t_+-q^2}+\sqrt{t_+-t_-}\right)^{3/2}
\left(\sqrt{t_+-q^2}+\sqrt{t_+}\right)^{-5}
\,\frac{t_+-q^2}{(t_+-t_0)^{1/4}}\,,
\end{split}
\end{gather}
where $t_-=(m_B-m_\pi)^2$, and $\chi_{1^-}(0)$ is the derivative of the transverse component of
the polarization function (i.e., the Fourier transform of the vector
two-point function) $\Pi_{\mu\nu}(q)$ at Euclidean momentum
$Q^2=-q^2=0$. It is computed perturbatively, using operator product
expansion techniques, by relating the $B\to\pi\ell\nu$ decay amplitude
to $\ell\nu\to B\pi$ inelastic scattering via crossing symmetry and
reproducing the correct value of the inclusive $\ell\nu\to X_b$ amplitude.
We will refer to the series parameterization with the outer function
in Eq.~(\ref{eq:comp_of}) as Boyd, Grinstein, and Lebed (BGL).  The
perturbative and OPE truncations imply that the bound is not strict,
and one should take it as
\begin{gather}
\sum_{n=0}^N a_n^{2} \lesssim 1\,,
\end{gather}
where this holds for any choice of $N$.  Since the values of $|z|$ in
the kinematical region of interest are well below~1 for judicious
choices of $t_0$, this provides a very stringent bound on systematic
uncertainties related to truncation for $N\geq 2$. On the other hand,
the outer function in Eq.~(\ref{eq:comp_of}) is somewhat unwieldy and,
more relevantly, spoils the correct large $q^2$ behaviour and induces
an unphysical singularity at the $B\pi$ threshold.

A simpler choice of outer function has been proposed by Bourrely,
Caprini and Lellouch (BCL) in Ref.~\cite{Bourrely:2008za}, which leads to a
parameterization of the form
\begin{gather}
\label{eq:bcl}
f_+(q^2)=\frac{1}{1-q^2/m_{B^*}^2}\,\sum_{n=0}^N a_n^{+}(t_0) z(q^2,t_0)^n\,.
\end{gather}
This satisfies all the basic properties of the form factor, at the price
of changing the expression for the bound to
\begin{gather}
\sum_{j,k=0}^N B_{jk}(t_0)a_j^{+}(t_0)a_k^{+}(t_0) \leq 1\,.
\end{gather}
The constants $B_{jk}$ can be computed and shown to be
$|B_{jk}|\lesssim \cO(10^{-2})$ for judicious choices of
$t_0$; therefore, one again finds that truncating at $N\geq 2$
provides sufficiently stringent bounds for the current level of
experimental and theoretical precision.  It is actually possible to
optimize the properties of the expansion by taking
\begin{gather}
t_0 = t_{\rm opt} = (m_B+m_\pi)(\sqrt{m_B}-\sqrt{m_\pi})^2\,,
\end{gather}
which for physical values of the masses results in the semileptonic
domain being mapped onto the symmetric interval $|z| \ltapprox 0.279$
(where this range differs slightly for the $B^{\pm}$ and $B^0$ decay
channels), minimizing the maximum truncation error.  If one also
imposes that the asymptotic behaviour ${\rm Im}\,f_+(q^2) \sim
(q^2-t_+)^{3/2}$ near threshold is satisfied, then the highest-order
coefficient is further constrained as
\begin{gather}
\label{eq:red_coeff}
a_N^{+}=-\,\frac{(-1)^N}{N}\,\sum_{n=0}^{N-1}(-1)^n\,n\,a_n^{+}\,.
\end{gather}
Substituting the above constraint on $a_N^{+}$ into Eq.~(\ref{eq:bcl})
leads to the constrained BCL parameterization
\begin{gather}
\label{eq:bcl_c}
f_+(q^2)=\frac{1}{1-q^2/m_{B^*}^2}\,\sum_{n=0}^{N-1} a_n^{+}\left[z^n-(-1)^{n-N}\,\frac{n}{N}\,z^N\right]\,,
\end{gather}
which is the standard implementation of the BCL parameterization used
in the literature.

Parameterizations of the BGL and BCL kind, to which we will refer
collectively as ``$z$-parameterizations'', have already been adopted
by the BaBar and Belle collaborations to report their results, and
also by the Heavy Flavour Averaging Group (HFAG, later renamed HFLAV).
Some lattice
collaborations, such as FNAL/MILC and ALPHA, have already started to
report their results for form factors in this way.  The emerging trend
is to use the BCL parameterization as a standard way of presenting
results for the $q^2$-dependence of semileptonic form factors. Our
policy will be to quote results for $z$-parameterizations when the
latter are provided in the paper (including the covariance matrix of
the fits); when this is not the case, but the published form factors
include the full correlation matrix for values at different $q^2$, we
will perform our own fit to the constrained BCL ansatz
in Eq.~(\ref{eq:bcl_c}); otherwise no fit will be quoted.
We however stress the importance of providing, apart from parameterization
coefficients, values for the form factors themselves (in the continuum limit
and at physical quark masses) for a number of values of $q^2$, so that
the results can be independently parameterized by the readers if so wished.

\paragraph{The scalar form factor for $B\to\pi\ell\nu$}

The discussion of the scalar $B\to \pi$ form factor is very similar. The main differences are the absence of a constraint analogue to Eq.~(\ref{eq:red_coeff}) and the choice of the overall pole function. In our fits we adopt the simple expansion:
\begin{gather}
\label{eq:bcl_f0}
f_0 (q^2) = \sum_{n=0}^{N-1} a_n^0 \; z^n \, .
\end{gather}
We do impose the exact kinematical constraint $f_+ (0) = f_0 (0)$ by expressing the $a_{N-1}^0$ coefficient in terms of all remaining $a_n^+$ and $a_n^0$ coefficients. This constraint introduces important correlations between the $a_n^+$ and $a_n^0$ coefficients; thus only lattice calculations that present the correlations between the vector and scalar form factors can be used in an average that takes into account the constraint at $q^2 = 0$. 

Finally we point out that we do not need to use the same number of parameters for the vector and scalar form factors. For instance, with $(N^+ = 3, N^0 = 3)$ we have $a_{0,1,2}^+$ and $a_{0,1}^0$, while with $(N^+ = 3, N^0 = 4)$ we have $a_{0,1,2}^+$ and $a_{0,1,2}^0$ as independent fit parameters. In our average we will choose the combination that optimizes uncertainties. 

\paragraph{Extension to other form factors}

The discussion above largely extends to form factors for other semileptonic transitions (e.g., $B_s\to K$ and $B_{(s)} \to D^{(*)}_{(s)}$,
and semileptonic $D$ and $K$ decays).
Details are discussed in the relevant sections.

A general discussion of semileptonic meson decay in this context can be found,
e.g., in Ref.~\cite{Hill:2006ub}. Extending what has been discussed above for
$B\to\pi$, the form factors for a generic $H \to L$
transition will display a cut starting at the production threshold $t_+$, and the optimal
value of $t_0$ required in $z$-parameterizations is $t_0=t_+(1-\sqrt{1-t_-/t_+})$
(where $t_\pm=(m_H\pm m_L)^2$).
For unitarity bounds to apply, the Blaschke factor has to include all sub-threshold
poles with the quantum numbers of the hadronic current --- i.e., vector (resp. scalar) resonances
in $B\pi$ scattering for the vector (resp. scalar) form factors of $B\to\pi$, $B_s\to K$,
or $\Lambda_b \to p$; and vector (resp. scalar) resonances
in $B_c\pi$ scattering for the vector (resp. scalar) form factors of $B\to D$
or $\Lambda_b \to \Lambda_c$.\footnote{A more complicated analytic structure
may arise in other cases, such as channels with vector mesons in the final state.
We will however not discuss form-factor parameterizations for any such process.}
Thus, as emphasized above, the control over systematic uncertainties brought in by using
$z$-parameterizations strongly depends on implementation details.
This has practical consequences, in particular, when the resonance spectrum
in a given channel is not sufficiently well-known. Caveats may also
apply for channels where resonances with a nonnegligible width appear.
A further issue is whether $t_+=(m_H+m_L)^2$ is the proper choice for the start of the cut in cases such as $B_s\to K\ell\nu$ and $B\to D\ell\nu$, where there are lighter two-particle states that project on the current ($B$,$\pi$ and $B_c$,$\pi$ for the two processes, respectively).\footnote{We are grateful
to G.~Herdo\'{\i}za, R.J.~Hill, A.~Kronfeld and A.~Szczepaniak for illuminating discussions
on this issue.}
In any such
situation, it is not clear a priori that a given $z$-parameterization will
satisfy strict bounds, as has been seen, e.g., in determinations of the proton charge radius
from electron-proton scattering~\cite{Hill:2010yb,Hill:2011wy,Epstein:2014zua}.

The HPQCD collaboration pioneered a variation on the $z$-parameterization
approach, which they refer to as a ``modified $z$-expansion,'' that
is used to simultaneously extrapolate their lattice simulation data
to the physical light-quark masses and the continuum limit, and to
interpolate/extrapolate their lattice data in $q^2$.  This entails
allowing the coefficients $a_n$ to depend on the light-quark masses,
squared lattice spacing, and, in some cases the charm-quark mass and
pion or kaon energy.  Because the modified $z$-expansion is not
derived from an underlying effective field theory, there are several
potential concerns with this approach that have yet to be studied.
The most significant is that there is no theoretical
derivation relating the coefficients of the modified $z$-expansion to
those of the physical coefficients measured in experiment; it
therefore introduces an unquantified model dependence in the
form-factor shape. As a result, the applicability of unitarity bounds has to be examined carefully.
Related to this, $z$-parameterization coefficients implicitly depend on quark masses,
and particular care should be taken in the event that some state can move
across the inelastic threshold as quark masses are changed (which would
in turn also affect the form of the Blaschke factor). Also, the lattice-spacing dependence of form factors provided by Symanzik effective theory
techniques may not extend trivially to $z$-parameterization coefficients.
The modified $z$-expansion is now being utilized by collaborations
other than HPQCD and for quantities other than $D \to \pi \ell \nu$
and $D \to K \ell \nu$, where it was originally employed.
We advise treating results that utilize the modified $z$-expansion to
obtain form-factor shapes and CKM matrix elements with caution,
however, since the systematics of this approach warrant further study.

\paragraph{Choice of form-factor basis for chiral-continuum extrapolations}

For pseudoscalar-to-pseudoscalar transitions $P_1\to P_2$ (such as $B\to \pi$ or $B_s \to K$), the chiral and continuum extrapolations
have often been performed in a different basis $f_\parallel$, $f_\perp$ given by \cite{El-Khadra:2001wco}
\begin{equation}
 \langle P_2 (p^\prime) | V^\mu | P_1(p) \rangle = \sqrt{2 M_1} [ v^\mu f_\parallel (E_2) + p^{\prime\mu}_\perp f_\perp(E_2) ]. \label{eq:fparallelfperp}
\end{equation}
Here, $v^\mu=p^\mu/M_1$ is the initial-meson four-velocity, $p^{\prime \mu}_\perp=p^{\prime\mu}-(v\cdot p^\prime)v^\mu$ is the projection of the final-meson
momentum in the direction perpendicular to $v^\mu$, and the form factors are taken to be functions of $E_2=v\cdot p^\prime $ (the energy of the final-state meson in the initial-meson rest frame). After the chiral and continuum extrapolations, the standard form factors are then constructed as the linear combinations
\begin{align}
f_0(q^2) &= \frac{\sqrt{2M_1}}{M^2_1-M^2_2}\big[ (M_1 - E_2)f_\parallel(E_2) + (E_2^2 - M^2_K)f_\perp(E_2)\big],\\
f_+(q^2) &= \frac{1}{\sqrt{2M_1}}\left[f_\parallel(E_2) + (M_1 -
  E_2)f_\perp(E_2)\right].
\end{align}
The decomposition (\ref{eq:fparallelfperp}) is motivated by heavy-meson chiral perturbation theory and is also convenient for the extraction of the form
factors from the correlation functions. For example, for $B\to \pi$, heavy-meson chiral perturbation theory predicts, at leading-order in both the chiral and the heavy-quark expansion,
\begin{align}
 f_\perp(E_\pi) &= \frac{1}{f_\pi}\frac{g_{B^*B\pi}}{E_\pi+\Delta_{B^*}}, \\
 f_\parallel(E_\pi) &= \frac{1}{f_\pi},
\end{align}
where $\Delta_{B^*}=M_{B^*}-M_B$. For a general transition $P_1\to P_2$, the chiral and continuum extrapolations were therefore commonly performed by fitting functions of the form
\begin{align}
 f_\perp(E_2) &= \frac{1}{E_2 + \Delta_\perp}\bigg[ ... \bigg]
\end{align}
and
\begin{align}
 f_\parallel(E_2) &= \frac{1}{E_2 + \Delta_\parallel}\bigg[ ... \bigg]\hspace{4ex}\text{or}\hspace{4ex} f_\parallel(E_2) = \bigg[ ... \bigg]
\end{align}
with $\Delta_\perp = M_{1^-}-M_1$ and $\Delta_\parallel = M_{0^+}-M_1$, where $M_{1^-}$ and $M_{0^+}$ denote the masses of the bound states with $J^P=1^-$ and $J^P=0^+$ that couple to the weak current, and the ellipsis in the brackets denote terms describing the remaining dependence on the quark masses, lattice spacing, and kinematics. The terms in front of the brackets introduce poles at $E_2=-\Delta$, which corresponds to $q^2\approx M_{J^P}^2$ for large $M_1$. Depending on the process, there may be no QCD-stable bound state with $J^P=0^+$, in which case this pole factor for $f_\parallel$ is usually omitted.

A problem with the above prescription is that, for finite heavy-quark mass, the $J^P$ quantum numbers of the poles appearing in the form factors are definite only in the helicity basis of the form factors, with $J^P=1^-$ for $f_+$ and $J^P=0$ for $f_0$. In particular, the form factor $f_\parallel$, being a linear combination of $f_+$ and $f_0$, also has a pole at the lower mass $M_{1^-}$ that is neglected when using the above functions. The alternative is to perform the chiral-continuum extrapolations for $f_+$ and $f_0$ using
\begin{align}
 f_+(E_2) &= \frac{1}{E_2 + \Delta_+}\bigg[ ... \bigg]
\end{align}
and
\begin{align}
 f_0(E_2) &= \frac{1}{E_2 + \Delta_0}\bigg[ ... \bigg]\hspace{4ex}\text{or}\hspace{4ex} f_0(E_2) = \bigg[ ... \bigg],
\end{align}
where $\Delta_+ = M_{1^-}-M_1$ and $\Delta_0 = M_{0^+}-M_1$ now truly correspond to the lowest pole in each form factor. The authors of Ref.~\cite{Flynn:2023nhi} found that this method (in the case of $B_s\to K$ form factors) yields significantly different results for the extrapolated $f_0$ when compared to extrapolating $f_\parallel$, $f_\perp$ and then reconstructing $f_+$ and $f_0$. Lattice determinations of the form factors based on extrapolations of $f_\parallel$, $f_\perp$ may therefore have an uncontrolled systematic error, and directly extrapolating $f_+$ and $f_0$ appears to be the better choice.

\subsection{Explicit parameterizations used in the form factor fits}
\label{sec:zparam_explicit}
In order to reconstruct the form factors from the results of fits performed using a $z$-parameterization it is necessary not only to use the correct version of the parameterization but also to adopt {\em exactly} the same numerical values for all ancillary quantities that enter the fit (e.g., location of poles). In particular, users must avoid utilizing the most updated numerical inputs for these quantities with $z$-coefficients extracted using older values. The purpose of this appendix is to eliminate all ambiguities in the implementation of the fit results presented in Secs.~\ref{sec:DDecays} and \ref{sec:BDecays}.

\subsubsection{$D\to K$ form factors}
\label{sec:app_D2K}

BCL parameterization:
\begin{align}
f_+(q^2) &= \frac{1}{1-q^2/m_{D_s^*}^2}\,\sum_{n=0}^{N^+-1} a_n^{+}\left[z^n-(-1)^{n-N^+}\,\frac{n}{N^+}\,z^N\right] \; , \\
f_0(q^2) &= \frac{1}{1-q^2/m_{D_s^*(0^+)}^2}\,\sum_{n=0}^{N^0-1} a_n^{0} z^n \; .  
\end{align}
The kinematical constraint $f_+(0)=f_0(0)$ is implemented expressing $a^0_{N^0-1}$ in terms of the other coefficients. We use $t_+=(m_D+m_K)^2$, $t_- = (m_D-m_K)^2$ and  $t_0 = t_+ - \sqrt{t_+ (t_+ - t_-)}$. The numerical inputs are: $m_D = 1.87265$ GeV, $m_{D_s^*} = 2.1122$ GeV, $m_{D_s^*(0^+)} = 2.317$ GeV, and $m_K = 0.495644$ GeV.

\subsubsection{$B\to \pi$ form factors}
\label{sec:app_B2pi}
BCL parameterization:
\begin{align}
f_+(q^2) &= \frac{1}{1-q^2/m_{B^*}^2}\,\sum_{n=0}^{N^+-1} a_n^{+}\left[z^n-(-1)^{n-N^+}\,\frac{n}{N^+}\,z^N\right] \; , \\
f_0(q^2) &= \sum_{n=0}^{N^0-1} a_n^{0} z^n \; .
\end{align}
The kinematical constraint $f_+(0)=f_0(0)$ is implemented expressing $a^0_{N^0-1}$ in terms of the other coefficients. We use $t_+=(m_B+m_\pi)^2$ and $t_0 = (m_B+m_\pi)(\sqrt{m_B}-\sqrt{m_\pi})$. The numerical inputs are: $m_{B^*} = 5.32471$ GeV, $m_B = 5.27934$ GeV and $m_\pi = 0.1349768$ GeV. 

Results for the form factor $f_T$ are taken directly from Ref.~\cite{Bailey:2015nbd} where we refer the reader for details on the parameterization.

\subsubsection{$B_s\to K$ form factors}
\label{sec:app_Bs2K}
BCL parameterization:
\begin{align}
f_+(q^2) &= \frac{1}{1-q^2/m_{B^*}^2}\,\sum_{n=0}^{N^+-1} a_n^{+}\left[z^n-(-1)^{n-N^+}\,\frac{n}{N^+}\,z^N\right] \; , \\
f_0(q^2) &= \frac{1}{1-q^2/m_{B^*(0^+)}^2}\,\sum_{n=0}^{N^0-1} a_n^{0} z^n \; .  
\end{align}
The kinematical constraint $f_+(0)=f_0(0)$ is implemented expressing $a^0_{N^0-1}$ in terms of the other coefficients. We use $t_+=(m_B+m_\pi)^2$, $t_- = (m_{B_s}-m_K)^2$ and  $t_0 = t_+ - \sqrt{t_+ (t_+ - t_-)}$. The numerical inputs are: $m_B = 5.27931$ GeV, $m_{B^*} = 5.3251$ GeV, $m_{B_s} = 5.36688$ GeV, $m_{B^*(0^+)} = 5.68$ GeV, $m_K = 0.493677$ GeV and $m_\pi = 0.1349766$ GeV. 

\subsubsection{$B\to K$ form factors}
\label{sec:app_B2K}
BCL parameterization:
\begin{align}
f_+(q^2) &= \frac{1}{1-q^2/m_{B_s^*}^2}\,\sum_{n=0}^{N^+-1} a_n^{+}\left[z^n-(-1)^{n-N^+}\,\frac{n}{N^+}\,z^N\right] \; , \\
f_0(q^2) &= \frac{1}{1-q^2/m_{B_s^*(0^+)}^2}\,\sum_{n=0}^{N^0-1} a_n^{0} z^n \; , \\  
f_T(q^2) &= \frac{1}{1-q^2/m_{B_s^*}^2}\,\sum_{n=0}^{N^T-1} a_n^{T} z^n \; .  
\end{align}
The kinematical constraint $f_+(0)=f_0(0)$ is implemented expressing $a^0_{N^0-1}$ in terms of the other coefficients. We use $t_+=(m_B+m_K)^2$ and $t_0 = (m_B+m_K)(\sqrt{m_B}-\sqrt{m_K})^2$. The numerical inputs are: $m_B = 5.27931$ GeV, $m_{B_s^*} = 5.4154$ GeV, $m_{B_s^*(0^+)} = 5.718$ GeV and $m_K = 0.493677$ GeV. 

\subsubsection{$B\to D$ form factors}
\label{sec:app_B2D}
BCL parameterization:
\begin{align}
f_+(q^2) &= \sum_{n=0}^{N^+-1} a_n^{+}\left[z^n-(-1)^{n-N^+}\,\frac{n}{N^+}\,z^N\right] \; , \\
f_0(q^2) &= \sum_{n=0}^{N^0-1} a_n^{0} z^n \; . 
\end{align}
The kinematical constraint $f_+(0)=f_0(0)$ is implemented expressing $a^0_{N^0-1}$ in terms of the other coefficients. We use $t_+=(m_B+m_D)^2$ and $t_0 = (m_B+m_D)(\sqrt{m_B}-\sqrt{m_D})$. The numerical inputs are: $m_B = 5.27931$ GeV and $m_D = (1.86483 + 1.86965)/2$ GeV. 

\subsubsection{$B_s\to D_s$ form factors}
\label{sec:app_Bs2Ds}
Results for the form factors are taken directly from Table VIII of Ref.~\cite{McLean:2019qcx} where we refer the reader for details on the parameterization.

\subsubsection{$B\to D^*$ form factors}
\label{sec:app_B2D*}
We adopt the BGL parameterization used in Ref.~\cite{FermilabLattice:2021cdg}: the form factors are given in Eqs.~(63) and (64), the poles for the Blaschke factors are given in Table~9, the four outer functions in Eqs.~(67)--(70) and the remaining numerical inputs in Table~10. We impose the kinematic constraints at zero and max recoil (see Eqs.(72) and (73) of Ref.~\cite{FermilabLattice:2021cdg}) by eliminating the coefficients $a_0^{F_1}$ and $a_0^{F_2}$. 

\subsubsection{$B_s\to D_s^*$ form factors}
\label{sec:app_Bs2Ds*}
We adopt the same BGL parameterization described in Sec.~\ref{sec:app_B2D*}. Both the outer functions and the location of the poles are identical to the $B\to D^*$ case, and the kinematical constraints are imposed in the same way. The only difference are the masses $m_{B_s} = 5.36688$ GeV and $m_{D_s^*} = 2.112$ GeV.

%% file: Glossary/app_qed.tex
\subsection{Inclusion of electromagnetic effects}
\label{app:qed}
Electromagnetism on a lattice can be formulated using a naive discretization of
the Maxwell action $S[A_{\mu}]=\frac{1}{4}\int d^4
x\,\sum_{\mu,\nu}[\partial_{\mu}A_{\nu}(x)-\partial_{\nu}A_{\mu}(x)]^2$. Even in
its noncompact form, the action remains gauge invariant. This is not the case
for non-Abelian theories for which one uses the traditional compact Wilson gauge
action (or an improved version of it). Compact actions for QED feature spurious
photon-photon interactions which vanish only in the
continuum limit. This is one of the main reason why the noncompact action is
the most popular so far. It was used in all the calculations presented in this
review. Gauge-fixing is necessary for noncompact actions because of the usual infinite measure of equivalent gauge orbits which contribute to the path integral. It was shown~\citep{Hansen:2018zre,Lucini:2015hfa} that gauge-fixing is not necessary with compact actions, including in the construction of interpolating operators for charged states. 

Although discretization is straightforward, simulating QED in a finite volume is
more challenging. Indeed, the long range nature of the interaction suggests
that important finite-size effects have to be expected. In the case of periodic
boundary conditions, the situation is even more critical: a naive implementation
of the theory features an isolated zero-mode singularity in the photon
propagator. It was first proposed in~\citep{Duncan:1996xy} to fix the global
zero-mode of the photon field $A_{\mu}(x)$ in order to remove it from the
dynamics. This modified theory is generally named $\mathrm{QED}_{\mathrm{TL}}$.
Although this procedure regularizes the theory and has the right classical
infinite-volume limit, it is nonlocal because of the zero-mode fixing. As
first discussed in~\citep{Borsanyi:2014jba}, the nonlocality in time of
$\mathrm{QED}_{\mathrm{TL}}$ prevents the existence of a transfer matrix, and
therefore a quantum-mechanical interpretation of the theory. Another
prescription named $\mathrm{QED}_{\mathrm{L}}$, proposed
in~\citep{Hayakawa:2008an}, is to remove the zero-mode of $A_{\mu}(x)$
independently for each time slice. This theory, although still
nonlocal in space, is local in time and has a well-defined transfer matrix.
Whether these nonlocalities constitute an issue to extract infinite-volume
physics from lattice-QCD+$\mathrm{QED}_{\mathrm{L}}$ simulations is, at the time
of this review, still an open question. However, it is known through analytical
calculations of electromagnetic finite-size effects at $\cO(\alpha)$ in hadron
masses~\citep{Hayakawa:2008an,deDivitiis:2013xla,Davoudi:2014qua,Borsanyi:2014jba,Fodor:2015pna,Lubicz:2016xro,Davoudi:2018qpl,DiCarlo:2021apt},
meson leptonic decays~\citep{Lubicz:2016xro,DiCarlo:2021apt}, and the hadronic vacuum
polarization~\citep{Bijnens:2019ejw} that $\mathrm{QED}_{\mathrm{L}}$ does not
suffer from a problematic (e.g., UV divergent) coupling of short- and
long-distance physics due to its nonlocality, and is likely safe to use for these quantities. Another strategy, first proposed
in~\citep{Gockeler:1989wj} and used by the QCDSF collaboration, is to bound the
zero-mode fluctuations to a finite range. Although more minimal, it is still
a nonlocal modification of the theory and so far finite-size effects for this
scheme have not been investigated. Two proposals for local
formulations of finite-volume QED emerged. The first one described
in~\citep{Endres:2015gda} proposes to use massive photons to regulate zero-mode
singularities, at the price of (softly) breaking gauge invariance. The second
one presented in~\citep{Lucini:2015hfa}, based on earlier works~\cite{Wiese:1991ku,Polley:1993bn}, avoids the zero-mode issue by using
anti-periodic boundary conditions for $A_{\mu}(x)$. In this approach, gauge
invariance requires the fermion field to undergo a charge conjugation
transformation over a period, breaking electric charge conservation. These local
approaches have the potential to constitute cleaner approaches to finite-volume
QED. They have led to first numerical studies at unphysical masses~\citep{Clark:2022wjy,RCstar:2022yjz}, but were not used in any calculation reviewed in this paper.

Once a finite-volume theory for QED is specified, there are various ways to
compute QED effects themselves on a given hadronic quantity. The most direct
approach, first used in~\citep{Duncan:1996xy}, is to include QED directly in the
lattice simulations and assemble correlation functions from charged quark
propagators. Another approach proposed in~\citep{deDivitiis:2013xla}, is to
exploit the perturbative nature of QED, and compute the leading-order
corrections directly in pure QCD as matrix elements of the electromagnetic
current. Both approaches have their advantages and disadvantages and as shown
in~\citep{Giusti:2017dmp}, are not mutually exclusive. A critical comparative
study can be found in~\citep{Boyle:2017gzv}.

Finally, most of the calculations presented here made the choice of computing
electromagnetic corrections in the electro-quenched approximation. In this
limit, one assumes that only valence quarks are charged, which is equivalent to
neglecting QED corrections to the fermionic determinant. This approximation reduces
dramatically the cost of lattice-QCD+QED calculations since it allows the reuse of previously generated QCD configurations. If QED is introduced pertubatively through current insertions, the electro-quenched approximation avoids computing disconnected contributions coming from the electromagnetic current in the vacuum, which are generally challenging to determine precisely.
The electromagnetic contributions from sea quarks to hadron-mass splittings are known to be flavour-SU(3) 
and large-$N_c$ suppressed, thus electro-quenched simulations are
expected to have an $\cO(10\%)$ accuracy for the leading electromagnetic effects.
This suppression is in principle rather weak and results obtained from
electro-quenched simulations might feature uncontrolled systematic errors. For
this reason, the use of the electro-quenched approximation constitutes the
difference between \good~and \soso~in the FLAG criterion for the inclusion of
isospin-breaking effects.

%% file: qmass/qmass_app.tex
\subsection{Notes to Sec.~\ref{sec:qmass} on quark masses}

\ifx\reducedapptables\undefined 
\begin{table}[!ht]
{\footnotesize
\begin{tabular*}{\textwidth}[l]{l @{\extracolsep{\fill}} c c c l}
\hline\hline \\[-1.0ex]
Collab. & Ref. & $\Nf$ & $a$ [fm] & Description
\\[1.0ex] \hline \hline \\[-1.0ex]
ETM 21A  & \cite{Alexandrou:2021gqw}  & 2+1+1 & \parbox[t]{3cm}{0.07, 0.08, 0.09}  & \parbox[t]{4.5cm}{Scale set from $w_0$.} \\	[2.0ex] 
\\
[4.0ex]\hline\\[-1.0ex]
	MILC 18  & \cite{Basak:2018yzz}  & 2+1+1 & \parbox[t]{3cm}{\centering 0.042, 0.06, 0.09, 0.12, 0.15\\}& \parbox[t]{4.5cm}{Scale set from $f_\pi$.} \\
[4.0ex] \hline \\
\end{tabular*}
\caption{Continuum extrapolations/estimation of lattice artifacts in
  determinations of $m_{ud}$, $m_s$ and, in some cases $m_u$ and
  $m_d$, with $\Nf=2+1+1$ quark flavours.}
}
\end{table}
\fi

\begin{table}[!ht]
{\footnotesize
\begin{tabular*}{\textwidth}[l]{l @{\extracolsep{\fill}} c c c l}
\hline\hline \\[-1.0ex]
Collab. & Ref. & $\Nf$ & $a$ [fm] & Description
\\[1.0ex] \hline \hline \\[-1.0ex]
CLQCD 23  & \cite{CLQCD:2023sdb} & 2+1&0.052, 0.077, 0.11 & \parbox[t]{5.0cm}{smeared Wilson-clover/Symanzik} \\
[2.0ex] %
\ifx\reducedapptables\undefined
\hline \\[-1.0ex]
ALPHA 19  & \cite{Bruno:2019vup} & 2+1 & 0.05, 0.064, 0.076, 0.086& \parbox[t]{5.0cm}{improved Wilson-clover/Symanzik}\\
[2.0ex] \hline
\\[-1.0ex]
{BMW 16A}& \cite{Fodor:2016bgu} & 2+1 & 0.05, 0.06, 0.07, 0.09, 0.11 &  \parbox[t]{4.5cm}{Scale set through $M_{\Omega}$ and $M_{\Xi}$.}\\
[2.0ex] \hline \\[-1.0ex]
{MILC 16}& \cite{Basak:2016jnn} & 2+1 & 0.045, 0.06, 0.09, 0.12 &  \parbox[t]{4.5cm}{Scale set through $f_\pi$.}
 \hline \\[-1.0ex]
{QCDSF/UKQCD 15}& \cite{Horsley:2015eaa} & 2+1 & 0.07 &  \parbox[t]{4.5cm}{Scale set through the gradient flow parameter $w_0$.}\\
[4.0ex] \hline \\[-1.0ex]
{RBC/UKQCD 14B}& \cite{Blum:2014tka} & 2+1 & 0.063, 0.084,\, 0.114,\, 0.144  &  \parbox[t]{4.5cm}{Scale set through $M_\Omega$.}\\
[2.0ex] \hline \\[-1.0ex]
{RBC/UKQCD 12}& \cite{Arthur:2012opa} & 2+1 & 0.085,\, 0.113,\, 0.144  &  \parbox[t]{4.5cm}{Scale set through $M_\Omega$.}\\
[2.0ex] \hline \\[-1.0ex]
PACS-CS 12 & \cite{Aoki:2012st} & 1+1+1 & 0.09
& \parbox[t]{4.5cm}{Reweighting of PACS-CS 08 $\Nf=2+1$ QCD
  configurations with e.m. and $m_u\ne m_d$.} \\
[7.0ex] \hline \\[-1.0ex]
Laiho 11 & \cite{Laiho:2011np} & 2+1 & 0.06, 0.09, 0.15
& \parbox[t]{4.5cm}{MILC staggered ensembles~\cite{Bazavov:2010yq}, scale set using
  $r_1$ determined by HPQCD with $\Upsilon$ splittings, pseudoscalar
  decay constants, through $r_1$~\cite{Davies:2009tsa}.} \\ [13.0ex]
\fi
\hline \hline
\\[-1.0ex]
\end{tabular*}
\caption{Continuum extrapolations/estimation of lattice artifacts in
  determinations of $m_{ud}$, $m_s$ and, in some cases $m_u$ and
  $m_d$, with $\Nf=2+1$ quark flavours.}
}
\end{table}

\ifx\reducedapptables\undefined
\begin{table}[!ht]
\addtocounter{table}{-1}
{\footnotesize
 \begin{tabular*}{\textwidth}[l]{l @{\extracolsep{\fill}} c c c l}
 \hline\hline \\[-1.0ex]
 Collab. & Ref. & $\Nf$ & $a$ [fm] & Description
 \\[1.0ex] \hline \hline \\[-1.0ex]
PACS-CS 10 & \cite{Aoki:2010wm} &  2+1 &  0.09
& \parbox[t]{4.5cm}{cf.~PACS-CS 08} \\
[1.0ex] \hline \\[-1.0ex]
{MILC 10A}& \cite{Bazavov:2010yq} & 2+1 & & cf.~MILC 09, 09A \\
[1.0ex] \hline \\[-1.0ex]
{BMW 10A, 10B}& \cite{Durr:2010vn,Durr:2010aw}
& 2+1 & \parbox[t]{2.3cm}{0.054, 0.065, 0.077, 0.093, 0.116} &  Scale set via
$M_\pi, M_K, M_\Omega$.\\
\end{tabular*}
 \caption{(cntd.) Continuum extrapolations/estimation of lattice artifacts in
   determinations of $m_{ud}$, $m_s$ and, in some cases $m_u$ and
   $m_d$, with $\Nf=2+1$ quark flavours.}
}
\end{table}
\newpage
\fi

\ifx\reducedapptables\undefined
\begin{table}[!ht]
{\footnotesize
\begin{tabular*}{\textwidth}[l]{l @{\extracolsep{\fill}} c c c l}
\hline \hline  \\[-1.0ex]
Collab. & Ref. & $\Nf$ & $M_{\pi,\text{min}}$ [MeV] & Description
\\[1.0ex] \hline \hline \\[-1.0ex]
ETM 21A  & \cite{Alexandrou:2021gqw}  & 2+1+1 &physical  & \parbox[t]{4.5cm}{} \\[1.0ex]
[4.0ex]\hline\\[-1.0ex]
MILC 18  & \cite{Basak:2018yzz}  & 2+1+1 & $130~\mathrm{MeV}$  & \parbox[t]{6.0cm}{$m_u-m_d$ obtained from $\hat{M}_{K^+}-\hat{M}_{K^0}$ extrapolated to physical masses. EM contributions are subtracted using $\epsilon$ from a dedicated $\Nf=2+1$ QCD+QED calculation.} \\
[9.0ex] \hline \\[-1.0ex]
FNAL/MILC/TUMQCD 18  & \cite{Bazavov:2018omf}  & 2+1+1 & $130~\mathrm{MeV}$  & \parbox[t]{6.0cm}{Mass extrapolation done through a general fit function which combines HQET and $\chi$PT terms.} \\
\end{tabular*}
\caption{Chiral extrapolation/minimum pion mass in determinations of
  $m_{ud}$, $m_s$ and, in some cases, $m_u$ and
  $m_d$,  with $\Nf=2+1+1$ quark flavours.}
}
\end{table}
\fi

\begin{table}[!ht]
{\footnotesize
%
\caption{Chiral extrapolation/minimum pion mass in determinations of
  $m_{ud}$, $m_s$, and in some cases $m_u$ and
  $m_d$,  with $\Nf=2+1$ quark flavours.}
}
\end{table}

\ifx\reducedapptables\undefined
\begin{table}[!ht]
\addtocounter{table}{-1}
{\footnotesize
\begin{tabular*}{\textwidth}[l]{l @{\extracolsep{\fill}} c c c l}
\hline \hline  \\[-1.0ex]
Collab. & Ref. & $\Nf$ & $M_{\pi,\text{min}}$ [MeV] & Description
\\[1.0ex] \hline \hline \\[-1.0ex]

{PACS-CS 09}& \cite{Aoki:2009ix}& 2+1 & 135 & \parbox[t]{5.5cm}{Physical point reached by
reweighting technique, no chiral extrapolation needed.}\\
[7.0ex] \hline \\[-1.0ex]
{HPQCD 09A, 10}&  \cite{Davies:2009ih,McNeile:2010ji}& 2+1 &  &  cf.~MILC 09\\
[1.0ex] \hline \\[-1.0ex]
{MILC 09A, 09} & \cite{Bazavov:2009fk,Bazavov:2009bb} & 2+1 & 177, 224 & \parbox[t]{5.5cm}{
		NLO SU(3) RS{\Ch}PT, continuum {\Ch}PT at NNLO and NNNLO and NNNNLO 
		analytic terms. The lightest Nambu-Goldstone mass
                is 177 MeV (09A) and 224 MeV (09)
		(at $a=$0.09fm) and the lightest RMS mass is 258MeV (at
                $a=$0.06fm).} \\ 
[18.0ex] \hline \\[-1.0ex]
PACS-CS 08 &\cite{Aoki:2008sm}& 2+1 &156& 
		\parbox[t]{5.5cm}{NLO SU(2) {\Ch}PT and SU(3)
                  (Wilson){\Ch}PT.}\\
[4.0ex] \hline \\[-1.0ex]
RBC/UKQCD 08	&\cite{Allton:2008pn}	&2+1& \parbox[t]{2cm}{242 (valence), 330 (sea)} & \parbox[t]{5.5cm}{SU(3) PQ{\Ch}PT 
and heavy kaon NLO SU(2) PQ{\Ch}PT fits.} \\
[4.0ex] \hline \\[-1.0ex]
{CP-PACS/JLQCD 07}&  \cite{Ishikawa:2007nn} & 2+1 & 620 & \parbox[t]{5.5cm}{NLO Wilson {\Ch}PT
fits to meson masses.} \\
[1.0ex] \hline \\[-1.0ex]
{HPQCD 05}& \cite{Mason:2005bj}& 2+1 & 240 & \parbox[t]{5.5cm}{PQ RS{\Ch}PT fits.} \\
[1.0ex] \hline \\[-1.0ex]
\parbox[t]{4.5cm}{HPQCD/MILC/UKQCD 04,\\ MILC 04}
& \cite{Aubin:2004ck,Aubin:2004fs}& 2+1 & 240 & \parbox[t]{5.5cm}{PQ RS{\Ch}PT fits.} \\
[4.0ex] \hline \hline \\[-1.0ex]
\end{tabular*}
\caption{(cntd.) Chiral extrapolation/minimum pion mass in determinations of
  $m_{ud}$, $m_s$ and, in some cases $m_u$ and
  $m_d$,  with $\Nf=2+1$ quark flavours.}
}
\end{table}
\newpage
\fi

\ifx\reducedapptables\undefined
\begin{table}
{\footnotesize
\begin{tabular*}{\textwidth}[l]{l @{\extracolsep{\fill}} c c c c l}
\hline\hline  \\[-1.0ex]
Collab. & Ref. & $\Nf$ & $L$ [fm] & $M_{\pi,\text{min}} L$ & Description
\\[1.0ex] \hline \hline \\[-1.0ex]
ETM 21A  & \cite{Alexandrou:2021gqw}  & 2+1+1 & 2.5 -- 5.6 & 3.6 & \parbox[t]{4.5cm}{} \\[1.0ex]
[4.0ex]\hline\\[-1.0ex]
MILC 18  & \cite{Basak:2018yzz}  & 2+1+1 & 4.83 -- 6.12  & 3.2 & \parbox[t]{4.3cm}{Universal $\mathrm{QED}_{\mathrm{TL}}$ FV effects subtracted and higher order model using S$\chi$PT for the $\Nf=2+1$ calculation of $\epsilon$.} \\
[9.0ex] \hline \\[-1.0ex]
HPQCD 18  & \cite{Lytle:2018evc}  & 2+1+1 & 2.88 -- 5.76  & 3.8 & \parbox[t]{4.3cm}{Finite-volume effects negligible for strange-quark mass.} \\
\end{tabular*}
\caption{Finite-volume effects in determinations of $m_{ud}$, $m_s$
  and, in some cases $m_u$ and $m_d$, with $\Nf=2+1+1$ quark flavours.}
}
\end{table}
\fi

\begin{table}
{\footnotesize
\begin{tabular*}{\textwidth}[l]{l @{\extracolsep{\fill}} c c c c l}
\hline\hline  \\[-1.0ex]
Collab. & Ref. & $\Nf$ & $L$ [fm] & $M_{\pi,\text{min}} L$ & Description
\\[1.0ex] \hline \hline \\[-1.0ex]
CLQCD 23  & \cite{CLQCD:2023sdb} & 2+1 & 2.5 -- 5.1 & 3.45   &\\
[2.0ex]
\ifx\reducedapptables\undefined
\hline \\[-1.0ex]
ALPHA 19  & \cite{Bruno:2019vup} & 2+1 & 2.4 -- 4.1 & $\gtrsim 4.0$ &\\ [1.0ex] 
\hline\\[-1.0ex]
{BMW 16A}& \cite{Fodor:2016bgu} & 2+1 &  $\gtrsim 5.0$ & $\gtrsim 4.0$ &  \parbox[t]{4.5cm}{Universal $\mathrm{QED}_{\mathrm{TL}}$ FV effects subtracted and $\cO(1/L^3)$ fitted.}\\
[4.0ex] \hline \\[-1.0ex]
{MILC 16}& \cite{Basak:2016jnn} & 2+1 & 1.4 -- 5.5 & 2.7 & \parbox[t]{4.5cm}{Universal $\mathrm{QED}_{\mathrm{TL}}$ FV effects subtracted and higher order model using S$\chi$PT.}\\[7.0ex] 
\hline \\[-1.0ex]
{QCDSF/UKQCD 15}& \cite{Horsley:2015eaa} & 2+1 & 1.7,\,2.2,\,3.4& & \parbox[t]{4.5cm}{Effective field theory used to extrapolate to infinite volume.} \\
[4.0ex] \hline \\[-1.0ex]
{RBC/UKQCD 14B}& \cite{Blum:2014tka} & 2+1 & 2.0,\,2.7,\,4.6,\,5.4 &  $3.8$ &  
\parbox[t]{4.5cm}{Uses FV chiral perturbation theory to estimate the error, which is deemed negligible and omitted.}\\
[10.0ex] \hline \\[-1.0ex]
{RBC/UKQCD 12}& \cite{Arthur:2012opa} & 2+1 & 2.7, 4.6 &  $\gtrsim 4.0$
& \parbox[t]{4.5cm}{Uses FV chiral perturbation theory to estimate the error.} \\
[4.0ex] \hline \\[-1.0ex]
PACS-CS 12 & \cite{Aoki:2012st} &  1+1+1 & & & cf.~PACS-CS 08 \\
[1.0ex] \hline \\[-1.0ex]
Laiho 11 & \cite{Laiho:2011np} & 2+1 &
\parbox[t]{2cm}{2.5, 2.9, 3.0,\\ 3.6, 3.8, 4.8}&\parbox[t]{1.5cm}{4.1 (val.)\\
4.1 (sea)}&  \parbox[t]{4.5cm}{Data corrected using NLO SU(3) $\chi$PT finite-V formulae.} \\
[4.0ex] 
\fi
\hline \hline\\[-1.0ex]
\end{tabular*}
\caption{Finite-volume effects in determinations of $m_{ud}$, $m_s$
  and, in some cases $m_u$ and $m_d$, with $\Nf=2+1$ quark flavours.}
}
\end{table}
 
\ifx\reducedapptables\undefined
\begin{table}
\addtocounter{table}{-1}
{\footnotesize
\begin{tabular*}{\textwidth}[l]{l @{\extracolsep{\fill}} c c c c l}
\hline\hline  \\[-1.0ex]
Collab. & Ref. & $\Nf$ & $L$ [fm] & $M_{\pi,\text{min}} L$ & Description
\\[1.0ex] \hline \hline \\[-1.0ex]
PACS-CS 10 & \cite{Aoki:2010wm} &  2+1 & & & cf.~PACS-CS 08 \\
[1.0ex] \hline \\[-1.0ex]
{MILC 10A}& \cite{Bazavov:2010yq} & 2+1 & & & cf.~MILC 09A, 09 \\
[1.0ex] \hline \\[-1.0ex]
{BMW 10A, 10B}& \cite{Durr:2010vn,Durr:2010aw}
& 2+1 &  $\gtrsim 5.0$ & $\gtrsim 4.0$ & \parbox[t]{3.7cm}{FV corrections
  below 5 per mil on the largest lattices.}\\
\end{tabular*}
\caption{(cntd.) Finite-volume effects in determinations of $m_{ud}$, $m_s$
  and, in some cases $m_u$ and $m_d$, with $\Nf=2+1$ quark flavours.}
}
\end{table}
\newpage
\fi

\ifx\reducedapptables\undefined
\begin{table}[!ht]
{\footnotesize
\begin{tabular*}{\textwidth}[l]{l @{\extracolsep{\fill}} c c l}
\hline\hline \\[-1.0ex]
Collab. & Ref. & $\Nf$  & Description
\\[1.0ex] \hline \hline \\[-1.0ex]
ETM 21A  & \cite{Alexandrou:2021gqw}  & 2+1+1 &  Nonperturbative (RI/MOM) \\[1.0ex]
[4.0ex]\hline\\[-1.0ex]
MILC 18  & \cite{Basak:2018yzz}  & 2+1+1 & \parbox[t]{4.7cm}{Renormalization not required for $m_u/m_{d}$.} \\
[4.0ex] \hline \\[-1.0ex]
FNAL/MILC/TUMQCD 18  & \cite{Bazavov:2018omf}  & 2+1+1 & \parbox[t]{4.7cm}{MRS scheme from~\citep{Brambilla:2017hcq}.} \\
[3.0ex] \hline \\[-1.0ex]
FNAL/MILC 17  & \cite{Bazavov:2017lyh}  & 2+1+1 & \parbox[t]{4.7cm}{Renormalization not required for quark mass ratios.}\\
[4.0ex] \hline \\[-1.0ex]
HPQCD 18  & \cite{Lytle:2018evc}  & 2+1+1 & \parbox[t]{4.7cm}{Nonperturbative renormalization (RI/SMOM).}\\
[4.0ex] \hline \\[-1.0ex]
RM123 17  & \cite{Giusti:2017dmp}  & 2+1+1 & \parbox[t]{4.7cm}{Nonperturbative renormalization (RI/MOM).}\\[4.0ex] 
\hline \\[-1.0ex]
HPQCD 14A & \cite{Chakraborty:2014aca} & 2+1+1 & \parbox[t]{4.7cm}{Renormalization not required through the use of the ratio $m_c/m_s$.} \\
[7.0ex] \hline \\[-1.0ex]
{FNAL/MILC 14A}& \cite{Bazavov:2014wgs} & 2+1+1 & \parbox[t]{4.7cm}{Renormalization not required for $m_s/m_{ud}$.} \\
[4.0ex] \hline \\[-1.0ex]
ETM 14& \cite{Carrasco:2014cwa} & 2+1+1 & \parbox[t]{4.7cm}{Nonperturbative renormalization (RI/MOM).} \\[4.0ex]
 \hline  \hline \\[-1.0ex]
\end{tabular*}
\caption{Renormalization in determinations of $m_{ud}$, $m_s$
  and, in some cases $m_u$ and $m_d$, with $\Nf=2+1+1$ quark flavours.}
}
\end{table}
\fi

\begin{table}[!ht]
{\footnotesize
%
\caption{Renormalization in determinations of $m_{ud}$, $m_s$
  and, in some cases $m_u$ and $m_d$, with $\Nf=2+1$ quark flavours.}
}
\end{table}
 

%% file: qmass/qmass_mc_app.tex
\clearpage
\ifx\reducedapptables\undefined
\begin{table}[!ht]
{\footnotesize
\begin{tabular*}{\textwidth}[l]{l @{\extracolsep{\fill}} c c c l}
\hline\hline \\[-1.0ex]
Collab. & Ref. & $\Nf$ & $a$ [fm] & Description
\\[1.0ex] \hline \hline \\[-1.0ex]
ETM 21A  & \cite{Alexandrou:2021gqw}  & 2+1+1 & \parbox[t]{3cm}{0.07, 0.08, 0.09}  & \parbox[t]{4.5cm}{Scale set from $w_0$. Twisted mass action for charm quarks.} \\
[4.0ex]\hline\\[-1.0ex]
HPQCD 20A  & \cite{Hatton:2020qhk}  & 2+1+1 & \parbox[t]{3cm}{0.03, 0.042,0.06, 0.09, 0.12}  & \parbox[t]{4.5cm}{Scale set from $w_0$ and $f_\pi$. HISQ action for charm quarks.} \\[4.0ex]
\hline\\[-1.0ex]
HPQCD 18  & \cite{Lytle:2018evc}  & 2+1+1 & \parbox[t]{3cm}{0.06, 0.09, 0.12}  & \parbox[t]{4.5cm}{Scale set from $w_0$ and $f_\pi$. HISQ action for charm quarks.} \\
[4.0ex]\hline\\[-1.0ex]
FNAL/MILC/TUMQCD 18  & \cite{Bazavov:2018omf}  & 2+1+1 & \parbox[t]{3cm}{0.03, 0.042, 0.06, 0.09, 0.12, 0.15}  & \parbox[t]{4.5cm}{Scale set from $f_\pi$. HISQ action for charm quarks.} \\
[4.0ex] 
\hline \\[-1.0ex]
HPQCD 14A & \cite{Chakraborty:2014aca} & 2+1+1 & 0.06, 0.09, 0.12, 0.15  & \parbox[t]{4.5cm}{Scale set through the Wilson flow parameter $w_0$.} \\
[4.0ex] \hline \\[-1.0ex]
ETM 14& \cite{Carrasco:2014cwa} & 2+1+1 & 0.062, 0.082, 0.089  & \parbox[t]{4.5cm}{Scale set through $F_\pi$.}\\
[1.0ex] \hline \\[-1.0ex]
ETM 14A & \cite{Alexandrou:2014sha} & 2+1+1 & 0.062, 0.082, 0.089 & \parbox[t]{4.5cm}{Scale set through the nucleon mass $M_N$.} \\
[4.0ex]
 \hline\hline \\[-1.0ex]
\end{tabular*}
\caption{Continuum extrapolations/estimation of lattice artifacts in the determinations of $m_c$ with $\Nf=2+1+1$ quark flavours.}
}
\end{table}
\fi

\begin{table}[!ht]
{\footnotesize
%
\caption{Continuum extrapolations/estimation of lattice artifacts in the determinations of $m_c$ with $\Nf=2+1$ quark flavours.}
}
\end{table}

\ifx\reducedapptables\undefined
\begin{table}[!ht]
{\footnotesize
\begin{tabular*}{\textwidth}[l]{l @{\extracolsep{\fill}} c c c l}
\hline \hline  \\[-1.0ex]
Collab. & Ref. & $\Nf$ & $M_{\pi,\text{min}}$ [MeV] & Description
\\[1.0ex] \hline \hline \\[-1.0ex]
ETM 21A  & \cite{Alexandrou:2021gqw}  & 2+1+1 &physical  & \parbox[t]{4.5cm}{} \\
[4.0ex]\hline\\[-1.0ex]
HPQCD 20A  & \cite{Hatton:2020qhk}  & 2+1+1 &  physical   & \parbox[t]{4.5cm}{} \\[2.0ex]
[4.0ex]\hline\\[-1.0ex]
HPQCD 18  & \cite{Lytle:2018evc}  & 2+1+1 & $130~\mathrm{MeV}$  & \parbox[t]{4.50cm}{Fits include sea quark mass tunings. $m_c$ set from $\eta_c$ meson.} \\[4.0ex]\hline\\[-1.0ex]
FNAL/MILC/TUMQCD 18  & \cite{Bazavov:2018omf} & 2+1+1 & $130$ ($173_{RMS}$)& \\
[1.0ex]
 \hline \\[-1.0ex]
HPQCD 14A & \cite{Chakraborty:2014aca} & 2+1+1 & $128$ ($173_{RMS}$)& \\
[1.0ex] \hline \\[-1.0ex]
ETM 14& \cite{Carrasco:2014cwa} & 2+1+1 & $180_{\pi^0}$ ($220_{\pi^\pm}$)  & \\
[1.0ex] \hline \\[-1.0ex]
ETM 14A & \cite{Alexandrou:2014sha} & 2+1+1 & 210 & cf.~ETM 14 \\
[1.0ex]
 \hline  \hline \\[-1.0ex]
\end{tabular*}
\caption{Chiral extrapolation/minimum pion mass in the determinations of $m_c$ with $\Nf=2+1+1$ quark flavours.}
}
\end{table}
\fi

\begin{table}[!ht]
{\footnotesize
\begin{tabular*}{\textwidth}[l]{l @{\extracolsep{\fill}} c c c l}
\hline \hline  \\[-1.0ex]
Collab. & Ref. & $\Nf$ & $M_{\pi,\text{min}}$ [MeV] & Description
\\[1.0ex] \hline \hline \\[-1.0ex]
  {ALPHA~23} &\cite{Bussone:2023kag} & 2+1 & 200    & \parbox[t]{4.5cm}{} \\[2.0ex]
\ifx\reducedapptables\undefined
\hline \\[-1.0ex]
ALPHA 21  & \cite{Heitger:2021apz} & 2+1 & 198 & \\
[2.0ex] \hline \\[-1.0ex]
\\[-2.0ex]
Petreczky 19  & \cite{Petreczky:2019ozv} & 2+1 &  161 &\\[2.0ex]
[3.0ex] \hline \\[-1.0ex]
\\[-1.0ex]
Maezawa 16  & \cite{Maezawa:2016vgv} & 2+1 & 160 &  \\
[1.0ex] \hline \\[-1.0ex]
JLQCD 16 & \cite{Nakayama:2016atf} & 2+1 & 230 &  \\
[1.0ex]
 \hline \\[-1.0ex]
$\chi$QCD 14 & \cite{Yang:2014sea} & 2+1 & 290 & \\
[3.0ex] \hline \\[-1.0ex]
HPQCD 10 & \cite{McNeile:2010ji} & 2+1 & 260 & \\
[1.0ex] \hline \\[-1.0ex]
HPQCD 08B & \cite{Allison:2008xk} & 2+1 &  & \\
[1.0ex]
\fi
 \hline  \hline \\[-1.0ex]
\end{tabular*}
\caption{Chiral extrapolation/minimum pion mass in the determinations of $m_c$ with $\Nf=2+1$ quark flavours.}
}
\end{table}

\ifx\reducedapptables\undefined
\begin{table}
{\footnotesize
\begin{tabular*}{\textwidth}[l]{l @{\extracolsep{\fill}} c c c c l}
\hline\hline  \\[-1.0ex]
Collab. & Ref. & $\Nf$ & $L$ [fm] & $M_{\pi,\text{min}} L$ & Description
\\[1.0ex] \hline \hline \\[-1.0ex]
ETM 21A  & \cite{Alexandrou:2021gqw}  & 2+1+1 & 2.5 -- 5.6 & 3.6 & \parbox[t]{4.5cm}{} \\
[2.0ex]\hline\\[-1.0ex]
HPQCD 20A  & \cite{Hatton:2020qhk}  & 2+1+1 & 1.9 -- 5.76 & 3.8 & \parbox[t]{4.5cm}{} \\
[1.0ex]
[4.0ex]\hline\\[-1.0ex]
HPQCD 18  & \cite{Lytle:2018evc}  & 2+1+1 & 2.88 -- 5.76  & 3.8 & \parbox[t]{4.3cm}{Finite-volume effects negligible for charm-quark mass.} \\
[4.0ex] \hline \\[-1.0ex]
FNAL/MILC/TUMQCD 18  & \cite{Bazavov:2018omf} & 2+1+1 & 2.89 -- 6.12  & 3.7 & \\
 \hline \\[-1.0ex]
HPQCD 14A & \cite{Chakraborty:2014aca} & 2+1+1 & 2.5 -- 5.8  & 3.7 & \\
[3.0ex] \hline \\[-1.0ex]
ETM 14& \cite{Carrasco:2014cwa} & 2+1+1 & 2.0 -- 3.0  & $2.7_{\pi^0}$ ($3.3_{\pi^\pm}$) & \\
[3.0ex] \hline \\[-1.0ex]
ETM 14A & \cite{Alexandrou:2014sha} & 2+1+1 & 2.0 -- 3.0 & $2.7_{\pi^0}$ ($3.3_{\pi^\pm}$) & \\
[1.0ex]
 \hline  \hline \\[-1.0ex]
\end{tabular*}
\caption{Finite-volume effects in the determinations of $m_c$ with $\Nf=2+1+1$ quark flavours.}
}
\end{table}
\fi

\begin{table}
{\footnotesize
\begin{tabular*}{\textwidth}[l]{l @{\extracolsep{\fill}} c c c c l}
\hline\hline  \\[-1.0ex]
Collab. & Ref. & $\Nf$ & $L$ [fm] & $M_{\pi,\text{min}} L$ & Description
\\[1.0ex] \hline \hline \\[-1.0ex]
  {ALPHA~23} &\cite{Bussone:2023kag} & 2+1 & \parbox[t]{1.75cm}{$2.7,\, 2.4,\,3.1/4.1,$\\ 2.4/3.1} &  \parbox[t]{1.6cm}{$3.9,\,5.1,\,4.2,$ 4.1} & No explicit discussion of FSE.\\ [3.0ex]
\ifx\reducedapptables\undefined
\hline \\[-1.0ex]
ALPHA 21  & \cite{Heitger:2021apz} & 2+1 & 2.5 -- 4.0 & 4.2  & \\
[2.0ex] \hline \\[-1.0ex]
\\[-2.0ex]
Petreczky 19  & \cite{Petreczky:2019ozv} & 2+1 & 1.6 -- 5.2 & 4.2  &\\
[1.0ex]
[3.0ex] \hline \\[-1.0ex]
\\[-1.0ex]
Maezawa 16  & \cite{Maezawa:2016vgv} & 2+1 & 2.6 -- 5.2 & 4.2 & \\ 
[1.0ex] \hline \\[-1.0ex]
JLQCD 16 & \cite{Nakayama:2016atf}& 2+1 & 2.6 -- 2.8 & 4.4 & \\ 
 \hline \\[-1.0ex]
$\chi$QCD 14 & \cite{Yang:2014sea} &  2+1 &  2.8 & 4.1 & \\
[1.0ex] \hline \\[-1.0ex]
HPQCD 10 & \cite{McNeile:2010ji} & 2+1 & 2.3 -- 3.4 & 3.8 & \\
%
[1.0ex]
\fi
 \hline  \hline \\[-1.0ex]
\end{tabular*}
\caption{Finite-volume effects in the determinations of $m_c$ with $\Nf=2+1$ quark flavours.}
}
\end{table}

\ifx\reducedapptables\undefined
\begin{table}[!ht]
{\footnotesize
\begin{tabular*}{\textwidth}[l]{l @{\extracolsep{\fill}} c c l}
\hline\hline \\[-1.0ex]
Collab. & Ref. & $\Nf$  & Description
\\[1.0ex] \hline \hline \\[-1.0ex]
ETM 21A  & \cite{Alexandrou:2021gqw}  & 2+1+1 &  Nonperturbative (RI/MOM) \\
[2.0ex]\hline\\[-1.0ex]
HPQCD 20A  & \cite{Hatton:2020qhk}  & 2+1+1 & Nonperturbative (RI/SMOM) \\
[2.0ex]
[4.0ex]\hline\\[-1.0ex]
HPQCD 18  & \cite{Lytle:2018evc}  & 2+1+1 & {Nonperturbative (RI/SMOM).}\\
[1.0ex]\hline\\[-1.0ex]
FNAL/MILC/TUMQCD 18 & \cite{Bazavov:2018omf} & 2+1+1 & not required. \\
 \hline \\[-1.0ex]
HPQCD 14A & \cite{Chakraborty:2014aca} & 2+1+1 & Renormalization not required. \\
[1.0ex] \hline \\[-1.0ex]
ETM 14& \cite{Carrasco:2014cwa} & 2+1+1 & Nonperturbative renormalization (RI/MOM). \\
[1.0ex] \hline \\[-1.0ex]
ETM 14A & \cite{Alexandrou:2014sha} & 2+1+1 & Nonperturbative renormalization (RI/MOM). \\
[1.0ex]
 \hline  \hline \\[-1.0ex]
\end{tabular*}
\caption{Renormalization in the determinations of $m_c$ with $\Nf=2+1+1$ quark flavours.}
}
\end{table}
\fi

\begin{table}[!ht]
{\footnotesize
\begin{tabular*}{\textwidth}[l]{l @{\extracolsep{\fill}} c c l}
\hline\hline \\[-1.0ex]
Collab. & Ref. & $\Nf$  & Description
\\[1.0ex] \hline \hline \\[-1.0ex]
ALPHA 23  & \cite{Bussone:2023kag}& 2+1 & Schr\"odinger functional\\
[2.0ex]
\ifx\reducedapptables\undefined
\hline \\[-1.0ex]
\\[-2.0ex]
ALPHA 21  & \cite{Heitger:2021apz} & 2+1 & Schr\"odinger functional\\
[2.0ex] \hline \\[-1.0ex]
\\[-2.0ex]
Petreczky 19  & \cite{Petreczky:2019ozv} & 2+1  & not required \\
[2.0ex]
[3.0ex] \hline \\[-1.0ex]
\\[-1.0ex]
Maezawa 16  & \cite{Maezawa:2016vgv} & 2+1 & not required \\ [1.0ex] \hline
\\[-1.0ex]
JLQCD 16 & \cite{Nakayama:2016atf}& 2+1 &  not required \\  \hline
\\[-1.0ex]
$\chi$QCD 14 & \cite{Yang:2014sea} &  2+1 & Nonperturbative renormalization (RI/MOM). \\
[1.0ex] \hline \\[-1.0ex]
HPQCD 10 & \cite{McNeile:2010ji} & 2+1 &  Renormalization not required. \\
[1.0ex] \hline \\[-1.0ex]
HPQCD 08B & \cite{Allison:2008xk} & 2+1 & Renormalization not required. \\
[1.0ex]
\fi
 \hline  \hline \\[-1.0ex]
\end{tabular*}
\caption{Renormalization in the determinations of $m_c$ with $\Nf=2+1$ quark flavours.}
}
\end{table}


%% file: Vudus/VudVus_app.tex


\subsection{Notes to Sec.~\ref{sec:vusvud} on $|V_{ud}|$ and  $|V_{us}|$}
\label{app:VusVud}

\begin{table}[!h]
{\footnotesize
\begin{tabular*}{\textwidth}[l]{l @{\extracolsep{\fill}} c c c l}
\hline\hline \\[-1.0ex]
Collab. & Ref. & $\Nf$ & $a$ [fm] & Description
\\[1.0ex] \hline \hline \\[-2.0ex]

PACS 22 &\cite{Ishikawa:2022ulx} &  2+1 &  0.085, 0.063 &
\parbox[t]{5.9cm}{\vspace{-2mm} Nonperturbative $\cO(a)$ clover quark action.
Scale set from $\Xi$-baryon mass.}
\\
\\[-1.0ex] \hline \\[-2.5ex]

\ifx\reducedapptables\undefined

FNAL/MILC 18 &\cite{Bazavov:2018kjg} & 2+1+1 & 0.042, 0.06, 0.09, 0.12, 0.15&
\parbox[t]{5.9cm}{HISQ quark action. Relative scale through $r_1$.}
\\
[3.0ex] \hline \\[-2.5ex]

PACS 19 &\cite{Kakazu:2019ltq} & 2+1 & 0.085 &
\parbox[t]{5.9cm}{Nonperturbative $\cO(a)$ clover quark action.
Scale set from $\Xi$-baryon mass.}
\\
[3.0ex] \hline \\[-2.5ex]

{ETM 16}& \cite{Carrasco:2016kpy} & 2+1+1 & 0.062, 0.082, 0.089  &
		    \parbox[t]{5.9cm}{ 
		    Scale set through $f_\pi$. Automatic $\cO(a)$ improvement.} \\
[3.0ex] \hline  \\[-2.5ex]
FNAL/MILC 13E &\cite{Bazavov:2013maa}			&2+1+1&0.06, 0.09, 0.12, 0.15&
	\parbox[t]{5.9cm}{HISQ action for both sea and valence quarks. Relative scale through $r_1$, physical scale from 
	         pseudoscalar decay constants calculated with Asqtad fermions. The ensemble with $a \simeq 0.06$~fm 
	         is used only for cross-checking discretization effects.}\\
[17.0ex] \hline \\[-2.5ex]
%
%
JLQCD 17 &\cite{Aoki:2017spo}			&2+1&0.112&
	\parbox[t]{5.9cm}{Scale set through $\Omega$ mass.}\\
[0.0ex] \hline \\[-2.5ex]
RBC/UKQCD 15A	&\cite{Boyle:2015hfa}			&2+1&0.08, 0.11&
	\parbox[t]{5.9cm}{Scale set through $\Omega$ mass.}\\
[0.0ex] \hline \\[-2.5ex]
FNAL/MILC 12I		&\cite{Bazavov:2012cd}			&2+1&0.09, 0.12&
	\parbox[t]{5.9cm}{Relative scale $r_1$, physical scale determined from a 
		mixture of $f_\pi$, $f_K$, radial excitation of $\Upsilon$
		and $m_{D_s}-\frac 12 m_{ {\eta}_c}$.}\\
[6.0ex] \hline \\[-2.5ex]
%
%
%
%
%
%
ETM 09A		&\cite{Lubicz:2009ht}	&2  & 0.07, 0.09, 0.10&
		\parbox[t]{5.9cm}{
		Scale set through $f_\pi$. Automatic $\cO(a)$ impr., flavour symmetry
		breaking:\\ $(M_{PS}^{0})^2 - (M_{PS}^\pm)^2\sim \cO(a^2)$. \\
  		Three lattice spacings only for pion mass 470\,MeV.}\\
[12.0ex] 
\hline \\[-2.5ex]
%
%
%
%
\fi
\hline
%
\end{tabular*}
\caption{Continuum extrapolations/estimation of lattice artifacts in the
determinations of $f_+(0)$.}
}
\end{table}

\noindent
\begin{table}[!h]
{\footnotesize
\begin{tabular*}{\textwidth}[l]{l @{\extracolsep{\fill}} c c c l}
\hline \hline  \\[-1.0ex]
Collab. & Ref. & $\Nf$ & $M_{\pi,\text{min}}$ [MeV] & Description
\\[1.0ex] \hline \hline \\[-2.0ex]
PACS 22 & \cite{Ishikawa:2022ulx}	&  2+1&   135 &
\parbox[t]{6.3cm}{\vspace{-2mm} Physical point simulation at a single pion mass 135~MeV.}
\\
[3.0ex] \hline \\[-2.5ex]
\ifx\reducedapptables\undefined
FNAL/MILC 18 &\cite{Bazavov:2018kjg}	&2+1+1&$144_{\rm RMS}(135_{\rm \pi,5})$&
\parbox[t]{6.5cm}{Chiral interpolation through NLO SU(3) PQ staggered {\Ch}PT
                  with continuum {\Ch}PT at NNLO. Lightest Nambu-Goldstone 
	          mass is 135~MeV and lightest RMS mass is 144~MeV
		  at the same gauge ensemble with $a \simeq 0.06$~fm.}
\\
[12.0ex] \hline \\[-2.5ex]
PACS 19 & \cite{Kakazu:2019ltq}	&2+1&  135 &
\parbox[t]{6.3cm}{Physical point simulation at a single pion mass 135~MeV.}
\\
[3.0ex] \hline \\[-2.5ex]
{ETM 16}& \cite{Carrasco:2016kpy} & 2+1+1 & $180_{\rm \pi^0}(220_{\pi^\pm})$  & 
          \parbox[t]{6.3cm}{Chiral extrapolation performed through SU(2) or SU(3) {\Ch}PT.}\\
[3.0ex] \hline  \\[-2.5ex]
FNAL/MILC 13E &\cite{Bazavov:2013maa}			&2+1+1&$173_{\rm RMS}(128_{\rm \pi,5})$&
		\parbox[t]{6.3cm}{NLO SU(3) PQ staggered {\Ch}PT with
		        continuum {\Ch}PT at NNLO. Lightest Nambu-Goldstone 
		        mass is 128~MeV and lightest RMS mass is 173~MeV
			for the same gauge ensemble with $a \simeq 0.09$~fm.
			}\\
[12.0ex] \hline \\[-2.5ex]
%
%
JLQCD 17 &\cite{Aoki:2017spo}			&2+1&290&
		\parbox[t]{6.3cm}{NNLO SU(3) {\Ch}PT.}\\
[3.0ex] \hline \\[-2.5ex]
RBC/UKQCD 15A	&\cite{Boyle:2015hfa}			&2+1&140&
		\parbox[t]{6.3cm}{NLO SU(3) {\Ch}PT with phenomenological ansatz 
			for higher orders or polynomial models.}\\
[6.0ex] \hline \\[-2.5ex]
FNAL/MILC 12I		&\cite{Bazavov:2012cd}			&2+1&$378_{\rm RMS}(263_{\rm \pi,5})$&
		\parbox[t]{6.3cm}{NLO SU(3) PQ staggered {\Ch}PT with either 
			phenomenological NNLO ansatz or NNLO {\Ch}PT.
			Lightest Nambu-Goldstone mass is 263~MeV with 
			$a=0.12$~fm and lightest RMS mass is 378~MeV
			with $a=0.09$~fm.
			}\\
[12.0ex] \hline \\[-2.5ex]
%
%
%
%
%
%
ETM 09A		&\cite{Lubicz:2009ht}	&2  &$210_{\pi^0}(260_{\pi^\pm})$& 
		\parbox[t]{6.3cm}{NLO heavy kaon SU(2) {\Ch}PT and
		NLO SU(3) {\Ch}PT and phenomenological ansatz for higher orders.}\\
[6.0ex] 
\hline \\[-2.5ex]
%
%
%
%
\fi
\hline
%
\end{tabular*}
\caption{Chiral extrapolation/minimum pion mass in determinations of $f_+(0)$.
}
}
\end{table}

\noindent
\begin{table}[!h]
{\footnotesize
\begin{tabular*}{\textwidth}[l]{l @{\extracolsep{\fill}} c c c c l}
\hline \hline  \\[-1.0ex]
Collab. & Ref. & $\Nf$ &$L$ [fm]& $M_{\pi,\text{min}}L$ & Description
\\[1.0ex] \hline \hline \\[-1.5ex]
PACS 22 &\cite{Ishikawa:2022ulx}   & 2+1&  10.9 &  7.5 & \parbox[t]{5cm}{}
		\parbox[t]{5.2cm}{}\\
[1.0ex] \hline \\[-2.5ex]
\ifx\reducedapptables\undefined
FNAL/MILC 18 &\cite{Bazavov:2018kjg}   &2+1+1&2.6--5.8&$4.2_{\rm RMS}(3.9_{\pi,5})$& \parbox[t]{5cm}{The values correspond to $M_{\pi,\rm RMS}=144$~MeV and 
		$M_{\pi,5}=135$~MeV, respectively.}
		\parbox[t]{5.2cm}{}\\
[6.5ex] \hline \\[-2.5ex]
PACS 19 &\cite{Kakazu:2019ltq}   &2+1& 10.9 & 7.5 & \parbox[t]{5cm}{}
		\parbox[t]{5.2cm}{}\\
[1.0ex] \hline \\[-2.5ex]
{ETM 16}& \cite{Carrasco:2016kpy} & 2+1+1 & 2.0--3.0 & $2.7_{\pi^0}(3.3_{\pi^\pm})$ &
        \parbox[t]{5.2cm}{FSE observed only in the slopes of the vector and scalar form factors.} \\
[3.0ex]\hline  \\[-2.5ex]
FNAL/MILC 13E &\cite{Bazavov:2013maa} 		&2+1+1&2.9--5.8&$4.9_{\rm RMS}(3.6_{\pi,5})$& \parbox[t]{5cm}{The values correspond to $M_{\pi,\rm RMS}=173$~MeV and 
		$M_{\pi,5}=128$~MeV, respectively.}
		\parbox[t]{5.2cm}{}\\
[6.0ex] \hline \\[-2.5ex]
%
%
JLQCD 17 &\cite{Aoki:2017spo} 	&2+1&1.8, 2.7&4.1& 
		\parbox[t]{5.2cm}{}\\
[0.0ex] \hline \\[-2.5ex]
RBC/UKQCD 15A	&\cite{Boyle:2015hfa} 	&2+1&2.6, 5.2&3.9& 
		\parbox[t]{5.2cm}{}\\
[0.0ex] \hline \\[-2.5ex]
FNAL/MILC 12I		&\cite{Bazavov:2012cd} 		&2+1&2.4--3.4&$6.2_{\rm RMS}(3.8_{\pi,5})$& \parbox[t]{5cm}{The values correspond to $M_{\pi,\rm RMS}=378$~MeV and 
		$M_{\pi,5}=263$~MeV, respectively.}
		\parbox[t]{5.2cm}{}\\
[6.0ex] \hline \\[-2.5ex]
%
%
%
%
%
%
ETM 09A		&\cite{Lubicz:2009ht}	&2  &2.1, 2.8&$3.0_{\pi^0}(3.7_{\pi^\pm})$& 
		\parbox[t]{5.2cm}{Two volumes at $M_\pi=300$\,MeV and {\Ch}PT-motivated 
		estimate of the error due to FSE.}\\
[6.0ex] 
\hline \\[-2.5ex]
%
%
%
%
\fi
\hline
%
\end{tabular*}
\caption{Finite-volume effects in determinations of $f_+(0)$.
}
}
\end{table}

\begin{table}[!h]
{\footnotesize
\begin{tabular*}{\textwidth}[l]{l @{\extracolsep{\fill}} c c c l}
\hline \hline  \\[-1.0ex]
Collab. & Ref. & $\Nf$ & $a$ [fm] & Description
\\[1ex] \hline \hline \\[-2.0ex]
{ETM 21} & \cite{Alexandrou:2021bfr} & 2+1+1 & 0.07, 0.08, 0.09 &
\parbox[t]{5.2cm}{Wilson-clover twisted mass quark action. Relative scale through gradient flow scale $w_0$ and absolute scale through $f_\pi$.}\\
[9.0ex] \hline \\[-2.0ex]
{CLQCD 23} & \cite{CLQCD:2023sdb} & 2+1 & 0.105, 0.077, 0.052 &
\parbox[t]{5.2cm}{Tadpole-improved clover quark action with one-step stout smearing. Scale set through gradient-flow scale $w_0$.}\\
\ifx\reducedapptables\undefined
{CalLat 20} & \cite{Miller:2020xhy} & 2+1+1 & 0.06, 0.09, 0.12, 0.15&
\parbox[t]{5.2cm}{M\"obius domain-wall valence quarks on gradient-flowed HISQ ensembles.  Relative scale through the gradient-flow scale $w_0$.}\\
[9.0ex] \hline \\[-2.5ex]
{FNAL/MILC 17}& \cite{Bazavov:2017lyh} & 2+1+1 & 0.03, 0.042, 0.06,\,0.09,\,0.12,\,0.15  &  
	\parbox[t]{5.2cm}{ 
	HISQ action for both valence and sea quarks. Absolute scale through $f_\pi$.}\\
[3.0ex]  \hline  \\[-2.5ex]
{ETM 14E}& \cite{Carrasco:2014poa} & 2+1+1 & 0.062, 0.082, 0.089  &
		    \parbox[t]{5.2cm}{ 
		    Scale set through $f_\pi$. Automatic $\cO(a)$ improvement, flavour symmetry
		    breaking: $(M_{PS}^{0})^2-(M_{PS}^\pm)^2\sim \cO(a^2)$. Discretization and volume effects
		    due to the $\pi^0 - \pi^\pm$ mass splitting are taken into account through {\Ch}PT for
		    twisted-mass fermions.} \\
[17.0ex] \hline  \\[-2.5ex]
%
%
\hline  \\[-2.5ex]
%
%
%
\fi
\hline 
\end{tabular*}
\caption{Continuum extrapolations/estimation of lattice artifacts in
  determinations of $f_K/f_\pi$.} 
}
\end{table}

\ifx\reducedapptables\undefined
\begin{table}[!h]
{\footnotesize
\begin{tabular*}{\textwidth}[l]{l @{\extracolsep{\fill}} c c c l}
\hline \hline  \\[-1.0ex]
 
Collab. & Ref. & $\Nf$ & $a$ [fm] & Description
\\[1ex] \hline \hline \\[-2.5ex]
QCDSF/UKQCD 16  &\cite{Bornyakov:2016dzn}		&2+1&
		\parbox[t]{3.5cm}{\centering 0.059, 0.068, 0.074, 0.082}&
		\parbox[t]{5.7cm}{Scale set through $M_N$. Nonperturbative
		$\cO(a)$ clover improvement.}\\
[3.0ex] \hline \\[-2.5ex]
D\"urr 16  	&\cite{Durr:2016ulb,Scholz:2016kcr}		&2+1&
 		\parbox[t]{3.5cm}{\centering 0.054, 0.065, 0.077, \\ 0.092, 0.12}&
 		\parbox[t]{5.7cm}{Scale set through $M_\Omega$. Perturbative
 		$\cO(a)$-improvement.}\\
[3.0ex] \hline \\[-2.5ex]
 
{RBC/UKQCD 14B}& \cite{Blum:2014tka} & 2+1 & 0.063,\, 0.085,\, 0.114  &  
 	\parbox[t]{5.7cm}{ 
	Scale set through $m_\Omega$.}\\
\hline  \\[-2.5ex]

{RBC/UKQCD 12}& \cite{Arthur:2012opa} & 2+1 & 0.09,\, 0.11,\, 0.14  &  
	\parbox[t]{5.7cm}{ 
	Scale set through $m_\Omega$.}\\
\hline  \\[-2.5ex]

{Laiho 11}& \cite{Laiho:2011np} & 2+1 & 0.06, 0.09, 0.125 &  
	\parbox[t]{5.7cm}{Scale set through $r_1$ and $\Upsilon$ and 
	continuum extrapolation based on MA{\Ch}PT.}\\
[3.0ex] \hline  \\[-2.5ex]

{JLQCD/TWQCD 10}& \cite{Noaki:2010zz} & 2+1 & 0.112  & Scale set
through $M_\Omega$. \\
\hline  \\[-2.5ex]

{RBC/UKQCD 10A}& \cite{Aoki:2010dy} & 2+1 & 0.087, 0.114 & Scale set
through $M_\Omega$. \\
\hline  \\[-2.5ex]
 
MILC 10 & \cite{Bazavov:2010hj} & 2+1 & 0.045, 0.06, 0.09  & \parbox[t]{5.7cm}{3 lattice
spacings, continuum extrapolation by means of RS{\Ch}PT.}\\
[3.0ex] \hline  \\[-2.5ex]
BMW 10		&\cite{Durr:2010hr}		&2+1&
 		\parbox[t]{2.5cm}{\centering 0.07, 0.08,\\ 0.12}&
 		\parbox[t]{5.7cm}{Scale set through $M_{\Omega,\Xi}$. Perturbative
 		$\cO(a)$-improvement.}\\
[3.0ex] \hline \\[-2.5ex]
 
JLQCD/TWQCD 09A &\cite{Noaki:2008iy}	&2+1& 0.1184(3)(21) & 
		\parbox[t]{5.7cm}{Scale set through $F_\pi$. Automatic 
		$\cO(a)$-improvement due to chiral symmetry of action.}\\[6.0ex]
\hline \\[-2.5ex]

{PACS-CS 09}& \cite{Aoki:2009ix} & 2+1 & 0.0900(4)  & Scale set
through $M_\Omega$. \\
\hline  \\[-2.5ex]
 
MILC 09A	&\cite{Bazavov:2009fk}	&2+1& 0.045, 0.06, 0.09 & 
		\parbox[t]{5.7cm}{Scale set through $r_1$ and $\Upsilon$ and 
		continuum extrapolation based on RS{\Ch}PT.}\\
[3.0ex]  \hline \\[-2.5ex]
 
MILC 09		&\cite{Bazavov:2009bb}	&2+1& 	\parbox[t]{2.5cm}{\centering 0.045, 0.06, \\0.09, 0.12} & 
		\parbox[t]{5.7cm}{Scale set through $r_1$ and $\Upsilon$ and 
		continuum extrapolation based on RS{\Ch}PT.}\\
[3.0ex] \hline \\[-2.5ex]
 
Aubin 08	&\cite{Aubin:2008ie}	&2+1& 0.09, 0.12 & 	
		\parbox[t]{5.7cm}{Scale set through $r_1$ and $\Upsilon$ and 
		continuum extrapolation based on MA{\Ch}PT.}\\
[3.0ex]  \hline \\[-2.5ex]
 
PACS-CS 08, 08A &\cite{Aoki:2008sm,Kuramashi:2008tb}& 2+1 & 0.0907(13) & 
		\parbox[t]{5.7cm}{Scale set 
		through $M_\Omega$. Nonperturbatively $\cO(a)$-improved. }\\
[3.0ex]  \hline \\[-2.5ex]
 
HPQCD/UKQCD 07	&\cite{Follana:2007uv}	&2+1& 0.09, 0.12, 0.15 & 
 		\parbox[t]{5.7cm}{Scale set through $r_1$ and $\Upsilon$ and 
 		continuum extrapolation on continuum-{\Ch}PT motivated ansatz. Taste
 		breaking of sea quarks ignored.}\\
[9.0ex]  
\hline \\[-2.5ex]
 
RBC/UKQCD 08	&\cite{Allton:2008pn}	&2+1& 0.114(2) & 
		\parbox[t]{5.7cm}{Scale set through $M_\Omega$. Automatic 
		$\cO(a)$-improvement due to approximate chiral symmetry. 
		$(\Lambda_{\rm QCD}a)^2\approx 4\%$ systematic error
		due to lattice artifacts added.}\\
[9.0ex]  \hline \\[-2.5ex]
 
NPLQCD 06	&\cite{Beane:2006kx}	&2+1& 0.125 & 
		\parbox[t]{5.7cm}{Scale set through $r_0$ and $F_\pi$. Taste breaking
		of sea quarks ignored.}\\
[3.0ex]  \hline \\[-2.5ex]
 
MILC 04	&\cite{Aubin:2004fs}	 &2+1& 0.09, 0.12 & \parbox[t]{5.7cm}{Scale
   set through $r_1$ and $\Upsilon$ and continuum extrapolation based on RS{\Ch}PT.} \\
[3.0ex] %
\hline
\hline
\end{tabular*}
\caption{Continuum extrapolations/estimation of lattice artifacts in
  determinations of $f_K/f_\pi$ for $\Nf=2+1$ simulations.}
}
\end{table}
\fi

\ifx\reducedapptables\undefined
\begin{table}[!h]
{\footnotesize
\begin{tabular*}{\textwidth}[l]{l @{\extracolsep{\fill}} c c c l}
\hline \hline  \\[-1.0ex]
Collab. & Ref. & $\Nf$ & $a$ [fm] & Description
\\[1ex] \hline \hline \\[-1.5ex]
%
%
%
%
%
ETM 09		&\cite{Blossier:2009bx}		&2  & 0.07, 0.09, 0.10 & 
		\parbox[t]{5.7cm}{Scale set through $F_\pi$.
		Automatic $\cO(a)$ impr., flavour symmetry
		breaking: $(M_{PS}^{0})^2-(M_{PS}^\pm)^2\sim \cO(a^2)$. 
		  }\\
[6.0ex] 
\hline \\[-2.5ex]
%
%
\hline
%
\end{tabular*}
\caption{Continuum extrapolations/estimation of lattice artifacts in
  determinations of $f_K/f_\pi$ for $\Nf=2$ simulations.}
}
\end{table}
\fi

\vfill
\noindent
\begin{table}[!h]
{\footnotesize
\begin{tabular*}{\textwidth}[l]{l @{\extracolsep{\fill}} c c c l}
\hline \hline  \\[-1.0ex]
Collab. & Ref. & $\Nf$ & $M_{\pi,\text{min}}$ [MeV] & Description
\\[1ex] \hline \hline \\[-2.0ex]
{ETM 21}& \cite{Alexandrou:2021bfr} & 2+1+1 & 134 & 
          \parbox[t]{6.2cm}{ 
            Chiral extrapolation based on NLO SU(2) {\Ch}PT.
          }\\
[3.0ex]\hline  \\[-2.0ex]
{CLQCD 23}& \cite{CLQCD:2023sdb} & 2+1 & 135.5 & 
            \parbox[t]{6.2cm}{
              Two additional partially-quenched quark masses
              on each ensemble.
              Chiral extrapolation based on NLO SU(2) partially-quenched {\Ch}PT
              for $f_\pi$ and a polynomial extrapolation for $f_K$.
          }\\
[3.0ex]\hline  \\[-2.0ex]
\ifx\reducedapptables\undefined
{CalLat 20}& \cite{Miller:2020xhy} & 2+1+1 & 157 & 
          \parbox[t]{6.2cm}{ 
            Chiral extrapolation based on NNLO SU(3) {\Ch}PT.
            We quote the root-mean-square (RMS) mass of the valence and
            valence-sea pions as $M_{\pi,\text{min}}$.
            The smallest mass is 176~MeV for the HISQ sea pions,
            which do not enter until NNLO in the {\Ch}PT expression.
          }\\
[17.0ex]\hline  \\[-2.5ex]
{FNAL/MILC 17}& \cite{Bazavov:2017lyh} & 2+1+1 & $143_{\rm RMS}(128_{\pi,5})$  & 
          \parbox[t]{6.2cm}{ 
	Linear interpolation to physical point. The lightest RMS mass is from the 
	$a=0.06$~fm ensemble and the lightest Nambu-Goldstone mass is from the
	$a=0.09$~fm ensemble.}\\
[9.0ex]\hline  \\[-2.5ex]
{ETM 14E}& \cite{Carrasco:2014poa} & 2+1+1 & $180_{\rm \pi^0}(220_{\pi^\pm})$  & 
          \parbox[t]{6.2cm}{Chiral extrapolation performed through SU(2) {\Ch}PT or polynomial fit.}\\
[3.0ex]\hline  \\[-2.5ex]
%
%
{HPQCD 13A}& \cite{Dowdall:2013rya} & 2+1+1 &$173_{\rm RMS}(128_{\pi,5})$  &  
	\parbox[t]{6.2cm}{ 
	NLO {\Ch}PT supplemented by model for NNLO. Both the lightest RMS and 
	the lightest
	Nambu-Goldstone mass are from the $a=0.09$~fm ensemble.}\\
[9.0ex] 
\hline  \\[-2.5ex]
%
%
%
%
\fi
\hline
%
\end{tabular*}
\caption{Chiral extrapolation/minimum pion mass in determinations of $f_K / f_\pi$.
}
}
\end{table}

\ifx\reducedapptables\undefined
\noindent
\vspace{-3cm}
\begin{table}[!h]
{\footnotesize
\begin{tabular*}{\textwidth}[l]{l @{\extracolsep{\fill}} c c c l}
\hline \hline  \\[-1.0ex]
Collab. & Ref. & $\Nf$ & $M_{\pi,\text{min}}$ [MeV] & Description
\\[1ex] \hline \hline \\[-1.5ex]
QCDSF/UKQCD 16  &\cite{Bornyakov:2016dzn}		&2+1& 220 & 
		\parbox[t]{6.2cm}{Expansion around the SU(3) symmetric point 
		$m_u = m_d = m_s = (m_u + m_d + m_s)^{phys} / 3$.}\\
[3.0ex] \hline \\[-2.5ex]
D\"urr 16   	&\cite{Durr:2016ulb,Scholz:2016kcr}			&2+1& 130 & 
		\parbox[t]{6.2cm}{Comparison between SU(3) {\Ch}PT 
		and polynomial fit-ans\"atze. }\\
[3.0ex] \hline \\[-2.5ex]
{RBC/UKQCD 14B}& \cite{Blum:2014tka} & 2+1 & 139 &\parbox[t]{6.2cm}{
		NLO PQ SU(2) {\Ch}PT as well as analytic ans\"atze.}\\
[3.0ex] \hline \\[-2.5ex]
%
%
%
%
%
MILC 10       & \cite{Bazavov:2010hj}  & 2+1 & $258_{\rm RMS}(177_{\pi,5})$ &  \parbox[t]{6.2cm}{Lightest
Nambu-Goldstone mass is 177\,MeV (at 0.09\,fm) and
 lightest RMS mass is 258\,MeV (at 0.06\,fm). NLO rS{\Ch}PT and NNLO {\Ch}PT.} \\
[9ex]\hline \\[-2.5ex]
BMW 10		&\cite{Durr:2010hr}		&2+1& 190 & 
		\parbox[t]{6.2cm}{Comparison of various fit-ans\"atze:
		SU(3) {\Ch}PT, heavy kaon SU(2) {\Ch}PT, polynomial. }\\
[3.0ex] \hline \\[-2.5ex]
%
%
%
%
HPQCD/UKQCD 07	&\cite{Follana:2007uv}	&2+1&$375_{\rm RMS}(263_{\pi,5})$& 
		\parbox[t]{6.2cm}{NLO SU(3) chiral perturbation theory
		with NNLO and NNNLO analytic terms. 
		The lightest RMS mass is from the $a=0.09$~fm ensemble and 
		the lightest Nambu-Goldstone mass is from the 
		$a=0.12$~fm ensemble.}\\
[12.0ex] 
\hline \\[-2.5ex]
\hline\\
\end{tabular*}
\vskip -0.5cm
\caption{Chiral extrapolation/minimum pion mass in determinations of $f_K/f_\pi$ for $\Nf=2+1$ simulations. The subscripts RMS and $\pi,5$ in the case of staggered fermions indicate
the root-mean-square mass and the Nambu-Goldstone boson mass. In the case
of twisted-mass fermions $\pi^0$ and $\pi^\pm$ indicate the neutral and
charged pion mass and where applicable, ``val'' and ``sea'' indicate valence
and sea pion masses.}
}
\end{table}
\fi

\ifx\reducedapptables\undefined
\noindent
\vspace{-3cm}
\begin{table}[!h]
\addtocounter{table}{-1}
{\footnotesize
\begin{tabular*}{\textwidth}[l]{l @{\extracolsep{\fill}} c c c l}
\hline \hline  \\[-1.0ex]
Collab. & Ref. & $\Nf$ & $M_{\pi,\text{min}}$ [MeV] & Description
\\[1ex] \hline \hline \\[-2.5ex]
Aubin 08	&\cite{Aubin:2008ie}	&2+1&$329_{\rm RMS}(246_{\pi,5})$& 
		\parbox[t]{6.2cm}{NLO MA{\Ch}PT. According to \cite{Bazavov:2009fk}
		the lightest sea Nambu-Goldstone mass is 246\,MeV (at $a=0.09$ fm)
		and the lightest RMS mass is 329\,MeV (at $a=0.09$ fm).}\\
[9.0ex] \hline \\[-2.5ex]
 
PACS-CS 08, 08A &\cite{Aoki:2008sm,Kuramashi:2008tb}& 2+1 &156& 
		\parbox[t]{6.2cm}{NLO SU(2) {\Ch}PT and SU(3) (Wilson){\Ch}PT.}\\
\hline \\[-2.5ex]
 
HPQCD/UKQCD 07	&\cite{Follana:2007uv}	&2+1&$375_{\rm RMS}(263_{\pi,5})$& 
		\parbox[t]{6.2cm}{NLO SU(3) chiral perturbation theory
		with NNLO and NNNLO analytic terms. 
		The lightest RMS mass is from the $a=0.09$~fm ensemble and 
		the lightest Nambu-Goldstone mass is from the 
		$a=0.12$~fm ensemble.}\\
[12.0ex] \hline \\[-2.5ex]
RBC/UKQCD 08	&\cite{Allton:2008pn}	&2+1&$330_{\rm sea}$, $242_{\rm val}$& 
		\parbox[t]{6.2cm}{While SU(3) PQ{\Ch}PT fits were studied,
		final results are based on 
		heavy kaon NLO SU(2) PQ{\Ch}PT. }\\
[6.0ex] \hline \\[-2.5ex]
 
NPLQCD 06	&\cite{Beane:2006kx}	&2+1&300& 
		\parbox[t]{6.2cm}{NLO SU(3) {\Ch}PT and some NNLO terms. 
		The sea RMS mass for the employed lattices is heavier.}\\
[6.0ex] \hline \\[-2.5ex]
 
MILC 04	&\cite{Aubin:2004fs}	 &2+1& $400_{\rm RMS}(260_{\pi,5})$ & \parbox[t]{6.2cm}{PQ RS{\Ch}PT fits. The lightest sea Nambu-Goldstone mass is 260\,MeV
		(at $a=0.12$ fm) and the lightest RMS mass is 400\,MeV (at $a=0.09$ fm).} \\
[9ex]\hline\hline\\

\end{tabular*}
\vskip -0.5cm
\caption{(cntd.) Chiral extrapolation/minimum pion mass in determinations of $f_K/f_\pi$ for $\Nf=2+1$ simulations. The subscripts RMS and $\pi,5$ in the case of staggered fermions indicate
the root-mean-square mass and the Nambu-Goldstone boson mass. In the case
of twisted-mass fermions $\pi^0$ and $\pi^\pm$ indicate the neutral and
charged pion mass and where applicable, ``val'' and ``sea'' indicate valence
and sea pion masses.}
}
\end{table}
\fi

\ifx\reducedapptables\undefined
\begin{table}[!h]
{\footnotesize
\begin{tabular*}{\textwidth}[l]{l @{\extracolsep{\fill}} c c c l}
\hline \hline  \\[-1.0ex]
Collab. & Ref. & $\Nf$ & $M_{\pi,\text{min}}$ [MeV] & Description
\\[1ex] \hline \hline \\[-2.5ex]
%
\\
%
%
%
%
ETM 09		&\cite{Blossier:2009bx}		&2  &$210_{\pi^0}(260_{pi^\pm})$& 
		\parbox[t]{6.2cm}{NLO heavy meson SU(2) {\Ch}PT and NLO
		SU(3) {\Ch}PT.}\\
[4.0ex] \hline \\[-2.5ex]
%
%
\hline\\
\end{tabular*}
\vskip -0.5cm
\caption{Chiral extrapolation/minimum pion mass in determinations of $f_K/f_\pi$ for $\Nf=2$ simulations. The subscripts RMS and $\pi,5$ in the case of staggered fermions indicate
the root-mean-square mass and the Nambu-Goldstone boson mass. In the case
of twisted-mass fermions $\pi^0$ and $\pi^\pm$ indicate the neutral and
charged pion mass and where applicable, ``val'' and ``sea'' indicate valence
and sea pion masses.}
}
\end{table}
\fi

\noindent
\begin{table}[!h]
{\footnotesize
\begin{tabular*}{\textwidth}[l]{l @{\extracolsep{\fill}} c c c c l}
\hline \hline  \\[-1.0ex]
Collab. & Ref. & $\Nf$ &$L$ [fm]& $M_{\pi,\text{min}}L$ & Description
\\[1ex] \hline \hline \\[-2.0ex]
{ETM 21}& \cite{Alexandrou:2021bfr} & 2+1+1 &2.0--5.6 & 3.8 &  
\parbox[t]{4.7cm}{Three different volumes at $M_\pi=253$~MeV
and $a=0.08$~fm.
}\\
[3.0ex] \hline  \\[-2.0ex]
{CLQCD 23}& \cite{CLQCD:2023sdb} & 2+1 & 2.5--5.1 & 3.5 &  
\parbox[t]{4.7cm}{Two volumes
at four simulation points with $M_\pi=200$\,--\,300~MeV
on two coarser lattices.
}\\
[3.0ex] \hline  \\[-2.0ex]
\ifx\reducedapptables\undefined
{CalLat 20}& \cite{Miller:2020xhy} & 2+1+1 &2.4--7.2 & 3.8 &  
\parbox[t]{4.7cm}{Three different volumes at $M_\pi=220$~MeV
and $a=0.12$~fm.
}\\
[3.0ex] \hline  \\[-2.5ex]
{FNAL/MILC 17}& \cite{Bazavov:2017lyh} & 2+1+1 &2.4--6.1 & $3.9_{\rm RMS}(3.7_{\pi,5}) $ &  
	\parbox[t]{4.7cm}{ }\\
[0.0ex] \hline  \\[-2.5ex]
{ETM 14E}& \cite{Carrasco:2014poa} & 2+1+1 & 2.0 -- 3.0 & $2.7_{\pi^0}(3.3_{\pi^\pm})$ &
        \parbox[t]{4.7cm}{FSE for the pion is corrected through resummed NNLO {\Ch}PT for twisted-mass fermions,
        which takes into account the effects due to the $\pi^0 - \pi^\pm$ mass splitting.} \\
[12.0ex] \hline \\[-2.5ex]
%
	\parbox[t]{4.7cm}{ }\\
%
{HPQCD 13A}& \cite{Dowdall:2013rya} & 2+1+1 &2.5--5.8& $4.9_{\rm RMS}(3.7_{\pi,5})$  &  
	\parbox[t]{4.7cm}{ }\\
[1.0ex] \hline  \\[-2.5ex]
%
%
%
\fi
\hline\\
\end{tabular*}
\caption{Finite-volume effects in determinations of $f_K/f_\pi$.
}
}
\end{table}

\ifx\reducedapptables\undefined
\begin{table}[!h]
{\footnotesize
\begin{tabular*}{\textwidth}[l]{l @{\extracolsep{\fill}} c c c c l}
\hline \hline  \\[-1.0ex]
Collab. & Ref. & $\Nf$ &$L$ [fm]& $M_{\pi,\text{min}}L$ & Description
\\[1ex] \hline \hline \\[-2.5ex]
QCDSF/UKQCD 16  &\cite{Bornyakov:2016dzn}	&2+1& 
{2.0--2.8} &3.0& 
		\parbox[t]{4.5cm}{...}\\
[3.0ex] \hline \\[-2.5ex]
D\"urr 16  	&\cite{Durr:2016ulb,Scholz:2016kcr}		&2+1& 
{1.5--5.5} &3.85& 
		\parbox[t]{4.5cm}{Various volumes for comparison and
		corrections for FSE from NLO {\Ch}PT with re-fitted coefficients.}\\
[9.0ex] \hline \\[-2.5ex]
\ifx\reducedapptables\undefined
{RBC/UKQCD 14B}& \cite{Blum:2014tka} & 2+1 & 2.0,\,2.7,\,4.6,\,5.4 &  $3.8$ & 
\\
[0.0ex] \hline \\[-2.5ex]
%
%
%
%
%
MILC 10       & \cite{Bazavov:2010hj}                 & 2+1 & 2.5-3.8 &
$7.0_{\rm RMS}(4.0_{\pi,5})$  & \parbox[t]{4.5cm}{$L\!\geq\!2.9\,\fm$ for the lighter masses.}\\
[0.0ex] \hline \\[-2.5ex]
BMW 10		&\cite{Durr:2010hr}		&2+1& 
{2.0--5.3} &4.0& 
		\parbox[t]{4.5cm}{Various volumes for comparison and
		correction for FSE from {\Ch}PT using \cite{Colangelo:2005gd}.}\\
[6.0ex] \hline \\[-2.5ex]
%
%
%
%
%
%
%
HPQCD/UKQCD 07	&\cite{Follana:2007uv}	&2+1&
{2.4--2.9}&$4.1_{\rm RMS}(3.8_{\pi,5})$& 
		\parbox[t]{4.5cm}{Correction for FSE from {\Ch}PT using 
		\cite{Colangelo:2005gd}.}\\
[3.0ex] \hline \\[-2.5ex]
%
%
%
%
%
%
%
%
%
ETM 09		&\cite{Blossier:2009bx}		&2  &2.0--2.7&$3.0_{\pi^0}(3.7_{\pi^\pm})$& 
		\parbox[t]{4.5cm}{Correction for FSE from {\Ch}PT 
		\cite{Gasser:1986vb,Gasser:1987ah,Colangelo:2005gd}.}\\
[3.0ex] 
\hline \\[-2.5ex]
%
%
\fi
\hline
\end{tabular*}
\caption{Finite-volume effects in determinations of $f_K/f_\pi$ for
  $\Nf=2+1$ and $\Nf=2$.
The subscripts RMS and $\pi,5$ in the case of staggered fermions indicate
the root-mean-square mass and the Nambu-Goldstone boson mass. In the case
of twisted-mass fermions $\pi^0$ and $\pi^\pm$ indicate the neutral and
charged pion mass and where applicable, ``val'' and ``sea'' indicate valence
and sea pion masses.}
}
\end{table}
\fi


%% file: BK/BK_app.tex
\subsection{Notes to section \ref{sec:BK} on Kaon mixing}
\label{app-BK}


\subsubsection{Kaon $B$-parameter $B_K$}
\label{app:BKSM}


\begin{table}[!ht]

{\footnotesize
\begin{tabular*}{\textwidth}{l c c c l}
\hline\hline \\[-1.0ex]
Collab. & Ref. & $\Nf$ & $a$ [fm] & Description 
\\[1.0ex] \hline \hline \\[-1.0ex]
%
RBC/UKQCD~24 & \cite{Boyle:2024gge} & 2+1 & 0.114, 0.084, 0.073  &
\parbox[t]{6.0cm}{Combined continuum and chiral (NLO SU(2))  extrapolation fits. Assigned systematic error at the per-mille level}.
\\[7.0ex] 
\hline\hline
\end{tabular*}
\caption{Continuum extrapolations/estimation of lattice artifacts in
determinations of $B_K$.}
}
\end{table}


\begin{table}[!ht]
{\footnotesize
\begin{tabular*}{\textwidth}{l @{\extracolsep{\fill}} c c c l}
\hline\hline \\[-1.0ex]
{Collab.} & {Ref.} & {$\Nf$} & {$M_{\pi,\rm min}\,[\mev]$} & {Description}  
\\[1.0ex] \hline \hline \\[-1.0ex]
RBC/UKQCD~24 & \cite{Boyle:2024gge} & 2+1 & 
\parbox[t]{1.8cm}{139, \, 139, \, 232}
 & \parbox[t]{6cm}{Chiral extrapolations based on  SU(2)-$\chi$PT fits at NLO.
Systematic uncertainties  amount to less than half a per cent.}
\\[7.0ex] 
\hline\hline
\end{tabular*}
\caption{Chiral extrapolation/minimum pion mass in
  determinations of $B_K$.}
}
\end{table}


\begin{table}[!ht]
{\footnotesize
\begin{tabular*}{\textwidth}{l @{\extracolsep{\fill}} c c c c l}
\hline\hline \\[-1.0ex]
Collab. & Ref. & $\Nf$ & $L$ [fm] & ${M_{\pi,\rm min}}L$ & Description 
\\[1.0ex] \hline \hline \\
RBC/UKQCD~24 & \cite{Boyle:2024gge} & 2+1
& \parbox[t]{1.8cm}{5.5,  \, 5.4, \,  3.5, \,  2.6 } & $3.9, \, 3.8, \, 4.1$ 
& \parbox[t]{6cm}{Finite-volume effects are found to be negligible
  compared to  other systematic effects  and are thus omitted in the final
  error budget.} 
\\[11.0ex] 
\hline\hline
\end{tabular*}
\caption{Finite-volume effects in determinations of $B_K$.}
}
\end{table}


\begin{table}[!ht]
{\footnotesize
\begin{tabular*}{\textwidth}{l @{\extracolsep{\fill}} c c c c l}
\hline\hline \\[-1.0ex]
& & & & running & \\
\rb{Collab.} & \rb{Ref.} & \rb{$\Nf$} & \rb{Ren.} & match. & \rb{Description} 
\\[1.0ex] \hline \hline \\[-1.0ex]
%
RBC/UKQCD~24 & \cite{Boyle:2024gge} & 2+1 & RI & PT1$\ell$ &
  \parbox[t]{5cm}{Two different RI-SMOM schemes used to estimate  
a 1\% systematic error owing to the perturbative matching to $\msbar$.}
\\[11.0ex]
\hline\hline
\end{tabular*}
\caption{Running and matching in determinations of $B_K$.}
}
\end{table}

\clearpage

\subsubsection{Kaon BSM $B$-parameters}
\label{app-Bi}



\begin{table}[!ht]

{\footnotesize
\begin{tabular*}{\textwidth}{l c c c l}
\hline\hline \\[-1.0ex]
Collab. & Ref. & $\Nf$ & $a$ [fm] & Description 
\\[1.0ex] \hline \hline \\[-1.0ex]
RBC/UKQCD~24 & \cite{Boyle:2024gge} & 2+1 & 0.114, 0.084, 0.073  &
\parbox[t]{6.0cm}{Systematic uncertainties ranging from a minimum of 0.4\% (for the case of $B_2$) to 1.9\% (for the case of $B_3$). } 
\\[7.0ex]
\hline\hline
\end{tabular*}
\caption{Continuum extrapolations/estimation of lattice artifacts in
  determinations of the BSM $B_i$ parameters.}
}
\end{table}


\begin{table}[!ht]
{\footnotesize
\begin{tabular*}{\textwidth}{l @{\extracolsep{\fill}} c c c l}
\hline\hline \\[-1.0ex]
{Collab.} & {Ref.} & {$\Nf$} & {$M_{\pi,\rm min}\,[\mev]$} & {Description}  
\\[1.0ex] \hline \hline \\[-1.0ex]
%
RBC/UKQCD~24 & \cite{Boyle:2024gge} & 2+1 & 
\parbox[t]{1.8cm}{139, \, 139, \, 232}
& \parbox[t]{6cm}{Chiral extrapolations based on  SU(2)-$\chi$PT fits at NLO.
	Systematic uncertainties  amount to less than half a percent.}
\\[7.0ex] 

\hline \hline
\end{tabular*}
\caption{Chiral extrapolation/minimum pion mass in
  determinations of the BSM $B_i$ parameters.}
}

\end{table}


\begin{table}[!ht]
{\footnotesize
\begin{tabular*}{\textwidth}{l @{\extracolsep{\fill}} c c c c l}
\hline\hline \\[-1.0ex]
Collab. & Ref. & $\Nf$ & $L$ [fm] & ${M_{\pi,\rm min}}L$ & Description 
\\[1.0ex] \hline \hline \\[-1.0ex]
RBC/UKQCD~24 & \cite{Boyle:2024gge} & 2+1
& \parbox[t]{1.8cm}{5.5,  \, 5.4, \,  3.5 \,  2.6} & $3.9, \, 3.8, \, 4.1$ 
& \parbox[t]{5cm}{Finite-volume effects  are at most at the 2 per-mille level. They are negligible compared to other systematic effects and are therefore omitted in the error budget.} 
\\[13.0ex] 
\hline \hline
\end{tabular*}
\caption{Finite-volume effects in determinations of the BSM $B_i$
  parameters. } }
\end{table}


\begin{table}[!ht]
{\footnotesize
\begin{tabular*}{\textwidth}{l @{\extracolsep{\fill}} c c c c l}
\hline\hline \\[-1.0ex]
& & & & running & \\
\rb{Collab.} & \rb{Ref.} & \rb{$\Nf$} & \rb{Ren.} & match. & \rb{Description} 
\\[1.0ex] \hline \hline \\[-1.0ex]
RBC/UKQCD~24 & \cite{Boyle:2024gge} & 2+1 & RI & PT1$\ell$ &
\parbox[t]{5cm}{Two different RI-SMOM schemes used to estimate
the systematic error owing to the perturbative matching to $\msbar$; minimal value of about 0.7\% for the case of $B_2$ and maximal of 2.4\% for  $B_3$.}
\\ [15.0ex] 
\hline \hline

\end{tabular*}
\caption{Running and matching in determinations of the BSM $B_i$ parameters.}
}
\end{table}

\newpage

%% file: BK/Kpipi_app.tex
\subsubsection{$K \to \pi\pi$ decay amplitudes}
\label{app-Kpipi}


\begin{table}[!ht]

{\footnotesize
\begin{tabular*}{\textwidth}{l @{\extracolsep{\fill}} c c c l}
\hline\hline \\[-1.0ex]
Collab. & Ref. & $\Nf$ & $a$ [fm] & Description 
\\[1.0ex] \hline \hline \\[-1.0ex]
RBC/UKQCD~23A & \cite{Blum:2023mtn} & 2+1 & 0.193 &
\parbox[t]{7.5cm}{Single lattice spacing.}\\ 
[1.0ex] \hline \hline \\[-1.0ex]
\end{tabular*}
\caption{Continuum extrapolations/estimation of lattice artifacts in
  determinations of the $K \to \pi\pi$ decay amplitudes.}
}
\end{table}


\begin{table}[!ht]
{\footnotesize
\begin{tabular*}{\textwidth}{l @{\extracolsep{\fill}} c c c l}
\hline\hline \\[-1.0ex]
{Collab.} & {Ref.} & {$\Nf$} & {$M_{\pi,\rm min}\,[\mev]$} & {Description}  
\\[1.0ex] \hline \hline \\[-1.0ex]
RBC/UKQCD~23A & \cite{Blum:2023mtn} & 2+1 & 
\parbox[t]{1.8cm}{142.6}
& \parbox[t]{6cm}{Single pion mass value, close to the physical point.}\\
[4.0ex]\hline\hline
\end{tabular*}
\caption{Chiral extrapolation/minimum pion mass in
  determinations of the $K \to \pi\pi$ decay amplitudes.}
}
\end{table}


\begin{table}[!ht]
{\footnotesize
\begin{tabular*}{\textwidth}{l @{\extracolsep{\fill}} c c l c l}
\hline\hline \\[-1.0ex]
Collab. & Ref. & $\Nf$ & $L$ [fm] & ${M_{\pi,\rm min}}L$ & Description 
\\[1.0ex] \hline \hline \\[-1.0ex]
RBC/UKQCD~23A & \cite{Blum:2023mtn} & 2+1
& 4.6 & $3.3$ 
& \parbox[t]{5cm}{Finite-volume effects amount to a 7\% systematic error
	contribution to the final error budget of  $A_0$ and $A_2$.  }\\
[7.0ex]\hline\hline \\[-1.0ex]
\end{tabular*}
\caption{Finite-volume effects in determinations of  the $K \to \pi\pi$ decay amplitudes.}
}
\end{table}


\begin{table}[!ht]

  {\footnotesize
\begin{tabular*}{\textwidth}{l @{\extracolsep{\fill}} c c c c l}
\hline\hline \\[-1.0ex]
& & & & running & \\
\rb{Collab.} & \rb{Ref.} & \rb{$\Nf$} & \rb{Ren.} & match. & \rb{Description} 
\\[1.0ex] \hline \hline \\[-1.0ex]
RBC/UKQCD~23A & \cite{Blum:2023mtn} & 2+1 & RI & PT1$\ell$ &
\parbox[t]{5cm}{Two different RI-SMOM schemes are used.
	One of the two schemes is used for the final analysis. A systematic error ranging from 6\% to  16\%, depending on the considered case, is included based on the dispersion of other sets of intermediate scheme and scales. Systematic uncertainties 
	 arising from the computation of the Wilson coefficients
	in the $\msbar$ scheme amount to 12\%.}\\
\\[-1.0ex] \hline\hline \\[-1.0ex]
\end{tabular*}
\caption{Running and matching in determinations of  the $K \to \pi\pi$ decay amplitudes.}
}
\end{table}

%% file: HQ/HQ_app.tex
\subsection{Notes to Sec.~\ref{sec:DDecays} on $D$-meson decay constants and form factors}
\label{app:DDecays}



\input{HQ/AppendixTables/DLapp.tex}

\clearpage

\input{HQ/AppendixTables/DSLapp.tex}

\clearpage

\subsection{Notes to Sec.~\ref{sec:BDecays} on $B$-meson decay
  constants, mixing parameters and form factors}
\label{app:BDecays}



\input{HQ/AppendixTables/BLapp.tex}
\clearpage

\input{HQ/AppendixTables/BBapp.tex}
\clearpage


\input{HQ/AppendixTables/BSLapp.tex}

%% file: HQ/AppendixTables/DLapp.tex

\begin{table}[!htb]


{\footnotesize
\begin{tabular*}{\textwidth}{l @{\extracolsep{\fill}} c c c   l}
\hline\hline \\[-1.0ex]
{Collab.} & {Ref.} & {$\Nf$} & {$M_{\pi,\rm min}\,[\mev]$}  & {Description}  
\\[1.0ex] \hline \hline \\[-1.0ex]
%
\parbox[t]{2cm}{ETM~13F \\ ETM~14E \\ ETM~21B} & \cite{Dimopoulos:2013qfa,Carrasco:2014poa,Dimopoulos:2021qsf} & 2+1+1 &  \parbox[t]{2cm}{245, 239, 211\\~\\$167,\; 137,\; 134$}  & \parbox[t]{5.8cm}{$f_{D_s}\sqrt{m_{D_s}}$ in ETM~13F and  $f_{D_s}/m_{D_s}$ in ETM~14E are extrapolated
  using both a quadratic and a linear fit in $m_l$ plus $O(a^2)$ terms.
  In ETM~21B either $w_0$ or the $D_{(s)}$~meson mass are used as scaling variables in the
  chiral-continuum extrapolations.\\[-1.7ex]} 
\\[18.0ex]
\hline\hline
\end{tabular*}
\caption{Chiral extrapolation/minimum pion mass in $\Nf=2+1+1$ determinations of the
  $D$- and $D_s$-meson decay constants.  For actions with multiple
 species of pions, masses quoted are the RMS pion masses.    The
 different $M_{\pi,\rm min}$ entries correspond to the different lattice
 spacings.} 
}
\end{table}
\begin{table}[!htb]



{\footnotesize
\begin{tabular*}{\textwidth}{l @{\extracolsep{\fill}} c c c   l}
\hline\hline \\[-1.0ex]
{Collab.} & {Ref.} & {$\Nf$} & {$M_{\pi,\rm min}\,[\mev]$}  & {Description}  
\\[1.0ex] \hline \hline \\[-1.0ex]
  {RQCD/ALPHA~24} &\cite{Kuberski:2024pms}  & 2+1  & \parbox[t]{2cm}{335, 129, 155, 130, 175, 337}    & \parbox[t]{5.8cm}{The dependence on light and strange quark masses
    is described using fit ans\"atze inspired by NLO HM$\chi$PT.}\\[10.0ex]
  {ALPHA~23} &\cite{Bussone:2023kag} & 2+1 & \parbox[t]{2cm}{277, 415, 200, 257}    & \parbox[t]{5.8cm}{HM$\chi$PT expressions are used for the quantities
    $(8t_0)^{3/4}f_{D_{(s)}}\sqrt{m_{D_{(s)}}}$.}\\[5.0ex]
\ifx\reducedapptables\undefined
  \hline \\[-2.0ex]
{$\chi$QCD 20A} & \cite{Chen:2020qma} & 2+1  & 114 & \parbox[t]{5.8cm}{Simulations are performed very close to the physical point
and linear interpolations/extrapolations are used to correct for mismatches.}\\[10.0ex]
\hline \\[-2.0ex]
\parbox[t]{2cm}{RBC/UKQCD~17, RBC/UKQCD 18A} & \cite{Boyle:2017jwu, Boyle:2018knm} & 2+1 & \parbox{2cm}{139, 139, 232} & \parbox[t]{5.8cm}{
The lattice spacing, pion-mass and charm-quark mass dependences are fit simultaneously through a Taylor expansion in $a^2$, $(m_\pi^2-
m_\pi^{2 phys})$ and $1/m_H-1/m_{D_{(s)}}$.}\\[12.0ex]
\hline \\[-2.0ex]
$\chi$QCD~14& \cite{Yang:2014sea} & 2+1 & \parbox[t]{2cm}{334, 296} & \parbox[t]{5.8cm}{Chiral and continuum extrapolations
 are peformed simultaneously using linear fits in $m_l$ (and quadratic or including partially quenched chiral logs, in order to assess the systematic error)
plus terms up to O$(a^2)$ and O$(a^4m_c^4)$.}
\\[15.0ex] \hline \\[-2.0ex]
HPQCD 12A& \cite{Na:2012iu} & 2+1 & \parbox[t]{2cm}{460, 329} & \parbox[t]{5.8cm}{Chiral and continuum extrapolations
 are peformed simultaneously using PQHM$\chi$PT augmented by $a$-dependent terms: $c_0(am_c)^2 + c_1(am_c)^4$.}
\\[10.0ex] \hline \\[-2.0ex]
FNAL/MILC 11 & \cite{Bazavov:2011aa} & 2+1 & \parbox[t]{2cm}{570, 440, 320} &
\parbox[t]{5.8cm}{Chiral and continuum extrapolations are peformed 
simultaneously using HM$\chi$PT for rooted staggered quarks.
Effects of hyperfine and flavour splittings are also included.}
\\[13.0ex] \hline \\[-2.0ex]
PACS-CS 11 & \cite{Namekawa:2011wt} & 2+1 & \parbox[t]{2cm}{152} & \parbox[t]{5.8cm}{Simulations are  
reweighted in the light- and strange-quark masses to the physical point.}
\\[7.0ex] \hline \\[-2.0ex]
HPQCD 10A & \cite{Davies:2010ip} & 2+1 & \parbox[t]{2cm}{542, 460, 329, 258, 334} & \parbox[t]{5.8cm}{Chiral and continuum 
extrapolations are performed simultaneously. Polynomials up to  $\left( \frac{m_{q,sea}-m_{q,phys}}{m_{q,phys}}\right)^2$ for $q=s,l$ and
up to  $(am_{c})^8$ are kept.}
\\[12.0ex] \hline \\[-2.0ex]
HPQCD/UKQCD 07 & \cite{Follana:2007uv} & 2+1 & \parbox[t]{2cm}{542, 460, 329} & \parbox[t]{5.8cm}{Combined chiral and continuum extrapolations using
HM$\chi$PT at NLO augmented by second and third-order polynomial terms in $m_q$ and terms up to $a^4$.}
\\[10.0ex] \hline \\[-2.0ex]
FNAL/MILC 05 & \cite{Aubin:2005ar} & 2+1 &
\parbox[t]{2cm}{$>~440\;, 440\;, 400$} & \parbox[t]{5.8cm}{Chiral extrapolations are first performed at each lattice spacing
uisng NLO HM$\chi$PT for rooted staggered quarks. Lattice artifacts are then extrapolated linearly in $a^2$. }
\\[13.0ex] 
%
\fi
\hline\hline
\end{tabular*}
\caption{Chiral extrapolation/minimum pion mass in $\Nf=2+1$ determinations of the
  $D$- and $D_s$-meson decay constants.  For actions with multiple
 species of pions, masses quoted are the RMS pion masses.    The
 different $M_{\pi,\rm min}$ entries correspond to the different lattice
 spacings.} 
}
\end{table}

\begin{table}[!htb]


{\footnotesize
\begin{tabular*}{\textwidth}{l @{\extracolsep{\fill}} c c c c l}
\hline\hline \\[-1.0ex]
Collab. & Ref. & $\Nf$ & $L$ [fm] & ${M_{\pi,\rm min}}L$ & Description 
\\[1.0ex] \hline \hline \\[-1.0ex]
%
\parbox[t]{2.5cm}{ETM~13F \\ ETM~14E \\ ETM 21B} & \cite{Dimopoulos:2013qfa,Carrasco:2014poa,Dimopoulos:2021qsf} & 2+1+1 & \parbox[t]{1.75cm}{2.3/4.6, 2.6/5.2, 3.3/5.5}   & 
\parbox[t]{1.6cm}{$3.8, 3.6, 3.7 $}  &  \parbox[t]{4.2cm}{The comparison of two different volumes at the two largest lattice spacings indicates that FV effects
are below the statistical errors. No explicit discussion of FSE in ETM~21B.} 
\\[15.0ex]
\hline\hline
\end{tabular*}
\caption{Finite-volume effects in $\Nf=2+1+1$ determinations of the $D$- and $D_s$-meson
 decay constants.  Each $L$-entry corresponds to a different
 lattice spacing, with multiple spatial volumes at some lattice
 spacings. For actions with multiple species of pions, the lightest masses are
quoted.} 
}
\end{table}
\begin{table}[!htb]
{\footnotesize
\begin{tabular*}{\textwidth}{l @{\extracolsep{\fill}} c c c c l}
\hline\hline \\[-1.0ex]
Collab. & Ref. & $\Nf$ & $L$ [fm] & ${M_{\pi,\rm min}}L$ & Description 
\\[1.0ex] \hline \hline \\[-1.0ex]
  {RQCD/ALPHA~24} &\cite{Kuberski:2024pms}  & 2+1 &  \parbox[t]{1.75cm}{$2.34,\,2.05/5.46,\\2.4/4.8,\,2.0/6.1,\\2.4/4.8,\,2.48$} &  \parbox[t]{1.6cm}{$4,\,3.6,\\3.8,\,4.05,\\4.2,\,4.2$} &
 \parbox[t]{4.9cm}{By comparing different volumes (up to 5) at fixed pion mass, FSE are estimated to be negligible once the cut $m_\pi L \geq 3.5$ or $L\geq 2.3$ fm is imposed.}\\[14.5ex]
  {ALPHA~23} &\cite{Bussone:2023kag} & 2+1 & \parbox[t]{1.75cm}{$2.7,\, 2.4,\,3.1/4.1,$\\ 2.4/3.1} &  \parbox[t]{1.6cm}{$3.9,\,5.1,\,4.2,$ 4.1} & No explicit discussion of FSE.
\\[2.0ex]
\ifx\reducedapptables\undefined
       {$\chi$QCD 20A} & \cite{Chen:2020qma} & 2+1 &  \parbox[t]{1.75cm}{5.5} &  \parbox[t]{1.6cm}{3.2} & No explicit discussion of FSE.
\\[2.0ex] 
\hline \\[-2.0ex]
\parbox[t]{2.2cm}{RBC/UKQCD~17, RBC/UKQCD 18A} & \cite{Boyle:2017jwu, Boyle:2018knm} & 2+1 & 
 \parbox[t]{1.75cm}{5.5/2.7, 5.4/2.7, 3.4} & \parbox[t]{1.6cm}{$3.86, 3.78$, $4.11$}  &
\parbox[t]{4.9cm}{FV errors estimated to be at the permille level by either comparing values of $m_\pi L$
  to the study of FSE by MILC in~\cite{Bazavov:2014wgs} or by 1-loop HM$\chi$PT.} \\[12.0ex]
\hline \\[-1.0ex]
$\chi$QCD~14& \cite{Yang:2014sea} & 2+1 &
\parbox[t]{1.75cm}{2.7, 2.7}
& \parbox[t]{1.6cm}{4.6, 4.1}  &
\parbox[t]{4.7cm}{No explicit discussion of FV effects.}  
\\[5.0ex] \hline \\[-1.0ex]
HPQCD 12A& \cite{Na:2012iu} & 2+1 &
\parbox[t]{1.75cm}{2.4/2.8, 2.4/3.4}
& \parbox[t]{1.6cm}{3.8, 4.2}  &
\parbox[t]{4.7cm}{FV errors  estimated by comparing finite- and infinite-volume $\chi$PT.}  
\\[8.0ex] \hline \\[-1.0ex]
FNAL/MILC 11 & \cite{Bazavov:2011aa} & 2+1 & 
	     \parbox[t]{1.75cm}{2.4, 2.4/2.88, 2.52/3.6} & \parbox[t]{1.6cm}{3.9, 3.8, 4,2 }  &
\parbox[t]{4.7cm}{FV errors  estimated using  finite-volume $\chi$PT.}
\\[8.0ex] \hline \\[-1.0ex]
PACS-CS 11 & \cite{Namekawa:2011wt} & 2+1 &
	     \parbox[t]{1.75cm}{$2.88$} & \parbox[t]{1.6cm}{2.2 (before reweighting)}  &
\parbox[t]{4.7cm}{No discussion of FV effects.}
\\[8.0ex] \hline \\[-1.0ex]
HPQCD 10A & \cite{Davies:2010ip} & 2+1 &
             \parbox[t]{1.75cm}{2.4, 2.4/2.88/3.36, $2.52, 2.88$, 2.82} & \parbox[t]{1.6cm}{$3.9, 3.8$, $4.1, 4.5$, 4.6 }  &
\parbox[t]{4.7cm}{FV errors  estimated using finite- vs.~infinite-volume $\chi$PT.}
\\[10.0ex] \hline \\[-1.0ex]
HPQCD/UKQCD 07 & \cite{Follana:2007uv} & 2+1 & 
	     \parbox[t]{1.75cm}{2.4, 2.4/2.88, 2.52} & \parbox[t]{1.6cm}{3.9, 3.8, 4.1}  &
\parbox[t]{4.7cm}{FV errors  estimated using finite- vs.~infinite-volume $\chi$PT.}
\\[8.0ex] \hline \\[-1.0ex]
FNAL/MILC 05 & \cite{Aubin:2005ar} & 2+1 & \parbox[t]{1.75cm}{2.8, 2.9, 2.5} & \parbox[t]{1.6cm}{$3.8, 3.8,$ $4.1$ }  &
\parbox[t]{4.7cm}{FV errors estimated to be  $1.5\%$ or less from $\chi$PT.} 
\\[5.0ex] 
\fi
\hline\hline
\end{tabular*}
\caption{Finite-volume effects in $\Nf=2+1$ determinations of the $D$- and $D_s$-meson
 decay constants.  Each $L$-entry corresponds to a different
 lattice spacing, with multiple spatial volumes at some lattice
 spacings. For actions with multiple species of pions, the lightest masses are
quoted.} 
}
\end{table}
\begin{table}[!htb]


{\footnotesize
\begin{tabular*}{\textwidth}{l @{\extracolsep{\fill}} c c l l l}
\hline\hline \\[-1.65ex]
Collab. & Ref. & $\Nf$ & $a$ [fm]  & Continuum extrapolation & Scale Setting
\\[.2ex] \hline \hline \\[-1.0ex]
\parbox[t]{2cm}{ETM~13F \\ETM~14E \\ETM 21B} & \cite{Dimopoulos:2013qfa,Carrasco:2014poa,Dimopoulos:2021qsf} & 2+1+1 & \parbox[t]{.9cm}{0.095, 0.081, 0.069}  & \parbox[t]{3.8cm}{Chiral and continuum extrapolations  
performed simultaneously by adding an $O(a^2)$ term to the chiral fits.} &
\parbox[t]{3.5cm}{Relative scale set through $w_0$ or  $M_{c's'}$, the mass of a fictitious meson made of valence quarks of mass $r_0m_{s'}\!=0.22$ and
$r_0m_{c'}=2.4$. Absolute scale through $f_\pi^{iso}$.\\} 
\\[9.0ex]
\hline\hline
\end{tabular*}
\caption{Lattice spacings and description of actions used in $\Nf=2+1+1$  determinations
 of the $D$- and $D_s$-meson decay constants.} 
}
\end{table}
\begin{table}[!htb]


{\footnotesize
\begin{tabular*}{\textwidth}{l @{\extracolsep{\fill}} c c l l l}
\hline\hline \\[-1.65ex]
Collab. & Ref. & $\Nf$ & $a$ [fm]  & Continuum extrapolation & Scale Setting
\\[.2ex] \hline \hline \\[-1.65ex]
  {RQCD/ALPHA~24} &\cite{Kuberski:2024pms}  & 2+1 &
  \parbox[t]{.9cm}{0.098, 0.085, 0.075, 0.064, 0.049, 0.039} & \parbox[t]{3.8cm}{Terms up to $a^4$ (possibly
          with mass-dependent coefficients) are included in the chiral-continuum extrapolation.}
& \parbox[t]{3.4cm}{
   $t_0$ is used as intermediate scale
          with the physical scale set by a combination of $f_\pi$ and $f_K$ in the isosymmetric limit.}\\[15.0ex]
        {ALPHA~23} &\cite{Bussone:2023kag} & 2+1 & \parbox[t]{.9cm}{0.085, 0.075, 0.064, 0.049} & \parbox[t]{3.8cm}{${\mathcal{O}}(a^2)$ terms,
          with mass-dependent coefficients, are included in the chiral-continuum extrapolation.}
          & \parbox[t]{3.4cm}{$t_0$ is used as intermediate scale
          with the physical scale set by a combination of $f_\pi$ and $f_K$ in the isosymmetric limit.}\\[15.0ex]
\ifx\reducedapptables\undefined
{$\chi$QCD 20A} & \cite{Chen:2020qma} & 2+1  & \parbox[t]{.9cm}{0.115} & \parbox[t]{3.8cm}{Cutoff effects are estimated
to be 2\% by comparing the result for $f_{D_s}$ with the one obtained in~\cite{Yang:2014sea}.} & \parbox[t]{3.4cm}{$m_{\Omega}$ used for scale
    setting by RBC/UKQCD, which produced the ensemble}\\[10.0ex]
  \hline \\[-2.0ex]
RBC/UKQCD~17, \\ RBC/UKQCD 18A & \cite{Boyle:2017jwu, Boyle:2018knm} & 2+1 &   \parbox[t]{.9cm}{0.11, 0.08, 0.07} &
\parbox[t]{3.8cm}{The lattice spacing, pion-mass and charm-quark mass dependences are fit simultaneously through a Taylor 
expansion in $a^2$, $(m_\pi^2-m_\pi^{2 phys})$ and $1/m_H-1/m_{D_{(s)}}$.} &
\parbox[t]{3.4cm}{The lattice scale and physical light-quark masses have been determined using $m_\pi$, $m_K$ and 
  $m_\Omega$ as inputs.}
\\[18.0ex]

\hline \\[-1.65ex]
$\chi$QCD~14 & \cite{Yang:2014sea} & 2+1 & \parbox[t]{.9cm}{0.113, 0.085} &
\parbox[t]{3.8cm}{
Chiral and continuum extrapolations  performed in global fits
including linear terms in $m_l$ and terms up to O$(a^2)$ and O$(a^4m_c^4)$.} 
& \parbox[t]{3.4cm}{Relative scale set through $r_0$, fixed together with the charm and strange quark masses
using $m_{D_s}$, $m_{D_s^*}$ amd $m_{J/\psi}$ as inputs.}
\\[15.0ex] \hline \\[-1.65ex]
HPQCD 12A& \cite{Na:2012iu} & 2+1 & \parbox[t]{.9cm}{0.12, 0.09} &
\parbox[t]{3.8cm}{
Chiral and continuum extrapolations  peformed 
simultaneously  using PQHM$\chi$PT augmented by $a$-dependent terms: $c_0(am_c)^2 + c_1(am_c)^4$.
} & \parbox[t]{3.4cm}{Relative scale set through $r_1$; absolute scale from $f_\pi$, $f_{\rm K}$ and the $\Upsilon$ splitting.
Uncertainties from both $r_1$ and $r_1/a$  propagated.}
\\[15.0ex] \hline \\[-1.65ex]
FNAL/MILC 11 & \cite{Bazavov:2011aa} & 2+1 &
             \parbox[t]{.9cm}{0.15, 0.12, 0.09} &
\parbox[t]{3.8cm}{
Chiral and continuum extrapolations  performed 
simultaneously using 1-loop HM$\chi$PT for rooted staggered quarks. Effects of hyperfine and flavour splittings
are also included. 
} & \parbox[t]{3.4cm}{Relative scale set through $r_1=0.3117(22)$. The error in $r_1$
comes from the spread of different absolute scale determinations using $f_\pi$, $f_{\rm K}$ and the $\Upsilon$ splitting.}
\\[21.0ex] \hline \\[-1.65ex]
PACS-CS 11 & \cite{Namekawa:2011wt} & 2+1 &
             \parbox[t]{.9cm}{$0.09$} & \parbox[t]{3.8cm}{
Cutoff effects from the heavy-quark action estimated by naive power counting to be at the percent level.} 
& \parbox[t]{3.4cm}{Scale set through $m_\Omega$.}
\\[12.0ex] \hline \\[-2.1ex]
HPQCD 10A & \cite{Davies:2010ip} & 2+1 &
\parbox[t]{.9cm}{0.15, 0.12, 0.09, 0.06, 0.044} &
\parbox[t]{3.8cm}{
Chiral and continuum 
extrapolations  performed simultaneously. Polynomials 
up to $am_{c}^8$ are kept (even powers only).} & \parbox[t]{3.4cm}{See the discussion for HPQCD 12A.}
\\[12.0ex] \hline \\[-1.65ex]
HPQCD/UKQCD 07 & \cite{Follana:2007uv} & 2+1 & 
\parbox[t]{.9cm}{0.15, 0.12, 0.09} &
\parbox[t]{3.8cm}{
Combined chiral and continuum extrapolations using
HM$\chi$PT at NLO augmented by second and third-order polynomial terms in $m_q$ and terms up to $a^4$.
} & \parbox[t]{3.4cm}{Scale set through $r_1$ obtained from the $\Upsilon$ spectrum using the 
nonrelativistic QCD action for $b$ quarks. Uncertainty propagated among the systematics.}
\\[18.0ex] \hline \\[-1.65ex]
FNAL/MILC 05 & \cite{Aubin:2005ar} & 2+1 &
\parbox[t]{.9cm}{0.175, 0.121, 0.086} &
\parbox[t]{3.8cm}{Most light-quark cutoff effects are removed through
 NLO HM$\chi$PT for rooted staggered quarks. Continuum values are then obtained 
by averaging the $a \approx 0.12$ and $a \approx 0.09$ fm results.}
 &  \parbox[t]{3.4cm}{Scale set through $r_1$ obtained from the $\Upsilon$ spectrum using the 
nonrelativistic QCD action for $b$ quarks.}
\\[21.0ex] 
%
%
\fi
\hline\hline
\end{tabular*}
\caption{Lattice spacings and description of actions used in $\Nf=2+1$  determinations
 of the $D$- and $D_s$-meson decay constants.} 
}
\end{table}

\begin{table}[!htb]
{\footnotesize
\begin{tabular*}{\textwidth}{l @{\extracolsep{\fill}} c c c l}
\hline\hline \\[-1.0ex]
Collab. & Ref. & $\Nf$ &  Ren. & Description
\\[1.0ex] \hline \hline \\[-1.0ex]
\parbox[t]{2cm}{ETM~13F \\ ETM~14E \\ ETM 21B} & \cite{Dimopoulos:2013qfa,Carrasco:2014poa,Dimopoulos:2021qsf} & 2+1+1 & $-$ & \parbox[t]{6.9cm}{The axial current is absolutely normalized.} 
\\[6.0ex]
\hline\hline
\end{tabular*}
\caption{Operator renormalization in $\Nf=2+1+1$ determinations of the $D$- and $D_s$-meson decay constants.} 
}
\end{table}

\begin{table}[!htb]
{\footnotesize
\begin{tabular*}{\textwidth}{l @{\extracolsep{\fill}} c c c l}
\hline\hline \\[-1.0ex]
Collab. & Ref. & $\Nf$ &  Ren. & Description
\\[1.0ex] \hline \hline \\[-1.0ex]
  {RQCD/ALPHA~24} &\cite{Kuberski:2024pms}  & 2+1 & SF &
  \parbox[t]{6.9cm}{The axial current is nonperturbatively improved and renormalized.}
\\[3.0ex] \hline \\[-1.0ex]
  {ALPHA~23} &\cite{Bussone:2023kag} & 2+1 &$-$ &
\parbox[t]{6.9cm}{The axial current is absolutely normalized.}
\\[3.0ex]
\ifx\reducedapptables\undefined
$\chi$QCD 20A & \cite{Chen:2020qma} & 2+1 & 
RI &
\parbox[t]{6.9cm}{The decay constants are extracted from an exact lattice Ward identity or from NP renormalized operators.}
\\[7.0ex]
\hline \\[-2.0ex]
RBC/UKQCD~17, \\ RBC/UKQCD 18A & \cite{Boyle:2017jwu, Boyle:2018knm} & 2+1 & mNPR &
\parbox[t]{6.9cm}{The local current is renormalized nonperturbatively for the case of the unmixed action; however
in the actual computation the domain wall height is chosen differently in the valence than in the sea and the effect
of that on the renormalization constant is estimated to be 0.4\% through a study in the RI/SMOM scheme.\\}
\\[3.0ex] 
\hline \\[-1.0ex]
$\chi$QCD~14 & \cite{Yang:2014sea} & 2+1 & 
RI &
\parbox[t]{6.9cm}{The decay constant is extracted from an exact lattice Ward identity and from the NP renormalized axial
current.\\}
\\[3.0ex] \hline \\[-1.0ex]
HPQCD 12A& \cite{Na:2012iu} & 2+1 & 
$-$ &
\parbox[t]{6.9cm}{The axial current is absolutely normalized.}
\\[3.0ex] \hline \\[-1.0ex]
FNAL/MILC 11 & \cite{Bazavov:2011aa} & 2+1 &
mNPR &
\parbox[t]{6.9cm}{2-loop and higher-order perturbative truncation errors estimated to be the full size of the 1-loop term.\\}
\\[3.0ex] \hline \\[-1.0ex]
PACS-CS 11 & \cite{Namekawa:2011wt} & 2+1 &
PT1$\ell$+NP &
\parbox[t]{6.9cm}{Mass dependent part of the renormalization constant of the axial current computed at 1-loop; the NP contribution
is added in the chiral limit.\\}
\\[3.0ex] \hline \\[-1.ex]
HPQCD 10A & \cite{Davies:2010ip} & 2+1 &
$-$ &
\parbox[t]{6.9cm}{The axial current is absolutely normalized.}
\\[3.0ex] \hline \\[-1.0ex]
HPQCD/UKQCD 07 & \cite{Follana:2007uv} & 2+1 & 
$-$&
\parbox[t]{6.9cm}{The axial current is absolutely normalized.}
\\[3.0ex] \hline \\[-1.0ex]
FNAL/MILC 05 & \cite{Aubin:2005ar} & 2+1 &
mNPR &
\parbox[t]{6.9cm}{Errors due to higher order corrections in the perturbative part are estimated to be $1.3\%$.}
\\[3.0ex] \\[-1.0ex]
\fi
\hline\hline
\end{tabular*}
\caption{Operator renormalization in $\Nf=2+1$ determinations of the $D$- and $D_s$-meson decay constants.} 
}
\end{table}



\begin{table}[!htb]
{\footnotesize
\begin{tabular*}{\textwidth}{l @{\extracolsep{\fill}} c c c l}
\hline\hline \\[-1.2ex]
Collab. & Ref. & $\Nf$ &  Action & Description
\\[1.0ex] \hline \hline \\[-1.2ex]
%
\parbox[t]{2cm}{ETM~13F\\ ETM~14E \\ ETM 21B} & \cite{Dimopoulos:2013qfa,Carrasco:2014poa,Dimopoulos:2021qsf} & 2+1+1 & tmWil  &
\parbox[t]{3.9cm}{$0.15 \lsim am_h \lsim 0.28$.} 
\\[6.0ex]
\hline\hline
\end{tabular*}
\caption{Heavy-quark treatment in  $\Nf=2+1+1$ determinations of the $D$-and $D_s$-meson decay constants.} 
}
\end{table}
%

\begin{table}[!htb]
{\footnotesize
\begin{tabular*}{\textwidth}{l @{\extracolsep{\fill}} c c c l}
\hline\hline \\[-1.2ex]
Collab. & Ref. & $\Nf$ &  Action & Description
\\[1.0ex] \hline \hline \\[-1.2ex]
%
\parbox[t]{2.7cm}{RQCD/ALPHA~24} &\cite{Kuberski:2024pms}  & 2+1 & npSW &
  \parbox[t]{3.9cm}{$0.1\leq am_h\leq 0.3$.}
\\[3.0ex] \hline \\[-1.0ex]
\parbox[t]{2.7cm}{ALPHA~23} &\cite{Bussone:2023kag} & 2+1 & tmWil on npSW &
\parbox[t]{3.9cm}{$0.13 \leq am_h\leq 0.26$}
\\[3.0ex]
\ifx\reducedapptables\undefined
$\chi$QCD 20A & \cite{Chen:2020qma} & 2+1 & 
Overlap on DW &
\parbox[t]{3.9cm}{$am_c\approx 0.73$ set by using $m_{D_s}$.\\}
\\[4.0ex]
\hline \\[-1.2ex]
\parbox[t]{2cm}{RBC/UKQCD~17, RBC/UKQCD 18A} & \cite{Boyle:2017jwu,Boyle:2018knm} & 2+1 & 
\parbox[t]{2cm}{
M\"obius-DWF on  Shamir-DWF or M\"obius-DWF} &
\parbox[t]{3.9cm}{$0.18<am_h<0.4$. Charm discretization errors estimated using different ways to define the charm quark mass
(through $D$, $D_s$ or $\eta_c$) in the global fits.\\} 
\\[4.0ex]
\hline \\[-1.2ex]
$\chi$QCD~14 & \cite{Yang:2014sea} & 2+1 & 
Overlap on DW &
\parbox[t]{3.9cm}{$0.29<am_c<0.75$. Heavy-quark discretization errors estimated by including $(am_c)^2$ and $am_c)^4$ terms
in the chiral/continuum extrapolation.\\}
\\[4.0ex] \hline \\[-1.4ex]
HPQCD 12A& \cite{Na:2012iu} & 2+1 & 
HISQ &
\parbox[t]{3.9cm}{$0.41<am_c<0.62$. Heavy-quark discretization errors estimated using different fit ans\"atze to be $\approx~1.2\%$.\\}
\\[4.0ex] \hline \\[-1.4ex]
FNAL/MILC 11 & \cite{Bazavov:2011aa} & 2+1 &
Fermilab  &
\parbox[t]{3.9cm}{Discretization errors from charm quark estimated through a combination of Heavy-Quark and Symanzik Effective Theories to be around 3\% for $f_{D_{(s)}}$ and negligible for the ratio.\\}
\\[4.0ex] \hline \\[-1.4ex]
PACS-CS 11 & \cite{Namekawa:2011wt} & 2+1 &
Tsukuba &
\parbox[t]{3.9cm}{$am_c\approx 0.57$. Heavy-quark discretization errors estimated to be at the percent level by power counting.\\}
\\[4.0ex] \hline \\[-1.2ex]
HPQCD 10A & \cite{Davies:2010ip} & 2+1 &
HISQ &
\parbox[t]{3.9cm}{$0.193< am_c<0.825$. Heavy-quark discretization errors estimated by changing the fit-inputs to be $\approx~0.4\%$.\\}
\\[4.0ex] \hline \\[-1.2ex]
HPQCD/UKQCD 07 & \cite{Follana:2007uv} & 2+1 & 
HISQ &
\parbox[t]{3.9cm}{$0.43<am_c<0.85$.  Heavy-quark discretization errors estimated from the chiral/continuum fits to be $\approx~0.5\%$.\\
}
\\[4.0ex] \hline \\[-1.4ex]
FNAL/MILC 05 & \cite{Aubin:2005ar} & 2+1 &
Fermilab  &
\parbox[t]{3.9cm}{Discretization errors from charm quark estimated via heavy-quark power-counting at 4.2\% for $f_{D_{(s)}}$ and 0.5\% for the ratio.\\}
\\[4.0ex] 
%
\fi
\hline\hline
\end{tabular*}
\caption{Heavy-quark treatment in  $\Nf=2+1$ determinations of the $D$- and $D_s$-meson decay constants.} 
}
\end{table}


%% file: HQ/AppendixTables/DSLapp.tex
\FloatBarrier
\subsubsection{Form factors for semileptonic decays of charmed hadrons}
\label{app:DtoPi_Notes}
\FloatBarrier

\begin{table}[!ht]

{\footnotesize
\begin{tabular*}{\textwidth}{l @{\extracolsep{\fill}} c c c l l}
\hline\hline \\[-1.0ex]
Collab. & Ref. & $\Nf$ & $a$ [fm] & Continuum extrapolation & Scale setting 
\\[1.0ex] \hline \hline \\[-1.0ex]
FNAL/MILC 22 & \cite{FermilabLattice:2022gku} & 2+1+1 & \parbox[t]{1.5cm}{ 0.12, 0.088, 0.057, 0.042  }   & \parbox[t]{4.2cm}{ Combined chiral-continuum extrapolation using SU(2) heavy-meson rooted staggered chiral perturbation theory.  } & \parbox[t]{35mm}{ Scale setting using gradient flow $w_0$ with physical scale from $f_\pi$.  } 
\\[16.0ex] \hline \\[-2.0ex]
Meinel 21B & \cite{Meinel:2021mdj} & 2+1 & \parbox[t]{1.5cm}{0.0828(3), 0.1106(3)}   & \parbox[t]{4.2cm}{Combined chiral-continuum extrapolation as part of the expansion of form factor shape in powers of $w-1$. Systematics estimated by varying fit form.} & \parbox[t]{35mm}{Scale setting using $\Omega$ mass in Ref.~\cite{Blum:2014tka}.} 
\\[16.0ex] \hline \\[-2.0ex]
HPQCD 21A & \cite{Chakraborty:2021qav} & 2+1+1 & \parbox[t]{1.3cm}{0.042, 0.06, 0.09, 0.12, 0.15}  &
\parbox[t]{4.1cm}{Modified $z$-expansion fit combining the continuum and chiral extrapolations and the momentum-transfer dependence. Discretization effects assumed dominated by the charm scale. Discretization errors on form factors between 0.4\% and 1.2\% as a function of the momentum transfer.} & \parbox[t]{3.5cm}{Scale setting from $f_\pi$ via the flow quantity $w_0$~\cite{Dowdall:2013rya,Chakraborty:2014aca,Chakraborty:2016mwy}.}
\\[23.0ex]  \\ \hline
 \\[-2.0ex]
Zhang  21  & \cite{Zhang:2021oja} & 2+1 & \parbox[t]{1.3cm}{0.080, 0.11}   &
\parbox[t]{4.1cm}{Continuum extrapolation combined with fit to $q^2$-dependence of form factors in a ``modified'' $z$-expansion. Systematics estimated from difference between extrapolated results and results at smallest lattice spacing, and difference between two current renormalization methods.} & \parbox[t]{3.5cm}{Set from Wilson-flow quantity $w_0$.}
\\[25.0ex]  \\ \hline
 \\[-2.0ex]
HPQCD 20 & \cite{Cooper:2020wnj} & 2+1+1 & \parbox[t]{1.3cm}{0.06, 0.09, 0.12, 0.15}  &
\parbox[t]{4.1cm}{Modified $z$-expansion fit combining the continuum and chiral extrapolations and the momentum-transfer dependence, and, for the heavy-HISQ spectator $b$ quark, the dependence on $1/m_Q$. The analysis combines data with NRQCD $b$ quarks and data with HISQ heavy quarks.} & \parbox[t]{3.5cm}{Scale setting from $f_\pi$ via the flow quantity $w_0$~\cite{Dowdall:2013rya,Chakraborty:2014aca,Chakraborty:2016mwy}.}
\\[23.0ex]  \\ 
\hline\hline
\end{tabular*}
\caption{Continuum extrapolations/estimation of lattice artifacts in $\Nf=2+1+1$ determinations of form factors for semileptonic decays of charmed hadrons. For HPQCD 22, see Tab.~\protect\ref{tab:AppTableRareBCont}.}
}
\end{table}

\ifx\reducedapptables\undefined

\begin{table}[!ht]

{\footnotesize
\begin{tabular*}{\textwidth}{l @{\extracolsep{\fill}} c c c l l}
\hline\hline \\[-1.0ex]
Collab. & Ref. & $\Nf$ & $a$ [fm] & Continuum extrapolation & Scale setting 
\\[1.0ex] \hline \hline \\[-1.0ex]
ETM 17D, 18 & \cite{Lubicz:2017syv,Lubicz:2018rfs} & 2+1+1 & \parbox[t]{1.3cm}{0.062, 0.082, 0.089}  &
\parbox[t]{4.1cm}{Modified $z$-expansion fit combining the continuum and chiral extrapolations and the momentum-transfer dependence. Lattice-spacing dependence through $\mathcal{O}(a^2)$, with systematic uncertainty estimated by adding $\mathcal{O}(a^4)$ terms constrained by priors. Additional terms included to fit artifacts due to the breaking of rotational invariance. Meson momenta tuned to be constant with changing lattice spacing and volume.\\} & \parbox[t]{3.5cm}{Relative scale set through $M_{c's'}$, the mass of a fictitious meson made of valence quarks of mass $r_0m_{s'} = 0.22$ and $r_0m_{c'} = 2.4$. Absolute scale from the
experimental value of $f_\pi$.
}
\\[17.0ex] \hline \\[-2.0ex]
JLQCD 17B & \cite{Kaneko:2017xgg} & 2+1 & \parbox[t]{1.3cm}{0.044, 0.055, 0.080}   &
\parbox[t]{4.1cm}{Joint chiral-continuum extrapolation, with mass dependence based on hard-pion HQ$\chi$PT.} & \parbox[t]{3.5cm}{Set from $t_0$ by using the value in physical units provided in~\cite{Borsanyi:2012zs}.}
\\[10.0ex] \hline \\[-2.0ex]
Meinel 16, 17  & \cite{Meinel:2016dqj,Meinel:2017ggx} & 2+1 & \parbox[t]{1.3cm}{0.085, 0.11}   &
\parbox[t]{4.1cm}{Joint chiral-continuum extrapolation, combined with fit to $q^2$-dependence of form factors in a ``modified'' $z$-expansion. Systematics estimated by varying fit form.} & \parbox[t]{3.5cm}{Set from $\Upsilon(2S)$--$\Upsilon(1S)$ splitting, cf.~\cite{Meinel:2010pv}.}
\\[16.0ex] \hline \\[-2.0ex]
HPQCD 10B, 11 & \cite{Na:2010uf,Na:2011mc} & 2+1 & \parbox[t]{1.3cm}{0.09, 0.12}  &
\parbox[t]{4.1cm}{Modified $z$-expansion fit combining the continuum and chiral extrapolations and the momentum-transfer dependence. Leading discretization errors from $(am_c)^n$ charm-mass effects (see Tab.~\ref{tab:DtoPiKHQ}). 
    Subleading $(aE)^n$ discretization corrections 
     estimated to be 1.0\% for both 
    $D\to\pi$ and $D\to K$.} & \parbox[t]{3.5cm}{Relative scale $r_1/a$ set from the static-quark potential.  Absolute scale $r_1$ set from several quantities including $f_\pi$, $f_K$, and $\Upsilon$ $2S-1S$ splitting c.f. HPQCD~09B~\cite{Davies:2009tsa}.  Scale uncertainty estimated to be 0.7\% in $D\to\pi$ and and 0.2\% in $D\to K$.}
\\[30.0ex] \hline \\[-2.0ex]
\parbox[t]{1cm}{FNAL/MILC~04} & \cite{Aubin:2004ej} & 2+1 & 0.12  &
\parbox[t]{4.1cm}{Discretization effects from light-quark sector estimated to be 4\% by power counting. Discretization effects from final-state pion and kaon energies estimated to be 5\%.} & \parbox[t]{3.5cm}{Scale set through $\Upsilon$ $2S-1S$ splitting c.f. HPQCD 03~\cite{Wingate:2003gm}.  Error in $a^{-1}$ estimated to be 1.2\%, but scale error in dimensionless form factor negligible compared to other uncertainties.}
\\[20.0ex]
\hline\hline
\end{tabular*}
\caption{Continuum extrapolations/estimation of lattice artifacts in $\Nf=2+1$ determinations of form factors for semileptonic decays of charmed hadrons.}
}
\end{table}

\begin{table}[!ht]

{\footnotesize
\begin{tabular*}{\textwidth}{l @{\extracolsep{\fill}} c c c l l}
\hline\hline \\[-1.0ex]
Collab. & Ref. & $\Nf$ & $a$ [fm] & Continuum extrapolation & Scale setting 
\\[1.0ex] \hline \hline \\[-1.0ex]
ETM 11B & \cite{DiVita:2011py} & 2 & \parbox[t]{1.3cm}{0.068, 0.086, 0.102} &
\parbox[t]{4.1cm}{Discretization errors estimated to be 5\% for $D\to\pi$ and 3\% for $D\to K$ from comparison of results in the continuum limit to those at the finest lattice spacing.}
& \parbox[t]{3.5cm}{Scale set through $f_\pi$ c.f. ETM~07A~\cite{Boucaud:2007uk} and ETM~09C~\cite{Baron:2009wt}.}
\\[15.0ex]
\hline\hline
\end{tabular*}
\caption{Continuum extrapolations/estimation of lattice artifacts in $\Nf=2$ determinations of form factors for semileptonic decays of charmed hadrons.}
}
\end{table}
\fi

\begin{table}[!ht]
{\footnotesize
\begin{tabular*}{\textwidth}{l @{\extracolsep{\fill}} c  c c  l}
\hline\hline \\[-1.0ex]
{Collab.} & {Ref.} & {$\Nf$} & {$M_{\pi,\rm min}\,[\mev]$}  & {Description}  
\\[1.0ex] \hline \hline \\[-1.0ex]
FNAL/MILC 22 & \cite{FermilabLattice:2022gku} & 2+1+1 & \parbox[t]{2cm}{ 135, 130, 134, 308 }   & \parbox[t]{6cm}{ Combined chiral-continuum extrapolation using SU(2) heavy-meson rooted staggered chiral perturbation theory at NLO, including NNLO analytic terms.  } 
\\[12.0ex] \hline \\[-2.0ex]
Meinel 21B & \cite{Meinel:2021mdj} & 2+1 & \parbox[t]{2cm}{303, 340}
& \parbox[t]{6cm}{Combined chiral-continuum extrapolation as part of the expansion of form factor shape in powers of $w-1$. Systematic uncertainty estimated by repeating fit with added higher-order terms.}
\\[14.0ex]
\hline \\[-2.0ex]
HPQCD 21A & \cite{Chakraborty:2021qav} & 2+1+1 & \parbox[t]{2cm}{315, 329, 129, 132, 131}
 & \parbox[t]{6cm}{Modified $z$-expansion fit combining the continuum and chiral extrapolations and the momentum-transfer dependence. Polynomial dependence on quark masses, supplemented by a pion chiral logarithm. Fit result compared with alternative approach based on cubic splines in $q^2$.}
\\[19.0ex] \hline \\[-2.0ex]

Zhang  21  & \cite{Zhang:2021oja} & 2+1 & \parbox[t]{2cm}{300, 290}   &
\parbox[t]{6cm}{Dependence on pion mass neglected. No estimate of resulting systematic uncertainty.}
\\[7.0ex] \hline \\[-2.0ex]
HPQCD 20 & \cite{Cooper:2020wnj} & 2+1+1 & \parbox[t]{2cm}{329, 316, 132/305, 131/305}  &
\parbox[t]{6cm}{Modified $z$-expansion fit combining the continuum and chiral extrapolations and the momentum-transfer dependence, and, for the heavy-HISQ spectator $b$ quark, the dependence on $1/m_Q$. The analysis combines data with NRQCD $b$ quarks and data with HISQ heavy quarks.}
\\[18.0ex] 
\ifx\reducedapptables\undefined
 \hline \\[-2.0ex]
ETM 17D, 18 & \cite{Lubicz:2017syv,Lubicz:2018rfs} & 2+1+1 & \parbox[t]{2cm}{220, 258, 275}
 & \parbox[t]{6cm}{Modified $z$-expansion fit combining the continuum and chiral extrapolations and the momentum-transfer dependence. Chiral log term in vector and scalar form factors set to hard-pion $\chi$PT prediction~\cite{Bijnens:2010ws}. Systematic uncertainty in tensor form factor estimated by comparing fits with and without chiral log terms.}
\\[21.0ex] \hline \\[-2.0ex]
JLQCD 17B & \cite{Kaneko:2017xgg} & 2+1 & \parbox[t]{2cm}{284, 296, 226}
 & \parbox[t]{6cm}{Joint chiral-continuum extrapolation, with mass dependence based on hard-pion Q$\chi$PT.}
\\[7.0ex] \hline \\[-2.0ex]
Meinel 17 & \cite{Meinel:2017ggx} & 2+1 & \parbox[t]{2.5cm}{227, 245 \\ (valence-valence) \\ 263, 294 \\ (valence-sea)}
& \parbox[t]{6cm}{Joint chiral-continuum extrapolation, combined with fit to $q^2$-dependence of form factors in a ``modified'' $z$-expansion. Only analytic NLO terms $\propto (m_\pi^2-m_{\pi,{\rm phys}}^2)$ included in light-mass dependence. Systematic uncertainty estimated by repeating fit with added higher-order terms.}
\\[21.0ex] \hline \\[-2.0ex]
Meinel 16 & \cite{Meinel:2016dqj} & 2+1 & \parbox[t]{2cm}{295, 139}
 & \parbox[t]{6cm}{Modified $z$-expansion fit combining the continuum and chiral extrapolations and the momentum-transfer dependence. Analytic function in $m_\pi, m_{\eta_s}$ used for mass dependence. ($\eta_s$ stands for a nonsinglet meson with two mass-degenerate valence quarks of mass $m_s$, used to set the strange scale.} 
\\[21ex] 
\hline \\[-2.0ex]
HPQCD 10B, 11 & \cite{Na:2010uf, Na:2011mc} & 2+1 & \parbox[t]{2cm}{390, 390}
& \parbox[t]{6cm}{Modified $z$-expansion fit combining the continuum and chiral extrapolations and the momentum-transfer dependence. Contributions to error budget from light valence and sea-quark mass dependence estimated to be 2.0\% for $D \to\pi$ and 1.0\% for $D \to K$. }
\\[18.0ex] \hline \\[-2.0ex]
FNAL/MILC~04 & \cite{Aubin:2004ej} & 2+1 & \parbox[t]{2cm}{510}
 & \parbox[t]{6cm}{Fit to S$\chi$PT, combined with the Becirevic-Kaidalov ansatz for the momentum-transfer dependence of form factors. Error estimated to be 3\% for $D \to\pi$ and 2\% for $D \to K$ by comparing fits with and without one extra analytic term.}
\\[15.0ex] \hline \\[-2.0ex]
ETM 11B & \cite{DiVita:2011py} & 2 & \parbox[t]{2cm}{270}
& \parbox[t]{6cm}{SU(2) tmHM$\chi$PT plus Becirevic-Kaidalov ansatz for fits to the momentum-transfer dependence of form factors.  Fit uncertainty estimated to be 7\% for $D \to\pi$ and 5\% for $D \to K$
by considering fits with and without NNLO corrections of order $\cO(m_\pi^4)$ and/or higher-order terms through $E^5$, and by excluding data with $E \gtapprox 1$~GeV.}
\\[24.0ex]
\fi
\hline\hline
\end{tabular*}
\caption{Chiral extrapolation/minimum pion mass in
  determinations of form factors for semileptonic decays of charmed hadrons.  For actions with multiple species of pions, masses quoted are the RMS pion masses for $\Nf=2+1$ and the Goldstone mode mass for $\Nf=2+1+1$.  The different $M_{\pi,\rm min}$ entries correspond to the different lattice spacings.  For HPQCD 22, see Tab.~\protect\ref{tab:AppTableRareBChir}.}
}
\end{table}

\begin{table}[!ht]
{\footnotesize
\begin{tabular*}{\textwidth}{l @{\extracolsep{\fill}} c c  c c l}
\hline\hline \\[-1.0ex]
Collab. & Ref. & $\Nf$ & $L$ [fm] & ${M_{\pi,\rm min}}L$ & Description 
\\[1.0ex] \hline \hline \\[-1.0ex]
FNAL/MILC 22 & \cite{FermilabLattice:2022gku} & 2+1+1 & \parbox[t]{1.8cm}{  5.76, 4.22/5.63, 2.74/3.65/5.47, 2.69 }   & \parbox[t]{1.8cm}{3.95, 3.72, 3.72, 4.20}  & \parbox[t]{5cm}{Finite-volume effects removed by correction to chiral logs due to sums over discrete momenta; corrections are $\mathcal{O}(0.01)\%$ overall. Effect of frozen topological charge at finest lattice spacing also corrected using $\chi$PT and found to be $\lesssim0.03\%$.   } 
\\[20.0ex] \hline \\[-2.0ex]
Meinel 21B & \cite{Meinel:2021mdj} & 2+1 & \parbox[t]{1.8cm}{2.7, 2.7}  & \parbox[t]{1.8cm}{4.1, 4.6} & \parbox[t]{5cm}{Finite-volume effects not quantified. Effects from unstable $\Lambda^*(1520)$ not quantified.}
\\[10.0ex] \hline \\[-2.0ex]
HPQCD 21A & \cite{Chakraborty:2021qav} & 2+1+1 & \parbox[t]{1.8cm}{2.73, 2.72, 2.81/5.62, 2.93/5.87, 2.45/4.89} & $\gtrsim$ 3.7 & 
\parbox[t]{5cm}{Finite-volume correction included in chiral fit, claimed to be a negligible effect.
Effect of frozen topology in finest ensemble not discussed.}
\\[10.0ex] \hline \\[-1.0ex]
Zhang  21  & \cite{Zhang:2021oja} & 2+1+1 & \parbox[t]{1.8cm}{2.6, 2.6} & $\gtrsim$ 3.8 & 
\parbox[t]{5cm}{No discussion of finite-volume effects.}
\\[5.0ex] \hline \\[-1.0ex]
HPQCD 20 & \cite{Cooper:2020wnj} & 2+1+1 & \parbox[t]{1.8cm}{2.72, 2.81, 2.93/5.87, 2.45/4.89} & $\gtrsim$ 3.8 & 
\parbox[t]{5cm}{Physical point ensemble at $a \simeq 0.15~{\rm fm}$ has $m_\pi L=3.3$; the statement $m_\pi L \gtrsim 3.8$ applies to the other five ensembles.}
\\[10.0ex]
\ifx\reducedapptables\undefined
\hline \\[-1.0ex]
ETM 17D, 18 & \cite{Lubicz:2017syv,Lubicz:2018rfs} & 2+1+1 & \parbox[t]{1.8cm}{2.97, 1.96/2.61, 2.13/2.84} & 3.31, 3.42, 3.49 & 
\parbox[t]{5cm}{Extrapolation to infinite volume performed by including term $\propto e^{-M_\pi L}/(M_\pi L)$ in global fit.}
\\[8.0ex] \hline \\[-1.0ex]
JLQCD 17B & \cite{Kaneko:2017xgg} & 2+1 & \parbox[t]{1.8cm}{2.8, 2.6, 2.6/3.9} & 4.0, 3.9, 4.4 & 
\parbox[t]{5cm}{No discussion of finite-volume effects.}
\\[5.0ex] \hline \\[-1.0ex]
Meinel 17 & \cite{Meinel:2017ggx} & 2+1 & \parbox[t]{1.8cm}{2.7, 2.7} & \parbox[t]{1.8cm}{$\gtrsim 3.6$ \\ (valence-sea pion)}  &
\parbox[t]{5cm}{Finite-volume effect estimated at 3\% from experience on $\chi$PT estimates of finite-volume effects for heavy-baryon axial couplings.
}
\\[12.0ex] \hline \\[-1.0ex]
Meinel 16 & \cite{Meinel:2016dqj} & 2+1 & \parbox[t]{1.8cm}{2.7, 2.6/5.3} & 4.1, 3.7 & 
\parbox[t]{5cm}{Finite-volume effect estimated to be at 1.0\% level.}
\\[5.0ex] 
\hline \\[-1.0ex]
HPQCD 10B, 11 & \cite{Na:2010uf,Na:2011mc} & 2+1 & \parbox[t]{1.8cm}{2.4, 2.4/2.9} & $\gtrsim 3.8$ & 
\parbox[t]{5cm}{Finite-volume effects estimated to be 0.04\% for $D\to\pi$ and 0.01\% for $D\to K$ by comparing the ``$m_\pi^2 {\rm log} (m_\pi^2)$" term in infinite and finite volume.}
\\[12.0ex] \hline \\[-1.0ex]
FNAL/MILC 04 & \cite{Aubin:2004ej} & 2+1 & \parbox[t]{1.8cm}{2.4/2.9} & $\gtrsim 3.8$ & 
\parbox[t]{5cm}{No explicit estimate of finite-volume error, but expected to be small for simulation masses and volumes.}
\\[7.0ex] \hline \\[-1.0ex]
ETM 11B & \cite{DiVita:2011py} & 2 & \parbox[t]{1.8cm}{2.2, 2.1/2.8, 2.4} & $\gtrsim 3.7$ &
\parbox[t]{5cm}{Finite-volume uncertainty estimated to be at most 2\% by considering
fits with and without the lightest pion mass point at $m_\pi L \approx 3.7$.}
\\[12.0ex]
\fi
\hline\hline
\end{tabular*}
\caption{Finite-volume effects in determinations of form factors for semileptonic decays of charmed hadrons.  Each $L$-entry corresponds to a different lattice
spacing, with multiple spatial volumes at some lattice spacings.  For actions with multiple species of pions, the lightest pion masses are quoted.  For HPQCD 22, see Tab.~\protect\ref{tab:AppTableRareBFV}.}
}
\end{table}
\begin{table}[!ht]
{\footnotesize
\begin{tabular*}{\textwidth}{l @{\extracolsep{\fill}} c c c l}
\hline\hline \\[-1.0ex]
Collab. & Ref. & $\Nf$ & Ren. & Description 
\\[1.0ex] \hline \hline \\[-1.0ex]
FNAL/MILC 22 & \cite{FermilabLattice:2022gku} & 2+1+1 & NPR  & \parbox[t]{6cm}{ Nonperturbative renormalization by imposing the PCVC relation. } 
\\[5.0ex] \hline \\[-2.0ex]
Meinel 21B & \cite{Meinel:2021mdj} & 2+1 & mNPR  & \parbox[t]{6cm}{Residual matching factors $\rho$ computed at 1-loop for vector and axial-vector currents, but at tree-level only for tensor currents. A systematic uncertainty is assigned to $\rho_{T^{\mu\nu}}$ as the double of ${\rm max}(|\rho_{A^{\mu}}-1|,|\rho_{V^{\mu}}-1|)$.}
\\[16.0ex] \hline \\[-2.0ex]
HPQCD 21A & \cite{Chakraborty:2021qav} & 2+1+1 & NP &
\parbox[t]{6cm}{Vector current normalized by imposing Ward identity at zero recoil.}
\\[5.0ex] \hline \\[-1.0ex]
Zhang  21  & \cite{Zhang:2021oja} & 2+1 & NP & 
\parbox[t]{6cm}{Local vector current renormalized using ratio to conserved vector current. Axial current renormalized using ratio of off-shell quark matrix elements. }
\\[12.0ex] \hline \\[-1.0ex]
HPQCD 20 & \cite{Cooper:2020wnj} & 2+1+1 & NP &
\parbox[t]{6cm}{Vector current normalized by imposing Ward identity at zero recoil.}
\\[5.0ex]
\ifx\reducedapptables\undefined
\hline \\[-1.0ex]
ETM 17D, 18 & \cite{Lubicz:2017syv,Lubicz:2018rfs} & 2+1+1 & RI'-MOM &
\parbox[t]{6cm}{Vector current normalization obtained nonperturbatively by imposing
charge conservation in $D\to D$, $K \to K$, and $\pi \to \pi$ transitions. Tensor current renormalization factors computed in 
RI'-MOM. Scalar form factor is absolutely normalized from chiral symmetry. Renormalized tensor 
current matched to $\overline{\rm MS}$ using 2-loop perturbation theory.}
\\[24.0ex] \hline \\[-1.0ex]
JLQCD 17B & \cite{Kaneko:2017xgg} & 2+1 & mNPR &
\parbox[t]{6cm}{Nonperturbative renormalization of vector current.}
\\[5.0ex] \hline \\[-1.0ex]
Meinel 16, 17  & \cite{Meinel:2016dqj,Meinel:2017ggx} & 2+1 & mNPR &
\parbox[t]{6cm}{Nonperturbative renormalization of singlet currents, residual factor computed at 1-loop in tadpole-improved perturbation theory.}
\\[10.0ex] 
\hline \\[-1.0ex]
HPQCD 10B, 11 & \cite{Na:2010uf, Na:2011mc} & 2+1 & --- &
\parbox[t]{6cm}{Form factor extracted from absolutely normalized scalar-current matrix element then using kinematic constraint at zero momentum-transfer $f_+(0) = f_0(0)$.}
\\[10.0ex] \hline \\[-1.0ex]
FNAL/MILC~04 & \cite{Aubin:2004ej} & 2+1 & mNPR &
\parbox[t]{6cm}{Size of 2-loop correction to current renormalization factor assumed to be negligible.}
\\[5.0ex] \hline \\[-1.0ex]
ETM 11B & \cite{DiVita:2011py} & 2 & --- &
\parbox[t]{6cm}{Form factors extracted from double ratios insensitive to current normalization.}
\\[4.0ex]
\fi
\hline\hline
\end{tabular*}
\caption{Operator renormalization in determinations of form factors for semileptonic decays of charmed hadrons.  For HPQCD 22, see Tab.~\protect\ref{tab:AppTableRareBRen}.}
}
\end{table}
\begin{table}[!ht]
{\footnotesize
\begin{tabular*}{\textwidth}{l @{\extracolsep{\fill}} c c c l}
\hline\hline \\[-1.0ex]
Collab. & Ref. & $\Nf$ & Action & Description 
\\[1.0ex] \hline \hline \\[-1.0ex]
FNAL/MILC 22 & \cite{FermilabLattice:2022gku} & 2+1+1 & HISQ  & \parbox[t]{6cm}{ Valence heavy-quark masses range from $0.9$ to $2$ times the physical charm mass, with $0.164 \leq a m_h \leq 0.8935$} 
\\[8.0ex] \hline \\[-2.0ex]

Meinel 21B & \cite{Meinel:2021mdj} & 2+1 & \parbox[t]{30mm}{Columbia RHQ for both the $b$ and $c$ quarks.} & \parbox[t]{6cm}{Discretization errors discussed as part of combined chiral-continuum-$w$ fit. Higher-order fit also includes $\mathcal{O}(\alpha_s a |\mathbf{p}|)$ terms to account for missing radiative corrections to $\mathcal{O}(a)$ improvement of the currents.}
\\[12.0ex] \hline \\[-2.0ex]
HPQCD 21A & \cite{Chakraborty:2021qav} & 2+1+1 & HISQ &
\parbox[t]{6cm}{Bare charm-quark mass $0.194 \lesssim am_c \lesssim 0.8605$.}
\\[4.0ex] \hline \\[-1.0ex]
Zhang  21  & \cite{Zhang:2021oja} & 2+1+1 & SW &
\parbox[t]{6cm}{Bare charm-quark mass $0.235 \lesssim am_c \lesssim 0.485$. No $\mathcal{O}(a)$ improvement of currents.}
\\[6.0ex] \hline \\[-1.0ex]
HPQCD 20 & \cite{Cooper:2020wnj} & 2+1+1 & \parbox{2cm}{ Charm: HISQ \\ Bottom \\ (spectator): \\ HISQ and NRQCD} & 
\parbox[t]{6cm}{ Bare charm-quark HIQS mass $0.274 \lesssim am_c \lesssim 0.827$. \\  Bare bottom-quark HIQS mass $0.274 \lesssim am_b \lesssim 0.8$.}
\\[12.0ex]
\ifx\reducedapptables\undefined
\hline \\[-1.0ex]
ETM 17D, 18 & \cite{Lubicz:2017syv,Lubicz:2018rfs} & 2+1+1 & tmWil &
\parbox[t]{6cm}{Bare charm-quark mass $0.14 \lesssim am_c \lesssim 0.29$.}
\\[4.0ex] \hline \\[-1.0ex]
JLQCD 17B & \cite{Kaneko:2017xgg} & 2+1 & M\"obius DWF &
\parbox[t]{6cm}{Charm quark matched to its physical value.}
\\[4.0ex] \hline \\[-1.0ex]
Meinel 16, 17  & \cite{Meinel:2016dqj,Meinel:2017ggx} & 2+1 & Anisotropic SW. &
\parbox[t]{6cm}{Residual $\mathcal{O}(a)$ improvement coefficients in currents computed in 1-loop tadpole-improved perturbation theory.}
\\[7.0ex] 
\hline \\[-1.0ex]
HPQCD 10B, 11 & \cite{Na:2010uf,Na:2011mc} & 2+1 & HISQ &
\parbox[t]{6cm}{Bare charm-quark mass $am_c \sim 0.41$--0.63.  Errors of $(am_c)^n$ estimated within modified $z$-expansion to be 1.4\% for $D\to K$ and 2.0\% for $D \to\pi$.  Consistent with expected size of dominant 1-loop cutoff effects on the finest lattice spacing, $\cO(\alpha_S (am_c)^2 (v/c) )\sim 1.6$\%.}
\\[18.0ex] \hline \\[-1.0ex]
FNAL/MILC~04 & \cite{Aubin:2004ej} & 2+1 & Fermilab &
\parbox[t]{6cm}{Discretization errors from charm quark estimated via heavy-quark power-counting to be 7\%.}
\\[7.0ex] \hline \\[-1.0ex]
ETM 11B & \cite{DiVita:2011py} & 2 & tmWil &
\parbox[t]{6cm}{Bare charm-quark mass $am_c \sim 0.17$--0.30.   Expected size of $\cO((am_c)^2)$ cutoff effects on the finest lattice spacing consistent with quoted 5\% continuum-extrapolation uncertainty.}
\\[13.0ex]
\fi
\hline\hline
\end{tabular*}
\caption{Heavy-quark treatment in determinations of form factors for semileptonic decays of charmed hadrons.  For HPQCD 22, see Tab.~\protect\ref{tab:AppTableRareBHQ}. \label{tab:DtoPiKHQ} 
}}
\end{table}

%% file: HQ/AppendixTables/BLapp.tex
\subsubsection{$B_{(s)}$-meson decay constants}
\label{app:fB_Notes}
\begin{table}[!htb]
{\footnotesize
\begin{tabular*}{\textwidth}{l @{\extracolsep{\fill}} c c c   l}
\hline\hline \\[-1.0ex]
Collab. & \hspace{-0.4cm}Ref. & $\Nf$ & $M_{\pi,\rm min}\,[\mev]$  & Description 
\\[1.0ex] \hline \hline \\[-2.0ex]
\parbox[t]{2.8cm}{Frezzotti 24} 
& \parbox[t]{1.0cm}{\cite{Frezzotti:2024kqk} } & 2+1+1 
& \parbox[t]{1.3cm}{175, 140, \\137, 141} &
\parbox[t]{7.0cm}{One light-quark mass per lattice spacing. Chiral effects expected to be subdominant compared to other effects.}
\\[6.0ex] 
\ifx\reducedapptables\undefined
\hline \\[-2.0ex]
\parbox[t]{2.8cm}{FNAL/MILC 17 } 
& \parbox[t]{1.0cm}{\cite{Bazavov:2017lyh} } & 2+1+1
& \parbox[t]{1.3cm}{130, 133, 130, 135, 134, 309} &
\parbox[t]{7.0cm}{Multiple values of pion masses at each lattice spacing, except for the finest lattice.
Chiral extrapolation performed using the heavy-meson rooted all-staggered $\chi$PT~\cite{Bernard:2013qwa}.}
\\[10.0ex] \hline \\[-2.0ex]
\parbox[t]{2.8cm}{HPQCD 17A  \\ HPQCD 13} 
& \parbox[t]{1.0cm}{\cite{Hughes:2017spc} \\ \cite{Dowdall:2013tga} } & 2+1+1 
& \parbox[t]{1.3cm}{310, 294, 173} & 
\parbox[t]{7.0cm}{Two or three pion masses at each lattice spacing, one each with a physical mass GB pion. 
NLO (full QCD) HM$\chi$PT supplemented by generic $a^2$ and $a^4$ terms is used to extrapolate to the physical pion mass.}
\\[12.0ex] \hline \\[-2.0ex]
\parbox[t]{2.8cm}{ETM 16B  \\ ETM 13E }
& \parbox[t]{1.0cm}{\cite{Bussone:2016iua} \\ \cite{Carrasco:2013naa} } & 2+1+1 
& \parbox[t]{1.3cm}{245, 239, 211}
& \parbox[t]{7.0cm}{$M_{\pi,{\rm min}}$ refers to the charged pions.
Linear and NLO (full QCD) HM$\chi$PT formulae supplemented by an $a^2$ term are used for the chiral-continuum extrapolation.
In ETC 13, the chiral-fit error is estimated from the difference between the NLO HM$\chi$PT and linear fits with half the difference used as estimate of the systematic error. 
The ratio $z_s$ is fit using just linear HM$\chi$PT supplemented by an $a^2$ term. 
On the other hand, in ETC 16B, the systematic error is estimated by using data points with $M_{\pi} < 350$ MeV for the chiral-continuum fit. }
\\[29.0ex]
\fi
\hline\hline
\end{tabular*}
\caption{Chiral extrapolation/minimum pion mass in determinations of the
  $B$- and $B_s$-meson decay constants for $\Nf=2+1+1$ simulations.  
  The different $M_{\pi,\rm min}$ entries correspond to the different lattice
 spacings.} 
\ifx\reducedapptables\undefined
\caption{Chiral extrapolation/minimum pion mass in determinations of the
  $B$- and $B_s$-meson decay constants for $\Nf=2+1+1$ simulations.  
  For actions with multiple species of pions, masses quoted are the RMS pion masses.    
  The different $M_{\pi,\rm min}$ entries correspond to the different lattice
 spacings.} 
\fi
}
\end{table}


\begin{table}[!htb]
{\footnotesize
\begin{tabular*}{\textwidth}{l @{\extracolsep{\fill}} c c c   l}
\hline\hline \\[-1.0ex]
Collab. & \hspace{-0.4cm}Ref. & $\Nf$ & $M_{\pi,\rm min}\,[\mev]$  & Description
\\[1.0ex] \hline \hline \\[-2.0ex]
\parbox[t]{2.8cm}{QCDSF/UKQCD\\/CSSM 22}
& \parbox[t]{1.0cm}{\cite{Hollitt:2022exk} } & 2+1 
& \parbox[t]{1.3cm}{280, 155,\\226, 290} 
& \parbox[t]{7.0cm}{Between one and three light-quark masses per lattice spacing. 
Generic fits to $(M_\pi^2/X_\pi^2-1)^2$ and $a^2(M_\pi^2/X_\pi^2-1)$ in the combined chiral-continuum extrapolation, with systematic errors estimated to
be from $1.3\%$ in $f_{B_{s}}/f_B$.}
\\[12.0ex] \hline \\[-2.0ex] 
\parbox[t]{2.8cm}{RBC/UKQCD 22}
& \parbox[t]{1.0cm}{\cite{Black:2022eph} } & 2+1 
& \parbox[t]{1.3cm}{340, 302,\\267, 371} 
& \parbox[t]{7.0cm}{Between one and three light-quark masses per lattice spacing. 
Combined chiral-continuum extrapolation using NLO SU(2) Heavy-Meson $\chi$PT. No explicit estimate of systematic errors.}
\\[9.0ex] 
\ifx\reducedapptables\undefined
\parbox[t]{2.8cm}{RBC/UKQCD 18A}
& \parbox[t]{1.0cm}{\cite{Boyle:2018knm} } & 2+1 
& \parbox[t]{1.3cm}{139, 139, 232} 
& \parbox[t]{7.0cm}{Three or four light-quark masses per lattice spacing except for the finest lattice spacing. 
Generic fits to $m_\pi^2-(m_\pi^{\mathrm{phys}})^2$ and $a^2$ in the combined chiral-continuum extrapolation, with systematic errors estimated to
be from $0.3\%$ to $0.5\%$ in $f_{B_{s}}/f_B$.}
\\[15.0ex] \hline \\[-2.0ex]
\parbox[t]{2.8cm}{RBC/UKQCD~14  \\ RBC/UKQCD 13}
& \parbox[t]{1.0cm}{\cite{Christ:2014uea} \\ \cite{Witzel:2013sla} } & 2+1 
& \parbox[t]{1.3cm}{329, 289} 
& \parbox[t]{7.0cm}{Two or three light-quark masses per lattice spacing.
In RBC/UKQCD~14, three to four light valence-quark  masses that are heavier than the sea-quark masses are also employed to have partially-quenched points.
NLO SU(2) HM$\chi$PT is used.  
In RBC/UKQCD~14, the fit with only the unitary points is the central analysis procedure, and the systematic errors in the combined chiral-continuum extrapolation are estimated to be from $3.1\%$ to $5.9\%$ in the decay constants and the SU(3)-breaking ratios.}
\\[29.0ex] \hline \\[-2.0ex]
\parbox[t]{2.8cm}{RBC/UKQCD~14A}
& \parbox[t]{1.0cm}{\cite{Aoki:2014nga} } & 2+1 
& \parbox[t]{1.3cm}{327, 289} 
& \parbox[t]{7.0cm}{Two or three light-quark masses per lattice spacing. 
NLO SU(2) HM$\chi$PT is used in the combined chiral-continuum extrapolation. 
The systematic errors in this extrapolation are estimated to be $3.54\%$ for $f_{B}$, $1.98\%$ for $f_{B_{s}}$, and $2.66\%$ for $f_{B_{s}}/f_{B}$.}
\\[15.0ex] \hline \\[-2.0ex]
\parbox[t]{2.8cm}{HPQCD 12} 
& \parbox[t]{1.0cm}{\cite{Na:2012kp} } & 2+1 
& \parbox[t]{1.3cm}{390, 390} 
& \parbox[t]{7.0cm}{Two or three pion masses at each lattice spacing. 
NLO (full QCD) HM$\chi$PT supplemented by NNLO analytic terms and generic $a^2$ and $a^4$ terms is used. 
The systematic error is estimated by varying the fit Ansatz, in particular for the NNLO analytic terms and the $a^{2n}$ terms. }
\\[15.0ex] \hline \\[-2.0ex]
\parbox[t]{2.8cm}{HPQCD 11A} 
& \parbox[t]{1.0cm}{\cite{McNeile:2011ng} } & 2+1 
& \parbox[t]{1.3cm}{570, 450, 390, 330, 330} 
& \parbox[t]{7.0cm}{One light sea-quark mass only at each lattice spacing. 
The sea-quark mass dependence is assumed to be negligible, based on the calculation of $f_{D_s}$ in Ref.~\cite{Davies:2010ip}, where the sea-quark extrapolation error is estimated as $0.34\%$. } 
\\[13.0ex] \hline \\[-2.0ex]
\parbox[t]{2.8cm}{FNAL/MILC 11} 
& \parbox[t]{1.0cm}{\cite{Bazavov:2011aa} } & 2+1 
& \parbox[t]{1.3cm}{570, 440, 320} 
& \parbox[t]{7.0cm}{Three to five sea-quark masses per lattice spacing, and  $9-12$ valence light quark masses per ensemble. 
NLO partially quenched HMrS$\chi$PT including $1/m$ terms and  supplemented by NNLO analytic and $\alpha_s^2a^2$ terms is used. 
The systematic error is estimated by varying the fit Ansatz, in particular the NNLO analytic terms and the chiral scale. }
\\[18.0ex] \hline \\[-2.0ex]
\parbox[t]{2.8cm}{RBC/UKQCD 10C}
& \parbox[t]{1.0cm}{\cite{Albertus:2010nm} } & 2+1 
& \parbox[t]{1.3cm}{430} 
& \parbox[t]{7.0cm}{Three light-quark masses at one lattice spacing. 
NLO SU(2) $\chi$PT is used. The systematic error is estimated from the difference between NLO $\chi$PT and linear fits as $\sim 7\%$.}
\\[10.0ex] \hline \\[-2.0ex]
\parbox[t]{2.8cm}{HPQCD 09} & \parbox[t]{1.0cm}{\cite{Gamiz:2009ku} }
& 2+1 & \parbox[t]{1.3cm}{440, 400} & 
\parbox[t]{7.0cm}{Four or two pion masses per lattice spacing. NLO (full QCD) 
HMrS$\chi$PT supplemented by NNLO analytic terms and  $\alpha_s a^2, a^4$ terms is used. 
The chiral fit error is estimated by varying the fit Ansatz, in particular, by adding or removing NNLO and discretization terms. }
\\[15.0ex]
\fi
\hline\hline
\end{tabular*}
\caption{Chiral extrapolation/minimum pion mass in determinations of the
  $B$- and $B_s$-meson decay constants for $\Nf=2+1$ simulations. 
  The different $M_{\pi,\rm min}$ entries correspond to the different lattice
 spacings.} 
\ifx\reducedapptables\undefined
\caption{Chiral extrapolation/minimum pion mass in determinations of the
  $B$- and $B_s$-meson decay constants for $\Nf=2+1$ simulations.  
  For actions with multiple species of pions, masses quoted are the RMS pion masses.    
  The different $M_{\pi,\rm min}$ entries correspond to the different lattice
 spacings.} 
 \fi
}
\end{table}

\ifx\reducedapptables\undefined
\begin{table}[!htb]
{\footnotesize
\begin{tabular*}{\textwidth}{l @{\extracolsep{\fill}} c c c   l}
\hline\hline \\[-1.0ex]
Collab. & \hspace{-0.4cm}Ref. & $\Nf$ & $M_{\pi,\rm min}\,[\mev]$  & Description 
\\[1.0ex] \hline \hline \\[-2.0ex]
\parbox[t]{2.8cm}{Balasubramamian 19} 
& \parbox[t]{1.0cm}{\cite{Balasubramanian:2019net}} & 2 
& \parbox[t]{1.3cm}{282, 194, 269}
& \parbox[t]{7.0cm}{Two pion masses for the coarsest and finest lattice spacings, and four pion masses for the intermediate lattice spacing. 
Generic fits to $m_\pi^2$ and $a^2$ in the combined continuum-chiral extrapolation, with systematic effects obtained by including a generic NLO chiral term $m_\pi^2 \log m_\pi^2$.
Detailed error budget not provided, but total systematic uncertainty estimated as $1.1$-$1.2\%$.}
\\[24.0ex]
\hline \\[-2.0ex]
\parbox[t]{2.8cm}{ALPHA 14  \\ ALPHA 13 \\ ALPHA 12A } & 
\parbox[t]{1.0cm}{\cite{Bernardoni:2014fva} \\ \cite{Bernardoni:2013oda} \\ \cite{Bernardoni:2012ti} } & 2 
& \parbox[t]{1.3cm}{280, 190, 270}
& \parbox[t]{7.0cm}{LO and NLO HM$\chi$PT supplemented by a term linear in $a^2$ are used. 
In ALPHA 13 and ALPHA 12A, the final result is an average between LO and NLO with half the difference used as estimate of the systematic error.
In ALPHA 14, the NLO fit is used as the central analysis procedure, and the LO results are used to estimate the systematic errors ($0.9\%$ for $f_{B_{s}}$, $1.1\%$ for $f_{B}$, and $1.6\%$ for $f_{B_{s}}/f_{B}$).}
\\[24.0ex]\hline \\[-2.0ex]
\parbox[t]{2.8cm}{ETM 13B, 13C \\ ETM 12B  \\ ETM 11A}  & 
\parbox[t]{1.0cm}{\cite{Carrasco:2013zta,Carrasco:2013iba} \\ \cite{Carrasco:2012de} \\ \cite{Dimopoulos:2011gx} } & 2 
& \parbox[t]{1.3cm}{410, 275, 300, 270} &
\parbox[t]{7.0cm}{$M_{\pi,{\rm min}}$ refers to the charged pions. 
Linear and NLO (full QCD) HM$\chi$PT supplemented by an $a^2$ term is used.
The chiral fit error is estimated from the difference between the NLO HM$\chi$PT and linear fits with half the difference used as estimate of the systematic error. 
For the static limit calculation in ETM 11A, $\Phi_s^{\rm stat}$ is extrapolated assuming a constant in light-quark mass.
The ratio $\Phi_s^{\rm stat}/\Phi_{\ell}^{\rm stat}$ is fit using three different chiral fit forms (NLO HM$\chi$PT, linear, and quadratic) to estimate the chiral fir error. }
\\[29.0ex] \hline \\[-2.0ex]
\parbox[t]{2.8cm}{ALPHA 11}
& \parbox[t]{1.0cm}{\cite{Blossier:2011dk} }  & 2 
& \parbox[t]{1.3cm}{331, 268, 267}
& \parbox[t]{7.0cm}{Linear and NLO (full QCD) HM$\chi$PT supplemented by a term linear in $a^2$ are used. 
The final result is an average between linear and NLO fits with half the difference used as estimate of the systematic error. }
\\[12.0ex]\hline \\[-2.0ex]
\parbox[t]{2.8cm}{ETM 09D} & \parbox[t]{1.0cm}{\cite{Blossier:2009hg} } & 2 
& \parbox[t]{1.3cm}{410, 275, 300}
& \parbox[t]{7.0cm}{$M_{\pi,{\rm min}}$ refers to the charged pions.  
Linear and NLO (full QCD) HM$\chi$PT is used. 
The final result given by the average of NLO HM$\chi$PT and linear {\it Ans\"atze} $\pm$ half the difference).}
\\[9.0ex] 
\hline\hline
\end{tabular*}
\caption{Chiral extrapolation/minimum pion mass in determinations of the
  $B$- and $B_s$-meson decay constants for $\Nf=2$ simulations.  
  For actions with multiple species of pions, masses quoted are the RMS pion masses (where available).    
  The different $M_{\pi,\rm min}$ entries correspond to the different lattice
 spacings.} 
}
\end{table}
\fi



\begin{table}[!htb]
{\footnotesize
\begin{tabular*}{\textwidth}{l @{\extracolsep{\fill}} c c c c l}
\hline\hline \\[-1.0ex]
Collab. & \hspace{-0.4cm}Ref. & $\Nf$ & $L$ [fm] & ${M_{\pi,\rm min}}L$ & Description 
\\[1.0ex] \hline \hline \\[-2.0ex]
\parbox[t]{2.8cm}{Frezzotti 24}
& \parbox[t]{1.0cm}{\cite{Frezzotti:2024kqk} } & 2+1+1 
& \parbox[t]{1.8cm}{4.36, 5.09,\\5.46, 5.46} 
& \parbox[t]{1.5cm}{3.9, 3.6,\\3.8, 3.9 } 
& \parbox[t]{4.9cm}{Finite-volume effects estimated to be subdominant to other sources of uncertainty, based in part on calculations in a larger ensemble in~\cite{Frezzotti:2023ygt}}
\\[2.0ex] 
\ifx\reducedapptables\undefined
\hline \\[-2.0ex]
\parbox[t]{2.8cm}{FNAL/MILC 17 }
& \parbox[t]{1.0cm}{\cite{Bazavov:2017lyh}} & 
2+1+1 & \parbox[t]{1.8cm}{2.4$\sim$4.8/\\3.0$\sim$5.8/\\3.0$\sim$5.6/\\2.9$\sim$5.4/\\2.9, 6.1/\\3.3}  
& \parbox[t]{1.5cm}{3.2, 3.9, 3.7, 3.7, 4.2, $\mbox{ }\mbox{ }\mbox{ }$4.8 } 
& \parbox[t]{4.9cm}{ Finite-size effects estimated, using alternative EFT fits, to be $0.1\%$ for $f_{B^{0}}$, $0.07\%$ for $f_{B_{s}}$, and $0.03\%$ for $f_{B_{s}}/f_{B^{0}}$.}
\\[15.0ex] \hline \\[-2.0ex]
\parbox[t]{2.8cm}{HPQCD 17A  \\ HPQCD 13}
& \parbox[t]{1.0cm}{\cite{Hughes:2017spc} \\ \cite{Dowdall:2013tga} } & 2+1+1 
& \parbox[t]{1.8cm}{2.4/3.5/4.7, 2.9/3.8/5.8, 2.8/5.6}  
& \parbox[t]{1.5cm}{3.30, 3.88, 3.66} 
& \parbox[t]{4.9cm}{ The analysis uses finite-volume $\chi$PT.  No explicit estimation of the systematic error arising from finite-size effects. }
\\[10.0ex] \hline \\[-2.0ex]
\parbox[t]{2.8cm}{ETM 16B  \\ ETM 13E}
& \parbox[t]{1.0cm}{\cite{Bussone:2016iua} \\ \cite{Carrasco:2013naa} } & 2+1+1 
& \parbox[t]{1.8cm}{2.84/2.13, 2.61/1.96, 2.97} 
& \parbox[t]{1.5cm}{3.53, 3.16, 3.19} 
& \parbox[t]{4.9cm}{ The data show no statistically discernible finite-volume effects, and the related systematic error is not explicitly estimated. }
\\[10.0ex] 
\fi
\hline\hline
\end{tabular*}
\caption{Finite-volume effects in determinations of the $B$- and $B_s$-meson
 decay constants for $\Nf=2+1+1$ simulations.  Each $L$-entry corresponds to a different
 lattice spacing.} 
}
\ifx\reducedapptables\undefined
\caption{Finite-volume effects in determinations of the $B$- and $B_s$-meson
 decay constants for $\Nf=2+1+1$ simulations.  Each $L$-entry corresponds to a different
 lattice spacing, with multiple spatial volumes at some lattice
 spacings.} 
\fi
\end{table}
%

\begin{table}[!htb]
{\footnotesize
\begin{tabular*}{\textwidth}{l @{\extracolsep{\fill}} c c c c l}
\hline\hline \\[-1.0ex]
Collab. & \hspace{-0.4cm}Ref. & $\Nf$ & $L$ [fm] & ${M_{\pi,\rm min}}L$ & Description 
\\[1.0ex] \hline \hline \\[-2.0ex]
\parbox[t]{2.8cm}{QCDSF/UKQCD\\/CSSM 22}
& \parbox[t]{1.0cm}{\cite{Hollitt:2022exk} } & 2+1 
& \parbox[t]{1.8cm}{2.62, 2.36/3.55,\\ 3.26/4.35, 2.83} 
& \parbox[t]{1.5cm}{3.86, 3.10/4.07,\\ 4.37/3.42, 4.03} 
& \parbox[t]{4.9cm}{Final result for $f_{B_{s}}/f_B$ includes fits to ensembles with $M_\pi L > 4$. 
No explicit estimate of FV effects.}
\\[9.0ex]\hline \\[-2.0ex]
\parbox[t]{2.8cm}{RBC/UKQCD 22}
& \parbox[t]{1.0cm}{\cite{Black:2022eph} } & 2+1 
& \parbox[t]{1.8cm}{2.65, 2.65,\\ 3.40, 2.00} 
& \parbox[t]{1.5cm}{4.57, 4.06,\\ 4.60, 3.77} 
& \parbox[t]{4.9cm}{No explicit estimate of FV effects.}
\\[3.0ex]
\ifx\reducedapptables\undefined
\hline \\[-2.0ex]
\parbox[t]{2.8cm}{RBC/UKQCD 18A}
& \parbox[t]{1.0cm}{\cite{Boyle:2018knm} } & 2+1 
& \parbox[t]{1.8cm}{2.65/5.47, 2.65/5.35, 3.49} 
& \parbox[t]{1.5cm}{3.86, 3.78, 4.11} 
& \parbox[t]{4.9cm}{Finite-volume effects are estimated to be $0.18\%$ in $f_{B_{s}}/f_B$.}
\\[7.0ex] \hline \\[-2.0ex]
\parbox[t]{2.8cm}{RBC/UKQCD~14 \\ RBC/UKQCD 13} & 
\parbox[t]{1.0cm}{ \cite{Christ:2014uea}\\  \cite{Witzel:2013sla}} & 2+1 
& \parbox[t]{1.8cm}{2.64, 2.75} 
& \parbox[t]{1.5cm}{4.38, 4.03}
& \parbox[t]{4.9cm}{In RBC/UKQCD~14, finite-volume effects are estimated to be negligible for $f_{B_{s}}$, $0.4\%$ for $f_{B^{0}}$,
$0.5\%$ for $f_{B^{+}}$ and the SU(3)-breaking ratios.}
\\[13.0ex] \hline \\[-2.0ex]
\parbox[t]{2.8cm}{RBC/UKQCD~14A} 
& \parbox[t]{1.0cm}{\cite{Aoki:2014nga}} & 2+1 
& \parbox[t]{1.8cm}{2.74, 2.76} 
& \parbox[t]{1.8cm}{4.53, 4.05} 
& \parbox[t]{4.9cm}{Finite-volume effects are estimated to be negligible for $f_{B_{s}}$, $0.82\%$ for $f_{B}$, and $1\%$ for $f_{B_{s}}/f_{B}$.}
\\[7.0ex] \hline \\[-2.0ex]
\parbox[t]{2.8cm}{HPQCD 12} 
& \parbox[t]{1.0cm}{\cite{Na:2012kp}} & 2+1 
& \parbox[t]{1.8cm}{2.4/2.9, 2.5/3.6} 
& \parbox[t]{1.5cm}{3.84, 4.21} 
& \parbox[t]{4.9cm}{FV error is taken from Ref.~\cite{Follana:2007uv} for HPQCD's $D$ meson analysis, where it was estimated using finite-volume $\chi$PT .}
\\[9.0ex] \hline \\[-2.0ex]
\parbox[t]{2.8cm}{HPQCD 11A} 
& \parbox[t]{1.0cm}{\cite{McNeile:2011ng}} & 2+1 
& \parbox[t]{1.8cm}{2.4, 2.4, 2.5, 2.9, 2.9} 
& \parbox[t]{1.5cm}{3.93, 4.48, 4.14, 4.49, 4.54} 
& \parbox[t]{4.9cm}{FV error is assumed to negligible.}
\\[6.0ex] \hline \\[-2.0ex]
\parbox[t]{2.8cm}{FNAL/MILC 11} 
& \parbox[t]{1.0cm}{\cite{Bazavov:2011aa}} & 2+1 
& \parbox[t]{1.8cm}{2.4, 2.4/2.9, 2.5/3.6} 
& \parbox[t]{1.5cm}{3.93, 3.78, 4.14}  
& \parbox[t]{4.9cm}{FV error is estimated using  finite-volume $\chi$PT.}
\\[4.0ex] \hline \\[-2.0ex]
\parbox[t]{2.8cm}{RBC/UKQCD 10C} 
& \parbox[t]{1.0cm}{\cite{Albertus:2010nm}} & 2+1 
& \parbox[t]{1.8cm}{1.8} & $3.9$ 
& \parbox[t]{4.9cm}{FV error estimated using  finite-volume $\chi$PT to be $1\%$ for SU(3)-breaking ratios. }
\\[7.0ex] \hline \\[-2.0ex]
\parbox[t]{2.8cm}{HPQCD 09} 
& \parbox[t]{1.0cm}{\cite{Gamiz:2009ku}} & 2+1 
& \parbox[t]{1.8cm}{2.4/2.9, 2.5}
& \parbox[t]{1.5cm}{3.78, 4.14}
& \parbox[t]{4.9cm}{FV error is assumed to negligible.}
\\[1.0ex]
\fi
\hline\hline
\end{tabular*}
\caption{Finite-volume effects in determinations of the $B$- and $B_s$-meson
 decay constants for $\Nf=2+1$ simulations.  Each $L$-entry corresponds to a different
 lattice spacing, with multiple spatial volumes at some lattice
 spacings.} 
}
\end{table}


\ifx\reducedapptables\undefined
\begin{table}[!htb]
{\footnotesize
\begin{tabular*}{\textwidth}{l @{\extracolsep{\fill}} c c c c l}
\hline\hline \\[-1.0ex]
Collab. & \hspace{-0.4cm}Ref. & $\Nf$ & $L$ [fm] & ${M_{\pi,\rm min}}L$ & Description 
\\[1.0ex] \hline \hline \\[-2.0ex]
\parbox[t]{3.0cm}{Balasubramamian 19} & \parbox[t]{1.0cm}{\cite{Balasubramanian:2019net}} 
& 2 & \parbox[t]{1.8cm}{2.4/3.6, 2.1/3.1/4.2, 2.3/3.1} & \parbox[t]{1.5cm}{4.1, 4.1, 4.0} 
& \parbox[t]{4.9cm}{No explicit estimate of FV errors, but expected to be much smaller than other uncertainties.}
\\[6.5ex] \hline \\[-2.0ex]
\parbox[t]{2.8cm}{ALPHA 14 \\ALPHA 13 \\ALPHA 12A \\ALPHA 11} & 
\parbox[t]{1.0cm}{\cite{Bernardoni:2014fva}\\  \cite{Bernardoni:2013oda}\\ \cite{Bernardoni:2012ti}\\ \cite{Blossier:2011dk}} & 2 &
\parbox[t]{1.8cm}{2.4/3.6, 2.1/3.1/4.2, 2.3/3.1} & \parbox[t]{1.5cm}{5.2, 4.1, 4.2}  & 
\parbox[t]{4.9cm}{No explicit estimate of FV errors, but expected to be much smaller than other uncertainties.}
\\[9.0ex] \hline \\[-2.0ex]
\parbox[t]{2.8cm}{ETM 13B, 13C \\ETM 12B \\ETM 11A} & 
\parbox[t]{1.0cm}{ \cite{Carrasco:2013zta,Carrasco:2013iba}\\ \cite{Carrasco:2012de}\\ \cite{Dimopoulos:2011gx}} & 2 &
	     \parbox[t]{1.8cm}{2.4, 2.0/2.7, 2.1, 1.7/2.6} &
	      \parbox[t]{1.5cm}{5, 3.7, 3.3, 3.5}  &
\parbox[t]{4.9cm}{FV errors are found to be negligible by comparing results at $m_\pi L = 3.3$ and $m_\pi L=4.3$ for $m_\pi \simeq 310$ MeV.}
\\[9.0ex]
\hline\hline
\end{tabular*}
\caption{Finite-volume effects in determinations of the $B$- and $B_s$-meson
 decay constants for $\Nf=2$ simulations.  Each $L$-entry corresponds to a different
 lattice spacing, with multiple spatial volumes at some lattice
 spacings.} 
}
\end{table}
\fi


\begin{table}[!htb]
{\footnotesize
\begin{tabular*}{\textwidth}{l @{\extracolsep{\fill}} c c c l l}
\hline\hline \\[-1.0ex]
Collab. & \hspace{-0.4cm}Ref. & $\Nf$ & $a$ [fm]  & Continuum extrapolation & Scale setting
\\[1.0ex] \hline \hline \\[-2.0ex]
\parbox[t]{2.8cm}{Frezzotti 24}
& \parbox[t]{1.0cm}{\cite{Frezzotti:2024kqk} } & 2+1+1 
& \parbox[t]{0.9cm}{0.091, 0.080, 0.068, 0.057} 
& \parbox[t]{3.8cm}{Continuum extrapolation linear in $a^2$. }
& \parbox[t]{3.6cm}{ Scale set by $w_0$, with details described in Ref.~\cite{Alexandrou:2021gqw}. }
\\[9.0ex] 
\ifx\reducedapptables\undefined
\hline \\[-2.0ex]
\parbox[t]{2.8cm}{FNAL/MILC 17 }
& \parbox[t]{1.0cm}{\cite{Bazavov:2017lyh} } & 2+1+1 
& \parbox[t]{0.9cm}{0.15, 0.12, 0.09, 0.06, 0.042, 0.032}
& \parbox[t]{3.8cm}{Continuum extrapolation linear in $a^2$. }
& \parbox[t]{3.6cm}{ Scale set by $f_{\pi}$, with details described in Ref.~\cite{Bazavov:2014wgs}. }
\\[15.0ex] \hline \\[-1.0ex]
\parbox[t]{2.8cm}{HPQCD 17A  \\ HPQCD 13 }
& \parbox[t]{1.0cm}{\cite{Hughes:2017spc} \\ \cite{Dowdall:2013tga} } & 2+1+1 
& \parbox[t]{0.9cm}{0.15, 0.12, 0.09}
& \parbox[t]{3.8cm}{Combined  continuum and chiral extrapolation. 
Continuum extrapolation errors estimated to be $0.7\%$ in HPQCD 13, and $1.1\%$ in HPQCD 17A. }
& \parbox[t]{3.6cm}{Scale set from $\Upsilon$(2S-1S) splitting, see Ref.~\cite{Dowdall:2011wh}. 
Scale uncertainty included in statistical error. }
\\[15.0ex] \hline \\[-1.0ex]
\parbox[t]{2.8cm}{ETM 16B  \\ ETM 13E }
& \parbox[t]{1.0cm}{\cite{Bussone:2016iua} \\ \cite{Carrasco:2013naa} } & 2+1+1 
& \parbox[t]{0.9cm}{0.089, 0.082, 0.062}
& \parbox[t]{3.8cm}{Combined  continuum and chiral extrapolation, linear in $a^2$.}
& \parbox[t]{3.8cm}{Scale set from $f_\pi$. Scale setting uncertainty included in combined statistical and systematic error. 
Discretization effects are also estimated by removing data at the coarsest lattice spacing in the chiral-continuum fit.}
\\[23.0ex] 
\fi
\hline\hline
\end{tabular*}
\caption{Continuum extrapolations/estimation of lattice artifacts in determinations of 
the $B$- and $B_s$-meson decay constants for $\Nf=2+1+1$ simulations.} 
}
\end{table}


\begin{table}[!htb]
{\footnotesize
\begin{tabular*}{\textwidth}{l @{\extracolsep{\fill}} c c c l l}
\hline\hline \\[-1.0ex]
Collab. & \hspace{-0.4cm}Ref. & $\Nf$ & $a$ [fm]  & Continuum extrapolation & Scale setting
\\[1.0ex] \hline \hline \\[-2.0ex]
\parbox[t]{2.8cm}{QCDSF/UKQCD\\/CSSM 22}
& \parbox[t]{1.0cm}{\cite{Hollitt:2022exk} } & 2+1 
& \parbox[t]{0.9cm}{0.082, 0.074, 0.068, 0.059} 
& \parbox[t]{3.8cm}{Combined continuum and chiral extrapolation omits the term linear in $a^2$. 
Systematic errors associated with discretization effects subdominant in $f_{B_{s}}/f_B$.}
& \parbox[t]{3.8cm}{Scale setting procedure and scale uncertainty are not discussed.}
\\[15.0ex] \hline \\[-2.0ex]
\parbox[t]{2.8cm}{RBC/UKQCD 22}
& \parbox[t]{1.0cm}{\cite{Black:2022eph} } & 2+1 
& \parbox[t]{0.9cm}{0.11, 0.083, 0.071, 0.063} 
& \parbox[t]{3.8cm}{Combined continuum and chiral extrapolation includes term linear in $a^2$.  
No estimate of systematic errors associated with discretization effects.}
& \parbox[t]{3.8cm}{Scale setting procedure and scale uncertainty are not discussed.}
\\[15.0ex]
\ifx\reducedapptables\undefined
\hline \\[-2.0ex]
\parbox[t]{2.8cm}{RBC/UKQCD 18A}
& \parbox[t]{0.5cm}{\cite{Boyle:2018knm} } & 2+1 
& \parbox[t]{0.9cm}{0.11, 0.084, 0.073} 
& \parbox[t]{3.8cm}{Combined continuum and chiral extrapolation, linear in $a^2$.
Systematic errors estimated to be from $0.3\%$ to $0.5\%$ in $f_{B_{s}}/f_B$.}
& \parbox[t]{3.8cm}{Scale set by the $\Omega$ baryon mass. 
No estimate for scale uncertainty, but expected to be negligible in $f_{B_s}/f_B$.}
\\[12.0ex] \hline \\[-2.0ex]
\parbox[t]{2.8cm}{RBC/UKQCD~14  \\ RBC/UKQCD 13A}
& \parbox[t]{0.5cm}{\cite{Christ:2014uea} \\ \cite{Witzel:2013sla}} & 2+1 
& \parbox[t]{0.9cm}{0.11, 0.086}
& \parbox[t]{3.8cm}{Combined  continuum and chiral extrapolation with linear in $a^2$ term. 
In RBC/UKQCD 14,  the systematic errors from this procedure are estimated to be from $3.1\%$ to $5.9\%$ in the decay constants and the
SU(3)-breaking ratios.
}
& \parbox[t]{3.8cm}{Scale set by the $\Omega$ baryon mass. 
In RBC/UKQCD 14, scale uncertainty estimated to be $1.5\%$ in the decay constants, and $0.1\%$ in the SU(3)-breaking ratios.}
\\[21.0ex] \hline \\[-2.0ex]
\parbox[t]{2.8cm}{RBC/UKQCD~14A}
& \parbox[t]{0.5cm}{\cite{Aoki:2014nga} } & 2+1 
& \parbox[t]{0.9cm}{0.11, 0.086}
& \parbox[t]{3.8cm}{Chiral-continuum extrapolation with linear in $a^2$ term is employed, with the systematic errors estimated to be from $1.98\%$ to $3.54\%$ in the decay constants and $f_{B_{s}}/f_{B}$. 
Discretization errors at $\cO(\alpha_{s} a)$ in the static-light system are estimated to be $1\%$ in the decay constants, and $0.2\%$ in $f_{B_{s}}/f_{B}$. 
}
& \parbox[t]{3.8cm}{Scale set by the $\Omega$ baryon mass.}
\\[29.0ex] \hline \\[-2.0ex]
\parbox[t]{2.8cm}{HPQCD 12} 
& \parbox[t]{0.5cm}{\cite{Na:2012kp} } & 2+1 
& \parbox[t]{0.9cm}{0.12, 0.09}
& \parbox[t]{3.8cm}{Combined  continuum and chiral extrapolation. 
Continuum extrapolation errors estimated to be $0.9\%$. 
}
& \parbox[t]{3.8cm}{Relative scale $r_1/a$ from the static-quark potential.  
Absolute scale $r_1$ from $f_\pi$, $f_K$, and $\Upsilon$(2S-1S) splitting. 
Scale uncertainty estimated to be $1.1\%$.}
\\[15.0ex] \hline \\[-2.0ex]
\parbox[t]{2.8cm}{HPQCD 11A} 
& \parbox[t]{0.5cm}{\cite{McNeile:2011ng}} & 2+1 
& \parbox[t]{0.9cm}{0.15, 0.12, 0.09, 0.06, 0.045}
& \parbox[t]{3.8cm}{ $am_Q \approx 0.2 - 0.85$. Combined  continuum and HQET fit. 
Continuum extrapolation error estimated by varying the fit ansatz and the included data points to be $0.63\%$. Discretization errors appear to decrease with increasing heavy-meson mass. 
 } 
& \parbox[t]{3.8cm}{Relative scale $r_1/a$ from the static-quark potential.  
Absolute scale $r_1$ from $f_\pi$, $f_K$, and $\Upsilon$(2S-1S) splitting. 
Scale uncertainty estimated to be $0.74\%$.}
\\[23.0ex] \hline \\[-2.0ex]
\parbox[t]{2.8cm}{FNAL/MILC 11} 
& \parbox[t]{0.5cm}{\cite{Bazavov:2011aa} } & 2+1 
& \parbox[t]{0.9cm}{0.15, 0.12, 0.09}
& \parbox[t]{3.8cm}{Combined continuum and chiral extrapolation. Continuum extrapolation errors estimated to be 1.3\%.
}
& \parbox[t]{3.8cm}{Relative scale $r_1/a$ from the static-quark potential.  
Absolute scale $r_1$ from $f_\pi$, $f_K$, and $\Upsilon$(2S-1S) splitting.  
Scale uncertainty estimated to be 1~MeV.}
\\[15.0ex] \hline \\[-2.0ex]
\parbox[t]{2.8cm}{RBC/UKQCD 10C} 
& \parbox[t]{0.5cm}{\cite{Albertus:2010nm}} & 2+1 
& \parbox[t]{0.9cm}{0.11}
& \parbox[t]{3.8cm}{One lattice spacing with discretization errors estimated by power counting as $3\%$.}
& \parbox[t]{3.8cm}{Scale set by the $\Omega$ baryon mass. 
Combined scale and mass tuning uncertainties on $f_{B_s}/f_B$ estimated as 1\%}
\\[9.0ex] \hline \\[-2.0ex]
\parbox[t]{2.8cm}{HPQCD 09} 
& \parbox[t]{0.5cm}{\cite{Gamiz:2009ku} } & 2+1 
& \parbox[t]{0.9cm}{0.12, 0.09}
& \parbox[t]{3.8cm}{Combined continuum and chiral extrapolation. Continuum extrapolation errors estimated to be 3\%. 
}
& \parbox[t]{3.8cm}{Relative scale $r_1/a$ from the static-quark potential.  
Absolute scale $r_1$  from the $\Upsilon$(2S-1S) splitting. 
Scale uncertainty estimated to be $2.3\%$. }
\\[12.0ex] 
\fi
\hline\hline
\end{tabular*}
\caption{Continuum extrapolations/estimation of lattice artifacts in determinations of 
the $B$ and $B_s$ meson decay constants for $\Nf=2+1$ simulations.} 
}
\end{table}

\ifx\reducedapptables\undefined

\begin{table}[!htb]
{\footnotesize
\begin{tabular*}{\textwidth}{l @{\extracolsep{\fill}} l c c l l}
\hline\hline \\[-1.0ex]
Collab. & Ref. & $\Nf$ & $a$ [fm]  & Continuum extrapolation & Scale setting
\\[1.0ex] \hline \hline \\[-2.0ex]
\parbox[t]{3.0cm}{Balasubramamian 19} 
& \parbox[t]{0.5cm}{\cite{Balasubramanian:2019net}} & 2 
& \parbox[t]{0.9cm}{0.075, 0.065, 0.048}
& \parbox[t]{4.3cm}{Combined continuum and chiral extrapolation with linear in $a^2$ term. 
Detailed error budget not provided, but total systematic uncertainty estimated as $1.1$-$1.2\%$.}
& \parbox[t]{4.3cm}{Scale set from $f_\pi$. Scale setting uncertainty estimated as $0.9\%$.}
\\[24.0ex]\hline \\[-2.0ex]
\parbox[t]{2.0cm}{ALPHA 14  \\ ALPHA 13 \\ ALPHA 12A \\ ALPHA 11}
& \parbox[t]{0.5cm}{\cite{Bernardoni:2014fva} \\ \cite{Bernardoni:2013oda} \\ \cite{Bernardoni:2012ti} \\ \cite{Blossier:2011dk}} & 2 
& \parbox[t]{0.9cm}{0.075, 0.065, 0.048} 
& \parbox[t]{4.3cm}{Combined continuum and chiral extrapolation with linear in $a^2$ term. 
Continuum extrapolation errors estimated to be 5~MeV in ALPHA 11. 
The continuum extrapolation with a term linear in $a$ also investigated
in ALPHA 14, and within the statistical error no discernable difference was observed.
}
& \parbox[t]{4.3cm}{Relative scale set from $r_0$. 
Absolute scale set from $f_K$. 
Scale setting uncertainty included in combined statistical and extrapolation error.}
\\[29.0ex] \hline \\[-2.0ex]
\parbox[t]{2.4cm}{ETM 13B, 13C \\ ETM 12B  \\ ETM 11A} 
& \parbox[t]{0.8cm}{\cite{Carrasco:2013zta,Carrasco:2013iba} \\ \cite{Carrasco:2012de} \\ \cite{Dimopoulos:2011gx} }& 2 
& \parbox[t]{0.9cm}{0.098, 0.085, 0.067, 0.054} 
& \parbox[t]{4.3cm}{Combined continuum and chiral extrapolation, with a term linear in $a^2$.
ETM 12 and 13 include heavier masses than ETM 11A. 
Discretization error included in combined statistical and systematic error, estimated by dropping the data at the coarsest lattice spacing
as $\sim 0.5 - 1\%$. 
}
& \parbox[t]{4.3cm}{Scale set from $f_\pi$. 
Scale setting uncertainty included in combined statistical and systematic error. }
\\[26.0ex] \hline \\[-2.0ex]
\parbox[t]{2.0cm}{ETM 09D} 
& \parbox[t]{0.5cm}{\cite{Blossier:2009hg} } & 2 
& \parbox[t]{0.9cm}{0.098, 0.085, 0.067} 
& \parbox[t]{4.3cm}{Combined continuum and chiral extrapolation with a term linear in $a^2$.}
& \parbox[t]{4.3cm}{Scale set from $f_\pi$. 
Scale setting uncertainty included in combined statistical and systematic error.}
\\[9.0ex] 
\hline\hline
\end{tabular*}
\caption{Continuum extrapolations/estimation of lattice artifacts in determinations of 
the $B$- and $B_s$- meson decay constants for $\Nf=2$ simulations.} 
}
\end{table}
\fi

%
\begin{table}[!htb]
{\footnotesize
\begin{tabular*}{\textwidth}{l @{\extracolsep{\fill}} c c c l}
\hline\hline \\[-1.0ex]
Collab.  &  Ref. & $\Nf$ & Ren. & Description  
\\[1.0ex] \hline \hline \\[-2.0ex]
\parbox[t]{2.8cm}{Frezzotti 24}
& \parbox[t]{1.0cm}{\cite{Frezzotti:2024kqk} } & 2+1+1 
& -- 
& \parbox[t]{7.5cm}{ Nonperturbative operator renormalization provided by ETMC by private communication and unpublished at the time of this review.}
\\[6.0ex] 
\ifx\reducedapptables\undefined
\hline \\[-2.0ex]
\parbox[t]{2.8cm}{FNAL/MILC 17 }
& \parbox[t]{0.5cm}{\cite{Bazavov:2017lyh} } & 2+1+1 
& -- 
& \parbox[t]{7.5cm}{The current used for this work is absolutely normalized. }
\\[4.0ex] \hline \\[-2.0ex]
\parbox[t]{2.8cm}{HPQCD 17A  \\ HPQCD 13 }
& \parbox[t]{0.5cm}{\cite{Hughes:2017spc} \\ \cite{Dowdall:2013tga} }
& 2+1+1 & PT1$\ell$
& \parbox[t]{7.5cm}{ The NRQD effective current is matched through $O(1/m$) and renormalized using 1-loop PT. 
Included are all terms though $O(\alpha_s)$,  $O(\alpha_s \, a)$, $O(\Lambda_{\rm QCD}/M)$, $O(\alpha_s/a M)$ ,  $O(\alpha_s \, \Lambda_{\rm QCD}/M)$.
The dominant error is due to unknown $O(\alpha_s^2)$ contributions to the current renormalization. 
The error is estimated as  $\sim1.4$\% in HPQCD 13, and as $\sim2$\% in HPQCD 17A. }
\\[18.0ex]  \hline \\[-2.0ex]
\parbox[t]{2.8cm}{ETM 16B  \\ ETM 13E }
& \parbox[t]{0.5cm}{\cite{Bussone:2016iua} \\ \cite{Carrasco:2013naa} }
& 2+1+1 
& --, PT 
& \parbox[t]{7.5cm}{The current used for the relativistic decay constants is absolutely normalized. 
 The ratio is constructed from the relativistic decay constant data and the heavy-quark pole masses. 
 Ratios of pole-to-$\msbar$ mass conversion factors are included at N$^{3}$LO in continuum perturbation theory, and the matching between the continuum QCD and HQET currents is performed at N$^{2}$LO.}
\\[20.0ex]  \\[-2.0ex]
\fi
\hline\hline
\end{tabular*}
\caption{Description of the renormalization/matching procedure adopted
  in the determinations of the $B$- and $B_s$-meson decay constants for
  $\Nf  = 2 + 1+1$ simulations.} 
}
\end{table}



\begin{table}[!htb]
{\footnotesize
\begin{tabular*}{\textwidth}{l @{\extracolsep{\fill}} c c c l}
\hline\hline \\[-1.0ex]
\rb{Collab.}  &  \hspace{-0.4cm}\rb{Ref.} & \rb{$\Nf$} & \rb{Ren.} & \rb{Description}  
\\[1.0ex] \hline \hline \\[-2.0ex]
\parbox[t]{2.8cm}{QCDSF/UKQCD\\/CSSM 22}
& \parbox[t]{1.0cm}{\cite{Hollitt:2022exk} } & 2+1 
& \parbox[t]{1.3cm}{mNPR} 
& \parbox[t]{7.5cm}{Operator renormalization is calculated partially nonperturbatively as $Z_{B_q} = \rho_A^{bq}\sqrt{Z_V^{bb}Z_V^{qq}}$, with perturbative contribution neglected, $\rho_A^{bq}=1$.}
\\[7.0ex] \hline \\[-2.0ex]
\parbox[t]{2.8cm}{RBC/UKQCD 22}
& \parbox[t]{1.0cm}{\cite{Black:2022eph} } & 2+1 
& \parbox[t]{1.3cm}{mNPR} 
& \parbox[t]{7.5cm}{Operator renormalization is calculated partially nonperturbatively as $Z_{B_q} = \rho_A^{bq}\sqrt{Z_V^{bb}Z_V^{qq}}$.}
\\[4.0ex] 
\ifx\reducedapptables\undefined
\parbox[t]{2.8cm}{RBC/UKQCD 18A}
& \parbox[t]{1.3cm}{\cite{Boyle:2018knm} } & 2+1 
& \parbox[t]{1.3cm}{$-$} 
& \parbox[t]{7.5cm}{Mass-independent operator renormalization cancels in $f_{B_{s}}/f_B$.}
\\[3.0ex] \hline \\[-2.0ex]
\parbox[t]{2.8cm}{RBC/UKQCD~14 \\RBC/UKQCD 13}
& \parbox[t]{1.3cm}{ \cite{Christ:2014uea} \\ \cite{Witzel:2013sla}} & 2+1 
& mNPR 
& \parbox[t]{7.5cm}{In RBC/UKQCD~14, the error is dominated by the perturbative aspect, and is estimated to be $1.7\%$ for the decay constants by taking the full size of the 1-loop correction for the fine lattice. }
\\[9.0ex] \hline \\[-2.0ex]
\parbox[t]{2.8cm}{RBC/UKQCD~14A}
& \parbox[t]{1.3cm}{ \cite{Aoki:2014nga}} & 2+1 
& PT1$\ell$ 
& \parbox[t]{7.5cm}{A two-step matching procedure is employed, first from QCD to HQET in the continuum at $m_{b}$, then to HQET on the lattice at $a^{-1}$ with $O(pa)$ and $O(m_{q} a)$ errors included. 
Both matching steps are accurate to 1-loop, and the running between $m_{b}$ and $a^{-1}$ is performed at 2-loop accordingly. 
The error is estimated using a power-counting argument to be $6\%$ for the decay constants.}
\\[20.0ex] \hline \\[-2.0ex]
\parbox[t]{2.8cm}{HPQCD 12, 09} 
& \parbox[t]{1.3cm}{\cite{Na:2012kp,Gamiz:2009ku}} & 2+1 
& PT1$\ell$
& \parbox[t]{7.5cm}{The NRQD effective current is matched through $O(1/m$) and renormalized using 1-loop PT.  
Included are all terms though $O(\alpha_s)$,  $O(\alpha_s \, a)$, $O(\Lambda_{\rm QCD}/M)$, $O(\alpha_s/a M)$, $O(\alpha_s \, \Lambda_{\rm QCD}/M)$. 
The dominant error is due unknown $O(\alpha_s^2)$ contributions to the current renormalization. 
The authors take the perturbative error as $\sim 2  \rho_0 \, \alpha_s^2$, where $\rho_0$ is the coefficient of the 1-loop correction to the leading term, which yields an error of $\sim4$\%. }
\\[23.0ex] \hline \\[-2.0ex]
\parbox[t]{2.8cm}{HPQCD 11A} 
& \parbox[t]{1.2cm}{\cite{McNeile:2011ng}} & 2+1 
& -- 
& \parbox[t]{7.5cm}{This work uses PCAC together with an absolutely normalized current.}
\\[4.0ex] \hline \\[-2.0ex]
\parbox[t]{2.8cm}{FNAL/MILC 11} 
& \parbox[t]{1.3cm}{\cite{Bazavov:2011aa}} & 2+1 
& mNPR 
& \parbox[t]{7.5cm}{The authors' estimate of the perturbative errors is comparable in size to the actual 1-loop corrections. }
\\[4.0ex] \hline \\[-2.0ex]
\parbox[t]{2.8cm}{RBC/UKQCD 10C} & \parbox[t]{1.2cm}{\cite{Albertus:2010nm}} & 2+1 
& PT1$\ell$ 
& \parbox[t]{7.5cm}{The static-light current is matched through $O(\alpha_s a, \alpha_s)$ and renormalized using 1-loop tadpole improved PT. 
For massless light quarks, the renormalization factors cancel in the ratio of decay constants.}
\\[12.0ex] 
\fi
\hline\hline
\end{tabular*}
\caption{Description of the renormalization/matching procedure adopted in the determinations of the $B$- and $B_s$-meson decay constants for
  $\Nf  = 2+1$ simulations.} 
}
\end{table}

\ifx\reducedapptables\undefined
\begin{table}[!htb]
{\footnotesize
\begin{tabular*}{\textwidth}{l @{\extracolsep{\fill}} c c c l}
\hline\hline \\[-1.0ex]
\rb{Collab.}  &  \hspace{-0.4cm}\rb{Ref.} & \rb{$\Nf$} & \rb{Ren.} & \rb{Description}  
\\[1.0ex] \hline \hline \\[-2.0ex]
\parbox[t]{3.0cm}{Balasubramamian 19} 
& \parbox[t]{1.0cm}{\cite{Balasubramanian:2019net}} & 2 
& PT2$\ell$
& \parbox[t]{7.50cm}{Perturbative coefficient relating HQET and QCD currents included at NNLO in continuum perturbation theory. 
Systematic uncertainties estimated using fits including only NLO coefficients.}
\\[12.0ex] \hline \\[-2.0ex]
\parbox[t]{2.8cm}{ALPHA 14 \\ALPHA 13 \\ALPHA 12A \\ALPHA 11} & 
\parbox[t]{1.3cm}{ \cite{Bernardoni:2014fva}\\ \cite{Bernardoni:2013oda}\\ \cite{Bernardoni:2012ti}\\ \cite{Blossier:2011dk}} & 2 
& NPR 
& \parbox[t]{7.5cm}{The authors use the Schr\"odinger functional for the NP matching. }
\\[9.0ex] \hline \\[-2.0ex]
\parbox[t]{2.8cm}{ETM 13B, 13C \\ETM 12B \\ETM 11A} & 
\parbox[t]{1.3cm}{ \cite{Carrasco:2013zta,Carrasco:2013iba}\\ \cite{Carrasco:2012de}\\ \cite{Dimopoulos:2011gx}} & 2 
& --, PT1$\ell$ 
& \parbox[t]{7.5cm}{The current used for the relativistic decay constants is absolutely normalized. 
{\bf Interpolation method:} The static limit current renormalization is calculated in 1-loop mean field improved perturbation theory, there half the correction is used to estimate the  error.  
{\bf Ratio method:} The ratio is constructed from the relativistic decay constant data and the heavy-quark pole masses. 
Ratios of pole-to-$\msbar$ mass  conversion factors are included at NLO in continuum perturbation theory.}
\\[25.0ex]
\hline\hline
\end{tabular*}
\caption{Description of the renormalization/matching procedure adopted
  in the determinations of the $B$- and $B_s$-meson decay constants for
  $\Nf  = 2$ simulations.} 
}
\end{table}
\fi


%
\begin{table}[!htb]
{\footnotesize
\begin{tabular*}{\textwidth}{l @{\extracolsep{\fill}} c c c l}
\hline\hline \\[-2.0ex]
Collab. & \hspace{-0.4cm}Ref. & $\Nf$ &  Action  &  Description
\\[1.0ex] \hline \hline \\[-1.0ex]
\parbox[t]{2.8cm}{Frezzotti 24}
& \parbox[t]{1.0cm}{\cite{Frezzotti:2024kqk} } & 2+1+1 
& tmWil
& \parbox[t]{6.5cm}{ Heavy-strange meson extrapolated to physical $B_s$ mass using HQET scaling linear in $B/m_{H_s}$, with contributions from QCD-HQET current matching and HQET axial current anomalous dimension at 1-loop.}
\\[12.0ex] 
\ifx\reducedapptables\undefined
\hline  \\[-2.0ex]
\parbox[t]{2.8cm}{FNAL/MILC 17 }
& \parbox[t]{1.0cm}{\cite{Bazavov:2017lyh} } & 2+1+1 
& HISQ 
& \parbox[t]{7.5cm}{ The discretization effects are estimated by repeating the fit by either adding the coarsest ($a \approx 0.15$ fm)  ensembles or omitting
  the finest ($a \approx 0.03$ fm) ensembles, as well as changing the fit Ansatz in the combined EFT and continuum-limit analysis.}
\\[12.0ex] \hline  \\[-2.0ex]
\parbox[t]{2.8cm}{HPQCD 17A  \\ HPQCD 13 }
& \parbox[t]{1.0cm}{\cite{Hughes:2017spc} \\ \cite{Dowdall:2013tga} } & 2+1+1 
& NRQCD 
& \parbox[t]{7.5cm}{HQ truncation effects estimated as in HPQCD 09  to be $1.0\%$.}
\\[4.0ex] \hline  \\[-2.0ex]
\parbox[t]{2.8cm}{ETM 16B  \\ ETM 13E }
& \parbox[t]{1.0cm}{\cite{Bussone:2016iua} \\ \cite{Carrasco:2013naa} } & 2+1+1 
& tmWil 
& \parbox[t]{7.5cm}{The estimate of the discretization effects is described in the continuum table. 
Ratios of pole-to-$\msbar$ mass conversion factors are included at N$^{3}$LO in continuum perturbation theory, and the matching between the continuum QCD and HQET currents is performed at  N$^{2}$LO. 
The systematic error in this procedure is very small compared to other systematic effects.  }
\\[18.0ex] 
\fi
\hline\hline
\end{tabular*}
\caption{Heavy-quark treatment in determinations of the $B$- and $B_s$-meson decay constants for $\Nf  = 2+ 1 +1$ simulations.} 
}
\end{table}


\begin{table}[!htb]
{\footnotesize
\begin{tabular*}{\textwidth}{l @{\extracolsep{\fill}} c c c l}
\hline\hline \\[-2.0ex]
Collab. & \hspace{-0.4cm}Ref. & $\Nf$ &  Action  &  Description
\\[1.0ex] \hline \hline \\[-1.0ex]
\parbox[t]{2.8cm}{QCDSF/UKQCD\\/CSSM 22}
& \parbox[t]{1.0cm}{\cite{Hollitt:2022exk} } & 2+1 
& RHQ 
& \parbox[t]{7.5cm}{HQ tuning effects are estimated to be $0.06\%$ in$f_{B_{s}}/f_B$. 
HQ discretization effects not explicitly estimated, although the continuum-limit fits do not indicate a strong $a^2$ dependence.}
\\[9.0ex] \hline \\[-2.0ex]
\parbox[t]{2.8cm}{RBC/UKQCD 22}
& \parbox[t]{1.0cm}{\cite{Black:2022eph} } & 2+1 
& RHQ 
& \parbox[t]{7.5cm}{HQ tuning and discretization effects not explicitly estimated. HQ tuning of new finest ensemble ongoing.}
\\[4.0ex] 
\ifx\reducedapptables\undefined
\hline \\[-2.0ex]
\parbox[t]{2.8cm}{RBC/UKQCD 18A}
& \parbox[t]{1.0cm}{\cite{Boyle:2018knm} } & 2+1 
& DWF 
& \parbox[t]{7.5cm}{Linear extrapolation in inverse heavy-quark mass $1/M_H$, with systematic errors estimated, by excluding heaviest and lightest valence-quark masses, to be $0.47\%$ in $f_{B_{s}}/f_B$. 
HQ discretization effects are  estimated to be $0.01\%$ in$f_{B_{s}}/f_B$.}
\\[12.0ex] \hline \\[-2.0ex]
\parbox[t]{2.8cm}{RBC/UKQCD~14 \\RBC/UKQCD 13A} 
& \parbox[t]{1.2cm}{ \cite{Christ:2014uea} \\ \cite{Witzel:2013sla}} & 2+1 
& RHQ 
& \parbox[t]{7.5cm}{In RBC/UKQCD 14, the heavy-quark discretization errors are estimated to be $1.7\%$ in the decay constants, and $0.3\%$ in the SU(3)-breaking ratios.}
\\[6.0ex] \hline \\[-2.0ex]
\parbox[t]{2.8cm}{RBC/UKQCD~14A} 
& \parbox[t]{1.2cm}{\cite{Aoki:2014nga}} & 2+1 
& Static 
& \parbox[t]{7.5cm}{Static-limit computation, with $\cO(\Lambda_{{\rm QCD}}/m_{h})$ errors estimated to be $10\%$ for the decay constants, and $2.2\%$ for $f_{B_{s}}/f_{B}$.}
\\[6.0ex] \hline \\[-2.0ex]
\parbox[t]{2.8cm}{HPQCD 12} 
& \parbox[t]{1.2cm}{\cite{Na:2012kp}} & 2+1 
& NRQCD 
& \parbox[t]{7.5cm}{HQ truncation effects estimated as in HPQCD 09  to be $1.0\%$}
\\[3.0ex] \hline \\[-2.0ex]
\parbox[t]{2.8cm}{HPQCD 11A} 
& \parbox[t]{1.2cm}{\cite{McNeile:2011ng}} & 2+1 
& HISQ
& \parbox[t]{7.5cm}{ The analysis uses a combined continuum and $1/m$ extrapolation. }
\\[3.0ex] \hline \\[-2.0ex]
\parbox[t]{2.8cm}{FNAL/MILC 11} 
& \parbox[t]{1.2cm}{\cite{Bazavov:2011aa}} & 2+1 
& RHQ
& \parbox[t]{7.5cm}{HQ discretization effects are included in the combined chiral and continuum fits, and are estimated by varying the fit Ansatz and excluding the data at the coarsest lattice spacing to be $\sim 2\%$, consistent with simple power counting estimates but larger than the residual discretization errors observed
in the data.}
\\[15.0ex] \hline \\[-2.0ex]
\parbox[t]{2.8cm}{RBC/UKQCD 10C} 
& \parbox[t]{1.2cm}{\cite{Albertus:2010nm}} & 2+1 
& Static 
& \parbox[t]{7.5cm}{Truncation effects of $\cO(1/m_h)$ on the SU(3)-breaking ratios are estimated by power counting to be 2\%.}
\\[3.0ex] \hline \\[-2.0ex]
\parbox[t]{2.8cm}{HPQCD 09} 
& \parbox[t]{1.2cm}{\cite{Gamiz:2009ku}} & 2+1 
& NRQCD
& \parbox[t]{7.5cm}{The leading HQ truncation effects are of $\cO(\alpha_s \Lambda_{\rm QCD}/m_h)$ due to the tree-level coefficient of the ${\bm \sigma}\cdot{\rm\bf B}$ term. 
The error is estimated by calculating the $B^*-B$ hyperfine splitting and comparing with experiment as $1\%$.}
\\[12.0ex] 
\fi
\hline\hline
\end{tabular*}
\caption{Heavy-quark treatment in $\Nf=2+1$ determinations of the $B$-and $B_s$-meson decay constants.} 
}
\end{table}

\ifx\reducedapptables\undefined
\begin{table}[!htb]
{\footnotesize
\begin{tabular*}{\textwidth}{l @{\extracolsep{\fill}} c c c l}
\hline\hline \\[-2.0ex]
Collab. & \hspace{-0.4cm}Ref. & $\Nf$ &  Action  &  Description
\\[1.0ex] \hline \hline \\[-1.0ex]
\parbox[t]{3.0cm}{Balasubramamian 19} 
& \parbox[t]{1.0cm}{\cite{Balasubramanian:2019net}} & 2 
& Wilson
& \parbox[t]{7.0cm}{Expansion in inverse of pseudoscalar heavy-strange meson mass, $1/M_{H_{s}}$. 
HQ discretization effects included in generic fit linear in $a^2(aM_{H_{s}})^2$. 
Detailed error budget not provided, but systematic uncertainty estimated as $1.1$-$2.3\%$ in $f_{B_s}$.}
\\[12.0ex] \hline \\[-2.0ex]
\parbox[t]{2.8cm}{ALPHA 14 \\ALPHA 13 \\ALPHA 12A \\ALPHA 11} & 
\parbox[t]{1.2cm}{ \cite{Bernardoni:2014fva}\\ \cite{Bernardoni:2013oda}\\ \cite{Bernardoni:2012ti}\\ \cite{Blossier:2011dk}} & 2 
& HQET 
& \parbox[t]{7.5cm}{NP improved through $\cO(1/m_h)$. \
Truncation errors of $\cO\left [ (\Lambda_{\rm QCD}/m_h)^2 \right ]$ are not included. \\}
\\[9.0ex] \hline \\[-2.0ex]
\parbox[t]{2.8cm}{ETM 13B, 13C \\ETM 12B \\ETM 11A} & 
\parbox[t]{1.2cm}{ \cite{Carrasco:2013zta,Carrasco:2013iba}\\ \cite{Carrasco:2012de}\\ \cite{Dimopoulos:2011gx}} & 2 
& tmWil 
& \parbox[t]{7.5cm}{The estimate of the discretization effects is described in the continuum table. 
In both methods the relativistic data are matched to HQET using NLO continuum PT in an intermediate step, and converted back to QCD at the end. 
The error due to HQET matching (estimated by replacing the NLO expressions with LO) is a very small contribution to the systematic error due to the 
heavy-quark mass dependence. 
The variation observed from adding heavier masses to their data and/or including $1/m_h^3$ terms is $0.4-1.3\%$.}
\\[29.0ex]
\hline\hline
\end{tabular*}
\caption{Heavy-quark treatment in $\Nf=2$ determinations of the $B$-and $B_s$-meson decay constants.} 
}
\end{table}
\fi

%% file: HQ/AppendixTables/BBapp.tex
\subsubsection{$B_{(s)}$-meson mixing matrix elements}
\label{app:BMix_Notes}
\begin{table}[!ht]
{\footnotesize
\begin{tabular*}{\textwidth}{l @{\extracolsep{\fill}} c c c l l}
\hline\hline \\[-1.0ex]
Collab. & Ref. & $\Nf$ & $a$ [fm] & Continuum extrapolation & Scale setting 
\\[1.0ex] \hline \hline \\[-1.0ex]
HPQCD 19A & \cite{Dowdall:2019bea} & 2+1+1 & 
\parbox[t]{0.9cm}{0.15, 0.12, 0.09} & 
\parbox[t]{4.3cm}{Discretization errors start from $\alpha_sa^2$ and are included in the systematic error.
		 It is estimated as $1.8$\% for individual bag parameters.
		 Residual $\alpha_sa^2$ and $a^4$ errors from wrong-spin contributions are subtracted by including them in the chiral  fit.} &
\parbox[t]{4.2cm}{Scale setting done using $\Upsilon$ and $\Upsilon'$ mass splitting \cite{Dowdall:2011wh}.}
\\[24.0ex] 
\hline\hline
\end{tabular*}
\caption{Continuum extrapolations/estimation of lattice artifacts 
 in determinations of the neutral $B$-meson mixing matrix elements for
 $\Nf=2+1+1$ simulations.}
\label{tab:BBcontinum4}
}
\end{table}

\begin{table}[!ht]
{\footnotesize
\begin{tabular*}{\textwidth}{l @{\extracolsep{\fill}} c c c l l}
\hline\hline \\[-1.0ex]
Collab. & Ref. & $\Nf$ & $a$ [fm] & Continuum extrapolation & Scale setting 
\\[1.0ex] \hline \hline \\[-1.0ex]
RBC/UKQCD 18A & \cite{Boyle:2018knm} & 2+1 & 
\parbox[t]{0.9cm}{0.11, 0.08, 0.07} & 
\parbox[t]{4.3cm}{Combined continuum ($a^2$) and heavy quark ($1/m_H$)
		 extrapolation with the LO pion mass dependence ($m_\pi^2$)
		 in the global fit.} &
\parbox[t]{4.2cm}{Lattice scale and target quark masses are set using
		  $\Omega$, $K$ and $\pi$ masses \cite{Aoki:2010dy,Blum:2014tka,Boyle:2017jwu}.}
\\[12.0ex] 
\ifx\reducedapptables\undefined
\hline \\[-2.0ex]
FNAL/MILC 16& \cite{Bazavov:2016nty} & 2+1 & 
\parbox[t]{0.9cm}{0.12, 0.09, 0.06, 0.045} & 
\parbox[t]{4.3cm}{Combined continuum and chiral extrapolation with
		 NLO HMrS$\chi$PT and NNLO analytic terms as well as the
		 terms for the heavy quark
		 discretization errors up to $a^3$, heavy quark mass
		 mismatch, and renormalization error of $\alpha_s^2$.} &
\parbox[t]{4.2cm}{Relative scale $r_1/a$ is set via static-quark
		  potential. Absolute scale is set as $r_1=0.3117(22)$ fm.
		  See the description of FNAL/MILC 12 below.
		  The scale uncertainty on $\xi$, e.g., is estimated as 0.6\%.} 
\\[20.0ex] 
\hline \\[-2.0ex]
RBC/UKQCD 14A& \cite{Aoki:2014nga} & 2+1 & \parbox[t]{0.9cm}{0.11, 0.086} & 
\parbox[t]{4.3cm}{Combined continuum and chiral extrapolation with
		 SU(2) NLO HM$\chi$PT and linear in quark mass
		 both with $\cO(a^2)$ terms.
		 The combined continuum and chiral extrapolation
		 uncertainty is estimated as 2.55, 2.13 and 3.08\% for
		 $f_B\sqrt{B_B}$, $f_{B_s}\sqrt{B_{B_s}}$ and $\xi$
		 respectively.} &
\parbox[t]{4.2cm}{Scale is set using the $\Omega^-$ mass as input 
		 \cite{Aoki:2010dy}. The scale uncertainty is estimated
		 as 1.84, 1.86 and 0.05\% for $f_B\sqrt{B_B}$, 
		 $f_{B_s}\sqrt{B_{B_s}}$ and $\xi$ respectively.}
\\[26.0ex] \hline \\[-2.0ex]
FNAL/MILC 12 & \cite{Bazavov:2012zs} & 2+1 & \parbox[t]{0.9cm}{0.12, 0.09} & 
\parbox[t]{4.3cm}{Combined continuum and chiral extrapolation with
		 NLO rHMS$\chi$PT, NNLO analytic and generic 
		 $\cO(\alpha_s^2 a^2, a^4)$ terms. 
		 Combined statistical, chiral and light-quark discretization error 
		 is estimated, by examining the variation with different fit Ans\"atze
		 to be 3.7\% on $\xi$.} &
\parbox[t]{4.2cm}{Relative scale $r_1/a$ is set via static-quark
		  potential. Absolute scale $r_1=0.3117(22)$ fm is
		  determined \cite{Bazavov:2011aa} through averaging the
		  $f_\pi$ input and the estimate of HPQCD collaboration 
		  \cite{Davies:2009tsa}. The scale uncertainty on $\xi$ is estimated as 0.2\%.} 
\\[23.0ex] \hline \\[-2.0ex]
FNAL/MILC 11A & \cite{Bouchard:2011xj} & 2+1 & \parbox[t]{0.9cm}{0.12, 0.09, 0.06} & 
\parbox[t]{4.3cm}{Combined continuum and chiral extrapolation with
		 NLO rHMS$\chi$PT, NNLO analytic and and generic 
		 $\cO(\alpha_s^2 a^2, a^4)$ terms.} & 
\parbox[t]{4.2cm}{See above. The error in $r_1$ yields a 3\% 
		 uncertainty on $f^2_B B_B$.} 
\\[12.0ex] \hline \\[-2.0ex]
\parbox[t]{2.8cm}{RBC/UKQCD 10C} & \parbox[t]{0.5cm}{\cite{Albertus:2010nm}} & 2+1 & \parbox[t]{0.9cm}{0.11} &
\parbox[t]{4.3cm}{Only one lattice spacing is used. Discretization error
		 is estimated to be 4\% on $\xi$ by power counting.}&
\parbox[t]{4.2cm}{Scale is set using the $\Omega^-$ mass as input 
		 \cite{Allton:2008pn}.The error on $\xi$ due to the combined scale and 
		 light-quark mass uncertainties is estimated as 1\%.} 
\\[12.0ex] \hline \\[-2.0ex]
HPQCD 09 & \cite{Gamiz:2009ku} & 2+1 & \parbox[t]{1.1cm}{0.12, 0.09} &
\parbox[t]{4.3cm}{Combined continuum and chiral extrapolation with
		 NLO rHMS$\chi$PT and NNLO analytic terms. Light-quark
		 discretization error
		 is estimated as 3, 2 and 0.3\% for $f_B\sqrt{B_B}$,
		 $f_{B_s}\sqrt{B_{B_s}}$ and $\xi$ respectively.} &
\parbox[t]{4.2cm}{Relative scale $r_1/a$ is set via static-quark
		  potential. Absolute scale $r_1=0.321(5)$ fm is
		  determined through $\Upsilon$ mass \cite{Gray:2005ur}.
		  The error on $f_B\sqrt{B_B}$ due to the scale uncertainty
		  is estimated as 2.3\%.} 
\\[18.0ex] \hline \\[-2.0ex]
HPQCD 06A & \cite{Dalgic:2006gp} & 2+1 & \parbox[t]{1.1cm}{0.12} & 
\parbox[t]{4.3cm}{Only one lattice spacing is used. Light-quark discretization error
		 on $f_{B_s}^2B_{B_s}$ is estimated as 4\% 
		 by power counting.} &
\parbox[t]{4.2cm}{Scale is set using the $\Upsilon$ $2S-1S$ splitting as 
		 input \cite{Gray:2005ur}. The error on $f^2_B B_B$ due to the scale uncertainty
		  is estimated as 5\%.} 
\\[12.0ex]
\fi
\hline\hline
\end{tabular*}
\caption{Continuum extrapolations/estimation of lattice artifacts 
 in determinations of the neutral $B$-meson mixing matrix elements for
 $\Nf=2+1$ simulations.}
\label{tab:BBcontinum}
}
\end{table}

\ifx\reducedapptables\undefined

\begin{table}[!ht]
{\footnotesize
\begin{tabular*}{\textwidth}{l @{\extracolsep{\fill}}c c c l l}
\hline\hline \\[-1.0ex]
Collab. & Ref. & $\Nf$ & $a$ [fm] & Continuum extrapolation & Scale setting 
\\[1.0ex] \hline \hline \\[-1.0ex]
ETM 13B  & \cite{Carrasco:2013zta} & 2 & \parbox[t]{1.1cm}{0.098,
	     0.085, 0.067, 0.054} & 
\parbox[t]{4.3cm}{
	     Combined chiral and continuum extrapolation, with a term linear in $a^2$.
	     Discretization error is estimated by omitting the
	     coarsest lattice as 0.5, 1.7, 1.3 and 1.0 \% for 
	     $B_{B_s}$, $B_B$, $B_{B_s}/B_B$ and $\xi$ respectively.
	     The heavy-quark masses vary in the range $0.13 \lsim  a m_h \lsim 0.85$.} &
\parbox[t]{4.0cm}{See below.} 
\\[26.0ex] 
\hline \\[-2.0ex]
\parbox[t]{2.4cm}{ETM 12A, 12B}  & \cite{Carrasco:2012dd,Carrasco:2012de} & 2 & \parbox[t]{1.1cm}{0.098,
	     0.085, 0.067} & 
\parbox[t]{4.3cm}{Combined chiral and continuum extrapolation, with a term linear in $a^2$.
		Discretization error included in combined statistical, chiral and continuum extrapolation error
		and estimated as $4.5$\%. 
		The heavy-quark masses vary in the range $0.25 \lsim  a m_h \lsim 0.6$.} &
\parbox[t]{4.0cm}{Relative scale $r_0/a$ set from the static-quark
		     potential. Absolute scale set from $f_\pi$. Scale setting uncertainty 
		     included in combined statistical and systematic error.} 
\\[23.0ex] 
\hline\hline
\end{tabular*}
\caption{Continuum extrapolations/estimation of lattice artifacts 
 in determinations of the neutral $B$-meson mixing matrix elements for
 $\Nf=2$ simulations.}
\label{tab:BBcontinum2}
}
\end{table}

\fi

\begin{table}[!ht]
{\footnotesize
\begin{tabular*}{\textwidth}{l @{\extracolsep{\fill}} c c c  l}
\hline\hline \\[-1.0ex]
Collab. & Ref. & $\Nf$ & {$M_{\pi,\rm min}\,[\mev]$}  & {Description}
\\[1.0ex] \hline \hline \\[-1.0ex]
HPQCD 19A& \cite{Dowdall:2019bea} & 2+1+1 & \parbox[t]{2cm}{311, 241, --}
	     &  \parbox[t]{6cm}{Pion mass in the Goldstone channel is as small as 130 MeV
		 for two coarser lattices. NLO HMrS$\chi$PT is used with NNLO analytic terms and other
		 discretization errors. Staggered wrong-spin contributions are included.
		 }
\\[7.0ex] 
\hline \\[-2.0ex]
RBC/UKQCD 18A& \cite{Boyle:2018knm} & 2+1 & \parbox[t]{2cm}{139, 139, 234}
	     &  \parbox[t]{6cm}{Combined continuum ($a^2$) and heavy quark ($1/m_H$)
		 extrapolation with the LO pion mass dependence ($m_\pi^2$)
		 in the global fit.}
\\[9.0ex] 
\ifx\reducedapptables\undefined
\hline \\[-2.0ex]
FNAL/MILC 16& \cite{Bazavov:2016nty} & 2+1 & \parbox[t]{2cm}{464,
	     280, 257, 332}
	     &  \parbox[t]{6cm}{Combined continuum and chiral extrapolation with
		 NLO HMrS$\chi$PT, NNLO analytic terms and other
		 discretization errors.
		 See the entry in Tab.~\ref{tab:BBcontinum}.
		 The breakdown of the chiral error on $\xi$ is 0.4\% and
		 not the dominant one.}
\\[15.0ex] 
\hline \\[-2.0ex]
RBC/UKQCD 14A& \cite{Aoki:2014nga} & 2+1 & \parbox[t]{2cm}{327, 289}
	     &  \parbox[t]{6cm}{Combined continuum and chiral extrapolation with
		 SU(2) NLO HM$\chi$PT and linear in quark mass
		 both with $\cO(a^2)$ terms.
		 The chiral fit error is estimated from 
		 difference between the NLO HM$\chi$PT and linear fits,
		 and further from eliminating the heaviest $ud$ quark
		 mass point.}
\\[18.0ex] \hline \\[-2.0ex]
FNAL/MILC 12 & \cite{Bazavov:2012zs} & 2+1 & \parbox[t]{2cm}{440, 320}
	     &  \parbox[t]{6cm}{Combined continuum and chiral extrapolation with
		 NLO rHMS$\chi$PT and NNLO analytic terms.
		 See the entry in Tab.~\ref{tab:BBcontinum}.
		 The omission of wrong-spin contributions \cite{Bernard:2013dfa} 
		 in the HMrS$\chi$PT is treated as a systematic error and
		 estimated to be 3.2\% for $\xi$.}
\\[15.0ex] \hline \\[-2.0ex]
FNAL/MILC 11A & \cite{Bouchard:2011xj} & 2+1 & \parbox[t]{2cm}{440, 320, 250} & 
\parbox[t]{6cm}{Combined continuum and chiral extrapolation with
		NLO rHMS$\chi$PT and NNLO analytic terms.}
\\[6.0ex] \hline \\[-2.0ex]
RBC/UKQCD 10C & \cite{Albertus:2010nm} & 2+1 & \parbox[t]{2cm}{430} &
\parbox[t]{6cm}{Linear fit matched with SU(2) NLO HM$\chi$PT at the lightest
		$ud$ mass point is used as the preferred fit. Many different fit Ans\"atze
		are considered. The systematic
		 error is estimated from the difference between the SU(2) 
		 HM$\chi$PT fit described above and a linear fit. }
\\[18.0ex]\hline \\[-2.0ex] 
HPQCD 09 & \cite{Gamiz:2009ku} & 2+1 & \parbox[t]{2cm}{440, 400} &
\parbox[t]{6cm}{Combined continuum and chiral extrapolation with
		NLO rHMS$\chi$PT and NNLO analytic terms.} 
\\[6.0ex] \hline \\[-2.0ex]
HPQCD 06A & \cite{Dalgic:2006gp} & 2+1 & \parbox[t]{2cm}{510} & 
\parbox[t]{6cm}{Two sea $ud$ quark masses $m_{ud}/m_s=0.25$ and $0.5$
		 are used to calculate the matrix element for $B_s$
		 meson at the predetermined value of the strange-quark
		 mass. No significant sea-quark mass dependence is observed
		 and the value at the lighter sea $ud$ mass is taken as the result.} 
\\[18.0ex] \hline \\[-2.0ex]
\parbox[t]{2.2cm}{ETM 13B\\ ETM 12A, 12B}  & \parbox[t]{1.4cm}{\cite{Carrasco:2013zta}\\ \cite{Carrasco:2012dd,Carrasco:2012de}} & 2 & \parbox[t]{2cm}{410, 275, 300, 270} & 
\parbox[t]{6cm}{$M_{\pi,{\rm min}}$ refers to the charged pions, where 270 MeV
	on the finest lattice only included in ETM 13B. Linear and NLO (full
	QCD) HM$\chi$PT supplemented by an $a^2$ term is used. The chiral fit error is
	estimated from the difference between the NLO HM$\chi$PT and linear fits. }
\\[18.0ex] 
%
\fi
\hline\hline
\end{tabular*}
\caption{Chiral extrapolation/minimum pion mass
 in determinations of the neutral $B$-meson mixing matrix elements.
 For actions with multiple species of pions, masses quoted are the RMS
 pion masses (where available).  The different $M_{\pi,\rm min}$ entries correspond to the
 different lattice spacings.}
}
\end{table}

\begin{table}[!ht]
{\footnotesize
\begin{tabular*}{\textwidth}{l @{\extracolsep{\fill}} c c c c l}
\hline\hline \\[-1.0ex]
Collab. & Ref. & $\Nf$ & $L$ [fm] & ${M_{\pi,\rm min}}L$ & Description
\\[1.0ex] \hline \hline \\[-1.0ex]
HPQCD 19A & \cite{Dowdall:2019bea} & 2+1+1 & 
\parbox[t]{2.0cm}{2.4/3.5/4.6, 2.9/3.8/5.7, 2.8} & 
\parbox[t]{1.8cm}{7.3, 7.0, -} & 
\parbox[t]{5.1cm}{FV error is estimated to be negligible from FV HM$\chi$PT.} 
\\[7.0ex] 
\hline \\[-2.0ex]
RBC/UKQCD 18A & \cite{Boyle:2018knm} & 2+1 & 
\parbox[t]{2.0cm}{2.7/5.5, 2.6/5.3, 3.5} & 
\parbox[t]{1.8cm}{3.9, 3.8, 4.0} & 
\parbox[t]{5.1cm}{FV error is estimated to be less than $0.18\%$ for
		  SU(3)-breaking ratios from FV HM$\chi$PT.} 
\\[7.0ex] 
\ifx\reducedapptables\undefined
\hline \\[-2.0ex]
FNAL/MILC 16& \cite{Bazavov:2016nty} & 2+1 & 
\parbox[t]{2.0cm}{2.4/2.9, 2.5/2.9/3.6/5.8, 2.9/3.4/3.8, 2.9} & 
\parbox[t]{1.8cm}{6.8, 8.2, 5.0, 4.8} & 
\parbox[t]{5.1cm}{FV error is estimated to be less than $0.1\%$ for
		  SU(3)-breaking ratios from FV HMrS$\chi$PT.} 
\\[9.0ex] 
\hline \\[-2.0ex]
RBC/UKQCD 14A
& \cite{Aoki:2014nga} & 2+1 & 
\parbox[t]{1.8cm}{2.74, 2.76} & 4.5, 4.0 & 
\parbox[t]{5.1cm}{FV error is estimated from SU(2) $\chi$PT to be 
		  0.76, 0, 1.07\% for $f_B\sqrt{B_B}$,
		   $f_{B_s}\sqrt{B_{B_s}}$ and $\xi$ respectively.} 
\\[6.0ex] \hline \\[-2.0ex]
FNAL/MILC 12 & \cite{Bazavov:2012zs} & 2+1 & 
\parbox[t]{1.8cm}{2.4/2.9, 2.5} & 6.4, 5.1 & 
\parbox[t]{5.1cm}{FV error is estimated to be less than $0.1\%$ for
		     SU(3)-breaking ratios from FV HMrS$\chi$PT.} 
\\[6.0ex] \hline \\[-2.0ex]
FNAL/MILC 11A & \cite{Bouchard:2011xj} & 2+1 & 
\parbox[t]{1.8cm}{2.4/2.9, 2.5/2.9/3.6, 3.8} & 6.4, 5.8, 4.9 &
\parbox[t]{5.1cm}{FV error on $f_B\sqrt{B_B}$ is estimated to be less
		  than $1\%$, 
		  which is inferred from the study of the $B$-meson decay
		  constant using FV HM$\chi$PT \cite{Bazavov:2011aa}.} 
\\[9.0ex] \hline \\[-2.0ex]
RBC/UKQCD 10C & \cite{Albertus:2010nm} & 2+1 & 
\parbox[t]{1.8cm}{1.8} & 3.9 & 
\parbox[t]{5.1cm}{FV error estimated through FV HM$\chi$PT as $1\%$ for SU(3)-breaking ratios. }
\\[6.0ex]\hline \\[-2.0ex] 
HPQCD 09 & \cite{Gamiz:2009ku} & 2+1 & 
\parbox[t]{1.8cm}{2.4/2.9, 2.5} & 6.4, 5.1 &
\parbox[t]{5.1cm}{No explicit estimate of FV error, but expected to be much smaller than other uncertainties.}
\\[6.0ex] \hline \\[-2.0ex]
HPQCD 06A & \cite{Dalgic:2006gp} & 2+1 & 
\parbox[t]{1.8cm}{2.4} & 6.2 & 
\parbox[t]{5.1cm}{No explicit estimate of FV error, but expected to be much smaller than other uncertainties.}
\\[6.0ex] \hline \\[-2.0ex]
\parbox[t]{2.2cm}{ETM 13B\\ ETM 12A, 12B}  &
     \parbox[t]{1.4cm}{\cite{Carrasco:2013zta}\\ \cite{Carrasco:2012dd,Carrasco:2012de}} & 2 & 
\parbox[t]{1.8cm}{2.4, 2.0/2.7, 2.1, 1.7/2.6} & 
\parbox[t]{1.8cm}{5.0, 3.7, 3.3, 3.5} & 
\parbox[t]{5.1cm}{$L=1.7/2.6\, {\rm fm}$ only included in ETM~13B. FV error is assumed to be negligible based on the study of $D$-meson decay constants in 
Ref.~\cite{Blossier:2009bx}.}
\\[9.0ex] 
\fi
\hline\hline
\end{tabular*}
\caption{Finite-volume effects 
 in determinations of the neutral $B$-meson mixing matrix elements.
 Each $L$-entry corresponds to a different
 lattice spacing, with multiple spatial volumes at some lattice
 spacings. 
 For actions with multiple species of pions, masses quoted are the RMS
 pion masses (where available).
}
}
\end{table}

\begin{table}[!ht]
{\footnotesize
\begin{tabular*}{\textwidth}{l @{\extracolsep{\fill}} c c c l}
\hline\hline \\[-1.0ex]
Collab. & Ref. & $\Nf$ & Ren. & Description 
\\[1.0ex] \hline \hline \\[-1.0ex]
HPQCD 19A & \cite{Dowdall:2019bea} & 2+1+1 & PT1$\ell$  & 
\parbox[t]{6cm}{HISQ-NRQCD 4-quark operators are matched through $O(1/M)$ and renormalized to 1-loop:
		 included are those of $O(\alpha_s)$, $O(\Lambda_{\rm QCD}/M)$, $O(\alpha_s/a M)$, 
		 $O(\alpha_s \, \Lambda_{\rm QCD}/M)$. 
		 Remnant error is dominated by $O(\alpha_s \Lambda_{\rm QCD}/M)~2.9$\% 
		 and $O(\alpha_s^2)~2.1$\% for individual bag parameters.
		 Associated error for their SU(3) breaking ratio are negligible.}
\\[24.0ex] 
\hline \\[-2.0ex]
RBC/UKQCD 18A & \cite{Boyle:2018knm} & 2+1 & - & 
\parbox[t]{6cm}{Operators are renormalized multiplicatively 
		 due to chiral symmetry of DWF. 
		 No need to calculate the renormalization factor
		 since only the SU(3) breaking ratios are examined.}
\\[12.0ex] 
\ifx\reducedapptables\undefined
\hline \\[-2.0ex]
FNAL/MILC 16& \cite{Bazavov:2016nty}  & 2+1 & mNPR & 
\parbox[t]{6cm}{mNPR is used with 1-loop lattice perturbation theory
		to renormalize the four-quark operators
		with heavy quarks rotated to eliminate tree-level $\cO(a)$
		errors. The error from neglecting higher order corrections 
		is estimated to be $0.5$\% on $\xi$.} 
\\[15.0ex] 
\hline \\[-2.0ex]
RBC/UKQCD 14A & \cite{Aoki:2014nga} & 2+1 & PT$1l$ & 
\parbox[t]{6cm}{Static-light four-quark operators are 
		renormalized with 1-loop
		mean field improved PT.
		The errors due to neglected higher order effects 
		are estimated for purely $\cO\alpha_s^2$ to be
		$6$\% on the matrix elements 
		or $1.2$\% on $\xi$ and for $\cO\alpha_s^2 a^2$
		to be $1$\% or $0.2$\% respectively.}
\\[18.0ex] \hline \\[-2.0ex]
FNAL/MILC 12 & \cite{Bazavov:2012zs} & 2+1 & PT$1l$ & 
\parbox[t]{6cm}{1-loop mean-field improved PT
		 is used to renormalize the four-quark operators
		 with heavy quarks rotated to eliminate tree-level $\cO(a)$
		 errors. The error from neglecting higher order corrections 
		 is estimated to be $0.5$\% on $\xi$.} 
\\[15.0ex] \hline \\[-2.0ex]
FNAL/MILC 11A & \cite{Bouchard:2011xj} & 2+1 & PT$1l$ &
\parbox[t]{6cm}{1-loop mean-field improved PT is used to 
                renormalize the four-quark operators
		 with heavy quarks rotated to eliminate tree-level $\cO(a)$
		 errors. The error from neglected higher order corrections 
		 is estimated to be $4$\% on $f_B\sqrt{B_B}$.} 
\\[15.0ex] \hline \\[-2.0ex]
RBC/UKQCD 10C & \cite{Albertus:2010nm} & 2+1 & PT$1l$ & 
\parbox[t]{6cm}{Static-light four-quark operators are 
		 renormalized with 1-loop
		 mean field improved PT.
		 The error due to neglected higher order effects 
		 is estimated to be $2.2$\% on $\xi$.}
\\[9.0ex]\hline \\[-2.0ex]
HPQCD 09 & \cite{Gamiz:2009ku} & 2+1 & PT$1l$ &
\parbox[t]{6cm}{Four-quark operators in lattice NRQCD are matched to
		 QCD through order $\alpha_s$, $\Lambda_{\rm QCD}/M$ 
		 and  $\alpha_s/(aM)$ \cite{Gamiz:2008sk} using 1-loop PT. 
		 The error due to neglected higher order effects is estimated to 
		 be 4\% on $f_B\sqrt{B_B}$ and 0.7\% on $\xi$.}
\\[15.0ex] \hline \\[-2.0ex]
HPQCD 06A & \cite{Dalgic:2006gp} & 2+1 & PT$1l$ &
\parbox[t]{6cm}{Four-quark operators in lattice NRQCD are matched to
		 full QCD through order $\alpha_s$, $\Lambda_{\rm QCD}/M$ 
		 and  $\alpha_s/(aM)$ \cite{Gamiz:2008sk}. The error  
		 is estimated as $\sim 1\cdot \alpha_s^2$ to be 9\% on
		 $f_{B_s}^2 B_{B_s}$}
\\[9.0ex] \hline \\[-2.0ex]
ETM 13B, 12A, 12B & \cite{Carrasco:2013zta,Carrasco:2012dd,Carrasco:2012de} & 2 & NPR & 
\parbox[t]{6cm}{The bag parameters are nonperturbatively 
	         renormalized in the RI'-MOM scheme. They are calculated
	         as functions of the ($\msbar$) heavy-quark mass (renormalized
	         nonperturbatively in RI/MOM). }
\\[12.0ex] 
%
%
%
\fi
\hline\hline
\end{tabular*}
\caption{Operator renormalization 
 in determinations of the neutral $B$-meson mixing matrix elements.
}
\label{tab:BBmatching}
}
\end{table}

\begin{table}[!ht]
{\footnotesize
\begin{tabular*}{\textwidth}{l @{\extracolsep{\fill}} c c c l}
\hline\hline \\[-1.0ex]
Collab. & Ref. & $\Nf$ & Action & Description
\\[1.0ex] \hline \hline \\[-1.0ex]
HPQCD 19A & \cite{Dowdall:2019bea} & 2+1+1 & NRQCD &
\parbox[t]{8cm}{See the entry in Tab.~\ref{tab:BBmatching}.}
\\[4.0ex] 
\hline \\[-2.0ex]
RBC/UKQCD 18A & \cite{Boyle:2018knm} & 2+1 & DWF &
\parbox[t]{8cm}{Domain-wall fermion with 3 stout-smearing extends the reach to heavy mass,
		 allowing to simulate up to half of the b-quark mass.
		 Heavy mass errors on $\xi$ are estimated as 0.8\% from fitting range and 0.4\% 
		 from higher order ($1/M^2$) by power counting.}
\\[10.0ex] 
\ifx\reducedapptables\undefined
\hline \\[-2.0ex]
FNAL/MILC 16& \cite{Bazavov:2016nty} & 2+1 & Fermilab & 
\parbox[t]{8cm}{The heavy-quark discretization error is a dominant error 
		comparable to the statistical error. It reads 4.6\%,
		3.2\% or 0.7\% for the $B_d$, $B_s$ matrix element or $\xi$.
		}
\\[6.0ex] 
\hline \\[-2.0ex]
RBC/UKQCD 14A& \cite{Aoki:2014nga} & 2+1 & Static & 
\parbox[t]{8cm}{Two different static-quark actions with HYP1 and HYP2
		 smearings are used and the continuum extrapolation is
		 constrained so the two values converges in the limit.
		 The error due to the missing $1/m_b$ corrections is
		 estimated to be $12$\% for individual matrix
		 elements or $2.2$\% on $\xi$ using power-counting.}
\\[15.0ex] \hline \\[-2.0ex]
FNAL/MILC 12 & \cite{Bazavov:2012zs} & 2+1 & Fermilab & 
\parbox[t]{8cm}{The heavy-quark discretization error on $\xi$ is estimated
		to be 0.3 \%. The error on $\xi$ 
		 due to the uncertainty in the $b$-quark mass is
		 estimated to be 0.4 \%.} 
\\[6.0ex] \hline \\[-2.0ex]
FNAL/MILC 11A & \cite{Bouchard:2011xj} & 2+1 & Fermilab &
\parbox[t]{8cm}{The 
		 heavy-quark discretization error on $f_B\sqrt{B_B}$
		 is estimated as 4\% using power-counting.} 
\\[3.0ex] \hline \\[-2.0ex]
RBC/UKQCD 10C & \cite{Albertus:2010nm} & 2+1 & Static & 
\parbox[t]{8cm}{Two different static-quark actions with Ape and HYP
		 smearings are used. The discretization error
		 on $\xi$ is estimated as  $\sim 4$\% and  the error due to the missing
		 $1/m_b$ corrections  as
		 $\sim 2$\%, both using power-counting.}
\\[9.0ex]\hline \\[-2.0ex]
HPQCD 09 & \cite{Gamiz:2009ku} & 2+1 & NRQCD &
\parbox[t]{8cm}{Heavy-quark truncation errors due to relativistic corrections
		 are estimated to be 2.5, 2.5 and 0.4 \% for $f_B\sqrt{B_B}$,
		 $f_{B_s}\sqrt{B_{B_s}}$ and $\xi$ respectively.}
\\[6.0ex] \hline \\[-2.0ex]
HPQCD 06A & \cite{Dalgic:2006gp} & 2+1 & NRQCD &
\parbox[t]{8cm}{Heavy-quark truncation errors due to relativistic corrections
		 are estimated to be 3\% for $f_{B_s}^2{B_{B_s}}$.}
\\[3.0ex] \hline \\[-2.0ex]
\parbox[t]{2.2cm}{ETM 13B\\ ETM 12A,12B}  &
     \parbox[t]{1.4cm}{\cite{Carrasco:2013zta}\\ \cite{Carrasco:2012dd,Carrasco:2012de}} & 2 & 
	 tmWil & 
\parbox[t]{8cm}{The ratio method is used to perform an interpolation to
	the physical $b$ quark mass from the simulated
	heavy mass and the known static limit.
	In an intermediate step, the ratios include HQET matching
	factors calculated to tree-level, leading-log, 
	and next-to-leading-log (ETM 13B only)
	in continuum PT.
	The interpolation uses a polynomial up to
	quadratic in the inverse quark-mass. 
	The systematic errors added together with those of the chiral
	fit are estimated as $1.3-1.6$\% for bag parameters for ETM 13B,
	while they are estimated from
	changing the interpolating polynomial as $2$\% and from changing the order of HQET 
	matching factors as $3$\% for ETM 12A and 12B.}
\\[32.0ex]
\fi
\hline\hline
\end{tabular*}
\caption{Heavy-quark treatment
 in determinations of the neutral $B$-meson mixing matrix elements.
}}
\end{table}

%% file: HQ/AppendixTables/BSLapp.tex
\FloatBarrier
\subsubsection{Form factors entering determinations of $|V_{ub}|$ ($B \to \pi\ell\nu$, $B_s \to K\ell\nu$, $\Lambda_b\to p\ell\bar{\nu}$)}
\label{app:BtoPi_Notes}
\FloatBarrier



\begin{table}[!ht]

{\footnotesize
\begin{tabular*}{\textwidth}{l c c c l l}
\hline\hline \\[-1.0ex]
Collab. & Ref. & $\Nf$ & $a$ [fm] & Continuum extrapolation & Scale setting 
\\[1.0ex] \hline \hline \\[-1.0ex]
RBC/UKQCD 23 & \cite{Flynn:2023nhi} & 2+1 & \parbox[t]{20mm}{0.071, 0.083, 0.11}   &
\parbox[t]{4.2cm}{Joint chiral-continuum extrapolation using SU(2) hard-pion HM$\chi$PT.
Systematic uncertainty estimated by varying fit ansatz and form of coefficients, as well as implementing different
cuts on data.} & \parbox[t]{30mm}{Scale implicitly set in the light-quark sector using the $\Omega^-$ mass, cf.~\cite{Blum:2014tka,Boyle:2017jwu,Boyle:2018knm}.}
\\[18.0ex] \hline \\[-2.0ex]
\SLjlqcdBpi & \cite{Colquhoun:2022atw} & 2+1 & \parbox[t]{20mm}{0.044, 0.055, 0.080}   &
\parbox[t]{4.2cm}{Discretization effects treated using overall factors of $(1+C_{a^2}(\Lambda_{\mathrm{QCD}}a)^2+C_{(am_Q)^2}(am_Q)^2)$, with independent coefficients for the two form factors. Systematic uncertainties estimated by adding $C_{a^4}(\Lambda_{\mathrm{QCD}}a)^4$ or $C_{(am_Q)^4}(am_Q)^4)$ terms.
} & \parbox[t]{30mm}{Relative scale set using gradient-flow time $t_0^{1/2}/a$. Absolute scale $t_0^{1/2}$ taken from Ref.~\cite{Borsanyi:2012zs}.}
\\[26.0ex] \hline  \\[-2.0ex]
\SLfnalmilcBsK & \cite{Bazavov:2019aom} & 2+1 & \parbox[t]{20mm}{0.06, 0.09, 0.12}   &
\parbox[t]{4.2cm}{
HMrS$\chi$PT expansion used at next-to-leading
order in SU(2) and leading order in $1/M_B$, including next-to-next-to-leading-order
(NNLO) analytic and generic discretization terms. Hard kaons assumed to decouple. Systematic uncertainties
estimated by varying fit ansatz and data range. The (stat + chiral extrap + HQ discretization + $g_{\pi}$) uncertainty
dominates the error budget, ranging from 2--3\% at $q^2 \gtrsim 21~{\rm GeV}^2$ to up to 8-10\%
in the lower end of the accessed $q^2$ interval.} & \parbox[t]{30mm}{Relative scale $r_1/a$ set from the static-quark potential.  Absolute scale $r_1$, including related uncertainty estimates, taken from \cite{Bazavov:2011aa}.}
\\[26.0ex]
\ifx\reducedapptables\undefined
\hline \\[-2.0ex]
\SLfnalmilcBpi & \cite{Lattice:2015tia} & 2+1 & \parbox[t]{20mm}{0.045, 0.06, 0.09, 0.12}   &
\parbox[t]{4.2cm}{Fit to HMrS$\chi$PT to remove light-quark discretization errors. Residual heavy-quark discretization errors estimated with power-counting. Total (stat + chiral extrap + HQ discretization + $g_{B^*B\pi}$) error estimated to be 3.1\% for $f_+$ and 3.8\% for $f_0$ at $q^2=20~{\rm GeV}^2$.} & \parbox[t]{30mm}{Relative scale $r_1/a$ set from the static-quark potential.  Absolute scale $r_1$, including related uncertainty estimates, taken from \cite{Bazavov:2011aa}.}
\\[26.0ex] \hline \\[-2.0ex]
\parbox[t]{20mm}{\SLLambdabp} & \cite{Detmold:2015aaa} & 2+1 & \parbox[t]{20mm}{0.0849(12), 0.1119(17)}   &
\parbox[t]{4.2cm}{Joint chiral-continuum extrapolation, combined with fit to $q^2$-dependence of form factors in a ``modified'' $z$-expansion. Systematics estimated by varying fit form and $\cO(a)$ improvement parameter values.} & \parbox[t]{30mm}{Set from $\Upsilon(2S)$--$\Upsilon(1S)$ splitting, cf.~\cite{Meinel:2010pv}.}
\\[21.0ex] \hline \\[-2.0ex]
\SLrbcukqcdBpi & \cite{Flynn:2015mha} & 2+1 & \parbox[t]{20mm}{0.086,0.11}   &
\parbox[t]{4.2cm}{Joint chiral-continuum extrapolation using SU(2) hard-pion HM$\chi$PT.
Systematic uncertainty estimated by varying fit ansatz and form of coefficients, as well as implementing different
cuts on data; ranges from 5.0\% to 10.9\% for $B\to\pi$ form factors, and 2.5\% to 5.1\% for $B_s\to K$.
Light-quark and gluon discretization errors estimated at 1.1\% and 1.3\%, respectively.} & \parbox[t]{30mm}{Scale implicitly set in the light-quark sector using the $\Omega^-$ mass, cf.~\cite{Aoki:2010dy}.}
\\[35.0ex] \hline \\[-2.0ex]
\SLhpqcdBsK & \cite{Bouchard:2014ypa} & 2+1 & \parbox[t]{20mm}{0.09,0.12}   &
\parbox[t]{4.2cm}{Combined chiral-continuum extrapolation using hard-pion rHMS$\chi$PT. (No explicit estimate of discretization effects.)} & \parbox[t]{30mm}{Relative scale $r_1/a$ set from the static-quark potential. Absolute scale $r_1$
set to 0.3133(23)~fm.}
\\[12.0ex] \hline \\[-2.0ex]
HPQCD 06 & \cite{Dalgic:2006dt} & 2+1 & \parbox[t]{20mm}{0.09,0.12}  &
\parbox[t]{4.2cm}{Central values obtained from data at $a=0.12$~fm.  Discretization errors observed to be within the statistical error by comparison with data at $a=0.09$~fm.} &  \parbox[t]{30mm}{Relative scale $r_1/a$ set from the static-quark potential.  Absolute scale $r_1$ set through $\Upsilon$ $2S-1S$ splitting c.f.~HPQCD 05B~\cite{Gray:2005ur}.}
\\[18.0ex]
\fi
\hline\hline
\end{tabular*}
\caption{Continuum extrapolations/estimation of lattice artifacts in
  determinations of $B\to\pi\ell\nu$, $B_s\to K\ell\nu$, and $\Lambda_b\to p\ell\bar{\nu}$ form factors.}
}
\end{table}
\begin{table}[!ht]
{\footnotesize
\begin{tabular*}{\textwidth}{l @{\extracolsep{\fill}} c c c   l}
\hline\hline \\[-1.0ex]
{Collab.} & {Ref.} & {$\Nf$} & {$M_{\pi,\rm min}\,[\mev]$}  & {Description}  
\\[1.0ex] \hline \hline \\[-1.0ex]
RBC/UKQCD 23 & \cite{Flynn:2023nhi} & 2+1 & \parbox[t]{2cm}{268, 301, 340}
& \parbox[t]{5cm}{Joint chiral-continuum extrapolation using SU(2) hard-pion HM$\chi$PT.
Systematic uncertainty estimated by varying fit ansatz and form of coefficients, as well as implementing different
cuts on data.}
\\[18.0ex] \hline \\[-2.0ex]
\SLjlqcdBpi & \cite{Colquhoun:2022atw} & 2+1 & \parbox[t]{20mm}{300, 300, 230}   &
\parbox[t]{5cm}{Chiral extrapolation uses SU(2) hard-pion heavy-meson chiral perturbation theory at next-to-leading
order. Systematic uncertainty estimated by adding $M_\pi^4$ terms or by making the coefficients of the chiral logs fit parameters. }
\\[19.0ex] \hline  \\[-2.0ex]
\SLfnalmilcBsK & \cite{Bazavov:2019aom} & 2+1 & \parbox[t]{2cm}{255, 277, 456}
& \parbox[t]{5cm}{HMrS$\chi$PT expansion used at next-to-leading
order in SU(2) and leading order in $1/M_B$, including next-to-next-to-leading-order
(NNLO) analytic and generic discretization terms. Hard kaons assumed to decouple. Systematic uncertainties
estimated by varying fit ansatz and data range. }
\\[18.0ex]
\ifx\reducedapptables\undefined
\hline \\[-2.0ex]
\SLfnalmilcBpi & \cite{Lattice:2015tia} & 2+1 & \parbox[t]{2cm}{330, 260, 280, 470}
& \parbox[t]{5cm}{Simultaneous chiral-continuum extrapolation and $q^2$ interpolation using NNLO SU(2) hard-pion HMrS$\chi$PT.  Systematic error estimated by adding higher-order analytic terms and varying the $B^*$-$B$-$\pi$ coupling.}
\\[18.0ex] \hline \\[-2.0ex]
\parbox[t]{20mm}{\SLLambdabp} & \cite{Detmold:2015aaa} & 2+1 & \parbox[t]{2.5cm}{227, 245 \\ (valence-valence) \\ 263, 294 \\ (valence-sea)}
& \parbox[t]{5cm}{Joint chiral-continuum extrapolation, combined with fit to $q^2$-dependence of form factors in a ``modified'' $z$-expansion. Only analytic NLO terms $\propto (m_\pi^2-m_{\pi,{\rm phys}}^2)$ included in light-mass dependence. Systematic uncertainty estimated by repeating fit with added higher-order terms.}
\\[23.0ex] \hline \\[-2.0ex]
\SLrbcukqcdBpi & \cite{Flynn:2015mha} & 2+1 & \parbox[t]{2cm}{289, 329}
& \parbox[t]{5cm}{Joint chiral-continuum extrapolation using SU(2) hard-pion HM$\chi$PT.
Systematic uncertainty estimated by varying fit ansatz and form of coefficients, as well as implementing different
cuts on data; ranges from 5.0\% to 10.9\% for $B\to\pi$ form factors, and 2.5\% to 5.1\% for $B_s\to K$.}
\\[23.0ex] \hline \\[-2.0ex]
\SLhpqcdBsK & \cite{Bouchard:2014ypa} & 2+1 & \parbox[t]{2cm}{295, 260}
& \parbox[t]{5cm}{Combined chiral-continuum extrapolation using hard-pion rHMS$\chi$PT. (No explicit estimate of extrapolation systematics.)}
\\[9.0ex] \hline \\[-2.0ex]
%
%
HPQCD 06 & \cite{Dalgic:2006dt} & 2+1 & \parbox[t]{2cm}{400, 440}
& \parbox[t]{5cm}{First interpolate data at fixed quark mass to fiducial values of $E_\pi$ using the Becirevic-Kaidalov and Ball-Zwicky ans\"atze, then extrapolate data at fixed $E_\pi$ to physical quark masses using SU(3) rHMS$\chi$PT. Systematic error estimated by varying interpolation and extrapolation fit functions.}
\\[24.0ex]
%
\fi
\hline\hline
\end{tabular*}
\caption{Chiral extrapolation/minimum pion mass in
  determinations of $B\to\pi\ell\nu$, $B_s\to K\ell\nu$, and $\Lambda_b\to p\ell\bar{\nu}$ form factors.  For actions with multiple species of pions, masses quoted are the RMS pion masses.    The different $M_{\pi,\rm min}$ entries correspond to the different lattice spacings.}
}
\end{table}
\begin{table}[!ht]
{\footnotesize
\begin{tabular*}{\textwidth}{l @{\extracolsep{\fill}} c c c c l}
\hline\hline \\[-1.0ex]
Collab. & Ref. & $\Nf$ & $L$ [fm] & ${M_{\pi,\rm min}}L$ & Description 
\\[1.0ex] \hline \hline \\[-1.0ex]
RBC/UKQCD 23 & \cite{Flynn:2023nhi} & 2+1 & \parbox[t]{1.8cm}{3.4, 2.7, 2.6} & $4.6$, $4.0, 4.4$  &
\parbox[t]{5cm}{Finite-volume effects removed by correction to chiral logs due to sums over discrete momenta;
quoted maximum corrections are 0.13\% for $f_+$ and 0.06\% for $f_0$.
}
\\[12.0ex] \hline \\[-1.0ex]
\SLjlqcdBpi & \cite{Colquhoun:2022atw} & 2+1 & \parbox[t]{20mm}{2.6, 3.9}   & $\gtrsim 4.0$  & 
\parbox[t]{5cm}{Finite-volume effects in form factors deemed negligible. Bias in pion mass due to topology freezing at finest lattice spacing estimated to be $\sim0.1\%$.  }
\\[9.0ex] \hline  \\[-2.0ex]
\SLfnalmilcBsK & \cite{Bazavov:2019aom} & 2+1 & \parbox[t]{1.8cm}{3.8, 2.5/2.9/3.6/5.8, 2.9} & $\gtrsim 3.8$  &
\parbox[t]{5cm}{Finite-volume effects estimated by comparing infinite volume integrals with finite sums in HMrS$\chi$PT, found to be negligible.
}
\\[9.0ex]
\ifx\reducedapptables\undefined
\hline \\[-1.0ex]
\SLfnalmilcBpi & \cite{Lattice:2015tia} & 2+1 & \parbox[t]{1.8cm}{2.9, 2.9/3.4/3.8, 2.5/2.9/3.6/5.8, 2.4/2.9} & $\gtrsim 3.8$  &
\parbox[t]{5cm}{Finite-volume effects estimated by replacing infinite-volume chiral logs with sums over discrete momenta, found to be negligible.
}
\\[9.0ex] \hline \\[-1.0ex]
\parbox[t]{20mm}{\SLLambdabp} & \cite{Detmold:2015aaa} & 2+1 & \parbox[t]{1.8cm}{2.7, 2.7} & \parbox[t]{1.8cm}{$\gtrsim 3.6$ \\ (valence-sea pion)}  &
\parbox[t]{5cm}{Finite-volume effect estimated at 3\% from experience on $\chi$PT estimates of finite-volume effects for heavy-baryon axial couplings.
}
\\[9.0ex] \hline \\[-1.0ex]
\SLrbcukqcdBpi & \cite{Flynn:2015mha} & 2+1 & \parbox[t]{1.8cm}{2.8, 2.6} & $4.0, 4.4$  &
\parbox[t]{5cm}{Finite-volume effects estimated by correction to chiral logs due to sums over discrete momenta;
quoted 0.3-0.5\% for $f_+$ and 0.4-0.7\% for $f_0$ for $B\to\pi$, and 0.2\% for $f_+$ and 0.1-0.2\% for $f_0$ for $B_s\to K$.
}
\\[15.0ex] \hline \\[-1.0ex]
\SLhpqcdBsK & \cite{Bouchard:2014ypa} & 2+1 & \parbox[t]{1.8cm}{2.5, 2.4/2.9} & $\gtrsim 3.8$  &
\parbox[t]{5cm}{Finite-volume effects estimated by shift of pion log, found to be negligible.
}
\\[4.0ex] \hline \\[-1.0ex]
%
%
HPQCD 06 & \cite{Dalgic:2006dt} & 2+1 & \parbox[t]{1.8cm}{2.4/2.9} & $\gtrsim 3.8$ & 
\parbox[t]{5cm}{No explicit estimate of finite-volume error, but expected to be much smaller than other uncertainties.}
\\[7.0ex]
%
%
\fi
\hline\hline
\end{tabular*}
\caption{Finite-volume effects in determinations of $B\to\pi\ell\nu$, $B_s\to K\ell\nu$, and $\Lambda_b\to p\ell\bar{\nu}$ form factors.  Each $L$-entry corresponds to a different lattice
spacing, with multiple spatial volumes at some lattice spacings.  For actions with multiple species of pions, the lightest masses are quoted.}
}
\end{table}
\begin{table}[!ht]
{\footnotesize
\begin{tabular*}{\textwidth}{l @{\extracolsep{\fill}} c c c l}
\hline\hline \\[-1.0ex]
Collab. & Ref. & $\Nf$ & Ren. & Description 
\\[1.0ex] \hline \hline \\[-1.0ex]
RBC/UKQCD 23 & \cite{Flynn:2023nhi} & 2+1 & mNPR &
\parbox[t]{6cm}{Perturbative truncation error estimated as full size of $O(\alpha_s)$ correction at the 0.083 fm lattice spacing.}
\\[7.0ex] \hline \\[-1.0ex]
\SLjlqcdBpi & \cite{Colquhoun:2022atw} & 2+1 & NPR   &  
\parbox[t]{6cm}{$Z_{V_{qq}}$ obtained using position-space current-current correlators. For heavier quark masses, $\sqrt{Z_{V_{QQ}}  Z_{V_{qq}}}$ is used, where $Z_{V_{QQ}}$ is the renormalization factor of the flavour-conserving temporal vector current, determined using charge conservation.  }
\\[16.0ex] \hline  \\[-2.0ex]
\SLfnalmilcBsK & \cite{Bazavov:2019aom} & 2+1 & mNPR &
\parbox[t]{6cm}{Perturbative truncation error estimated at 1\% with size of 1-loop correction on next-to-finest ensemble.}
\\[7.0ex]
\ifx\reducedapptables\undefined
\hline \\[-1.0ex]
\SLfnalmilcBpi & \cite{Lattice:2015tia} & 2+1 & mNPR &
\parbox[t]{6cm}{Perturbative truncation error estimated at 1\% with size of 1-loop correction on next-to-finest ensemble.}
\\[7.0ex] \hline \\[-1.0ex]
\parbox[t]{20mm}{\SLLambdabp} & \cite{Detmold:2015aaa} & 2+1 & mNPR &
\parbox[t]{6cm}{Perturbative truncation error estimated at 1\% with size of 1-loop correction on next-to-finest ensemble.}
\\[7.0ex] \hline \\[-1.0ex]
\SLrbcukqcdBpi & \cite{Flynn:2015mha} & 2+1 & mNPR &
\parbox[t]{6cm}{Perturbative truncation error estimated as largest of power counting, effect from value of $\alpha_s$ used, numerical integration. Nonperturbative normalization of flavour-diagonal currents computed by fixing values of ratios of meson two-point functions to three-point functions with an extra current inversion, cf.~\cite{Christ:2014uea}}
\\[21.0ex] \hline \\[-1.0ex]
\SLhpqcdBsK & \cite{Bouchard:2014ypa} & 2+1 & mNPR &
\parbox[t]{6cm}{Currents matched using 1-loop HISQ lattice perturbation theory, omitting $\cO(\alpha_s\Lambda_{\rm QCD}/m_b$. Systematic uncertainty resulting from 1-loop matching and neglecting $\cO(\Lambda_{\rm QCD}^2/m_b^2$ terms estimated at 4\% from power counting.}
\\[16.0ex] \hline \\[-1.0ex]
%
%
HPQCD 06 & \cite{Dalgic:2006dt} & 2+1 & PT1$\ell$ &
\parbox[t]{6cm}{Currents included through $\cO(\alpha_S \Lambda_{\rm QCD}/M$, $\alpha_S/(aM)$, $\alpha_S\, a \Lambda_{\rm QCD})$. Perturbative truncation error estimated from power-counting.}
\\[10.0ex]
%
%
\fi
\hline\hline
\end{tabular*}
\caption{Operator renormalization in determinations of $B\to\pi\ell\nu$, $B_s\to K\ell\nu$, and $\Lambda_b\to p\ell\bar{\nu}$ form factors.}
}
\end{table}
\begin{table}[!ht]
{\footnotesize
\begin{tabular*}{\textwidth}{l @{\extracolsep{\fill}} c c c l}
\hline\hline \\[-1.0ex]
Collab. & Ref. & $\Nf$ & Action & Description 
\\[1.0ex] \hline \hline \\[-1.0ex]
RBC/UKQCD 23 & \cite{Flynn:2023nhi} & 2+1 & Columbia RHQ &
\parbox[t]{8cm}{Heavy-quark discretization errors estimated by power counting.}
\\[4.0ex] \hline \\[-1.0ex]
\SLjlqcdBpi & \cite{Colquhoun:2022atw} & 2+1 & DWF   &  
\parbox[t]{8cm}{ Bare heavy-quark masses satisfy $a m_Q<0.7$ and reach from the charm mass up to 2.44 times the charm mass. Form factors extrapolated linearly in $1/m_Q$ to the bottom mass.}
\\[10.0ex] \hline  \\[-2.0ex]
\SLfnalmilcBsK & \cite{Bazavov:2019aom} & 2+1 & Fermilab &
\parbox[t]{8cm}{(See comments for continuum limit extrapolation.)}
\\[3.0ex]
\ifx\reducedapptables\undefined
\hline \\[-1.0ex]
\SLfnalmilcBpi & \cite{Lattice:2015tia} & 2+1 & Fermilab &
\parbox[t]{8cm}{Total statistical + chiral extrapolation + heavy-quark discretization + $g_{B^*B\pi}$ error estimated to be 3.1\% for $f_+$ and 3.8\% for $f_0$ at $q^2=20~{\rm GeV}^2$.}
\\[7.0ex] \hline \\[-1.0ex]
\parbox[t]{20mm}{\SLLambdabp} & \cite{Detmold:2015aaa} & 2+1 & Columbia RHQ &
\parbox[t]{8cm}{Discretization errors discussed as part of combined chiral-continuum-$q^2$ fit, stemming from $a^2|\mathbf{p}|^2$ terms.}
\\[4.0ex] \hline \\[-1.0ex]
\SLrbcukqcdBpi & \cite{Flynn:2015mha} & 2+1 & Columbia RHQ &
\parbox[t]{8cm}{Discretization errors estimated by power counting to be 1.8\% for $f_+$ and 1.7\% for $f_0$.}
\\[4.0ex] \hline \\[-1.0ex]
\SLhpqcdBsK & \cite{Bouchard:2014ypa} & 2+1 & NRQCD &
\parbox[t]{8cm}{Currents matched using 1-loop HISQ lattice perturbation theory, omitting $\cO(\alpha_s\Lambda_{\rm QCD}/m_b$. Systematic uncertainty resulting from 1-loop matching and neglecting $\cO(\Lambda_{\rm QCD}^2/m_b^2$ terms estimated at 4\% from power counting.}
\\[12.0ex] \hline \\[-1.0ex]
%
%
HPQCD 06 & \cite{Dalgic:2006dt} & 2+1 & NRQCD &
\parbox[t]{8cm}{Discretization errors in $f_+(q^2)$ estimated to be $\cO (\alpha_s (a \Lambda_{\rm QCD})^2) \sim  3\%$.  Relativistic errors estimated to be $\cO((\Lambda_{\rm QCD}/M)^2) \sim 1\%$.}
\\[7.0ex]
%
%
\fi
\hline\hline
\end{tabular*}
\caption{Heavy-quark treatment in determinations of $B\to\pi\ell\nu$, $B_s\to K\ell\nu$, and $\Lambda_b\to p\ell\bar{\nu}$ form factors.}
}
\end{table}

\FloatBarrier
\subsubsection{Form factors for rare decays of beauty hadrons}
\label{app:BtoK_Notes}
\FloatBarrier

\begin{table}[!ht]

{\footnotesize
\begin{tabular*}{\textwidth}{l c c c l l}
\hline\hline \\[-1.0ex]
Collab. & Ref. & $\Nf$ & $a$ [fm] & Continuum extrapolation & Scale setting 
\\[1.0ex] \hline \hline \\[-1.0ex]
HPQCD 22 & \cite{Parrott:2022rgu} & 2+1+1 & \parbox[t]{20mm}{ 0.15, 0.12, 0.090, 0.088, 0.059, 0.044 }   & \parbox[t]{4.2cm}{Combined extrapolation in lattice spacing, light-quark mass, strange-quark mass, heavy-quark mass, and momentum transfer using modified $z$ expansion. Stability tested by varying fit form, changing prior widths, and removing data subsets.} & \parbox[t]{30mm}{Scale setting using gradient flow $w_0$ with physical scale from $f_\pi$. } 
\\[24.0ex]  \hline \\[-2.0ex]
\parbox[t]{2.0cm}{Meinel 20, Meinel 21B} & \cite{Meinel:2020owd,Meinel:2021mdj} & 2+1 & \parbox[t]{20mm}{0.0828(3), 0.1106(3)}   & \parbox[t]{4.2cm}{Combined chiral-continuum extrapolation as part of the expansion of form factor shape in powers of $w-1$. Systematic uncertainty estimated by repeating fit with added higher-order terms.} & \parbox[t]{30mm}{Scale setting using $\Omega$ mass in Ref.~\cite{Blum:2014tka}.} 
\\[18.0ex]
\ifx\reducedapptables\undefined
Detmold 16 & \cite{Detmold:2016pkz} & 2+1 & \parbox[t]{20mm}{0.0849(12), 0.1119(17)}   &
\parbox[t]{4.2cm}{Joint chiral-continuum extrapolation, combined with fit to $q^2$-dependence of form factors in a ``modified'' $z$-expansion. Systematics estimated by varying fit form and $\cO(a)$ improvement parameter values. Contrary to Detmold 15 $\Lambda_b \rightarrow p$, no odd powers of spatial momenta are included in the fit ansatz, based on cubic symmetry.} & \parbox[t]{30mm}{Set from $\Upsilon(2S)$--$\Upsilon(1S)$ splitting, cf.~\cite{Meinel:2010pv}.}
\\[32.0ex] \hline \\[-2.0ex]
FNAL/MILC 15E & \cite{Bailey:2015nbd} & 2+1 & \parbox[t]{20mm}{0.045, 0.06, 0.09, 0.12}   &
\parbox[t]{4.2cm}{Fit to SU(2) HMrS$\chi$PT for the combined chiral-continuum limit extrapolation.
Combined stat + chiral extrap + HQ discretization + $g_{B^*B\pi}$ error provided as a function of $q^2$ for each form factor; representative impact on $f_T$ quoted as 3.8\% at $q^2=20~{\rm GeV}^2$.} & \parbox[t]{30mm}{Relative scale $r_1/a$ set from the static-quark potential. Absolute scale $r_1$, including related uncertainty estimates, taken from \cite{Bazavov:2011aa}.}
\\[26.0ex] \hline \\[-2.0ex]
\fi
\ifx\reducedapptables\undefined
 \hline \\[-2.0ex]
 FNAL/MILC 15D & \cite{Bailey:2015dka} & 2+1 & \parbox[t]{20mm}{0.045, 0.06, 0.09, 0.12}   &
\parbox[t]{4.2cm}{Fit to SU(2) HMrS$\chi$PT for the combined chiral-continuum limit extrapolation.
Combined stat + chiral extrap + HQ discretization + $g_{B^*B\pi}$ error provided as a function of $q^2$ for each form factor, ranging between $\sim 1.4$\% and $\sim 2.8$\%.} & \parbox[t]{30mm}{Relative scale $r_1/a$ set from the static-quark potential. Absolute scale $r_1$, including related uncertainty estimates, taken from \cite{Bazavov:2011aa}.}
\\[23.0ex] \hline \\[-2.0ex]
HPQCD 13E & \cite{Bouchard:2013mia} & 2+1 & \parbox[t]{20mm}{0.09,0.12}   &
\parbox[t]{4.2cm}{Combined chiral-continuum extrapolation using rHMS$\chi$PT. Errors provided as a function of $q^2$, combined total ranging from $\sim 3$\% to $\sim 5$\% in data region.} & \parbox[t]{30mm}{Relative scale $r_1/a$ set from the static-quark potential. Absolute scale $r_1$
set to 0.3133(23)~fm.}
\\[15.0ex]
\fi
\hline\hline
\end{tabular*}
\caption{Continuum extrapolations/estimation of lattice artifacts in
  determinations of form factors for rare decays of beauty hadrons.\label{tab:AppTableRareBCont}}
}
\end{table}
\begin{table}[!ht]
{\footnotesize
\begin{tabular*}{\textwidth}{l @{\extracolsep{\fill}} c c c   l}
\hline\hline \\[-1.0ex]
{Collab.} & {Ref.} & {$\Nf$} & {$M_{\pi,\rm min}\,[\mev]$}  & {Description}  
\\[1.0ex] \hline \hline \\[-1.0ex]
HPQCD 22 & \cite{Parrott:2022rgu} & 2+1+1 & \parbox[t]{20mm}{ 131, 132, 313, 128, 325, 308 }   & \parbox[t]{5cm}{Combined extrapolation in lattice spacing, light-quark mass, strange-quark mass, heavy-quark mass, and momentum transfer using modified $z$ expansion. Logarithms from hard-pion $\chi$PT included. Stability tested by varying fit form, changing prior widths, and removing data subsets.}
\\[24.0ex]  \hline \\[-2.0ex]
\parbox[t]{2.0cm}{Meinel 20, Meinel 21B} & \cite{Meinel:2020owd,Meinel:2021mdj} & 2+1 &  \parbox[t]{2cm}{303, 340}
& \parbox[t]{5cm}{Combined chiral-continuum extrapolation as part of the expansion of form factor shape in powers of $w-1$. Systematic uncertainty estimated by repeating fit with added higher-order terms.}
\\[16.0ex]
\ifx\reducedapptables\undefined
\hline \\[-2.0ex]
FNAL/MILC 15D & \cite{Bailey:2015dka} & 2+1 & \parbox[t]{2cm}{330, 260, 280, 470}
& \parbox[t]{5cm}{Simultaneous chiral-continuum extrapolation and $q^2$ interpolation using SU(2) HMrS$\chi$PT,
with a hard-kaon $\chi$PT treatment of high-energy kaons. Combined stat + chiral extrap + HQ discretization + $g_{B^*B\pi}$ error provided as a function of $q^2$ for each form factor, ranging between $\sim 1.4$\% and $\sim 2.8$\%.}
\\[23.0ex] \hline \\[-2.0ex]
HPQCD 13E & \cite{Bouchard:2013mia} & 2+1 & \parbox[t]{2cm}{295, 260}
& \parbox[t]{5cm}{Combined chiral-continuum extrapolation using rHMS$\chi$PT. Errors provided as a function of $q^2$, combined total ranging from $\sim 3$\% to $\sim 5$\% in data region.}
\\[12.0ex]
\hline \\[-2.0ex]
Detmold 16 & \cite{Detmold:2016pkz} & 2+1 &  \parbox[t]{2.5cm}{227, 245 \\ (valence-valence) \\ 263, 294 \\ (valence-sea)}
& \parbox[t]{5cm}{Joint chiral-continuum extrapolation, combined with fit to $q^2$-dependence of form factors in a ``modified'' $z$-expansion. Only analytic NLO terms $\propto (m_\pi^2-m_{\pi,{\rm phys}}^2)$ included in light-mass dependence. Systematic uncertainty estimated by repeating fit with added higher-order terms.}
\\[23.0ex] \hline \\[-2.0ex]
FNAL/MILC 15E & \cite{Bailey:2015nbd} & 2+1 & \parbox[t]{2cm}{330, 260, 280, 470}
& \parbox[t]{5cm}{Simultaneous chiral-continuum extrapolation and $q^2$ interpolation using SU(2) HMrS$\chi$PT,
with a hard-pion $\chi$PT treatment of high-energy pions.}
\\[12.0ex] \hline \\[-2.0ex]
\fi
\hline\hline
\end{tabular*}
\caption{Chiral extrapolation/minimum pion mass in
  determinations of form factors for rare decays of beauty hadrons. For actions with multiple species of pions, masses quoted are the RMS pion masses for $\Nf=2+1$ and the Goldstone mode mass for $\Nf=2+1+1$. The different $M_{\pi,\rm min}$ entries correspond to the different lattice spacings. \label{tab:AppTableRareBChir}}
}
\end{table}
\begin{table}[!ht]
{\footnotesize
\begin{tabular*}{\textwidth}{l @{\extracolsep{\fill}} c c c c l}
\hline\hline \\[-1.0ex]
Collab. & Ref. & $\Nf$ & $L$ [fm] & ${M_{\pi,\rm min}}L$ & Description 
\\[1.0ex] \hline \hline \\[-1.0ex]
HPQCD 22 & \cite{Parrott:2022rgu} & 2+1+1 &\parbox[t]{1.8cm}{2.4/4.8, 2.88/5.76, 2.88, 5.63, 2.83, 2.82 }   & \parbox[t]{1.8cm}{3.19, 3.86, 4.57, 3.66, 4.67, 4.41}  & \parbox[t]{4.8cm}{Finite-volume effects included in fit by replacing infinite-volume chiral logs with sums over discrete momenta. }
\\[14.0ex]  \hline \\[-2.0ex]
\parbox[t]{2.0cm}{Meinel 20, Meinel 21B} & \cite{Meinel:2020owd,Meinel:2021mdj} & 2+1 & \parbox[t]{1.8cm}{2.7, 2.7}  & \parbox[t]{1.8cm}{4.1, 4.6} & \parbox[t]{4.8cm}{Finite-volume effects not quantified. Effects from unstable $\Lambda^*(1520)$ not quantified.}
\\[8.0ex]
\ifx\reducedapptables\undefined
\hline \\[-1.0ex]
FNAL/MILC 15D & \cite{Bailey:2015dka} & 2+1 & \parbox[t]{1.8cm}{2.9, 2.9/3.8, 2.5/2.9/3.6/5.8, 2.4/2.9} & $\gtrsim 3.8$  &
\parbox[t]{4.8cm}{Finite-volume effects estimated by replacing infinite-volume chiral logs with sums over discrete momenta, found to be negligible.
}
\\[9.0ex] \hline \\[-1.0ex]
HPQCD 13E & \cite{Bouchard:2013mia} & 2+1 & \parbox[t]{1.8cm}{2.5, 2.4/2.9} & $\gtrsim 3.8$  &
\parbox[t]{4.8cm}{Finite-volume effects included in combined chiral-continuum extrapolation.
}
\\[4.0ex] 
\hline \\[-2.0ex]
Detmold 16 & \cite{Detmold:2016pkz} & 2+1 & \parbox[t]{1.8cm}{2.7, 2.7} & \parbox[t]{1.8cm}{$\gtrsim 3.6$ \\ (valence-sea pion)}  &
\parbox[t]{4.8cm}{Finite-volume effect estimated at 3\% from experience on $\chi$PT estimates of finite-volume effects for heavy-baryon axial couplings.
}
\\[9.0ex] \hline \\[-1.0ex]
\parbox[t]{2.7cm}{FNAL/MILC 15E} & \cite{Bailey:2015nbd} & 2+1 & \parbox[t]{1.8cm}{2.9, 2.9/3.8, 2.5/2.9/3.6/5.8, 2.4/2.9} & $\gtrsim 3.8$  &
\parbox[t]{4.8cm}{Finite-volume effects estimated by replacing infinite-volume chiral logs with sums over discrete momenta, found to be negligible.
}
\\[9.0ex] \hline \\[-1.0ex]
\fi
\hline\hline
\end{tabular*}
\caption{Finite-volume effects in determinations of form factors for rare decays of beauty hadrons.  Each $L$-entry corresponds to a different lattice
spacing, with multiple spatial volumes at some lattice spacings. For actions with multiple species of pions, masses quoted are the RMS pion masses for $\Nf=2+1$ and the Goldstone mode mass for $\Nf=2+1+1$. \label{tab:AppTableRareBFV}}
}
\end{table}
\begin{table}[!ht]
{\footnotesize
\begin{tabular*}{\textwidth}{l @{\extracolsep{\fill}} c c c l}
\hline\hline \\[-1.0ex]
Collab. & Ref. & $\Nf$ & Ren. & Description 
\\[1.0ex] \hline \hline \\[-1.0ex]
HPQCD 22 & \cite{Parrott:2022rgu} & 2+1+1 &  NPR    & \parbox[t]{6cm}{ $Z_V$ and $Z_A$ obtained from Ward identities. $Z_T$ determined using RI-SMOM.}
\\[10.0ex]  \hline \\[-2.0ex]
\parbox[t]{2.0cm}{Meinel 20, Meinel 21B} & \cite{Meinel:2020owd,Meinel:2021mdj} & 2+1 & mNPR  & \parbox[t]{6cm}{Residual matching factors $\rho$ computed at 1-loop for vector and axial-vector currents, but at tree-level only for tensor currents. A systematic uncertainty is assigned to $\rho_{T^{\mu\nu}}$ as the double of ${\rm max}(|\rho_{A^{\mu}}-1|,|\rho_{V^{\mu}}-1|)$.}
\\[16.0ex]
\ifx\reducedapptables\undefined
\hline \\[-1.0ex]
FNAL/MILC 15D & \cite{Bailey:2015dka} & 2+1 & mNPR &
\parbox[t]{6cm}{Perturbative truncation error estimated at 1\% for $f_+$ and $f_0$ and 2\% for $f_T$, using size of 1-loop correction on next-to-finest ensemble.}
\\[10.0ex] \hline \\[-1.0ex]
HPQCD 13E & \cite{Bouchard:2013mia} & 2+1 & mNPR &
\parbox[t]{6cm}{Currents matched using 1-loop massless-HISQ lattice perturbation theory. Associated systematic uncertainty dominates quoted 4\% uncertainty from matching, charm quenching, and electromagnetic and isospin-breaking effects.}
\\[16.0ex]
\hline \\[-2.0ex]
Detmold 16 & \cite{Detmold:2016pkz} & 2+1 & mNPR &
\parbox[t]{6cm}{Same procedure as in Detmold 15 $\Lambda_p \rightarrow p$ for vector and axial currents. For tensor currents, the residual renormalization factor is set to its tree-level value $\rho_{T^{\mu\nu}}=1$. A systematic uncertainty is assigned as the double of ${\rm max}[1-\rho_{A^{\mu}},1-\rho_{V^{\mu}}]$, using the known 1-loop values of the residual matchings for the vector and axial currents.}
\\[21.0ex] \hline \\[-1.0ex]
FNAL/MILC 15E & \cite{Bailey:2015nbd} & 2+1 & mNPR &
\parbox[t]{6cm}{Perturbative truncation error on $f_T$ estimated at 0.7\% from the ratio of singlet renormalization constants and 2.0\% from the residual renormalization.}
\\[10.0ex] \hline \\[-1.0ex]
\fi
\hline\hline
\end{tabular*}
\caption{Operator renormalization in determinations of form factors for rare decays of beauty hadrons. \label{tab:AppTableRareBRen}}
}
\end{table}
\begin{table}[!ht]
{\footnotesize
\begin{tabular*}{\textwidth}{l @{\extracolsep{\fill}} c c c l}
\hline\hline \\[-1.0ex]
Collab. & Ref. & $\Nf$ & Action & Description 
\\[1.0ex] \hline \hline \\[-1.0ex]
HPQCD 22 & \cite{Parrott:2022rgu} & 2+1+1 &  HISQ    & \parbox[t]{6cm}{Extrapolation to the physical $b$-quark mass using terms with powers and logarithms of the inverse heavy-meson mass in the modified $z$-expansion fit. Heavy-quark masses in lattice units satisfy $am_h \leq 0.9$. The heavy-light pseudoscalar meson mass reaches $\approx 0.94\: M_{B,{\rm phys.}}$.}
\\[18.0ex]  \hline \\[-2.0ex]
\parbox[t]{2.0cm}{Meinel 20, Meinel 21B} & \cite{Meinel:2020owd,Meinel:2021mdj} & 2+1 & Columbia RHQ & \parbox[t]{6cm}{Discretization errors discussed as part of combined chiral-continuum-$w$ fit. Higher-order fit also includes $\mathcal{O}(\alpha_s a |\mathbf{p}|)$ terms to account for missing radiative corrections to $\mathcal{O}(a)$ improvement of the currents.}
\\[12.0ex]
\ifx\reducedapptables\undefined
\hline \\[-1.0ex]
FNAL/MILC 15D & \cite{Bailey:2015dka} & 2+1 & Fermilab &
\parbox[t]{7.8cm}{Combined stat + chiral extrap + HQ discretization + $g_{B^*B\pi}$ error provided as a function of $q^2$ for each form factor, ranging between $\sim 1.4$\% and $\sim 2.8$\%.}
\\[7.0ex] \hline \\[-1.0ex]
HPQCD 13E & \cite{Bouchard:2013mia} & 2+1 & NRQCD &
\parbox[t]{7.8cm}{Currents matched using 1-loop massless-HISQ lattice perturbation theory. Associated systematic uncertainty dominates quoted 4\% uncertainty from matching, charm quenching, and electromagnetic and isospin-breaking effects.}
\\[12.0ex]
\hline \\[-2.0ex]
Detmold 16 & \cite{Detmold:2016pkz} & 2+1 & Columbia RHQ &
\parbox[t]{7cm}{Discretization errors discussed as part of combined chiral-continuum-$q^2$ fit, stemming from $a^2|\mathbf{p}|^2$ terms.}
\\[4.0ex] \hline \\[-1.0ex]
\parbox[t]{2.7cm}{FNAL/MILC 15E} & \cite{Bailey:2015nbd} & 2+1 & Fermilab &
\parbox[t]{7.8cm}{Combined stat + chiral extrap + HQ discretization + $g_{B^*B\pi}$ error provided as a function of $q^2$ for each form factor; representative impact on $f_T$ quoted as 3.8\% at $q^2=20~{\rm GeV}^2$.}
\\[10.0ex] \hline \\[-1.0ex]
\fi
\hline\hline
\end{tabular*}
\caption{Heavy-quark treatment in determinations of form factors for rare decays of beauty hadrons. \label{tab:AppTableRareBHQ}}
}
\end{table}

\clearpage
\subsubsection{Form factors entering determinations of $|V_{cb}|$ ($B_{(s)} \to D_{(s)}^{(*)}\ell\nu$,  $\Lambda_b \to \Lambda_c^{(*)} \ell \bar{\nu}$) and $R(D_{(s)})$}
\label{app:BtoD_Notes}
\vspace{-0.48cm}
\begin{table}[!ht]
{\footnotesize
\begin{tabular*}{\textwidth}{l @{\extracolsep{\fill}} c c c l l}
\hline\hline \\[-1.0ex]
Collab. & Ref. & $\Nf$ & $a$ [fm] & Continuum extrapolation & Scale setting 
\\[1.0ex] \hline \hline \\[-1.0ex]
\parbox[t]{20mm}{HPQCD 23}                & \cite{Harrison:2023dzh}              & 2+1+1 & \parbox[t]{13mm}{ 0.044, 0.058, 0.088} 
                                          & \parbox[t]{40mm}{Combined chiral-continuum and heavy-quark extrapolations using HMrS$\chi$PT. The recoil dependence in powers of $(w-1)$ is fitted using BGL-inspired coefficients. Zero-recoil uncertainty negligible compared to other sources of error.}
                                          & \parbox[t]{30mm}{Scale setting from Wilson flow, fixing the slope $\left.t\tfrac{d}{dt}\left\{t^2\left\langle E(t)\right\rangle\right\}\right|_{t=w_0^2} = 0.3$, with $w_0$ taken from~\cite{Dowdall:2013rya}. Uncertainty related to scale setting estimated at $\approx 0.5\%$.}
\\[24.5ex] \hline \\[-2.0ex]
\parbox[t]{20mm}{JLQCD 23}                & \cite{Aoki:2023qpa}                  & 2+1   & \parbox[t]{15mm}{ 0.044, 0.055, 0.080}                     
                                          & \parbox[t]{40mm}{Combined chiral-continuum and heavy-quark extrapolations using HM$\chi$PT. Each form factor is extrapolated separately. Zero-recoil uncertainty estimated at $\approx 0.9\%$.}
                                          & \parbox[t]{30mm}{Scale setting from Wilson flow, fixing the slope $\left.t\tfrac{d}{dt}\left\{t^2\left\langle E(t)\right\rangle\right\}\right|_{t=w_0^2} = 0.3$, with $w_0$ taken from~\cite{Borsanyi:2012zs}. Uncertainty related to scale setting estimated at $\approx 1.7\%$.} 
\\[24.5ex] \hline \\[-2.0ex]
\parbox[t]{23mm}{\SLfnalmilcBDstar}       & \cite{FermilabLattice:2021cdg}       & 2+1   & \parbox[t]{15mm}{ 0.045, 0.06, 0.09, 0.12, 0.15}
                                          & \parbox[t]{40mm}{Combined chiral-continuum extrapolation using HMrS$\chi$PT. Total uncertainty quoted at 0.7\%.}
                                          & \parbox[t]{30mm}{Relative scale $r_1/a$ set from the static-quark potential.  Absolute scale $r_1$, including related uncertainty estimates, taken from \cite{Bazavov:2011aa}. Uncertainty related to scale setting estimated at less than 0.1\%.} 
\\[29.5ex] \hline \\[-2.0ex]
\parbox[t]{20mm}{Meinel 21, Meinel 21B} & \cite{Meinel:2021rbm,Meinel:2021mdj} & 2+1   & \parbox[t]{15mm}{0.0828(3), 0.1106(3)}
                                          & \parbox[t]{40mm}{Combined chiral-continuum extrapolation as part of the expansion of form factor shape in powers of $w-1$. Systematics estimated by varying fit form.}
                                          & \parbox[t]{30mm}{Scale setting using $\Omega$ mass in Ref.~\cite{Blum:2014tka}.} 
\\ [15.5ex] 
\ifx\reducedapptables\undefined
\hline
\parbox[t]{20mm}{HPQCD 19} & \cite{McLean:2019qcx} & 2+1+1 & \parbox[t]{1.5cm}{0.044, 0.059, 0.09} &
\parbox[t]{40mm}{Combined chiral-continuum extrapolation. Fractional error from continuum limit and
extrapolation to physical $b$ mass at zero recoil quoted as 1.20\%.}
& \parbox[t]{30mm}{Scale setting from $f_\pi$ via the flow quantity $w_0$~\cite{Dowdall:2013rya,Chakraborty:2014aca,Chakraborty:2016mwy}.} 
\\[14.0ex] \hline \\[-2.0ex]
\parbox[t]{20mm}{HPQCD 19B} & \cite{McLean:2019sds} & 2+1+1 & \parbox[t]{1.5cm}{0.044, 0.059, 0.09} &
\parbox[t]{40mm}{Combined chiral-continuum extrapolation. Fractional error from continuum limit and extrapolation to physical $b$ mass at zero recoil quoted as 0.73\% and 0.69\%, respectively.}
& \parbox[t]{30mm}{Scale setting from $f_\pi$ via the flow quantity $w_0$~\cite{Dowdall:2013rya,Chakraborty:2014aca,Chakraborty:2016mwy}.} 
\\[20.0ex]
\hline
\parbox[t]{2.0cm}{HPQCD 17B} & \cite{Harrison:2017fmw} & 2+1+1 & \parbox[t]{1.5cm}{0.09, 0.12, 0.15} &
\parbox[t]{4.5cm}{Combined chiral-continuum extrapolation. $\mathcal{O}(a^2)$ uncertainty on the two relevant form factors at zero recoil estimated to be 0.7\% and 1.4\%.}
& \parbox[t]{35mm}{Determined from $\Upsilon(2S-1S)$ splitting in~\cite{Dowdall:2011wh}.} 
\\[12ex]  \hline
\parbox[t]{2.6cm}{HPQCD 17} & \cite{Monahan:2017uby}
 & 2+1 & \parbox[t]{1.5cm}{0.09, 0.12} &
\parbox[t]{4.2cm}{Combined chiral-continuum extrapolation as part of modified $z$-expansion of form factors, which also includes uncertainty related to matching of NRQCD and relativistic currents.}
& \parbox[t]{30mm}{Implicitly set from $r_1$.} 
\\[18ex] \hline \\[-2.0ex]
\parbox[t]{26mm}{Datta 17} & \cite{Datta:2017aue} & 2+1 & \parbox[t]{1.5cm}{0.0849(12), 0.1119(17)}   &
\parbox[t]{4.2cm}{Joint chiral-continuum extrapolation, combined with fit to $q^2$-dependence of form factors in a ``modified'' $z$-expansion. Systematics estimated by varying fit form and $\cO(a)$ improvement parameter values.} & \parbox[t]{30mm}{Set from $\Upsilon(2S)$--$\Upsilon(1S)$ splitting, cf.~\cite{Meinel:2010pv}.}
\\[21.0ex] 
\hline \\[-2.0ex]
\parbox[t]{26mm}{\SLfnalmilcBD} & \cite{Lattice:2015rga} & 2+1 & \parbox[t]{1.5cm}{ 0.045, 0.06, 0.09, 0.12} &
\parbox[t]{4.2cm}{Combined chiral-continuum extrapolation using HMrS$\chi$PT. Form factors fitted to NLO $\chi$PT, with chiral logs taken from staggered version of the Chow-Wise result, modified to include taste-breaking terms. $\cO(a^2)$ terms introduced based on power-counting arguments. Total uncertainty estimated at 0.6\% for $f_+$ and 0.5\% for $f_0$ for the largest recoil.}
& \parbox[t]{30mm}{Relative scale $r_1/a$ set from the static-quark potential.  Absolute scale $r_1$, including related uncertainty estimates, taken from \cite{Bazavov:2011aa}. Uncertainty related to scale setting estimated at 0.2\%.} 
\\[35ex] \hline \\[-2.0ex]
\parbox[t]{26mm}{\SLLambdabc} & \cite{Detmold:2015aaa} & 2+1 & \parbox[t]{1.5cm}{0.0849(12), 0.1119(17)}   &
\parbox[t]{4.2cm}{Joint chiral-continuum extrapolation, combined with fit to $q^2$-dependence of form factors in a ``modified'' $z$-expansion. Systematics estimated by varying fit form and $\cO(a)$ improvement parameter values.} & \parbox[t]{30mm}{Set from $\Upsilon(2S)$--$\Upsilon(1S)$ splitting, cf.~\cite{Meinel:2010pv}.}
\\[21.0ex] \hline \\[-2.0ex]
\fi
\hline\hline
\end{tabular*}
\caption{Continuum extrapolations/estimation of lattice artifacts in
    $\Nf=2+1$ determinations of $B_{(s)} \to D_{(s)}^{(*)}\ell\nu$ and $\Lambda_b \to \Lambda_c^{(*)} \ell \bar{\nu}$ form factors, and of $R(D_{(s)})$.}
}
\end{table}

\ifx\reducedapptables\undefined

\begin{table}[!ht]
{\footnotesize
\begin{tabular*}{\textwidth}{l @{\extracolsep{\fill}} c c c l l}
\hline\hline \\[-1.0ex]
Collab. & Ref. & $\Nf$ & $a$ [fm] & Continuum extrapolation & Scale setting 
\\[1.0ex] \hline \hline \\[-1.0ex]
\parbox[t]{26mm}{Atoui  13} & \cite{Atoui:2013zza} & 2& \parbox[t]{20mm}{0.054, 0.067, 0.085, 0.098} &
\parbox[t]{4.2cm}{Combined continuum and chiral extrapolation, with linear terms in $a^2$ and $m_{\rm sea}$.
No dependence on $a$ or $m_{\rm sea}$ observed within errors. Stability of results vs.~fits with no $m_{\rm sea}$-dependence checked.}
& \parbox[t]{30mm}{Scale set through $F_\pi$.} 
\\[18ex]
\hline\hline
\end{tabular*}
\caption{Continuum extrapolations/estimation of lattice artifacts in
    $\Nf=2$ determinations of $B_{(s)} \to D_{(s)}^{(*)}\ell\nu$ and $\Lambda_b \to \Lambda_c^{(*)} \ell \bar{\nu}$ form factors, and of $R(D_{(s)})$.}
}
\end{table}

\fi

\begin{table}[!ht]
{\footnotesize
\begin{tabular*}{\textwidth}{l @{\extracolsep{\fill}} c c c  l}
\hline\hline \\[-1.0ex]
{Collab.} & {Ref.} & {$\Nf$} & {$M_{\pi,\rm min}\,[\mev]$}  & {Description}  
\\[1.0ex] \hline \hline \\[-1.0ex]
\parbox[t]{20mm}{HPQCD 23}                & \cite{Harrison:2023dzh}              & 2+1+1 & \parbox[t]{20mm}{315, 135, 129}
                                          & \parbox[t]{50mm}{Combined chiral-continuum and heavy-quark extrapolations using HMrS$\chi$PT. The recoil dependence in powers of $(w-1)$ is fitted using BGL-inspired coefficients. Zero-recoil uncertainty negligible compared to other sources of error.}
\\[19.0ex] \hline \\[-2.0ex]
\parbox[t]{20mm}{JLQCD 23}                & \cite{Aoki:2023qpa}                  & 2+1   & \parbox[t]{20mm}{284, 300, 226}
                                          & \parbox[t]{50mm}{Combined chiral-continuum and heavy-quark extrapolations using HM$\chi$PT. Each form factor is extrapolated separately. Zero-recoil uncertainty estimated at $\approx 0.9\%$.}
\\[13.5ex] \hline \\[-2.0ex]
\parbox[t]{23mm}{\SLfnalmilcBDstar}       & \cite{FermilabLattice:2021cdg}       & 2+1   & \parbox[t]{20mm}{320, 220, 180, 270, 340}
                                          & \parbox[t]{50mm}{Combined chiral-continuum extrapolation using HMrS$\chi$PT.  Systematic errors estimated by adding higher-order analytic terms and varying the $D^*$-$D$-$\pi$ coupling. Total uncertainty quoted at 0.7\%.}
\\[15.5ex] \hline \\[-2.0ex]
\parbox[t]{20mm}{Meinel 21, Meinel 21B} & \cite{Meinel:2021rbm,Meinel:2021mdj} & 2+1   & \parbox[t]{2cm}{303, 340}
                                          & \parbox[t]{50mm}{Combined chiral-continuum extrapolation as part of the expansion of form factor shape in powers of $w-1$. Systematic uncertainty estimated by repeating fit with added higher-order terms.}
\\[16.0ex]
\ifx\reducedapptables\undefined
\hline
\parbox[t]{26mm}{HPQCD 19} & \cite{McLean:2019qcx} & 2+1+1 & \parbox[t]{2cm}{315, 329, 129 \cite{Bazavov:2017lyh}}
 & \parbox[t]{5cm}{Combined chiral-continuum extrapolation using rS$\chi$PT. No specific uncertainty coming from chiral extrapolation quoted.}
\\[11.0ex]
\hline \\[-2.0ex]
\parbox[t]{26mm}{HPQCD 19B} & \cite{McLean:2019sds} & 2+1+1 & \parbox[t]{2cm}{315, 329, 129 \cite{Bazavov:2017lyh}}
 & \parbox[t]{5cm}{Combined chiral-continuum extrapolation using rS$\chi$PT. No specific uncertainty coming from chiral extrapolation quoted.}
\\[10.0ex]
\hline \\[-2.0ex]
\parbox[t]{26mm}{HPQCD 17B} & \cite{Harrison:2017fmw} & 2+1+1 & \parbox[t]{2cm}{130, 133, 130}
 & \parbox[t]{5cm}{Combined chiral-continuum extrapolation using rS$\chi$PT. No specific uncertainty coming from chiral extrapolation quoted.}
\\[10.0ex]
\hline \\[-2.0ex]
\parbox[t]{26mm}{HPQCD 17} & \cite{Monahan:2017uby}  & 2+1 & \parbox[t]{2cm}{295, 260}
 & \parbox[t]{5cm}{Combined chiral-continuum extrapolation as part of modified $z$-expansion of form factors. Systematic uncertainties at $q^2=0$ estimated to 0.80\% ($B_s \to D_s$) and 1.14\% ($B \to D$, with input from hard-pion $\chi$PT).}
\\[18.0ex] \hline \\[-2.0ex]
\parbox[t]{26mm}{Datta 17} & \cite{Datta:2017aue} & 2+1 & \parbox[t]{2.5cm}{227, 245 \\ (valence-valence) \\ 263, 294 \\ (valence-sea)}
& \parbox[t]{5cm}{Joint chiral-continuum extrapolation, combined with fit to $q^2$-dependence of form factors in a ``modified'' $z$-expansion. Only analytic $NLO$ terms $\propto (m_\pi^2-m_{\pi,{\rm phys}}^2)$ included in light-mass dependence. Systematic uncertainty estimated by repeating fit with added higher-order terms.}
\\[23.0ex] 
\hline \\[-2.0ex]
\parbox[t]{26mm}{\SLfnalmilcBD} & \cite{Lattice:2015rga} & 2+1 & \parbox[t]{2cm}{330, 260, 280, 470}
 & \parbox[t]{5cm}{Combined chiral-continuum extrapolation using HMrS$\chi$PT. Form factors fitted to NLO $\chi$PT, with chiral logs taken from staggered version of the Chow-Wise result, modified to include taste-breaking terms. $\cO(a^2)$ terms introduced based on power-counting arguments. Total uncertainty estimated at 0.6\% for $f_+$ and 0.5\% for $f_0$ for the largest recoil.}
\\[29.0ex] \hline \\[-2.0ex]
\parbox[t]{26mm}{\SLLambdabc} & \cite{Detmold:2015aaa} & 2+1 & \parbox[t]{2.5cm}{227, 245 \\ (valence-valence) \\ 263, 294 \\ (valence-sea)}
& \parbox[t]{5cm}{Joint chiral-continuum extrapolation, combined with fit to $q^2$-dependence of form factors in a ``modified'' $z$-expansion. Only analytic $NLO$ terms $\propto (m_\pi^2-m_{\pi,{\rm phys}}^2)$ included in light-mass dependence. Systematic uncertainty estimated by repeating fit with added higher-order terms.}
\\[23.0ex] \hline \\[-2.0ex]
\fi
\hline\hline
\end{tabular*}
\caption{Chiral extrapolation/minimum pion mass in $\Nf=2+1$
  determinations of $B_{(s)} \to D_{(s)}^{(*)}\ell\nu$ and $\Lambda_b \to \Lambda_c^{(*)} \ell \bar{\nu}$ form factors, and of $R(D_{(s)})$.  For actions with multiple species of pions, masses quoted are the RMS pion masses for $\Nf=2+1$ and the Goldstone mode mass for $\Nf=2+1+1$.  The different $M_{\pi,\rm min}$ entries correspond to the different lattice spacings.}
}
\end{table}

\ifx\reducedapptables\undefined

\begin{table}[!ht]
{\footnotesize
\begin{tabular*}{\textwidth}{l @{\extracolsep{\fill}} c c c  l}
\hline\hline \\[-1.0ex]
{Collab.} & {Ref.} & {$\Nf$} & {$M_{\pi,\rm min}\,[\mev]$}  & {Description}  
\\[1.0ex] \hline \hline \\[-1.0ex]
Atoui  13 & \cite{Atoui:2013zza} & 2 & \parbox[t]{2cm}{270, 300, 270, 410}
 & \parbox[t]{5cm}{Combined continuum and chiral extrapolation, with linear terms in $a^2$ and $m_{\rm sea}$.
No dependence on $a$ or $m_{\rm sea}$ observed within errors. Stability of results vs.~fits with no $m_{\rm sea}$
dependence checked.}
\\[15.0ex]
\hline\hline
\end{tabular*}
\caption{Chiral extrapolation/minimum pion mass in $\Nf=2$
  determinations of $B_{(s)} \to D_{(s)}^{(*)}\ell\nu$ and $\Lambda_b \to \Lambda_c^{(*)} \ell \bar{\nu}$ form factors, and of $R(D_{(s)})$.  For actions with multiple species of pions, masses quoted are the RMS pion masses.  The different $M_{\pi,\rm min}$ entries correspond to the different lattice spacings.}
}
\end{table}

\fi
\begin{table}[!ht]
{\footnotesize
\begin{tabular*}{\textwidth}{l @{\extracolsep{\fill}} c c c c l}
\hline\hline \\[-1.0ex]
Collab. & Ref. & $\Nf$ & $L$ [fm] & ${M_{\pi,\rm min}}L$ & Description 
\\[1.0ex] \hline \hline \\[-1.0ex]
\parbox[t]{20mm}{HPQCD 23}                & \cite{Harrison:2023dzh}              & 2+1+1 & \parbox[t]{18mm}{2.8, 2.8--5.5, 2.8--5.6}                & \parbox[t]{18mm}{4.5, 3.7, 3.7}
                                          & \parbox[t]{35mm}{Finite-volume effects expected to be negligible, including the effect of frozen topology in the finest ensemble, according to~\cite{FermilabLattice:2022gku}.}
\\[12.0ex] \hline \\[-2.0ex]
\parbox[t]{20mm}{JLQCD 23}                & \cite{Aoki:2023qpa}                  & 2+1   & \parbox[t]{18mm}{2.8, 2.6, 2.6--3.9}                     & \parbox[t]{18mm}{4.0, 4.0, 4.4}
                                          & \parbox[t]{35mm}{Study effects of topology freezing and compare ensembles with similar parameters but different volumes.}
\\[12.0ex] \hline \\[-2.0ex]
\parbox[t]{23mm}{\SLfnalmilcBDstar}       & \cite{FermilabLattice:2021cdg}       & 2+1   & \parbox[t]{18mm}{4.6, 4.3--6.3, 4.1--5.8, 3.8--6.2, 3.9} & $\gtrsim 3.8$
                                          & \parbox[t]{35mm}{Finite-volume error estimated to be negligible at zero recoil using HMrS$\chi$PT. Given the values $m_\pi L\gtrsim 3.7$ and the smallness of the chiral logs, expectations are that finite-volume errors remain negligible in the whole recoil range.}
\\[24.5ex] \hline \\[-2.0ex]
\parbox[t]{20mm}{Meinel 21, Meinel 21B} & \cite{Meinel:2021rbm,Meinel:2021mdj} & 2+1   & \parbox[t]{18mm}{2.7, 2.7}                               & \parbox[t]{18mm}{4.1, 4.6}
                                          & \parbox[t]{35mm}{Finite-volume effects not quantified. Effects from unstable $\Lambda_c^*$ not quantified.}
\\[7.0ex]
\ifx\reducedapptables\undefined
\hline \\[-1.0ex]
\parbox[t]{26mm}{HPQCD 19} & \cite{McLean:2019qcx} & 2+1+1 & \parbox[t]{1.8cm}{2.73, 2.72, 2.81/5.62} & \parbox[t]{1.8cm}{4.3, 4.5, 3.7/4.5, 3.9/4.5, 3.3/3.8}  &
\parbox[t]{3.5cm}{Finite-volume effects neglected. This includes the effect of frozen topology in the finest ensemble.}
\\[10.0ex] \hline \\[-2.0ex]
\parbox[t]{26mm}{HPQCD 19B} & \cite{McLean:2019sds} & 2+1+1 & \parbox[t]{1.8cm}{2.73, 2.72, 2.81/5.62} & \parbox[t]{1.8cm}{4.3, 4.5, 3.7/4.5, 3.9/4.5, 3.3/3.8}  &
\parbox[t]{3.5cm}{Finite-volume effects neglected. This includes the effect of frozen topology in the finest ensemble.}
\\[10.0ex] \hline
\parbox[t]{26mm}{HPQCD 17B} & \cite{Harrison:2017fmw} & 2+1+1 & \parbox[t]{1.8cm}{2.4/3.7/4.8, 3.0/3.9/5.8, 3.0/5.6} & $\gtrsim 3.2$  &
\parbox[t]{3.5cm}{Finite-volume effects estimated and subtracted using rs$\chi$PT formulae. No associated uncertainty quoted.}
\\[10.0ex] \hline \\[-1.0ex]
\parbox[t]{26mm}{HPQCD 17} & \cite{Monahan:2017uby} & 2+1 & \parbox[t]{1.8cm}{2.5, 2.4/2.9} & $\gtrsim 3.8$  &
\parbox[t]{3.5cm}{Finite-volume effects estimated to be below $0.01\%$.}
\\[4.0ex] \hline \\[-1.0ex]
\parbox[t]{26mm}{Datta 17} & \cite{Datta:2017aue} & 2+1 & \parbox[t]{1.8cm}{2.7, 2.7} & \parbox[t]{1.8cm}{$\gtrsim 3.6$ \\ (valence-sea pion)}  &
\parbox[t]{3.5cm}{No explicit discussion of finite-volume effects. Analysis identical to that of vector and axial form factors in \SLLambdabc.}
\\[15.0ex] 
\hline \\[-1.0ex]
\parbox[t]{26mm}{\SLfnalmilcBD} & \cite{Lattice:2015rga} & 2+1 & \parbox[t]{1.8cm}{2.9, 2.9--3.8, 2.5--5.8, 2.4/2.9} & $\gtrsim 3.8$ & 
\parbox[t]{3.5cm}{ Finite-volume error estimated to be negligible in \cite{Bailey:2012rr}.}
\\[7.0ex] \hline \\[-1.0ex]
\parbox[t]{26mm}{\SLLambdabc} & \cite{Detmold:2015aaa} & 2+1 & \parbox[t]{1.8cm}{2.7, 2.7} & \parbox[t]{1.8cm}{$\gtrsim 3.6$ \\ (valence-sea pion)}  &
\parbox[t]{3.5cm}{Finite-volume effect estimated at 1.5\% from experience on $\chi$PT estimates of finite-volume effects for heavy-baryon axial couplings.
}
\\[13.0ex] \hline \\[-1.0ex]
Atoui  13 & \cite{Atoui:2013zza} & 2 & \parbox[t]{1.8cm}{1.7/2.6, 2.1, 2.0/2.7, 2.4} & $\gtrsim 3.6$ & 
\parbox[t]{3.5cm}{No volume dependence observed within errors.}
\\[4.0ex]
\fi
\hline\hline
\end{tabular*}
\caption{Finite-volume effects in determinations of $B_{(s)} \to D_{(s)}^{(*)}\ell\nu$ and $\Lambda_b \to \Lambda_c^{(*)} \ell \bar{\nu}$ form factors, and of $R(D_{(s)})$.  Each $L$-entry corresponds to a different lattice spacing, with multiple spatial volumes at some lattice spacings.  For actions with multiple species of pions, the lightest pion masses are quoted.}
}
\end{table}
\begin{table}[!ht]
{\footnotesize
\begin{tabular*}{\textwidth}{l @{\extracolsep{\fill}} c c c l}
\hline\hline \\[-1.0ex]
Collab. & Ref. & $\Nf$ & Ren. & Description 
\\[1.0ex] \hline \hline \\[-1.0ex]
\parbox[t]{20mm}{HPQCD 23}                & \cite{Harrison:2023dzh}              & 2+1+1 & NPR
                                          & \parbox[t]{60mm}{Vector (axial) currents renormalized nonperturbatively using the PCVC (PCAC) relation.}
\\[6.5ex] \hline \\[-2.0ex]
\parbox[t]{20mm}{JLQCD 23}                & \cite{Aoki:2023qpa}                  & 2+1   & mNPR
                                          & \parbox[t]{60mm}{Majority of current renormalization factor cancels in ratio of lattice correlation functions. Remaining correction expected to behave better than $O(a)$, and vanishes in the continuum limit.}
\\[13.0ex] \hline \\[-2.0ex]
\parbox[t]{23mm}{\SLfnalmilcBDstar}       & \cite{FermilabLattice:2021cdg}       & 2+1   & mNPR
                                          & \parbox[t]{60mm}{Majority of current renormalization factor cancels in double ratio of lattice correlation functions.  Remaining correction calculated with 1-loop tadpole-improved lattice perturbation theory. Systematic uncertainty estimated at 0.1\%.}
\\[15.0ex] \hline \\[-2.0ex]
\parbox[t]{20mm}{Meinel 21, Meinel 21B} & \cite{Meinel:2021rbm,Meinel:2021mdj} & 2+1   & mNPR  
                                          & \parbox[t]{60mm}{Residual matching factors $\rho$ computed at 1-loop for vector and axial-vector currents, but at tree-level only for tensor currents. A systematic uncertainty is assigned to $\rho_{T^{\mu\nu}}$ as the double of ${\rm max}(|\rho_{A^{\mu}}-1|,|\rho_{V^{\mu}}-1|)$.}
\\[16.0ex]
\ifx\reducedapptables\undefined
\hline \\[-1.0ex]
\parbox[t]{26mm}{HPQCD 19} & \cite{McLean:2019qcx} & 2+1+1 & NP &
\parbox[t]{6cm}{Currents renormalized nonperturbatively using PCVC relation.}
\\[11.0ex] \hline \\[-2.0ex]
\parbox[t]{26mm}{HPQCD 19B} & \cite{McLean:2019sds} & 2+1+1 & NP &
\parbox[t]{6cm}{Currents renormalized nonperturbatively using PCAC relation.}
\\[5.0ex] \hline \\[-2.0ex]
\parbox[t]{26mm}{HPQCD 17B} & \cite{Harrison:2017fmw} & 2+1+1 & 1-loop &
\parbox[t]{6cm}{1-loop matching of currents taken from~\cite{Monahan:2012dq}.}
\\[4.0ex] \hline \\[-1.0ex]
\parbox[t]{26mm}{HPQCD 17} & \cite{Monahan:2017uby} & 2+1 & 1-loop &
\parbox[t]{6cm}{1-loop matching of currents taken from~\cite{Monahan:2012dq}.}
\\[4.0ex] \hline \\[-1.0ex]
\parbox[t]{26mm}{Datta 17} & \cite{Datta:2017aue} & 2+1 & (incomplete) &
\parbox[t]{6cm}{mNPR procedure followed, fixing the residual renormalization factor to its tree-level value $\rho_{T^{\mu\nu}}=1$. A systematic uncertainty is assigned as the double of ${\rm max}[1-\rho_{A^{\mu}},1-\rho_{V^{\mu}}]$, using the known 1-loop values of the residual matchings for the vector and axial currents.}
\\[18.0ex] 
\hline \\[-1.0ex]
\parbox[t]{26mm}{\SLfnalmilcBD} & \cite{Lattice:2015rga} & 2+1 & mNPR &
\parbox[t]{6cm}{Form factors extracted from ratios of correlators that renormalize with
ratios of current normalizations, computed at 1-loop in perturbation theory.
Dependence of renormalization factor on recoil parameter $w$ neglected.
Systematic uncertainty due to perturbative truncation and $w$-dependence estimated
by power counting to 0.7\%.}
\\[21.0ex] \hline \\[-1.0ex]
\parbox[t]{26mm}{\SLLambdabc} & \cite{Detmold:2015aaa} & 2+1 & mNPR &
\parbox[t]{6cm}{Perturbative truncation error estimated at 1\% with size of 1-loop correction on next-to-finest ensemble.}
\\[7.0ex] \hline \\[-1.0ex]
Atoui  13 & \cite{Atoui:2013zza} & 2 & --- &
\parbox[t]{6cm}{Observables obtained from ratios that do not require renormalization. Checks performed by comparing with results coming from currents that are renormalized separately with nonperturbative $Z_{\rm V}$.}
\\[12.0ex]
\fi
\hline\hline
\end{tabular*}
\caption{Operator renormalization in determinations of $B_{(s)} \to D_{(s)}^{(*)}\ell\nu$ and $\Lambda_b \to \Lambda_c^{(*)} \ell \bar{\nu}$ form factors, and of $R(D_{(s)})$.}
}
\end{table}
\begin{table}[!ht]
{\footnotesize
\begin{tabular*}{\textwidth}{l @{\extracolsep{\fill}} c c c l}
\hline\hline \\[-1.0ex]
Collab. & Ref. & $\Nf$ & Action & Description 
\\[1.0ex] \hline \hline \\[-1.0ex]
\parbox[t]{20mm}{HPQCD 23}                & \cite{Harrison:2023dzh}              & 2+1+1 & \parbox[t]{30mm}{HISQ for both the $b$ and $c$ quarks.}
                                          & \parbox[t]{60mm}{Values of bare heavy-quark masses up to $am_h=0.8$. The error from continuum limit and extrapolation to physical $b$ mass at zero recoil is quite small, but it becomes dominant at mid-recoil.}
\\[12.0ex] \hline \\[-2.0ex]
\parbox[t]{20mm}{JLQCD 23}                & \cite{Aoki:2023qpa}                  & 2+1   & \parbox[t]{30mm}{M\"obius Domain-Wall for both the $b$ and $c$ quarks.} 
                                          & \parbox[t]{60mm}{Values of bare heavy-quark masses up to $am_h=0.69$. The systematics associated to the extrapolation to physical $b$ mass stays under 4\% for all form factors in the whole recoil range.}
\\[12.0ex] \hline \\[-2.0ex]
\parbox[t]{23mm}{\SLfnalmilcBDstar}       & \cite{FermilabLattice:2021cdg}       & 2+1   & \parbox[t]{30mm}{Fermilab RHQ for both the $b$ and $c$ quarks.}
                                          & \parbox[t]{60mm}{Discretization errors discussed as part of combined chiral-continuum stemming from $\alpha_s a$, $a^2$ and $a^3$ terms.}
\\[6.5ex] \hline \\[-2.0ex]
\parbox[t]{20mm}{Meinel 21, Meinel 21B} & \cite{Meinel:2021rbm,Meinel:2021mdj} & 2+1   & \parbox[t]{30mm}{Columbia RHQ for both the $b$ and $c$ quarks.}
                                          & \parbox[t]{60mm}{Discretization errors discussed as part of combined chiral-continuum-$w$ fit. Higher-order fit also includes $\mathcal{O}(\alpha_s a |\mathbf{p}|)$ terms to account for missing radiative corrections to $\mathcal{O}(a)$ improvement of the currents.}
\\[12.5ex]
\ifx\reducedapptables\undefined
\hline \\[-1.0ex]
\parbox[t]{26mm}{HPQCD 19} & \cite{McLean:2019qcx} & 2+1+1 & \parbox[t]{30mm}{HISQ for both the $b$ and $c$ quarks.} &
\parbox[t]{6cm}{Values of bare heavy-quark masses up to $am_h=0.8$. Fractional error from continuum limit and
extrapolation to physical $b$ mass at zero recoil quoted as 1.20\%.}
\\[10.0ex] \hline \\[-2.0ex]
\parbox[t]{26mm}{HPQCD 19B} & \cite{McLean:2019sds} & 2+1+1 & \parbox[t]{30mm}{HISQ for both the $b$ and $c$ quarks.} &
\parbox[t]{6cm}{Values of bare heavy-quark masses up to $am_h=0.8$. Fractional error from
extrapolation to physical $b$ mass quoted as 0.69\%.}
\\[7.0ex] \hline \\[-2.0ex]
\parbox[t]{26mm}{HPQCD 17B} & \cite{Harrison:2017fmw} & 2+1+1 & \parbox[t]{30mm}{NRQCD for b quark,\\ HISQ for c quark} &
\parbox[t]{6cm}{$\mathcal{O}(\alpha_s^2)$ uncertainty quoted as 2.1\% and 2.5\% for the two
relevant form factors; $\mathcal{O}(\alpha_s\Lambda_{\rm QCD}/m_b,(\Lambda_{\rm QCD}/m_b)^2)$ errors quoted as 0.9\% and 0.8\% in both cases.}
\\[10.0ex] \hline \\[-1.0ex]
\parbox[t]{26mm}{HPQCD 17} & \cite{Monahan:2017uby} & 2+1 & \parbox[t]{30mm}{NRQCD for b quark,\\ HISQ for c quark} &
\parbox[t]{6cm}{Discretization errors estimated 
via power counting to be 2.47\% ($B_s \to D_s$) and 2.59\% ($B \to D$) at $q^2=0$.}
\\[7.0ex] \hline \\[-1.0ex]
\parbox[t]{26mm}{Datta 17} & \cite{Datta:2017aue} & 2+1 & Columbia RHQ &
\parbox[t]{6cm}{Discretization errors discussed as part of combined chiral-continuum-$q^2$ fit, stemming from $a^2|\mathbf{p}|^2$ terms.}
\\[7.0ex] 
\hline \\[-1.0ex]
\parbox[t]{26mm}{\SLfnalmilcBD} & \cite{Lattice:2015rga} & 2+1 & Fermilab &
\parbox[t]{6cm}{Discretization errors of form factors estimated 
via power counting to be 0.4\%.}
\\[4.0ex] \hline \\[-1.0ex]
\parbox[t]{26mm}{\SLLambdabc} & \cite{Detmold:2015aaa} & 2+1 & Columbia RHQ &
\parbox[t]{6cm}{Discretization errors discussed as part of combined chiral-continuum-$q^2$ fit, stemming from $a^2|\mathbf{p}|^2$ terms.}
\\[7.0ex] \hline \\[-1.0ex]
Atoui  13 & \cite{Atoui:2013zza} & 2 & tmWil &
\parbox[t]{6cm}{Results obtained from step-scaling in heavy-quark mass via the ratio method. Separate continuum limit extrapolations with mild $a^2$-dependence carried out for each mass point separately. Result at physical value of $m_b$ obtained by interpolation between data region and known exact HQET limit.}
\\[21.0ex]
\fi
\hline\hline
\end{tabular*}
\caption{Heavy-quark treatment in determinations of $B_{(s)} \to D_{(s)}^{(*)}\ell\nu$ and $\Lambda_b \to \Lambda_c^{(*)} \ell \bar{\nu}$ form factors, and of $R(D_{(s)})$.}
}
\end{table}

%% file: Alpha_s/Alpha_s_app.tex
%
%
%
%
%
%


\subsection{Notes to Sec.~\ref{sec:alpha_s} on the strong coupling $\alpha_{\rm s}$}


\subsubsection{Renormalization scale and perturbative behaviour}



\begin{table}[!htb]
   \footnotesize
\begin{tabular*}{\textwidth}[l]{l@{\extracolsep{\fill}}rlllllllll}
\hline\hline \\[-1.0ex]
      Collab. & Ref. & $\Nf$ & $\alpha_{\rm eff}$ & $n_{\mathrm{l}}$ & Description \\
      & & & &  \\
      \hline
      \hline
      & & & &  \\[-0.1cm]

\parbox[t]{2.1cm}{Hasenfratz~23} & \cite{Hasenfratz:2023bok} & 0 &  $0.095 < \alpha_{\rm eff} < 1.26$ & 2
                  & \parbox[t]{0.555\textwidth}
                    {GF scheme does not reach perturbative asymptotics.} \\      
\\[-0.1cm] \hline\\[-1.5ex] 
      
Wong~23    & \cite{Wong:2023jvr}   & 0 &  $0.095 < \alpha_{\rm eff} < 1.26$ & 2
                  & \parbox[t]{0.555\textwidth}
                    {GF scheme does not reach perturbative asymptotics.} \\
\hline
   \hline
\end{tabular*}
\caption{Renormalization scale and perturbative behaviour
         of $\alpha_s$ determinations for $\Nf=0$. }
\label{tab_Nf=0_renormalization_a}
\end{table}


\ifx\reducedapptables\undefined

\begin{table}[!htb]
\addtocounter{table}{-1}
   \footnotesize
   \begin{tabular*}{\textwidth}[l]{l@{\extracolsep{\fill}}rlllllllll}
\hline\hline \\[-1.0ex]
      Collab. & Ref. & $\Nf$ & $\alpha_{\rm eff}$ & $n_{\mathrm{l}}$ & Description \\
      & & & &  \\
      \hline
      \hline
      & & & &  \\[-0.1cm]

\parbox[t]{0.14\textwidth}{Boucaud 00A} & \cite{Boucaud:2000ey}         &  0       &
  \parbox[t]{0.09\textwidth}{0.35-0.55, 0.25-0.45, 0.22-0.28, 0.18-0.22} & 2
& \parbox[t]{0.555\textwidth}{$\alpha_{\widetilde{\rm MOM}}$ with
  $p =$ 2--10\,GeV. Fitted to $n_{\mathrm{l}} = 2$ loop perturbation theory
  with power correction.
 ($\beta$ = 6.0, 6.2, 6.4, 6.8.)} \\
\\[-0.1cm] \hline\\[-1.5ex] 

\parbox[t]{0.14\textwidth}{Boucaud 00B} & \cite{Boucaud:2000nd}         &  0       &
  \parbox[t]{0.09\textwidth}{0.35--0.55, 0.25--0.45, 0.22--0.28, 0.18--0.22} & 2
& \parbox[t]{0.555\textwidth}{$\alpha_{\rm MOM}$ with
  $2 \leq \mu \leq 10$\, GeV. Consistency check of $n_{\mathrm{l}} = 2$ loop
  perturbation formula with gluon condensate.
  $\beta_2^{MOM}=1.5\times \beta_2^{\widetilde{MOM}}$ is needed.
  ($\beta$ = 6.0, 6.2, 6.4, 6.8.)} \\
\hline
   \hline
\end{tabular*}
\caption{(cntd.) Renormalization scale and perturbative behaviour
         of $\alpha_s$ determinations for $\Nf=0$.}
\label{tab_Nf=0_renormalization_b}
\end{table}

\fi


\ifx\reducedapptables\undefined

\begin{table}[!htb]
   \footnotesize
   \begin{tabular*}{\textwidth}[l]{l@{\extracolsep{\fill}}rlllllllll}
\hline\hline \\[-1.0ex]
      Collab. & Ref. & $\Nf$ & $\alpha_{\rm eff}$ & $n_{\mathrm{l}}$ & Description \\
      &&&&&&& & \\[-0.1cm]
      \hline
      \hline
      &&&&&&&&& \\[-0.1cm]

\parbox[t]{2.0cm}{Karbstein 18} & \cite{Karbstein:2018mzo} & 2 
    & 0.28 - 0.41 & 3
    & \parbox[t]{0.555\textwidth}{$\alpha_V(p)$ for momentum 
                                $1.5 < p < 3.0\,\mbox{GeV}$.
                                Values computed from the 
                                quoted $\Lambda$ parameter with the 2-loop
                                $\beta$ function; larger values (0.32--0.62)
                                are obtained with 3-loop running. 
                                As with ETM 11C
                                central values are taken from 
                                $a = 0.042\,\mbox{fm}$ lattice
                                with $L = 1.3\,\mbox{fm}$ and 
                                $m_\pi = 350\,\mbox{MeV}$.}
\\ \\[-0.1cm] 
\hline\\[-1.5ex] 
ALPHA 12 & \cite{Fritzsch:2012wq} & 2  & 
\parbox[t]{0.1\textwidth}{see \\ ALPHA 04} & 2
    & \parbox[t]{0.5550\textwidth}{Determination of $\Lambda_\msbar / f_K$ using ALPHA 04}
\\
\hline\hline
\end{tabular*}
\caption{Renormalization scale and perturbative behaviour 
of $\alpha_s$ determinations for $\Nf=2$.}
\label{tab_Nf=2_renormalization}
\end{table}

\fi


\begin{table}[!htb]
   \footnotesize
   %
\caption{Renormalization scale and perturbative behaviour of $\alpha_s$
         determinations for $\Nf=3$.}
\label{tab_Nf=3_renormalization_a}
\end{table}


\ifx\reducedapptables\undefined

\begin{table}[!htb]
\addtocounter{table}{-1}
   \footnotesize
   \begin{tabular*}{\textwidth}[l]{l@{\extracolsep{\fill}}rlllllllll}
\hline\hline \\[-1.0ex]
      Collab. & Ref. & $\Nf$ & $\alpha_{\rm eff}$ & $n_{\mathrm{l}}$ & Description \\
      &&&&& \\[-0.1cm]
      \hline
      \hline
      &&&&& \\[-0.1cm]

HPQCD 08B    & \cite{Allison:2008xk}  & 2+1 &  
0.38  & 2 &\parbox[t]{0.54\textwidth}{Fit of the ratios to PT
at the charm mass
including $(a m)^{2i}$ terms with $i\leq 2\ldots 4$; coefficients constrained by priors. } 
\\ \\[-0.1cm] \hline\\[-1.5ex]
 {HPQCD 08A}
   & \cite{Davies:2008sw}
   & 2+1
   & \parbox[t]{0.1\textwidth}
        {0.15-0.4}
   & 2
   & \parbox[t]{0.54\textwidth}
        {$\alphav(q^*)$ for a variety of short-distance quantities,
         using same method as in HPQCD 05A.}
\\ \\[-0.1cm] \hline\\[-1.5ex]
 {Maltman 08}
   & \cite{Maltman:2008bx}
   & 2+1
   &  & 2
   & \parbox[t]{0.54\textwidth}
        {Re-analysis of HPQCD 05A for a restricted set of
         short-distance quantities with similar results.}
\\ \\[-0.1cm] \hline\\[-1.5ex]
 {HPQCD 05A}
   & \cite{Mason:2005zx}
   & 2+1
   & \parbox[t]{0.1\textwidth}
        {0.2-0.4}
   & 2
   & \parbox[t]{0.54\textwidth}
        {$\alphav(q^*)$ for a variety of short-distance quantities.
        }
\\
      &&&&& \\[-0.1cm]
      \hline
      \hline
\end{tabular*}
\caption{(cntd.) Renormalization scale and perturbative behaviour of $\alpha_s$
         determinations for $\Nf=3$.}
\label{tab_Nf=3_renormalization_b}
\end{table}

\fi


\ifx\reducedapptables\undefined

\begin{table}[!htb]
   \vspace{3.0cm}
   \footnotesize
   \begin{tabular*}{\textwidth}[l]{l@{\extracolsep{\fill}}rlllllllll}
\hline\hline \\[-1.0ex]
      Collab. & Ref. & $\Nf$ & $\alpha_{\rm eff}$ & $n_{\mathrm{l}}$ & Description \\
      &&&&& \\[-0.1cm]
      \hline
      \hline
      &&&&& \\[-0.1cm]

 HPQCD 14A   &  \cite{Chakraborty:2014aca} & 2+1+1 &  
 0.11-0.33 & 2 
   &  \parbox[t]{0.54\textwidth}{
           Range given for $\alpha_\mathrm{eff}$ from $\tilde R_4$. 
           Fit of ratios $R_n\;n=4\ldots10 $ to perturbation theory
           including $(a m)^{2i}$ terms with $i\leq $10--20 and 
           higher-order perturbative terms;
           coefficients constrained by priors.}
 \\ \\[-0.1cm] \hline\\[-1.5ex]
 { {ETM 13D}}       &  \cite{Blossier:2013ioa} & 2+1+1 &  0.26-0.7    & 3 & 
	\parbox[t]{0.54\textwidth}{$\alpha_\mathrm{T}(p)$ for $p=$1.6--6.5\,GeV. 
	Update of \cite{Blossier:2012ef}
                                   with improved power law determination.}
 \\ \\[-0.1cm] \hline\\[-1.5ex]
 { ETM 12C}       & \cite{Blossier:2012ef}   & 2+1+1 &
0.24-0.38 & 3 &
	\parbox[t]{0.54\textwidth}{$\alpha_\mathrm{T}(p)$ for $p=$1.7--6.8\,GeV. 
	Fit to PT with gluon condensate correction or higher power.}
	 \\ \\[-0.1cm] \hline\\[-1.5ex]
      ALPHA 10A & \cite{Tekin:2010mm} & 4                  & 0.07-0.28 & 2
        & \parbox[t]{0.54\textwidth}{$\alpha_\mathrm{SF}(1/L)$.
        Comparison to PT with 2-, 3-loop $\beta$-function.}
 \\ \\[-0.1cm] \hline\\[-1.5ex]
 {ETM 11D}      & \cite{Blossier:2011tf}   & 2+1+1 & 
	0.24-0.4& 3 &
	\parbox[t]{0.54\textwidth}{$\alpha_\mathrm{T}(p)$ for $p=$3.8--7.1\,GeV
        with H(4)-procedure. Fit to PT with gluon condensate correction.}
 \\ \\[-0.1cm] \hline\\[-1.5ex]
      Perez 10 & \cite{PerezRubio:2010ke} & 4           & 0.06-0.28 & 2
        & \parbox[t]{0.54\textwidth}{$\alpha_\mathrm{SF}(1/L)$.
          Comparison with 1-, 2-, 3-loop $\beta$-function.}\\
      &&&&&&\\[-0.1cm]
      \hline
      \hline
\end{tabular*}
\caption{Renormalization scale and perturbative behaviour of $\alpha_s$
         determinations for $\Nf=4$.}
\label{tab_Nf=4_renormalization}
\end{table}

\fi

\clearpage


\subsubsection{Continuum limit}


\begin{table}[!htb]
   \footnotesize
   \begin{tabular*}{\textwidth}[l]{l@{\extracolsep{\fill}}rlllllllll}
\hline\hline \\[-1.0ex]
      Collab. & Ref. & $\Nf$ & $a\,\mu$  & Description\\
      & & & &  \\
      \hline
      \hline
      & & & &  \\[-0.1cm]

Hasenfratz~23 & \cite{Hasenfratz:2023bok} & 0 &  $0.158 < a/\sqrt{8t} < 0.29$ &
\parbox[t]{0.39\textwidth}{GF scheme, infinite-volume extrapolation, \\ direct determination of the $\beta$-function.}\\

\\[-0.1cm] \hline\\[-1.5ex] 
      
Wong~23    & \cite{Wong:2023jvr} & 0 &  $0.16 < a/\sqrt{8t} < 0.28$ &
\parbox[t]{0.39\textwidth}{GF scheme, infinite-volume extrapolation, \\ direct determination of the $\beta$-function.}\\
\hline
      \hline
\end{tabular*}
\caption{Continuum limit for $\alpha_s$ determinations with $\Nf=0$.}
\label{tab_Nf=0_continuumlimit_a}
\end{table}


\ifx\reducedapptables\undefined

\begin{table}[!htb]
\addtocounter{table}{-1}
   \footnotesize
   \begin{tabular*}{\textwidth}[l]{l@{\extracolsep{\fill}
   }rlllllllll}
\hline\hline \\[-1.0ex]
      Collaboration & Ref. & $\Nf$ & $a\,\mu$  & Description\\
      & & & &  \\
      \hline
      \hline
      & & & &  \\[-0.1cm]

    {Boucaud 00A}      &\cite{Boucaud:2000ey}           &  0    & 
    {$a =$ 0.1, 0.07, 0.05, 0.03\,fm}  &
    {At $\alpha_s \leq 0.3$ $ap = $1.9, 1.4, 1.0, 0.6.}\\
\\[-0.1cm] \hline\\[-1.5ex]
    {Boucaud 00B}   & \cite{Boucaud:2000nd}      & 0         & 
    {$a =$ 0.1, 0.07, 0.05, 0.03\,fm} & 
    {At $\alpha_s \leq 0.3$ $ap =$ 1.9, 1.4, 1.0, 0.6.}\\
\hline
      \hline
\end{tabular*}
\caption{(cntd.) Continuum limit for $\alpha_s$ determinations with $\Nf=0$.}
\label{tab_Nf=0_continuumlimit_b}
\end{table}

\fi


\ifx\reducedapptables\undefined

\begin{table}[!htb]
   \vspace{0.0cm}
   \footnotesize
   \begin{tabular*}{\textwidth}[l]{l@{\extracolsep{\fill}}rlllllllll}
\hline\hline \\[-1.0ex]
      Collab. & Ref. & $\Nf$ &  $a\,\mu$ & Description   \\
      &&&&&&& & \\[-0.1cm]
      \hline
      \hline
      &&&&&&&&& \\[-0.1cm]
Karbstein 14 & \cite{Karbstein:2014bsa} & 2  & 
\parbox[t]{0.1\textwidth}{0.32 -- 1.19 \\ 0.63 -- 1.19 }
    & 
\parbox[t]{0.555\textwidth}{at $p=1.5$\,GeV \\
at $p=3$\,GeV, roughly coincides with $\alpha_s=0.3$.}
\\ \\[-0.1cm] \hline\\[-1.5ex] 
ALPHA 12 & \cite{Fritzsch:2012wq} & 2 &  
\parbox[t]{0.25\textwidth}{$a=$ 0.049, 0.066, 0.076 fm from $f_K$} & 
\parbox[t]{0.46\textwidth}{2-loop $\cO(a)$ boundary improvement, 
linear extrapolation of $L_\mathrm{max} f_\mathrm{K}$ in $a^2$.}
 \\
\hline
      \hline
\end{tabular*}
\caption{Continuum limit for $\alpha_s$ determinations with $\Nf=2$.}
\label{tab_Nf=2_continuumlimit}
\end{table}

\fi


\begin{table}[!htb]
   \vspace{0.0cm}
   \footnotesize
   %
\caption{Continuum limit for $\alpha_s$ determinations with $\Nf=3$.}
\label{tab_Nf=3_continuumlimit_a}
\end{table}


\ifx\reducedapptables\undefined

\begin{table}[!htb]
   \vspace{0.0cm}
   \addtocounter{table}{-1}
   \footnotesize
   \begin{tabular*}{\textwidth}[l]{l@{\extracolsep{\fill}}rlllllllll}
\hline\hline \\[-1.0ex]
      Collab. & Ref. & $\Nf$ & $a\,\mu$ & Description \\
      &&&&&&& & \\[-0.1cm]
      \hline
      \hline
      &&&&&&& \\[-0.1cm]

PACS-CS 09A& \cite{Aoki:2009tf} & 2+1 & 
$a/L=1/8,1/6,1/4$ & 
\parbox[t]{0.42\textwidth}{Tree-level
$\cO(a)$ boundary improvement, which has been seen to behave better
than 1-loop in simulations \cite{Takeda:2004xha}; 
weighted average of $a/L=1/8,1/6$ for step-scaling function
which agrees with a
linear extrapolation in $a/L$ of all data points of the 
SSF. Linear extrapolation  in $a/L$ of $L_\mathrm{max}m_\rho$ with 
$a/L_\mathrm{max} = 1/8, 1/6, 1/4$.}
 \\ \\[-0.1cm] \hline\\[-1.5ex]

{HPQCD 08B}   & \cite{Allison:2008xk}  & 2+1 &  
\parbox[t]{0.25\textwidth}{ $a\mu=2 a\bar m_c$ = 0.8,\, 1.2,\, 1.7,\, 2.1} & 
\parbox[t]{0.42\textwidth}{4 lattice spacings with heavy-quark mass
approximately the charm mass, where $\alpha_{R_4}(\mu) = 0.38$.}
 \\ \\[-0.1cm] \hline\\[-1.5ex]

{HPQCD 08A}
   & \cite{Davies:2008sw}
   & 2+1
   & \parbox[t]{0.25\textwidth}
        {6 lattice spacings with $a = 0.18$ - $0.045\,\mbox{fm}$}
   & \parbox[t]{0.555\textwidth}
        {$r_1$ using $\Upsilon$ spectrum splitting.}
 \\ \\[-0.1cm] \hline\\[-1.5ex]

{Maltman 08}
   & \cite{Maltman:2008bx}
   & 2+1
   & \parbox[t]{0.25\textwidth}
        {5 lattice spacings with $a = 0.18$ - $0.06\,\mbox{fm}$}
   & \parbox[t]{0.42\textwidth}
        {Re-analysis of HPQCD 05A with additional lattice
         spacings $a = 0.06$, $0.15\,\mbox{fm}$.}
 \\ \\[-0.1cm] \hline\\[-1.5ex]

{HPQCD 05A}
   & \cite{Mason:2005zx}
   & 2+1
   & \parbox[t]{0.25\textwidth}
        {3 lattice spacings with $a = 0.18$ - $0.09\,\mbox{fm}$}
   & \parbox[t]{0.555\textwidth}
        {$r_1$ using $\Upsilon$ spectrum splitting.}
\\ 
        &&&&&&&&& \\[-0.1cm]
   \hline
      \hline
\end{tabular*}
\caption{(cntd.) Continuum limit for $\alpha_s$ determinations with $\Nf=3$.}
\label{tab_Nf=3_continuumlimit_b}
\end{table}

\fi


\ifx\reducedapptables\undefined

\begin{table}[!htb]
   \vspace{3.0cm}
   \footnotesize
   \begin{tabular*}{\textwidth}[l]{l@{\extracolsep{\fill}}rlllllllll}
\hline\hline \\[-1.0ex]
      Collab. & Ref. & $\Nf$ & $a\,\mu$ & Description \\
      &&&& \\[-0.1cm]
      \hline
      \hline
      &&&& \\[-0.1cm]
HPQCD 14A   &  \cite{Chakraborty:2014aca} & 2+1+1 & 
\parbox[t]{0.25\textwidth}{$a\mu=2 a\bar m_c = 0.78 - 2.09$}& 
\parbox[t]{0.42\textwidth}{4 lattice spacings;
             2 lattice spacings with $ a\mu  \leq 1.5$ 
             and one more lattice spacing with $ a\mu  \lsim 1.6$ 
             when $\alpha_{R_4}(\mu) \leq 0.3$.}
 \\ \\[-0.1cm] \hline\\[-1.5ex]
ETM 13D         &  \cite{Blossier:2013ioa} & 2+1+1 &    
  $a=0.060, 0.068$ fm from $f_\pi$          &For $\alpha_s \leq 0.3$, $ap=1.5,1.7$.
	\parbox[t]{0.555\textwidth}{Update of \cite{Blossier:2012ef}.}
\\ \\[-0.1cm] \hline\\[-1.5ex]      
{ETM 12C}       & \cite{Blossier:2012ef}   & 2+1+1 &  
      $a=0.061, 0.078$ fm from $f_\pi$ & 
      \parbox[t]{0.42\textwidth}{Global fit with $(ap)^2$ discretization effects.
       For $\alpha_s \leq 0.3$, $ap = 1.5, 2.2$.}       
\\ \\[-0.1cm] \hline\\[-1.5ex]      
{ETM 11D}      & \cite{Blossier:2011tf}   & 2+1+1 & 
      $a = 0.061, 0,078$\,fm & {For $\alpha_s \leq 0.3$, $ap = 1.5, 2.0$.}
\\ \\[-0.1cm] \hline\\[-1.5ex]      
      ALPHA 10A & \cite{Tekin:2010mm} & 4                  & 
      $a/L = 1/4, 1/6, 1/8$ & Constant or global linear fit in $(a/L)^2$.
\\ \\[-0.1cm] \hline\\[-1.5ex]      
      Perez 10 & \cite{PerezRubio:2010ke} & 4 & $a/L=1/4, 1/6, 1/8$ &
      \parbox[t]{0.42\textwidth}{Linear extrapolation in $(a/L)^2$.
      1-loop improvement at the boundary.}
\\  &&&&&&&&& \\[-0.1cm]
   \hline
   \hline
\end{tabular*}
\caption{Continuum limit for $\alpha_s$ determinations with $\Nf=4$.}
\label{tab_Nf=4_continuumlimit}
\end{table}

\fi

%
%

%% file: NME/NME_app.tex
\newpage
\subsection{Notes to Sec.~\ref{sec:NME} on nucleon matrix elements}
\label{subsec:Notes to NME}

\input{NME/Tables/app_gast_isovector}

\clearpage
\clearpage
\input{NME/Tables/app_sigmaud_s}
\clearpage
\input{NME/Tables/app_sigmaud_fh}

 \clearpage
 \clearpage
 \input{NME/Tables/app_x_isovector}

%% file: NME/Tables/app_gast_isovector.tex
\begin{table}[!ht]
{\footnotesize
\begin{tabular*}{\textwidth}{l @{\extracolsep{\fill}} c c c l}
\hline\hline \\[-1.0ex]
Collab. & Ref. & $\Nf$ & $a$ [fm] & Description 
\\[.3ex]\\[-2.ex] \hline \hline \\[-1.0ex]
%
ETM 23 & \cite{Alexandrou:2023qbg} & 2+1+1 & 0.057, 0.069, 0.080
& \parbox[t]{6.3cm}{Extrapolation via a fit which is linear in
$a^2$.}
\\[0.5ex]\\[0.5ex]  \hline\\[-0.5ex]
PNDME 23 & \cite{Jang:2023zts} & 2+1+1 & 0.06, 0.09, 0.12, 0.15
& \parbox[t]{6.3cm}{Physical-point
extrapolations performed simultaneously, keeping only the
leading-order terms in the various expansion parameters.}
\\[0.5ex]\\[0.5ex]\hline\\[-0.5ex]
ETM 22 & \cite{Alexandrou:2022dtc} & 2+1+1 & 0.057, 0.069, 0.080
& \parbox[t]{6.3cm}{Extrapolation via a fit which is linear in
$a^2$.}
\\[0.5ex]\\[0.5ex]
\ifx\reducedapptables\undefined
\hline \\[-0.5ex]
CalLat 19 & \cite{Walker-Loud:2019cif} & 2+1+1 & 0.15,0.12,0.09
& \parbox[t]{6.3cm}{Extrapolation to the physical point via
  simultaneous fit in the lattice spacing, $M_\pi$ and $M_\pi L$,
  including terms of order $a^2$ and $a^4$.}
\\[0.5ex]\\[0.5ex]  \hline\\[-0.5ex]
ETM 19 & \cite{Alexandrou:2019brg} & 2+1+1 & 0.08
& \parbox[t]{6.3cm}{Single lattice spacing.}
\\[0.5ex]\\[0.5ex]
PNDME 18 & \cite{Gupta:2018qil} & 2+1+1 & 0.15,0.12,0.09,0.06
& \parbox[t]{6.3cm}{Extrapolation performed including a linear term in
  $a$ as part of a simultaneous fit in $a$, $M_\pi$ and $M_\pi L$.} 
\\[0.5ex]\\[0.5ex]  \hline\\[-0.5ex]
CalLat 18 & \cite{Chang:2018uxx} & 2+1+1 & 0.15,0.12,0.09
& \parbox[t]{6.3cm}{Extrapolation to the physical point via
  simultaneous fit in the lattice spacing, $M_\pi$ and $M_\pi L$,
  including terms of order $a^2$ and $a^4$.}
\\[0.5ex]\\[0.5ex]  \hline\\[-0.5ex]
CalLat 17 & \cite{Berkowitz:2017gql} & 2+1+1 & 0.15,0.12,0.09
& \parbox[t]{6.3cm}{Extrapolation to the physical point via
  simultaneous fit in the lattice spacing, $M_\pi$ and $M_\pi L$,
  including terms of order $a^2$ and $a^4$.}
\\[0.5ex]\\[0.5ex]  \hline\\[-0.5ex]
PNDME 16 & \cite{Bhattacharya:2016zcn} & 2+1+1 & 0.12,0.09,0.06
& \parbox[t]{6.3cm}{ Extrapolation performed including a linear term
  in $a$ as part of a simultaneous fit in $a$, $M_\pi$ and $M_\pi L$.}
\\[0.5ex]\\[0.5ex]  \hline\\[-0.5ex]
PNDME 15 & \cite{Bhattacharya:2015wna,Bhattacharya:2015esa} & 2+1+1
& 0.12,0.09,0.06 & \parbox[t]{6.3cm}{Extrapolation performed
  including a linear term in $a$ as part of a simultaneous fit in $a$,
  $M_\pi$ and $M_\pi L$.}
\\[0.5ex]\\[0.5ex]  \hline\\[-0.5ex]
PNDME 13 & \cite{Bhattacharya:2013ehc} & 2+1+1 & 0.12
& \parbox[t]{6.3cm}{Single lattice spacing.}
\\[0.5ex]\\[0.5ex]
\fi
\hline \hline \\[-1.0ex]
\end{tabular*}
\caption{Continuum extrapolations/estimation of lattice artifacts in
  determinations of the isovector axial, scalar and tensor charges
  with $\Nf=2+1+1$ quark flavours.}
}
\end{table}

\begin{table}[!ht]
{\footnotesize
\begin{tabular*}{\textwidth}{l @{\extracolsep{\fill}} c c c l}
\hline\hline \\[-1.0ex]
Collab. & Ref. & $\Nf$ & $a$ [fm] & Description 
\\[.3ex]\\[-2.ex] \hline \hline \\[-1.0ex]
%
Mainz 24 & \cite{Djukanovic:2024krw} & 2+1 &0.049,0.064,0.076,0.086
& \parbox[t]{6.3cm}{Extrapolation performed as part of a simultaneous
fit in $a$, $M_\pi$ and $M_\pi L$.}
\\[0.5ex]\\[0.5ex]  \hline\\[-0.5ex]
PACS 23 & \cite{Tsuji:2023llh} & 2+1 & 0.063, 0.085
& \parbox[t]{6.3cm}{Discretization effects estimated by difference between two ensembles.}
\\[0.5ex]\\[0.5ex]  \hline\\[-0.5ex]
RQCD 23 & \cite{Bali:2023sdi} & 2+1 & \parbox[t]{4.0cm}{0.039, 0.049, 0.064, 0.076, 0.086, 0.098}
& \parbox[t]{6.3cm}{Extrapolation performed using terms up to $a^2$ in the lattice spacing.}
\\[0.5ex]\\[0.5ex]  \hline\\[-0.5ex]
\parbox[t]{1.3cm}{QCDSF/ \par UKQCD/ \par CSSM 23} & \cite{QCDSFUKQCDCSSM:2023qlx} & 2+1 & 0.052, 0.059, 0.068, 0.074, 0.082
& \parbox[t]{6.3cm}{Extrapolation performed including leading discretization effects.}
\\[0.5ex]\\[0.5ex]  \hline\\[-0.5ex]
PACS 22B & \cite{Tsuji:2022ric} & 2+1 & 0.085
& \parbox[t]{6.3cm}{Single lattice spacing.}
\\[0.5ex]\\[0.5ex]  \hline\\[-0.5ex]
Mainz~22 & \cite{Djukanovic:2022wru} & 2+1 & 0.049,0.064,0.076,0.086 &
\parbox[t]{6.1cm}{Extrapolation performed as part of a simultaneous
fit in $a$, $M_\pi$ and $M_\pi L$.}
\\[0.5ex]\\[0.5ex]
\ifx\reducedapptables\undefined
\hline \\[-0.5ex]
$\chi$QCD 21 & \cite{Liu:2021irg} & 2+1 & 0.06,0.08,0.11,0.14
& \parbox[t]{6.3cm}{Extrapolation performed using a quadratic term
  in $a$ as part of a simultaneous fit in $a$, $M_\pi$ and $M_\pi L$.}
\\[0.5ex]\\[0.5ex]  \hline\\[-0.5ex]
NME 21 & \cite{Park:2021ypf} & 2+1 & 0.07,0.09,0.13
& \parbox[t]{6.3cm}{Extrapolation performed using a linear term
  in $a$ as part of a simultaneous fit in $a$, $M_\pi$ and $M_\pi L$.}
\\[0.5ex]\\[0.5ex]  \hline\\[-0.5ex]
RBC/UKQCD 19 & \cite{Abramczyk:2019fnf} & 2+1 & 0.14
& \parbox[t]{6.3cm}{Single lattice spacing.}
\\[0.5ex]\\[0.5ex]  \hline\\[-0.5ex]
LHPC 19 & \cite{Hasan:2019noy} & 2+1 & 0.09,0.12
& \parbox[t]{6.3cm}{No extrapolation performed; final result taken from finer lattice with an enlarged error bar to account for result at coarser $a$.}
\\[0.5ex]\\[0.5ex]  \hline\\[-0.5ex]
Mainz 19 & \cite{Harris:2019bih} & 2+1 & 0.05,0.06,0.08,0.09
& \parbox[t]{6.3cm}{Extrapolation performed as part of a simultaneous
fit in $a$, $M_\pi$ and $M_\pi L$.}
\\[0.5ex]\\[0.5ex]  \hline\\[-0.5ex]
PACS 18A & \cite{Shintani:2018ozy} & 2+1 & 0.085 &
\parbox[t]{6.3cm}{Single lattice spacing.}
\\[0.5ex]\\[0.5ex]
PACS 18 & \cite{Ishikawa:2018rew} & 2+1 & 0.09 & \parbox[t]{6.3cm}{Single lattice spacing.}
\\[0.5ex]\\[0.5ex]  \hline\\[-0.5ex]
$\chi$QCD 18 & \cite{Liang:2018pis} & 2+1 & 0.14,0.11,0.08
& \parbox[t]{6.3cm}{Extrapolation to the physical point via
  simultaneous fit in the lattice spacing, $M_\pi$ and $M_\pi L$,
  including terms of order $a^2$.}
\\[0.5ex]\\[0.5ex]  \hline\\[-0.5ex]
JLQCD 18 & \cite{Yamanaka:2018uud} & 2+1 & 0.11 & \parbox[t]{6.3cm}{Single lattice spacing.}
\\[0.5ex]\\[0.5ex]  \hline\\[-0.5ex]
LHPC 12 & \cite{Green:2012ej} & 2+1 & 0.12,0.09
&  \parbox[t]{6.3cm}{No extrapolation or uncertainty
  estimated. Comparison between two lattice spacings at a single pion
  mass shows no deviation within statistical errors.} 
\\[0.5ex]\\[0.5ex]  \hline\\[-0.5ex]
LHPC 12A & \cite{Green:2012ud} & 2+1 & 0.12,0.09 
& \parbox[t]{6.3cm}{No statistically significant discretization effects
  observed. Results assumed to be constant in $a$.}
\\[0.5ex]\\[0.5ex] 
\fi
\hline \hline \\[-1.0ex]
\end{tabular*}
\caption{Continuum extrapolations/estimation of lattice artifacts in
  determinations of the isovector axial, scalar and tensor charges
  with $\Nf=2+1$ quark flavours.}
}
\end{table}

\ifx\reducedapptables\undefined
\begin{table}[!ht]
\addtocounter{table}{-1}
{\footnotesize
\begin{tabular*}{\textwidth}{l @{\extracolsep{\fill}} c c c l}
\hline\hline \\[-1.0ex]
Collab. & Ref. & $\Nf$ & $a$ [fm] & Description 
\\[.3ex]\\[-2.ex] \hline \hline \\[-1.0ex]
LHPC 10 & \cite{Bratt:2010jn} & 2+1 & 0.12 & \parbox[t]{6.3cm}{Single lattice spacing.}
\\[0.5ex]\\[0.5ex]  \hline\\[-0.5ex]
RBC/UKQCD 10D & \cite{Aoki:2010xg} & 2+1 & 0.11 &  \parbox[t]{6.3cm}{Single lattice spacing.}
\\[0.5ex]\\[0.5ex]  \hline\\[-0.5ex]
RBC/UKQCD 09B & \cite{Yamazaki:2009zq} & 2+1 & 0.11 & \parbox[t]{6.3cm}{Single lattice spacing.}
\\[0.5ex]\\[0.5ex]  \hline\\[-0.5ex]
RBC/UKQCD 08B & \cite{Yamazaki:2008py} & 2+1 & 0.11 & \parbox[t]{6.3cm}{Single lattice spacing.}
\\[0.5ex]\\[0.5ex]  \hline\\[-0.5ex]
LHPC 05 & \cite{Edwards:2005ym} & 2+1 & 0.12 & \parbox[t]{6.3cm}{Single lattice spacing.}
\\[0.5ex]\\[0.5ex]
\hline \hline \\[-1.0ex]
\end{tabular*}
\caption{(cntd.) Continuum extrapolations/estimation of lattice artifacts in
  determinations of the isovector axial, scalar and tensor charges
  with $\Nf=2+1$ quark flavours.}
}
\end{table}

\begin{table}[!ht]
{\footnotesize
\begin{tabular*}{\textwidth}{l @{\extracolsep{\fill}} c c c l}
\hline\hline \\[-1.0ex]
Collab. & Ref. & $\Nf$ & $a$ [fm] & Description 
\\[.3ex]\\[-2.ex] \hline \hline \\[-1.0ex]
%
Mainz 17 & \cite{Capitani:2017qpc} & 2 & 0.08,0.06,0.05
& \parbox[t]{6.3cm}{No statistically significant discretization effects
  observed. Results assumed to be constant in $a$.}
\\[0.5ex]\\[0.5ex]  \hline\\[-0.5ex]
ETM 17 & \cite{Alexandrou:2017qyt}  & 2 & 0.09  & \parbox[t]{6.3cm}{Single lattice spacing.}
\\[0.5ex]\\[0.5ex]  \hline\\[-0.5ex]
ETM 17B & \cite{Alexandrou:2017hac}  & 2 & 0.09 & \parbox[t]{6.3cm}{Single lattice spacing.}
\\[0.5ex]\\[0.5ex]  \hline\\[-0.5ex]
ETM 15D & \cite{Abdel-Rehim:2015owa} & 2 & 0.09 & \parbox[t]{6.3cm}{Single lattice spacing.}
\\[0.5ex]\\[0.5ex]  \hline\\[-0.5ex]
RQCD 14 & \cite{Bali:2014nma} & 2 & 0.08,0.07,0.06
& \parbox[t]{6.3cm}{No statistically significant discretization effects
  observed. No extrapolation performed, but residual systematic error
  due to discretization effects is estimated.}
\\[0.5ex]\\[0.5ex]  \hline\\[-0.5ex]
QCDSF 13 & \cite{Horsley:2013ayv} & 2 & 0.08,0.07,0.06
& \parbox[t]{6.3cm}{No statistically significant discretization
  effects observed. Results assumed to be constant in $a$.}
\\[0.5ex]\\[0.5ex]  \hline\\[-0.5ex]
Mainz 12 & \cite{Capitani:2012gj} & 2 & 0.08,0.06,0.05
& \parbox[t]{6.3cm}{No statistically significant discretization
  effects observed. Results assumed to be constant in $a$.} 
\\[0.5ex]\\[0.5ex]  \hline\\[-0.5ex]
RBC 08 & \cite{Lin:2008uz} & 2 & 0.11 & \parbox[t]{6.3cm}{Single lattice spacing.}
\\[0.5ex]\\[0.5ex]  \hline\\[-0.5ex]
QCDSF 06 & \cite{Khan:2006de} & 2 & 0.09,0.08,0.07,0.06
& \parbox[t]{6.3cm}{No statistically significant discretization
  effects observed. Results assumed to be constant in $a$.}
\\[0.5ex]\\[0.5ex]
\hline \hline \\[-1.0ex]
\end{tabular*}
\caption{Continuum extrapolations/estimation of lattice artifacts in
  determinations of the isovector axial, scalar and tensor charges
  with $\Nf=2$ quark flavours.}
}
\end{table}
\fi

\begin{table}[!ht]
{\footnotesize
\begin{tabular*}{\textwidth}{l @{\extracolsep{\fill}} c c c l}
\hline\hline \\[-1.0ex]
{Collab.} & {Ref.} & {$\Nf$} & {$M_{\pi,\rm min}\,[\mev]$} & {Description}  
\\[.3ex]\\[-2.ex] \hline \hline \\[-1.0ex]
%
ETM 23 & \cite{Alexandrou:2023qbg} & 2+1+1 & 140, 138, 141
& \parbox[t]{6.3cm}{Three pion masses within 3\% of the physical value.}
\\[0.5ex]\\[0.5ex]  \hline\\[-0.5ex]
PNDME 23 & \cite{Jang:2023zts} & 2+1+1 & 321, 228, 138, 136
& \parbox[t]{6.3cm}{Fit performed including leading-order pion-mass dependence.}
\\[0.5ex]\\[0.5ex] \hline\\[-0.5ex]
ETM 22 & \cite{Alexandrou:2022dtc} & 2+1+1 & 140, 138, 141
& \parbox[t]{6.3cm}{Three pion masses within 3\% of the physical value.}
\\[0.5ex]\\[0.5ex]
\ifx\reducedapptables\undefined
\hline \\[-0.5ex]
CalLat 19 & \cite{Walker-Loud:2019cif} & 2+1+1 & 135,130,220
& \parbox[t]{6.3cm}{Fit performed including analytic and nonanalytic
  terms in $M_\pi$ up to order $M_\pi^4$.}
\\[0.5ex]\\[0.5ex]  \hline\\[-0.5ex]
ETM 19 & \cite{Alexandrou:2019brg} & 2+1+1 & 139
& \parbox[t]{6.3cm}{Single pion mass within 3\% of the physical value.}
\\[0.5ex]\\[0.5ex]
PNDME 18 & \cite{Gupta:2018qil} & 2+1+1 & 321,225,138,136
& \parbox[t]{6.3cm}{Fit performed including a linear term in $M_\pi^2$
  and investigating a logarithmic term as part of a simultaneous fit
  in $a$, $M_\pi$ and $M_\pi L$.}
\\[0.5ex]\\[0.5ex]  \hline\\[-0.5ex]
CalLat 18 & \cite{Chang:2018uxx} & 2+1+1 & 130,130,220
& \parbox[t]{6.3cm}{Fit performed including analytic and nonanalytic
  terms in $M_\pi$ up to order $M_\pi^4$.}
\\[0.5ex]\\[0.5ex]  \hline\\[-0.5ex]
CalLat 17 & \cite{Berkowitz:2017gql} & 2+1+1 &
130,220,310 & \parbox[t]{6.3cm}{Fit performed including analytic and
  nonanalytic terms in $M_\pi$ up to order $M_\pi^4$.}
\\[0.5ex]\\[0.5ex]  \hline\\[-0.5ex]
PNDME 16 & \cite{Bhattacharya:2016zcn} & 2+1+1 & 225,138,235
& \parbox[t]{6.3cm}{Fit performed including linear and logarithmic
  terms in $M_\pi^2$ as part of a simultaneous fit in $a$, $M_\pi$ and
  $M_\pi L$.}
\\[0.5ex]\\[0.5ex]  \hline\\[-0.5ex]
PNDME 15 & \cite{Bhattacharya:2015wna,Bhattacharya:2015esa} & 2+1+1
& 225,138,235 & \parbox[t]{6.3cm}{Fit performed including a linear
  term in $M_\pi^2$ as part of a simultaneous fit in $a$, $M_\pi$ and
  $M_\pi L$.}
\\[0.5ex]\\[0.5ex]  \hline\\[-0.5ex]
PNDME 13 & \cite{Bhattacharya:2013ehc} & 2+1+1 & 228
&  \parbox[t]{6.3cm}{Fit performed with a linear term in $M_\pi^2$.}
\\[0.5ex]\\[0.5ex]
\fi
\hline \hline \\[-1.0ex]
\end{tabular*}
\caption{Chiral extrapolation/minimum pion mass in determinations of
  the isovector axial, scalar and tensor charges with $\Nf=2+1+1$
  quark flavours.} 
}
\end{table}

\begin{table}[!ht]
{\footnotesize
\begin{tabular*}{\textwidth}{l @{\extracolsep{\fill}} c c c l}
\hline\hline \\[-1.0ex]
{Collab.} & {Ref.} & {$\Nf$} & {$M_{\pi,\rm min}\,[\mev]$} & {Description}  
\\[.3ex]\\[-2.ex] \hline \hline \\[-1.0ex]
%
Mainz 24 & \cite{Djukanovic:2024krw} & 2+1 & 176, 130,218,228
& \parbox[t]{6.3cm}{Physical-point extrapolations
were performed simultaneously in the lattice spacing, pion mass, and
volume.}
\\[0.5ex]\\[0.5ex]  \hline\\[-0.5ex]
PACS 23 & \cite{Tsuji:2023llh} & 2+1 & 138, 135
& \parbox[t]{6.3cm}{Three near-physical pion masses.}
\\[0.5ex]\\[0.5ex]  \hline\\[-0.5ex]
RQCD 23 & \cite{Bali:2023sdi} & 2+1 & 336,176, 131,156,127,338
& \parbox[t]{6.3cm}{Extrapolations performed using leading-order chiral expressions for the
pion mass.}
\\[0.5ex]\\[0.5ex]  \hline\\[-0.5ex]
\parbox[t]{1.3cm}{QCDSF/ \par UKQCD/ \par CSSM 23} & \cite{QCDSFUKQCDCSSM:2023qlx} & 2+1 & 290,315,270,220,280,347
&  \parbox[t]{6.3cm}{Combined pion-mass, lattice-spacing, 
and volume extrapolations, performed around chiral SU(3) point.}
\\[0.5ex]\\[0.5ex]  \hline\\[-0.5ex]
PACS 22B & \cite{Tsuji:2022ric} & 2+1 & 135
& \parbox[t]{6.3cm}{Two near-physical pion masses.}
\\[0.5ex]\\[0.5ex]  \hline\\[-0.5ex]
Mainz~22 & \cite{Djukanovic:2022wru} & 2+1 & 176, 130,218,228
& \parbox[t]{6.3cm}{Physical-point extrapolations
were performed simultaneously in the lattice spacing, pion mass, and
volume.}
\\[0.5ex]\\[0.5ex]
\ifx\reducedapptables\undefined
\hline \\[-0.5ex]
$\chi$QCD 21 & \cite{Liu:2021irg} & 2+1 & 139, 171, 302, 337, 371
& \parbox[t]{6.3cm}{Extrapolation performed using quadratic terms
  in $M_{\pi}$ from partially quenched chiral perturbation theory as part of a simultaneous fit in $a$, $M_\pi$ and $M_\pi L$.}
\\[0.5ex]\\[0.5ex]  \hline\\[-0.5ex]
NME 21 & \cite{Park:2021ypf} & 2+1 & 170,170,270,285
& \parbox[t]{6.3cm}{Extrapolation performed using a quadratic term
  in $M_{\pi}$ as part of a simultaneous fit in $a$, $M_\pi$ and $M_\pi L$.}
\\[0.5ex]\\[0.5ex]  \hline\\[-0.5ex]
RBC/UKQCD 19 & \cite{Abramczyk:2019fnf} & 2+1 & 170,250
& \parbox[t]{6.3cm}{Does not quote extrapolated value, but shows low mass dependence.}
\\[0.5ex]\\[0.5ex]  \hline\\[-0.5ex]
LHPC 19 & \cite{Hasan:2019noy} & 2+1 & 133,137
&  \parbox[t]{6.3cm}{Does not perform extrapolation; comparison of results at two near-physical pion masses.}
\\[0.5ex]\\[0.5ex]  \hline\\[-0.5ex]
Mainz 19 & \cite{Harris:2019bih} & 2+1 & 220, 290, 200, 260
& \parbox[t]{6.3cm}{Extrapolation performed using logarithmic and
quadratic terms in $M_\pi$ as part of a simultaneous fit in $a$,
$M_\pi$ and $M_\pi L$.}
\\[0.5ex]\\[0.5ex]  \hline\\[-0.5ex]
PACS 18A & \cite{Shintani:2018ozy} & 2+1 & 135
& \parbox[t]{6.3cm}{Single, physical pion mass.}
\\[0.5ex]\\[0.5ex]
PACS 18 & \cite{Ishikawa:2018rew} & 2+1 & 146
& \parbox[t]{6.3cm}{Single, nearly physical pion mass.}
\\[0.5ex]\\[0.5ex]  \hline\\[-0.5ex]
$\chi$QCD 18 & \cite{Liang:2018pis} & 2+1 & 171,337,302
& \parbox[t]{6.3cm}{Linear fit in $M_\pi^2$, including terms
  containing $M_\pi L$.}
\\[0.5ex]\\[0.5ex]  \hline\\[-0.5ex]
JLQCD 18 & \cite{Yamanaka:2018uud} & 2+1 & 293
& \parbox[t]{6.3cm}{Fit performed with linear and quadratic terms in
  $M_\pi^2$.}
\\[0.5ex]\\[0.5ex] 
\fi
\hline \hline \\[-1.0ex]
\end{tabular*}
\caption{Chiral extrapolation/minimum pion mass in determinations of
  the isovector axial, scalar and tensor charges with $\Nf=2+1$
  quark flavours.} 
}
\end{table}

\ifx\reducedapptables\undefined
\begin{table}[!ht]
\addtocounter{table}{-1}
{\footnotesize
\begin{tabular*}{\textwidth}{l @{\extracolsep{\fill}} c c c l}
\hline\hline \\[-1.0ex]
{Collab.} & {Ref.} & {$\Nf$} & {$M_{\pi}\,[\mev]$} & {Description}  
\\[.3ex]\\[-2.ex] \hline \hline \\[-1.0ex]
%
LHPC 12 & \cite{Green:2012ej} & 2+1 & 149,317
&  \parbox[t]{6.3cm}{Chiral fit performed with $g_A$ and $F_{\pi}$
  held fixed to NNLO in SU(2) HBChPT. Several other $M_\pi$
  investigated.} 
\\[0.5ex]\\[0.5ex]  \hline\\[-0.5ex]
LHPC 12A & \cite{Green:2012ud} & 2+1 & 149,317
& \parbox[t]{6.3cm}{Chiral fit formula based on the ``small scale
  expansion'' to order $\epsilon^3$ with some coefficients fixed.}
\\[0.5ex]\\[0.5ex]  \hline\\[-0.5ex]
LHPC 10 & \cite{Bratt:2010jn} & 2+1 & 293 & \parbox[t]{6.3cm}{Chiral
  fit formula based on the ``small scale expansion'' to order
  $\epsilon^3$ with some coefficients fixed.}
\\[0.5ex]\\[0.5ex]  \hline\\[-0.5ex]
RBC/UKQCD 10D & \cite{Aoki:2010xg} & 2+1 & 329,416,555,668
&  \parbox[t]{6.3cm}{Constant fit to heaviest three and linear fit to
  lightest two pion masses gives the quoted range.}
\\[0.5ex]\\[0.5ex]  \hline\\[-0.5ex]
RBC/UKQCD 09B & \cite{Yamazaki:2009zq} & 2+1 & 329
& \parbox[t]{6.3cm}{Linear fit in $M_\pi^2$, supplemented by a term in
  $e^{-M_\pi L}$ to describe finite-volume effects.}
\\[0.5ex]\\[0.5ex]  \hline\\[-0.5ex]
RBC/UKQCD 08B & \cite{Yamazaki:2008py} & 2+1 & 329
& \parbox[t]{6.3cm}{Linear fit in $M_\pi^2$, supplemented by a term in 
  $e^{-M_\pi L}$ to describe finite-volume effects.}
\\[0.5ex]\\[0.5ex]  \hline\\[-0.5ex]
LHPC 05 & \cite{Edwards:2005ym} & 2+1 & 353 & \parbox[t]{6.3cm}{Chiral
  fit formula based on the ``small scale expansion'' with some
  coefficients fixed.}
\\[0.5ex]\\[0.5ex]
\hline \hline \\[-1.0ex]
\end{tabular*}
\caption{(cntd.) Chiral extrapolation/minimum pion mass in determinations of
  the isovector axial, scalar and tensor charges with $\Nf=2+1$
  quark flavours.} 
}
\end{table}

\begin{table}[!ht]
{\footnotesize
\begin{tabular*}{\textwidth}{l @{\extracolsep{\fill}} c c c l}
\hline\hline \\[-1.0ex]
{Collab.} & {Ref.} & {$\Nf$} & {$M_{\pi,\rm min}\,[\mev]$} & {Description}  
\\[.3ex]\\[-2.ex] \hline \hline \\[-1.0ex]
%
Mainz 17 & \cite{Capitani:2017qpc} & 2 & 268,193,261
& \parbox[t]{6.3cm}{Main result from linear fit in $M_\pi^2$ with mass
  cut of $M_\pi\leq 300$\,MeV. Systematic error estimated from
  considering fits containing logarithmic terms.}
\\[0.5ex]\\[0.5ex]  \hline\\[-0.5ex]
ETM 17 & \cite{Alexandrou:2017qyt} & 2 & 130
& \parbox[t]{6.3cm}{Single, nearly physical, pion mass.}
\\[0.5ex]\\[0.5ex]  \hline\\[-0.5ex]
ETM 17B & \cite{Alexandrou:2017hac} & 2 & 130
& \parbox[t]{6.3cm}{Single, nearly physical, pion mass.}
\\[0.5ex]\\[0.5ex]  \hline\\[-0.5ex]
ETM 15D & \cite{Abdel-Rehim:2015owa} & 2 & 130
& \parbox[t]{6.3cm}{Single, nearly physical, pion mass.}
\\[0.5ex]\\[0.5ex]  \hline\\[-0.5ex]
RQCD 14 & \cite{Bali:2014nma} & 2 & 280,150,260
& \parbox[t]{6.3cm}{Linear fit in $M_\pi^2$ with pion mass cut
  ($M_\pi^2<0.1$~GeV$^2$).} 
\\[0.5ex]\\[0.5ex]  \hline\\[-0.5ex]
QCDSF 13 & \cite{Horsley:2013ayv} & 2 & 259,158,262
& \parbox[t]{6.3cm}{Chiral fit of $g_A/f_\pi$ including linear,
  quadratic and logarithmic terms in $M_\pi^2$.}
\\[0.5ex]\\[0.5ex]  \hline\\[-0.5ex]
Mainz 12 & \cite{Capitani:2012gj} & 2 & 312,277,430
& \parbox[t]{6.3cm}{Main result from linear fit in
  $M_\pi^2$. Systematic error estimated from considering additional
  terms containing $M_\pi^4$ and $M_\pi L$.}
\\[0.5ex]\\[0.5ex]  \hline\\[-0.5ex]
RBC 08 & \cite{Lin:2008uz} & 2 & 493 &  \parbox[t]{6.3cm}{Linear fit
  in $M_\pi^2$.}
\\[0.5ex]\\[0.5ex]  \hline\\[-0.5ex]
QCDSF 06 & \cite{Khan:2006de} & 2 & 617,664,666,726
& \parbox[t]{6.3cm}{Chiral fit formula based on the ``small scale 
  expansion'' with some fixed coefficients and finite-volume
  corrections subtracted.}
\\[0.5ex]\\[0.5ex] 
\hline \hline \\[-1.0ex]
\end{tabular*}
\caption{Chiral extrapolation/minimum pion mass in determinations of
  the isovector axial, scalar and tensor charges with $\Nf=2$
  quark flavours.} 
}
\end{table}
\fi

\begin{table}[!ht]
{\footnotesize
\begin{tabular*}{\textwidth}{l @{\extracolsep{\fill}} c c c c l}
\hline\hline \\[-1.0ex]
Collab. & Ref. & $\Nf$ & $L$ [fm] & ${M_{\pi,\rm min}}L$ & Description 
\\[.3ex]\\[-2.ex] \hline \hline \\[-1.0ex]
%
ETM 23 & \cite{Alexandrou:2023qbg} & 2+1+1 & \parbox[t]{1.4cm}{5.1,\\5.4,\\5.5}
& \parbox[t]{1.cm}{3.6,\\3.8,\\3.9} & \parbox[t]{5.9cm}{No extrapolation performed.}
\\[0.5ex]\\[0.5ex]  \hline\\[-0.5ex]
PNDME 23 & \cite{Jang:2023zts} & 2+1+1 & \parbox[t]{1.4cm}{2.4,\\2.9-4.8,\\2.9-5.8,\\2.9-5.8}
& \parbox[t]{1.cm}{3.62,\\2.98 } & \parbox[t]{5.9cm}{Physical-point
extrapolations performed simultaneously, using the
leading-order terms in the various expansion parameters.}
\\[0.5ex]\\[0.5ex] \hline\\[-0.5ex]
ETM 22 & \cite{Alexandrou:2022dtc} & 2+1+1 &  \parbox[t]{1.4cm}{5.1,\\5.4,\\5.5}
& \parbox[t]{1.cm}{3.62,\\2.98 } & \parbox[t]{5.9cm}{No extrapolation performed.}
\\[0.5ex]\\[0.5ex]
\ifx\reducedapptables\undefined
\hline \\[-0.5ex]
CalLat 19 & \cite{Walker-Loud:2019cif} & 2+1+1
& \parbox[t]{1.4cm}{2.4--7.2,\\2.9--5.8,\\2.9--4.2}
& \parbox[t]{1.cm}{4.9,\\3.8,\\4.7} & \parbox[t]{5.9cm}{Fit
  performed including a term of the form
  ${M_\pi^2}e^{-M_\pi L}/{\sqrt{M_{\pi}L}}$ as part of a
  simultaneous fit in $a^2$, $M_\pi$ and $M_\pi L$.}
\\[0.5ex]\\[0.5ex]  \hline\\[-0.5ex]
ETM 19 & \cite{Alexandrou:2019brg} & 2+1+1
& \parbox[t]{1.4cm}{5.12,\\4.5-6.0}
& \parbox[t]{1.cm}{3.62,\\2.98 } & \parbox[t]{5.9cm}{Final result quoted from the $M_\pi L = 3.62$ ensemble only.}
\\[0.5ex]\\[0.5ex]
PNDME 18 & \cite{Gupta:2018qil} & 2+1+1
& \parbox[t]{1.4cm}{2.4,\\2.9--4.8,\\2.8--5.6,\\2.8--5.5}
& \parbox[t]{1.cm}{3.9,\\5.5,\\3.9,\\3.7} & \parbox[t]{5.9cm}{Fit
  performed including a term of the form $M_\pi^2e^{-M_\pi L}$ as
  part of a simultaneous fit in $a$, $M_\pi$ and $M_\pi L$.}
\\[0.5ex]\\[0.5ex]  \hline\\[-0.5ex]
CalLat 18 & \cite{Chang:2018uxx} & 2+1+1
& \parbox[t]{1.4cm}{2.4--4.8,\\2.9--5.8,\\2.9--4.2}
& \parbox[t]{1.cm}{3.2,\\3.9,\\4.7} & \parbox[t]{5.9cm}{Fit
  performed including a term of the form
  ${M_\pi^2}e^{-M_\pi L}/{\sqrt{M_{\pi}L}}$ as part of a
  simultaneous fit in $a^2$, $M_\pi$ and $M_\pi L$.}
\\[0.5ex]\\[0.5ex]  \hline\\[-0.5ex]
CalLat 17 & \cite{Berkowitz:2017gql} & 2+1+1
& \parbox[t]{1.4cm}{2.4--4.8,\\2.9--4.8,\\2.9}
& \parbox[t]{1.cm}{3.2,\\5.4,\\4.5} & \parbox[t]{5.9cm}{Fit
  performed including a term of the form
  ${M_\pi^2}e^{-M_\pi L}/{\sqrt{M_{\pi}L}}$ as part of a
  simultaneous fit in $a^2$, $M_\pi$ and $M_\pi L$.}
\\[0.5ex]\\[0.5ex]  \hline\\[-0.5ex]
PNDME 16 & \cite{Bhattacharya:2016zcn} & 2+1+1
& \parbox[t]{1.4cm}{2.9--4.8,\\2.8--5.6,\\2.8--3.7}
& \parbox[t]{1.cm}{5.5,\\3.9,\\4.4} & \parbox[t]{5.9cm}{Fit performed
  including a term of the form $M_\pi^2e^{-M_\pi L}$ as part of a
  simultaneous fit in $a$, $M_\pi$ and $M_\pi L$.} 
\\[0.5ex]\\[0.5ex]  \hline\\[-0.5ex]
PNDME 15 & \cite{Bhattacharya:2015wna,Bhattacharya:2015esa} & 2+1+1
& \parbox[t]{1.4cm}{2.9--4.8,\\2.8--5.6,\\2.8--3.7}
& \parbox[t]{1.cm}{5.5,\\3.9,\\4.4} & \parbox[t]{5.9cm}{Fit performed
  including a term of the form $\exp\{-M_\pi L\}$ as part of a
  simultaneous fit in $a$, $M_\pi$ and $M_\pi L$.} 
\\[0.5ex]\\[0.5ex]  \hline\\[-0.5ex]
PNDME 13 & \cite{Bhattacharya:2013ehc} & 2+1+1
& \parbox[t]{1.4cm}{2.9--3.8} & \parbox[t]{1.cm}{4.4}
& \parbox[t]{6.3cm}{Finite-volume effects not estimated.}
\\[0.5ex]\\[0.5ex]
\fi
\hline\hline  \\[-1.0ex]
\end{tabular*}
\caption{Finite-volume effects in determinations of the isovector
  axial, scalar and tensor charges with $\Nf=2+1+1$ quark flavours.}
}
\end{table}

\begin{table}[!ht]
{\footnotesize
\begin{tabular*}{\textwidth}{l @{\extracolsep{\fill}} c c c c l}
\hline\hline \\[-1.0ex]
Collab. & Ref. & $\Nf$ & $L$ [fm] & ${M_{\pi,\rm min}}L$ & Description 
\\[.3ex]\\[-2.ex] \hline \hline \\[-1.0ex]
%
Mainz 24 & \cite{Djukanovic:2024krw} & 2+1 & \parbox[t]{1.4cm}{4.1\\4.8,\\6.1,\\4.7} & \parbox[t]{1.4cm}{4.7,\\5.4,\\4.0,\\4.2}
& \parbox[t]{6.3cm}{Extrapolation performed including a term of the form
$M_\pi^2e^{-M_\pi L}/\sqrt{M_\pi L}$  as part of a simultaneous fit in $a$, $M_\pi$ and $M_\pi L$.}
\\[0.5ex]\\[0.5ex]  \hline\\[-0.5ex]
PACS 23 & \cite{Tsuji:2023llh} & 2+1 & \parbox[t]{1.4cm}{5.5-10.9,\\10.9} & \parbox[t]{1.4cm}{3.8-7.5,\\7.5}
& \parbox[t]{6.3cm}{Negligible finite-volume effects seen between volumes.}
\\[0.5ex]\\[0.5ex]  \hline\\[-0.5ex]
RQCD 23 & \cite{Bali:2023sdi} & 2+1 & \parbox[t]{1.4cm}{4.8,\\2.75-11,\\4.9-9.7,\\4.1-12.3,\\6.3-9.4,\\7.5} & \parbox[t]{1.4cm}{4.0,\\3.5,\\3.8,\\4.1,\\4.3,\\4.3}
& \parbox[t]{6.3cm}{Leading asymptotic form for finite-volume corrections used for extrapolation.}
\\[0.5ex]\\[0.5ex]  \hline\\[-0.5ex]
\parbox[t]{1.3cm}{QCDSF/ \par UKQCD/ \par CSSM 23} & \cite{QCDSFUKQCDCSSM:2023qlx} & 2+1 & \parbox[t]{1.4cm}{2.6,\\2.4,\\3.3-4.4,\\2.8,\\2.5} & \parbox[t]{1.4cm}{4.0,\\3.9}
& \parbox[t]{6.3cm}{Combined pion-mass, lattice-spacing, 
and volume extrapolations performed.}
\\[0.5ex]\\[0.5ex]  \hline\\[-0.5ex]
PACS 22B & \cite{Tsuji:2022ric} & 2+1 &
\parbox[t]{1.4cm}{5.5-10.9} &
\parbox[t]{1.4cm}{3.8-7.5}
& \parbox[t]{6.3cm}{Negligible finite-volume effects seen between volumes.}
\\[0.5ex]\\[0.5ex]  \hline\\[-0.5ex]
Mainz~22 & \cite{Djukanovic:2022wru} & 2+1 &  \parbox[t]{1.4cm}{4.1\\4.8,\\6.1,\\4.7} & \parbox[t]{1.4cm}{4.7,\\5.4,\\4.0,\\4.2}& \parbox[t]{5.9cm}{Extrapolation performed including a term of the form
 $M_\pi^2e^{-M_\pi L}/\sqrt{M_\pi L}$ as part of a simultaneous fit in $a$, $M_\pi$ and $M_\pi L$.}
\\[0.5ex]\\[0.5ex] 
\ifx\reducedapptables\undefined
\hline \\[-0.5ex]
$\chi$QCD 21 & \cite{Liu:2021irg} & 2+1 & \parbox[t]{1.4cm}{4.5\\5.3,\\2.6,\\2.6,\\1.92} & \parbox[t]{1.4cm}{3.7,\\3.9,\\3.6,\\3.8,\\3.6}
& \parbox[t]{6.3cm}{Extrapolation performed including a term of the form
  $ e^{-M_\pi L}$ as part of a simultaneous fit in $a$, $M_\pi$ and $M_\pi L$.}
\\[0.5ex]\\[0.5ex]  \hline\\[-0.5ex]
NME 21 & \cite{Park:2021ypf} & 2+1 & \parbox[t]{1.4cm}{4.2,\\2.9-4.3,\\4.3-5.8,\\3.4-5.0} & \parbox[t]{1.4cm}{5.87,\\4.09,\\3.75,\\4.28}
& \parbox[t]{6.3cm}{Extrapolation performed including a term of the form
  ${M_\pi^2}e^{-M_\pi L}/{\sqrt{M_{\pi}L}}$ as part of a
  simultaneous fit in $a^2$, $M_\pi$ and $M_\pi L$.}
\\[0.5ex]\\[0.5ex]  \hline\\[-0.5ex]
RBC/UKQCD 19 & \cite{Abramczyk:2019fnf} & 2+1 & \parbox[t]{1.4cm}{4.5} & \parbox[t]{1.4cm}{4.0}
& \parbox[t]{6.3cm}{Does not quote extrapolated value.}
\\[0.5ex]\\[0.5ex]  \hline\\[-0.5ex]
LHPC 19 & \cite{Hasan:2019noy} & 2+1 &	\parbox[t]{1.4cm}{5.9,\\5.6} & \parbox[t]{1.4cm}{4.0,\\3.9}
& \parbox[t]{6.3cm}{Does not quote extrapolated value.}
\\[0.5ex]\\[0.5ex]  \hline\\[-0.5ex]
Mainz 19 & \cite{Harris:2019bih} & 2+1 &
\parbox[t]{1.4cm}{2.8-4.1,\\2.4-3.6,\\2.1-4.1,2.4-3.2} &
\parbox[t]{1.4cm}{4.7,\\5.3,\\4.2,\\4.3}
& \parbox[t]{6.3cm}{Extrapolation performed including a term of the
form ${M_\pi^2}e^{-M_\pi L}/{\sqrt{M_{\pi}L}}$ as part of a
  simultaneous fit in $a^2$, $M_\pi$ and $M_\pi L$.}
\\[0.5ex]\\[0.5ex]  \hline\\[-0.5ex]
PACS 18A & \cite{Shintani:2018ozy} & 2+1 & \parbox[t]{1.4cm}{10.8}
& \parbox[t]{1.cm}{7.4} & \parbox[t]{5.9cm}{Single spatial volume of
  10.8\,fm at physical pion mass $M_\pi L=7.4$.}
\\[0.5ex]\\[0.5ex] 
PACS 18 & \cite{Ishikawa:2018rew} & 2+1 & \parbox[t]{1.4cm}{8.2}
& \parbox[t]{1.cm}{6.0} & \parbox[t]{5.9cm}{Single spatial volume of
  8.2\,fm at near-physical pion mass $M_\pi L=6$.}
\\[0.5ex]\\[0.5ex] 
\fi
\hline\hline  \\[-1.0ex]
\end{tabular*}
\caption{Finite-volume effects in determinations of the isovector
  axial, scalar and tensor charges with $\Nf=2+1$ quark flavours.}
}
\end{table}

\ifx\reducedapptables\undefined
\begin{table}[!ht]
\addtocounter{table}{-1}
{\footnotesize
\begin{tabular*}{\textwidth}{l @{\extracolsep{\fill}} c c c c l}
\hline\hline \\[-1.0ex]
Collab. & Ref. & $\Nf$ & $L$ [fm] & ${M_{\pi,\rm min}}L$ & Description 
\\[.3ex]\\[-2.ex] \hline \hline \\[-1.0ex]
$\chi$QCD 18 & \cite{Liang:2018pis} & 2+1
& \parbox[t]{1.4cm}{4.6,\\2.7,\\2.6}
& \parbox[t]{1.cm}{4.0,\\4.5,\\4.1} & \parbox[t]{5.9cm}{Fit performed
  including a term of the form $e^{-M_\pi L}$ as part of a
  simultaneous fit in $a$, $M_\pi$ and $M_\pi L$.}
\\[0.5ex]\\[0.5ex]  \hline\\[-0.5ex]
JLQCD 18 & \cite{Yamanaka:2018uud} & 2+1
& \parbox[t]{1.4cm}{1.8--2.7} & \parbox[t]{1.cm}{4.0}
& \parbox[t]{5.9cm}{Finite-volume effects not estimated.}
\\[0.5ex]\\[0.5ex]  \hline\\[-0.5ex]
LHPC 12 & \cite{Green:2012ej} & 2+1
& \parbox[t]{1.4cm}{2.8--5.6,\\2.9} & \parbox[t]{1.cm}{4.2,\\4.6}
& \parbox[t]{6.3cm}{No uncertainty estimated. Comparison between two
  volumes at a single pion mass shows no deviation within statistical
  errors.}
\\[0.5ex]\\[0.5ex]  \hline\\[-0.5ex]
LHPC 12A & \cite{Green:2012ud} & 2+1
& \parbox[t]{1.4cm}{2.8--5.6,\\2.9} & \parbox[t]{1.cm}{4.2,\\4.6}
& \parbox[t]{5.9cm}{Finite-volume effects investigated and found to be
  negligible.}
\\[0.5ex]\\[0.5ex]  \hline\\[-0.5ex]
LHPC 10 & \cite{Bratt:2010jn} & 2+1 & \parbox[t]{1.4cm}{2.5--3.5}
& \parbox[t]{1.cm}{3.7} & \parbox[t]{5.9cm}{Finite-volume effects 
  included in chiral fit formula and found to be negligible.}
\\[0.5ex]\\[0.5ex]  \hline\\[-0.5ex]
RBC/UKQCD 10D & \cite{Aoki:2010xg} & 2+1 & \parbox[t]{1.4cm}{2.7}
& \parbox[t]{1.cm}{4.6} & \parbox[t]{5.9cm}{No uncertainty
  estimated. Comparison between two volumes for the three heaviest 
  pion masses shows no deviation within statistical
  errors.}
\\[0.5ex]\\[0.5ex]  \hline\\[-0.5ex]
RBC/UKQCD 09B & \cite{Yamazaki:2009zq} & 2+1 & \parbox[t]{1.4cm}{2.7}
& \parbox[t]{1.cm}{4.6} & \parbox[t]{5.9cm}{Chiral fit ansatz contains
  term in $e^{-M_\pi L}$ to describe finite-volume effects.}
\\[0.5ex]\\[0.5ex]  \hline\\[-0.5ex]
RBC/UKQCD 08B & \cite{Yamazaki:2008py} & 2+1 & \parbox[t]{1.4cm}{2.7}
& \parbox[t]{1.cm}{4.6} & \parbox[t]{5.9cm}{Chiral fit ansatz contains
  term in $e^{-M_\pi L}$ to describe finite-volume effects.}
\\[0.5ex]\\[0.5ex]  \hline\\[-0.5ex]
LHPC 05 & \cite{Edwards:2005ym} & 2+1 & \parbox[t]{1.4cm}{2.5--3.5}
& \parbox[t]{1.cm}{6.2} & \parbox[t]{5.9cm}{Finite-volume effects
  included in chiral fit formula and found to be negligible.}
\\[0.5ex]\\[0.5ex]
\hline\hline  \\[-1.0ex]
\end{tabular*}
\caption{(cntd.) Finite-volume effects in determinations of the isovector
  axial, scalar and tensor charges with $\Nf=2+1$ quark flavours.}
}
\end{table}

\begin{table}[!ht]
{\footnotesize
\begin{tabular*}{\textwidth}{l @{\extracolsep{\fill}} c c c c l}
\hline\hline \\[-1.0ex]
Collab. & Ref. & $\Nf$ & $L$ [fm] & ${M_{\pi,\rm min}}L$ & Description 
\\[.3ex]\\[-2.ex] \hline \hline \\[-1.0ex]
%
Mainz 17 & \cite{Capitani:2017qpc} & 2
& \parbox[t]{1.4cm}{2.5--3.8,\\2.0--4.0,\\2.4--3.2}
& \parbox[t]{1.cm}{5.0,\\4.0,\\4.4} & \parbox[t]{5.9cm}{Finite-volume
  effects not estimated.}
\\[0.5ex]\\[0.5ex]  \hline\\[-0.5ex]
ETM 17 & \cite{Alexandrou:2017qyt} & 2 & \parbox[t]{1.4cm}{4.5}
& \parbox[t]{1.cm}{3.0} & \parbox[t]{6.3cm}{Finite-volume effects not
  estimated.}
\\[0.5ex]\\[0.5ex]  \hline\\[-0.5ex]
ETM 17B & \cite{Alexandrou:2017hac} & 2 & \parbox[t]{1.4cm}{4.5}
& \parbox[t]{1.cm}{3.0} & \parbox[t]{5.9cm}{Finite-volume effects not
  estimated.}
\\[0.5ex]\\[0.5ex]  \hline\\[-0.5ex]
ETM 15D & \cite{Abdel-Rehim:2015owa} & 2 & \parbox[t]{1.4cm}{4.5}
& \parbox[t]{1.cm}{3.0} & \parbox[t]{5.9cm}{Finite-volume effects not
  estimated.}
\\[0.5ex]\\[0.5ex]  \hline\\[-0.5ex]
RQCD 14 & \cite{Bali:2014nma} & 2
& \parbox[t]{1.4cm}{2.6,\\1.7--4.5,\\1.9--2.9}
& \parbox[t]{1.cm}{3.7,\\3.5,\\3.8} & \parbox[t]{5.9cm}{No uncertainty
  estimated. Comparison between volumes for two pion masses, including
  the lightest, shows no deviation within statistical errors.} 
\\[0.5ex]\\[0.5ex]  \hline\\[-0.5ex]
QCDSF 13 & \cite{Horsley:2013ayv} & 2
& \parbox[t]{1.4cm}{1.2--2.4,\\0.9--3.4,\\1.4--2.9}
& \parbox[t]{1.cm}{3.2,\\2.7,\\3.8} & \parbox[t]{5.9cm}{Finite-volume
  effects assumed to be negligible in ratio $g_A/f_\pi$.}
\\[0.5ex]\\[0.5ex]  \hline\\[-0.5ex]
Mainz 12 & \cite{Capitani:2012gj} & 2
& \parbox[t]{1.4cm}{2.5,\\2.0--3.0,\\2.4}
& \parbox[t]{1.cm}{4.0,\\4.3,\\5.2} & \parbox[t]{5.9cm}{Finite-volume
  corrections estimated in $\chi$PT and subtracted before chiral fit.}
\\[0.5ex]\\[0.5ex]  \hline\\[-0.5ex]
RBC 08 & \cite{Lin:2008uz} & 2 & \parbox[t]{1.4cm}{1.8}
& \parbox[t]{1.cm}{4.6} & \parbox[t]{5.9cm}{Finite-volume effects
  investigated, but not included in error budget. All volumes below 2\,fm.}
\\[0.5ex]\\[0.5ex]  \hline\\[-0.5ex]
QCDSF 06 & \cite{Khan:2006de} & 2
& \parbox[t]{1.4cm}{1.5,\\1.2--1.8,\\0.9--1.7,\\1.4}
& \parbox[t]{1.cm}{4.7,\\6.1,\\5.7,\\5.3}
& \parbox[t]{5.9cm}{Finite-volume effects included in fit formula
  based on ``small scale expansion''. All spatial volumes below
  2\,fm.}
\\[0.5ex]\\[0.5ex]
\hline\hline  \\[-1.0ex]
\end{tabular*}
\caption{Finite-volume effects in determinations of the isovector
  axial, scalar and tensor charges with $\Nf=2$ quark flavours.}
}
\end{table}
\fi

\begin{table}[!ht]
{\footnotesize
\begin{tabular*}{\textwidth}{l @{\extracolsep{\fill}} c c c l}
\hline\hline \\[-1.0ex]
Collab. & Ref. & $\Nf$ & Ren.
\\[.3ex]\\[-2.ex] \hline \hline \\[-1.0ex]
%
ETM 23 & \cite{Alexandrou:2023qbg} & 2+1+1 & RI'-MOM  \\[0.5ex]\\[0.5ex]
\hline\\[-0.5ex]
PNDME 23 & \cite{Jang:2023zts} & 2+1+1 & RI-SMOM  \\[0.5ex]\\[0.5ex]
\hline\\[-0.5ex]
ETM 22 & \cite{Alexandrou:2022dtc} & 2+1+1 & RI'-MOM  \\[0.5ex]\\[0.5ex]
\ifx\reducedapptables\undefined
\hline \\[-0.5ex]
CalLat 19 & \cite{Walker-Loud:2019cif} & 2+1+1 & RI-MOM  \\[0.5ex]\\[0.5ex]
\hline\\[-0.5ex]
ETM 19 & \cite{Alexandrou:2019brg} & 2+1+1 & RI-MOM  \\[0.5ex]\\[0.5ex]
\hline\\[-0.5ex]
PNDME 18 & \cite{Gupta:2018qil} & 2+1+1 & RI-SMOM  \\[0.5ex]\\[0.5ex]
\hline\\[-0.5ex]
CalLat 18 & \cite{Chang:2018uxx} & 2+1+1 & RI-MOM  \\[0.5ex]\\[0.5ex]
\hline\\[-0.5ex]
CalLat 17 & \cite{Berkowitz:2017gql} & 2+1+1 & RI-MOM  \\[0.5ex]\\[0.5ex]
\hline\\[-0.5ex]
PNDME 16 & \cite{Bhattacharya:2016zcn} & 2+1+1 & RI-SMOM  \\[0.5ex]\\[0.5ex]
\hline\\[-0.5ex]
PNDME 15 & \cite{Bhattacharya:2015wna,Bhattacharya:2015esa} & 2+1+1 & RI-SMOM    \\[0.5ex]\\[0.5ex]
\hline\\[-0.5ex]
PNDME 13 & \cite{Bhattacharya:2013ehc} & 2+1+1 & RI-SMOM    \\[0.5ex]\\[0.5ex]
\fi
 \hline \hline \\[-1.0ex]
\end{tabular*}
\caption{Renormalization in determinations of the isovector axial,
  scalar and tensor charges with $\Nf=2+1+1$ quark
  flavours.}
}
\end{table}

\begin{table}[!ht]
{\footnotesize
\begin{tabular*}{\textwidth}{l @{\extracolsep{\fill}} c c c l}
\hline\hline \\[-1.0ex]
Collab. & Ref. & $\Nf$ & Ren.
\\[.3ex]\\[-2.ex] \hline \hline \\[-1.0ex]
Mainz 24 & \cite{Djukanovic:2024krw} & 2+1 & RI-SMOM  \\[0.5ex]\\[0.5ex]
\hline\\[-0.5ex]
PACS 23 & \cite{Tsuji:2023llh} & 2+1 & RI-SMOM  \\[0.5ex]\\[0.5ex]
\hline\\[-0.5ex]
RQCD 23 & \cite{Bali:2023sdi} & 2+1 & RI'-SMOM  \\[0.5ex]\\[0.5ex]
\hline\\[-0.5ex]
\parbox[t]{1.3cm}{QCDSF/ \par UKQCD/ \par CSSM 23} & \cite{QCDSFUKQCDCSSM:2023qlx} & 2+1 & RI'-MOM  \\[0.5ex]\\[0.5ex]
\hline\\[-0.5ex]
PACS 22B & \cite{Tsuji:2022ric} & 2+1 & RI-SMOM
\\[0.5ex]\\[0.5ex]
\hline\\[-0.5ex]
Mainz~22 & \cite{Djukanovic:2022wru} & 2+1 & RI-SMOM  \\[0.5ex]\\[0.5ex]
\ifx\reducedapptables\undefined
\hline \\[-0.5ex]
$\chi$QCD 21 & \cite{Liu:2021irg} & 2+1 & RI-MOM  \\[0.5ex]\\[0.5ex]
\hline\\[-0.5ex]
NME 21 & \cite{Park:2021ypf} & 2+1 & RI-SMOM  \\[0.5ex]\\[0.5ex]
\hline\\[-0.5ex]
RBC/UKQCD 19 & \cite{Abramczyk:2019fnf} & 2+1 & RI-SMOM  \\[0.5ex]\\[0.5ex]
\hline\\[-0.5ex]
LHPC 19 & \cite{Hasan:2019noy} & 2+1 & RI-MOM, RI-SMOM  \\[0.5ex]\\[0.5ex]
\hline\\[-0.5ex]
Mainz 19 & \cite{Harris:2019bih} & 2+1 & RI-MOM, SF
\\[0.5ex]\\[0.5ex]
\hline\\[-0.5ex]
PACS 18A & \cite{Shintani:2018ozy} & 2+1 & SF  \\[0.5ex]\\[0.5ex]
\hline\\[-0.5ex]
PACS 18 & \cite{Ishikawa:2018rew} & 2+1 & SF  \\[0.5ex]\\[0.5ex]
\hline\\[-0.5ex]
$\chi$QCD 18 & \cite{Liang:2018pis} & 2+1 & RI-MOM  \\[0.5ex]\\[0.5ex]
\hline\\[-0.5ex]
JLQCD 18 & \cite{Yamanaka:2018uud} & 2+1 & RI-MOM  \\[0.5ex]\\[0.5ex]
\hline\\[-0.5ex]
LHPC 12 & \cite{Green:2012ej} & 2+1 & RI-MOM  \\[0.5ex]\\[0.5ex]
\hline\\[-0.5ex]
LHPC 12A & \cite{Green:2012ud} & 2+1 & RI-MOM  \\[0.5ex]\\[0.5ex]
\hline\\[-0.5ex]
LHPC 10 & \cite{Bratt:2010jn} & 2+1 & RI-MOM  \\[0.5ex]\\[0.5ex]
\hline\\[-0.5ex]
RBC/UKQCD 10D & \cite{Aoki:2010xg} & 2+1 & RI-MOM    \\[0.5ex]\\[0.5ex]
\hline\\[-0.5ex]
RBC/UKQCD 09B & \cite{Yamazaki:2009zq} & 2+1 & $g_A/g_V$  \\[0.5ex]\\[0.5ex]
\hline\\[-0.5ex]
RBC/UKQCD 08B & \cite{Yamazaki:2008py} & 2+1 & $g_A/g_V$  \\[0.5ex]\\[0.5ex]
\hline\\[-0.5ex]
LHPC 05 & \cite{Edwards:2005ym} & 2+1 & $g_A/g_V$  \\[0.5ex]\\[0.5ex]
\fi
 \hline \hline \\[-1.0ex]
\end{tabular*}
\caption{Renormalization in determinations of the isovector axial,
  scalar and tensor charges with $2+1$ quark
  flavours.}
}
\end{table}

\ifx\reducedapptables\undefined
\begin{table}[!ht]
{\footnotesize
\begin{tabular*}{\textwidth}{l @{\extracolsep{\fill}} c c c l}
\hline\hline \\[-1.0ex]
Collab. & Ref. & $\Nf$ & Ren.
\\[.3ex]\\[-2.ex] \hline \hline \\[-1.0ex]
%
Mainz 17 & \cite{Capitani:2017qpc} & 2 & SF  \\[0.5ex]\\[0.5ex]
\hline\\[-0.5ex]
ETM 17B & \cite{Alexandrou:2017hac} & 2 & RI'-MOM  \\[0.5ex]\\[0.5ex]
\hline\\[-0.5ex]
ETM 15D & \cite{Abdel-Rehim:2015owa} & 2 & RI'-MOM  \\[0.5ex]\\[0.5ex]
\hline\\[-0.5ex]
RQCD 14 & \cite{Bali:2014nma} & 2 & RI'-MOM  \\[0.5ex]\\[0.5ex]
\hline\\[-0.5ex]
QCDSF 13 & \cite{Horsley:2013ayv} & 2 & $g_A/f_\pi$  \\[0.5ex]\\[0.5ex]
\hline\\[-0.5ex]
Mainz 12 & \cite{Capitani:2012gj} & 2 & SF  \\[0.5ex]\\[0.5ex]
\hline\\[-0.5ex]
RBC 08 & \cite{Lin:2008uz} & 2 & $g_A/g_V$  \\[0.5ex]\\[0.5ex]
\hline\\[-0.5ex]
QCDSF 06 & \cite{Khan:2006de} & 2 & RI'-MOM  \\[0.5ex]\\[0.5ex]
 \hline \hline \\[-1.0ex]
\end{tabular*}
\caption{Renormalization in determinations of the isovector axial,
  scalar and tensor charges with $\Nf=2$ quark
  flavours.}
}
\end{table}
\fi

\begin{table}[!ht]
{\footnotesize
\begin{tabular*}{\textwidth}{l @{\extracolsep{\fill}} c c c l}
\hline\hline \\[-1.0ex]
Collab. & Ref. & $\Nf$ & $\tau$ [fm] & Description 
\\[.3ex]\\[-2.ex]\hline \hline \\[-1.0ex]
%
ETM 23 & \cite{Alexandrou:2023qbg} & 2+1+1 &\parbox[t]{3.5cm}{ [0.6--1.6] \par[0.6--1.5] \par [0.5--1.2]}
& \parbox[t]{5.4cm}{Compared results
from the plateau, summation method and two-state fits.}
\\[0.5ex]\\[0.5ex]  \hline\\[-0.5ex]
PNDME 23 & \cite{Jang:2023zts} & 2+1+1 & \parbox[t]{3.5cm}{[0.8--1.4]\par[1--1.4,1--1.4,1--1.4,1--1.7]\par[0.9--1.4,0.9--1.4,0.7--1.4]\par[1--1.4,1.1--1.4,1--1.4,1.1--1.4,1--1.3]}
& \parbox[t]{5.4cm}{Several strategies to remove excited-state contributions, including removing $N\pi$ contributions.}
\\[0.5ex]\\[0.5ex]  \hline\\[-0.5ex]
ETM 22 & \cite{Alexandrou:2022dtc} & 2+1+1 & \parbox[t]{3.5cm}{ [0.6--1.6] \par[0.6--1.5] }
& \parbox[t]{5.4cm}{Compared results
from the plateau, summation method and two-state fits.}
\\[0.5ex]\\[0.5ex]
\ifx\reducedapptables\undefined
\hline \\[-0.5ex]
CalLat 19 & \cite{Walker-Loud:2019cif} & 2+1+1 & all
& \parbox[t]{5.4cm}{Two-state fits to the $\tau$-dependence of summed
  operator insertion for $\tau\geq0.3$\,fm.}
\\[0.5ex]\\[0.5ex]  \hline\\[-0.5ex]
ETM 19 & \cite{Alexandrou:2019brg} & 2+1+1
& \parbox[t]{3.5cm}{[0.64--1.6]}
& \parbox[t]{5.4cm}{Fits to the $\tau$- and $t$-dependence of three-point
  correlators using multiple methods (plateau, two- and three-state fits, summation), with the two-state fits quoted for the final result.}
\\[0.5ex]\\[0.5ex]
PNDME 18 & \cite{Gupta:2018qil} & 2+1+1
& \parbox[t]{3.5cm}{[0.8--1.4]\par [1.0--1.4,1.0--1.4]\par
  [0.9--1.2,0.9--1.2,0.7--1.4]\par [0.9--1.4,0.9--1.4,0.9--1.2]}
& \parbox[t]{5.4cm}{Fits to the $\tau$- and $t$-dependence of three-point
  correlators using two or three lowest-lying states.}
\\[0.5ex]\\[0.5ex]  \hline\\[-0.5ex]
CalLat 18 & \cite{Chang:2018uxx} & 2+1+1 & all
& \parbox[t]{5.4cm}{Two-state fits to the $\tau$-dependence of summed
  operator insertion for $\tau\geq0.3$\,fm.}
\\[0.5ex]\\[0.5ex]  \hline\\[-0.5ex]
CalLat 17 & \cite{Berkowitz:2017gql} & 2+1+1 & all
& \parbox[t]{5.4cm}{Two-state fits to the $\tau$-dependence of summed
  operator insertion for $\tau\geq0.3$\,fm.}
\\[0.5ex]\\[0.5ex]  \hline\\[-0.5ex]
PNDME 16 & \cite{Bhattacharya:2016zcn} & 2+1+1
& \parbox[t]{3.5cm}{[1.0--1.4,1.0--1.4]\par
  [0.9--1.2,0.9--1.2,0.9--1.2]\par [0.9--1.4,0.9--1.4]}
& \parbox[t]{5.4cm}{Fits to the $\tau$ and $t$-dependence of three-point
  correlators using two or three lowest-lying states.}
\\[0.5ex]\\[0.5ex]  \hline\\[-0.5ex]
PNDME 15 & \cite{Bhattacharya:2015wna,Bhattacharya:2015esa} & 2+1+1
& \parbox[t]{3.5cm}{[1.0--1.4,1.0--1.4]\par
  [0.9--1.2,0.9--1.2,0.9--1.2]\par [0.9--1.4,0.9--1.4]}
& \parbox[t]{5.4cm}{Fits including up to two states are
  investigated.} \\[0.5ex]\\[0.5ex] 
\hline\\[-0.5ex]
PNDME 13 & \cite{Bhattacharya:2013ehc} & 2+1+1
& \parbox[t]{3.5cm}{[1.0--1.4,0.9--1.4]} &   \parbox[t]{5.4cm}{Fits
  including up to two states are investigated.}
\\[0.5ex]\\[0.5ex]
\fi
\hline \hline \\[-1.0ex]
\end{tabular*}
\caption{Control of excited-state contamination in determinations of
  the isovector axial, scalar and tensor charges with $\Nf=2+1+1$
  quark flavours. The comma-separated list of numbers in square brackets denote the range of source-sink separations $\tau$ (in fermi) at each value of the bare coupling.}
}
\end{table}

\begin{table}[!ht]
{\footnotesize
\begin{tabular*}{\textwidth}{l @{\extracolsep{\fill}} c c c l}
\hline\hline \\[-1.0ex]
Collab. & Ref. & $\Nf$ & $\tau$ [fm] & Description 
\\[.3ex]\\[-2.ex]\hline \hline \\[-1.0ex]
%
Mainz 24 & \cite{Djukanovic:2024krw} & 2+1 & \parbox[t]{3.5cm}{[0.4--1.5,0.4--1.5,0.4--1.5,0.4--1.5]\par [1--1.2]\par[0.3--1.5,0.3--1.5,0.3--1.5]\par[0.3--1.4,0.3--1.4,0.3--1.4,0.3--1.4,0.3--1.4]\par[0.2--1.4,0.2--1.4,0.2--1.4]\par[0.3--1.4]\par[0.9--1.5]}
& \parbox[t]{5.4cm}{Two-state fits to the summation method.}
\\[0.5ex]\\[0.5ex]  \hline\\[-0.5ex]
PACS 23 & \cite{Tsuji:2023llh} & 2+1 & \parbox[t]{3.5cm}{[0.8--1.1]\par [0.75--1.3,0.75--1.7]\par
  [0.72--1.46,0.72--1.46]\par [0.8--1.4,0.9--1.5]}
& \parbox[t]{5.4cm}{Excited-state contributions estimated using different time separations and smearings.}
\\[0.5ex]\\[0.5ex]  \hline\\[-0.5ex]
RQCD 23 & \cite{Bali:2023sdi} & 2+1 &  \parbox[t]{3.5cm}{[0.7--1.3,0.7--1.3,0.7--1.3]\par[0.5--1.2,0.5--1.2,0.5--1.2,0.5--1.2,0.5--1.2,0.5--1.2,0.5--1.2,0.5--1.2,0.5--1.2,0.5--1.2,0.5--1.2,0.5--1.2,0.5--1.2,0.5--1.2,0.5--1.2]\par[0.7--1.2,0.7--1.2,0.7--1.2,0.7--1.2,0.7--1.2,0.7--1.2,0.7--1.2,0.7--1.2,0.7--1.2,0.7--1.2,0.7--1.2]\par[0.7--1.2,0.7--1.2,0.7--1.2,0.7--1.2,0.7--1.2,0.7--1.2,0.7--1.2,0.7--1.2,0.7--1.2,0.7--1.2,0.7--1.2]\par[0.7--1.2,0.7--1.2,0.7--1.2,0.7--1.2,0.7--1.2,0.7--1.2]\par[0.7--1.3,0.7--1.3]}
& \parbox[t]{5.4cm}{Simultaneous two- and three-state fits of up to four different
observables using four time separations.}
\\[0.5ex]\\[0.5ex]  \hline\\[-0.5ex]
\parbox[t]{1.3cm}{QCDSF/ \par UKQCD/ \par CSSM 23} & \cite{QCDSFUKQCDCSSM:2023qlx} & 2+1 & \parbox[t]{3.5cm}{all}
& \parbox[t]{5.4cm}{Energies from fits to
two-point correlation functions, where a weighted average is taken of
the results obtained when varying the fitting range.}
\\[0.5ex]\\[0.5ex] 
\hline\hline\\[-0.5ex]
\end{tabular*}
\caption{Control of excited-state contamination in determinations of
  the isovector axial, scalar and tensor charges with $\Nf=2+1$
  quark flavours. The comma-separated list of numbers in square brackets denote the range of source-sink separations $\tau$ (in fermi) at each value of the bare coupling.}
}
\end{table}

\begin{table}[!ht]
\addtocounter{table}{-1}
{\footnotesize
\begin{tabular*}{\textwidth}{l @{\extracolsep{\fill}} c c c l}
\hline\hline \\[-1.0ex]
Collab. & Ref. & $\Nf$ & $\tau$ [fm] & Description 
\\[.3ex]\\[-2.ex]\hline \hline \\[-1.0ex]
%
PACS 22B & \cite{Tsuji:2022ric} & 2+1 & \parbox[t]{3.5cm}{[0.9--1.4]} & \parbox[t]{5.4cm}{Excited-state contributions estimated using different time separations and smearings.}
\\[0.5ex]\\[0.5ex]  \hline\\[-0.5ex]
Mainz~22 & \cite{Djukanovic:2022wru} & 2+1 & \parbox[t]{3.5cm}{[0.4--1.5,0.4--1.5,0.4--1.5,0.4--1.5]\par [1--1.2]\par[0.3--1.5,0.3--1.5,0.3--1.5]\par[0.3--1.4,0.3--1.4,0.3--1.4,0.3--1.4,0.3--1.4]\par[0.2--1.4,0.2--1.4,0.2--1.4]\par[0.3--1.4]\par[0.9--1.5]}
& \parbox[t]{5.4cm}{Two-state fits to the summation method.}
\\[0.5ex]\\[0.5ex]
\ifx\reducedapptables\undefined
\hline \\[-0.5ex]
$\chi$QCD 21 & \cite{Liu:2021irg} &  2+1
& \parbox[t]{3.5cm}{[0.9--1.1]\par [1--1.2]\par[0.9--1.3,0.9--1.3]\par[1--1.5]}
& \parbox[t]{5.4cm}{Fits to the $\tau$- and $t$-dependence of three-point
  correlators using up to three lowest-lying states.}
\\[0.5ex]\\[0.5ex]  \hline\\[-0.5ex]
NME 21 & \cite{Park:2021ypf} & 2+1
& \parbox[t]{3.5cm}{[1.0--1.8]\par [0.75--1.3,0.75--1.7]\par
  [0.72--1.46,0.72--1.46]\par [0.8--1.4,0.9--1.5]}
& \parbox[t]{5.4cm}{Fits to the $\tau$- and $t$-dependence of three-point
  correlators using three lowest-lying states.}
\\[0.5ex]\\[0.5ex]  \hline\\[-0.5ex]
RBC/UKQCD 19 & \cite{Abramczyk:2019fnf} & 2+1 &  \parbox[t]{3.5cm}{[1.0,1.3]}
& \parbox[t]{5.4cm}{Result quoted from $t$=1.3~fm point only; consistency in results from $t$=1.0~fm is demonstrated.}
\\[0.5ex]\\[0.5ex]  \hline\\[-0.5ex]
LHPC 19 & \cite{Hasan:2019noy} & 2+1 &\parbox[t]{3.5cm}{[0.4--1.4]\par [0.9-1.5]}
& \parbox[t]{5.4cm}{Multiple analysis methods used (plateau, summation, and two-state fits to each), and combined to produce final result.}
\\[0.5ex]\\[0.5ex]  \hline\\[-0.5ex]
Mainz 19 & \cite{Harris:2019bih} & 2+1
& \parbox[t]{3.5cm}{[1.0--1.4,1.0--1.4,1.0--1.4]\par
  [1.0--1.5,1.0--1.5]\par [1.0--1.4,1.0--1.4,1.0--1.4,1.0--1.4]\par
  [1.0--1.4,1.0--1.3]} & \parbox[t]{5.4cm}{Fits to the $\tau$- and $t$-dependence of correlator ratios using the two lowest-lying states.}
\\[0.5ex]\\[0.5ex]  \hline\\[-0.5ex]
PACS 18A & \cite{Shintani:2018ozy} & 2+1 &
\parbox[t]{3.5cm}{[0.9--1.4]}
& \parbox[t]{5.4cm}{Average of plateau values at the three largest source-sink separations.}
\\[0.5ex]\\[0.5ex]
PACS 18 & \cite{Ishikawa:2018rew} & 2+1 & \parbox[t]{3.5cm}{[1.3]}
& \parbox[t]{5.4cm}{Plateau fits of correlator ratio at $\tau=1.3$\,fm.}
\\[0.5ex]\\[0.5ex]  \hline\\[-0.5ex]
$\chi$QCD 18 & \cite{Liang:2018pis} & 2+1 & \parbox[t]{3.5cm}{[1.0--1.6]\par [0.9--1.3]\par [1.0--1.2]} &  \\[0.5ex]\\[0.5ex]
\hline\\[-0.5ex]
JLQCD 18 & \cite{Yamanaka:2018uud} & 2+1 & all
& \parbox[t]{5.4cm}{Fits to the leading (ground state) $\tau$- and $t$-dependence.}
\\[0.5ex]\\[0.5ex]
\fi
\hline\hline\\[-0.5ex]
\end{tabular*}
\caption{(cntd.) Control of excited-state contamination in determinations of
  the isovector axial, scalar and tensor charges with $\Nf=2+1$
  quark flavours. The comma-separated list of numbers in square brackets denote the range of source-sink separations $\tau$ (in fermi) at each value of the bare coupling.}
}
\end{table}

\ifx\reducedapptables\undefined
\begin{table}[!ht]
\addtocounter{table}{-1}
{\footnotesize
\begin{tabular*}{\textwidth}{l @{\extracolsep{\fill}} c c c l}
\hline\hline \\[-1.0ex]
Collab. & Ref. & $\Nf$ & $\tau$ [fm] & Description 
\\[.3ex]\\[-2.ex]\hline \hline \\[-1.0ex]
%
LHPC 12 & \cite{Green:2012ej} & 2+1
& \parbox[t]{3.5cm}{[0.9--1.4,0.9--1.4,0.9--1.4,0.9--1.4,0.9--1.4]\par
  [0.9--1.4]} &  \parbox[t]{5.3cm}{Result is given by the average
  over central points in plateau, and a fitting uncertainty is
  estimated from the variation over two to three points.}
\\[0.5ex]\\[0.5ex]  \hline\\[-0.5ex]
LHPC 12A & \cite{Green:2012ud} & 2+1
& \parbox[t]{3.5cm}{[0.9--1.4,0.9--1.4,0.9--1.4,0.9--1.4,0.9--1.4]\par
  [0.9--1.4]}
& \parbox[t]{5.4cm}{Fits to the leading (ground state) $\tau$-dependence of summed correlator ratios.}
\\[0.5ex]\\[0.5ex]  \hline\\[-0.5ex]
LHPC 10 & \cite{Bratt:2010jn} & 2+1
& \parbox[t]{3.5cm}{[1.1,1.1,1.1,1.1,1.1--1.2,1.1]}
& \parbox[t]{5.4cm}{Plateau fits of correlator ratio at
  $\tau=1.1$\,fm. Larger source-sink separation on one ensemble as
  cross check.} 
\\[0.5ex]\\[0.5ex] \hline\\[-0.5ex]
RBC/UKQCD 10D & \cite{Aoki:2010xg} & 2+1
& \parbox[t]{3.5cm}{[1.4,1.4,1.4,1.4]} & \parbox[t]{5.4cm}{Single
  source-sink separation considered.}
\\[0.5ex]\\[0.5ex]  \hline\\[-0.5ex]
RBC/UKQCD 09B & \cite{Yamazaki:2009zq} & 2+1
& \parbox[t]{3.5cm}{[1.4,1.4,1.4,1.4]} & \parbox[t]{5.4cm}{Plateau
  fits to correlator ratios for $\tau=1.4$\,fm.}
\\[0.5ex]\\[0.5ex]  \hline\\[-0.5ex]
RBC/UKQCD 08B & \cite{Yamazaki:2008py} & 2+1
& \parbox[t]{3.5cm}{[1.4,1.4,1.4,1.4]} & \parbox[t]{5.4cm}{Plateau
  fits to correlator ratios for $\tau=1.4$\,fm.}
\\[0.5ex]\\[0.5ex]  \hline\\[-0.5ex]
LHPC 05 & \cite{Edwards:2005ym} & 2+1
& \parbox[t]{3.5cm}{[1.1]} & \parbox[t]{5.4cm}{Plateau
  fits of correlator ratio at single value of $\tau$.}
\\[0.5ex]\\[0.5ex]  \hline \hline \\[-1.0ex]
\end{tabular*}
\caption{(cntd.) Control of excited-state contamination in determinations of
  the isovector axial, scalar and tensor charges with $\Nf=2+1$
  quark flavours. The comma-separated list of numbers in square brackets denote the range of source-sink separations $\tau$ (in fermi) at each value of the bare coupling.}
}
\end{table}

\begin{table}[!ht]
{\footnotesize
\begin{tabular*}{\textwidth}{l @{\extracolsep{\fill}} c c c l}
\hline\hline \\[-1.0ex]
Collab. & Ref. & $\Nf$ & $\tau$ [fm] & Description 
\\[.3ex]\\[-2.ex]\hline \hline \\[-1.0ex]
%
Mainz 17 & \cite{Capitani:2017qpc} & 2
& \parbox[t]{3.5cm}{[0.8--1.3,0.8--1.3,0.8--1.3,0.8--1.3]\par
  [0.7--1.1,0.7--1.1,0.7--1.1,0.7--1.1]\par
  [0.7--1.1,0.7--1.1,0.7--1.1]} & \parbox[t]{7.3cm}{Two-state fits of
  correlator ratio with fixed energy gaps. Summation method consistent
  within the total quoted error.}
\\[0.5ex]\\[0.5ex]  \hline\\[-0.5ex]
ETM 17 & \cite{Alexandrou:2017qyt} & 2
& \parbox[t]{3.5cm}{[0.9--1.7]} & \parbox[t]{7.3cm}{Fits including up
  to two states and summation method are considered. Result taken from
  the largest source-sink separation and error estimated from
  comparison with two-state fits.} 
\\[0.5ex]\\[0.5ex]  \hline\\[-0.5ex]
ETM 17B & \cite{Alexandrou:2017hac} & 2
& \parbox[t]{3.5cm}{[0.9--1.3]} & \parbox[t]{7.3cm}{Fits including up
  to two states and the summation method are considered. No excited-state contamination seen within statistical errors. Result taken
  from the smallest source-sink separation at which plateau fits and
  two-state fits agree. Systematic error estimated as the difference
  with the summation method.}
\\[0.5ex]\\[0.5ex]  \hline\\[-0.5ex]
ETM 15D & \cite{Abdel-Rehim:2015owa} & 2
& \parbox[t]{3.5cm}{[0.9--1.3]} & \parbox[t]{7.3cm}{Fits including up
  to two states and the summation method are considered. No excited-state contamination seen within statistical errors. Result taken
  from the largest source-sink separation and no error estimated.}
\\[0.5ex]\\[0.5ex]  \hline\\[-0.5ex]
RQCD 14 & \cite{Bali:2014nma} & 2 & \parbox[t]{3.5cm}{[1.1]\par
  [1.1--1.2,0.5--1.2,0.6--1.1]\par [1.1,1.1,1.1]}
& \parbox[t]{7.3cm}{Multiple source-sink separations produced on two
  ensembles to determine a single source-sink separation to be used
  for all other ensembles.}
\\[0.5ex]\\[0.5ex]  \hline\\[-0.5ex]
QCDSF 13 & \cite{Horsley:2013ayv} & 2
& \parbox[t]{3.5cm}{[1.0,1.0,1.0,1.0,1.0]\par
  [1.0,1.0,1.0,0.8--1.3,1.0,1.0,1.0]\par [1.0,1.0,1.0,1.0,1.0,1.0]}
& \parbox[t]{7.3cm}{Plateau fits to $g_A/f_\pi$ for
  $\tau=1.0$\,fm. Multiple source-sink separations investigated on one
  ensemble as cross check.}
\\[0.5ex]\\[0.5ex]  \hline \hline \\[-1.0ex]
\end{tabular*}
\caption{Control of excited-state contamination in determinations of
  the isovector axial, scalar and tensor charges with $\Nf=2$
  quark flavours. The comma-separated list of numbers in square brackets denote the range of source-sink separations $\tau$ (in fermi) at each value of the bare coupling.}
}
\end{table}

\begin{table}[!ht]
\addtocounter{table}{-1}
{\footnotesize
\begin{tabular*}{\textwidth}{l @{\extracolsep{\fill}} c c c l}
\hline\hline \\[-1.0ex]
Collab. & Ref. & $\Nf$ & $\tau$ [fm] & Description 
\\[.3ex]\\[-2.ex]\hline \hline \\[-1.0ex]
%
Mainz 12 & \cite{Capitani:2012gj} & 2
& \parbox[t]{3.5cm}{[0.8--1.3,0.8--1.3,0.8--1.3,0.8--1.3]\par
  [0.7--1.1,0.7--1.1,0.7--1.1,0.7--1.1,0.7--1.1]\par
  [0.7--1.1,0.7--1.1]} & \parbox[t]{7.3cm}{Fits to the leading (ground
  state) $\tau$-dependence of summed correlator ratios.}
\\[0.5ex]\\[0.5ex]  \hline\\[-0.5ex]
RBC 08 & \cite{Lin:2008uz} & 2
& \parbox[t]{3.5cm}{[1.1,1.1,1.1--1.4]} & \parbox[t]{7.3cm}{Two
  source-sink separations produced for lightest pion mass to determine
  a single source-sink separation to be used for all other ensembles.}
\\[0.5ex]\\[0.5ex]  \hline\\[-0.5ex]
QCDSF 06 & \cite{Khan:2006de} & 2
& \parbox[t]{3.5cm}{[1.2,1.2,1.2]\par [1.2,1.2,1.2]\par
  [1.2,1.2,1.2,1.2]\par [1.2,1.2,1.2]} & \parbox[t]{7.3cm}{Ground
  state fits performed. No investigation of excited-state effects.}
\\[0.5ex]\\[0.5ex]  \hline \hline \\[-1.0ex]
\end{tabular*}
\caption{(cntd.) Control of excited-state contamination in determinations of
  the isovector axial, scalar and tensor charges with $\Nf=2$
  quark flavours. The comma-separated list of numbers in square brackets denote the range of source-sink separations $\tau$ (in fermi) at each value of the bare coupling.}
}
\end{table}
\fi

%% file: NME/Tables/app_sigmaud_s.tex
\begin{table}[!ht]

{\footnotesize
\begin{tabular*}{\textwidth}{l @{\extracolsep{\fill}} c c c l}
\hline\hline \\[-1.0ex]
Collab. & Ref. & $\Nf$ & $a$ [fm] & Description 
\\[.3ex]\\[-2.ex] \hline \hline \\[-1.0ex]
%
PNDME 21 & \cite{Gupta:2021ahb} & 2+1+1 & 0.12,0.09,0.06 & \parbox[t]{6.3cm}{Joint continuum and chiral fit includes a $aM_\pi^2$ term.}
\\[0.5ex]\\[0.5ex]
\hline \\[-0.5ex]
\ifx\reducedapptables\undefined
ETM 19 & \cite{Alexandrou:2019brg} & 2+1+1 & 0.08 & \parbox[t]{6.3cm}{Not estimated. }
\\[0.5ex]\\[0.5ex]
\hline \\[-0.5ex]
\fi
Mainz 23& \cite{Agadjanov:2023efe} & 2+1 & 0.08,0.07,0.06,0.05 & \parbox[t]{6.3cm}{Joint continuum, chiral and finite-volume (correlated) fit of $\sigma_{\pi N}$ and $\sigma_s$ includes a $aM_{\pi,K}^2$ term. Fits are performed including and excluding this term.}
\\[0.5ex]\\[0.5ex]
\ifx\reducedapptables\undefined
JLQCD 18 & \cite{Yamanaka:2018uud} & 2+1 & 0.11 &  \parbox[t]{6.3cm}{  Discretization effects are estimated to be $\cO(\Lambda^2_{QCD}a^2)\sim 8\%$ with $\Lambda_{QCD}\sim 500$~MeV. This error is neglected for $\sigma_s$ as it is much smaller than the overall uncertainty.} \\[0.5ex]\\[0.5ex]
\hline\\[-0.5ex]
$\chi$QCD 15A & \cite{Yang:2015uis} & 2+1 & 0.11,0.08 &  \parbox[t]{6.3cm}{  Joint continuum, chiral and volume fit with a linear term in $a^2$.} \\[0.5ex]\\[0.5ex]
\hline\\[-0.5ex]
$\chi$QCD 13A & \cite{Gong:2013vja} & 2+1 & 0.11 &  \parbox[t]{6.3cm}{Not estimated. } \\[0.5ex]\\[0.5ex]
\hline\\[-0.5ex]
JLQCD 12A & \cite{Oksuzian:2012rzb} & 2+1 & 0.11 &  \parbox[t]{6.3cm}{  Discretization effects are estimated at $9\%$ taking $\cO(\Lambda^2_{QCD}a^2)$ with $\Lambda_{QCD}\approx 500$~MeV. This error is neglected as this is much smaller than the statistical accuracy.} \\[0.5ex]\\[0.5ex]
\hline\\[-0.5ex]
Engelhardt 12 & \cite{Engelhardt:2012gd} & 2+1 & 0.12 &  \parbox[t]{6.3cm}{  Estimate $3\%$ discretization error based on a study of the nucleon mass in Ref.~\cite{Walker-Loud:2008rui}.} \\[0.5ex]\\[0.5ex]
\hline\\[-0.5ex]
ETM 16A & \cite{Abdel-Rehim:2016won} & 2 & 0.09 &  \parbox[t]{6.3cm}{ Uncertainty in the fixing the lattice spacing is given based on the analysis of Ref.~\cite{Abdel-Rehim:2015pwa}.  } \\[0.5ex]\\[0.5ex]
\hline\\[-0.5ex]
RQCD 16 & \cite{Bali:2016lvx} & 2 & 0.08,0.07,0.06 &  \parbox[t]{6.3cm}{  No significant discretization effects observed.} \\[0.5ex]\\[0.5ex]
\hline\hline\\[-0.5ex]
MILC 12C & \cite{Freeman:2012ry} & 2+1+1 & 0.15,0.12,0.09,0.06 &  \parbox[t]{6.3cm}{  Combined chiral and continuum fit with an $a^2$ term. Coefficient constrained with a Bayesian prior with a 1 standard deviation width corresponding to a $10\%$ effect at $a=0.12$~fm.} \\[0.5ex]\\[0.5ex]
\hline\\[-0.5ex]
MILC 12C & \cite{Freeman:2012ry} & 2+1 & 0.12,0.09,0.06 &  \parbox[t]{6.3cm}{Analysis and estimation
of discretisation effects as for the $2+1+1$ result in the same publication~(see above).} \\[0.5ex]\\[0.5ex]
\hline\\[-0.5ex]
MILC 09D & \cite{Toussaint:2009pz} & 2+1 & 0.12,0.09,0.06 &  \parbox[t]{6.3cm}{Analysis and estimation         
of discretisation effects as in Ref.~\cite{Freeman:2012ry}~(see above).} \\[0.5ex]\\[0.5ex]
\fi
\hline \hline \\[-1.0ex]
\end{tabular*}
\caption{
\ifx\reducedapptables\undefined
Continuum extrapolation/estimation of lattice artifacts in the determinations of $\sigma_{\pi N}$ and $\sigma_s$. 
The calculations of $\sigma_s$ by MILC 12C and MILC 09D employ a hybrid approach, while all other results were obtained 
from a direct calculation.
\else
  Continuum extrapolation/estimation of lattice artifacts in direct determinations of $\sigma_{\pi N}$ and $\sigma_s$. 
\fi
}
}
\end{table}

\begin{table}[!ht]
{\footnotesize
\begin{tabular*}{\textwidth}{l @{\extracolsep{\fill}} c c c l}
\hline\hline \\[-1.0ex]
{Collab.} & {Ref.} & {$\Nf$} & {$M_{\pi,\rm min}\,[\mev]$} & {Description}  
\\[.3ex]\\[-2.ex] \hline \hline \\[-1.0ex]
%
PNDME 21 & \cite{Gupta:2021ahb} & 2+1+1 & 228,138,235 & \parbox[t]{6.3cm}{Joint continuum and chiral fit including the SU(2) NNLO baryon $\chi$PT~\cite{Hoferichter:2015hva} terms.}
\\[0.5ex]\\[0.5ex]
\hline \\[-0.5ex]
\ifx\reducedapptables\undefined
ETM 19 & \cite{Alexandrou:2019brg} & 2+1+1 & 139  & \parbox[t]{6.3cm}{ Simulate close to $M_\pi^{phys}$. }
\\[0.5ex]\\[0.5ex]
\hline \\[-0.5ex]
JLQCD 18 & \cite{Yamanaka:2018uud} & 2+1 & 293 & \parbox[t]{7.1cm}{$\sigma_{\pi N}$:  Fit including linear and quadratic terms in $M_\pi^2$. Difference with a linear fit taken as the systematic error. $\sigma_s$:  Linear fit in $M_\pi^2$. Difference with a constant fit taken as the systematic error.}  \\[0.5ex]\\[0.5ex]
\hline\\[-0.5ex]
\fi
Mainz 23& \cite{Agadjanov:2023efe} & 2+1 & 219, 154, 128, 174 & \parbox[t]{7.1cm}{Joint continuum, chiral and finite-volume (correlated) fit of $\sigma_{\pi N}$ and $\sigma_s$ utilizing SU(3) NNLO covariant baryon $\chi$PT with the EOMS loop regularization scheme~\cite{Gegelia:1999gf,Fuchs:2003qc,Lehnhart:2004vi}. Cuts on the pion mass excluding ensembles with $M_\pi>220$, $285$ or $360$~MeV are made. }
\\[0.5ex]\\[0.5ex]
\ifx\reducedapptables\undefined
$\chi$QCD 15A & \cite{Yang:2015uis} & 2+1 & 139,300 & \parbox[t]{7.1cm}{  Joint continuum, chiral and volume fit. Partially quenched analysis with $M_\pi \in [114,400]$~MeV. $\sigma_{\pi N}$: employ partially quenched SU(2) $\chi$PT~\cite{Beane:2002vq,Tiburzi:2005is,Walker-Loud:2008rui}. $\sigma_s$: linear terms in the sea and valence $M_\pi^2$ are included. Unitary $M_{\pi}$ given.}  \\[0.5ex]\\[0.5ex]
\hline\\[-0.5ex]
$\chi$QCD 13A & \cite{Gong:2013vja} & 2+1 & 331 & \parbox[t]{7.1cm}{  Partially quenched study with a valence $M_{\pi,min}=250$~MeV. Combined chiral and excited-state fit. Linear fit in
the valence light-quark mass~(and $m_s$ to interpolate to the physical strange-quark mass). Unitary $M_{\pi}$ given.}  \\[0.5ex]\\[0.5ex]
\hline\\[-0.5ex]
JLQCD 12A & \cite{Oksuzian:2012rzb} & 2+1 & 300 & \parbox[t]{7.1cm}{  Linear fit in $m_{ud}$~(and $m_s$ to interpolate to the physical strange quark mass). Difference with SU(3) HB$\chi$PT~\cite{WalkerLoud:2004hf} fit taken as the systematic error.}  \\[0.5ex]\\[0.5ex]
\hline\\[-0.5ex]
Engelhardt 12 & \cite{Engelhardt:2012gd} & 2+1 & 293 & \parbox[t]{7.1cm}{  Linear fit in $M_\pi^2$. Systematic error of $6\%$ estimated from comparing with a constant fit.}  \\[0.5ex]\\[0.5ex]
\hline\\[-0.5ex]
ETM 16A & \cite{Abdel-Rehim:2016won} & 2 & 130 & \parbox[t]{7.1cm}{  Simulate close to $M_\pi^{phys}$.}  \\[0.5ex]\\[0.5ex]
\hline\\[-0.5ex]
RQCD 16 & \cite{Bali:2016lvx} & 2 & 280,150,260 & \parbox[t]{7.1cm}{$\sigma_{\pi N}$:  Rescaling with $M_\pi^2$ of the result at lightest $M_\pi$. Result is consistent with~(but the error more conservative than) performing a baryon $\chi$PT~\cite{Steininger:1998ya,Becher:1999he} fit to $M_\pi\le 420$~MeV. $\sigma_s$: Linear fit in $M_\pi^2$.}  \\[0.5ex]\\[0.5ex]
\fi
\hline\hline\\[-0.5ex]
\end{tabular*}
\caption{
Chiral extrapolation/minimum pion mass in
direct determinations of $\sigma_{\pi N}$ and $\sigma_s$.
}
}
\end{table}

\ifx\reducedapptables\undefined
\begin{table}[!ht]
{\footnotesize
\begin{tabular*}{\textwidth}{l @{\extracolsep{\fill}} c c c l}
\hline\hline \\[-1.0ex]
{Collab.} & {Ref.} & {$\Nf$} & {$M_{\pi,\rm min}\,[\mev]$} & {Description}  
\\[.3ex]\\[-2.ex] \hline \hline \\[-1.0ex]
MILC 12C & \cite{Freeman:2012ry} & 2+1+1 & 131,216,221,329 & \parbox[t]{7.1cm}{  Combined chiral and continuum fit with a linear term in $m_{ud}$. Bayesian prior used to constrain the coefficient with the central value taken from the $\Nf=2+1$ analysis and a width equal to the error of the slope. A $7\%$ error is added to account for omitted higher order terms in $\chi$PT.}  \\[0.5ex]\\[0.5ex]
\hline\\[-0.5ex]
MILC 12C & \cite{Freeman:2012ry} & 2+1 & 269,247,224 & \parbox[t]{7.1cm}{  Combined chiral and continuum fit with a linear term in $m_{ud}$. A $7\%$ error is added to account for omitted higher order terms in $\chi$PT.}  \\[0.5ex]\\[0.5ex]
\hline\\[-0.5ex]
MILC 09D & \cite{Toussaint:2009pz} & 2+1 & 269,247,224 & \parbox[t]{7.1cm}{  Combined chiral and continuum fit with a linear term in $m_{ud}$. A $7\%$ error is added to account for omitted higher order terms in $\chi$PT.}  \\[0.5ex]\\[0.5ex]
\hline\hline
\end{tabular*}
\caption{Chiral extrapolation/minimum pion mass in
hybrid calculations of $\sigma_s$.}
}
\end{table}
\fi

\begin{table}[!ht]
{\footnotesize
\begin{tabular*}{\textwidth}{l @{\extracolsep{\fill}} c c c c l}
\hline\hline \\[-1.0ex]
Collab. & Ref. & $\Nf$ & $L$ [fm] & ${M_{\pi,\rm min}}L$ & Description 
\\[.3ex]\\[-2.ex] \hline \hline \\[-1.0ex]
%
PNDME 21 & \cite{Gupta:2021ahb} & 2+1+1 & \parbox[t]{1.4cm}{2.9--3.8\\2.8--5.6\\2.8--3.7} & \parbox[t]{1.cm}{4.4,\\3.9\\4.4}  &  \parbox[t]{6.3cm}{Finite-volume terms are not included in the joint continuum-chiral extrapolation.}
\\[0.5ex]\\[0.5ex]
\hline \\[-0.5ex]
\ifx\reducedapptables\undefined
ETM 19 & \cite{Alexandrou:2019brg} & 2+1+1 & \parbox[t]{1.4cm}{5.1} &
\parbox[t]{1.cm}{3.6} & \parbox[t]{6.3cm}{FVE are found to be small from an investigation using two
near physical point $\Nf=2$ ensembles with $M_\pi L=3.0$ and $4.0$.}
\\[0.5ex]\\[0.5ex]
\hline \\[-0.5ex]
\fi
Mainz 23& \cite{Agadjanov:2023efe} & 2+1 &  \parbox[t]{1.4cm}{2.7--4.1\\2.4--4.8\\2.0--6.1\\2.4--4.7} & \parbox[t]{1.cm}{4.6\\3.8\\4.0\\4.2} & \parbox[t]{5.7cm}{Joint continuum, chiral and finite-volume (correlated) fit of $\sigma_{\pi N}$ and $\sigma_s$ includes a term derived from the SU(2) finite-volume expression for the nucleon mass in Ref.~\cite{Beane:2004tw}. Fits are performed including and excluding this term.  }
\\[0.5ex]\\[0.5ex]
\ifx\reducedapptables\undefined
JLQCD 18 & \cite{Yamanaka:2018uud} & 2+1 & \parbox[t]{1.4cm}{1.8--2.7} & \parbox[t]{1.cm}{4.0} & \parbox[t]{5.7cm}{  FVE expected to be small for $M_\pi L\gtrsim 4$.}  \\[0.5ex]\\[0.5ex]
\hline\\[-0.5ex]
$\chi$QCD 15A & \cite{Yang:2015uis} & 2+1 & \parbox[t]{1.4cm}{2.7--5.5,\\2.6} & \parbox[t]{1.cm}{3.9,\\4.0} & \parbox[t]{5.7cm}{  Joint continuum, chiral and volume fit with 
leading order FV terms included for $\sigma_{\pi N}$ and a $e^{-M_\pi L}$ term included for $\sigma_s$. Unitary $M_{\pi,min}L$ given.}  \\[0.5ex]\\[0.5ex]
\hline\\[-0.5ex]
$\chi$QCD 13A & \cite{Gong:2013vja} & 2+1 & \parbox[t]{1.4cm}{2.7} & \parbox[t]{1.cm}{4.6} & \parbox[t]{5.7cm}{Not estimated.  Unitary $M_{\pi,min}L$ given.}  \\[0.5ex]\\[0.5ex]
\hline\\[-0.5ex]
JLQCD 12A & \cite{Oksuzian:2012rzb} & 2+1 & \parbox[t]{1.4cm}{1.8--2.7} & \parbox[t]{1.cm}{4.1} & \parbox[t]{5.7cm}{  No significant effects observed when comparing $L/a=16$ and 24 for the two lightest $M_\pi$.}  \\[0.5ex]\\[0.5ex]
\hline\\[-0.5ex]
Engelhardt 12 & \cite{Engelhardt:2012gd} & 2+1 & \parbox[t]{1.4cm}{2.5} & \parbox[t]{1.cm}{3.7} & \parbox[t]{5.7cm}{  A generic $10\%$ FVE is added based on possible FVE in $g_A$.}  \\[0.5ex]\\[0.5ex]
\hline\\[-0.5ex]
ETM 16A & \cite{Abdel-Rehim:2016won} & 2 & \parbox[t]{1.4cm}{4.5} & \parbox[t]{1.cm}{3.0} & \parbox[t]{5.7cm}{ FVE from baryon $\chi$PT following Ref.~\cite{QCDSF-UKQCD:2003hmh} estimated to lead to a $5\%$ increase in $\sigma_{\pi N}$, included as a systematic. This increase is taken as an upper bound on FVE for $\sigma_s$.}  \\[0.5ex]\\[0.5ex]
\hline\\[-0.5ex]
RQCD 16 & \cite{Bali:2016lvx} & 2 & \parbox[t]{1.4cm}{2.6,\\1.7--4.5,\\1.9--2.9} & \parbox[t]{1.cm}{3.7,\\3.5,\\3.8} & \parbox[t]{5.7cm}{  No significant FVE seen for $LM_{\pi}=3.4-6.7$ at $M_\pi=290$~MeV.}  \\[0.5ex]\\[0.5ex]
\hline\hline\\[-0.5ex]
MILC 12C & \cite{Freeman:2012ry} & 2+1+1 & \parbox[t]{1.4cm}{2.5--4.9,\\2.9--4.9,\\2.8--4.2,\\2.7} & \parbox[t]{1.cm}{3.3,\\5.4,\\4.7,\\4.5} & \parbox[t]{5.7cm}{  Estimated to be $2\%$, less than for the $\Nf=2+1$ analysis, since the volumes are larger.}  \\[0.5ex]\\[0.5ex]
\hline\\[-0.5ex]
MILC 12C & \cite{Freeman:2012ry} & 2+1 & \parbox[t]{1.4cm}{2.3--2.8,\\2.3--3.4,\\2.8--3.8} & \parbox[t]{1.cm}{3.8,\\4.2,\\4.3} & \parbox[t]{5.7cm}{  Estimated to be $3\%$, based on FVE observed in the nucleon mass.}  \\[0.5ex]\\[0.5ex]
\hline\\[-0.5ex]
MILC 09D & \cite{Toussaint:2009pz} & 2+1 & \parbox[t]{1.4cm}{2.3--2.8,\\2.3--3.4,\\2.8--3.8} & \parbox[t]{1.cm}{3.8,\\4.2,\\4.3} & \parbox[t]{5.7cm}{  Estimated to be $3\%$, based on FVE observed in the nucleon mass.}  \\[0.5ex]\\[0.5ex]
\fi
\hline\hline
\end{tabular*}
\caption{
\ifx\reducedapptables\undefined
Finite-volume effects in the determinations of $\sigma_{\pi N}$ and  $\sigma_s$. 
The calculations of $\sigma_s$ by MILC 12C and MILC 09D employ a hybrid approach, while all other results were obtained 
from a direct calculation.
\else
Finite-volume effects in direct determinations of $\sigma_{\pi N}$ and  $\sigma_s$. 
\fi
}
}
\end{table}

\begin{table}[!ht]
{\footnotesize
\begin{tabular*}{\textwidth}{l @{\extracolsep{\fill}} c c c l}
\hline\hline \\[-1.0ex]
Collab. & Ref. & $\Nf$ & Ren. &  Description
\\[.3ex]\\[-2.ex] \hline \hline \\[-1.0ex]
%
PNDME 21 & \cite{Gupta:2021ahb} & 2+1+1 & -/- & \parbox[t]{6.3cm}{Flavour mixing occurs due to breaking of chiral symmetry. The ratio of $Z^{ns}/Z^{s}$ is estimated to be close to 1 and the mixing is neglected.}
\\[0.5ex]\\[0.5ex]
\hline \\[-0.5ex]
\ifx\reducedapptables\undefined
ETM 19 & \cite{Alexandrou:2019brg} & 2+1+1 & na/na   &  \parbox[t]{6.3cm}{ } \\[0.5ex]\\[0.5ex]
\hline \\[-0.5ex]
\fi
Mainz 23& \cite{Agadjanov:2023efe} & 2+1 & NP/NP &\parbox[t]{6.3cm}{Flavour mixing occurs due to breaking of chiral symmetry. The mixing is implemented using ratios of the light- and strange-quark masses rather than utilising $Z^{ns}/Z^{s}$. }
\\[0.5ex]\\[0.5ex]
\ifx\reducedapptables\undefined
JLQCD 18 & \cite{Yamanaka:2018uud} & 2+1 & na/na  &  \parbox[t]{6.3cm}{ } \\[0.5ex]\\[0.5ex]
\hline\\[-0.5ex]
$\chi$QCD 15A & \cite{Yang:2015uis} & 2+1 & na/na   &  \parbox[t]{6.3cm}{ } \\[0.5ex]\\[0.5ex]
\hline\\[-0.5ex]
$\chi$QCD 13A & \cite{Gong:2013vja} & 2+1 & $-$/na   &  \parbox[t]{6.3cm}{ } \\[0.5ex]\\[0.5ex]
\hline\\[-0.5ex]
JLQCD 12A & \cite{Oksuzian:2012rzb} & 2+1 & $-$/na   &  \parbox[t]{6.3cm}{ } \\[0.5ex]\\[0.5ex]
\hline\\[-0.5ex]
Engelhardt 12 & \cite{Engelhardt:2012gd} & 2+1 &$-$/na   &  \parbox[t]{6.3cm}{  Flavour mixing due to residual chiral symmetry breaking for a DW action estimated to be $1\%$ and neglected.} \\[0.5ex]\\[0.5ex]
\hline\\[-0.5ex]
ETM 16A & \cite{Abdel-Rehim:2016won} & 2 & na/na   &  \parbox[t]{6.3cm}{ } \\[0.5ex]\\[0.5ex]
\hline\\[-0.5ex]
RQCD 16 & \cite{Bali:2016lvx} & 2 & na/NP   &  \parbox[t]{6.3cm}{  Flavour mixing occurs due to breaking of chiral symmetry when evaluating $\sigma_s$. The ratio of $Z^{ns}/Z^{s}$ is computed nonperturbatively.} \\[0.5ex]\\[0.5ex]
\hline\hline\\[-0.5ex]
MILC 12C & \cite{Freeman:2012ry} & 2+1+1 & 2-loop   &  \parbox[t]{6.3cm}{  $\langle N|s\bar{s}|N\rangle$ computed. $Z_m$ for Asqtad action applied. The 1-loop factors for Asqtad and HISQ are very similar, cf Refs.~\cite{McNeile:2012xh} and~\cite{Mason:2005bj}.} \\[0.5ex]\\[0.5ex]
\hline\\[-0.5ex]
MILC 12C & \cite{Freeman:2012ry} & 2+1 & 2-loop   &  \parbox[t]{6.3cm}{  $\langle N|s\bar{s}|N\rangle$ computed. $Z_m$ from Ref.~\cite{Mason:2005bj}.} \\[0.5ex]\\[0.5ex]
\hline\\[-0.5ex]
MILC 09D & \cite{Toussaint:2009pz} & 2+1 & na   &  \parbox[t]{6.3cm}{ } \\[0.5ex]\\[0.5ex]
\fi
 \hline \hline \\[-1.0ex]
\end{tabular*}
\caption{
\ifx\reducedapptables\undefined
Renormalization for determinations of $\sigma_{\pi N}$ and $\sigma_s$.
 The calculations of $\sigma_s$ by MILC 12C and MILC 09D employ a hybrid approach.
 For the remaining direct determinations, the 
\else
  Renormalization for direct determinations of $\sigma_{\pi N}$ and $\sigma_s$.
The 
  \fi
type of
  renormalization~(Ren.) is given for $\sigma_{\pi N}$ first and
  $\sigma_s$ second. The label 'na' indicates that no renormalization is required.  
}

}
\end{table}
\begin{table}[ht!]
  {\footnotesize
\begin{tabular*}{\textwidth}{l @{\extracolsep{\fill}} c c c l}
\hline\hline \\[-1.0ex]
Collab. & Ref. & $\Nf$ & $\tau$ [fm] & Description 
\\[.3ex]\\[-2.ex]\hline \hline \\[-1.0ex]
%
PNDME 21 & \cite{Gupta:2021ahb} & 2+1+1 & \parbox[t]{3.5cm}{[1.0--1.7,1.0--1.7]/all\par [0.9--1.4,0.9--1.4]/all\par [1.1--1.4,1.1--1.4]/all} &  \parbox[t]{6.3cm}{The two- and three-point functions are fitted simultaneously including four and three states, respectively. The final results are obtained using a narrow-width prior to set the first excited-state energy to that of the lowest $N\pi$ state.}
\\[0.5ex]\\[0.5ex]
\hline \\[-0.5ex]
\ifx\reducedapptables\undefined
ETM 19 & \cite{Alexandrou:2019brg} & 2+1+1 &  \parbox[t]{3.5cm}{[0.6--1.6]/all}  &  \parbox[t]{6.3cm}{Two-state fit to all $\tau$. A comparison is made with three-state fits, plateau fits and the summation method.} \\[0.5ex]\\[0.5ex]
\hline \\[-0.5ex]
\fi
Mainz 23& \cite{Agadjanov:2023efe} & 2+1 & \parbox[t]{3.5cm}{[0.3--1.5]/all\par [0.3--1.5]/all\par [0.3--1.4]/all\par [0.2--1.4]/all} &\parbox[t]{6.3cm}{Summation method including the ground-state terms and ratio fits including an excited state with the energy fixed with a prior to the lowest $N\pi$ energy are considered. The final result combines results from both fit types. }
\\[0.5ex]\\[0.5ex]
\ifx\reducedapptables\undefined
JLQCD 18 & \cite{Yamanaka:2018uud} & 2+1 & all/all &  \parbox[t]{6.3cm}{  Simultaneous plateau fit for $\tau/a\gtrsim 11$.} \\[0.5ex]\\[0.5ex]
\hline\\[-0.5ex]
$\chi$QCD 15A & \cite{Yang:2015uis} & 2+1 & \parbox[t]{3.5cm}{[0.9--1.3,0.9--1.4]\par [1.0--1.2]/all} &  \parbox[t]{6.3cm}{  Two-state fit with $t\in[0.2,\tau-0.2]$~fm.} \\[0.5ex]\\[0.5ex]
\hline\\[-0.5ex]
$\chi$QCD 13A & \cite{Gong:2013vja} & 2+1 & $-$/all &  \parbox[t]{6.3cm}{  Combined chiral and excited-state fit using the summation method with $\tau/a=7-14$.} \\[0.5ex]\\[0.5ex]
\hline\\[-0.5ex]
JLQCD 12A & \cite{Oksuzian:2012rzb} & 2+1 & $-$/all &  \parbox[t]{6.3cm}{  Simultaneous plateau fit to $\tau/a\in [12,23]$ with $t/a\in [5,\tau-5]$.} \\[0.5ex]\\[0.5ex]
\hline\\[-0.5ex]
Engelhardt 12 & \cite{Engelhardt:2012gd} & 2+1 & $-$/all &  \parbox[t]{6.3cm}{  Ratio of three-point to two-point function averaged over $t/a\in [3,7]$ for $\tau/a=10$.} \\[0.5ex]\\[0.5ex]
\hline\\[-0.5ex]
ETM 16A & \cite{Abdel-Rehim:2016won} & 2 & \parbox[t]{3.5cm}{[0.9--1.7]/all} &  \parbox[t]{6.3cm}{  Plateau fit to $\tau=1.7$~fm. Systematic error estimated from plateau fits for $\tau \ge 1.5$~fm. Comparison made with 2-state fits and the summation method.} \\[0.5ex]\\[0.5ex]
\hline\\[-0.5ex]
RQCD 16 & \cite{Bali:2016lvx} & 2 & \parbox[t]{3.5cm}{[1.1]\par [1.1--1.2,0.5--1.2,0.6--1.1]\par [1.1,1.1,1.1]/all} &  \parbox[t]{6.3cm}{  Two state~(simultaneous plateau) fit for the connected~(disconnected) terms.} \\[0.5ex]\\[0.5ex]
\fi
\hline \hline \\[-1.0ex]
\end{tabular*}
\caption{
  Control of excited-state contamination in
  direct determinations of $\sigma_{\pi N}$ and $\sigma_s$. 
  The comma-separated list of numbers in
  square brackets denote the range of source-sink separations $\tau$
  (in fermi) at each value of the bare coupling. The range of $\tau$ for the connected~(disconnected)
  contributions to the three-point correlation functions is given
  first~(second). If a wide range of $\tau$ values is available this
  is indicated by ``all'' in the table. } }
\end{table}

\ifx\reducedapptables\undefined
\begin{table}[ht!]
{\footnotesize
\begin{tabular*}{\textwidth}{l @{\extracolsep{\fill}} c c c l}
\hline\hline \\[-1.0ex]
Collab. & Ref. & $\Nf$ & $\tau$ [fm] & Description 
\\[.3ex]\\[-2.ex]\hline \hline \\[-1.0ex]
%
MILC 12C & \cite{Freeman:2012ry} & 2+1+1 & all &  \parbox[t]{6.3cm}{A modified nucleon correlator is fitted including a positive and negative parity state starting from $\approx 0.6$~fm. A $5\%$ error for possible excited-state contamination is included.} \\[0.5ex]\\[0.5ex]
\hline\\[-0.5ex]
MILC 12C & \cite{Freeman:2012ry} & 2+1 & all &  \parbox[t]{6.3cm}{Analysis and estimation
of excited-state contamination as for the $2+1+1$ result in the same publication~(see above).} \\[0.5ex]\\[0.5ex]
\hline\\[-0.5ex]
MILC 09D & \cite{Toussaint:2009pz} & 2+1 & all &  \parbox[t]{6.3cm}{A modified nucleon correlator is fitted including a positive and negative parity state starting from $\approx 0.6$~fm. A $10\%$ error for possible excited-state contamination is included.} \\[0.5ex]\\[0.5ex]
\hline \hline \\[-1.0ex]
\end{tabular*}
\caption{Control of excited-state contamination in
 determinations of $\sigma_s$ using a hybrid approach. 
If a wide range of $\tau$ values is available this
  is indicated by ``all'' in the table. } }
\end{table}
\fi

%% file: NME/Tables/app_sigmaud_fh.tex
\begin{table}[!ht]

{\footnotesize
\begin{tabular*}{\textwidth}{l @{\extracolsep{\fill}} c c c l}
\hline\hline \\[-1.0ex]
Collab. & Ref. & $\Nf$ & $a$ [fm] & Description 
\\[.3ex]\\[-2.ex] \hline \hline \\[-1.0ex]
%
\ifx\reducedapptables\undefined
BMW 20A & \cite{Borsanyi:2020bpd} & 1+1+1+1 & 0.10,0.09,0.08,0.06 &  \parbox[t]{6.8cm}{ Combined continuum, chiral and volume fit within an extended frequentist method. $\cO(\alpha a)$ or $\cO(a^2)$ terms are included. $M_N$ used to fix the lattice spacing. } \\[0.5ex]\\[0.5ex]
\hline \\[-0.5ex]
ETM 14A & \cite{Alexandrou:2014sha} & 2+1+1 & 0.09,0.08,0.06 &  \parbox[t]{6.8cm}{  No significant discretization effects are observed. $M_N$ used to fix the lattice spacing.} \\[0.5ex]\\[0.5ex]
\hline\\[-0.5ex]
\fi
RQCD 22& \cite{RQCD:2022xux} & 2+1 &  \parbox[t]{2.5cm}{0.10,0.09,0.08,\\0.06,0.05,0.04}  & \parbox[t]{6.8cm}{Combined continuum, chiral and volume fit to the baryon octet. Leading $O(a^2)$ terms are included in the parameterisation. Fits are performed also excluding the coarsest lattice spacing.}
\\[0.5ex]\\[0.5ex]
\ifx\reducedapptables\undefined
BMW 15 & \cite{Durr:2015dna} & 2+1 & \parbox[t]{2.5cm}{0.12,0.09,0.08,\\0.07,0.05} &  \parbox[t]{6.8cm}{  Combined continuum, chiral and volume fit to $M_{N,\Omega,\pi,K}$ within an extended frequentist method. For $M_N$, $\cO(\alpha a)$ or $\cO(a^2)$ errors on the physical limit are included. Systematics include shift when adding discretization terms to the slope of $M_N$ with $m_{ud,s}$, fixing $M_N$ to expt..} \\[0.5ex]\\[0.5ex]
\hline\\[-0.5ex]
Junnarkar 13 & \cite{Junnarkar:2013ac} & 2+1 & 0.12,0.09 &  \parbox[t]{6.8cm}{Discretization effects are not resolved.} \\[0.5ex]\\[0.5ex]
\hline\\[-0.5ex]
Shanahan 12 & \cite{Shanahan:2012wh} & 2+1 & 0.09 &  \parbox[t]{6.8cm}{  Two results quoted, differing on setting the lattice spacing per ensemble or in the chiral limit.} \\[0.5ex]\\[0.5ex]
\hline\\[-0.5ex]
JLQCD 12A & \cite{Oksuzian:2012rzb} & 2+1 & 0.11 &  \parbox[t]{6.8cm}{  Discretization effects are estimated to be $\cO(\Lambda^2_{QCD}a^2)\sim 9\%$ with $\Lambda_{QCD}\sim 500$~MeV and neglected as the statistical error is much larger.} \\[0.5ex]\\[0.5ex]
\hline\\[-0.5ex]
QCDSF 11 & \cite{Horsley:2011wr} & 2+1 & 0.08 &  \parbox[t]{6.8cm}{  Not estimated.} \\[0.5ex]\\[0.5ex]
\hline\\[-0.5ex]
BMW 11A & \cite{Durr:2011mp} & 2+1 & 0.12,0.08,0.06 &  \parbox[t]{6.8cm}{  Joint continuum and chiral extrapolation of octet baryons masses with $\cO(a)$, $\cO(a^2)$ terms or no discretization terms. Final result from consideration of all fit combinations weighted with the fit quality.} \\[0.5ex]\\[0.5ex]
\hline\\[-0.5ex]
Martin~Camalich 10 & \cite{MartinCamalich:2010fp} & 2+1 & 0.09 &  \parbox[t]{6.8cm}{  Not estimated.} \\[0.5ex]\\[0.5ex]
\hline\\[-0.5ex]
PACS-CS 09 & \cite{Ishikawa:2009vc} & 2+1 & 0.09 &  \parbox[t]{6.3cm}{  Not estimated.} \\[0.5ex]\\[0.5ex]
\hline\hline
\end{tabular*}
\caption{Finite-volume effects in determinations of $\sigma_{\pi N}$  and $\sigma_s$  from the Feynman-Hellmann method.}
}
\end{table}
\fi
\ifx\reducedapptables\undefined
\begin{table}[!ht]
\addtocounter{table}{-1}
{\footnotesize
\begin{tabular*}{\textwidth}{l @{\extracolsep{\fill}} c c c l}
\hline\hline \\[-1.0ex]
Collab. & Ref. & $\Nf$ & $a$ [fm] & Description 
\\[.3ex]\\[-2.ex] \hline \hline \\[-1.0ex]
%
Walker-Loud 08 & \cite{Walker-Loud:2008rui} & 2+1 & 0.12 &  \parbox[t]{6.3cm}{Not estimated. } \\[0.5ex]\\[0.5ex]
\hline\\[-0.5ex]
QCDSF 12 & \cite{Bali:2012qs} & 2 & 0.08,0.07,0.06 &  \parbox[t]{6.3cm}{  No significant discretization effects observed. $M_N$ is used to fix the lattice spacing.} \\[0.5ex]\\[0.5ex]
\hline\\[-0.5ex]
JLQCD 08B & \cite{Ohki:2008ff} & 2 & 0.12 &  \parbox[t]{6.3cm}{  Not estimated.} \\[0.5ex]\\[0.5ex]
\fi
\hline \hline \\[-1.0ex]
\end{tabular*}
\ifx\reducedapptables\undefined
\caption{(cntd) Continuum extrapolations/estimation of lattice artifacts in
  determinations of $\sigma_{\pi N}$ and $\sigma_s$ from the Feynman-Hellmann method.}
\else
\caption{Continuum extrapolations/estimation of lattice artifacts in
  determinations of $\sigma_{\pi N}$ and $\sigma_s$ from the Feynman-Hellmann method.}
\fi
}
\end{table}

\begin{table}[!ht]
{\footnotesize
\begin{tabular*}{\textwidth}{l @{\extracolsep{\fill}} c c c l}
\hline\hline \\[-1.0ex]
{Collab.} & {Ref.} & {$\Nf$} & {$M_{\pi,\rm min}\,[\mev]$} & {Description}  
\\[.3ex]\\[-2.ex] \hline \hline \\[-1.0ex]
%
\ifx\reducedapptables\undefined
BMW 20A & \cite{Borsanyi:2020bpd} & 1+1+1+1 & 238,200,219,199 &  \parbox[t]{6.8cm}{ Combined continuum, chiral and volume fit within an extended frequentist method. Terms with $M_\pi^{2,3,4}$ and $M_K^2$ are included and cuts of $M_\pi < 420$~MeV and $M_\pi < 360$~MeV are made.} \\[0.5ex]\\[0.5ex]
\hline \\[-0.5ex]
Walker-Loud 08 & \cite{Walker-Loud:2008rui} & 2+1 & 294 &  \parbox[t]{6.3cm}{Fit using NNLO SU(2) HB$\chi$PT~\cite{Tiburzi:2005na} with a zero and nonzero value for $g_{\Delta N}$. Errors on the LECs that are not free parameters determine the systematic uncertainty. } \\[0.5ex]\\[0.5ex]
\fi
\ifx\reducedapptables\undefined
ETM 14A & \cite{Alexandrou:2014sha} & 2+1+1 & 261,256,213 & \parbox[t]{6.3cm}{  Fit using SU(2) HB$\chi$PT to $\cO(p^3)$~\cite{Gasser:1987rb}. Error assigned to difference with fit using HB$\chi$PT in SSE to $\cO(p^4)$~\cite{Procura:2006bj}.}  \\[0.3ex]\\
\hline\\[-0.5ex]
\fi
RQCD 22& \cite{RQCD:2022xux} & 2+1 & \parbox[t]{2.5cm}{338,127,216\\131,176,336}   & \parbox[t]{6.3cm}{  Combined continuum, chiral and volume fit to the baryon octet. Fits utilizing SU(3) NNLO covariant baryon $\chi$PT with the EOMS loop regularization scheme~\cite{Gegelia:1999gf,Fuchs:2003qc,Lehnhart:2004vi} are performed. Cuts on the flavour average meson mass squared are made.}
\\[0.5ex]\\[0.5ex]
\ifx\reducedapptables\undefined
BMW 15 & \cite{Durr:2015dna} & 2+1 & 136,131,120,182,219 & \parbox[t]{6.3cm}{  Combined continuum, chiral and volume fit to $M_{N,\Omega,\pi,K}$ within an extended frequentist method~\cite{Durr:2008zz}. Terms linear in $m_{ud,s}$ are included and cuts of $M_\pi\le 480$~MeV and $M_\pi \le 320$~MeV are made. Higher order terms in the fit are also considered.}  \\[0.3ex]\\
\hline\\[-0.5ex]
Junnarkar 13 & \cite{Junnarkar:2013ac} & 2+1 & 380,238 & \parbox[t]{6.3cm}{  $\sigma_s$ from $M_N$ with one $m_s$ value above and one below the physical $m_s$. Weighted average is made of fits to $\sigma_s$~(including a $M_\pi^2$ term) and $f_{T_s}$~(including a $M_\pi$ or $M_\pi^2$ term). Correlations are taken into account.}  \\[0.3ex]\\
\hline\\[-0.5ex]
Shanahan 12 & \cite{Shanahan:2012wh} & 2+1 & 296 & \parbox[t]{6.3cm}{  Fit to baryon octet masses using finite-range regularization of baryon $\chi$PT~\cite{Young:2002cj}. Coefficients held fixed in the fit are varied by $10\%$.}  \\[0.3ex]\\
\hline\\[-0.5ex]
JLQCD 12A & \cite{Oksuzian:2012rzb} & 2+1 & 300 & \parbox[t]{6.3cm}{  Only $\sigma_s$ computed. Reweighting used to vary $M_N$ in a region $m_s^\prime\in[m_s-25\text{\,MeV},m_s+25\text{\,MeV}]$. Linear fit to extract the slope at physical $m_s$ for two different sea $m_s$. Linear fit in $m_{ud}$ and $m_s$. 1-loop SU(3) HB$\chi$PT~\cite{WalkerLoud:2004hf} used to estimate the systematic error.}  \\[0.3ex]\\
\hline\\[-0.5ex]
QCDSF 11 & \cite{Horsley:2011wr} & 2+1 & 328 & \parbox[t]{6.3cm}{SU(3) flavour expansion along simulation trajectory with average quark mass held fixed~\cite{Bietenholz:2011qq}. Fit with linear flavour breaking terms, with systematics from next order terms included in the error.}  \\[0.3ex]\\
\hline\\[-0.5ex]
BMW 11A & \cite{Durr:2011mp} & 2+1 & 273,197,321 & \parbox[t]{6.3cm}{  Joint continuum and chiral extrapolation of octet baryon masses. Fits involving Taylor and Pad\'{e} expansions and SU(3) baryon $\chi$PT~\cite{Oller:2006yh,Frink:2006hx} and cuts of $M_\pi<410$~MeV and $M_\pi<550$~MeV. Final result from consideration of all fit combinations weighted with the fit quality.}  \\[0.3ex]\\
\fi
\hline\hline
\end{tabular*}
\caption{Chiral extrapolation/minimum pion mass in
  determinations of $\sigma_{\pi N}$  and $\sigma_s$ from the Feynman-Hellmann method.}
}
\end{table}

\ifx\reducedapptables\undefined
\begin{table}[!ht]
\addtocounter{table}{-1}
{\footnotesize
\begin{tabular*}{\textwidth}{l @{\extracolsep{\fill}} c c c l}
\hline\hline \\[-1.0ex]
{Collab.} & {Ref.} & {$\Nf$} & {$M_{\pi,\rm min}\,[\mev]$} & {Description}  
\\[.3ex]\\[-2.ex] \hline \hline \\[-1.0ex]
%
Martin~Camalich 10 & \cite{MartinCamalich:2010fp} & 2+1 & 156 & \parbox[t]{6.3cm}{  Fit to baryon octet masses using NLO covariant baryon $\chi$PT in the EOMS scheme~\cite{Gegelia:1999gf,Fuchs:2003qc}. Uncertainty from omitted higher order terms estimated as half the difference between LO and NLO.}  \\[0.3ex]\\
\hline\\[-0.5ex]
PACS-CS 09 & \cite{Ishikawa:2009vc} & 2+1 & 156 & \parbox[t]{6.3cm}{  Fit $M_N$ with $\cO(p^3)$ SU(2) HB$\chi$PT~\cite{Gasser:1987rb}.}  \\[0.5ex]\\[0.5ex]
\hline\\[-0.5ex]
Walker-Loud 08 & \cite{Walker-Loud:2008rui} & 2+1 & 294 &  \parbox[t]{6.3cm}{Fit using NNLO SU(2) HB$\chi$PT~\cite{Tiburzi:2005na} with a zero and nonzero value for $g_{\Delta N}$. Errors on the LECs that are not free parameters determine the systematic uncertainty. } \\[0.5ex]\\[0.5ex]
\hline\\[-0.5ex]
QCDSF 12 & \cite{Bali:2012qs} & 2 & 478,158,262 & \parbox[t]{6.3cm}{  Fit $M_N$ to $\cO(p^4)$ baryon $\chi$PT~\cite{Procura:2003ig}. Slope at $M_\pi=290$~MeV is fixed from a direct determination of $\sigma_{\pi N}$. Results consistent with $\cO(p^2)$ and $\cO(p^3)$ fits.}  \\[0.5ex]\\[0.5ex]
\hline\\[-0.5ex]
JLQCD 08B & \cite{Ohki:2008ff} & 2 & 288 & \parbox[t]{6.3cm}{  Fit $M_N$ using a reduced form of covariant baryon $\chi$PT~\cite{Procura:2003ig} with the systematic error determined from $\cO(p^3)$ and $\cO(p^4)$ fits.}  \\[0.5ex]\\[0.5ex]
\hline\hline
\end{tabular*}
\caption{(cntd.) Chiral extrapolation/minimum pion mass in
  determinations of $\sigma_{\pi N}$  and $\sigma_s$ from the Feynman-Hellmann method.}
}
\end{table}
\fi

\begin{table}[!ht]
{\footnotesize
\begin{tabular*}{\textwidth}{l @{\extracolsep{\fill}} c c c c l}
\hline\hline \\[-1.0ex]
Collab. & Ref. & $\Nf$ & $L$ [fm] & ${M_{\pi,\rm min}}L$ & Description 
\\[.3ex]\\[-2.ex] \hline \hline \\[-1.0ex]
%
\ifx\reducedapptables\undefined
BMW 20A & \cite{Borsanyi:2020bpd} & 1+1+1+1 & \parbox[t]{1.4cm}{3.2,\\2.9--4.3,\\2.6--3.8,\\1.9--3.8} & \parbox[t]{1.cm}{4.0,\\4.3,\\4.1,\\4.1}  &  \parbox[t]{5.3cm}{ Combined continuum, chiral and volume fit within an extended frequentist method. FV term $M_\pi^{1/2}L^{-3/2}e^{-M_\pi L}$ added.} \\[0.5ex]\\[0.5ex]
\hline \\[-0.5ex]
ETM 14A & \cite{Alexandrou:2014sha} & 2+1+1 & \parbox[t]{1.4cm}{1.9--3.0,\\2.0--2.6,\\2.1--3.1} & \parbox[t]{1.cm}{4.0,\\3.4,\\3.4} & \parbox[t]{5.3cm}{  No significant finite-volume effects are observed.}  \\[0.5ex]\\[0.5ex]
\hline\\[-0.5ex]
\fi
RQCD 22& \cite{RQCD:2022xux} & 2+1 &  \parbox[t]{1.4cm}{2.3,\\2.0--5.5,\\2.4--4.8,\\2.0--6.1,\\2.4--4.7\\2.5} & \parbox[t]{1.cm}{4.0,\\3.5,\\5.3,\\4.0,\\4.2,\\4.2}   &  \parbox[t]{5.3cm}{  Combined continuum, chiral and volume fit to the baryon octet. Finite-volume terms from NNLO covariant baryon $\chi$PT (with no new fit parameters) are included in the fit~\cite{Ren:2012aj,Procura:2006bj,QCDSF-UKQCD:2003hmh}. Ensembles with $L<2.3$~fm are excluded from the analysis.}
\\[0.5ex]\\[0.5ex]
\ifx\reducedapptables\undefined
BMW 15 & \cite{Durr:2015dna} & 2+1 & \parbox[t]{1.4cm}{1.9--5.6,\\1.5--5.9,\\2.5--4.9,\\2.1--4.2,\\1.7--3.4} & \parbox[t]{1.cm}{3.9,\\3.9,\\3.0,\\3.9,\\3.8} & \parbox[t]{5.3cm}{  Combined continuum, chiral and volume fit to $M_{N,\Omega,\pi,K}$ within an extended frequentist method with finite-volume corrections following Ref.~\cite{Colangelo:2005cg}}  \\[0.5ex]\\[0.5ex]
\hline\\[-0.5ex]
Junnarkar 13 & \cite{Junnarkar:2013ac} & 2+1 & \parbox[t]{1.4cm}{2.4,\\3.6} & \parbox[t]{1.cm}{4.4,\\4.1} & \parbox[t]{5.3cm}{  Not estimated.}  \\[0.5ex]\\[0.5ex]
\hline\\[-0.5ex]
Shanahan 12 & \cite{Shanahan:2012wh} & 2+1 & \parbox[t]{1.4cm}{2.9} & \parbox[t]{1.cm}{4.3} & \parbox[t]{5.3cm}{  Baryons masses are finite-volume corrected~\cite{QCDSF-UKQCD:2003hmh}.}  \\[0.5ex]\\[0.5ex]
\hline\\[-0.5ex]
JLQCD 12A & \cite{Oksuzian:2012rzb} & 2+1 & \parbox[t]{1.4cm}{1.8--2.7} & \parbox[t]{1.cm}{4.1} & \parbox[t]{5.3cm}{  No significant effects observed when comparing $L/a=16$ and 24 for the two lightest $M_\pi$.}  \\[0.5ex]\\[0.5ex]
\hline\\[-0.5ex]
QCDSF 11 & \cite{Horsley:2011wr} & 2+1 & \parbox[t]{1.4cm}{1.8--2.5} & \parbox[t]{1.cm}{4.1} & \parbox[t]{5.3cm}{  Not estimated.}  \\[0.5ex]\\[0.5ex]
\hline\\[-0.5ex]
BMW 11A & \cite{Durr:2011mp} & 2+1 & \parbox[t]{1.4cm}{2.0--3.9,\\2.0--3.9,\\2.0--2.5} & \parbox[t]{1.cm}{4.1,\\3.9,\\4.1} & \parbox[t]{5.3cm}{  FVE found to be small in Ref.~\cite{Durr:2008zz}.}  \\[0.5ex]\\[0.5ex]
\hline\\[-0.5ex]
Martin~Camalich 10 & \cite{MartinCamalich:2010fp} & 2+1 & \parbox[t]{1.4cm}{2.9} & \parbox[t]{1.cm}{2.3} & \parbox[t]{5.3cm}{  Baryons masses are finite-volume corrected using Ref.~\cite{Beane:2004tw}.}  \\[0.5ex]\\[0.5ex]
\hline\\[-0.5ex]
PACS-CS 09 & \cite{Ishikawa:2009vc} & 2+1 & \parbox[t]{1.4cm}{2.9} & \parbox[t]{1.cm}{2.3} & \parbox[t]{5.3cm}{  Estimated to be less than $1\%$ in $M_N$ using Ref.~\cite{Beane:2004tw}.}  \\[0.5ex]\\[0.5ex]
\hline\hline
\end{tabular*}
\caption{Finite-volume effects in determinations of $\sigma_{\pi N}$  and $\sigma_s$  from the Feynman-Hellmann method.}
}
\end{table}
\fi
\ifx\reducedapptables\undefined
\begin{table}[!ht]
\addtocounter{table}{-1}
{\footnotesize
\begin{tabular*}{\textwidth}{l @{\extracolsep{\fill}} c c c c l}
\hline\hline \\[-1.0ex]
Collab. & Ref. & $\Nf$ & $L$ [fm] & ${M_{\pi,\rm min}}L$ & Description 
\\[.3ex]\\[-2.ex] \hline \hline \\[-1.0ex]
%

Walker-Loud 08 & \cite{Walker-Loud:2008rui} & 2+1 & \parbox[t]{1.4cm}{2.5} & \parbox[t\
]{1.cm}{3.7} &\parbox[t]{5.3cm}{FV effects expected to be $1\%$ or less for the lightest pion mass based on~\cite{Beane:2004tw}. } \\[0.5ex]\\[0.5ex]
\hline\\[-0.5ex]
QCDSF 12 & \cite{Bali:2012qs} & 2 & \parbox[t]{1.4cm}{1.8,\\1.7--3.4,\\1.9--2.9} & \parbox[t]{1.cm}{4.4,\\2.7,\\3.8} & \parbox[t]{5.3cm}{  Finite-volume corrections are applied~\cite{QCDSF-UKQCD:2003hmh}.}  \\[0.5ex]\\[0.5ex]
\hline\\[-0.5ex]
JLQCD 08B & \cite{Ohki:2008ff} & 2 & \parbox[t]{1.4cm}{1.9} & \parbox[t]{1.cm}{2.8} & \parbox[t]{5.3cm}{ Systematic error based on fits with and without FVE of Ref.~\cite{QCDSF-UKQCD:2003hmh}.}  \\[0.5ex]\\[0.5ex]
\fi
\hline\hline
\end{tabular*}
\ifx\reducedapptables\undefined
\caption{(cntd.) Finite-volume effects in determinations of $\sigma_{\pi N}$  and $\sigma_s$  from the Feynman-Hellmann method.}
}
\else
\caption{Finite-volume effects in determinations of $\sigma_{\pi N}$  and $\sigma_s$  from the Feynman-Hellmann method.}
}
\fi
\end{table}

%% file: NME/Tables/app_x_isovector.tex
\begin{table}[!ht]
{\footnotesize
\begin{tabular*}{\textwidth}{l @{\extracolsep{\fill}} c c c l}
\hline\hline \\[-1.0ex]
Collab. & Ref. & $\Nf$ & $a$ [fm] & Description 
\\[.3ex]\\[-2.ex] \hline \hline \\[-1.0ex]
%
ETM 22       & \cite{Alexandrou:2022dtc} & 2+1+1 & 0.08,0.07,0.06 & \parbox[t]{6.3cm}{Linear extrapolation in $a^2$.}\\[0.5ex]\\[0.5ex] \hline\\[-0.5ex]
PNDME 20A     & \cite{Mondal:2020cmt} & 2+1+1 & 0.15,0.12,0.09,0.06 & \parbox[t]{6.3cm}{Extrapolation performed using a linear term in $a$ as part of a simultaneous fit in
$a$, $M_\pi$ and $M_\pi L$.}\\[0.5ex]\\[0.5ex]\hline\\[-0.5ex]
ETM 20C       & \cite{Alexandrou:2020sml} & 2+1+1 & 0.08 & \parbox[t]{6.3cm}{Single lattice spacing.}\\[0.5ex]\\[0.5ex]\hline\\[-0.5ex]
ETM 19A       & \cite{Alexandrou:2019ali} & 2+1+1 & 0.08 & \parbox[t]{6.3cm}{Single lattice spacing.}\\[0.5ex]\\[0.5ex]
\hline\hline\\[-0.5ex]
\end{tabular*}
\caption{Continuum extrapolations/estimation of lattice artifacts in
  determinations of the isovector unpolarised, helicity and
  transversity second moments with $\Nf=2+1+1$ quark flavours.}  }
\end{table}

\begin{table}[!ht]
{\footnotesize
\begin{tabular*}{\textwidth}{l @{\extracolsep{\fill}} c c c l}
\hline\hline \\[-1.0ex]
Collab. & Ref. & $\Nf$ & $a$ [fm] & Description 
\\[.3ex]\\[-2.ex] \hline \hline \\[-1.0ex]
%
Mainz 24     &\cite{Djukanovic:2024krw}  & 2+1 & 0.09,0.08,0.06,0.05 & \parbox[t]{6.3cm}{A number of simultaneous $a$, $M_\pi$ and $M_\pi L$ fits are performed using a linear term in $a$. The final results are obtained from a weighted average.} \\[0.5ex]\\[0.5ex]\hline\\[-0.5ex]
LHPC  24     & \cite{Rodekamp:2023wpe}   & 2+1 & 0.12,0.09 & \parbox[t]{6.3cm}{A Bayesian fit is performed including an $a$ and an $a^2$ term.}\\[0.5ex]\\[0.5ex]\hline\\[-0.5ex]
NME 21A       & \cite{Mondal:2021oot} & 2+1 & 0.127, 0,09, 0.07 & \parbox[t]{6.3cm}{Extrapolation performed using a linear term in $a$ as part of a simultaneous fit in
$a$, $M_\pi$ and $M_\pi L$.}\\[0.5ex]\\[0.5ex]\hline\\[-0.5ex]
NME 20       & \cite{Mondal:2020ela} & 2+1 & 0.127, 0,09, 0.07  & \parbox[t]{6.3cm}{Extrapolation performed using a linear term in $a$ as part of a simultaneous fit in
$a$, $M_\pi$ and $M_\pi L$.}\\[0.5ex]\\[0.5ex]\hline\\[-0.5ex]
Mainz 19        &\cite{Harris:2019bih}  & 2+1  & 0.05,0.06,0.08,0.09 & \parbox[t]{6.3cm}{Extrapolation performed as part of a simultaneous fit in $a$, $M_\pi$ and $M_\pi L$.}\\[0.5ex]\\[0.5ex]\hline\\[-0.5ex]
$\chi$QCD 18A & \cite{Yang:2018nqn}   & 2+1   & 0.143, 0.11, 0.114, 0.083 & \parbox[t]{6.3cm}{Partially quenched calculation. Extrapolation performed as part of a simultaneous fit in $a^2$, $M_\pi$ and $M_\pi L$ using expression from HBChPT.} \\[0.5ex]\\[0.5ex]\hline\\[-0.5ex]
LHPC 12A        & \cite{Green:2012ud}   & 2+1  &  0.12,0.09  & \parbox[t]{6.3cm}{No statistically significant discretization effects   observed. Results assumed to be constant in $a$.}\\[0.5ex]\\[0.5ex]\hline\\[-0.5ex]
LHPC 10         & \cite{Bratt:2010jn}   & 2+1 & 0.12 & \parbox[t]{6.3cm}{Single lattice spacing.}\\[0.5ex]\\[0.5ex]\hline\\[-0.5ex]
RBC/UKQCD 10D   & \cite{Aoki:2010xg}    & 2+1 & 0.11 & \parbox[t]{6.3cm}{Single lattice spacing.}\\[0.5ex]\\[0.5ex]
\hline\hline\\[-0.5ex]
\end{tabular*}
\caption{Continuum extrapolations/estimation of lattice artifacts in
  determinations of the isovector unpolarised, helicity and
  transversity second moments with $\Nf=2+1$ quark flavours.}  }
\end{table}

\begin{table}[!ht]
{\footnotesize
\begin{tabular*}{\textwidth}{l @{\extracolsep{\fill}} c c c l}
\hline\hline \\[-1.0ex]
Collab. & Ref. & $\Nf$ & $a$ [fm] & Description 
\\[.3ex]\\[-2.ex] \hline \hline \\[-1.0ex]
%
RQCD 18   & \cite{Bali:2018zgl}        & 2 & 0.081, 0.071, 0.060 & No significant $O(a)$ effects observed.  \\[0.5ex]\\[0.5ex]\hline\\[-0.5ex]
ETM 17C   & \cite{Alexandrou:2017oeh}  & 2 & 0.0938 & Single ensemble.  \\[0.5ex]\\[0.5ex]\hline\\[-0.5ex]
ETM 15D   & \cite{Abdel-Rehim:2015owa} & 2 & 0.093  & Single ensemble. \\[0.5ex]\\[0.5ex]\hline\\[-0.5ex]
RQCD 14A  & \cite{Bali:2014gha}        & 2 & 0.081, 0.071, 0.060& Analysis not conclusive.  \\[0.5ex]\\[-0.5ex]
\hline\hline\\[-0.5ex]
\end{tabular*}
\caption{Continuum extrapolations/estimation of lattice artifacts in
  determinations of the isovector unpolarised, helicity and
  transversity second moments with $\Nf=2$ quark flavours.}  }
\end{table}

\begin{table}[!ht]
{\footnotesize
\begin{tabular*}{\textwidth}{l @{\extracolsep{\fill}} c c c l}
\hline\hline \\[-1.0ex]
{Collab.} & {Ref.} & {$\Nf$} & {$M_{\pi,\rm min}\,[\mev]$} & {Description}  
\\[.3ex]\\[-2.ex] \hline \hline \\[-1.0ex]
%
ETM 22       & \cite{Alexandrou:2022dtc} & 2+1+1 & 140,137,141 & \parbox[t]{6.3cm}{Simulate close to $M_\pi^{phys}$.}\\[0.5ex]\\[0.5ex]\hline\\[-0.5ex]
PNDME 20A     & \cite{Mondal:2020cmt} & 2+1+1 & 321,228,138,136 & \parbox[t]{6.3cm}{Extrapolation performed using a linear term in $M_\pi^2$ as part of a simultaneous fit in
$a$, $M_\pi$ and $M_\pi L$.}\\[0.5ex]\\[0.5ex]\hline\\[-0.5ex]
ETM 20C       & \cite{Alexandrou:2020sml} & 2+1+1 & 139 & \parbox[t]{6.3cm}{Single pion mass within 3\% of the physical value.}\\[0.5ex]\\[0.5ex]\hline\\[-0.5ex]
ETM 19A       & \cite{Alexandrou:2019ali} & 2+1+1 &  139& \parbox[t]{6.3cm}{Single pion mass within 3\% of the physical value.}\\[0.5ex]\\[0.5ex]
\hline \hline \\[-1.0ex]
\end{tabular*}
\caption{Chiral extrapolation/minimum pion mass in determinations of
  the isovector unpolarised, helicity and
  transversity second moments  with $\Nf=2+1+1$
  quark flavours.} 
}
\end{table}

\begin{table}[!ht]
{\footnotesize
\begin{tabular*}{\textwidth}{l @{\extracolsep{\fill}} c c c l}
\hline\hline \\[-1.0ex]
{Collab.} & {Ref.} & {$\Nf$} & {$M_{\pi,\rm min}\,[\mev]$} & {Description}  
\\[.3ex]\\[-2.ex] \hline \hline \\[-1.0ex]
%
Mainz 24     &\cite{Djukanovic:2024krw}  & 2+1   & 228,218,130,176& \parbox[t]{6.3cm}{A number of simultaneous $a$, $M_\pi$ and $M_\pi L$ fits are performed using terms which appear in the SU(2) NNLO ChPT expression for the axial charge and (alternatively) a linear term in $M_\pi^2$. The final results are obtained from a weighted average.} \\[0.5ex]\\[0.5ex]\hline\\[-0.5ex]
LHPC  24     & \cite{Rodekamp:2023wpe}   & 2+1 & 136,133 & \parbox[t]{6.3cm}{Simulate close to $M_\pi^{phys}$.}\\[0.5ex]\\[0.5ex]\hline\\[-0.5ex]
NME 21A       & \cite{Mondal:2021oot} & 2+1 & 285, 169, 167 & \parbox[t]{6.3cm}{Extrapolation performed using a linear term in $M_\pi^2$ as part of a simultaneous fit in
$a$, $M_\pi$ and $M_\pi L$.}\\[0.5ex]\\[0.5ex]\hline\\[-0.5ex]
NME 20       & \cite{Mondal:2020ela} & 2+1 &  285, 169, 167 & \parbox[t]{6.3cm}{Extrapolation performed using a linear term in $M_\pi^2$ as part of a simultaneous fit in
$a$, $M_\pi$ and $M_\pi L$.}\\[0.5ex]\\[0.5ex]\hline\\[-0.5ex]
Mainz 19        &\cite{Harris:2019bih}  & 2+1  & 290, 200, 260 & \parbox[t]{6.3cm}{Extrapolation performed using logarithmic and quadratic terms in $M_\pi$ as part of a simultaneous fit in $a$,
$M_\pi$ and $M_\pi L$.}\\[0.5ex]\\[0.5ex]\hline\\[-0.5ex]
$\chi$QCD 18A & \cite{Yang:2018nqn}   & 2+1   &171,330,139,300  & \parbox[t]{6.3cm}{Partially quenched calculation. Extrapolation performed as part of a simultaneous fit in $a^2$, $M_\pi$ and $M_\pi L$ using expression from HBChPT.} \\[0.5ex]\\[0.5ex]\hline\\[-0.5ex]
LHPC 12A        & \cite{Green:2012ud}   & 2+1  & 149,317 & \parbox[t]{6.3cm}{Chiral fit formula based on the ``small-scale   expansion'' to order $\epsilon^3$ with some coefficients fixed.}\\[0.5ex]\\[0.5ex]\hline\\[-0.5ex]
LHPC 10         & \cite{Bratt:2010jn}   & 2+1 &  293 & \parbox[t]{6.3cm}{Chiral
  fit formula based on the ``small-scale expansion'' to order
  $\epsilon^3$ with some coefficients fixed.}\\[0.5ex]\\[0.5ex]\hline\\[-0.5ex]
RBC/UKQCD 10D   & \cite{Aoki:2010xg}    & 2+1 & 329,416,555,668 & \parbox[t]{6.3cm}{Constant fit to heaviest three and linear fit to lightest two pion masses gives the quoted range.}\\[0.5ex]\\[0.5ex]
\hline \hline \\[-1.0ex]
\end{tabular*}
\caption{Chiral extrapolation/minimum pion mass in determinations of
  the isovector unpolarised, helicity and
  transversity second moments  with $\Nf=2+1$
  quark flavours.} 
}
\end{table}

\begin{table}[!ht]
{\footnotesize
\begin{tabular*}{\textwidth}{l @{\extracolsep{\fill}} c c c l}
\hline\hline \\[-1.0ex]
{Collab.} & {Ref.} & {$\Nf$} & {$M_{\pi,\rm min}\,[\mev]$} & {Description}  
\\[.3ex]\\[-2.ex] \hline \hline \\[-1.0ex]
%
RQCD 18   & \cite{Bali:2018zgl}        & 2 & 280, 150, 260 & Chiral fit using BChPT. Kept terms up to $O(M_\pi^3)$. \\[0.5ex]\\[0.5ex]\hline\\[-0.5ex]
ETM 17C   & \cite{Alexandrou:2017oeh}  & 2 & $\approx$135 & Single ensemble within 3\% of physical pion mass. \\[0.5ex]\\[0.5ex]\hline\\[-0.5ex]
ETM 15D   & \cite{Abdel-Rehim:2015owa} & 2 & 131 & Single ensemble within 3\% of physical pion mass. \\[0.5ex]\\[0.5ex]\hline\\[-0.5ex]
RQCD 14A  & \cite{Bali:2014gha}        & 2 & 280, 150, 260 & No significant dependence on pion mass observed. \\[0.5ex]\\[0.5ex]
\hline \hline \\[-1.0ex]
\end{tabular*}
\caption{Chiral extrapolation/minimum pion mass in determinations of
  the isovector unpolarised, helicity and
  transversity second moments  with $\Nf=2$
  quark flavours.} 
}
\end{table}

\begin{table}[!ht]
{\footnotesize
\begin{tabular*}{\textwidth}{l @{\extracolsep{\fill}} c c c c l}
\hline\hline \\[-1.0ex]
Collab. & Ref. & $\Nf$ & $L$ [fm] & ${M_{\pi,\rm min}}L$ & Description 
\\[.3ex]\\[-2.ex] \hline \hline \\[-1.0ex]
%
ETM 22       & \cite{Alexandrou:2022dtc} & 2+1+1 & \parbox[t]{1.4cm}{5.1,\\5.5,\\5.5}
& \parbox[t]{1.cm}{3.6,\\3.8,\\3.9} & \parbox[t]{5.9cm}{Finite-volume effects not estimated.}\\[0.5ex]\\[0.5ex]\hline\\[-0.5ex]
PNDME 20A     & \cite{Mondal:2020cmt} & 2+1+1 &  \parbox[t]{1.4cm}{2.4,\\4.8,\\5.6,\\5.5}
& \parbox[t]{1.cm}{3.9,\\5.5,\\3.9,\\3.7} & \parbox[t]{5.9cm}{Fit performed using a
  term of the form $M_\pi^2e^{-M_\pi L}/\sqrt{M_\pi L}$ as part of a simultaneous fit in
$a$, $M_\pi$ and $M_\pi L$.}\\[0.5ex]\\[0.5ex]\hline\\[-0.5ex]
ETM 20C       & \cite{Alexandrou:2020sml} & 2+1+1 &   \parbox[t]{1.4cm}{5.1}
& \parbox[t]{1.cm}{3.6} & \parbox[t]{5.9cm}{Finite-volume effects not estimated.}\\[0.5ex]\\[0.5ex]\hline\\[-0.5ex]
ETM 19A       & \cite{Alexandrou:2019ali} & 2+1+1 &   \parbox[t]{1.4cm}{5.1}
& \parbox[t]{1.cm}{3.6} & \parbox[t]{5.9cm}{Finite-volume effects not estimated.}\\[0.5ex]\\[0.5ex]
\hline\hline  \\[-1.0ex]
\end{tabular*}
\caption{Finite-volume effects in determinations of the isovector
  unpolarised, helicity and transversity second moments with
  $\Nf=2+1+1$ quark flavours.}  }
\end{table}

\begin{table}[!ht]
{\footnotesize
\begin{tabular*}{\textwidth}{l @{\extracolsep{\fill}} c c c c l}
\hline\hline \\[-1.0ex]
Collab. & Ref. & $\Nf$ & $L$ [fm] & ${M_{\pi,\rm min}}L$ & Description 
\\[.3ex]\\[-2.ex] \hline \hline \\[-1.0ex]
%
Mainz 24     &\cite{Djukanovic:2024krw}  & 2+1   &  \parbox[t]{1.4cm}{4.1,\\4.8,\\6.1,\\4.7}
& \parbox[t]{1.cm}{4.7,\\5.4,\\4.0,\\4.2} & \parbox[t]{5.9cm}{A number of simultaneous $a$, $M_\pi$ and $M_\pi L$ fits are performed using term of the form $M_\pi^2e^{-M_\pi L}/\sqrt{M_\pi L}$. The final results are obtained from a weighted average.} \\[0.5ex]\\[0.5ex]\hline\\[-0.5ex]
LHPC  24     & \cite{Rodekamp:2023wpe}   & 2+1 &  \parbox[t]{1.4cm}{5.6,\\5.9}
& \parbox[t]{1.cm}{3.9,\\4.0} & \parbox[t]{5.9cm}{Finite-volume effects are not estimated.}\\[0.5ex]\\[0.5ex]\hline\\[-0.5ex]
NME 21A       & \cite{Mondal:2021oot} & 2+1 &  \parbox[t]{1.4cm}{4.1,\\3.0--5.8,\\3.5--5.1}
& \parbox[t]{1.cm}{5.9,\\5.1,\\4.3} & \parbox[t]{5.9cm}{Fit performed using a
  term of the form $M_\pi^2e^{-M_\pi L}/\sqrt{M_\pi L}$ as part of a simultaneous fit in
$a$, $M_\pi$ and $M_\pi L$.}\\[0.5ex]\\[0.5ex]\hline\\[-0.5ex]
NME 20       & \cite{Mondal:2020ela} & 2+1 &  \parbox[t]{1.4cm}{4.1,\\3.0--5.8,\\3.5--5.1}
& \parbox[t]{1.cm}{5.9,\\5.1,\\4.3} & \parbox[t]{5.9cm}{Fit performed using a
  term of the form $M_\pi^2e^{-M_\pi L}/\sqrt{M_\pi L}$ as part of a simultaneous fit in
$a$, $M_\pi$ and $M_\pi L$.}\\[0.5ex]\\[0.5ex]\hline\\[-0.5ex]
Mainz 19        &\cite{Harris:2019bih}  & 2+1  &  \parbox[t]{1.5cm}{2.8-4.1,\\2.4-3.6,\\2.1-4.1, 2.4-3.2} &
\parbox[t]{1.cm}{4.7,\\5.3,\\4.2,\\4.3} & \parbox[t]{5.9cm}{Extrapolation performed including a term of the
form ${M_\pi^2}e^{-M_\pi L}/{\sqrt{M_{\pi}L}}$ as part of a
  simultaneous fit in $a^2$, $M_\pi$ and $M_\pi L$.}\\[0.5ex]\\[0.5ex]\hline\\[-0.5ex]
$\chi$QCD 18A & \cite{Yang:2018nqn}   & 2+1   &  \parbox[t]{1.4cm}{4.6,\\2.7,\\5.5,\\2.6}
& \parbox[t]{1.cm}{4.0,\\4.4,\\3.9,\\4.0} & \parbox[t]{5.9cm}{Extrapolation performed including a term of the
form $e^{-M_\pi L}$ as part of a simultaneous fit in $a^2$, $M_\pi$ and $M_\pi L$.} \\[0.5ex]\\[0.5ex]\hline\\[-0.5ex]
LHPC 12A        & \cite{Green:2012ud}   & 2+1  & \parbox[t]{1.4cm}{2.8--5.6,\\2.9} & \parbox[t]{1.cm}{4.2,\\4.6} & \parbox[t]{5.9cm}{Finite-volume effects investigated and found to be   negligible.}\\[0.5ex]\\[0.5ex]\hline\\[-0.5ex]
LHPC 10         & \cite{Bratt:2010jn}   & 2+1 &  \parbox[t]{1.4cm}{2.5--3.5}
& \parbox[t]{1.cm}{3.7} & \parbox[t]{5.9cm}{Finite-volume effects 
  included in chiral fit formula and found to be negligible.}\\[0.5ex]\\[0.5ex]\hline\\[-0.5ex]
RBC/UKQCD 10D   & \cite{Aoki:2010xg}    & 2+1 & \parbox[t]{1.4cm}{2.7}
& \parbox[t]{1.cm}{4.6} & \parbox[t]{5.9cm}{No uncertainty
  estimated. Comparison between two volumes for the three heaviest 
  pion masses shows no deviation within statistical
  errors.}\\[0.5ex]\\[0.5ex]
\hline\hline  \\[-1.0ex]
\end{tabular*}
\caption{Finite-volume effects in determinations of the isovector
  unpolarised, helicity and transversity second moments with
  $\Nf=2+1$ quark flavours.}  }
\end{table}

\begin{table}[!ht]
{\footnotesize
\begin{tabular*}{\textwidth}{l @{\extracolsep{\fill}} c c c c l}
\hline\hline \\[-1.0ex]
Collab. & Ref. & $\Nf$ & $L$ [fm] & ${M_{\pi,\rm min}}L$ & Description 
\\[.3ex]\\[-2.ex] \hline \hline \\[-1.0ex]
%
RQCD 18   & \cite{Bali:2018zgl}        & 2 & \parbox[t]{1.4cm}{2.6,\\1.7-4.6,\\1.9-2.9}  & \parbox[t]{1.cm}{3.7,\\ 2.8,\\ 3.8} & \parbox[t]{3.9cm}{No significant finite-volume effects observed.}\\[0.5ex]\\[0.5ex]\hline\\[-0.5ex]
ETM 17C   & \cite{Alexandrou:2017oeh}  & 2 & \parbox[t]{1.4cm}{4.5}  & \parbox[t]{1.cm}{3.1}  & Single ensemble  \\[0.5ex]\\[0.5ex]\hline\\[-0.5ex]
ETM 15D   & \cite{Abdel-Rehim:2015owa} & 2 & \parbox[t]{1.4cm}{4.5} & \parbox[t]{1.cm}{3.0}  & Single ensemble \\[0.5ex]\\[0.5ex]\hline\\[-0.5ex]
RQCD 14A  & \cite{Bali:2014gha}        & 2 & \parbox[t]{1.4cm}{2.6,\\1.7-4.6,\\1.9-2.9}  & \parbox[t]{1.cm}{3.7,\\2.8,\\3.8} & \parbox[t]{3.9cm}{No significant finite-volume effects observed.}\\[0.5ex]\\[0.5ex]
\hline\hline  \\[-1.0ex]
\end{tabular*}
\caption{Finite-volume effects in determinations of the isovector
  unpolarised, helicity and transversity second moments with
  $\Nf=2$ quark flavours.}  }
\end{table}

\begin{table}[!ht]
{\footnotesize
\begin{tabular*}{\textwidth}{l @{\extracolsep{\fill}} c c c l}
\hline\hline \\[-1.0ex]
Collab. & Ref. & $\Nf$ & Ren.
\\[.3ex]\\[-2.ex] \hline \hline \\[-1.0ex]
%
ETM 22       & \cite{Alexandrou:2022dtc} & 2+1+1 &  RI$^\prime$-MOM \\[0.2ex]\\[0.2ex]\hline\\[-1.0ex]
PNDME 20A     & \cite{Mondal:2020cmt} & 2+1+1 & RI$^\prime$-MOM\\[0.2ex]\\[0.2ex]\hline\\[-1.0ex]
ETM 20C       & \cite{Alexandrou:2020sml} & 2+1+1 & RI$^\prime$-MOM\\[0.2ex]\\[0.2ex]\hline\\[-1.0ex]
ETM 19A       & \cite{Alexandrou:2019ali} & 2+1+1 & RI$^\prime$-MOM\\[0.2ex]\\[0.2ex]
\hline\\[-1.0ex]
Mainz 24     &\cite{Djukanovic:2024krw}  & 2+1   & RI-MOM \\[0.2ex]\\[0.2ex]\hline\\[-1.0ex]
LHPC  24     & \cite{Rodekamp:2023wpe}   & 2+1 & RI$^\prime$-MOM/RI-SMOM \\[0.2ex]\\[0.2ex]\hline\\[-1.0ex]
NME 21A       & \cite{Mondal:2021oot}    & 2+1 & RI$^\prime$-MOM \\[0.2ex]\\[0.2ex]\hline\\[-1.0ex]
NME 20       & \cite{Mondal:2020ela}     & 2+1 & RI$^\prime$-MOM  \\[0.2ex]\\[0.2ex]\hline\\[-1.0ex]
Mainz 19        &\cite{Harris:2019bih}  & 2+1  & RI-MOM \\[0.2ex]\\[0.2ex]\hline\\[-1.0ex]
$\chi$QCD 18A & \cite{Yang:2018nqn}     & 2+1  & RI-MOM  \\[0.2ex]\\[0.2ex]\hline\\[-1.0ex]
LHPC 12A        & \cite{Green:2012ud}   & 2+1  & RI-MOM  \\[0.2ex]\\[0.2ex]\hline\\[-1.0ex]
LHPC 10         & \cite{Bratt:2010jn}   & 2+1 &  RI-MOM \\[0.2ex]\\[0.2ex]\hline\\[-1.0ex]
RBC/UKQCD 10D   & \cite{Aoki:2010xg}    & 2+1 &  RI-MOM\\[0.2ex]\\[0.2ex]
\hline\\[-1.0ex]
RQCD 18      & \cite{Bali:2018zgl}        & 2 & RI$^\prime$-MOM  \\[0.2ex]\\[0.2ex]\hline\\[-1.0ex]
ETM 17C      & \cite{Alexandrou:2017oeh}  & 2 & RI$^\prime$-MOM  \\[0.2ex]\\[0.2ex]\hline\\[-1.0ex]
ETM 15D      & \cite{Abdel-Rehim:2015owa} & 2 & RI$^\prime$-MOM  \\[0.2ex]\\[0.2ex]\hline\\[-1.0ex]
RQCD 14A     & \cite{Bali:2014gha}        & 2 & RI$^\prime$-MOM  \\[0.2ex]\\[0.2ex]
 \hline \hline \\[-1.0ex]
\end{tabular*}
\caption{Renormalization in determinations of the isovector
  unpolarised, helicity and transversity second moments with
  $\Nf=2+1+1$, $\Nf=2+1$ and $\Nf=2$  quark flavours.}  }
\end{table}

\begin{table}[!ht]
{\footnotesize
\begin{tabular*}{\textwidth}{l @{\extracolsep{\fill}} c c c l}
\hline\hline \\[-1.0ex]
Collab. & Ref. & $\Nf$ & $\tau$ [fm] & Description 
\\[.3ex]\\[-2.ex]\hline \hline \\[-1.0ex]
%
ETM 22       & \cite{Alexandrou:2022dtc} & 2+1+1 &
\parbox[t]{3.5cm}{[0.6--1.6]\par [0.6--1.5]\par
  [0.5--1.2]} &\parbox[t]{5.4cm}{Two-state fit to all $\tau$. A comparison is made
with plateau fits and the summation method.} \\[0.5ex]\\[0.5ex]\hline\\[-0.5ex]
PNDME 20A     & \cite{Mondal:2020cmt} & 2+1+1 & \parbox[t]{3.5cm}{[0.8--1.4]\par [1.0--1.7,1.0--1.7]\par
  [0.9--1.4,0.9--1.4,0.9--1.4]\par [1.0--1.4,0.9--1.3]} & \parbox[t]{5.4cm}{Fits to the $\tau$- and $t$-dependence of three-point correlators using two or three lowest-lying states.}  \\[0.5ex]\\[0.5ex]\hline\\[-0.5ex]
ETM 20C       & \cite{Alexandrou:2020sml} & 2+1+1 &\parbox[t]{3.5cm}{[0.6--1.6]} & \parbox[t]{5.4cm}{Two-state fit to all $\tau$. A comparison is made
with plateau fits and the summation method.} \\[0.5ex]\\[0.5ex]\hline\\[-0.5ex]
ETM 19A       & \cite{Alexandrou:2019ali} & 2+1+1 &\parbox[t]{3.5cm}{[0.6--1.6]} & \parbox[t]{5.4cm}{Two-state fit to all $\tau$. A comparison is made
with plateau fits and the summation method.} \\[0.5ex]\\[0.5ex]
\hline \hline \\[-1.0ex]
\end{tabular*}
\caption{Control of excited-state contamination in determinations of
  the isovector unpolarised, helicity and transversity second moments
  with $\Nf=2+1+1$ quark flavours. The comma-separated list of numbers
  in square brackets denote the range of source-sink separations
  $\tau$ (in fermi) at each value of the bare coupling.}  }
\end{table}

\begin{table}[!ht]
{\footnotesize
\begin{tabular*}{\textwidth}{l @{\extracolsep{\fill}} c c c l}
\hline\hline \\[-1.0ex]
Collab. & Ref. & $\Nf$ & $\tau$ [fm] & Description 
\\[.3ex]\\[-2.ex]\hline \hline \\[-1.0ex]
%
Mainz 24     &\cite{Djukanovic:2024krw}  & 2+1   & \parbox[t]{3.5cm}{[0.4--1.5,0.4--1.5,0.4--1.5,0.4--1.5]\par [0.3--1.5,0.3--1.5,0.3--1.5]\par [0.3--1.4,0.3--1.4,0.3--1.4,0.3--1.4,0.3--1.4]\par [0.2--1.4,0.2--1.4,0.2--1.4]} & \parbox[t]{5.4cm}{Two-state summation method.} \\[0.5ex]\\[0.5ex]\hline\\[-0.5ex]
LHPC  24     & \cite{Rodekamp:2023wpe}   & 2+1 & \parbox[t]{3.5cm}{[0.3--1.4]\par [0.9--1.5]} & \parbox[t]{5.4cm}{Two-state fits to the three-point function and the summation method are utilised, with the final results obtained from a weighted average.}\\[0.5ex]\\[0.5ex]\hline\\[-0.5ex]
NME 21A       & \cite{Mondal:2021oot} & 2+1 & \parbox[t]{3.8cm}{[1.3--1.8]\par [1.1--1.5,1.3--1.7,1.1--1.5,1.1--1.5]\par [1.1--1.4,1.2--1.5]} & \parbox[t]{5.4cm}{Fits to the $\tau$- and $t$-dependence of three-point correlators using two or three lowest-lying states.}\\[0.5ex]\\[0.5ex]\hline\\[-0.5ex]
NME 20       & \cite{Mondal:2020ela} & 2+1 &  \parbox[t]{3.8cm}{[1.3--1.8]\par [1.1--1.5,1.3--1.7,1.1--1.5,1.1--1.5]\par [1.1--1.4,1.2--1.5]} & \parbox[t]{5.4cm}{Fits to the $\tau$- and $t$-dependence of three-point correlators using two or three lowest-lying states.}\\[0.5ex]\\[0.5ex]\hline\\[-0.5ex]
Mainz 19        &\cite{Harris:2019bih}  & 2+1  & \parbox[t]{3.5cm}{[1.0--1.4,1.0--1.4,1.0--1.4]\par
  [1.0--1.5,1.0--1.5]\par [1.0--1.4,1.0--1.4,1.0--1.4,1.0--1.4]\par
  [1.0--1.4,1.0--1.3]} & \parbox[t]{5.4cm}{Fits to the $\tau$- and $t$-dependence of correlator ratios using the two lowest-lying states.}\\[0.5ex]\\[0.5ex]\hline\\[-0.5ex]
$\chi$QCD 18A & \cite{Yang:2018nqn}   & 2+1   & \parbox[t]{3.5cm}{[0.7--1.5]}  & \parbox[t]{5.4cm}{two-state fits to the three-point function.} \\[0.5ex]\\[0.5ex]\hline\\[-0.5ex]
LHPC 12A        & \cite{Green:2012ud}   & 2+1  &\parbox[t]{3.5cm}{[0.9--1.4,0.9--1.4,0.9--1.4,0.9--1.4,0.9--1.4]\par [0.9--1.4]} & \parbox[t]{5.4cm}{Fits to the leading (ground state) $\tau$-dependence of summed correlator ratios.}\\[0.5ex]\\[0.5ex]\hline\\[-0.5ex]
LHPC 10         & \cite{Bratt:2010jn}   & 2+1 & \parbox[t]{3.5cm}{[1.1,1.1,1.1,1.1,1.1--1.2,1.1]} & \parbox[t]{5.4cm}{Plateau fits of correlator ratio at
  $\tau=1.1$\,fm. Larger source-sink separation on one ensemble as
  cross check.}\\[0.5ex]\\[0.5ex]\hline\\[-0.5ex]
RBC/UKQCD 10D   & \cite{Aoki:2010xg}    & 2+1 & \parbox[t]{3.5cm}{[1.4,1.4,1.4,1.4]} & \parbox[t]{5.4cm}{Single
  source-sink separation considered.}\\[0.5ex]\\[0.5ex]
\hline \hline \\[-1.0ex]
\end{tabular*}
\caption{Control of excited-state contamination in determinations of
  the isovector unpolarised, helicity and transversity second moments
  with $\Nf=2+1$ quark flavours. The comma-separated list of numbers
  in square brackets denote the range of source-sink separations
  $\tau$ (in fermi) at each value of the bare coupling.}  }
\end{table}

\begin{table}[!ht]
{\footnotesize
\begin{tabular*}{\textwidth}{l @{\extracolsep{\fill}} c c c l}
\hline\hline \\[-1.0ex]
Collab. & Ref. & $\Nf$ & $\tau$ [fm] & Description 
\\[.3ex]\\[-2.ex]\hline \hline \\[-1.0ex]
%
RQCD 18    & \cite{Bali:2018zgl}        & 2 & \parbox[t]{1.5cm}{Not given} & Analysis limited by statistics\\[0.5ex]\\[0.5ex]\hline\\[-0.5ex]
ETM 17C    & \cite{Alexandrou:2017oeh}  & 2 & \parbox[t]{1.5cm}{[0.9--1.5]}  & \parbox[t]{5.4cm}{Result from plateau method with $\tau/a=14$. Consistent with estimates from summation and two-state fit methods.} \\[0.5ex]\\[0.5ex]\hline\\[-0.5ex]
ETM 15D    & \cite{Abdel-Rehim:2015owa} & 2 & \parbox[t]{1.5cm}{[0.9--1.3]}  & \parbox[t]{5.4cm}{Result from plateau method with $\tau/a=14$. Consistent with estimate from the summation method.} \\[0.5ex]\\[0.5ex]\hline\\[-0.5ex]
RQCD 14A   & \cite{Bali:2014gha}        & 2 & \parbox[t]{1.5cm}{Not given} & \parbox[t]{5.4cm}{Plateau value at larger $\tau/a$ consistent with two-state fit. }\\[0.5ex]\\[0.5ex]
\hline \hline \\[-1.0ex]
\end{tabular*}
\caption{Control of excited-state contamination in determinations of
  the isovector unpolarised, helicity and transversity second moments
  with $\Nf=2$ quark flavours. The comma-separated list of numbers
  in square brackets denote the range of source-sink separations
  $\tau$ (in fermi) at each value of the bare coupling.}  }
\end{table}

%% file: ScaleSetting/scalesetting_app.tex
\subsection{Notes to Sec.~\ref{sec:scalesetting} on scale setting}

\begin{table}[!ht]
{\footnotesize
\begin{tabular*}{\textwidth}[l]{l @{\extracolsep{\fill}} c c c l}
\hline\hline \\[-1.0ex]
Collab. & Ref. & $\Nf$ & $a$ [fm] & Description
\\[1.0ex] \hline \hline \\[-1.0ex]
TUMQCD 22 & \cite{Brambilla:2022het} & 2+1+1 & \parbox[t]{3.5cm}{\centering 0.15294, 0.12224, 0.08786, 0.05662,  0.0426, 0.03216} &  \parbox[t]{5.5cm}{MILC  ensembles
  from on-shell Symanzik-improved gauge action and rooted HISQ fermions}\\
[7.0ex] \hline \\[-1.0ex]
ETM 21 & \cite{Alexandrou:2021bfr} & 2+1+1 & 0.069, 0.079, 0.097 & \\
[1.0ex]
\ifx\reducedapptables\undefined
\hline \\[-1.0ex]
CalLat 20A & \cite{Miller:2020evg} & 2+1+1 & \parbox[t]{3.5cm}{$[0.056, 0.082, 0.11, 0.13]$, $[0.057, 0.087, 0.12, 0.15]$}   & \parbox[t]{5.5cm}{M\"obius Domain-wall valence quarks on HISQ /  MILC sea; lattice spacings depending on scheme.} \\
[7.0ex] \hline \\[-1.0ex]
 BMW 20   &  \cite{Borsanyi:2020mff} & 1+1+1+1 & \parbox[t]{3.5cm}{\centering 0.0640, 0.0787, 0.0952, 0.1116, 0.1191, 0.1315} &  \parbox[t]{5.5cm}{Staggered fermion computation with isospin breaking and QED.}  \\[4.0ex] 
\hline\\[-1.0ex]
  ETM 20 & \cite{Dimopoulos:2020eqd} & 2+1+1 & 0.069, 0.079, 0.097 & \parbox[t]{5.5cm}{Wilson TM
  fermions at maximal twist including clover term. Note that  lattice spacings are not explicitely given in \cite{Dimopoulos:2020eqd}.}\\
  [10.0ex] \hline \\[-1.0ex]
ETM 18A & \cite{Alexandrou:2018egz} & 2+1+1 & 0.08 & \parbox[t]{5.5cm}{Wilson TM fermions at maximal twist including clover term.}\\
  [4.0ex] \hline \\[-1.0ex]
FNAL/MILC 17 &  \cite{Bazavov:2017lyh} & 2+1+1 &  \parbox[t]{3.5cm}{\centering 0.03, 0.042, 0.06, 0.09, 0.12, 0.15} & Determination of $f_{p4s}$.\\
[4.0ex]
\hline \\[-1.0ex]
MILC 15   & \cite{Bazavov:2015yea}  & 2+1+1 & \parbox[t]{3.5cm}{\centering 0.06, 0.09, 0.12, 0.15\\}& \parbox[t]{5.5cm}{Highly improved staggered quarks.} \\
[2.0ex] \hline \\[-1.0ex]
ETM 14 & \cite{Carrasco:2014cwa} & 2+1+1 & 0.089, 0.082, 0.062 & \parbox[t]{5.5cm}{Wilson TM fermions at maximal twist.}\\
[2.0ex] \hline \\[-1.0ex]
FNAL/MILC 14A &  \cite{Bazavov:2014wgs} & 2+1+1 & 0.06, 0.09, 0.12, 0.15  & \\
[2.0ex]
\hline \\[-1.0ex]
HPQCD 13A &  \cite{Dowdall:2013rya} & 2+1+1 &  0.15, 0.12, 0.09 & \\
[2.0ex] \hline \\[-1.0ex]
HPQCD 11B &  \cite{Dowdall:2011wh} & 2+1+1 & 0.09, 0.12, 0.15 &
                                                                \parbox[t]{5.5cm}{NRQCD for $b$ quark and HISQ for valence light quarks and MILC sea.} \\
[4.0ex]
\fi
\hline \hline\\[-1.0ex]
\end{tabular*}
\caption{Continuum extrapolations/estimation of lattice artifacts in scale
  determinations with $\Nf=2+1+1$ quark flavours.}
}
\end{table}

\begin{table}[!ht]
{\footnotesize
\begin{tabular*}{\textwidth}[l]{l @{\extracolsep{\fill}} c c c l}
\hline\hline \\[-1.0ex]
Collab. & Ref. & $\Nf$ & $a$ [fm] & Description
\\[1.0ex] \hline \hline \\[-1.0ex]
Hudspith 24 & \cite{Hudspith:2024kzk}    & 2+1 &  \parbox[t]{3.5cm}{\centering 0.098, 0.085, 0.075, 0.064, 0.049, 0.039} &  \parbox[t]{5.5cm}{NP ${\cal O}(a)$-improved Wilson fermions with tree-level Symanzik improved gauge action.} \\
[7.0ex] \hline 
\\[-1.0ex] 
Asmussen 23 & \cite{Asmussen:2023pia}    & 2+1 &  \parbox[t]{3.5cm}{\centering  0.085, 0.075, 0.064, 0.049, 0.039} &  \parbox[t]{5.5cm}{NP ${\cal O}(a)$-improved Wilson fermions with tree-level Symanzik improved gauge action.} \\
[7.0ex] \hline \\[-1.0ex]
RQCD 22 & \cite{RQCD:2022xux}    & 2+1 &  \parbox[t]{3.5cm}{\centering 0.098, 0.085, 0.075, 0.064, 0.049, 0.039} &  \parbox[t]{5.5cm}{NP ${\cal O}(a)$-improved Wilson fermions with tree-level Symanzik improved gauge action.} \\
[7.0ex] \hline \\[-1.0ex]
RBC/Bielefeld 07 & \cite{Cheng:2007jq} & 2+1 & \parbox[t]{3.5cm}{\centering 0.3 - 0.05} & \parbox[t]{5.5cm}{Simulations with improved staggered fermions (p4fat3-action: smeared 1-link term and bent 3-link terms) at 27 different values of $\beta$.}\\
[10.0ex]
\ifx\reducedapptables\undefined
\hline \\[-1.0ex]
CLS 21 & \cite{Strassberger:2021tsu} & 2+1 & \parbox[t]{3.5cm}{\centering 0.038, 0.049, 0.063, 0.075 0.085}  &  \parbox[t]{5.5cm}{NP ${\cal O}(a)$-improved Wilson fermions with tree-level Symanzik improved gauge action.}   \\
[7.0ex] \hline \\[-1.0ex]
CLS 16 & \cite{Bruno:2016plf} & 2+1 & 0.085, 0.065, 0.05 &
                                                              \parbox[t]{5.5cm}{NP ${\cal O}(a)$-improved Wilson fermions with LW gauge action.} \\
[4.0ex] \hline \\[-1.0ex]
QCDSF/UKQCD 15B & \cite{Bornyakov:2015eaa} & 2+1 &  \parbox[t]{3.5cm}{\centering 0.059, 0.068, 0.074, 0.082} &  $(a_\text{max}/a_\text{min})^2=1.94$ \\
[4.0ex]\hline \\[-1.0ex]
$\chi$QCD 14& \cite{Yang:2014sea} & 2+1 & 0.084, 0.112 & \parbox[t]{5.5cm}{Valence overlap fermions on domain wall fermion gauge configurations.}\\
[4.0ex] \hline \\[-1.0ex]
HotQCD 14 & \cite{Bazavov:2014pvz} & 2+1 & $[0.04,0.25]$ & \parbox[t]{5.5cm}{HISQ and tree-level improved Symanzik gauge action. Several ensembles with $a$ in the quoted range and with a single $M_\pi\approx 160\MeV$.} \\[10.0ex]
\hline\\ [-1.0ex]
RBC/UKQCD 14  & \cite{Blum:2014tka}  & 2+1 & \parbox[t]{3.5cm}{\centering 0.06, 0.08, 0.11, 0.14\\}& \parbox[t]{5.5cm}{Two versions of Domain Wall Fermions combined each with the Iwasaki gauge action.} \\[7.0ex]
\hline\\ [-1.0ex]
BMW 12A &  \cite{Borsanyi:2012zs} & 2+1 & \parbox[t]{3.5cm}{\centering 0.053, 0.065, 0.077, 0.093} & \parbox[t]{5.5cm}{ Wilson fermion computation.} \\
[4.0ex]\hline \\[-1.0ex]
HotQCD 11 & \cite{Bazavov:2011nk} & 2+1 & \parbox[t]{3.5cm}{\centering $[0.066,0.25]$\\}& \parbox[t]{5.5cm}{HISQ and tree-level improved Symanzik gauge action. Several ensembles with $a$ in the quoted range.} \\[7.0ex]
\hline \\[-1.0ex]
RBC/UKQCD 10A &  \cite{Aoki:2010dy} & 2+1 & 0.087, 0.114 & \\
[2.0ex]
\hline \\[-1.0ex]
MILC 10 & \cite{Bazavov:2010hj} & 2+1 & 0.045, 0.06, 0.09 & Asqtad staggered quarks.
\\[2.0ex]
\fi
\hline \hline \\[-1.0ex]
\end{tabular*}
\caption{Continuum extrapolations/estimation of lattice artifacts in scale
  determinations with $\Nf=2+1$ quark flavours.}
}
\end{table}

\ifx\reducedapptables\undefined
\begin{table}[!ht]
\addtocounter{table}{-1}
{\footnotesize
\begin{tabular*}{\textwidth}[l]{l @{\extracolsep{\fill}} c c c l}
\hline\hline \\[-1.0ex]
Collab. & Ref. & $\Nf$ & $a$ [fm] & Description
\\[1.0ex] \hline \hline \\[-1.0ex]
HPQCD 09B &  \cite{Davies:2009tsa} & 2+1 &
                                           \parbox[t]{3.5cm}{\centering 0.044, 0.059, 0.085, 0.12, 0.15} & \parbox[t]{5.5cm}{NRQCD for $b$ quark and HISQ for valence light quarks on MILC asqtad sea.}
 \\
[4.0ex]\hline \\[-1.0ex]
MILC 09 & \parbox[t]{1.5cm}{\cite{Bazavov:2009bb,Bazavov:2009fk,Bazavov:2009tw} }& 2+1 & \parbox[t]{3.5cm}{\centering $[0.045,0.18]$\\}& \parbox[t]{5.5cm}{Asqtad staggered quarks, six different lattice spacings in the quoted range.} \\
[4.0ex]\hline \\[-1.0ex]
PACS-CS 08 & \cite{Aoki:2008sm} & 2+1 & \parbox[t]{3.5cm}{\centering $0.09$\\}& \parbox[t]{5.5cm}{Several ensembles of Wilson clover quarks and Iwasaki gauge action at a single lattice spacing.} \\
[7.0ex] \hline \\[-1.0ex]
HPQCD 05B &  \cite{Gray:2005ur} & 2+1 & 0.09, 0.12, 0.17  & \parbox[t]{5.5cm}{NRQCD with tree level tadpole improved couplings on MILC asqtad sea.} \\
[4.0ex]\hline \\[-1.0ex]
Aubin 04 & \cite{Aubin:2004wf} & 2+1 & \parbox[t]{3.5cm}{\centering 0.09, 0.12\\}& \parbox[t]{5.5cm}{Asqtad staggered quarks.}\\ [2.0ex]
\hline \hline \\[-1.0ex]
\end{tabular*}
\caption{(cntd.) Continuum extrapolations/estimation of lattice artifacts in scale
  determinations with $\Nf=2+1$ quark flavours.}
}
\end{table}
\fi

\begin{table}
{\footnotesize
\begin{tabular*}{\textwidth}[l]{l @{\extracolsep{\fill}} c c c c l}
\hline \hline  \\[-1.0ex]
Collab. & Ref. & $\Nf$ & $M_{\pi,\text{min}}$ [MeV] &  $M_{\pi} L$  & Description
\\[1.0ex] \hline \hline \\[-1.0ex]

TUMQCD 22 & \cite{Brambilla:2022het} & 2+1+1 & 129 & 3.25 $-$ 4.17 & At the physical point.\\
[1.0ex] \hline \\[-1.0ex]
  ETM 21 & \cite{Alexandrou:2021bfr} & 2+1+1 & 134.2 & 3.78 & \\
  [1.0ex]
\ifx\reducedapptables\undefined
\hline \\[-1.0ex]
CalLat 20A & \cite{Miller:2020evg}    & 2+1+1 &  130 & 3.9 &  \\
[2.0ex] \hline \\[-1.0ex]
BMW 20   &  \cite{Borsanyi:2020mff} & 1+1+1+1 &   \parbox[t]{2.75cm}{Several pion masses within $\pm$3\% of physical value.}  &  3.0   &       \\
[7.0ex] \hline\\[-1.0ex]
ETM 20 & \cite{Dimopoulos:2020eqd} & 2+1+1 &  135 & 3.5 & \parbox[t]{5.5cm}{Only  indirectly inferred.}\\
[2.0ex] \hline \\[-1.0ex]
  ETM 18A & \cite{Alexandrou:2018egz} & 2+1+1 &  139.8 & 3.6 & One ensemble only.\\
[2.0ex] \hline \\[-1.0ex]
FNAL/MILC 17 &  \cite{Bazavov:2014wgs} & 2+1+1 & 129 & 3.7  & Determination of $f_{4ps}, M_{4ps}$.\\
[2.0ex]
\hline \\[-1.0ex]
MILC 15  & \cite{Bazavov:2015yea} & 2+1+1 & 125 & 3.7
&  \parbox[t]{5.5cm}{Four ensembles at physical point.}\\
[2.0ex] \hline \\[-1.0ex]
ETM 14 & \cite{Carrasco:2014cwa} & 2+1+1 & 211 & 3.19 &\\
[2.0ex] \hline \\[-1.0ex]
FNAL/MILC 14A &  \cite{Bazavov:2014wgs} & 2+1+1 & 130 & 3.7 & Determination of $f_{4ps}, M_{4ps}$.\\
[2.0ex]
\hline \\[-1.0ex]
HPQCD 13A &  \cite{Dowdall:2013rya} & 2+1+1 & 128 & 3.7 & \\
[2.0ex] \hline \\[-1.0ex]
HPQCD 11B & \cite{Dowdall:2011wh} & 2+1+1 & 211 & 4.0 & \parbox[t]{5.5cm}{NRQCD $\Upsilon$ 2s-1s splitting and $\eta_s$ as input.}
\\[4.0ex]
\fi
\hline\hline \\[-1.0ex]
\end{tabular*}
\caption{Chiral extrapolation and finite-volume effects in scale determinations 
  with $\Nf=2+1+1$ quark flavours. We list the minimum pion mass $M_{\pi,\text{min}}$ and $M_{\pi} L \equiv M_{\pi,\text{min}} [L(M_{\pi,\text{min}})]_\text{max}$ is evaluated at the maximum value of $L$ available at $M_{\pi} =M_{\pi,\text{min}}$.  }
}
\end{table}

\begin{table}
{\footnotesize
\begin{tabular*}{\textwidth}[l]{l @{\extracolsep{\fill}} c c c c l}
\hline \hline  \\[-1.0ex]
Collab. & Ref. & $\Nf$ & $M_{\pi,\text{min}}$ [MeV] &  $M_{\pi} L$  & Description
\\[1.0ex] \hline \hline \\[-1.0ex]
Hudspith 24 & \cite{Hudspith:2024kzk}   & 2+1 & 127/131 & 3.51/4.05 & \parbox[t]{5.0cm}{At $m = m_\text{symm}$.} \\
[1.0ex] \hline \\[-1.0ex]
Asmussen 23 & \cite{Asmussen:2023pia}    & 2+1 & 127/131 & 3.51/4.05 & \parbox[t]{5.0cm}{At $m = m_\text{symm}$.} \\
[1.0ex] \hline \\[-1.0ex]
RQCD 22 & \cite{RQCD:2022xux}    & 2+1 & 127/131 & 3.51/4.05 & \parbox[t]{5.0cm}{At $m = m_\text{symm}$.} \\
           &                        &     & 200 & 4.14 & \parbox[t]{5.0cm}{At $\tilde m_s = \tilde m_{s,\text{phys}}$} \\
[1.0ex]\hline \\[-1.0ex]
RBC/Bielefeld 07 & \cite{Cheng:2007jq} & 2+1 & 220 &  5.456 & \\
[2.0ex]
\ifx\reducedapptables\undefined
\hline \\[-1.0ex]
CLS 21 & \cite{Strassberger:2021tsu} & 2+1 & 134  & -- & No further detailed information given. \\
[4.0ex] \hline \\[-1.0ex]
CLS 16 & \cite{Bruno:2016plf} & 2+1 & 200 & 4.2 & \parbox[t]{5.0cm}{Extrapolation along line $m_u+m_d+m_s=\mathrm{const.}$} \\
[4.0ex] 
\hline\\[-1.0ex]
QCDSF/UKQCD 15B & \cite{Bornyakov:2015eaa} & 2+1 & 228 & 4.1 & \\
[2.0ex]
\hline\\[-1.0ex]
$\chi$QCD 14& \cite{Yang:2014sea} & 2+1 & 290 & 4.0 & \parbox[t]{5.0cm}{}\\
[2.0ex] \hline \\[-1.0ex]
HotQCD 14 & \cite{Bazavov:2014pvz} & 2+1 & 160 & 4.8 & \parbox[t]{5.0cm}{HISQ staggered quarks, the pion mass quoted is $M_\pi$. Results on 20 ensembles with volumes in the range $[2.6,6.1]$~fm, we quote $M_\pi^{min} L^{max}$.}\\ 
[12.0ex]\hline \\[-1.0ex]
RBC/UKQCD 14  & \cite{Blum:2014tka}  & 2+1 &  $139$ & $3.9$ &  \\
[2.0ex] 
\hline\\[-1.0ex]
BMW 12A &  \cite{Borsanyi:2012zs} & 2+1 &  131 and 120 & 3.92 and 3.0 & \\
[2.0ex]\hline \\[-1.0ex]
HotQCD 11 & \cite{Bazavov:2011nk}    & 2+1 & $160$  & $4.8$ &  \parbox[t]{5.0cm}{HISQ staggered quarks, the pion mass quoted is $M_\pi$. Results on 20 ensembles with volumes in the range $[3.2,6.1]$~fm, we quote $M_\pi^{min}L^{max}$. }\\
[12.0ex]
\hline \\[-1.0ex]
RBC/UKQCD 10A &  \cite{Aoki:2010dy} & 2+1 & 290 & 4.0 & \\
[2.0ex]
\hline \\[-1.0ex]
MILC 10 & \cite{Bazavov:2010hj} & 2+1 & 255 & 4.8 & \parbox[t]{5.0cm}{Determination from global fit, scale from $f_\pi$.}
\\[4.0ex]
\fi
\hline  \hline \\[-1.0ex]
\end{tabular*}
\caption{Chiral extrapolation and finite-volume effects in scale determinations 
  with $\Nf=2+1$ quark flavours. We list the minimum pion mass $M_{\pi,\text{min}}$ and $M_{\pi} L \equiv M_{\pi,\text{min}} [L(M_{\pi,\text{min}})]_\text{max}$ is evaluated at the maximum value of $L$ available at $M_{\pi} =M_{\pi,\text{min}}$.  }
}
\end{table}

\ifx\reducedapptables\undefined
\begin{table}
\addtocounter{table}{-1}
{\footnotesize
\begin{tabular*}{\textwidth}[l]{l @{\extracolsep{\fill}} c c c c l}
\hline \hline  \\[-1.0ex]
Collab. & Ref. & $\Nf$ & $M_{\pi,\text{min}}$ [MeV] &  $M_{\pi} L$  & Description
\\[1.0ex] \hline \hline \\[-1.0ex]
HPQCD 09B&  \cite{Davies:2009tsa} & 2+1 & 211 & 4.0 & Extension of HPQCD 05B  \cite{Gray:2005ur}.\\
[2.0ex]\hline \\[-1.0ex]
MILC 09 & \parbox[t]{.5cm}{\cite{Bazavov:2009bb,Bazavov:2009fk,Bazavov:2009tw}}    & 2+1 & 258  & $4.3$ &  \parbox[t]{5.0cm}{Asqtad staggered quarks, the pion mass quoted is $M_\pi^{RMS}$.}\\
[7.0ex]\hline \\[-1.0ex]
PACS-CS 08 & \cite{Aoki:2008sm} & 2+1 & 156  & $2.3$ &  \parbox[t]{5.0cm}{Clover quarks, several pion masses in the range $[156,702]$~MeV, single lattice spacing.}\\
[4.0ex]\hline \\[-1.0ex]
HPQCD 05B & \cite{Gray:2005ur} & 2+1 & 270 & 3.7 & NRQCD $\Upsilon$ 2s-1s splitting as input.\\
[2.0ex]\hline \\[-1.0ex]
Aubin 04 & \cite{Aubin:2004wf}    & 2+1 & 253  & $3.8$ &  \parbox[t]{5.0cm}{Asqtad staggered quarks, the pion mass quoted is the Goldstone mass.}\\
[4.0ex]
\hline \hline \\[-1.0ex]
\end{tabular*}
\caption{(cntd.) Chiral extrapolation and finite-volume effects in scale determinations 
  with $\Nf=2+1$ quark flavours. We list the minimum pion mass $M_{\pi,\text{min}}$ and $M_{\pi} L \equiv M_{\pi,\text{min}} [L(M_{\pi,\text{min}})]_\text{max}$ is evaluated at the maximum value of $L$ available at $M_{\pi} =M_{\pi,\text{min}}$.  }
}
\end{table}
 \fi

%% file: FLAG.bib
@article{Cheng:2007jq,
    author = "{[RBC/Bielefeld 07] M. Cheng} and others",
    title = "{The QCD equation of state with almost physical quark masses}",
    eprint = "0710.0354",
    archivePrefix = "arXiv",
    primaryClass = "hep-lat",
    reportNumber = "BNL-NT-07-38, BI-TP-2007-20, CU-TP-1181",
    doi = "10.1103/PhysRevD.77.014511",
    journal = "Phys. Rev. D",
    volume = "77",
    pages = "014511",
    year = "2008"
}

@article{Baikov:2014qja,
    author = {Baikov, P. A. and Chetyrkin, K. G. and K\"uhn, J. H.},
    title = "{Quark Mass and Field Anomalous Dimensions to ${\cal O}(\alpha_s^5)$}",
    eprint = "1402.6611",
    archivePrefix = "arXiv",
    primaryClass = "hep-ph",
    reportNumber = "SFB-CPP-14-16, TTP14-007",
    doi = "10.1007/JHEP10(2014)076",
    journal = "JHEP",
    volume = "10",
    pages = "076",
    year = "2014"
}

@article{Boyle:2021kqn,
    author = {Boyle, Peter and Erben, Felix and J\"uttner, Andreas and Kaneko, Takashi and Marshall, Michael and Portelli, Antonin and Witzel, Oliver and Del Debbio, Luigi and Tsang, Justus Tobias},
    title = "{BSM $B - \bar{B}$ mixing on JLQCD and RBC/UKQCD $N_f=2+1$ DWF ensembles}",
    eprint = "2111.11287",
    archivePrefix = "arXiv",
    primaryClass = "hep-lat",
    doi = "10.22323/1.396.0224",
    journal = "PoS",
    volume = "LATTICE2021",
    pages = "224",
    year = "2022"
}

@article{Beneke:2001at,
    author = "Beneke, M. and Feldmann, T. and Seidel, D.",
    title = "{Systematic approach to exclusive $B \to  V l^+ l^-$, $V \gamma$ decays}",
    eprint = "hep-ph/0106067",
    archivePrefix = "arXiv",
    reportNumber = "PITHA-01-05",
    doi = "10.1016/S0550-3213(01)00366-2",
    journal = "Nucl. Phys. B",
    volume = "612",
    pages = "25--58",
    year = "2001"
}

@article{Grinstein:2004vb,
    author = "Grinstein, Benjamin and Pirjol, Dan",
    title = "{Exclusive rare $B \to K^*\ell^+\ell^-$ decays at low recoil: Controlling the long-distance effects}",
    eprint = "hep-ph/0404250",
    archivePrefix = "arXiv",
    reportNumber = "CTP-MIT-3490",
    doi = "10.1103/PhysRevD.70.114005",
    journal = "Phys. Rev. D",
    volume = "70",
    pages = "114005",
    year = "2004"
}

@article{Beylich:2011aq,
    author = "Beylich, M. and Buchalla, G. and Feldmann, T.",
    title = "{Theory of $B \to K^{(*)}\ell^+ \ell^-$ decays at high $q^2$: OPE and quark-hadron duality}",
    eprint = "1101.5118",
    archivePrefix = "arXiv",
    primaryClass = "hep-ph",
    reportNumber = "CERN-PH-TH-2010-305",
    doi = "10.1140/epjc/s10052-011-1635-0",
    journal = "Eur. Phys. J. C",
    volume = "71",
    pages = "1635",
    year = "2011"
}

@article{Korner:1989qb,
    author = "Korner, J. G. and Schuler, G. A.",
    title = "{Exclusive Semileptonic Heavy Meson Decays Including Lepton Mass Effects}",
    reportNumber = "DESY-89-122, MZ-TH-88-14",
    doi = "10.1007/BF02440838",
    journal = "Z. Phys. C",
    volume = "46",
    pages = "93",
    year = "1990"
}

@article{UTfit:2022hsi,
    author = "Bona, Marcella and others",
    collaboration = "UTfit",
    title = "{New UTfit Analysis of the Unitarity Triangle in the Cabibbo-Kobayashi-Maskawa scheme}",
    eprint = "2212.03894",
    archivePrefix = "arXiv",
    primaryClass = "hep-ph",
    reportNumber = "YITP-SB-2022-40",
    doi = "10.1007/s12210-023-01137-5",
    journal = "Rend. Lincei Sci. Fis. Nat.",
    volume = "34",
    pages = "37--57",
    year = "2023"
}

@article{Harrison:2020nrv,
    author = "{[HPQCD 20C] J.~Harrison} and Davies, Christine T. H. and Lytle, Andrew",
    title = "{$R(J/\psi)$ and $B_c^- \rightarrow J/\psi \ell^-\bar{\nu}_\ell$ Lepton Flavor Universality Violating Observables from Lattice QCD}",
    eprint = "2007.06956",
    archivePrefix = "arXiv",
    primaryClass = "hep-lat",
    doi = "10.1103/PhysRevLett.125.222003",
    journal = "Phys. Rev. Lett.",
    volume = "125",
    number = "22",
    pages = "222003",
    year = "2020"
}

@article{Belle-II:2022fsw,
    author = "Abudin\'en, F. and others",
    collaboration = "Belle-II",
    title = "{Reconstruction of $B \to \rho \ell \nu_{\ell}$ decays identified using hadronic decays of the recoil $B$ meson in 2019 -- 2021 Belle II data}",
    eprint = "2211.15270",
    archivePrefix = "arXiv",
    primaryClass = "hep-ex",
    reportNumber = "BELLE2-CONF-PH-2022-029",
    month = "11",
    year = "2022"
}

@article{Leskovec:2022ubd,
    author = "Leskovec, Luka and Meinel, Stefan and Petschlies, Marcus and Negele, John and Paul, Srijit and Pochinsky, Andrew and Rendon, Gumaro",
    title = "{A lattice QCD study of the $B\to\pi\pi\ell\bar{\nu}$ transition}",
    eprint = "2212.08833",
    archivePrefix = "arXiv",
    primaryClass = "hep-lat",
    doi = "10.22323/1.430.0416",
    journal = "PoS",
    volume = "LATTICE2022",
    pages = "416",
    year = "2023"
}

@inproceedings{Leskovec:2024sfx,
    author = "Leskovec, Luka and Meinel, Stefan and Petschlies, Marcus and Negele, John and Paul, Srijit and Pochinsky, Andrew and Rendon, Gumaro",
    title = "{Lattice outlook on $B\to\rho\ell\bar{\nu}$ and $B\to K^\star \ell \ell$}",
    booktitle = "{12th International Workshop on the CKM Unitarity Triangle}",
    eprint = "2403.19543",
    archivePrefix = "arXiv",
    primaryClass = "hep-lat",
    month = "3",
    year = "2024"
}

@article{Bowler:2004zb,
    author = "{[UKQCD 04] K.~C.~Bowler} and Gill, J. F. and Maynard, C. M. and Flynn, J. M.",
    title = "{$B\to\rho\ell\nu$ form-factors in lattice QCD}",
    eprint = "hep-lat/0402023",
    archivePrefix = "arXiv",
    reportNumber = "EDINBURGH-2004-04, SHEP-0404",
    doi = "10.1088/1126-6708/2004/05/035",
    journal = "JHEP",
    volume = "05",
    pages = "035",
    year = "2004"
}

@article{Flynn:2008zr,
    author = "Flynn, Jonathan M. and Nakagawa, Yoshiyuki and Nieves, Juan and Toki, Hiroshi",
    title = "{$|V_{ub}|$ from Exclusive Semileptonic $B \to \rho$ Decays}",
    eprint = "0812.2795",
    archivePrefix = "arXiv",
    primaryClass = "hep-ph",
    doi = "10.1016/j.physletb.2009.04.037",
    journal = "Phys. Lett. B",
    volume = "675",
    pages = "326--331",
    year = "2009"
}

@article{Jeong:2024glv,
    author = "Jeong, Hwancheol and DeTar, Carleton and El-Khadra, Aida X. and G\'amiz, Elvira and Gelzer, Zechariah and Gottlieb, Steven and Jay, William and Kronfeld, Andreas and Lytle, Andrew and Vaquero, Alejandro",
    title = "{Form factors for semileptonic B-decays with HISQ light quarks and clover b-quarks in Fermilab interpretation}",
    eprint = "2402.14924",
    archivePrefix = "arXiv",
    primaryClass = "hep-lat",
    reportNumber = "FERMILAB-CONF-24-0077-T",
    doi = "10.22323/1.453.0253",
    journal = "PoS",
    volume = "LATTICE2023",
    pages = "253",
    year = "2024"
}

@article{Mohanta:2024hzi,
    author = "{[JLQCD 24] P.~Mohanta} and Kaneko, Takashi and Hashimoto, Shoji",
    title = "{$B_s \to K\ell\nu$ form factors from lattice QCD with domain-wall heavy quarks.}",
    eprint = "2401.01570",
    archivePrefix = "arXiv",
    primaryClass = "hep-lat",
    doi = "10.22323/1.453.0267",
    journal = "PoS",
    volume = "LATTICE2023",
    pages = "267",
    year = "2024"
}

@article{Flynn:2021ttz,
    author = "Flynn, Jonathan and Hill, Ryan and Juettner, Andreas and Soni, Amarjit and Tsang, Justus Tobias and Witzel, Oliver",
    title = "{Form factors for semileptonic $B\to \pi\ , B_s\to K$ and $B_s\to D_s$ decays}",
    eprint = "2112.10580",
    archivePrefix = "arXiv",
    primaryClass = "hep-lat",
    reportNumber = "Siegen SI-HEP-2021-36",
    doi = "10.22323/1.396.0306",
    journal = "PoS",
    volume = "LATTICE2021",
    pages = "306",
    year = "2022"
}

@article{Marshall:2022xbz,
    author = {Marshall, Michael and Boyle, Peter and Del Debbio, Luigi and Erben, Felix and Flynn, Jonathan and J\"uttner, Andreas and Portelli, Antonin and Tsang, Justus Tobias and Witzel, Oliver},
    title = "{Semileptonic $D \rightarrow \pi \ell \nu$, $D \rightarrow K \ell \nu$ and $D_s \rightarrow K \ell \nu$ decays with 2+1f domain wall fermions}",
    eprint = "2201.02680",
    archivePrefix = "arXiv",
    primaryClass = "hep-lat",
    doi = "10.22323/1.396.0416",
    journal = "PoS",
    volume = "LATTICE2021",
    pages = "416",
    year = "2022"
}

@article{Belle-II:2020dyp,
    author = "Abudin\'en, F. and others",
    collaboration = "Belle-II",
    title = "{Measurement of the semileptonic $\bar{B}^0 \to D^{*+} \ell^{-} \nu_{\ell}$ branching fraction with fully reconstructed $B$ meson decays and 34.6 fb$^{-1}$of Belle II data}",
    eprint = "2008.10299",
    archivePrefix = "arXiv",
    primaryClass = "hep-ex",
    reportNumber = "BELLE2-CONF-PH-2020-009",
    month = "8",
    year = "2020"
}

@article{BESIII:2018xre,
    author = "Ablikim, M. and others",
    collaboration = "BESIII",
    title = "{First Measurement of the Form Factors in $D^+_{s}\rightarrow K^0 e^+\nu_e$ and $D^+_{s}\rightarrow K^{*0} e^+\nu_e$ Decays}",
    eprint = "1811.02911",
    archivePrefix = "arXiv",
    primaryClass = "hep-ex",
    doi = "10.1103/PhysRevLett.122.061801",
    journal = "Phys. Rev. Lett.",
    volume = "122",
    number = "6",
    pages = "061801",
    year = "2019"
}

@article{Belle-II:2022ggx,
    author = "Abudin\'en, F. and others",
    collaboration = "Belle-II",
    title = "{Measurement of the $\Lambda_c^+$ Lifetime}",
    eprint = "2206.15227",
    archivePrefix = "arXiv",
    primaryClass = "hep-ex",
    reportNumber = "Belle II Preprint 2022-003, KEK Preprint 2022-13",
    doi = "10.1103/PhysRevLett.130.071802",
    journal = "Phys. Rev. Lett.",
    volume = "130",
    number = "7",
    pages = "071802",
    year = "2023"
}

@article{BESIII:2022ysa,
    author = "Ablikim, M. and others",
    collaboration = "BESIII",
    title = "{Study of the Semileptonic Decay $\Lambda_c^+\to\Lambda e^+\nu e$}",
    eprint = "2207.14149",
    archivePrefix = "arXiv",
    primaryClass = "hep-ex",
    doi = "10.1103/PhysRevLett.129.231803",
    journal = "Phys. Rev. Lett.",
    volume = "129",
    number = "23",
    pages = "231803",
    year = "2022"
}

@article{BESIII:2023jxv,
    author = "Ablikim, M. and others",
    collaboration = "BESIII",
    title = "{Study of $\Lambda_c^+ \rightarrow \Lambda \mu^+\nu_{\mu}$ and test of lepton flavor universality with $\Lambda_c^+ \rightarrow \Lambda \ell^+\nu_{\ell}$ decays}",
    eprint = "2306.02624",
    archivePrefix = "arXiv",
    primaryClass = "hep-ex",
    doi = "10.1103/PhysRevD.108.L031105",
    journal = "Phys. Rev. D",
    volume = "108",
    number = "3",
    pages = "L031105",
    year = "2023"
}

@article{FermilabLattice:2022gku,
    author = "{[FNAL/MILC 22] A. Bazavov} and others",
    title = "{D-meson semileptonic decays to pseudoscalars from four-flavor lattice QCD}",
    eprint = "2212.12648",
    archivePrefix = "arXiv",
    primaryClass = "hep-lat",
    reportNumber = "MIT-CTP/5513, FERMILAB-PUB-22-943-T",
    doi = "10.1103/PhysRevD.107.094516",
    journal = "Phys. Rev. D",
    volume = "107",
    number = "9",
    pages = "094516",
    year = "2023"
}

@article{Parrott:2022zte,
    author = "{[HPQCD 22A] W.~G.~Parrott} and Bouchard, C. and Davies, C. T. H.",
    title = "{Standard Model predictions for $B\to K\ell^+\ell^-$, $B\to K\ell_1^-\ell_2^+$ and $B\to K\nu\bar{\nu}$ using form factors from Nf=2+1+1 lattice QCD}",
    eprint = "2207.13371",
    archivePrefix = "arXiv",
    primaryClass = "hep-ph",
    doi = "10.1103/PhysRevD.107.014511",
    journal = "Phys. Rev. D",
    volume = "107",
    number = "1",
    pages = "014511",
    year = "2023",
    note = "[Erratum: Phys.Rev.D 107, 119903 (2023)]"
}

@article{Parrott:2022rgu,
    author = "{[HPQCD 22] W.~G.~Parrott} and Bouchard, C. and Davies, C. T. H.",
    title = "{$B\to K$ and $D\to K$ form factors from fully relativistic lattice QCD}",
    eprint = "2207.12468",
    archivePrefix = "arXiv",
    primaryClass = "hep-lat",
    doi = "10.1103/PhysRevD.107.014510",
    journal = "Phys. Rev. D",
    volume = "107",
    number = "1",
    pages = "014510",
    year = "2023"
}

@article{Meinel:2023wyg,
    author = "Meinel, Stefan",
    title = "{Status of next-generation $\Lambda_b \to p, \Lambda, \Lambda_c$ form-factor calculations}",
    eprint = "2309.01821",
    archivePrefix = "arXiv",
    primaryClass = "hep-lat",
    doi = "10.22323/1.453.0275",
    journal = "PoS",
    volume = "LATTICE2023",
    pages = "275",
    year = "2024"
}

@article{Farrell:2023vnm,
    author = "Farrell, Callum and Meinel, Stefan",
    title = "{Form factors for the charm-baryon semileptonic decay $\Xi_c\to \Xi \ell \nu$ from domain-wall lattice QCD}",
    eprint = "2309.08107",
    archivePrefix = "arXiv",
    primaryClass = "hep-lat",
    doi = "10.22323/1.453.0210",
    journal = "PoS",
    volume = "LATTICE2023",
    pages = "210",
    year = "2024"
}

@article{DiRisi:2023npw,
    author = "Di Risi, Vigilante and Iacobacci, Davide and Sannino, Francesco",
    title = "{$\Lambda_b\to\Lambda_c^*$ at $1/m_c^2$ heavy quark mass order}",
    eprint = "2309.03553",
    archivePrefix = "arXiv",
    primaryClass = "hep-ph",
    doi = "10.1103/PhysRevD.109.036021",
    journal = "Phys. Rev. D",
    volume = "109",
    number = "3",
    pages = "036021",
    year = "2024"
}

@article{Papucci:2021pmj,
    author = "Papucci, Michele and Robinson, Dean J.",
    title = "{Form factor counting and HQET matching for new physics in $\Lambda_b\to\Lambda_c^*l\nu$}",
    eprint = "2105.09330",
    archivePrefix = "arXiv",
    primaryClass = "hep-ph",
    reportNumber = "CALT-2021-020",
    doi = "10.1103/PhysRevD.105.016027",
    journal = "Phys. Rev. D",
    volume = "105",
    number = "1",
    pages = "016027",
    year = "2022"
}

@article{Meinel:2021mdj,
    author = "{S. Meinel} and Rendon, Gumaro",
    title = "{$\Lambda_c \to \Lambda^*(1520)$ form factors from lattice QCD and improved analysis of the $\Lambda_b \to \Lambda^*(1520)$ and $\Lambda_b \to \Lambda_c^*(2595,2625)$ form factors}",
    eprint = "2107.13140",
    archivePrefix = "arXiv",
    primaryClass = "hep-lat",
    doi = "10.1103/PhysRevD.105.054511",
    journal = "Phys. Rev. D",
    volume = "105",
    number = "5",
    pages = "054511",
    year = "2022"
}

@article{ATLAS:2018cur,
    author = "Aaboud, Morad and others",
    collaboration = "ATLAS",
    title = "{Study of the rare decays of $B^0_s$ and $B^0$ mesons into muon pairs using data collected during 2015 and 2016 with the ATLAS detector}",
    eprint = "1812.03017",
    archivePrefix = "arXiv",
    primaryClass = "hep-ex",
    reportNumber = "CERN-EP-2018-291",
    doi = "10.1007/JHEP04(2019)098",
    journal = "JHEP",
    volume = "04",
    pages = "098",
    year = "2019"
}

@article{CMS:2019bbr,
    author = "Sirunyan, Albert M and others",
    collaboration = "CMS",
    title = "{Measurement of properties of B$^0_\mathrm{s}\to\mu^+\mu^-$ decays and search for B$^0\to\mu^+\mu^-$ with the CMS experiment}",
    eprint = "1910.12127",
    archivePrefix = "arXiv",
    primaryClass = "hep-ex",
    reportNumber = "CMS-BPH-16-004, CERN-EP-2019-215",
    doi = "10.1007/JHEP04(2020)188",
    journal = "JHEP",
    volume = "04",
    pages = "188",
    year = "2020"
}

@article{LHCb:2021awg,
    author = "Aaij, Roel and others",
    collaboration = "LHCb",
    title = "{Measurement of the $B^0_s\to\mu^+\mu^-$ decay properties and search for the $B^0\to\mu^+\mu^-$ and $B^0_s\to\mu^+\mu^-\gamma$ decays}",
    eprint = "2108.09283",
    archivePrefix = "arXiv",
    primaryClass = "hep-ex",
    reportNumber = "CERN-EP-2021-133, LHCb-PAPER-2021-008",
    doi = "10.1103/PhysRevD.105.012010",
    journal = "Phys. Rev. D",
    volume = "105",
    number = "1",
    pages = "012010",
    year = "2022"
}

@article{LHCb:2021vsc,
    author = "Aaij, R. and others",
    collaboration = "LHCb",
    title = "{Analysis of Neutral B-Meson Decays into Two Muons}",
    eprint = "2108.09284",
    archivePrefix = "arXiv",
    primaryClass = "hep-ex",
    reportNumber = "CERN-CERN-EP-2021-132, LHCb-PAPER-2021-007",
    doi = "10.1103/PhysRevLett.128.041801",
    journal = "Phys. Rev. Lett.",
    volume = "128",
    number = "4",
    pages = "041801",
    year = "2022"
}

@article{Janowski:2021yvz,
    author = "Janowski, Tadeusz and Pullin, Ben and Zwicky, Roman",
    title = "{Charged and neutral $ {\overline{B}}_{u,d,s} $ $\to \gamma$ form factors from light cone sum rules at NLO}",
    eprint = "2106.13616",
    archivePrefix = "arXiv",
    primaryClass = "hep-ph",
    reportNumber = "CP3-Origins-2020-14 DNRF90",
    doi = "10.1007/JHEP12(2021)008",
    journal = "JHEP",
    volume = "12",
    pages = "008",
    year = "2021"
}

@article{Kozachuk:2017mdk,
    author = "Kozachuk, Anastasiia and Melikhov, Dmitri and Nikitin, Nikolai",
    title = "{Rare FCNC radiative leptonic $B_{s,d}\to \gamma l^+l^-$ decays in the standard model}",
    eprint = "1712.07926",
    archivePrefix = "arXiv",
    primaryClass = "hep-ph",
    doi = "10.1103/PhysRevD.97.053007",
    journal = "Phys. Rev. D",
    volume = "97",
    number = "5",
    pages = "053007",
    year = "2018"
}

@article{Belov:2024vkv,
    author = "Belov, Ilia and Berezhnoy, Alexander and Melikhov, Dmitri",
    title = "{Nonfactorizable charming-loop contribution to FCNC $B_s\to \gamma l^+l^-$ decay}",
   eprint = "2404.01222",
    archivePrefix = "arXiv",
    primaryClass = "hep-ph",
    doi = "10.1103/PhysRevD.109.114012",
    journal = "Phys. Rev. D",
    volume = "109",
    number = "11",
    pages = "114012",
    year = "2024"
}

@article{Guadagnoli:2023zym,
    author = "Guadagnoli, Diego and Normand, Camille and Simula, Silvano and Vittorio, Ludovico",
    title = "{From $D_{s}\rightarrow\gamma$ in lattice QCD to $B_{s}\rightarrow\mu\mu\gamma$ at high $q^{2}$}",
    eprint = "2303.02174",
    archivePrefix = "arXiv",
    primaryClass = "hep-ph",
    reportNumber = "LAPTH-006/23",
    doi = "10.1007/JHEP07(2023)112",
    journal = "JHEP",
    volume = "07",
    pages = "112",
    year = "2023"
}

@article{Frezzotti:2024kqk,
    author = "{R. Frezzotti} and Gagliardi, G. and Lubicz, V. and Martinelli, G. and Sachrajda, C. T. and Sanfilippo, F. and Simula, S. and Tantalo, N.",
    title = "{The $B_{s}\to \mu^{+}\mu^{-}\gamma$ decay rate at large $q^{2}$ from lattice QCD}",
    eprint = "2402.03262",
    archivePrefix = "arXiv",
    primaryClass = "hep-lat",
    doi = "10.1103/PhysRevD.109.114506",
    journal = "Phys. Rev. D",
    volume = "109",
    number = "11",
    pages = "114506",
    year = "2024"
}

@article{Hollitt:2022exk,
    author = "{[QCDSF/UKQCD/CSSM 22] S. Hollitt} and others",
    title = "{Measurements of $SU(3)_f$ symmetry breaking in B meson decay constants}",
    eprint = "2201.10779",
    archivePrefix = "arXiv",
    primaryClass = "hep-lat",
    reportNumber = "ADP-22-3/T1174, DESY-22-011, LTH 1293",
    doi = "10.22323/1.396.0549",
    journal = "PoS",
    volume = "LATTICE2021",
    pages = "549",
    year = "2022"
}

@article{Black:2022eph,
    author = "{[RBC/UKQCD 22] M.~Black} and Witzel, Oliver",
    title = "{B Meson Decay Constants Using Relativistic Heavy Quarks}",
    eprint = "2212.10125",
    archivePrefix = "arXiv",
    primaryClass = "hep-lat",
    reportNumber = "SI-HEP-2022-37, P3H-22-125",
    doi = "10.22323/1.430.0405",
    journal = "PoS",
    volume = "LATTICE2022",
    pages = "405",
    year = "2023"
}

@article{Akaike:1974vps,
    author = "Akaike, H.",
    title = "{A new look at the statistical model identification}",
    doi = "10.1109/TAC.1974.1100705",
    journal = "IEEE Trans. Automatic Control",
    volume = "19",
    number = "6",
    pages = "716--723",
    year = "1974"
}

@article{Parisi:1983ae,
    author = "Parisi, G.",
    editor = "Itzykson, C. and Pomeau, Y. and Sourlas, N.",
    title = "{The Strategy for Computing the Hadronic Mass Spectrum}",
    reportNumber = "LNF-83-36-P",
    doi = "10.1016/0370-1573(84)90081-4",
    journal = "Phys. Rept.",
    volume = "103",
    pages = "203--211",
    year = "1984"
}

@inproceedings{Lepage:1989hd,
    author = "Lepage, G. Peter",
    title = "{The Analysis of Algorithms for Lattice Field Theory}",
    booktitle = "{Theoretical Advanced Study Institute in Elementary Particle Physics}",
    reportNumber = "CLNS-89-971",
    month = "6",
    year = "1989"
}

@article{Frezzotti:2023ygt,
    author = "Frezzotti, R. and Tantalo, N. and Gagliardi, G. and Sanfilippo, F. and Simula, S. and Lubicz, V. and Mazzetti, F. and Martinelli, G. and Sachrajda, C. T.",
    title = "{Lattice calculation of the Ds meson radiative form factors over the full kinematical range}",
    eprint = "2306.05904",
    archivePrefix = "arXiv",
    primaryClass = "hep-lat",
    doi = "10.1103/PhysRevD.108.074505",
    journal = "Phys. Rev. D",
    volume = "108",
    number = "7",
    pages = "074505",
    year = "2023"
}

@article{Giusti:2023pot,
    author = "Giusti, Davide and Kane, Christopher F. and Lehner, Christoph and Meinel, Stefan and Soni, Amarjit",
    title = "{Methods for high-precision determinations of radiative-leptonic decay form factors using lattice QCD}",
    eprint = "2302.01298",
    archivePrefix = "arXiv",
    primaryClass = "hep-lat",
    doi = "10.1103/PhysRevD.107.074507",
    journal = "Phys. Rev. D",
    volume = "107",
    number = "7",
    pages = "074507",
    year = "2023"
}

@article{Dimopoulos:2021qsf,
    author = "{[ETM 21B] P.~Dimopoulos} and Frezzotti, Roberto and Garofalo, Marco and Simula, Silvano",
    title = "{$K$- and $D_{(s)}$-meson leptonic decay constants with physical light, strange and charm quarks by ETMC}",
    eprint = "2110.01294",
    archivePrefix = "arXiv",
    primaryClass = "hep-lat",
    doi = "10.22323/1.396.0472",
    journal = "PoS",
    volume = "LATTICE2021",
    pages = "472",
    year = "2021"
}

@article{Bali:2021qem,
    author = {{[RQCD 21] G.~S.~Bali} and Braun, Vladimir and Collins, Sara and Sch\"afer, Andreas and Simeth, Jakob},
    title = "{Masses and decay constants of the \ensuremath{\eta} and \ensuremath{\eta}' mesons from lattice QCD}",
    eprint = "2106.05398",
    archivePrefix = "arXiv",
    primaryClass = "hep-lat",
    doi = "10.1007/JHEP08(2021)137",
    journal = "JHEP",
    volume = "08",
    pages = "137",
    year = "2021"
}

@article{Kuberski:2024pms,
    author = {{[RQCD/ALPHA 24] S. Kuberski} and Joswig, Fabian and Collins, Sara and Heitger, Jochen and S\"oldner, Wolfgang},
    title = "{$\mathrm{D}$ and $\mathrm{D_s}$ decay constants in $N_{\rm f}=2+1$ QCD with Wilson fermions}",
    eprint = "2405.04506",
    archivePrefix = "arXiv",
    primaryClass = "hep-lat",
    reportNumber = "CERN-TH-2024-052, MITP-24-047, MS-TP-24-09",
    doi = "10.1007/JHEP07(2024)090",
    journal = "JHEP",
    volume = "07",
    pages = "090",
    year = "2024"
}

@article{Bussone:2023kag,
    author = "{[ALPHA 23] A. Bussone} and Conigli, Alessandro and Frison, Julien and Herdo\'\i{}za, Gregorio and Pena, Carlos and Preti, David and S\'aez, Alejandro and Ugarrio, Javier",
    title = "{Hadronic physics from a Wilson fermion mixed-action approach: Charm quark mass and $D_{(s)}$ meson decay constants}",
    eprint = "2309.14154",
    archivePrefix = "arXiv",
    primaryClass = "hep-lat",
    reportNumber = "IFT-UAM/CSIC-23-114",
    doi = "10.1140/epjc/s10052-024-12816-4",
    journal = "Eur. Phys. J. C",
    volume = "84",
    number = "5",
    pages = "506",
    year = "2024"
}

@article{ParticleDataGroup:2022pth,
    author = "Workman, R. L. and others",
    collaboration = "Particle Data Group",
    title = "{Review of Particle Physics}",
    doi = "10.1093/ptep/ptac097",
    journal = "PTEP",
    volume = "2022",
    pages = "083C01",
    year = "2022"
}

@article{Bordone:2021oof,
    author = "Bordone, Marzia and Capdevila, Bernat and Gambino, Paolo",
    title = "{Three loop calculations and inclusive $V_{cb}$}",
    eprint = "2107.00604",
    archivePrefix = "arXiv",
    primaryClass = "hep-ph",
    doi = "10.1016/j.physletb.2021.136679",
    journal = "Phys. Lett. B",
    volume = "822",
    pages = "136679",
    year = "2021"
}

@article{Belle-II:2023okj,
    author = "Adachi, I. and others",
    collaboration = "Belle-II",
    title = "{Determination of $|V_{cb}|$ using $\bar B^0\to D^{*,+}\ell^-\bar\nu_\ell$ decays with Belle II}",
    eprint = "2310.01170",
    archivePrefix = "arXiv",
    primaryClass = "hep-ex",
    reportNumber = "Belle II Preprint 2023-014, KEK Preprint 2023-28",
    doi = "10.1103/PhysRevD.108.092013",
    journal = "Phys. Rev. D",
    volume = "108",
    number = "9",
    pages = "092013",
    year = "2023"
}

@article{Belle:2023bwv,
    author = "Prim, M. T. and others",
    collaboration = "Belle",
    title = "{Measurement of differential distributions of $B\to D^\ast\ell\bar\nu_\ell$ and implications on $|V_{cb}|$}",
    eprint = "2301.07529",
    archivePrefix = "arXiv",
    primaryClass = "hep-ex",
    reportNumber = "Belle Preprint 2022-34; KEK Preprint 2022-47",
    doi = "10.1103/PhysRevD.108.012002",
    journal = "Phys. Rev. D",
    volume = "108",
    number = "1",
    pages = "012002",
    year = "2023"
}

@article{Aoki:2023qpa,
    author = "{[JLQCD 23] Y.~Aoki} and Colquhoun, B. and Fukaya, H. and Hashimoto, S. and Kaneko, T. and Kellermann, R. and Koponen, J. and Kou, E.",
    title = {{$B\to D^*\ell \nu_\ell$ semileptonic form factors from lattice QCD with M\"obius domain-wall quarks}},
    eprint = "2306.05657",
    archivePrefix = "arXiv",
    primaryClass = "hep-lat",
    reportNumber = "KEK-CP-393, OU-HET-1186",
    doi = "10.1103/PhysRevD.109.074503",
    journal = "Phys. Rev. D",
    volume = "109",
    number = "7",
    pages = "074503",
    year = "2024"
}

@article{Harrison:2023dzh,
    author = "{[HPQCD 23] J.~Harrison} and Davies, Christine T. H.",
    title = "{$B\to D^*$ and $B_s\to D_s^*$ vector, axial-vector and tensor form factors for the full $q^2$ range from lattice QCD}",
    eprint = "2304.03137",
    archivePrefix = "arXiv",
    primaryClass = "hep-lat",
    doi = "10.1103/PhysRevD.109.094515",
    journal = "Phys. Rev. D",
    volume = "109",
    number = "9",
    pages = "094515",
    year = "2024"
}

@article{Lytle:2024zfr,
    author = "Lytle, Andrew and DeTar, Carleton and G\'amiz, Elvira and Gottlieb, Steven and Jay, William and El-Khadra, Aida X. and Kronfeld, Andreas and Laiho, Jack and Simone, James N. and Vaquero, Alejandro",
    title = "{B-meson semileptonic decays from highly improved staggered quarks}",
    eprint = "2403.03959",
    archivePrefix = "arXiv",
    primaryClass = "hep-lat",
    reportNumber = "FERMILAB-CONF-24-0032-T, MIT-CTP/5674",
    doi = "10.22323/1.453.0240",
    journal = "PoS",
    volume = "LATTICE2023",
    pages = "240",
    year = "2024"
}

@article{Kaneko:2021tlw,
    author = "Kaneko, Takashi and Aoki, Y. and Colquhoun, B. and Faur, M. and Fukaya, H. and Hashimoto, S. and Koponen, J. and Kou, E.",
    title = "{$B \to D^{(*)}\ell\nu$ semileptonic decays in lattice QCD with domain-wall heavy quarks}",
    eprint = "2112.13775",
    archivePrefix = "arXiv",
    primaryClass = "hep-lat",
    reportNumber = "KEK-CP-387",
    doi = "10.22323/1.396.0561",
    journal = "PoS",
    volume = "LATTICE2021",
    pages = "561",
    year = "2022"
}

@article{Colquhoun:2022atw,
    author = "{[JLQCD 22] B.~Colquhoun} and Hashimoto, Shoji and Kaneko, Takashi and Koponen, Jonna",
    title = {{Form factors of $B\to \pi\ell\nu$ and a determination of $|V_{ub}|$ with M\"obius domain-wall fermions}},
    eprint = "2203.04938",
    archivePrefix = "arXiv",
    primaryClass = "hep-lat",
    reportNumber = "KEK-CP-382",
    doi = "10.1103/PhysRevD.106.054502",
    journal = "Phys. Rev. D",
    volume = "106",
    number = "5",
    pages = "054502",
    year = "2022"
}

@article{Bruno:2019xed,
    author = "{[ALPHA 19B] M.~Bruno} and Campos, I. and Koponen, J. and Pena, Carlos and Preti, David and Ramos, Alberto and Vladikas, Anastassios",
    title = "{Light and strange quark masses from $N_f=2+1$ simulations with Wilson fermions}",
    eprint = "1903.04094",
    archivePrefix = "arXiv",
    primaryClass = "hep-lat",
    doi = "10.22323/1.334.0220",
    journal = "PoS",
    volume = "LATTICE2018",
    pages = "220",
    year = "2019"
}

@article{Bruno:2019vup,
    author = "{[ALPHA 19] M.~Bruno} and Campos, Isabel and Fritzsch, Patrick and Koponen, Jonna and Pena, Carlos and Preti, David and Ramos, Alberto and Vladikas, Anastassios",
    title = "{Light quark masses in ${N_\mathrm{f}=2+1}$ lattice QCD with Wilson fermions}",
    eprint = "1911.08025",
    archivePrefix = "arXiv",
    primaryClass = "hep-lat",
    reportNumber = "CERN-TH-2019-174, IFT-UAM/CSIC-19-151, KEK-CP-372",
    doi = "10.1140/epjc/s10052-020-7698-z",
    journal = "Eur. Phys. J. C",
    volume = "80",
    number = "2",
    pages = "169",
    year = "2020"
}

@article{Hatton:2020qhk,
    author = "{[HPQCD 20A] D.~Hatton} and Davies, C. T. H. and Galloway, B. and Koponen, J. and Lepage, G. P. and Lytle, A. T.",
    title = "{Charmonium properties from lattice $QCD$+QED : Hyperfine splitting, $J/\psi$ leptonic width, charm quark mass, and $a^c_\mu$}",
    eprint = "2005.01845",
    archivePrefix = "arXiv",
    primaryClass = "hep-lat",
    doi = "10.1103/PhysRevD.102.054511",
    journal = "Phys. Rev. D",
    volume = "102",
    number = "5",
    pages = "054511",
    year = "2020"
}

@article{Hatton:2020vzp,
    author = "{[HPQCD 20D] D.~Hatton} and Davies, C. T. H. and Lepage, G. P. and Lytle, A. T.",
    title = "{Renormalization of the tensor current in lattice QCD and the $J/\psi$ tensor decay constant}",
    eprint = "2008.02024",
    archivePrefix = "arXiv",
    primaryClass = "hep-lat",
    doi = "10.1103/PhysRevD.102.094509",
    journal = "Phys. Rev. D",
    volume = "102",
    number = "9",
    pages = "094509",
    year = "2020"
}

@article{Hatton:2021syc,
    author = "{[HPQCD 21] D.~Hatton} and Davies, C. T. H. and Koponen, J. and Lepage, G. P. and Lytle, A. T.",
    title = "{Determination of $\overline{m}_b/\overline{m}_c$ and $\overline{m}_b$ from $n_f=4$ lattice QCD$+$QED}",
    eprint = "2102.09609",
    archivePrefix = "arXiv",
    primaryClass = "hep-lat",
    doi = "10.1103/PhysRevD.103.114508",
    journal = "Phys. Rev. D",
    volume = "103",
    number = "11",
    pages = "114508",
    year = "2021"
}

@article{Heitger:2021apz,
    author = "{[ALPHA 21] J.~Heitger} and Joswig, Fabian and Kuberski, Simon",
    title = "{Determination of the charm quark mass in lattice QCD with $2+1$ flavours on fine lattices}",
    eprint = "2101.02694",
    archivePrefix = "arXiv",
    primaryClass = "hep-lat",
    reportNumber = "MS-TP-21-01",
    doi = "10.1007/JHEP05(2021)288",
    journal = "JHEP",
    volume = "05",
    pages = "288",
    year = "2021"
}

@ARTICLE{1100705,
author={H. Akaike},
journal={IEEE Transactions on Automatic Control},
title={A new look at the statistical model identification},
year={1974},
volume={19},
number={6},
pages={716-723},
doi={10.1109/TAC.1974.1100705},
ISSN={0018-9286},
month={Dec},}

@article{Aaij:2012nna,
      author         = "Aaij, R and others",
      title          = "{First evidence for the decay $B_s \to  \mu^+ \mu^-$}",
      collaboration  = "LHCb",
      journal        = "Phys.Rev.Lett.",
      volume         = "110",
      pages          = "021801",
      year           = "2013",
      eprint         = "1211.2674",
      archivePrefix  = "arXiv",
      primaryClass   = "Unknown",
      reportNumber   = "CERN-PH-EP-2012-335, LHCB-PAPER-2012-043",
      SLACcitation   = "%%CITATION = ARXIV:1211.2674;%%",
}

@article{LHCbRICHGroup:2012mgd,
    author = "Adinolfi, M. and others",
    collaboration = "LHCb RICH Group",
    title = "{Performance of the LHCb RICH detector at the LHC}",
    eprint = "1211.6759",
    archivePrefix = "arXiv",
    primaryClass = "physics.ins-det",
    reportNumber = "CERN-LHCB-DP-2012-003, LHCB-DP-2012-003",
    doi = "10.1140/epjc/s10052-013-2431-9",
    journal = "Eur. Phys. J. C",
    volume = "73",
    pages = "2431",
    year = "2013"
}

@article{Aaij:2015bfa,
      author         = "Aaij, Roel and others",
      title          = "{Determination of the quark coupling strength $|V_{ub}|$
                        using baryonic decays}",
      collaboration  = "LHCb",
      journal        = "Nature Phys.",
      volume         = "11",
      year           = "2015",
      pages          = "743-747",
      doi            = "10.1038/nphys3415",
      eprint         = "1504.01568",
      archivePrefix  = "arXiv",
      primaryClass   = "hep-ex",
      reportNumber   = "CERN-PH-EP-2015-084, LHCB-PAPER-2015-013",
      SLACcitation   = "%%CITATION = ARXIV:1504.01568;%%"
}

@article{Aaij:2015nea,
      author         = "Aaij, Roel and others",
      title          = "{First measurement of the differential branching fraction
                        and $C\!P$ asymmetry of the $B^\pm\to\pi^\pm\mu^+\mu^-$
                        decay}",
      collaboration  = "LHCb",
      journal        = "JHEP",
      volume         = "10",
      year           = "2015",
      pages          = "034",
      doi            = "10.1007/JHEP10(2015)034",
      eprint         = "1509.00414",
      archivePrefix  = "arXiv",
      primaryClass   = "hep-ex",
      reportNumber   = "LHCB-PAPER-2015-035, CERN-PH-EP-2015-219",
      SLACcitation   = "%%CITATION = ARXIV:1509.00414;%%"
}

@article{Aaij:2017vad,
      author         = "Aaij, Roel and others",
      title          = "{Measurement of the $B^0_s\to\mu^+\mu^-$ branching
                        fraction and effective lifetime and search for
                        $B^0\to\mu^+\mu^-$ decays}",
      collaboration  = "LHCb",
      journal        = "Phys. Rev. Lett.",
      volume         = "118",
      year           = "2017",
      number         = "19",
      pages          = "191801",
      doi            = "10.1103/PhysRevLett.118.191801",
      eprint         = "1703.05747",
      archivePrefix  = "arXiv",
      primaryClass   = "hep-ex",
      reportNumber   = "CERN-EP-2017-041, LHCB-PAPER-2017-001",
      SLACcitation   = "%%CITATION = ARXIV:1703.05747;%%"
}

@article{Abdel-Rehim:2015owa,
      author         = "{[ETM 15D] A. Abdel-Rehim} and others",
      title          = "{Nucleon and pion structure with lattice QCD simulations
                        at physical value of the pion mass}",
      journal        = "Phys. Rev.",
      volume         = "D92",
      year           = "2015",
      number         = "11",
      pages          = "114513",
      doi            = "10.1103/PhysRevD.92.114513, 10.1103/PhysRevD.93.039904",
      note           = "[Erratum: Phys. Rev.D93,no.3,039904(2016)]",
      eprint         = "1507.04936",
      archivePrefix  = "arXiv",
      primaryClass   = "hep-lat",
      SLACcitation   = "%%CITATION = ARXIV:1507.04936;%%"
}

@article{Abdel-Rehim:2015pwa,
      author         = "{[ETM 15A] A. Abdel-Rehim} and others",
      title          = "{Simulating QCD at the physical point with $N_f=2$ Wilson
                        twisted mass fermions at maximal twist}",
      year           = "2015",
      journal        = "Phys. Rev.",
      volume         = "D95",
      number         = "9",
      pages          = "094515",
      doi            = "10.1103/PhysRevD.95.094515",
      eprint         = "1507.05068",
      archivePrefix  = "arXiv",
      primaryClass   = "hep-lat",
      reportNumber   = "DESY-15-121",
      SLACcitation   = "%%CITATION = ARXIV:1507.05068;%%"
}

@article{Abdel-Rehim:2016won,
      author         = "{[ETM 16A] A. Abdel-Rehim} and Alexandrou, C. and Constantinou, M.
                        and Hadjiyiannakou, K. and Jansen, K. and Kallidonis, Ch.
                        and Koutsou, G. and Vaquero Aviles-Casco, A.",
      title          = "{Direct Evaluation of the Quark Content of Nucleons from
                        Lattice QCD at the Physical Point}",
      journal        = "Phys. Rev. Lett.",
      volume         = "116",
      year           = "2016",
      number         = "25",
      pages          = "252001",
      doi            = "10.1103/PhysRevLett.116.252001",
      eprint         = "1601.01624",
      archivePrefix  = "arXiv",
      primaryClass   = "hep-lat",
      SLACcitation   = "%%CITATION = ARXIV:1601.01624;%%"
}

@article{Belle:2018ezy,
    author = "Waheed, E. and others",
    collaboration = "Belle",
    title = "{Measurement of the CKM matrix element $|V_{cb}|$ from $B^0\to D^{*-}\ell^ {+} \nu_\ell$ at Belle}",
    eprint = "1809.03290",
    archivePrefix = "arXiv",
    primaryClass = "hep-ex",
    doi = "10.1103/PhysRevD.100.052007",
    journal = "Phys. Rev. D",
    volume = "100",
    number = "5",
    pages = "052007",
    note = "[Erratum: Phys.Rev.D 103, 079901 (2021)]",
    year = "2019",
}

@article{Abe:2005sh,
      author         = "Abe, K. and others",
      title          = "{Measurement of $D^0 \to \pi l \nu (K l \nu)$ and their
                        form-factors}",
      collaboration  = "Belle",
      year           = "2005",
      eprint         = "hep-ex/0510003",
      archivePrefix  = "arXiv",
      primaryClass   = "hep-ex",
      reportNumber   = "BELLE-CONF-519, LP2-5-154, EPS05-491, BELLE-CONF-0519",
      SLACcitation   = "%%CITATION = HEP-EX/0510003;%%",
}

@article{Ablikim:2015prg,
      author         = "Ablikim, M. and others",
      title          = "{Measurement of the absolute branching fraction for
                        $\Lambda^+_{c}\to \Lambda e^+\nu_e$}",
      collaboration  = "BESIII",
      journal        = "Phys. Rev. Lett.",
      volume         = "115",
      year           = "2015",
      number         = "22",
      pages          = "221805",
      doi            = "10.1103/PhysRevLett.115.221805",
      eprint         = "1510.02610",
      archivePrefix  = "arXiv",
      primaryClass   = "hep-ex",
      SLACcitation   = "%%CITATION = ARXIV:1510.02610;%%"
}

@article{Ablikim:2016vqd,
       author         = "Ablikim, Medina and others",
       title          = "{Measurement of the absolute branching fraction for
                         $\Lambda_c^+\rightarrow \Lambda \mu^+\nu_{\mu}$}",
       collaboration  = "BESIII",
       journal        = "Phys. Lett.",
       volume         = "B767",
       year           = "2017",
       pages          = "42-47",
       doi            = "10.1016/j.physletb.2017.01.047",
       eprint         = "1611.04382",
       archivePrefix  = "arXiv",
       primaryClass   = "hep-ex",
       SLACcitation   = "%%CITATION = ARXIV:1611.04382;%%"
}

@article{Ablikim:2017oaf,
      author         = "Ablikim, M. and others",
      title          = "{Measurement of $e^{+}e^{-}\rightarrow
                        \pi^{+}\pi^{-}\psi(3686)$ from 4.008 to 4.600~GeV and
                        observation of a charged structure in the
                        $\pi^{\pm}\psi(3686)$ mass spectrum}",
      collaboration  = "BESIII",
      journal        = "Phys. Rev.",
      volume         = "D96",
      year           = "2017",
      number         = "3",
      pages          = "032004",
      doi            = "10.1103/PhysRevD.96.032004",
      eprint         = "1703.08787",
      archivePrefix  = "arXiv",
      primaryClass   = "hep-ex",
      SLACcitation   = "%%CITATION = ARXIV:1703.08787;%%"
}

@article{Ablikim:2018frk,
      author         = "Ablikim, M. and others",
      title          = "{Measurement of the branching fraction for the
                        semi-leptonic decay $D^{0(+)}\to \pi^{-(0)}\mu^+\nu_\mu$
                        and test of lepton universality}",
      collaboration  = "BESIII",
      journal        = "Phys. Rev. Lett.",
      volume         = "121",
      year           = "2018",
      number         = "17",
      pages          = "171803",
      doi            = "10.1103/PhysRevLett.121.171803",
      eprint         = "1802.05492",
      archivePrefix  = "arXiv",
      primaryClass   = "hep-ex",
      SLACcitation   = "%%CITATION = ARXIV:1802.05492;%%"
}

@article{Abramczyk:2019fnf,
    author = "{[RBC/UKQCD 19] M. Abramczyk} and Blum, Thomas and Izubuchi, Taku and Jung, Chulwoo and Lin, Meifeng and Lytle, Andrew and Ohta, Shigemi and Shintani, Eigo",
    title = "{Nucleon mass and isovector couplings in 2+1-flavor dynamical domain-wall lattice QCD near physical mass}",
    eprint = "1911.03524",
    archivePrefix = "arXiv",
    primaryClass = "hep-lat",
    reportNumber = "KEK-TH-2167, RBRC-1320",
    doi = "10.1103/PhysRevD.101.034510",
    journal = "Phys. Rev. D",
    volume = "101",
    number = "3",
    pages = "034510",
    year = "2020"
}

@Article{Accardi:2016ndt,
  author         = "Accardi, A. and others",
  title          = "{A critical appraisal and evaluation of modern PDFs}",
  journal        = "Eur. Phys. J.",
  volume         = "C76",
  year           = "2016",
  number         = "8",
  pages          = "471",
  doi            = "10.1140/epjc/s10052-016-4285-4",
  eprint         = "1603.08906",
  archivePrefix  = "arXiv",
  primaryClass   = "hep-ph",
  reportNumber   = "DESY-16-041, DO-TH-16-05, JLAB-THY-16-2231, LTH-1081",
  SLACcitation   = "%%CITATION = ARXIV:1603.08906;%%"
}

@article{Adachi:2012mm,
     author         = "Adachi, I. and others",
     title          = "{Measurement of $B^- \to \tau^- \bar{\nu}_\tau$ with a
                       hadronic tagging method using the full data sample of
                       Belle}",
     collaboration  = "Belle",
     journal        = "Phys. Rev. Lett.",
     volume         = "110",
     pages          = "131801",
     doi            = "10.1103/PhysRevLett.110.131801",
     year           = "2013",
     eprint         = "1208.4678",
     archivePrefix  = "arXiv",
     primaryClass   = "hep-ex",
     reportNumber   = "BELLE-CONF-1201, KEK-PREPRINT-2012-28,
                       BELLE-PREPRINT-2012-26",
     SLACcitation   = "%%CITATION = ARXIV:1208.4678;%%",
}

@Article{Adams:2008db,
     author    = "Adams, David H.",
     title     = "{The rooting issue for a lattice fermion formulation
                  similar to staggered fermions but without taste mixing}",
     journal   = "Phys. Rev.",
     volume    = "D77",
     year      = "2008",
     pages     = "105024",
     eprint    = "0802.3029",
     archivePrefix = "arXiv",
     primaryClass  =  "hep-lat",
     doi       = "10.1103/PhysRevD.77.105024",
     SLACcitation  = "%%CITATION = 0802.3029;%%"
}

@article{Adams:2013qkq,
      author         = "Adams, C. and others",
      title          = "{Scientific Opportunities with the Long-Baseline Neutrino
                        Experiment}",
      collaboration  = "LBNE",
      year           = "2013",
      eprint         = "1307.7335",
      archivePrefix  = "arXiv",
      primaryClass   = "hep-ex",
      reportNumber   = "FERMILAB-CONF-13-300, BNL-101354-2013-JA",
      SLACcitation   = "%%CITATION = ARXIV:1307.7335;%%",
}

@article{Ademollo:1964sr,
      author         = "Ademollo, M. and Gatto, Raoul",
      title          = "{Nonrenormalization Theorem for the Strangeness Violating
                        Vector Currents}",
      journal        = "Phys.Rev.Lett.",
      volume         = "13",
      pages          = "264-265",
      doi            = "10.1103/PhysRevLett.13.264",
      year           = "1964",
}

@Article{Ademollo_Gatto,
     author    = "Ademollo, M. and Gatto, Raoul",
     title     = "{Nonrenormalization theorem for the strangeness violating
                  vector currents}",
     journal   = "Phys. Rev. Lett.",
     volume    = "13",
     year      = "1964",
     pages     = "264-265",
     doi       = "10.1103/PhysRevLett.13.264",
     SLACcitation  = "%%CITATION = PRLTA,13,264;%%"
}

@article{Agashe:2014kda,
      author         = "{K. A. Olive} and others",
      title          = "{Review of Particle Physics}",
      collaboration  = "Particle Data Group",
      journal        = "Chin. Phys.",
      volume         = "C38",
      year           = "2014",
      pages          = "090001 and 2015 update",
      doi            = "10.1088/1674-1137/38/9/090001",
      SLACcitation   = "%%CITATION = CHPHD,C38,090001;%%"
}

@article{Akhoury:1993uw,
      author         = "Akhoury, R. and Sterman, George F. and Yao, Y.P.",
      title          = "{Exclusive semileptonic decays of $B$ mesons into light
                        mesons}",
      journal        = "Phys.Rev.",
      volume         = "D50",
      pages          = "358-372",
      doi            = "10.1103/PhysRevD.50.358",
      year           = "1994",
      SLACcitation   = "%%CITATION = PHRVA,D50,358;%%",
}

@article{Alarcon:2011zs,
       author         = "Alarcon, J. M. and Martin Camalich, J. and Oller, J. A.",
       title          = "{The chiral representation of the $\pi N$ scattering
                         amplitude and the pion-nucleon sigma term}",
       journal        = "Phys. Rev.",
       volume         = "D85",
       year           = "2012",
       pages          = "051503",
       doi            = "10.1103/PhysRevD.85.051503",
       eprint         = "1110.3797",
       archivePrefix  = "arXiv",
       primaryClass   = "hep-ph",
       SLACcitation   = "%%CITATION = ARXIV:1110.3797;%%"
 }

@article{Albertus:2010nm,
      author         = "{[RBC/UKQCD 10C] C. Albertus} and others",
      title          = "{Neutral B-meson mixing from unquenched lattice QCD with
                        domain-wall light quarks and static b-quarks}",
      journal        = "Phys.Rev.",
      volume         = "D82",
      pages          = "014505",
      doi            = "10.1103/PhysRevD.82.014505",
      year           = "2010",
      eprint         = "1001.2023",
      archivePrefix  = "arXiv",
      primaryClass   = "hep-lat",
      reportNumber   = "CU-TP-1192, EDINBURGH-2010-1, RBRC-827, SHEP-0928",
      SLACcitation   = "%%CITATION = ARXIV:1001.2023;%%",
}

@article{Alexandrou:1990dq,
      author         = "Alexandrou, C. and Jegerlehner, F. and Gusken, S. and
                        Schilling, K. and Sommer, R.",
      title          = "{B meson properties from lattice QCD}",
      journal        = "Phys. Lett.",
      volume         = "B256",
      year           = "1991",
      pages          = "60-67",
      doi            = "10.1016/0370-2693(91)90219-G",
      reportNumber   = "WU-B-90-18-REV, PSI-PR-90-28-REV",
      SLACcitation   = "%%CITATION = PHLTA,B256,60;%%"
}

@article{Alexandrou:2014sha,
      author         = "{[ETM 14A] C. Alexandrou} and Drach, V. and Jansen, K. and
                        Kallidonis, C. and Koutsou, G.",
      title          = "{Baryon spectrum with $N_f=2+1+1$ twisted mass fermions}",
      journal        = "Phys. Rev.",
      volume         = "D90",
      year           = "2014",
      number         = "7",
      pages          = "074501",
      doi            = "10.1103/PhysRevD.90.074501",
      eprint         = "1406.4310",
      archivePrefix  = "arXiv",
      primaryClass   = "hep-lat",
      reportNumber   = "DESY-14-096",
      SLACcitation   = "%%CITATION = ARXIV:1406.4310;%%"
}

@article{Alexandrou:2017hac,
      author         = "{[ETM 17B] C. Alexandrou}  and Constantinou, Martha and
                        Hadjiyiannakou, Kyriakos and Jansen, Karl and Kallidonis,
                        Christos and Koutsou, Giannis and Vaquero Aviles-Casco,
                        Alejandro",
      title          = "{Nucleon axial form factors using $N_f$ = 2 twisted mass
                        fermions with a physical value of the pion mass}",
      journal        = "Phys. Rev.",
      volume         = "D96",
      year           = "2017",
      number         = "5",
      pages          = "054507",
      doi            = "10.1103/PhysRevD.96.054507",
      eprint         = "1705.03399",
      archivePrefix  = "arXiv",
      primaryClass   = "hep-lat",
      SLACcitation   = "%%CITATION = ARXIV:1705.03399;%%"
}

@article{Alexandrou:2017oeh,
      author         = "{[ETM 17C] C. Alexandrou} and Constantinou, M. and Hadjiyiannakou,
                        K. and Jansen, K. and Kallidonis, C. and Koutsou, G. and
                        Vaquero Avilés-Casco, A. and Wiese, C.",
      title          = "{Nucleon Spin and Momentum Decomposition Using Lattice
                        QCD Simulations}",
      journal        = "Phys. Rev. Lett.",
      volume         = "119",
      year           = "2017",
      number         = "14",
      pages          = "142002",
      doi            = "10.1103/PhysRevLett.119.142002",
      eprint         = "1706.02973",
      archivePrefix  = "arXiv",
      primaryClass   = "hep-lat",
      reportNumber   = "DESY-17-086",
      SLACcitation   = "%%CITATION = ARXIV:1706.02973;%%"
}

@article{Alexandrou:2017qyt,
      author         = "{[ETM 17] C. Alexandrou} and others",
      title          = "{Nucleon scalar and tensor charges using lattice QCD
                        simulations at the physical value of the pion mass}",
      journal        = "Phys. Rev.",
      volume         = "D95",
      year           = "2017",
      number         = "11",
      pages          = "114514",
      doi            = "10.1103/PhysRevD.96.099906, 10.1103/PhysRevD.95.114514",
      note           = "[Erratum: Phys. Rev.D96,no.9,099906(2017)]",
      eprint         = "1703.08788",
      archivePrefix  = "arXiv",
      primaryClass   = "hep-lat",
      SLACcitation   = "%%CITATION = ARXIV:1703.08788;%%"
}

@article{Alexandrou:2017xwd,
      author         = "{[ETM 17A] C. Alexandrou}  and Kallidonis, Christos",
      title          = "{Low-lying baryon masses using $N_f=2$ twisted mass
                        clover-improved fermions directly at the physical pion
                        mass}",
      journal        = "Phys. Rev.",
      volume         = "D96",
      year           = "2017",
      number         = "3",
      pages          = "034511",
      doi            = "10.1103/PhysRevD.96.034511",
      eprint         = "1704.02647",
      archivePrefix  = "arXiv",
      primaryClass   = "hep-lat",
      SLACcitation   = "%%CITATION = ARXIV:1704.02647;%%"
}

@article{Alexandrou:2018egz,
      author         = "{[ETM 18A] C.~Alexandrou} and others",
      title          = "{Simulating twisted mass fermions at physical light,
                        strange and charm quark masses}",
      journal        = "Phys. Rev.",
      volume         = "D98",
      year           = "2018",
      number         = "5",
      pages          = "054518",
      doi            = "10.1103/PhysRevD.98.054518",
      eprint         = "1807.00495",
      archivePrefix  = "arXiv",
      primaryClass   = "hep-lat",
      SLACcitation   = "%%CITATION = ARXIV:1807.00495;%%"
}

@article{Alexandrou:2019brg,
    author = "{[ETM 19] C. Alexandrou} and Bacchio, S. and Constantinou, M. and Finkenrath, J. and Hadjiyiannakou, K. and Jansen, K. and Koutsou, G. and Vaquero Aviles-Casco, A.",
    title = "{Nucleon axial, tensor, and scalar charges and $\sigma$-terms in lattice QCD}",
    eprint = "1909.00485",
    archivePrefix = "arXiv",
    primaryClass = "hep-lat",
    doi = "10.1103/PhysRevD.102.054517",
    journal = "Phys. Rev. D",
    volume = "102",
    number = "5",
    pages = "054517",
    year = "2020"
}

@Article{Alford:1995hw,
     author    = "Alford, Mark G. and Dimm, W. and Lepage, G. P. and Hockney,
                  G. and Mackenzie, P. B.",
     title     = "{Lattice QCD on small computers}",
     journal   = "Phys. Lett.",
     volume    = "B361",
     year      = "1995",
     pages     = "87-94",
     eprint    = "hep-lat/9507010",
     archivePrefix = "arXiv",
     doi       = "10.1016/0370-2693(95)01131-9",
     SLACcitation  = "%%CITATION = HEP-LAT/9507010;%%"
}

@Article{AliKhan:2001tx,
     author    = "{[CP-PACS 01] A. Ali Khan} and others",
     title     = "{Light hadron spectroscopy with two flavors of dynamical
                  quarks on the lattice}",
     journal   = "Phys. Rev.",
     volume    = "D65",
     year      = "2002",
     pages     = "054505",
     note  = "{Erratum: {\it Phys. Rev.} {\bf D66} (2003) 059901}",
     eprint    = "hep-lat/0105015",
     archivePrefix = "arXiv",
     doi       = "10.1103/PhysRevD.65.054505",
     SLACcitation  = "%%CITATION = HEP-LAT/0105015;%%" 
}

@Article{Allison:2008xk,
     author    = "{[HPQCD 08B] I. Allison} and others",
     title     = "{High-precision charm-quark mass from current-current
                  correlators in lattice and continuum QCD}",
     journal   = "Phys. Rev.",
     volume    = "D78",
     year      = "2008",
     pages     = "054513",
     eprint    = "0805.2999",
     archivePrefix = "arXiv",
     primaryClass  =  "hep-lat",
     doi       = "10.1103/PhysRevD.78.054513",
     SLACcitation  = "%%CITATION = 0805.2999;%%"
}

@article{Allton:1998sm,
      author         = "Allton, C. R. and Conti, L. and Donini, A. and Gimenez,
                        V. and Giusti, Leonardo and Martinelli, G. and Talevi, M.
                        and Vladikas, A.",
      title          = "{B parameters for Delta S = 2 supersymmetric operators}",
      journal        = "Phys. Lett.",
      volume         = "B453",
      year           = "1999",
      pages          = "30-39",
      doi            = "10.1016/S0370-2693(99)00283-X",
      eprint         = "hep-lat/9806016",
      archivePrefix  = "arXiv",
      primaryClass   = "hep-lat",
      reportNumber   = "EDINBURGH-98-8, FTUV-98-40, IFIC-98-41, ROME1-1209-98,
                        ROM2F-98-16, SNS-PH-1998-011, SWAT-190",
      SLACcitation   = "%%CITATION = HEP-LAT/9806016;%%"
}

@Article{Allton:2008pn,
     author    = "{[RBC/UKQCD 08] C. Allton} and others",
     title     = "{Physical results from 2+1 flavor domain wall QCD and SU(2)
                  chiral perturbation theory}",
     journal   = "Phys. Rev.",
     volume    = "D78",
     year      = "2008",
     pages     = "114509",
     eprint    = "0804.0473",
     archivePrefix = "arXiv",
     primaryClass  =  "hep-lat",
     doi       = "10.1103/PhysRevD.78.114509",
     SLACcitation  = "%%CITATION = 0804.0473;%%"
}

@article{Amhis:2019ckw,
    author = "{[HFLAV 18] Y.~Amhis} and others",
    title = "{Averages of $b$-hadron, $c$-hadron, and $\tau$-lepton properties as of 2018}",
    eprint = "1909.12524",
    archivePrefix = "arXiv",
    primaryClass = "hep-ex",
    doi = "10.1140/epjc/s10052-020-8156-7",
    journal = "Eur. Phys. J. C",
    volume = "81",
    number = "3",
    pages = "226",
    year = "2021"
}

@Article{Amoros:2001cp,
     author    = "{G. Amoros, J. Bijnens} and Talavera, P.",
     title     = "{QCD isospin breaking in meson masses, decay constants and
                  quark mass  ratios}",
     journal   = "Nucl. Phys.",
     volume    = "B602",
     year      = "2001",
     pages     = "87-108",
     eprint    = "hep-ph/0101127",
     archivePrefix = "arXiv",
     doi       = "10.1016/S0550-3213(01)00121-3",
     SLACcitation  = "%%CITATION = HEP-PH/0101127;%%"
}

@Article{Anikeev:2001rk,
     author    = "Anikeev, K. and others",
     title     = "{$B$ physics at the Tevatron: Run II and beyond}",
     year      = "2001",
     eprint    = "hep-ph/0201071",
     archivePrefix = "arXiv",
     SLACcitation  = "%%CITATION = HEP-PH/0201071;%%"
}

@article{Antonelli:2009ws,
      author         = "Antonelli, Mario and others",
      title          = "{Flavor physics in the quark sector}",
      journal        = "Phys.Rept.",
      volume         = "494",
      pages          = "197-414",
      doi            = "10.1016/j.physrep.2010.05.003",
      year           = "2010",
      eprint         = "0907.5386",
      archivePrefix  = "arXiv",
      primaryClass   = "hep-ph",
      reportNumber   = "BNL-90299-2009-BC, CERN-PH-TH-2009-112,
                        FERMILAB-PUB-09-323-T, LAL-09-111, MPP-2009-88,
                        MZ-TH-09-22, MKPH-T-09-14, SLAC-R-926, TUM-HEP-728-09,
                        WSU-HEP-0902",
      SLACcitation   = "%%CITATION = ARXIV:0907.5386;%%",
}

@Article{Antonelli:2010yf,
     author    = "Antonelli, M. and others",
     title     = "{An evaluation of $|V_{us}|$ and precise tests of the Standard
                  Model from world data on leptonic and semileptonic kaon
                  decays}", 
     journal   = "Eur. Phys. J.",
     volume    = "C69",
     year      = "2010",
     pages     = "399-424",
     eprint    = "1005.2323",
     archivePrefix = "arXiv",
     primaryClass  =  "hep-ph",
     doi       = "10.1140/epjc/s10052-010-1406-3",
     SLACcitation  = "%%CITATION = 1005.2323;%%"
}

@Article{Antonio:2007pb,
     author    = "{[RBC/UKQCD 07A] D. J. Antonio} and others",
     title     = "{Neutral kaon mixing from 2+1 flavor domain wall QCD}",
     journal   = "Phys. Rev. Lett.",
     volume    = "100",
     year      = "2008",
     pages     = "032001",
     eprint    = "hep-ph/0702042",
     archivePrefix = "arXiv",
     doi       = "10.1103/PhysRevLett.100.032001",
     SLACcitation  = "%%CITATION = HEP-PH/0702042;%%"
}

@Article{Antonio:2008zz,
     author    = "{[RBC 07A] D. J. Antonio} and others",
     title     = "{Localization and chiral symmetry in 3 flavor domain wall
                  QCD}",
     journal   = "Phys. Rev.",
     volume    = "D77",
     year      = "2008",
     pages     = "014509",
     eprint    = "0705.2340",
     archivePrefix = "arXiv",
     primaryClass  =  "hep-lat",
     doi       = "10.1103/PhysRevD.77.014509",
     SLACcitation  = "%%CITATION = 0705.2340;%%"
}

@article{Anzai:2009tm,
      author         = "Anzai, C. and Kiyo, Y. and Sumino, Y.",
      title          = "{Static QCD potential at three-loop order}",
      journal        = "Phys.Rev.Lett.",
      volume         = "104",
      pages          = "112003",
      doi            = "10.1103/PhysRevLett.104.112003",
      year           = "2010",
      eprint         = "0911.4335",
      archivePrefix  = "arXiv",
      primaryClass   = "hep-ph",
      reportNumber   = "TU-856, KEK-TH-1339",
      SLACcitation   = "%%CITATION = ARXIV:0911.4335;%%",
}

@article{Aoki:1994pc,
      author         = "{S. Aoki} and Fukugita, M. and Hashimoto, S.
                        and Ishizuka, N. and Mino, H. and others",
      title          = "{Manifestation of sea quark effects in the strong
                        coupling constant in lattice QCD}",
      journal        = "Phys.Rev.Lett.",
      volume         = "74",
      pages          = "22-25",
      doi            = "10.1103/PhysRevLett.74.22",
      year           = "1995",
      eprint         = "hep-lat/9407015",
      archivePrefix  = "arXiv",
      primaryClass   = "hep-lat",
      reportNumber   = "UTHEP-280",
      comment = "Aoki 94",
      SLACcitation   = "%%CITATION = HEP-LAT/9407015;%%",
}

@article{Aoki:2001ra,
      author         = "Aoki, Sinya and Kuramashi, Yoshinobu and Tominaga,
                        Shin-ichi",
      title          = "{Relativistic heavy quarks on the lattice}",
      journal        = "Prog.Theor.Phys.",
      volume         = "109",
      pages          = "383-413",
      doi            = "10.1143/PTP.109.383",
      year           = "2003",
      eprint         = "hep-lat/0107009",
      archivePrefix  = "arXiv",
      primaryClass   = "hep-lat",
      reportNumber   = "KEK-CP-110",
      SLACcitation   = "%%CITATION = HEP-LAT/0107009;%%",
}

@Article{Aoki:2002uc,
     author    = "{[JLQCD 02] S. Aoki} and others",
     title     = "{Light hadron spectroscopy with two flavors of $O(a)$-
                  improved dynamical quarks}",
     journal   = "Phys. Rev.",
     volume    = "D68",
     year      = "2003",
     pages     = "054502",
     eprint    = "hep-lat/0212039",
     archivePrefix = "arXiv",
     doi       = "10.1103/PhysRevD.68.054502",
     SLACcitation  = "%%CITATION = HEP-LAT/0212039;%%"
}

@article{Aoki:2003dg,
      author         = "Aoki, Sinya and Kayaba, Yasuhisa and Kuramashi,
                        Yoshinobu",
      title          = "{A perturbative determination of mass dependent O(a)
                        improvement coefficients in a relativistic heavy quark
                        action}",
      journal        = "Nucl.Phys.",
      volume         = "B697",
      pages          = "271-301",
      doi            = "10.1016/j.nuclphysb.2004.07.017",
      year           = "2004",
      eprint         = "hep-lat/0309161",
      archivePrefix  = "arXiv",
      primaryClass   = "hep-lat",
      SLACcitation   = "%%CITATION = HEP-LAT/0309161;%%",
}

@Article{Aoki:2003yv,
     author    = "Aoki, Sinya",
     title     = "{Chiral perturbation theory with Wilson-type fermions
                  including $a^2$  effects: $N_f = 2$ degenerate case}",
     journal   = "Phys. Rev.",
     volume    = "D68",
     year      = "2003",
     pages     = "054508",
     eprint    = "hep-lat/0306027",
     archivePrefix = "arXiv",
     doi       = "10.1103/PhysRevD.68.054508",
     SLACcitation  = "%%CITATION = HEP-LAT/0306027;%%"
}

@Article{Aoki:2004ht,
     author    = "{[RBC 04] Y. Aoki} and others",
     title     = "{Lattice QCD with two dynamical flavors of domain wall
                  fermions}",
     journal   = "Phys. Rev.",
     volume    = "D72",
     year      = "2005",
     pages     = "114505",
     eprint    = "hep-lat/0411006",
     archivePrefix = "arXiv",
     doi       = "10.1103/PhysRevD.72.114505",
     SLACcitation  = "%%CITATION = HEP-LAT/0411006;%%"
}

@Article{Aoki:2004ta,
     author    = "Aoki, Sinya and {B\"ar}, Oliver",
     title     = "{Twisted-mass QCD, O(a) improvement and Wilson chiral
                  perturbation  theory}",
     journal   = "Phys. Rev.",
     volume    = "D70",
     year      = "2004",
     pages     = "116011",
     eprint    = "hep-lat/0409006",
     archivePrefix = "arXiv",
     doi       = "10.1103/PhysRevD.70.116011",
     SLACcitation  = "%%CITATION = HEP-LAT/0409006;%%"
}

@article{Aoki:2004th,
      author         = "Aoki, Sinya and Kayaba, Yasuhisa and Kuramashi,
                        Yoshinobu",
      title          = "{Perturbative determination of mass dependent O(a)
                        improvement coefficients for the vector and axial vector
                        currents with a relativistic heavy quark action}",
      journal        = "Nucl.Phys.",
      volume         = "B689",
      pages          = "127-156",
      doi            = "10.1016/j.nuclphysb.2004.04.009",
      year           = "2004",
      eprint         = "hep-lat/0401030",
      archivePrefix  = "arXiv",
      primaryClass   = "hep-lat",
      SLACcitation   = "%%CITATION = HEP-LAT/0401030;%%",
}

@article{Aoki:2005et,
      author         = "{[CP-PACS/JLQCD 05] S. Aoki} and others",
      title          = "{Nonperturbative O(a) improvement of the Wilson quark
                        action with the RG-improved gauge action using the
                        Schr{\"o}dinger functional method}",
      journal        = "Phys.Rev.",
      volume         = "D73",
      pages          = "034501",
      doi            = "10.1103/PhysRevD.73.034501",
      year           = "2006",
      eprint         = "hep-lat/0508031",
      archivePrefix  = "arXiv",
      primaryClass   = "hep-lat",
      reportNumber   = "KEK-CP-163, UTHEP-509, UTCCS-P-15, HUPD-0506",
      SLACcitation   = "%%CITATION = HEP-LAT/0508031;%%",
}

@article{Aoki:2005vt,
      author         = "Aoki, Y. and Fodor, Z. and Katz, S.D. and Szabo, K.K.",
      title          = "{The equation of state in lattice QCD: with physical
                        quark masses towards the continuum limit}",
      journal        = "JHEP",
      volume         = "0601",
      pages          = "089",
      doi            = "10.1088/1126-6708/2006/01/089",
      year           = "2006",
      eprint         = "hep-lat/0510084",
      archivePrefix  = "arXiv",
      primaryClass   = "hep-lat",
      SLACcitation   = "%%CITATION = HEP-LAT/0510084;%%",
}

@Article{Aoki:2007ka,
     author    = "Aoki, Sinya and Fukaya, Hidenori and Hashimoto, Shoji and
                  Onogi, Tetsuya",
     title     = "{Finite volume QCD at fixed topological charge}",
     journal   = "Phys. Rev.",
     volume    = "D76",
     year      = "2007",
     pages     = "054508",
     eprint    = "0707.0396",
     archivePrefix = "arXiv",
     primaryClass  =  "hep-lat",
     doi       = "10.1103/PhysRevD.76.054508",
     SLACcitation  = "%%CITATION = 0707.0396;%%"
}

@Article{Aoki:2008sm,
     author    = "{[PACS-CS 08] S. Aoki} and others",
     title     = "{2+1 flavor lattice QCD toward the physical point}",
     journal   = "Phys. Rev.",
     volume    = "D79",
     year      = "2009",
     pages     = "034503",
     eprint    = "0807.1661",
     archivePrefix = "arXiv",
     primaryClass  =  "hep-lat",
     doi       = "10.1103/PhysRevD.79.034503",
     SLACcitation  = "%%CITATION = 0807.1661;%%"
}

@Article{Aoki:2008ss,
     author    = "{[JLQCD 08] S. Aoki} and others",
     title     = "{$B_K$ with two flavors of dynamical overlap fermions}",
     journal   = "Phys. Rev.",
     volume    = "D77",
     year      = "2008",
     pages     = "094503",
     eprint    = "0801.4186",
     archivePrefix = "arXiv",
     primaryClass  =  "hep-lat",
     doi       = "10.1103/PhysRevD.77.094503",
     SLACcitation  = "%%CITATION = 0801.4186;%%"
}

@Article{Aoki:2009ix,
     author    = "{[PACS-CS 09] S. Aoki} and others",
     title     = "{Physical point simulation in 2+1 flavor lattice QCD}",
      journal   = "Phys. Rev.",
     volume    = "D81",
     year      = "2010",
     pages     = "074503",
     eprint    = "0911.2561",
     archivePrefix = "arXiv",
     primaryClass  =  "hep-lat",
     doi       = "10.1103/PhysRevD.81.074503",
     SLACcitation  = "%%CITATION = 0911.2561;%%"
}

@article{Aoki:2009tf,
      author         = "{[PACS-CS 09A] S. Aoki} and others",
      title          = "{Precise determination of the strong coupling constant
                         in $N_f = 2+1$ lattice QCD with the Schr\"odinger
                         functional scheme}",
      journal        = "JHEP",
      volume         = "0910",
      pages          = "053",
      doi            = "10.1088/1126-6708/2009/10/053",
      year           = "2009",
      eprint         = "0906.3906",
      archivePrefix  = "arXiv",
      primaryClass   = "hep-lat",
      reportNumber   = "UTHEP-584, UTCCS-P-54",
      SLACcitation   = "%%CITATION = ARXIV:0906.3906;%%",
}

@article{Aoki:2010dy,
      author         = "{[RBC/UKQCD 10A] Y. Aoki} and others",
      title          = "{Continuum limit physics from 2+1 flavor domain wall
                        QCD}",
      journal        = "Phys.Rev.",
      volume         = "D83",
      pages          = "074508",
      doi            = "10.1103/PhysRevD.83.074508",
      year           = "2011",
      eprint         = "1011.0892",
      archivePrefix  = "arXiv",
      primaryClass   = "hep-lat",
      SLACcitation   = "%%CITATION = ARXIV:1011.0892;%%",
}

@Article{Aoki:2010pe,
     author    = "{[RBC/UKQCD 10B] Y. Aoki} and others",
     title     = "{Continuum limit of $B_K$ from 2+1 flavor domain wall
                  QCD}",
      journal        = "Phys.Rev.",
      volume         = "D84",
      pages          = "014503",
      doi            = "10.1103/PhysRevD.84.014503",
      year           = "2011",
      eprint         = "1012.4178",
      archivePrefix  = "arXiv",
      primaryClass   = "hep-lat",
      reportNumber   = "CU-TP-1196, EDINBURGH-2010-12, KEK-TH-1366, RBRC-843,
                        SHEP-1016, MPP-2010-172",
      SLACcitation   = "%%CITATION = ARXIV:1012.4178;%%",
}

@article{Aoki:2010wm,
      author         = "{[PACS-CS 10] S. Aoki} and others",
      title          = "{Non-perturbative renormalization of quark mass in $N_f =
                        2+1$ QCD with the Schr{\"o}dinger functional scheme}",
      journal        = "JHEP",
      volume         = "1008",
      pages          = "101",
      doi            = "10.1007/JHEP08(2010)101",
      year           = "2010",
      eprint         = "1006.1164",
      archivePrefix  = "arXiv",
      primaryClass   = "hep-lat",
      reportNumber   = "UTHEP-609, UTCCS-P-59",
      SLACcitation   = "%%CITATION = ARXIV:1006.1164;%%",
}

@article{Aoki:2010xg,
      author         = "{[RBC/UKQCD 10D] Y. Aoki} and Blum, Tom and Lin, Huey-Wen and Ohta,
                        Shigemi and Sasaki, Shoichi and Tweedie, Robert and
                        Zanotti, James and Yamazaki, Takeshi",
      title          = "{Nucleon isovector structure functions in (2+1)-flavor
                        QCD with domain wall fermions}",
      journal        = "Phys. Rev.",
      volume         = "D82",
      year           = "2010",
      pages          = "014501",
      doi            = "10.1103/PhysRevD.82.014501",
      eprint         = "1003.3387",
      archivePrefix  = "arXiv",
      primaryClass   = "hep-lat",
      SLACcitation   = "%%CITATION = ARXIV:1003.3387;%%"
}

@article{Aoki:2012st,
      author         = "{[PACS-CS 12] S. Aoki} and Ishikawa, K.-I. and Ishizuka, N. and Kanaya,
                        K. and Kuramashi, Y. and others",
      title          = "{1+1+1 flavor QCD + QED simulation at the physical point}", 
      journal        = "Phys.Rev.",
      volume         = "D86",
      pages          = "034507",
      doi            = "10.1103/PhysRevD.86.034507",
      year           = "2012",
      eprint         = "1205.2961",
      archivePrefix  = "arXiv",
      primaryClass   = "hep-lat",
      reportNumber   = "UTCCS-P-66",
      SLACcitation   = "%%CITATION = ARXIV:1205.2961;%%",
}

@article{Aoki:2012xaa,
      author         = "{[RBC/UKQCD 12A] Y. Aoki} and others",
      title          = "{Nonperturbative tuning of an improved relativistic
                        heavy-quark action with application to bottom
                        spectroscopy}",
      journal        = "Phys.Rev.",
      volume         = "D86",
      pages          = "116003",
      doi            = "10.1103/PhysRevD.86.116003",
      year           = "2012",
      eprint         = "1206.2554",
      archivePrefix  = "arXiv",
      primaryClass   = "hep-lat",
      SLACcitation   = "%%CITATION = ARXIV:1206.2554;%%",
}

@article{Aoki:2013ldr,
      author         = "{[FLAG 13] S. Aoki} and Aoki, Yasumichi and Bernard, Claude and
                        Blum, Tom and Colangelo, Gilberto and others",
      title          = "{Review of lattice results concerning low-energy particle
                        physics}",
      journal        = "Eur.Phys.J.",
      volume         = "C74",
      pages          = "2890",
      doi            = "10.1140/epjc/s10052-014-2890-7",
      year           = "2014",
      eprint         = "1310.8555",
      archivePrefix  = "arXiv",
      primaryClass   = "hep-lat",
      reportNumber   = "CP3-ORIGINS-2013-040-DNRF90, DIAS-2013-40,
                        FERMILAB-PUB-13-484-T, FTUAM-13-28, IFIC-13-76,
                        IFT-UAM-CSIC-13-106, MITP-13-067, YITP-13-114",
      SLACcitation   = "%%CITATION = ARXIV:1310.8555;%%",
}

@article{Aoki:2014nga,
      author         = "{[RBC/UKQCD 14A] Y. Aoki} and Ishikawa, Tomomi and Izubuchi, Taku
                        and Lehner, Christoph and Soni, Amarjit",
      title          = "{Neutral $B$ meson mixings and $B$ meson decay constants
                        with static heavy and domain-wall light quarks}",
      journal        = "Phys. Rev.",
      volume         = "D91",
      year           = "2015",
      number         = "11",
      pages          = "114505",
      doi            = "10.1103/PhysRevD.91.114505",
      eprint         = "1406.6192",
      archivePrefix  = "arXiv",
      primaryClass   = "hep-lat",
      reportNumber   = "RBRC-1080",
      SLACcitation   = "%%CITATION = ARXIV:1406.6192;%%"
}

@article{Aoki:2015pba,
      author         = "{[JLQCD 15A] S. Aoki} and Cossu, G. and Feng, X. and Hashimoto, S. and
                        Kaneko, T. and Noaki, J. and Onogi, T.",
      title          = "{Light meson electromagnetic form factors from
                        three-flavor lattice QCD with exact chiral symmetry}",
      journal        = "Phys. Rev.",
      volume         = "D93",
      year           = "2016",
      number         = "3",
      pages          = "034504",
      doi            = "10.1103/PhysRevD.93.034504",
      eprint         = "1510.06470",
      archivePrefix  = "arXiv",
      primaryClass   = "hep-lat",
      reportNumber   = "KEK-CP-325, YITP-15-70",
      SLACcitation   = "%%CITATION = ARXIV:1510.06470;%%"
}

@article{Aoki:2016frl,
      author         = "{[FLAG 16] S. Aoki} and others",
      title          = "{Review of lattice results concerning low-energy particle
                        physics}",
      journal        = "Eur. Phys. J.",
      volume         = "C77",
      year           = "2017",
      number         = "2",
      pages          = "112",
      doi            = "10.1140/epjc/s10052-016-4509-7",
      eprint         = "1607.00299",
      archivePrefix  = "arXiv",
      primaryClass   = "hep-lat",
      reportNumber   = "CP3-Origins-2016-023-DNRF90, DESY-16-111, DIAS-2016-23,
                        Edinburgh-2016-11, FTUAM-16-23, HIM-2016-02,
                        IFT-UAM-CSIC-16-057, LPT-Orsay-16-47, MITP-16-059,
                        RM3-TH-16-7, ROM2F-2016-05, YITP-16-77",
      SLACcitation   = "%%CITATION = ARXIV:1607.00299;%%"
}

@article{Aoki:2017paw,
      author         = "{[JLQCD 17A] S. Aoki} and Cossu, G. and Fukaya, H. and Hashimoto, S.
                        and Kaneko, T.",
      title          = "{Topological susceptibility of QCD with dynamical
                        M{\"o}bius domain wall fermions}",
      journal        = "PTEP",
      volume         = "2018",
      year           = "2018",
      number         = "4",
      pages          = "043B07",
      doi            = "10.1093/ptep/pty041",
      eprint         = "1705.10906",
      archivePrefix  = "arXiv",
      primaryClass   = "hep-lat",
      reportNumber   = "OU-HET-937, KEK-CP-359, YITP-17-51",
      SLACcitation   = "%%CITATION = ARXIV:1705.10906;%%"
}

@article{Aoki:2017spo,
      author         = "{[JLQCD 17] S. Aoki} and Cossu, G. and Feng, X. and Fukaya, H. and
                        Hashimoto, S. and Kaneko, T. and Noaki, J. and Onogi, T.",
      title          = "{Chiral behavior of $K \to \pi l \nu$ decay form factors
                        in lattice QCD with exact chiral symmetry}",
      journal        = "Phys. Rev.",
      volume         = "D96",
      year           = "2017",
      number         = "3",
      pages          = "034501",
      doi            = "10.1103/PhysRevD.96.034501",
      eprint         = "1705.00884",
      archivePrefix  = "arXiv",
      primaryClass   = "hep-lat",
      reportNumber   = "KEK-CP-357, OU-HET-928",
      SLACcitation   = "%%CITATION = ARXIV:1705.00884;%%"
}

@article{FlavourLatticeAveragingGroupFLAG:2021npn,
    author = "{[FLAG 21] Y. Aoki} and others",
    title = "{FLAG Review 2021}",
    eprint = "2111.09849",
    archivePrefix = "arXiv",
    primaryClass = "hep-lat",
    reportNumber = "CERN-TH-2021-191, JLAB-THY-21-3528, FERMILAB-PUB-21-620-SCD-T",
    doi = "10.1140/epjc/s10052-022-10536-1",
    journal = "Eur. Phys. J. C",
    volume = "82",
    number = "10",
    pages = "869",
    year = "2022"
}

@article{Arnesen:2005ez,
      author         = "Arnesen, M. Christian and Grinstein, Benjamin and
                        Rothstein, Ira Z. and Stewart, Iain W.",
      title          = "{A precision model independent determination of $|V_{ub}|$
                        from $B \to  \pi e \nu$}",
      journal        = "Phys.Rev.Lett.",
      volume         = "95",
      pages          = "071802",
      doi            = "10.1103/PhysRevLett.95.071802",
      year           = "2005",
      eprint         = "hep-ph/0504209",
      archivePrefix  = "arXiv",
      primaryClass   = "hep-ph",
      reportNumber   = "MIT-CTP-3620, CMU-HEP-0504, UCSD-PTH-05-04",
      SLACcitation   = "%%CITATION = HEP-PH/0504209;%%",
}

@Article{Arthur:2010ht,
     author    = "{[RBC 10] R. Arthur} and Boyle, P. A.",
     title     = "{Step scaling with off-shell renormalisation}",
      journal        = "Phys.Rev.",
      volume         = "D83",
      pages          = "114511",
      doi            = "10.1103/PhysRevD.83.114511",
      year           = "2011",
      eprint         = "1006.0422",
      archivePrefix  = "arXiv",
      primaryClass   = "hep-lat",
      SLACcitation   = "%%CITATION = ARXIV:1006.0422;%%",
}

@article{Arthur:2012opa,
      author         = "{[RBC/UKQCD 12] R. Arthur} and others",
      title          = "{Domain wall QCD with near-physical pions}",
      journal        = "Phys.Rev.",
      volume         = "D87",
      pages          = "094514",
      doi            = "10.1103/PhysRevD.87.094514",
      year           = "2013",
      eprint         = "1208.4412",
      archivePrefix  = "arXiv",
      primaryClass   = "hep-lat",
      SLACcitation   = "%%CITATION = ARXIV:1208.4412;%%",
}

@article{Asakawa:2015vta,
      author         = "{[FlowQCD 15] M. Asakawa} and Iritani, Takumi
                        and Kitazawa, Masakiyo and Suzuki, Hiroshi",
      title          = "{Determination of Reference Scales for Wilson Gauge
                        Action from Yang--Mills Gradient Flow}",
      year           = "2015",
      eprint         = "1503.06516",
      archivePrefix  = "arXiv",
      primaryClass   = "hep-lat",
      reportNumber   = "KYUSHU-HET-152, YITP-15-19",
      SLACcitation   = "%%CITATION = ARXIV:1503.06516;%%"
}

@article{Ashman:1987hv,
      author         = "Ashman, J. and others",
      title          = "{A Measurement of the Spin Asymmetry and Determination of
                        the Structure Function g(1) in Deep Inelastic Muon-Proton
                        Scattering}",
      booktitle      = "{Internal spin structure of the nucleon. Proceedings,
                        Symposium, SMC Meeting, New Haven, USA, January 5-6,
                        1994}",
      collaboration  = "European Muon",
      journal        = "Phys. Lett.",
      volume         = "B206",
      year           = "1988",
      pages          = "364",
      doi            = "10.1016/0370-2693(88)91523-7",
      reportNumber   = "CERN-EP-87-230",
      SLACcitation   = "%%CITATION = PHLTA,B206,364;%%"
}

@article{Athenodorou:2018wpk,
      author         = "Athenodorou, Andreas and Finkenrath, Jacob and Knechtli,
                        Francesco and Korzec, Tomasz and Leder, Bj{\"o}rn and
                        Marinkovic, Marina Krstic and Sommer, Rainer",
      title          = "{How perturbative are heavy sea quarks?}",
      journal        = "Nucl. Phys.",
      volume         = "B943",
      year           = "2019",
      pages          = "114612",
      doi            = "10.1016/j.nuclphysb.2019.114612",
      eprint         = "1809.03383",
      archivePrefix  = "arXiv",
      primaryClass   = "hep-lat",
      reportNumber   = "DESY 18-134, HU-EP-18/28, WUB/18-03, DESY-18-134,
                        HU-EP-18-28, WUB-18-03",
      SLACcitation   = "%%CITATION = ARXIV:1809.03383;%%"}

@article{Atoui:2013zza,
      author         = "Atoui, Mariam and Morenas, Vincent and Becirevic, Damir
                        and Sanfilippo, Francesco",
      Title = {$B_s \to D_s \ell \nu_\ell$ near zero recoil in and beyond the Standard Model},
      journal        = "Eur. Phys. J.",
      volume         = "C74",
      year           = "2014",
      number         = "5",
      pages          = "2861",
      doi            = "10.1140/epjc/s10052-014-2861-z",
      eprint         = "1310.5238",
      archivePrefix  = "arXiv",
      primaryClass   = "hep-lat",
      reportNumber   = "LPT-13-74",
      SLACcitation   = "%%CITATION = ARXIV:1310.5238;%%"
}

@article{Aubert:2009ac,
      author         = "Aubert, Bernard and others",
      title          = "{Measurement of $|V(cb)|$ and the Form-Factor Slope in
                        $\overline{B}\rightarrow D \ell^- \overline{\nu}_\ell$ Decays in Events Tagged by a
                        Fully Reconstructed $B$ Meson}",
      collaboration  = "BaBar",
      journal        = "Phys. Rev. Lett.",
      volume         = "104",
      year           = "2010",
      pages          = "011802",
      doi            = "10.1103/PhysRevLett.104.011802",
      eprint         = "0904.4063",
      archivePrefix  = "arXiv",
      primaryClass   = "hep-ex",
      reportNumber   = "BABAR-PUB-09-009, SLAC-PUB-13580",
      SLACcitation   = "%%CITATION = ARXIV:0904.4063;%%"
}

@article{Aubert:2009wt,
      author         = "Aubert, Bernard and others",
      title          = "{A search for $B^+ \to \ell^+ \nu_{\ell}$ recoiling
                        against $B^{-}\to D^{0} \ell^{-}\bar{\nu} X$}",
      collaboration  = "Babar",
      journal        = "Phys.Rev.",
      volume         = "D81",
      pages          = "051101",
      doi            = "10.1103/PhysRevD.81.051101",
      year           = "2010",
      eprint         = "0912.2453",
      archivePrefix  = "arXiv",
      primaryClass   = "hep-ex",
      reportNumber   = "SLAC-PUB-13866, BABAR-09-029, BABAR-PUB-09-029",
      SLACcitation   = "%%CITATION = ARXIV:0912.2453;%%",
}

@Article{Aubin:2003mg,
     author    = "Aubin, C. and Bernard, C.",
     title     = "{Pion and kaon masses in staggered chiral perturbation
                  theory}",
     journal   = "Phys. Rev.",
     volume    = "D68",
     year      = "2003",
     pages     = "034014",
     eprint    = "hep-lat/0304014",
     archivePrefix = "arXiv",
     doi       = "10.1103/PhysRevD.68.034014",
     SLACcitation  = "%%CITATION = HEP-LAT/0304014;%%"
}

@Article{Aubin:2003uc,
     author    = "Aubin, C. and Bernard, C.",
     title     = "{Pseudoscalar decay constants in staggered chiral
                  perturbation theory}",
     journal   = "Phys. Rev.",
     volume    = "D68",
     year      = "2003",
     pages     = "074011",
     eprint    = "hep-lat/0306026",
     archivePrefix = "arXiv",
     doi       = "10.1103/PhysRevD.68.074011",
     SLACcitation  = "%%CITATION = HEP-LAT/0306026;%%"
}

@Article{Aubin:2004ck,
     author    = "{[HPQCD/MILC/UKQCD 04] C. Aubin} and others",
     title     = "{First determination of the strange and light quark masses
                  from full  lattice QCD}",
     journal   = "Phys. Rev.",
     volume    = "D70",
     year      = "2004",
     pages     = "031504",
     eprint    = "hep-lat/0405022",
     archivePrefix = "arXiv",
     doi       = "10.1103/PhysRevD.70.031504",
     SLACcitation  = "%%CITATION = HEP-LAT/0405022;%%"
}

@article{Aubin:2004ej,
      author         = "{[FNAL/MILC 04] C. Aubin} and others",
      title          = "{Semileptonic decays of D mesons in three-flavor lattice
                        QCD}",
      journal        = "Phys.Rev.Lett.",
      volume         = "94",
      pages          = "011601",
      doi            = "10.1103/PhysRevLett.94.011601",
      year           = "2005",
      eprint         = "hep-ph/0408306",
      archivePrefix  = "arXiv",
      primaryClass   = "hep-ph",
      reportNumber   = "FERMILAB-PUB-04-195-T",
      SLACcitation   = "%%CITATION = HEP-PH/0408306;%%",
}

@Article{Aubin:2004fs,
     author    = "{[MILC 04] C. Aubin} and others",
     title     = "{Light pseudoscalar decay constants, quark masses and low
                  energy  constants from three-flavor lattice QCD}",
     journal   = "Phys. Rev.",
     volume    = "D70",
     year      = "2004",
     pages     = "114501",
     eprint    = "hep-lat/0407028",
     archivePrefix = "arXiv",
     doi       = "10.1103/PhysRevD.70.114501",
     SLACcitation  = "%%CITATION = HEP-LAT/0407028;%%"
}

@Article{Aubin:2004wf,
     author    = "Aubin, C. and others",
     title     = "{Light hadrons with improved staggered quarks: Approaching
                  the continuum  limit}",
     journal   = "Phys. Rev.",
     volume    = "D70",
     year      = "2004",
     pages     = "094505",
     eprint    = "hep-lat/0402030",
     archivePrefix = "arXiv",
     doi       = "10.1103/PhysRevD.70.094505",
     SLACcitation  = "%%CITATION = HEP-LAT/0402030;%%"
}

@Article{Aubin:2005aq,
     author    = "Aubin, C. and Bernard, C.",
     title     = "{Staggered chiral perturbation theory for heavy-light
                  mesons}",
     journal   = "Phys. Rev.",
     volume    = "D73",
     year      = "2006",
     pages     = "014515",
     eprint    = "hep-lat/0510088",
     archivePrefix = "arXiv",
     doi       = "10.1103/PhysRevD.73.014515",
     SLACcitation  = "%%CITATION = HEP-LAT/0510088;%%"
}

@article{Aubin:2005ar,
      author         = "{[FNAL/MILC 05] C. Aubin} and Bernard, C. and DeTar, Carleton E. and Di
                        Pierro, M. and Freeland, Elizabeth Dawn and others",
      title          = "{Charmed meson decay constants in three-flavor lattice
                        QCD}",
      journal        = "Phys.Rev.Lett.",
      volume         = "95",
      pages          = "122002",
      doi            = "10.1103/PhysRevLett.95.122002",
      year           = "2005",
      eprint         = "hep-lat/0506030",
      archivePrefix  = "arXiv",
      primaryClass   = "hep-lat",
      reportNumber   = "FERMILAB-PUB-05-257-T",
      SLACcitation   = "%%CITATION = HEP-LAT/0506030;%%",
}

@Article{Aubin:2008ie,
     author    = "{C. Aubin} and Laiho, Jack and Van de Water, Ruth S.",
     title     = "{Light pseudoscalar meson masses and decay constants from
                  mixed action lattice QCD}",
     journal   = "PoS",
     volume    = "LAT2008",
     year      = "2008",
     pages     = "105",
     eprint    = "0810.4328",
     archivePrefix = "arXiv",
     primaryClass  =  "hep-lat",
     comment = "Aubin 08",
     SLACcitation  = "%%CITATION = 0810.4328;%%"
}

@Article{Aubin:2008wk,
     author    = "Aubin, C. and Laiho, Jack and Van de Water, Ruth S.",
     title     = "{Discretization effects and the scalar meson correlator in
                  mixed-action lattice simulations}",
     journal   = "Phys. Rev.",
     volume    = "D77",
     year      = "2008",
     pages     = "114501",
     eprint    = "0803.0129",
     archivePrefix = "arXiv",
     primaryClass  =  "hep-lat",
     doi       = "10.1103/PhysRevD.77.114501",
     SLACcitation  = "%%CITATION = 0803.0129;%%"
}

@Article{Aubin:2009jh,
     author    = "{C. Aubin} and Laiho, Jack and Van de Water, Ruth S.",
     title     = "{The neutral kaon mixing parameter $B_K$ from unquenched
                  mixed-action lattice QCD}",
     journal   = "Phys. Rev.",
     volume    = "D81",
     year      = "2010",
     pages     = "014507",
     eprint    = "0905.3947",
     archivePrefix = "arXiv",
     primaryClass  =  "hep-lat",
     doi       = "10.1103/PhysRevD.81.014507",
     comment = "Aubin 09",
     SLACcitation  = "%%CITATION = 0905.3947;%%"
}

@Article{Auerbach:2008ut,
     author    = "Auerbach, N.",
     title     = "{Coulomb corrections to superallowed $\beta$ decay in
                  nuclei}",
     journal   = "Phys. Rev.",
     volume    = "C79",
     year      = "2009",
     pages     = "035502",
     eprint    = "0811.4742",
     archivePrefix = "arXiv",
     primaryClass  =  "nucl-th",
     doi       = "10.1103/PhysRevC.79.035502",
     SLACcitation  = "%%CITATION = 0811.4742;%%"
}

@article{Babich:2006bh,
      author         = "Babich, Ronald and Garron, Nicolas and Hoelbling,
                        Christian and Howard, Joseph and Lellouch, Laurent and
                        Rebbi, Claudio",
      title          = "{K0 -- anti-K0 mixing beyond the standard model and
                        CP-violating electroweak penguins in quenched QCD with
                        exact chiral symmetry}",
      journal        = "Phys. Rev.",
      volume         = "D74",
      year           = "2006",
      pages          = "073009",
      doi            = "10.1103/PhysRevD.74.073009",
      eprint         = "hep-lat/0605016",
      archivePrefix  = "arXiv",
      primaryClass   = "hep-lat",
      reportNumber   = "CPT-P07-2006",
      SLACcitation   = "%%CITATION = HEP-LAT/0605016;%%"
}

@article{Babich:2010qb,
      author         = "Babich, R. and Brannick, J. and Brower, R. C. and Clark,
                        M. A. and Manteuffel, T. A. and McCormick, S. F. and
                        Osborn, J. C. and Rebbi, C.",
      title          = "{Adaptive multigrid algorithm for the lattice
                        Wilson-Dirac operator}",
      journal        = "Phys. Rev. Lett.",
      volume         = "105",
      year           = "2010",
      pages          = "201602",
      doi            = "10.1103/PhysRevLett.105.201602",
      eprint         = "1005.3043",
      archivePrefix  = "arXiv",
      primaryClass   = "hep-lat",
      SLACcitation   = "%%CITATION = ARXIV:1005.3043;%%"
}

@Article{Bae:2010ki,
     author    = "{[SWME 10] T. Bae} and others",
     title     = "{$B_K$ using HYP-smeared staggered fermions in $N_f=2+1$
                  unquenched QCD}",
     journal   = "Phys. Rev.",
     volume    = "D82",
     year      = "2010",
     pages     = "114509",
     eprint    = "1008.5179",
     archivePrefix = "arXiv",
     primaryClass  =  "hep-lat",
     doi       = "10.1103/PhysRevD.82.114509",
     SLACcitation  = "%%CITATION = 1008.5179;%%"
}

@article{Bae:2011ff,
      author         = "{[SWME 11A] T. Bae} and others",
      title          = "{Kaon $B$-parameter from improved staggered fermions in
                        $N_f=2+1$ QCD}",
      journal        = "Phys.Rev.Lett.",
      volume         = "109",
      pages          = "041601",
      doi            = "10.1103/PhysRevLett.109.041601",
      year           = "2012",
      eprint         = "1111.5698",
      archivePrefix  = "arXiv",
      primaryClass   = "hep-lat",
      SLACcitation   = "%%CITATION = ARXIV:1111.5698;%%",
}

@article{Bae:2013lja,
     author    = "{[SWME 13] T. Bae} and others",
      title          = "{Update on $B_K$ and $\varepsilon_K$ with staggered
                        quarks}",
      journal        = "PoS",
      volume         = "LATTICE2013",
      pages          = "476",
      year           = "2013",
      eprint         = "1310.7319",
      archivePrefix  = "arXiv",
      primaryClass   = "hep-lat",
      SLACcitation   = "%%CITATION = ARXIV:1310.7319;%%",
}

@article{Bae:2013tca,
      author         = "{[SWME 13A] T. Bae} and others",
      title          = "{Neutral kaon mixing from new physics: matrix elements in
                        $N_f=2+1$ lattice QCD}",
      journal        = "Phys. Rev.",
      volume         = "D88",
      year           = "2013",
      number         = "7",
      pages          = "071503",
      doi            = "10.1103/PhysRevD.88.071503",
      eprint         = "1309.2040",
      archivePrefix  = "arXiv",
      primaryClass   = "hep-lat",
      reportNumber   = "LA-UR-13-26958",
      SLACcitation   = "%%CITATION = ARXIV:1309.2040;%%"
}

@article{Bae:2014sja,
      author         = "{[SWME 14] T. Bae} and others",
      title          = "{Improved determination of $B_K$ with staggered quarks}",
      journal        = "Phys. Rev.",
      volume         = "D89",
      year           = "2014",
      number         = "7",
      pages          = "074504",
      doi            = "10.1103/PhysRevD.89.074504",
      eprint         = "1402.0048",
      archivePrefix  = "arXiv",
      primaryClass   = "hep-lat",
      SLACcitation   = "%%CITATION = ARXIV:1402.0048;%%"
}

@inproceedings{Bahr:2014iqa,
      author         = "{[ALPHA 14B] F. Bahr} and Bernardoni, Fabio and Bulava, John and
                        Joseph, Anosh and Ramos, Alberto and Simma, Hubert and
                        Sommer, Rainer",
      title          = "{Form factors for $\mathrm B_\mathrm s \to \mathrm K \ell
                        \nu$ decays in Lattice QCD}",
      booktitle      = "{8th International Workshop on the CKM Unitarity Triangle
                        (CKM2014) Vienna, Austria, September 8-12, 2014}",
      url            = "https://inspirehep.net/record/1328088/files/arXiv:1411.3916.pdf",
      year           = "2014",
      eprint         = "1411.3916",
      archivePrefix  = "arXiv",
      primaryClass   = "hep-lat",
      reportNumber   = "DESY-14-217, SFB-CPP-14-90",
      SLACcitation   = "%%CITATION = ARXIV:1411.3916;%%"
}

@article{Bai:2014cva,
      author         = "Bai, Z. and Christ, N. H. and Izubuchi, T. and Sachrajda,
                        C. T. and Soni, A. and Yu, J.",
      title          = "{$K_L-K_S$ Mass Difference from Lattice QCD}",
      journal        = "Phys. Rev. Lett.",
      volume         = "113",
      year           = "2014",
      pages          = "112003",
      doi            = "10.1103/PhysRevLett.113.112003",
      eprint         = "1406.0916",
      archivePrefix  = "arXiv",
      primaryClass   = "hep-lat",
      SLACcitation   = "%%CITATION = ARXIV:1406.0916;%%"
}

@article{Bai:2015nea,
      author         = "{[RBC/UKQCD 15G] Z.~Bai} and others",
      title          = "{Standard Model Prediction for Direct CP Violation in
                        {$K \to \pi \pi$} Decay}",
      journal        = "Phys. Rev. Lett.",
      volume         = "115",
      year           = "2015",
      number         = "21",
      pages          = "212001",
      doi            = "10.1103/PhysRevLett.115.212001",
      eprint         = "1505.07863",
      archivePrefix  = "arXiv",
      primaryClass   = "hep-lat",
      reportNumber   = "RBRC-1141",
      SLACcitation   = "%%CITATION = ARXIV:1505.07863;%%"
}

@article{Bai:2016gzv,
      author         = "Bai, Ziyuan",
      title          = "{Long distance part of $\epsilon_K$ from lattice QCD}",
      booktitle      = "{Proceedings, 34th International Symposium on Lattice
                        Field Theory (Lattice 2016): Southampton, UK, July 24-30,
                        2016}",
      journal        = "PoS",
      volume         = "LATTICE2016",
      year           = "2017",
      pages          = "309",
      doi            = "10.22323/1.256.0309",
      eprint         = "1611.06601",
      archivePrefix  = "arXiv",
      primaryClass   = "hep-lat",
      SLACcitation   = "%%CITATION = ARXIV:1611.06601;%%"
}

@article{Baikov:2008jh,
      author         = "Baikov, P. A. and Chetyrkin, K. G. and Kuhn, Johann H.",
      title          = "{Order $\alpha_s^4$ QCD Corrections to Z and tau Decays}",
      journal        = "Phys. Rev. Lett.",
      volume         = "101",
      year           = "2008",
      pages          = "012002",
      doi            = "10.1103/PhysRevLett.101.012002",
      eprint         = "0801.1821",
      archivePrefix  = "arXiv",
      primaryClass   = "hep-ph",
      reportNumber   = "SFB-CPP-08-04, TTP08-01",
      SLACcitation   = "%%CITATION = ARXIV:0801.1821;%%"
}

@article{Baikov:2016tgj,
      author         = "Baikov, P. A. and Chetyrkin, K. G. and Kuhn, J. H.",
      title          = "{Five-Loop Running of the QCD coupling constant}",
      journal        = "Phys. Rev. Lett.",
      volume         = "118",
      year           = "2017",
      number         = "8",
      pages          = "082002",
      doi            = "10.1103/PhysRevLett.118.082002",
      eprint         = "1606.08659",
      archivePrefix  = "arXiv",
      primaryClass   = "hep-ph",
      reportNumber   = "TTP16-026",
      SLACcitation   = "%%CITATION = ARXIV:1606.08659;%%"
}

@article{Bailey:2007iq,
      author         = "Bailey, Jon A.",
      title          = "{Staggered heavy baryon chiral perturbation theory}",
      journal        = "Phys.Rev.",
      volume         = "D77",
      pages          = "054504",
      doi            = "10.1103/PhysRevD.77.054504",
      year           = "2008",
      eprint         = "0704.1490",
      archivePrefix  = "arXiv",
      primaryClass   = "hep-lat",
      SLACcitation   = "%%CITATION = ARXIV:0704.1490;%%",
}

@article{Bailey:2008wp,
      author         = "{[FNAL/MILC 08A] J. A. Bailey} and others",
      title          = "{The $B \to \pi \ell \nu$ semileptonic form factor from
                        three-flavor lattice QCD: a model-independent
                        determination of $|V_{ub}|$}",
      journal        = "Phys.Rev.",
      volume         = "D79",
      pages          = "054507",
      doi            = "10.1103/PhysRevD.79.054507",
      year           = "2009",
      eprint         = "0811.3640",
      archivePrefix  = "arXiv",
      primaryClass   = "hep-lat",
      reportNumber   = "FERMILAB-PUB-08-541-T",
      SLACcitation   = "%%CITATION = ARXIV:0811.3640;%%",
}

@article{Bailey:2010gb,
      author         = "{[FNAL/MILC 10] J. A. Bailey} and others",
      title          = "{$B \to D^* \ell \nu$ at zero recoil: an update}",
      journal        = "PoS",
      volume         = "LAT2010",
      pages          = "311",
      year           = "2010",
      eprint         = "1011.2166",
      archivePrefix  = "arXiv",
      primaryClass   = "hep-lat",
      reportNumber   = "FERMILAB-CONF-10-450-T",
      SLACcitation   = "%%CITATION = ARXIV:1011.2166;%%",
}

@article{Bailey:2012rr,
      author         = "{[FNAL/MILC 12C] J. A. Bailey} and others",
      title          = "{$B_s\to D_s/B\to D$ semileptonic form-factor ratios and
                        their application to BR($B^0_s\to \mu^+\mu^-$)}",
      journal        = "Phys.Rev.",
      volume         = "D85",
      pages          = "114502",
      doi            = "10.1103/PhysRevD.85.114502, 10.1103/PhysRevD.86.039904",
      year           = "2012",
      eprint         = "1202.6346",
      archivePrefix  = "arXiv",
      primaryClass   = "hep-lat",
      reportNumber   = "FERMILAB-PUB-12-047-T",
      SLACcitation   = "%%CITATION = ARXIV:1202.6346;%%",
}

@article{Bailey:2012sa,
      author         = "{[FNAL/MILC 12G] J. A. Bailey} and others",
      title          = "{Charm semileptonic decays and $|V_{cs(d)}|$ from heavy
                        clover quarks and 2+1 flavor asqtad staggered ensembles}",
      journal        = "PoS",
      volume         = "LAT2012",
      pages          = "272",
      year           = "2012",
      eprint         = "1211.4964",
      archivePrefix  = "arXiv",
      primaryClass   = "hep-lat",
      reportNumber   = "FERMILAB-CONF-12-610-T",
      SLACcitation   = "%%CITATION = ARXIV:1211.4964;%%",
}

@article{Bailey:2012wb,
      author         = "Bailey, Jon A. and Kim, Hyung-Jin and Lee, Weonjong and
                        Sharpe, Stephen R.",
      title          = "{Kaon mixing matrix elements from
                        beyond-the-Standard-Model operators in staggered chiral
                        perturbation theory}",
      journal        = "Phys. Rev.",
      volume         = "D85",
      year           = "2012",
      pages          = "074507",
      doi            = "10.1103/PhysRevD.85.074507",
      eprint         = "1202.1570",
      archivePrefix  = "arXiv",
      primaryClass   = "hep-lat",
      SLACcitation   = "%%CITATION = ARXIV:1202.1570;%%"
}

@article{Bailey:2014tva,
      author         = "{[FNAL/MILC 14] J. A. Bailey} and others",
      title          = "{Update of $|V_{cb}|$ from the $\bar{B}\to
                        D^*\ell\bar{\nu}$ form factor at zero recoil with
                        three-flavor lattice QCD}",
      journal        = "Phys. Rev.",
      volume         = "D89",
      year           = "2014",
      number         = "11",
      pages          = "114504",
      doi            = "10.1103/PhysRevD.89.114504",
      eprint         = "1403.0635",
      archivePrefix  = "arXiv",
      primaryClass   = "hep-lat",
      reportNumber   = "FERMILAB-PUB-14-039-T",
      SLACcitation   = "%%CITATION = ARXIV:1403.0635;%%"
}

@article{Bailey:2015dka,
      author         = "{[FNAL/MILC 15D] J. A. Bailey} and others",
      title          = "{$B\to Kl^+l^-$ decay form factors from three-flavor
                        lattice QCD}",
      journal        = "Phys. Rev.",
      volume         = "D93",
      year           = "2016",
      number         = "2",
      pages          = "025026",
      doi            = "10.1103/PhysRevD.93.025026",
      eprint         = "1509.06235",
      archivePrefix  = "arXiv",
      primaryClass   = "hep-lat",
      reportNumber   = "FERMILAB-PUB-15-403-T",
      SLACcitation   = "%%CITATION = ARXIV:1509.06235;%%"
}

@article{Bailey:2015nbd,
      author         = "{[FNAL/MILC 15E] J. A. Bailey} and others",
      title          = "{$B\to\pi\ell\ell$ form factors for new-physics searches
                        from lattice QCD}",
      journal        = "Phys. Rev. Lett.",
      volume         = "115",
      year           = "2015",
      number         = "15",
      pages          = "152002",
      doi            = "10.1103/PhysRevLett.115.152002",
      eprint         = "1507.01618",
      archivePrefix  = "arXiv",
      primaryClass   = "hep-ph",
      reportNumber   = "FERMILAB-PUB-15-288-T",
      SLACcitation   = "%%CITATION = ARXIV:1507.01618;%%"
}

@article{Bailey:2015tba,
      author         = "{[SWME 15B] J. A. Bailey} and Jang, Yong-Chull and Lee, Weonjong and
                        Park, Sungwoo",
      title          = "{Standard Model evaluation of $\varepsilon_K$ using
                        lattice QCD inputs for $\hat{B}_K$ and $V_{cb}$}",
      journal        = "Phys. Rev.",
      volume         = "D92",
      year           = "2015",
      number         = "3",
      pages          = "034510",
      doi            = "10.1103/PhysRevD.92.034510",
      eprint         = "1503.05388",
      archivePrefix  = "arXiv",
      primaryClass   = "hep-lat",
      SLACcitation   = "%%CITATION = ARXIV:1503.05388;%%"
}

@article{Bailey:2018feb,
      author         = "{[SWME 18] J.A. Bailey} and others", 
      title          = "{Updated evaluation of $\epsilon_K$ in the standard model
                        with lattice QCD inputs}",
      journal        = "Phys. Rev.",
      volume         = "D98",
      year           = "2018",
      number         = "9",
      pages          = "094505",
      doi            = "10.1103/PhysRevD.98.094505",
      eprint         = "1808.09657",
      archivePrefix  = "arXiv",
      primaryClass   = "hep-lat",
      SLACcitation   = "%%CITATION = ARXIV:1808.09657;%%"
}

@article{Baker:2006ts,
      author         = "Baker, C.A. and Doyle, D.D. and Geltenbort, P. and Green,
                        K. and van der Grinten, M.G.D. and others",
      title          = "{An Improved experimental limit on the electric dipole
                        moment of the neutron}",
      journal        = "Phys.Rev.Lett.",
      volume         = "97",
      pages          = "131801",
      doi            = "10.1103/PhysRevLett.97.131801",
      year           = "2006",
      eprint         = "hep-ex/0602020",
      archivePrefix  = "arXiv",
      primaryClass   = "hep-ex",
      SLACcitation   = "%%CITATION = HEP-EX/0602020;%%",
}

@article{Abel:2020pzs,
    author = "Abel, C. and others",
    title = "{Measurement of the Permanent Electric Dipole Moment of the Neutron}",
    eprint = "2001.11966",
    archivePrefix = "arXiv",
    primaryClass = "hep-ex",
    doi = "10.1103/PhysRevLett.124.081803",
    journal = "Phys. Rev. Lett.",
    volume = "124",
    number = "8",
    pages = "081803",
    year = "2020"
}

@article{Bali:1992ru,
      author         = "{G. S. Bali} and Schilling, Klaus",
      title          = "{Running coupling and the $\Lambda$-parameter from $SU(3)$
                        lattice simulations}",
      journal        = "Phys.Rev.",
      volume         = "D47",
      pages          = "661-672",
      doi            = "10.1103/PhysRevD.47.661",
      year           = "1993",
      eprint         = "hep-lat/9208028",
      archivePrefix  = "arXiv",
      primaryClass   = "hep-lat",
      reportNumber   = "WUB-92-29",
      comment = "Bali 92",
      SLACcitation   = "%%CITATION = HEP-LAT/9208028;%%",

}

@article{Bali:2009hu,
      author         = "Bali, Gunnar S. and Collins, Sara and Schafer, Andreas",
      title          = "{Effective noise reduction techniques for disconnected
                        loops in Lattice QCD}",
      journal        = "Comput. Phys. Commun.",
      volume         = "181",
      year           = "2010",
      pages          = "1570-1583",
      doi            = "10.1016/j.cpc.2010.05.008",
      eprint         = "0910.3970",
      archivePrefix  = "arXiv",
      primaryClass   = "hep-lat",
      SLACcitation   = "%%CITATION = ARXIV:0910.3970;%%"
}

@article{Bali:2011ks,
      author         = "{[QCDSF 11] G. S. Bali} and others",
      title          = "{The strange and light quark contributions to the nucleon
                        mass from Lattice QCD}",
      journal        = "Phys. Rev.",
      volume         = "D85",
      year           = "2012",
      pages          = "054502",
      doi            = "10.1103/PhysRevD.85.054502",
      eprint         = "1111.1600",
      archivePrefix  = "arXiv",
      primaryClass   = "hep-lat",
      reportNumber   = "ADELAIDE-ADP-11-33-T755, EDINBURGH-2011-33,
                        LIVERPOOL-LTH-930, DESY-11-245",
      SLACcitation   = "%%CITATION = ARXIV:1111.1600;%%"
}

@article{Bali:2012qs,
      author         = "{[QCDSF 12] G. Bali} and Bruns, P.C. and Collins, S. and Deka, M.
                        and Glasle, B. and others",
      title          = "{Nucleon mass and sigma term from lattice QCD with two
                        light fermion flavors}",
      journal        = "Nucl.Phys.",
      volume         = "B866",
      pages          = "1-25",
      doi            = "10.1016/j.nuclphysb.2012.08.009",
      year           = "2013",
      eprint         = "1206.7034",
      archivePrefix  = "arXiv",
      primaryClass   = "hep-lat",
      reportNumber   = "ADP-12-29-T796, DESY-12-105, EDINBURGH-2012-11",
      SLACcitation   = "%%CITATION = ARXIV:1206.7034;%%",
}

@article{Bali:2014gha,
      author         = "{[RQCD 14A] G. Bali} and Collins, Sara and Gläßle, Benjamin
                        and G{\"o}ckeler, Meinulf and Najjar, Johannes and R{\"o}dl,
                        Rudolf H. and Schäfer, Andreas and Schiel, Rainer W. and
                        Sternbeck, André and Söldner, Wolfgang",
      title          = "{The moment $\langle x\rangle_{u-d}$ of the nucleon from
                        $N_f=2$ lattice QCD down to nearly physical quark masses}",
      journal        = "Phys. Rev.",
      volume         = "D90",
      year           = "2014",
      number         = "7",
      pages          = "074510",
      doi            = "10.1103/PhysRevD.90.074510",
      eprint         = "1408.6850",
      archivePrefix  = "arXiv",
      primaryClass   = "hep-lat",
      SLACcitation   = "%%CITATION = ARXIV:1408.6850;%%"
}

@article{Bali:2014nma,
      author         = "{[RQCD 14] G.~S. Bali} and Collins, Sara and Gl{\"a}ssle, Benjamin
                        and G{\"o}ckeler, Meinulf and Najjar, Johannes and R{\"o}dl,
                        Rudolf H. and Sch{\"a}fer, Andreas and Schiel, Rainer W. and
                        S{\"o}ldner, Wolfgang and Sternbeck, Andr\'e",
      title          = "{Nucleon isovector couplings from $N_f=2$ lattice QCD}",
      journal        = "Phys. Rev.",
      volume         = "D91",
      year           = "2015",
      number         = "5",
      pages          = "054501",
      doi            = "10.1103/PhysRevD.91.054501",
      eprint         = "1412.7336",
      archivePrefix  = "arXiv",
      primaryClass   = "hep-lat",
      SLACcitation   = "%%CITATION = ARXIV:1412.7336;%%"
}

@article{Bali:2016lvx,
      author         = "{[RQCD 16] G.~S. Bali} and Collins, Sara and Richtmann, Daniel
                        and Sch{\"a}fer, Andreas and S{\"o}ldner, Wolfgang and
                        Sternbeck, Andr\'e",
      title          = "{Direct determinations of the nucleon and pion $\sigma$
                        terms at nearly physical quark masses}",
      journal        = "Phys. Rev.",
      volume         = "D93",
      year           = "2016",
      number         = "9",
      pages          = "094504",
      doi            = "10.1103/PhysRevD.93.094504",
      eprint         = "1603.00827",
      archivePrefix  = "arXiv",
      primaryClass   = "hep-lat",
      SLACcitation   = "%%CITATION = ARXIV:1603.00827;%%"
}

@article{Bali:2016umi,
      author         = "{[RQCD 16A] G. S. Bali} and Scholz, Enno E. and Simeth, Jakob and
                        Söldner, Wolfgang",
      title          = "{Lattice simulations with $N_f=2+1$ improved Wilson
                        fermions at a fixed strange quark mass}",
      journal        = "Phys. Rev.",
      volume         = "D94",
      year           = "2016",
      number         = "7",
      pages          = "074501",
      doi            = "10.1103/PhysRevD.94.074501",
      eprint         = "1606.09039",
      archivePrefix  = "arXiv",
      primaryClass   = "hep-lat",
      SLACcitation   = "%%CITATION = ARXIV:1606.09039;%%"
}

@article{Bali:2017jyw,
      author         = "Bali, G. S. and Collins, S. and G{\"o}ckeler, M. and
                        Piemonte, S. and Sternbeck, A.",
      title          = "{Non-perturbative renormalization of flavor singlet quark
                        bilinear operators in lattice QCD}",
      booktitle      = "{Proceedings, 34th International Symposium on Lattice
                        Field Theory (Lattice 2016): Southampton, UK, July 24-30,
                        2016}",
      journal        = "PoS",
      volume         = "LATTICE2016",
      year           = "2016",
      pages          = "187",
      doi            = "10.22323/1.256.0187",
      eprint         = "1703.03745",
      archivePrefix  = "arXiv",
      primaryClass   = "hep-lat",
      SLACcitation   = "%%CITATION = ARXIV:1703.03745;%%"
}

@article{Balitsky:1993ki,
      author         = "Balitsky, I.I. and Beneke, M. and Braun, Vladimir M.",
      title          = "{Instanton contributions to the $\tau$ decay widths}",
      journal        = "Phys.Lett.",
      volume         = "B318",
      pages          = "371-381",
      doi            = "10.1016/0370-2693(93)90142-5",
      year           = "1993",
      eprint         = "hep-ph/9309217",
      archivePrefix  = "arXiv",
      primaryClass   = "hep-ph",
      reportNumber   = "MPI-PH-93-62, PSU-TH-130",
      SLACcitation   = "%%CITATION = HEP-PH/9309217;%%",
}

@article{Ball:2004ye,
      author         = "Ball, Patricia and Zwicky, Roman",
      title          = "{New results on $B \to \pi, K, \eta$ decay form factors
                        from light-cone sum rules}",
      journal        = "Phys.Rev.",
      volume         = "D71",
      pages          = "014015",
      doi            = "10.1103/PhysRevD.71.014015",
      year           = "2005",
      eprint         = "hep-ph/0406232",
      archivePrefix  = "arXiv",
      primaryClass   = "hep-ph",
      reportNumber   = "IPPP-04-23, DCPT-04-46, TPI-MINN-04-25",
      SLACcitation   = "%%CITATION = HEP-PH/0406232;%%",
}

@Article{Banks:1975gq,
     author    = "Banks, Tom and Susskind, Leonard and Kogut, John B.",
     title     = "{Strong coupling calculations of lattice gauge theories:
                  (1+1)-dimensional exercises}",
     journal   = "Phys. Rev.",
     volume    = "D13",
     year      = "1976",
     pages     = "1043",
     doi       = "10.1103/PhysRevD.13.1043",
     SLACcitation  = "%%CITATION = PHRVA,D13,1043;%%"
}

@Article{Bar:2002nr,
     author    = "{B\"ar}, Oliver and Rupak, Gautam and Shoresh, Noam",
     title     = "{Simulations with different lattice Dirac operators for
                  valence and sea quarks}",
     journal   = "Phys. Rev.",
     volume    = "D67",
     year      = "2003",
     pages     = "114505",
     eprint    = "hep-lat/0210050",
     archivePrefix = "arXiv",
     doi       = "10.1103/PhysRevD.67.114505",
     SLACcitation  = "%%CITATION = HEP-LAT/0210050;%%"
}

@Article{Bar:2003mh,
     author    = "{B\"ar}, Oliver and Rupak, Gautam and Shoresh, Noam",
     title     = "{Chiral perturbation theory at $O(a^2)$ for lattice QCD}",
     journal   = "Phys. Rev.",
     volume    = "D70",
     year      = "2004",
     pages     = "034508",
     eprint    = "hep-lat/0306021",
     archivePrefix = "arXiv",
     doi       = "10.1103/PhysRevD.70.034508",
     SLACcitation  = "%%CITATION = HEP-LAT/0306021;%%"
}

@Article{Bar:2005tu,
     author    = "{B\"ar}, Oliver and Bernard, Claude and Rupak, Gautam and
                  Shoresh, Noam",
     title     = "{Chiral perturbation theory for staggered sea quarks and
                  Ginsparg-Wilson  valence quarks}",
     journal   = "Phys. Rev.",
     volume    = "D72",
     year      = "2005",
     pages     = "054502",
     eprint    = "hep-lat/0503009",
     archivePrefix = "arXiv",
     doi       = "10.1103/PhysRevD.72.054502",
     SLACcitation  = "%%CITATION = HEP-LAT/0503009;%%"
}

@Article{Bar:2010jk,
     author    = "{B\"ar}, Oliver",
     title     = "{Chiral logs in twisted mass lattice QCD with large isospin
                  breaking}",
      journal        = "Phys.Rev.",
      volume         = "D82",
      pages          = "094505",
      doi            = "10.1103/PhysRevD.82.094505",
      year           = "2010",
      eprint         = "1008.0784",
      archivePrefix  = "arXiv",
      primaryClass   = "hep-lat",
      reportNumber   = "HU-EP-10-44, SFB-CCP-10-72",
      SLACcitation   = "%%CITATION = ARXIV:1008.0784;%%",
}

@article{Bar:2013ora,
      author         = "B{\"a}r, Oliver and Golterman, Maarten",
      title          = "{Chiral perturbation theory for gradient flow
                        observables}",
      journal        = "Phys. Rev.",
      volume         = "D89",
      year           = "2014",
      number         = "3",
      pages          = "034505",
      doi            = "10.1103/PhysRevD.89.099905, 10.1103/PhysRevD.89.034505",
      note           = "[Erratum: {\it Phys. Rev.} {\bf D89} (2014) 099905]",
      eprint         = "1312.4999",
      archivePrefix  = "arXiv",
      primaryClass   = "hep-lat",
      SLACcitation   = "%%CITATION = ARXIV:1312.4999;%%"
}

@article{Bar:2016jof,
      author         = "B{\"a}r, Oliver",
      title          = "{Nucleon-pion-state contribution in lattice calculations
                        of moments of parton distribution functions}",
      journal        = "Phys. Rev.",
      volume         = "D95",
      year           = "2017",
      number         = "3",
      pages          = "034506",
      doi            = "10.1103/PhysRevD.95.034506",
      eprint         = "1612.08336",
      archivePrefix  = "arXiv",
      primaryClass   = "hep-lat",
      reportNumber   = "YITP-16-143",
      SLACcitation   = "%%CITATION = ARXIV:1612.08336;%%"
}

@article{Bar:2016uoj,
      author         = "B{\"a}r, Oliver",
      title          = "{Nucleon-pion-state contribution in lattice calculations
                        of the nucleon charges $g_A,g_T$ and $g_S$}",
      journal        = "Phys. Rev.",
      volume         = "D94",
      year           = "2016",
      number         = "5",
      pages          = "054505",
      doi            = "10.1103/PhysRevD.94.054505",
      eprint         = "1606.09385",
      archivePrefix  = "arXiv",
      primaryClass   = "hep-lat",
      reportNumber   = "YITP-16-81",
      SLACcitation   = "%%CITATION = ARXIV:1606.09385;%%"
}

@article{Bar:2017gqh,
      author         = "B{\"a}r, Oliver",
      title          = "{Multi-hadron-state contamination in nucleon observables
                        from chiral perturbation theory}",
      booktitle      = "{Proceedings, 35th International Symposium on Lattice
                        Field Theory (Lattice 2017): Granada, Spain, June 18-24,
                        2017}",
      journal        = "EPJ Web Conf.",
      volume         = "175",
      year           = "2018",
      pages          = "01007",
      doi            = "10.1051/epjconf/201817501007",
      eprint         = "1708.00380",
      archivePrefix  = "arXiv",
      primaryClass   = "hep-lat",
      SLACcitation   = "%%CITATION = ARXIV:1708.00380;%%"
}

@Article{Baron:2009wt,
     author    = "{[ETM 09C] R. Baron} and others",
     title     = "{Light meson physics from maximally twisted mass lattice
                  QCD}",
    journal   = "JHEP",
     volume    = "08",
     year      = "2010",
     pages     = "097",
     eprint    = "0911.5061",
     archivePrefix = "arXiv",
     primaryClass  =  "hep-lat",
     doi       = "10.1007/JHEP08(2010)097",
     SLACcitation  = "%%CITATION = 0911.5061;%%"
}

@article{Baron:2010bv,
      author         = "{[ETM 10] R. Baron} and others",
      title          = "{Light hadrons from lattice QCD with light (u,d), strange
                        and charm dynamical quarks}",
      journal        = "JHEP",
      volume         = "1006",
      pages          = "111",
      doi            = "10.1007/JHEP06(2010)111",
      year           = "2010",
      eprint         = "1004.5284",
      archivePrefix  = "arXiv",
      primaryClass   = "hep-lat",
      reportNumber   = "DESY-10-054, HU-EP-10-18, IFIC-10-11, SFB-CPP-10-29,
                        LPT-ORSAY-10-28, LTH873, LPSC1042, MS-TP-10-09,
                        ROM2F-2010-08",
      SLACcitation   = "%%CITATION = ARXIV:1004.5284;%%",
}

@Article{Baron:2011sf,
     author    = "{[ETM 11] R. Baron} and others",
     title     = "{Light hadrons from $N_f=2+1+1$ dynamical twisted mass
                  fermions}",
     journal   = "PoS",
     volume    = "LAT2010",
     year      = "2010",
     pages     = "123",
     eprint    = "1101.0518",
     archivePrefix = "arXiv",
     primaryClass  =  "hep-lat",
     SLACcitation  = "%%CITATION = 1101.0518;%%"
}

@article{Barone:2023tbl,
    author = {Barone, Alessandro and Hashimoto, Shoji and J\"uttner, Andreas and Kaneko, Takashi and Kellermann, Ryan},
    title = "{Approaches to inclusive semileptonic B$_{(s)}$-meson decays from Lattice QCD}",
    eprint = "2305.14092",
    archivePrefix = "arXiv",
    primaryClass = "hep-lat",
    reportNumber = "KEK-CP-0394, CERN-TH-2023-087",
    doi = "10.1007/JHEP07(2023)145",
    journal = "JHEP",
    volume = "07",
    pages = "145",
    year = "2023"
}

@Article{Basak:2008na,
     author    = "{[MILC 08] S. Basak} and others",
     title     = "{Electromagnetic splittings of hadrons from improved
                  staggered quarks in full QCD}",
     journal   = "PoS",
     volume    = "LAT2008",
     year      = "2008",
     pages     = "127",
     eprint    = "0812.4486",
     archivePrefix = "arXiv",
     primaryClass  =  "hep-lat",
     SLACcitation  = "%%CITATION = 0812.4486;%%"
}

@article{Basak:2016jnn,
      author         = "{[MILC 16] S. Basak} and others",
      title          = "{Electromagnetic effects on the light pseudoscalar mesons
                        and determination of $m_u/m_d$}",
      booktitle      = "{Proceedings, 33rd International Symposium on Lattice
                        Field Theory (Lattice 2015): Kobe, Japan, July 14-18,
                        2015}",
      journal        = "PoS",
      volume         = "LATTICE2015",
      year           = "2016",
      pages          = "259",
      eprint         = "1606.01228",
      archivePrefix  = "arXiv",
      primaryClass   = "hep-lat",
      SLACcitation   = "%%CITATION = ARXIV:1606.01228;%%"
}

@article{Basak:2018yzz,
      author         = "{[MILC 18] S.~Basak} and others",
      title          = "{Lattice computation of the electromagnetic contributions
                        to kaon and pion masses}",
      journal        = "Phys. Rev.",
      volume         = "D99",
      year           = "2019",
      pages          = "034503",
      doi            = "10.1103/PhysRevD.99.034503",
      eprint         = "1807.05556",
      archivePrefix  = "arXiv",
      primaryClass   = "hep-lat",
      reportNumber   = "FERMILAB-PUB-18-341-T",
      SLACcitation   = "%%CITATION = ARXIV:1807.05556;%%"
}

@article{Bautista:2015yza,
      author         = "Bautista, Irais and Bietenholz, Wolfgang and Dromard,
                        Arthur and Gerber, Urs and Gonglach, Lukas and Hofmann,
                        Christoph P. and Mej\'ia-D\'iaz, H\'ector and Wagner, Marc",
      title          = "{Measuring the Topological Susceptibility in a Fixed
                        Sector}",
      journal        = "Phys. Rev.",
      volume         = "D92",
      year           = "2015",
      number         = "11",
      pages          = "114510",
      doi            = "10.1103/PhysRevD.92.114510",
      eprint         = "1503.06853",
      archivePrefix  = "arXiv",
      primaryClass   = "hep-lat",
      SLACcitation   = "%%CITATION = ARXIV:1503.06853;%%"
}

@Article{Bazavov:2009bb,
     author    = "{[MILC 09] A. Bazavov} and others",
     title     = "{Full nonperturbative QCD simulations with 2+1 flavors of
                  improved staggered quarks}",
     journal   = "Rev. Mod. Phys.",
     volume    = "82",
     year      = "2010",
     pages     = "1349-1417",
     eprint    = "0903.3598",
     archivePrefix = "arXiv",
     primaryClass  =  "hep-lat",
     SLACcitation  = "%%CITATION = 0903.3598;%%"
}

@Article{Bazavov:2009fk,
     author    = "{[MILC 09A] A. Bazavov} and others",
     title     = "{MILC results for light pseudoscalars}",
     journal   = "PoS",
     volume    = "CD09",
     year      = "2009",
     pages     = "007",
     eprint    = "0910.2966",
     archivePrefix = "arXiv",
     primaryClass  =  "hep-ph",
     SLACcitation  = "%%CITATION = 0910.2966;%%"
}

@Article{Bazavov:2009tw,
     author    = "{[MILC 09B] A. Bazavov} and others",
     title     = "{Results from the MILC collaboration's SU(3) chiral
                  perturbation theory analysis}",
     journal   = "PoS",
     volume    = "LAT2009",
     year      = "2009",
     pages     = "079",
     eprint    = "0910.3618",
     archivePrefix = "arXiv",
     primaryClass  =  "hep-lat",
     SLACcitation  = "%%CITATION = 0910.3618;%%"
}

@Article{Bazavov:2010hj,
     author    = "{[MILC 10] A. Bazavov}  and others",
     title     = "{Results for light pseudoscalar mesons}",
     journal   = "PoS",
     volume    = "LAT2010",
     year      = "2010",
     pages     = "074",
     eprint    = "1012.0868",
     archivePrefix = "arXiv",
     primaryClass  =  "hep-lat",
     SLACcitation  = "%%CITATION = 1012.0868;%%"
}

@Article{Bazavov:2010xr,
     author    = "{[MILC 10] A. Bazavov} and others", 
     title     = "{Topological susceptibility with the asqtad action}",
     journal   = "Phys. Rev.",
     volume    = "D81",
     year      = "2010",
     pages     = "114501",
     eprint    = "1003.5695",
     archivePrefix = "arXiv",
     primaryClass  =  "hep-lat",
     doi       = "10.1103/PhysRevD.81.114501",
     SLACcitation  = "%%CITATION = 1003.5695;%%"
}

@Article{Bazavov:2010yq,
     author    = "{[MILC 10A] A. Bazavov} and others",
     title     = "{Staggered chiral perturbation theory in the two-flavor
                  case and SU(2) analysis of the MILC data}",
     journal   = "PoS",
     volume    = "LAT2010",
     year      = "2010",
     pages     = "083",
     eprint    = "1011.1792",
     archivePrefix = "arXiv",
     primaryClass  =  "hep-lat",
     SLACcitation  = "%%CITATION = 1011.1792;%%"
}

@article{Bazavov:2011aa,
      author         = "{[FNAL/MILC 11] A. Bazavov} and others",
      title          = "{$B$- and $D$-meson decay constants from three-flavor lattice
                        QCD}",
      journal        = "Phys.Rev.",
      volume         = "D85",
      pages          = "114506",
      doi            = "10.1103/PhysRevD.85.114506",
      year           = "2012",
      eprint         = "1112.3051",
      archivePrefix  = "arXiv",
      primaryClass   = "hep-lat",
      SLACcitation   = "%%CITATION = ARXIV:1112.3051;%%",
}

@article{Bazavov:2011fh,
      author         = "{[MILC 11] A. Bazavov} and others",
      title          = "{Properties of light pseudoscalars from lattice QCD with
                        HISQ ensembles}",
      journal        = "PoS",
      volume         = "LAT2011",
      pages          = "107",
      year           = "2011",
      eprint         = "1111.4314",
      archivePrefix  = "arXiv",
      primaryClass   = "hep-lat",
      SLACcitation   = "%%CITATION = ARXIV:1111.4314;%%",
}

@article{Bazavov:2011nk,
      author         = "{[HotQCD 11] A. Bazavov} and Bhattacharya, T. and Cheng, M.
                        and DeTar, C. and Ding, H.T. and others",
      title          = "{The chiral and deconfinement aspects of the QCD
                        transition}",
      journal        = "Phys.Rev.",
      volume         = "D85",
      pages          = "054503",
      doi            = "10.1103/PhysRevD.85.054503",
      year           = "2012",
      eprint         = "1111.1710",
      archivePrefix  = "arXiv",
      primaryClass   = "hep-lat",
      SLACcitation   = "%%CITATION = ARXIV:1111.1710;%%",
}

@article{Bazavov:2012cd,
      author         = "{[FNAL/MILC 12I] A. Bazavov} and Bernard, C. and Bouchard, C.M. and DeTar,
                        C. and Du, Daping and others",
      title          = "{Kaon semileptonic vector form factor and determination
                        of $|V_{us}|$ using staggered fermions}",
      journal        = "Phys.Rev.",
      volume         = "D87",
      pages          = "073012",
      doi            = "10.1103/PhysRevD.87.073012",
      year           = "2013",
      eprint         = "1212.4993",
      archivePrefix  = "arXiv",
      primaryClass   = "hep-lat",
      SLACcitation   = "%%CITATION = ARXIV:1212.4993;%%",
}

@article{Bazavov:2012dg,
      author         = "{[FNAL/MILC 12B] A. Bazavov} and others",
      title          = "{Pseudoscalar meson physics with four dynamical quarks}",
      journal        = "PoS",
      volume         = "LAT2012",
      pages          = "159",
      year           = "2012",
      eprint         = "1210.8431",
      archivePrefix  = "arXiv",
      primaryClass   = "hep-lat",
      reportNumber   = "FERMILAB-CONF-12-601-T",
      SLACcitation   = "%%CITATION = ARXIV:1210.8431;%%",
}

@article{Bazavov:2012ka,
      author         = "{A. Bazavov} and Brambilla, Nora and Garcia i Tormo,
                        Xavier and Petreczky, Peter and Soto, Joan and others",
      title          = "{Determination of $\alpha_s$ from the QCD static
                         energy}",
      journal        = "Phys.Rev.",
      volume         = "D86",
      pages          = "114031",
      doi            = "10.1103/PhysRevD.86.114031",
      year           = "2012",
      eprint         = "1205.6155",
      archivePrefix  = "arXiv",
      primaryClass   = "hep-ph",
      reportNumber   = "TUM-EFT-31-12, UB-ECM-PF-11-71, ICCUB-12-122",
      comment = "Bazavov 12",
      SLACcitation   = "%%CITATION = ARXIV:1205.6155;%%",
}

@article{Bazavov:2012xda,
      author         = "{[MILC 12B] A. Bazavov} and others",
      title          = "{Lattice QCD ensembles with four flavors of highly
                        improved staggered quarks}",      
      journal        = "Phys.Rev.",
      volume         = "D87",
      pages          = "054505",
      doi            = "10.1103/PhysRevD.87.054505",
      year           = "2013",
      eprint         = "1212.4768",
      archivePrefix  = "arXiv",
      primaryClass   = "hep-lat",
      reportNumber   = "FERMILAB-PUB-12-796-T",
      SLACcitation   = "%%CITATION = ARXIV:1212.4768;%%",
}

@article{Bazavov:2012zs,
      author         = "{[FNAL/MILC 12] A. Bazavov} and Bernard, C. and Bouchard, C.M. and DeTar,
                        C. and Di Pierro, M. and others",
      title          = "{Neutral B-meson mixing from three-flavor lattice QCD:
                        determination of the SU(3)-breaking ratio $\xi$}",
      journal        = "Phys.Rev.",
      volume         = "D86",
      pages          = "034503",
      doi            = "10.1103/PhysRevD.86.034503",
      year           = "2012",
      eprint         = "1205.7013",
      archivePrefix  = "arXiv",
      primaryClass   = "hep-lat",
      reportNumber   = "FERMILAB-PUB-12-258-PPD",
      SLACcitation   = "%%CITATION = ARXIV:1205.7013;%%",
}

@article{Bazavov:2013cp,
      author         = "{[MILC 13A] A. Bazavov} and Bernard, C. and DeTar, C. and Foley, J.
                        and Freeman, W. and others",
      title          = "{Leptonic decay-constant ratio {$f_{K^+}/f_{\pi^+}$} from
                        lattice QCD with physical light quarks}",
      journal        = "Phys.Rev.Lett.",
      volume         = "110",
      pages          = "172003",
      doi            = "10.1103/PhysRevLett.110.172003",
      year           = "2013",
      eprint         = "1301.5855",
      archivePrefix  = "arXiv",
      primaryClass   = "hep-ph",
      SLACcitation   = "%%CITATION = ARXIV:1301.5855;%%",
}

@article{Bazavov:2013maa,
      author         = "{[FNAL/MILC 13E] A. Bazavov} and others",
      title          = "{Determination of $|V_{us}|$ from a lattice-QCD
                        calculation of the $K\to\pi\ell\nu$ semileptonic form
                        factor with physical quark masses}",
      journal        = "Phys. Rev. Lett.",
      volume         = "112",
      year           = "2014",
      number         = "11",
      pages          = "112001",
      doi            = "10.1103/PhysRevLett.112.112001",
      eprint         = "1312.1228",
      archivePrefix  = "arXiv",
      primaryClass   = "hep-ph",
      reportNumber   = "FERMILAB-PUB-13-504-T",
      SLACcitation   = "%%CITATION = ARXIV:1312.1228;%%"
}

@article{Bazavov:2013nfa,
      author         = "{[FNAL/MILC 13] A. Bazavov} and others",
      title          = "{Charmed and strange pseudoscalar meson decay constants
                        from HISQ simulations}",
      booktitle      = "{Proceedings, 31st International Symposium on Lattice
                        Field Theory (Lattice 2013)}",
      journal        = "PoS",
      volume         = "LATTICE2013",
      year           = "2014",
      pages          = "405",
      eprint         = "1312.0149",
      archivePrefix  = "arXiv",
      primaryClass   = "hep-lat",
      reportNumber   = "FERMILAB-CONF-13-562-T",
      SLACcitation   = "%%CITATION = ARXIV:1312.0149;%%"
}

@article{Bazavov:2014pvz,
      author         = "{[HotQCD 14] A. Bazavov} and others",
      title          = "{Equation of state in (2+1 )-flavor QCD}",
      journal        = "Phys.Rev.",
      number         = "9",
      volume         = "D90",
      pages          = "094503",
      doi            = "10.1103/PhysRevD.90.094503",
      year           = "2014",
      eprint         = "1407.6387",
      archivePrefix  = "arXiv",
      primaryClass   = "hep-lat",
      reportNumber   = "BNL-105928-2014-JA",
      SLACcitation   = "%%CITATION = ARXIV:1407.6387;%%",
}

@article{Bazavov:2014soa,
      author         = "{A. Bazavov} and Brambilla, N. and
                        Garcia i Tormo, X. and Petreczky, P. and Soto J. and Vairo, A.",
      title          = "{Determination of $\alpha_s$ from the QCD static energy:
                        An update}",
      journal        = "Phys.Rev.",
      number         = "7",
      volume         = "D90",
      pages          = "074038",
      doi            = "10.1103/PhysRevD.90.074038",
      year           = "2014",
      eprint         = "1407.8437",
      archivePrefix  = "arXiv",
      primaryClass   = "hep-ph",
      reportNumber   = "TUM-EFT-47-14, UB-ECM-PF-14-81, ICCUB-14-055",
      SLACcitation   = "%%CITATION = ARXIV:1407.8437;%%",
}

@article{Bazavov:2014wgs,
      author         = "{[FNAL/MILC 14A] A. Bazavov} and others",
      title          = "{Charmed and light pseudoscalar meson decay constants
                        from four-flavor lattice QCD with physical light quarks}",
      journal        = "Phys.Rev.",
      number         = "7",
      volume         = "D90",
      pages          = "074509",
      doi            = "10.1103/PhysRevD.90.074509",
      year           = "2014",
      eprint         = "1407.3772",
      archivePrefix  = "arXiv",
      primaryClass   = "hep-lat",
      reportNumber   = "FERMILAB-PUB-14-230-T",
      SLACcitation   = "%%CITATION = ARXIV:1407.3772;%%",
}

@article{Bazavov:2015yea,
      author         = "{[MILC 15] A. Bazavov} and others",
      title          = "{Gradient flow and scale setting on MILC HISQ ensembles}",
      journal        = "Phys. Rev.",
      volume         = "D93",
      year           = "2016",
      number         = "9",
      pages          = "094510",
      doi            = "10.1103/PhysRevD.93.094510",
      eprint         = "1503.02769",
      archivePrefix  = "arXiv",
      primaryClass   = "hep-lat",
      reportNumber   = "FERMILAB-PUB-15-284-T",
      SLACcitation   = "%%CITATION = ARXIV:1503.02769;%%"
}

@article{Bazavov:2016nty,
      author         = "{[FNAL/MILC 16] A. Bazavov} and others",
      title          = "{$B^0_{(s)}$-mixing matrix elements from lattice QCD for
                        the Standard Model and beyond}",
      journal        = "Phys. Rev.",
      volume         = "D93",
      year           = "2016",
      number         = "11",
      pages          = "113016",
      doi            = "10.1103/PhysRevD.93.113016",
      eprint         = "1602.03560",
      archivePrefix  = "arXiv",
      primaryClass   = "hep-lat",
      reportNumber   = "FERMILAB-PUB-16-030-T",
      SLACcitation   = "%%CITATION = ARXIV:1602.03560;%%"
}

@article{Bazavov:2017lyh,
      author         = "{[FNAL/MILC 17] A. Bazavov} and others",
      title          = "{$B$- and $D$-meson leptonic decay constants from
                        four-flavor lattice QCD}",
      journal        = "Phys. Rev.",
      volume         = "D98",
      year           = "2018",
      number         = "7",
      pages          = "074512",
      doi            = "10.1103/PhysRevD.98.074512",
      eprint         = "1712.09262",
      archivePrefix  = "arXiv",
      primaryClass   = "hep-lat",
      reportNumber   = "FERMILAB-PUB-17/491-T, FERMILAB-PUB-17-491-T",
      SLACcitation   = "%%CITATION = ARXIV:1712.09262;%%"
}

@article{Bazavov:2018kjg,
      author         = "{[FNAL/MILC 18] A. Bazavov} and others",
      title          = "{$|V_{us}|$ from $K_{\ell 3}$ decay and four-flavor
                        lattice QCD}",
      journal        = "Phys. Rev.",
      volume         = "D99",
      year           = "2019",
      number         = "11",
      pages          = "114509",
      doi            = "10.1103/PhysRevD.99.114509",
      eprint         = "1809.02827",
      archivePrefix  = "arXiv",
      primaryClass   = "hep-lat",
      reportNumber   = "FERMILAB-PUB-18-439-T",
      SLACcitation   = "%%CITATION = ARXIV:1809.02827;%%"}

@article{Bazavov:2018omf,
      author         = "{[FNAL/MILC/TUMQCD 18] A. Bazavov} and others",
      title          = "{Up-, down-, strange-, charm-, and bottom-quark masses
                        from four-flavor lattice QCD}",
      journal        = "Phys. Rev.",
      volume         = "D98",
      year           = "2018",
      number         = "5",
      pages          = "054517",
      doi            = "10.1103/PhysRevD.98.054517",
      eprint         = "1802.04248",
      archivePrefix  = "arXiv",
      primaryClass   = "hep-lat",
      reportNumber   = "FERMILAB-PUB-17-492-T, TUM-EFT-107-18,
                        FERMILAB-PUB-17/492-T, TUM-EFT 107/18",
      SLACcitation   = "%%CITATION = ARXIV:1802.04248;%%"}

@article{Bazavov:2019aom,
    author = "{[FNAL/MILC 19] A. Bazavov} and others",
    title = "{$B_s\to K\ell\nu$ decay from lattice QCD}",
    eprint = "1901.02561",
    archivePrefix = "arXiv",
    primaryClass = "hep-lat",
    reportNumber = "FERMILAB-PUB-19-005-T",
    doi = "10.1103/PhysRevD.100.034501",
    journal = "Phys. Rev. D",
    volume = "100",
    number = "3",
    pages = "034501",
    year = "2019"
}

@article{Beane:2002vq,
      author         = "Beane, Silas R. and Savage, Martin J.",
      title          = "{Nucleons in two flavor partially quenched chiral
                        perturbation theory}",
      journal        = "Nucl. Phys.",
      volume         = "A709",
      year           = "2002",
      pages          = "319-344",
      doi            = "10.1016/S0375-9474(02)01086-2",
      eprint         = "hep-lat/0203003",
      archivePrefix  = "arXiv",
      primaryClass   = "hep-lat",
      reportNumber   = "NT-UW-02-004",
      SLACcitation   = "%%CITATION = HEP-LAT/0203003;%%"
}

@article{Beane:2004rf,
      author         = "Beane, Silas R. and Savage, Martin J.",
      title          = "{Baryon axial charge in a finite volume}",
      journal        = "Phys. Rev.",
      volume         = "D70",
      year           = "2004",
      pages          = "074029",
      doi            = "10.1103/PhysRevD.70.074029",
      eprint         = "hep-ph/0404131",
      archivePrefix  = "arXiv",
      primaryClass   = "hep-ph",
      reportNumber   = "UNH-04-04, NT-UW-04-08, JLAB-THY-04-14",
      SLACcitation   = "%%CITATION = HEP-PH/0404131;%%"
}

@article{Beane:2004tw,
      author         = "Beane, Silas R.",
      title          = "{Nucleon masses and magnetic moments in a finite volume}",
      journal        = "Phys. Rev.",
      volume         = "D70",
      year           = "2004",
      pages          = "034507",
      doi            = "10.1103/PhysRevD.70.034507",
      eprint         = "hep-lat/0403015",
      archivePrefix  = "arXiv",
      primaryClass   = "hep-lat",
      reportNumber   = "UNH-04-01, JLAB-THY-04-20",
      SLACcitation   = "%%CITATION = HEP-LAT/0403015;%%"
}

@Article{Beane:2006fk,
     author    = "Beane, Silas R. and Orginos, Kostas and Savage, Martin J.",
     title     = "{Strong-isospin violation in the neutron proton mass
                  difference from  fully-dynamical lattice QCD and PQQCD}",
     journal   = "Nucl. Phys.",
     volume    = "B768",
     year      = "2007",
     pages     = "38-50",
     eprint    = "hep-lat/0605014",
     archivePrefix = "arXiv",
     doi       = "10.1016/j.nuclphysb.2006.12.023",
     SLACcitation  = "%%CITATION = HEP-LAT/0605014;%%"
}

@Article{Beane:2006kx,
     author    = "{[NPLQCD 06] S. R. Beane} and Bedaque, P. F. and Orginos, K. and Savage,
                  M. J.",
     title     = "{$f_K/f_\pi$ in full QCD with domain wall valence quarks}",
     journal   = "Phys. Rev.",
     volume    = "D75",
     year      = "2007",
     pages     = "094501",
     eprint    = "hep-lat/0606023",
     archivePrefix = "arXiv",
     doi       = "10.1103/PhysRevD.75.094501",
     SLACcitation  = "%%CITATION = HEP-LAT/0606023;%%"
}

@article{Becher:1999he,
      author         = "Becher, Thomas and Leutwyler, H.",
      title          = "{Baryon chiral perturbation theory in manifestly Lorentz
                        invariant form}",
      journal        = "Eur. Phys. J.",
      volume         = "C9",
      year           = "1999",
      pages          = "643-671",
      doi            = "10.1007/PL00021673",
      eprint         = "hep-ph/9901384",
      archivePrefix  = "arXiv",
      primaryClass   = "hep-ph",
      reportNumber   = "BUTP-99-1",
      SLACcitation   = "%%CITATION = HEP-PH/9901384;%%"
}

@article{Becher:2005bg,
      author         = "Becher, Thomas and Hill, Richard J.",
      title          = "{Comment on form-factor shape and extraction of $|V_{ub}|$
                        from $B \to\pi l \nu$}",
      journal        = "Phys.Lett.",
      volume         = "B633",
      pages          = "61-69",
      doi            = "10.1016/j.physletb.2005.11.063",
      year           = "2006",
      eprint         = "hep-ph/0509090",
      archivePrefix  = "arXiv",
      primaryClass   = "hep-ph",
      reportNumber   = "FERMILAB-PUB-05-385-T, SLAC-PUB-11468",
      SLACcitation   = "%%CITATION = HEP-PH/0509090;%%",
}

@article{Becirevic:1999kt,
      author         = "Be{\'c}irevi{\'c}, Damir and Kaidalov, Alexei B.",
      title          = "{Comment on the heavy $\to$ light form-factors}",
      journal        = "Phys.Lett.",
      volume         = "B478",
      pages          = "417-423",
      doi            = "10.1016/S0370-2693(00)00290-2",
      year           = "2000",
      eprint         = "hep-ph/9904490",
      archivePrefix  = "arXiv",
      primaryClass   = "hep-ph",
      reportNumber   = "LPT-ORSAY-99-32, ROMA-99-1248",
      SLACcitation   = "%%CITATION = HEP-PH/9904490;%%",
}

@Article{Becirevic:2000cy,
     author    = "Be{\'c}irevi{\'c}, D. and others",
     title     = "{$K^0 \bar{K}^0$ mixing with Wilson fermions without
                  subtractions}",
     journal   = "Phys. Lett.",
     volume    = "B487",
     year      = "2000",
     pages     = "74-80",
     eprint    = "hep-lat/0005013",
     archivePrefix = "arXiv",
     doi       = "10.1016/S0370-2693(00)00797-8",
     SLACcitation  = "%%CITATION = HEP-LAT/0005013;%%"
}

@Article{Becirevic:2004ya,
     author    = "{[SPQcdR 04] D. Be{\'c}irevi{\'c}} and others",
     title     = "{The $K \to\pi$ vector form factor at zero momentum transfer
                  on the  lattice}",
     journal   = "Nucl. Phys.",
     volume    = "B705",
     year      = "2005",
     pages     = "339-362",
     eprint    = "hep-ph/0403217",
     archivePrefix = "arXiv",
     doi       = "10.1016/j.nuclphysb.2004.11.017",
     SLACcitation  = "%%CITATION = HEP-PH/0403217;%%"
}

@Article{Becirevic:2005ta,
     author    = "{[SPQcdR 05] D. Be{\'c}irevi{\'c}} and others",
     title     = "{Non-perturbatively renormalised light quark masses from a
                  lattice  simulation with $N_f= 2$}",
     journal   = "Nucl. Phys.",
     volume    = "B734",
     year      = "2006",
     pages     = "138-155",
     eprint    = "hep-lat/0510014",
     archivePrefix = "arXiv",
     doi       = "10.1016/j.nuclphysb.2005.11.014",
     SLACcitation  = "%%CITATION = HEP-LAT/0510014;%%"
}

@article{Becirevic:2014kaa,
      author         = "Becirevic, Damir and Yaouanc, Alain Le and Oyanguren,
                        Arantza and Roudeau, Patrick and Sanfilippo, Francesco",
      title          = "{Insight into $D/B\to \pi \ell \nu_\ell$ decay using the
                        pole models}",
      year           = "2014",
      eprint         = "1407.1019",
      archivePrefix  = "arXiv",
      primaryClass   = "hep-ph",
      reportNumber   = "LPT-14-58",
      SLACcitation   = "%%CITATION = ARXIV:1407.1019;%%"
}

@article{Bedaque:2004kc,
      author         = "Bedaque, Paulo F.",
      title          = "{Aharonov-Bohm effect and nucleon nucleon phase shifts on
                        the lattice}",
      journal        = "Phys.Lett.",
      volume         = "B593",
      pages          = "82-88",
      doi            = "10.1016/j.physletb.2004.04.045",
      year           = "2004",
      eprint         = "nucl-th/0402051",
      archivePrefix  = "arXiv",
      primaryClass   = "nucl-th",
      reportNumber   = "LBNL-54557",
      SLACcitation   = "%%CITATION = NUCL-TH/0402051;%%",
}

@article{Beneke:1996gn,
      author         = "Beneke, M. and Buchalla, G. and Dunietz, I.",
      title          = "{Width difference in the $B_s-\bar{B_s}$ system}",
      journal        = "Phys.Rev.",
      volume         = "D54",
      pages          = "4419-4431",
      doi            = "10.1103/PhysRevD.54.4419, 10.1103/PhysRevD.83.119902",
      year           = "1996",
      eprint         = "hep-ph/9605259",
      archivePrefix  = "arXiv",
      primaryClass   = "hep-ph",
      reportNumber   = "SLAC-PUB-7165, FERMILAB-PUB-96-095-T",
      SLACcitation   = "%%CITATION = HEP-PH/9605259;%%",
}

@article{Berkowitz:2017gql,
      author         = "{[CalLat 17] E. Berkowitz} and others",
      title          = "{An accurate calculation of the nucleon axial charge with
                        lattice QCD}",
      year           = "2017",
      eprint         = "1704.01114",
      archivePrefix  = "arXiv",
      primaryClass   = "hep-lat",
      SLACcitation   = "%%CITATION = ARXIV:1704.01114;%%"
}

@article{Bernard:1992qa,
      author         = "Bernard, Veronique and Kaiser, Norbert and Kambor,
                        Joachim and Meissner, Ulf G.",
      title          = "{Chiral structure of the nucleon}",
      journal        = "Nucl. Phys.",
      volume         = "B388",
      year           = "1992",
      pages          = "315-345",
      doi            = "10.1016/0550-3213(92)90615-I",
      reportNumber   = "BUTP-92-15, CRN-92-24, TUM-T31-28-92",
      SLACcitation   = "%%CITATION = NUPHA,B388,315;%%"
}

@Article{Bernard:1993sv,
     author    = "Bernard, Claude W. and Golterman, Maarten F. L.",
     title     = "{Partially quenched gauge theories and an application to
                  staggered fermions}",
     journal   = "Phys. Rev.",
     volume    = "D49",
     year      = "1994",
     pages     = "486-494",
     eprint    = "hep-lat/9306005",
     archivePrefix = "arXiv",
     doi       = "10.1103/PhysRevD.49.486",
     SLACcitation  = "%%CITATION = HEP-LAT/9306005;%%"
}

@Article{Bernard:2000gd,
     author    = "Bernard, Claude W. and others",
     title     = "{The static quark potential in three flavor QCD}",
     journal   = "Phys. Rev.",
     volume    = "D62",
     year      = "2000",
     pages     = "034503",
     eprint    = "hep-lat/0002028",
     archivePrefix = "arXiv",
     doi       = "10.1103/PhysRevD.62.034503",
     SLACcitation  = "%%CITATION = HEP-LAT/0002028;%%"
}

@article{Bernard:2003gq,
    author         = "Bernard, Claude and others",
    title          = "{Topological susceptibility with the improved Asqtad
                      action}",
    journal        = "Phys. Rev.",
    volume         = "D68",
    year           = "2003",
    pages          = "114501",
    doi            = "10.1103/PhysRevD.68.114501",
    eprint         = "hep-lat/0308019",
    archivePrefix  = "arXiv",
    primaryClass   = "hep-lat",
    reportNumber   = "EDINBURGH-2003-11",
    SLACcitation   = "%%CITATION = HEP-LAT/0308019;%%"
}

@Article{Bernard:2006ee,
     author    = "Bernard, Claude and Golterman, Maarten and Shamir, Yigal",
     title     = "{Observations on staggered fermions at non-zero lattice
                  spacing}",
     journal   = "Phys. Rev.",
     volume    = "D73",
     year      = "2006",
     pages     = "114511",
     eprint    = "hep-lat/0604017",
     archivePrefix = "arXiv",
     doi       = "10.1103/PhysRevD.73.114511",
     SLACcitation  = "%%CITATION = HEP-LAT/0604017;%%"
}

@Article{Bernard:2006vv,
     author    = "Bernard, Claude and Golterman, Maarten and Shamir, Yigal
                  and Sharpe, Stephen R.",
     title     = "{Comment on 'chiral anomalies and rooted staggered
                  fermions'}",
     journal   = "Phys. Lett.",
     volume    = "B649",
     year      = "2007",
     pages     = "235-240",
     eprint    = "hep-lat/0603027",
     archivePrefix = "arXiv",
     doi       = "10.1016/j.physletb.2007.04.018",
     SLACcitation  = "%%CITATION = HEP-LAT/0603027;%%"
}

@Article{Bernard:2006zw,
     author    = "Bernard, C.",
     title     = "{Staggered chiral perturbation theory and the fourth-root
                  trick}",
     journal   = "Phys. Rev.",
     volume    = "D73",
     year      = "2006",
     pages     = "114503",
     eprint    = "hep-lat/0603011",
     archivePrefix = "arXiv",
     doi       = "10.1103/PhysRevD.73.114503",
     SLACcitation  = "%%CITATION = HEP-LAT/0603011;%%"
}

@Article{Bernard:2007eh,
     author    = "Bernard, Claude and Golterman, Maarten and Shamir, Yigal
                  and Sharpe, Stephen R.",
     title     = "{'t Hooft vertices, partial quenching, and rooted staggered
                  QCD}",
     journal   = "Phys. Rev.",
     volume    = "D77",
     year      = "2008",
     pages     = "114504",
     eprint    = "0711.0696",
     archivePrefix = "arXiv",
     primaryClass  =  "hep-lat",
     doi       = "10.1103/PhysRevD.77.114504",
     SLACcitation  = "%%CITATION = 0711.0696;%%"
}

@Article{Bernard:2007ma,
     author    = "Bernard, Claude and Golterman, Maarten and Shamir, Yigal",
     title     = "{Effective field theories for QCD with rooted staggered
                  fermions}",
     journal   = "Phys. Rev.",
     volume    = "D77",
     year      = "2008",
     pages     = "074505",
     eprint    = "0712.2560",
     archivePrefix = "arXiv",
     primaryClass  =  "hep-lat",
     doi       = "10.1103/PhysRevD.77.074505",
     SLACcitation  = "%%CITATION = 0712.2560;%%"
}

@article{Bernard:2007ps,
      author         = "{[MILC 07] C. Bernard} and others",
      title          = "{Status of the MILC light pseudoscalar meson project}",
      journal        = "PoS",
      volume         = "LAT2007",
      pages          = "090",
      year           = "2007",
      eprint         = "0710.1118",
      archivePrefix  = "arXiv",
      primaryClass   = "hep-lat",
      SLACcitation   = "%%CITATION = ARXIV:0710.1118;%%",
}

@Article{Bernard:2007qf,
     author    = "Bernard, Claude and DeTar, Carleton E. and Fu, Ziwen and
                  Prelovsek, Sasa",
     title     = "{Scalar meson spectroscopy with lattice staggered
                  fermions}",
     journal   = "Phys. Rev.",
     volume    = "D76",
     year      = "2007",
     pages     = "094504",
     eprint    = "0707.2402",
     archivePrefix = "arXiv",
     primaryClass  =  "hep-lat",
     doi       = "10.1103/PhysRevD.76.094504",
     SLACcitation  = "%%CITATION = 0707.2402;%%"
}

@article{Bernard:2007tk,
      author         = "Bernard, Veronique and Passemar, Emilie",
      title          = "{Matching chiral perturbation theory and the dispersive
                        representation of the scalar K pi form-factor}",
      journal        = "Phys. Lett.",
      volume         = "B661",
      year           = "2008",
      pages          = "95-102",
      doi            = "10.1016/j.physletb.2008.02.004",
      eprint         = "0711.3450",
      archivePrefix  = "arXiv",
      primaryClass   = "hep-ph",
      SLACcitation   = "%%CITATION = ARXIV:0711.3450;%%"
}

@article{Bernard:2008dn,
      author         = "{[FNAL/MILC 08] C. Bernard} and others",
      title          = "{The $\bar{B} \to D^{*} \ell \bar{\nu}$ form factor at
                        zero recoil from three-flavor lattice QCD: a model
                        independent determination of $|V_{cb}|$}",
      journal        = "Phys.Rev.",
      volume         = "D79",
      pages          = "014506",
      doi            = "10.1103/PhysRevD.79.014506",
      year           = "2009",
      eprint         = "0808.2519",
      archivePrefix  = "arXiv",
      primaryClass   = "hep-lat",
      reportNumber   = "FERMILAB-PUB-08-316-T",
      SLACcitation   = "%%CITATION = ARXIV:0808.2519;%%",
}

@Article{Bernard:2008gr,
     author    = "Bernard, Claude and Golterman, Maarten and Shamir, Yigal
                  and Sharpe, Stephen R.",
     title     = "{Reply to: Comment on 't Hooft vertices, partial
                  quenching, and rooted staggered QCD}",
     journal   = "Phys. Rev.",
     volume    = "D78",
     year      = "2008",
     pages     = "078502",
     eprint    = "0808.2056",
     archivePrefix = "arXiv",
     primaryClass  =  "hep-lat",
     doi       = "10.1103/PhysRevD.78.078502",
     SLACcitation  = "%%CITATION = 0808.2056;%%"
}

@article{Bernard:2010fp,
      author         = "Bernard, V. and Lage, M. and Meissner, U. -G. and
                        Rusetsky, A.",
      title          = "{Scalar mesons in a finite volume}",
      journal        = "JHEP",
      volume         = "01",
      year           = "2011",
      pages          = "019",
      doi            = "10.1007/JHEP01(2011)019",
      eprint         = "1010.6018",
      archivePrefix  = "arXiv",
      primaryClass   = "hep-lat",
      reportNumber   = "PREPRINT-HISKP-TH-10-21, FZJ-IKP(TH)-2010-18",
      SLACcitation   = "%%CITATION = ARXIV:1010.6018;%%"
}

@article{Bernard:2013dfa,
      author         = "{[MILC 13B] C.~Bernard}",
      title          = "{Neutral B mixing in staggered chiral perturbation
                        theory}",
      journal        = "Phys.Rev.",
      volume         = "D87",
      pages          = "114503",
      doi            = "10.1103/PhysRevD.87.114503",
      year           = "2013",
      eprint         = "1303.0435",
      archivePrefix  = "arXiv",
      primaryClass   = "hep-lat",
      SLACcitation   = "%%CITATION = ARXIV:1303.0435;%%",
}

@article{Bernard:2013eya,
      author         = "Bernard, C. and Bijnens, J. and Gamiz, E.",
      title          = "{Semileptonic kaon decay in staggered chiral perturbation
                        theory}",
      journal        = "Phys. Rev.",
      volume         = "D89",
      year           = "2014",
      number         = "5",
      pages          = "054510",
      doi            = "10.1103/PhysRevD.89.054510",
      eprint         = "1311.7511",
      archivePrefix  = "arXiv",
      primaryClass   = "hep-lat",
      SLACcitation   = "%%CITATION = ARXIV:1311.7511;%%"
}

@article{Bernard:2013qwa,
      author         = "Bernard, Claude and Komijani, Javad",
      title          = "{Chiral Perturbation Theory for All-Staggered Heavy-Light
                        Mesons}",
      journal        = "Phys.Rev.",
      volume         = "D88",
      pages          = "094017",
      doi            = "10.1103/PhysRevD.88.094017",
      year           = "2013",
      eprint         = "1309.4533",
      archivePrefix  = "arXiv",
      primaryClass   = "hep-lat",
      SLACcitation   = "%%CITATION = ARXIV:1309.4533;%%",
}

@article{Bernard:2017npd,
      author         = "Bernard, C. and Toussaint, D.",
      title          = "{Effects of nonequilibrated topological charge
                        distributions on pseudoscalar meson masses and decay
                        constants}",
      journal        = "Phys. Rev.",
      volume         = "D97",
      year           = "2018",
      number         = "7",
      pages          = "074502",
      doi            = "10.1103/PhysRevD.97.074502",
      eprint         = "1707.05430",
      archivePrefix  = "arXiv",
      primaryClass   = "hep-lat",
      SLACcitation   = "%%CITATION = ARXIV:1707.05430;%%"
}

@article{Bernard:2017scg,
      author         = "Bernard, Claude and Bijnens, Johan and G{\'a}miz, Elvira and
                        Relefors, Johan",
      title          = "{Twisted finite-volume corrections to $K_{l3}$ decays
                        with partially-quenched and rooted-staggered quarks}",
      journal        = "JHEP",
      volume         = "03",
      year           = "2017",
      pages          = "120",
      doi            = "10.1007/JHEP03(2017)120",
      eprint         = "1702.03416",
      archivePrefix  = "arXiv",
      primaryClass   = "hep-lat",
      reportNumber   = "LU-TP-16-50",
      SLACcitation   = "%%CITATION = ARXIV:1702.03416;%%"
}

@article{Bernardoni:2010nf,
      author         = "{F. Bernardoni} and Hernandez, Pilar and Garron,
                        Nicolas and Necco, Silvia and Pena, Carlos",
      title          = "{Probing the chiral regime of $N_{f}$= 2 QCD with mixed
                        actions}",
     journal   = "Phys. Rev.",
     volume    = "D83",
     year      = "2011",
     pages     = "054503",
     eprint    = "1008.1870",
     archivePrefix = "arXiv",
     primaryClass  =  "hep-lat",
     doi       = "10.1103/PhysRevD.83.054503",
     comment = "Bernardoni 10",
     SLACcitation  = "%%CITATION = 1008.1870;%%"
}

@article{Bernardoni:2011kd,
      author         = "{F. Bernardoni} and Garron, Nicolas and Hernandez,
                        Pilar and Necco, Silvia and Pena, Carlos",
      title          = "{Light quark correlators in a mixed-action setup}",
      journal        = "PoS",
      volume         = "LAT2011",
      pages          = "109",
      year           = "2011",
      eprint         = "1110.0922",
      archivePrefix  = "arXiv",
      primaryClass   = "hep-lat",
      reportNumber   = "CERN-PH-TH-2011-229, IFT-UAM-CSIC-11-68,
                        EDINBURGH-2011-28, FTUAM-11-55, DESY-11-171, IFIC-11-54",
      comment = "Bernardoni 11",
      SLACcitation   = "%%CITATION = ARXIV:1110.0922;%%",
}

@article{Bernardoni:2012ti,
      author         = "{[ALPHA 12A] F. Bernardoni} and Blossier, B. and Bulava, J. and Della
                        Morte, M. and Fritzsch, P. and others",
      title          = "{B-physics from HQET in two-flavour lattice QCD}",
      journal        = "PoS",
      volume         = "LAT2012",
      pages          = "273",
      year           = "2012",
      eprint         = "1210.7932",
      archivePrefix  = "arXiv",
      primaryClass   = "hep-lat",
      reportNumber   = "DESY-12-188, CERN-PH-TH-2012-289, HU-EP-12-35,
                        LPT-ORSAY-12-107, MS-TP-12-14, SFB-CPP-12-78,
                        TCDMATH-12-09",
      SLACcitation   = "%%CITATION = ARXIV:1210.7932;%%",
}

@article{Bernardoni:2013oda,
      author 	     = "{[ALPHA 13] F. Bernardoni} and Blossier, B. and Bulava, J. and Della Morte, M. and Fritzsch, P. and others",
      title          = "{B-physics with $N_f=2$ Wilson fermions}",
      booktitle      = "{Proceedings, 31st International Symposium on Lattice
                        Field Theory (Lattice 2013)}",
      journal        = "PoS",
      volume         = "LATTICE2013",
      year           = "2014",
      pages          = "381",
      eprint         = "1309.1074",
      archivePrefix  = "arXiv",
      primaryClass   = "hep-lat",
      reportNumber   = "DESY-13-154, TCD-MATH-13-10, HU-EP-13-41, MS-TP-13-22,
                        IFIC-13-58, SFB-CPP-13-61",
      SLACcitation   = "%%CITATION = ARXIV:1309.1074;%%"
}

@article{Bernardoni:2013xba,
      author         = "{[ALPHA 13C] F. Bernardoni} and others",
      title          = "{The b-quark mass from non-perturbative $N_f=2$ Heavy
                        Quark Effective Theory at $O(1/m_h)$}",
      journal        = "Phys. Lett.",
      volume         = "B730",
      year           = "2014",
      pages          = "171-177",
      doi            = "10.1016/j.physletb.2014.01.046",
      eprint         = "1311.5498",
      archivePrefix  = "arXiv",
      primaryClass   = "hep-lat",
      reportNumber   = "DESY-13-224, HU-EP-13-66, IFIC-13-82, LPT-ORSAY-13-85,
                        MITP-13-066, MS-TP-13-30, SFB-CPP-13-96, TCD-MATH-13-14,
                        --MS-TP-13-30",
      SLACcitation   = "%%CITATION = ARXIV:1311.5498;%%"
}

@article{Bernardoni:2014fva,
      author         = "{[ALPHA 14] F. Bernardoni} and others",
      title          = "{Decay constants of B-mesons from non-perturbative HQET
                        with two light dynamical quarks}",
      journal        = "Phys.Lett.",
      volume         = "B735",
      pages          = "349-356",
      doi            = "10.1016/j.physletb.2014.06.051",
      year           = "2014",
      eprint         = "1404.3590",
      archivePrefix  = "arXiv",
      primaryClass   = "hep-lat",
      reportNumber   = "DESY-14-048, HU-EP-14-14, IFIC-14-24,
                        CP3-ORIGINS-2014-011-DNRF90, DIAS-2014-11,
                        LPT-ORSAY-14-19, MITP-14-026, MS-TP-14-18, SFB-CPP-14-20,
                        TCD-14-03",
      SLACcitation   = "%%CITATION = ARXIV:1404.3590;%%",
}

@article{Bernardson:1993he,
      author         = "Bernardson, S. and McCarty, P. and Thron, C.",
      title          = "{Monte Carlo methods for estimating linear combinations
                        of inverse matrix entries in lattice QCD}",
      journal        = "Comput. Phys. Commun.",
      volume         = "78",
      year           = "1993",
      pages          = "256-264",
      doi            = "10.1016/0010-4655(94)90004-3",
      SLACcitation   = "%%CITATION = CPHCB,78,256;%%"
}

@article{Bernreuther:1981sg,
      author         = "Bernreuther, Werner and Wetzel, Werner",
      title          = "{Decoupling of heavy quarks in the minimal subtraction
                        scheme}",
      journal        = "Nucl.Phys.",
      volume         = "B197",
      pages          = "228",
      doi            = "10.1016/0550-3213(82)90288-7",
      year           = "1982",
      reportNumber   = "CLNS-81-500",
      SLACcitation   = "%%CITATION = NUPHA,B197,228;%%",

}

@article{Bertone:2012cu,
      author         = "{[ETM 12D] V. Bertone} and others",
      title          = "{Kaon Mixing Beyond the SM from N$_{f}$=2 tmQCD and model
                        independent constraints from the UTA}",
      journal        = "JHEP",
      volume         = "03",
      year           = "2013",
      pages          = "089",
      doi            = "10.1007/JHEP07(2013)143, 10.1007/JHEP03(2013)089",
      note           = "[Erratum: JHEP07,143(2013)]",
      eprint         = "1207.1287",
      archivePrefix  = "arXiv",
      primaryClass   = "hep-lat",
      SLACcitation   = "%%CITATION = ARXIV:1207.1287;%%"
}

@article{Besson:2009uv,
      author         = "Besson, D. and others",
      title          = "{Improved measurements of {$D$} meson semileptonic decays to
                        $\pi$ and $K$ mesons}",
      collaboration  = "CLEO",
      journal        = "Phys. Rev.",
      volume         = "D80",
      year           = "2009",
      pages          = "032005",
      doi            = "10.1103/PhysRevD.80.032005",
      eprint         = "0906.2983",
      archivePrefix  = "arXiv",
      primaryClass   = "hep-ex",
      reportNumber   = "CLNS-09-2049, CLEO-09-02",
      SLACcitation   = "%%CITATION = ARXIV:0906.2983;%%"
}

@article{Bethke:2011tr,
      author         = "Bethke, Siegfried and Hoang, Andre H. and Kluth, Stefan
                        and Schieck, Jochen and Stewart, Iain W. and others",
      title          = "{Workshop on Precision Measurements of $\alpha_s$}",
      year           = "2011",
      eprint         = "1110.0016",
      archivePrefix  = "arXiv",
      primaryClass   = "hep-ph",
      reportNumber   = "FERMILAB-CONF-11-611-T, MIT-CTP-4301",
      SLACcitation   = "%%CITATION = ARXIV:1110.0016;%%",
}

@article{Bhattacharya:2005rb,
      author         = "Bhattacharya, Tanmoy and Gupta, Rajan and Lee, Weonjong
                        and Sharpe, Stephen R. and Wu, Jackson M. S.",
      title          = "{Improved bilinears in lattice QCD with non-degenerate
                        quarks}",
      journal        = "Phys. Rev.",
      volume         = "D73",
      year           = "2006",
      pages          = "034504",
      doi            = "10.1103/PhysRevD.73.034504",
      eprint         = "hep-lat/0511014",
      archivePrefix  = "arXiv",
      primaryClass   = "hep-lat",
      reportNumber   = "LA-UR-05-8131",
      SLACcitation   = "%%CITATION = HEP-LAT/0511014;%%"
}

@article{Bhattacharya:2011qm,
      author         = "Bhattacharya, Tanmoy and Cirigliano, Vincenzo and Cohen,
                        Saul D. and Filipuzzi, Alberto and Gonzalez-Alonso, Martin
                        and others",
      title          = "{Probing Novel Scalar and Tensor Interactions from
                        (Ultra)Cold Neutrons to the LHC}",
      journal        = "Phys.Rev.",
      volume         = "D85",
      pages          = "054512",
      doi            = "10.1103/PhysRevD.85.054512",
      year           = "2012",
      eprint         = "1110.6448",
      archivePrefix  = "arXiv",
      primaryClass   = "hep-ph",
      reportNumber   = "LA-UR-11-11460, NT@UW-11-16, IFIC-11-57, FTUV-11-1007,
                        NPAC-11-14",
      SLACcitation   = "%%CITATION = ARXIV:1110.6448;%%",                                                                                                  
}

@article{Bhattacharya:2013ehc,
      author         = "{[PNDME 13] T. Bhattacharya} and Cohen, Saul D. and Gupta, Rajan
                        and Joseph, Anosh and Lin, Huey-Wen and Yoon, Boram",
      title          = "{Nucleon Charges and Electromagnetic Form Factors from
                        2+1+1-Flavor Lattice QCD}",
      journal        = "Phys. Rev.",
      volume         = "D89",
      year           = "2014",
      number         = "9",
      pages          = "094502",
      doi            = "10.1103/PhysRevD.89.094502",
      eprint         = "1306.5435",
      archivePrefix  = "arXiv",
      primaryClass   = "hep-lat",
      reportNumber   = "LA-UR-13-24606, NT@UW-13-23",
      SLACcitation   = "%%CITATION = ARXIV:1306.5435;%%"
}

@article{Bhattacharya:2015esa,
      author         = "{[PNDME 15] T. Bhattacharya} and Cirigliano, Vincenzo and Gupta,
                        Rajan and Lin, Huey-Wen and Yoon, Boram",
      title          = "{Neutron Electric Dipole Moment and Tensor Charges from
                        Lattice QCD}",
      journal        = "Phys. Rev. Lett.",
      volume         = "115",
      year           = "2015",
      number         = "21",
      pages          = "212002",
      doi            = "10.1103/PhysRevLett.115.212002",
      eprint         = "1506.04196",
      archivePrefix  = "arXiv",
      primaryClass   = "hep-lat",
      reportNumber   = "LA-UR-15-24210",
      SLACcitation   = "%%CITATION = ARXIV:1506.04196;%%"
}

@article{Bhattacharya:2015wna,
      author         = "{[PNDME 15A] T. Bhattacharya} and Cirigliano, Vincenzo and Cohen,
                        Saul and Gupta, Rajan and Joseph, Anosh and Lin, Huey-Wen
                        and Yoon, Boram",
      title          = "{Iso-vector and Iso-scalar Tensor Charges of the Nucleon
                        from Lattice QCD}",
      journal        = "Phys. Rev.",
      volume         = "D92",
      year           = "2015",
      number         = "9",
      pages          = "094511",
      doi            = "10.1103/PhysRevD.92.094511",
      eprint         = "1506.06411",
      archivePrefix  = "arXiv",
      primaryClass   = "hep-lat",
      reportNumber   = "DESY-15-128, LA-UR-15-23801",
      SLACcitation   = "%%CITATION = ARXIV:1506.06411;%%"
}

@article{Bhattacharya:2016zcn,
      author         = "{[PNDME 16] T. Bhattacharya} and Cirigliano, Vincenzo and Cohen,
                        Saul and Gupta, Rajan and Lin, Huey-Wen and Yoon, Boram",
      title          = "{Axial, Scalar and Tensor Charges of the Nucleon from
                        2+1+1-flavor Lattice QCD}",
      journal        = "Phys. Rev.",
      volume         = "D94",
      year           = "2016",
      number         = "5",
      pages          = "054508",
      doi            = "10.1103/PhysRevD.94.054508",
      eprint         = "1606.07049",
      archivePrefix  = "arXiv",
      primaryClass   = "hep-lat",
      reportNumber   = "LA-UR-16-20522",
      SLACcitation   = "%%CITATION = ARXIV:1606.07049;%%"
}

@article{Bietenholz:1995cy,
      author         = "Bietenholz, Wolfgang and Wiese, U.J.",
      title          = "{Perfect lattice actions for quarks and gluons}",
      journal        = "Nucl.Phys.",
      volume         = "B464",
      pages          = "319-352",
      doi            = "10.1016/0550-3213(95)00678-8",
      year           = "1996",
      eprint         = "hep-lat/9510026",
      archivePrefix  = "arXiv",
      primaryClass   = "hep-lat",
      reportNumber   = "MIT-CTP-2475",
      SLACcitation   = "%%CITATION = HEP-LAT/9510026;%%",
}

@Article{Bietenholz:2010jr,
     author    = "{[QCDSF/UKQCD 10] W. Bietenholz} and others",
     title     = "{Tuning the strange quark mass in lattice simulations}",
     journal   = "Phys. Lett.",
     volume    = "B690",
     year      = "2010",
     pages     = "436-441",
     eprint    = "1003.1114",
     archivePrefix = "arXiv",
     primaryClass  =  "hep-lat",
     doi       = "10.1016/j.physletb.2010.05.067",
     SLACcitation  = "%%CITATION = 1003.1114;%%"
}

@article{Bietenholz:2011qq,
      author         = "Bietenholz, W. and others",
      title          = "{Flavour blindness and patterns of flavour symmetry
                        breaking in lattice simulations of up, down and strange
                        quarks}",
      journal        = "Phys. Rev.",
      volume         = "D84",
      year           = "2011",
      pages          = "054509",
      doi            = "10.1103/PhysRevD.84.054509",
      eprint         = "1102.5300",
      archivePrefix  = "arXiv",
      primaryClass   = "hep-lat",
      reportNumber   = "DESY-11-030, EDINBURGH-2011-09, LTH-909",
      SLACcitation   = "%%CITATION = ARXIV:1102.5300;%%"
}

@article{Bietenholz:2016ymo,
      author         = "Bietenholz, Wolfgang and Czaban, Christopher and Dromard,
                        Arthur and Gerber, Urs and Hofmann, Christoph P. and
                        Mej\'ia-D\'iaz, H\'ector and Wagner, Marc",
      title          = "{Interpreting Numerical Measurements in Fixed Topological Sectors}",
      journal        = "Phys. Rev.",
      volume         = "D93",
      year           = "2016",
      number         = "11",
      pages          = "114516",
      doi            = "10.1103/PhysRevD.93.114516",
      eprint         = "1603.05630",
      archivePrefix  = "arXiv",
      primaryClass   = "hep-lat",
      SLACcitation   = "%%CITATION = ARXIV:1603.05630;%%"
}

@article{Bigi:2016mdz,
      author         = "Bigi, Dante and Gambino, Paolo",
      title          = "{Revisiting $B\to D \ell \nu$}",
      journal        = "Phys. Rev.",
      volume         = "D94",
      year           = "2016",
      number         = "9",
      pages          = "094008",
      doi            = "10.1103/PhysRevD.94.094008",
      eprint         = "1606.08030",
      archivePrefix  = "arXiv",
      primaryClass   = "hep-ph",
      SLACcitation   = "%%CITATION = ARXIV:1606.08030;%%"
}

@article{Bigi:2017njr,
      author         = "Bigi, Dante and Gambino, Paolo and Schacht, Stefan",
      title          = "{A fresh look at the determination of $|V_{cb}|$ from
                        $B\to D^{*} \ell \nu$}",
      journal        = "Phys. Lett.",
      volume         = "B769",
      year           = "2017",
      pages          = "441-445",
      doi            = "10.1016/j.physletb.2017.04.022",
      eprint         = "1703.06124",
      archivePrefix  = "arXiv",
      primaryClass   = "hep-ph",
      SLACcitation   = "%%CITATION = ARXIV:1703.06124;%%"
}
